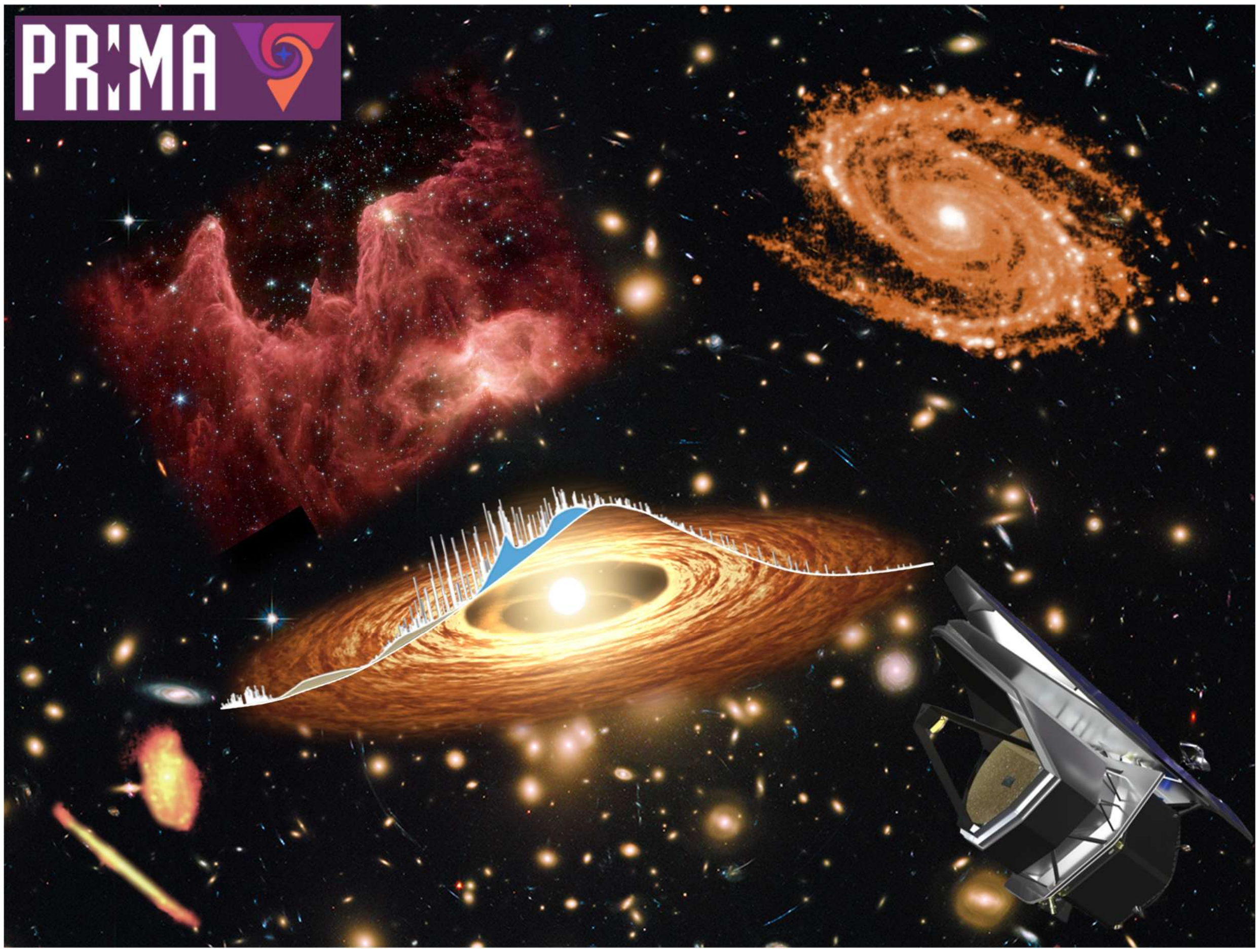

# PRIMA General Observer Science Book Volume 2

Editors: A. Moullet (NRAO) , D. Burgarella (Laboratoire d'Astrophysique de Marseille), T. Kataria (Jet Propulsion Laboratory, California Institute of Technology), H. Beuther (Max Planck Institute for Astronomy), C. Battersby (University of Connecticut), M. Cheng (UC Irvine), T. Essinger-Hileman (NASA Goddard Space Flight Center), H. Inami (Hiroshima University), E. Mills (University of Kansas), T. Nagao (Ehime University), S. Unwin (Jet Propulsion Laboratory, California Institute of Technology) (Eds.)

## Cover Image Credits:

PRIMA telescope: https://prima.ipac.caltech.edu/page/phase-a-selection

PRIMA spectrum: https://prima.ipac.caltech.edu/page/pi-science

Pillars of creation: https://www.jpl.nasa.gov/news/spitzer-captures-cosmic-mountains-of-creation/

Planetary disk: https://esawebb.org/images/weic2416a/

M81 Spitzer: https://www.spitzer.caltech.edu/image/ssc2003-06d3-24-micron-image-of-the-spiral-galaxy-messier-81

Spitzer galaxies: https://www.jb.man.ac.uk/news/2012/Spitzer/

# Table of Contents

# Highlights at a Glance

| |
|---|
| In **5000 h**, PRIMA can build an unprecedented all sky dust map, improving on Infrared Astronomical Satellite IRAS's all sky survey with an angular resolution >30 times finer and a sensitivity >60 times better. This legacy dataset would uncover troves of new cold objects (debris disks, circumstellar disks, brown dwarfs), new populations of high-z objects, and serve as the community reference for decades of studies in dust extinction, dust polarization and multi-scale magnetic fields (section 92; Saydjari et al.) |
| In **60 h**, PRIMA can measure the elemental C:O ratio in circumstellar gas in a statistically significant sample of 35 debris disks, allowing to examine the relationship between the elemental composition of primary circumstellar material and that of evolved exoplanets atmospheres (section 80; MacGregor et al.) |
| In **200 h**, PRIMA can provide a detailed inventory of dust mineralogy in about 1200 post-main sequence stars in the Milky Way and the Local Group, to compare to ISM dust mineralogy so as to understand how dust is processed after ejection from stars (section 108, Kemper et al.) |
| In **8 h**, PRIMA can map the strongest lensing galaxy cluster fields to reveal the faint end of contributions to the Cosmic Infrared Background, resolving the remaining unexplained fraction of the CIB. (section 5, Hatsukade et al.) |
| In **360 h**, PRIMA can detect for the first time cold dust in debris belts around a sample of 50 white dwarfs, witnessing the evolution paths of planetary systems at the end of their stars' life (section 105, Bonsor et al. ) |
| In **26 h**, PRIMA can achieve deep integrations on recently discovery 'little red dots' to determine the presence or absence of warm dust, providing more evidence towards the nature of these enigmatic sources (section 40, McKinney et al.) |
| In **30 h**, PRIMA can search for the presence of warm intergalactic hydrogen gas in a local galaxy cluster, shedding light on the mechanisms for gas heating in rarefied environments (section 32, Kaneko et al.) |
| In **70 h**, PRIMA can survey the temperature and density structures in a sample of 25 planetary nebulae, offering new insights into the mass loss history of their progenitors and the processes driving nebular evolution (section 119, Ueta et al. ) |
| In **200 h**, PRIMA can monitor the polarization of three bright blazars over half a year, verifying the coevolution of magnetic field and particle acceleration in potential sources of extragalactic cosmic rays and neutrinos. (section 4, Zhang et al.) |
| In **1000 h**, PRIMA can perform intensity mapping over a 1 square degree area, enabling three-dimensional mapping of PAH emission over cosmic time up to z~5 (section 17, Cheng et al.) |
| In **40 h**, PRIMA can conduct an unprecedented far-IR surface spectrum survey on a sample of 30 Kuiper Belt Objects, enabling new detections of silicate and organic features,<br>and measurements of the dust-to-ice ratio to constrain the thermal and chemical history of these bodies (section 58, Pinilla Alonso et al.) |
| In **1050 h**, PRIMA can map 45 square degrees in the Large and Small Magellanic Clouds and resolve magnetic fields on molecular cloud scales at better than 10 pc resolution, helping to understand the interplay of cloud and galactic magnetic fields (section 66, Fissel et al.) |

PRIMA (The PRobe for Infrared Mission for Astrophysics) is a far-infrared (FIR) observatory concept responding to NASA's APEX opportunity. A cryogenically cooled 1.8-m telescope feeds two complementary instruments that together provide unprecedented instantaneous spectral coverage across 24–235 µm. PRIMAger delivers ultra-sensitive, multi-band spectrophotometric imaging with R=8 from 24–80 µm and *polarimetric* imaging in four broad bands from 80–235 µm. FIRESS offers versatile spectroscopy from R=100 up to about 4400 x 112 / $\lambda$ across the same wavelength range. PRIMA's mapping speed surpasses that of Herschel and Spitzer by 2–4 orders of magnitude, enabling true survey-mode science as well as targeted programs. Because PRIMA is defined withing the APEX framework, about 25 % of the mission time will be used to carry out a well-defined Principal Investigator (PI) observing program. This program is organized to provide the international community with scientific breakthroughs focused on three topics previously identified by the Astro2020 Decadal survey as priority areas for a far-IR probe mission:

- Origins of Planetary Atmospheres: tracing the astrochemical signatures of planet formation,
- Buildup of Dust and Metals: measuring the increase of heavy elements and interstellar dust from early galaxies to today,
- Evolution of Galactic Systems: probing the co-evolution of galaxies and their supermassive black holes across cosmic time.

In addition to this PI program, about 75 % of the mission time is allocated to a General Observer (GO) program that will be offered to the international community. With >80% observing efficiency, the nominal five-year mission yields more than 26,000 hours for community-led science.

The PRIMA General Observer Science Book, Volume 2 (GO book Vol. 2) is a collection of 120 contributions authored by community members, each articulating a scientific case that is enabled by PRIMA's tremendous sensitivity advance, broad spectral coverage and polarization capability (See the Highlights at a Glance table.). GO Book Vol 2 is the follow-up of the GO Book Vol. 1 (Moullet et al., 2023 2023arXiv231020572M) issued in November 2023. This second volume reflects the strong development of the community interest, awareness and involvement in PRIMA, following the selection of PRIMA for Phase A study in October 2024. In particular, a large number of interested community members is now engaged in 10 field-specific PRIMA working groups (https://prima.ipac.caltech.edu/page/working-groups-step2).

Collectively, science cases in Volume 2 cover a much broader range of topics than the Volume 1 cases, and than the PI program. Cases address more than 90% of the scientific questions and discovery areas in Astro 2020 (see Table 1) and leverage all the scientific opportunities offered by PRIMA's observing modes, including:

- Exploit full-band coverage to secure continuum and diagnostic lines (e.g., fine-structure cooling lines) with consistent calibration across 24–235 µm, minimizing cross-instrument systematics.
- Perform survey-class mapping speeds to propose both ultra-wide to all-sky surveys and resolved studies in large nearby galaxies and the ISM.

- Activate FIR polarimetry at wavelengths longer than 80 µm to trace magnetic fields, dust physics, and scattering geometries at scales from protoplanetary disks to galactic environments
- Incorporate time-domain strategies: in particular multi-epoch monitoring of variable accretion and dust formation are enabled by PRIMA's agility.
- Set the base for coordinated campaigns with optical/near-IR on the ground but also in space and radio/submm facilities. This multi-wavelength approach uses PRIMA's FIR anchor to close the energy budget and break degeneracies in SED and line-ratio diagnostics.

In aggregate, the time requested to perform the science cases of this GO Book Volume 2 amounts to 50.400 hours, more than the nominal GO time availability. This is an unambiguous and clear evidence of strong community pull. The portfolio presented in the Volume 2 provides a snapshot of the community's collected propositions, includes both legacy-scale surveys and focused, high-impact programs, ensuring decades of archival value (See the Highlights at a Glance table). Where appropriate, the PRIMA team lightly edited the contributed cases and harmonized assumptions (backgrounds, calibration, confusion regimes) to enable fair comparisons.

# What is New in GO Book Volume 2

GO Book Vol. 2 is based on the latest characterization of the expected performance of PRIMA and its instruments, which is accessible through updated documentation and an exposure time calculator (https://prima.ipac.caltech.edu/page/etc-calc). Compared to Vol.1, the PRIMA team has refined the end-to-end performance picture in specific areas that matter directly for program feasibility and yield:

- Updated sensitivity curves and mapping-speed projections that fold in refined thermal backgrounds, detector noise spectra, and scan-map efficiencies for both PRIMAger and FIRESS.
- Bandpass edge and throughput validation, clarifying line placement near filter edges and the effective R for narrow-band imaging, low-R and high-R spectroscopy.
- Confusion-noise characterization, revisiting confusion limits versus depth/area at shortward (<80 µm) and longward (>80 µm) regimes, with guidance on when polarization or higher-R spectroscopy de-confuses blended fields.
- Spectroscopic performance verification, consolidating line-sensitivity predictions across R ~ 75 (low-R) – 13,000 (high-R)

Note that 26 cases in this book correspond to updated Volume 1 cases, using this new information to reevaluate integration times compared to the initial estimates in Volume 1.

Following several public events (April 2025 workshop in Marseille, France, June 2025 EAS session in Ireland, May 2025 workshop in Pasadena, June 2025 workshop in Tokyo), the team sustained broad engagement through an efficient communication and iterative release of fact sheets and simulator aligned to the maturing design. PRIMA's working groups helped engaging the widest community and to provide well-defined and high-impact science cases.

**Table 1.** Science Panel Questions and discovery areas as defined in the Astro2020 Decadal Survey (Tables 2.1 and 2.2, p 2-49), and the number of PRIMA GO Book Vol 2 Science Cases addressing them. Some science cases address multiple questions.

| Panel | Decadal Question | Citation | Number of Cases in GO Book | Example case |
|---|---|---|---|---|
| Compact Objects and Energetic Phenomena | What are the mass and spin distributions of neutron stars and stellar mass black holes | B-Q1 | 1 | "Probing particle acceleration in blazars using far-infrared polarimetry monitoring with PRIMA", Zhang et al., section 4 |
| | What powers the diversity of explosive phenomena across the electromagnetic spectrum? | B-Q2 | 6 | "Probing GRB reverse shocks in the far-IR", Ho et al., section 2 |
| | Why do some compact objects eject material in nearly light-speed jets, and what is that material made of? | B-Q3 | 1 | "PRIMA systematic study of particle acceleration in hot spots of radio galaxies", Isobe et al., section 29 |
| | What seeds supermassive black holes and how do they grow? | B-Q4 | 14 | "Ghosts of disruptions past: Revealing the growth of supermassive black holes via stellar capture with PRIMA", De et al., section 21 |
| Cosmology | What are the properties of dark matter and the dark sector? | C-Q2 | 2 | "PRIMAx, Axion like dark matter searches with PRIMA" Basu-Thakur et al., section 7 |
| | What physics drives the expansion and large-scale evolution of the Universe? | C-Q3 | 3 | "Evolution of ISM properties of galaxies along large-scale structures probed with multi-object optical fiber spectroscopy and PRIMA IR spectroscopy" Kubo et al., section 6 |
| Galaxies | How did the intergalactic medium and the first sources of radiation evolve from cosmic dawn through the epoch of reionization? | D-Q1 | 7 | "Far-infrared line properties of Green Pea galaxies, the best local analogues of galaxies in the epoch of reionization era" Hashimoto et al., section 26 |
| | How do gas, metals, and dust flow into, through, and out of galaxies? | D-Q2 | 42 | "M81 y sus Primas con PRIMA: Spectroscopic Mapping of the M81 Group with FIRESS" Levy et al., section 37 |
| | How do supermassive black holes form and how is their growth coupled to the evolution of their host galaxies? | D-Q3 | 33 | "Probing the Dust-Obscured Universe: A Blind Spectroscopic Survey with PRIMA to Trace Star Formation and Black Hole Growth at Cosmic Noon" Spignolio et al., section 45 |
| | How do the histories of galaxies and their dark matter halos shape their observable properties? | D-Q4 | 17 | "The impact of morphology, environment and AGN on ISM conditions across the Hubble Sequence" Davis et al., section 20 |

| Panel | Decadal Question | Citation | Number of Cases in GO Book | Example case |
|---|---|---|---|---|
| | Mapping the Circumgalactic Medium and Intergalactic Medium in Emission | D-DA | 2 | "A Survey of Intergalactic Warm Molecular Gas with PRIMA", Kaneko et al., section 32 |
| Exoplanets, Astrobiologyand the Solar System | What is the range of planetary system architectures, and is the configuration of the solar system common? | E-Q1 | 16 | "Locating water snowline positions in protoplanetary disks by analyzing Keplerian rotation profiles of water lines" Notsu et al., section 85 |
| | What are the properties of individual planets, and which processes lead to planetary diversity? | E-Q2 | 14 | "From Dust to Planetesimals/Comets: Probing Radial Mixing and Mineral Evolution in Planet-Forming Disks with PRIMA" Honda et al., section 72 |
| | How do habitable environments arise and evolve within the context of their planetary systems? | E-Q3 | 13 | "Measurement of water isotopologue ratios: a probe of planet formation cosmochemistry" Salyk et al., section 91 |
| | How can signs of habitable life be identified and interpreted in the context of their planetary environments? | E-Q4 | 1 | "Comparing the Carbon-to-Oxygen Ratio of Debris Disks to Exoplanets" MacGregor et al., section 80 |
| ISM and Star and Planet Formation | How do star-forming structures arise from, and interact with, the diffuse ISM? | F-Q1 | 33 | "Characterizing the interplay between Galactic star formation and ionization feedback with PRIMA" Zavagno et al., section 100 |
| | What regulates the structure and motions within molecular clouds? | F-Q2 | 16 | "Mapping Magnetic Field Strength from Molecular Cloud Envelopes to Dense Cores and Filaments" Lis et al., section 77 |
| | How does gas flow from parsec scales down to protostars and their disks? | F-Q3 | 11 | "A Baseline Spectroscopic Survey of Protostars within 500 pc: Accretion Driven Feedback in Evolving and Outbursting Protostars" Megeath et al., section 83 |
| | Is planet formation fast or slow? | F-Q4 | 11 | "Variability and Transient Debris Dust with PRIMA" Marshall et al., section 81 |
| Stars, the Sun and Stellar Populations | What are the most extreme stars and stellar populations? | G-Q1 | 13 | "Investigating the formation of anomalous chemistry and complex structures in the circumstellar envelopes surrounding extreme AGB stars" Alonso Hernandez et al., section 107 |
| | How does multiplicity affect the way a star lives and dies? | G-Q2 | 5 | "PRIMA's view of Clumping in Massive Star Winds" Najarro et al., section 111 |
| | What would stars look like if we view them like we do the Sun? | G-Q3 | 3 | "Tracing mass-loss, wind acceleration and dust formation with FIRESS" Scicluna et al., section 117 |
| | How do the Sun and other stars create space weather? | G-Q4 | 4 | "Exploring the Properties of Stellar Flares in the Far-Infrared 109 PRIMA" MacGregor et al., section 109 |

# Compact Objects and Energetic Phenomena

# 1. PRIMA Follow-up Observations of Transients Discovered by mm/submm Surveys

Anna Ho (Cornell University), Dave Clements (Imperial College London), Dr Mike Peel (Imperial College London)

Today, the far-IR/mm time-variable sky is largely unexplored. By the time PRIMA flies, ground-based CMB experiments, including the Simons Observatory, will be monitoring large areas of the sky for transient mm sources as a byproduct of their CMB observations and will produce regular alerts for newly discovered mm transients. The first results from the current generation of such experiments have shown that they are capable of detecting a wide variety of transient sources at mm wavelengths, ranging from stars in our own galaxy to extragalactic sources likely associated with AGN. These results, together with theoretical predictions for other classes of sources, indicate that future facilities may be able to detect many interesting classes of sources, including gamma-ray bursts (GRBs), tidal disruption events (TDEs), interacting supernovae (SNe), and fast blue optical transients (FBOTs). There is also the possibility that these CMB observatories will uncover new classes of mm-variable sources. For such transients, observing the far-IR portion of their spectral energy distributions would significantly improve our understanding of their radiation mechanisms and the physics of the underlying source. PRIMA follow-up of representative examples of various transients discovered by mm surveys could thus have a powerful impact on our understanding of a wide range of astrophysical phenomena.

## Science Justification

Transient phenomena have been studied very little in the far-IR, although a wide range of Galactic and extragalactic sources are expected to show transient activity at these wavelengths. In particular, it has become increasingly common to detect transients at slightly shorter wavelengths, i.e., in the submillimeter band. By repeatedly mapping the sky, cosmic microwave background (CMB) experiments have serendipitously detected a large number of stellar flares (Naess et al. 2021, Guns et al. 2021, Li et al. 2023, Tandoi et al. 2024), as well as several transients likely associated with active galactic nuclei (AGN; Guns et al. 2021, Biermann et al. 2024) and one classical nova (Biermann et al. 2024). Monitoring of star-forming regions has uncovered protostellar sources that vary on year-long timescales (e.g., Lee et al., 2021). Surprisingly bright submm-wavelength emission has also been observed during targeted follow-up observations of a wide variety of transients discovered at higher energies, including gamma-ray bursts (e.g., de Ugarte Postigo et al. 2011), tidal disruption events (e.g., Andreoni et al. 2022), and fast blue optical transients ("FBOTs"; e.g., Ho et al. 2023). The prototype mm-bright FBOT was the transient AT2018cow, which was 50 mJy at 230 GHz for 50 days and was even detected with ALMA's Band 9 receiver at nearly 1 THz (Ho et al. 2019).

Today's CMB experiments, such as the Simons Observatory—which will be operating while PRIMA is flying—represent a significant enhancement to our ability to detect mm/submm transients because they will cover large areas of the sky with a roughly daily cadence. The wide coverage and high cadence enable far more efficient selection of transients with bright submm emission (currently, the follow-up detection rate is low). Recent calculations (Eftekhari et al., 2022), along with extrapolations from the current discovery rate with the South Pole Telescope, suggest that these future CMB experiments could detect a significant number of transient sources over their lifetimes that would be worth following up with PRIMA.

The addition of far-IR observations to the study of mm/submm transients can provide a range of physical insights into these diverse sources. For the most luminous sources—including GRBs, FBOTs, and jetted TDEs—the principal emission mechanism is synchrotron radiation. In these cases, observations in the far-IR can help determine the turnover frequency of the synchrotron emission and thus the parameters of the shock responsible for it. This will provide insights into the medium surrounding the transient source, including its density and magnetic field, as well as a better estimate of the shock's overall energy budget (e.g., Chevalier 1998, Kulkarni et al. 1998). The peak of the synchrotron spectral energy distribution (SED) remains in the far-IR/submm band for hours to weeks (Perley et al. 2014, Laskar et al. 2019, Ho et al. 2019, Andreoni et al. 2022), so rapid-turnaround observations may be necessary. Such observations can also help determine the energy distribution of the underlying emitting electrons (e.g., Margalit & Quataert 2021, 2024, Ho et al. 2022), which is interesting for better understanding the shock-acceleration process.

In addition to synchrotron radiation, far-IR observations could probe newly formed dust. An increasing number of TDEs are being discovered in the IR band using WISE data (e.g., Masterson et al. 2024). These TDEs had been missed by surveys at higher energies (optical) because they take place in heavily dust-obscured regions (e.g., Mattila et al. 2018). This leads to dust heating and much of the energy of the TDE being reradiated into the infrared. The infrared emission of a dust-obscured TDE can last for years. For such events, the addition of short-wavelength far-IR emission, in PRIMA's high frequency bands, will add new constraints to the energetics of such objects and our understanding of the circumnuclear dust heated by them.

PRIMA observations of Galactic submm transients would also be valuable. For example, accreting protostars that vary on month-to-year timescales may be the subject of specific monitoring programs. Still, it may also appear as targets of opportunity in the case of significant events. The known mm/submm protostar transients are all quite bright (Lee et al., 2021), with 850-micron fluxes > 5 Jy, so there may be numerous less luminous systems that will be identified as transients by future observations by CMB experiments and others. With their long-term variability, these sources can be easily dealt with as targets of opportunity or as part of specific monitoring projects. Classical novae are some of the brightest submm transients in the sky (e.g., Chomiuk et a. 2014), and there is significant interest in obtaining far-IR to cm observations because the basic emission mechanism at these wavelengths is unknown. Multi-frequency observations could help determine the relative contributions of different emission mechanisms, which, in turn, could help study the connection between shocks in novae and newly formed dust.

Finally, there is the possibility of novel or unusual sources that may be high-priority targets. An example of this type of source is AT2018cow (Ho et al., 2019)—an extreme FBOT with very luminous mm emission also detected in the X-ray. It is thought to be the result of a fast sub-relativistic shock in a dense environment, but the origin of this shock remains unclear. Submm observations proved crucial to understanding the nature of this source, identifying the frequency of the spectral turnover, but it may well be that other extreme sources have turnovers at still higher, far-IR frequencies accessible only by PRIMA.

## Instruments and Modes Used

PRIMAger PPI and PHI mapping.

## Approximate Integration Time

Varies significantly depending on source type due to large expected variations in brightness (e.g., classical novae can be in the Jy range, while most extragalactic transients will be at the detection threshold). For a given source, we request 3 epochs of follow-up PRIMAger observations on a logarithmic timescale, with a cadence that depends on the light-curve evolution. At peak brightness, a source would have to be at least 35 mJy for a 5-sigma detection by a CMB survey (e.g., the SO single-epoch 1-sigma RMS is 7 mJy at 95 GHz). Assuming a roughly 0.3 mJy RMS for PRIMAger in 10 minutes of observations (including overheads, from the PRIMAger exposure time calculator), we could track such a source as it fades by an order of magnitude in brightness. To be conservative, we estimate that we would require 3 epochs for 5 sources per year, for a total of 75 epochs over the five years of the mission, totaling 12.5 hours of observations.

## Special Capabilities Needed

Monitoring over days to years timescales, depending on the nature of the transient.

## Synergies with Other Facilities

Synergy directly with CMB experiments that will provide event triggers, and with other monitoring projects such as VRO-LSST

## Description of Observations

Single-source photometric observations in all PRIMAGER bands with 10-minute observing time per epoch, including overheads. Observations to be repeated on timescales determined to be appropriate for each specific source by other observatories that can characterize the timescale of variation.

# 2. Probing GRB Reverse Shocks in the Far-IR

Anna Ho (Cornell), Smaranika Banerjee (OKC), Tanmoy Laskar (Utah), Daniel Perley (LJMU)

Gamma ray bursts (GRBs) are some of the most energetic explosive events in the Universe. Modeling the afterglow of GRBs—synchrotron radiation from the collision of the jet with the ambient medium—provides insight into the structure, velocity, and magnetization of the relativistic outflow launched by the collapsing star, which in turn constraints the nature of the compact object "engine" and the progenitor star, as well as the physics of the outflow. Detailed multifrequency ALMA observations of GRBs have yielded significant progress in early-afterglow studies by enabling distinct components of the afterglow emission to be separated and have also revealed heretofore unexpected temporal and spectral behaviors indicative of new astrophysics in the hydrodynamics or radiative processes involved. Models currently available in the literature suggest that the peak of GRB afterglow SEDs is located at far-IR wavelengths (e.g., about 100 microns), a few minutes to hours post-explosion. Therefore, quick (within 10 minutes) follow-up and monitoring observations of GRB afterglows at far-IR wavelengths, with careful characterization of the SEDs, could enable significant progress in our understanding of GRBs.

## Science Justification

Gamma-ray bursts (GRBs) are among the most energetic explosive events in the Universe: their enormous luminosities mean they can be detected at high redshifts and thus serve as probes of the early phases of cosmic history. While significant progress has been made both theoretically and observationally over the past decades, a number of major questions about the progenitors and the physics of GRBs remain elusive (for reviews, see Piran 2004, Kumar & Zhang 2015).

In the basic long-duration GRB model, in the core collapse of a massive star, accretion or rotation of a newly formed compact object launches an ultra-relativistic outflow that tunnels through the star and breaks out into interstellar space. The afterglow of GRBs—synchrotron radiation from the collision of the jet with the ambient medium—is a key diagnostic of the progenitor system. Modeling the afterglow spectral energy distribution (SED) as it evolves with time provides insight into the structure, velocity, and magnetization of the relativistic outflow launched by the collapsing star, which in turn constrains the nature of the compact object "engine" and the progenitor star, as well as the physics of the outflow. Afterglow emission can show very complex behavior, particularly at early times. Detailed multifrequency ALMA observations of GRBs have yielded significant progress in early-afterglow studies by enabling distinct components in the afterglow emission to be separated (e.g, Laskar et al. 2018, 2019; Perley et al. 2014). However, they have also revealed unexpected temporal behaviors (e.g., slowly evolving peaks and rebrightenings) and unexpected spectral components, indicative of new astrophysics between either the hydrodynamics or radiative processes involved (e.g., azimuthally or radially structured outflows, multiple episodes of engine activity, incomplete particle acceleration).

Models currently available in the literature suggest that the peak of GRB afterglow SEDs is located at far-IR wavelengths (e.g., about 100 microns) a few minutes to hours post-explosion, and that it shifts to longer wavelengths over time (see Figures 1 and 2 below). The time evolution of SEDs is closely coupled to the environment and explosion properties, and prompt multiwavelength observations are needed to break the parameter degeneracy. Therefore, quick (within 10 minutes) follow-up and monitoring observations of GRB afterglows, particularly at far-IR wavelengths, with careful characterization of the SEDs, will be key. Despite such importance, there is currently no GRB afterglow observation at far-IR wavelengths. PRIMA has the potential to be a game-changer on this topic.

Thus, far-IR observations being conducted by PRIMA can advance our understanding of relativistic outflows from stellar explosions and address the question raised by the recently released Astro2020 decadal survey, "B-Q2. WHAT POWERS THE DIVERSITY OF EXPLOSIVE PHENOMENA ACROSS THE ELECTROMAGNETIC SPECTRUM?"

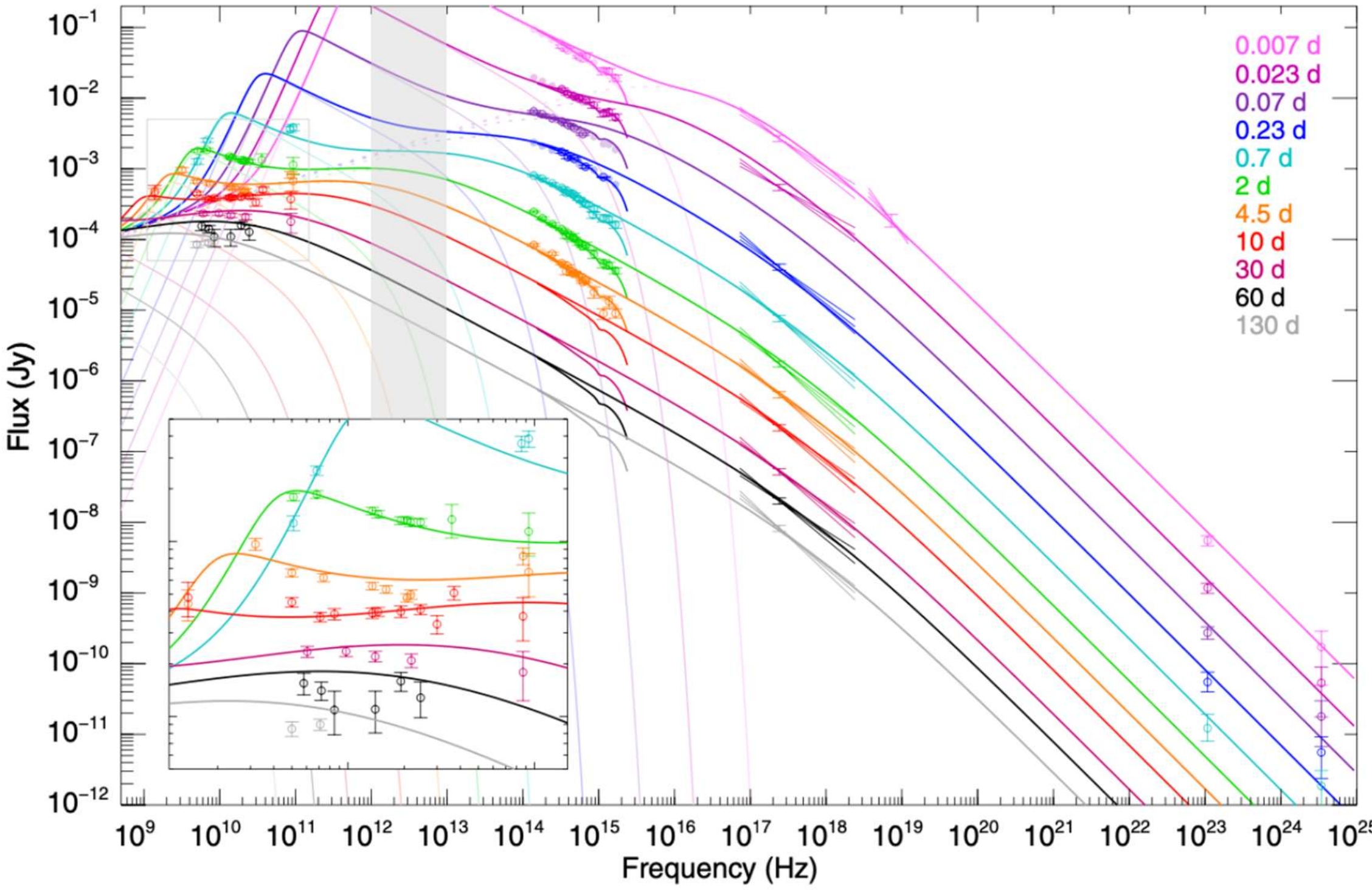

**Figure 1.** Observations of the afterglow of GRB 130427A. The SEDs originating from reverse shocks take the peak values at far-IR wavelengths (the PRIMA band is shaded: 10^12-10^13 Hz, corresponding to 300-30 micron). Figure adapted from Perley et al. 2014.

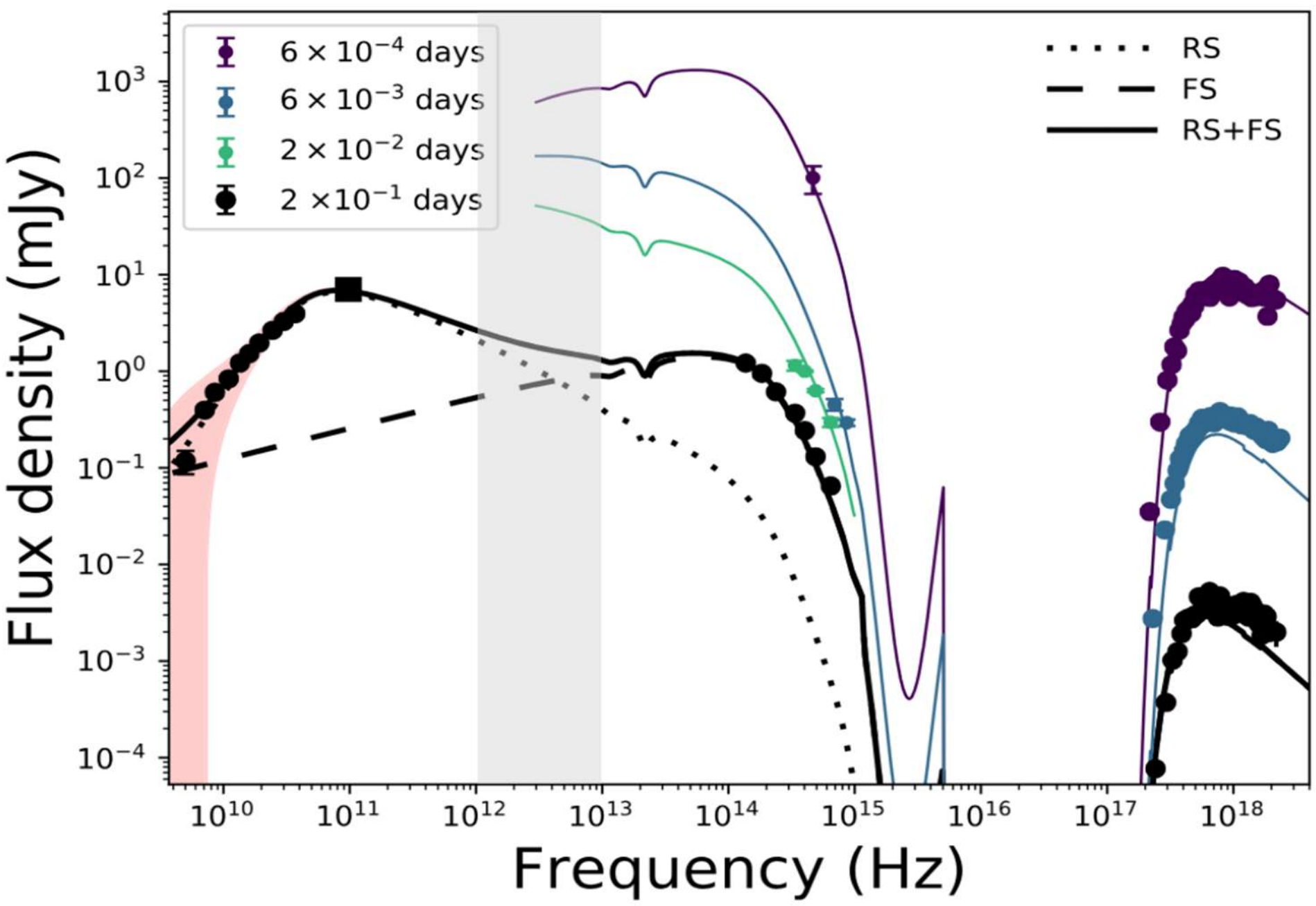

**Figure 2.** Observations of the afterglow of GRB 190114C. As in Figure 1, a key missing range of observations is in the far-IR at early times, when it is critical to decompose the afterglow into the forward and reverse shock components. The PRIMA band is shaded. Figure adapted from Laskar et al. 2019.

## Instruments and Modes Used

FIRESS pointed low-res mode.

## Approximate Integration Time

Integrating for 0.5 hours (including overheads, from the exposure time calculator) gives a 5-sigma point-like sensitivity of 860 micro Jy in Band 2 (100-200 microns, i.e., few x 10^12 Hz).

## Special Capabilities Needed

None

## Synergies with Other Facilities

GRB instruments (e.g., Fermi, Swift), submillimeter (ALMA), radio (VLA, ngVLA)

## Description of Observations

For GRBs we use observations of GRB 130427A (z=0.340; Figure 1) as a model, since its physical parameters (e.g., jet energy and opening angle) are considered to be relatively typical for GRBs (Perley et al. 2014). We request four observations on a logarithmic timescale (times given post GRB): 10min, 20min, 40min, and 80min. An integration time of 0.5 hr (including overheads) would enable us to detect a source as bright as ~1 mJy. At the distance of GRB 130427A, the flux densities at these times would be 1000 mJy, 200 mJy, 60 mJy, and 30 mJy. Therefore, we could detect a burst like GRB 130427A a factor of five as distant, or out to z=1.3. In practice, most GRBs have an unknown redshift and poor localization (square degrees), and not all of them have bright reverse shocks (the fraction is not well constrained). Therefore, we would focus on well-localized GRBs that are known to have bright afterglows in the wavelength range relevant to PRIMA. We estimate that we would trigger on ~10 GRBs over the course of the five-year mission (~2/year), and that we would additionally obtain deeper (~2 hrs total including overheads) observations for a subset (say 2-3). The total request is therefore 10 x 0.5 hr + 3 x 2 hr = 11 hr.

# 3. Probing Jet Evolution and Morphology Using Infrared Polarization Map with PRIMA

Haocheng Zhang (University of Maryland Baltimore County/NASA Goddard Space Flight Center)

Relativistic plasma jets from supermassive black holes are crucial to the understanding of jet launching, propagation, and interaction with surrounding medium, which can feedback to the co-evolution of the central black hole and host galaxy. General relativistic magnetohydrodynamic simulations have revealed the importance of magnetic fields in jet evolution and morphology. However, so far we can only study jet magnetic field with radio polarization maps. PRIMA will bring the novel infrared polarization map to this research area. This permits us to improve our knowledge of the jet magnetic field morphology and evolution, as well as how the jet energy dissipation works during its interaction with the surrounding medium.

## Science Justification

Radio galaxies possess powerful relativistic plasma jets that are launched from the central supermassive black holes. They are closely related to many important physical questions in high-energy and time-domain multi-messenger astronomy, such as the jet launching, black hole growth, feedback, and cosmic ray acceleration. While radio galaxy jets are mostly studied in the radio band, recent observations have found that they can emit in X-ray and gamma-ray bands, in particular, there is evidence that the X-ray emission can be variable at kpc distance from the central engine [1]. This suggests efficient particle acceleration very far from the black hole, which may be sites for very high energy cosmic ray acceleration and neutrino emission.

To address these questions, we need to understand the jet propagation, evolution, and its interaction with the surrounding medium. Recent global general relativistic magnetohydrodynamic simulations have found that the magnetic field morphology and evolution play an important role in jet dynamics [2]. However, since most of the observations are in radio bands, which are dominated by emission from relatively low-energy electrons, we can only study the global magnetic field morphology rather than potential sites with magnetic evolution, energy dissipation and active particle acceleration. The variable X-ray knots at kpc scale indeed shed light on the particle acceleration, but there is no polarization information, thus the magnetic field morphology and evolution in these particle acceleration sites remain unclear.

The infrared polarimeter PRIMAger onboard PRIMA can open a new infrared polarimetry window for the study of radio galaxies. PRIMAger can map the polarized emission for nearby bright radio galaxies, such as Cen A. The THz band of PRIMAger can map synchrotron emission from higher-energy electrons, better probing the energy dissipation and particle acceleration sites. If any differences are found between PRIMA polarized emission map and radio counterparts, they may hint on the energy stratification in these radio jets, and help us to investigate the physical conditions and particle acceleration mechanisms in these radio galaxies.

The multi-wavelength (radio to infrared) polarized emission maps can be compared with global hybrid magnetohydrodynamic and particle transport simulations. Global general relativistic magnetohydrodynamic simulations have found that the jet can suffer from strong kink instabilities beyond the Bondi radius, due to the feedback of jet interactions with the surrounding medium to the black hole accretion (Figure 1) [2]. Such unstable regions can trigger magnetic reconnection and turbulence that lead to efficient particle acceleration. They can be promising interpretation for the variable X-ray knots at kpc scale, and potential origins for extragalactic cosmic rays and neutrinos. Although we still need coupled particle transport simulations to examine how the particle acceleration works, preliminary polarized radiation transfer simulations have hinted that these sites can make variable high-energy emission [3]. These in-development simulation results can be compared with radio to PRIMA polarized emission maps and shed light on the jet feedback and particle acceleration in radio galaxies.

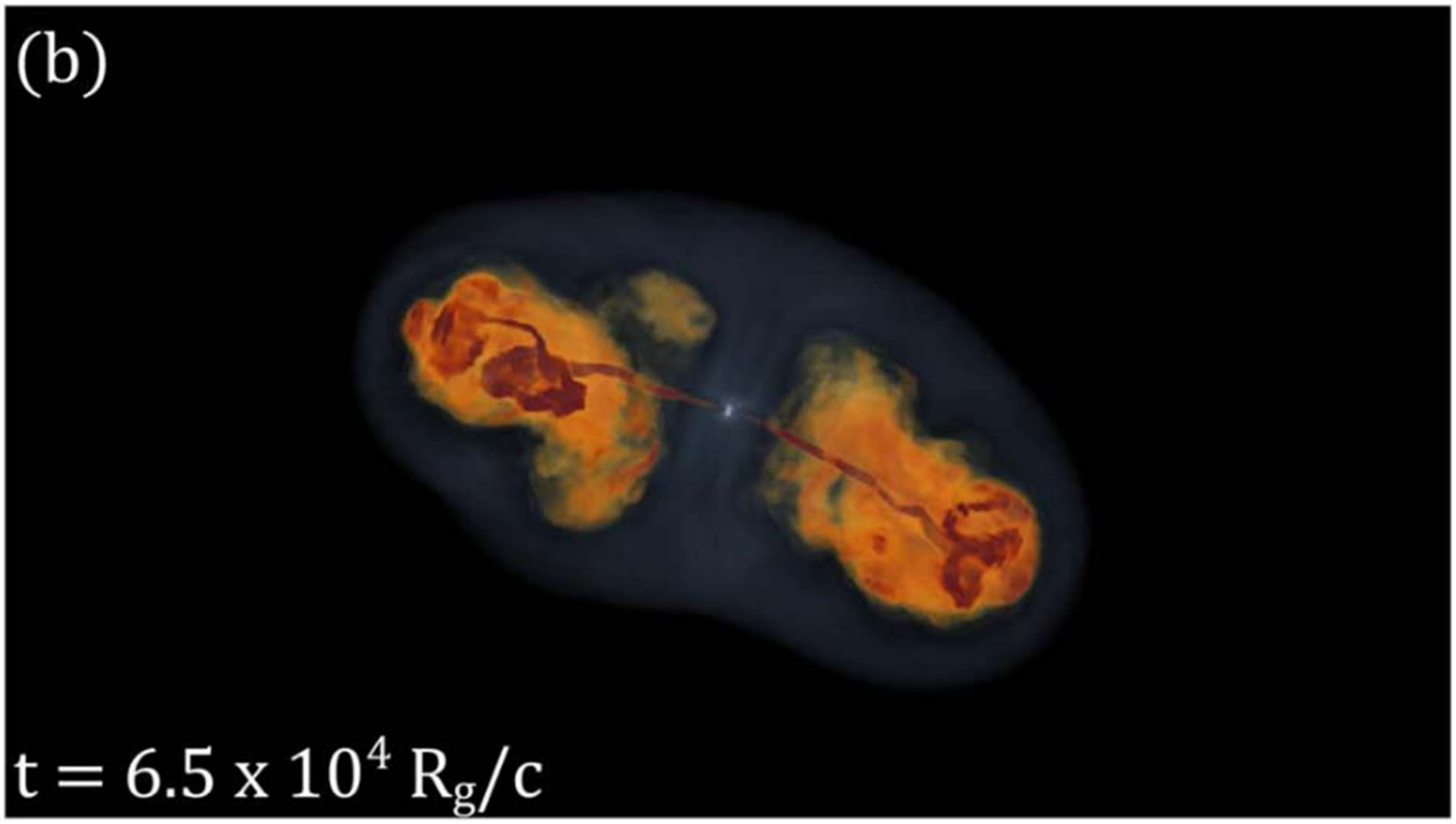

**Figure 1.** A snapshot of global general relativistic magnetohydrodynamic simulations of a kink-unstable jet beyond the Bondi radius [2].

## Instruments and Modes Used

PRIMAGer mapping in the polarimetry band of a 20x20 arc min area (galaxy and jet)

## Approximate Integration Time

Cen A surface brightness is on the order of 0.3 mJy/arcsec$^2$ at 250 microns [4], which we estimate corresponds to about 2 mJy/beam at PRIMager's longest wavelengths. To obtain a minimal detectable polarization of 1%, PRIMA will take roughly 13 hours to map the jet and the galaxy (400 arcmin$^2$).

## Special Capabilities Needed

N/A

## Synergies with Other Facilities

VLA and Chandra images are already available.

## Description of Observations

PRIMAger can obtain the infrared polarized emission map for Cen A in about half a day . Since the Cen A jet is rather steady in time, it is not necessary to perform the observation continuously. If possible, it will be helpful to have roughly contemporary radio and X-ray images of the Cen A jet with PRIMA observation.

# 4. Probing Particle Acceleration in Blazars using Far-Infrared Polarimetry Monitoring with PRIMA

Haocheng Zhang (University of Maryland, Baltimore County/NASA Goddard Space Flight Center)

Blazars are among the most powerful cosmic particle accelerators in our Universe. Optical polarimetry and the recently available X-ray polarimetry have revealed that polarization signatures can vary significantly during blazar flares, indicating the co-evolution of magnetic fields and particle acceleration in high-synchrotron-peaked blazars. With the far-infrared polarimetry capability of PRIMA, we can expand the multi-wavelength polarimetry study of blazars. In particular, this will put unprecedented constraints on the particle acceleration and magnetic field evolution in the low-synchrotron-peaked blazars, which are the most energetic type of blazars and potential sources of extragalactic cosmic rays and neutrinos.

## Science Justification

Blazars are relativistic plasma jets launched from actively accreting supermassive black holes that point very close to our line of sight. They exhibit highly variable, nonthermal-dominated emission across the entire electromagnetic spectrum on all time scales, ranging from minutes to years. Recent IceCube detection of a very high-energy neutrino event in coincidence with a flare in the blazar TXS 0506+056 suggests that blazars can be potential neutrino emitters [1]. Additionally, observations suggest potential quasi-periodic oscillations in a few blazars, suggesting that they may host binary supermassive black holes at the center [2]. These findings make blazars top targets for time-domain and multi-messenger astronomy. Among all blazars, low-synchrotron-peaked blazars, whose primary electron synchrotron spectrum peaks in the infrared bands, are believed to be the most powerful. They typically have a higher bulk Lorentz factor and magnetic field strength in the flaring region, ideal for accelerating very high-energy cosmic rays. A comprehensive understanding of the particle acceleration and physical conditions in these blazar flaring sites is thus crucial.

Despite decades of theoretical and observational efforts, the particle acceleration mechanism and physical conditions of these blazars remain elusive. This is because the three frequently considered particle acceleration mechanisms, namely shock, magnetic reconnection, and turbulence, all can reproduce similar spectra and light curves as observed [3]. Recently, the X-ray polarimetry IXPE and optical polarimeters have found a higher X-ray polarization degree than the optical counterparts in high-synchrotron-peaked blazars, suggesting an energy-stratified jet [4]. This provides the first clean constraint on the jet physical conditions, promoting follow-up studies on the particle acceleration mechanisms. However, the above only applies to high-synchrotron-peaked blazars, because the synchrotron spectrum of low-synchrotron-peaked blazars only extends to the optical band. To achieve similar insights on particle acceleration and jet physical conditions for low-synchrotron-peaked blazars, we need infrared polarimetry.

PRIMA has exactly the necessary polarimetry capability. The PRIMAger instrument can perform polarimetry from 92-235 micron. Given the high flux (on the order of Jy) of bright low-synchrotron-peaked blazars, PRIMA can obtain minimal detectable polarization down to 1% in less than an hour, ideal for tracking the highly variable blazar emission. Combined with optical polarization data, PRIMA will deliver unprecedented constraints on particle acceleration and jet physical conditions in low-synchrotron-peaked blazars.

To fully understand the blazar flaring mechanism, we need to self-consistently study the plasma dynamics and particle transport in the blazar zone. In the recent decade, combined particle-in-cell and radiation transfer simulations have matured. Previous works have shown that multi-wavelength polarization, in particular, the time-dependent patterns can distinguish magnetic reconnection and turbulence [5]. The following figure shows a comparison between time-dependent infrared and optical polarization degree and angle for magnetic reconnection and turbulence [6]. We can see that not only are the infrared and optical evolution different, but that difference is also different between reconnection and turbulence. We plan to perform a parameter study for both mechanisms and statistically evaluate the differences in time-dependent polarization signatures. When PRIMA launches, these results and future simulations for the shock scenario can be tested by both long-term unbiased daily monitoring and ToO observations when blazars are flaring. We expect to unambiguously distinguish shock, magnetic reconnection, and turbulence and shed light on the physical conditions of the low-synchrotron-peaked blazars as well as the maximal energy these blazars can achieve for accelerating cosmic rays.

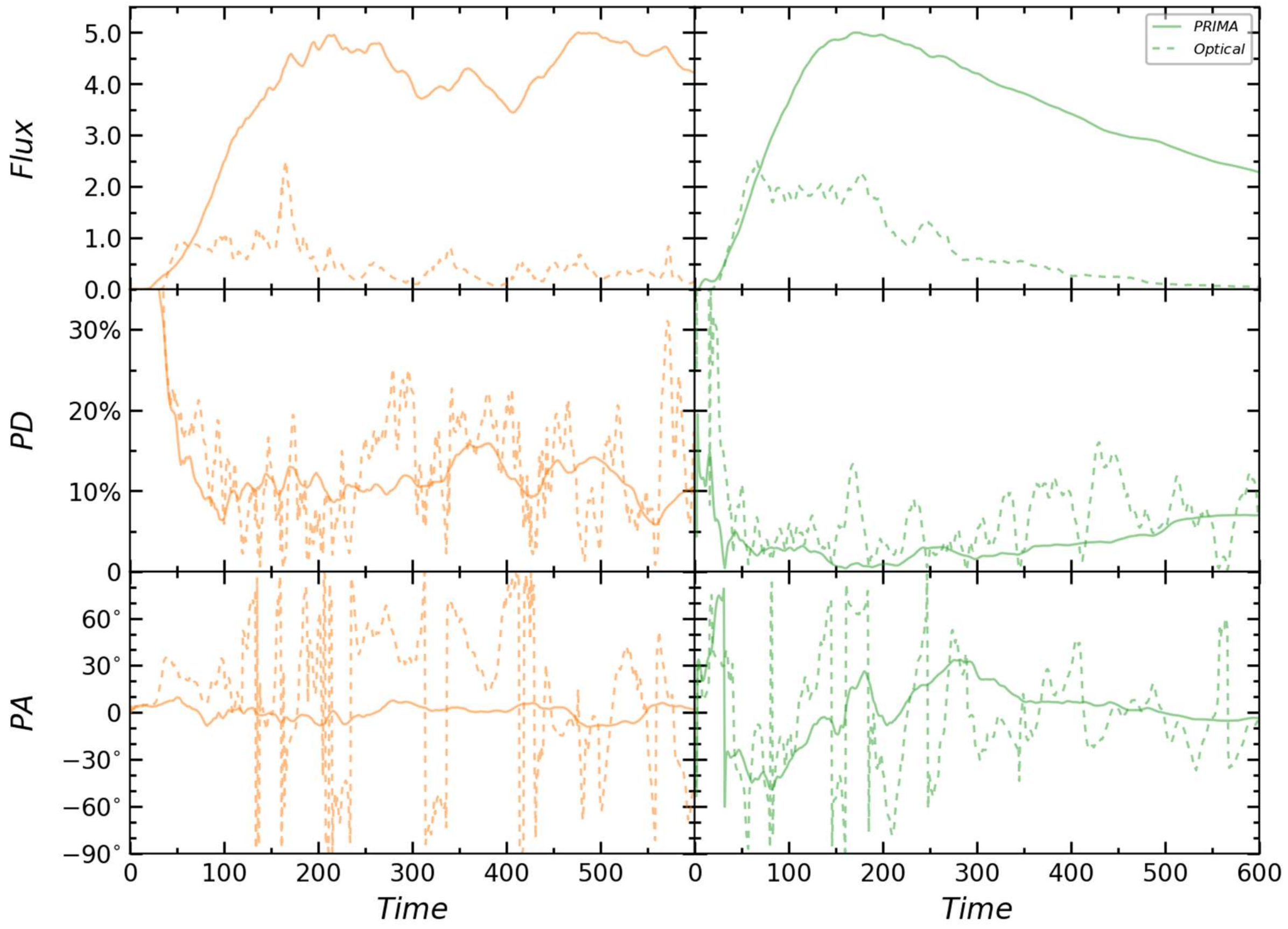

**Figure 1.** PRIMA and optical light curves, polarization degree, and angle for magnetic reconnection (left panels) and turbulence (right panels).

## Instruments and Modes Used

The PRIMAger imager in its polarimeter mode for all bands will be used.

## Approximate Integration Time

We plan to perform a half-year daily polarimetry monitoring of a few selected bright blazars and a few ToO observations with hourly polarization data when some of these blazars show flaring activities. Bright low-synchrotron-peaked blazars can reach ~Jy level flux in the PRIMAger bands. We need to reach a minimal detectable polarization of ~1%, which takes about 10 minutes. We will select three bright blazars for daily monitoring with the 96 micron band for half a year, which can take at most 190 hours. We plan to also do ToO for three bright flares for all four bands with as continuous data as possible. Typical blazar flares last for a few days to one week; they may take in total about 200 hours.

## Special Capabilities Needed

N/A

## Synergies with Other Facilities

To obtain the best scientific return, we will need multi-wavelength supports, such as VLBA, optical polarimetry at Steward Observatory, COSI, and/or Fermi-LAT.

## Description of Observations

We will use PRIMAger to perform daily monitoring for three bright blazars. A sample list can be 3C 454.3, BL Lac, and 3C 279. We will perform this monitoring with the 96 micron band for half a year, and perform optical polarization monitoring, COSI, or Fermi-LAT monitoring for the same blazars with one-day cadence as well, regardless of whether these blazars are flaring or not. We will then study the statistical properties of this unbiased dataset and examine whether it matches better with the magnetic reconnection or turbulence simulation predictions.

We will also use PRIMAger to perform ToO observations for a few bright blazar flares. They will be triggered based on a specific flux level. We will do this with all four bands of the polarimeter and obtain hourly flux and polarization variations during flares. Results will be compared with simulations to check how they match the simulated variability patterns.

# Cosmology

# 5. Resolving the Cosmic Infrared Background with PRIMAger through Gravitational Lensing Clusters

Bunyo Hatsukade (National Astronomical Observatory of Japan)

We propose deep 250 µm observations with PRIMAger targeting galaxy lensing cluster fields to probe the faint end of number counts contributing to the Cosmic Infrared Background (CIB). While Herschel surveys have resolved only ~15–20% of the CIB into individual sources due to severe source confusion, recent developments in Wiener filtering and deblending techniques allow PRIMAger to push significantly below the classical confusion limit. By leveraging strong gravitational lensing (magnification factors > 10), we aim to detect intrinsically faint sources down to ~1–2 mJy, essential for resolving the remaining, unexplained fraction of the CIB. We estimate that surveying ~13 square degrees across multiple high-magnification cluster regions will enable statistically robust measurements of the faint end of number counts, potentially uncovering new populations of faint, dusty galaxies. Our proposed observations will provide critical insights into the obscured star formation and galaxy evolution of the universe.

## Science Justification

The Cosmic Infrared Background (CIB) represents the integrated infrared light emitted throughout the history of the universe, primarily originating from the activity of star formation and active galactic nuclei (AGNs) that have been absorbed and reradiated by dust (Lagache et al. 2005). The CIB encodes the cumulative history of star formation and black hole accretion, making its detailed characterization critical for understanding the complete picture of cosmic activity. In the far-infrared, the dominant sources of the CIB are believed to be dust-obscured star-forming galaxies.

Deep surveys with the Herschel Space Observatory have attempted to resolve the sources contributing to the CIB. However, at 250 µm—where Herschel's bands overlap with the PRIMA wavelength range—only about 15–20% of the CIB was resolved into discrete sources (Oliver et al. 2010; Béthermin et al. 2012). This limitation arises primarily from source confusion, where multiple faint objects blend within a single telescope beam due to finite angular resolution. To circumvent this, stacking analyses have been applied using known source catalogs, which allowed ~70% of the CIB to be resolved into point sources (Béthermin et al. 2012). However, stacking analysis inherently relies on existing catalogs, meaning that any sources absent from these catalogs remain unaccounted for. In addition, such analyses require careful corrections for clustering effects and contamination from nearby bright sources. An alternative approach is using P(D) fluctuation analysis, which statistically infers the contribution of sources below the confusion limit (e.g., Glenn et al. 2010). Recently, Varnish et al. (2025) applied P(D) analysis to the Herschel-SPIRE Dark Field (Pearson et al. 2025) and reported evidence of previously unidentified bumps and turnovers in the number counts below 2 mJy, hinting at the possible existence of an unknown population of faint sources. However, the P(D) method is limited by its dependence on prior assumptions and the difficulty of interpreting the derived physical results.

In this study, we aim to probe the faint end of number counts at 250 µm through deep observations with PRIMAger. Although PRIMA's 1.8 m aperture faces significant confusion limits, Donnellan et al. (2024) have demonstrated through simulations that the application of Wiener filtering techniques and XID+ deblending tools can reliably extract fluxes for sources below the classical confusion limit. Without relying on priors, Wiener-filtered blind catalogs can achieve 95% purity down to depths of ~18 mJy at 235 µm—about 2.5 times deeper than the classical confusion limit.

However, theoretical models and past observations indicate that to fully resolve the 250 µm CIB, detections of sources fainter than ~2 mJy are required (Béthermin et al. 2012). This is where gravitational lensing by galaxy clusters becomes a key observational strategy. Thanks to strong lensing, intrinsically faint background galaxies can be magnified well above the nominal detection limit.

This approach has proven successful with Herschel cluster surveys (e.g., Egami et al. 2010) and in the (sub-)millimeter using ALMA. For example, the ALMA Lensing Cluster Survey (ALCS; Fujimoto et al. 2024) surveyed 33 massive lensing clusters, placing tight constraints on the faint-end number counts at 1 mm and resolving a significant fraction of the CIB.

By targeting cluster fields with magnification factors exceeding ~10, PRIMAger will be capable of detecting sources intrinsically fainter than 2 mJy at 250 µm, offering a unique window into the unresolved components of the CIB.

## Instruments and Modes Used

PRIMAger PPI Band 4 (235 µm).

## Approximate Integration Time

According to the Exposure Time Calculator, reaching a survey depth (5σ) of 18 mJy in PRIMAger 235 µm (band 4) over a total survey area of 13 square degrees would require ~8.5 hours of integration time.

## Special Capabilities Needed

N/A

## Synergies with Other Facilities

To investigate the nature of extremely faint sources, sensitive multi-wavelength observations are essential. Given the limited spatial resolution of PRIMA, follow-up observations with higher spatial resolution are necessary. By constructing spectral energy distributions (SEDs) using multi-wavelength data from facilities such as JWST and ALMA, we aim to elucidate the physical properties of the sources that contribute significantly to the CIB.

## Description of Observations

According to the 250 µm number count models of Béthermin et al. (2012), ~2 mJy sources are expected at a surface density of ~2 sources per square arcminute, while ~1 mJy sources are expected at ~4 sources per square arcminute. We aim to detect this number of sources five times to achieve statistically meaningful constraints on the faint end of the number counts.

We estimate the area of high-magnification regions ($\mu \gtrsim 10$) based on the estimates from Fujimoto et al. (2024), where the ALCS surveyed the high-magnification regions of 33 lensing clusters. Out of a total surveyed area of 133 square arcminutes, ~ 1 square arcminute had a magnification factor of ~10. Therefore, to increase the number of faint sources detected by a factor of five, we need to survey ~5 times this area. The construction of an optimal target cluster sample remains an open task. There is also room for adjustment regarding how much each galaxy cluster region should be covered. The Herschel Lensing Survey (HLS) used the Large Map mode of SPIRE, covering 17' x 17' per cluster (Egami et al. 2010). Based on this, our proposed survey would cover a total area of ~13 square degrees.

According to the PRIMA Exposure Calculator, achieving a 5σ survey depth of 18 mJy over 13 square degrees would require ~ 8.5 hours of observing time.

# 6. Evolution of ISM Properties of Galaxies Along Large-Scale Structures Probed with Multi-Object Optical Fiber Spectroscopy and PRIMA IR Spectroscopy

Mariko Kubo (Kwansei Gakuin University), Hideki Umehata (Nagoya University), Tohru Nagao (Ehime University), Hanae Inami (Hiroshima University)

Galaxies have formed and evolved along the filamentary large-scale structures and clusters of galaxies at their nodes, woven by baryons and dark matter. At high redshift, cold gas inflows along filaments promote star formation, while as the cluster core is virialized, hot intra-cluster medium (ICM) halts cold gas accretion onto galaxies. In the meantime, outflows from galaxies eject metals to enrich the intergalactic medium. To understand the origin of mass and environmental dependence on galaxies today, and the evolution of intergalactic gas, it is essential to resolve such an ecosystem of matter on a large scale. The optical/NIR wide field imagers like Subaru/HSC, Euclid, Rubin, and Roman (will) explore large-scale structures of galaxies at high redshifts. Still, it is hard to observe diffuse filamentary intergalactic gas distribution directly. Alternatively, the impression of gas inflows and outflows can be seen as the mass and environmental dependence of star formation and gas-phase metallicity determined by the balance of inflows, outflows, and star formation. Now, Subaru/PFS and VLT/MOON, wide-field multi-object fiber spectrographs in the optical, enable us to explore the ISM properties of galaxies statistically down to $M^* \sim 1E+9$ $M_{sun}$ or less along large-scale structures at $z < 1.5$. However, only the metallicities of the unattenuated parts of galaxies can be measured with the optical line diagnostics. Here, we propose PRIMA FIRESS spectroscopy for z = 1-1.5 large-scale structures, synergistic with optical wide-field spectroscopic observations, to perform the first panchromatic view of the mass and environmental dependence of gas-phase metallicities.

## Science Justification

### Broader Context

Galaxies are formed at the intersections of filamentary networks of matter woven by baryons and dark matter, called the cosmic web. From a broad perspective, galaxies are distributed along large-scale structures and in clusters at the intersections. The physical properties are tightly correlated with their mass and environment (Peng et al. 2010), and how matter flows along large-scale structures to promote galaxy formation is one of the key questions of modern cosmology. At early epochs, efficient cold gas accretion along filaments promotes star formation in galaxies, whereas at later epochs, as cluster halos grow, the hot intra-cluster medium (ICM) appears to suppress cold gas accretion onto galaxies. Moreover, ICM can remove the cold gas from infalling galaxies through ram-pressure stripping to halt their star formation. On the other hand, outflows from galaxies eject metals into their surroundings to form metal-enriched ICM of modern clusters, but dense surrounding gas can prevent outflows, confining metals within the interstellar medium (ISM) of galaxies. Thus, resolving such a large-scale ecosystem of galaxies and intergalactic gas is essential to understanding the evolution of galaxies.

Large-scale structures at z > 0.5 have been explored by wide-field optical/NIR imagers such as Hyper Suprime-Cam (HSC) on the Subaru Telescope (e.g., Hayashi et al. 2018), and now Euclid, Rubin, and Roman will further explore galaxy distributions at high redshift. However, it is challenging to observe the neutral intergalactic medium (IGM) along the cosmic web. Instead, the metallicities and star formation of galaxies should reflect cold gas inflows and confinement of outflows via their gaseous environments (e.g., Pérez-Martínez et al. 2024). Now, the multi-object wide-field fiber spectrographs, Subaru/PFS and VLT/MOONS, enable us to effectively characterize the large-scale structures and ISM properties of galaxies, including those with low masses ($M^* < 1E+9\ M_{sun}$), down to z < 1.5 using optical/NIR emission line diagnostics (Figure 1). However, it only informs about the unobscured part of the ionized gas in galaxies. To obtain a panchromatic picture of galaxies and the intergalactic gas ecosystem, we need far-infrared emission lines that are observable for heavily obscured galaxies.

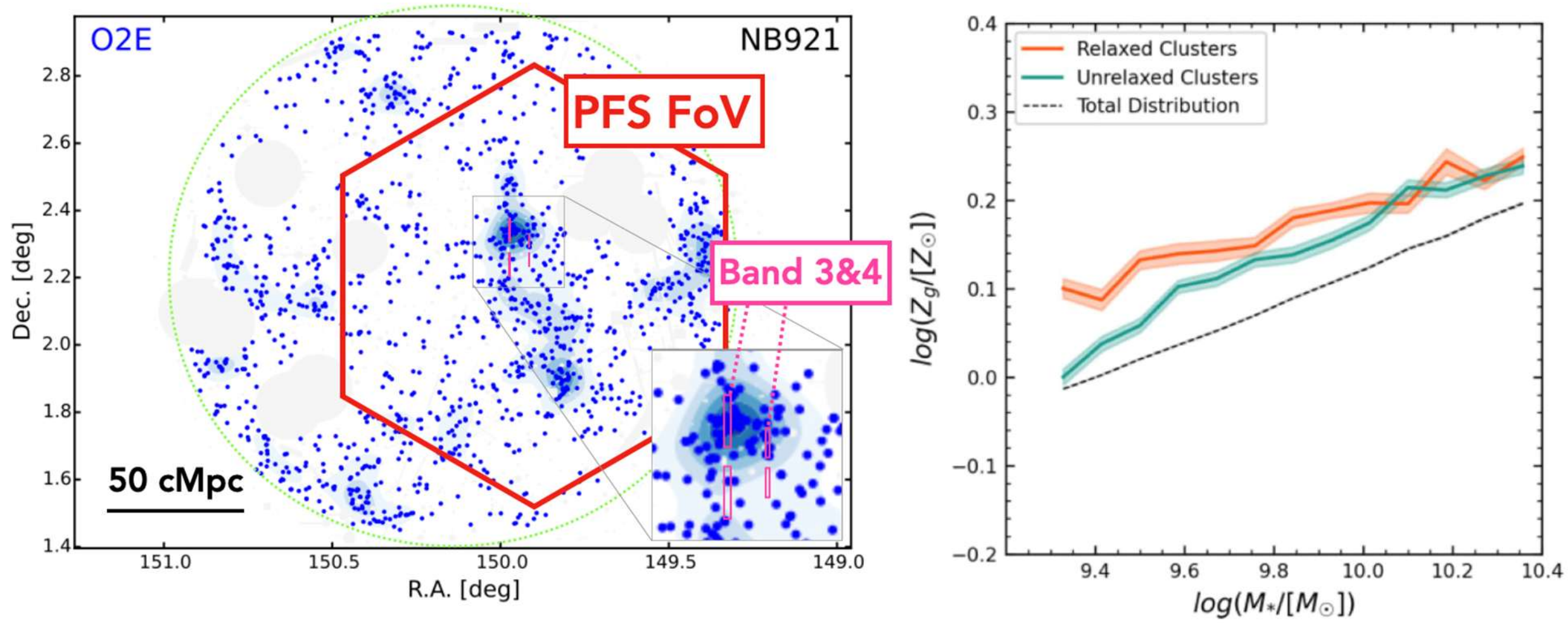

**Figure 1.** (*Left*) The emission line galaxies at z ∼ 1.5 selected from the HSC-SSP survey (Hayashi et al. 2018) in the COSMOS field compared to Subaru/PFS and PRIMA band 3 & 4 FoVs. (*Right*) Mass dependence of metallicities in clusters, unrelaxed clusters, and the field at z ∼ 0.6 from Horizon simulation (Rowntree et al. 2024).

## Science Questions

How environmental effects have processed galaxies, and the ISM has evolved in line with the environment.

## Need for PRIMA

First, a high sensitivity in the far-infrared is needed to perform emission line diagnostics of galaxies at high redshift. A high survey speed is also essential for characterizing the environments of galaxies and performing their statistical study. The PRIMA is the only observatory to achieve a deep and wide infrared survey with angular resolution and sensitivity enough for high-redshift galaxies.

## Interpretation Methods

The metallicity of the gas is constrained with [O III]51.80+[O III]88.33/[N III]57.21 line diagnostics (Nagao et al. 2011; Peng et al. 2021).

## Link to Testable Hypotheses

The metallicities of galaxies measured with PRIMA will be compared to those measured by multi-object and wide-field fiber spectroscopy in the optical/NIR by Subaru/PFS and VLT/MOONS, as well as slit-less spectroscopy by Euclid and Roman. The results will be directly compared to cosmological hydrodynamical simulations (e.g., Rowntree et al. 2024 in Figure 1)

## Instruments and Modes Used

The FIRESS spectrometer. In its low-resolution mode will be used.

## Approximate Integration Time

We request the sensitivity to perform [O III]51.81, [O III]88.36, and [N III] 57.32 line diagnostics (Nagao et al. 2011) to constrain metallicities for galaxies with L > 1E+12 Lsun at z ~ 1. According to Spinoglio et al. (2012), [O III]51.81 and [O III]88.36 intensities are ~4E-19 and 6E-19 Wm-2, respectively. We request to measure [N III]57.32 down to [O III]51.81 / [N III]57.32 ~ 5 (Z ~0.2 Zsun, Nagao et al. 2011). Thus, the requested 5σ sensitivity is 1E-19 Wm-2, which needs 3.6 hours of integration per pointing. Along with large-scale structures, several luminous infrared sources will be simultaneously observed with the FIRESS wide FoV (Figure 1, right). Only Bands 3 and 4 are needed for the requested line diagnostics, but the band 1 and 2 spectroscopy at higher resolution is helpful to reduce source deblending at dense cluster core regions. Then we request 15 pointings (~100 sources). In total, ~100 hours are required.

## Special Capabilities Needed

None

## Synergies with Other Facilities

We target the large-scale structures at z = 1–1.5 in the COSMOS field probed by narrow-band surveys using the HSC on the Subaru Telescope. For the structures in the COSMOS field, metallicities based on the rest-frame optical line diagnostics will be available for a few 1000 objects ($M^*$>1E+9 $M_{sun}$) using the PFS on the Subaru Telescope. By measuring molecular gas mass using ALMA, we can discuss mass loading factors (outflow/SFR) in the context of the gas regulator model (Peng & Maiolino 2014).

## Description of Observations

### Immediate Goal

We aim to measure the metallicities of a statistical number of dusty galaxies along large-scale structures to discuss the environmental dependence on the ISM properties.

### Targets

We plan to observe large-scale structures at z = 1–1.5 because both optical emission line diagnostics with state-of-the-art multi-object fiber spectrographs, such as Subaru/PFS and VLT/MOONS, and far-infrared emission line diagnostics with PRIMA are available. The structures in the COSMOS field are feasible because an extensive PFS spectroscopic project for emission-line galaxies (M*> 1E+9 $M_{sun}$) has begun (S25A: PI MK). The PRIMA will target L > 1E+12 Lsun sources at z∼1 (L>2E+12Lsun at z ∼ 1.5), which can be selected from the multi-wavelength photometric catalog in the case of the COSMOS field.

### Instruments and Observational Strategy

We request FIRESS low-res pointing observations to perform [O III]51.80+[O III]88.33/[N III]57.21 line diagnostics. The requested sensitivity is 1E-19Wm-2, achieved with 3.6 hours of integration. We request 30 pointings to observe 100∼200 sources along large-scale structures. In total, ∼100 hours are required.

# 7. PRIMAx, Axion-Like Dark Matter Searches with PRIMA

Ritoban Basu Thakur (JPL / Caltech), Ryan Plestid (Caltech / CERN), Elena Pinetti (CCA Flatiron)

The nature of dark matter remains a provocative scientific mystery. Outstanding questions in particle physics have motivated several very light (and consequently wave-like) dark matter models including axions and dark photons. These particles couple feebly to Standard Model particles (i.e., ordinary matter) producing faint, but ``smoking gun'', signatures such as nearly mono-energetic photon lines. Detection of these excess photons in halos would enable the first optical observation of dark matter and supply a measurement of its mass. While the particle physics community has been developing direct detection measurements, astrophysical searches – especially using space telescope data continue to provide some of the most sensitive probes of dark matter interactions. Dark matter masses between $10^{-3}$ – 1 eV/$c^2$ are essentially unexplored at present. PRIMA is the only proposed astrophysics mission that can probe this parameter space and (as we discuss) could test well-motivated axion models that address the strong-CP problem. Besides discovery potential, PRIMA will be crucial for cross-verification of any terrestrial detection that awaits us. PRIMAx sets the case for synergistically using PRIMA data from various surveys to probe new parameter space for Axion-like dark matter.

## Science Justification

### Scientific Question

Astro2020 [1] identifies "the unknown physical nature of dark matter [as] an outstanding grand [challenge] in both physics and astronomy." Wave-like dark matter (DM), with masses <$10^{-1}$ eV and de Broglie $\lambda > 10$ µm, are frontrunner candidates. These are closely linked to the Quantum Chromodynamics (QCD) axion, which was theorized to solve the strong-CP problem (the anomalously small electric dipole moment of the neutron). In models of wave-like dark matter, an interaction term often connects the dark sector to the electromagnetic sector of the Standard Model of particle physics. This interaction allows dark matter to convert to regular photons in the presence of a background magnetic field, in a low-rate process. Alternatively, daughter photons from dark matter decay may be searched for using telescope data. For decaying dark matter, the photon energy is one-half that of the dark matter mass. Today, combining laboratory and astrophysical searches, a significant observational gap exists for 250 µm $> \lambda >$ 25 µm or for dark matter masses of 5-50 meV/$c^2$. The question PRIMAx will answer is this: Are there any hints of such photons in the astrophysical context, where we have a good probe of the dark matter density?

### Broader Context

Fig. 1 shows a survey of current and future (next decade) status of axion dark matter searches [2]. This is expressed as the coupling constant between axions and photons vs the axion mass.

Past astrophysical studies constrain the QCD axion quite well above 100 meV/c$^2$, however explorations are lacking at lower masses. Several direct detection experiments are proposed and their reach in the next 10 years is shown by the dashed blue line. One immediately notices that an indirect detection search using PRIMA is competitive with frontier direct detection searches and may therefore could discover QCD axion dark matter.

Indirect detection searches (such as the one outlined herein) offer a complimentary probe of axion-like dark matter and can be cross correlated with direct detection experiments. If a signal is seen in a laboratory experiment corresponding to a specific axion mass, a dedicated analysis to test the compatibility of the signal with decaying axion dark matter can be carried out. The converse is also possible if a discovery is first made using PRIMA's data, and this is especially relevant for laboratory techniques that rely on resonance scanning; external data that constrain the axion's mass is invaluable. An indirect detection search using PRIMA's data is naturally broadband and offers discover potential over a wide range of masses.

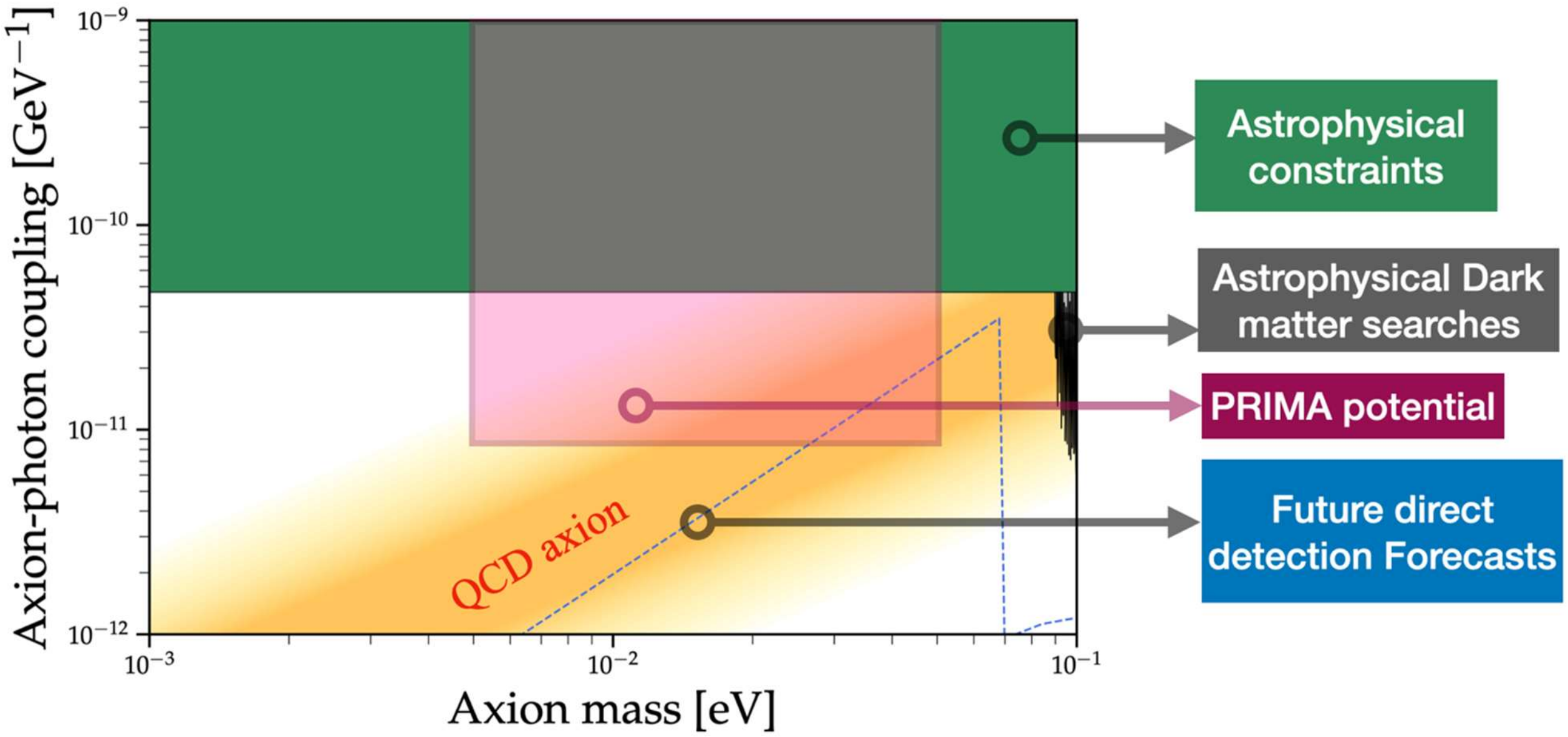

**Fig.1.** Axion-photon coupling landscape shows the ability of PRIMA with modest observations to probe new dark matter parameter space. The QCD axion (gold) band is one theoretical target region. There are currently no real searches < 0.1 eV, and future direct detection experiments plan to reach the blue dashed boundary in a decade or so. Fig adapted from [2], credit Ciaran O'Hare.

## Need for PRIMA

There is an absence of astrophysical constraints in Fig. 1, because there has not been deep space telescope data in the far infrared / millimeter-wave to search for potential signals of decaying dark matter. Terrestrial telescopes at millimeter wavelengths are poorly suited for these searches because of excessive opacity/ photon noise. A recent analysis of near infrared data from JWST has yielded new and improved sensitivity for dark matter in the ~1 eV/c$^2$ mass range [3, 4]. PRIMA covers 25-250 µm, corresponding to a 5-50 meV/c$^2$ dark matter mass window. This opens several opportunities for dark matter searches. Modest precision measurements from PRIMA data can provide initial sensitivity to 5-50 meV/c$^2$ dark matter and therefore offer the opportunity to discover axion-like dark matter (even probing part of the QCD axion band).

## Methods and Hypothesis Testing

PRIMAx will deploy methods ranging from simple flux or photon counting limits and spatial and spectral cross correlation searches to stacked searches, which have been developed by the dark matter community [4-9]. In Fig. 1, we use the continuum model idea from [4] to make an order-of-magnitude forecast. This model searches for narrow ($\frac{\lambda}{\Delta\lambda} \sim 10^3$) line emissions on top of a smooth background to set a limit on the dark matter mass. We emphasize that this methodology is robust and conservative (more detailed analyses are expected to yield even stronger limits).

For a given coupling constant and photon frequency (i.e., dark matter mass), we can estimate the photon flux. The magnitude of the coupling constant is shown in Fig.1, with some upper limits from various searches. This is compared to observations, and if a null result is observed (no excess photons), one can set an upper limit on the size of the photon-axion coupling strength. Crucially, even limited survey sizes can yield valuable sensitivity. By way of contrast, knowledge of the dark matter density in the line of sight and integration times is crucial. For e.g., Fig. 2 shows what two patches/pointings may be. The exact locations are not very important; the contrast in the dark matter density (grayscale) is more important.

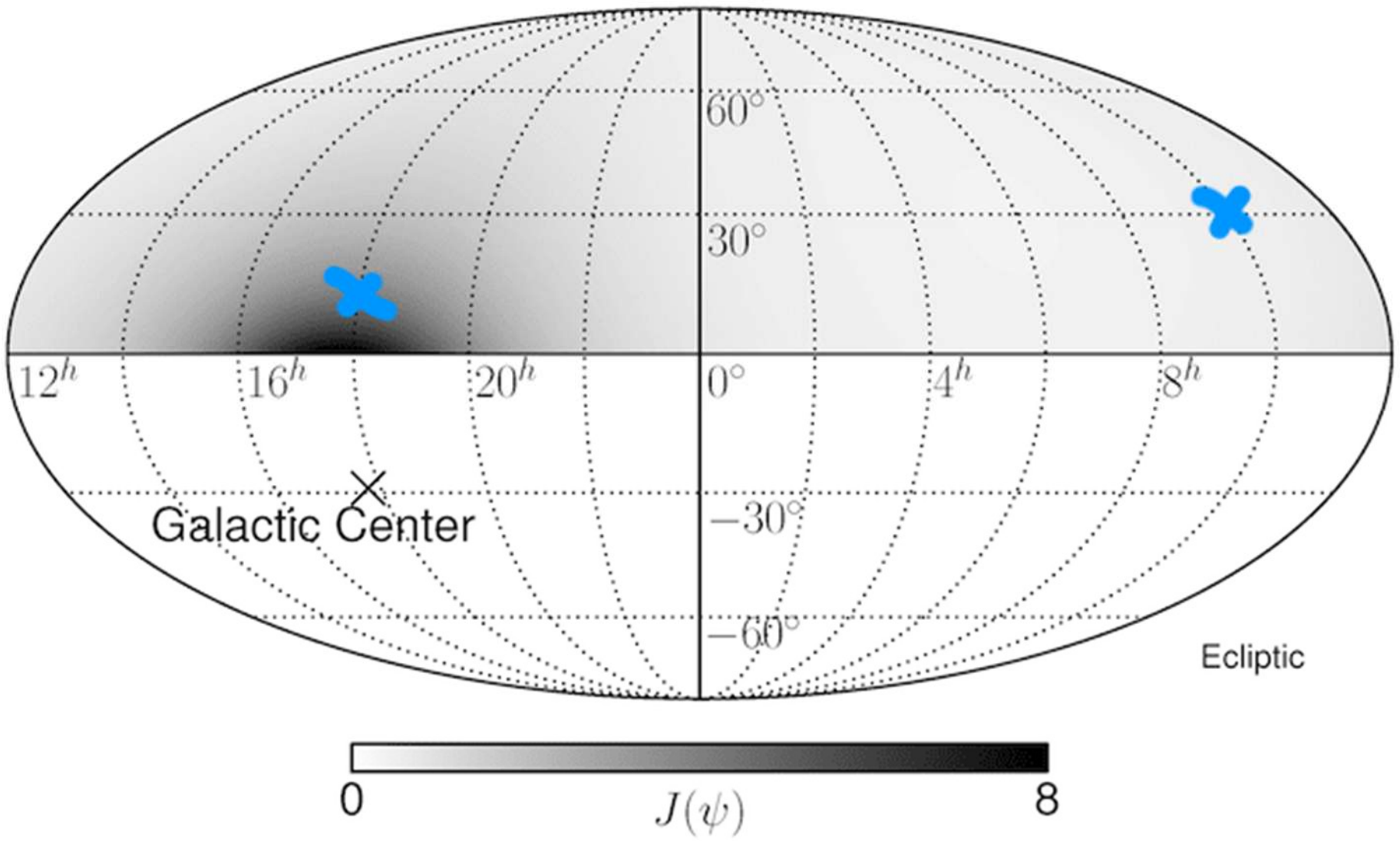

**Fig. 2.** Line-of-sight integral of dark matter density for the Northern Hemisphere in equatorial coordinates, from [5]. Example pointings (blue crosses) probe different line emission strengths by probing regions with different dark matter densities. The exact location or map size of these pointings are not very relevant, simply that different (grayscale) dark matter densities are captured.

For the estimate in Fig. 1, we assume a measurement of ~0.1 MJy/sr flux limit. Real observations will inevitably contain spectral variations, and so we expect that the actual limit will be centered around coupling ~ $10^{-11}$ with fluctuations due to noise in the data.

The process described above is akin to bump-hunting in particle physics and can be further improved with Bayesian methods if one has data from multiple pointings/ small patches [10-12].

Consider (spectral) data from two sky patches with different known dark matter densities. In the absence of any line detection, the lowest flux patch sets the limit. However, if there is a detection, joint probability estimators can be used to provide much deeper constraints than naïve counting limits.

The spectral resolution, $R = \frac{\lambda}{\Delta\lambda}$, for such dark matter searches, is set by the Maxwell-Boltzmann velocity distribution of dark matter and roughly maps to $R \sim 10^3$. Thus, if any unknown signal is seen, it is possible to follow up in high-*R* mode and understand if the line shape matches what is expected from dark matter astrophysics. In addition to strengthening the discovery potential of the search, this methodology could be used to disentangle axion dark matter spectral features from noise, thereby improving sensitivity even in the absence of a dark matter signal. Furthermore, as shown by [6], with half a dozen pointings with high-*R* measurements, one can disentangle the indirect galactic dark matter signal via velocity spectroscopy.

The above discussion is in the context of probing our galactic dark matter halo. It is also possible to invoke weighted filters by stacking data from selected targets where we have measurements of large dark matter densities, e.g., dwarf galaxies, [7-9]. These methods may further enhance sensitivity; however, a more detailed study is required before projections can be made.

## Instruments and Modes Used

Details on observation strategy (later section) discuss that this effort can be realized with existing observation plans (those of the deep variety with sensitivities < 0.1MJy/sr) from other GO programs. Primary study only needs the low-resolution data, and should signals be discovered, high-resolution follow-up may be pursued to understand the origin of excess flux at a specific wavelength. To that end, we will use PRIMAGER (both bands) first. And FIRESS FTM, high-res mode matched to $R \sim 10^3$ can be used for follow-up studies, if and as required.

As will be discussed in the observations section, our proposed work is synergistic with other GO programs. To achieve a first round of dark matter search results, we will not need specific dedicated surveys and can use data from other GO programs. Nonetheless, we quote our needs in a stand-alone manner here.

Our minimal requirements are with PRIMAGER for a low-resolution line search.

Potential FIRESS surveys may be required, especially aimed at understanding line emissions.

## Approximate Integration Time

Following the overhead included sensitivity available at https://prima.ipac.caltech.edu/page/instruments, we essentially need 5x data sets matching the table entries, for 10x more time, which are our minimal needs for standalone PRIMAGER based studies. This implies a total time of 500 hours (5 pointings, 100 hours each).

Upon discovery of signals, high-resolution data from FIRESS will be needed, as detailed in the observation section. The exact integration time will depend on the line strength and frequency. However, with FTM 1 hour 5 sigma flux sensitivity of $7 \times 10^{-19}$ W/m$^2$, we estimate a few thousand hours for one pointing (wherever the line emission was captured in coarse PRIMAGER data).

## Special Capabilities Needed

None

## Synergies with Other Facilities

As mentioned, results from PRIMAx will complement terrestrial direct searches of dark matter. In the case of any detection, these results will guide the design of future terrestrial experiments. Thus, there is a strong synergy with a slew of different experiments, primarily BREAD [2, and references therein]. The first author Basu Thakur is involved as JPL-PI with the BREAD collaboration and will develop joint analyses for space-based and ground-based dark matter searches.

## Description of Observations

Our proposed PRIMAx effort does not need specific observation targets and can be pursued synergistically with other planned GO programs, e.g., PRIM(All): A PRIMA All-Sky, Polarized, Dust Map and Point Source Catalog [First author Saydjari, and Basu Thakur is a co-author]. Dark matter search studies as proposed here can be done on most observations that are sufficiently deep (~0.1 MJy/sr). The main challenge in such searches is the foreground modeling; thus, it is programmatically beneficial to run the analysis as a knock-on process to specific astrophysics studies that the other GO programs will naturally pursue. Nonetheless, ideal observations for dark matter searches can be categorized as (i) blank sky observations, which will be done for PRIM(All), (ii) targeted observations of objects that are rich in dark matter and identified already, such as lensed objects, dwarf galaxies, globular clusters etc. Several such objects will be studied in detail by other GO programs, and with their astrophysics teams, we co-analyze the data for such indirect dark matter searches.

As dark matter searches are not as specific as traditional astronomy surveys, at this time, we can't specify a list of targets, though initial targets (pointings) can be follow-ups to [4]. Our initial plan, akin to [3,4] is to perform a fast cadence of excess-photons like line searches. We intend to do this with a variety of data sets, collaborating with other GO programs. Irrespective of this shared-effort philosophy, we have quoted stand-alone metrics in the instrumentation and modes section. We first quote metrics for a coarse spectral scan with PRIMAGER. Should PRIMAx discover lines in the spectrum which are not readily explained by astrophysical modeling, we will request time on FIRESS high-resolution mode as a follow-up to pursue velocity spectroscopy studies, as we expect dark matter engendered signals to have $R{\sim}10^3$.

## Acknowledgments

This research was carried out at the Jet Propulsion Laboratory, California Institute of Technology, under a contract with the National Aeronautics and Space Administration (80NM0018D0004).

# Galaxies

# 8. The Metallicity-PAH Relationship from the Local Universe to Cosmic Noon

Leonor Arriscado (Centro de Astrobiología, Spain), Irene Shivaei (Centro de Astrobiología, Spain), Leindert Boogaard (Leiden University, Netherlands), Tanio Díaz-Santos (IA-FORTH, Greece), Jed McKinney (University of Texas, Austin, USA)

Dust is a key component of galaxies. Although interstellar dust grains comprise less than 1% of the interstellar medium (ISM) mass, they play a critical role by absorbing and scattering up to 99% of a galaxy's ultraviolet (UV) light and re-emitting it in the infrared (IR) [1,2,20]. Among these dust grains, polycyclic aromatic hydrocarbon (PAH) molecules—small carbonaceous particles—dominate the mid-IR emission in most galaxies [3,22,23,24].

Previous studies have explored the relationship between PAH emission and gas-phase metallicity, both in the local universe [4,16,17,18,25,26,27] and at cosmic noon [5,28], revealing a correlation between these properties from redshift z ~ 0 to z ~ 2. However, most of the metallicity measurements used in these works are either obtained indirectly (e.g., derived from the fundamental metallicity relation, FMR) or based on optical emission lines, which are subject to dust attenuation uncertainties.

With the unprecedented capabilities of PRIMA, we can now probe metallicity diagnostic lines in the far-infrared (FIR) with FIRESS, which is unaffected by dust attenuation. In synergy with JWST/MIRI, this will enable us to trace the relationship between PAH emission and metallicity from the nearby universe (z ~ 0) to cosmic noon (z ~ 2) for a large sample of galaxies, potentially establishing PAH luminosity as a reliable tracer of metallicity. We aim to conduct a FIRESS low-res survey of 100 arcmin$^2$ overlapping the MIRI multi-band imaging fields, with depth of $1.8\times10^{-23}$ W/cm$^2$ at 3σ, sufficient to detect all the required FIR line metallicity diagnostics at z~0-2 for galaxies with IR luminosities brighter than $10^{12}$ L$_\odot$ to $10^{13}$ L$_\odot$ across the redshift range. The survey will take about 220 hours.

## Science Justification

Understanding the life cycle of dust in galaxies is fundamental to tracing how baryonic matter evolves over cosmic time. PAHs are crucial in this effort, as they dominate the mid-IR emission, providing insight into both dust content and stellar populations in galaxies. PAHs contribute significantly to gas heating through the photoelectric effect ([7,8,9,21]). At the same time, they serve as efficient "coolers" of radiation from young and massive stars, limiting the penetration of UV light into surrounding molecular clouds. This shielding effect reduces the dissociation of $H_2$ and potentially decreases the volume of gas available for future star formation. Various studies have shown a correlation between PAH emission strength or mass fraction and gas-phase metallicity from the local universe to cosmic noon, reflecting the production, growth, and destruction mechanisms of these very small grains in the ISM [e.g., 4,5,16,17,18,19,25,26,27,28].

These studies predominantly rely on indirect metallicity estimates, such as those inferred from the FMR, or on optical emission-line diagnostics. The optical line diagnostics are more reliable but affected by dust attenuation and electron temperature variations (see Fig.1). PRIMA's FIR spectroscopic capabilities (via FIRESS) will allow us to trace FIR emission lines (e.g., [O III] 88 µm, [N III] 57 µm), which provide a direct and dust-insensitive probe of gas-phase metallicity (Fig.1) [10,15,29,30,31,32,33,34].

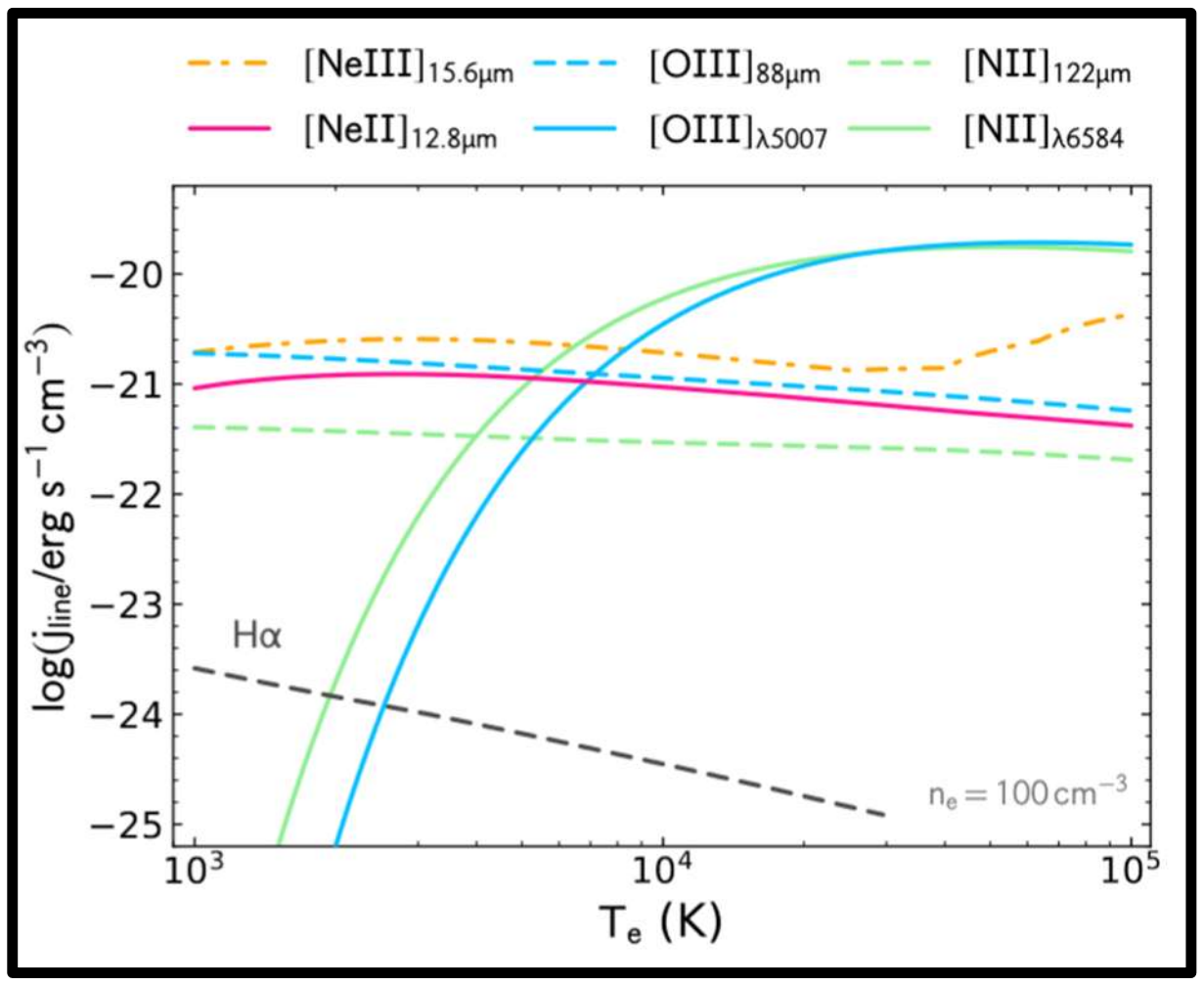

**Fig.1**. Emissivity of fine-line structures ($j_{line}$) vs. electronic temperature ($T_e$) (figure reproduced from [10]). For lines in the optical range ($[OIII]_{\lambda5007}$ and $[NII]_{\lambda6584}$), emissivity is strongly dependent on electronic temperature, in contrast with the IR lines.

PRIMA's FIR spectrometer (FIRESS) combined with JWST MIRI photometry offers an unprecedented opportunity to simultaneously measure robust and dust-unbiased metallicities and PAH fractions, to study PAH formation and destruction processes. This enables a holistic view of the dust-metallicity relationship in galaxies over a broad redshift range, bridging the critical wavelength gap between MIRI and ALMA that is currently inaccessible and essential for studying dusty and high-redshift galaxies (see Fig.2).

This synergy between FIRESS and MIRI will offer new insights into the relationship between PAHs and gas-phase metallicity. Firstly, with robust FIR-based metallicity estimates, we can revisit the PAH-metallicity relation. Secondly, we can test whether PAH luminosity - or, more accurately, PAH luminosity normalized to total dust emission or to the continuum mid-IR emission at rest-frame ~20µm - can serve as a tracer of metallicity, independent of galaxy type or redshift. This would provide a powerful new tool for estimating metallicities in dusty galaxies and enable metallicity mapping in large-area surveys.

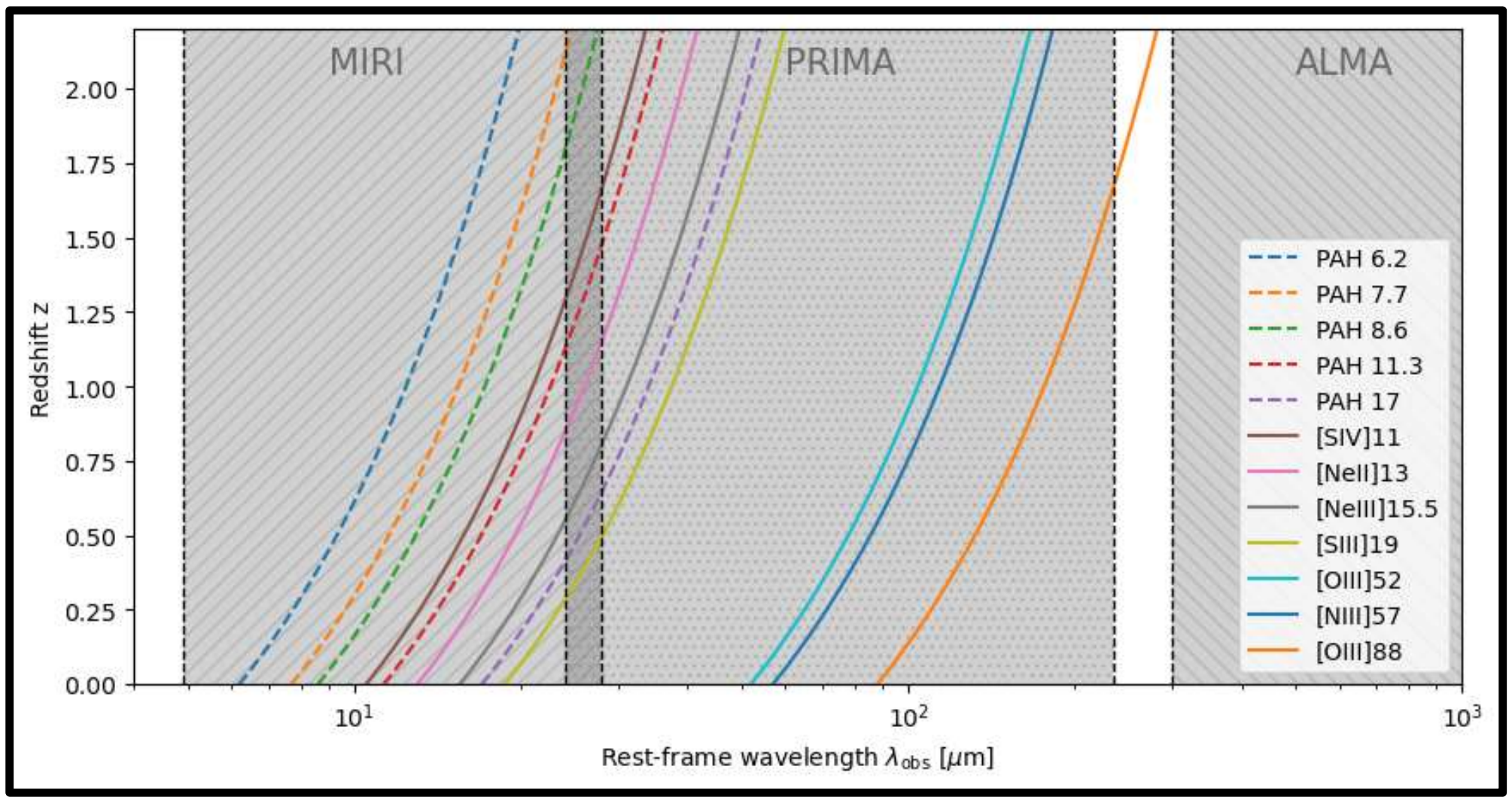

**Fig. 2**. Wavelength coverage of key FIR spectral features (PAHs, emission lines) accessible with MIRI, PRIMA, and ALMA, from the local universe (z ~ 0) to cosmic noon (z ~ 2).

The goal of this proposal is to map the MIRI multi-band surveys using FIRESS low-resolution mapping mode to detect the necessary FIR lines for metallicity estimates. As of Cycle 4 of JWST, there is ~1100 arcmin$^2$ coverage of multiband MIRI photometry in extragalactic deep fields with sufficient depths to detect PAH emission down to the LIRG regime at z<2. The largest surveys include MINERVA, COSMOS-3D, PRIMER, and MEOW. We propose a 100 arcmin$^2$ FIRESS survey, overlapping with a representative subset of the MIRI fields.

The FIR emission lines used in this study are selected based on [11,12], highlighting the flux ratios ([OIII]λ52μm+88μm)/[NIII]57μm and ([NeII]λ12.8μm/[NeIII]λ15.5μm)/([SIV]11μm+[SIII]18μm) as robust metallicity tracers.

**This study, enabled by PRIMA, will allow us to:**

- Quantify the relationship between PAH emission strength and FIR-derived gas-phase metallicity from the local universe to cosmic noon.
- Assess whether the scatter in the PAH-metallicity relation is significantly reduced when using dust-unbiased FIR metallicities compared to traditional optical estimates.
- This flux-limited FIR map will enable various other projects and science cases, including stacking to reach lower line fluxes. By targeting well-studied legacy deep fields such as CANDELS, we maximize scientific return by building on extensive multiwavelength datasets from HST, Spitzer, Herschel, and Chandra. The broad sky coverage of these earlier surveys is ideally matched to PRIMA's wide-field mapping capabilities. The combined FIR spectroscopy and mid-IR imaging are essential to fully constrain the PAH SEDs and trace gas-phase metallicities. The resulting dataset will be of lasting value, supporting a broad range of extragalactic studies with physically consistent FIR data.

## Instruments and Modes Used

This science case will use FIRESS R~100 mapping over 100 arcmin$^2$ (1 map).

## Approximate Integration Time

Under 220 hours on FIRESS low-resolution mode for a 100 arcmin$^2$ map (based on ETC. This provides a sensitivity of 1.2x$10^{-23}$ W/cm$^2$ at 3σ (calculated at 129μm), sufficient to detect the metallicity lines at the targeted redshift range (z~0-2) (see Fig.3).

## Special Capabilities Needed

None

## Synergies with Other Facilities

- HST, Spitzer, Herschel, Chandra: Observations overlap with the CANDELS field
- MIRI: Provides PAH coverage for the targeted redshift range

## Description of Observations

We use the PRIMA Exposure Time Calculator (ETC) to estimate integration times, assuming a galaxy infrared luminosity of $L_{IR}$=($10^{12}$-$10^{13}$) $L_{\odot}$. Line luminosities are derived using the $L_{line}$-$L_{IR}$ calibrations presented in [13]. Based on these estimates, we calculate the expected line fluxes across our target redshift range for the diagnostic lines of interest (see Fig. 3).

The metallicity-sensitive line ratios we adopt are ([OIII]λ52μm+88μm)/[NIII]57μm at lower redshifts (z<1.67) [14], and ([NeII] 12.8 μm+[NeIII] 15.5 μm)/([SIV] 11 μm+[SIII] 18 μm) at higher redshifts (z>1.18) [15]. For a 100 arcmin$^2$ map, and assuming a 3σ depth of 1.2x$10^{-23}$ W/cm2 or 5σ depth of 3.0x$10^{-23}$ W/m$^2$ at 129 μm, the total integration time estimated with the ETC is under 220 hours.

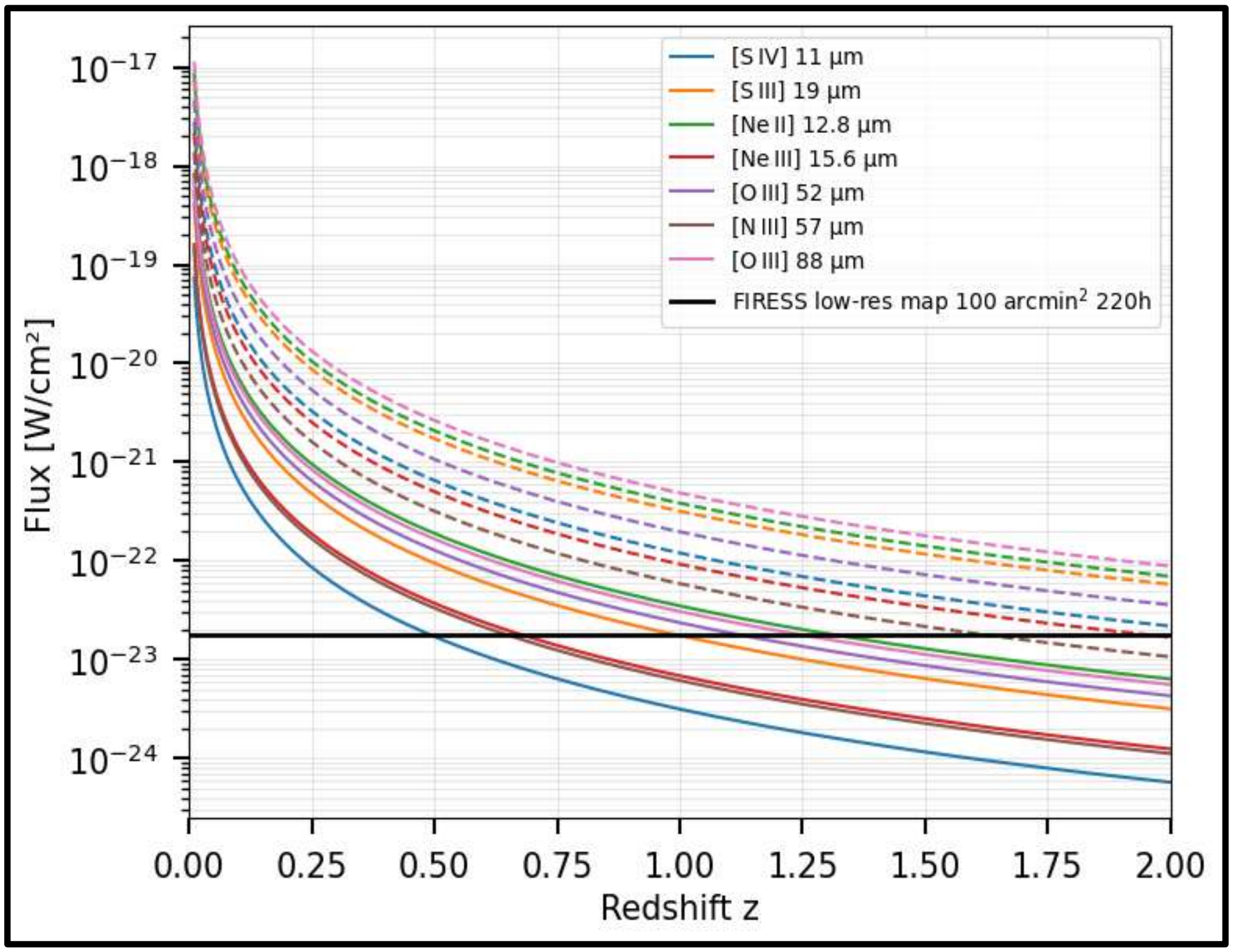

**Fig.3.** Line emission fluxes across the targeted redshift range for the FIR lines used in the metallicity diagnostics, assuming a galaxy infrared luminosity of $L_{IR}=10^{12}\ L_{\odot}$ (solid line) and $L_{IR}=10^{13}\ L_{\odot}$ (dashed line). The horizontal black line represents the sensitivity limit of FIRESS for a 100 arcmin$^2$ map, assuming a 3σ depth of $1.2\times10^{-23}$ W/cm$^2$ at 129µm.

# 9. A New Hope for Obscured AGN: The PRIMA-NewAthena Alliance

Luigi Barchiesi (University of Cape Town, South Africa), F. J. Carrera (IFCA, CSIC-University of Cantabria, Spain), C. Vignali (University of Bologna, Italy), F. Pozzi (University of Bologna, Italy), L. Marchetti (University of Cape Town, South Africa), C. Gruppioni (INAF Bologna, South Africa), I. Delvecchio (INAF Bologna, South Africa), L. Bisigello (INAF Padova, South Africa), F. Calura (INAF Bologna, South Africa), K. Caputi (Kapteyn Astronomical Institute, U. of Groningen, The Netherlands), M. Vaccari (University of Cape Town, South Africa)

Understanding the Active Galactic Nuclei (AGN)-galaxy co-evolution, feedback processes, and the evolution of Black Hole Accretion rate Density (BHAD) requires accurately estimating the contribution from obscured AGN. Recent works show that the fraction of obscured AGN increases with redshift (up to ~80% at z~6) and that at high-z, due to the host galaxies being compact and gas-rich, the ISM can contribute significantly to the AGN obscuration. The key role played by obscured AGN is also revealed by X-ray background (XRB) synthesis models, which predict that a large fraction of the yet-unresolved XRB is due to the most obscured AGN (Compton thick, CT: $N_H > 10^{24} cm^{-2}$). However, detecting these sources is challenging due to significant extinction at the wavelengths typically used to trace their emission.

We propose to perform a deep survey using all the PRIMAger bands (24-261 µm) for the same fields that will be observed with the ESA NewAthena X-ray Observatory. We predict that the combination of these facilities is a powerful tool for selecting and characterizing all types of AGN. While NewAthena is particularly effective at detecting the most luminous, the unobscured, and the moderately obscured AGN, PRIMA excels at identifying heavily obscured sources, including Compton-thick AGN (of which we expect 7,500 detections per $deg^2$). PRIMA will detect ~60 times more sources than *Herschel* over the same area and will allow us to accurately measure the BHAD evolution up to $z$~8, better than any current IR or X-ray survey, finally revealing the true contribution of Compton-thick AGN to the BHAD evolution.

## Science Justification

One of the key open issues in astrophysics is the role of super-massive black holes (SMBHs) in shaping galaxies and the influence of the active galactic nuclei (AGN) on the star formation (SF) properties. While it is now accepted that the growth in BH mass of the AGN and stellar mass ($M*$) of the host galaxy are coupled, the exact mechanism, timescales, and how these influence each other are still a matter of debate. Therefore, their study is fundamental for understanding the evolution of both galaxy and AGN.

### Obscured Accretion Phase

Despite the differences between the various co-evolution models, for example, regarding the co-evolution triggering mechanism and the involved timescales, all scenarios agree that a key phase is that of obscured accretion. In this phase, most of the BH accretion is expected to take place; however, the high quantity of material fueling the SMBH growth also has the effect of hiding the core of the galaxy. Thus, this first phase of growth takes place in very obscured conditions, making its study extremely challenging.

The presence of a population of very obscured AGN is also fundamental to explaining the gap between the measured BHAD and what is expected from simulations (see Fig. 1). At z>3, our measurements of the BHAD rely mostly on X-ray surveys and are at least one order of magnitude lower than what is expected from cosmological simulations. This gap could be bridged if we assume that a significant fraction of the BH accretion happens in very obscured conditions and is missed by X-ray surveys. We are seeing hints of this population with IR surveys (e.g., Delvecchio et al. 2014), although in the high-z Universe (z>3), we must rely almost uniquely on the UV and X-ray bands, which are prone to miss the highly obscured AGN. Recently, the uncertainty on the high-z evolution of the BHAD has been exacerbated by the JWST discovery of "Little Red Dots" (LRDs), which, if we assume to be dust-reddened broad-line AGN, provide BHAD measures more than one magnitude higher than the previous ones (Yang et al. 2023).

These "missing" AGN are also extremely important for explaining the X-ray background (XRB), as the current spectrum of the XRB can only be reproduced by assuming the presence of a population of very obscured AGN that we are currently able to efficiently detect only at low redshift (e.g., Gilli et al. 2007).

A breakthrough in the selection and identification of AGN is expected with the new ESA X-ray observatory NewAthena, due for launch in the 2030s. NewAthena, with its combination of angular resolution, field of view and collecting area at ~1 keV, is the ideal instrument to perform X-ray surveys and will be more than two orders of magnitude faster than Chandra and XMM-Newton (Nandra et al. 2013). Unfortunately, even this new generation instrument will not be able to completely probe the AGN parameter space: most of the CT-AGN will be detected only at very low-z, and the detection of low-luminosity AGN at high-z will be challenging. However, the detection of these elusive sources is well within the capabilities of the PRIMA mission. Deep PRIMAger photometric surveys will allow us to detect these sources up to and beyond z~6, and via the combined use of the two observatories, we will be able to overcome the degeneracies due to single-band studies.

### Simulations

We used XRB synthesis models to predict the total number of AGN as a function of intrinsic X-ray luminosity ($L_X$), amount of obscuration ($N_H$) and redshift (see Barchiesi et al. 2021, 2025). The predictions of the AGN emission are based on the SEDs of more than 500 AGN from the COSMOS field with both X-ray spectra and optical-to-FIR SED-fitting. For each $L_X$-$N_H$-z bin, we estimated the fraction of AGN that will be detected by NewAthena, taking into account the source density and the instrument capabilities. Regarding the IR side of our simulations, we used the above-mentioned SED and randomly extracted 20 of them from the corresponding bins. We measured their total (AGN+host) flux densities in all the PRIMAger bands. If the SED had a higher flux than the survey sensitivity (filter-wise), we considered that source as detected. We assumed a 1000 hr, 1 $deg^2$ deep survey for PRIMA (similar to those envisioned in Donnellan et al. 2024 and Bisigello et al. 2024; see justification below) and compared it with the deepest 1 $deg^2$ of the NewAthena survey (which will comprise 30 300 ks-fields for a total of 12$deg^2$ and 2,500 hr).

To investigate the capabilities of PRIMA in constraining the BHAD evolution, we simulated PRIMA observations at ~98µm, used those to measure the AGN bolometric luminosity function (LF), and consequently recovered the BHAD evolution. In particular, the AGN bolometric LF was measured via the $1/V_{max}$ method and fitted with modified Schechter functions via a non-linear least square fitting algorithm. We fixed the slopes of the bright- and faint-end of the LF to the value found by Delvecchio et al. 2014. By integrating the LFs, with an assumed radiative efficiency of the SMBH of $\epsilon$=0.1 (Delvecchio et al. 2018), we were able to estimate the capabilities of PRIMA in measuring the BHAD evolution. We estimated the associated uncertainties by performing the SED extraction and LF fitting 100 times.

### PRIMA and NewAthena Capabilities

Our simulations show that more than 17,000 AGN (70% of all the AGN at z<10) will be detected by at least one of the PRIMAger filters, while ~7,500 (30% of all AGN) will be detected with all the PRIMAger filters. Moreover, ~6,500 AGN (26%) will also have a detection in the X-rays with NewAthena. The combination of PRIMA and NewAthena will be a powerful tool to completely sample the AGN population up to very high-z, regardless of the source obscuration. NewAthena will excel in detecting the high-luminosity obscured sources and almost all the unobscured AGN, while PRIMA will be very effective in recovering the heavily obscured and CT-AGN population that NewAthena will miss.

A deep PRIMAger survey will provide us with enough detections to allow us to measure the bolometric LFs up to z=8. Thus, we anticipate achieving precise measurements of the BHAD up to z~8 (Fig. 1), enabling us to determine definitively whether the discrepancy between X-ray-derived estimates and theoretical predictions arises from a population of heavily obscured AGN. Furthermore, PRIMA will provide an unprecedented opportunity to study the evolution of AGN and galaxies, probing back to an age of the Universe of ~ 700 Myr. While the BHAD values we obtained depend on the assumed evolution of the total number of AGN, the associated uncertainties do not and effectively illustrate PRIMA's potential in measuring the high-redshift evolution of the BHAD free from obscuration biases. Indeed, it has been shown that starting from IR-derived BHADs (which predict an intrinsically higher BHAD at high-z), we can effectively recover the BHAD with even greater accuracy (Barchiesi et al. 2025).

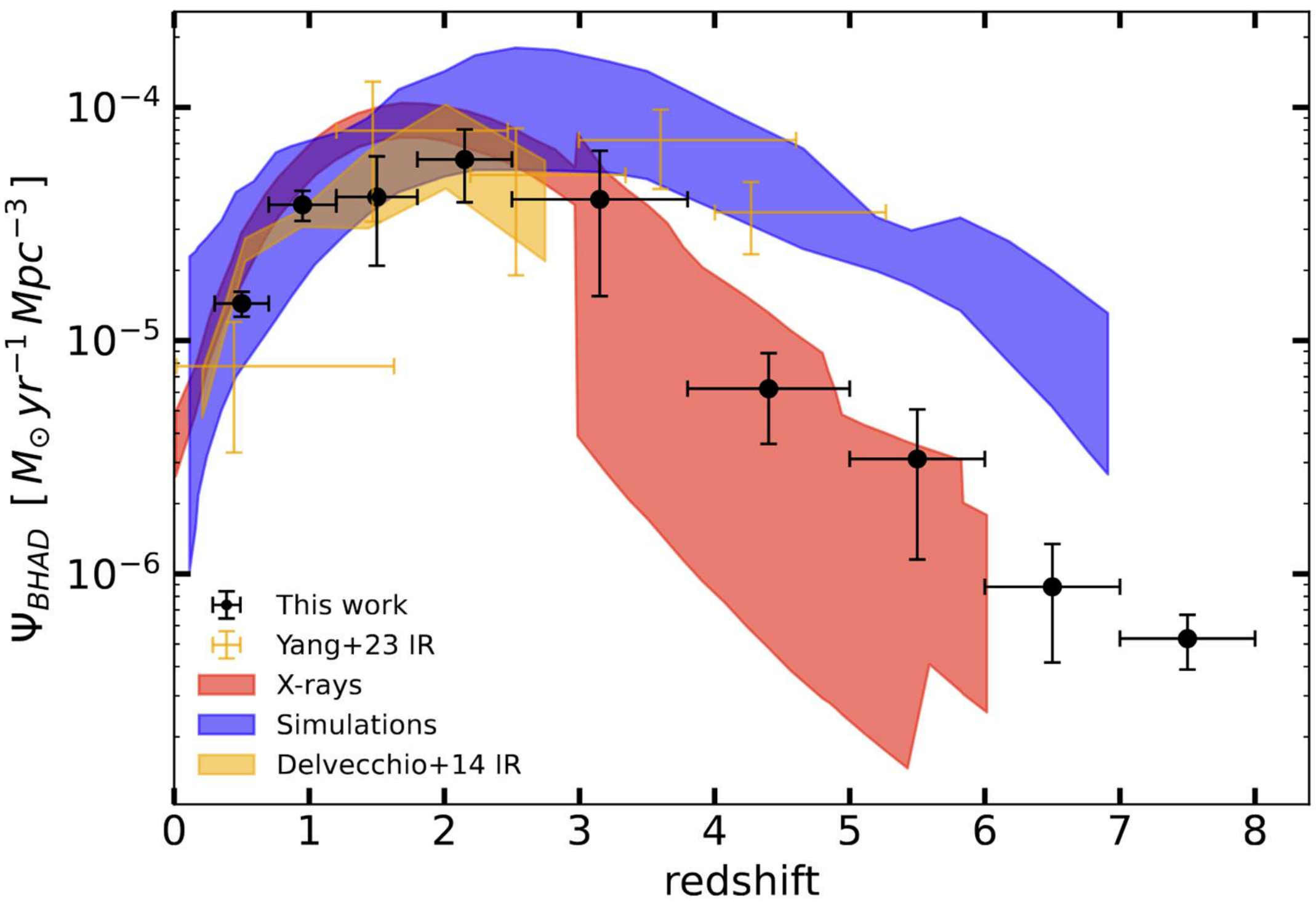

**Figure 1.** Prediction of PRIMA-derived measurements of the BHAD. The black points represent the predictions of the BHAD obtainable with a PRIMA Deep survey with PPI1 (at 98μm, Barchiesi et al. 2025). For comparison, we report the BHAD measured from X-ray surveys (red; Ueda et al. 2014, Vito et al. 2014, Aird et al. 2015, Vito et al. 2018, Pouliasis et al. 2024), and IR surveys (orange; Delvecchio et al. 2014, Yang et al. 2023). In blue, BHAD predictions from simulations (Shankar et al. 2013, Sijacki et al. 2015, Volonteri et al. 2016) are shown. Our points follow the BHAD measured from the X-rays, as our simulations started from X-ray background modeling. PRIMA will be extremely effective in probing the z>3 regime, which shows a gap between the BHAD measurements and the predictions from simulations. This gap could be due to a population of heavily obscured AGN missed by current surveys (mostly X-rays) that PRIMA will be able to reveal and characterize.

## Instruments and Modes Used

The proposed observing program requires ~36 maps of size 60’x60’ each in both the hyperspectral and polarimeter bands of PRIMAger. Polarimetry information is not required.

## Approximate Integration Time

We estimate a total approximate integration time of 1000 hr for this survey.

We took into consideration various survey sizes and depths. As we are most interested in the detection of the sources that will be missed by the NewAthena survey (CT-AGN with $L_x < 10^{43}$ erg/s), we focused on a deep 1 deg$^2$ survey, matched to the deepest part of the NewAthena survey strategy. A ∼1000 hr deep survey will provide 5σ sensitivities between 92μJy (at ∼25 μm) and 229μJy (at ∼235 μm) computed using the latest PRIMAger characteristics available here (Barchiesi et al. 2025). The first seven filters (i.e., the entirety of PHI-A and the first part of PHI-B) will be above the classical confusion limit, and it has been demonstrated that using those as priors, the flux in the confused bands can be recovered within a ∼20% accuracy (Donnellan et al. 2024).

Such a survey will allow us to detect almost all the AGN up to redshift 2. For $L_X$>$10^{43}$erg/s, we also expect to have detections at z>4. As most of the AGN lie at z ≤ 3, we expect PRIMA to detect ~30% of all AGN in all the 16 bands and more than 70% in at least one band. Moreover, thanks to the continuous coverage at mid- and far-IR wavelengths, SED-fitting will allow us to recover (within 0.1 dex of the true value) the bolometric luminosity for 75% of our sources and 63% of those at z>3. Focusing on sources at 6<z<8, it has been shown that PRIMA will not only be able to detect the most luminous AGN, but also measure their bolometric luminosity with reasonable accuracy, enabling us to reconstruct the bolometric LFs and BHAD even at these high redshifts (Barchiesi et al. 2025).

## Special Capabilities Needed

N/A

## Synergies with Other Facilities

The synergies with the ESA X-ray observatory NewAthena are extremely important for two reasons. On the one hand, the combined use of both observatories allows us to cover the whole AGN parameter space. NewAthena will excel in detecting the unobscured AGN and the high-luminosity obscured ones up to very high-z. PRIMA will be extremely good in detecting and characterizing AGN regardless of their obscuration, practically filling in the gaps of NewAthena.

On the other hand, exploiting both wavelengths helps in recognizing these sources as AGN. At z>2, an X-ray photometric detection with NewAthena corresponds to a luminosity of $L_X$>$10^{42}$erg/s, usually associated with AGN activity. Moreover, NewAthena will provide X-ray spectra alongside the detection, allowing us to constrain at least the AGN intrinsic luminosity and obscuration. In addition, once the X-ray-derived intrinsic luminosity has been obtained, it can be used in the SED-fitting to disentangle the AGN and host-galaxy contributions, thus achieving better constraints on stellar mass, SFR, IR luminosity, and AGN bolometric luminosity. For CT-AGN, for which the X-ray spectral analysis suffers from the strong degeneracy between AGN intrinsic luminosity and obscuration, it has been shown that the AGN bolometric luminosity derived from IR SED-fitting can be used as prior in the X-ray spectral analysis to solve the degeneracy and put stronger constraints on the AGN properties (Laloux et al. 2022).

Finally, having near-IR coverage, such as the one provided by the JWST COSMOS-Web survey, will help in delivering priors to deblend the sources, thus allowing us to make the best of the deepest PRIMAger observations.

## Description of Observations

To exploit the synergies between X-ray and IR, PRIMA will need to target the same fields as NewAthena. At the moment, this is not defined yet, although it is known that it will be one of the deep fields with multi-wavelength coverage (e.g., COSMOS, CDFS, EUCLID fields, Rubin-LSST Deep Drilling fields, or those of the Roman NIR). Targeting an already extensively studied field will provide further benefits, such as having already at our disposal a complete multi-λ photometric coverage that will allow us to perform reliable SED-fitting, separating the AGN and host-galaxy contribution, thus allowing us to distinguish between dusty SF-galaxies and obscured AGN. In particular, it has been shown that exploiting PRIMAger photometry and optical and NIR coverages similar to those in the COSMOS field, we will be able to constrain (within 0.1 dex) the bolometric luminosity for 75% of our simulated sources and 63% of those at z>3, a significant improvement over the ∼20% of the sources without mid-IR and far-IR coverage. We found similar performances regarding the constraining of the SFR and stellar mass of the host galaxy (Barchiesi et al. 2025).

Having accurate redshifts will be essential for the characterization of newly discovered sources. The PRIMA surveys could be designed to target fields that will have comprehensive deep spectroscopic coverage by the time of their launch. For instance, in the coming years, fields such as COSMOS, GOODS, and the Euclid Deep Fields will receive spectroscopic coverage from 4MOST. Furthermore, large-scale surveys like the Wide-Area VISTA Extragalactic Survey (WAVES) and the Optical, Radio Continuum, and HI Deep Spectroscopic Survey (ORCHIDSS) will target over 180,000 galaxies in these fields. Additionally, PRIMA could also benefit from upcoming facilities such as the Wide Field Spectroscopic telescope (WFS), which is expected to provide multi-object spectroscopy for approximately 250 million sources (Bacon et al. 2024).

# 10. Radio-Selected NIR-Dark Sources

Meriem Behiri (SISSA), Alberto Traina (INAF OAS Bologna)

In this study, we aim to observe the FIR emission of dust-obscured star-forming galaxies selected in the radio bands. These objects, elusive by definition in the optical and near-infrared, often remain undetected even in the far-infrared due to sensitivity and confusion limits of current observatories, making it difficult to reconstruct their full Spectral Energy Distribution (SED) and estimate accurate redshifts. This observational gap hampers our understanding of their nature and cosmic role. Addressing this challenge with future facilities like PRIMA is therefore essential. But why focus on the FIR emission of radio-selected NIR-dark galaxies? These systems are likely among the most dust-enshrouded and actively star-forming galaxies in the Universe. Their contribution to the cosmic star formation rate density, metal production, and their potential role as progenitors of massive quiescent galaxies make them crucial to models of galaxy formation and evolution. We propose a study of radio-selected sources in COSMOS with PRIMA, leveraging deep radio surveys, like COSMOS VLA 3 GHz, to extract faint FIR fluxes below the classical confusion limit. Approaching the FIR using the radio, instead of infrared and optical, allows us to target the most obscured sources. Radio observations are, in fact, not affected by dust extinction, and thanks to interferometers like VLA and SKAO in the future, it is possible to reach sub-arcsec resolutions. Using the well-established FIR–radio correlation (FIRRC), we will estimate FIR fluxes for NIR-dark radio sources, calibrate the relation with those detected in Herschel data, and predict the sensitivity needed for PRIMA to detect the most obscured star-forming galaxies at high redshift. This approach provides a model-independent roadmap for maximising the scientific return of FIR surveys. This proposal addresses the Astro 2020 Decadal survey key question: How do gas, metals, and dust flow into, through, and out of galaxies? RS-NIRdark are likely undergoing intense star formation hidden behind heavy dust columns. Their far-infrared emission provides a direct tracer of dust content and obscured star formation activity, key to understanding how baryons cycle through galaxies.

## Science Justification

The Far-Infrared Radio Correlation (FIRRC) is a well-established empirical relation observed in star-forming galaxies (SFGs), linking their FIR and radio emissions (e.g., Delhaize et al., 2017; Ivison et al., 2010). This correlation is rooted in the processes associated with star formation (Condon et al., 1992; Murphy et al., 2009). One of the main channels of dust production is through the ejecta of supernovae (Gall et al., 2011). Once formed, this dust plays a dual role in star formation regulation: it shields molecular gas from high-energy heating radiation, promoting cooling and, thus, collapse, while also reprocessing ultraviolet radiation into the far-infrared (Narayanan et al., 2018). In SFGs, this process goes along with the radio emission in the GHz regime, produced mainly by synchrotron radiation from SNe explosions, offering a proxy for recent massive star formation (Condon, 1992). Thus, since both FIR and radio emissions are by-products of star formation, a strong correlation is observed between these two spectral regions in star-forming galaxies (Bressan et al., 2002; Murphy et al., 2009). In this proposal, we want to use PRIMA to complete our picture of radio-selected opt/NIR-dark sources (RS-NIRdark), fully exploiting the correlation FIRRC, fundamental in these cases where no hint of optical or NIR emission is given. To do so, an excellent starting point to match state-of-the-art data is constituted by the sources from the COSMOS VLA 3GHz survey (Smolcic et al., 2017), in particular NIR-dark objects, defined as those lacking counterparts in COSMOS2020 (Weaver et al., 2022). To minimise AGN contamination, we exclude X-ray detections (Civano et al., 2016). These likely represent high-redshift, dust-obscured galaxies invisible to optical/NIR surveys. Our approach, in line with Talia et al. (2021), Behiri et al. (2023) and Gentile et al. (2025), leverages the multi-wavelength coverage of COSMOS, from X-rays to radio. Studying the FIR emission of NIR-dark galaxies is key for two reasons. First, these galaxies lack optical and NIR data, making redshift and stellar mass estimation challenging if using just MIR and radio data. The dust grey-body is a prominent SED feature, allowing reliable redshift constraints and improving stellar mass estimation via energy balance methods. Using the FIRRC, we can probe populations of dust-obscured star-forming galaxies that are undetected or faint in existing FIR surveys due to sensitivity limits or confusion noise. This hinders the possibility of a complete knowledge of the role the RS-NIRdarks in galaxy evolution: these are among the best candidates as missing progenitors of massive early-type galaxies and might explain a missing contribution to the high-z star-formation rate density (e.g. Talia+21, Behiri+23). To quantify the FIRRC, we use the parameter $q_{FIR}$, which is defined as (Yun et al., 2001; Magnelli et al., 2015): $q_{FIR} = \log 10\left(\frac{L_{FIR}}{3.75 \times 1012\, W}\right) - \log 10\left(\frac{L_{1.4GHz}}{WHz^{-1}}\right)$, where $L_{FIR}$ is the total infrared luminosity (integrated over 8–1000 µm) and $L_{1.4GHz}$ is the radio luminosity at 1.4 GHz. To reach our goal, we first cross-match the radio-selected NIR-dark sources with the Herschel Super-deblended catalogue (Jin et al., 2018) to identify FIR counterparts. For sources with reliable redshifts and at least three FIR detections in Herschel, we calibrate the FIRRC, excluding AGN, to avoid bias (Figure 1). It is no surprise that the FIRRC is normalised towards higher values: these sources were selected to be among the dustiest objects, as they should be completely obscured star-forming galaxies. The complementary NIRbright sample would lower the normalisation.

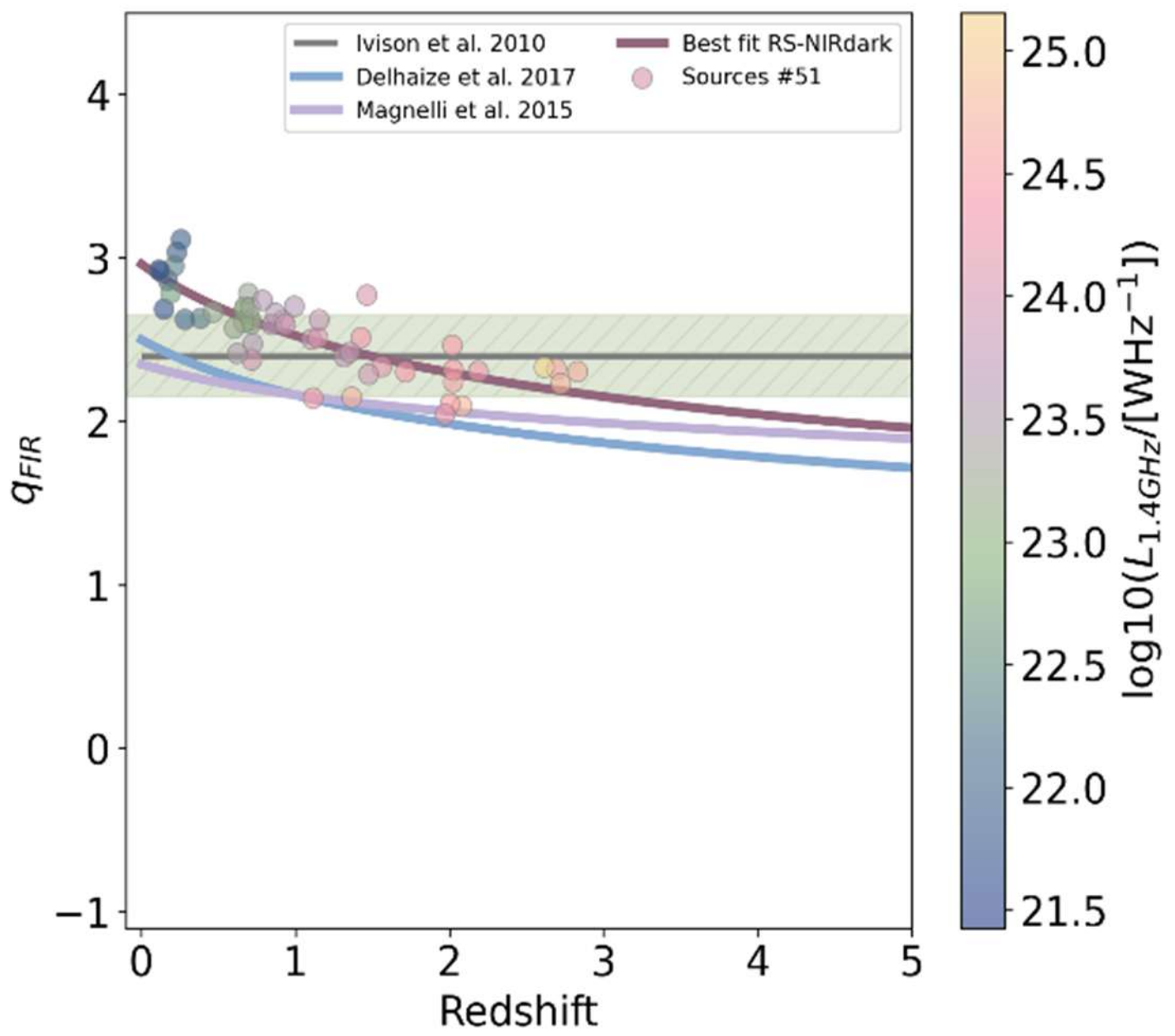

**Figure 1.** This plot shows the $q_{FIR}$ vs redshift for the RS-NIRdarks in COSMOS. The shaded area represents the 1σ dispersion around the median value of $q_{FIR}$.

We then apply this relation to the remaining radio sources to estimate their expected FIR fluxes and derive the sensitivity required for their detection. We restrict the study to sources with $L_{1.4GHz}$ < $10^{24}$ $WHz^{-1}$, to further minimise the appearance of AGN contamination. Unlike classical SED fitting, which struggles without FIR or optical/NIR data, our approach uses the FIRRC to derive model-independent FIR flux fair estimates. The preliminary highlights the presence of selected NIR-dark radio sources in COSMOS falling below the Herschel detection threshold (Jin et al., 2018), with predicted FIR fluxes around 5.6 mJy at 235μm (Figure 2). While this is below the SPIRE confusion limit, such sources are accessible to PRIMA thanks to its higher resolution and advanced deblending, enabling the discovery of FIR-faint, dust-obscured galaxies previously out of reach. *Caveat:* This remains a conservative estimate based on a sub-sample of well-behaved sources, while other studies (e.g., Behiri et al., 2023) indicate that nearly half of radio-selected NIR-dark sources go undetected by Herschel. Further, most of these FIR-faint sources are now detected in COSMOS-Webb with JWST, confirming their dusty nature and reinforcing the need for deeper FIR observations to constrain their SEDs and dust properties. Their detection with COSMOS-Webb will provide robust priors for PRIMA deblending, enabling recovery of sources below the classical confusion limit. According to the official PRIMA Exposure Time Calculator (ETC), the required integration times to reach a 5σ sensitivity of 0.15 mJy vary significantly across PRIMAger polarimeter bands: ~1392 h/deg² at 96 µm, ~2913 h/deg² at 126 µm, ~5108 h/deg² at 172 µm, and ~9364 h/deg² at 235 µm. Therefore, the previously quoted value of "~1500 h/deg²" applies only to the shortest wavelength and does not represent the full range. This is in the absence of confusion. Classical confusion noise limits blind extractions (up to 46 mJy at 235μm), advanced deblending tools like XID+ successfully applied in COSMOS with Herschel (Jin et al., 2018; Wang et al., 2019) can recover sources up to 0.7- 7.0 mJy at 5σ when guided by priors. In our study, the main priors would come from radio observations. A 2 deg² PRIMA survey reaching 5σ depth of 5.6 mJy at 235 µm requires only ~13.4 hours, according to the PRIMA ETC. This confirms the feasibility of the proposed program. It is possible to relax the SNR required for counterparts in case of multiple reliable counterparts from different surveys.

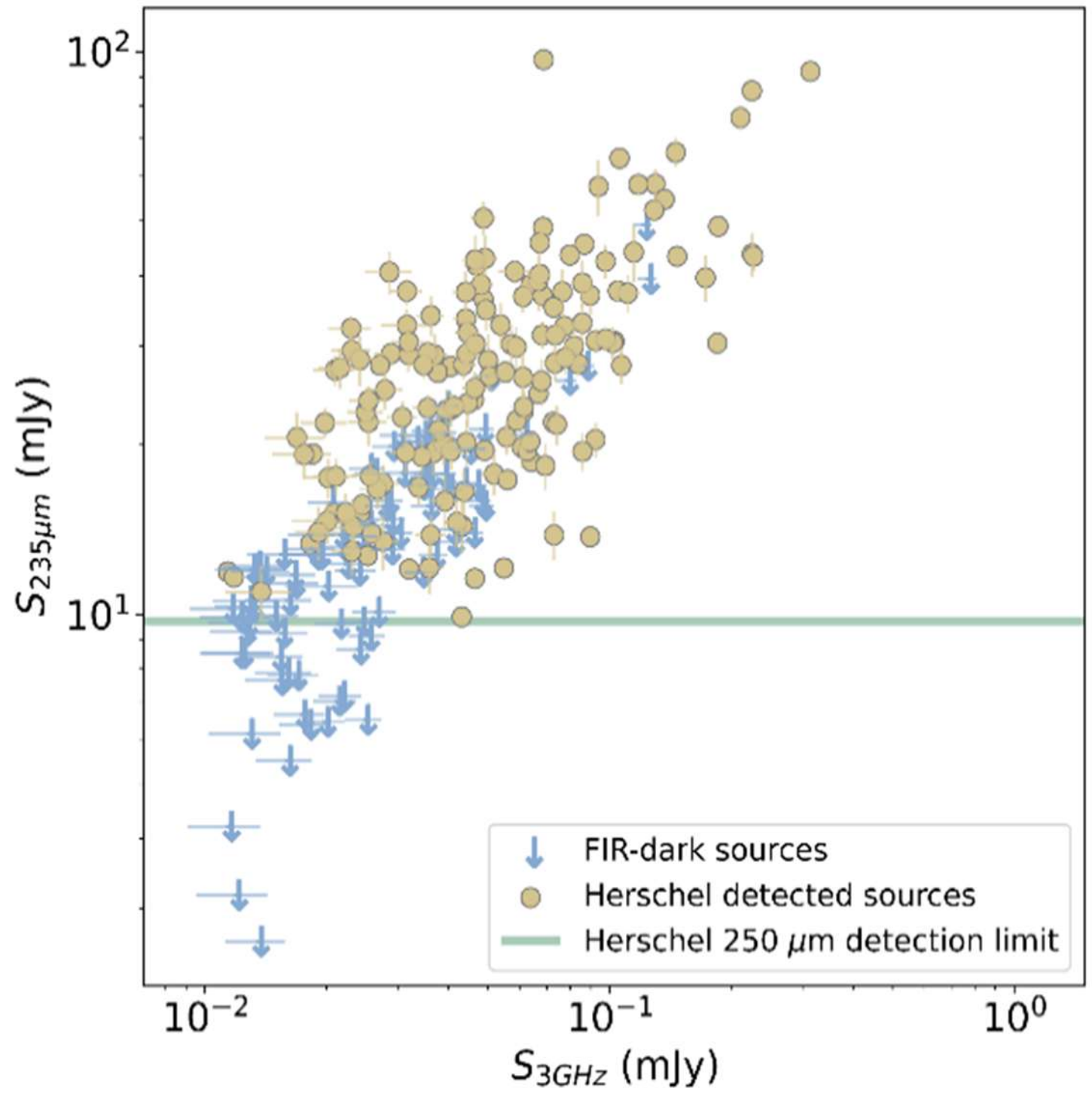

**Figure 2.** This plot shows the FIR-dark sources (in blue) and the Herschel-detected sources (in yellow) at 235 microns. The solid line represents the Herschel detection limit at 250 microns.

## Instruments and Modes Used

PRIMAger: Deep FIR continuum imaging in the **three longer Polarimeter bands** (126, 172, and 235 µm), selected to sample the FIR SED of dusty, high-redshift radio-selected NIR-dark galaxies for an 85’x85’ field. The 96 µm band is excluded due to the extremely low expected fluxes and prohibitive integration times. Expected fluxes are derived from the FIR–radio correlation and modeled greybody SEDs (T = 35 K, β = 1.8, z ≈ 3.5). The polarimetry information is not necessary for our science case, but it could be a useful byproduct.

## Approximate Integration Time

Our requested program includes targeted integration over 2 deg² to detect radio-selected NIR-dark galaxies at high redshift, with predicted fluxes of:

- ~5.6 mJy at 235 µm
- ~2.6 mJy at 172 µm
- ~0.43 mJy at 126 µm

Integration times (for 5.6σ detection over 2 deg², from PRIMA ETC):

- ~13.4 h at 235 µm

- ~33.5 h at 172 µm
- ~709 h at 126 µm

The program focuses on full detection at 235 and 172 µm, while shallower depth at 126 µm is accepted for SED fitting. The **total time request is ~60–80 hours**, prioritising the two longest bands. The 96 µm band is not included: expected fluxes for our targets are <30 µJy, implying integration times >50,000 hours/deg².

## Special Capabilities Needed

- High-resolution FIR imaging to mitigate confusion noise and isolate faint counterparts.
- Multi-wavelength catalogues including radio, optical/NIR or ALMA (sub)mm to enable accurate cross-matching.

## Synergies with Other Facilities

The program leverages the multi-wavelength coverage of the COSMOS field, including VLA 3GHz radio imaging, JWST MIRI data (COSMOS-webb), and (sub)mm observations (e.g., CHAMPS, ALMA WSU). The FIR data will complement these by directly probing obscured star formation and dust properties, providing a comprehensive picture of galaxy evolution obscured by dust. Looking ahead, radio and FIR polarimetry offer promising complementary diagnostics: radio polarisation can trace magnetic field structures and synchrotron emission properties, while FIR polarimetry reveals dust grain alignment and magnetic fields in cold ISM phases. The synergy between radio and FIR polarisation studies, especially with next-generation facilities like SKAO and future FIR polarimeters, will open new windows into the role of magnetic fields in galaxy evolution and dust physics. This will enhance not only our understanding of SFGs, but also of radio-quiet AGN, often as faint as SFGs in the radio but with clear AGN features in the dust size and temperature distribution, and presenting polarisation hints.

## Description of Observations

We propose deep FIR imaging with PRIMAger covering the COSMOS field tiles overlapping with the VLA 3GHz survey footprint. The targeted sensitivities are based on predicted fluxes from the FIRRC calibrated on radio-selected NIR-dark sources with Herschel counterparts. Observations will be performed in multiple bands to constrain dust temperatures and masses. Polarimetric measurements may be included to investigate magnetic field structures related to star formation feedback and dust alignment.

# 11. Dust Heating in High-Redshift Quasars with PRIMA

Manuela Bischetti (Università di Trieste), Francesco Salvestrini (INAF OATs)

Luminous quasars at high redshift (z>4), powered by rapidly growing black holes with masses $M_{BH} \gtrsim 10^8\ M_\odot$, offer a unique window into the early growth phases of the first massive galaxies and into the role of AGN feedback in shaping their evolution during the Universe's first billion years. Although many observational campaigns have explored the role of AGN-driven outflows using ionized, atomic, and molecular gas tracers, the impact of feedback on the warm (T>70-100 K) dust component remains largely uncharted and is limited to a handful of sources. This warm dust is mostly located in the inner kiloparsec of high-redshift galaxies and, although it represents only a small fraction of the total dust mass, it can dominate the mid-to-far infrared emission in luminous quasars (e.g., Fernandez-Aranda+2025; Meyer+25; Di Mascia+23). Here, we propose a PRIMAger survey of the warm dust emission in a representative sample of 20 luminous 4.5 < z < 6.5 quasars to reveal the impact of AGN emission on the dust component.

The sample will be a reference for future studies and the improvement of the current modeling of the quasar emission involving radiative transfer models, which are limited to a handful of sources.

## Science Justification

By transferring energy and momentum to the surrounding dusty environment, black holes in luminous quasars can substantially affect the conditions of the interstellar and circumgalactic medium of their host galaxies. UV radiation heats the dust, leading grains to re-emit in the far-IR. Moreover, radiation pressure on dust grains may drive powerful outflows (e.g., Ishibashi et al. 2018) that push away the gas surrounding the BH, clearing up a few lines of sight (e.g., Costa et al. 2022) and possibly influencing the host galaxy growth (e.g., Bischetti et al. 2019, 2024). Signatures of such interplay between black hole/stellar radiation and dust grains are encoded in the rest-frame UV-to-FIR spectral energy distribution (SED) of galaxies.

An essential ingredient for understanding the impact of feedback from supermassive black holes on the star formation in high-redshift massive galaxies and, in turn, how black hole–galaxy co-evolution proceeds in the distant Universe, is measuring the black hole's contribution to dust heating on galactic scales. Indeed, this is crucial to provide unbiased measurements of the star formation rate (SFR) in galaxies hosting luminous quasars during the first 3 billion years of cosmic history (z ~ 2 – 6).

Most SFR measurements in high-redshift quasar host galaxies are based on the far-infrared (FIR) luminosity and rely on the assumption that the latter is mainly due to reprocessed emission by dust heated by stars. The FIR photometry is generally fitted with a modified black body (MBB) function (e.g., Salvestrini et al. 2025, dashed curve in Fig. 1), which depends on the dust mass and, crucially, on the dust temperature ($T_{dust}$). The total FIR luminosity, $L_{FIR}$, is computed by integrating the MBB function in the wavelength range 8−1000 µm, and a SFR−$L_{FIR}$ calibration (e.g., Kennicutt & Evans 2012) is used to infer the SFR in the host galaxy. However, this computation is very sensitive to the value of $T_{dust}$, which is often assumed because of the lack of photometric samplings close to the peak of the dust emission. This results in uncertainties on the SFR up to one order of magnitude. Direct measurements of $T_{dust}$ in quasar host galaxies at z>2 are available only for a relatively small number of targets with Herschel observations (e.g., Duras et al. 2017) or with ALMA observations in the highest frequency bands (e.g., Tripodi et al. 2024), which can only be carried out under rare atmospheric conditions.

Radiative transfer (RT) studies assuming relatively simple dust distributions in z~2-6 quasar host galaxies reported a quasar contribution to $L_{FIR}$ in the range 30-60% (e.g., Duras et al. 2017, Tsukui et al. 2023). Recent applications of RT to hydrodynamic simulations of z~6 quasar halos have shown that SFR can be overestimated by a factor of ≈ 3 (10) for quasar bolometric luminosities $L_{bol} \approx 10^{12}\ L_\odot$ ($10^{13}\ L_\odot$), implying that the SFRs of $z \sim 6$ quasars can be overestimated by over an order of magnitude (Di Mascia et al. 2021, 2023).

To systematically investigate the contribution from quasars to dust-heating at galactic scales, we propose a campaign of a sample of 20 quasars at 4.5<z<6.5 probing the 24-84 µm spectral range covered by the PRIMA Hyperspectral Imager with three different pointings, for a total integration time of 600 hours. PRIMA spectral coverage and sensitivity will be crucial to reveal the emission from warm dust in a statistically sound sample of luminous ($L_{FIR}>10^{13}\ L_\odot$) quasars.

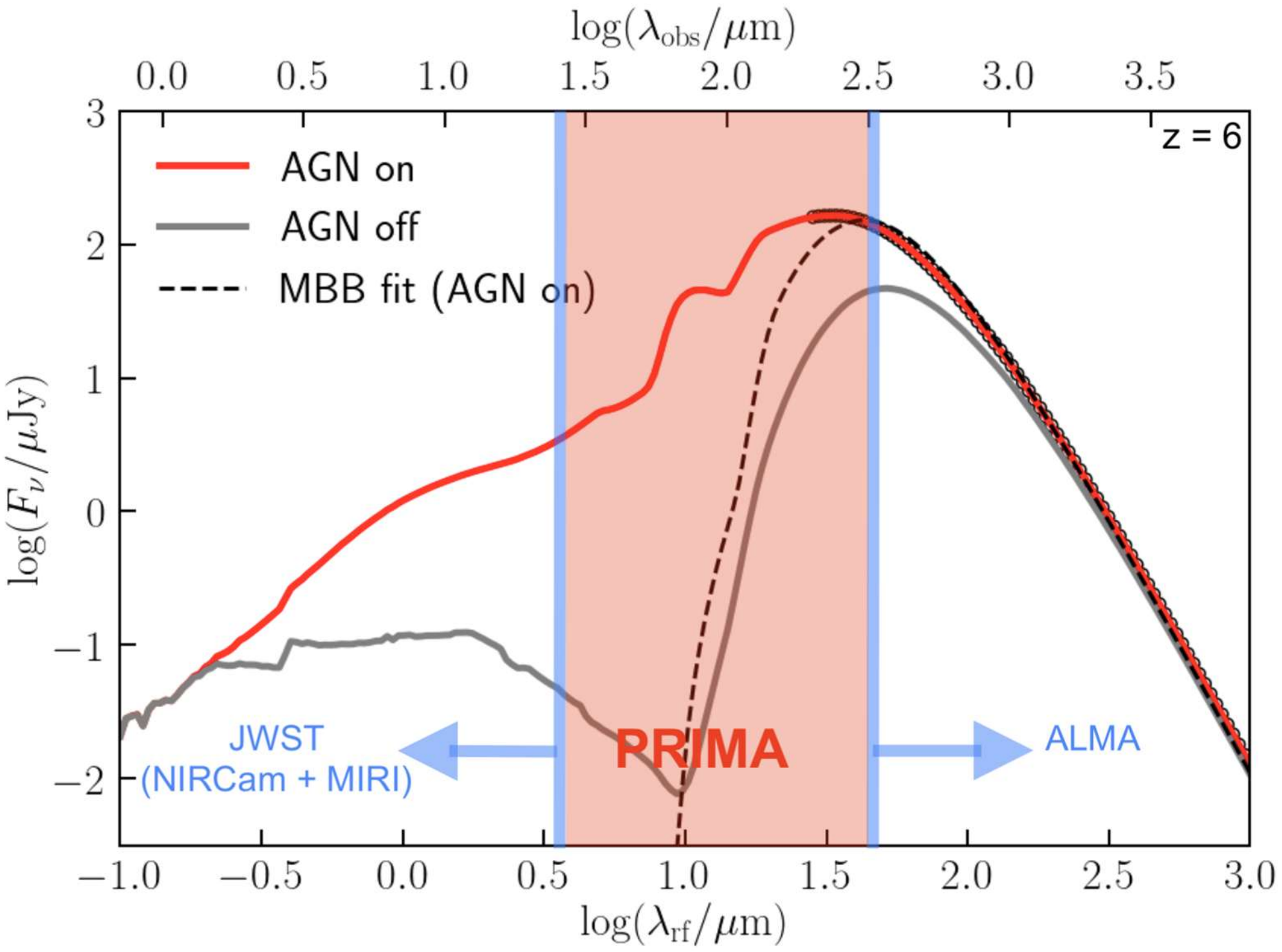

**Figure 1.** Adapted from Di Mascia et al. 2022. SED of a simulated massive galaxy at z=6, massive with no active black hole (grey curve). The solid red line shows the flux increase obtained by including the contribution from a luminous quasar out to 6 kpc via radiative transfer (red curve). Adapted from Di Mascia et al. (2022). The black dashed line indicates the SED fit of the millimeter part of the SED performed according to the modified black-body function (e.g., Salvestrini et al. 2025), from which the vast majority of SFR measurements in high-redshift galaxies are inferred. Blue arrows show the spectral range sampled by JWST and ALMA, which fall short in sampling the peak of the galaxy-scale dust emission in the hosts of luminous quasars. PRIMA will be key to robustly detecting the mid-IR band emission and providing a robust measurement of quasar contribution to dust heating in massive host galaxies in the first 3 Gyr of cosmic time, breaking down systematics on SFR measurements at these epochs.

## Instruments and Modes Used

This observing program requires 2.2'x2.2' maps in the PRIMAger hyperspectral bands (PHI1 and PHI2) towards 20 targets, listed in the Description of Observations section.

## Approximate Integration Time

We expect to conduct the survey in a total time of ~600 hours: ~10 hr per pointing per target, considering 3 pointings for each target. Three pointings are needed for each target to obtain three photometric points (roughly 30, 55, and 80 microns) with the hyperspectral mode to cover the requested field of view (~2.2x2.2 squared arcmin, the same as that of JWST/NIRCam) in each photometric band. The requested 5sigma sensitivities at 30, 55, and 80 microns are 100, 200, and 300 microJy, respectively. The sensitivities were estimated based on the available modeling for the spectral energy distribution (SED) of quasars at similar redshifts and luminosities.

## Special Capabilities Needed

N/A

## Synergies with Other Facilities

Precisely quantifying quasar-driven dust heating on galactic scales demands fully sampled SEDs from the rest-frame UV/optical through the FIR. As Fig. 1 illustrates, achieving this requires:

- Deep near- and mid-infrared imaging and spectroscopy (NIR/MIR) with both space-based (e.g., JWST, WISE) and ground-based facilities (e.g., VLT/X-shooter, ELT) to constrain the stellar and AGN continuum, hot dust;
- High-sensitivity sub-millimeter interferometry with ALMA over multiple bands.

JWST/NIRCam observations in the near-infrared and ALMA ones in the far-infrared will be crucial to identify potential contaminants in the field of view and to run deconvolution methods to investigate the environment of high-z luminous quasars.

## Description of Observations

Three pointings of the hyperspectral imager are needed to fully sample the SED of each target with three photometric points at 30, 55, and 80 microns. Our goal field of view is 2.2x2.2 arcmin$^2$, the same as that of JWST/NIRCam, enough to investigate the environment of the high-redshift quasars (~700-900 kpc$^2$ at the redshift of the targets). The requested 5σ sensitivities at 30, 55, and 80 microns are 100, 200, and 300 µJy, respectively. These sensitivities will allow us to detect each target with at least S/N>10-15 in each band.

Although the required sensitivities may approach the confusion limit in some bands, we do not expect significant contamination because our targets are intrinsically bright (LFIR>5E12 $L_\odot$). When studying the surrounding environment, confusion is mitigated by NIRCam imaging that can help identify faint targets by using NIRCam detections as priors for source deconvolution techniques. Moreover, in the noisier cases, we can enhance the signal-to-noise ratio by combining the two PRIMAger bands (≈ 24–45 µm and 45–84 µm) in two single snapshots.

### Targets

| | |
|---|---|
| ATLASJ029-36 | z=6.0199 |
| SDSSJ0100+2802 | z=6.3 |
| VDESJ0224-4711 | z=6.526 |
| PSOJ036.5+03 | z=6.5329 |
| PSO231-20 | z=6.5869 |
| PSOJ011+09 | z=6.444 |
| PSOJ083+11 | z=6.346 |
| PJ308-21 | z=6.2342 |
| PSOJ009-10 | z=5.938 |
| PSOJ023-02 | z=5.817 |
| PSOJ025-11 | z=5.816 |
| PSOJ065-26 | z=6.179 |
| PSOJ089-15 | z=5.972 |
| PSOJ108+08 | z=5.955 |
| J2054-0005 | z=6.039 |
| W2246−0526 | z=4.601 |
| J0331−0741 | z=4.737 |
| J1341+0141 | z = 4.700 |
| J1511+0408 | z=4.679 |
| J1433+0227 | z = 4.727 |

# 12. Spectroscopy of Extremely Luminous Galaxies: AGNs at Cosmic Noon, PAH Emission, Silicate Absorption and Detailed Dust SEDs.

Andrew Blain (Leicester, UK), Tanio Dias-Santos (FORTH, Greece)

The most luminous dusty galaxies ever discovered are almost certainly powered by a blend of AGN and star-formation activity, with complex internal structure (e.g., Fernandez-Arana et al. 2024). Furthermore, they have hot interstellar mediums (ISMs) with well-measured fine-structure lines from ALMA observations. However, they do not have detailed spectral observations in the rest-frame 7-40 micron window, which is important for determining the overall shape of the dust SED, the degree of silicate absorption, and the suite of PAH features, including aliphatic and aromatic carbon molecular emission/absorption. PRIMA-FIRESS is in a unique position to address these questions for unusually hot extreme galaxies. While related to Case 16 in PRIMA GO Book, Vol. 1, this is a substantially revised case based on additional information that has come to light about the conditions in the ISM of WISE-selected extreme galaxies (hot dust-obscured galaxies or HotDOGS). There are approximately 500 well-studied HotDOGs, and a statistical survey of their most powerful emission in the PRIMA-FIRESS range would be a very important baseline for ISM astrophysics under the most extreme conditions. Based on the FIRESS low-resolution exposure time calculator, a typical 10mJy 22-micron continuum source can be dissected into a spectrum with signal-to-noise 100, in R=100 channels in about 0.25 hr. Subsequent high-resolution inspection of specific key features identified could then also be conducted in a reasonable time. A 200-hour large program could thus reveal an array of spectral richness in this least-explored window, where these most remarkable objects are most luminous.

## Science Justification

The growth of galaxies, including their associated supermassive black holes, stellar mass, and enrichment of metals, is a key persistent question as new observing facilities become available. The most luminous objects (Tsai et al. 2015; Fernandez-Arana et al. 2024) include perhaps the most useful clues about the wider evolution of galaxies. The most extreme objects of all are found to have very powerful mid-infrared (IR) emission, ideally matched by the observing capabilities of PRIMA, especially FIRESS.

By investigating the ISM of these galaxies, the processes by which star formation and active galactic nuclei fueling are generated, feedback takes place, and energy is transformed between different wavebands can be determined. The presence of absorption against bright central regions can reveal size information otherwise beyond the resolution limit of PRIMA, while the complete spectral coverage of small grain/PAH emission can reveal the intensity of radiation illuminating the ISM. More pronounced emission is likely associated with less intense radiation and a lack of grain destruction/evaporation, while smooth emission would indicate the destruction of material responsible for spectral features. While spatial resolution must be obtained using ALMA at present, only a space-borne instrument can probe the hot mid-IR ISM. PRIMA-FIRESS has the speed to determine an accurate, high signal-to-noise mid-IR spectrum for a significant fraction of the 2000 most-luminous galaxies emitting in the mid-IR waveband over the whole sky, catalogued from the WISE survey (the HotDOGs), and subsequently the subject of systematic study by the proposers and others.

The shape of the broad continuum SED provides information about the relative masses of dust at different temperatures and the opacity conditions in the ISM. Previously covered in only broad-band WISE and Herschel data, the much greater sensitivity and spectral resolution of PRIMA will allow details of the abundance and illumination of dust to be determined for the first time at these luminosities and redshifts, underlying narrower spectral features

Furthermore, compared with other high-redshift galaxies that have significant dust content and UV/optical absorption, the WISE HotDOGs are extremely powerful in the mid-IR, rest-frame 5-50 microns, as compared with the longer wavelength far-IR bands probed by ALMA. This previously largely unexplored spectral region is almost ideally matched to the PRIMA-FIRESS wavelength coverage when redshifted from cosmic noon, and the HotDOGs thus represent the most accessible galaxies with intense emission that PRIMA can investigate.

Broad silicate grain absorption features at restframe wavelengths of 9.7 microns can be used to determine the column density of silicate dust, and provide limits on the extent of the hottest emission regions around an enshrouded AGN, providing insight into spatial scales too fine to probe using PRIMA directly.

Broad PAH features can be inspected to reveal information about carbon chemistry in the ISM, and the illumination conditions of small grains, in comparison with local ISM conditions, to highlight internal structure, opacity, and constituents of the ISM of the target sample. Narrow line features found can be followed up using the high-resolution FIRESS mode to resolve line widths, multiple dynamical components, and other features that can further help to elucidate conditions in the ISM.

Without high-signal-to-noise spectra from PRIMA, there can be no detailed insight into the hot bulk ISM of the most luminous galaxies in the Universe. Previous observations have lacked the sensitivity and spectral resolution to answer these questions, and PRIMA will make a unique contribution to putting them in the context of other well-studied systems at lower redshifts and luminosities. Comparison with ALMA information at higher redshifts will also allow us to determine whether the HotDOGs are an entirely new phenomenon emerging after recombination and growing in significance towards cosmic noon, or show a continuation of the astrophysical conditions being revealed in the unexpectedly large number of high-redshift galaxies during and prior to reionization that are being found by JWST.

The great sensitivity of PRIMA-FIRESS will allow a comprehensive sample to be obtained, revealing the true range of properties in these extreme mid-IR selected galaxies, in which the most intense star-formation and AGN fueling is taking place. PRIMA is uniquely able to investigate the most luminous spectral region of the most luminous galaxies that have ever existed.

## Instruments and Modes Used

FIRESS pointed low-resolution and high-resolution modes.

## Approximate Integration Time

The PRIMA-FIRESS ETC estimates that a 10-mJy 25-micron source can be measured at a signal-to-noise ratio of 100 in each channel of a R=100 spectrum in 0.1 hours, corresponding to a 5-sigma detectable line flux of 6 x $10^{-19}$ $Wm^{-2}$. Covering all 4 FIRESS bands, that is, with two instrument settings, would thus require about 0.2 hours per object, for a total time of about 100 hours for 500 targets selected from the total 2000-strong sample.

In a high-resolution-mode follow-up observation with FIRESS, potentially able to isolate the velocity width of any identified features, a 3.1-hr observation would be required to reach the same line flux to the same signal-to-noise ratio in a 10-mJy continuum source. Assuming the 20 sources have such interesting features, a total time of order 60 hours would be required to obtain high-resolution spectra of chosen lines.

## Special Capabilities Needed

Not applicable

## Synergies with Other Facilities

ALMA spectral imaging has already been carried out for the most luminous galaxies (e.g. Fernandez-Arana et al. 2024). As ALMA's capability expands, joint analysis of the PRIMA-FIRESS spectral range and the SED of these galaxies in ALMA atmospheric windows will be important. The arrival of SKA/ngVLA will enable high-resolution images of this sample, with appropriate redshifted molecular lines to be investigated. Our targets are the most powerful sample of galaxies at the most active epoch in the Universe's history, and investigating the properties of their ISM will be interesting whenever other new facilities become available.

## Description of Observations

Spectral observations of the mid-IR continuum shape and spectral features for a large sample of the most extreme galaxies at redshifts of 2-5. A significant fraction of the 2000 known mid-IR selected galaxies with known redshifts and ancillary follow-up data, in addition to the initial detection measurements using WISE, can be included. Low-resolution pointed observations using FIRESS would be conducted for 500 targets initially. As outlined above, two settings of the FIRESS low-resolution mode are required, for about 0.1 hr each, to obtain spectra with resolution R=100 and a signal-to-noise ratio of 100. This would require 100 hours of observing time. Any targets that show interesting narrow lines in the results can be investigated using the high-resolution FIRESS mode to reveal dynamics via line shapes and structure in 3.1 hours each. Twenty such targets are anticipated, requiring a further 60 hours of follow-up observations.

# 13. The π-IR Survey: A 10,000 deg$^2$ Far-IR Legacy Survey with PRIMAger

Denis Burgarella (Aix Marseille Univ, CNRS, CNES, LAM, Marseille, France), Alessandro Boselli (LAM), Véronique Buat (LAM, France), Dave Clements (Imperial College, UK), Stephen Eales (Cardiff Univ., UK), Carlotta Gruppioni (INAF, Italy), Takuya Hashimoto (University of Tsukuba, Japan), Hanae Inami (Hiroshima Univ., Japan), Guilaine Lagache (LAM, France), Arielle Moullet (NRAO, USA), Chris Pearson (RAL Space, UK), Giulia Rodighiero (Univ. Padova, Italy), Marc Sauvage (CEA Saclay, France), Steve Serjeant (Open University), Rachel Somerville (Flatiron Institute, USA), et al.

The π-IR survey is a community-driven, legacy-scale far-infrared mapping program designed to cover approximately π steradians (~25%) of the sky using the *PRIMAger* instrument aboard the *PRIMA* mission. Operating across 24–260 µm, *PRIMAger* combines hyperspectral imaging, polarimetry, and exceptional mapping speed to deliver a transformative dataset for extragalactic and cosmological science. The survey will provide confusion-limited photometry in PHI (R ~ 8) and confusion-limited photometry in PPI (R ~ 4), and polarimetry enabling detailed investigations of galaxy evolution, dust-obscured star formation, interstellar dust, and large-scale structure out to redshift z ~ 4.

Leveraging *PRIMAger*'s high survey efficiency, π-IR will cover wide with both spectral and polarimetric modes. It will detect over 16 million galaxies, offering the statistical power needed to study cosmic environments and structure formation across time. The survey is designed to work in concert with NASA's *Roman Space Telescope* and ESA's *Euclid* mission, providing far-infrared counterparts to their high-resolution near-infrared imaging and spectroscopy. This synergy will enable robust multi-wavelength studies of stellar mass assembly, dust physics, and the evolution of galaxies at cosmic noon and beyond.

While extragalactic and cosmological science are the core drivers, the π-IR dataset will also support diverse astrophysical investigations, including Solar System small bodies and proto-planetary disks. Crucially, the π-IR survey is structured as an open, community-oriented initiative: all raw and value-added data products will be released to the public immediately, maximizing accessibility, scientific return, and engagement across the astronomical community.

## Science Justification

We propose the π-IR Survey, a community-oriented, legacy-scale far-infrared (FIR) mapping project that will capitalize on the unprecedented capabilities of the *PRIMAger* instrument aboard NASA's *PRIMA* Probe-class mission. Covering approximately π steradians (~25%) of the sky across the 24–260 µm range, the survey will combine hyperspectral imaging, polarimetry, and exceptional mapping speed to deliver a transformative dataset for extragalactic and cosmological science. All raw and value-added data products will be immediately and publicly released, ensuring broad scientific return and accessibility for the full community.

### Scientific Objectives

The π-IR survey is designed to address key questions at the intersection of galaxy evolution and cosmology:

- How does dust-obscured star formation evolve across cosmic time?
- What is the role of dust, metallicity, and environment in shaping galaxy growth?
- How do cosmic structures and galaxy clustering evolve from z ~ 0 to z ~ 4?
- What is the contribution of dusty AGN to the cosmic infrared background?

By achieving photometry (R ~ 8) and polarimetry (R ~ 4), the survey will enable precise measurements of the obscured star formation rate density (SFRD), ISM dust properties, and magnetic field structures over cosmic history. With a projected detection of ~16 million galaxies, π-IR will allow robust statistical studies across a wide dynamic range in mass, luminosity, and environment.

### Survey Design and Legacy Impact

*PRIMAger*'s high mapping speed is a key enabler of the π-IR program, allowing wide-area coverage in both hyperspectral and polarimetric modes. These capabilities are otherwise mutually exclusive in most space-based infrared instruments.

Figure 1 presents the predicted redshift distribution of detected sources, based on simulations from [26]. The left axis shows expected detections in a 2 deg² field, while the right axis extrapolates to the full π-IR survey area. Gaussian approximations to the redshift histograms illustrate the relative contributions across cosmic epochs. The survey will have strong constraining power from the nearby universe out to the epoch of reionization.

### Synergy with Euclid and Roman

A key strength of π-IR is its strategic synergy with NASA's *Roman* and ESA's *Euclid* missions, which are delivering near-infrared imaging and spectroscopy over overlapping sky areas. Roman and Euclid will provide accurate redshifts, morphologies, and stellar mass estimates, while π-IR supplies the FIR emission tracing dust and obscured star formation.

Combined datasets will enable:

- Robust measurements of total SFRs (UV+FIR) and dust attenuation
- Characterization of dusty AGN and merger-driven starbursts
- Environmental studies of galaxy evolution using joint clustering and weak lensing
- Cross-correlation analyses to probe the cosmic infrared background and matter distribution

This coordinated multi-wavelength approach directly supports NASA’s long-term astrophysics goals and the recommendations of the Astro2020 Decadal Survey.

## Broader Scientific Utility

Although optimized for extragalactic and cosmological science, the π-IR survey will produce a high-quality dataset applicable to a wide range of astrophysical domains, including:

- Solar System: Detection and mapping of small bodies and the Zodiacal dust cloud
- Galactic star Formation: Characterization of proto-planetary disks, cores, and embedded protostars
- Milky Way ISM: Polarimetric maps of magnetic fields and dust filaments at high Galactic latitude

By offering full-sky access to FIR hyperspectral and polarimetric data, π-IR will become a critical resource for the astronomical community for decades.

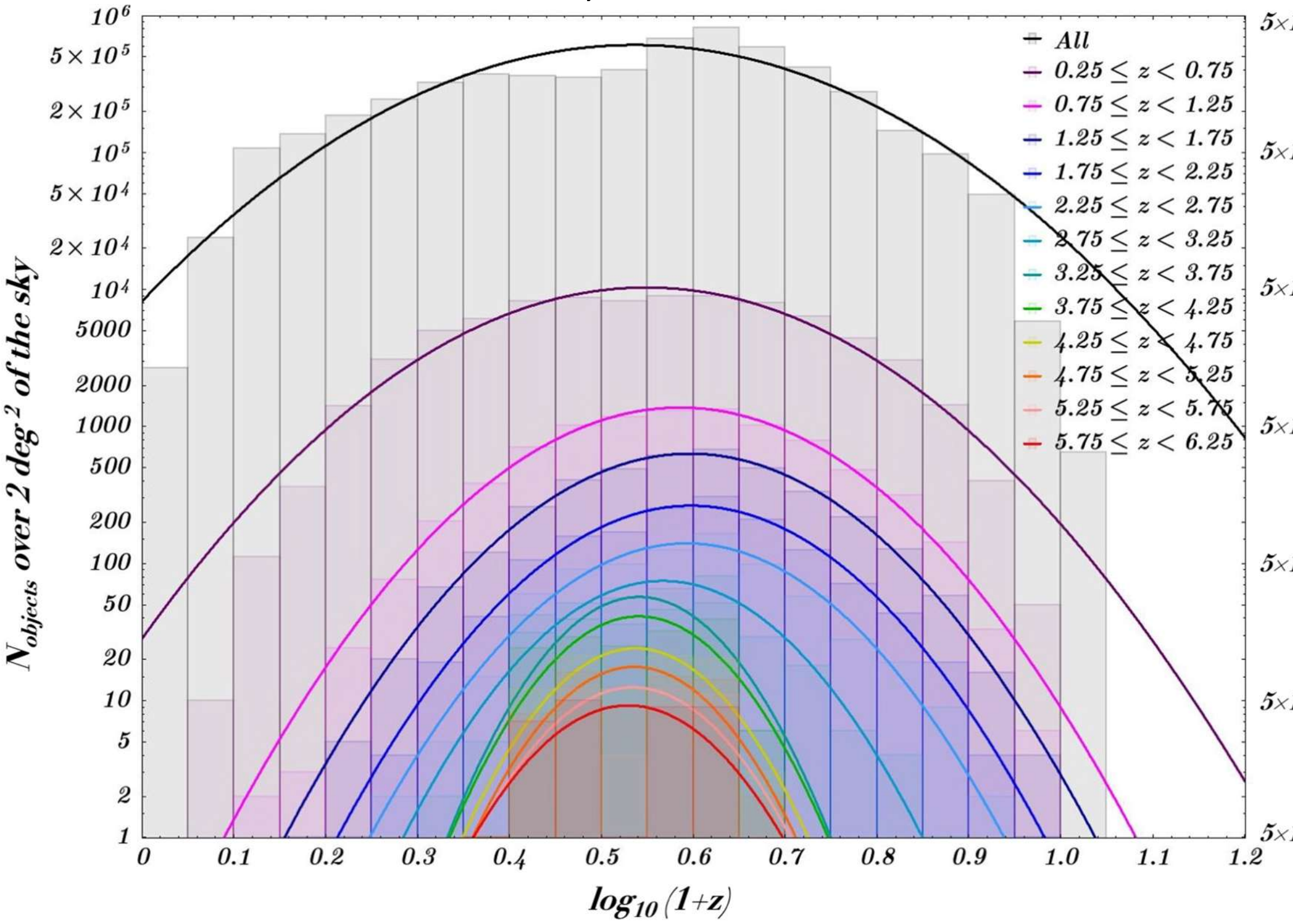

**Figure 1.** *Predicted redshift distribution of detected sources in the π-IR survey.* The histogram (left axis) shows the number of objects expected in a representative 2 deg² field, while the right axis scales to the full π-IR survey area (~10,300 deg²). Predictions are based on simulations from the SIDES model [26]. Overlaid Gaussian curves approximate the redshift distributions and provide an estimate of the relative contributions from different cosmic epochs. The figure highlights π-IR's ability to probe galaxy populations from the local universe out to z > 4, with strong statistical power across all redshifts.

## Instrument and Modes Used

PRIMAGer maps at all bands. ~10,313 deg² area for wide-area component; targeted **10′×10′ tiles** for deep fields. Polarimetry required. **Estimated Integration Time**: ~1800 hours to achieve **SNR = 5** across all bands

## Approximate Integration Time

To estimate the sensitivities and time of the π-IR survey, we use a Python script from Marc Sauvage (private communication). The most constraining parameter is the scan speed, and we have tested three options. The normal scan speed is 60 arcsec/s, and the maximum scan speed is 252 s/sec. We assume we will use the maximum scan speed, which would lead to a total duration of ~1800 hours over π steradians.

## Special Capabilities Needed

The success of the π-IR survey depends on a combination of instrument-level capabilities, optimized operations, and scalable data processing infrastructure. These special requirements are necessary to fully exploit *PRIMAger*'s unique strengths and deliver high-value, legacy-class science to the community.

### Observing Capabilities

**Continuous Wide-Area Mapping**: The survey requires support for efficient, large-scale sky mapping over approximately 10313 deg². This includes optimized scan strategies, overlap control, and minimal overheads to ensure uniform sensitivity and full spectral/polarimetric coverage.

**Survey Tiling and Scheduling**: Execution of the survey in a structured tiling scheme is essential. We will require tiling capabilities at multiple scales, π steradians for wide-area coverage and 10′×10′ tiles for deep fields, alongside scheduling flexibility to accommodate Galactic avoidance, calibration targets, and complementary mission windows.

**High-Precision Polarimetric Calibration**: Polarimetry is essential to the survey's main science goals. Accurate calibration of instrumental polarization and cross-band systematics across the four polarimetric bands (96, 126, 172, and 235 μm) is required.

### Pipeline and Data Handling

The scope and complexity of the π-IR dataset necessitate scalable and efficient data processing workflows**.** We anticipate assessing the need for:

**Dedicated Pipeline Modules**: We will assess whether specific components of the *PRIMA* mission's pipeline can be adapted or expanded to support the π-IR survey's volume and modes, including hyperspectral data cubes, polarimetric Stokes maps, and large-area mosaics.

**Automated Quality Control**: Real-time or near-real-time quality checks will be critical to flag data anomalies, verify calibration integrity, and track survey progress.

**Survey-Oriented Data Products**: Pipeline outputs will be structured to facilitate community access and scientific usability, including source catalogs, background-subtracted maps, and calibration metadata.

### Value-Added Data Products

Beyond the standard mission deliverables, the π-IR team will produce a suite of value-added science products**,** including:

**Spectral Energy Distribution (SED) Fitting with a dedicated CIGALE[1] code**: Using the π-IR photometry, supplemented with *FIRESS* spectroscopic data and multi-wavelength ancillary datasets (e.g., Roman, Euclid, WISE, Herschel), we will perform consistent SED fitting across the survey.

**Physical Parameter Catalogs**: Publicly released catalogs will include derived quantities such as:

- Dust-obscured star formation rates
- Stellar masses and dust temperatures and masses
- AGN diagnostics and estimation of the AGN fractions
- Polarization fractions and position angles
    - Emission line properties (where FIRESS data is available)

**Multi-Tiered Product Access**: We will ensure all products are delivered in standard formats with full documentation and versioning, suitable for both expert users and the broader community.

## Synergies with Other Facilities

Given its wide sky coverage and wavelength range, the π-IR survey is uniquely positioned to build upon and enhance the legacy of major multi-wavelength, wide-area surveys**.** The full value of π-IR will emerge through coordinated analysis with overlapping datasets from NASA and international facilities. Below, we outline key synergies with current and legacy missions that will enable comprehensive, multi-dimensional studies of galaxy evolution, star formation, and cosmic structure.

### Key Synergies

**GALEX (UV All-Sky Imaging)**

- Enables measurement of unobscured star formation when combined with π-IR dust-obscured FIR data.
- Enhances SED fitting for galaxies by extending wavelength coverage into the UV, improving constraints on star formation histories and dust attenuation.
- Legacy GALEX fields overlap with Euclid and Herschel, enabling rich multi-wavelength datasets for π-IR sources.

[1] CIGALE is (probably) the fastest spectrophotometric fitting code (ref). CIGALE is physically oriented : https :cigale/lam.fr

**Euclid (Optical/NIR Imaging and Spectroscopy)**

- Provides photometric and spectroscopic redshifts, stellar masses, and morphologies over 15000 deg².
- Offers critical environmental context, enabling studies of how galaxy evolution correlates with large-scale structure (e.g., clusters, filaments, voids).
- Weak lensing maps from Euclid enhance structural and clustering analyses of π-IR-selected galaxies.

**Roman Space Telescope (Deep NIR Imaging and Grism Spectroscopy)**

- Offers high-sensitivity imaging and spectroscopy ideal for identifying and characterizing faint, high-z counterparts to π-IR sources.
- Combined with Euclid, delivers a full optical-to-NIR view of π-IR galaxies, enabling complete stellar and dust component separation.

**WISE (NIR + MIR All-Sky Imaging)**

- Serves as a bridge between NIR (Euclid/Roman) and FIR (π-IR), crucial for SED modeling and AGN selection.
- Provides calibration for bright sources and contributes to identifying infrared-bright AGN and starbursts.

**Planck (Sub-mm to Microwave All-Sky Maps)**

- Enables absolute calibration and zero-point anchoring of diffuse FIR background.
- Supports foreground modeling, especially for Galactic cirrus and zodiacal emission subtraction in π-IR fields.

**eROSITA (X-ray All-Sky Survey)**

- Identifies AGN and galaxy clusters, enabling π-IR studies of dust and star formation in X-ray-selected environments.
- Facilitates cross-correlation of star-forming galaxies and AGN with large-scale X-ray structure, enhancing cosmological studies.

### Multi-Facility Integration for π-IR Science

The π-IR survey is designed with intentional overlap with the footprints of Euclid**,** Roman**,** WISE**,** GALEX, and Planck**,** creating a powerful multi-wavelength dataset. This will enable:

- Robust SED fitting from UV to FIR for millions of galaxies
- Separation of obscured and unobscured star formation
- Characterization of galaxy populations across cosmic environments
- Calibration and modeling of dust emission, ISM physics, and magnetic fields

- Joint studies of the cosmic infrared and X-ray backgrounds

These synergies ensure that π-IR not only stands as a high-impact FIR survey but also becomes a foundational pillar of NASA's broader multi-mission astrophysics program.

**Table 1.** Multi-Wavelength Survey Landscape (Area >5,000 deg²)

| Facility | Wavelength Range [μm] | Sensitivity (5σ) | Survey Area |
|---|---|---|---|
| GALEX | 0.135–0.28 | $m_{AB}$ ~ 20.5 (FUV, NUV) | All-sky (41,253 deg²) |
| Euclid | 0.55–2.0 | VIS: $I_E$=26.2; NISP: $Y_E$=24.3, $J_E$=24.5, $H_E$=24.4 | 15,000 deg² |
| Roman HLS | 0.5–2.3 | $m_{AB}$ > 26.2 in Y, J, H | ~5,000 deg² |
| WISE | 3.4–22 | 0.08–6 mJy (W1–W4) | All-sky |
| π-IR (PRIMAger) | 24–260 | 0.8–3 mJy | ~10,300 deg² |
| Planck | 350–3000 | 0.5–2 MJy/sr | All-sky |
| eROSITA | 0.00012–0.012 (X-ray) | $1\times10^{-14}$ erg/s/cm² (0.5–2 keV) | All-sky |

## Description of Observations

The π-IR survey will tile approximately π steradians (~10000 deg²) of the sky using the *PRIMAger* instrument aboard the *PRIMA* mission. The survey will be conducted in both of *PRIMAger*'s key observing modes: simultaneous hyperspectral imaging (24–84 μm, R ~ 8) and simultaneous polarimetric mapping (96, 126, 172, and 235 μm, R ~ 4). These modes will operate concurrently throughout the survey to maximize observational efficiency and scientific return.

Observations will be performed in a structured tiling strategy. It will consist of contiguous 10′×10′ tiles, mosaicked to ensure uniform coverage across the full survey region. A nominal integration time of ~1800 hours is planned to reach a 5σ detection limit of 0.8–3 mJy, depending on wavelength.

The survey footprint has been optimized to maximize overlap with existing and upcoming multi-wavelength surveys, including Euclid**,** Roman**,** WISE, and eROSITA, while avoiding regions of high Galactic confusion. The expected yield includes detections of ~16 million galaxies, with robust photometry and polarimetry enabling detailed spectral energy distributions and magnetic field measurements across a wide range of redshifts and environments.

A wide-area survey spanning π steradians (¼ of the sky) is particularly well-matched to the statistical nature of the core science goals, including galaxy evolution, large-scale structure, and dust-obscured star formation. This scale allows for the detection of ~16 million galaxies in mid- and far-infrared, mitigating both cosmic variance and observational biases while preserving sensitivity to rare populations.

A dedicated calibration plan will ensure high-fidelity photometric and polarimetric performance. Calibration observations will include both internal references and external astrophysical standards to maintain stable throughput and minimize instrumental systematics.

The observing strategy is designed to exploit *PRIMAger*'s exceptional mapping speed and multi-mode efficiency. It balances wide-area coverage with targeted depth, enabling the π-IR survey to address its primary extragalactic and cosmology goals while generating high-value data products for broader astrophysical use.

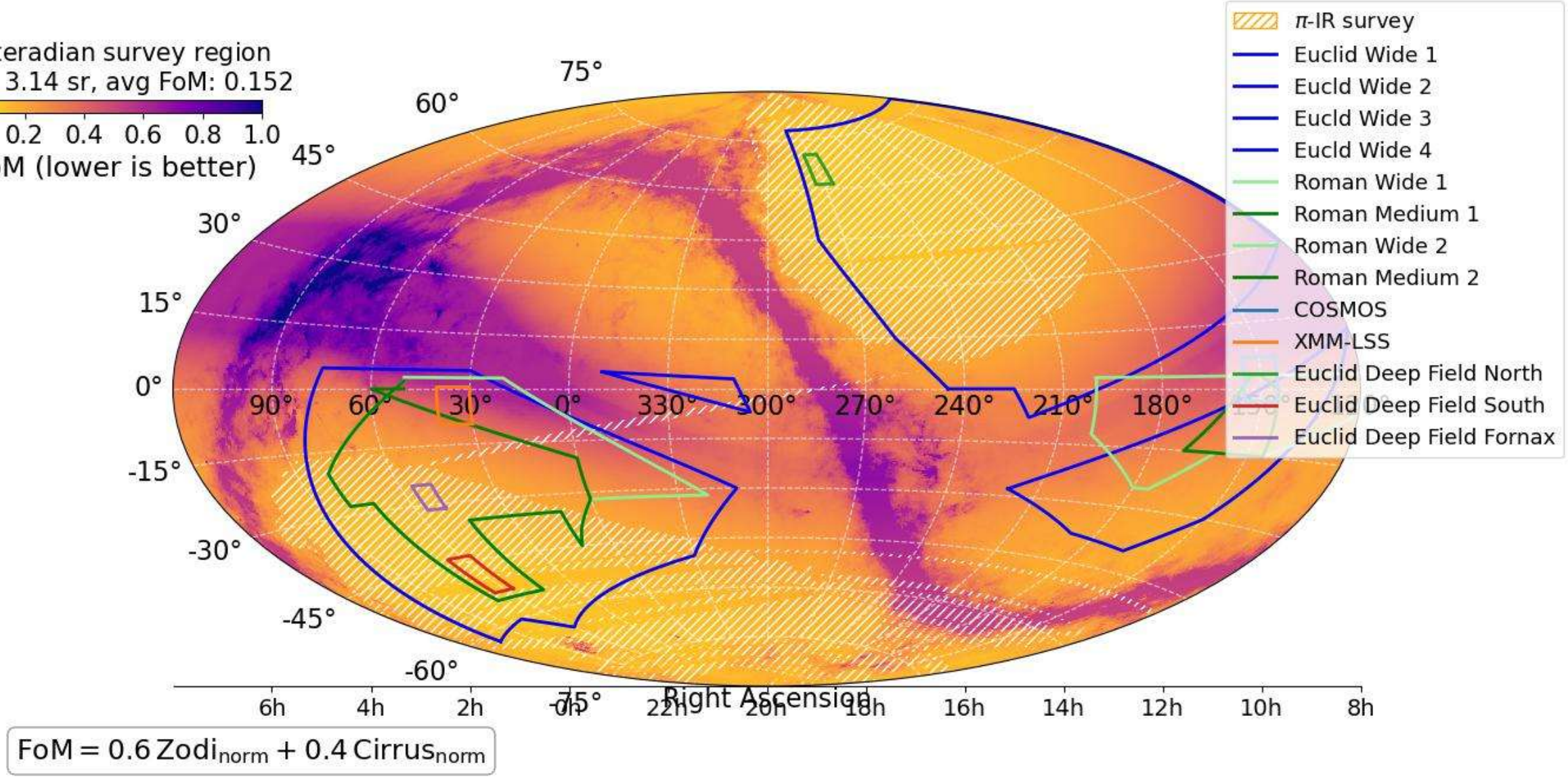

**Figure 2.** *π-IR Survey coverage and Figure of Merit (FoM) map.* The hatched region shows the proposed wide-area π-IR survey footprint (~10,300 deg²), optimized for low foregrounds and overlap with major multi-wavelength surveys (Euclid, Roman, WISE, eROSITA). Deep fields near the ecliptic poles are marked with solid boxes and will receive enhanced integration. The background FoM map reflects a combination of sky visibility, foreground contamination, and ancillary dataset overlap, guiding the field selection to maximize scientific return across extragalactic and cosmological objectives.

# 14. Warm Dust in the First Few Billion Years: Low-Metallicity Galaxies, Dust-Obscured AGN and their Role in the Activity History of the Universe at z>4.

Karina I. Caputi (Kapteyn Astronomical Institute, U. of Groningen, The Netherlands), Manuela Bischetti (INAF – OA Trieste; Italy), Jianwei Lyu (Steward Observatory, US), Francesca Pozzi (U. of Bologna; Italy), Pierluigi Rinaldi (Space Telescope Science Institute,US), Francesco Salvestrini (INAF – OA Trieste, Italy), Irene Shivaei (CAB, CSIC-INTA, Spain), Tsutomu Takeuchi (Nagoya University/ISM, Japan), Luigi Barchiesi (U. of Cape Town, Southafrica), Edoardo Iani (ISTA Vienna, Austria)

Numerous pieces of evidence indicate that dust was warmer in galaxies at high redshifts than in the local Universe. The identification of warm dust in high-z galaxies is, thus, key to obtaining an unbiased view of the roles of dusty star-formation and nuclear activity in the global activity history of the Universe in the first few billion years of cosmic time. We propose to follow up the main JWST deep survey fields with PRIMAger at 24-84 microns to constrain the presence of warm dust in galaxies at z>4. Our main science goals are to: 1) obtain a complete census of the most luminous IR galaxies with warm dust at high z, reaching below the 10^12 $L_{sun}$ regime; 2) disentangle whether the warm dust in these sources is heated by star formation at low metallicities or by the presence of nuclear activity (or both), via the joint analysis of the PRIMA images and ancillary JWST data; 3) determine the roles of these sources to the cosmic star-formation and black-hole accretion densities in the first few billion years of cosmic time.

## Science Justification

Understanding how galaxies formed and evolved through cosmic time is essential to explaining the Universe that we see today. While at present most galaxies have little dust content, it is well known that star formation in the past was, instead, substantially affected by dust attenuation. This is evident from the increasing number density of luminous and ultra-luminous infrared (IR) galaxies with redshift (e.g. Le Floc'h et al. 2005; Takeuchi et al. 2005; Caputi et al. 2007).

Indeed, these dusty galaxies had a major contribution, and even dominated, the cosmic star-formation rate density (SFRD) at redshifts z~1-4 (e.g., Gruppioni et al. 2020; Zavala et al. 2021). At higher redshifts, the picture is much less clear, as IR galaxy surveys derive dusty-galaxy contributions to the total cosmic SFRD differing in more than 1 dex, and generally higher than the values predicted by galaxy formation models (e.g., Pillepich et al. 2018). Interestingly, the shorter the wavelength of the IR surveys, the higher the inferred contribution to the CSFRD, which suggests that **tracing warm dust is crucial to obtain a complete picture of the obscured star-formation activity in the first few billion years of cosmic time.**

At z>4 most galaxies have sub-solar metallicities, although these abundances are still high enough to allow for the presence of small dust grains, which are typically heated by hard radiation fields. Dusty low-metallicity galaxies, such as the prototype Haro11 object (e.g., Cormier et al. 2012), are known in the local Universe and at z~2 (Shivaei et al. 2022).

In parallel, the cosmic black-hole accretion density (BHAD) derived from X-ray surveys is substantially lower than the theoretical predictions, especially at z>3. On the one hand, this is due to the fact that typical photometric surveys only select high-z AGN with little or no dust obscuration (e.g., Bischetti et al. 2023), and a significant fraction of high-z active galactic nuclei (AGN) are dust-obscured, as shown with recent JWST Mid Infrared Instrument (MIRI) observations (e.g., Lyu et al. 2024). Indeed, the IR-selected AGN account for a substantial fraction of the BHAD at cosmic noon, and the total BHAD value does not seem to decline up to at least z=4 (Yang et al. 2023). On the other hand, the vast majority of the newly JWST-discovered AGN at high z are not detected in X-rays (e.g., Yue et al. 2024; Maiolino et al. 2025). This, in turn, may be the consequence of an evolution in the AGN X-ray spectral shape at high z (Zappacosta et al. 2023) and/or the lack of a coronal component in metal-poor systems (Simmonds et al. 2016). These sources are expected to have mostly dust-free broad-line regions (Maiolino et al. 2025); still, it is unclear whether dust could be present farther away from the central engine, either in a circumnuclear region or the more extended host interstellar medium.

**A systematic census of the sources of warm dust at mid-IR wavelengths is key to quantify the importance of classical, dust-obscured AGN, as well as constrain the properties of the newly discovered, lower luminosity, X-ray-weak AGN, at high z. A coherent study of these two independent AGN populations, via the analysis of IR surveys, will shed light on their roles in the cosmic nuclear activity within the first two billion years after the Big Bang.**

While sub-millimetre telescopes like ALMA can detect luminous dusty sources at z>4, they are biased towards the detection of galaxies with mainly cold dust (Td=30-40 K). Both theoretical galaxy models (e.g., Sommovigo et al. 2022) and different observational pieces of evidence (e.g., Bakx et al. 2020) suggest that warm dust (Td >80 K) could be dominant in star-forming galaxies at high z. In the case of dust heated by the winds of a central black-hole, the temperature can easily reach several hundred and up to ~1000-1500 K, depending on the distance of the dust clouds to the central engine. Thus, the peak of the IR SEDs of these sources typically falls below the ALMA (most sensitive) wavelength coverage. In addition, ALMA's survey power is limited, given its small field of view.

Obtaining an unbiased view of the dusty IR Universe in the first few billion years requires a telescope capable of surveying relatively large areas at mid-/far-IR wavelengths. Particularly, tracing the warm dust emission at rest-frame wavelengths 3-15 microns at z>4 implies observing beyond the wavelength coverage of JWST and below the wavelengths that ALMA can probe. **PRIMA will be the only telescope capable of obtaining IR maps in the ~24-84-micron wavelength regime,** which is fundamental to quantify the importance of warm dust-obscured star formation and black-hole activity in the early Universe.

The warm dust emission at rest-frame wavelengths 3-15 microns produced by low-metallicity dusty galaxies and hot dusty tori of intermediate-luminosity AGN have different spectral energy distributions (SEDs). Although both follow a power-law shape, the low-metallicity dusty galaxy

SEDs are typically much steeper (Fig. 1). In composite systems, where the IR SED difference is less obvious, the joint analysis of multi-wavelength data, including PRIMAger data, will be fundamental to reveal the nature of the dusty sources present in the Universe only 1.5 billion years after the Big Bang.

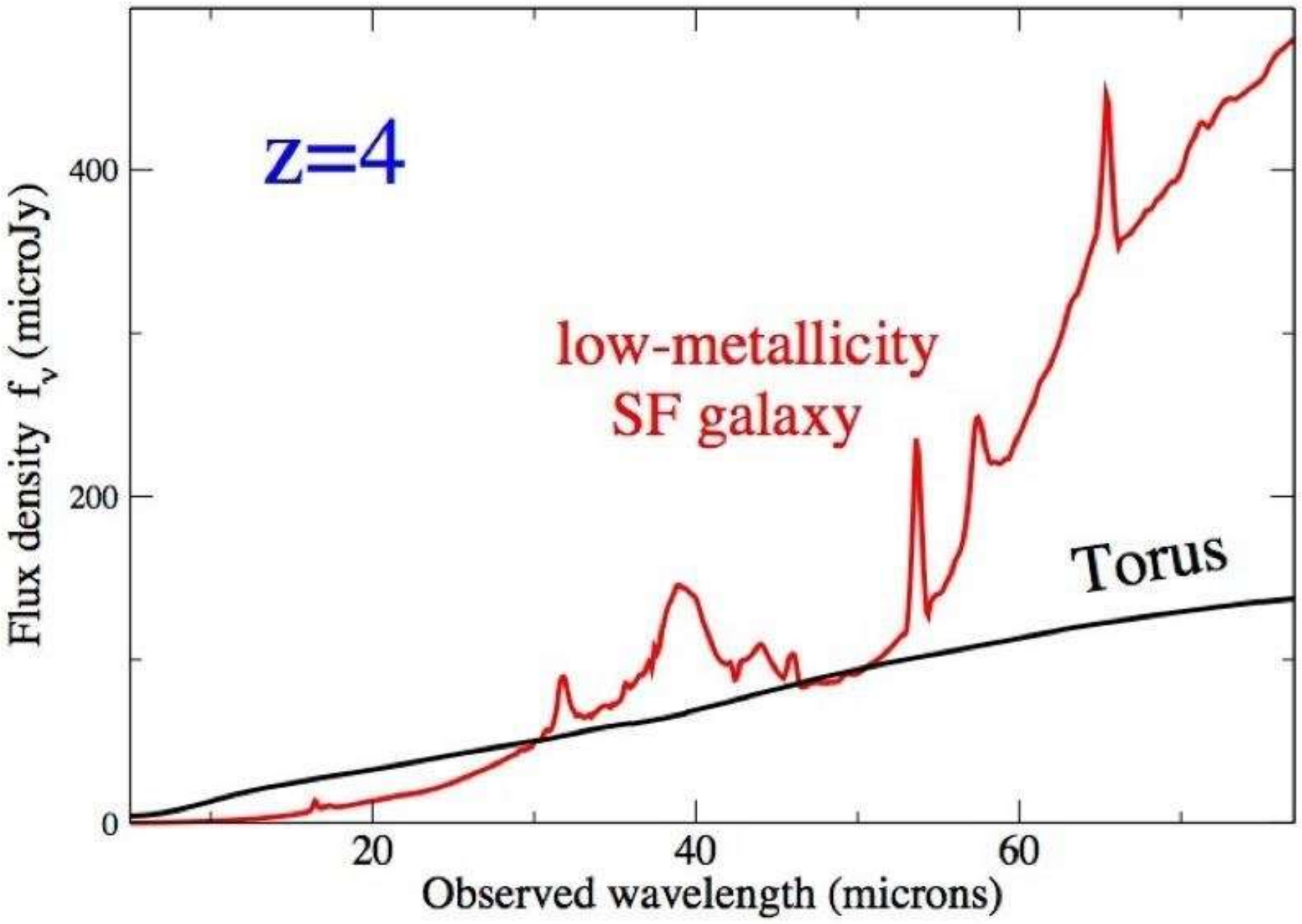

**Figure 1**. Mid-IR SEDs of a HARO11-like low-metallicity galaxy and an AGN hot dusty torus (Lyu+24), redshifted to z=4. In this example, the separation of both SEDs is very clear. In other, less evident cases, the joint analysis of PRIMA and JWST data will allow us to reveal the nature of the warm-dust sources.

We propose to conduct a blank survey of about 0.4 sq. deg. with the PRIMA hyperspectral Imager (PRIMAger) to **a depth of about 30 microJy (3 sigma) in two integrated PRIMAger bands**, namely about 24-45 and 45-84 microns, over a number of JWST blank fields, such as the CEERS, JADES and PRIMER fields (Fig. 2). This will allow us to detect 1000-1500 **luminous and ultra-luminous IR galaxies at z>4** (based on Béthermin et al. 2017, 2024).

Our main scientific goal is to determine the nature of the warm dusty sources at z>4 and understand their role in the activity history of the Universe. Current analysis based on JWST data finds that some low-metallicity galaxies with warm dust can mimic the properties of AGN (Hainline et al. 2016). The PRIMA mid-IR imaging at 24-84 microns, analysed in conjunction with the JWST ancillary data, will help us to disentangle low-metallicity star-forming galaxies from AGN-dominated sources, and conduct a general study of composite systems with a multi-wavelength approach. Simultaneously, our proposed survey will allow us to obtain a complete census of the most luminous IR galaxies with warm dust at high z, reaching luminosities below the 10^12 Lsun regime. Finally, the PRIMAger collected data will also allow for the detection of ~15000-20000 sources at z<4, enabling legacy science, which is beyond the scope of this Proposal.

This science program will be conducted with **approximately 1250 h of PRIMA observing time** (according to the PRIMA ETC; see section Approx. Integration Time below).

## Instruments and Modes Used

This observing program requests observations using PRIMAger using both bands in hyperspectral mode, with about 10-15 maps of variable size between 5’x5’ and ~15’x15’.

## Approximate Integration Time

We request a sensitivity of 50 microJy (3 sigma) @30microns in each of the 6 PRIMAger sub-bands, in order to **get a depth of 30 microJy (3 sigma) in two integrated PRIMAger bands**, namely 24-45 and 45-84 microns. This will allow us to detect about 1000-1500 **luminous and ultra-luminous dusty galaxies at z>4**. According to the PRIMA ETC, achieving such sensitivity for 10-15 maps of variable size (between 25 and 225 sq. arcmin) requires ~1250 h of observing time.

## Special Capabilities Needed

None.

## Synergies with Other Facilities

Our proposal is a clear example of PRIMA/JWST synergies. Our scientific goals will require the joint analysis of data from both observatories.

## Description of Observations

We propose to conduct our survey with the PRIMAger hyperspectral mode to a depth of about 50 microJy (3 sigma), over a number of JWST deep blank fields (e.g., CEERS, JADES, PRIMER fields; see Fig. 2). This requested sensitivity refers to each of the 6 PRIMAger sub-bands, implying that **we will get a depth of about 30 microJy (3 sigma) in two integrated PRIMAger bands**, namely 24-45 and 45-84 microns. This will allow us to detect **about 1000-1500 luminous and ultra-luminous dusty galaxies at z>4** (based on Béthermin et al. 2017, 2024). With two mid-IR data points, we will determine the mid-IR colours of all sources and, for the brightest ones, we will make use of the PRIMAger hyperspectral power to determine the mid-IR SED shapes in more detail.

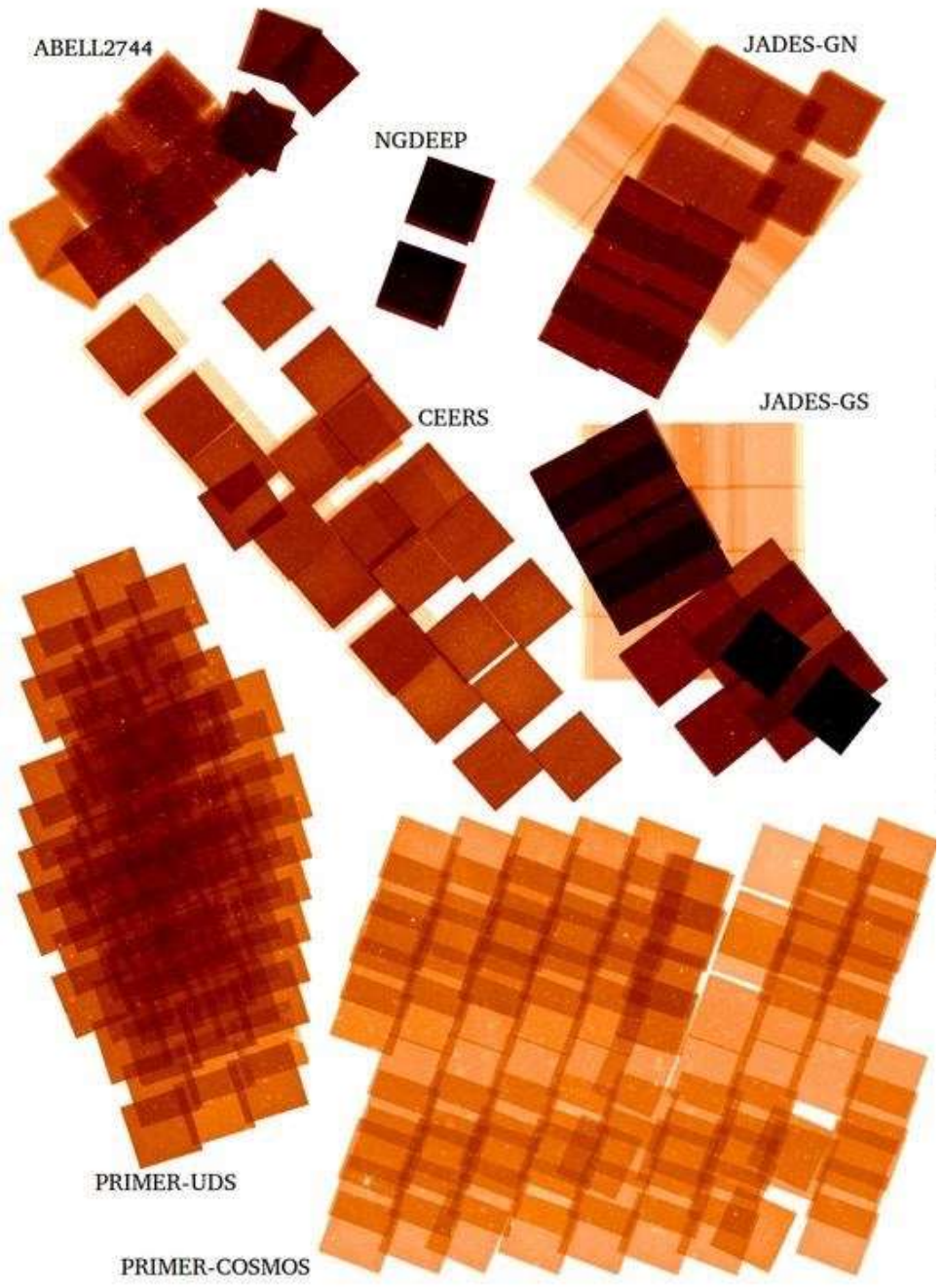

**Figure 2.** Overview of current JWST deep fields covering a total of ~0.2 sq. deg. The area covered with JWST imaging will increase in the coming years and should at least be double by the time of PRIMA launch
Figure Credit: ASTRODEEP (Merlin +24)

Given the requested sensitivity, our PRIMA survey will be about twice as deep as any of the GTO PRIMAger surveys and will be below the confusion limit (~50 microJy at 35 microns; Béthermin et al. 2024; Donnellan et al. 2024). We will mitigate confusion by making use of the deep near- and mid-IR imaging from JWST: we will consider JWST IR sources as priors in algorithms for flux deconvolution, such as XID+ (Donnellan et al. 2024) or TPHOT (Merlin et al. 2016), in order to recover their PRIMA fluxes.

# 15. Warm Dust Component in Nearby Galaxies

Viviana Casasola (INAF-IRA Bologna, Italy), Simone Bianchi (INAF-OAA Firenze, Italy), Francesco Calura (INAF-OAS Bologna, Italy), Golshan Ejlali (IPM, Tehran, Iran), Jacopo Fritz (IRyA, UNAM Morelia, Mexico), Hanae Inami (Hiroshima University), Logan Jones (Space Telescope Science Institute, Baltimore, USA), Evangelos Dimitrios Paspaliaris (INAF-OAA Firenze, Italy), Francesca Pozzi (University of Bologna, Italy), Francesco Salvestrini (INAF-OATs, Italy), Vidhi Tailor (INAF-IRA Bologna, Italy), Tsutomu T. Takeuchi (Nagoya University/ISM, Japan)

We propose a PRIMAger survey to characterize the warm dust component ($T \gtrsim 30$ K) in a representative sample of 18 nearby galaxies from the DustPedia project. This component, largely missed by Herschel, serves as a key tracer of recent star formation, AGN activity, and stochastic heating processes in grains. Our goal is to map the distribution of warm dust across galactic disks and explore its correlation with cold dust, gas, stars, and other main galaxy properties. PRIMA will be crucial for accurately modeling the warm dust component, addressing significant uncertainties in current data, and providing essential constraints to better understand the warm dust in nearby galaxies.

We will perform spectro-photometric mapping with PRIMA's Hyperspectral (24–84 µm) and Polarimetric (92–235 µm) Imagers, covering areas of 2 × optical size per target (0.4–0.9 deg²) to trace emission from galaxy centers to outskirts. Total requested time is ∼260 hours, ensuring 5σ sensitivity to faint emission in all PRIMAger bands. Using spatially resolved UV-to-mm spectral energy distribution fitting and ancillary multi-wavelength data, we will constrain dust temperature, mass, and heating sources, enabling the first comprehensive view of warm dust in the Local Universe. The proposed survey will benefit from synergies with ALMA (molecular gas), JWST (PAHs, stellar populations), and MeerKAT (atomic gas), allowing for a comprehensive multi-phase characterization of the ISM and main galaxy properties.

## Science Justification

Herschel has transformed our understanding of the properties of the interstellar medium (ISM), particularly concerning dust. Its observations provided dust emission maps with a resolution of up to 36′′ across the wavelength range of 70–500 µm, enabling detailed analyses of dust emission, masses, and temperatures within galaxies. Herschel has been crucial for investigating the cold dust component (T∼15–20 K; Galliano et al. 2018), which does not emit significantly at wavelengths <50 µm (e.g., Draine 2003). A prime example of cold dust characterization in nearby galaxies, thanks to Herschel, is the DustPedia project (Davies et al. 2017). DustPedia is based on a sample of 875 extended ($D_{25} \gtrsim 8'$) galaxies of all morphological types, located within <40 Mpc and observed by Herschel, along with a database of multi-wavelength (UV/ sub-mm) imagery and photometry that greatly exceeds the scope, in terms of the number of galaxies and wavelength coverage, of any similar survey (Clark et al. 2018). Thanks to DustPedia, we were able to characterize the radial distribution of dust and other properties in spiral galaxies, revealing that dust has scale lengths roughly twice as large as those of stars. We also explored an unprecedented set of integrated and spatially resolved (sub-kpc/kpc) scaling relations between dust, cold gas (CO and HI), and other galaxy properties (e.g., stellar mass, SFR; Casasola et al. 2022). However, Herschel missed the warmer dust component, with emission peaking in the MIR.

While the cold dust component in galaxies has been extensively studied, a comprehensive characterization of the warm dust component ($T \gtrsim 30$ K) in the Local Universe remains lacking. Warm dust, predominantly heated by radiation from young stellar populations and active galactic nuclei or originating from the emission of small grains subject to stochastic heating, is generally found in star-forming regions and within the diffuse ISM. Due to its heating mechanisms and spatial distribution, this component is likely to exhibit a stronger correlation with recent star formation activity and molecular gas reservoirs than its colder dust counterpart. Addressing this gap is crucial for a complete understanding of the interplay between dust, gas, and star formation in nearby galaxies.

We propose to carry out a PRIMAger survey focused on characterizing the warm dust component in a representative and well-studied subset of nearby galaxies selected from the DustPedia sample. This study aims to address the following key scientific questions:

- What is the spatial distribution of warm dust within galactic disks?
- How does warm dust distribution compare with cold dust, stars, and cold gas?
- How do the properties of warm dust vary as a function of galaxy environment and properties, such as morphology, SFR, AGN activity, and metallicity?

By leveraging the extensive ancillary data from the DustPedia database, we will perform both global and spatially resolved (pixel-by-pixel) UV-to-mm spectral energy distribution (SED) fitting. This approach will allow us to precisely constrain dust and stellar properties across the sample, including the key contribution of warm dust emission that PRIMA will probe. PRIMA will be crucial for accurately modeling the warm dust component, addressing significant uncertainties in current data. For instance, as shown in Fig. 1, the SED of a typical nearby galaxy (NGC 2146, Ejlali et al. 2024), modeled with a two-component modified blackbody (2MBB), reveals substantial uncertainty in the flux modeled in the 30–60 μm range using Herschel and Spitzer data, highlighting the essential role of PRIMA in better constraining the warm dust component.

We aim to exploit the capabilities of the Hyperspectral Imager (PHI; 24–84 μm) and the Polarimetric Imager (PPI; 92–235 μm) onboard the PRIMA facility to carry out spectro-photometric observations of a well-characterized sample of nearby galaxies. PHI provides a unique opportunity to investigate the warm dust component ($T \approx 30$–$120$ K), as well as the emission from small grains undergoing stochastic heating. The 60–90 μm range is crucial for probing dust at intermediate temperatures, marking the transition between warm and cold dust emission regimes. The PPI bands play a key role in refining our understanding of the cold dust component mapped by Herschel.

PRIMA represents a significant advancement in both sensitivity and angular resolution compared to previous IR missions. While IRAS offered angular resolutions of ∼0.5′-2′ with limited sensitivity, PRIMA improves on both by over an order of magnitude. WISE achieved ∼6″ resolution at 22 μm; PRIMA delivers up to ∼1.2 times better resolution in the MIR and up to ∼2 times better resolution at longer wavelengths, where WISE lacks coverage. Compared to Spitzer, PRIMA improves angular resolution by up to ∼1.2× in the MIR and ∼3.6× at FIR wavelengths, with sensitivity gains of several factors. Relative to Herschel, PRIMA offers up to ∼2× better angular resolution in the 60–200 μm range and ∼3–5× higher sensitivity. These advancements make PRIMA uniquely suited to detect fainter, more diffuse dust structures, enabling unprecedented mapping of dust in the outskirts of galaxies within the Local Universe. It will resolve features at physical scales ranging from tens of parsecs to a few kiloparsecs, depending on the wavelength and the galaxy distance.

We propose to use PRIMAger to map the DustPedia sub-sample examined in Casasola et al. (2017, 2022) and Tailor et al. (2025). This sample includes 18 nearby (distances = 3–20 Mpc), large (D25 = 8′-24′, corresponding to ∼11–62 kpc), and face-on galaxies, all well-resolved in the 500μm-SPIRE-Herschel band (FWHM = 36′′). For this targeted galaxy sample, thanks to Herschel, we have already characterized the radial distribution of cold dust emission, dust mass, and temperature of the cold dust component, as well as the interstellar radiation field, young and old stars, cold gas, and SFR. This sample allows us to explore a valuable parameter space, including morphology (Hubble type T = 2–7), stellar mass ($2 \times 10^{9}$–$1 \times 10^{11}$ M⊙), SFR (0.2–13.0 M⊙ yr$^{-1}$ ), and the presence or absence of structural features such as stellar bars and central AGN.

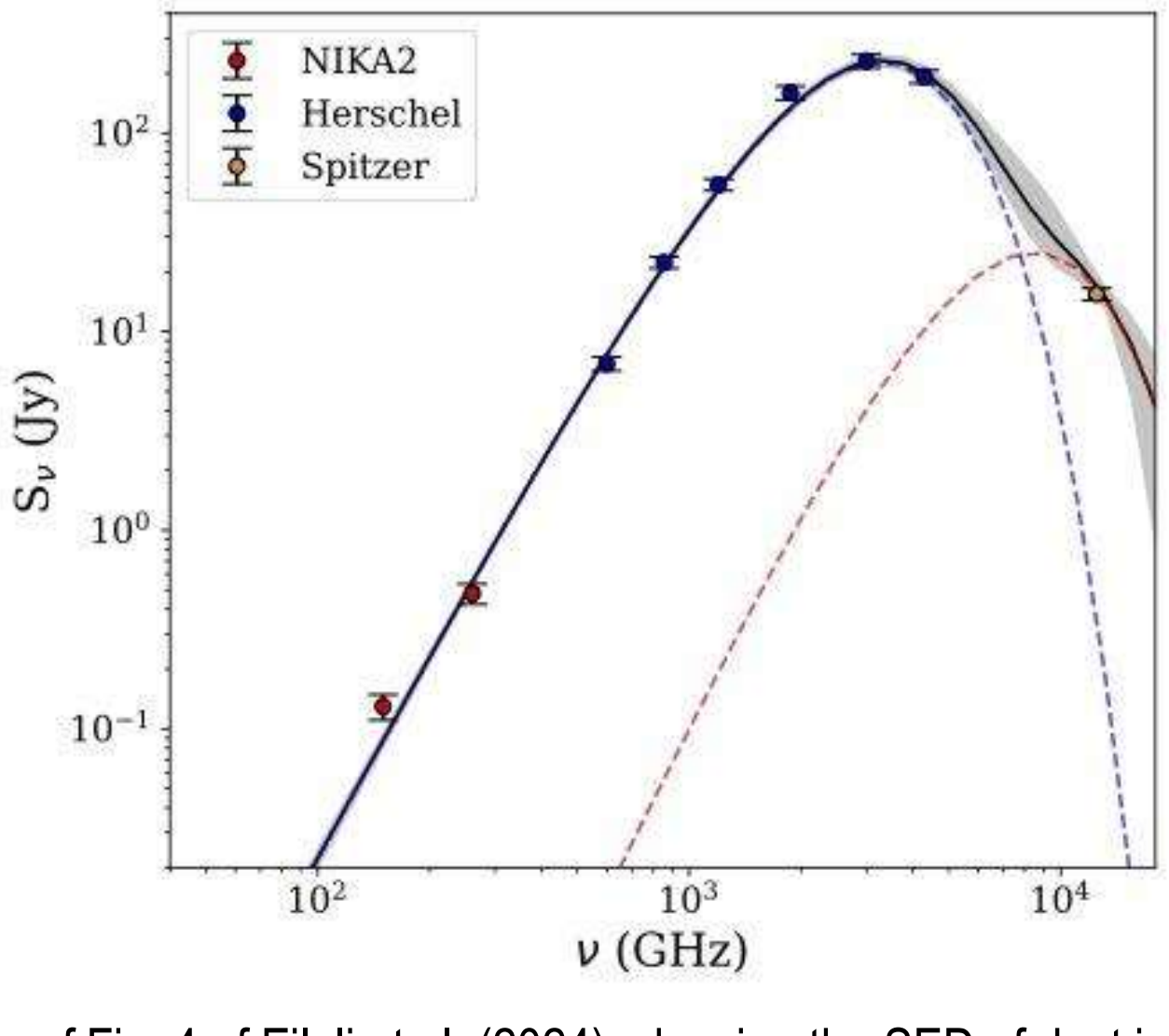

**Fig. 1.** Reproduced version of Fig. 4 of Ejlali et al. (2024), showing the SED of dust in the starburst galaxy NGC 2146. The dashed blue and red lines show the cold and warm dust components, respectively, and the black line shows the total emission. The grey area shows the uncertainty in the modeled flux. The uncertainty in the warm component of dust is large, despite the data point from Spitzer MIPS at 24 μm. This uncertainty is even larger in galaxies with lower SFR than this starburst galaxy, resulting in a warm dust component that is not well constrained.

## Instruments and Modes Used

This proposed survey uses PRIMAger data for all the hyperspectral and polarimeter bands as the polarimetric information is not needed. The map size depends on the angular extent of each target. The targets of this survey are 18 nearby galaxies: NGC 300, NGC 628, NGC 925, NGC 1097, NGC 1365, IC 342, NGC 2403, NGC 3031, NGC 3521, NGC 3621, NGC 4725, NGC 4736, NGC 5055, NGC 5194, NGC 5236, NGC 5457, NGC 6946, and NGC 7793.

## Approximate Integration Time

The requested observing time has been calculated using the PRIMA Exposure Time Calculator (ETC) and the sensitivity values provided on the official PRIMA website (see Table "Required PRIMAger Sensitivity"). We request a total of ∼224 hours to map the entire sample across all PRIMAger bands (Hyperspectral Imager and Polarimetry Imager). The map sizes range from 0.4 to 0.9 $deg^2$, depending on the angular extent of each galaxy, with individual map integration times between 9.4 and 18.5 hours per target. Based on Herschel observations, this integration time will enable 5σ detection of faint emission in the outermost regions of the galaxies' optical disks. For instance, it corresponds to a point-source sensitivity of ∼2.2 mJy (equivalent to a surface brightness of ∼0.74 MJy/sr) at 65 µm, approximately the central wavelength of the PHI2 band.

## Special Capabilities Needed

None

## Synergies with Other Facilities

The synergy with both space- and ground-based facilities is crucial. All proposed galaxies have been thoroughly mapped across the UV to submillimeter wavelengths (e.g., Herschel, HST, and GALEX; see the DustPedia database). They are also prime targets of major observatories probing other critical ISM components and galaxy properties intimately linked to the dust features that PRIMA will investigate. Specifically, ALMA provides detailed maps of molecular gas (e.g., see PHANGS data; Leroy et al. 2021); JWST offers insights into polycyclic aromatic hydrocarbons (PAHs), warm dust emission, ionized gas regions, IR molecular lines, and stellar populations; MeerKAT (soon complemented by SKA) traces the atomic gas (HI) component.

## Description of Observations

We propose to map the sample of 18 nearby galaxies simultaneously across all PRIMAger bands (Hyperspectral Imager, Polarimetry Imager). For each galaxy, we will cover an area extending to 2 × D25, ensuring complete spatial coverage of both the main disk and the outer regions. This strategy maximizes observational efficiency and minimizes total observing time.

# 16. The Radial Variation of the Silicate-to-Carbon Ratio in M31 Probed by PRIMA

Jérémy Chastenet (Sterrenkundig Observatorium, Universiteit Gent, Krijgslaan 281-S9, 9000 Gent, Belgium), Ilse De Looze (Sterrenkundig Observatorium, Universiteit Gent, Krijgslaan 281-S9, 9000 Gent, Belgium), Maarten Baes (Sterrenkundig Observatorium, Universiteit Gent, Krijgslaan 281-S9, 9000 Gent, Belgium), Simone Bianchi (INAF – Osservatorio Astrofisico di Arcetri, Largo E. Fermi 5, 50125 Firenze, Italy), Viviana Casasola (INAF - Istituto di Radioastronomia, Via Piero Gobetti 101, 40129 Bologna, Italy), Laure Ciesla (Aix Marseille Univ, CNRS, CNES, LAM, Marseille, France), Stephen Eales (School of Physics & Astronomy, Cardiff University, The Parade, Cardiff CF24 3AA, UK), Jacopo Fritz (Instituto de Radioastronomía y Astrofísica, Universidad Nacional Autónoma de México, Morelia, Michoacán 58089, Mexico), Frédéric Galliano (Université Paris-Saclay, Université Paris Cité, CEA, CNRS, AIM, 91191, Gif-sur-Yvette, France), Suzanne C. Madden (Université Paris-Saclay, Université Paris Cité, CEA, CNRS, AIM, 91191, Gif-sur-Yvette, France), Angelos Nersesian (STAR Institute, Quartier Agora – Allée du six Août, 19c B-4000 Liège, Belgium), Monica Relaño (Dept. Fisica Teorica y del Cosmos, E-18071 Granada, Spain; Instituto Universitario Carlos I de Fisica Teorica y Computacional, Universidad de Granada, E-18071 Granada, Spain), Matthew W. L. Smith (School of Physics & Astronomy, Cardiff University, The Parade, Cardiff CF24 3AA, UK), Stefan van der Giessen (Sterrenkundig Observatorium, Universiteit Gent, Krijgslaan 281-S9, 9000 Gent, Belgium), Emmanuel Xilouris (Institute for Astronomy, Astrophysics, Space Applications & Remote Sensing, National Observatory of Athens, P. Penteli, 15236 Athens, Greece)

The properties of interstellar dust grains are being scrutinized more than ever before, with the advent of large facilities. Infrared emission from dust grains is a powerful asset that can help constrain their physical and chemical properties. Among these, the relative ratio of carbon-rich to silicate-rich grains remains one that has not yet been investigated thoroughly, due to the lack of dedicated instruments and modeling limitations.

With this proposal, we aim to quantify the modeling degeneracies inherent in constraining the far-infrared (far-IR) slope of the dust emission spectral energy distribution. Used as a proxy for the silicate-to-carbon ratio, we find that recovering the far-IR slope is affected by the estimate of the local radiation field, and the input abundances of different grain species. We show that PRIMA's hyperspectral imaging will lead to better constrained local radiation fields, which will aid—together with PRIMA's polarization capabilities—to better constrain the silicate-to-carbon ratio in M31, and how it spatially varies within the galaxy.

## Science Justification

Dust in the interstellar medium (ISM) of galaxies is a key component in regulating the energy distribution in its local environment. Dust grains absorb ultraviolet (UV) and optical light from young stellar populations, which increases their temperature, and re-emit that light in lower energy wavelengths, in the infrared (IR), as they cool down. This leads to (i) a significant decrease in the photons' energy, making dust grains act as a coolant, and (ii) a nuisance when it comes to estimating stellar-related quantities, such as stellar mass or star formation rate (De Looze et al. 2014, Nersesian et al. 2020a, Nersesian et al. 2020b, Verstocken et al. 2020, Viaene et al. 2020), as a major part of that light can be extinguished and reddened by dust grains (Viaene et al. 2016, Bianchi et al. 2018). Dust grains also participate in several physicochemical processes in the ISM. They act as a catalyst for the formation of $H_2$ (Le Bourlot et al. 2012, Bron et al. 2014), leading to the formation of molecular clouds necessary for star formation, and the smallest grains account for most of the photoelectric effect, heating the gas (Wolfire et al. 1995). For these reasons, understanding the life cycle, formation, and destruction of dust grains is critical so that we can not only correct their effect on other galactic properties but also better understand galactic evolution as a whole.

To do so, we often measure dust properties by comparing observations with models that are meant to encompass physical properties of interstellar dust: size distribution, composition, abundance, etc. Here we use emission measurements, from mid-infrared (mid-IR) to sub-millimeter (sub-mm) wavelengths. Emission models can be split between rather simplistic analytical models, e.g., modified blackbody, reproducing well the emission from large grains, in thermal equilibrium (Gordon et al. 2014, Meisner & Finkbeiner 2015) and physically-motivated, either observationally- or laboratory-based models (Zubko et al. 2004, Draine & Li 2007, Jones et al. 2013, Jones et al. 2017).

A substantial amount of work in the past few decades has led to some solid knowledge of interstellar dust: it is made up of grains ranging from a few nm to about 1 µm in size; it shows features associated to the presence of amorphous silicate-rich material, with metallic nano-inclusions to account for depletions, and features arising from vibrational bonds of aromatic-rich material, and a population of amorphous carbon-rich grains; grains are also likely not spherical but elongated or irregular in shape, as suggested by the detection of polarized light, caused by preferential alignment of non-spherical grains along the magnetic fields. Spatially integrated and resolved studies of nearby galaxies have shown a wide range of variations between and within galaxies, proving that interstellar dust is an evolving component (Muñoz-Mateos et al. 2009, Boselli et al. 2010, Ciesla et al. 2014, Rémy-Ruyer et al. 2014, Hunt et al. 2015, Davies et al. 2017, Aniano et al. 2020, Galliano et al. 2021, Abdurro'uf et al. 2022, Casasola et al. 2022, Dale et al. 2023, Chastenet et al. 2025).

Using physical models, we investigate the possibility of recovering a silicate-to-carbon ratio for large grain populations. Being able to disentangle silicate and carbon dust content would inform us of their relative spatial distribution, which could be linked to formation (O-rich vs C-rich Asymptotic Giant Branch stars; Gail et al. 2009, Goldman et al. 2022) and destruction (in supernova shocks; Bocchio et al. 2014, Slavin et al. 2015, Hu et al. 2019, Kirchschlager et al. 2024) processes. Understanding the differential evolution of carbon- vs silicate-rich dust grains would give us great insights into the grain life cycle in the ISM. This approach is heavily motivated by observational and experimental evidence of a varying far-IR slope (Bot et al. 2010, Galliano et al. 2011, Demyk et al. 2012, Galametz et al. 2014, Gordon et al. 2014, Paradis et al. 2019, Demyk et al. 2022).

In M31, in particular, Smith et al. (2012) provide resolved maps of the dust temperature, dust mass surface density, and spectral index, β (the slope of the far-IR in blackbody models) at 36" resolution. Their work shows radial trends of the β parameter, increasing from the center to a radius of about 3 kpc and decreasing to the outskirts (see also Tabatabaei et al. 2014, Whitworth et al. 2019, Clark et al. 2023). Assuming that temperature variations (correlated with β) should be limited to the diffuse ISM, we argue that the β variation could be predominantly driven by a varying silicate-to-carbon ratio, and we investigate how to recover this parameter. In Chastenet et al. (2017), the silicate-to-carbon ratio of large grains is tentatively inferred in the SMC and LMC but is met with severe limitations due to model degeneracies and limited spectral coverage.

This science case is built upon previous work, and leverages more recent data such as JCMT, longward of the visible break in the far-IR—critical to estimate the silicate-to-carbon ratio—and updated physical model (THEMIS2; Ysard et al. 2024) to capitalize on theoretical work on dust polarization. Using The Heterogeneous dust Evolution Model for Interstellar Solids (THEMIS, Jones et al. 2017)—a mixture of amorphous silicates and hydrocarbon grains, both with aromatic-rich mantles—a suite of broadband photometry of M31—WISE, Herschel, and JCMT data—we find that new observations from PRIMA can help alleviate model degeneracies and yield stronger estimates of the silicate-to-carbon ratio. Specifically, the Hyperspectral Imager (PHI) can improve the recovery of radiation field parameters by at least 30–40% by adding measurements in the 24–84 μm regime. The $I_{pol}/I_{tot}$ spectrum can be very informative. For example, considering carbonaceous grains polarization alters the SED, from a monotonous raise to the presence of a bump. The PPI measurements will sample the IR regime sufficiently to be confronted with models varying the silicate-to-carbon ratio, assuming polarizing efficiencies from laboratory measurements.

With PRIMA, we will derive the radial variations of the silicate-to-carbon ratio across a large area of M31 and reveal more details about the life cycle of dust grains in the ISM.

## Instruments and Modes Used

This observation is performed with PRIMAger, and all photometric data of hyperspectral and polarimeter bands are needed to map 25'' x 82'' area in total. Polarimetry information is also needed.

## Approximate Integration Time

We measure the emission in 3 key regions in M31: a region in one of the main bright arms, a diffuse inter-arm region, and a region in an arm in the outskirts. We use these average surface brightnesses and target signal-to-noise ratios to calculate the exposure time following prescriptions from the PRIMA team.

For a $0.6^{\circ 2}$ mapped area:

- For the total intensity in the Polarimetric Imager, we aim for a S/N = 5 detection in the faintest region (inter-arm) with 12.1, 13.5, 10.5, and 6.1 MJy/sr (PPI1 through PPI4), which leads to short integration times within a few minutes.
- For the polarized intensity, given the much lower sensitivity in that mode, we aim for S/N = 5 in the bright central arm for a conservative 2% polarization, with fluxes of 0.7, 1.1, 1.3, and 1.0 MJy/sr, leading to about 7 hours of integration.
- We aim for a 1.25 MJy/sr detection at S/N = 3 for the Hyperspectral Imager, which leads to a maximum of about 4 hours of integration.

## Special Capabilities Needed

There is a limited window for the position angle: the proposal aims to sample several environments in M31, which requires including the bulge of the galaxy, spiral arms, and diffuse interarm regions.

## Synergies with Other Facilities

N/A

## Description of Observations

For the best outcome of this study, a nearby target is required for better physical resolution, and galaxies in the Local Group are obvious candidates. The Magellanic Clouds are different enough from the Milky Way that the interpretation and applicability of our method would be more difficult. The other candidate, M33, is simply smaller and more distant, making it less ideal. For distance, size, metallicity, and overall similarity to the Milky Way, M31 stands as the best target for this project. It has also been more studied and has a greater wealth of ancillary data, and we believe it represents a good case for a first solid application of this program: observed by Herschel in its entirety (Fritz et al. 2012), and with the JCMT (Smith et al. 2021), covering the far-IR spectrum and all infrared bands relevant for our science goals; in the optical, there is the stunning HST imagery of M31 from the PHAT (Dalcanton et al. 2012) and PHAST (Chen et al. 2025) programs. Stellar population variations inferred from HST could be checked against the silicate-to-carbon ratio spatial variations to link dust production and processing with stellar environment; there is also high velocity-resolution H I mapping of the northern half of M31 (Koch et al. 2021), and CO mapping available as well (Caldú-Primo & Schruba 2016). All of these data can help separate the ISM into different phases and help inform the model to be used (e.g., ‘dense’ vs ‘diffuse’ THEMIS) to track the silicate-to-carbon ratio variations.

As a Milky Way-like galaxy in an external environment, with an exceptional closeness that allows for highly resolved data, M31 is an ideal laboratory to probe the variations of dust grain properties in extragalactic regions.

To build upon these works and focus on radial variations and other known properties of dust in M31, we want to sample as many ISM environments as possible. To keep exposure time reasonable, we select a 0.6°$^2$ area to be observed with PRIMA's instruments, with two main goals:

- New radiation field estimates. The 24–84 µm measurements from PRIMA would not directly constrain the silicate-to-carbon ratio, but will help estimate the radiation field and γ parameter. This, in turn, is expected to alleviate degeneracies in the far-IR. In our program, we include the bulge of the galaxy, which is known to have a softer radiation field (Draine et al. 2014). This specific environment will be an excellent region to see the effect of a better radiation field estimate between the older bulge and the more diffuse ISM.
- Radial variations of the silicate-to-carbon ratio. The polarimeter will provide high spatial resolution maps of polarized far-IR emission, directly bringing constraints on the dust species abundances. The selected area covers radii from the center to the outskirts of the galaxy. This will allow us to compare radial trends of the target silicate-to-carbon ratio with previous work, namely that of the mentioned spectral index, β.

# 17. Probing PAH from All Galaxies Across Redshifts—PAH Intensity Mapping with PRIMA

Yun-Ting Cheng (Caltech/JPL), Brandon S. Hensley (JPL), Thomas S.-Y. Lai (IPAC)

Intensity mapping is a promising technique for probing galaxy evolution and large-scale structure by measuring spatial fluctuations in the integrated emission from all sources. We propose to conduct intensity mapping of polycyclic aromatic hydrocarbons (PAHs) across redshift using PRIMA. PAHs are key tracers of star formation activity and small grains in the interstellar medium of galaxies. Leveraging a recently developed technique, feature intensity mapping, we can constrain the total PAH intensity as a function of redshift by cross-correlating large-scale spectral imaging maps across different PRIMA channels. This method extends the established line intensity mapping formalism to broad spectral features, enabling three-dimensional mapping of PAH emission over cosmic time. We propose a 1000-hour observation with the PRIMAger Hyperspectral Instrument (PHI) to map a 1 deg² area for PAH intensity mapping. According to our forecast, this survey will achieve high-sensitivity detections of the large-scale power spectra of several PAH features out to z~5. Given that individual detections in such a PHI survey are limited to only the brightest galaxies at higher redshift (z>3), intensity mapping offers access to the otherwise inaccessible bulk population of high-redshift galaxies. This provides a complementary probe to help build a more complete understanding of galaxy formation and evolution in the early universe.

## Science Justification

Intensity mapping is an emerging technique that probes galaxy formation and evolution, as well as the cosmic large-scale structure (LSS) throughout cosmic time, by mapping the aggregate emission field from all emitting sources in the universe, including faint or diffuse galaxies that remain unresolved in individual detections [1,2,3]. Line intensity mapping (LIM) performs intensity mapping using a specific spectral line in a large-scale multi-wavelength imaging survey, typically with moderate spectral resolution. With the line-of-sight distance inferred from the redshift–frequency relation, LIM datasets directly map the 3D LSS of the universe traced by the spectral line emission [4,5]. Furthermore, the statistical properties of LIM maps, such as the clustering power of the intensity field, inform the properties of the emitting source population. As galaxy surveys typically face challenges in detecting a sufficient number of galaxies at high redshift (z >3) for mapping the 3D LSS and for obtaining representative samples to understand galaxy evolution, LIM promises to fill this gap in the high-redshift regime. PRIMA's far-infrared (FIR) coverage will open a new window for LIM using a large number of FIR spectral lines that trace various physical processes in galaxies (e.g., star formation, black hole accretion, etc.). The potential of LIM with PRIMA has been explored in detail in the PRIMA GO Book-1, Case #9 [6].

In addition to spectral lines, one of the most prominent features in the infrared spectral energy distribution (SED) of galaxies is the emission bands from polycyclic aromatic hydrocarbons (PAHs), which can contribute up to 20% of the total infrared luminosity [7]. PAHs serve as important tracers of star formation and dust composition in galaxies, and thus are of significant interest and represent a promising target for intensity mapping [8]. However, PAHs are broad spectral features that span multiple channels in PRIMA, making the typical LIM formalism inapplicable for direct use with PAHs. To address this challenge, the proposal team has recently developed a technique called "feature intensity mapping" (FIM; [9]). FIM is an extension of the LIM formalism that accounts for the spectral profile of the emission feature and the instrument's channel passband. By cross-correlating intensity maps across all spectral channels, FIM enables forward modeling to reconstruct the bias-weighted intensity, b(z)*I(z), for multiple PAH features as a function of redshifts (z), where I(z) is the mean intensity of each PAH feature and b(z) is the large-scale bias factor.

We forecast the sensitivity for detecting the PAH intensity mapping signal with PRIMA. Specifically, we consider a 1 deg² survey with PHI and a 0.1 deg² survey with FIRESS, each with 1000 hours of integration time. We fit the bias-weighted intensity as a function of redshift for 9 PAH features spanning 6.2 to 17 µm. Previous studies have shown that while the spectral shapes of individual PAH features exhibit negligible variation, the relative strengths between different features vary across galaxies. Therefore, we fix the central wavelengths and widths of the PAH features based on templates from previous studies, while allowing the amplitude of each feature as a function of redshift to vary as free parameters.

The results of our sensitivity forecast are shown in Figure 1. With PHI, we find that 8 out of the 9 PAH features considered can be detected with >5σ significance at z <3. Furthermore, the two brightest PAH features (7.7 and 11.3 µm) can achieve >5σ detection even at redshifts as high as z=5.

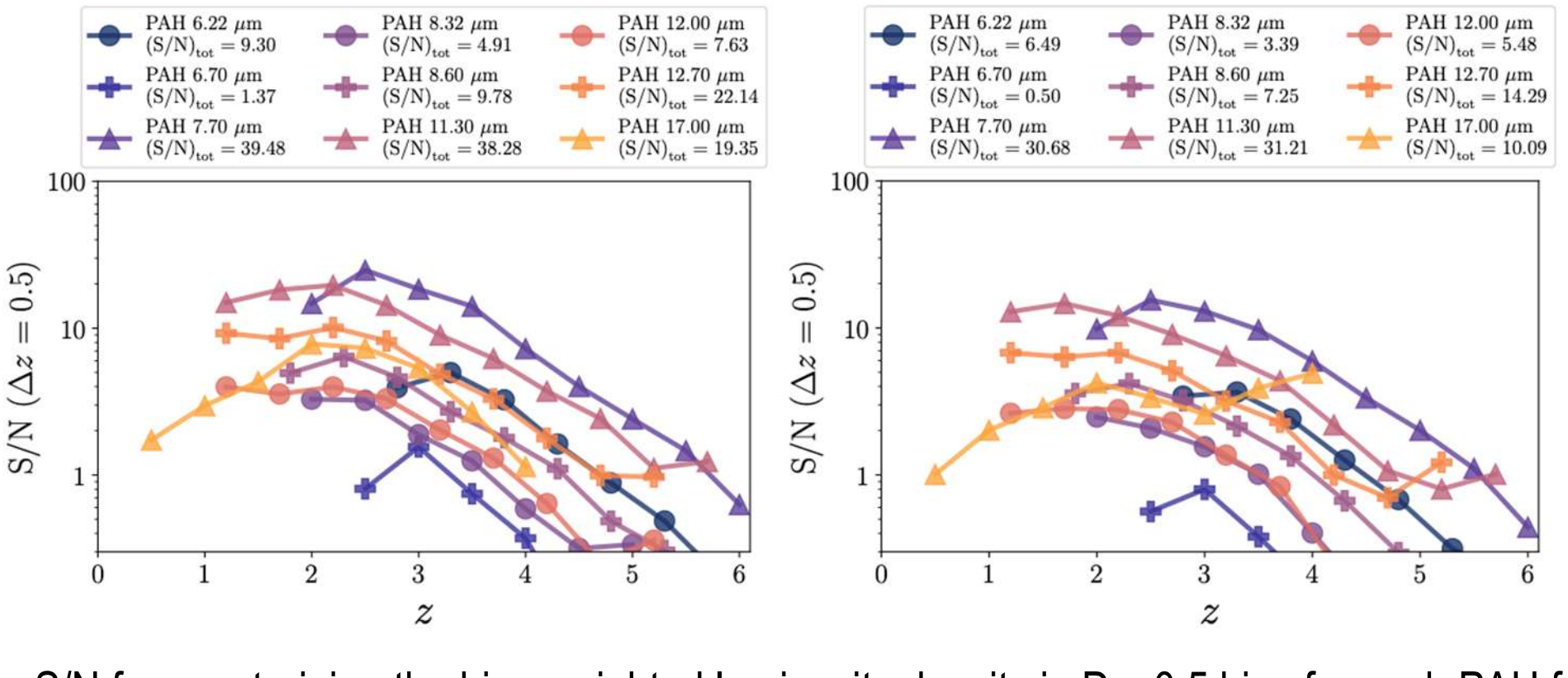

**Figure 1.** The S/N for constraining the bias-weighted luminosity density in Dz=0.5 bins for each PAH feature in the PHI (left) and FIRESS (right). The total S/N of each feature, summed over all redshift bins, is provided in the legend.

We find that PHI achieves better sensitivity for PAH intensity mapping than FIRESS. Despite PHI's lower spectral resolution (R~10), it is sufficient to resolve PAH features. Its lower noise level and larger survey area together result in higher sensitivity in the FIM measurements. We have verified that our chosen survey areas are optimal for 1000-hour integrations for both PHI and FIRESS.

Based on these findings, we propose a 1000-hour PHI observation to map a 1 deg$^2$ field in an extragalactic sky region.

The individual source detection limit for high-redshift galaxies with PRIMA is constrained by both sensitivity and confusion noise. [10] studied the galaxy populations detectable in a wide (10 deg$^2$) and a deep (1 deg$^2$) PHI survey, each with 1500 hours of integration. The deep survey in their study corresponds closely to our proposed GO case. Using simulations based on a semi-analytical model, they found that in the deep survey, only galaxies with star formation rates > 100 M_sun/yr are detectable at z > 4. This corresponds to an exceptionally luminous but unrepresentative subset of the high-redshift galaxy population. Our intensity mapping technique, however, enables measurements of the bulk population of high-redshift galaxies that remain beyond the reach of individual detection. Therefore, PAH intensity mapping will offer a crucial probe to advance our understanding of galaxy evolution at high redshift, helping to answer key questions about star formation and the dust and metal buildup in the bulk population of high-redshift galaxies.

## Instruments and Modes Used

This observing program requests a 1 deg$^2$ field using PRIMAger's hyperspectral mode with a total integration time of 1000 hours.

## Approximate Integration Time

1000 Hrs

## Special Capabilities Needed

None

## Synergies with Other Facilities

None

## Description of Observations

Uniformly scan a 1 deg$^2$ field with a total integration time of 1000 hours. The observing field will be selected in an extragalactic area, preferably a field overlapping with other deep galaxy and/or intensity mapping surveys to facilitate future opportunities for synergy.

## Acknowledgments

This research was carried out at the Jet Propulsion Laboratory, California Institute of Technology, under a contract with the National Aeronautics and Space Administration (80NM0018D0004).

# 18. A PRIMA Survey of the Deepest Herschel SPIRE Field

Dave Clements (Imperial College London), Chris Pearson (RAL Space)

The deepest Herschel image at 250, 350, and 500 microns, the SPIRE Dark Field (SDF), suggests the presence of a population of sub-mJy sources that are not accounted for in current far-IR/submm galaxy evolution models (see Pearson et al., 2025; Varnish et al., 2025). These results are based on P(D) analysis of a region observed repeatedly by Herschel-SPIRE for calibration purposes. There are a number of other anomalies in this field, including an apparent overdensity of SCUBA2 850 micron sources in part of this field that are not detected by SPIRE, so-called SPIRE dropouts (Parmar et al., in prep). These are candidate z>6 dusty star-forming galaxies (DSFGs) based on the stacking of their far-IR fluxes. Further observations of this field, including SCUBA2 observations to cover it completely, are planned, alongside existing deep IRAC and MIPS data. These will provide multiwavelength complementary data running from the X-ray to the radio, making the field an ideal target for a deep PRIMAger survey. We here propose detailed observations of this 0.25 sq. deg. field with both the PPI and PHI to determine the nature of the sources contributing to these counts and thus to allow us to finally understand the nature and evolution of the sources contributing to the CIB.

## Science Justification

The Cosmic Infrared Background (CIB) contains the integrated history of star formation in the universe. Comparison to the integrated UV-Opt-IR background shows that roughly half of the energy generated in the universe through star formation is obscured by dust and re-radiated at far-IR wavelengths. A full understanding of the history of galaxy and star formation over cosmic time thus requires observations at these long wavelengths. Observations by Herschel and other mid-IR-to-submm facilities have made significant advances in probing this population over the last 20 years, with up to 70% or more of the CIB resolved into individual sources brighter than the Herschel confusion limit of ~10s of mJy at 250-500 microns. However, analysis of the deepest Herschel field available, the SPIRE Dark Field (SDF) (Pearson et al., 2025) using P(D) analysis to probe below the confusion limit (Varnish et al., 2025, Fig 1) provides evidence of a second bump in the source number counts at sub-mJy fluxes, indicating the presence of an unexpected population of far-IR sources at flux levels of a few 100 microJy. At 250 microns, the counts of this new population peak at fluxes around 400 microJy. At still fainter fluxes, the counts drop away, indicating that this sub-mJy population is the last significant source of far-IR emission contributing to the CIB. The nature of this sub-mJy population, however, remains to be determined. There are also a number of other oddities in the SDF. SCUBA2 observations of the central 12x12 arcminute deepest SPIRE region, for example, have revealed an over-density of 850 micron sources lacking SPIRE detections – so-called SPIRE dropouts – compared to the density of such sources in other fields (Greenslade et al., 2019; Parmar et al., in prep.). SPIRE-dropouts are candidate high-redshift dusty star-forming galaxies (DSFGs). A stacking analysis of the SDF SPIRE dropouts suggests that they lie at z ~ 6.5 and make a significant contribution to the star formation rate density at that epoch (Parmar et al., in prep).

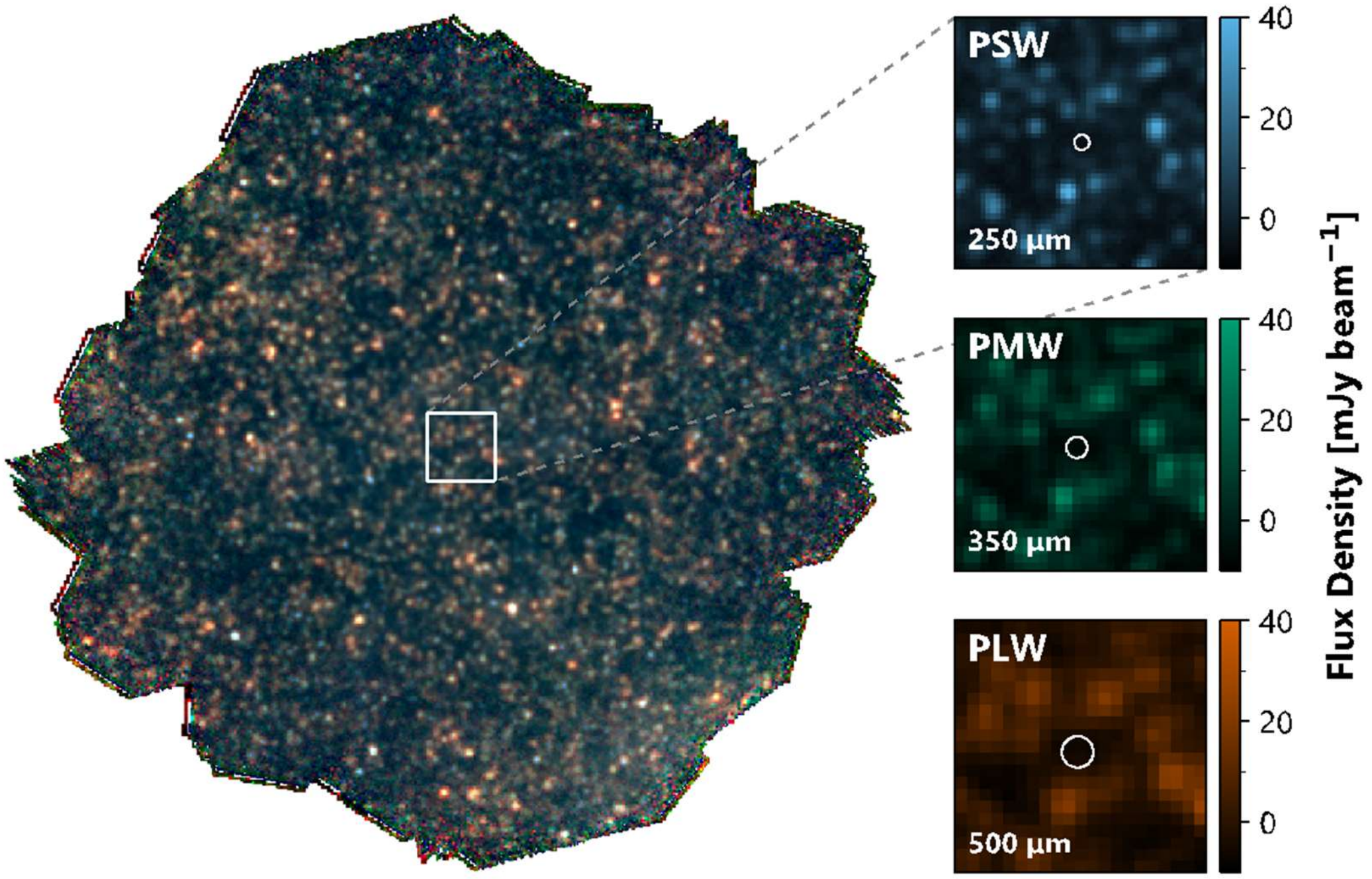

**Figure 1.** Left: 3 colour image of 250, 350, and 500 micron (blue, green, red, respectively) of the SPIRE Deep field. The image is roughly 30' across with the deepest part of the field lying in the central 12'. Right: blow-ups of part of the deepest central region.

Given the unique depth of the SDF in the SPIRE bands, the unexpected number count bump revealed by P(D) analysis, and other factors, the SDF is clearly a tempting target for further observations at a range of wavelengths. We are already in the process of obtaining deeper optical data for this field from Subaru, as well as coverage of the full 30x30 arcmin region with SCUBA2. Further observations at multiple wavelengths will be obtained before the launch of PRIMA. However, the key mid-to-far-IR band needed to study sources likely to be heavily affected by dust extinction can only be provided by PRIMA.

We here propose observations of the SDF with both halves of PRIMA – the PPI, to provide coverage of longer wavelengths and specifically to provide sub-confusion depths, comparable to the SPIRE depth, at 235 microns and in its shorter bands. The scientific goal for these PPI observations is:

- To confirm the P(D) results from SPIRE, that there is an unexpected bump at sub-mJy flux levels in the counts at 250 microns
- To explore the sub-confusion counts at shorter wavelengths in the other PPI bands. This will provide additional information on the nature of the sources responsible for the sub-mJy bump, testing, for example, whether this bump becomes more significant at shorter wavelengths, potentially setting redshift constraints on the population responsible.

- To allow cross-correlation of the SPIRE and PPI images at different frequencies, power spectrum analysis, and other statistical probes of the SPIRE SDF to further explore the nature of the sub-mJy population.
- Once theoretical models of the sub-mJy-bump population are available, forward modeling techniques will be used to test them against the existing SDF observations and the proposed observations with PRIMA PPI.

The PRIMA exposure time calculator indicates that the 0.25 sq. deg. area of the SDF can be covered to a 5 sigma depth of 400uJy, which is well matched to the 250 micron counts bump, in 330 hours.

We also propose PHI observations with PRIMAger, surveying this field to depths appropriate for detecting the rest frame mid-IR emission from DSFGs with fluxes of a few mJy at 250 microns. This will allow several investigations:

- Reveal the individual rest frame near -mid-IR SEDs of DSFGs detected in this field, allowing DSFGs a factor of 10 fainter than those detected by Herschel to be individually studied. The role of PAHs, AGNs, etc., as well as basic parameters such as star formation rate, dust temperature, dust mass, etc., will thus be reliably calculated.
- Allow guided source extraction methods such as XID+SED to be used to cross-match and extract PHI, PPI, and SPIRE sources at fluxes well below the confusion limit, providing the first insights into the properties of such objects.
- Since PHI observations are not sensitive enough to reach the fluxes needed to probe individual sources in the 400 microJy bump to be studied, we will use stacking and other statistical techniques to extract the population rest-frame near-to-mid-IR properties of the bump sources using the PHI data.
- All of these analyses will be tested against forward modeling of the sub-mJy DSFG population.

Existing and future optical/near-IR data from Spitzer, HST, Subaru, JWST, submm data from JCMT, and other observations will provide additional multiwavelength data to help understand the faint DSFG population and to test models of its content and evolution. In this way, the unique legacy of the SPIRE Dark Field will provide a Rosetta stone for understanding the sub-mJy DSFG population.

## Instruments and Modes Used

PRIMAger imaging of a 30’ x 30’ region with all of the hyperspectral and polarimeter bands. No polarimetry information needed.

## Approximate Integration Time

To reach a 235 micron depth of 400 microJy 5 sigma over this region, the exposure time calculator indicates an integration time of 330 hours is needed.

Since the PHI is offset from the PPI by 16 arcmin, separate dedicated time with PHI will be needed to cover the same field. A depth of 850 microJy at 84 microns is required to detect a 4mJy 250 micron source at z~2, assuming an ALESS-like SED. This will take 58.2 hours, so we request this time to probe the ~mJy DSFG population in this unique field.

## Special Capabilities Needed

The ability to choose the spacecraft rotation angle at the time of observation would be useful so that flanking fields can be optimally aligned.

## Synergies with Other Facilities

There are clearly strong synergies with the Herschel legacy in the SDF. This field also encompasses the Spitzer IRAC dark field, as well as MIPS, SCUBA2, AKARI, HST, and JWST observations. Though these do not all encompass the entire SDF at this time, further observations are being proposed (e.g., HST, JCMT, Subaru in 2025B), and we expect a thorough multiwavelength dataset to be available for the SDF by the time PRIMA launches.

## Description of Observations

The SDF will be observed in area mapping mode with PPI and, separately, with PHI, allowing the full region to be covered to the desired depth. Flanking fields observed in parallel will reach comparable, or, for PHI during the PPI observations, deeper depths, but the details of these will depend on the spacecraft rotation angle at the time of the observations. The ability to set requirements on rotation angle would allow these parallel observations to be optimized for scientific output.

# 19. The Sub-mJy Far-IR Population

Dave Clements (Imperial College London), Chris Pearson (RAL Space)

The deepest Herschel images at 250, 350, and 500 microns, the Herschel Dark Field, suggest the presence of a population of sub-mJy sources that are not accounted for in current far-IR/submm galaxy evolution models (Pearson et al., 2025; Varnish et al., 2025). These results are based on P(D) (probability of deflection) analysis of a region observed repeatedly by Herschel-SPIRE for calibration purposes. However, only this one small field over the entire sky has observations to this depth, so these results, which have the potential to fundamentally change our view of dusty galaxy evolution, are potentially subject to uncertainties due to cosmic variance, large-scale structure, and even galactic foreground contamination. The most significant divergence from current number count and galaxy evolution models in the Dark Field is at 250 microns, where a second peak in the Euclidean normalized counts appears at a flux of 400 microJy. The 250 micron band is well matched to PRIMAger's longest wavelength, or 235 microns, so we propose to survey multiple regions of the sky to a 5 sigma sensitivity at 235 microns of 400 microJy.

## Science Justification

### Introduction

The Cosmic Infrared Background (CIB) contains the integrated history of star formation in the universe. Comparison to the integrated UV-Opt-IR shows that roughly half of the energy generated in the universe through star formation is obscured by dust and re-radiated at far-IR wavelengths. A full understanding of the history of galaxy and star formation over cosmic time thus requires observations at these long wavelengths. Observations by Herschel and other mid-IR-to-submm facilities have made significant advances in probing this population over the last 20 years, with up to 70% or more of the CIB resolved into individual sources brighter than the Herschel confusion limit of ~10s of mJy at 250-500 microns. However, analysis of the deepest Herschel field available, SPIRE Dark Field (SDF) (Pearson et al., 2025) using P(D) analysis which analyses the pixel flux distributions to probe below the confusion limit (Varnish et al., 2025, Fig 1; Scheuer, 1957) provides evidence of a second bump in the number counts, indicating the presence of an unexpected population of far-IR sources at flux levels of a few 100 microJy. At 250 microns, the counts of this new population peak at fluxes around 400 microJy. At still fainter fluxes, the counts drop away, indicating that this sub-mJy population is the last significant source of far-IR emission contributing to the CIB. The nature of this sub-mJy population, however, remains to be determined.

While individual sources responsible for this population might be uncovered by other observatories (Bisigello et al., 2023) a full understanding of this population requires further deep observations in the far-IR. PRIMA observations using both the long-wavelength PRIMAger channels are thus required.

The PRIMAger long wavelength channels between 235 and 92 microns will be able to both confirm the presence of this bump in the 250 micron counts and provide much-needed additional constraints on the nature of the sources contributing to it. By observing four additional 0.25 sq. deg. fields, scattered around the sky in low galactic foreground fields, we will be able to test whether the SDF counts bump is a local or global phenomenon. The PRIMA exposure time calculator indicates that each 0.25 sq. deg. region will require 330 hours of observations to reach a 5 sigma sensitivity of 400 microJy, sufficient to detect the bump in the 250 micron counts seen in the SDF. These same observations will reach deeper sensitivities at shorter wavelengths: ~300, 220, and 160 microJy at 172, 126, and 92 microns, respectively. These shorter wavelength data will place significant new constraints on the far-IR counts through P(D) and other analyses, allowing us to estimate the redshifts of the major populations contributing to these counts. In this way, the origin of the SDF counts bump will finally be understood, and the populations contributing to the CIB will have been fully identified. PRIMA is the only facility planned that will be able to conduct these unprecedentedly deep far-IR observations.

## Interpretation Methods

Since these observations will reach well below the PRIMA confusion limit, methods necessary to deal with extreme confusion are needed. Our first step will be to use well-established P(D) codes, such as pofdaffine (Glenn et al., 2010), to repeat the SDF analysis in the new PRIMA survey regions. Power spectrum and cross-correlation analysis between the different PRIMA bands will then place constraints on the CIB sources responsible for the counts using a number of approaches, including forward modeling from galaxy evolution models. Individually detected sources will be analysed using SED fitting codes. We target fields are likely to be well-studied regions of the extragalactic sky, so higher resolution shorter wavelength data from e.g., Spitzer, Euclid, Roman, or PRIMA. HIS imaging will likely be available. We will thus use advanced prior-based cross-matching tools to cross-identify and determine the fluxes of as many of the confused sources as possible. Even in the absence of data from other facilities, this will be possible using the higher resolution, shorter wavelength PRIMAger bands, e.g., 92 microns. At this stage, the most advanced such tool is XID+SED (Hurley et al., 2017; Pearson et al., 2017), but we expect more advanced tools to be developed by the time of launch, with some potentially being specifically designed for PRIMAger use.

We will also use SED fitting tools to cross-match fluxes from UV to radio, and including X-ray if available, to determine the physical properties of the sources, including stellar mass, star formation rate, AGN content, photometric redshift, star formation history, and more. Our current tool of choice for this is CIGALE (Boquien et al., 2019), but we expect more advanced tools, potentially optimized for PRIMAger, to be available at the time of launch.

Follow-up spectroscopy for specific sources of interest or for statistically relevant sub-samples will be proposed subsequent to the completion of the imaging survey. These observations will validate the photometric-based methods for determining redshift and other properties, and provide additional information, such as emission line diagnostics, to further our understanding of this population.

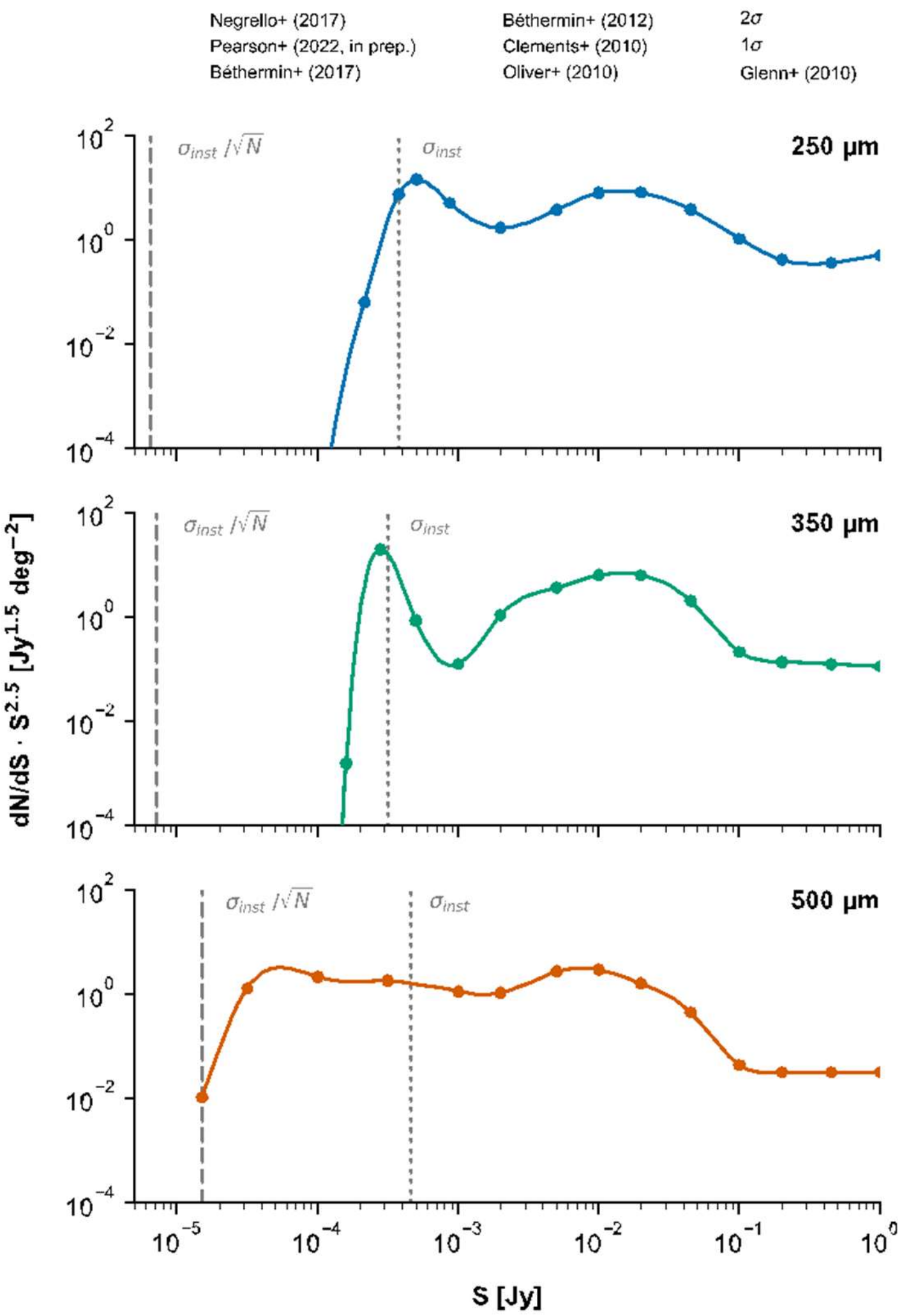

**Figure 1**. Result of P(D) analysis applied to the SPIRE Deep field observations. The analysis agrees with discrete number counts and existing models at brighter fluxes, but shows an additional population of sub-mJy sources at 250 microns and plausible evidence for similar populations at 350 and potentially at 500 microns. The dark grey band indicates the 68% confidence internal of the P(D) derived counts, the light grey region the 95% confidence interval.

## Link to Testable Hypotheses

Determine whether the SDF counts bump is a feature of the population and not a result of LSS or other foreground effects.

Determine the nature of the sources contributing to the faintest far-IR counts.

Test far-IR galaxy number count models by comparing the observations to models using forward modeling techniques.

## Instruments and Modes Used

PRIMAGer mapping in the PPI bands.

## Program Size

Large

## Approximate Integration Time

We aim to reach a 5 sigma sensitivity of 400 microJy in the 235 micron PPI band over 4 different 0.25 sq. deg. regions spread over the sky. These observations require a total of 330 hours according to the PRIMA exposure time calculator.

## Special capabilities needed

None

## Synergies with Other Facilities

Selection of fields with existing (or future) deep imaging from UV, and ideally X-ray, through optical to submm and radio will be beneficial though not required.

## Description of Observations

These observations require deep far-IR images of the fields using all bands of the PRIMAger PPI instrument. Deep observations of four 0.25 sq. deg. regions, each amounting to 330 hours total integration, are needed.

# 20. The Impact of Morphology, Environment and AGN on ISM Conditions Across the Hubble Sequence

Dr Timothy A. Davis (Cardiff University, UK)

We propose a comprehensive far-infrared (FIR) spectroscopic survey of 180 nearby galaxies using FIRESS on PRIMA to investigate the physical conditions of the interstellar medium (ISM) across a diverse range of environments, morphologies, and nuclear activity. By targeting key FIR fine-structure lines ([CII] 158 µm, [OI] 63 µm, [NII] 122/205 µm, [OIII] 88 µm), we aim to characterize the heating and cooling balance of the ISM, which is fundamental to understanding star formation and galaxy evolution. This survey addresses three critical science goals: (1) exploring how ISM conditions vary with galaxy morphology, particularly in systems below the star-forming main sequence which were understudied by Herschel; (2) assessing environmental impacts on the ISM by studying galaxies across the different environments present in the nearby Virgo and Fornax clusters; and (3) probing the role of AGN in shaping ISM properties across a wide luminosity range. FIRESS's sensitivity and resolution will allow us to map velocity-resolved ISM diagnostics at kiloparsec scales, enabling comparison with existing (e.g., ALMA, MeerKAT, MUSE) and future state-of-the-art multi-wavelength datasets. This legacy dataset will significantly advance our understanding of the ISM and star formation regulation in the local universe, while providing a crucial reference for interpreting high-redshift observations.

## Science Justification

The atomic and ionic fine-structure lines [CII] (158 µm) and [OI] (63µm) are the dominant cooling lines of the neutral interstellar gas, and therefore they are important for understanding the heating and cooling balance in this phase of the interstellar medium (ISM). The [NII] (122, 205µm) and [OIII] (88µm) lines are also prominent far-IR (FIR) tracers of ionized gas. These lines provide diagnostics for inferring physical conditions of the gas, such as temperatures, densities, and radiation field strengths, by comparing with models of photodissociation regions (e.g., Kaufman et al. 2006; Pound & Wolfire 2008). They are also used as tracers of the star-formation rate, similar in spirit to the optical nebular lines, but with lower extinction (Zhao et al. 2016).

Furthermore, because the FIR lines are convenient for high-resolution millimeter interferometry at redshifts, they are quickly becoming workhorses for the study of high-redshift galaxy evolution. In the context of understanding the high-redshift universe, it is *crucial* to understand the behavior of these lines in as many different types of local galaxies as possible. If we want to understand how the ISM and its cooling/heating balance, we need to cover multiple bases:

**The Impact of Morphology**: as galaxies transition away from the main sequence, the neutral ISM becomes compressed, and is typically found at high densities deep within the galaxy bulge. This environment is very different from classic spiral galaxy discs, with a harsh radiation environment, high cosmic ray fluxes, and extreme midplane pressures. These systems have low star-formation efficiencies (e.g., Davis et al. 14), perhaps in part due to changes in the ISM conditions. Existing observations of galaxies beyond the main sequence with Herschel in these key atomic diagnostics were limited, with small samples of systems covered (e.g. Lapham et al. 17). These found high [N II]/[C II] and low [C II]/FIR line ratios as galaxies become more early-type, but the true cause of these correlations is unknown.

**The Impact of Environment**: Galaxies that live in galaxy clusters experience a range of phenomena that their cousins in the field do not. Ram pressure and stripping, strangulation, and hard radiation environments all clearly affect the ISM. Galaxies within a cluster typically have very high molecular to atomic gas ratios as their gas is compressed and the atomic medium removed. Large ALMA and MeerKAT/VLA surveys have mapped the impact of clusters on the molecular and atomic medium, but the crucial information on the energy balance within the ISM that FIR lines give is missing. Herschel did not significantly contribute to this question, as the majority of its cluster surveys were photometric, leaving a huge discovery space where PRIMA can capitalize.

**The Impact of AGN:** Active galactic nuclei are thought to be hugely important in regulating the evolution of galaxies. Bright AGN are thought to do a large amount of damage to the ISM, expelling it from galaxies. Lower luminosity AGN, however, are also likely to be important. Heating the ISM locally, driving turbulence, and increasing cosmic ray fluxes. Tracing how the conditions within the ISM change as a function of AGN luminosity requires large samples of AGN systems in the local universe to be observed, not just those with high luminosity.

In all these cases, studying nearby galaxies gives us the opportunity to resolve the system along the FIRESS slit, both spatially and spectrally, using the high-resolution mode. This will allow spectrophotometric study of different regions within each system, allowing the physics at work to be probed in detail.

Together, the science cases above argue the need for a large FIR spectroscopic survey of nearby galaxies, in all environments, concentrating not just on massive spirals but spanning the full range of galaxy morphology.

## Instruments and Modes Used

In this program, the FIRESS spectrometer will be used in the high-resolution mode. The full spectral range should be covered for 180 targets.

## Approximate Integration Time

To reach a 5 sigma line flux of 5e-18 W m-2 (an order of magnitude deeper than archival Herschel PACS observations of similar sources) against a continuum flux of 1 Jy for the [CII] line requires only 0.1 hours per source. Observing the full four spectral modules with dithering to place them along the galaxy's major axis, we require two tunings per source. Thus, the total FIRESS time required on-source (without overheads) is 36 hours.

## Special Capabilities Needed

Position angle—the slits should be aligned along the galaxy's major axes.

## Synergies with Other Facilities

This case would be hugely synergistic with other large facilities. It would add value to existing ALMA and MeerKAT programs, and provide prime targets and input for next-generation facilities like IFU instruments on the ELTs, the SKA, with HWO and UVEX.

## Description of Observations

FIRESS's high-resolution spectroscopy allows the key FIR cooling and ionized gas lines to be observed and velocity resolved, allowing mapping of the ISM conditions along the FIRESS slits, which should be positioned along the major axis of these systems. The line fluxes can be compared with PDR, XDR, and CRDR models, and with tracers at other wavelengths (e.g., from ALMA, MeerKAT/SKA/VLA/nGVLA, JWST, HST, and ground-based optical IFUs). PRIMA is sensitive enough to probe 10x deeper than Herschel PACS typical spectroscopic sensitivities in only ~15 minutes per source, allowing large samples to be covered.

Here we suggest a three-tiered approach to cover the key science goals above:

**A cluster sample,** consisting of ~100 galaxies of all spectral types detected in cold-ISM tracers across the Virgo and Fornax clusters. These clusters are optimal as they are nearby, providing kiloparsec resolution along the FIRESS slit, and have a huge suite of ancillary information (e.g., large programs on ALMA, and MeerKAT revealing the molecular gas, MUSE large programs revealing the stellar and ionized properties, etc).

**A morphology-matched sample.** In addition to the cluster samples, we need to ensure FIR spectroscopic information is not limited to massive field spirals. These will be used in concert with the cluster galaxies to disentangle the effects of morphology from the effects of the cluster environment. Combining data from other proposed PRIMA programs will provide the massive spirals, but a comparison sample of fifty galaxies below the main sequence will be selected from existing surveys with large amounts of ancillary data, including ATLAS3D, PHANGS-extension, and WISDOM to fully sample the mass-size plane.

Together, these will cover a large range in AGN properties at low luminosity, but to ensure the local high-luminosity AGN population (which is most comparable to the high-z population covered in PI science cases) we will include observations of an additional 30 AGN selected based on their ancillary data and luminosity (e.g., from the MUSE and ALMA large programs such as CARS etc).

In total, this results in a putative sample of 180 galaxies observed that will allow trends with all key galaxy properties to be revealed and explored, making a huge leap forward over the information present from Herschel.

We would observe the full four spectral modules with dithering to place each of them along the galaxy's major axis.

# 21. Ghosts of Disruptions Past: Revealing the Growth of Supermassive Black Holes via Stellar Capture with PRIMA

Kishalay De (Columbia University/Flatiron Institute), Megan Masterson (MIT), Jacob Jencson (Caltech/IPAC), Christos Panagiotou (MIT), Ryan Lau (NSF/NOIRLab)

The formation and growth of supermassive black holes (SMBHs) form one of the cornerstones of modern astrophysics. While gas accretion has been unequivocally shown to be the dominant mechanism in growing the most massive SMBHs, the growth mechanism for the lightest SMBHs has long been suggested to be significantly influenced by stellar capture and tidal disruption. With the advent of panchromatic time domain surveys, we can now routinely find the resulting electromagnetic flares as Tidal Disruption Events (TDEs); however, directly measuring the bolometric energy output in TDEs as a tracer of the total accreted mass remains difficult owing to the natural far-ultraviolet-dominated emission of SMBH accretion disks.

The most viable method relies on the ubiquitous presence of circumnuclear dust in galactic nuclei, which have been shown to produce luminous dust echoes in the infrared bands. We propose a far-IR spectroscopy campaign using PRIMA/FIRESS in the low-resolution (R∼10 binned) mode of the luminous infrared echoes of nearby TDEs (∼ 150 events at < 400 Mpc) at phases of 10–20 years after eruption. By observing at late phases when the echo spectrum becomes dominated by far-infrared emission, the proposed observations would uniquely reveal i) the properties of the dust surrounding nearby SMBHs at spatial scales of tens of parsecs and ii) the total energy radiated in these events, which may be dominated by accretion at late phases.

## Science Justification

### Introduction

Ubiquitously residing at the center of nearly every galaxy, supermassive black holes (SMBHs) regulate the evolution of their host galaxies, inject energy into the intergalactic medium, and play a pivotal role in shaping entire Galactic ecosystems over cosmic time. While most of the cosmic SMBH mass density is primarily attributed to growth via gas accretion, the formation of both the highest and lowest ends of the SMBH mass function remains elusive (Volonteri et al. 2021). Unlike their massive counterparts, the formation of the likely most common lowest mass SMBHs ($< 10^6$ $M_\odot$; including one in the Milky Way) has long been argued to be influenced by mechanisms involving the tidal disruption of nearby stars onto (initially) lower mass seed black holes (Freitag & Benz 2002, Stone et al. 2017).

The disruption of a star by an SMBH will lead to a temporary burst of accretion and an associated flare of emission across the electromagnetic spectrum, commonly known as a tidal disruption event (TDE). Although the first TDEs were discovered in the soft X-rays, the discovery of several dozen TDEs in ongoing optical time-domain surveys (which will grow to thousands in the Rubin era) is now enabling novel TDE population studies (see Gezari et al. 2021 for a recent review). Combining the (largely uncertain) TDE rate of approximately once every $10^5$ to $10^3$ years (Sazonov et a. 2021, Yao et al. 2023) with theoretical results that find a few 10% of the stellar mass (~1 $M_\odot$) being accreted onto the BH (Jiang et al. 2014), TDEs potentially contribute substantially to low mass SMBH growth over a Hubble time. Despite existing theoretical work, direct observational measurements constraining the role of TDEs in SMBH growth remain inaccessible due to limitations posed by the fact that most of their energy is radiated in the opaque EUV band.

### Luminous dust echoes identified from infrared surveys.

Early work using the NEOWISE survey showed that dust clouds surrounding the SMBH can produce luminous infrared (IR) echoes (van Velzen et al. 2016). These IR flares not only provide a powerful probe of the flare energetics, as the reprocessed dust emission acts as a bolometer for the radiated energy from the soft X-ray to optical bands, but also serve as a way to discover TDEs that would otherwise be hidden by dust obscuration. Although historically limited, novel data processing techniques on archival data from the NEOWISE survey (Mainzer et al. 2014) has enabled the identification of dozens of nearby TDEs with luminous IR echoes, including events which have no counterparts at other wavebands (Panagiotou et al. 2023, Masterson et al. 2024).

The identification of these luminous IR echoes in the local universe (< 200 Mpc) has i) shown that IR TDEs occur at a rate comparable to those discovered in optical and X-ray bands, despite being overlooked for decades, and ii) recently enabled the first mid-infrared spectroscopic measurements of the dust emission using JWST/MIRI. The MIRI data both confirm the presence of coronal emission lines powered by SMBH accretion in these events and provide strong constraints on the dust-emitting region that powers the observed IR flare via the continuum emission. Although unparalleled in its ability to identify the nature of the flares, the JWST data reveal tantalizing clues about the long-term evolution of accretion in TDE flares that can only be addressed with follow-up in the far-IR band. With targeted follow-up of a volume-limited sample of TDE dust echoes decades after eruption, PRIMA/FIRESS low-resolution spectroscopy will be able to address the two largest unknowns: i) What is the contribution of very late-time accretion in mass accumulation during TDEs? and ii) what is the bolometric IR energy output in TDEs?

## Science Questions

### What is the bolometric IR energy output in TDEs?

With dust being completely opaque to intrinsic TDE radiation, the total bolometric energy output in the IR emission from the dust provides a powerful route to measuring the total energy output of the central engine (tracing the total accreted mass). However, as speculated in the early theoretical models for dust echoes of TDEs (e.g., Lu et al. 2016), the complete IR light curve is dominated by light travel-time effects as the TDE light front is gradually absorbed by the surrounding dust and re-radiated in the infrared as parabolic light fronts for a distant observer. The resulting dust spectrum therefore evolves from being dominated by hot dust emission (peaking in the WISE and JWST bands) to cool dust emission (in the far-IR/PRIMA band) at late times as the light front is dominated by dust at large radii. This has been confirmed with the first

JWST observations of TDE dust echoes, demonstrating striking consistency with simple dust echo models incorporating light travel-time effects (Masterson et al. 2025).

As the echo spectrum becomes dominated by the cooler dust at larger radii, its spectrum becomes dominated by far-IR emission, peaking at ~ 30–50 microns 1–2 decades after the disruption. Importantly, the dust emission becomes nearly unobservable at shorter mid-IR wavelengths (< 30 micron) after the intrinsic emission shuts off due to the steep cut-off below the characteristic peak wavelength of the dust spectrum (Figure 1). Depending on the dust distribution and light curve evolution, the evolution of this late-time far-IR component is critical to computing the total energy output in the dust echo. Given its smooth spectrum, PRIMA/FIRESS low-resolution observations would be perfectly suited to measuring the late-time dust echo in the far-IR bands for echoes that have been identified over the duration of NEOWISE (~ 10–20 years prior to PRIMA).

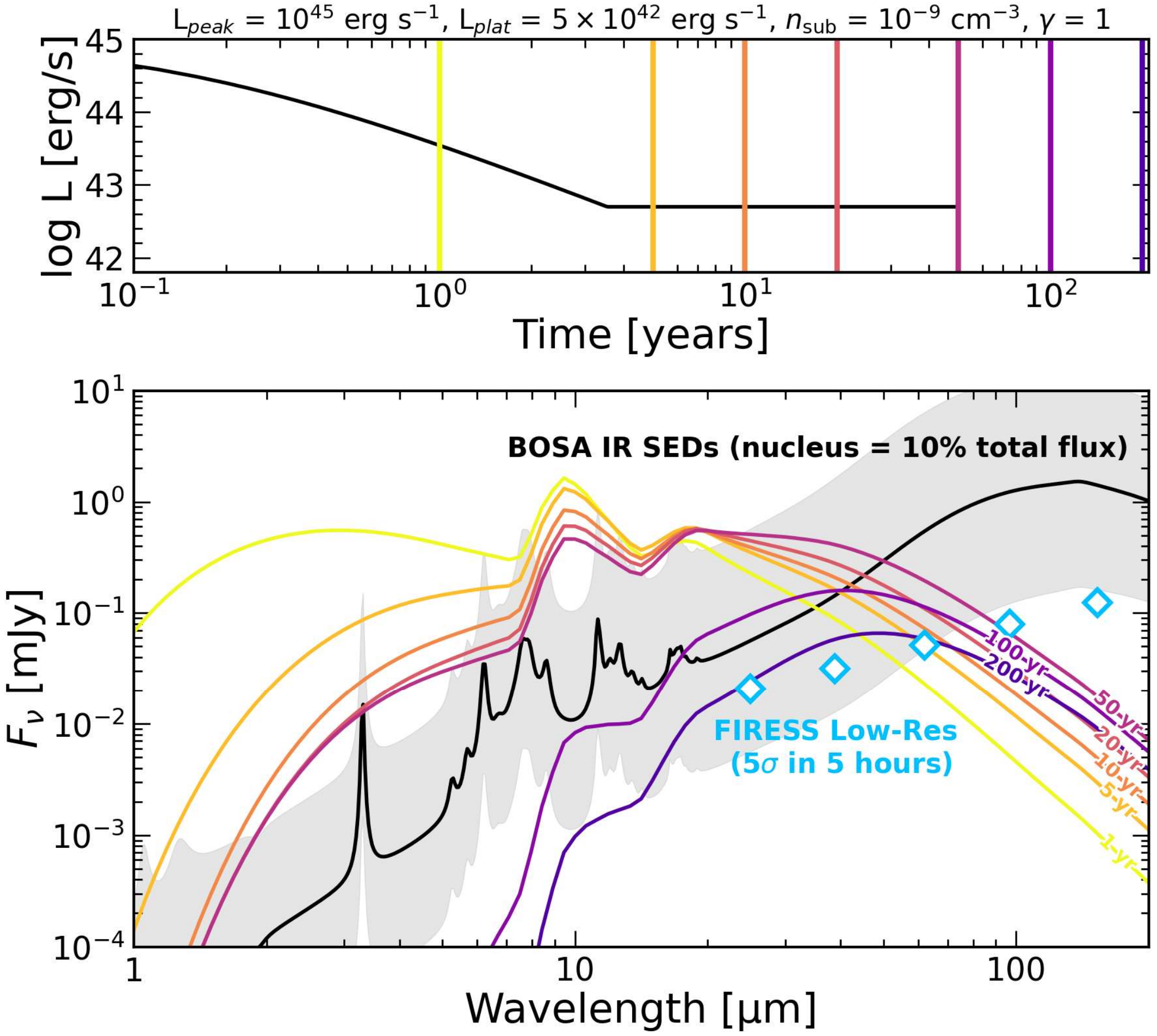

**Figure 1.** Top: Simulated bolometric light curve of a TDE following an initial power-law decay, followed by a late-time plateau lasting 50 years post-disruption. Bottom: Simulated spectra of the resulting dust echo using parameters inferred from recent JWST data (Masterson et al. 2025) for a dust echo at a distance of 400 Mpc. The colored lines show the temporal evolution of the echo spectrum from 1–200 years after the eruption, as marked in the top panel. The gray and black lines show a range and mean of spectra for the nucleus of the host galaxy (over a range of star formation rates/total IR luminosity) -- assumed to be 10% of the total galaxy luminosity. The blue diamonds show the estimated FIRESS sensitivity with R~10 spectroscopy for a 5-hour exposure.

### What is the contribution of very late-time accretion in mass accumulation during TDEs?

Although optical TDEs are identified and followed up during the peak of their optical emission, long-term follow-up of the nearest and brightest TDEs reveals that late-time plateaus in the optical and UV bands are common (Mummery et al. 2024). These plateaus are thought to result from the viscous spreading of the newly formed accretion flow and indicate ongoing mass accretion at very late phases. While directly detecting this UV emission is challenging due to the high opacity of the intergalactic medium, evidence for the same late-time plateaus has now emerged from JWST follow-up of the luminous dust echoes (Masterson et al. 2025). The mid-IR spectra show clear evidence for an excess from hotter dust emission (producing spectral excess at < 10 micron initially), which continues to be a dominant contribution more than a decade after the event. As the overall light front for the echo evolves to longer wavelengths at late phases, this late-time plateau emission can continue to be a large power source for the far-IR component and produce a long-lived far-IR excess that lasts for many decades after the eruption, even as the shorter wavelength mid-IR emission completely fades away (Figure 1). PRIMA/FIRESS low-resolution follow-up of decades-old TDE dust echoes will be uniquely suited to quantifying the contribution of the late-time plateaus that may even dominate the overall mass accretion budget in TDEs (Mummery & van Velzen 2024).

## Instruments and Modes Used

FIRESS pointed observations, binned to R=10.

## Approximate Integration Time

As shown in Figure 1, 5h exposure times with FIRESS will be sensitive enough to detect luminous dust echoes of TDEs out to 400 Mpc. Taking the rate of infrared TDE echoes from NEOWISE data (Masterson et al. 2024), we estimate about 150 TDEs to be identified within a volume of 400 Mpc over the duration of NEOWISE. This is consistent with ongoing follow-up efforts for luminous nuclear transients in archival NEOWISE data. As the majority of these events will be at the further end of the volume, we estimate a total integration time of about 600 h.

## Special Capabilities Needed

None

## Synergies with Other Facilities

This effort has very strong synergies with past, ongoing, and upcoming infrared surveys. The nominal parent sample is identified from NEOWISE data, which concluded operations in 2024. We expect more such events to be identified from ongoing IR surveys like WINTER (Lourie et al. 2020) and PRIME (Yama et al. 2023). The upcoming NEO-Surveyor mission will provide a recent stream of TDE dust echoes that will enable the earliest time far-IR follow-up with PRIMA.

## Description of Observations

We propose to observe the locations of a sample of IR-selected TDEs identified from prior surveys. The data will be acquired with a single pointing FIRESS low-resolution spectrum (R∼10 bin). The targets will be at the center of nearby Galactic nuclei, where there will be contamination from the host galaxy light. Therefore, we estimate the host galaxy contribution with SED models from Boquien & Salim (2021) and show them in Figure 1. As shown, although the host galaxy may dominate the spectrum in heavily star-forming galaxies, the time variable dust echo spectrum can either be directly detected as a distinct excess over the galaxy SED, or via repeated observations to detect temporal evolution if needed. Observations in this mode provide the most sensitive configuration for obtaining the far-IR (∼25–200 μm) SED from cool dust emission. The resulting spectra can be modeled using techniques incorporating constant time delay dust emission surfaces, as done in Masterson et al. 2025.

# 22. The AGN Nature of Little Red Dots via PRIMA Stacking

Ivan Delvecchio (INAF-OAS, Italy), Benjamin Magnelli (CEA-Saclay, France), Emanuele Daddi (CEA-Saclay, France), Sirio Belli (Univ. Bologna, Italy), Alberto Traina (INAF-OAS, Italy), Carlotta Gruppioni (INAF-OAS, Italy), Kohei Ichikawa (Univ. Waseda, Japan), Kohei Inayoshi (KIAA, PKU, China), Masafusa Onoue (Univ. Waseda, Japan), Dave Clements (Imperial College London, UK), Jed McKinney (Univ. Texas Austin, US), David Setton (Univ. Princeton, US), Fabrizio Gentile (CEA-Saclay, France), Caitlin M. Casey (UC Santa Barbara, US), Hollis B. Akins (Univ. Texas Austin, US), Maximilien Franco (CEA-Saclay, France), Takumi Tanaka (Univ. Tokyo / Kavli IPMU, Japan).

Our science case aims at elucidating the highly debated nature of the enigmatic population of Little Red Dots (LRDs) found at z>4, owing to the unique rest-frame mid-IR coverage of PRIMA. Specifically, our main goal is constraining the median rest-frame 5-10µm spectral energy distribution (SED) of LRDs by exploiting PRIMAger image stacking. After performing median stacking of all the photometrically-selected LRDs, as obtained from available JWST/NIRCam, JWST/MIRI, and ALMA data, we find hints of a rising mid-IR slope up to rest-frame ~3µm (MIRI-21µm at z~6), possibly suggesting AGN-heated dust emission from a torus. We argue that PRIMAger imaging at longer wavelengths (i.e., 5-10µm rest-frame at z~6) is key for confirming this hypothesis and distinguishing between starburst and AGN (torus) emission. To this end, we propose an ultra-deep (confusion-limited) PRIMAger survey with PHI1 (and PHI2) over a COSMOS-Web-like area (0.5 deg$^2$), from which LRDs would be observed as "bonus" targets.

## Science Justification

### Scientific Context

The James Webb Space Telescope (JWST) has recently revealed a puzzling population of compact and extremely red galaxies at z>4, dubbed "Little Red Dots" (LRDs). These are characterized by broad emission lines, compact morphology, V-shaped optical spectra, and extremely red colors (Kocevski et al. 2023; Harikane et al. 2023; Labbè et al. 2023; Matthee et al. 2024; Maiolino et al. 2024). Despite the growing literature on these elusive objects, whether their emission is powered by Active Galactic Nuclei (AGN) or extreme star-formation (SF) activity remains strongly debated. If LRDs are AGN, several studies suggest that the number densities of these AGN candidates are more than an order of magnitude higher than expected from extrapolating quasar luminosity functions (e.g., Kokorev et al. 2024; Kocevski et al. 2024, Pizzati et al. 2025), and their black holes are overmassive relative to local BH-galaxy scaling relations (e.g., Kokorev et al. 2023; Greene et al. 2024). Recently, Naidu et al. (2025) interpreted the strong Balmer absorption features seen in LRDs with a model labeled "black hole star" (BH*), which is, an early black hole embedded in a dense gas forming a dust-free "atmosphere". Therefore, these sources might be tracing an early phase of black hole growth. Alternatively, it has been argued that LRDs could be compact massive galaxies with extreme starburst activity (e.g., Baggen et al. 2024; Pérez-González et al. 2024). Nonetheless, none of these LRDs have been detected by ALMA, neither individually nor in stacking, providing loose limits on the contribution of obscured star formation and cold dust emission (Akins et al. 2024; Casey et al. 2024; 2025).

## Are LRDs dust-obscured AGN?

**Our primary goal is constraining the median rest-frame 5-10µm SED of LRDs.**

We selected the latest compilation of about 500 (photometrically-selected) LRDs from Gloudemans et al. (2025) over the COSMOS and GOODS-South fields. First, we performed UV-stacking of archival ALMA-Band6 (1.1mm) from the combined A3COSMOS+A3GOODSS surveys (Liu et al. 2019; Adscheid et al. 2024; Magnelli et al. 2024) at the positions of LRDs. This yielded a non-detection, though 5x deeper than previously found (Akins et al. 2024), similar to that presented in Casey et al. (2025). Then we also stacked all available NIRCam and MIRI data in the image plane over the COSMOS-Web (Casey et al. 2023), PRIMER-COSMOS, and SMILES (GOODS-S) surveys (Alberts et al. 2024; Rieke et al. 2024).

The top panel of **Fig.1** (top) displays the median stacked cutouts, while the bottom panel shows the corresponding stacked fluxes (or 5σ upper limits) overlaid to the best-fit SED from Casey et al. (2024) at the average redshift of the sample ($<z> = 6.2$). We further note that *median* stacking should be less sensitive to blending issues from nearby sources, contrary to *mean* stacking. Despite the sample selection being heterogeneous, we find hints of a rising mid-IR slope up to rest-frame ~3µm (MIRI-21µm at z~6)—albeit with only upper limits longwards—possibly suggesting AGN-heated dust emission. The underlying redshift distribution is broad (nearly all sources are in the range $4<z<7.5$). We note that about 30% of LRDs are spectroscopically confirmed: our analysis adopted the best available redshift. Multi-band fitting of the median SED, comparing the runs with vs without AGN templates, will elucidate whether an AGN torus is significantly required at a population level. We stress that detecting hot dust continuum from AGN would strongly challenge the BH* model, which predicts a dust-free atmosphere of dense gas around the BH. Hence, testing this scenario is crucial for advancing our knowledge of this elusive population.

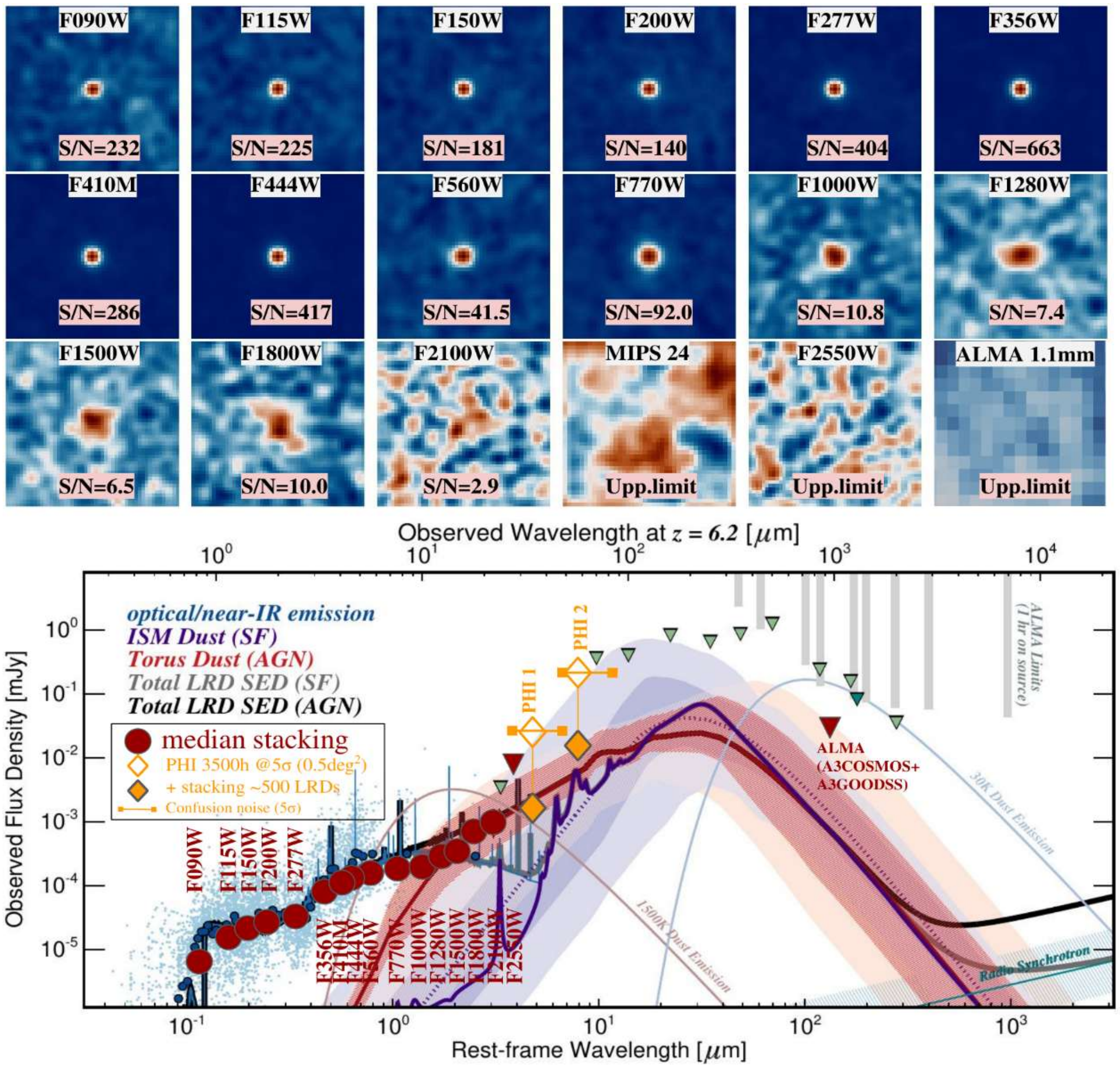

**Fig.1.** *(Top)* Median stacked cutouts at each band, reporting the corresponding signal-to-noise ratio, or upper limit in case of non-detection. *(Bottom)* Adapted from Casey et al. (2024). Median stacked spectral energy distribution of LRDs in all available NIRCam and MIRI bands over multiple JWST Legacy Fields (COSMOS-Web, PRIMER-COSMOS, and SMILES). Red circles and downward arrows indicate median-stacked detections and (5σ) upper limits, respectively. Best-fit templates are overlaid from Casey et al. (2024, not a fit to our current stacking). Open diamonds mark PRIMAger PHI1 (30.9μm) and PHI2 (52.3μm) 5σ detection limit, as reached with 3.500 hours net exposure over 0.5 deg$^2$ (i.e., confusion-limited). Assuming to stack ~500 LRDs in the same area, filled diamonds mark the expected 5σ stacked flux limits with PHI1 and PHI2.

## The need for PRIMA

PRIMAger imaging at longer wavelengths (5-10μm rest-frame at z~6) is uniquely sensitive to host dust continuum, and thus key for confirming our hypothesis and distinguishing between starburst and AGN (torus) dust emission. This science case is meant to be a "bonus" from an ultra-deep (confusion-limited) imaging survey covering a 0.5 deg$^2$ area.

## Instruments and Modes Used

This observing program requires the hyperspectral band of PRIMAger (PHI1 and PHI2) over a total area of 0.5 $deg^2$, consisting of contiguous pointings.

## Approximate Integration Time

We base our required integration time on the expected emission level of the AGN torus from Casey et al. (2024), which our NIRCam and MIRI stacked fluxes nicely match up to MIRI-21µm. The expected AGN flux at 30µm is about 2µJy: assuming ~500 LRDs in the same area, hence a S/N boost of x20, the required depth of each pointing must be around 40µJy. Because the 5σ confusion limit at 30.9 µm is 37 µJy (Béthermin et al. 2024), we foresee a confusion-limited survey in PHI1. A detection from PRIMAger stacking will strongly favor the AGN torus model. Similarly, a non-detection will add a constraining upper limit to AGN-heated emission. If we integrate the signal over the full PRIMAger Hyperspectral PHI1 filter (i.e., 6 sub-bands), we can divide the exposure time obtained from the PRIMA ETC (https://prima.ipac.caltech.edu/page/etc-calc) by sqrt(6). Considering a total of 0.5 $deg^2$ area, the global integration time is about 3500 hours. Simultaneous PHI2 imaging will come for free (see "Description of Observations").

## Special Capabilities Needed

None.

## Synergies with Other Facilities

PRIMA will bridge the wavelength gap between JWST and ALMA, which is paramount for elucidating the AGN nature of LRDs and testing recent BH models (e.g., Naidu et al. 2025). JWST already provides large samples of LRDs with accurate photometry and positional priors. The fraction of spectroscopic redshifts (now close to ~30%) will soon dramatically increase in the COSMOS-Web area, thanks to the COSMOS-3D treasury program (Kakiichi et al. 2024), which will provide NIRCam slitless spectroscopy over most of the COSMOS-Web area. Our proposed PRIMAger observations will also capitalize on the ALMA constraints obtained from archival A3COSMOS+A3GOODSS continuum imaging. Particularly, SED-fitting, including PRIMAger stacking, will strongly benefit from ALMA data, which currently places a constraining upper limit to the cold (T<100 K) dust content originating from galaxy star formation. Therefore, PRIMA and ALMA will probe different complementary ranges of dust temperature and emission mechanisms. In summary, PRIMA will be highly synergic with both JWST and ALMA, providing a unique far-infrared view of the dusty Universe that will remain unsurpassed even in the next decades.

## Description of Observations

The observational strategy foresees an imaging survey of a total area of 0.5 $deg^2$, consisting of contiguous pointings with the PRIMAger Hyperspectral band PHI1 (and simultaneously in PHI2), matching (for reference) the COSMOS-Web NIRCam coverage. These observations will be confusion-limited in PHI1 (30.9 µm), pushing PRIMAger to its deepest level. The 5σ confusion limit is 37 µJy (Béthermin et al. 2024); hence, the total observing time is about 3500 hours.

Simultaneous imaging will be obtained in PHI2: for comparison, at 52.3 μm, the 5σ confusion limit (0.25 mJy) will be reached in about 12 hours (but possibly overcome as described in Donnellan et al. (2024), based on PHI1 and NIRCam/MIRI priors).

# 23. Far-IR Emission Exploration in Local Analogues of Reionization-Era Galaxies

Yoshinobu Fudamoto (Chiba University)

Over the past decade, sensitive mm/submm interferometers have provided an unprecedented view of the interstellar medium (ISM) of high-redshift galaxies (e.g., at z>6). These observations have revealed the presence of a massive cold gas reservoir, total star-formation activities, and ionized/neutral gas properties using far-infrared (FIR) emission lines, such as [CII]158um and [OIII]88um. These studies started to show the diversity of ISM properties of galaxies at high redshift. However, observational constraints on low mass, numerous populations of galaxies are mostly untouched, due to their faintness and the required prohibitive amount of observation time. Thus, our understanding of conditions of star formation and ISM structure in early galaxies remains incomplete. To bridge this gap, observations of local analogs that replicate the physical conditions of high-redshift galaxies offer a powerful and efficient alternative. Leveraging the far-infrared spectroscopic capabilities of PRIMA/FIRESS, we propose to target a carefully selected sample of local galaxies to measure key diagnostic emission lines such as [OI] 63/146 µm, [C II] 158 µm, [O III] 52/88 µm, and [N II] 122/205 µm. These observations will enable multi-phase ISM characterization, offering insights into the ionized, atomic, and molecular components of the ISM. PRIMA/FIRESS will allow us to efficiently build a statistically robust sample and establish relations between emission line properties with ISM conditions, metallicity, and star formation properties in galaxies analogous to those at z > 6. This will provide critical constraints for interpreting high-redshift observations and for calibrating models of early galaxy evolution.

## Science Justification

The interstellar medium (ISM) is an essential component of galaxies, serving as the source for star formation, the site of metal enrichment from stellar feedback, and the birthplace of subsequent generations of stars. Thus, characterizing ISM properties, such as metallicity, gas density, ionized and neutral gas masses, is essential to understanding how galaxies form and evolve over cosmic time. The importance of understanding ISM properties becomes even more critical at high redshift (z > 6), during the epoch of reionization, when ionizing radiation from young, metal-poor galaxies played a central role in reionizing the Universe (e.g., Atek+24). In recent years, sensitive mm/submm interferometers such as ALMA and NOEMA have opened a new window into the ISM of such early galaxies by detecting key far-infrared (FIR) emission lines, including [C II] 158 µm and [O III] 88 µm (e.g., Inoue+16, Le Fèvre+20, Bouwens+21). These observations have yielded critical insights, including scaling relations between FIR line luminosities and galaxy properties, and emission-line diagnostics that enable us to probe ISM conditions (e.g., Schaerer+20, Witstok+22). However, studies using existing interferometers have been limited to the most luminous, massive galaxies, such as galaxies with high star formation rates (SFR > 30 $M_{\odot}$/yr, Faisst+20) that are bright enough to be observed in reasonable integration times. In contrast, the lower-mass, more numerous populations of star-forming galaxies, likely the primary drivers of cosmic reionization, remain largely untouched by the current observations. As these galaxies typically have modest SFRs ($\lesssim$10 $M_{\odot}$/yr), building statistically meaningful samples (e.g., N > 50) of such faint systems remains a major challenge with current ground-based facilities. As a result, key questions about the ISM conditions in typical reionization-era galaxies, such as neutral gas masses, neutral gas densities, and gas covering fractions, remain unanswered.

To overcome these limitations, observations of carefully selected local analogs that replicate the physical conditions of early galaxies offer a powerful and efficient alternative. Compact, highly star-forming dwarf galaxies, such as Green Pea and Blueberry galaxies (e.g., Cardamone et al. 2009; Izotov et al. 2011; Yang et al. 2017), have the relevant properties of high-redshift galaxies, including low metallicities, high ionization parameters, and high specific star formation rates. These systems provide ideal laboratories for probing ISM physics in environments analogous to those of the early universe. Leveraging the unprecedented far-infrared spectroscopic capabilities of PRIMA/FIRESS, we aim to address several key science questions using local analogues of reionization-era galaxies.

## Science Question

**(1) Scaling relation with FIR emission lines with galaxy properties:**

Interferometer observations of high redshift galaxies have revealed that fine structure tightly correlates with galaxy properties, such as SFR and neutral gas masses (Schaerer+20, Zanella+18). In particular, [CII] 158um emission lines have been employed as a tracer of SFR and the mass of neutral gas, while [OIII]88um has been shown to correlate with SFR (de Looze+14). However, these observations are predominantly derived from observations of massive metal-rich galaxies, which introduce significant biases when extrapolated to metal-poor galaxies. To investigate their properties and establish a scaling relation, we propose to investigate the FIR line properties of local analogs of reionization-era galaxies. This will provide a foundation for interpreting FIR observations of high-redshift galaxies, where [C II] is often the only accessible tracer of neutral gas, and for constructing physically motivated models of galaxy evolution in the early universe (Left panel of Figure 1).

**(2) FIR emission line diagnostics to explore ISM properties of galaxies:**

Combinations of emission line ratios provide key diagnostics of several important ISM parameters, such as gas density, local UV strength, and electron density (Fudamoto+25, Harikane+25). Especially, fine structure lines in FIR are free from dust attenuation. Their low critical densities allow them to trace diffuse, low-temperature phases of the ISM, complementing UV/optical lines that probe only dense ionized gas. PRIMA/FIRESS observations of Green Pea galaxies will enable multi-phase ISM characterization and refine diagnostics applicable to high-redshift systems (Right panel of Figure 1).

## Need for PRIMA

PRIMA is the only instrument that can perform observations of these fine structure lines of local analogues of reionization-era galaxies. PRIMA/FIRESS's extremely high sensitivity enables us to detect FIR emission lines of the local, star-forming dwarf galaxies in a very feasible amount of time (e.g., ~0.3-1hr in the low-resolution mode for [CII] 158µm lines of SFR~10Msun/yr galaxies in 5σ). These observations are only possible with PRIMA's capability. Thus, we need PRIMA to achieve these goals.

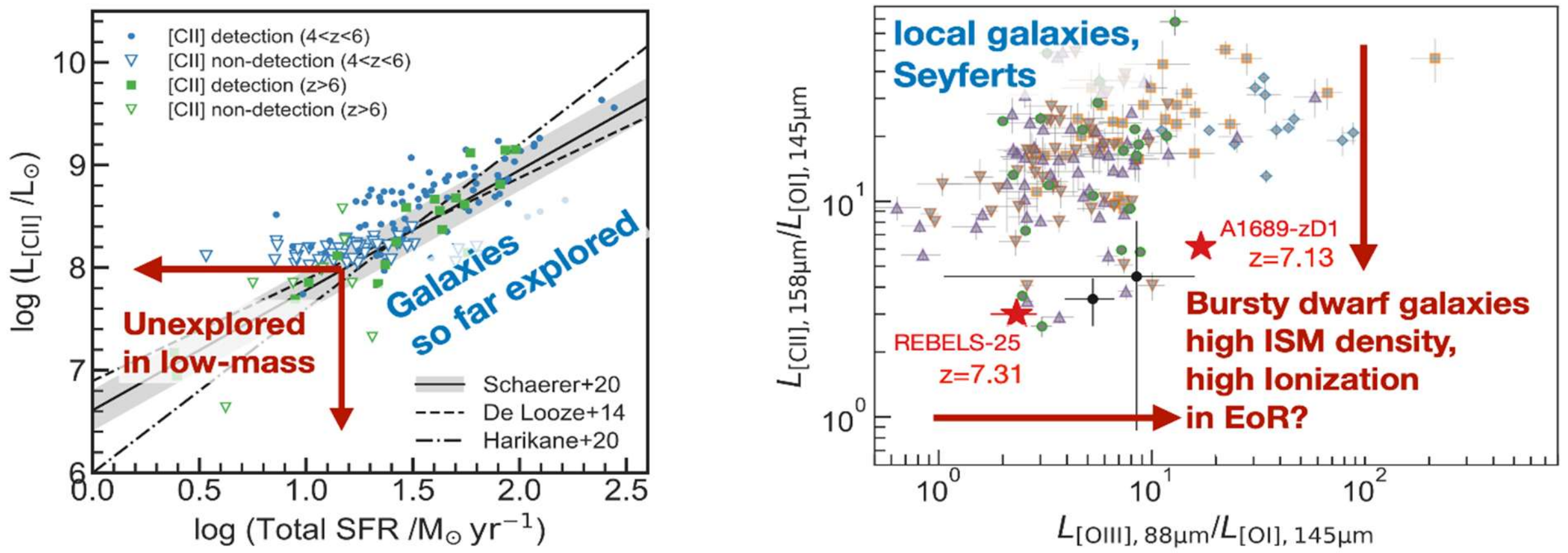

**Figure 1.** (Left Panel) Correlation between SFR and [CII]158µm line luminosity for high redshift galaxies. (Right Panel) Emission line luminosity ratio between [OI]146µm, [CII]158µm, [OIII]88µm lines. PRIMA observation of local analogues of the reionization-era galaxies enables us to understand ISM properties of these important populations of galaxies.

## Instruments and Modes Used

This science case requires the observation of 100 sources with FIRESS in the R~1 low-resolution mode.

## Approximate Integration Time

The primary target of this proposed observation is local, star-forming, luminous compact galaxies such as green peas and blueberries at z<0.1, where all key emission lines, up to [NII]205µm, are accessible for PRIMA/FIRESS. We base our request integration time scaled based on the luminosity of targets. For these sources, we first estimate [CII]158µm emission line luminosity using the already known correlation between SFR-$L_{[CII]}$. Based on these estimates, a galaxy at z=0.1 with SFR =3M⊙/yr has an estimated [CII] flux of $2.2\times10^{-18}$W/m$^2$. Thus, 5σ detection of the [CII] line can be achieved in <0.1hrs. Other luminous lines such as [OIII]52µm/88µm, [OI]63µm, [NII]122µm can be observed in a similar time scale. Other fainter emission lines, such as [OI]146µm and [NII]205µm, can be achieved in 0.75hours for each, assuming [CII]158µm/[OI]146µm=10 and [CII]158µm/[NII]205µm=10. Thus, 2 hours per source for the total of 7 emission lines. Therefore, the approximate integration time in total would be 200 hours for 100 sources.

## Synergies with Other Facilities

Deep radio observations can trace the hydrogen line at 21 cm (Kanekar+21). These observations. Additionally, mm/submm observations to search for molecular gas (Cormier+14) complement multi-phase observations of local analogue galaxies.

## Description of Observations

We propose to observe a sample of compact, low-metallicity, star-forming galaxies at z<0.1, such as Green Pea and Blueberry galaxies, using FIRESS onboard PRIMA, targeting key fine-structure emission lines including [C II] 158 μm, [O III] 52/88 μm, [N II] 122/205 μm, and [O I] 63/146 μm. These lines have been shown to provide sensitive diagnostics of ISM properties such as gas density, ionization state, and radiation field strength. The target sample consists of ~50 such galaxies selected from the previous UV/optical studies (Cardamone et al. 2009; Izotov et al. 2011; Yang et al. 2017). These galaxies are characterized by low metallicity (Z ~ 0.1–0.3 Zsun), high specific star formation rates, and compact morphologies—conditions analogous to those in reionization-era galaxies. The selected targets span a range of stellar masses (~$10^8$–$10^9$ Msun) and SFRs (~3–10 Msun yr$^{-1}$), providing a representative sample of low-mass, intensely star-forming systems. Each target will be observed in pointed mode with FIRESS's low-resolution spectroscopic setting (R ~100). Due to their compact sizes (r<0".2; e.g., Kim+21), single observations for each target can capture total fluxes of these emission lines. Integration times will be adjusted based on the emission line. In particular, each [CII] 158μm / [OIII] 52/88μm / [OI] 63 μm / [NII]122 μm emission line will require <0.1hours per line per target. Fainter lines, such as [OI]146 μm / [NII]205 μm, require longer integration time with ~0.75 hours.

# 24. Disentangling the Effects of Metallicity and Star Formation Activity on the Dust Properties

Frédéric Galliano (DAp, CEA Paris-Saclay, France), Maarten Baes (Ghent University, Belgium), Léo Belloir (DAp, CEA Paris-Saclay, France), Simone Bianchi (INAF, Italy), Caroline Bot (Observatoire astronomique de Strasbourg, France), Viviana Casasola (INAF, Italy), Jérémy Chastenet (Ghent University, Belgium), Christopher Clark (STScI, USA), Ilse DE Looze (Ghent University, Belgium), Golshan Ejlali (IPM, Tehran, Iran), Vianney Lebouteiller (DAp, CEA Paris-Saclay, France), Arianna Long (University of Washington, USA), Suzanne Madden (DAp, CEA Paris-Saclay, France), Takashi Onaka (Tokyo University, Japan), Lara Pantoni (Ghent University, Belgium), Evangelos Paspaliaris (INAF Arcetri, Italy), Monica Relaño Pastor (Granada University, Spain), Marc Sauvage (DAp, CEA Paris-Saclay, France), Matthew Smith (Cardiff University, UK), Tsutomu Takeuchi (Nagoya University, Japan) Manolis Xilouris (Athens Observatory, Greece)

The exceptional sensitivity of PRIMA will allow us to detect, for the first time in the infrared (IR), the population of low-surface brightness (quiescent), low-metallicity galaxies. These objects are the key to separating the effects of metallicity and star formation activity on the dust properties that observations of star-forming low-metallicity galaxies cannot. PRIMAger observations of the SED of these objects will thus be a novelty and important for understanding the early stages of dust evolution.

## Science Justification

Our understanding of dust evolution relies on comparing the grain properties in regions (Galactic or extragalactic) exhibiting different physical conditions (e.g., Galliano et al. 2018). Different objects are then considered as snapshots of galaxy evolution at different stages. The most used environmental parameter to quantify the evolution of a galaxy is its metallicity, *Z*, as it traces the cumulated elemental enrichment of its InterStellar Medium (ISM). However, different Star Formation Histories (SFH), and thus different evolutionary paths, can lead to the same *Z* at a given time. Consequently, we often face a degeneracy between the effects of metal enrichment and star formation activity when attempting to interpret dust evolution trends. This is illustrated by Fig. 1, showing the evolution of the fraction of small amorphous carbon grains as a function of *Z* and average starlight intensity, $\langle U \rangle$, in individual galaxies (Galliano *et al.* 2021). It ambiguously suggests that the evolution of these small grains could be driven either by *Z* or by the *specific Star Formation Rate* (sSFR). This is because the low-*Z* galaxies detected with *Herschel* are primarily actively star-forming. This selection effect therefore, results in a correlation between these two parameters in our sample, and we are unable to understand which one is fundamental. This degeneracy is encountered when looking at other quantities, too.

There is however, a population of low sSFR, low-*Z* galaxies, distinct from those in our sample (*e.g.* Lara-López *et al.* 2013). Deriving the dust properties in these objects would allow us to break this type of degeneracy, as these sources would necessarily appear as a distinct branch in one of the two panels of Fig. 1 (solution 1 or solution 2). In addition, learning more about these local galaxies is interesting for the interpretation of deep surveys, as they probe the faint end of the luminosity function. These objects are also local analogs to distant primordial galaxies. *Herschel* was successful in detecting a few very low-*Z* star-forming objects (essentially I Zw 18 and SBS 0335-052). PRIMA could observe a sample of 100 nearby very low-*Z* galaxies (1/10 to 1/30 $Z_\odot$; typical size 2'×2'). For instance, the ALFALFA $HI_{21cm}$ survey (Haynes *et al.* 2018) contains hundreds of galaxies with $M$(HI)<$10^8$ $M_\odot$ and dozens with $M$(HI)<$10^7$ $M_\odot$. Among them, several have been well observed in stellar and ionized gas tracers: the 12 sources from the SHIELD sample (Cannon *et al.*2011); the very low surface density, nearby object, Leo P (Giovanelli *et al.* 2013); the most metal-poor gas-rich galaxy known to date, AG 198691 (Hirschauer *et al.* 2016). We also need more statistics in the low-*Z*/high-sSFR branch. We thus need to explore the whole sSFR range at very low *Z*.

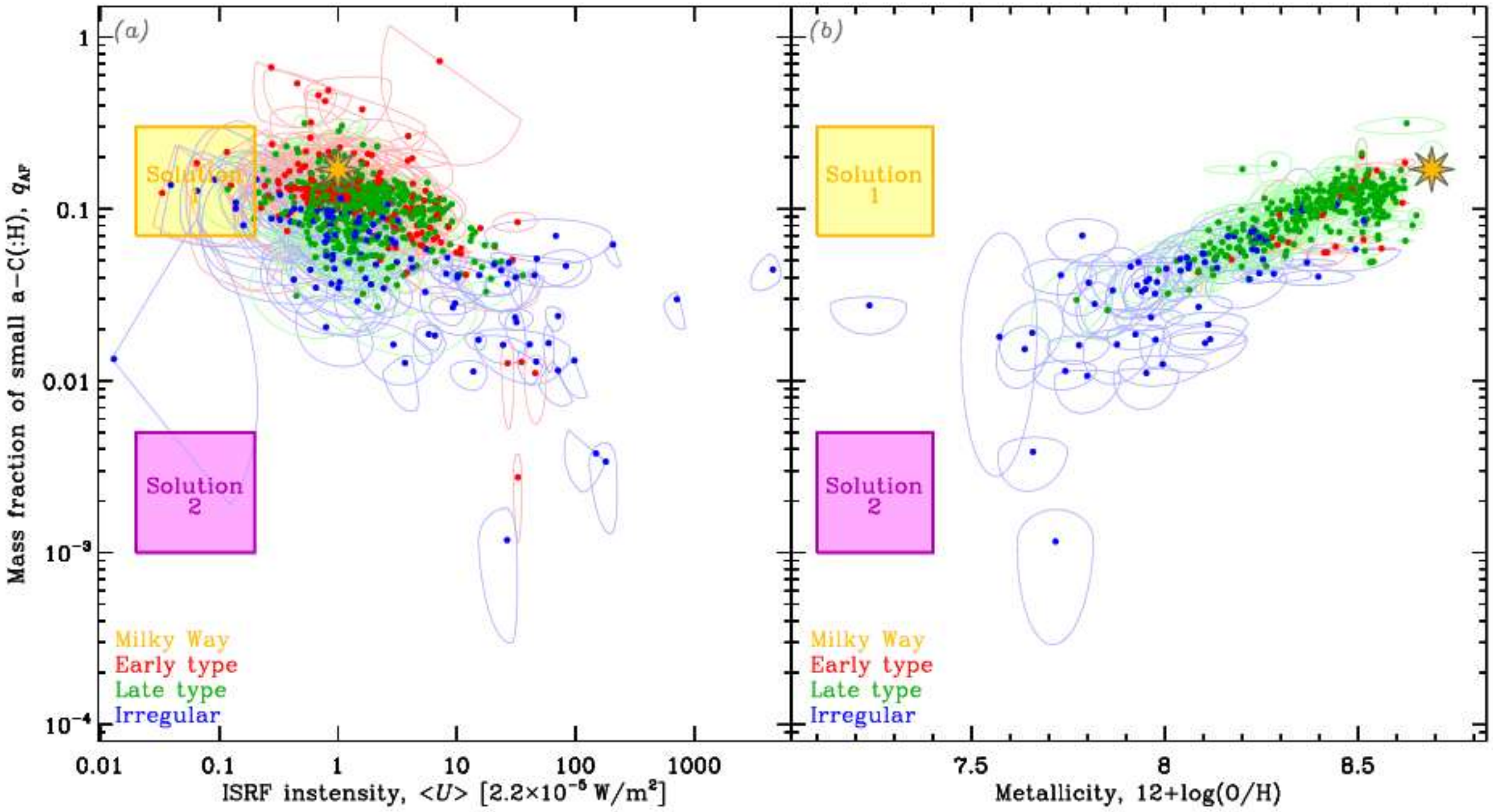

**Figure 1.** SEO Figures: The potential of quiescent very low-Z galaxies. We show the mass fraction of small amorphous carbon grains as a function of Z and starlight intensity. <U>. Each point corresponds to one galaxy of the Galliano et al. (2021) sample. We have added two hypothetical observations of a quiescent very-low-Z galaxy (solutions 1 and 2). Such observations would break the degeneracy between Z and <U>, as they cannot be consistent with both trends.

## Instruments and Modes Used

The proposed observations Request 100 2'x2' maps in FIRESS low-res mode, PRIMAger hyperspectral bands, and PRIMAger polarimeter bands (polarimetry information not needed).

## Special Capabilities

None.

## Approximate Integration Time

Low-resolution FIRESS spectroscopy of 100 2'×2' maps down to $\nu F_\nu=6.9\times10^{-19}$ W/m$^2$/beam (5σ) at 32 µm is about 6 hours per target or 600 hours total.

PRIMAger Hyperspectral band maps (minimum 10'x10' map size in current ETC) down to $\nu F_\nu=6.94\times10^{-17}$ W/m$^2$/beam (5σ) at 35 µm corresponds to 1.2 hours per target or 120 hours total. PRIMAger Polarimeter band maps can be done simultaneously in a total time of about 20 hours to the following sensitivity limits: Band 1: 100 2'×2' maps down to $\nu F_\nu=6.94\times10^{-17}$ W/m$^2$/beam (5σ), Band 2: 100 2'×2' maps down to $\nu F_\nu=1.83\times10^{-16}$ W/m$^2$/beam (5σ), Band 3: 100 2'×2' maps down to $\nu F_\nu=2.76\times10^{-16}$ W/m$^2$/beam (5σ), and Band 4: 100 2'×2' maps down to $\nu F_\nu=3.05\times10^{-16}$ W/m$^2$/beam (5σ).

Therefore, observations of all 100 sources would take a total of 600 hours in FIRESS low-resolution spectroscopy and about 140 hours with PRIMAger hyperspectral and polarimeter bands.

The flux sensitivity has been estimated using the dust model of Jones *et al.* (2017) with a radiation field intensity of $U$=1 and for $N$(HI)=$2\times10^{20}$ H/cm$^2$. At first approximation, the emission and the column density both scale with $Z$. We should thus scale the emission of the model by $Z/Z_\odot$. However, we are interested in the emission of an optically thin medium ($A$(V)≈0.1). This $A$(V) will be reached at a $Z_\odot/Z$ times higher column density than in the Milky Way. $Z$ therefore cancels out in this (rough) estimate. We do not know much about the ISM properties of low surface brightness galaxies. Most of these objects have been detected only through their stellar emission or $HI_{21cm}$. Refer to Galliano et al. (2025) for more details on the sensitivity estimates.

## Description of Observations

The unprecedented sensitivity of PRIMA should allow us to detect these objects at all IR wavelengths. Obtaining the well-sampled IR *Spectral Energy Distribution* (SED) of these objects would allow us to understand the role of $Z$ and sSFR in shaping the dust properties at early stages.

Ideally, we want to map this galaxy sample with all PRIMAger bands. We also need low-resolution spectral maps, with PRIMA-FIRESS, to better constrain the shape of the continuum in different regions.

# 25. Self-Consistently Probing the Dust Properties and the ISM Structure of Nearby Galaxies

Frédéric Galliano (DAp, CEA Paris-Saclay, France), Maarten Baes (Ghent University, Belgium), Léo Belloir (DAp, CEA Paris-Saclay, France), Simone Bianchi (INAF, Italy), Caroline Bot (Observatoire astronomique de Strasbourg, France), Viviana Casasola (INAF, Italy), Jérémy Chastenet (Ghent University, Belgium), Christopher Clark (STScI, USA), Stavroula Katsioli (Athens Observatory, Greece), Ilse De Looze (Ghent University, Belgium), Golshan Ejlali (IPM, Tehran, Iran), Mika Juvela (Helsinki University, Finland), Vianney Lebouteiller (DAp, CEA Paris-Saclay, France), Suzanne Madden (DAp, CEA Paris-Saclay, France), Takashi Onaka (Tokyo University, Japan), Lara Pantoni (Ghent University, Belgium), Evangelos Paspaliaris (INAF Arcetri, Italy), Monica Relaño Pastor (Granada University, Spain), Matthew Smith (Cardiff University, UK), Tsutomu Takeuchi (Nagoya University, Japan) Manolis Xilouris (Athens Observatory, Greece)

The idea is to exploit the fact that PRIMA will allow us to self-consistently obtain, at the same time, the dust continuum and FIR lines in nearby galaxies. In a sample of nearby galaxies, this would allow us to solve some of the degeneracies one often encounters when modeling dust SEDs. The information provided by the FIR lines can indeed be used to constrain the ISM/star topology. With this knowledge, we can thus solve the radiative transfer at subpixel scales, using the information provided by the gas lines to more accurately model the dust.

## Science Justification

One of the main challenges when interpreting the Spectral Energy Distribution (SED) of an object is to separate the effects due to the variation of the microscopic grain properties (composition, size distribution) and those due to the macroscopic distribution of the InterStellar Medium (ISM). This problem has been addressed, concerning the gas, by state-of-the-art models, where a large number of lines can be used to constrain both the physical conditions of the ISM and its topology in galaxies (*e.g.* Cormier *et al.* 2012). This has been possible because atomic physics is more precisely known than dust physics. The fact that different ions have a wide range of critical densities allows us to use a few well-chosen lines to characterize the main phases of the ISM. The knowledge of the average structure of the ISM and of the stellar distribution, provided by the lines, could thus be used as an *a priori* to model the dust properties. Yet, obtaining at the same time a well-sampled SED and the intensity of the main IR gas lines is observationally challenging and has been achieved in only a handful of sources, most of the time for a single pointing.

The combined Mid-InfraRed to Far-InfraRed (MIR-to-FIR) photometric and spectroscopic capability of PRIMA opens the window to obtaining consistent spatially-resolved maps of the total dust and multiphase gas properties in extragalactic regions. Such spectra were only obtained in the past by combining ISOSWS and ISOLWS (e.g., Peeters et al. 2002), but with poor sensitivity, poor spatial resolution, and stitching problems. If we have only a few broadbands, scattered over the whole IR domain, as is usually the case, we are unable to solve the degeneracy discussed above (e.g., Galliano et al. 2018, for a review). The unique feature brought by PRIMA will be consistent maps of the brightest far-IR lines, with the well-sampled dust continuum.

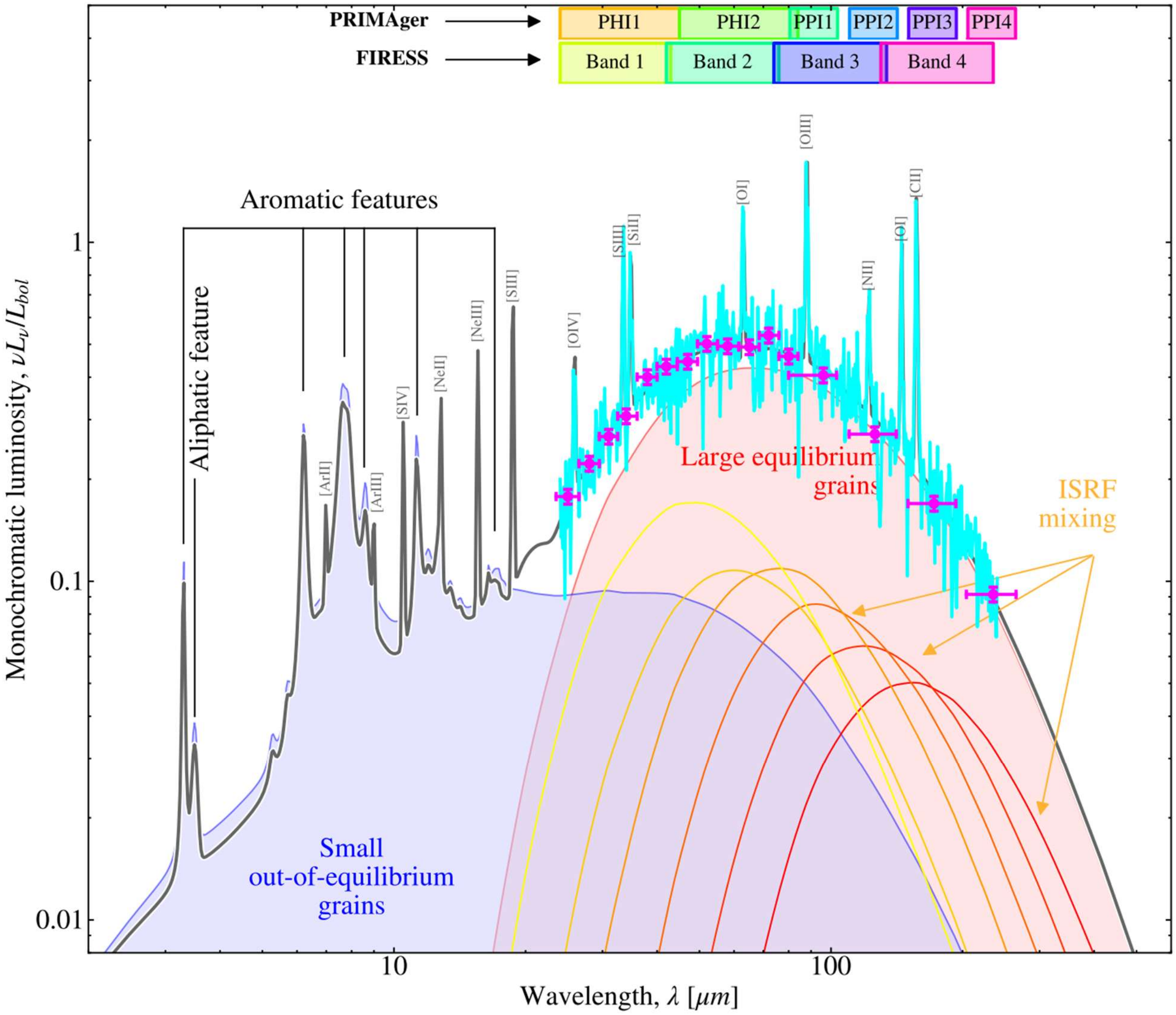

**Figure 1.** Typical SED of a star-forming region with the brightest gas lines. The magenta error bars correspond to broad-band spectrophotometry with R≈3.5 (10σ), and the cyan line is a R≈200 spectrum (5σ)

## Instruments and Modes Used

This science case 22 hours of FIRESS R~100 mapping and 29 hours of PRIMAger, for a grand total of about 50 hours.

## Description of Observations and Approximate Integration Time

We will observe a sample of nearby galaxies, both in narrow-band imaging and low-resolution spectroscopy. We do not need to reach the lowest emission of these objects, but we would need to have contiguous maps in order to understand the spatial variations. Such combined dust-gas maps could be modeled using Cloudy (Ferland *et al.* 2017) and the Meudon PDR code (Le Petit *et al.* 2006) for the gas, a dust evolution model such as THEMIS (Jones *et al.* 2017) for the dust, and a radiative transfer code such as SOC (Juvela 2019). We could observe 30 disk galaxies (20'×20') and 100 low-metallicity galaxies (2'×2'; 1/10 to 1/3 $Z_\odot$) in order to build an evolutionary sequence of the dust properties.

We will observe this galaxy sample with all PRIMAger bands, and with PRIMA-S at low spectral resolution.

- The continuum intensity is based on the Jones *et al.* (2017) model, at *U*=3, and $N$(HI)=$5\times10^{20}$ H/cm$^2$. This corresponds to the typical extended emission of disk galaxies found in the DustPedia sample. This is, however, not the most diffuse emission of the galaxy, but going down to $N$(HI)=$5\times10^{20}$ H/cm$^2$ is sufficient for the present science case.
- These fine-structure lines are all bright. In addition, we do not need to spectrally resolve them. Low-resolution spectroscopy should thus be sufficient to measure their intensity. We have estimated the line intensity by assuming they are proportional to the total IR luminosity. The proportionality factor is the average of the sample in Cormier *et al.* (2019). Sensitivity is not an issue for this science case, as we are not aiming at the faintest regions.

This program could be observed in about 50 hours.

# 26. Far-infrared Line Properties of Green Pea Galaxies, the Best Local Analogues of Galaxies in the Epoch of Reionization Era

Takuya Hashimoto (University of Tsukuba, Japan), Tohru Nagao (Ehime U., Japan), Hanae Inami (Hiroshima U., Japan), Yoichi Tamura (Nagoya U., Japan), Yoshi Fudamoto (Chiba U., Japan), Yurina Nakazato (U. Tokyo, Japan), Tsutomu T. Takeuchi (Naoya U., Japan), Yuri Nishimura (U. Tokyo, Uapan), Logan Jones (STScI, USA)

ALMA has detected the far-infrared (FIR) fine-structure emission lines in galaxies in the epoch of reionization (EoR) up to a redshift of z ~ 14. These EoR galaxies show strong [OIII] 88 μm, 2-10 times brighter than [CII] 158 μm in luminosity. Cosmological hydrodynamic simulations suggest that such high FIR line luminosity ratios may indicate highly-ionized gas or a low covering fraction of neutral gas in the interstellar medium (ISM), both facilitating the escape of LyC photons. However, only two LyC emitters in the local Universe, Haro 11 and Mrk 54, have been observed with FIR lines to date, hampering us from obtaining a census on the relation between the LyC escape and FIR line properties.

Here we propose to perform PRIMA/FIRESS observations of Green Pea galaxies (GPs), the best local analogues of galaxies in the EoR. GPs are low-mass, compact, and bursty star-forming galaxies with significant LyC escape fractions. The high sensitivity of FIRESS can immediately detect [NIII] 57 μm, [OI] 63 μm, [OIII] 52/88 μm, and [CII] 158 μm, as well as the OH 119 μm absorption line within 1-hr observation time per source. The data allows us to (i) systematically examine the relation between LyC escape and FIR line properties, (ii) obtain ISM properties such as electron density, N/O abundance ratio from the FIR lines, (iii) examine presence of multi-phase ISM based on the combination of optical and FIR lines, (iii) study the multi-phase gas outflow, a key to understand the LyC escape. These datasets will provide a golden reference for future radio astronomy projects that plan to study the FIR properties of distant galaxies.

## Science Justification

### Background

Far-infrared (FIR) fine-structure emission lines are major coolants of the interstellar medium (ISM), and provide key information on star formation rates (SFRs: de Looze et al. 2014) and properties of ISM (the gas density, metallicity, ionization parameter, e.g., Pereira-Santaella et al. 2017, Pound & Wolfire 2023). The FIR lines become increasingly important in the observations of galaxies in the epoch of reionization (EoR). ALMA has successfully detected [CII] 158 μm up to z = 8.3 (Bakx et al. 2020) and [OIII] 88 μm beyond z > 9 (Hashimoto et al. 2018). These EoR galaxies show strong [OIII] 88 μm, 2-10 times brighter than [CII] 158 μm in luminosity (Harikane et al.

2020). Cosmological hydrodynamic simulations suggest that such high FIR line luminosity ratios may indicate highly-ionized gas or a low covering fraction of neutral gas in the interstellar medium (ISM), both facilitating the escape of LyC photons (e.g., Katz et al. 2021; see also Figure 1). However, only two LyC emitters (LCEs) in the local Universe, Haro 11 and Mrk 54, have been observed with FIR lines to date (Cormier et al. 2012; Ura & Hashimoto et al. 2023; see also Figure 1), hampering us from obtaining a census on the relation between the LyC escape and FIR line properties. To overcome these, we propose to perform FIRESS observations for Green Pea galaxies (GPs), the best local analogues of galaxies in the EoR with significant LyC escape (e.g., Flurry et al. 2022).

## Science Goals

❶ We statistically test the correlations between $f_{esc}$ and FIR line luminosity ratios such as [OIII] 88 µm/[CII] 158 µm as proposed by Katz et al. (2021). If we could establish a link between them, we could place the FIR line observations into the context of the study of cosmic reionization.

❷ The proposed FIRESS observations will provide us with a reference FIR spectrum of GPs. Combined with the rest-frame UV and optical spectra of GPs, the data will be a golden reference for high-z galaxy studies, including those based on JWST+ALMA. The FIR line luminosity ratios are key to measuring the gas density, ionization parameter, and gas-phase metallicity (Pereira-Santaella et al. 2017). We also combine optical spectrum (e.g., [OII] doublet, [OIII] 4364, 5008 A) with FIR [OIII] 52/88 µm. This will allow us to examine if there exists a multi-phase ISM in GPs (Harikane et al. 2025).

❸ We will examine the presence and velocity of outflow in GPs. We will stack the spectra of GPs to achieve high S/N ratios in emission lines originating from the neutral phase ([OI] 63 µm and [CII] 158 µm) and ionized phase ([OIII] 88 µm). We will also examine the blueshift of the OH 119 µm absorption line to examine the velocity of molecular-phase outflow. These will be crucial to test if a strong outflow facilitates the LyC escape.

### Need for PRIMA

As shown in Figure 1, only two LCEs, Haro 11 and Mrk 54, were observed in FIR lines with the Herschel Space Observatory and SOFIA, respectively (Cormier et al. 2012; Ura et al. 2023). This is because most LCEs were discovered after the mission of Herschel Space Observatory. With PRIMA's superb sensitivity and wide coverage of wavelengths at a time, we can, for the first time, conduct an efficient FIR line survey for the GPs.

### Link to Testable Hypotheses

Theoretical studies based on cosmological hydrodynamic simulations suggest that the [OIII] 88 µm/[CII] 158 µm line luminosity ratio offers insights into the ISM (e.g., Katz et al. 2021; Vallini et al. 2021). As shown in Figure 1, the line ratios are expected to be linked with the LyC escape fraction, one of the key parameters to understand the process of cosmic reionization. The degeneracy mentioned above can be solved because the proposed FIRESS observations provide information on density, metallicity, and abundance ratio. Thus, we can test the hypothesis of the link between LyC vs. FIR line properties. We can also examine the presence and the velocity of outflow in GPs, which facilitate the LyC escape.

| Name | Redshift | $f_{esc}$ [%] | Reference | FIR obs. |
|---|---|---|---|---|
| **Haro11** | **0.021** | **3.2** | **Leitet+11** | **Cormier+12** |
| **Mrk 54** | **0.045** | **2.5** | **Leitherer+16** | **Ura & Hashimoto+23** |
| Tololo 1247-232 | 0.048 | 4.5 | Leitherer+16 | N/A |
| J0921+4509 | 0.235 | 1.0 | Borthakur+14 | N/A |
| **N > 50 Green Pea Galaxies** | **0.30-0.43** | **1 – 72** | **Izotov+16a,b, Izotov+18a,b Flury+22a,b** | **Proposed FIRESS study** |

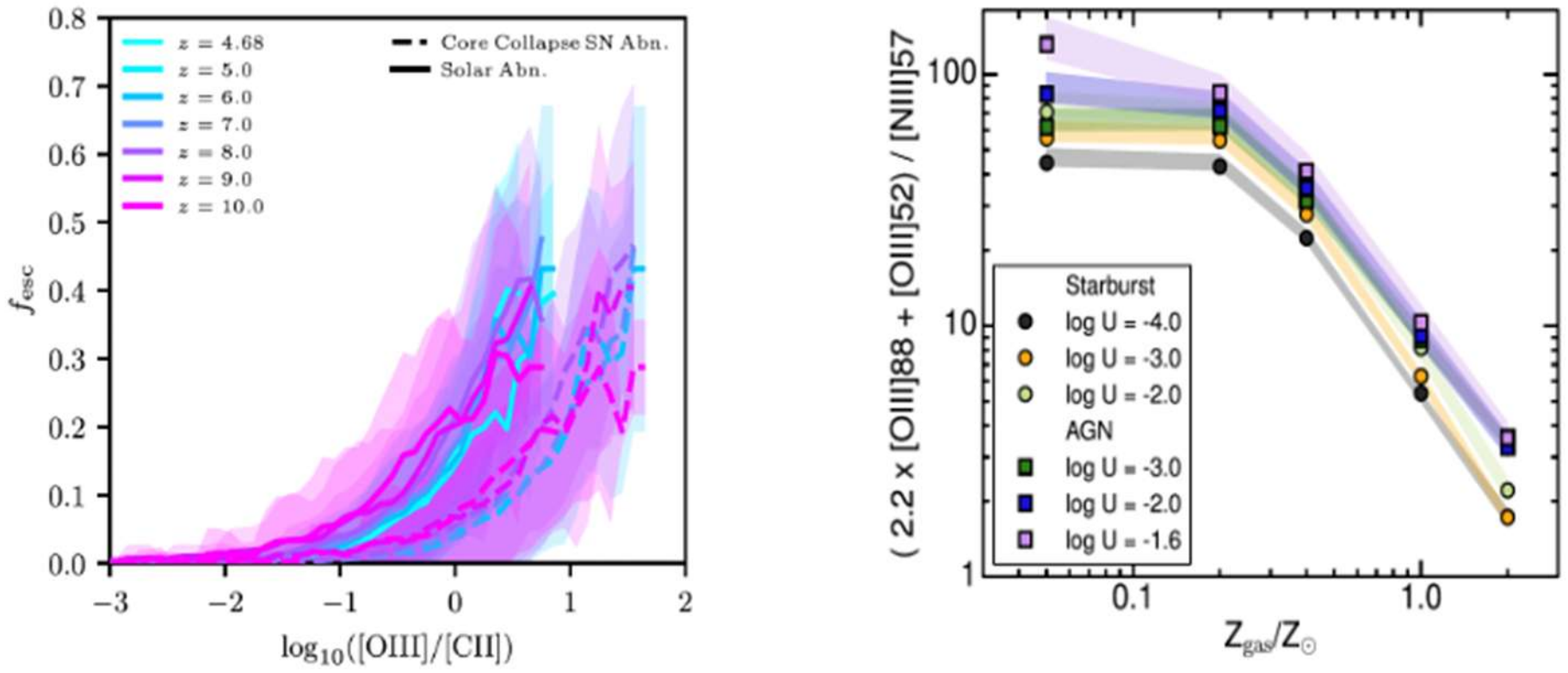

**Figure 1.** (**Top**) List of LCEs. The proposed study will increase the number of LCEs with FIR line observations by a factor of 10. (**Bottom left**) The positive correlation between $f_{esc}$ and [OIII]88/[CII]158 ratios based on simulations of Katz et al. (2021). The proposed study aims to observationally calibrate the relation, which will allow us to place the FIR line observations into the context of the study of reionization. (**Bottom right**) An example figure of how we derive ISM properties of galaxies based on FIR line ratios (Pereira-Santaella et al. 2017). We will measure SFRs, derive electron densities, and metallicities of GPs.

## Instruments and Modes Used

FIRESS pointed high-resolution observations, 1 pointing per target (N=40)

## Approximate Integration Time

Estimation can be based on PRIMA ETC, or another method (please specify). Do not include overheads (slews, calibration, etc.)

For the sensitivity calculation, we have used the PRIMA Exposure Time Calculator (ETC), FIRESS's High-resolution point source mode. Because we nominally aim to target N = 40 GPs, we assume a typical redshift (z = 0.26, corresponding to a luminosity distance of ~ 1300 Mpc), star formation rate of 10 Msun/yr. Under the assumption that half of the star formation is obscured by dust, the total infrared luminosity (LIR) integrated over the wavelength range of 8 – 1000 µm is $2.9 \times 10^{10}\ L_{\odot}$ if we adopt the dust temperature of 30 K and the emissivity index of 1.5. In the ETC, the continuum flux density values at relevant wavelengths are then read from the dust continuum spectral energy distribution.

We assume a relation between LIR and the line luminosities for the local dwarf galaxies in de Looze et al. (2014) for [OI] 63 μm, [OIII] 88 μm, and [CII] 158 μm. For fainter lines, we conservatively assume the line luminosity ratios of [OIII] 52 um/[OIII] 88 um = 0.5 (i.e., low-density limit of ne = 10 cm-3), [NII] 122 um/[OIII] 88 um = 0.1 (i.e., logU = -3 and Z = 0.2 Zsun) and [NIII] 57 um/(2.2[OIII] 88 um + [OIII] 52 um) = 1/50 (i.e., logU = -3 and Z = 0.2 Zsun) based on Pereira-Santaella et al. (2017), and [OI] 145 um/[OI] 63 um = 0.1. The resulting luminosity, flux, and required observation time are summarized in Table 1. In OH 119 um, we assume a 20% absorption against the continuum level (10 mJy), 2 mJy. If we aim to achieve 5 $\sigma$ detection per spectral resolution (~ 70 km/s), we need a sensitivity of 0.4 mJy per 70 km/s, corresponding to the effective flux of 0.028 Jy km/s (7.5E-19 W/m^2).

With 1-hr observation time per source, we can detect [OIII] 52/88 μm, [OI] 63 μm, [CII] 158 μm, and OH 119 μm at 5-sigma. The [NIII] 57 μm and [OI] 145 μm require 10-hr and 20-hr observation time, respectively. By stacking of N = 20 galaxies (T = 20 hrs), we can detect these fainter lines and can examine the presence of a broad wing in high S/N spectrum of brighter lines ([OI] 63 μm, [OIII] 52/88 μm, and [CII] 158 μm).

To statistically examine the LyC escape fraction and the FIR properties based on N = 20 LCEs and another N = 20 non-LCEs as a control sample, we request 40-hrs of observation.

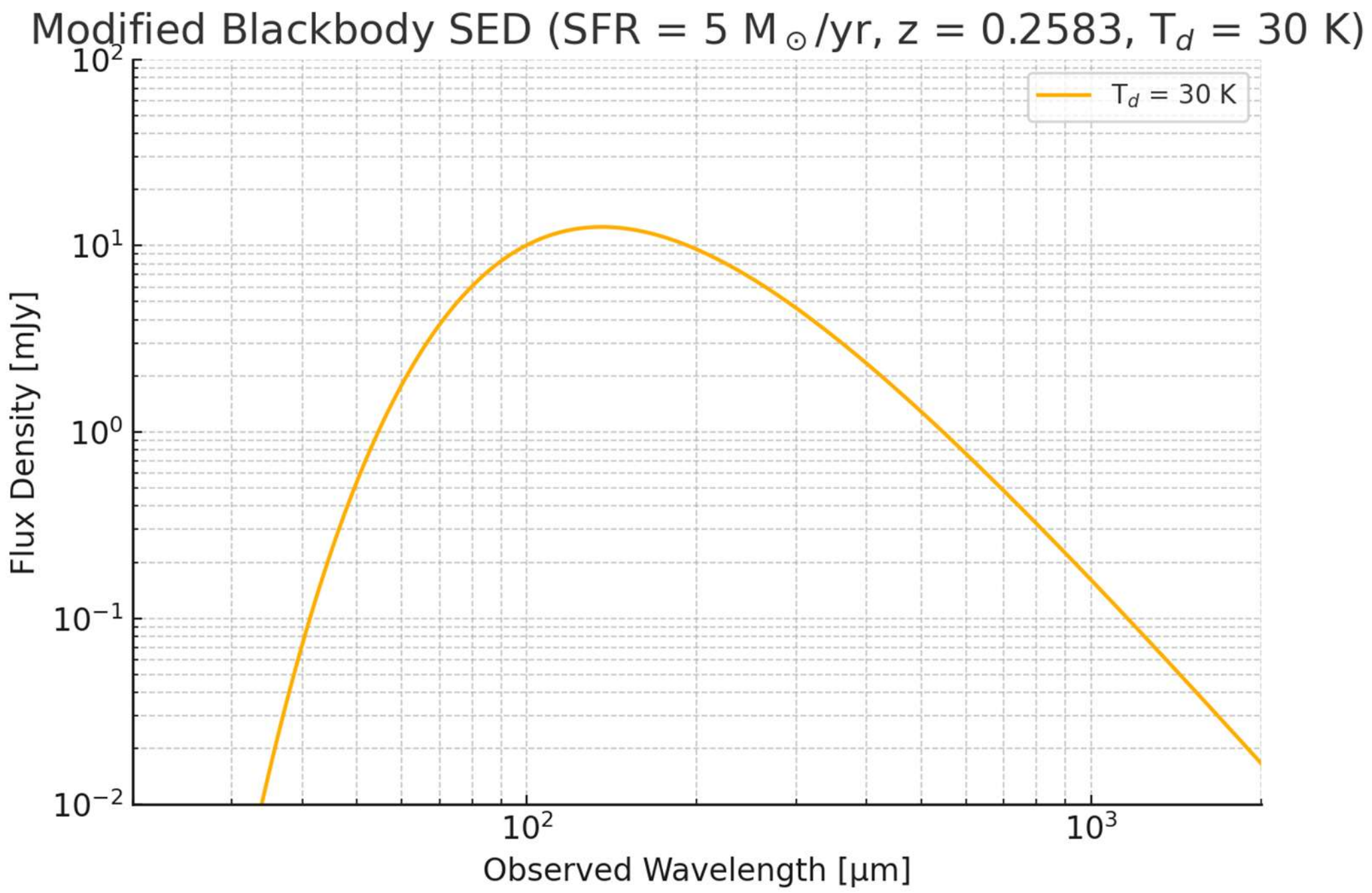

| Line | logL [Lsun] | Flux [W/m^2] | Rest λ [μm] | Obs λ [μm] | Continuum [Jy] | Time [hr] (5σ) |
|---|---|---|---|---|---|---|
| [OIII] 52μm | 8.05 | 2.2E-18 | 51.815 | 65 | 0.001 | 0.1 |
| [NIII] 57μm | 7.07 | 2.2E-19 | 57.317 | 72 | 0.002 | 10.5 |
| [OI] 63μm | 7.92 | 1.5E-18 | 63.1837 | 80 | 0.008 | 0.2 |
| [OIII] 88μm | 8.35 | 4.2E-18 | 88.356 | 111 | 0.011 | < 0.1 |
| [NII] 122μm | 7.35 | 4.2E-19 | 121.898 | 153 | 0.012 | 3.0 |
| [OI] 145μm | 6.92 | 1.5E-19 | 145.525 | 183 | 0.013 | 22.9 |
| [CII] 158μm | 8.34 | 4E-18 | 157.741 | 198 | 0.013 | < 0.1 |
| OH 119 μm | - | 7.5E-19 | 119 | 150 | 0.012 | 0.9 |

## Special Capabilities Needed

N/A.

## Synergies with Other Facilities

The proposed FIRESS observations provide synergy with various existing and facilities as well as future radio telescope projects as follows. ❶ The GPs have been extensively observed in the rest-frame UV (HST-COS), optical (e.g., Keck, VLT, Subaru), radio telescopes (e.g., GMRT, VLA, GBT), and X-ray (e.g., Chandra). Combined with these legacy data sets, the proposed FIRESS observations will for the first time provide a complete and multi-wavelength picture of GPs ranging from UV to far-infrared and beyond. We also note that the potential PRIMAger observations of GPs targeting FIR continuum emission are also highly complementary to the proposed FIRESS observations. ❷ As discussed in the Scientific Justification, the proposed FIRESS observations are motivated by and thus have strong synergies with previous FIR observations of two local LCEs with Herschel Space Observatory and SOFIA. ❸ Because GPs are the best local analogues of galaxies in the EoR, the proposed FIRESS data will be crucial reference to interpret the ongoing ALMA observations of high-z galaxies, as well as the combined analyses of JWST+ALMA. ❹ These data will be a golden reference for future radio astronomy projects such as ALMA2, ALMA3, and LST/AtLAST whose important goals include observations of distant galaxies with FIR lines.

## Description of Observations

We propose to perform FIRESS spectroscopy of GPs. Because the sky coordinates of GPs are located across the sky, we plan to target GPs one by one. The point-source mode is suitable for our purpose: the typical GPs have rest-frame UV continuum emission size of 0.2 – 1.0 kpc, well below 1 arcsecond. We do not aim to spatially resolve the GPS but intend to obtain the total line fluxes of each targeted line. To examine the line profiles based on a stacking analysis as well as the OH 119 µm absorption line, we aim to use the high-spectral resolution mode (R ~ 4400). This is sufficient to study the potential broad wing component of FIR lines as Herschel Space Observatory data (R ~ 1200) have demonstrated in local dwarf galaxies (Romano et al. 2023).

In order to statistically examine the relation between the LyC escape and FIR line properties, we nominally aim to target N = 20 LCEs (e.g., Flurry et al. 2021) and another N = 20 non-LCEs as a control sample. The high dynamic range of LyC escape fraction in GPs is a key in this regard.

# 27. PRIMA/PRIMAGER will Reveal Whether Little Red Dots are Dust-Reddened AGN or Naked Accretion System

Kohei Ichikawa (Waseda University, Japan), Kohei Inayoshi (KIAA, PKU, China), Masafusa Onoue (Waseda U., Japan), Tohru Nagao (Ehime U., Japan), Takuya Hashimoto (Tsukuba U., Japan), Hanae Inami (Hiroshima U., Japan), Ivan Delvecchio (INAF OAS, Italy), Takumi Tanaka (U. Tokyo/Kavli IPMU, Japan), Jacco van Loon (Keele U., UK)

Little Red Dots (LRDs) are population of active galactic nuclei (AGN) newly found by JWST at z>=3 (now expanding down to z~2). These objects are characterized by their compact morphology and red optical continuum, often interpreted as evidence of dust extinction with Av=3-5 mag. However, current JWST/MIRI shows very flatter SEDs and ALMA dust continuum follow-up observations show non-detections, suggesting a lower dust content of Av<3 mag. We propose to constrain the rest-frame mid-IR (MIR;5-15 μm) SEDs of LRDs at z~4-6 (and possibly at lower-z as well) using PRIMA's supreme sensitivity. With its expected performance, PRIMA will enable dust extinction constraints down to Av<0.6 mag for the brightest known LRD, A2744-45924. This approach can be extended to similarly luminous LRDs for several sources in JWST/COSMOS-Web Survey. If LRDs indeed show such low Av values, this provides strong impact in two aspects; one is to AGN luminosity and $M_{BH}$ estimations, reconsidering the total BH growth with current Av~3 mag estimates. Second, it would challenge the current understanding of their SEDs, suggesting that red optical continuum is intrinsic rather than due to dust extinction, potentially indicating a truncated accretion disk and a clear departure from the SED shape of known quasars. Readers are also referred to related studies in this GO book, including PRIMA stacking analysis of LRDs (Delvecchio et al.) and detailed MIR to far-IR SEDs of low-z analogs (Tanaka et al.).

## Science Justification

### Background: Discovery and enigmatic nature of “Little Red Dots”

The advent of JWST has revealed a previously unknown abundant population of compact, red sources called Little Red Dots (LRDs) at redshifts z>3 (e.g., Matthee et al. 2024). These objects show extremely compact morphologies and steep red continua in the rest-frame optical, previously attributed to dust extinction of Av~3-5 mag. Both galaxy and active galactic nucleus (AGN) dominated scenario remain viable (e.g., Akins et al. 2024), spectroscopy shows that over 60% of LRDs feature broad Balmer emission lines, confirming the presence of AGN. However, ALMA follow-ups have consistently reported non-detections of dust continuum emission, and JWST/MIRI photometry shows unusually flat SEDs at rest-frame 1-3 μm (e.g., Williams et al. 2024; Perez-Gonzalez et al. 2024), inconsistent with the picture of dust reddened AGN. This tension raises a critical question: are LRDs truly dust-reddened AGN, or do they represent a fundamentally new AGN or SMBH populations in early Universe (e.g., Naidu et al. 2025; Rusakov et al. 2025)?

## Science Question

### Are LRDs dust-reddened AGN or unusual AGN population?

We aim to obtain the intrinsic rest-frame 5-15 µm SEDs of LRDs at z=4-6, using PRIMAGER hyper-spectral model (R~10). If LRDs host dust-reddened AGN ($A_V$ ~1-3mag), their reprocessed MIR dust emission should be detectable. Our primary field of study is the JWST/COSMOS-Web area, where ~500 LRDs have been identified. In particular, we focus on the MIRI footprint region where MIRI, NIRCam, and NIRSpec are covered (0.2 deg$^2$, Akins et al. 2024; Lin et al. 2025).

Figure 1 shows that, for the brightest known LRD (A2744-45924 at z=4.47; Labbe et al. 2024), PRIMA can constrain Av down to Av < 0.6 mag. If PRIMA shows non-detections in luminous LRDs, we can rule out dust-reddened AGN scenario, and it would provide strong impact their AGN, accretion rate, and $M_{BH}$ estimates and their nature of red optical continuum.

## Science Goals

### 1. Dust Geometries and Dusty Torus Structure for LRDs

The Hyperspectral mode (R~10) by PRIMAGER enables robust constraints to dust SED of LRDs (Figure 1). Dust SED shapes will provide us about dust geometry (e.g., torus density profile and dust emission size), including constraints from the 9.7 µm silicate feature. Emission or absorption features can indicate whether we observe the source with face-on or edge-on view (Hao et al. 2007; Ichikawa et al. 2012, 2015).

### 2. Revised Estimates of AGN luminosity and $M_{BH}$

If current extinction assumption (Av~3 mag) are over-estimates, current intrinsic AGN luminosities and $M_{BH}$ are also over-estimated by factors of 2-4 (Chen et al. 2025; Greene & Ho 2005). PRIMAGER's MIR observations provide Av constraints, which will critically inform the total black hole accretion history and will reduce the currently reported tension to the Soltan's argument (Inayoshi & Ichikawa 2024), considering the abundance of LRDs (Greene et al. 2024; Kokorev et al. 2024).

### 3. Origin of red Optical Continuum in LRDs

A low Av value implies that red optical slope in LRDs is intrinsic, not extinction driven. This could indicate a truncated accretion disk with a large inner cavity, lacking hot accretion disk emission. If true, LRDs would present a unique accretion state (and SMBH population) not observed in typical quasars. In addition, since the lack of dust emission, we are able to observe for the first time the intrinsic accretion disk spectra longer than the dust sublimation wavelength ($\lambda$>2 µm), which was partially reported through polarization observations (e.g., Kishimoto et al. 2008)

## Need for PRIMA

JWST/MIRI and ALMA cannot probe the rest-frame 5-15 µm band for LRDs at z~4-6. PRIMA/PRIMAGER Hyperspectral mode (both for PHI1+PHI2) covers the unique wavelength range to detect dust-reprocessed emission and to distinguish between low- and high- dust obscuration scenarios. Its capability to conduct wide-area surveys of COSMOS-Web MIRI footprint (0.2 deg$^2$) enables statistically meaningful constraints on the LRD population (see Delvecchio et al. for stacking analysis).

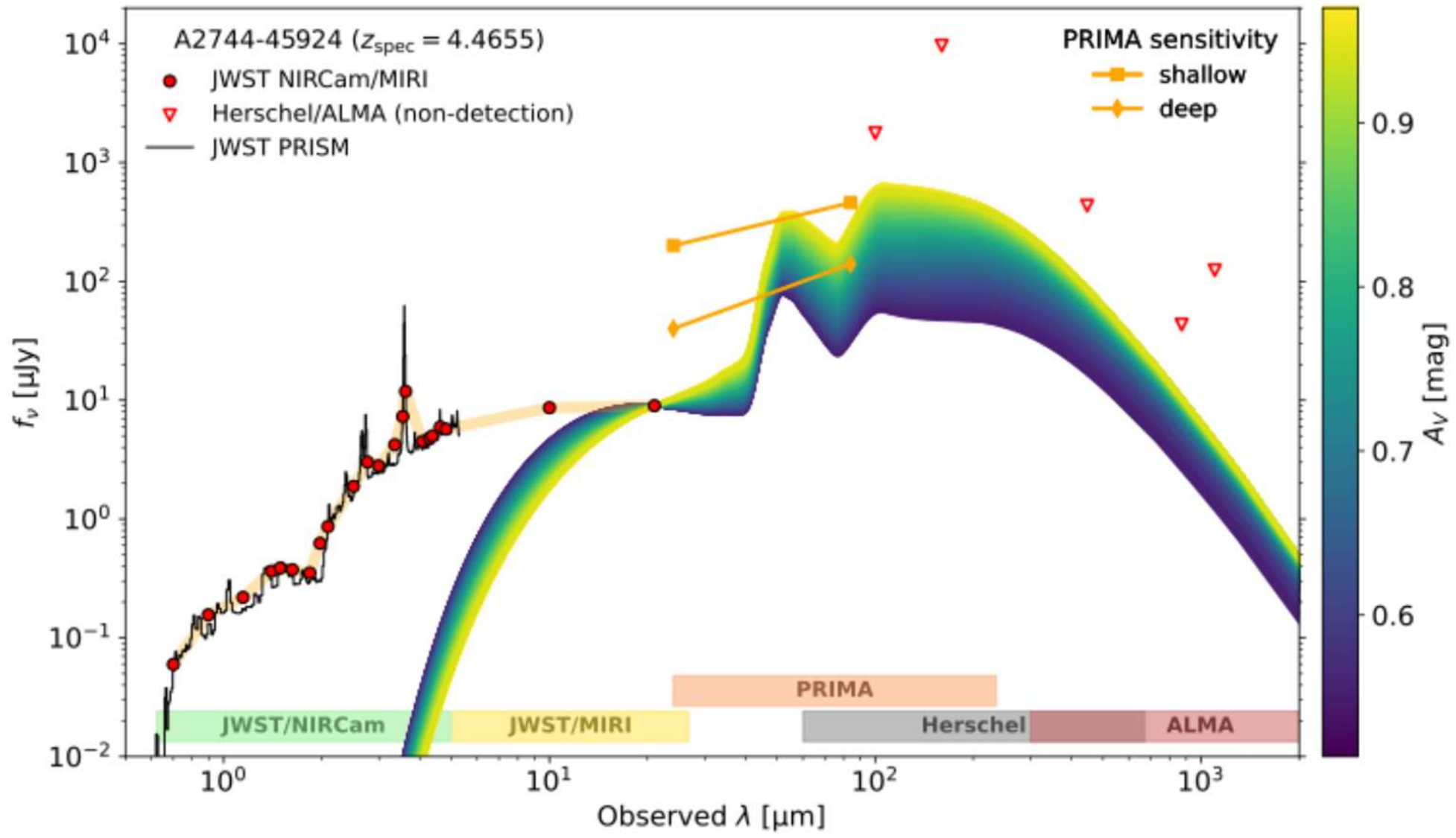

Figure 1: Multiwavelength SED of the brightest known LRD, A2744-45924 (z = 4.47), showing JWST/NIRCam and MIRI photometric detections (orange circles), and upper limits from Herschel and ALMA (triangles). The gray solid line shows the JWST/NIRSpec prism spectrum (Labbe et al. 2024). Colored curves indicate model dust SEDs with varying dust extinction (0.6 < Av < 1.0 mag; based on Chen et al. 2025 and Li et al. 2025). The orange solid lines represent the expected sensitivity curve of PRIMA/PRIMAGER PHI1+PHI2 for the proposed shallow (aquare) and deep (triangle) surveys.

## Instruments and Modes Used

This science case will map 0.2 deg$^2$ of the COSMOS-Web MIRI footprint with PRIMAger.

## Approximate Integration Time

To obtain meaningful constraints on dust extinction (Av) for LRDs, we use the SED of the currently brightest known LRD, A2744-45925 at z = 4.47, as a test case (Labbe et al. 2024). Although A2744-45925 is not located within the JWST-COSMOS field, we expect future identifications of spectroscopically confirmed LRDs with similar brightness levels in that field. We adopt JWST/COSMOS-Web MIRI footprint as our survey region, as it provides the widest area (0.2 deg$^2$) with full of NIRCam, NIRSpec, and MIRI coverage (Akins et al. 2024, Lin et al. 2025), maximizing the likelihood of detecting luminous LRDs.

### We define two survey depth goals

**Shallow Survey Goal (to constrain Av < 1 mag)**

This requires a flux limit of $f_{nu}$ < 200 uJy at 24 µm. The estimated survey time for 0.2 deg2 is 69.6 hrs based on PRIMA ETC. Under this depth, we expect to detect several (<10) LRDs if they have moderate extinction (Av = 1 mag) or above.

**Deep Survey Goal (to constrain Av < 0.6 mag)**

This requires a deeper flux limit of $f_{nu}$ < 40 uJy at 24 µm. According to PRIMA ETC, the estimated survey time increases to 1740 hrs. This depth would allow the detection of more than 10-20 LRDs, enabling statistical analysis of their dust extinction properties (Akins et al. 2024; Casey et al. 2024).

## Special Capabilities Needed

It is not applicable to our science case.

## Synergies with Other Facilities

Our science case has strong synergies with JWST, Euclid, and the Subaru/HSC SSP Deep fields in two key aspects. First, the legacy value of our PRIMA/PRIMAGER survey enhances these well-studied fields. Since LRDs have been identified in areas as abundant LRDs at z >= 4 by JWST (COSMOS-Web), and rarer but brighter LRDs at z <= 3 by Euclid and Subaru/HSC+NIR deep surveys (see Tanaka et al. in the same GO book for detailed LRD studies in these fields). Thus, the PRIMA/PRIMAGER survey will naturally provide unique multi-band MIR photometry that was either unavailable or too shallow in these datasets. This will significantly improve our ability to constrain dust SEDs not only for LRDs, but also for a broader population of AGNs and galaxies in the field.

Second, the PRIMA/PRIMAGER survey will inherently generate a transient MIR sky map for the survey regions (e.g., COSMOS field). COSMOS was previously covered by Spitzer/MIPS observations nearly 20 years ago (e.g., Sanders et al. 2007). With PRIMA's deeper sensitivity, all sources detected by Spitzer/MIPS should be re-detected by PRIMAGER. Therefore, any sources absent in PRIMAGER images are excellent candidates for MIR transients, such as tidal disruption events (TDEs) in dusty environments, which are often missed in optical surveys.

## Description of Observations

We propose to carry out a dedicated PRIMA/PRIMAGER survey of the JWST/COSMOS-Web field using the Hyperspectral Imaging mode (PHI1 + PHI2). The COSMOS-Web MIRI footprint, which spans 0.2 $deg^2$, is currently the largest contiguous area with deep JWST/NIRCam, NIRSpec, and MIRI coverage. This field hosts ~200-500 photometrically identified LRDs and multiple spectroscopically confirmed candidates. It represents the optimal field for a wide-area MIR follow-up survey. Our proposed mapping strategy will fully cover the JWST/MIRI coverage region, maximizing overlap with existing JWST datasets and enhancing the scientific legacy of the field.

**We adopt a dual-depth survey strategy**

- The shallow tier aims for a 24 µm flux density limit of fv=200 uJy, which corresponds to constraining Av < 1 mag for the brightest LRDs. The required total integration time for this tier is 69.6 hrs.
- The deep tier aims to reach a fainter flux limit of fv = 40 uJy, enabling constraints on Av < 0.6 mag. This depth would allow the detection of fainter or less-obscured LRDs and provide statistical studies on the dust content of LRDs. The deep tier requires up to 1740 hrs of integration time.

We will conduct mosaicked observations using multiple PRIMAGER pointings, each covering a 10'×10' field-of-view per pointing. The number of pointings is optimized to fully map the 0.2 $deg^2$ area while maintaining uniform depth. The hyperspectral mode ($R \approx 10$) will provide continuous wavelength coverage essential for modeling the dust SEDs and identifying key spectral features such as the 9.7 μm silicate absorption/emission.

No special calibration or timing constraints are required for this program. The survey design ensures flexibility in scheduling and allows for future time-domain comparisons with archival Spitzer/MIPS imaging, further increasing the long-term legacy value of the dataset.

# 28. A Complete View of the ISM at the Era of Peak Galaxy Growth

Hanae Inami (Hiroshima University, Japan), Amit Vishwas (Cornell University, USA), Luigi Spinoglio (IAPS INAF, Rome, Italy), Juan Antonio Fernández-Ontiveros (CEFCA, Teruel, Spain), Tohru Nagao (Ehime University, Japan), Hideki Umehata (Nagoya University, Japan), Jason Surace (California Institute of Technology, USA)

Determining the physical and chemical conditions of gas and dust in galaxies is vital for understanding galaxy evolution. At the peak of galaxy growth near z~2-3, active galaxies are deeply enshrouded in dust, and thus mid- and far-infrared observations are indispensable. PRIMA will bridge the gap between the wavelength range of JWST and ALMA, providing coverage of a suite of polycyclic aromatic hydrocarbon (PAH) features, fine-structure lines, and molecular hydrogen lines, allowing us to explore the interstellar medium (ISM) embedded in the dust. In this PRIMA program we propose to use FIRESS to observe active galaxies detected with ALMA and/or JWST in the distant Universe in order to directly trace the ISM conditions at dust embedded sites of galaxy activity. The redshift range from $3 < z < 4.5$ is ideal for PRIMA as it allows coverage of rest-frame wavelengths from 5 to 80μm, where the peak dust emission and rich spectral features reside. For galaxies at $z < 4$, PRIMA will efficiently detect PAH features and emission lines from neon, sulfur, argon, oxygen and silicon for dwarf-like main-sequence galaxies, allowing measurements of properties of small dust grains and warm dust, the star formation rate, gas ionization, density, and metallicity. At $z > 4$, it will be more feasible to observe luminous galaxies (e.g., starbursts) to obtain the same set of spectral features to explore their ISM. Simultaneous observations of multiple PAH features, multiple fine-structure lines, and molecular hydrogen lines, which are only available with PRIMA, are the only way to fully reveal the underlying condition of the ISM and the physical and chemical imprints of the assembly of active galaxies in the early Universe.

## Science Justification

### Broader Context

The physical and chemical conditions of gas and dust are fundamental to understanding galaxy evolution. In particular, emission lines play important roles in constraining properties such as star formation rate, star formation age, ionization, density, metallicity and dust properties. Recent ground- and space-based observatories have enabled observations of rest-frame optical emission lines at $z > 1$ to investigate these key physical parameters for early galaxies. However, due to the dustier nature of active galaxies at high redshifts of at least up to ~3 (e.g., Gruppioni et al. 2013, Magnelli et al. 2013) or possibly even higher (e.g., Gruppioni et al. 2020, Algera et al. 2023, Rodighiero et al. 2023, Barrufet et al. 2023), it is likely that we have not yet probed the extent and nature of star formation activity deeply enshrouded in dust. In addition, an unexpectedly large population of heavily obscured AGN have been discovered by ALMA and JWST in recent years (Gilli et al. 2022, Yang et al. 2023).

Although JWST is pushing this limit to higher redshift, its wavelength cutoff at 28µm still leaves important dust features and coolants of the ISM inaccessible. Several PAH bands are sprinkled throughout the rest-frame mid-infrared (mid-IR) bands at 3-20µm and can directly trace the properties of dust (e.g., Galliano et al. 2008, Stierwalt et al. 2014, Lai et al. 2020). Fine-structure lines are sensitive tracers of the physical conditions of the ISM, like the ionization parameter (e.g., Cormier et al. 2015, Inami et al. 2013). In particular, primary elements like neon, sulfur, argon, oxygen, and silicon are available in the mid-IR bands, which are less sensitive to large dust columns than optical lines. They provide powerful diagnostics for ongoing energetic processes that accompany star formation, supermassive black hole growth, and feedback mechanisms that regulate the growth of galaxies (e.g., Spinoglio et al. 2017). PRIMA covers wavelengths of ~25-250µm, which are otherwise inaccessible from Earth, and traces the rest-frame mid-to-far-IR spectra of galaxies. This is essential to seeing past the dusty veil and understanding the key processes that drive the assembly of massive galaxies in the early Universe.

### Science Questions

- What are the ISM properties of active galaxies in the early Universe?
- What role does dust play in our understanding of galaxy evolution?
- What is the fractional importance of AGN in active galaxies at high redshift?

### Need for PRIMA

For at least the next decade, PRIMA will be the *only* observatory that fully bridges the wavelength range between JWST and ALMA. This wavelength range covers many key spectral features that trace the physical conditions of the gas and dust. The dust-penetrating power of mid- and far-IR spectroscopy of PRIMA is essential because most active galaxies are known to be dusty at least up to z~3 and likely beyond. With ground-based rest-frame optical observations, the evolution of ISM properties from z~3 to z=0 has already been suggested (Shapley et al. 2015, Steidel et al. 2016) and now JWST is probing even farther into the distant Universe (e.g., Isobe et al. 2023, Sanders et al. 2024). However, these features are likely from less obscured regions, outside the active regions of galaxies. PRIMA will be able to reveal the hidden aspects of galaxies where most stars are being made and the vicinity of supermassive black holes.

The redshift range of $3 < z < 4.5$ will be the sweet spot for PRIMA to follow up galaxies that are detected with ALMA and JWST (Figure 1). The two strongest cooling lines, [C II]158 and [O III]88, are covered by ALMA Bands 7/8 and Bands 9/10, respectively, in this redshift range. On the other hand, JWST can find dusty active galaxies with imaging but its spectroscopy will miss most fine-structure lines, molecular hydrogen lines, and PAH emission for studying the ISM embedded in dust.

### Interpretation Methods

The wavelength range between ~5-80µm in a galaxy's rest-frame spectrum is filled with astrophysically important tracers. The most prominent features in this spectral range arise from PAH emission, especially at 6.2μm, 7.7μm, 8.6μm, 11.2μm, and 12.7μm (Tielens 2008). These PAH features can have large equivalent widths (EW≳0.5μm) when star formation is the dominant energy source; conversely when their EWs are < 0.1μm it is likely that an AGN dominates (Armus et al. 2007, Petric et al. 2011). When multiple PAH features are detected, their flux ratios can be used to determine the dust grain size and charge (e.g., Draine & Li 2001).

In addition, PRIMA will cover all of the following fine structure lines for galaxies from 3 < z < 4.5: [Ar II] 7μm, [Ar III] 8.99, [S IV] 10.5, [Ne II] 12.8, [Ne V] 14.3, [Ne III] 15.6, [S III] 18.7, [Ar III] 22, [Ne V] 24.3, [Fe I] 24, [O IV] 25.9, [Fe II] 26.0, [S III] 33.5, [Si II] 34.8, [Fe I] 35, [Fe II] 35, and [Ne III] 36. Among these fine structure lines, [Ne II] and [Ne III] are often the strongest and their line flux ratio is the most sensitive tracer of the star formation rate (Ho & Keto 2007, Mordini et al. 2021) and the hardness of the radiation field (e.g., Brandl et al. 2006, Bernard-Salas et al. 2009). This line ratio will facilitate breaking the degeneracy between gas ionization, density, and metallicity when combined with other line ratios, particularly strong lines such as [S III]33.5/[Si II]34.8 (Figure 2 left) that are detectable with PRIMA (cf. Snijder et al. 2007, Inami et al. 2013). Furthermore, when emission lines are present from the same species and the same excitation level, their ratios can be used for density diagnostics. In particular, [S III]33.5 and [S III]18.7 are expected to be as strong as [Ne III], and thus their ratio will be a powerful density tool. Although much weaker (at most ~0.1% of the total IR luminosity, $L_{IR}$), the coronal line [Ne V] at 14.3μm and 24.3μm is a robust indicator of the presence of an AGN. Thus, [Ne V] can be used to search for an AGN and [Ne V]/[Ne II] can be used to calculate the AGN contribution (or its upper limit) to the ISM (Figure 2 right). In addition, the [O IV] line, which is often brighter than [Ne V], also has a relatively high ionization potential, so [O IV]/[Ne II] can be used to identify an AGN (e.g., Spinoglio et al. 2022, Stone et al. 2022). If shocked gas is present, $H_2$ emission lines may be prominent and detectable with PRIMA (Ogle et al. 2010, Guillard et al. 2012). Mergers are more prominent at high redshifts (e.g., Ventou et al. 2017, Duncan et al. 2019) and thus merger-induced shocks traced by $H_2$ observations of them will open a new window to explore their nature. Combinations of multiple H2 lines will provide estimates of the temperature and mass of warm molecular gas (Higdon et al. 2006, Roussel et al. 2007).

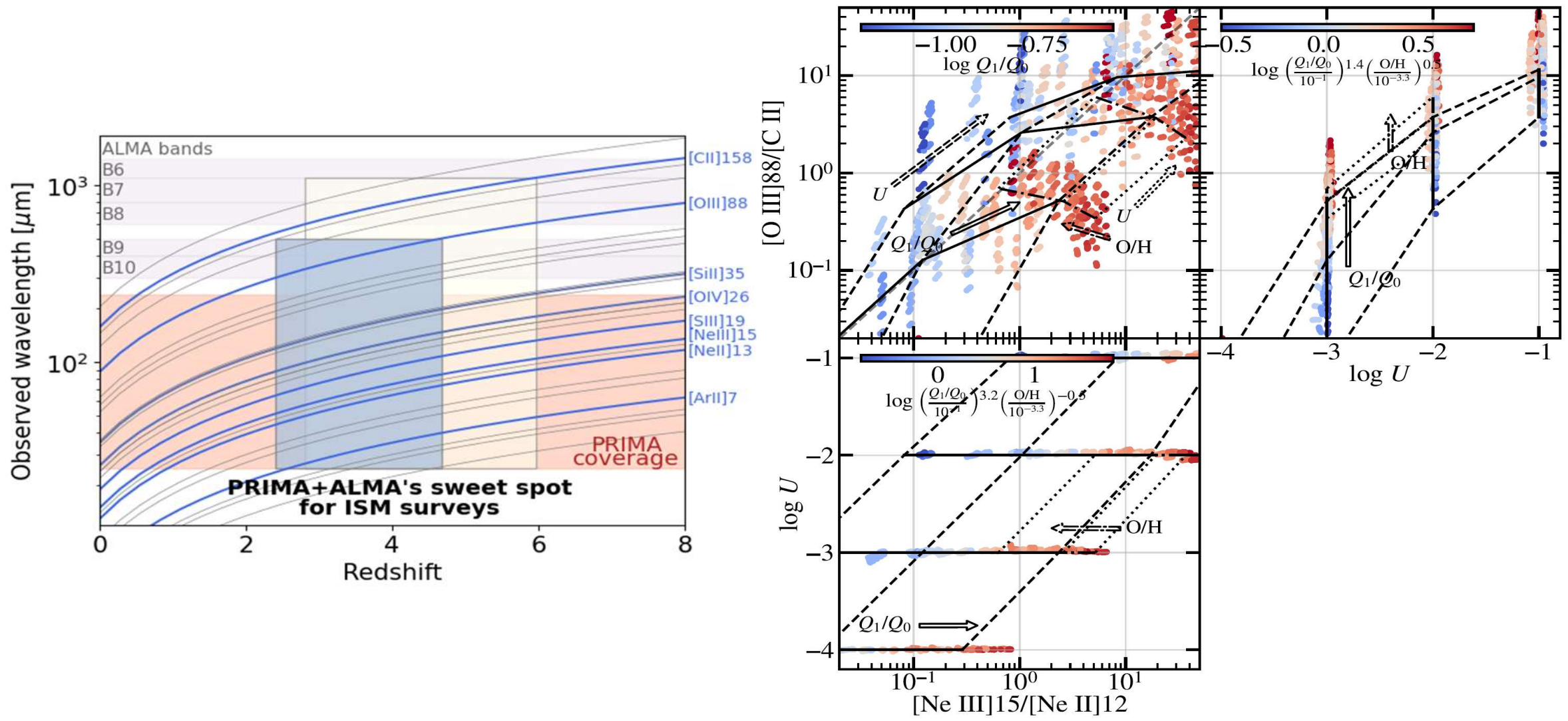

**Figure 1.** *Left*: The redshift range that is ideal for PRIMA to follow up ALMA- and JWST-selected samples. *Right*: A [Ne III]15/[Ne II]12 vs. [O III]88/[C II] diagnostics using Cloudy simulations (Peng et al. 2025ab). (top left) The model data points comparing the two line ratios, color-coded by Q1/Q0. The 2nd grid shows the effect of increasing Q1/Q0 (log Q1/Q0 = -1.4, -1.0, to -0.6) along the solid lines, and increasing U along the dashed lines. At log U = -3 and -2, a subgrid of O/H (dash-dotted lines at log (O/H) = -3.8, -3.4, -3.0) and U (dotted line) is also plotted. The diagonal gray dashed thick line is a visual aid for highlighting a linear trend. (top right) Illustration of the dependence of [O iii]88/[C ii] on log U, and the additional impact by Q1/Q0 and O/H. (bottom left) The dependence on Q1/Q0 and O/H for [Ne iii]15/[Ne ii]12. The plot indicates that interpretations using the mid-IR Neon lines are very useful, and easier to interpret for radiations field diagnostics.

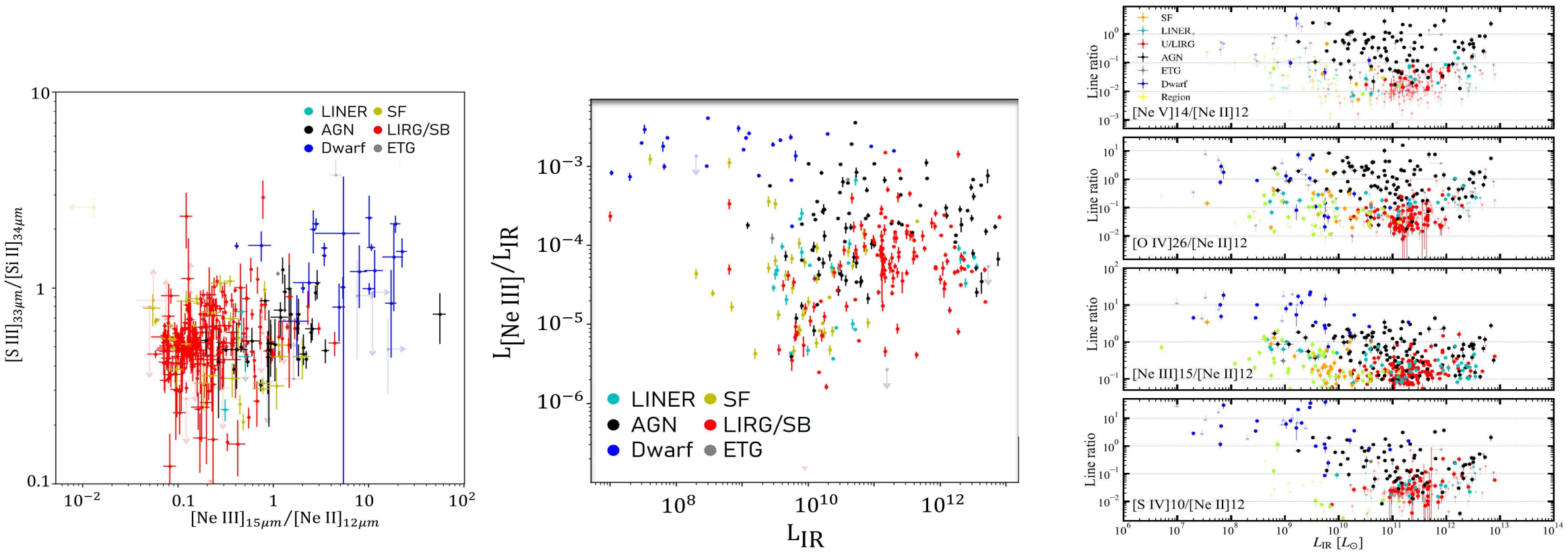

Figure 2. Parameter space of emission lines in the mid-IR that PRIMA can obtain with reliable detections. The data points from an exhaustive literature and archival search of data from space telescopes, mostly limited to local galaxies (e.g., Peng et al. 2025a, Inami et al. 2013, Cormier et al. 2015). *Left*: Line ratio-ratio diagram of the four prominent emission lines that PRIMA can easily detect, [Si II]33, [Si II]34, [Ne III]15, [Ne II]12. *Middle*: $L_{[Ne\ III]}/L_{IR}$ vs. $L_{IR}$, *Right*: Line ratios vs. $L_{IR}$ that enable quantifying the AGN fraction.

## Instruments and Modes Used

FIRESS low res pointing. 1 pointing and 2 spectral setups per source

## Approximate Integration Time

The observation depths are set by the need to detect the most important fine-structure lines for studying the ISM. At these depths, all the PAH features will be well detected. Based on observations of local galaxies near z=0, the median luminosity ratio of [Ne III] to total IR luminosity is 5e-4 for the whole sample, or 1e-3 for dwarf-like galaxies (Figure 2 middle). The [Ne II] line flux is 0.1-10x of [Ne III] and the [S III]18,33 lines and the [Si II] line are ~0.1-1x of [Ne III] depending on the age and metallicity. Given the FIRESS low-resolution point source sensitivity of 2e-19 W/m2 (1 hour, 5-sigma), in Table 1, we summarize the IR luminosity of dwarf-like galaxies that can be detected with all of these five important lines for the ISM diagnostics with 1, 3, 10, and 20 hour integrations. For more evolved galaxies, ~4x deeper observations will be needed. Given the sensitivity limits, we would attempt to primarily select unique targets with existing [CII] and/or [OIII] measurements to ensure high reliability detections.

To gain statistics in understanding of the ISM properties, 20 galaxies are proposed to be targeted in Δz=0.5 between z=3.0 and 5.0 (in total 80 galaxies). We assume that 1.8, 3.4, 6, and 10 hours (median) are needed to reach a 5-sigma detection for each galaxy in the redshift ranges of z=3.0-3.5, z=3.5-4.0, z=4.0-4.5, and z=4.5-5.0, respectively, for dwarf-like (LIR ~ 4e12 L⊙) or massive galaxies (LIR ~ 8e12 L⊙). This results in the integration time of 424 hours per pointing for the proposed 80 targets (20 in each redshift bin). Therefore, the total integration time is 848 hours.

| Integration time [hours] | 1 | 3 | 10 | 20 | LIR of the MS galaxies with M* = 1010 M⊙ (1011 M⊙) |
|---|---|---|---|---|---|
| z=1 | 2.9e+11 L⊙ | 1.7e+11 L⊙ | 9.2e+10 L⊙ | 6.5e+10 L⊙ | 3.9e+10 L⊙ (1.9e+11 L⊙) |
| z=3 | 4.2e+12 L⊙ | 2.4e+12 L⊙ | 1.3e+12 L⊙ | 9.4e+11 L⊙ | 1.4e+11 L⊙ (8.6e+11 L⊙) |
| z=5 | 1.4e+13 L⊙ | 8.1e+12 L⊙ | 4.4e+12 L⊙ | 3.1e+12 L⊙ | 2.0e+11 L⊙ (1.3+12 L⊙) |

Table 1. IR luminosities of galaxies at z=1, 3, and 5 to detect at least [Ne II], [Ne III], [S III]18, [S III]33, and [Si II] (in addition to all PAH features) with the given integration time of 1, 3, 10, and 20 hours for . For reference, IR luminosity converted from SFR of main-sequence galaxies with a stellar mass of $10^{10}$ M⊙ ($10^{11}$ M⊙) is shown in the rightmost column (Speagle et al. 2014). Galaxies above the MS relation will be detected with shorter integration times.

## Special Capabilities Needed

None

## Synergies with Other Facilities

ALMA, JWST, LMT, SKA, and AtLAST/LST (if realized) will provide the parent sample of targets to be followed up with PRIMA.

At longer wavelengths, ALMA observations increasingly detect the [C II]158µm and [O III]88µm lines (and their local dust continuum) for galaxies at z > 4 (e.g., Bethermin et al. 2020, Bouwens et al. 2022). These lines have been studied in galaxies in the local Universe with Herschel, but they are limited to z~0 due to a lack of sensitivity of the onboard spectrometers (e.g., Kennicutt et al. 2011, Diaz-Santos et al. 2017). At 6 < z < 10, the [O III]/[C II] ratios of currently known ``normal ’‘star-forming galaxies appear to be higher than those of local galaxies. The cause of this difference is still under debate (e.g., Harikane et al. 2020, Algera et al. 2024) due to a lack of observations of other fine-structure lines for these galaxies to interpret the findings. PRIMA, together with ALMA, can bridge this gap between z=0 and z=6-10, where observing a large sample of [O III] and [C II] lines is either entirely impossible or in sparse redshift windows (around z~3) due to telluric transmission (Figure 1).

Furthermore, LMT/TolTEC is expected to build up a large sample of distant galaxies, selected based on mm colors, in the near future (Bryan et al. 2018). Optically dark galaxies of this sample, in particular, will benefit from followup by PRIMA. If AtLAST/LST become available in the 2030s (Kawabe et al. 2016, Klaassen et al. 2020), their submm-selected high-z galaxies would be perfect candidates for PRIMA. Their spectroscopic capability may also provide large samples of [O III]- and/or [C II]-emitters to follow up.

At shorter wavelengths, JWST will also provide a source of good candidates for PRIMA followup based on imaging surveys. For example, the highest redshifts where the 3.3µm and 6.2µm PAH features can be detected with JWST are z=7 and 3.5, respectively, by MIRI (c.f., Spilker et al. 2023). PRIMA will be sensitive enough to detect the other PAH features at longer wavelengths in order to investigate dust grain properties. Moreover, JWST has started to identify NIRCam-dropout galaxies due to their dusty nature and/or large distance (e.g., McKinney et al. 2023, Perez-Gonzalez et al. 2024). These sources are likely at z > 4 and are obvious targets to follow up with ALMA. However, only a small number of spectral features are accessible with ALMA, preventing a full picture of this interesting and potentially dominant population of galaxies which build up most of the mass in the Universe. PRIMA will be the only observatory that can access their mid-IR spectra.

## Description of Observations

FIRESS will be used to target galaxies that have already been observed with far-infrared emission lines (e.g., [O III]88, [C II]158) using, for example, ALMA, JWST, LMT/TolTEC, AtLAST/LST (and PRIMA imaging). Pointed observations with the low-resolution mode of FIRESS will be needed. It is essential to cover the full 25-250µm observed wavelength range to simultaneously study multiple spectral features. Thus, two spectral settings are required per pointing to cover all four spectral bands.

Using the baseline FIRESS sensitivity of 2e-19 W/m$^2$ (5-sigma, 1 hour) across the band, a practical strategy will be selecting relatively luminous galaxies at z > 4, while observing ``normal ’’galaxies on the star-formation main-sequence at z < 4. At z > 4, luminous galaxies ($L_{IR} \gtrsim$ 1e13 $L_\odot$) are perfect targets for PRIMA (< 1-3 hours for each spectral setting to obtain at least the five primary lines). For galaxies at z < 4, on average 10-30 of the normal galaxies can be observed within 50-80 hours in total (~1-5 hours for each spectral setting).

With the low-resolution mode, we note that the 12.7μm PAH feature can overlap in coverage with the spectral line of [Ne II] at 12.8 μm. In addition, [Ne V]14.3 can be blended with the [Cl II] line at 14.4 μm, as well as [O IV]25.9 and [Fe II]26.0. We can either use [Ne V]24.3, which will not suffer from blending, to identify an AGN, or use [Ne V]14.3 and [O IV] in conjunction with PAH emission to avoid misidentifying [Cl II] as [Ne V]14.3 and [Fe II] as [O IV]. The observations undertaken by the Spitzer/IRS spectrometer have provided a significant baseline of extragalactic observations in the local Universe useful for studying the relative strengths of the PAH features and spectral lines, and which will aid in simultaneously fitting these features and spectral lines in the low-resolution mode (Diaz-Santos et al. 2025). Significant advances in these techniques will be made in the near future with the JWST NIRSpec and MIRI IFUs, which offer a higher spectral and spatial resolution than Spitzer, allowing us to refine our existing understanding from IRS spectra, and we will be able to apply these lessons to programs designed for FIRESS (R~100).

Our program could be spread over 2-3 cycles (~200 hours each), where we could collaboratively exploit the observation scheduling and observing efficiency benefits using newly discovered sources in the PRIMAger surveys that could make the spectroscopic redshift and luminosity cut.

# 29. PRIMA Systematic Study of Particle Acceleration in Hot Spots of Radio Galaxies

Naoki Isobe (ISAS/JAXA), Takao Nakagawa (Tokyo City University), Motoki Kino (Kogakuin University of Technology & Engineering), Hiroshi Nagai (National Astronomical Observatory of Japan), Shunsuke Baba (ISAS/JAXA), Makoto Tashiro (Saitama University)

The first systematic far-infrared study with PRIMAger is proposed of hot spots associated with Fanaroff-Riley type-II radio galaxies. The hot spots are widely regarded as jet-terminal strong shocks where particles are efficiently energized via the standard diffusive shock acceleration. Several important roles played by magnetic field within the acceleration process are widely recognized. Therefore, in order to specify the acceleration condition in the hot spots, it is indispensable to precisely evaluate their magnetic field strength. Previously, a comparison of inverse-Compton X-ray flux density to synchrotron radio one has long been a standard tool for the magnetic field estimation. However, the method is known to be significantly subjected to the assumption on the synchrotron and inverse-Compton spectra, for which a simple power-law form is implicitly utilized. Instead, the present study employs as a more reliable magnetic-field indicator a synchrotron spectral feature called the cooling break, which is expected to be created by mutual balance between the adiabatic and synchrotron losses. For typical hot spots, the cooling break is predicted to be within the far-infrared range as $10^{11} - 10^{13}$ Hz. Thanks to its unprecedented sensitivity in $(1.1 - 12.5) \times 10^{12}$ Hz, PRIMAger is regarded as an ideal instrument to gauge the magnetic field of the hot spots through the cooling break. The study tries to conduct 5′ x 5′ and 10′ x 10′ mappings with the Hyperspectral and Polarimeter channels, respectively, onto well-studied hot spots. For this purpose, a total exposure of about 330 hours is requested.

## Science Justification

Compact hot spots, hosted by Fanaroff-Riley type-II (FR-II) radio galaxies, are widely associated with strong shocks at the terminal of their relativistic jets [1]. In the standard picture, the hot spots are regarded as an on-going particle-acceleration site via the diffusive shock acceleration [2]. Thus, they are proposed as one of the promising candidates for the origin of ultra-high energy cosmic rays with an energy of $\gg 10^{18}$ eV. An anisotropy in the ultra-high-energy-cosmic-ray arrival direction [3] is typically ascribed to radio galaxies [4]. However, their detailed acceleration conditions have remained under debate.

Within the particle acceleration process, magnetic fields are theoretically predicted to play several key roles. Namely, they support energy transfer from the plasma flow to individual particles, confine the particles within the energy-transfer area, and radiatively cool the accelerated particles. As a result, the relation between the magnetic field and size is frequently utilized to investigate astrophysical particle accelerators [5, 6]. Therefore, it is of prime importance to evaluate the magnetic field strength $\boldsymbol{B}$ in the hot spots.

For more than two decades after the advent of the Chandra observatory, inverse Compton (IC) X-ray observations have been a conventional tool to evaluate the magnetic field in the hot spots. The synchrotron radio intensity is proportional to the product between the electron and magnetic-field energy densities, while the IC X-ray intensity is determined by the product between the energy densities of the electrons and seed photons. Because the seed photons in the hot spots are typically dominated by the synchrotron radiation itself (the synchrotron-self-Compton process [7]), the radio-to-X-ray intensity ratio is regarded as a good indicator of the magnetic field. This method was widely applied to X-ray-detected hot spots [8, 9], and a significant energy dominance of the electrons over the magnetic fields has been suggested in some objects.

The magnetic-field evaluation by the above method implicitly assumes a simple power-law like spectrum. However, the synchrotron and IC spectra from the standard diffusive shock acceleration under the continuous energy injection condition are theoretically predicted to become a broken power-law form [10]. Therefore, the method is inevitably subjected to errors due to this spectral simplification. It is strongly requested to evaluate the magnetic field in the hot spots through different approaches.

The present study proposes the break of the synchrotron spectrum, usually known as the cooling break, as another indicator of the magnetic field. The cooling break is determined by a mutual balance between electron radiative cooling and adiabatic losses [11], and its frequency is described as $\boldsymbol{\nu_b} = \frac{27\pi e m_e v^2 c}{\sigma_T^2} \boldsymbol{B}^{-3} \boldsymbol{L}^{-2}$, where $e$ is the electron charge, $m_e$ is the electron mass, $v$ is the downstream flow velocity in the shock frame, $c$ is the speed of light, $\sigma_T$ is the Thomson cross section, and $L$ is the source size. Because the size of the hot spots is possibly measured from high resolution interferometric radio images, an observational constraint on the break frequency yields a reliable estimate on the magnetic field.

Figure 1 shows the magnetic-field dependence of the synchrotron spectrum from the hot spots. The cooling break is calculated from the above equation for the median size and radio flux of the well-studied hot spots [8], $L = 2$ kpc and $S_\nu(5\ \mathrm{GHz}) = 0.1$ Jy respectively. Because the FR-II jets are thought to be highly relativistic, the downstream flow velocity for the relativistic shock (i.e., $v = \frac{1}{3}c$ [12]) is adopted. The strong shock condition ($\alpha = 0.5$) is assumed for the spectrum below the break. Following the continuous energy injection condition, the spectral index is supposed to change by $\Delta\alpha = 0.5$ at the break [10]. The synchrotron spectra displayed in Figure 1 take into account the typical cut-off frequency for hot spots with good quality mid-infrared data, $\nu_c = 4 \times 10^{14}$ Hz [13].

The well-studied hot spots are implied to exhibit the magnetic field in the range of $B = 100 - 300$ μG [9]. Thus, the cooling break method was pioneeringly applied to the hot spot D of the prototypal FR-II radio galaxy Cygnus A, by utilizing the Herschel far-infrared data [14]. As a result, the magnetic field in this object was precisely measured as $B = 120 - 150$ μG. This field strength was found to be by a factor of two weaker than that from the radio-to-X-ray intensity ratio alone.

However, due to its insufficient sensitivity, the Herschel study has been limited to only a few hot spots [14, 15].

The present study [16] aims at the first systematic search for the cooling break from the hot spots of FR-II radio galaxies to measure their magnetic field, by making most of the far-infrared photometric capability of PRIMAger. A combination of its Polarimetric and Hyperspectral bands ensures a wide far-infrared spectral coverage in the frequency range of $\nu = (1.1 - 12.5) \times 10^{12}$ Hz (or the wavelength one of $\lambda = 261 - 24$ μm). As indicated in Figure 1, the cooling break of the typical hot spots ($B = 100 - 300$ μG) is expected to be located in or just below the PRIMAger frequency range. Figure 1 also plots the sensitivity in the Polarimetric and Hyperspectral bands to be achieved in the nominal 1-degree$^2$ survey with a total exposure time of 10 hours. By optimizing the survey area and exposure time to improve the sensitivity by several factors, PRIMAger is inferred to detect the far-infrared emission from the typical hot spots (as detailed below). Thus, these properties make PRIMAger the ideal instrument for the proposed investigation.

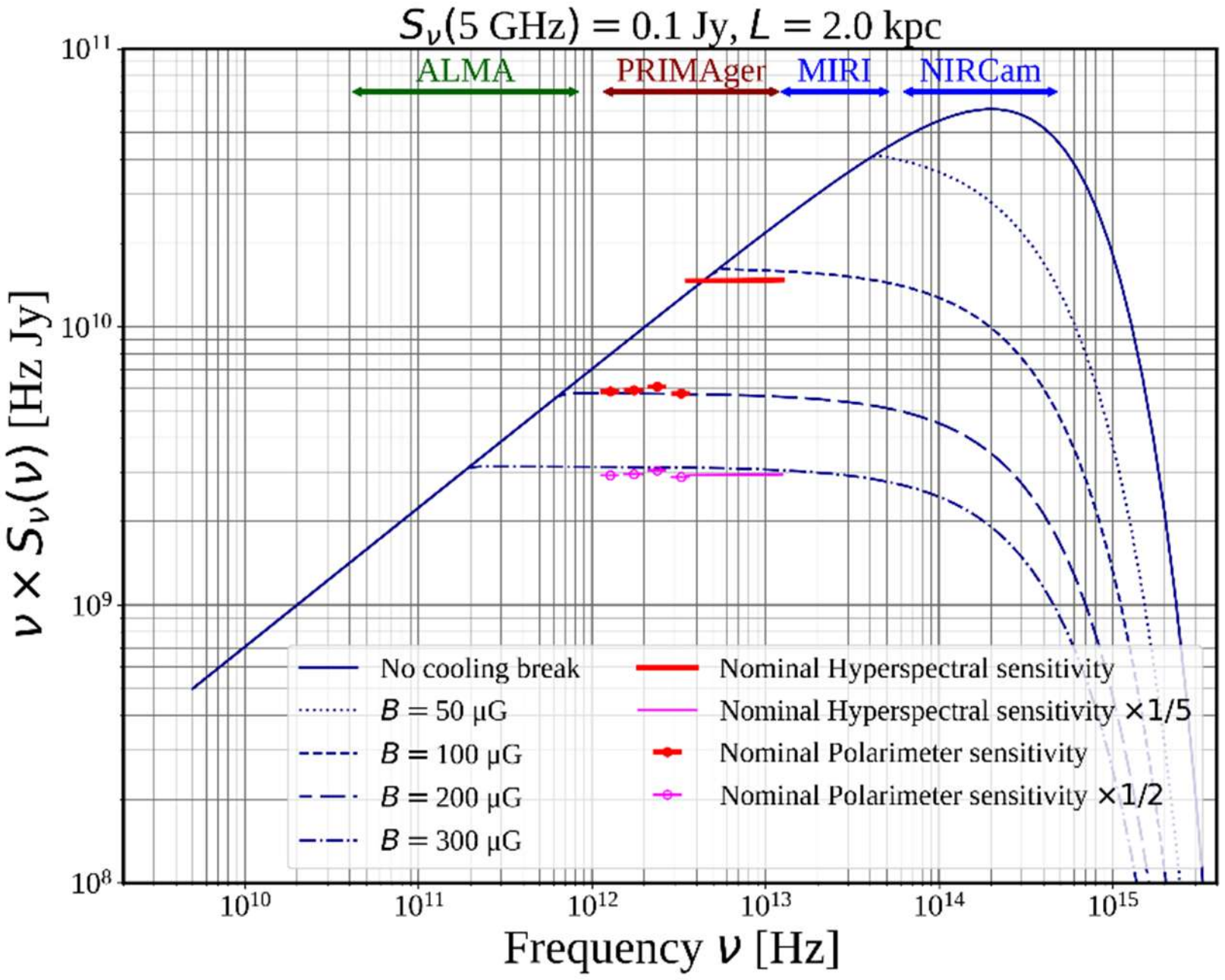

**Figure 1.** Synchrotron spectrum of hot spots, as a function of the magnetic field. The cooling break frequency is calculated by adopting $L = 2$ kpc and $v = \frac{1}{3}c$ as typical values. The radio flux density of $S_\nu(5\ \mathrm{GHz}) = 0.1$ Jy is employed.

## Instruments and Modes Used

PRIMAger mapping; 5'x5' Hyperspectral mapping on 49 objects and 10'x10' Polarimeter mapping (all 4 bands) on 39 objects. Polarimetric measurements are also helpful.

## Approximate Integration Time

The total PRIMAger integration time of 331 hours is requested, which is decomposed into 202 hours for the Hyperspectral band and 129 hours for the Polarimeter band. Among the proposed integration time, a higher priority is put on 62 hours to observe bright hot spots (32 hours for the Hyperspectral band and 29 hours for the Polarimeter one).

## Special Capabilities Needed

A small-area mapping mode for PRIMAger (e.g., 5'x5' and 10'x10' for the Hyperspectral and Polarimeter channels, respectively).

## Synergies with Other Facilities

In order to evaluate the cooling break frequency, radio data (especially interferometric ones) below the break is necessary. As the first step, the present study targets hot spots well-studied in the radio band. For systematic application of the cooling-break technique to a larger number of hot spots, a wide continuous spectral coverage for nearly 4 orders of magnitude enabled by a combination between ALMA, PRIMAger and JWST (see the arrows drawn in the top of Figure 1) is useful.

A combination of the cooling break method and conventional radio-to-X-ray ration one is expected to significantly improve the accuracy of the magnetic field estimation. This technique was demonstrated in the Herschel study of the hot spot of the FR-II radio galaxy Cygnus A [14]. In principle, by combining these two methods, it is possible to constrain another physical parameter additional to the magnetic field strength. One of the most important observationally unresolved parameters related to the cooling break is the downstream flow velocity $v$. Measurement of the flow velocity through the proposed method is expected to shed light on one of the very essential questions in the modern astrophysics, i.e., whether the FR-II jets are relativistic at the position of their terminal hot spots. Therefore, X-ray data with past, ongoing, and planned X-ray observatories will enhance the power of the proposed method [16].

## Description of Observations

The radio flux density, $S_\nu(5\ \mathrm{GHz})$, of candidate hot spots [8, 9, 13, 17, 18, 19] is plotted against their separation from the nucleus in Figure 2. The nucleus of FR-II radio galaxies tends to be brighter than their hot spots in the far-infrared range [14, 15]. In order to resolve them securely, their separation is requested to be larger than the PRIMAger beam size indicated with the thick vertical solid lines in Figure 2. In addition, for a precise evaluation of the contamination from the nucleus onto the hot spots, it is necessary to map them simultaneously. By considering the nuclear separation of the candidate hot spots ($\preccurlyeq 4.9$ arcmin), the smallest mapping area allowed for the Hyperspectral and Polarimeter bands (5' x 5' and 10' x 10', respectively, as of the submission of this document) is adopted.

Figure 1 indicates that the cooling break frequency is predicted to be lower, and correspondingly, the far-infrared flux density tends to be lower, as the magnetic field gets stronger. Hence, the upper-end value of the magnetic field among the well-studied hot spots ($B = 300\ \mu\mathrm{G}$ [9]) is adopted for the feasibility estimate, since the value is thought to give a reasonable lower limit on the far-infrared flux for individual objects. This means that the spectral shape displayed with the dash-dotted line in Figure 1 is utilized as the spectral template. This spectral template yields a radio-to-far-infrared flux conversion ratio as $S_\nu(92\ \mu\mathrm{m})/S_\nu(5\ \mathrm{GHz}) = 9.6 \times 10^{-3}$ and $S_\nu(24\ \mu\text{m})/S_\nu(5\ \text{GHz}) = 2.4 \times 10^{-3}$. By using the PRIMA Exposure Time Calculator, the far-infrared flux density detected with PRIMAger in 1 and 10 hours is estimated, and the corresponding radio flux density is plotted in Figure 2 with the thin and thick horizontal dashed lines, respectively.

For the hot spots located above the thin dashed lines in Figure 2 (32 and 29 ones for the Hyperspectral and Polarimeter bands, respectively), a 1-hour mapping is proposed. Those between the thin and thick lines (17 and 10 objects) require a mapping time of 10 hours.

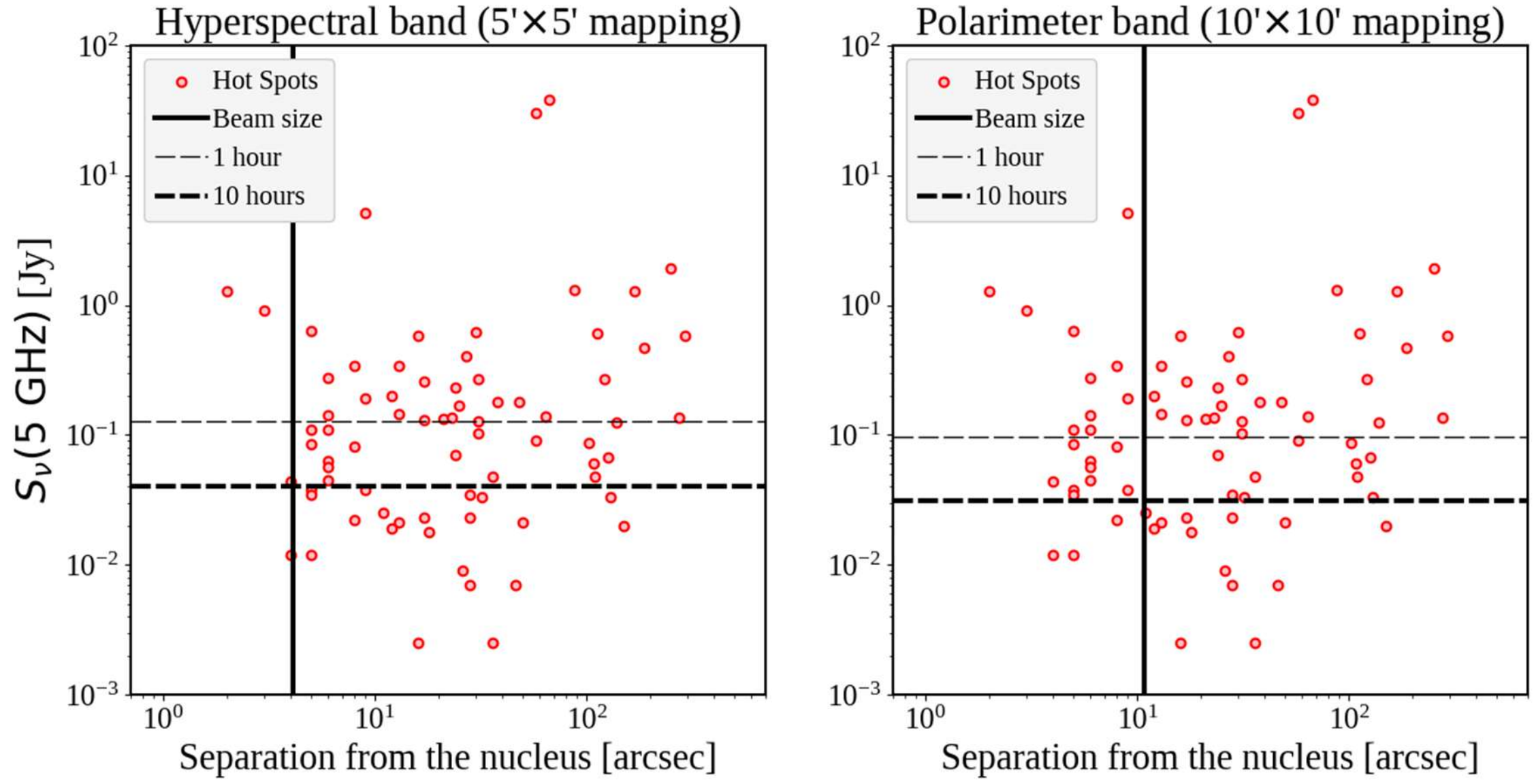

**Figure 2.** Relation between the radio flux density $S_\nu(5\ \text{GHz})$ of the candidate hot spots and their angular separation from the nucleus. The thick vertical solid line in the left and right panels, respectively, indicates the Hyperspectral and Polarimeter beam size at their shortest wavelength channels. For the adopted spectral template (the dash-dotted line Figure 1; $B = 300\ \mu\text{G}$), the radio flux density of $S_\nu(5\ \text{GHz}) = 0.1$ Jy corresponds to the far-infrared flux density of $S_\nu(24\ \mu\text{m}) = 0.24$ mJy and$S_\nu(92\ \mu\text{m}) = 0.96$ mJy at the Hyperspectral and Polarimeter channels, respectively. The thin and thick horizontal dashed lines in the left panel show the radio flux of which the corresponding far-infrared flux density is detectable with the 5' x 5' Hyperspectral mapping in 1 and 10 hours, respectively. Similar information for the 10' x 10' Polarimeter mapping is shown in the right panel.

# 30. A Far-Infrared Time-domain Survey of Nearby Galaxies with PRIMAger

Jacob Jencson (Caltech/IPAC), Kishalay De (Columbia), Viraj Karambelkar (Caltech), Mansi Kasliwal (Caltech), Ryan Lau (NOIRLab), Sam Rose (Caltech), Tom Steinmetz (CAMK Torun), Samaporn Tinyanont (NARIT), Charles Kilpatrick (Northwestern)

The infrared (IR) sky is replete with numerous classes of dust-forming stellar eruptions and explosions that occur over timescales from <days to centuries. These events play a critical role in the evolution of massive stars, and may contribute substantially to the build-up of dust and metals in the interstellar medium of galaxies (PRIMA Core Science Theme 3). The coldest and most heavily dust-obscured transients and variable stars are accessible only in the far-IR (>25 µm), but this region of the electromagnetic-spectrum remains almost entirely unexplored in the time-domain.

To uncover these events and quantify their contributions to galaxy dust budgets, we propose a far-IR time-series monitoring campaign of a sample of 35 nearby (<15 Mpc) star-forming galaxies with the PRIMAger Hyperspectral Imager (PHI; 24–84 µm). The proposed strategy utilizes the sensitivity and mapping efficiency of PHI and builds on the existing archives of Spitzer and Herschel imaging at these wavelengths. This provides immediate access with PRIMA to the 20–30 year timescales necessary to detect and characterize numerous classes of IR transients and variable stars.

## Science Justification

### Introduction

The life cycles of massive stars are remarkably dynamic, resulting in dramatic eruptions and explosions that play an outsized role in the chemical enrichment of galaxies. Often obscured by copious dust formation, time-domain infrared (IR) observations are vital to uncover these events. Despite recent developments in near- to mid-IR transient searches, the dynamic far-IR sky remains unexplored. Here, we propose a survey that capitalizes on the far-IR sensitivity and mapping capabilities of PRIMA to monitor nearby galaxies for densely obscured transients and variables. This search will reveal previously inaccessible phases of massive star evolution and enable a direct accounting of their contributions to the build-up of dust in galaxies (PRIMA Core Science Theme 3).

### Science Questions

*Dust in Core-collapse Supernovae: Core-collapse supernovae (CC SNe),* the explosive deaths of massive ($M \gtrsim 8\ M_\odot$) stars, are responsible for returning large reservoirs of energy and matter to the interstellar medium (ISM). Infrared observations that trace dust in SN explosions are particularly important to quantify the contribution of CC SNe to galaxy dust budgets (Dwek et al.

2007) and reveal the mass-loss histories of their progenitor stars via the detection of pre-existing dust in their circumstellar environments (Fox et al. 2011).

CC SNe have been considered promising sites for dust production for decades, especially considering the large quantities of dust seen in high-redshift galaxies in the early Universe (e.g., Cernuschi & Codina 1967). Theoretical work has indicated that significant quantities of dust (0.1–1 $M_{\odot}$) may condense in CC SN ejecta (e.g., P. Todini & A. Ferrara 2001; T. Nozawa et al. 2003, 2009). While early-time (<5 years) warm dust yields from Spitzer in the mid-IR were often orders of magnitude lower (e.g., Kotak 2009, Meikle et al. 2011), far-IR/sub-mm observations of SN 1987 and Galactic SN remnants revealed large reservoirs of cold dust formed on timescales of decades to centuries (e.g., Barlow et al. 2010, Matsuura et al. 2011, Temim et al. 2017).

Pre-existing dust around SNe may be radiatively or collisionally heated by the interaction of the SN shock with the circumstellar material (CSM) produced via the end-stage mass loss of the progenitor. These interactions can range from luminous Type IIn SNe with dense CSM from eruptive mass loss (Fox et al. 2011) to normal Type II SNe from red supergiant winds (Shahbandeh et al. 2023). Recent mid-IR observations have added to the growing classes of stripped-envelope SNe showing delayed interactions with massive, detached CSM shells (Meyers et al. 2024) likely produced via binary mass-transfer.

Efforts to disentangle the various origins and heating mechanisms of dust in SNe have been reinvigorated with the James Webb Space Telescope (JWST; Shahbandeh et al. 2025, Tinyanont et al. 2025, Zsíros et al. 2024). Still, far-IR observations on long timescales will be necessary to reveal cold (< 100 K) or optically thick dust formation that remains hidden shortward of 25 μm, and to probe distant CSM dust that maps further back (≳102-103 years) in the complex and diverse mass-loss histories of SN progenitor stars.

Dusty, Self-obscured Massive Variables and Transients: While many examples of luminous, dust-enshrouded variables stars are known (e.g., Humphreys et al. 1987, Jones et al. 1993, Thompson et al. 2009), the varied processes by which these systems eject enough material to encase themselves in massive CSM—and become the progenitors to strongly interacting, dust-forming CC SNe—remain poorly understood. Among the most famous examples is the exceptional η Carninae (η Car), a ~100 $M_{\odot}$ binary system that ejected ~10 $M_{\odot}$ during its explosive 19th century "Great Eruption" lasting ~20 years (Robinson et al. 1973, Smith et al. 2003). Optically obscured in the aftermath, η Car became the brightest extrasolar object in the IR sky (Westphal & Neugebauer 1969). Multiple physical mechanisms for the eruption have been proposed, including instabilities near the Eddington-limit (Humphreys & Davidson 1984) or a violent stellar collision (Smith 2011). A handful of analogous stars have been identified in nearby galaxies out to ~8 Mpc in Spitzer and Herschel 3–160 μm photometry, suggesting massive stars may enter these obscured states at a rate of 10–50% that of CC SNe (Khan et al. 2015a,b).

Numerous classes of self-obscuring events have been caught in action by recent transient searches in the near- to mid-IR (e.g., Kasliwal et al. 2017, Jencson et al. 2019b, De et al. 2020), including the luminous red novae (LRNe) associated with stellar mergers (Smith et al. 2016, Blagorodnova et al. 2017) and the intermediate-luminosity red transients (ILRTs) of debated origins—electron-capture SNe of super-AGB stars or non-terminal eruptions akin to η Car and the "SN impostors" (Jencson et al. 2019a, Rose et al. 2025). Far-IR observations beyond 25 μm over >decade timescales are necessary to reveal the most densely obscured members of these classes (e.g., the mergers of giant stars with enhanced mass-loss during an extended inspiral phase; Macleod et al. 2022, Karambelkar et al. 2025, Steinmetz et al. 2025), and to distinguish stars entering deeply embedded phases of their evolution from terminal fates, i.e., the direct collapse of massive stars into a black holes (Beasor et al. 2024, De et al. 2025).

### A Unique Role for PRIMA in IR Time-Domain Science

The efficient mapping capabilities and far-IR sensitivity of PRIMA open an entirely new regime of time-domain exploration. In particular, the science cases highlighted above require decades-long baselines at wavelengths >25 μm to probe the late-phases of dust-formation and reveal the cold, distant CSM dust in CC SNe, and to observe the most extreme processes by which massive stellar systems enshroud themselves in real time. Leveraging the archives of IR imaging of nearby galaxies with Spitzer and Herschel, these 20–30 year timescales will be immediately and uniquely accessible to PRIMA.

As shown in Figure 1 (and see Description of Observations), PRIMAger can efficiently map a sample of 35 nearby (< 15 Mpc) galaxies with available archival IR images at sufficient depth to detect a range of cold, dusty transients and variables. Specifically, these galaxies have hosted 23 known CC SNe in the last 30 years, providing a guaranteed science return of a large sample of late-time, cold dust constraints to quantify the dust yields of CC SNe and map their progenitor mass-loss histories. Similarly, this survey would observe an estimated 2–10 massive variables entering deeply embedded phases to quantify their rates and constrain the timescales and physical mechanisms of these transitions. A galaxy-targeted search also opens the potential for unanticipated discoveries, such as previously unknown decades- or centuries-old CC SNe entering new phases of interaction with distant CSM or entirely new classes of far-IR dominated transients not predicted by current stellar evolution models.

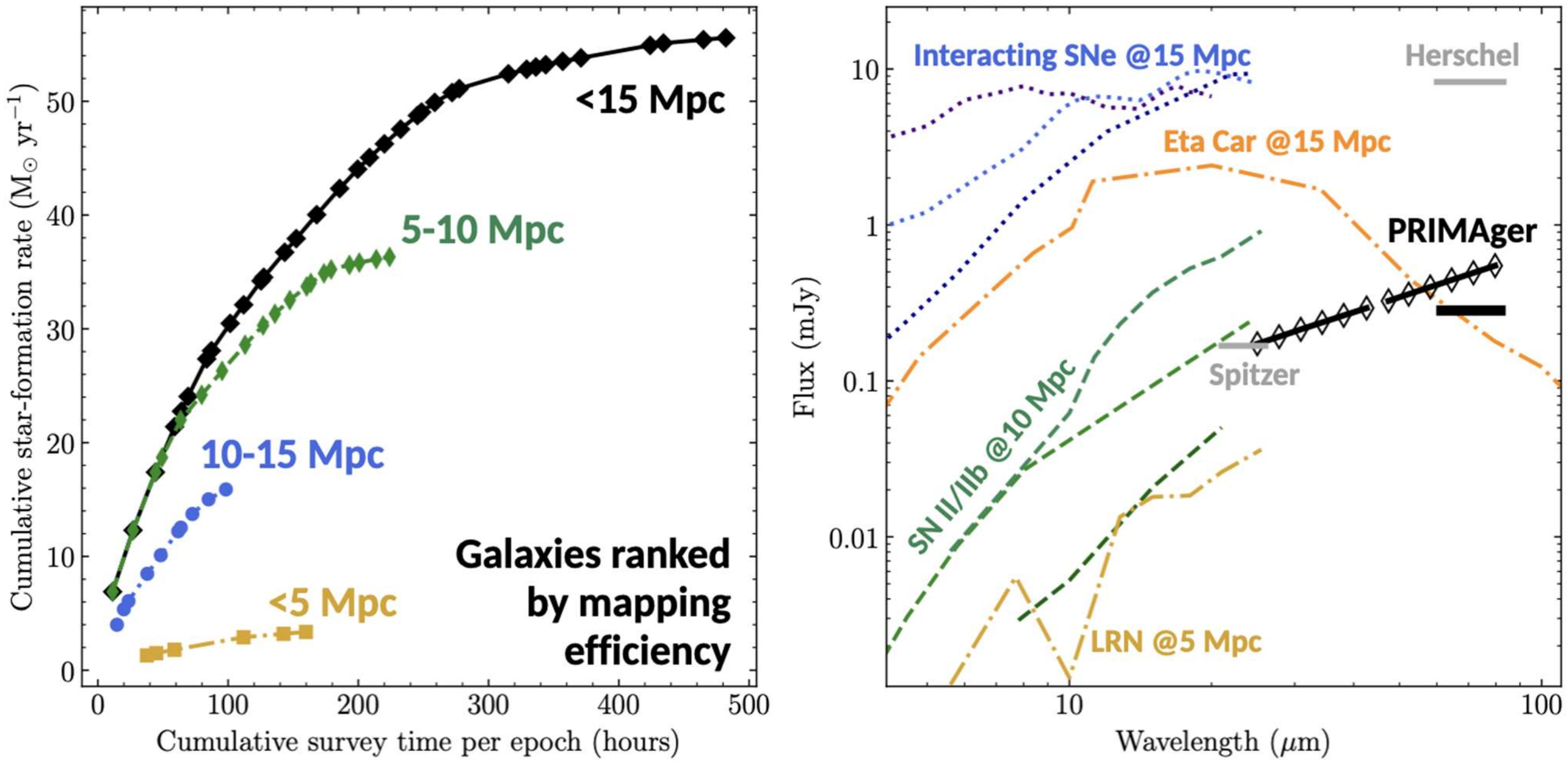

**Figure 1.** *Left*: Cumulative SFR probed in a survey of 35 galaxies within 15 Mpc (black diamonds) as a function of the time required to survey the sample. Galaxies are rank-ordered by their mapping efficiency (SFR/mapping time). Subsets of the sample within <5 Mpc, between 5–10 Mpc, and between 10–15 Mpc are shown as yellow squares, green thin diamonds, and blue circles, respectively. *Right*: Proposed 5σ depth of the PRIMAger PHI survey (black line with open diamonds) per resolution element, compared the available Spitzer/MIPS (24 μm) and Herschel/PACS (70 μm) imaging for this sample of galaxies. The PHI sensitivity binned to a comparable 70 μm bandpass to PACS is shown as the thick black line. Comparison SEDs of classes of transients and variables detectable at a range of relevant distances are shown as the multi-color curves (see Science Justification).

## Instruments and Modes Used

PRIMAGer mapping, hyperspectral band. One map for each of the 35 targets in the sample, observed on 1–3 epochs. Each map is ∼380 sq arcmin wide.

## Approximate Integration Time

480 hours per epoch for a sample of 35 galaxies.

## Special Capabilities Needed

For a cadenced survey, multi-epoch observations of each galaxy should be spaced approximately evenly throughout the 5 year mission duration. For example, for a 3 epoch survey, one observation per galaxy in each of years 1, 3, and 5.

## Synergies with Other Facilities

Long, decades-timescale baselines are provided by available Spitzer/MIPS 24 μm and Herschel/PACS 70 μm imaging for every galaxy in the sample. Additional IR time coverage may be available for a significant fraction of galaxies from JWST/MIRI and/or NIRCam imaging. Similarly, high-resolution archival HST imaging will provide valuable contextual information on the environments, progenitors, binary companions, and optical/UV counterparts of any transients and variables discovered by PRIMAger. Far-IR follow-up of select transients/variables may be conducted with PRIMA/FIRESS.

## Description of Observations

### Galaxy Sample

We propose to observe the sample of 35 star-forming within 15 Mpc with Spitzer/MIPS 24 µm and Herschel/PACS 70 µm imaging maps from the SINGS and KINGFISH projects (Kennicutt et al. 2003, 2011). The available IR maps provide the necessary initial reference epoch with 20–30 year time baselines to detect the slow evolution of the classes of IR transients and variable stars described above in the Science Justification. We focus on star-forming spiral galaxies to maximize the efficiency for discovering these transients and variables arising in massive stellar systems.

### Choice of Instrument/Mode and Required Sensitivity

We opt to use the PRIMAger Hyperspectral Imager (PHI) to take advantage of its efficient mapping capabilities and maximize our sensitivity to continuum sources at the wavelengths of interest for the transient search (24 and 70 µm). For a small 10' x 10' map, we can reach a 5σ sensitivity of 0.17 mJy at 24 µm to match the point-source depth of the available MIPS maps from SINGS with 14 hours of integration time. In the same integration time, we can reach a depth of 0.5 mJy per resolution element at 70 µm. And with spectral binning over 3 resolution elements, we can reach 0.28 mJy at 70 µm, ~30 times deeper than the available PACS imaging in a comparable bandpass. As a comparison, we estimate that reaching the equivalent line fluxes with FIRESS (binned to $\mathfrak{R}$=10) of 6.3–6.8 x $10^{-19}$ W $m^{-2}$ would require 150–180 hours for a 100 sq. arcmin map. We illustrate the proposed survey depths in comparison to the SEDs of several classes of transients and variables in Fig. 1 (right).

### Mapping Strategy

The galaxies in this sample span a range of sizes on the sky (~1–380 sq. arcmin), some of which are smaller than the minimum recommended map size of 10'x10'. To account for the reduced efficiency of these smaller maps (with one or both dimensions <10') in our integration time calculations with the ETC, we add 3.9' to the smaller dimension to account for the length of the PHI LVF detectors that must be scanned over the galaxy, and we add 1.8' to the larger dimension two account for gap between the two PHI detectors. We then scale the integration time from our fiducial calculation (14 h for 10'x10') with the area of the resulting map for each galaxy.

### Survey Time and Number of Epochs

As shown in Fig. 1 (left), the total time to survey all 35 galaxies is ~480 hours per epoch. Even with a single epoch, the availability of Spitzer and Herschel archival imaging for every galaxy in the sample immediately enables the decades-scale time-domain science that is the focus of this program. An additional 1–2 epochs over the 5 year prime mission duration would enable characterization of the ongoing evolution of detected sources, and maximize the potential to reveal unanticipated transients occurring on short timescales.

# 31. Contribution of Extremely Massive Black Holes with M > $10^{10}$ $M_\odot$ to Black Hole Accretion Rate at Cosmic Noon

Woong-Seob Jeong (Korea Astronomy and Space Science Institute [KASI], Korea), Seongjae Kim (KASI, Korea), Minjin Kim (Yonsei University, Korea), Alexandra Pope (UMASS, USA)

The co-evolution of supermassive black holes (SMBHs) and their host galaxies is a fundamental aspect of galaxy evolution, yet the growth mechanisms of the most massive SMBHs ($M_{BH}$ > $10^{10}$ $M_\odot$) at the cosmic noon remain poorly constrained. Hyperluminous Dust-Obscured Galaxies (DOGs) at this epoch are promising candidates for hosting these exceptionally massive, rapidly accreting SMBHs, likely fueled by gas-rich mergers. However, quantifying the contribution of these rare, extremely massive black holes to the total black hole accretion rate density (BHAD) at cosmic noon has been challenging due to their obscuration and the lack of comprehensive all-sky far-infrared data. The recently launched SPHEREx mission, with its all-sky near-infrared (NIR) spectral survey (0.75-5.0 μm), is expected to identify approximately 4,000 such hyperluminous DOGs by their distinctive red SEDs and broad emission lines (e.g., H$\alpha$, H$\beta$). Our initial study suggests that these extreme sources could account for over 15% of the BHAR at cosmic noon. While SPHEREx will provide an unprecedented catalog of such candidates, follow-up observations at the FIR range are crucial to confirm their obscured nature and accurately measure their bolometric luminosities and black hole accretion rates. This study aims to leverage SPHEREx-identified candidates to systematically investigate these extreme SMBHs through PRIMA follow-ups, ultimately quantifying their contribution to the BHAD at cosmic noon, a critical period for both galaxy and SMBH assembly.

## Science Justification

The co-evolution of supermassive black holes (SMBHs) and their host galaxies is a cornerstone of modern astrophysics, with strong observational evidence, such as the $M_{BH}$ -$\sigma_*$ relation, supporting their linked growth (Kormendy & Ho 2013). However, the detailed mechanisms governing SMBH formation and their vigorous growth phases, particularly for the most massive SMBHs, remain key open questions, even as studies span from cosmic dawn to cosmic noon. While SMBHs with masses in the range of $10^7$-$10^9$ $M_\odot$ have been extensively cataloged, for instance by the Sloan Digital Sky Survey (SDSS), the most massive black holes ($M_{BH}$ > $10^{10}$ $M_\odot$) are exceptionally rare and their evolution remains poorly understood at the epoch of "cosmic noon" (z ~ 1-3), a critical period when both the star formation rate density (SFRD) and SMBH accretion activity reached their peak (Madau & Dickinson 2014; Delvecchio et al. 2014).

### Hyperluminous Dust-Obscured Galaxies hosting Massive Black Holes

Dust-obscured galaxies (DOGs; Dey et al. 2008) residing around cosmic noon are considered as highly obscured galaxies with red [24 μm]/R-band color over ∼1,000. Given that [24 μm]-bright

DOGs show a high AGN fraction (often identified as power-law DOGs), these infrared-luminous DOGs are prime candidates for hosting highly obscured AGN, whose central engines are hidden by significant amounts of dust and gas. Our recent discoveries of hyperluminous ($L_{IR} > 10^{13} L_{\odot}$) DOGs hosting exceptionally massive black holes (e.g., $M_{BH} > 10^9 M_{\odot}$) indicate that such systems are effective tracers for identifying SMBHs with masses exceeding $10^{10} M_{\odot}$ (left panel of Figure 1). This aligns with models where rapid SMBH growth is fueled by gas-rich major mergers (Hopkins et al. 2008), conditions prevalent during the dense environment of cosmic noon. However, the contribution of these exceptionally massive black holes to the total black hole accretion rate density (BHAD) at cosmic noon remains unquantified. This is primarily due to their intrinsic rarity and the lack of comprehensive all-sky far-infrared (FIR) data necessary to pierce through the obscuring dust and characterize their accretion.

### Pre-selected Massive Black Holes with SPHEREx

The recently launched SPHEREx mission, with its all-sky near-infrared (NIR) spectral survey (0.75-5.0 μm), is poised to identify a significant population of candidates for such hyperluminous, obscured AGN. SPHEREx will detect these sources through their characteristic red SEDs and, where present, broad emission lines (such as H$\alpha$ and H$\beta$, redshifted into the SPHEREx bandpass at cosmic noon) indicative of AGN activity. As we have already found three hyperluminous DOGs hosting $M_{BH} > 10^{10} M_{\odot}$ in 12 deg$^2$, we expect the total number of those hyperluminous DOGs will be approximately 4,000 in all-sky NIR spectral survey. It also suggests that those extreme sample could explain more than 10% of BHAR at cosmic noon (right panel of Figure 1). While SPHEREx's 102 NIR spectral bands will be powerful for identifying candidates for these dust-reddened AGN, definitively confirming the level of obscuration and characterizing their full energy output (which peaks in the FIR for heavily obscured systems) requires complementary observations. This is where PRIMA's unique capabilities become critical. With superior sensitivity at far-infrared, PRIMA can follow up SPHEREx-selected obscured massive black hole candidates. Therefore, by leveraging SPHEREx as a "pathfinder" to create an unprecedented catalog of candidates, our scientific goal is to use PRIMA's superior far-infrared sensitivity to follow up these SPHEREx-selected obscured massive black hole candidates. This will enable us to systematically measure their black hole accretion rates (BHAR) and thereby quantify the contribution of these extreme SMBHs ($M_{BH} > 10^9 M_{\odot}$) to the total black hole accretion rate density (BHAD) during the peak epoch of cosmic activity.

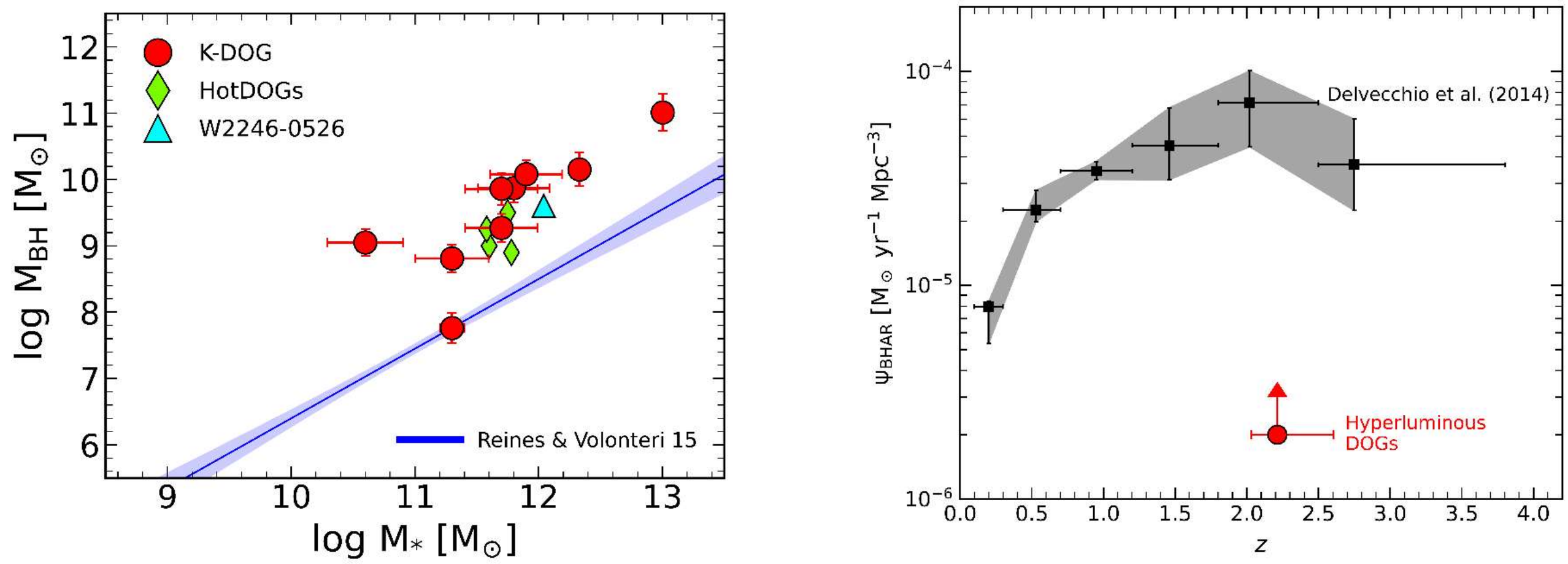

**Figure 1.** (Left) Black hole mass as a function of stellar mass from our hyperluminous DOG samples in red circles (Jeong et al., in prep.; Kim et al., submitted) and Hot DOGs in green diamonds (Assef et al. 2015; Wu et al. 2018). W2246-0526 is the most luminous Hot DOG target with bolometric luminosity of $\sim 4\times10^{15}$ $\mathbf{L}_{\odot}$ (Tsai et al. 2018). Most of hyperluminous DOGs and Hot DOGs locate over the scaling relation of local AGNs from Reines & Volonteri (2015), suggesting the overmassive evolutionary track of SMBHs at least hyperluminous galaxies. (Right) Black hole accretion rate density (BHAD) as a function of redshift. Black squares represent BHAD traced with Herschel from Delvecchio et al. (2014). Hyperluminous DOGs from our sample, a red circle, contributes at least ~10% to IR BHAD. From synergies with SPHEREx and PRIMA, we will conclude the accurate BHAD contribution of extremely massive black holes.

## Instruments and Modes Used

This science case will use observe 100 targets with PRIMAger.

## Approximate Integration Time

Estimation can be based on PRIMA ETC, or other method (please specify). Do not include overheads (slews, calibration, etc.)

We estimated the approximate number of hyperluminous DOGs hosting $M_{BH} > 10^{10}$ $M_{\odot}$ in all sky is 4,000 excluding galactic plane within |b| < 20 deg. According to Li et al. (2024) ($\phi_{BH}$ ~$2\times10^{-8}$ $Mpc^{-3}$ $dex^{-1}$ at z=4), the expected number of massive BHs with $M_{BH}$ ~ $10^{10} M_{\odot}$ is ~1.6 at a corresponding volume, when we consider the black hole mass function (BHMF). Considering the cosmic variance of the rare population like our extreme mass of BHs, we used the conservative number of massive BHs as ~2,500. From the PRIMA Exposure Time Calculator with polarimetric image mode, we set survey area as the minimum map size of polarimetric imager as 100 $arcmin^2$ and 5σ depth as 2.5 mJy, 3.5 mJy, 5.0 mJy, and 7.0 mJy for polarimetric imager 92 μm, 126 μm, 172 μm, and 235 μm, respectively. It yields 0.1 hours for each target. Thus, the total exposure time is 250 hours (2,500 targets × 0.1 hours). Similarly, for BHs with a mass of ~$10^{10.5} M_{\odot}$, the expected number is ~0.25 ($\phi_{BH}$ ~$1\times10^{-9}$ $Mpc^{-3}$ $dex^{-1}$ at z=4) within the same survey volume as the $10^{10} M_{\odot}$ case, implying a total of a few hundred (N~270). Since the pointing direction between polarimetric imager and hyperspectral imager are expected to differ within the minimum map size, we also perform hyperspectral imaging to obtain complete spectral energy distribution (SED) for ~100 targets, most massive or most luminous galaxies. These highly reliable sources will be selected from the parent sample of N ~270 based on redshift and SED shape, thereby characterizing these extreme population. The minimum map size is 25 arcmin$^2$ and 24μm 5σ depth as 0.6 mJy. As each exposure takes 0.3 hours, the total exposure time is 30 hours for hyperspectral imager (100 targets × 0.3 hours). Thus, we request 280 hours for the extremely massive BHs to investigate their contribution to BHAD at cosmic noon (250 hours for polarimetric imager & 30 hours for hyperspectral imager).

## Special Capabilities Needed

None

## Synergies with Other Facilities

### SPHEREx, WISE

As most of hyperluminous infrared galaxies were detected in WISE, we can fill up the gap between SPHEREx and PRIMA using ALLWISE data. PRIMA will provide deep targeted follow-up for SPHEREx-selected obscured AGNs and hyperluminous galaxies, measuring black hole accretion rates of ultra-massive black holes across cosmic noon.

## Description of Observations

We will carry out targeted observations using PRIMAger in both polarimetric imaging mode and hyperspectral imaging mode to follow up SPHEREx-selected hyperluminous dust-obscured galaxies (DOGs) hosting supermassive black holes (SMBHs) with masses exceeding $10^{10}$ $M_{\odot}$. Our goal is to constrain their contribution to the total black hole accretion rate density (BHAD) during cosmic noon (z ~ 1–3).

For the polarimetric imaging mode, we will observe approximately 2,500 targets, selected from SPHEREx, covering the minimum map size of 100 arcmin². We require a 5σ depth of 2.5 mJy (92 $\mu$m), 3.5 mJy (126 μm), 5.0 mJy (172 μm), and 7.0 mJy (235 μm) per field, which corresponds to 0.1 hours of integration time per source. This yields a total of 250 hours of observing time in this mode. These observations will allow us to detect the FIR emission from dust-obscured accreting SMBHs and robustly estimate their BHAR. To construct full IR SEDs, we will conduct hyperspectral imaging on a subset of 100 extreme objects (most massive or most luminous), using the minimum map size of 25 arcmin². We target a 5σ depth of 0.6 mJy at 24 μm, which corresponds to 0.3 hours per source, resulting in a total of 30 hours for this mode. These data will provide crucial diagnostics of dust properties, SED shapes, and enable more accurate BHAR estimates via multi-band SED fitting. Our requested depth for hyperspectral band (24 μm) and polarimetric band in Figure 2.

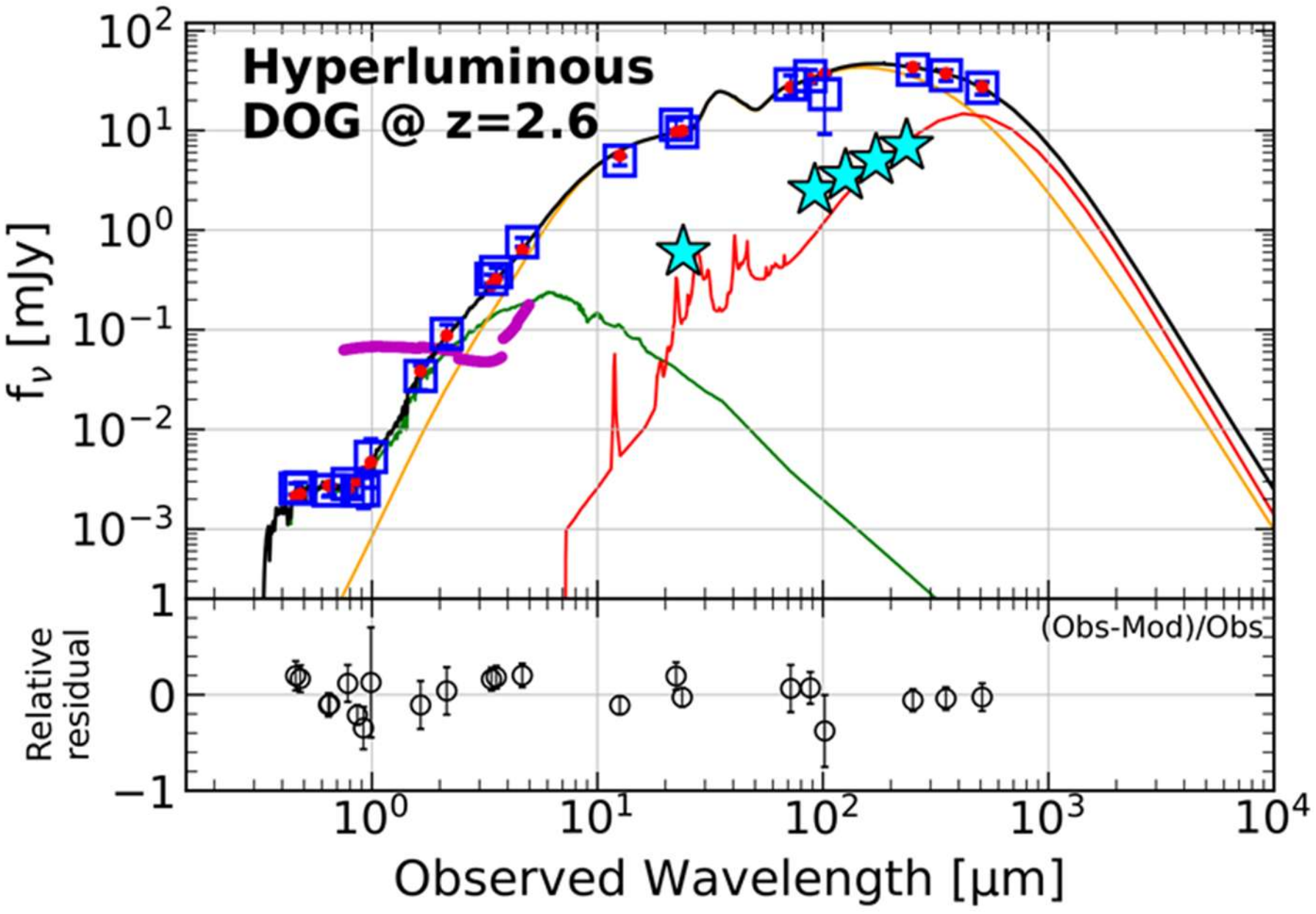

Figure 2. An example SED of a hyperluminous dust-obscured galaxy at z=2.6 (Kim et al, 2025, submitted). Blue squares denote the observed points, while violet points and cyan stars represent the expected and the requested depths with SPHEREx and PRIMA, respectively. Note that the emission lines used to identify extreme BHs from SPHEREx (e.g., H$\alpha$) are located at $\lambda \gtrsim 2\mu$m. Green, orange, and red SEDs are stellar, AGN, and dust components, respectively

# 32. A Survey of Intergalactic Warm Molecular Gas with PRIMA

Hiroyuki Kaneko (Niigata University, Japan)

Not only within a galaxy, but also in the intergalactic space of nearby galaxy groups and interacting galaxies, there is warm molecular hydrogen gas at temperatures of about 1000 K. The warm molecular gas is thought to be ejected from the galaxies and compressed by galaxy interactions. These results suggest that the warm molecular gas in intergalactic space may be gas ejected from galaxies and compressed by intergalactic interactions. However, there are a few examples of spatially resolved observations of warm molecular gas in intergalactic space. As a result, our understanding of its physical environment, including its mass, temperature, and distribution, is poor. Many questions remain unanswered, such as how it coexists with the high-temperature plasma in intergalactic space and by what mechanism it is heated.

By taking advantage of the wide field and high sensitivity slit spectroscopy of PRIMA /FIRESS Band 1 to image and observe $H_2$(0-0)S(0) emission lines (rest wavelength: 28.2 μm) in different environments in the local Universe at z<0.15, we will quantitatively understand how warm molecular gas affects galaxy evolution.

## Science Justification

Galaxy clusters are known to be filled with high-temperature gas plasma. High-temperature intergalactic gas prevents the selective stripping of HI gas and the accretion of cold gas, thereby suppressing star formation in clusters compared to field galaxies. These differences in galaxy environments are also reflected in the morphology of galaxies, which has long been known to evolve in high-density environments such as clusters and groups, leading to an increase in the number of elliptical galaxies with low gas content in the number of elliptical galaxies with low gas content (Dressler 1980). This suggests that intergalactic gas plays an important role in the evolution of clusters and galaxies.

X-ray observations have shown that the intergalactic high-temperature gas contains oxygen, nitrogen, and iron. This suggests that elements synthesized by supernova explosions in the galaxy have been ejected into intergalactic space. Possible mechanisms for the supply of such gas include outflows from galaxies, AGN jets, and galactic collisions. In fact, cold and hot gas associated with such structures has been observed.

On the other hand, the photoionization regions formed in starbursts, which are also the driving force behind these phenomena, shocks associated with galaxy collisions, or high-temperature environments can excite molecular hydrogen. Although molecular hydrogen is the most abundant molecule in the Universe, it is usually not directly observable at low temperatures due to its lack of a permanent dipole moment. However, in these extreme environments, pure rotational transitions of molecular hydrogen can be observed in the form of thermal or fluorescence emission; using PRIMA/FIRESS, the rotational emission line $H_2$(0-0)S(0) (28.2 µm, excitation temperature of 510K) can be observed from sources at z=0–0.175.

The first observations of $H_2$ rotation emission lines for galaxies were detected from NGC 6946 using ISO/SWS (Valentijnet al. 1996). After this, most observations continued to come from single-point observations of galaxies (Ogle et al. 2007; Egami et al. 2006; Ogle et al. 2012), with imaging observations coming from Stephan's Quintet (compact galaxy group: Appleton et al. 2017) and NGC 4258 (Seyfert galaxy: Ogle et al. 2014), Taffy I (colliding galaxies: Peterson et al. 2012), and few others.

These observations show that warm $H_2$ molecules are indeed present in intergalactic regions, along with outflows, AGN jets, galaxy collisions, and other galactic activity. In addition, these warm $H_2$ gas coexist with low-temperature molecular gas and star-forming regions. The physics that makes this coexistence possible, and the quantitative study of how much of the gas in galaxies is ejected into intergalactic space, providing high-temperature gas and plasma, are still poorly understood.

We propose to use PRIMA/FIRESS Band 1 to perform spatially resolved spectroscopic imaging of rotationally excited emission lines of warm $H_2$ molecules, which have been rarely observed in the past, for various environments ranging from galaxy clusters to galaxy groups to explore the spatial distribution of their physical states and their relation to galaxy morphology and star formation. Galaxy activity, such as AGN jets and galaxy collisions, also depends on the galaxy environment. For example, in high-density environments such as compact galaxy groups, the effects of galaxy collisions are expected to be stronger. On the other hand, in a galaxy cluster, the effect of the suppression of star formation activity in the environment due to the stripping away of cold HI gas by ram pressure is stronger than the increase in star formation activity due to frequent galaxy collisions. Understanding whether warm $H_2$ molecules exist in such different environments, and if so, how they affect their surroundings, is intrinsically important in understanding galaxy evolution.

The amount and distribution of warm $H_2$ gas in galaxies and intergalactic space, for which only a limited sample exists, can be spatially resolved by observations with PRIMA/FIRESS at realistic times. In combination with JWST/MIRI imaging observations of highly excited warm $H_2$ emission lines, the heating mechanism (temperature and velocity of shocks) can be identified, although model-dependent, since $H_2$ emission lines have a very short cooling time and require in situ heating. Further analysis of similar observations in different high-density regions (galaxy clusters and groups) could provide insight into the evolution of warm intergalactic material in intergalactic space and thus into the evolution of galaxies.

## Instruments and Modes Used

This proposed observation will be performed by using the FIRESS spectrometer, in the low-resolution mapping mode. The typical size of targets (galaxy groups and clusters at z<0.175) is 5’ x 5’.

## Approximate Integration Time

For observing $H_2$(0-0)S(0) emission (rest wavelength of 28.2 μm), we select low-resolution mapping mode with FIRESS Band 1. Assuming a survey area of 25 $arcmin^2$ and a mapping sensitivity of 8 x $10^{-19}$ ($W/m^2$, 5σ), the required observation time is 28 hours per source based on PRIMA ETC.

## Special Capabilities Needed

None

## Synergies with Other Facilities

This proposal aims to clarify the heating mechanism of molecular gas in intergalactic space by taking advantage of PRIMA's imaging capabilities to map warm $H_2$. However, synergies with other facilities can provide a comprehensive understanding of the physics of the multi-phase interstellar/intergalactic medium.

The distribution of warm molecular gas, cold molecular gas, and cold atomic gas will be investigated by HI observations (SKA/ngVLA) and multi-J CO observations (ALMA). We can compare the results with models such as CLOUDY and MAPPING III, as well as other models. It will greatly improve our understanding of shocks and energetics in mergers, galaxy groups, and galaxy clusters.

## Description of Observations

We propose low-resolution mapping observations with FIRESS Band 1 of $H_2$(0-0)S(0) emission lines in intergalactic space from relatively nearby to around z = 0.15, especially for galaxy clusters to galaxy clusters in medium to high density environments. Nearby galaxy clusters typically have a size of 5 arcmin x 5 arcmin. Since the source is spread out, the observational limit is determined solely by the areal luminosity sensitivity: according to Appleton et al. (2017), the $H_2$(0-0)S(0) emission line intensity at Stephan's Quintet is >8 x $10^{-19}$ ($W/m^2$) per 7.4” x 7.4”. Assuming a survey area of 25 $arcmin^2$ and a mapping sensitivity of 8 x $10^{-19}$ ($W/m^2$, 5σ), the required observation time is 28 hours per source. For example, if we set the total observing time limit to 1000 hours, 35 galaxy groups can be observed with this setting, which can be discussed in terms of the statistical features of warm $H_2$ gas in galaxy groups.

# 33. Crystals in the Interstellar Medium of Star-forming Galaxies at Cosmic Noon

Francisca Kemper (ICE-CSIC / ICREA / IEEC), Frédéric Galliano (CEA-Saclay), Janet Bowey (Cardiff), Howard A. Smith (Harvard-Smithsonian CfA), Hendrik Linz (MPIA), Henrik Spoon (Cornell), Paul van der Werf (Leiden Observatory), Yiqing Song (ESO-Chile / JAO)

Silicates are the most common form of dust in the universe. In the interstellar medium of our Milky Way they are found to be amorphous, while stellar ejecta contain significant fractions of crystalline silicates. Crystallization occurs through thermal annealing, while cosmic ray hits are thought to cause the amorphization of crystalline material. Nearby starbursting galaxies are reported to contain crystalline silicates, presumably recently formed by massive, rapidly evolving, stars. We thus hypothesize that measuring the crystalline fraction of silicates in starbursting galaxies helps to constrain galactic properties such as the star formation rate, the recent star formation history and the cosmic ray fluence. With its spectral coverage from 24 to 235 micron, the FIRESS spectrograph on PRIMA is uniquely capable of observing redshifted crystalline silicate features, which can be found at a rest wavelength from 8 to 72 micron. We propose to measure the crystalline silicate content of a sample of 230 starforming galaxies from $1.6<z<4$, bracketing the peak of cosmic star formation, also known as Cosmic Noon, in order to characterize the dust formation and evolution process in these galaxies, related to their star formation activity.

## Science Justification

Silicates are the main constituent of the interstellar dust in the Milky Way and most other galaxies. In the Milky Way, they are found to be almost completely amorphous, with the crystallinity fraction lower than the detection limit of 2% of the silicate mass (Kemper et al., 2004), while stellar ejecta may contain considerable amounts of crystalline silicates. It is thought that amorphitization occurs upon impact of cosmic ray hits, while crystalline silicates are formed through thermal annealing of amorphous silicates, or direct condensation at high temperatures, typically in the dust formation zone around mass-losing stars. Detecting crystallinity thus traces the balance between dust injection by stars in the ISM, the cosmic ray fluence, and dust destruction in supernova shocks, and is a useful tracer of starburst activity (Kemper et al., 2011). Indeed, Spitzer has revealed that many actively starforming galaxies host detectable amounts of crystalline silicates (e.g. Spoon et al., 2006, 2022).

Amorphous silicates show two broad resonances in the infrared at 9.7 and 18 micron due to Si-O stretching and O-Si-O bending modes in the silicate tetraedra. In the spectra of crystalline silicates these modes are split up in a number of narrow and sharply peaked resonances, while resonances at longer wavelengths appear due to lattice (phonon) modes. The longest wavelength feature due to crystalline silicates detected in astronomical objects is the 69 micron forsterite ($Mg_2SiO_4$) feature, which shifts to longer wavelengths (~72 micron) with the replacement of a small fraction of the Mg by Fe (Molster et al. 2002b). The laboratory measurements of crystalline silicates show a width of the spectral features of the order of 1-2 micron, and thus the spectral resolution of R~90-130 will resolve and detect a 1 micron wide 69 micron feature.

In their recent study, Spoon et al. (2022) catalogued the presence of extragalactic crystalline silicates in more than 700 relatively nearby galaxies using the Spitzer archive. As of yet, only very few extragalactic crystalline silicate detections have been reported in JWST observations (Rich et al. 2023, Ramos-Almeida et al. 2025).

JWST has a wavelength coverage up to 28 micron, and is not suitable to detect crystalline silicate features at redshifts z>1.5, despite its superb sensitivity in the mid-infrared. With a wavelength coverage of up to 38 micron for Spitzer, the situation was only marginally better for that observatory. Indeed, even at the rest wavelength, there are several important crystalline silicate features present between 28 (or 38) micron, including the strong 33 micron forsterite and 43 micron enstatite features and the diagnostic 69 micron forsterite feature (Molster et al. 2002a). Thus, discovery space for PRIMA, with FIRESS operating from 24 to 235 micron, can be found in exploring the crystallinity of silicates in galaxies in the era of peak star formation ($1 \lesssim z \lesssim 3$; Cosmic Noon).

Since crystalline silicate features are intrinsically much stronger than the features due to amorphous silicates, small amounts, of the order of a few percent, can be detected whenever the resonances due to amorphous silicates are visible. The dedicated survey proposed here will complement the data already available in the Spitzer archive. Spoon et al., (2022) have analysed 3335 Spitzer spectra of galaxies, of which 350 have a redshift $1<z<2$ and 204 have a redshift $z>2$. In order to complement this dataset in a meaningful way, and assuming a detection rate of crystalline silicates of 20%, we want to add 230 sources with a redshift of $1.5 \lesssim z \lesssim 4$, to bracket Cosmic Noon, but at the same time to not have overlap with what is already in the Spitzer archive.

The observations will allow us to measure the feature strength of crystalline silicates, with several features being present over the wavelength range covered. We can thus measure the occurrence and degree of crystallinity at Cosmic Noon, during the peak of star formation, study the mineralogical composition, and relate the measured parameters to characteristics of the galaxies, such as star formation rate, and cosmic ray fluence.

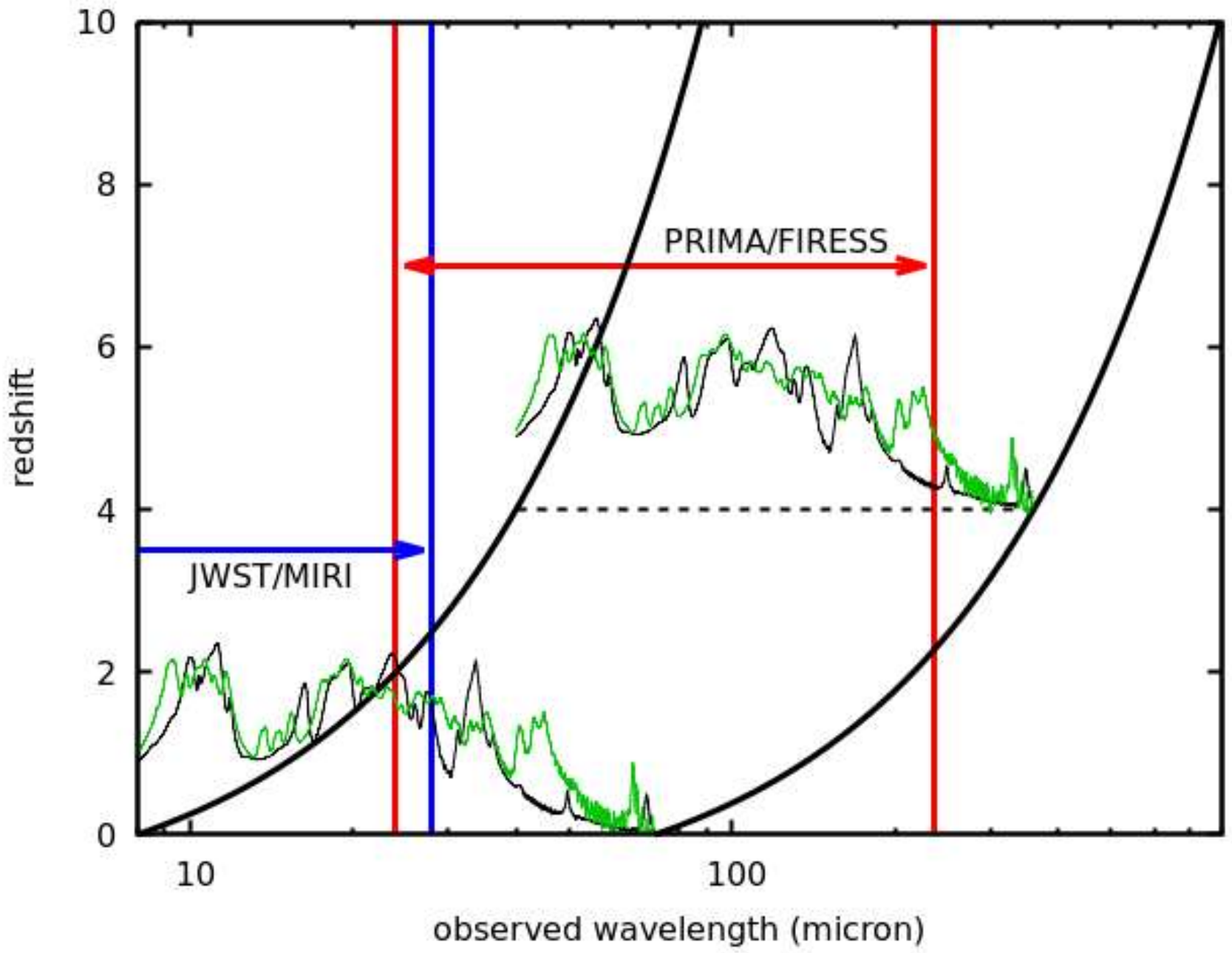

**Figure 1.** Discovery space of the FIRESS instrument on PRIMA when surveying extragalactic crystalline silicates. As a function of redshift, the wavelength range over which crystalline silicate features can be observed (demarcated by the two black curves, representing rest wavelengths of 8 and 72 micron) is shown. The two red lines show the observable range for the FIRESS instrument, and the blue line shows the long-wavelength limit of the JWST-MIRI instrument. From this plot is clear that the FIRESS wavelength coverage is ideally suited to detect crystalline silicates at Cosmic Noon (around z=2-3), and that it is uniquely capable of detecting the crystalline silicate features at wavelengths longer than 28 micron at any redshift. Laboratory spectra of crystalline silicate species forsterite (black; Koike et al. 1999) and enstatite (green; Koike et al. 1999) are shown for reference at the bottom of the plot at z=0 and in the middle of the plot at z=4.

## Instruments and Modes Used

This science case will make use of FIRESS' pointed low-resolution mode.

## Description of Observations

Using model spectra calculated by Dale & Helou (2002) it is estimated that a $10^{12}$ L⊙ normal star-forming galaxy placed at z=1.6 will have a flux density of $2 \times 10^{-19}$ W/m2 at 100 micron, corresponding to a rest wavelength of 38.5 micron (Figure 12, Farrah et al. 2019). This is a good tradeoff between the rising spectral energy distribution of thermal dust emission, and the strength of the crystalline silicate features compared to the opacity in the amorphous component. We aim to detect crystalline silicate features that are 10% above continuum, thus requiring the detection of $1 \times 10^{-19}$ W/m2. According to the PRIMA exposure time calculator, the integration time needed to obtain a full spectrum is ~7 hours. For brighter galaxies with luminosities of at least $10^{13}$ L⊙, integration times become significantly shorter, and a full spectrum can be obtained in 10 minutes at z=1.6. A z=4 galaxy is a factor of 3 more distant than a z=1.6 galaxy (in luminosity distance, assuming ΛCDM; Wright 2006), requiring 0.8 hours per $10^{13}$ L⊙ galaxy.

Assuming a typical value of 0.5 hours for the required integration time over the 1.6<z<4 redshift range, 200 galaxies with a luminosity of $10^{13}$ $L_\odot$ galaxies can be done in 100 hours. Additionally, we will also add smaller number (30) of $10^{12}$ $L_\odot$ galaxies with redshifts z<2, yielding an additional 200 hours of observing time, for a total of 300 hours of observing time.

# 34. A PRIMAger Survey to Investigate the Evolution of AGN Contribution to the Total Infrared Luminosity up to z ~ 4

Ece Kilerci, Department of Astronomy and Space Sciences, Science Faculty, Istanbul University, Beyazit 34119, Istanbul, Turkiye, Tomo Goto (National Tsing Hua University), Seong Jin Kim (National Tsing Hua University), Tetsuya Hashimoto (National Tsing Hua University)

The overall luminosity and properties of galaxies change over time due to the linked evolution of Active Galactic Nuclei (AGN) and star formation activity. Therefore, the contribution of AGN to the total infrared luminosity ($L_{\rm IR}$) is a tracer for galaxy evolution stages at different redshifts. An increasing AGN contribution trend with redshift is reported in the pre-JWST era. We propose a PRIMAger survey to investigate the evolution of AGN contribution to $L_{\rm IR}$ up to redshift 4. The proposed survey area includes more than 200 robust candidates from the most complete JWST-identified AGN sample to date at z ∼ 0–6 range. With PRIMAger PPI1 band detections down to 0.04 mJy, we estimate a total of 114 PRIMA AGN detections, 87 of which will be new, 3 times more than previous Herschel detections. PRIMA data will allow us to constrain the dust spectral energy distribution and measure accurate $L_{\rm IR}$ and AGN contribution up to z ~ 4. As a result of this survey, we will be able to examine the evolution of AGN contribution up to a higher z limit, for a larger sample with accurate measurements, and reach a firm conclusion.

## Science Justification

Active Galactic Nuclei (AGN) activity and star formation are fundamental processes that influence the co-evolution between Super Massive Black Holes (SMBHs) and their host galaxies. The interplay between AGN and star formation activity over time influences the evolution of galaxies. The evolutionary stages of galaxies across different cosmic epochs can be inferred by the contribution of AGN to the total far-infrared (FIR) luminosity (e.g., Symeonidis & Page, 2021).

The AGN contribution to the total IR luminosity (measured between 8-1000 micron), $f_{\rm AGN}$, is the ratio between AGN luminosity ($L_{\rm AGN}$) and total infrared luminosity ($L_{\rm IR}$). Although an increasing trend of $f_{\rm AGN}$ with z is reported in the literature (e.g., Wang et al. 2020), recent studies covering the lower-luminosity AGN discovered by JWST, do not fully support this trend (Chien et al. 2024). Since $L_{\rm IR}$ is measured from the spectral energy distribution (SED) analysis based on the extrapolation of the near-IR and mid-IR fluxes at much shorter wavelengths, without FIR detection (near the dust peak), $L_{\rm IR}$ measurements have a large uncertainty (e.g., Zavala, APJ, 2023). Therefore, FIR measurements around 100 $\mu$m are crucial to measure accurate $L_{\rm IR}$ and $f_{\rm AGN}$. While including AGN with a large range of IR luminosities is an important improvement, as shown by Chien et al. (2024), small sample size and lack of sensitive enough FIR measurements bring a large limitation to determining the trend of $f_{\rm AGN}$ as a function of redshift.

To overcome this limitation and investigate the $f_{\mathrm{AGN}}$ evolution up to a higher z (~ 4) with a larger sample of AGN with accurate $L_{\mathrm{IR}}$ and $f_{\mathrm{AGN}}$ measurements, we design a PRIMAger survey over a 35 arcmin² field with more than 200 robust JWST-detected AGN. The PRIMAger survey will be in the Systematic Mid-infrared Instrument Legacy Extragalactic Survey (SMILES; Alberts et al. 2024; Rieke et al. 2024) field that has the most complete AGN sample to date at z ∼ 0–6 range (Lyu et al. 2024) based on JWST observations. PRIMAgers' unique photometric coverage between 24-235 $\mu$m and rapid survey capabilities will open a new discovery space to measure $L_{\mathrm{IR}}$ and $f_{\mathrm{AGN}}$. In Figure 1, we show an example SED of a JWST-detected AGN that we plan to observe in the requested PRIMAger survey. PRIMAger photometric bands are shown with the red dots at longer wavelengths than 25 $\mu$m. As inferred from the shown SED analysis with CIGALE, we expect to detect 92 $\mu$m PPI1 band detection of this source close to our survey limit of 0.04 mJy.

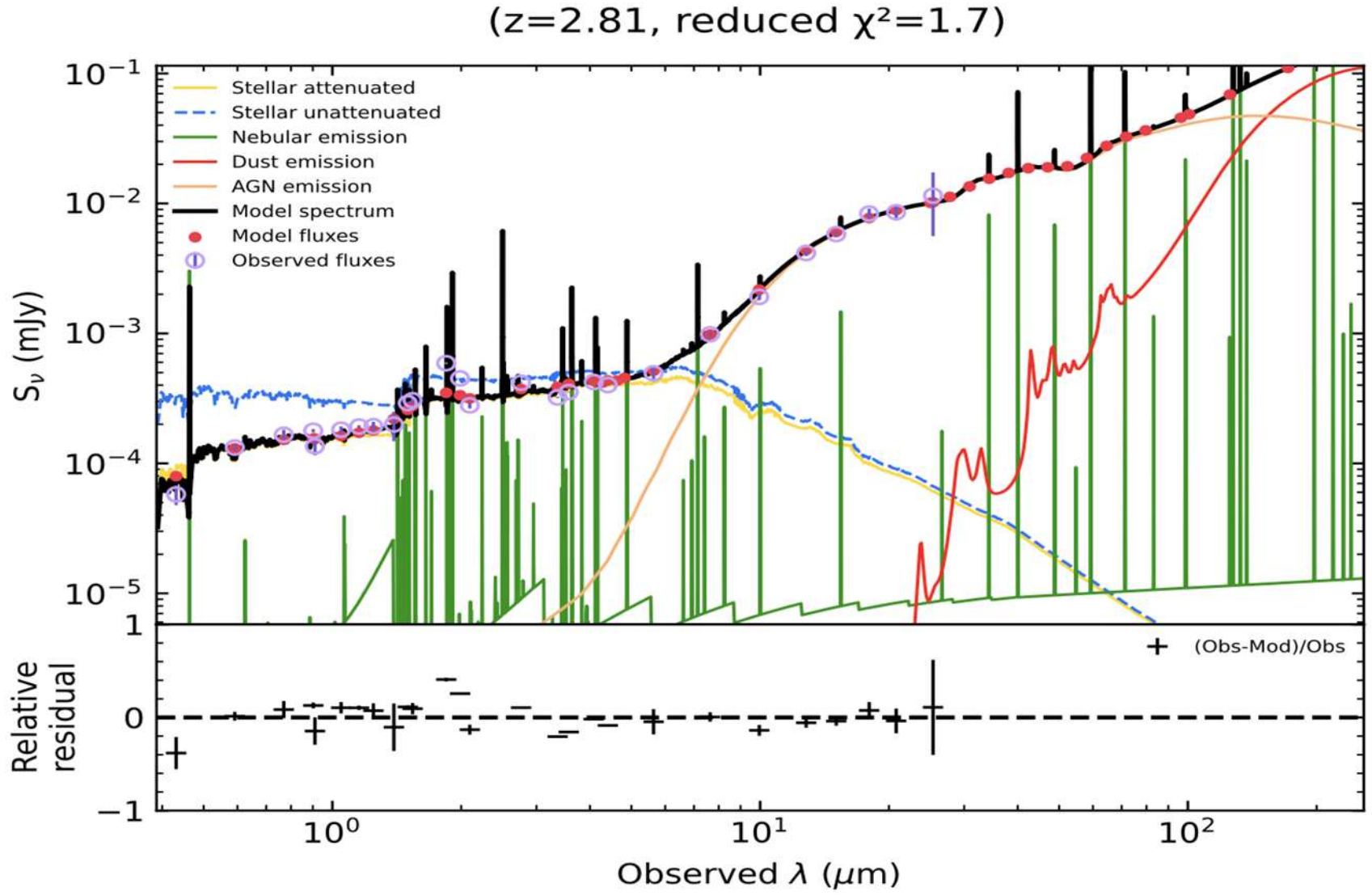

**Figure 1.** An example SED of an AGN from the JWST AGN sample of Lyu et al. (2024). Available MIRI and NIRCam data are combined with the PRIMAger filters.

Based on the SED analysis of 209 AGN, we expect to detect 114 AGN's 92 $\mu$m emission above our survey limit (0.04 mJy). In the previous deep Herschel survey in this field only 27 of the 209 AGN had 100 $\mu$m detection. We expect 3 times more AGN detection with PRIMA compared to Herschel.

As a result of the SED analysis, we measure $L_{\mathrm{IR}}$ and $f_{\mathrm{AGN}}$ for 209 AGN. Figure 2 shows $L_{\mathrm{IR}}$ versus $z$. The upper histogram along the x-axis shows the $z$ distribution. The bottom color bar shows $f_{\mathrm{AGN}}$ value. As shown in Figure 2, PRIMA will increase the lower luminosity AGN up to $z \sim 4$.

As a result of the requested PRIMA survey, we will have data which is 10 times deeper than the deepest available Herschel data. The FIR detected AGN number will increase 3 times with PRIMA (from 27 to 114). With this deep PRIMA imaging survey, we will have a volume limited 92 µm complete AGN sample that will allow us to find out the evolution of AGN contribution as function of redshift and $L_{\mathrm{IR}}$. Including lower luminosity AGN up to $z \sim 6$ will open a discovery space for the coevolution of AGNs and their host galaxies.

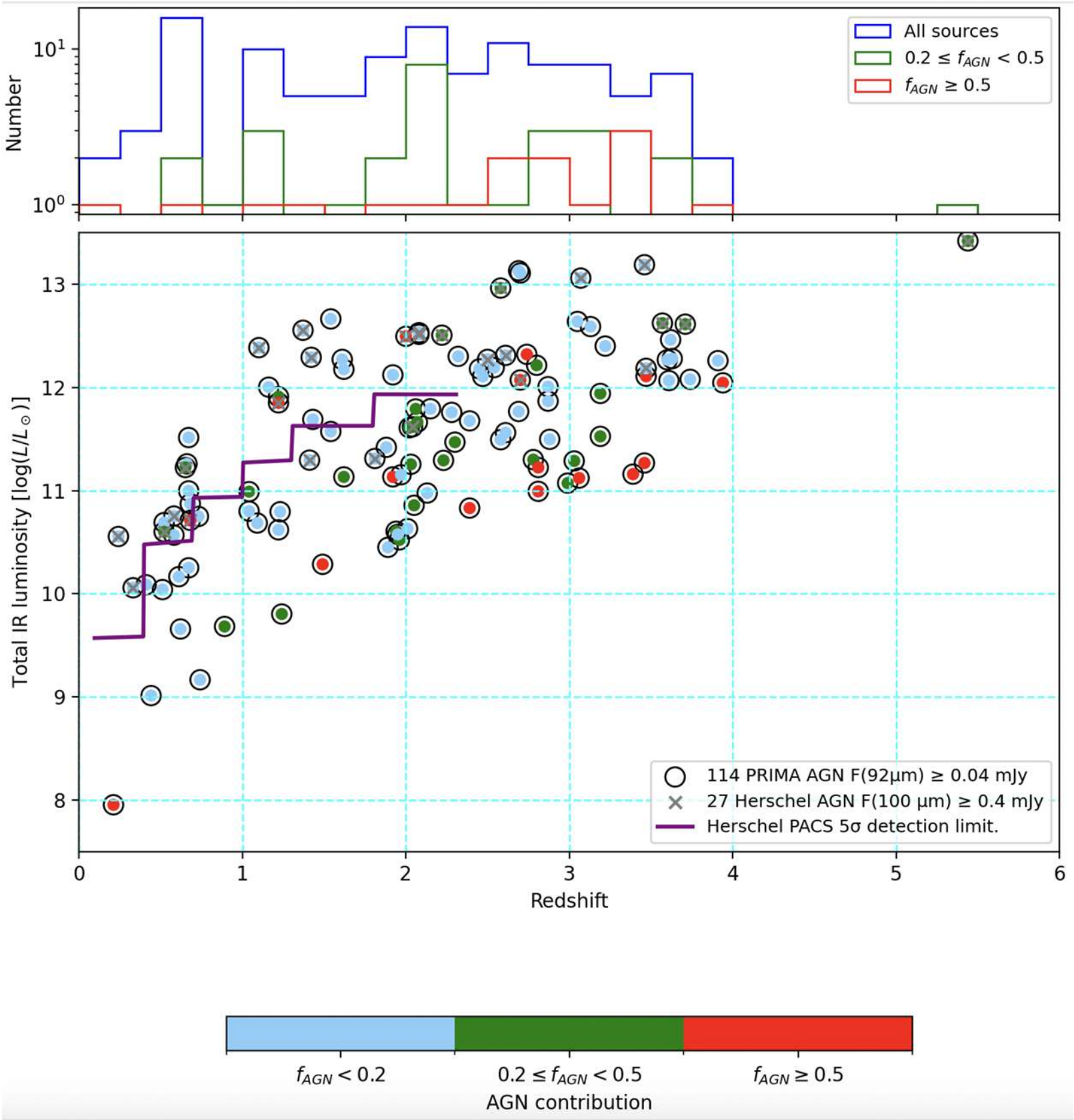

**Figure 2.** Total IR luminosity measured from SED analysis versus redshift. AGN fraction measured from the SED analysis is color coded as shown in the bottom bar. The purple line shows the Herschel completeness limits from Magnelli et al. (2013). The discovery space enabled by PRIMA is a 3 times increased number of AGN with FIR detection at cosmic noon and higher z. With PRIMA the FIR emission of these lower luminosity AGN will be measured for the first time.

## Instruments and Modes Used

This observing program requires PRIMAger small maps in polarimeter and hyperspectral bands. The program requires 2 maps of 4.2'x4.2' each in the PPI bands and 5 maps of 3.9'x1.3' each, in the PHI bands. Polarimetry information is not required.

## Approximate Integration Time

To map a 35 arcmin square area in polarimeter bands, 2 maps are needed. Based on PRIMA ETC 98 hours can reach 0.04 mJy at PPI1 band over 18 arcmin square area, therefore 196 hours is estimated for PPI1 mapping.

Since PPI and PHI observations cannot be obtained at the same time, seven 5.07 arcmin squared PHI maps are needed to cover 35 arcmin square area. To reach 0.1 mJy at 60 micron each 5.07 arcmin squared map requires 12.1 hours (based on the PRIMA ETC). So, PHI mapping requires 88.9 hours over 35armcin squared area. In total 284.9 hours is the requested PRIMA observation time.

## Special Capabilities Needed

None

## Synergies with Other Facilities

The survey area has ancillary data from Herschel, Hubble, JWST (MIRI and NIRCam).

## Description of Observations

PRIMAger polarimetric imager will map 35 sq. arcmin to 5σ = 40 µJy at PPI1 (92 µm) in 2 pointings (each with 4.2'x4.2' coverage). Since polarimetric and hyperspectral imager cannot observe at the same time, in order to have the photometric coverage of PHI1 and PHI2 bands PRIMAger hyperspectral imager will map 35 sq. arcmin to 5σ = 100 µJy at PHI2 (60 µm) in 7 pointings (each with 3.9'x1.3' coverage). The total time requested for the PRIMAger hyperspectral and polarimeter imager surveys is 284.9 hours.

# 35. PRIMA Photometric and Spectral Follow-up of Dusty Star-Forming Galaxies at Cosmic Noon

Ece Kilerci (Istanbul University)

Due to their heavily obscured intense star formation, dusty star-forming galaxies (DSFGs) significantly contribute to the cosmic star formation rate density at high redshifts. Therefore, they play a critical role in our understanding of galaxy evolution. Polycyclic aromatic hydrocarbon (PAH) molecules are important components of the interstellar medium (ISM) in DSFGs that trace star formation rate (SFR) and infrared luminosity. We design a PRIMAger survey to have full photometric coverage of a rare DSFG sample detected by ALMA. Additionally, we request deep FIRESS follow-up observations to measure PAHs in DSFGs at z = 2.0−4.6 range. PRIMA will provide a unique opportunity to measure the physical properties of DSFGs at cosmic noon.

## Science Justification

Dusty Star-Forming Galaxies (DSFGs; Casey et al. 2014) have an intense star formation rate (SFR) reaching 100-1000 Msolar/yr with high ultraviolet (UV) and optical radiation that is obscured by their interstellar dust. The dust re-emits the absorbed radiation in the FIR and submillimeter wavelengths. The peak of this thermal cold dust (20–50 K) emission is in the FIR- submillimeter wavelength (∼100 μm) range. DSFGs can be detected at high−z by their strong FIR/submillimeter emission. Most DSFGs are detected at the z ∼ 1−4 range called cosmic noon when star formation in the Universe reaches its highest value. The Atacama Large Millime er/submillimeter Array (ALMA) can measure the Rayleigh-Jeans tail of thermal dust emission of coldest dust grains in DSFGs at z = 2–3 (e.g., Casey et al. 2018). Therefore, DSFGs can be efficiently selected by deep 2−3 mm observations. The most extreme DSFGs detected at sub millimeter wavelengths (e.g., 850 μm) with ALMA or SCUBA are called Submillimeter Galaxies (SMG). Luminous and Ultra-Luminous Infrared Galaxies (LIRGs and ULIRGs) are also merger /interaction-triggered DSFG types.

The mid-IR spectra of DSFGs is crucial to measure their dust content, SFR, $M_{star}$, metallicity and AGN activity. The polycyclic aromatic hydrocarbon (PAH) important tracers of SFR. PAH luminosities correlate with SFR (e.g., Shipley et al. 2016) and IR luminosity (e.g., Farrah et al. 2008).

The mid-IR spectra of DSFGs at cosmic noon are very limited. To obtain the full mid-IR spectra of DSFGs at cosmic noon and at higher-z we design a FIRESS follow-up survey for a sample of rare DSFGs known as the MORA sample (Casey et al. 2021). MORA is a 2 mm ALMA survey in the Cosmic Evolution Survey (COSMOS) field. The updated Ex-MORA Survey covers 0.2 deg$^2$ and extends the number of robust 2 mm-bright sources to 33 within the 0.7 ≤ z ≤ 5.85 range. Our goal is to add PRIMA data to the well observed COSMOS-MORA Survey field that already have HST, JWST, Herschel, SCUBA, ALMA observations (Casey et al. 2021).

PRIMAGER's continuous photometric coverage is important to obtain mid- and far-infrared spectral energy distributions (SED). The goal of the requested PRIMAGER survey is to fill in the wavelength gap between the longest JWST band at 25 micron and the mm data. As shown in the example SEDs of two DSFGs in the MORA sample, PRIMA observations will have a significant advancement in the SEDs. Here all available HST/WFC3, Spitzer, Herschel, SCUBA and ALMA photometric data (Casey et al. 2021, Long et al. 2024) are combined with 16 PRIMA filters. As a result of our SED analysis, we find that most of the PRIMAger bands can be detected above the band-dependent sensitivity limit between ~0.2–0.5 mJy. In a 230 arcmin$^2$ field PRIMAger can reach this sensitivity level in ~16 hours.

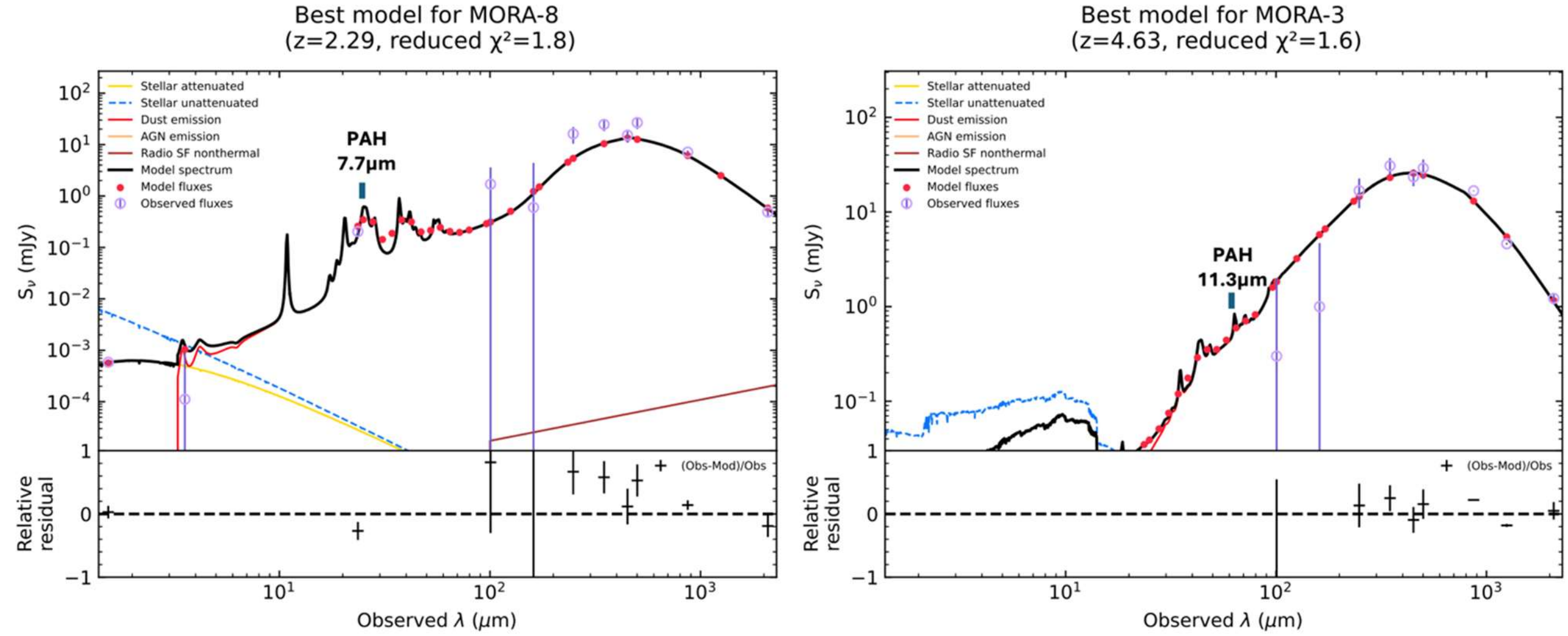

**Figure 1.** Best-fitting models of two DSFGs at cosmic noon showing the unique wavelength range of PRIMA to probe between 25–235 micron. The violet open circles and the red circles (filled) represent the observed fluxes and the model fluxes, respectively. The dashed blue lines show the unattenuated stellar emission, and the dust emission is shown by the solid red lines. Important PAH lines that are aimed to measure with the FIRESS survey are marked.

With PRIMA observations, the SEDs of DSFGs will be obtained for a large z range with additional PRIMA photometric data. These improved SEDs will give well-constrained physical parameters such as the SFR, IR luminosity, and dust luminosity.

From the SED analysis we estimate the expected PAH peak luminosities of these two sources. Based on MORA-8's SED we measure the peak of the PAH feature at 7.7 μm as 0.7 mJy (~ 7.2 $10^{-17}$ W m$^{-2}$). The peak of the PAH feature at 11.3 μm is measured as 0.9 mJy (~3.2 $10^{-17}$ W m$^{-2}$) for MORA-3. Based on these measurements, we estimate 1.5 $10^{-19}$ W m$^{-2}$ as a feasible sensitivity to reach our science goals.

PRIMA FIRESS low resolution point source observations reach 1.5 $10^{-19}$ W m$^{-2}$ sensitivity to reveal the physical parameters of DSFGs at cosmic noon. With FIRESS, for the first time, we will be able to investigate the combined mid-IR and FIR spectra of DSFGs at cosmic noon.

## Instruments and Modes Used

- PRIMAger mapping with both hyperspectral and polarimeter bands to cover a continuous area of 230 sq.arcmin. No polarimetry information is needed.

- FIRESS low-res pointed observations using all 4 bands for 33 targets.

## Approximate Integration Time

- For PRIMAger survey over 230 arcmin$^2$ with 0.2 mJy sensitivity required time is 137.2 hours for hyperspectral mapping at 60 µm; and 50 hours for polarimetric imaging with Band 1.
- For a single FIRESS low-resolution point source observation it takes 1.60 hours to detect the minimum line flux of 1.5 $10^{-19}$ W $m^{-2}$. Since we have 33 sources the requested time will be 52.8 hours for FIRESS observations.

## Special Capabilities Needed

None

## Synergies with Other Facilities

COSMOS-Web JWST and ALMA observations are available in this field.

## Description of Observations

PRIMAger will observe the 230 arcmin$^2$ COSMOS Web MORA Survey with hyperspectral and polarimeter bands. FIRESS follow-up observations of 33 known sources in low-resolution point source mode.

# 36. Panchromatic view of Protoclusters to Understand Environment-Dependent Galaxy and BH Growth Histories

Mariko Kubo (Kwansei Gakuin University), Hideki Umehata (Nagoya University), Tohru Nagao (Ehime University), Hanae Inami (Hiroshima University), Alberto Traina (alberto.traina@inaf.it)

The properties of galaxies depend on mass and environment in the local Universe. Protoclusters, overdense regions at high redshift, are plausible progenitors of modern clusters and important laboratories for investigating how the environment has shaped galaxies. The Atacama Large Millimeter/submillimeter Array (ALMA) telescope has explored heavily obscured star formation and active galactic nuclei (AGN) activities at high redshift, which are essential to characterize the galaxy formation and supermassive black hole (SMBH) accretion histories. However, ground-based facilities cannot observe rest-frame mid to far-infrared emission of protoclusters, which is the most essential bandpass that includes several emission lines critical to constrain dust-attenuated star formation, AGN, and interstellar mediam (ISM) properties. Furthermore, protoclusters have large spatial extents of several arcmin or larger and a large scatter in properties even at the same redshifts. Thus, we need statistical studies in the infrared to obtain the general picture of galaxy and super-massive black hole (SMBH) growth histories in protoclusters.

Here we propose a systematic and statistical study of heavily obscured galaxies in protoclusters. We perform PRIMAger imaging and FIRESS spectroscopy for a statistical number of protoclusters to show the environmental dependence on star formation and SMBH accretion histories. By resolving the AGN fraction along with the physical properties for each galaxy in protoclusters, the roles of AGNs in the emergence of the red sequence will be tested.

## Science Justification

### Broader Context

The physical properties of galaxies are tightly correlated with their mass and environments, such as groups, clusters, large-scale structures, and voids. However, when and how the dependence has arisen is not yet fully understood. To resolve the early stage of cluster evolution, protoclusters, plausible progenitors of clusters of galaxies today at cosmic noon or higher redshifts, have been discovered by the optical/near-infrared (near-IR) observations (Overzier et al. 2016). Moreover, Euclid, Rubin, and Roman will find and characterize a statistical number of protoclusters in the optical/near-IR; on the other hand, the sub-millimeter observations have shown significant overdensities of dusty starburst galaxies in protoclusters (e.g., Tamura et al. 2009; Dannerbauer et al. 2014; Oteo et al. 2018; Miller et al. 2018). Furthermore, deep X-ray observations using Chandra show that active galactic nuclei (AGNs) in protoclusters are heavily attenuated by gas and dust (Umehata et al. 2019; Vito et al. 2020; 2024, Figure 1 left). By stacking the low-resolution images of protoclusters at z~4 from the IR all-sky surveys such as Planck, the average of the total emission and IR spectral energy distribution (SED) of a protocluster is characterized (Kubo et al. 2019, Figure 1 right, see also Alberts et al. 2021). It suggests that roughly two-thirds of the total infrared emission of protoclusters likely comes from obscured sources, and the hot SED (~100 mJy in total at 100 μm) suggests notable warm infrared sources like AGNs or young starburst galaxies. Moreover, dusty starbursting protocluster cores, which have star formation densities several times higher than those of protoclusters, have been discovered (Oteo et al. 2018; Miller et al. 2018). These studies suggest that a large part of star formation and SMBH growths in protoclusters are heavily attenuated by dust and is hardly studied in the optical/near-IR.

Thus, to resolve the most dramatic episodes of galaxy evolution in protoclusters, i.e., the formation of giant ellipticals in clusters, we need to inspect obscured starbursts and AGNs by observing the rest-frame mid-to-far-IR regime. First, to constrain the total IR emission, measurements on the far-infrared SED are required, while only the Rayleigh-Jeans tails of SEDs can be probed by ground-based facilities. Second, mid to far-IR emission lines can characterize the ISM properties and AGNs in heavily obscured galaxies, which are hardly observed in the optical/near-IR. The JWST has now characterized dusty galaxies at high redshift through the Paschen series (Shimakawa et al. 2024). However, wide-field observation is necessary to characterize environments; each protocluster has a spatial extent of several arcmin or larger, and the properties of protoclusters range widely even at the same redshift. A uniform and statistical study is needed to describe the general picture of environmental effects. According to the AGN and galaxy co-evolution scenario (e.g., Hopkins et al. 2008), star formation is triggered by mergers, followed by a dusty quasi-stellar object (QSO) phase, and star formation is quenched via AGN feedback. Thus, resolving the AGN and star formation activities in each galaxy, which should reflect their evolutionary stages, together with their host properties, is crucial for testing the roles of AGNs in the emergence of massive quiescent galaxies.

### Science Question

- Environmental dependence on galaxy growth and BH accretion histories
- How the environmental effects have processed galaxies.

## Need for PRIMA

First, a high sensitivity at mid-to-far-IR wavelengths is required to constrain the infrared SEDs, including the peak dust emission, and perform emission line diagnostics for heavily obscured galaxies. A high survey speed is also essential for mapping the environments of galaxies at high redshift and performing their statistical study. PRIMA is the only observatory to achieve a deep and wide infrared survey with angular resolution and sensitivity enough for high-redshift galaxies.

We will conduct a PRIMAger systematic and statistical survey of the IR-luminous population in protoclusters for the first time. The IR luminosity function informs the evolution of star formation and SMBH accretion histories. In addition to the SEDs from PRIMAger, we will constrain the AGN prevalence in each galaxy robustly by emission line diagnostics using FIRESS. Together with the host properties characterized in the optical/near-IR, the roles of AGNs in the emergence of the red sequence will be tested. FIRESS will also probe the gas-phase metallicities, which can inform the balance between inflows, outflows, and star formation, as this balance is expected to depend on the mass and environment; enhanced accretion of cold gas at high redshift and confinement of outflows by the dense intracluster medium at low redshift are anticipated in overdense regions.

## Interpretation Methods

Dust masses, star formation rates (SFRs), and AGN luminosities of galaxies are measured by fitting the continuum SEDs with empirical and model templates. AGNs can also be identified through the PAH equivalent width (EW) (e.g., Smith et al. 2007). The BH accretion rate will be measured based on [OIV] 25.18 μm. The [O III] 51.80 μm / [N III] 57.21 μm line diagnostics infer gas-phase metallicities at z < 4 (Peng et al. 2021).

## Link to Testable Hypotheses

By stacking analyses of low-resolution all-sky infrared surveys, the average total dust emission of a protocluster has been constrained (Kubo et al. 2019; Alberts et al. 2021). This unresolved measurement can be compared to the resolved measurements by PRIMA. On the other hand, the AGN fraction in a galaxy is expected to depend on its evolutionary stage (e.g., Hopkins et al. 2008), and it is crucial to characterize the AGN fraction for each galaxy to test the roles of AGNs in the emergence of massive quiescent galaxies. The large volume hydrodynamical simulations will be directly compared to the results.

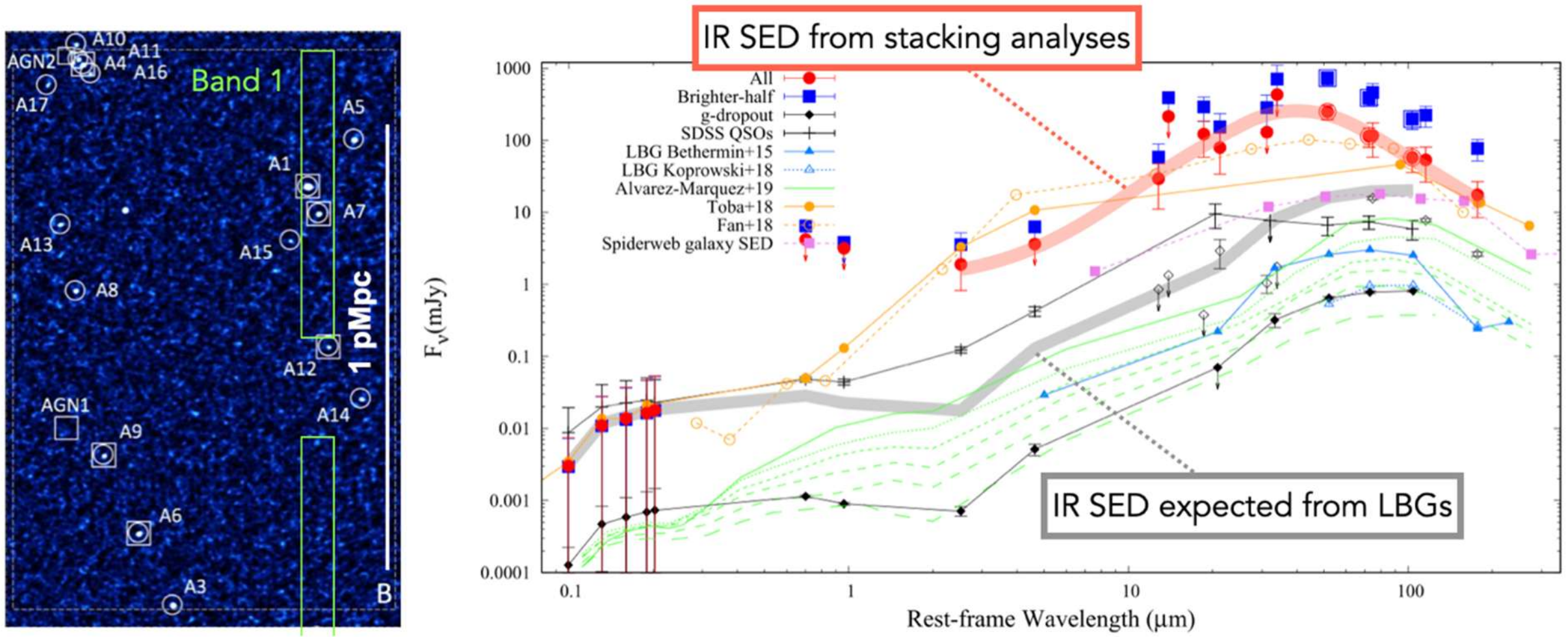

**Figure 1.** (Left) Background is sub-mm mapping for 2' x 3' at protocluster core at z = 3.09 in Umehata et al. (2019). White circles and squares show sub-millimeter (SMGs) and X-ray AGNs. The green rectangles show the FIRESS Band 1 field of view. (Right) The average of the total SED of a protocluster at z ~ 4 in Kubo et al. (2019).

## Instruments and Modes Used

This science case will use 64 FIRESS R~100 pointings and 36 PRIMAger pointings.

## Approximate Integration Time

The tables below present the approximate integration time for a protocluster. The required sensitivities for PRIMAger continuum are measured based on the SED templates for star-forming galaxies in Kirkpatrick et al. (2012). The required sensitivities for emission lines are calculated using the emission lines to infrared continuum luminosity ratios from Spinoglio et al. (2012). We will use FIRESS' low-resolution point-source mode. Several infrared luminous galaxies in a protocluster will be observed simultaneously within a field of view of FIRESS (Figure 1 left), and 3 pointings will cover 10-30 members. We request to observe 3 (or more) protoclusters / $\Delta z$=1. PRIMAger and FIRESS, ~60 hours and ~210 hours, respectively, are required in total.

**Table .:** Integration time for PRIMager

| Redshift | Luminosity ($L_{\odot}$) | Sensitivity (5 $\sigma$ at 126 µm in mJy) | N † | Integration time (h) |
|---|---|---|---|---|
| **1** | >3E+11 | 3.0 | 3 | 0.2 |
| **2** | >1E+12 | 1.2 | 3 | 1.1 |
| **3** | >2E+12 | 0.5 | 3 | 7.2 |
| **4** | >5E+12 | 0.4 | 3 | 10.3 |

The total exposure time for 3 protoclusters is 3*18.8 = 56.4 => ~60 hours.

**Table 2.** Integration time for FIRESS.

| redshift | Strong lines | L ($L_{\odot}$) | Sensitivity ($Wm^{-2}$) | Integration time per pointing | N † | Integration time (h)‡ |
|---|---|---|---|---|---|---|
| **1** | PAH, [Ne III] 15.55, [OIV]25.89, [O III] 51.81, 88.36, [O I] 63.18 | >1E+12 | 5E-19 | 0.14 | 5 | 1.4 |
| **2** | PAH, [Ne III] 15.55, [OIV]25.89, [O III] 51.81, 88.36, [O I] 63.18 | >2E+12 | 2E-19 | 0.9 | 5 | 9.0 |
| **3** | PAH, [Ne III] 15.55, [OIV]25.89, [O III] 51.81, 88.36, [O I]63.18 | >3E+12 | 1E-19 | 3.6 | 3 | 21.6 |
| **4** | PAH, [Ne III] 15.55, [OIV]25.89 | >1E+13 | 1E-19 | 3.6 | 3 | 21.6 |

† N is the number of pointings for a protocluster. ‡ A factor of 2 is needed to observe the full wavelength range with FIRESS. The total exposure time is 1.4+9.0+21.6+21.6 = 53.6 per cluster. For 4 protocluster, we thus have => 214.4 hours => ~210 hours.

## Special Capabilities Needed

None

## Synergies with Other Facilities

Target protoclusters will be selected from the deep and wide optical/near-IR surveys, namely Euclid/Rubin/Roman. To characterize the full infrared SEDs, including the Rayleigh-Jeans tail, follow-up observations with ALMA are also needed.

## Description of Observations

### Immediate Goal

Using PRIMAger, we will systematically study the number density of IR sources in protoclusters. We also aim to constrain their rest-frame mid-to-far-IR SEDs to determine photometric redshifts and whether the infrared emission originates from star formation and/or AGNs. We will also perform emission line diagnostics using the FIRESS for candidate protocluster members detected with PRIMAger (L>3E+12 $L_\odot$ at z ~ 3). We primarily aim for PAH, [Ne III] 15.55, [OIV]25.89, [O III] 51.81, 88.36, and [O I]63.18 emission lines. We will characterize ISM properties and star formation via [Ne III] 15.55, [O III] 51.81, 88.36, and [O I]63.18, AGN activity through PAH EW, and BH accretion rate through [OIV]25.89.

### Targets

We aim to observe 12 or more clusters/protoclusters (3 pcl / $\Delta$z = 1) at z = 1 - 4. The clusters/protoclusters are selected from the known protoclusters and those newly found through the Euclid/Rubin/Roman telescopes. The known protoclusters with intensive ALMA observations have higher priority. We also include systematically surveyed protoclusters from the Euclid/Rubin/Roman to discuss the variation of protoclusters. We note that a few protoclusters per 1 sq. deg. are expected at cosmic noon according to the cosmological numerical simulations and observations (Toshikawa et al. 2018). Then, a part of this program can be achieved by the wide-field survey in the PI program.

### Instruments and Observational Strategy

First, we request PRIMAger observations for protoclusters. The minimum usable map size for the Polarimeter band of 10’ x 10’ can cover protocluster and connecting large-scale structures. We request the observing time to detect galaxies with L>1E+12-1E+13 $L_\odot$ at 5σ significance. Table 1 and 2 shows the required time for each redshift bin. To observe three pcl/ $\Delta$z = 1, ~ 60 h is required. Second, we request FIRESS spectroscopic follow-up for the candidate protocluster members detected with the PRIMAger and the existing ground-based facilities. The requested sensitivity is 1-5E-19 $Wm^{-2}$, which is achieved with 0.1-3.6 hours exposure per pointing. As presented in Figure 1 left, FIRESS can effectively cover SMGs at a protocluster core by a few pointings. In total, 200 hours are required for FIRES spectroscopy.

# 37. M81 y sus Primas con PRIMA: Spectroscopic Mapping of the M81 Group with FIRESS

Rebecca Levy (Space Telescope Science Institute (STScI)), Eric Koch (NRAO), Serena Cronin (U. Maryland), Adam Smercina (STScI), Nicole Melso (RIT)

The far-infrared (FIR) holds the key to understanding the energy balance in the interstellar medium (ISM) and circumgalactic medium (CGM), as FIR spectral lines govern gas heating and cooling. The M81 group is the nearest example of an interacting galaxy system (d=3.6 Mpc) and is expected to merge in the next 1-2 Gyr. Previous multiwavelength imaging has revealed tidal streams of gas, dust, and stars connecting the three major group members M81, M82, and NGC3077. Already, this tidal interaction has resulted in AGN activity in M81, a unique and vigorous starburst in M82, and the perplexing morphology of NGC3077. This archetypal system has incredibly rich multiwavelength data and is ideally poised for deep spectroscopic mapping with PRIMA. We propose to map a 100 kpc × 100 kpc (95.5’ × 95.5’) region of the M81 group using the high spectral resolution mode of FIRESS from 24-235µm (R~2000-20000, ΔV~15-150 km/s). This deep survey will ensure robust detections of critical FIR spectral lines in the member galaxies, tidal streams, mysterious features (such as the “Cap”), and halo overall, yielding an unprecedented view of gas heating and cooling over a range of environments. With this large program, PRIMA will deliver an unparalleled legacy dataset.

## Science Justification

### Background

The far-infrared (FIR) holds the key to understanding the energy balance in the interstellar medium (ISM) and circumgalactic medium (CGM), as FIR spectral lines govern gas heating and cooling. While previous FIR missions have made substantial progress towards quantifying the energy balance in the Milky Way’s ISM and in regions of nearby galaxies, we know little about the properties of gas in galaxy halos, CGMs, and over a range of galaxy properties (mass, metallicity, star formation rate (SFR), etc). With its unparalleled mapping speed, sensitivity, and wavelength coverage, PRIMA is the key to unlock this door.

The M81 group is the ideal testbed for PRIMA to investigate gas heating and cooling in a range of environments. This nearby group, at a distance of only 3.6 Mpc, consists of ~30 (mostly dwarf) galaxies (Freedman et al. 1994; Karachentsev 2005). The main members of this group are M81, M82, and NGC3077 which are highlighted in Figure 1. The main galaxy, M81, has a modest SFR (~0.3-0.9 $M_\odot$ $yr^{-1}$; Gordon et al. 2004), roughly Milky Way mass (≈$6\times10^{10}$ $M_\odot$; Querejeta et al. 2015), and hosts a central low-luminosity active galactic nucleus (AGN) with a black hole mass of $7\times10^7$ $M_\odot$ (Devereux et al. 2003). The next largest member is M82, the archetypal starburst galaxy in the local Universe and the subject of hundreds of astrophysical studies and observations. M82 is small, with a dynamical radius of only 4 kpc (Greco et al. 2012) and stellar mass ~3×10$^{10}$ M$_\odot$ (Querejeta et al. 2015). Despite its small size, the disk is undergoing a vigorous starburst (SFR~10 M$_\odot$ yr$^{-1}$) which peaked 5 and 10 Myr ago (Förster Schreiber et al. 2003). This intense, concentrated star formation has resulted in a massive, multiphase superwind along the minor axis of M82, which has been studied across the electromagnetic spectrum. The existence of the "Cap" –a clump of gas ~12 kpc along the minor axis of M82—suggests past outflow activity from M82 (Tsuru et al. 2007). NGC3077 is a small irregular galaxy with an LMC-like stellar mass of ~2×10$^9$ M$_\odot$ (Querejeta et al. 2015). While M81 and M82 have ~solar metallicities, NGC3077 has a metallicity ~60% solar (Oparin et al. 2020).

Perhaps the most striking feature of the M81 group is due to the tidal interactions between the three main members. Atomic gas traced by HI clearly shows streams of material between M81, M82, and NGC3077 (Figure 1; Yun et al. 1994, de Blok et al. 2018). It is thought that this interaction is the trigger for M81's AGN (e.g., Yun et al. 1999), M82's starburst (Förster Schreiber et al. 2003), and NGC3077's peculiar morphology (Okamoto et al. 2023). In addition to HI, these streams contain stars (e.g., Smercina et al. 2020, Rao et al. 2025), molecular gas (e.g., Heithausen & Walter 2000), and dust (e.g., Walter et al. 2011). Moreover, recent deep wide-field Hα mapping from Dragonfly has also uncovered a tidal dwarf galaxy in formation near M82 (Pasha et al. 2021). The halo surrounding M81 is extremely metal poor (~6% solar; Smercina et al. 2020). It is expected that these three galaxies will merge in 1-2 Gyr (Oehm et al. 2017). A complete, wide-field and multi-wavelength study is clearly needed to understand this intriguing merging group.

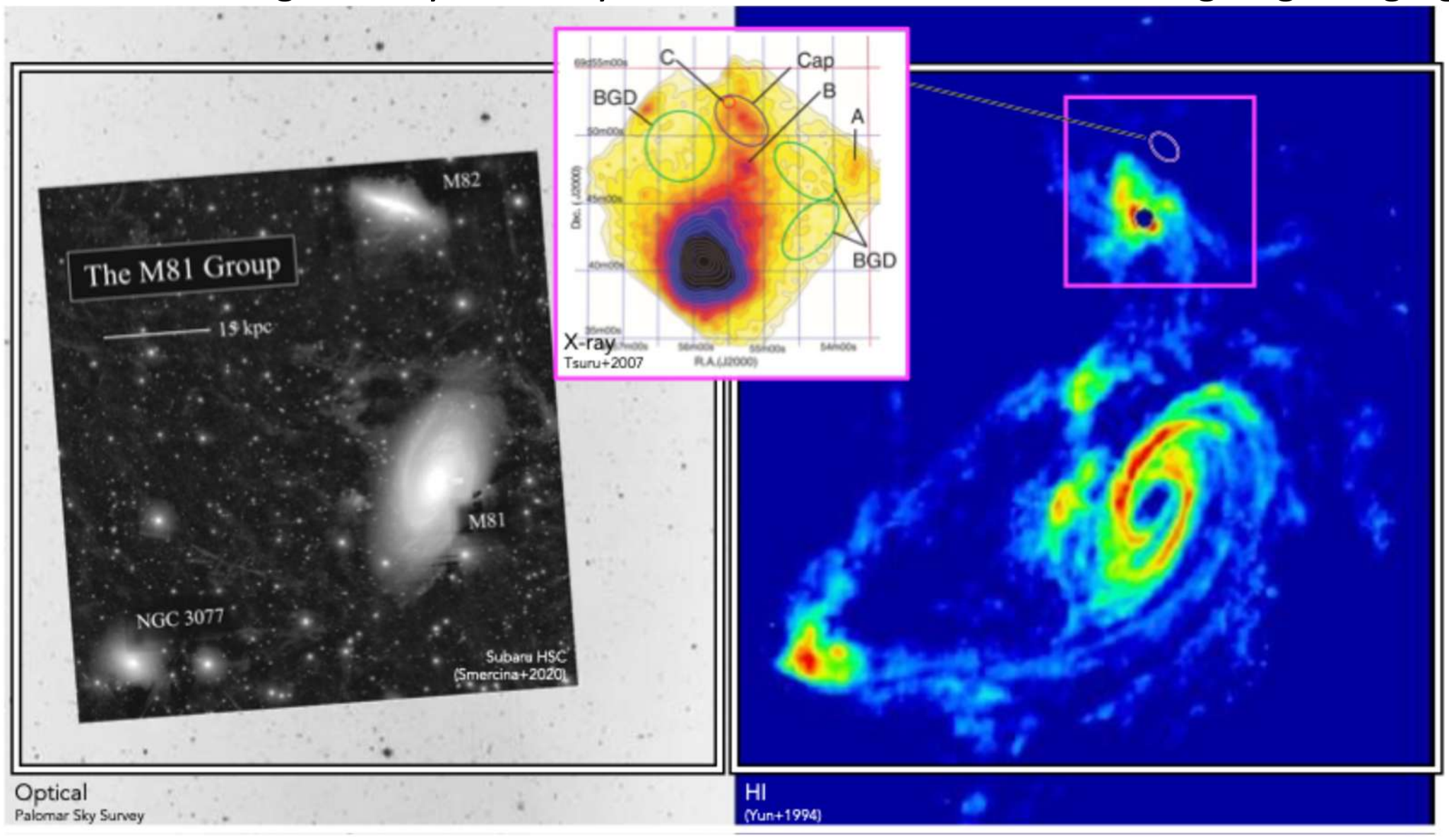

**Figure 1.** The M81 Group seen at optical wavelengths with Palomar and Subaru HSC (inset; Smercina et al. 2020) and in HI from the VLA (Yun et al. 1994). The center inset shows X-ray emission from the "Cap" (Tsuru et al. 2007). The white boxes show the proposed 100 kpc x 100 kpc (95.5' x 95.5') region to be mapped with PRIMA FIRESS FTM.

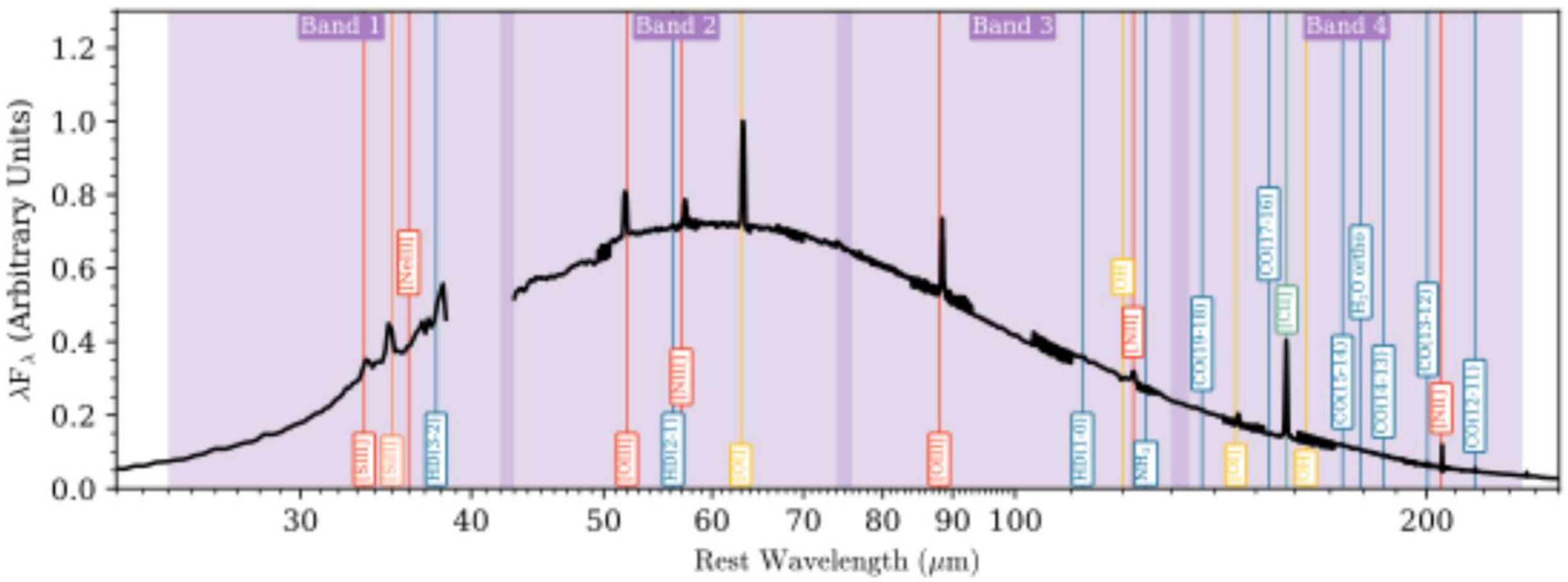

**Figure 2.** The wavelength coverage of FIRESS (purple shaded) compared to an SED of M82 (Kennicutt+2003; Brauher+2008; Kamenetzky+2012). Key diagnostic lines are shown, tracing the ionized (red), ionized/neutral (orange), neutral (yellow), ionized/neutral/molecular (green), and molecular (blue) ISM phases. Figure adapted from Levy+2024.

## Science Enabled by PRIMA

With PRIMA, we propose to use the high spectral resolution (FTM) mode of FIRESS to map the entire M81 group (100 kpc × 100 kpc = 95.5′ × 95.5′; Figure 1) from 24-235µm to a depth that ensures robust detections of key spectral lines in the galaxies and tidal streams. Some of these key lines are shown in Figure 2. FIRESS's high spectral resolution (R ~ 2,000-20,000, ΔV ~15-150 km/s) is required to 1) measure the complex gas kinematics, 2) robustly measure linewidths, which give insight into the turbulent state of the gas, and 3) remove any contribution from foreground galactic cirrus. PRIMA's ability to map high spectral resolution over such a large field of view is unprecedented and, for many of these key spectral lines, will provide the first spatially and spectrally resolved view of an on-going major merger, broadly informing unresolved observations for more distant mergers.

FIR transitions provide key diagnostics on ISM cooling, radiation field, temperature, density, and shocks. The 158µm transition of [CII] is the brightest line in the FIR, a key tracer of the cold neutral gas, and the major cooling channel in the ISM. High signal-to-noise observations of this emission in the halo and streams in this group will yield information about the bulk cold material and where gas is cooling (especially compared to the young stars in the streams). [OI] 63µm is another major cooling channel. Other key lines include [SiII] 35µm, [OI] 63, 145µm, [OIII] 52, 88µm, and [NII] 122, 205µm, whose strengths and ratios are diagnostics of the radiation field, temperature, density, and shocks. The [OIII] lines are clean tracers of ionized gas in dense star-forming regions, whereas the [NII] lines trace ionized gas in diffuse material. These lines are included in the MAPPINGS shock ionization models, which will aid in the interpretation of the observed line properties (Allen et al. 2008). Therefore, this complete FIRESS map gives us access to the major heating and cooling channels for gas in many phases and locations in the M81 group, providing a unique view of the ISM/CGM properties in the nearest external merging group of galaxies.

The FIR lines also provide key insight into the large-scale impact of feedback. For example, we will fully trace the full multiphase extent and structure of the outflow from M82. Typically, observations do not trace the full region between M82 and the "Cap" because of their large (11') separation. However, with PRIMA's mapping capabilities, we will be able to trace the multiphase and kinematic connections between the "Cap" and M82, and with full context within the larger M81 group.

## Instruments and Modes Used

This observing program requests a 1.6° × 1.6° map in all four bands of the FIRESS high-resolution mapping mode.

## Approximate Integration Time

We estimate the necessary observing time to map a 100 kpc × 100 kpc (95.5' × 95.5' = 1.6° × 1.6°) region of the M81 group using FIRESS's high-resolution mapping mode (FTM) in all 4 bands. Currently, this mode is not available in the ETC. Based on guidance from the PRIMA instrument team, FTM can be used to map 4.8 sq arcmin / 1 hour in 2 bands simultaneously to the equivalent sensitivity depth that one would obtain with 150s integration on a single pointing.

We estimate our desired sensitivity depth from high spectral resolution observations of the [CII] 158µm line with SOFIA/upGREAT (Levy et al. 2023). [CII] was detected 2 kpc away from the disk of M82 with a rms of ~10mK in 10 km/s channels. From HI imaging, the HI column density in the tidal streams is lower by a factor of ~10 than surrounding M82 (Yun et al. 1994). Using this ratio, we expect [CII] detections across the M81 group for a sensitivity of 1mK or ≈$7\times10^{-11}$ W $m^{-2}$ in 10 km/s channels. For the FTM, the velocity resolution at 158µm is ~100 km/s, so the channel width can likely be increased from this calculation; however, this channel width is applicable at shorter wavelengths.

[CII] is the brightest spectral line in the FIR, and we would also like to detect weaker lines across the entire FIRESS band. We therefore increase our desired sensitivity by a factor of 1000 or $7\times10^{-14}$ W $m^{-2}$, to ensure robust spectroscopic measurements of these weaker lines. Taking this as our 5σ sensitivity, the ETC returns that the necessary integration time is less than the minimum scan time of 150s.

Instead, we find that a 5σ sensitivity of $1\times10^{-18}$ W $m^{-2}$ corresponds to an integration time of 1800s. Scaling this to our required sensitivity yields an integration time of 6.8s ($=1800\text{s}\cdot[1\text{E-}18/7\text{E-}14]^{1/2}$). Therefore, we can use the FTM to map a 4.8 sq arcmin region to this depth in 2 bands in 163s. Scaling to all 4 bands (×2) and our desired area (×1900) yields a **total integration time of 172 hours**. With this large investment of time, PRIMA can deliver an unparalleled legacy dataset on FIR lines in star-forming, dwarf, and starbursting galaxies, their CGM, and tidal debris that will inform observations of mergers across the Universe.

## Special Capabilities Needed

N/A

## Synergies with Other Facilities

M82 is one of the most well-studied galaxies and has a myriad of ancillary data across the electromagnetic spectrum. FIR data exist from Herschel (Contursi et al. 2013; Herrera Camus et al. 2018) and SOFIA (e.g., Jones et al. 2019; Spinoglio et al. 2022; Lopez-Rodriguez et al. 2022; Levy et al. 2023). At shorter wavelengths, JWST Programs 1701 and 5145 include NIRCam and MIRI multi-filter imaging of M82, its outflow, and the halo. Within the next 2-3 years, SPHEREx will provide a complete map in the NIR. Very deep optical imaging with Subaru Hyper Suprime-Cam (HSC) exist covering the entire M81 group (Smercina et al. 2020; Figure 1), and deep narrowband coverage is now available thanks to Dragonfly (e.g., Lokhorst et al. 2022). M82 will almost certainly be an early target for Roman and the ELTs. At longer wavelengths, deep, resolved VLA radio continuum and HI 21cm maps of M82 and the M81 group exist (Yun et al. 1994; Martini et al. 2018; de Blok et al. 2018; Figure 1), and M82 will certainly be a prime target for the ngVLA. As the M81 group is in the north (+69°), it is not accessible by ALMA, but deep multi-line molecular gas tracers exist from NOEMA (Krieger et al. 2021) and the SMA (2024B-S019, 2022B-S034, 2018B-S059, 2018B-S063).

## Description of Observations

We will use FIRESS's FTM mapping mode to spectroscopically map the entire M81 group in all 4 bands (24-235μm) at high spectral resolution (R~2000-20000, ΔV~15-150 km/s). As shown in Figure 1, a 100 kpc × 100 kpc region is sufficient to capture the main group members and tidal streams; this corresponds to 95.5′ × 95.5′ at the distance of the M81 group. In mapping mode, the FTM can observe in two bands simultaneously, so the map will need to be repeated twice to obtain the full wavelength coverage needed to achieve the science goals (Figure 2). To achieve our required depth, we estimate needing 172 hours of on-source integration (see **Approximate integration time** section). There are no constraints on timing between observations. Given the pixel scale in each band, the spatial resolution of these observations corresponds to 132, 132, 222, and 400 pc in Bands 1-4 (noting that the corresponding diffraction-limited PSF sizes are 67, 118, 208, 365 pc). We defer to the instrument team in terms of the best mapping strategy and pattern for these observations.

# 38. Characterizing the Evolution of z=0.5 – 1.67 Galaxies with a PRIMA Spectroscopic Survey

Lun-Jun Liu (California Institute of Technology), Mathilde Van Cuyck (University of Illinois Urbana-Champaign), Charles M. Bradford (Jet Propulsion Laboratory, California Institute of Technology), Ryan P. Keenan (Max Planck Institute for Astronomy), TIM Consortium

The cosmic star formation rate has plummeted since its peak ~8 Gyr ago (z~2, cosmic noon) after the steady rise following the onset of galaxies in the epoch of reionization (z~6). We aim to study the processes underlying this decline through the aggregate measurements of star formation rate (SFR) and interstellar medium (ISM) conditions in galaxy populations since cosmic noon. Far-infrared gas-phase spectral lines offer unique opportunities to diagnose the multi-phase ISM and measure SFR in galaxies as they evolve, because the far-IR emission does not suffer dust obscuration. The PRobe far-Infrared Mission for Astrophysics (PRIMA), features a spectrometer FIRESS offering R~100 spectroscopy across the entire 24 to 235 μm band with unprecedented sensitivity. We propose a 1000-hour PRIMA-FIRESS spectroscopic survey on a 0.2 square degree field enclosing GOODS-South and examine yields in [OI]63μm and [OIII]88μm that encode star formation and gas cooling processes. We essentially perform spectral stacking analysis based on Euclid-like mock catalog, which offers enhanced signal-to-noise and provides aggregation over the population. We demonstrate (a) that for [OI]63μm and [OIII]88μm lines (two major coolants of the ISM) the z=0.5–1.67 line-SFR relation will be calibrated, and (b) the potential for the line-ratio analysis which constrains ISM diagnostics such as neutral and ionized contents.

## Science Justification

### Broader Context

Understanding the history of cosmic star formation and its connection to galaxy evolution is one of the most important challenges in modern astrophysics, which has been identified as one of the three science priority areas by the Astro2020 Decadal Survey. Over the past decade, extensive data have revealed a dramatic decline in the total star formation rate density and a shift in the nature of star-forming galaxies—from luminous, dust-obscured star-forming galaxies that once dominated cosmic star formation to a present-day population where such galaxies are nearly absent [1]. To fully understand galaxy evolution, observations in the far-infrared (IR) are essential, as about half of the energy from cosmic star formation is absorbed by dust and re-emitted in this regime [2,3]. This reprocessed energy also excites bright far-IR emission lines that remain largely unaffected by dust extinction. These lines provide critical diagnostics of star formation rates, the ionizing radiation field, and metallicity [4,5]. However, their study is made difficult by strong atmospheric absorptions.

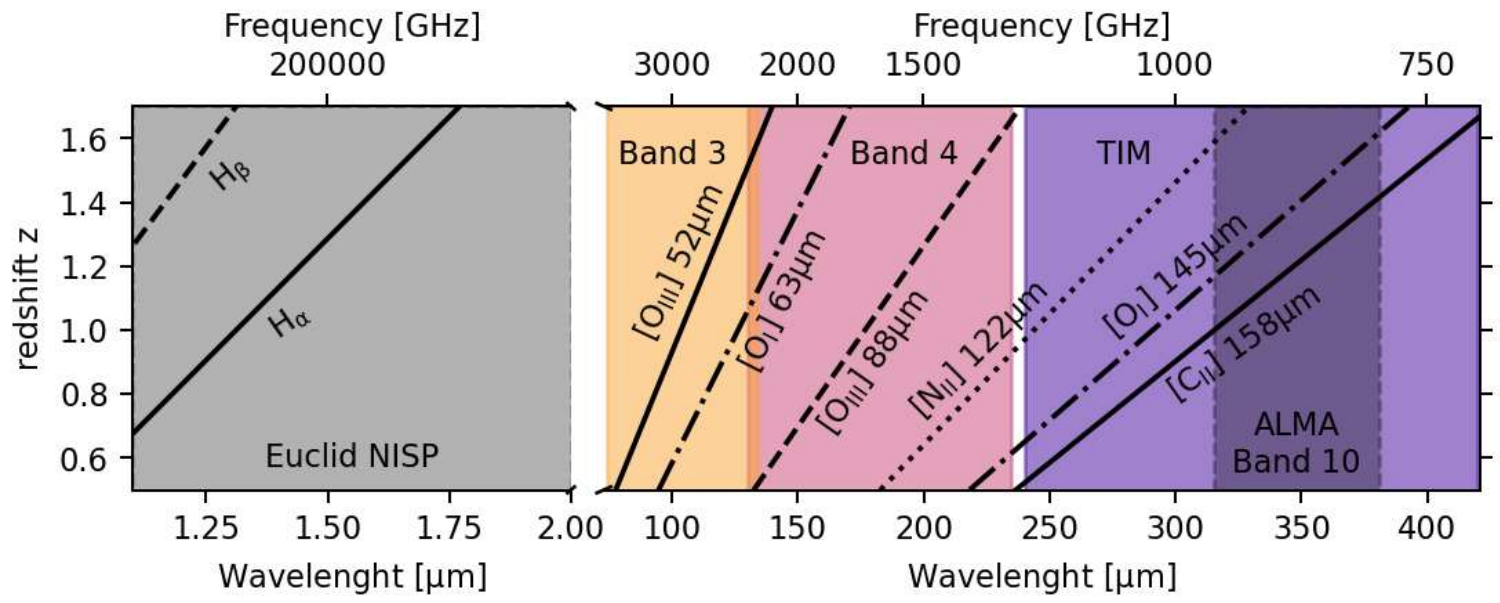

**Fig.1.** Redshift relation of relevant lines in the proposal.

Among the far-IR lines accessible with FIRESS, [OI]63 μm and [OIII]88 μm are major coolants and excellent proxies for the neutral and ionized gas reservoirs in galaxies.

### Science Questions

What drives the evolution of the star-forming galaxies after the cosmic Noon? How does their ionized and neutral content evolve with cosmic time and galaxy properties?

### A Unique Role for PRIMA in IR Time-Domain Science

To address these science questions, diagnostics of the multi-phase ISM are required and are unfeasible without far-IR emission lines that are immune to dust obscuration. In the (post-)cosmic noon regime, there are a suite of far-IR emission lines that can be observed by PRIMA, some of which fall in the PRIMA-FIRESS Band 3 and 4, as shown in Fig.1, and other lines can potentially be measured by the FIRESS Band 1 and 2.

We propose a blind spectroscopic survey which studies evolving galaxies through their far-IR fine-structure line emission in 3 ways: i) direct detections of lines in individual galaxies, ii) stacked aggregate line emission in samples to reach below the individual detection limit, and iii) intensity mapping to reveal mean intensities from all sources of emission. This proposal focuses on the second goal: stacking. We baseline Euclid-selected sources as priors, but other surveys with galaxy redshifts (e.g., Roman) could also be used. This proposed blind spectroscopic survey will sweep a 0.2-deg$^2$ field enclosing the GOODS-South field, a wealth of ancillary data that includes the ASTRODEEP [7] spec-z catalog. Furthermore, this proposed field is located in the 10-deg$^2$ Euclid Deep Field Fornax (EDF-F), which will obtain abundant galaxies spectroscopically confirmed by the Hα line. This will provide the spectroscopic prior for the stacking analysis of this proposal.

## Interpretation Methods and Link to Testable Hypotheses

Fig.2 showcases binning strategies that can perform high signal-to-noise measurement of the [OI]63μm and [OIII]88μm lines' stacked spectra and can extract the line flux ratio. The top 2 panels show binning the prior catalog in SFR and redshift, respectively. Notably, as shown on the top left panel the stacking will enable the characterization of the line-SFR relation to SFR ∼ 1 $M_{sun}$/yr (SNR > 5 in the 0.5<log(SFR)<1 bin), which will be a substantial improvement compared to the estimated direct detection.

We can further categorize the 1<log(SFR)<1.5 prior galaxies with redshift, aiming to trace the transition of average galaxy properties from cosmic noon (z > 1) to z < 1. As demonstrated in the right column of Fig. 2, all the stacked spectra are forecasted to be detected with SNR > 10. These 2 binning strategies will enable studying the evolution of these properties over the course of cosmic time and with respect to various galaxy properties (such as SFR).

Using this stacking strategy, we can probe below direct detection limits with high SNR to study the evolution of [OIII]88µm (ionized gas [4] and potentially metallicity [8]), [OI]63µm (neutral gas density [4]), and their ratio (tracing the neutral/ionized ISM volume ratio [9]). Including data from NASA's TIM balloon enables analysis of [OI]63µm/[CII]158µm to constrain PDR/XDR contributions fraction. Results can be compared to predictions from PDR models (PDR Toolbox, [12]) and photoionization models (CLOUDY [13]).

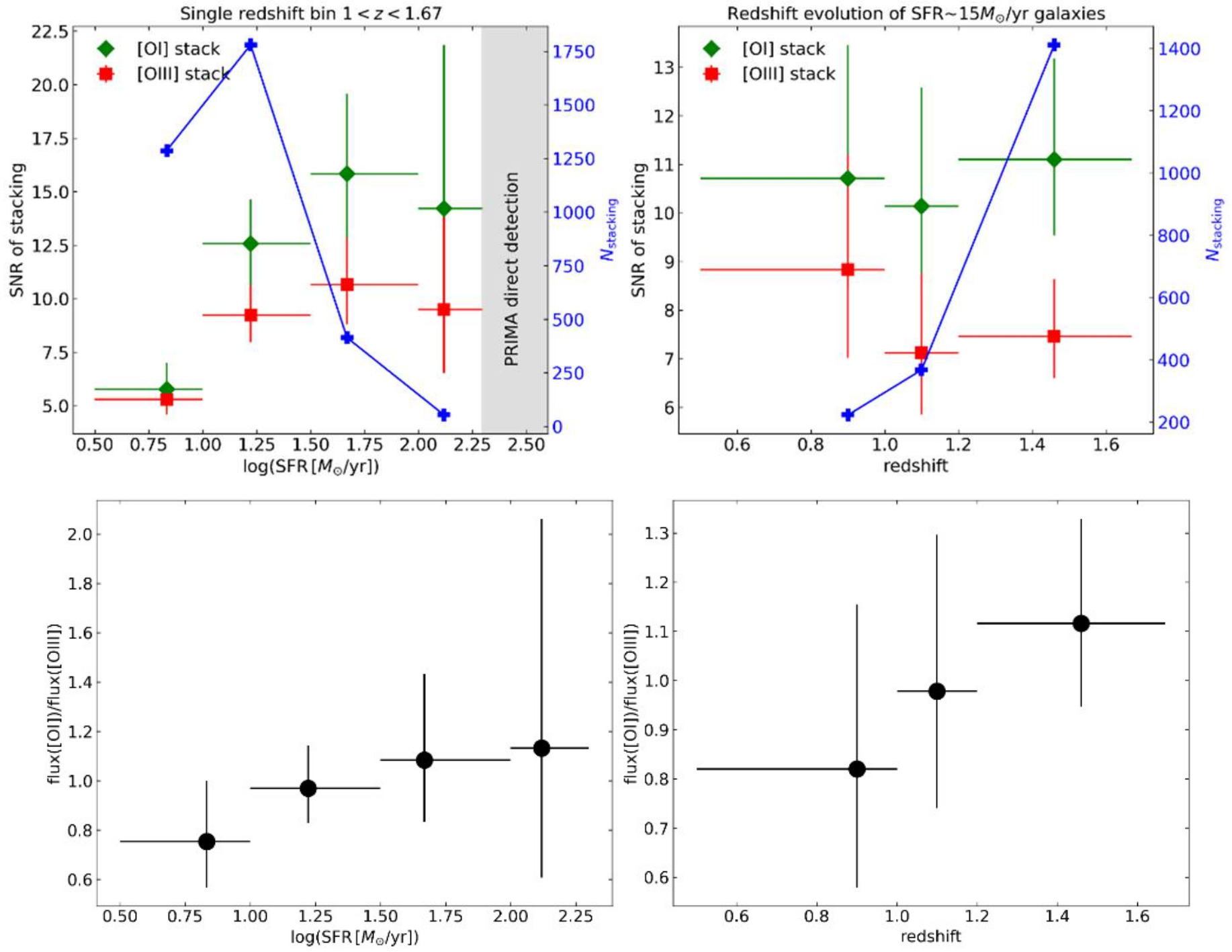

**Fig.2.** Example binning strategies and preliminary forecasts of the SNR of stacked results for 1000 Monte Carlo simulation runs. From the left to right column, the prior catalog is binned in SFR and further bin the 1<log(SFR)<1.5 galaxies in redshift, respectively, and then stack on the line emissions. *Top row*: The color-coded points show the bin range and the 1-sigma range of the stacked SNR in the 1000 runs. The grey region in the top left graph indicates the preliminary capability of the PRIMA direct detections. *Bottom row*: The anticipated line flux ratio [OI]63µm/[OIII]88µm based on their stacked spectra.

## Instruments and Modes Used

To carry out the scientific objectives of this proposal, PRIMA would use the FIRESS low-resolution mapping mode (R~100) to observe one 1° × 0.2° region centered on the GOODS-South field in bands 3 and 4. This field contains rich ancillary data in the GOODS-S field, and it lies within the ~10-deg$^2$ Euclid Deep Field Fornax (EDF-F) which will soon carry out a spec-z catalog and will also be covered by NASA's [CII] balloon experiment (Terahertz Intensity Mapper; TIM [10]).

## Approximate Integration Time

We proposed a 1000h PRIMA-FIRESS blind spectroscopic survey in the aforementioned field. The anticipated survey depth will be 2.5e-19 W/m$^2$–5e-19 W/m$^2$ in the wavelength range of 95 µm–235 µm, which ensures the full coverage of the [OI] and [OIII] line emission in z=0.5–1.67. Assuming a mock catalog of the EDF-F, and LIR-line luminosity relation from [14] for the [OI]63µm and [OIII]88µm, it would result in a significant detection of the stacked signal, in regimes below the direct detection regime of PRIMA.

As stated above, more far-IR lines are available within FIRESS bands, carrying valuable information for different science cases (i.e., metallicity, PDR fractions, strength of radiation field…) and we leave the prediction of their detection significance using Euclid stacking to future work.

## Special Capabilities Needed

N/A

## Synergies with Other Facilities

Euclid will measure the Hα line in z=0.5–1.67, the proposed survey field is enclosed by the 10 deg$^2$ Euclid Deep Field Fornax (EDF-F). This will provide an abundant number of galaxies for the investigation.

NASA's scientific balloon TIM (Terahertz Intensity Mapper, [10]) will measure the [CII]158 µm line in z=0.5–1.67 and target a 0.2 deg$^2$ field enclosing the GOODS-S field. The multi-line study will be enabled by, e.g., the line ratio between [CII] and [OIII] to infer the gas density of the photodissociation region in z=0.5–1.67 galaxies.

## Description of Observations

Blind spectroscopic survey using the PRIMA-FIRESS low-res mode R~130. This survey will uniformly scan a 0.2 deg^2 field with a total integration time of 1000 hours.

The observing field will be selected to enclose the GOODS-S field which has an abundant number of spectroscopically confirmed galaxies and to overlap with current and future galaxy surveys such as Euclid [6], enabling joint analysis.

## Acknowledgement

The research was carried out at the Jet Propulsion Laboratory, California Institute of Technology, under a contract with the National Aeronautics and Space Administration (80NM0018D0004).

# 39. Dust in High Redshift Quiescent Galaxies

Arianna Long (University of Washington), Jed McKinney (University of Texas at Austin), Erini Lambrides (NASA Goddard), Olivia Cooper (University of Texas), Sinclaire Manning (University of Massachusetts at Amherst), Darko Donevski (National Centre For Nuclear Research), Mariko Kubo (Tohoku University), Dave Clements (Imperial College London)

The abundant existence of massive quiescent galaxies in the early cosmos presents a challenge towards our understanding of rapid stellar mass assembly and subsequent star formation cessation (i.e. quenching). Understanding this phenomenon requires studying the conditions for star formation in the interstellar medium (ISM). Historically, studies focused on the stellar component owing to the accessibility and brightness of quiescent galaxies at rest-frame optical wavelengths. However, studies on the gaseous and dusty components of the ISM remain lacking. Observations and simulations of the nearby cosmos reveal non-negligible fractions of dust in the ISM of quenched galaxies. The nature of that dust—including its production mechanisms, chemical makeup, grain distributions, destruction / depletion mechanisms, and relationship with cold molecular gas—remains unknown. This knowledge gap permeates our attempts to discern a quiescent galaxy's future capacity for star formation. PRIMA FIRESS has the capability to observe the mid-to-far-infrared spectra of quiescent galaxies at z = 2.5-4, with the help of gravitational lensing. With a total of 20h per source using all four FIRESS bands, we can resolve polycyclic aromatic hydrocarbon (PAH) features, measure dust grain size distributions, constrain mass-weighted dust temperatures, and explore scenarios for dust production, destruction, expulsion, and depletion. All of these analyses combined will paint a detailed picture on how ISM conditions evolve from star forming through quiescence in early massive galaxies.

## Science Justification

Quiescent galaxies (QGs) are defined by their lack of star formation. The most massive QGs ($M_\star \gtrsim 10^{10.5}$ $M_\odot$) likely formed in the early cosmos (z > 2) then remained quenched for the majority of cosmic time [1–3]. Understanding the primary drivers of this quenching, along with the mechanisms that maintain quenching over several Gyr, requires in dept studies on their interstellar medium (ISM) content.

For years, the general consensus was that early massive quiescent galaxies stopped forming stars due to a lack of cold molecular gas [not warm atomic gas, 4]. Some studies on massive quiescent galaxies at $z \sim 1-2$ find low gas mass reservoirs ($f_{gas} < 5\%$), suggesting quenching happens rapidly via starvation and/or depletion [5, 6]; while others instead find significant reservoirs of molecular gas ($f_{gas} \sim 10-30\%$) indicating that quenching occurs slowly due to a reduction in overall star formation efficiency [7–9]. One of the primary drivers for so much uncertainty is a lack of knowledge on the dust content and properties in QGs. For example, accurate dust temperatures are vital for measuring dust masses from rest-frame far-infrared (IR) continuum data, which can then be used to estimate cold molecular gas masses [10, 11]. However, there is tension over whether $T_{dust}$ for quiescent galaxies is universal across cosmic time, or whether it evolves [6, 12]. There is also a severe gap in knowledge surrounding potentially exotic gas-to-dust ratios in QGs [13] due to rapid dust depletion and/or expulsion. Combined, these unknowns make it impossible to confidently understand the evolving interplay between cold dust and molecular gas in high-z QGs.

Over the last decade, several studies hinted at potentially significant dust reservoirs in QGs beyond the local Universe [7, 8, 14–16]. The origins of this dust is highly debated (in-situ production vs. ex-situ accretion from mergers), as are the timescales for dust production, destruction, expulsion and/or depletion. In contrast to the cold dust probed at rest-frame far-IR wavelengths, very little attention has been paid to the warm/hot dust in the mid-IR spectra of QGs beyond the local Universe [17–19]. Lying within the mid-IR spectra of these galaxies are likely polycyclic aromatic hydrocarbon (PAH) features [e.g., 20, 21] and even potential H2 emission lines. Decades of simulations and studies in local galaxies [22–26] show that the ratio of PAH features, such as the 7.7 μm and 11.3 μm bands, can be used to constrain a variety of physical ISM conditions, including: the dust grain size distribution, the strength of the local radiation field, and—when paired with far-IR data—depletion cycles for different chemical elements. Furthermore, H2 emission lines can be combined with the multi-wavelength IR data to directly constrain gas-to-dust ratios. All of this information is critical towards understanding the relationship between ISM conditions and galaxy quenching.

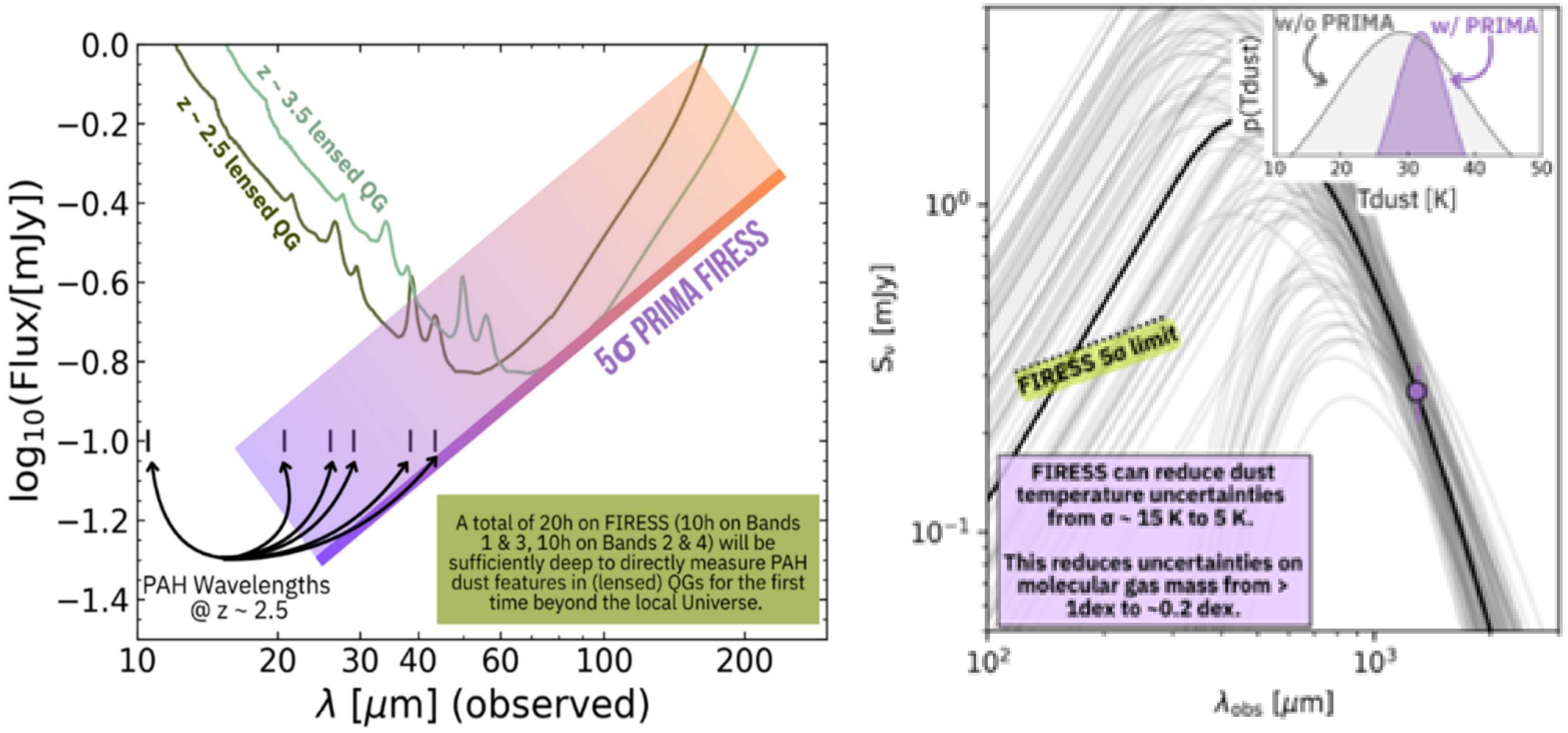

**Figure 1.** Left: a zoom in observed-frame mid-to-far-IR spectra of lensed dusty quiescent galaxies at z = 2.5 and 3.5. Overlaid are the predicted 5σ FIRESS depths assuming 10 hours each on Bands 1+3 and Bands 2+4, which are sufficiently deep to detect and resolve PAH features. Right: A model prediction of the far-IR dust blackbody using a lone ALMA 1.3mm continuum detection. Without constraints on the Wien's side of the dust blackbody, dust temperatures are highly uncertain. These drive significant uncertainties (>1 dex) on cold gas mass measurements, which are classically used to test quenching scenarios at high-z.

PRIMA will be able to explore and constrain all of these phenomena for the first time in QGs beyond the local Universe, specifically over a sample of gravitationally lensed z = 2 – 5 QGs from surveys such as REQUIEM-2D [27] or UNCOVER [28]. Gravitational lensing provides significant flux boosting for the typically faint IR spectra of QGs at z > 1 that would otherwise be undetectable. Using Band 1 & 2 on the Far-Infrared Enhanced Survey Spectrometer (FIRESS) with the native R = 100 resolving power, a suite of PAH features (including the 7.7 µm and 11.3 µm bands) can be studied from z ∼ 2 – 6 assuming lensing magnification factors of µ ∼ 5 − 20. Given simultaneous observations between Bands 1 & 3 and 2 & 4, this also effectively gives us constraints on the Wien's side of the far-IR dust blackbody for free. These measurements will significantly tighten unknowns on dust masses and temperatures in high-z QGs. Lastly, thanks to gravitational lensing, the coarser pixel resolutions and larger slits will not introduce significant contamination by nearby galaxies assuming they are well aligned along the extended image plane of lensed arcs.

With PRIMA, we can finally begin to answer questions such as: How does the dust-to-gas ratio evolve from extreme starbursts through to quiescence? What fraction of the dust is locked up in PAHs ($q_{PAH}$)? Are small dust grains preferentially destroyed during this process and, if so, what does this tell us about the dominant radiation field/ISM temperature in high-z QGs? What is the dominant dust production mechanism for quiescent galaxies, and how is that balanced with depletion to the gas-phase? How rapidly do these processes take hold and reflect in QG spectra, and are they coincident with the primary quenching event?

## Instruments and Modes Used

This observing program requests observations using FIRESS in low-resolution point-source mode, with 2 observations per galaxy (1 for each band configuration).

## Approximate Integration Time

Based on our ETC calculations, we estimate requiring 20 hours of time per source to achieve the needed sensitivity. For five sources, this is about 100 hours, assuming conservative instrumental specifications.

## Special Capabilities Needed

None

## Synergies with Other Facilities

This project will make significant strides in understanding dust and ISM evolution of QGs if paired with existing JWST and ALMA data. This includes imaging and spectroscopy as linking the results to galaxy ages, metallicities, carbon monoxide excitation, and potential AGN features will all provide additional crucial context. JWST/MIRI will also provide information on the 3.3 µm feature, as seen in e.g. [20]. The next generation VLA (ngVLA) will also prove significant by achieving deep sensitivities on cold gas and dust in high-z QGs, as will upcoming the Square Kilometer Array in studying 21 cm hydrogen emission out to z ∼ 8.

## Description of Observations

Our sensitivity limits are set by scaling low-z quiescent galaxy templates that include mid and far-IR dust [6, 19] to the known near-infrared flux densities of the lensed sample of quiescent galaxies from surveys like UNCOVER and REQUIEM-2D. Specifically, we aim to target the 5 brightest sources with near-IR continuum flux densities of 100 − 300 μJy at 1.25 μm. We also test for our desired sensitivities by scaling to known far-IR fluxes of the REQUIEM-2D sample [5], then using the modified blackbody SED fitting code [mcirsed, 29] to quantify the improvement on dust temperature measurements with the proposed long-wavelength depths in Band 3 & 4. We use the most up-to-date version of the PRIMA ETC to estimate the integration times needed to achieve the aforementioned science goals. We estimate an integration of time of 20 hours per source using FIRESS in low-resolution point-source mode, split between 10 hours of Bands 1/3 and another 10 hours of Bands 2/4. In this time we can achieve a 5σ RMS in the native R = 100 resolution of 50 μJy at 25 μm, 125, μJy at 62 μm, 300 μJy at 152 μm, and 470 μJy at 235 μm. Re-binning to R = 10 we can reach 16 μJy at 25 μm, 40, μJy at 62 μm, 100 μJy at 152 μm, and 150 μJy at 235 μm. These limits are sufficient to probe the rest-frame mid-to-far-infrared SEDs of lensed quiescent galaxies. Should these galaxies harbor significant PAH emission, the R = 100 spectra will enable us to decompose the PAHs and the continuum to infer PAH line ratios.

In fields like UNCOVER we will use the deep JWST and ALMA data to identify all bright infrared sources in the FoV of our PRIMA observations. This will be key for mitigating the effects of confusion on extracting flux densities from our targets.

# 40. Constraining the Total Infrared Luminosities of the Most Luminous Little Red Dots

Jed McKinney (UT Austin, NASA Hubble Fellow), David Setton (Princeton, Brinson Prize Fellow), Laia Barrufet (Edinburgh), Irene Shivaei (CAB), Ece Kilerci (Sabancı University), Tanmoy Bhowmik (SUST, Sylhet), Ivan Delvecchio (INAF), Justin Spilker (TAMU), Erini Lambrides (NASA Goddard), Keijan Chen (Peking University), Zhengrong Li (Peking University), Kohei Ichikawa (University of Waseda)

PRIMA will deliver orders-of-magnitude improvement in sensitivity at mid- through far-infrared wavelengths relative to previous space-based telescope missions. This enables unique constraint on the infrared spectral energy distribution of sources that are traditionally too faint to detect even by ALMA. One such category of sources are the enigmatic Little Red Dots (LRDs) discovered by the *James Webb Space Telescope (JWST)*. In this program we envision a pointed, direct-detection FIRESS experiment towards two of the brightest known LRDs, A2744-45924 and RUBIES-BLAGN-1. Deep ~26 hour observations of these LRDs could confirm or rule out the presence of hot (T~200 K) dust emission that is predicted by obscured AGN models. Combined with the existing MIRI and ALMA constraints, continuum confirmation of a lack of warm dust emission by FIRESS would significantly disfavor any significant dust attenuation in the AGN that power the most luminous LRDs.

## Science Justification

One of the persistent mysteries of the first few years of JWST's operation has been the nature of Little Red Dots (LRDs)—point sources characterized by their UV/optical V-shaped spectral energy distribution and broad Balmer emission (e.g., Greene et al. 2024). LRDs are highly numerous, with typical number densities of $10^{-5}$ cMpc$^{-3}$ at z ~ 5 (e.g., Kocevski et al. 2024), and as such, understanding their properties is crucial to understanding the demographics of the luminous high-z universe. Because an AGN is a natural explanation for broad lines, the prevailing interpretation has been that their red colors are the result of a highly dust-attenuated AGN, with the UV-optical dominated by an evolved stellar population that produces a Balmer break. However, the relative contribution of galaxy and AGN light in the rest-optical is subject to orders-of-magnitude uncertainty (Wang et al. 2024). Additionally, this model presents a number of problems. Even in stacks, LRDs show no evidence of the characteristic X-ray emission from a corona and mid-IR emission from a hot dust torus that is typically seen in massive AGN, and similarly, they are not detected in the far-IR where cold dust is luminous in highly attenuated sources (e.g., Akins et al. 2024, Lambrides et al. 2024). Thus, the bolometric luminosity of the broad population of LRDs is currently constrained only by upper limits.

Complementing this stacking analysis, Setton et al. 2025 presented deep panchromatic SED constraints of A2744-45924 (Labbe et al. 2024) and RUBIES-BLAGN-1 (Wang et al. 2024), two of the most luminous LRDs discovered to date in near-infrared JWST surveys, finding no evidence of emission from hot dust in MIRI detections or cold dust in deep ALMA non-detections. Together, these placed a hard cap on the total IR luminosity ~ 10^12 L⨀, disfavoring the aforementioned models that invoked a dust-attenuated AGN which predicted such high luminosities (see also Chen et al. 2025). These IR luminosity limits have motivated the exploration of exotic black hole models, such as a "Black Hole Star" that invokes hydrogen opacity rather than dust opacity to produce the characteristic LRD Balmer breaks and red continuum (e.g., Inayoshi et al. 2025). However, different implementations of this model have predicted a wide range of dust attenuation, ranging from none (e.g., Naidu et al. 2025) to Av ~ 2 (e.g., Ji et al. 2025), and current LIR constraints cannot distinguish between these models, or more typical galaxy+AGN models that require only modest dust attenuation.

PRIMA offers an opportunity to significantly improve these constraints. The current best constraints at rest-frame ~ 5−50 μm (observed-frame ~ 30−275 μm) come from shallow Herschel/PACs non-detections. A deep PRIMA drill with FIRESS could transform these constraints. ~ 26 hours of on-source time in the Band 1/Band 3 configuration would reach a continuum 5σ RMS of 10 μJy (38 μJy) at 24 μm (74 μm) after binning to R=10, deep enough to detect the extrapolation at rest-frame 5 μm of MIRI detections and over an order of magnitude deeper than the maximal IR luminosity (see Figure 1). Even non-detections would be extremely scientifically valuable; the bolometric luminosity of Little Red Dots has tremendous consequences for their implied black hole and stellar masses, and PRIMA is the only facility that could achieve these constraints. Additionally, the R=100 native resolution of FIRESS would be sensitive to emission in the rest-frame MIR, allowing for the first true constraints on PAHs in LRDs, in addition to star formation and AGN-sensitive fine structure lines.

## Instruments and Modes Used

FIRESS in low-resolution point source mode. Native R=100 resolution sufficient assuming dusty AGN torus scenario. Re-binning to R=10 needed to constraint alternative physical scenarios.

## Approximate Integration Time

26 hours per source (52 total) using FIRESS in low-resolution point-source mode (Estimates are based on the PRIMA ETC).

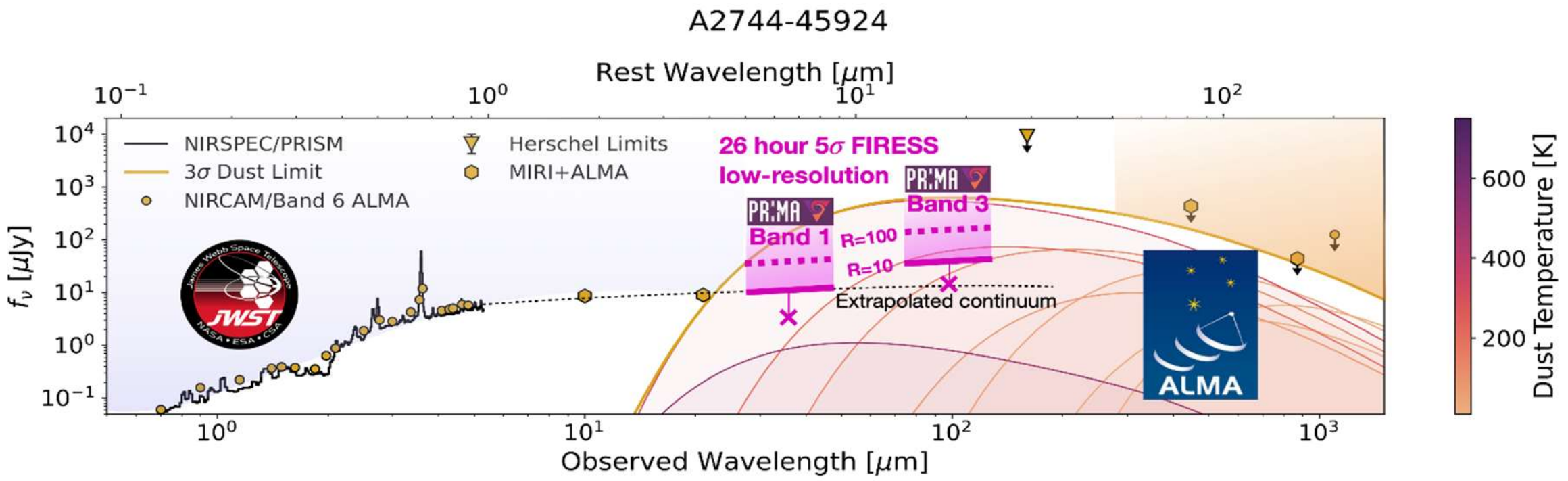

**Figure 1**. JWST+ALMA observations of the LRD A2744-45924 from Setton et al., 2025. With 26 hours using FIRESS in low-resolution point source mode we can simultaneously collect FIRESS Band 1 (24 − 43 µm) and Band 3 (74–134 µm) down to a 5σ, R = 10 continuum sensitivity of 10 µJy and 38 µJy respectively (solid purple lines). In FIRESS Band 1 this is sufficient to detect the continuum linearly extrapolated from JWST/MIRI F2100W (dashed black line) at 5σ. Should there be significant warm (Td ∼ 200) K dust emission, we could detect the continuum at > 5σ in the native R = 100 spectrum (dashed purple line). Purple × symbols show the depth reached by collapsing the full band to a single continuum point. PRIMA's unprecedented constraining power in the mid- through far-infrared is needed to ascertain the nature of LRDs

## Special Capabilities Needed

None

## Synergies with Other Facilities

This project will exploit the capabilities of both JWST and ALMA data. Both LRDs have already been observed with NIRSpec, yielding a reliable spectroscopic redshifts (z=4.4 and z=3.1). Both sources have been observed with JWST/MIRI, and show no evidence of a power law AGN torus at rest-frame 3 microns. In addition, both sources are undetected in ALMA dust-continuum at rest-frame 80-200 microns, providing key upper limits on their cold dust content.

## Description of Observations

We use the PRIMA Exposure Time Calculator (ETC) to make integration time estimates, leveraging the JWST and ALMA observations of Setton et al., 2025 shown in Figure 1 to anchor the RMS requirements. Our primary goal is to constrain the shape of the continuum emission beyond observed-frame 21 µm (λrest > 5 µm). Based on the modeling of Setton et al., 2025, the maximum possible far-infrared flux at these wavelengths is ∼ 50µJy at 24 µm and 300 µJy at 45 µm. However, the near-IR SEDs of LRDs do not exhibit rising continuum. Therefore, we adopt the known JWST/MIRI F2100W continuum flux density of 10 µJy as our 5σ RMS so that the observations are sensitive to a flat mid-infrared SED, allowable by models that do not invoke dust to explain the rest-frame UV/optical spectra of LRDs. FIRESS in low-resolution point-source mode, re-binning to R = 10, can achieve a 5σ RMS of 10 µJy at 24 µm in 26 hours. Simultaneously we would take a 5σ = 38 µJy R = 10 spectrum in FIRESS Band 3 (74 − 134 µm). This is sufficient to discriminate between a flat/falling spectrum vs. a rising dust SED.

Using the deep ALMA observations in the UNCOVER field from Fujimoto et al., 2023, we find that the nearest sub-mm source to A2744-45924 are 11 arcsec (22 µJy source, also 100 µJy at 21 µm) and 50 arcsec (150 μJy source) away. There is also a F2100W= 400 μJy source 13 arcsec away. Combined with the rich JWST photometric and spectroscopic coverage in UNCOVER we will mitigate the impact of blending (confusion) on our measurement because (1) the positions of the bright IR sources will be known from ALMA and JWST, and (2) the redshifts of all nearby galaxies are known from either deep, multi-band photometric coverage with JWST and/or spectroscopy.

# 41. Chemical Evolution and Feedback in Dust-Obscured Galaxies: Targeted PRIMA Spectroscopy of Metallicity and Outflows at Cosmic Noon

Tohru Nagao (Ehime University, Japan), J.A. Fernández Ontiveros (CEFCA, Spain), L. Spinoglio (INAF-IAPS, Italy), T. Hashimoto (U. of Tsukuba, Japan), K. Ichikawa (Tohoku U., Japan), H. Inami (Hiroshima U., Japan), T. Izumi (NAOJ, Japan), B. Pérez-Díaz (IAA-CSIC, Spain), E. Pérez-Montero (IAA-CSIC, Spain), T. T. Takeuchi (Nagoya U., Japan), Y. Tamura (Nagoya U., Japan), H. Umehata (Nagoya U., Japan), J. Vílchez (IAA-CSIC, Spain), T. Wada (NAOJ, Japan)

We propose a focused spectroscopic program using PRIMA's FIRESS instrument to investigate chemical enrichment and feedback mechanisms in a comprehensive sample of galaxies from cosmic noon ($1 < z < 3.1$) to the reionisation epoch. Building on existing JWST, ALMA, and ground-based datasets, this program will deliver low- and high-resolution spectra of ~75 dusty star-forming galaxies (DSFGs) and AGN, as well as ~20 high-redshift sources, providing an unprecedented view of galaxy evolution during critical epochs. Our primary goals include measuring nitrogen-to-oxygen (N/O) and oxygen-to-hydrogen (O/H) abundance ratios using robust mid-IR diagnostics—primarily [NIII]57 µm and [OIII]52,88 µm lines—and characterizing AGN-driven and starburst-driven outflows via high-resolution spectroscopy of [OIV]25.9 µm emission and molecular OH absorption lines. Additionally, we will obtain the first rest-frame mid-IR spectra of luminous and gravitationally lensed galaxies at the reionisation epoch ($z > 6$), exploring their chemical properties through key high-ionisation transitions ([NeIII]15.6 µm, [NeV]14.3 µm, [SIII]18.7 µm, and [SIV]10.5 µm). PRIMA's unique capabilities in sensitivity, wavelength coverage, and spectral resolution will enable these critical observations, probing deeply into obscured star-forming and accreting regions inaccessible by other facilities in the coming decade. This program will thus establish crucial empirical benchmarks for chemical evolution models and elucidate the physical mechanisms driving galaxy transformation and quenching across cosmic history.

## Science Justification

Chemical enrichment and feedback processes are central to the evolution of galaxies, governing their growth, structure, and eventual quenching. Heavy elements (metals) are synthesized in stars and redistributed to the interstellar medium (ISM) via stellar winds, supernovae, and active galactic nuclei (AGN) feedback. These processes shape galaxy scaling relations, such as the mass-metallicity relation and fundamental metallicity relation, and drive gas flows that regulate star formation. However, our current understanding of these mechanisms is largely based on optical diagnostics, which are unreliable in the heavily dust-obscured environments that dominate during cosmic noon (z ~ 1–3).

Mid- and far-infrared (IR) spectroscopy offers the best avenue for accessing these hidden regions. While Herschel and SOFIA provided crucial early insights, their sensitivity was insufficient to build large samples of high-z galaxies with robust metallicity measurements and outflow diagnostics. PRIMA/FIRESS offers the opportunity to overcome these limitations, enabling metallicity and feedback studies in representative dusty galaxies during the Universe's peak epoch of star formation and black hole growth.

This program addresses two interconnected questions:

1. What are the nitrogen-to-oxygen (N/O) and oxygen-to-hydrogen (O/H) abundance ratios in dusty galaxies at cosmic noon, and how do these trace chemical enrichment and gas accretion histories?
2. How do AGN and starburst-driven outflows manifest in obscured galaxies, and what are their roles in quenching star formation and regulating galaxy evolution?

PRIMA's FIRESS spectrometer is uniquely capable of addressing these questions by providing access to the full suite of key IR fine-structure lines:

- Metallicity diagnostics: [NIII]57 µm, [OIII]52 µm, and [OIII]88 µm enable robust N/O measurements in dusty environments inaccessible to optical tracers. Unlike optical [NII] lines, the [NIII]57 µm line is less affected by contamination from low-ionisation diffuse gas and provides a cleaner probe of star-forming regions. The combination of these lines with some other emission lines will give constraints on the O/H abundance ratio independently from the N/O abundance ratio.
- Feedback diagnostics: High-resolution spectroscopy of [OIV]25.9 µm and molecular OH (79, 119 µm) lines reveals outflow signatures via emission-line velocity profiles and P-Cygni profiles, tracing both ionized and molecular phases of galactic winds.

No other existing or planned facility offers PRIMA's combination of sensitivity, wavelength coverage (24–235 µm), and high-resolution capability (R > 2000), making it indispensable for this program.

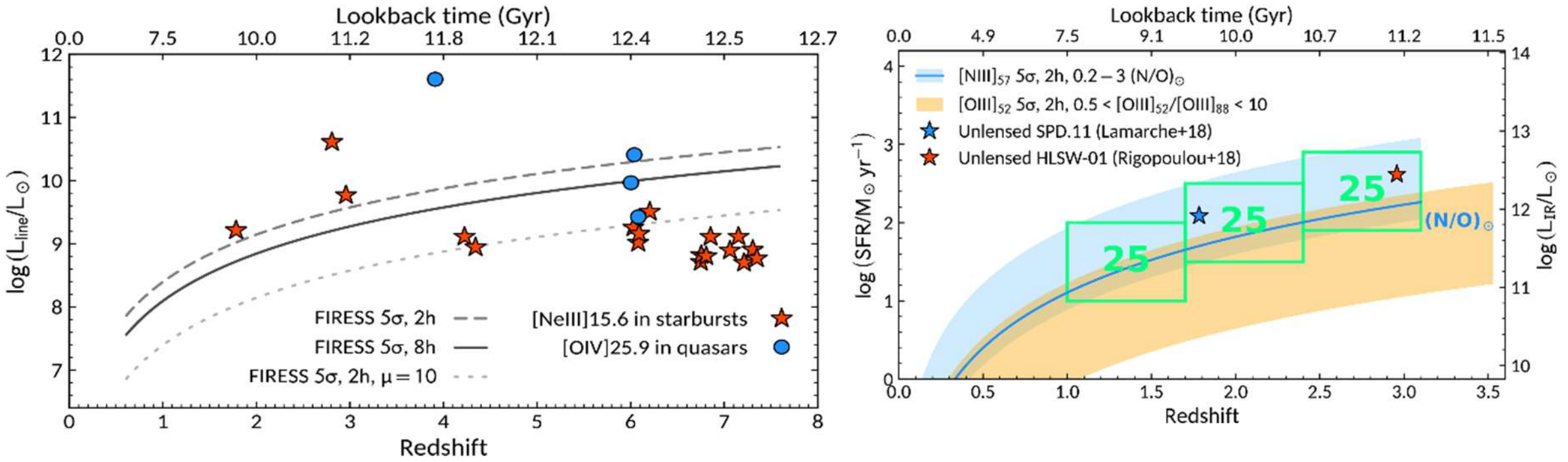

**Fig. 1. Left:** FIRESS detection limits (5**σ**, 2 h, low-res.) for [NIII]57 µm and [OIII]52 µm on the SFR-redshift plane. The blue-shaded area shows the [NIII]57 µm sensitivity based on the $L_{IR}$–[OIII]88 µm correlation (Mordini et al. 2021) for N/O = 0.2–3 (N/O)⊙ (Spinoglio et al. 2022), with the blue line indicating solar N/O = -0.86 dex (Asplund et al. 2021). The orange-shaded area shows the [OIII]52 µm sensitivity for density-dependent [OIII]52/[OIII]88 ratios (0.5–10; Fernández-Ontiveros et al. 2016, Spinoglio et al. 2022). Stars show the magnification-corrected luminosities of lensed galaxies with [NIII]57 detections (Lamarche et al. 2018, Rigopoulou et al. 2018). Green boxes indicate the proposed observational program coverage. **Right:** Expected [NeIII]15.6 µm fluxes in star-forming galaxies (red stars) and [OIV]25.9 µm in quasars (blue circles) based on high-z [OIII]88 µm detections (Ferkinhoff et al. 2010, Lamarche et al. 2018, Rigopoulou et al. 2018, Walter et al. 2018, De Breuck et al. 2019, Hashimoto et al. 2019a, 2019b, Harikane et al. 2020, Witstok et al. 2022, Ren et al. 2023, Algera et al. 2024), assuming typical line ratios from local galaxies (Cormier et al. 2015, 2019, Fernández-Ontiveros et al. 2016). Gray lines show FIRESS sensitivity limits for different integration times and magnifications (5**σ**, 2 h dashed; 8 h solid, 2 h with μ = 10 dotted).

FIRESS low-resolution mode (R > 85) will detect [NIII]57 µm and [OIII]52 µm lines in main-sequence galaxies from the local Universe to cosmic noon (z < 3.1). Fig. 1 (left panel) shows the expected FIRESS sensitivity (5**σ**, 2 h, low-res. full-band spectrum) for [NIII]57 µm emission (blue-shaded area), based on the log$L_{IR}$–[OIII]88 µm correlation for low-metallicity galaxies (Mordini et al. 2021) and N/O abundances of 0.2–3 (N/O)⊙ (Spinoglio et al. 2021). Blue and red stars indicate magnification-corrected luminosities for the two lensed galaxies with very marginal [NIII]57 µm detections (Lamarche et al. 2018, Rigopoulou et al. 2018). A proposed observing program could target 75 dusty star-forming galaxies across three redshift bins: 25 galaxies with $L_{IR} > 10^{10.8}$ L⊙ at 1 < z < 1.6, 25 with $L_{IR} > 10^{11.3}$ L⊙ at 1.6 < z < 2.4, and 25 with $L_{IR} > 10^{11.7}$ L⊙ at 2.4 < z < 3.1. Assuming a sensitivity of $1.9 \times 10^{-19}$ W $m^{-2}$ (5**σ**, 1 h per spectral setting), the [NIII]57 µm line can be detected in main-sequence cosmic noon galaxies with ~ 1–2 h integration per target. The total program needs ~ 360 hours, with two exposures per target covering the full spectral range.

Beyond N/O ratios, O/H abundances can be obtained using FIRESS's suite of nebular IR lines. Local Universe observations demonstrate that robust metallicities for both star-forming galaxies and AGN can be derived using the brightest IR lines (Fernández-Ontiveros et al. 2021, Pérez-Díaz et al. 2022). While N/O measurements extend to cosmic noon (z < 3.1), O/H abundances can be determined for luminous or gravitationally-lensed galaxies up to the reionization epoch. The right panel in Fig. 1 shows expected fluxes for [NeIII]15.6 µm in star-forming galaxies (red stars) and [OIV]25.9 µm in quasars (blue circles), based on high-z [OIII]88 µm detections (Ferkinhoff et al. 2010, Lamarche et al. 2018, Rigopoulou et al. 2018, Walter et al. 2018, De Breuck et al. 2019, Hashimoto et al. 2019a, 2019b, Harikane et al. 2020, Witstok et al. 2022, Ren et al. 2023, Algera et al. 2024) and typical line ratios ([NeIII]15.6/[OIII]88 ~ 0.4 and [OIV]25.9/[OIII]88 ~ 4) observed in local low-metallicity star-forming (Cormier et al. 2015, 2019) and Seyfert galaxies (Fernández-Ontiveros et al. 2016). Metallicity estimates can be obtained with, e.g. HII-CHI-mistry-IR (Pérez-Montero et al. 2014, Fernández-Ontiveros et al. 2021, Pérez-Diaz et al. 2022), which analyzes multiple emission lines ([NeII]12.8 μm, [NeIII]15.6 μm, [SIII]18.7, 33.5 μm, [SIV]10.5 μm, [OIII]52,88 μm, and [NIII]57 μm) while accounting for ionization parameter and N/O ratios. Though some lines fall outside FIRESS coverage at specific redshifts (e.g., [SIV]10.5 μm at z < 1.3, [OIII]88 μm at z > 1.6, [OIII]52 μm and [NIII]57 μm at z > 3.1), the code flexibly uses available lines for best-effort metallicity determinations.

FIRESS will also play a pivotal role in studying the feedback mechanisms that regulate galaxy evolution. Feedback episodes often manifest as multiphase gas outflows, with the molecular component carrying most of the mass, directly affecting the star-forming reservoir. These outflows can be identified in high-resolution spectra through P-Cygni profiles and absorption blueshifted wings in molecular lines such as the OH doublets at 119 and 79 μm (González-Alfonso et al. 2012, 2017b, Spoon et al. 2013), and through emission line wings of highly ionized emission lines such as [OIV]25.9 μm and [NeV]14.3,24.3 μm. These measurements will provide the mass, energetics, and kinematics of outflows across different galaxy evolutionary stages. By observing a sample of galaxies with high star-formation rates and AGN activity, FIRESS will disentangle the roles of starburst-driven winds and AGN-driven outflows in quenching star formation. Combining these results with observations from facilities like JWST and ALMA will enable a comprehensive multiphase characterization of feedback processes, revealing how galaxies transition from gas-rich, star-forming systems to passive, red-and-dead states.

## Instruments and Modes Used

This program requires 20 pointings in FIRESS high-res mode and 75+20 pointings x 2 spectral settings in FIRESS low-res mode.

## Approximate Integration Time

We will select targets of ~75 galaxies across three redshift bins:

- 25 galaxies with L_IR > $10^{10.8}$ $L_\odot$ at 1 < z < 1.6,
- 25 galaxies with L_IR > $10^{11.3}$ $L_\odot$ at 1.6 < z < 2.4,
- 25 galaxies with L_IR > $10^{11.7}$ $L_\odot$ at 2.4 < z < 3.1.

For each target, we will obtain:

- Low-resolution spectroscopy (R > 85) for metallicity diagnostics: Integration time of ~1-2 hours per target, enabling 5σ detections of the [NIII]57 μm line for main-sequence galaxies in the specified $L_{IR}$ ranges.

For a subset of ~20 galaxies showing strong AGN/starburst signatures:

- High-resolution spectroscopy (R > 2000) for outflow diagnostics: Integration time of ~3 hours per target to achieve S/N sufficient to resolve kinematic structures in [OIV]25.9 μm and OH lines.

In addition, we will observe ~20 high-z galaxies:

- Low-resolution spectroscopy (R > 85) for metallicity diagnostics: Integration time of ~2 hours per target, enabling 5σ detections of the [NeIII]15.6 μm line for bright and/or lensed high-z sources with $L_{Line}$ / μ > $10^9$ $L_\odot$, where μ is the lens magnification factor.

Total time request:

- Low-resolution component: ~75 targets × 1-2 h × 2 spectral sets ~ 200 hours.
- High-resolution component: ~20 targets × 3 h = 60 hours.
- High-z component: ~20 targets × 2 h × 2 spectral settings = 80 hours.
- Total estimated integration time: ~200 + 60 + 80 hours = 340 hours (excluding overheads).

Estimation method: Integration times are based on FIRESS sensitivity estimates (e.g., 5σ line sensitivity of ~$1.9 \times 10^{-19}$ W $m^{-2}$ in 1 h per spectral setting), scaled from performance simulations presented in the PRIMA JATIS paper.

## Special Capabilities Needed

N/A

## Synergies with Other Facilities

This program is designed to maximize complementarity with existing and forthcoming multi-wavelength datasets:

- JWST: Provides rest-frame optical and near-IR spectroscopy for the target sample, offering ancillary metallicity and SFR measurements to compare with IR-based diagnostics, as well as high-resolution imaging for precise source morphology and size.
- ALMA: Supplies sub-mm spectroscopy (e.g., [CII]158 μm, [NII]122 μm, [NII]205 μm) and continuum measurements, allowing us to cross-check metallicity estimates and constrain dust and gas masses.
- Ground-based optical/NIR surveys (e.g., VLT, Keck, Gemini, Subaru, and future ELTs): Deliver stellar mass, redshift, and environmental data for sample selection and interpretation.
- X-ray facilities (e.g., Chandra, XMM-Newton): Identify AGN components and provide complementary BHAR estimates, supporting interpretation of [O IV]25.9 μm measurements.

By targeting sources with rich ancillary data, we ensure robust multi-phase characterization of chemical enrichment and feedback processes, enabling powerful joint analyses that extend beyond the capabilities of PRIMA alone.

## Description of Observations

We will conduct pointed spectroscopic observations of ~75 known galaxies with redshifts between 1 and 3.1, selected from JWST, ALMA, and other well-characterized samples. Targets are chosen to span a wide range of stellar masses, IR luminosities, and AGN/starburst activity levels, ensuring diversity in metallicity and feedback properties. We will also observe ~20 additional galaxies at much higher redshift selected among bright and lensed systems detected at the reionisation epoch (Fig. 1, right).

For each galaxy:

- Low-resolution spectroscopy (R > 85):

  Two exposures of 1 hour each will cover the full 24–235 µm range, capturing key metallicity diagnostics ([NIII]57 µm, [OIII]52,88 µm) and supporting lines (e.g., [NeIII]15.6 µm, [SIII]33.5 µm). These observations will provide robust N/O and O/H measurements for all targets.

- High-resolution spectroscopy (R > 2000) (subset of ~20 targets):

  Single exposures of ~3 hours will focus on feedback tracers, including [OIV]25.9 µm and OH doublets (79, 119 µm), to resolve kinematic structures indicative of outflows.

- High-z galaxy observations with low-resolution spectroscopy (R > 85) (~20 targets):

  Two exposures of ~2 hours each to cover the full 24–235 µm range, to determine chemical abundances in galaxies at cosmic dawn from redshifted mid-IR nebular lines (e.g., [NeIII]15.6 µm, [NeV]14.3 µm, [SIII]18.7 µm, [SIV]10.5 µm). These observations will provide robust O/H measurements for all targets.

Observations will follow standard FIRESS templates. Targets will be prioritized for scheduling based on existing redshifts and visibility windows, with no special timing constraints. Calibration will use routine procedures to ensure high-fidelity flux and wavelength accuracy.

This strategy allows us to systematically probe chemical enrichment across cosmic noon and detect feedback signatures, linking metallicity buildup to outflow processes in dusty, high-z galaxies.

# 42. Rapid Carbonaceous Dust Evolution in EoR Galaxies: Synergies Among JWST, PRIMA, and ALMA

Yurina Nakazato (The University of Tokyo), Kosei Matsumoto (Ghent University), Tohru Nagao (Ehime University), Takuya Hashimoto (Tsukuba University), Hanae Inami (Hiroshima University), Daniel Ceverino (Universidad Autonoma de Madrid)

Recent JWST observations have revealed the existence of UV bumps at z = 4-7 galaxies for the time, which implies the presence of small carbon grains much earlier than expected. Subsequent theoretical studies insight that the bump may be caused by Polycyclic Aromatic Hydrocarbons (PAHs), although their formation pathways in the early Universe remain unclear. Currently, JWST can detect UV bumps but cannot directly observe the mid-infrared PAH emission bands. Similarly, ALMA is limited to observing longer-wavelength dust continuum emission and cannot probe the distinct PAH vibrational features. Only PRIMA's FIRESS instrument can directly detect PAH emission lines, thus uniquely bridging this observational gap. We propose FIRESS low-resolution spectroscopy targeting UV bump-detected galaxies observed by JWST and dusty galaxies identified by ALMA. Based on the 3D dust radiative transfer calculation in cosmological simulations, an integration time of 90-100 hours per galaxy will allow the detection of PAH features with S/N =5. This approach will directly test whether PAHs were already abundant at Cosmic Dawn ($z \gtrsim 6$). PAHs are potential carriers of the observed UV bump and their presence would imply rapid dust enrichment. Combining PRIMA data with JWST and ALMA observations will thus clarify dust formation and chemical evolution in the early Universe.

## Science Justification

### Broader Context

Recent results from the JWST /JADES survey show that ~20 % of spectroscopically confirmed galaxies at z > 4 (10 out of 49) exhibit a pronounced UV extinction bump (Witstok+23, Markov+23, 24, Ormerod+25, Fisher+25). Such a UV bump has been well known for over 60 years since as the 2175 Å extinction feature observed in the Milky Way and the Large Magellanic Cloud (Strecher+65), while notably absent in low-metallicity environments such as the Small Magellanic Cloud (SMC). The origin of the local UV bump is commonly attributed to small carbonaceous dust species such as Polycyclic Aromatic Hydrocarbons (PAHs), and graphite. PAHs are primarily produced by asymptotic giant branch (AGB) stars (Cherchneff+92). However, the typical evolution timescale for low-mass star ($M^* \lesssim 2.5\ M_{sun}$) to reach the AGB phase is ~ 1 Gyr (Dell'Agli+19), which exceeds the cosmic age at z~ 6-7 (i.e., $t_{age}$ = 760–920 Myr). Alternative dust formation pathways have been proposed by recent theoretical studies, such as grain shattering (Matsumoto+2024, Narayanan+2023). In this scenario, larger grains initially produced by Type-II supernovae can fragment into smaller grains. However, grain shattering also operates on relatively long timescales of several hundred million years (Hirashita+20), which still poses challenges in explaining dust formation in such young, high-redshift galaxies.

Moreover, the observed high-redshift UV bumps are systematically shifted toward longer wavelengths (2236 Å) and exhibit narrower widths compared to the well-known local UV bump at 2175 Å. While both graphite and PAHs adequately explain the local bump Draine+11, recent theoretical studies indicate that graphite alone cannot reproduce the longer wavelength position and narrower width observed in high-redshift UV bumps (Li+24). Lin+25 demonstrate that a mixture of PAH molecules with various sizes (carbon atom numbers) can successfully reproduce these observed features. This implies a surprising result that PAH molecules already existed within the first billion years after the Big Bang.

To obtain definitive evidence for the presence of small carbonaceous dust grains (PAHs) in high-redshift galaxies, it is crucial to directly detect their characteristic rest-frame mid-infrared vibrational emission bands, which can be achievable only through the proposed PRIMA survey. As a pilot study, we aim to detect at least the brightest PAH emission feature from as many galaxies at $z \gtrsim 6$ as possible.

### Science Questions

The main science questions addressed by this study are :

- "Is the JWST-detected high-redshift UV bump truly caused by PAHs?"
- "If confirmed, how common are PAHs in high-redshift galaxies (z>6)?"

These questions are closely related to one of PRIMA's key science themes, "Buildup of Dust and Metals".

### Need for PRIMA

PAH molecules exhibit distinctive vibrational emission features at rest-frame mid-infrared wavelengths: primarily at 3.3, 6.2, 7.7, 8.6, 11.3, and 12.7 µm. For galaxies in the Epoch of Reionization (z > 6), these spectral features shift into wavelengths (approximatively 46–62 µm) that are not accessible by JWST or ALMA. JWST detects only the rest-frame UV bump, whereas ALMA can detect only far-infrared/submillimeter dust continuum emission from larger dust grains. Therefore, only PRIMA can directly detect PAH emission features, and conclusively test the origin of the high-redshift UV bump and clarify the enigmatic process of dust formation at cosmic dawn.

## Interpretation Methods

We calculate PAH emission luminosity of high-z galaxies using the zoom-in cosmological simulations, FirstLight (Ceverino+17). We focus on a galaxy at z= 7.2 with a stellar mass of 5.39e9 $M_{sun}$. To calculate dust attenuation and subsequent infrared re-emission, we perform three-dimensional Monte Carlo radiative transfer simulations using the publicly available SKIRT code (Baes+11). Since the original FirstLight simulations do not include dust evolution or dust grain physics, we assume that the spatial distribution of dust traces that of metals, with a fixed dust-to-metal ratio (DTM) of 0.4. For the dust grain composition, we adopt dust models of MW from Weingartner & Draine 2001. Our calculations include important physical effects such as stochastic heating from small grains, and self-absorption in the calculation, which can increase the mid-IR/ FIR luminosity (Behrens+18). The details will be described in Nakazato in prep.

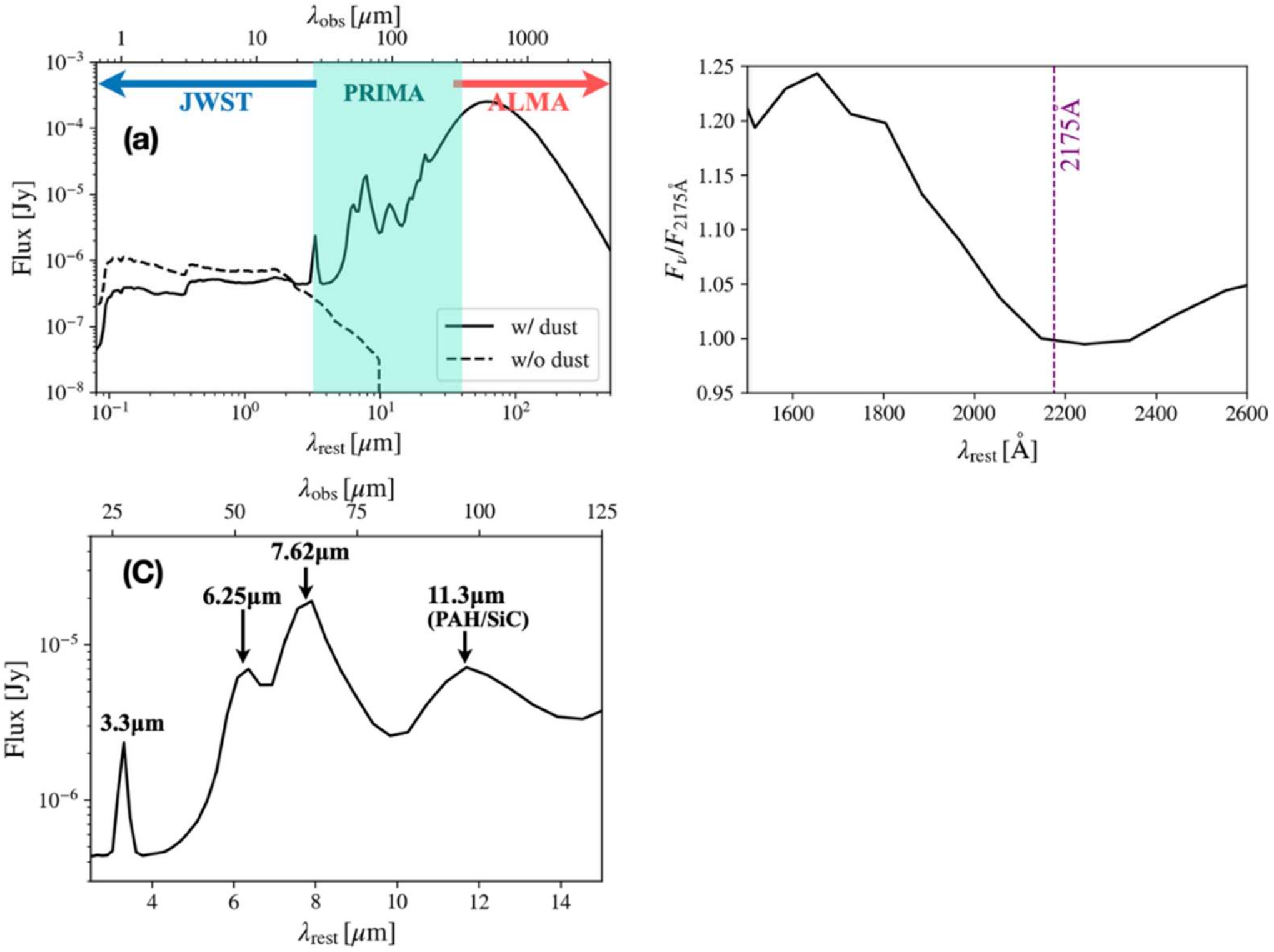

Figure 1 shows the SEDs for a z=7.2 simulated galaxy. The total IR luminosity is LIR = 2.93e12 $L_{sun}$, which dust continuum can be observed by ALMA (e.g., REBELS-12; Algera+24). Panel (a) highlights the wavelength coverage accessible by JWST, PRIMA, and ALMA, clearly showing that these facilities provide complementary information across UV-to-FIR wavelengths. Panel (b) illustrates the rest-frame UV continuum normalized at 2175Å, exhibiting a UV-bump at $\lambda_{rest}$ = 2100–2300Å. Panel (c) focuses on the rest-frame mid-IR emission from PAH thermal features, which are in the range PRIMA covers. Among these PAH emissions, the 7.62 μm feature is the brightest, therefore we focus our feasibility analysis specifically on detecting this feature.

Even if only the brightest PAH emission feature (7.7 μm) is detected, the measured flux can still provide a valuable estimate of the total PAH mass (MPAH) using luminosity-mass conversion factor obtained in the above simulation results. Such constraints are essential for testing whether PAHs alone can fully account for the observed UV bump. If the total PAH mass inferred from the detected emission and the derived upper limits are insufficient to explain the UV bump, it would strongly suggest the presence of additional carbonaceous dust components, providing critical constraints on alternative dust formation mechanisms in high-redshift galaxies (also see Section *Synergies with other facilities*).

## Instruments and Modes Used

FIRESS: pointed low-resolution, R~80-130, 1 pointing per target

## Approximate Integration Time

Table 1 shows the estimated fluxes of PAH 7.62 μm emission and corresponding integration timed derived using PRIMA Exposure Time Calculator. We estimate the expected PAH fluxes by scaling the reference PAH luminosity from the simulated galaxy shown in Figure 1, assuming that the PAH luminosity scales linearly with stellar mass. We then derive the flux at each redshift based on this scaling. Since our primary observational goal is to detect at least one PAH emission feature per galaxy, integration times are computed specifically for the strongest emission feature. If multiple PAH features are targeted for a single galaxy, additional integration time is required accordingly, as illustrated in Figure 1(c). Measuring the flux ratios among different PAH emission lines allows us to constrain the grain-size distribution of PAHs and ionization fractions of PAHs (e.g., McKinney+21, Chastenet+23).

We plan to observe a total of 15 galaxies at z $\gtrsim$ 6: REBELS-08, -12, -25, GNWY-7379420231, JADES-GS-z6-0, s00717, two additional sources from Witstok+23, and five additional sources from Markov+25. We thus request a total integration time of approximately 90–100 hours.

Tablel: Estimation of PAH 7.62 μm emission flux and corresponding integration time for UV-bump detected galaxies. (※1) There are no stellar mass information for 1433 3989 and 2750_449 in Markov+24, we roughly estimated the mass by scaling [0111]5007 flux of GNWY-7379420231.(※2)Ten galaxies at z=4-7 are detected with UV bumps. (※3) We roughly estimated the median redshift of 10 samples at z=4-8 in Figure 4 of Markov+25.

| Name | redshift | Ms [Msun] | F(7.62um) [W/m^2] | integration time [hr] | reference |
|---|---|---|---|---|---|
| FL958 | 7.2 | 5.35E+09 | 1.06E-18 | <5min | FiesrLight simulation |
| REBELS-08 | 6.749 | 2.04E+09 | 4.66E-19 | 0.17 | Fisher+25 |
| REBELS-15 | 6.875 | 2.51E+09 | 5.50E-19 | 0.12 | Fisher+25 |
| REBELS-25 | 7.307 | 1.17E+09 | 2.23E-19 | 0.73 | Fisher+25 |
| GNWY-7379420231 | 7.11 | 2.53E+08 | 5.13E-20 | 13.72 | Ormerod+25 |
| JADES-GS-z6-0 | 6.7 | 1.00E+08 | 2.32E-20 | 67.07 | Witstok+23 |
| median of UV-bump galaxies in Witstok+23 | 5.25 (※2) | 1.80E+08 | 7.38E-20 | 6.63 | Witstok+23 |
| s00717 | 6.932 | 4.90E+08 | 1.30E-19 | 2.14 | Markov+23 |
| 1433_3989 | 6.14 | ~2.5e7(※1) | 7.11E-21 | > 100 | Markov+25 |
| 2750_449 | 7.55 | ~2.5e7(※1) | 4.41E-21 | >100 | Markov+25 |
| median of UV-bump galaxies in Markov+25 | ~5 (※3) | 5.62E+08 | 2.58E-19 | 0.54 | Markov+25 |

## Special Capabilities Needed

None

## Synergies with Other Facilities

Our proposed PRIMA observations will offer powerful synergies with JWST and ALMA. JWST has already spectroscopically identified UV bumps in galaxies at high redshifts ($z > 4$) using NIRSpec and PRISM ($R \sim 100$). ALMA has detected rest-frame FIR dust continuum from some high-redshift galaxies with large stellar masses of $M^* \gtrsim 1e9\ M_{sun}$ (e.g., A1689-zD1, Big Three Dragons). However, JWST had not previously confirmed whether most of these ALMA-detected dusty galaxies exhibit UV bumps. Only very recently, 3 out of 12 REBELS galaxies were reported to show evidence of UV bumps at 4σ significance. Still, the presence of UV bumps remains uncertain for many other ALMA-detected dusty galaxies. PRIMA observations will therefore be essential for clarifying the presence and universality of PAH emission in such dusty galaxies. Even non-detections of PAH features will provide critical constraints by excluding PAH-based dust attenuation curves.

Table 2 summarizes the observational strategy and synergetic studies enabled by combining PRIMA, JWST, and ALMA. JWST can estimate metallicities and constrain attenuation curves, thereby testing for the presence of small grains, although their abundance remains uncertain. ALMA observations detect rest-frame FIR dust emission, but their limited wavelength coverage makes it difficult to accurately determine the peak wavelength of the gray-body emission. This limitation introduces large uncertainties in dust temperature estimates and consequently in the inferred total dust mass (Choban et al. 2024). PRIMA complements these by observing mid-IR dust emission, specifically small grains and PAH features. Combining JWST, ALMA, and PRIMA observations will enable accurate constraints on total dust mass, dust temperature distributions, and PAH mass fractions through comprehensive SED fitting. Such multi-wavelength analyses will significantly deepen our understanding of PAH and small-grain formation scenarios in the early Universe.

Table 2: Summary of our synergetic observation with JWST, ALMA, and PRIMA.

| **Detection combination** | UV bump+PAH emission | PAH emission + IR emission |
|---|---|---|
| **observation combination** | JWST+PRIMA | PRIMA+ALMA |
| **physical information** | PAH abundance/ attenuation curve | PAH abundance/ dust temperature (and dust mass) |
| **targets** | already detected UV-bump galaxies | ALMA-detected dust continuum emission in LBGs |

## Description of Observations

We propose to use PRIMA FIRESS in low-resolution mode ($R \sim 80$-130) to observe the PAH 7.62 µm emission feature from galaxies identified by JWST to show UV bumps and from dusty galaxies observed by ALMA. At the targeted redshifts, the PAH 7.62 µm feature shifts into the observed wavelength range of approximately 46–62 µm. We exclude targets requiring estimated exposure times exceeding 100 hours (see Table 1).

# 43. Recovering the Temperatures of the Warm and Cold Dust Components with PRIMA, in Synergy with Sub-mm Facilities, up to z ∼ 5

Francesca Pozzi (UniBo, Bologna, Italy), Alberto Traina (INAF-OAS, Bologna, Italy), Francesco Calura (INAF-OAS, Bologna, Italy), Michele Costa (UniBo, Bologna, Italy), Laura Bisigello (INAF-OAPd, Padova, Italy), Carlotta Gruppioni (INAF-OAS, Bologna, Italy), Luigi Barchiesi (UCT, Cape Town, South Africa), Ivan Delvecchio (INAF-OAS, Bologna, Italy), Livia Vallini (INAF-OAS, Bologna, Italy), Cristian Vignali (UniBo, Bologna, Italy), Viviana Casasola (INAF-IRA, Bologna, Italy), Laura Sommovigo (Flatiron Institute, NY), Karina Caputi (Kapteyn Astronomical Institute, Groningen, The Netherlands)

A key property of interstellar dust is its temperature ($T_{dust}$), which influences infrared luminosity, star formation rate estimates, and dust mass determinations. While detailed dust models have been developed for the Milky Way and the nearest galaxies, the knowledge of $T_{dust}$ in distant galaxies remains limited, especially at high redshift z, where a single temperature Modified Black Body (MBB) is often assumed. Existing studies suggest an evolution of $T_{dust}$ with z, but conflicting results beyond z ≈ 3 highlight the need for better observational constraints. We propose to utilize the PRIMAger instrument aboard the PRIMA satellite (24–235 μm) to significantly advance our understanding of dust temperature evolution across cosmic time.

We consider a 1000-hour, 1 deg² survey, corresponding to a 5σ depth of 0.3 mJy at 60 μm. At these flux limits, simulations using the SPRITZ phenomenological model predict detections of ∼ 35000 galaxies out to z ∼ 4 ∼ 5, with AGN and composite systems dominating at z > 2. With its photometric coverage, PRIMAger will enable robust modeling of two-temperature dust components for galaxies down to $M_{star} \sim 10^{7.5}\ M_{\odot}$ and $10^{10}\ M_{\odot}$ at z = 0 and z = 3, respectively. Synergy with sub-mm facilities (e.g., ALMA, AtLAST) will be crucial at z > 1.5 to complement PRIMAger data and fully resolve the FIR peak, with PRIMAger sampling the Wien regime and sub-mm facilities the Rayleigh-Jeans one.

## Science Justification

Dust obscures the light linked to star formation and black hole (BH) accretion processes, acts as a coolant for the ISM, and serves as a catalyst for chemical reactions that produce molecular gas. Consequently, understanding its physical properties is essential for building a comprehensive picture of the star formation and BH accretion processes in galaxies and how they evolve as a function across cosmic time. One of the key properties of the interstellar dust in galaxies is its temperature, $T_{dust}$. Determining this parameter is crucial not only to constrain the thermal condition of the ISM, but also for deriving the IR luminosity, the star-formation rates, and the dust mass budget (e.g., Gruppioni+2013, Pozzi et al. 2021, Sommovigo+22, Traina+24)—all fundamental properties in galaxy evolution. Interstellar dust emits in the mid/far-IR regime. The PRIMA satellite, operating in the 24–235 micron range, is therefore well-suited to studying dust emission.

In the Milky Way, the availability of data at high spatial resolution and sensitivity has allowed the development of dust models that account for a wide range of dust temperatures. These temperatures are physically linked to the grain size distribution, composition, and the intensity of the surrounding radiation field (Draine &Li 2007; Jones+13). These models allow for the successful reproduction of both the observed dust emission and the optical extinction curves. Outside our Galaxy, the knowledge of the dust properties is much more limited. Only in the nearest galaxies—primarily the Magellanic Clouds, where extinction curves have been measured along different sightlines—has detailed modeling been possible. For more distant galaxies, the understanding of the dust physical properties remains significantly uncertain, and simplified approaches have been followed.

In the Local Universe, thanks to data from the Spitzer and Herschel satellites, in the mid-IR and far-IR regimes, respectively, a two-temperature dust model has been constrained (e.g., Orellana+17). Under the approximation of thermal equilibrium, the emission of the two components is approximated by two Modified Black Bodies (MBB). The model includes a warm component (with $20 < T_{dust} < 70$ K), typically associated with photodissociation regions (PDRs), and a cold one (with $T_{dust} < 30$ K) that is primarily linked to the diffuse interstellar medium (ISM) (e.g., Draine & Li 2007).

At higher redshifts, a single dust component is typically assumed due to the lack of points sampling the Wien part of the dust emission. An evolution of $T_{dust}$ as a function of z has been observed, thanks to Herschel data up to $z \sim 2$–3 (e.g., Magnelli+2014; Bethermin+2015) and, in recent years, with ALMA, up to $z \sim 7$–8 (Sommovigo+22). The emerging picture is that the dust temperature of galaxies is, on average, warmer going towards higher redshift. However, while up to $z \sim 3$ there is a global agreement of a direct correlation of $T_{dust}$ with z, at higher there are discrepant results, with some authors confirming the direct evolution of $T_{dust}$ with z (e.g., Mitsuhashi+24) and others suggesting a flattening (e.g., Algera+24). The lack of a firm conclusion is mainly due to the limited number of rest-frame FIR points to constrain the dust peak, and the different selection bias affecting the observed samples (e.g., Sommovigo+22).

We propose to explore the capabilities of the PRIMAger instrument on board the PRIMA satellite (PI: GJ. Glenn) to achieve a significant progress in the understanding of $T_{dust}$ as a function of redshift z. With its unprecedented number of photometric points in the 24–235 μm range, PRIMAger will:

i) allow to constrain a two-temperature dust model up to $z \sim 4$–5;

ii) allow to significantly constrain the $T_{dust}$ evolution as a function of the redshift z up to $z \sim 4$–5, thanks to its ability in performing blind surveys in the FIR regime.

These goals can be achieved using PRIMA data alone up to z ~ 1–2. At higher redshift, synergy with sub-mm data, such as ALMA or future sub-mm facilities like AtLAST (able to perform large area surveys) will be essential.

We estimate the potential of PRIMAger to reach the science goals by considering the PI SCIENCE survey of 1000hr on 1deg2, which translates into a 5⍰ flux limit at 60 μm of 0.3 mJy. We simulate the expected number of detectable galaxies and their SED with the spectrophotometric realizations of IR-selected targets at all-z, SPRITZ (Bisigello+21). The SPRITZ model includes different population of IR active galaxies, as spiral and starburst galaxies, AGN, composite systems along with elliptical and irregular galaxies. We decompose the dust emission far-IR peak using a warm and a cold dust component.

Based on our simulation, we expect to detect galaxies up to z ~ 4–5, with AGN or composite system dominating the source counts at z > 2.

The two-temperatures components will be constrained for MS-galaxies at z=0 down to $M_{star} \sim 10^{7.5}$ M⊙ and down to $M_{star} \sim 10^{10}$ M⊙ at z ~ 3. At z = 0, the temperature uncertainties will be reduced by a factor of 2 and up to a factor of 10 for the cold and warm components, respectively, in comparison to the pre-PRIMA studies. At z = 3, thanks to the synergy with sub-mm facilities, the 1σ temperature uncertainties will be of the order of 1 % (Fig. 1). At very high-z, we note that at 60 μm the PAH emission must be considered together with the continuum warm dust emission.

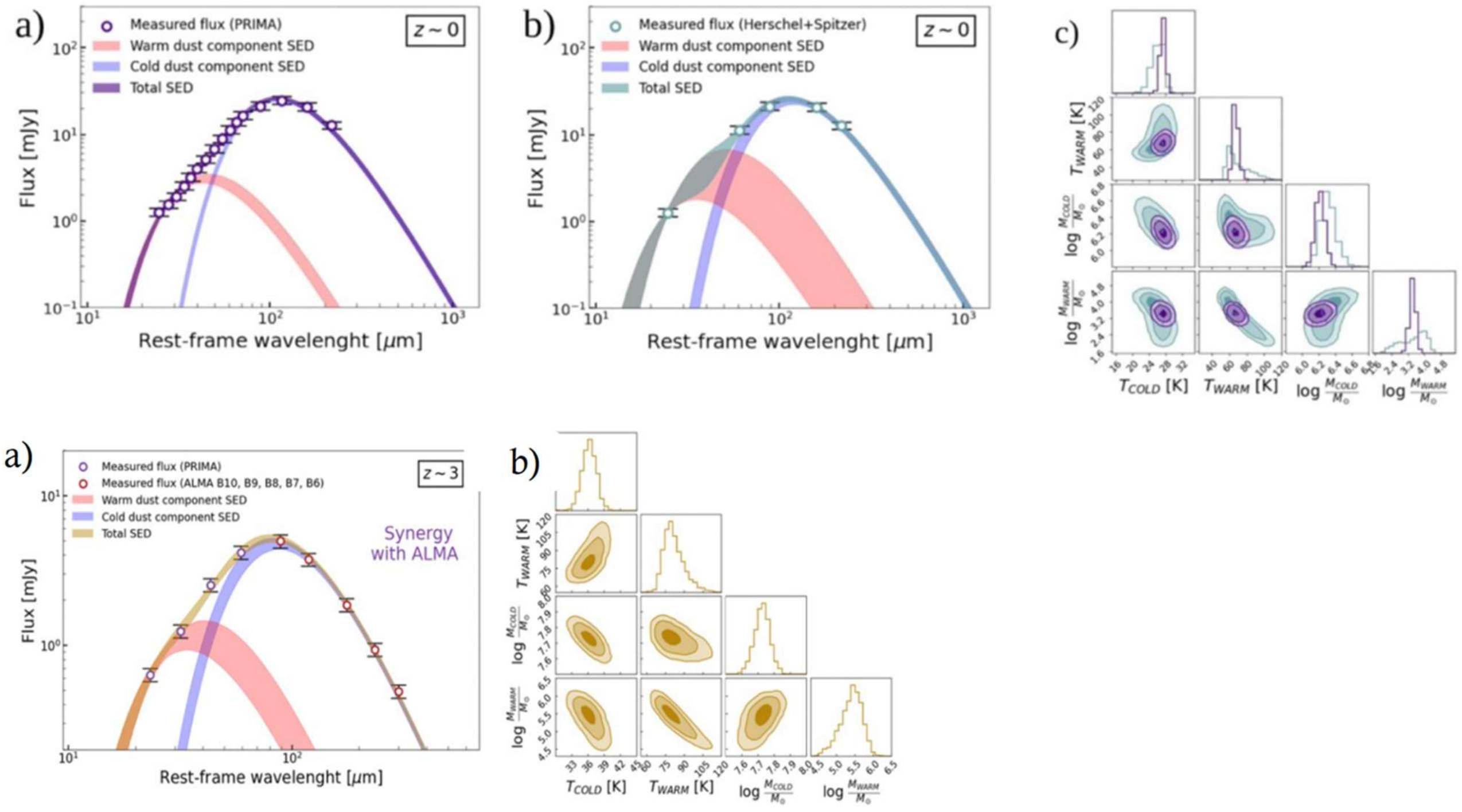

**Figure 1.** Simulated photometry from the SPRITZ realization (Bisigello+21) with overplotted MCMC best-fitting modelling of two MS-galaxies at z = 0,3 of $10^{10}$ ,$10^{10.6}$ M⊙, respectively. Top row: panels a), b) display the PRIMA and pre-PRIMA photometry, respectively for the z = 0 galaxy; panel c) the posterior probabilities of the dust parameters using the pre-PRIMA and PRIMA photometry (blue and violet contours, respectively). Bottom row: galaxy at z = 3 with the PRIMA points down to their detection limit (i.e., 0.3 mJy at 60 μm).

## Instruments and Modes Used

This science case will be carried out by using the PRIMAger data that will be obtained in the PI science survey. The data should cover 1 deg2 that is submivided into 36 contiguous pointins (10' x 10' each). All hyperspectral and polarimetric bands will be used in this science case but the polarimetric information is not needed.

## Approximate Integration Time

1000 hr

## Special Capabilities Needed

None

## Synergies with Other Facilities

PRIMA will be able to probe both parts of the dust peak (i.e., Wien and RJ-part) up to $z \sim 1$–$1.5$, with the caveat that the R-J fluxes probed by the PPI3 and PPI4 bands may be significantly affected by confusion. For this reason, especially at higher redshift, the synergy of PRIMA with current and future sub-mm and mm facilities (e.g., ALMA, JCMT/SCUBA-2, LMT, AtLAST) will be of key importance.

## Description of Observations

We plan to estimate $T_{dust}$ using a “reference” survey (i.e., the PI SCIENCE survey of 1000 hr on 1 $deg^2$). The exposure will allow us to reach a limiting flux at 60 μm of 0.3 mJy , computed using the on-line time exposure calculator. We use the 60 μm as ‘detection’ filter since it provides the highest observed IR flux, without reaching the confusion limit considering deblending tools (e.g., Donnellan+24) and the benefit of the rapid rise of the IR SED. We plan to use the 60 μm positions as priors to deblend the longer PRIMAger fluxes. Up to $z \sim 1$–$1.5$ PRIMager alone will be able to sample the dust SED, allowing to characterize the dust temperature of each galaxy. At higher redshift, the synergy with sub-mm/mm facilities will be key: PRIMAger will sample the Wien SED part, while the sub-mm facilities will sample the RJ-part.

From the SPRITZ simulation, we expect to detect $\sim$ 35000 galaxies above the detection limit (and above the confusion at 60 μm) at $0.5 < z < 5$.

# 44. Mid-Infrared AGN Diagnostics at Cosmic Noon

David Rupke (Rhodes College), R. Scott Barrows (Colorado), Ismael García-Bernete (CAB/ESAC), Daniel Stern (JPL/Caltech), Vivian U (Caltech/IPAC), Sylvain Veilleux (Maryland), Dominika Wylezalek (Heidelberg)

Luminous AGN and star formation (SF) peaked at Cosmic Noon ($z \simeq 1.5$–3). The dust properties of AGN at this epoch remain largely unexplored, but are key to disentangling the co-evolution of stellar mass buildup and black hole accretion. The mid-infrared spectral region (5–30 $\mu$m), explored in detail in the nearby Universe with the Spitzer Infrared Spectrograph and now JWST, contains a rich array of spectral diagnostics of the properties of AGN and star formation. The spectral range of FIRESS is well-matched to measuring these diagnostics in quasars at z = 2 and above, and its low-resolution mode is optimized to survey the broader dust features (PAH emission, silicate emission and absorption, continuum shapes) that dominate the mid-IR. We propose a pointed, low-resolution spectroscopic survey of 1000 luminous, WISE-selected AGN across the sky. We will sample a broad range of AGN ($L_{bol}$, obscuration, radio luminosity) and host ($M_*$, SFR) properties. The goal is to probe the dust properties of luminous AGN at Cosmic Noon to assess the co-evolution of AGN and SF, linked via feedback. We will follow-up promising targets with deep, high-resolution FIRESS-FTM observations to study narrower atomic and molecular features. These are a window to the details of feedback, particularly in obscured sources. The proposed survey will link to ongoing and future rest-frame mid-IR observations with JWST at z < 2, to bridge the current epoch with Cosmic Noon.

## Science Justification

The energy output from rapidly growing supermassive black holes at the centers of galaxies is now widely considered to be the main actor in regulating the evolution of massive galaxies in the Universe[1,2]. It is the epoch of Cosmic Noon, the peak of both star formation[3] and quasar activity[4] at $z \simeq 1.5$–3, in which powerful AGN (quasars, $L_{bol} > 10^{45}$ erg/s) played their starring role. A key evolutionary process connecting the two phenomena (SF and AGN) is the evacuation of dust by powerful, galaxy-scale winds in luminous quasars, and any corresponding cessation of star formation (negative feedback).

The rest-frame mid-IR—approximately 5–30 $\mu$m—provides a rich laboratory of discrete features for investigating the dust, molecules, and gas that obscure and surround the central engine. These include features that arise from the obscuring torus, from circumnuclear star formation, and from AGN feedback in the form of outflows. The diversity of AGN spectral properties, as seen in observations of AGN activity across luminosity in the nearby Universe[5,6,7], illustrate the mixing of AGN and star formation activity, as well as stages of AGN evolution and orientation.

These mid-IR AGN diagnostics have not been applied broadly outside of the local universe due to the absence of data. While Spitzer-IRS and JWST-MIRI have been optimized to observe 5–25 $\mu$m in local galaxies, they cannot reach the key features beyond z $\simeq$ 1.5. PRIMA's wavelength coverage and sensitivity provides an opportunity to apply these diagnostics out to the peak of AGN and SF activity at z > 2. The low-resolution mode of FIRESS—similar to the low-res modes of Spitzer-IRS and JWST-MIRI/LRS—provides a good balance between sensitivity and resolution, and is suitable for the broader dust, aromatic, and ice bands that dominate the mid-IR. The much higher-resolution FIRESS FTM is ideal for following up some of the most promising targets to detect and probe the kinematics of discrete atomic and molecular features.

We propose a FIRESS tiered survey of pointed observations of a large (approx. 1000 galaxy) sample of luminous AGN at z = 2–3, spanning luminosities from $10^{45}$ to $10^{48}$ erg/s, stellar masses $10^{9.5}$ to $10^{14}$ M$_\odot$, and star formation rates 10–1000 M$_\odot$/yr. We will also explore the diversity of obscuration—Type 1 vs. Type 2—and radio properties to understand how feedback impacts the dust and obscuration. In particular, FIRESS will be able to characterize deeply-embedded sources at these redshifts[13] using techniques pioneered at low z[14].

The key mid-IR diagnostics of AGN activity and feedback include PAH strengths, silicates in emission and absorption, hydrocarbon and ice absorption, emission-line ratios, and thermal continuum shapes. Figure 1 illustrates the diversity of mid-IR spectral shapes in two samples that reflect the wide range of properties we might encounter at Cosmic Noon. These include local IR-selected galaxies with a wide range of SFR and AGN contribution; and local optically-selected QSOs with a range of infrared properties[5,6,11]. Most easily visible in these spectra are the changing strengths of (a) PAH features at 6.2, 7.7, 8.6, 11.3, 12.7, and 15–18 $\mu$m[12]; (b) silicate emission and absorption at 9.8 and 18 $\mu$m[11]; and (c) the spectral slope above 10–20 $\mu$m[6]. Less visible, but equally diagnostic, are (d) the fine structure lines like [NeII] 12.81, [NeIII] 15.55, [SIII] 18.71, [NeV] 14.32 and 24.32, and [OIV] 25.89 $\mu$m[15]; (e) molecular hydrogen lines[16]; and (f) molecular absorption bands[17]. These all reflect some aspect of the changing mix of star formation and AGN activity in galaxies[6].

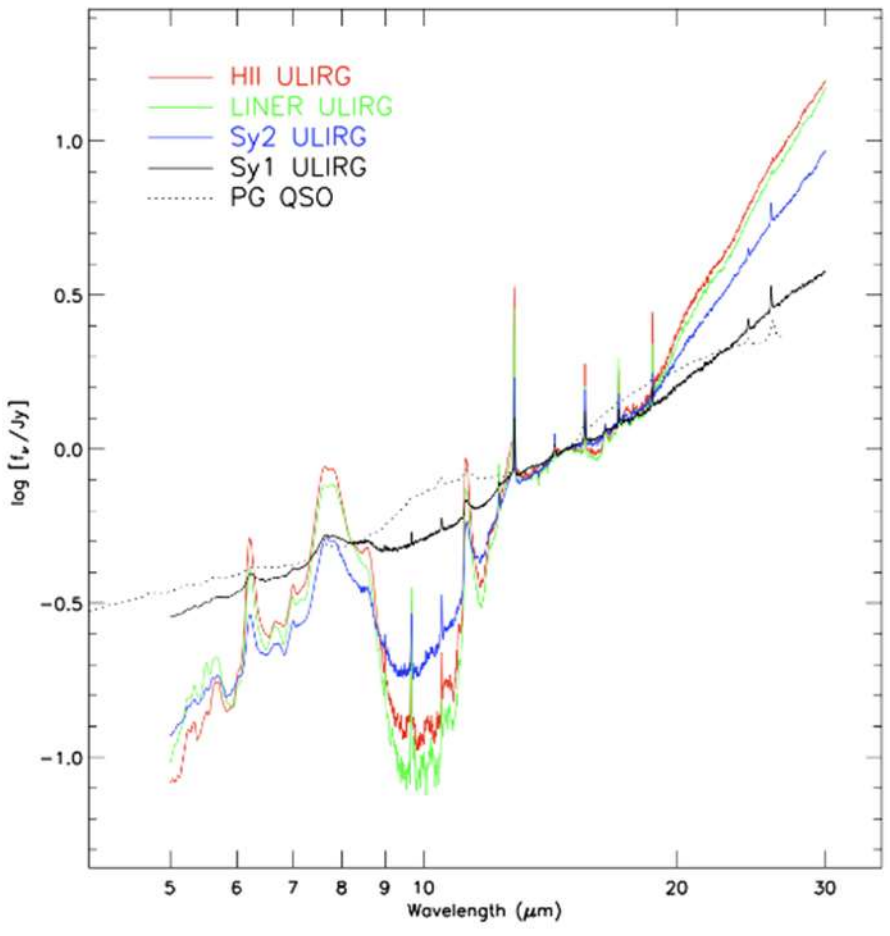

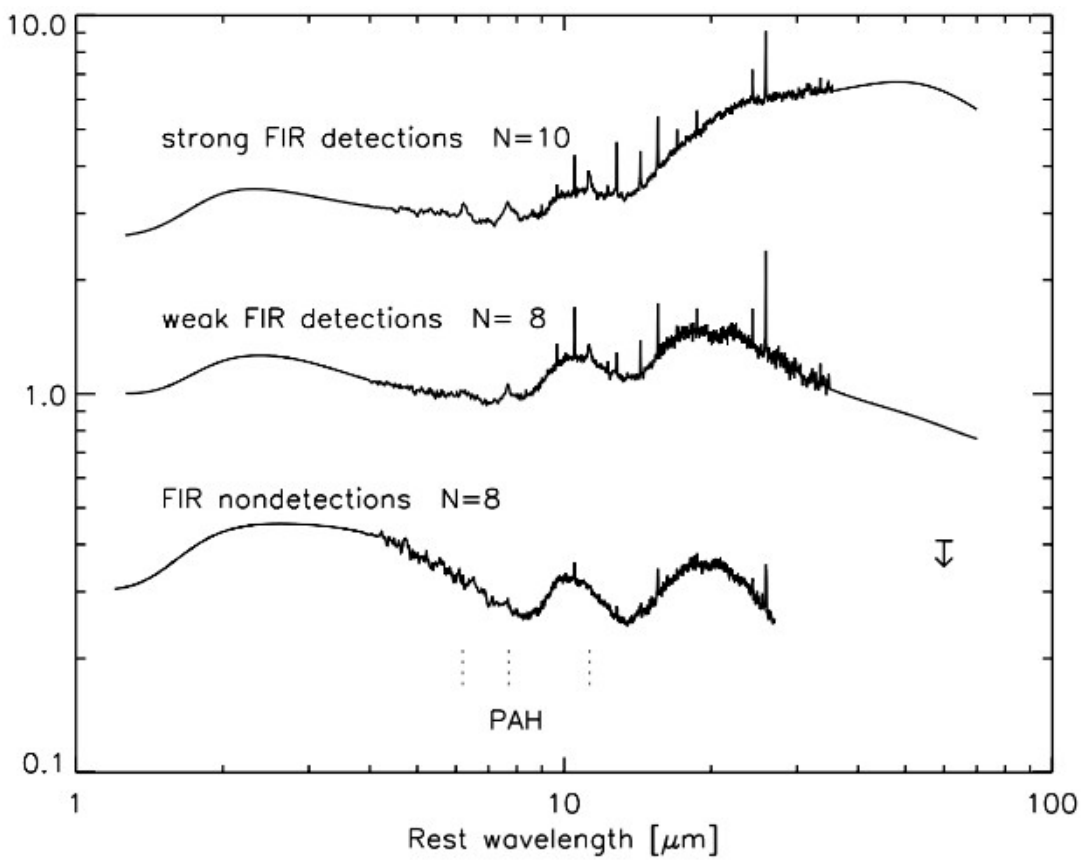

**Figure 1.** (left) The diversity of mid-IR spectral shapes in local IR-selected galaxies[6]. The colors represent different optical spectral types, which in turn reflect the relative contribution of SFR and AGN activity to the IR light. (right) The correlation of mid-IR and far-IR spectral shapes in local optically-selected quasars[5,11].

Each of these individually can serve as a diagnostic of the "AGN contribution" to a particular galaxy's bolometric luminosity. More generally, when combined with multi-wavelength diagnostics of the properties of AGN (blue vs. red quasars, radio-loud vs. radio-quiet, etc.), they may serve as a way to refine our understanding of how these different populations relate to one another and impact galaxy evolution. In particular, how do the dust properties connect to the obscuration? How does feedback impact SF? How do the IR properties of AGN at Cosmic Noon differ from those in the nearby Universe? The break in the luminosity function of AGN and the fraction of obscured AGN both rise with increasing z[4, 8]. Other factors like the evolution of cosmic metal content, which impact PAH and dust features[9], and the increase of SF and its molecular fuel with z[4,10], are likely to change the nature of the mid-IR spectra of Cosmic Noon quasars.

## Instruments and Modes Used

We will use the low-resolution mode of the FIRESS spectrometer, with R $\simeq$ 100, to survey 1000 targets. Pointed, spatially-unresolved observations allow for a survey of a broad range of AGN across the sky. We will follow up 15–20 targets at high resolution using FIRESS-FTM.

## Approximate Integration Time

The 5$\sigma$ point-source sensitivity of the FIRESS low-res mode is 2e-19 W m$^{-2}$ in one hour. We translate this to an average flux density over an unresolved line of 5e-18 erg s$^{-1}$ cm$^{-2}$ Å$^{-1}$, or 0.3 mJy. The typical W4 (22 $\mu$m) flux in the proposed sample (see Figure 2) is 3 mJy, which could be achieved in 3600s/100 = 36s. As a benchmark, S/N=15 in the continuum would allow robust fitting of silicate absorption features, which increases the integration time to 324s. To cover both wavelength settings requires twice the integration time, meaning a median exposure time of 10min per source. We propose to observe of order 1000 sources, which can then be achieved in 150–200 hours.

We will then use the FTM mode to follow up the most interesting sources. As a baseline, the fluxes of high-ionization fine structure lines in z $\simeq$ 0.2 quasars[5] reach 3e-17 W m$^{-2}$. At the lowest redshifts in the sample (z $\simeq$ 2), the flux of a line of the same luminosity is 1.2e-19 W m$^{-2}$. The 5$\sigma$ line sensitivity of the FIRESS FTM is 3e-19 W m$^{-2}$ in 6.3 hours, or 12.6 hours for the full wavelength range. The spectral resolution of FIRESS FTM is R = 4400 x (112/λ) = 12 320 at 40 $\mu$m, which is an average wavelength for bright fine structure lines when redshifted to z $\simeq$ 2. The quasars are expected to have broad lines due to outflowing as, so we can then bin to a resolution of R $\simeq$ 1800. This in turn allows us to reach the desired 1.2e-19 W m$^{-2}$ for 5$\sigma$ line detection and characterization in these quasars. We propose to observe of order 15 sources selected from the low-res survey for 15 hours each, for a total of 225 more hours.

The total time requested in both the large, low-resolution survey and targeted, high-resolution survey is then of order 400 hours.

## Special Capabilities Needed

None.

## Synergies with Other Facilities

JWST is poised to deliver constraints on the low wavelength end of the mid-IR up to and including Cosmic Dawn. This will connect the low-z universe up to z = 2 with the observations proposed here, at similar rest-frame wavelengths. At z>2, JWST will provide sensitive observations at lower rest-frame wavelengths than those available with PRIMA at the same redshifts through various programs (e.g., MILAN in Cycle 4; Wylezalek et al.) that will extend the diagnostic possibilities. We also envision proposing ALMA observations of a subset of the sample to constrain the spatially-resolved cold dust and molecular gas properties.

## Description of Observations

We will survey a broad range of luminous AGN at Cosmic Noon, subsampled by properties such as bolometric luminosity, host properties (stellar mass, SFR), radio properties, and obscuration. We select these as sources in the WISE AGN Catalog[18] for which AGN and host galaxy properties have been determined from broadband SED fitting[19]. This means that the observed-frame mid-IR fluxes are known from W4 measurements. We select z = 2 as the lower cutoff, as this means that the bottom end of the 9.8 $\mu$m silicate feature and the strongest PAH features near 8 $\mu$m are redshifted to the lower end of the FIRESS spectral range, while the upper mid-IR at rest-frame 30 $\mu$m is redshifted to 90 $\mu$m. We are able to access AGN up to z = 3.5 from the WISE AGN catalog (Figure 2), which spans the peak epoch of quasar activity in the Universe. At z = 3.5, the 8–30 $\mu$m range is redshifted to 36–145 $\mu$m. Thus, the FIRESS wavelength range is well-matched to probing the mid-IR at Cosmic Noon.

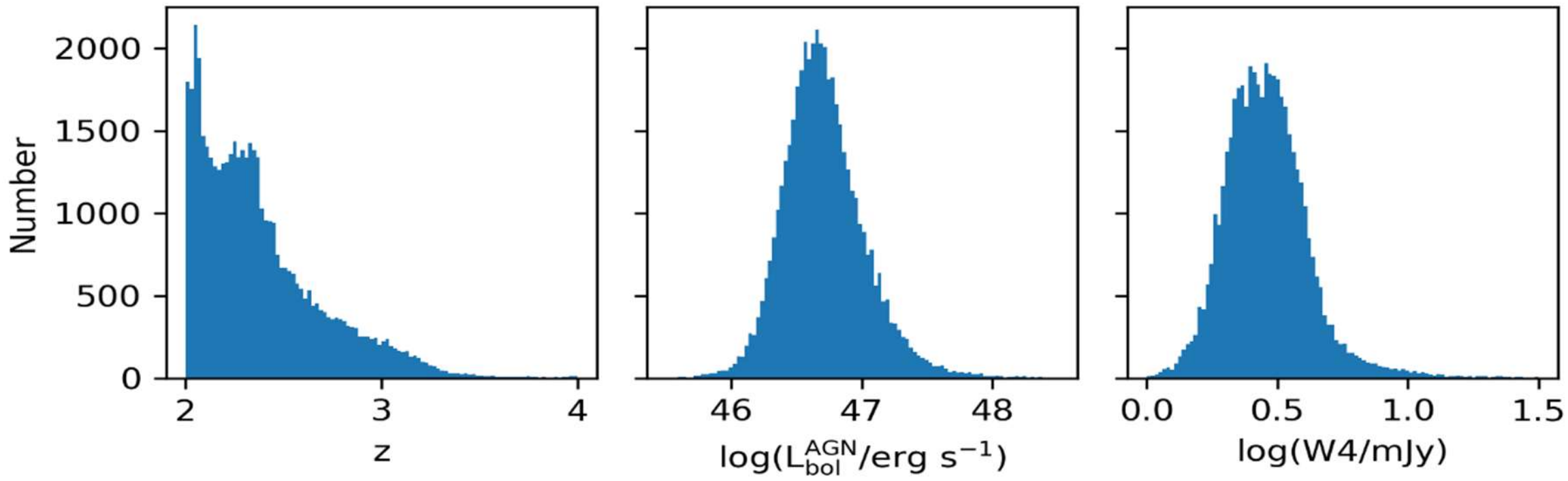

**Figure 2.** Distribution of z > 2 WISE-selected quasars with AGN and host properties from SED fitting[19].

The obscured (Type 2) fraction of AGN at z > 2 is relatively unexplored. However, techniques are emerging which will allow us to select obscured subsamples of the WISE quasars[20,21]. A large population of Extremely Red Quasars (ERQs) exists at these redshifts, as well, a subsample of which we will prioritize[22]. Finally, we will also target a set of radio-loud systems observed with Spitzer imaging[23] to ensure that the high-luminosity radio end is well-represented.

# 45. Probing the Dust-Obscured Universe: A Blind Spectroscopic Survey with PRIMA to Trace Star Formation and Black Hole Growth at Cosmic Noon

Luigi Spinoglio (INAF-IAPS, Italy), J. A. Fernández Ontiveros (CEFCA, Spain), T. Nagao (Ehime University, Japan), Hanae Inami, (Hiroshima University, Japan), Matt Malkan (UCLA, USA), Amit Wishwas (Cornell University, USA)

We propose a blind spectroscopic survey covering 200 arcmin² with PRIMA's Far-IR Enhanced Survey Spectrometer (FIRESS) to provide an unbiased census of dust-obscured star formation and black hole accretion across $0 < z < 4$. Our primary goal is to trace the co-evolution of galaxies and active galactic nuclei (AGN) during cosmic noon (z ~ 1–3), when both processes peaked. This survey will exploit the full spectral coverage of FIRESS (24–235 µm) in low-resolution mapping mode to detect key diagnostic features: polycyclic aromatic hydrocarbon (PAH) emission bands and nebular emission lines such as [OIII]52 µm, to quantify star formation rates (SFRs), and high-ionization lines such as [OIV]25.9 µm to measure black hole accretion rates (BHARs). We aim to detect ~1000 galaxies, including ~600 through the 11.3 µm PAH feature and ~900 via the [OIII]52 µm line, providing robust SFR measurements even in highly obscured environments. AGN signatures are expected in ~20–50% of these sources, offering critical constraints on the interplay between star formation and nuclear activity. PRIMA's unique spectral range and sensitivity will fill the gap between JWST and ALMA capabilities, enabling precise redshifts and source classification for dusty galaxies that are often missed in optical/UV surveys. This survey will provide a statistical foundation for testing key hypotheses about galaxy evolution and feedback effects during the Universe's most active epoch.

## Science Justification

Understanding galaxy evolution requires tracing both star formation and supermassive black hole accretion across cosmic time. These two fundamental processes peak during cosmic noon (z ~ 1–3), shaping the present-day galaxy population through complex interactions, including feedback mechanisms that regulate star formation and metal enrichment. However, approximately 90% of the UV/optical photons emitted by young stars and AGN during this epoch are absorbed by dust and reradiated in the infrared. This heavy obscuration hampers studies based on rest-frame UV and optical diagnostics, leading to incomplete and biased pictures of galaxy growth and AGN activity. Far-infrared (far-IR) spectroscopy provides the most direct window into these hidden processes. Selecting galaxies and AGN directly in the IR with FIRESS will provide a completely unbiased sample, because the mid-IR selects galaxies according to their bolometric flux (see, e.g. Spinoglio & Malkan 1989, Spinoglio et al. 2024). Previous facilities such as Herschel and SOFIA opened new avenues for IR spectroscopy but lacked the sensitivity and mapping speed needed to conduct deep, blind spectroscopic surveys of representative galaxy populations across cosmic noon. PRIMA, with its cryogenically-cooled 1.8 m mirror and the FIRESS spectrometer, offers a transformative leap forward, filling a critical observational gap between JWST and ALMA.

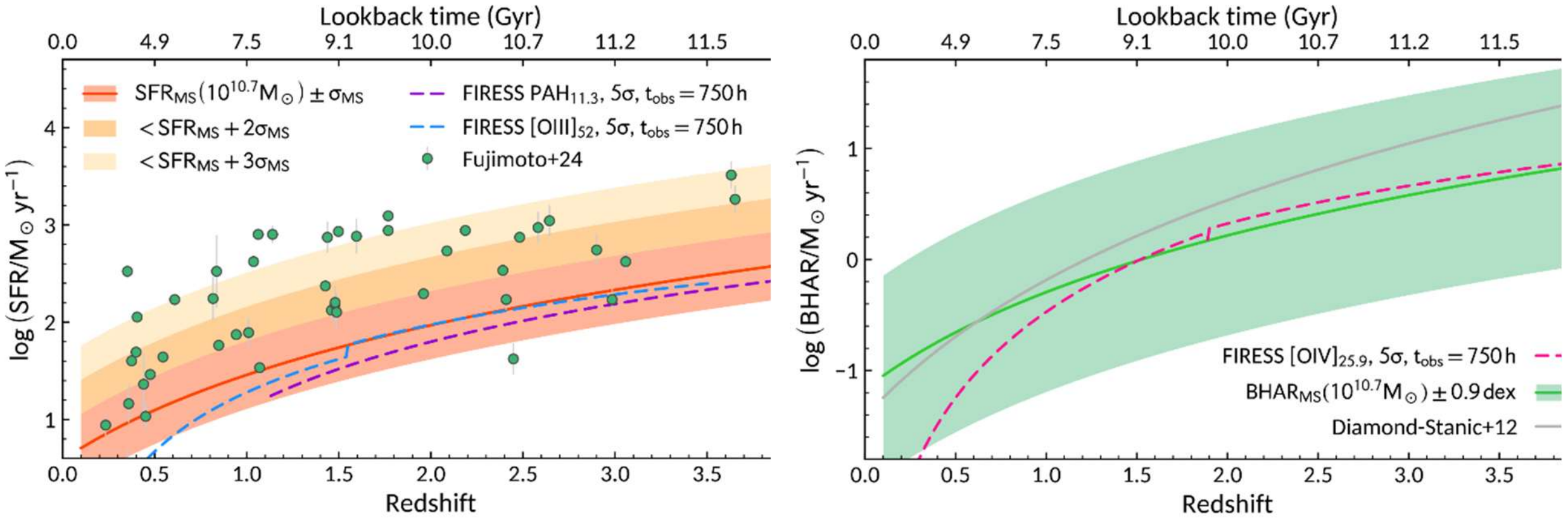

**Fig. 1. Left:** Star Formation Rate (left y-axis, SFR in $M_\odot$ $yr^{-1}$) and total IR luminosity (right y-axis) as a function of redshift for a $10^{10.7}$ $M_\odot$ galaxy in the Main Sequence (MS, red solid line; Scoville et al. 2017). The red-shaded area shows the 0.35 dex intrinsic scatter around the MS (Schreiber et al. 2015), while the dark and light orange-shaded areas indicate the location of galaxies whose SFR is +2$\sigma$ and +3$\sigma$ above the MS, respectively. The blue dashed line shows the SFR (left y-axis) and infrared luminosity (right y-axis) of a low-metallicity galaxy detectable through the $[OIII]_{52}$ line at 5$\sigma$ for a 200 arcmin$^2$ survey with 640 h of exposure time (750 h total observing time) with FIRESS low-resolution full-band mapping mode. The purple dashed line shows the same with the 11.3 µm PAH emission feature for a dusty star-forming galaxy. The green circles indicate the magnification-corrected infrared luminosities of galaxies detected in the ALCS survey (Fujimoto et al. 2024) with a measured spectroscopic redshift, which show that most galaxies detected in this survey can be spectroscopically detected by FIRESS. **Right:** Black hole accretion rate (BHAR; left axis) and total IR luminosity (right axis) versus redshift for a $10^{10.7}$ $M_\odot$ main-sequence galaxy. The green line and shaded area show the BHAR derived from a SFR-AGN luminosity correlation obtained for X-ray selected AGN, at redshift z = 1.2–1.7, using SED decomposition, with 0.9 dex scatter (Setoguchi et al. 2021). The grey solid line shows the estimate based on the BHAR-SFR relation obtained from $[OIV]_{25.9}$ Spitzer observations for local Seyfert galaxies in the revised Shapley-Ames catalog (Diamond-Stanic et al. 2012). The pink dashed line indicates the 5$\sigma$ detection limit for FIRESS in low-resolution full-band mapping mode for a sensitivity of 3.5 × $10^{-19}$ W $m^{-2}$ (5$\sigma$) at 24 µm, using $[OIV]_{25.9}$. FIRESS will measure BHAR through $[OIV]_{25.9}$ emission up to z = 4.

The proposed survey will answer the key question: What is the true distribution of star formation rate and black hole accretion rate in galaxies and AGN during cosmic noon, and how do these processes co-evolve? This includes testing the link between star formation and black hole growth and understanding the influence feedback on galaxy evolution.

PRIMA's unique capability to provide full spectral coverage between 24–235 µm at unprecedented sensitivity enables direct detection of key diagnostics:

- Star formation tracers: PAH emission bands (e.g., 11.3 µm), and nebular fine-structure lines such as [NeII]12.8 µm and [SIII]18.7,33.5 µm, bright in low-ionization environments characteristic of galaxies with solar-like metallicities, and mid-ionization nebular lines such as [OIII]52 µm, [NeIII]15.6 µm, [SIV]10.5 µm, which are bright in low-metallicity environments.

- Black hole accretion tracer: the [OIV]25.9 µm line, whose high ionization potential (~55 eV) just above the second ionization edge of helium ensures negligible contamination by star formation.

Only PRIMA/FIRESS can map these diagnostics over a relatively large area, reaching depths sufficient to detect main-sequence galaxies and AGN up to z ~ 4. Its multiplexing capability and fast mapping speed make it uniquely suited for blind spectroscopic surveys that are impractical with other present or future facilities.

For the galaxies detected in the 200 arcmin$^2$ blind survey, we will derive star formation rates using established calibrations based on PAH and nebular line luminosities (e.g. Mordini et al. 2021). Black hole accretion rates will be obtained from the [OIV]25.9 µm luminosity, leveraging well-tested local calibrations (e.g. Spinoglio et al. 2022). To assess FIRESS's capabilities for studying galaxy evolution, we compare its sensitivity limits with predicted SFR and BHAR –the two key parameters driving galaxy evolution– across cosmic time. The left panel in Fig. 1 shows the SFR (left axis) as a function of redshift for a galaxy in the so-called *main sequence* (Rodighiero et al. 2011, Elbaz et al. 2011). We have assumed the stellar mass of a Milky Way-like galaxy ($10^{10.7}$ $M_\odot$), and associated the values in luminosity (right axis) from the SFR--$L_{IR}$ relation (Scoville et al. 2017). Line-luminosity calibrations for galaxies in the local Universe have shown that the brightest SFR tracers are the PAH emission band at 11.3 µm, for dusty star-forming galaxies, and the [OIII]52 µm fine structure line for low-metallicity emission-line galaxies (Mordini et al. 2021). For the case assessment we also refer to the preparation of the Space Infrared telescope for Cosmology and Astrophysics (SPICA) galaxy evolution science program (Spinoglio et al. 2017, Spinoglio et al. 2021).

The right panel of Fig. 1 shows the evolution of BHAR with redshift. We estimate this evolution using two approaches based on the known SFR evolution (left panel), assuming that the parallel evolution of SFR and BHAR, observed on a global scale (Madau & Dickinson 2014), applies also to individual main-sequence galaxies. First, using the SFR–AGN luminosity correlation derived from X-ray selected galaxies at z = 1.2–1.7, through SED decomposition fitting (Setoguchi et al. 2021), ($L_{AGN}$ / erg s$^{-1}$) ~ $10^{44}$ × (SFR/$M_\odot$ yr$^{-1}$) with 0.9 dex scatter (green line and shaded area). Second, deriving the BHAR for a $10^{10.7}$ $M_\odot$ main-sequence galaxy using the SFR–BHAR correlation from Spitzer [OIV]25.9 µm observations of Seyfert galaxies in the local Universe (Diamond-Stanic et al. 2012; gray line).

Using an established correlation between the BHAR and the [OIV]25.9 µm emission-line luminosity (Mordini et al. 2021), we estimate FIRESS flux detection limits using the PRIMA Exposure Time Calculator (ETC) for a 3.5 × $10^{-19}$ W m$^{-2}$, 5$\sigma$ sensitivity at 24 µm (pink-dashed line). The proposed blind spectroscopic survey with FIRESS will detect AGN above the estimated knee of the luminosity function ($L^*_{IR}$) up to z ~ 4 using [OIV]25.9 µm, as shown by Fig. 1 for positive (Diamond-Stanic et al. 2012) and conservative (Setoguchi et al. 2021) predictions. Note that these BHARs represent instantaneous accretion rates during the AGN phase of these galaxies, not cosmic time-averaged values. Regarding the estimated number of source detections, if the parallel evolution of star formation and black hole accretion across cosmic time originates from the interplay between these two phenomena, this suggests that the number density of AGN and AGN-star formation composite sources is expected to be similar – within an order of magnitude

– to the observed number density of star forming galaxies (Fujimoto et al. 2024). Thus, we expect a considerable fraction of sources (~ 20–50%) to host an active nucleus and be detectable through the [OIV]25.9 µm line.

## Instruments and Modes Used

The project needs the FIRESS spectrometer in its low-resolution mapping mode.

## Approximate Integration Time

We propose to observe a 200 arcmin² field using the low-resolution full-band mapping mode of PRIMA's FIRESS spectrometer (R > 85). Based on the estimated FIRESS sensitivity in low-resolution map mode, provided by PRIMA ETC, we require $3.5 \times 10^{-19}$ W $m^{-2}$ (5σ) at 24 µm, assuming a spatial binning factor of ~1.7 to match the pixel scale of the bluest channels (7.′′6/px) to that of the reddest channel (22.′′9/px) and ensure an homogeneous spatial resolution across the full spectral range. To cover a common area of 20 × 10 $arcmin^2$ with all four channels, a 23.3 × 10 $arcmin^2$ field should be mapped (see Description of the observations). This requires a total observing time of 750 h, with 640 hours of integration time within the common 200 $arcmin^2$ field. This survey will enable detection of:

- ~600 galaxies via the 11.3 µm PAH band, redshifted into the PRIMA bands.
- ~900 galaxies via the [OIII]52 µm line.
- A significant number (~20-50%) of AGN through [O IV]25.9 µm emission.

No high-resolution mode is planned for this survey, as the focus is on blind mapping and census-building rather than kinematic studies.

### Estimation Method

We adopt the FIRESS sensitivity estimates provided by PRIMA ETC, presented in Fernández-Ontiveros et al. 2025 (JATIS accepted), which are anchored in expected instrument performance parameters and scaled to match the field size and total exposure time proposed.

## Special capabilities needed

No special capabilities are required for this programme. All targets are distant, extragalactic sources with no non-sidereal tracking, timing constraints, or specific position angle requirements. Standard calibration procedures are sufficient for the planned observations.

## Synergies with Other Facilities

- The blind spectroscopic survey will complement and enhance existing and forthcoming datasets from major multi-wavelength facilities:
- JWST: While JWST excels at mid-IR spectroscopy up to ~28 µm, PRIMA's coverage extends to 235 µm, filling the spectral gap and enabling joint analysis of PAH features and high-ionization fine-structure lines inaccessible to JWST at z > 1.

- ALMA: ALMA provides sub-mm line and continuum data, particularly tracing cold dust and molecular gas. PRIMA will add rest-frame mid- and far-IR spectroscopic diagnostics, enabling full multi-phase ISM characterisation and facilitating combined studies of dust-obscured star formation and AGN activity.
- Upcoming optical/IR surveys (e.g., Euclid, Rubin Observatory): These will provide deep imaging and photometric redshifts for source identification and cross-matching, allowing us to relate far-IR properties to optical rest-frame morphologies and environments.
- X-ray facilities (e.g., Chandra, XMM-Newton, Athena): PRIMA will detect Compton-thick AGN missed in X-rays, while X-ray data will confirm AGN classifications and offer independent BHAR constraints.

Together, these synergies will establish a comprehensive multi-wavelength view of galaxy and black hole co-evolution across cosmic noon and beyond.

## Description of Observations

We plan a blind spectroscopic survey covering 200 arcmin² using PRIMA's FIRESS low-resolution full-band mapping mode (R > 85). The survey field will be selected for its wealth of ancillary data across multiple wavelengths (e.g., GOODS-N, COSMOS), enabling rich multi-wavelength follow-up.

The observations will consist of contiguous mapping of the field to obtain full spectral coverage from 24–235 µm. The mapping strategy will prioritise uniform sensitivity across the area, using raster scanning with overlapping footprints to ensure consistent depth. Each map will be built up through multiple passes, allowing for effective cosmic ray rejection and calibration.

Given the ~3.3 arcmin separation on the sky between the slits of channels 1/3 and 2/4, a total coverage of 233 arcmin$^2$ would be required to map a 23.3 × 10 arcmin$^2$ field, ensuring a common area of 20 × 10 arcmin$^2$ covered by all four channels. This configuration requires a total observing time of 750 h, with an integration time of 640 hours over the common 200 arcmin$^2$ field. The corresponding sensitivity at 24 µm is $3.5 \times 10^{-19}$ W m$^{-2}$ (5σ). This configuration will allow us to detect:

- PAH 11.3 µm emission in dusty star-forming galaxies up to z ~ 3–4.
- [OIII]52 µm in low-metallicity galaxies, up to redshift z ~ 3.5.
- [OIV]25.9 µm in AGN, probing black hole accretion out to z ~ 4.

We will use standard FIRESS observing templates and calibration procedures, with no special instrument modes or configurations required. Data reduction will follow the standard FIRESS pipeline, and quality control will focus on ensuring flux calibration accuracy and spectral fidelity across the broad wavelength range.

This survey strategy maximises the statistical power of the dataset by providing a blind, unbiased census of galaxies and AGN across cosmic noon, crucial for deriving robust population properties and testing co-evolution hypotheses.

# 46. Probing the Missing FIR SED of Euclid-Discovered Bright Little Red Dots at z∼2 with PRIMA/FIRESS

Takumi S. Tanaka (the University of Tokyo/Kavli IPMU, Japan), Kohei Ichikawa (Waseda University, Japan), Tohru Nagao (Ehime U., Japan), Takuya Hashimoto (Tsukuba U., Japan), Hanae Inami (Hiroshima U., Japan)

Little Red Dots (LRDs) are a newly discovered population in the high-redshift Universe identified by JWST. LRDs are considered AGNs due to their compact morphology and Balmer broad lines. Their non-detections of X-ray emission, MIR hot torus emission, and significant variability all leave the nature of LRDs fundamentally different from that of local AGNs. The MIR-FIR SEDs of LRDs remain largely unconstrained and may represent the final key piece in understanding their nature. We propose deep PRIMA/FIRESS observations of extremely bright ($m_{5500}$=18) LRDs at z∼2, which are expected to be identified in the Euclid wide-field survey. Thanks to their lower redshifts than the typical LRDs, FIRESS can detect MIR-FIR emission even in cases of very low dust mass ($M_{dust}$=$10^2$ $M_\odot$) and can place robust constraints on the bolometric luminosity of LRDs. Furthermore, fully characterizing their SED shapes will provide key insights into the interpretation of LRDs.

## Science Justification

### Background

Thanks to its deep observations in the NIR-MIR wavelengths, James Webb Space Telescope (JWST) have been discovered low-luminosity active galactic nuclei (AGN) even at z∼6–8 (Harikane et al. 2024). This brings us closer to the early stages of black hole formation. Among these, a particularly notable population newly discovered by JWST is little red dots (LRDs, e.g., Matthee et al. 2024). LRDs are abundant at the high-z (z∼6–8) Universe characterized by a V-shaped SED from rest-UV to optical wavelengths, compact morphology, and Balmer broad lines suggestive of AGNs. That said, the non-detections of X-ray emission (e.g., Maiolino et al. 2025), MIR hot torus emission (Pérez-González et al. 2024), and variability (e.g., Kokubo & Harikane 2024) all suggest a fundamentally different nature from typical AGNs. Additionally, their number density sharply declines toward lower redshift (e.g., Kocevski et al. 2025), implying LRDs may be associated with primitive SMBHs unique to the high-z Universe, such as seed BHs in their very first accretion phases (Inayoshi 2025), super-Eddington accretion (Inayoshi & Maiolino 2025), or the BH-star model in which the early SMBH is embedded in neutral gas (Naidu et al. 2025). Although several models have been proposed, a definitive interpretation of LRDs remains elusive even three years after its first discovery (Labbé et al. 2023).

**This proposal:** Recent studies using wide-field surveys such as Subaru/HSC (Ma et al. 2025) and Euclid (Euclid Collaboration et al. 2025) have begun identifying low-z ($z \lesssim 4$) LRDs. According to the luminosity function of z∼2 LRDs, Euclid could identify hundreds of LRDs as bright as rest-frame 0.55 μm magnitude of $m_{5500}$=18. Thanks to their redshift, the peak of warm dust emission falls within the PRIMA/FIRESS Band 3 coverage. Moreover, their lower apparent magnitude compared to high-z LRDs makes them easier to detect. Thus, we propose deep PRIMA/FIRESS Band 1 and 3 observations of these bright z∼2 LRDs to test different SED scenarios. If the LRD exhibits dust emission consistent with the current upper limits from JWST/MIRI and ALMA (Casey et al. 2025), it could be significantly detected in both FIRESS Bands 1 and 3 (gray dotted line in Fig. 1). Even if the dust emission is weaker than the predictions by Casey et al. (2024) due to low dust masses ($M_{dust}$∼$10^2$ $M_\odot$, gray dash-dot line in Fig. 1), a detection would still be feasible. In these cases, follow-up observations with Bands 2 and 4 would enable us to fully constrain the SED and accurately estimate the dust temperature and dust mass. Moreover, even if the SED is dominated by blackbody emission, we expect a detection in FIRESS Band 1 (gray dashed line in Fig. 1).

By combining FIRESS data with NIR (JWST/NIRCam, Euclid) and JWST/MIRI observations, we can fully uncover the MIR-FIR SED. Through comparisons with models, this will enable a complete interpretation of the SED and uncover the physical nature of LRDs. Completing their SED is also crucial for constraining their bolometric luminosity ($L_{bol}$), which in turn provides an estimate of $M_{BH}$ under an assumed Eddington ratio. This approach does not rely on the single-epoch method, which is currently used to estimate $M_{BH}$ of LRDs but is calibrated with low-redshift AGNs and may not be directly applicable to LRDs.

### Need for PRIMA

Need for PRIMA: Constraining the shape of the dust emission requires identifying its peak, which in turn determines the dust temperature. The peak of dust emission is expected at rest-frame tens of microns, which fall into FIRESS Band 3 for z∼2 targets. PRIMA is uniquely capable of deep spectroscopy in this range.

Understanding the SEDs of LRDs is essential for interpreting their nature. Some recent studies suggest that the rest-frame optical red slope can be explained by blackbody radiation with T∼5000 K (e.g., Inayoshi et al. 2025; Kido et al. 2025; Liu et al. 2025), not by a strong dust-reddening. Others argue that the lack of JWST/MIRI and ALMA detections does not rule out the presence of warm (T∼100–500 K) dust emission (e.g., Setton et al. 2025). These interpretations are interdependent: if the red slope is caused by dust extinction, the absorbed energy should be re-emitted in the MIR-FIR. If such re-emission is absent, the red slope is likely intrinsic, supporting the blackbody scenario. However, their MIR-FIR SED remain poorly constrained, except for a few cases in the local Universe (Lin et al. 2025), and may hold the final key to completing our understanding of LRDs.

## Instruments and Modes Used

FIRESS pointed low resolution mode, on each target

## Approximate Integration Time

The integration time is estimated to detect blackbody emission (gray dashed line in Fig.1) from an extremely bright ($m_{5500}$=18) z∼2 LRDs in FIRESS Band 1 (30 microns). Using the PRIMA ETC, we confirmed that a 10-hour observation allows continuum detection of blackbody component with T=5000 K and $A_V$=1 over S/N>5 in R∼10 synthesized bins. If a warm dust component with a temperature of T∼100 K exists, which is undetectable with MIRI or ALMA, its SED would peak in Band 3, and it can be detected even with low dust masses of $10^2$ $M_\odot$ (gray dash-dot line in Fig. 1). We estimate that for a sample of 10 objects, a total of ∼50 hr is needed.

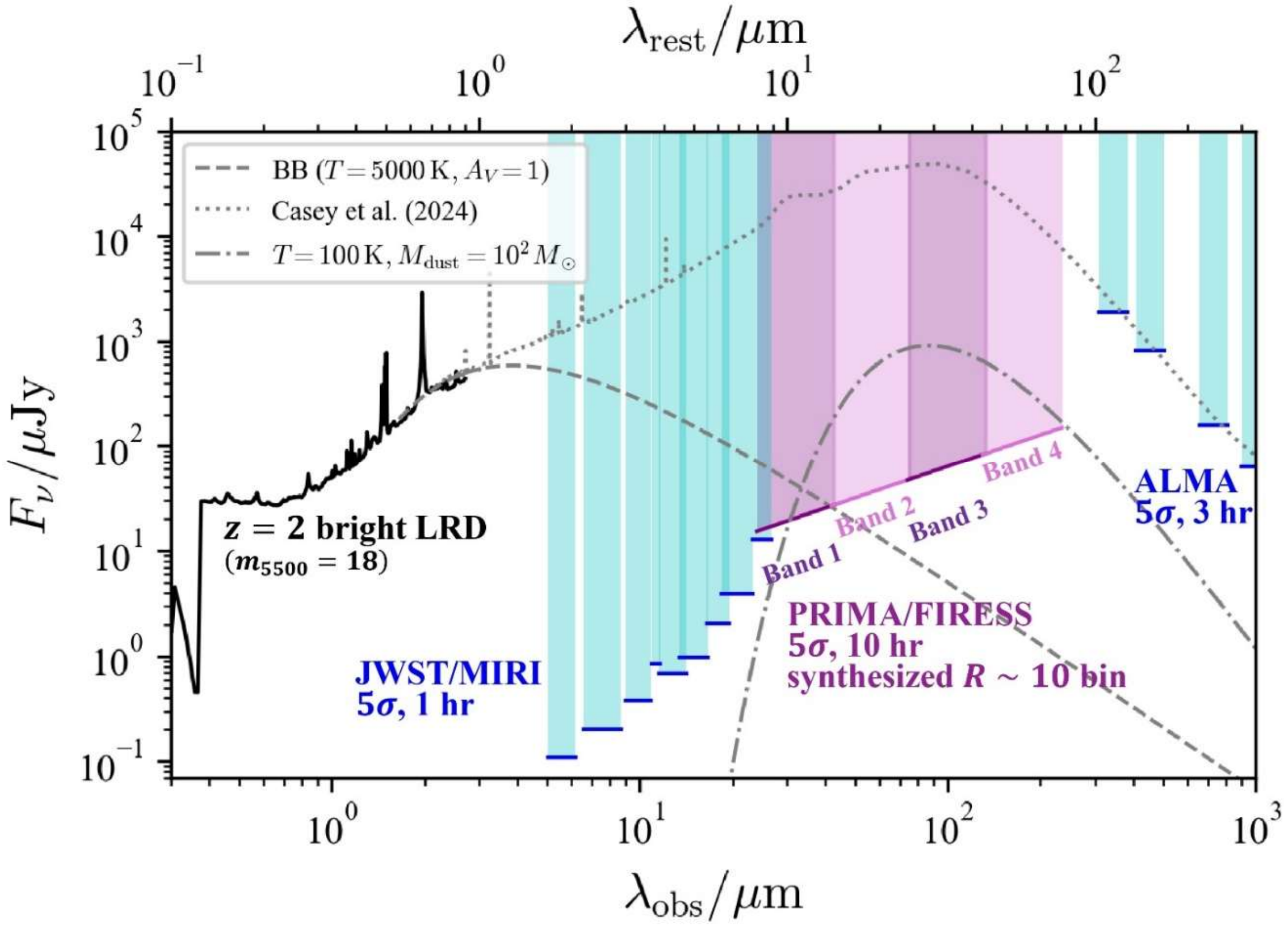

**Figure 1.** Predicted SED of a bright LRD at z∼2. The black solid line shows the stacked LRD SED from Akins et al. (2024). The gray dotted, dashed, and dash-dot lines represent a model SED consistent with the current ALMA and JWST/MIRI upper limits (Casey et al. 2024), a blackbody with T=5000 K and $A_V$=1, and a single dust component with T=100 K and $M_{dust}$=$10^2$ $M_\odot$ (assumed a modified blackbody with β=2 following Casey et al. 2024), respectively. All model SEDs are scaled to $m_{5500}$=18. Purple and blue rectangles indicate the 5σ depths of PRIMA/FIRESS (10 hr, synthesized R∼10 bins) and other facilities (JWST/MIRI with 1 hr and ALMA with 3 hr), respectively.

## Special Capabilities Needed

None

## Synergies with Other Facilities

In this science idea, we aim to target low-z bright LRDs expected to be discovered through multi-wavelength wide-field surveys that combine Euclid with ground-based observatories (e.g., UNIONS and DES). Such rare objects can only be identified thanks to their wide area coverage, making this a synergistic program in which PRIMA can deeply follow up on these unique objects. JWST/MIRI will also play an important role. Given the brightness of these LRDs, their blackbody components are likely detectable with MIRI. Once such bright LRDs are identified, MIRI observations prior to PRIMA's launch would allow us to establish a clearer prospect of their blackbody components in advance. On the other hand, as discussed earlier, PRIMA will have the capability to detect warm dust components with temperatures around 100 K, which remain out of reach for both MIRI and ALMA. With these complementary datasets of UNIONS, DES, Euclid, JWST, PRIMA, and ALMA, we will be able to complete the SEDs of LRDs in a fully synergistic manner from rest-frame UV to FIR.

## Description of Observations

We propose 10-hour PRIMA/FIRESS observations per target in Band 1 and Band 3 targeting bright LRDs at z∼2. The total expected number of such low-z bright LRDs found in Euclid and ground-based survey data is at most a few hundred. Considering the band center FWHM of 12.5'' in Band 3, we need to select isolated sources without nearby companions that could cause contamination, resulting in a final sample of roughly 1–10 LRDs, much less than the total expected number of such bright z∼2 LRDs

If warm dust emission is detected at levels exceeding the blackbody prediction (dash-dot line in Fig. 1), we will conduct follow-up observations in Band 2 and 4. Since the expected flux density of the dust component will already be constrained at that point, these additional observations may require much shorter integration times than the initial 10 hours. Combining all of Band 1 to 4 results will allow us to robustly determine the shape of the dust SED in the MIR-FIR regime and accurately constrain both the dust temperature and dust mass.

# 47. Probing Dust Porosity in Protoplanetary Disks with Far-IR Polarization Observations

Ryo Tazaki (University of Tokyo), Mitsuhiko Honda (Okayama University of Science), Yao-Lun Yang (RIKEN), B. Sargent (STScI/JHU)

Dust grains in protoplanetary disks coagulate into porous dust aggregates. Dust porosity strongly influences their subsequent dust evolution and planetesimal formation. However, actual values of dust porosity are still not yet convincingly constrained from observations. Recent polarimetric studies with ALMA have shown that self-scattering polarization is sensitive to dust porosity, and it could be used to constrain dust porosity. However, the limited wavelength coverage of ALMA polarimetry suffers from parameter degeneracy in porosity values. In this science case, we aim to constrain dust porosity through far-infrared self-scattering polarization observations for protoplanetary disks with PRIMA/PRIMAger. A wide wavelength coverage provided by PRIMA and ALMA will be sufficient to break parameter degeneracy in dust porosity. Our targets will be bright, inclined Class II disks in which ALMA has already detected self-scattering polarization. In the self-scattering mechanism, the polarization vector tends to align with the disk minor axis for inclined sources, and hence we expect to measure a net polarization signal even in spatially unresolved PRIMA observations. When combined with ALMA (sub)millimeter polarimetry, these data will be modeled with 3D radiative transfer simulations of both porous and compact grain populations. This strategy leverages the synergy between PRIMA and ALMA, yielding constraints on dust porosity in planet-forming environments.

## Science Justification

### Background

Planets form in protoplanetary disks. To initiate this process, dust particles within these disks must grow through collisions, eventually forming pebbles and kilometer-sized planetesimals. Despite decades of study, the pathway from submicron-sized grains to gravitationally bound planetesimals remains poorly understood. One of the most uncertain factors in dust evolution models is the internal structure—or porosity—of dust aggregates. The primary goal of this observation is to determine the porosity of dust aggregates in protoplanetary disks by detecting self-scattering-induced polarization at far-infrared wavelengths (PRIMAger/Bands 1 and 4).

### Why is dust Porosity Important?

Porosity influences dust evolution, planetesimal formation, and the interpretation of disk observations in three key ways. First, porosity governs the collisional outcomes of dust aggregates (Blum 2018). Bouncing and fragmentation are often the main barriers that inhibit grain growth, but these processes depend strongly on the internal porosity of aggregates (Johansen et al. 2014). Second, porosity modifies the gas drag experienced by aggregates, directly affecting their radial drift and vertical settling. Unlike compact grains, highly porous aggregates may grow directly into planetesimals without the need for the streaming instability (Okuzumi et al. 2012). Third, porosity significantly affects the interpretation of dust continuum observations: it can lower the opacity at millimeter wavelengths. If the dust in disks is indeed highly porous, current estimates of disk dust mass may be underestimated by at least a factor of about 6 (Liu et al. 2024). This may help resolve the so-called "mass budget problem", where inferred disk masses appear too small to account for the observed exoplanet population. In these ways, porosity not only directly impacts dust growth and planetesimal formation but also critically affects our interpretation of disk observations and our broader understanding of disk evolution.

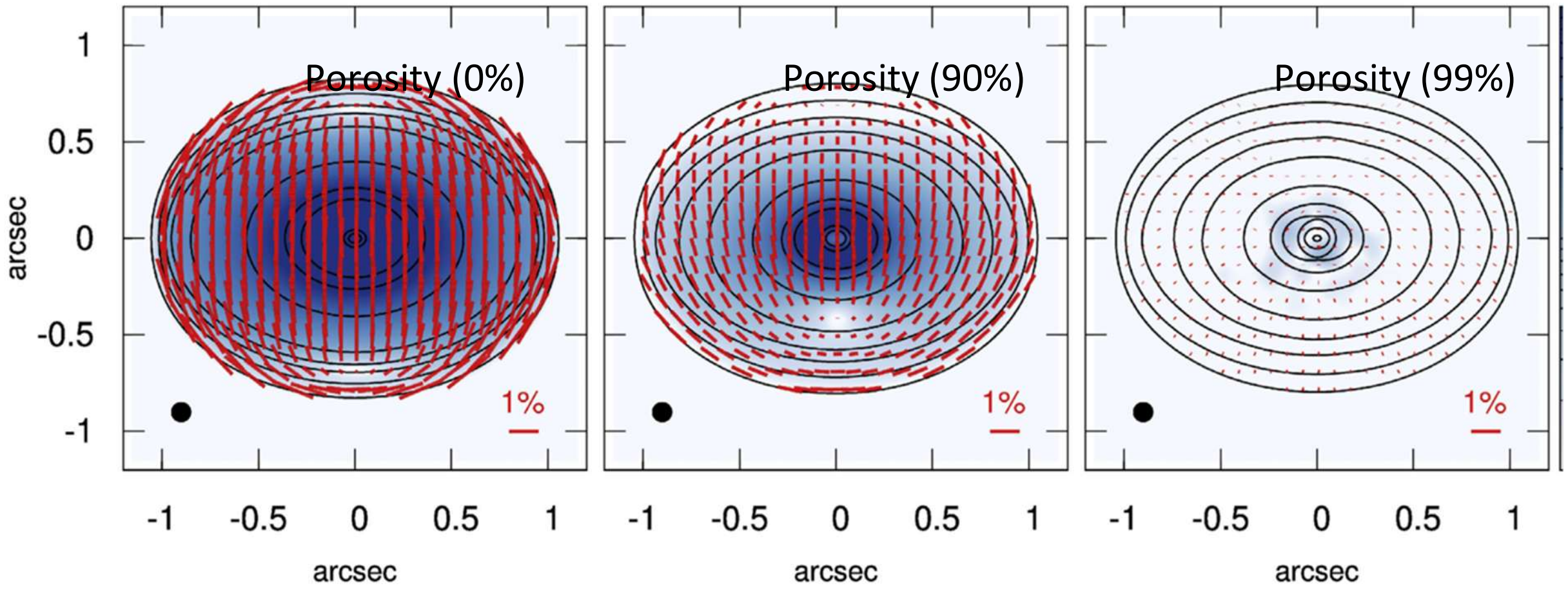

**Figure 1.** Self-scattering polarization patterns (contour: total intensity, color: polarized intensity, red bars: polarization vectors) and the effect of porosity on scattering polarization (non-porous grain to porous grains from left to right panels) (taken from Tazaki et al. 2019). Polarization orientations are uniformly aligned along the disk minor axis (e.g., Yang et al. 2016). This property prevents polarization cancellation even if the disk is unresolved.

### Why polarimetry?

ALMA has opened a new window for constraining dust porosity through its polarimetric capabilities at (sub)millimeter wavelengths. At these wavelengths, dust grains with sizes comparable to the observing wavelength can produce a phenomenon known as self-scattering, in which thermal emission from dust is scattered by other grains (Figure 1). Since self-scattering polarization is highly sensitive to dust porosity (Tazaki et al. 2019), polarimetric modeling of ALMA data has been used to constrain porosity (Zhang et al. 2023; Ueda et al. 2024). In particular, porosity changes the wavelength dependence of the polarization fraction (see Figure 2), and therefore, polarimetry covering a wide wavelength range is crucial for constraining dust porosity

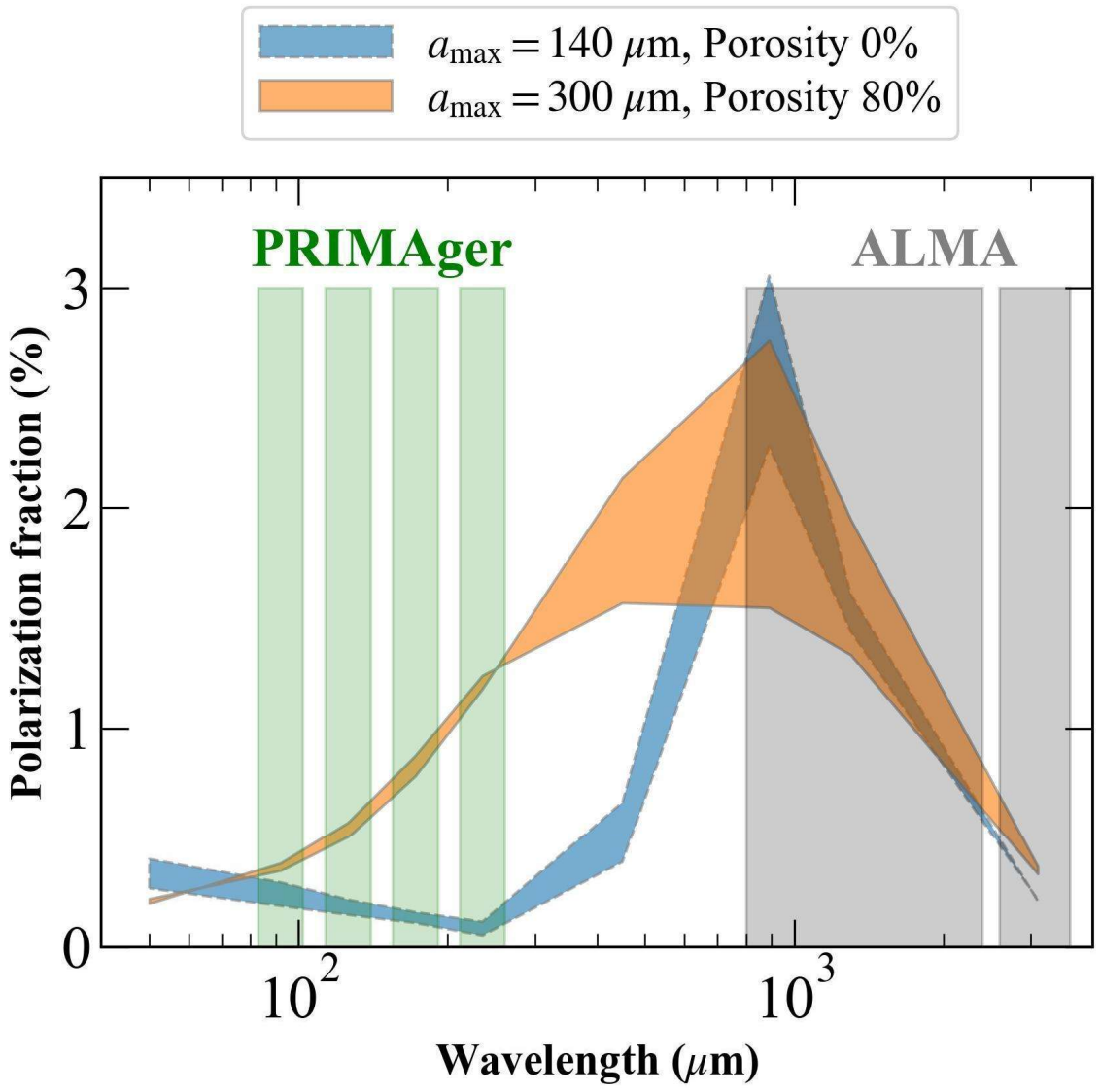

**Figure 2.** Radiative transfer model predictions for far-IR and mm wavelength polarization from self-scattering by non-porous (blue dashed) and porous (orange solid) models. The uncertainty range is due to different light scattering theory for computing optical properties of dust grains (Mie theory and DHS model) using OpTool (Dominik et al. 2021). The disk inclination is assumed to be 60 degrees. The porosity affects the wavelength dependence of the polarization profile.

### Why PRIMA?

As illustrated in Figure 2, far-infrared (far-IR) polarization, when combined with ALMA polarimetric measurements, can be used to test whether dust in disks is porous or not. SOFIA/HAWC+ detected far-IR polarization in HL Tau (Lietzow-Sinjen et al. 2025), but its limited sensitivity has prevented detailed investigation across a broader sample of sources. While ALMA does provide polarization measurements, its capabilities are restricted to Band 7 (0.9 mm) and longer wavelengths. The absence of shorter wavelength polarization data leads to parameter degeneracy in porosity, as shown in Figure 2. Far-IR polarization capability of PRIMA will provide a truly unique opportunity to reveal this missing polarization constraint.

### Interpretation Method

We will apply the same radiative-transfer frameworks that have already been successfully used to model ALMA polarimetry results (Ueda et al. 2024). Figure 2 compares radiative transfer simulations using RADMC-3D with a full polarization treatment (Dullemond et al. 2012) for two dust populations—highly porous dust grains (80 % porosity) versus compact grains—assuming an axisymmetric disk.

## Link to Testable Hypotheses

A measured polarization fraction at PRIMA wavelengths will determine the porosity of dust aggregates, which are then compared with predictions of dust evolution models (e.g., Michoulier et al. 2024). Even if PRIMA fails to detect polarization above its sensitivity limit, this will indicate that grains are predominantly compact. Such low-porosity grains may result from collisional fragmentation of larger bodies, suggesting that planetesimals have already formed in the disk (Tatsuuma et al. 2024).

## Instruments and Modes Used

We will use PRIMAger Band 4 (235 microns) to observe the polarized flux from disks.

## Approximate Integration Time

Since the expected integrated polarization fraction is on the order of ~1%, a total flux of 1 Jy would correspond to a polarized flux of ~10 mJy. To detect this polarization signal with a signal-to-noise ratio of 10, a noise level of 1 mJy is required. Because it's polarized intensity, we need to take a factor of sqrt(2), which leads to a noise level of 0.7 mJy. By inputting a 10'×10' (i.e., 100 square arcminutes) field of view and a 3.5 mJy as 5-sigma survey depth into the PRIMA Exposure Time Calculator (ETC) at PRIMAger/Band 4 (235 μm), we obtain 0.5 hours for one source. Thus, we estimate the approximate integration time of a single-band observation to be 10 hours for 20 sources, excluding overheads. We also request to observe the other three Bands to confirm the possible wavelength dependence within a far-IR wavelength domain. Because the four bands will be observed simultaneously, the total time will be 10 hours.

## Special Capabilities Needed

N/A

## Synergies with Other Facilities

These observations offer strong synergy with ALMA, as summarized in Figure 2. While ALMA's current polarimetric capabilities alone are insufficient to constrain dust porosity models, combining them with PRIMA's far-infrared polarization measurements allows us to break the existing parameter degeneracies.

## Description of Observations

We propose to observe protoplanetary disks using PRIMAger/Band 1 to 4. Multi-band polarimetry is crucial for establishing a wavelength dependence of the polarization fraction at far-IR wavelengths.

To make a statistical argument possible, it is desirable to detect at least 20 protoplanetary disks (ideally 10 T Tauri stars and 10 Herbig stars). According to the literature photometry, we can readily find >40 disks that are brighter than 1 Jy at 250 μm (close to the PRIMAger/Band 4) in various star-forming regions (e.g., Ribas et al. 2019, Stapper et al. 2022). We will further apply the following criteria to those samples: (i) moderate or higher inclinations (>35°) to avoid significant polarization cancellation; (ii) existing detections of self-scattering polarization with ALMA; and (iii) cleared envelopes to avoid contamination from cloud polarization, which may complicate the interpretation. Currently, the number of disks that meet all the criteria is 4 (2 Herbig stars and 2 T Tauri stars). However, it is worth noting that many sources have incomplete ALMA polarization studies or lack constraints on disk geometry (i.e., inclination). Future studies will increase the sample size further.

# 48. A Complete Census of Dust Properties in AGN: A Synergy with Subaru, JWST, and PRIMA

Yoshiki Toba (Ritsumeikan University), Hanae Inami (Hiroshima University), Tohru Nagao (Ehime University), Takuya Hashimoto (Tsukuba University)

Understanding the properties of cosmic dust in active galactic nuclei (AGN) is crucial for elucidating the co-evolution of galaxies and supermassive black holes (SMBHs). This study proposes a comprehensive survey of dust in AGN utilizing the forthcoming PRIMA FIR space telescope, in conjunction with observations from the Subaru Telescope and the James Webb Space Telescope (JWST). By employing PRIMAger's intensive (1 deg²) and broad (10 deg²) surveys over a cumulative 1500 observing hours, we aim to detect far-infrared (FIR) emissions from high-redshift quasars and AGNs. Through spectral energy distribution (SED) fitting with CIGALE for archival JWST data, we estimate physical dust properties, including mass and temperature, demonstrating that PRIMA can detect FIR emissions from up to 60% of JWST sources. This facilitates robust constraints on the dust mass function and dust mass density throughout cosmic time. Furthermore, the synergy with Subaru's Hyper Suprime-Cam (HSC) enables FIR detection of low-luminosity quasars at $z > 6$, and the Prime Focus Spectrograph (PFS) enables spectroscopic follow-up of faint PRIMA sources. The project also emphasizes complementary observations with GREX-PLUS for investigating optically dark, infrared-bright galaxies. Overall, the proposed survey will significantly enhance our understanding of the dusty Universe and the role of dust in the evolution of galaxies and AGN.

## Science Justification

Cosmic dust is a critical component for comprehending the formation and evolution of stars and galaxies. It is also significant for active galactic nuclei (AGN), as it can serve as a mass supply source for supermassive black holes (SMBHs), playing a crucial role in the growth of SMBHs in the early Universe (e.g., Ishibashi 2021; Park et al. 2022). Furthermore, dust associated with AGN hosts is one of the key elements controlling the star formation history and chemical evolution of the host galaxy, which is indispensable for understanding the co-evolution of galaxies and SMBHs.

One of the pioneering studies on high-redshift quasars is Leipski (2014), who investigated the broadband spectral energy distribution (SED) for SDSS quasars at $z > 5$, using optical to far-infrared data. However, they reported that only 17% of quasars were detected by SPIRE, indicating that the sensitivity of Herschel was insufficient to capture far-infrared photons from these quasars. Additionally, only 8% of quasars were successfully measured for their dust temperatures. Otherwise, the dust temperature was assumed to be 47 K to estimate dust mass, for example. Recent SCUBA-2 and ALMA observations, however, suggest that the dust temperature of quasars varies by up to 100 K in extreme cases. Furthermore, far-infrared (FIR) emission may not always originate from dust in host galaxies, and AGN emission could contribute to the far-infrared spectral energy distribution (SED), which affects the measurement of the star-formation rate. Therefore, deep and multi-wavelength infrared data are crucial to resolving this issue.

PRIMAger is an ideal instrument for addressing this issue. In this science case, we assume two survey modes: the PRIMA-wide (10 $deg^2$) and deep (1 $deg^2$) surveys, both of which have a total observing time of 1500 hours. Bisigello et al. (2024)[2] reported both the conservative expected survey depths (conservative) and the guaranteed payload requirement depths (payload). Figure 1 shows the SEDs of SDSS quasars with expected sensitivities by PRIMAger, which demonstrate that PRIMA can detect FIR emission from SDSS-class, luminous quasars at $z > 5$.

[2] It should be noted that expected sensitivities in the latest factsheet are slightly worse by a factor than what is assuming this work.

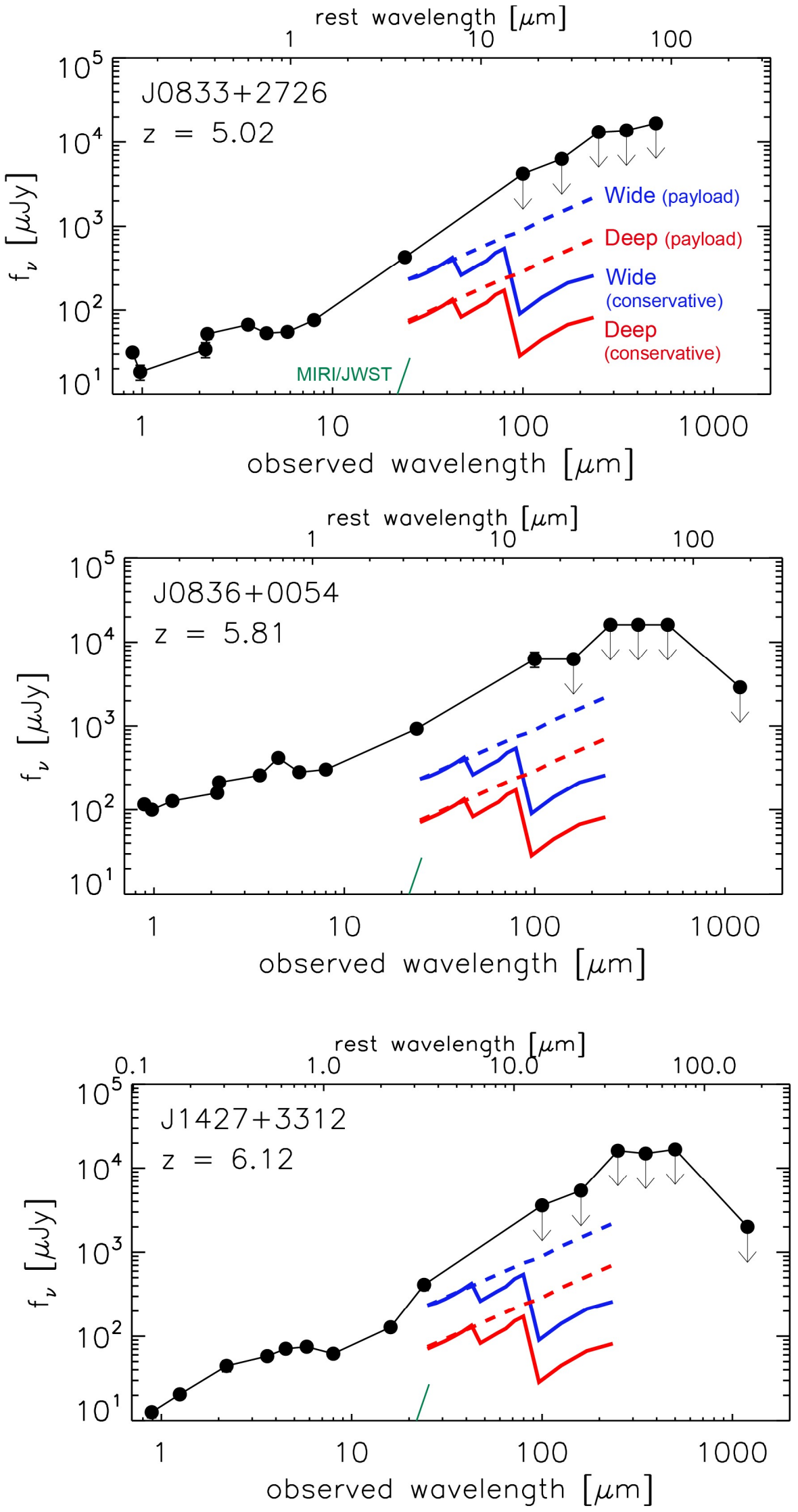

**Figure 1.** Examples of SEDs for SDSS quasars taken from Leipski et al. (2014). The red lines represent the anticipated sensitivity for a deep survey in both payload (dashed) and conservative (solid) modes, whereas the blue lines depict the expected sensitivity for a wide survey under the same modes, as documented in Bisigello et al. (2024).

In addition to studies on SDSS quasars, it is noteworthy that recent advancements facilitated by the JWST have identified numerous AGN candidates exhibiting a broad-line component in their spectra (e.g., Harikane et al. 2023; Matthee et al. 2024). A pertinent question arises: what proportion of JWST galaxies/AGNs can be detected by PRIMAger? To address this inquiry, we employed archival data from the DAWN JWST archive (Valentino et al. 2023), which encompasses 76,637 and 140,778 sources within the CEERS and JADES survey fields, respectively. To estimate the anticipated FIR flux of JWST sources, we conducted SED fitting using CIGALE (Boquien et al. 2019) with available datasets spanning from the optical to the FIR. It is important to note that deep PACS and SPIRE data are accessible in these fields, serving as upper limits for the SED fitting. Figure 2 presents examples of the SED fitting for JWST sources.

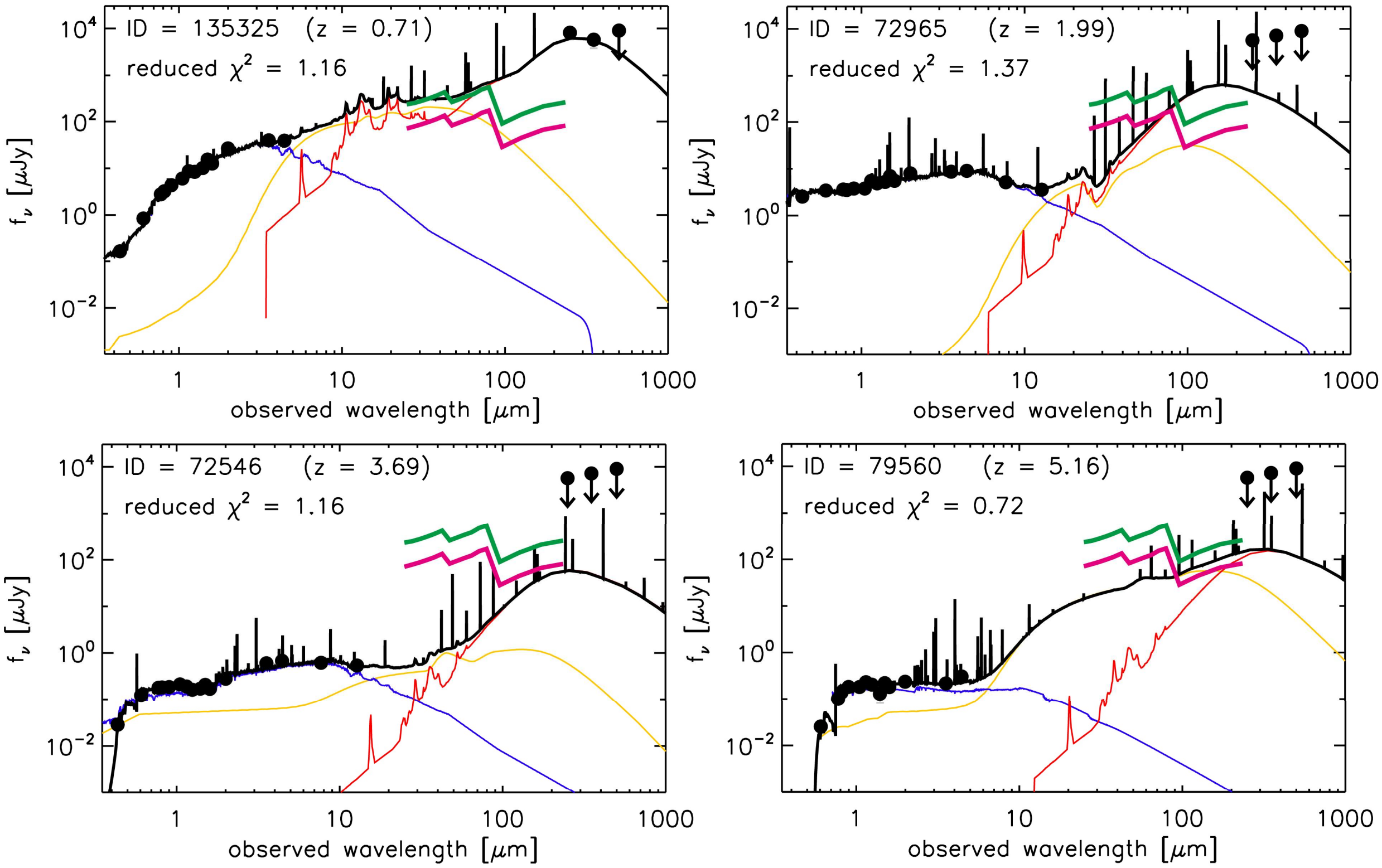

**Figure 2.** Examples of SEDs for JWST sources. The contributions from the stellar, AGN, and SF components to the total SED are shown as blue, yellow, and red lines, respectively. The black solid line represents the resultant best-fit SED. The green and magenta lines are the anticipated sensitivity for a wide and deep survey.

We subsequently estimated the proportion of JWST objects that are detectable by PRIMAger. Figure 3a illustrates the fraction of JWST objects detectable at 235 μm for each survey mode. If a deep survey with conservative sensitivity is implemented, approximately 60% of JWST sources, corresponding to over 100,000 sources, can be detected. The SED fitting also yields physical quantities of JWST sources. Figure 3b presents the mean dust mass estimated by CIGALE within each redshift bin. Notably, for the high-redshift Universe, the less massive end of the dust mass function has been inadequately constrained (e.g., Popping et al. 2017), a limitation that PRIMA can address. Furthermore, comprehending the dust mass density across cosmic time is a critical area of study.

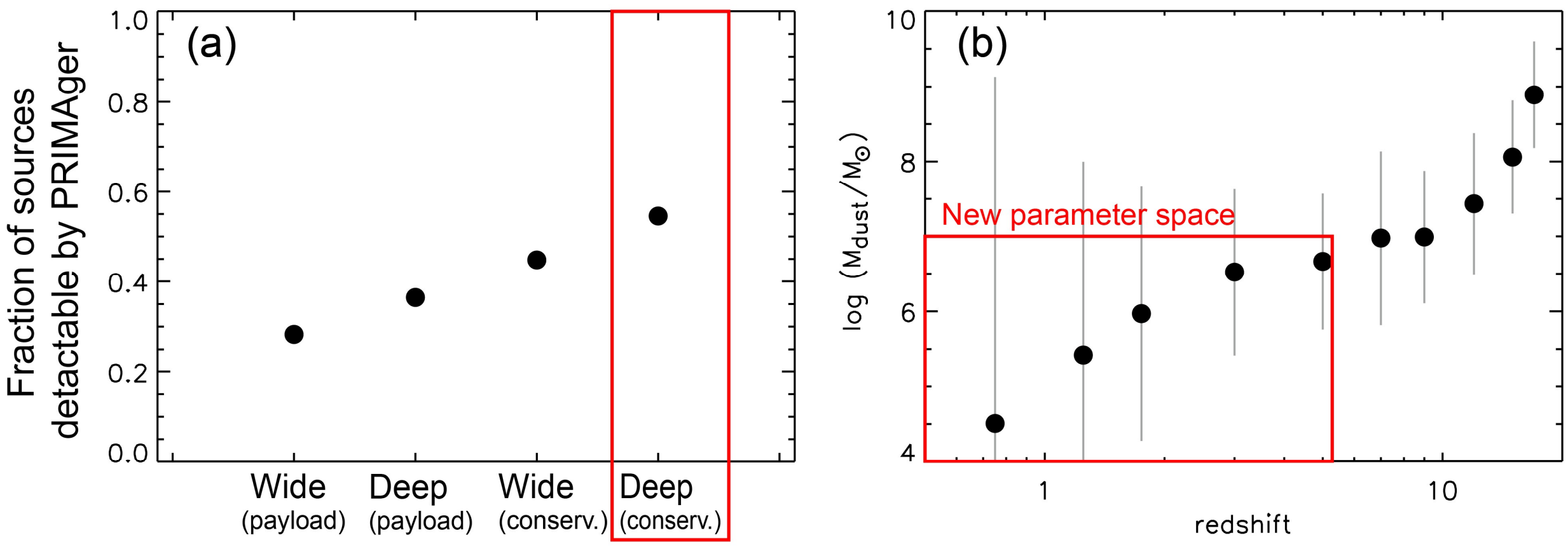

**Figure 3.** (a) Expected fraction of JWST sources detectable by PRIMA at 235 µm assuming PRIMAger wide and deep surveys. (b) Expected dust mass of JWST sources as a function of redshift.

## Instruments and Modes Used

This science case requires PRIMAger's mapping mode.

## Approximate Integration Time

1500 hours of exposure are necessary for both a wide and a deep survey to achieve the science goals mentioned above.

## Special Capabilities Needed

N/A

## Synergies with Other Facilities

### Synergy with Subaru Hyper Suprime-Cam (HSC) and Prime Focus Spectrograph (PFS)

The Subaru Hyper Suprime-Cam (HSC: Miyazaki et al. 2018) is an extensive mosaic CCD camera with a 1.5 degree diameter field of view, positioned at the prime focus of the Subaru Telescope. Initiated in 2014, the HSC legacy survey, as part of a Subaru strategic program (HSC-SSP: Aihara et al. 2018), facilitated the identification of over 100 quasars at $z > 6$ across 1200 deg$^2$ (e.g., Matsuoka et al. 2016). Despite the ongoing uncertainty regarding the dust properties of quasars, PRIMA is capable of detecting host dust emission, as illustrated in Figure 4.

The Subaru Prime Focus Spectrograph (PFS: Takada et al. 2014) is a multi-fiber spectrograph with a broad field of view, allowing for the simultaneous observation of up to 2400 objects with a wavelength coverage extending to 1.2 µm. The PFS-SSP commenced in March 2025, offering a diverse range of spectra for galaxies and AGNs (Greene et al. 2021). A systematic spectroscopic follow-up observation with PFS for PRIMA sources would create a beneficial synergy. We confirm that emission lines are detectable even for faint PRIMA quasars (imag ~24) at z ~ 4 through deep exposures (>5 hours).

### Synergy with Galaxy Reionization EXplorer and PLanetary Universe Spectrometer (GREX-PLUS)

The GREX-PLUS spacecraft is equipped with a telescope featuring a 1 m primary mirror aperture, which is cooled to 50 K. It will carry two scientific instruments: a wide-field camera operating in the 2-8 μm band and a high-resolution spectrometer with a wavelength resolution of 30,000 in the 10-18 μm band (see Inoue et al. 2023, for more details). Notably, the imaging capabilities of GREX-PLUS offer a means to investigate stellar emissions, even within optically dark galaxies. Figure 5 illustrates the anticipated spectral energy distribution (SED) of infrared-bright, optically dark galaxies and active galactic nuclei (AGNs) as reported in Toba et al. (2020). This highlights the synergy between GREX-PLUS and PRIMA in accurately determining the infrared SED for these highly dusty sources, even at redshifts greater than 4.

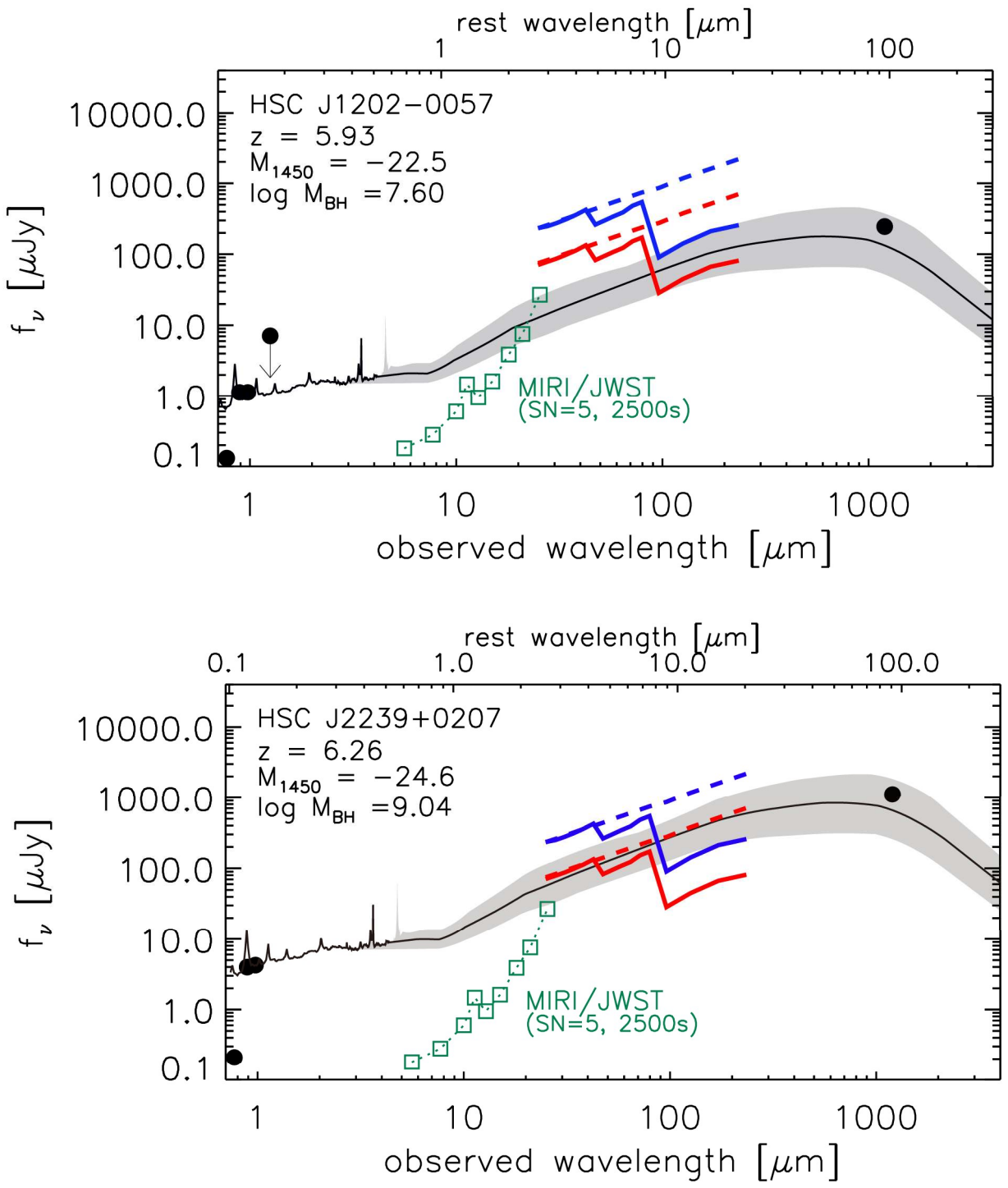

**Figure 4.** Examples of the SED for low-luminosity HSC quasars at z > 6 (e.g., Matsuoka et al. 2016). The red and blue lines are the same as Figure 1.

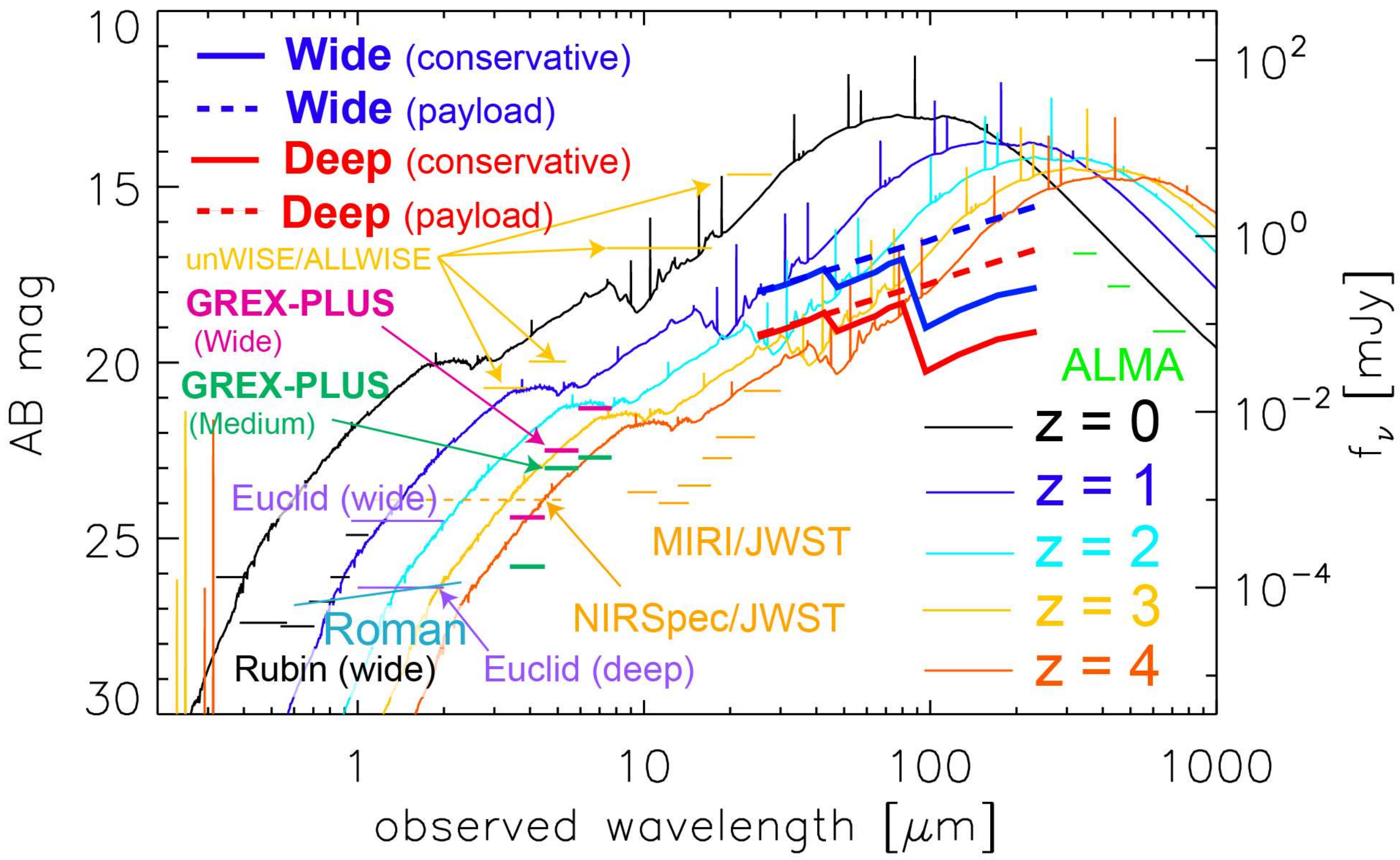

**Figure 5.** Typical SEDs of optically dark IR galaxies found in the AKARI NEP (Toba et al. 2020) as a function of redshift up to z = 4. The detection limits of ongoing and forthcoming missions are overplotted. The red and blue lines are the same as Figure 1.

## Description of Observations

We adhere to the methodologies outlined by Bisigello et al. (2024) (refer to Section 2.1), who delineated the proposed survey strategy utilizing PRIMAger, the photometric imager aboard the anticipated PRIMA FIR space telescope. The strategy encompasses two complementary surveys: a Deep Survey, covering 1 deg² over 1500 hours, aimed at detecting faint, high-redshift galaxies, and a Wide Survey, spanning 10 deg² over 1500 hours, designed to investigate brighter sources across a broader area. Galaxies are selected based on their FIR signal-to-noise ratio to ensure reliable spectral energy distribution (SED) measurements. These surveys are expected to yield a comprehensive dataset for examining galaxy and AGN co-evolution and for identifying targets for FIRESS spectroscopic follow-up.

# 49. Recovering the Dust Mass Budget and temperature with PRIMA

Alberto Traina (INAF-OAS, Bologna, Italy), Francesca Pozzi (UniBo, Bologna, Italy), Francesco Calura (INAF-OAS, Bologna, Italy), Michele Costa (UniBo, Bologna, Italy), Laura Bisigello (INAF-OAPd, Padova, Italy), Carlotta Gruppioni (INAF-OAS, Bologna, Italy), Luigi Barchiesi (UCT, Cape Town, South Africa), Ivan Delvecchio (INAF-OAS, Bologna, Italy), Livia Vallini (INAF-OAS, Bologna, Italy), Cristian Vignali (UniBo, Bologna, Italy), Viviana Casasola (INAF-IRA, Bologna, Italy), Golshan Ejlali (IPM, Tehran, Iran)

Achieving a complete picture of galaxy evolution is a primary goal of extragalactic astrophysics. To accomplish this ambitious task, many multi-wavelength surveys have been devoted to assessing the cosmic evolution of the cold gas and of the stellar mass across cosmic time. In this cosmic census, one elusive component is represented by interstellar dust. In this work, we exploit the archival value of the data provided by the deep survey planned in the PI science time, aimed at estimating the still poorly known dust mass function (DMF) at $z \sim 0.5 - 5$. We consider the spectro-photometric realization of the SPRITZ simulation and we compute the dust masses using single temperature Modified Grey Body functions. We show how PRIMA alone, thanks to its unprecedented sensitivities, will constrain the DMF at $z < 1.5$, in terms of mass and faint-end slope. At $z > 1.5$, we stress the key synergy with current or future sub-millimeter facilities, such as the JCMT/SCUBA-2, AtLAST, LMT and ALMA telescopes, that will allow us to probe the R−J regime of PRIMA selected galaxies.

## Science Justification

One major goal of extragalactic astronomy is to trace in the most complete possible way the evolution of baryonic matter across cosmic time. One fundamental component of galaxies is interstellar dust, representing the solid component of the interstellar medium and affecting the spectral properties of galaxies over a wide range of wavelengths, ranging from the far-infrared to the ultraviolet domain (e.g., Draine 2009), playing a key role in shaping their evolution (Cazaux & Tielens 2002, McKee & Ostriker 2007, Hopkins et al. 2012). How did the dust mass budget evolve through cosmic time? Addressing this question is of the outmost importance to constrain one significant component of the cold mass fraction in galaxies (e.g., Li 2004, Santini et al. 2014, Casasola et al. 2017), to access obscured star formation (e.g., Blain & Longair 1996, Gruppioni et al. 2017, Algera et al. 2023) and the fraction of heavy elements removed from the gas phase and incorporated into solid grains (e.g., Calura et al. 2008, Jenkins 2009, Vladilo et al. 2011).

A fundamental quantity suited for this purpose is the dust mass function (DMF). A thorough estimate of the evolution of the DMF will allow us to reconstruct how the production and destruction of interstellar dust have changed over time in galaxies of various masses (see Calura 2025 in prep. for a review). Thus far, the local DMF has been recovered by various authors (e.g., Vlahakis et al. 2005, Clements et al. 2010, Beeston et al. 2018), while the evolution of the DMF, based on FIR/sum-mm selection, has been subject of a handful of studies. These studies (e.g., Dunne et al. 2011) show a sharp increase in the bright end of the DMF at $0 < z < 0.5$ and it has been recently confirmed by Beeston et al. (2024) using a sample with an order of magnitude more galaxies than used in previous analyses.

In Pozzi et al. (2020) the first study of the DMF evolution from $z \sim 0.2$ up to $z \sim 2.5$ has been presented, based on Herschel-selected catalogue in the COSMOS field. The sample of Pozzi et al. (2020) consisted of $\sim$ 5500 sources with flux density > 16 mJy and estimated spectroscopic or photometric redshift. For each of these systems, the dust mass was derived from the observed flux by assuming a modified black-body relation, valid for a single-temperature dust component in thermal equilibrium and in the standard optically thin regime (Bianchi et al. 2013, Hunt et al. 2019). In absence of photometric points sampling both side of the dust emission (Wien and RJ regime), this estimate requires the assumption of a temperature value, which comes from an empirical relation between dust temperature, SFR and redshift out to $z \sim 2$ (Magnelli et al. 2014).

However, the observations performed with Herschel suffer from severe limitations due mostly to its poor sensitivity and resolution, which caused significant confusion noise and allowed one to probe only the bright-end (MDUST > 109 M⊙) of the DMF at $1 < z < 2$. At $z > 3$, there is only one ALMA-based exploration, from Traina et al. (2024). In this work, the authors used ~200 serendipitously detected galaxies from the ALMA A3COSMOS database (Liu et al. 2019, Adscheid et al. 2024) to sample the DMF at $z = 0.5 - 6$. However, in each redshift bin they were able to sample only the larger dust masses (MDUST = 109 − 1010 M⊙). Thus, a significant component of the DMF (i.e., the faint-end, MDUST ≲ 108 M⊙ ) is currently missing in high-redshift estimates. The present poor theoretical understanding of the evolution of the dust mass budget outlines the need for pursuing more surveys at high redshift, to better constrain the shape of the DMF and achieve new estimates at the so-called "Cosmic Noon", i.e. at $1 < z < 3$, where the measured peak of cosmic star formation lies.

We propose to explore the capabilities of The PRobe far-Infrared Mission for Astrophysics, PRIMA (PI: GJ. Glenn) to achieve a significant progress in the understanding of the dust mass budget evolution as a function of cosmic time. With its unprecedented features, PRIMA will: i) allow us to improve considerably the characterization of the DMF and the basic parameters that define its shape, enabling a better estimate of the faint end and the sampling of the characteristic dust mass across a wide redshift range; ii) extend the redshift range where the dust emission in individual galaxies can be probed, allowing us to derive the DMF, with PRIMA alone data, up to $z = 1.5$ and, in synergy with ALMA, up to $z = 5$; iii) allow us to significantly constrain dust production in galaxy formation models, improving our poor theoretical knowledge of this process and filling the gap in our understanding of dust-obscured galaxies. The prediction, derived from the SPRITZ, is shown in Figure 1.

To compute the dust masses, we will rely on the strength of PRIMA in determining the dust temperature and on the synergy with current or future sub-mm facilities (e.g., LMT, AtLAST) to measure a flux in the R-J regime of the SED (needed in the assumption of optically thin medium in the galaxy).

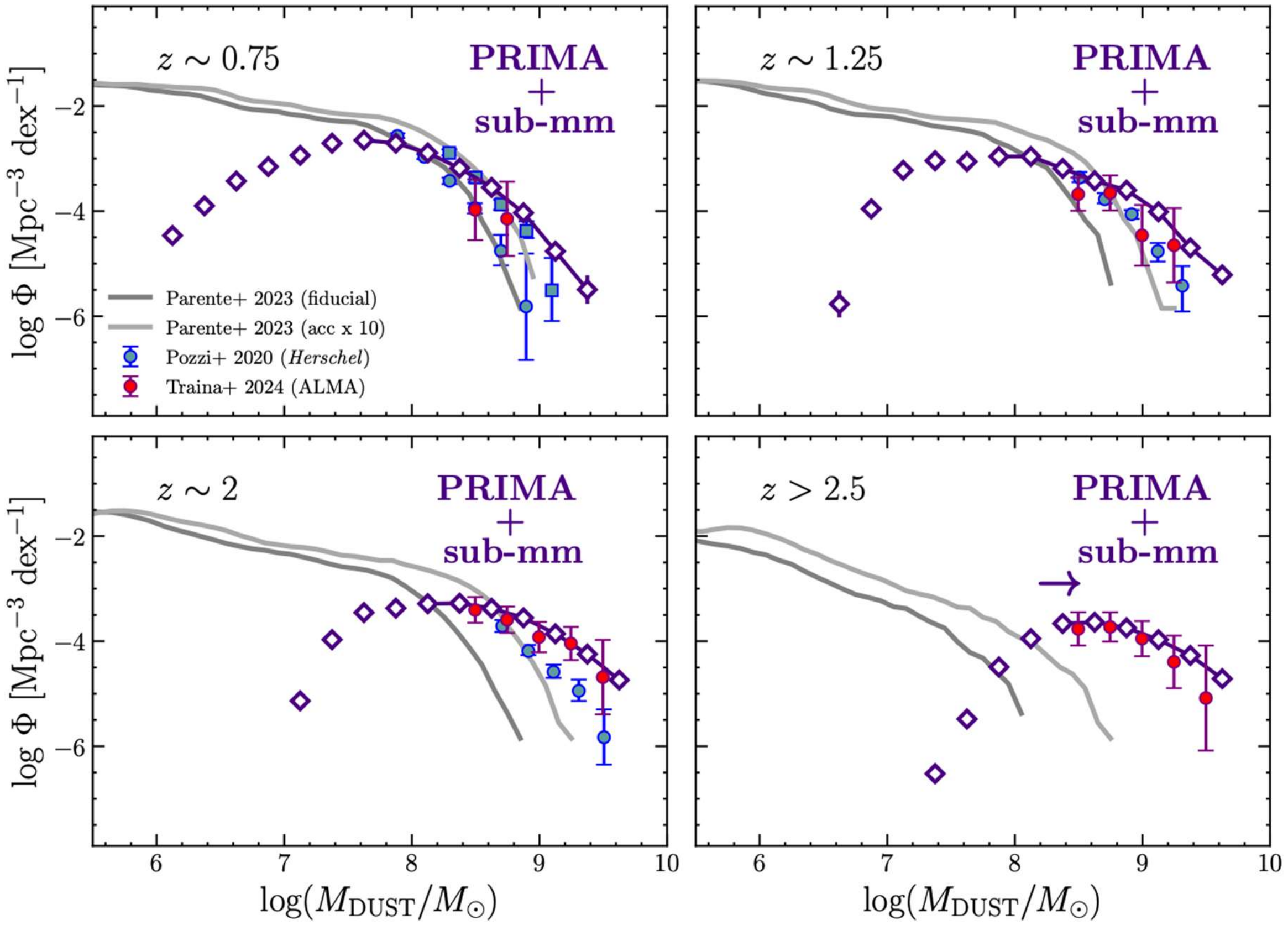

**Figure 1.** Dust mass functions at z ~ 0.75, z ~ 1.25, z ~ 2 and z > 2.5 from models and literature data, compared with the simulated DMF probed by PRIMA. The dark and light grey lines are two different realization of the SAM by Parente et al. (2022). Data from Pozzi et al. (2020) and Traina et al. (2024) are shown as blue and red points, respectively. Blue squares and circles refer to different redshift bins. The PRIMA predicted DMF is plotted as white diamonds and purple lines (complete regions).

## Instruments and Modes Used

This observing program uses all the hyperspectral and polarimeter bands of PRIMAger to map a 1x1 degree area. The polarimetry information is not required.

## Approximate Integration Time

This is a GI program that plans to use the data of the PI 1000hr survey.

## Special Capabilities Needed

N/A

## Synergies with Other Facilities

PRIMA will be able to probe the R-J part of the SED up to z ∼ 1.5, although the R-J flux in the last PRIMA band may be significantly affected by confusion. For this reason, especially at higher redshift, together with the lack of coverage of the R-J, the synergy of PRIMA with current of future sub-mm and mm facilities (e.g., JCMT/SCUBA-2, LMT, AtLAST) will be of key importance.

## Description of Observations

We plan to estimate the dust properties using a “reference” survey (i.e., the PI SCIENCE survey of 1000hr on 1deg2). We expect to have data in all the 12 HyperSpectral bands and in the 4 Polarimetry bands of the PRIMAger instrument. This coverage is necessary to characterize the dust temperature of each galaxy, which is required to derive the dust masses. Moreover, sampling the full dust emission is fundamental to properly estimate the dust luminosity. The exposure will allow us to reach a limiting flux at 60 µm of 0.3 mJy (which translates in going ∼ 10 times deeper than Herschel at 160 µm assuming the SEDs from the SPRITZ). An additional strength would be to have this survey in a field rich in ancillary data (i.e., the COSMOS field), allowing for a better determination of the galaxies’ physical properties.

From the SPRITZ simulation, we expect to detect ∼ 35000 galaxies above the detection limit (and above the confusion at 60 µm).

# 50. PRIMA Investigation of Obscured Active Galactic Nuclei in Star-Forming Galaxies at Cosmic Noon

Yoshihiro Ueda (Kyoto University), Satoshi Yamada (Tohoku University), Yoshiki Toba (Ritsumeikan University), Tohru Nagao (Ehime University)

It is a fundamental question in modern astronomy how supermassive black holes (SMBHs) in galactic centers formed and co-evolved with their host galaxies over the history of the Universe. A key process for understanding the co-evolution is a major merger, which triggers violent star formation and intense mass accretion onto SMBHs deeply obscured by surrounding gas and dust. Here, we propose to identify heavily obscured active galactic nuclei (AGNs) in rapidly star-forming galaxies at cosmic noon ($z \sim 1$–$3$) with PRIMA/FIRESS (24–235 μm), by detecting emission lines from ions with high ionization potential, such as [Ne V] 14.32 μm and [O IV] 25.89 μm. We select promising candidates on the basis of spectral energy distribution analysis from ultra-luminous infrared galaxies detected in imaging surveys with PRIMA and/or ALMA. We utilize our results to construct the luminosity function of the most heavily obscured AGN population, which is missing from the current census.

## Science Justification

Nearly all galaxies in the present universe contain supermassive black holes (SMBHs) in their centers, whose masses correlate well with those of galactic bulges over 4 orders of magnitude (e.g., see Kormendy & Ho 2013 for a review). This fact indicates that the formation process of SMBHs and that of galaxies are strongly coupled to each other. Thus, it is one of the most fundamental questions in modern astronomy how these SMBHs in galactic centers formed and "co-evolved" with their host galaxies over the history of the Universe. Since Active Galactic Nuclei (AGNs) are the very processes whereby the SMBHs grow by mass accretion, a complete census of AGNs is indispensable to elucidate their growth history. Previous X-ray surveys, performed at energies below 10 keV, have revealed the overall cosmological evolution of unobscured or moderately obscured AGNs, which shows cosmic "down-sizing" (e.g., Ueda et al. 2014; Fig. 1). These X-ray results may not be the whole story, however.

A key process for understanding the co-evolution of galaxies and SMBHs is a galactic merger. According to the current best understanding, major mergers of galaxies trigger violent star formation and intense mass accretion onto SMBHs deeply obscured by surrounding gas and dust, where the line-of-sight hydrogen column densities in the AGNs may far exceed $\log NH/cm^{-2} > 24$ (i.e., Compton thick). This means that a key aspect of the growth of SMBHs, the merging process, is largely hidden at most electromagnetic wavelengths, including soft X-rays below 10 keV.

Hard X-ray observations above 10 keV are useful for detecting Compton-thick AGNs (CTAGNs) thanks to their strong penetrating power against obscuration as far as $\log N_H/cm^{-2} \lesssim 25$. NuSTAR has shown >50% of late-stage AGN mergers in the local Universe are Compton-thick (Ricci et al.

2017; Yamada et al. 2021), indicating that these CTAGNs, unique tracers of mergers, may be a distinct population from normal AGNs, harboring much missing information about merger growth. However, the sensitivities of current missions are not sufficient to make a complete survey of this population covering cosmic noon, when the number density of AGNs is expected to be the highest. Hence, the basic questions on CTAGNs remain open: **(1) if their cosmological evolution is different from that of normal AGN populations and (2) what is the contribution of CTAGNs, largely missing from the current census, to the total growth of SMBHs.**

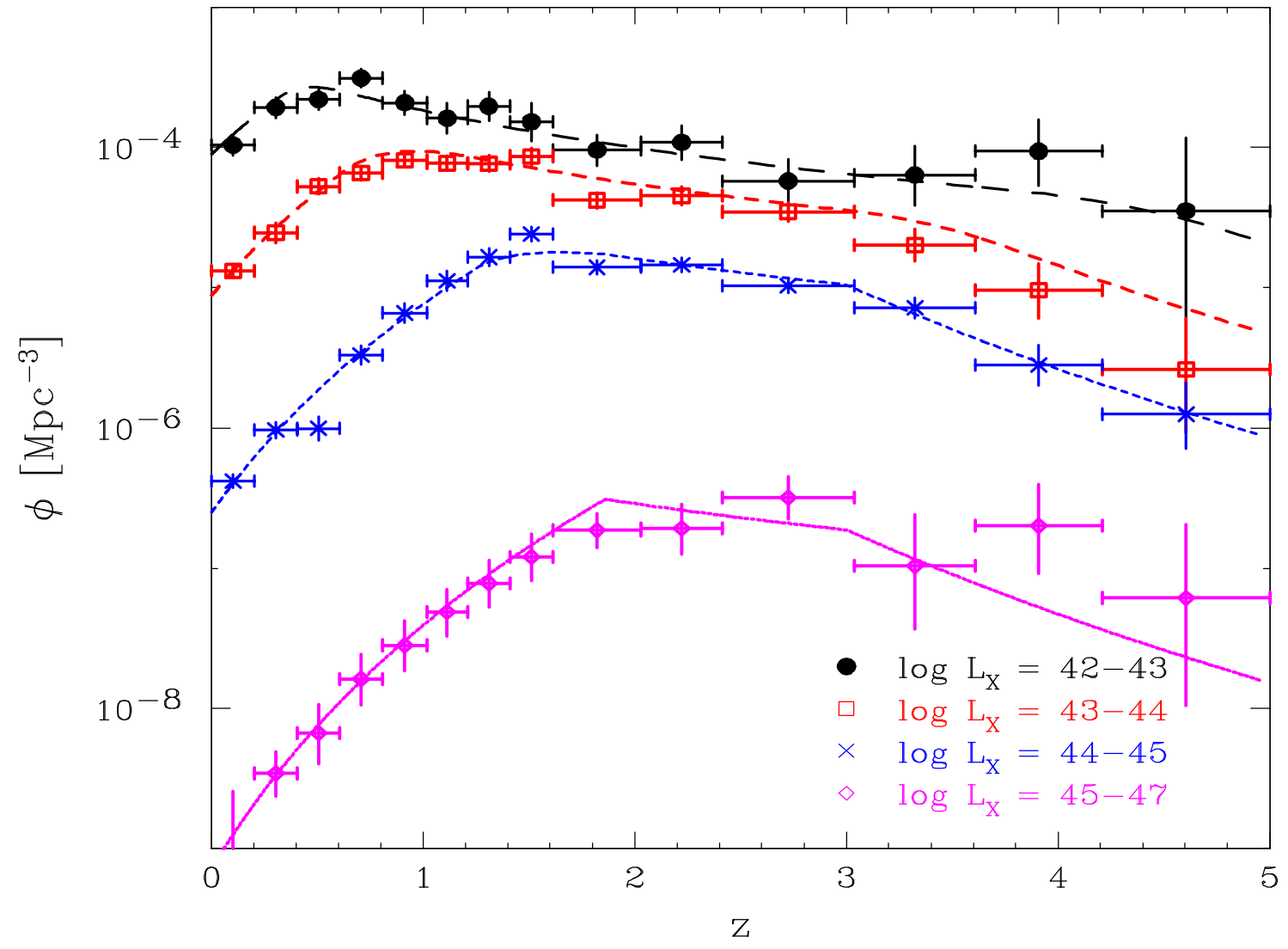

**Figure 1.** Comoving number density of AGNs plotted against redshift in different X-ray luminosity bins. The curves are the best-fit model, and the data points are calculated from either soft- or hard-band AGNs detected in X-ray surveys
(Ueda et al. 2014).

Infrared surveys have revealed that ultra-luminous infrared galaxies (ULIRGs), galaxies with ongoing intense star formation, exhibit a rapid number-density evolution from z = 0 toward z > 1, with a slope much steeper than those of any known AGN populations detectable below 10 keV (e.g., Goto et al. 2011). Many observations imply that a major fraction of ULIRGs contain deeply buried AGNs, although its estimate, except for the local universe, is highly uncertain. If most ULIRGs indeed contain buried AGNs with moderate luminosities, their number density could far exceed that of known AGNs at z ~ 1.5, leading to a transformational change in our understanding of the evolution of the whole SMBH population. Currently, however, along with current population synthesis models of the cosmic X-ray background (Ueda et al. 2014; Ananna et al. 2019), nearly all theoretical models of cosmological galaxy/SMBH formation, based on hydrodynamical simulations (e.g., Habouzit et al. 2022) or semi-analytic approach (e.g., Fanidakis et al. 2012; Shirakata et al. 2019), ignore the buried AGN population because their number-density evolution is observationally undetermined.

Attempts have been made to identify deeply obscured AGNs in rapidly star-forming galaxies at cosmic noon. Since AGN-heated dust emits mid-infrared (MIR) radiation, galaxies containing deeply obscured AGNs may show spectral energy distribution (SED) characterized by hot dust emission. For example, a systematic SED analysis of ALMA-detected sub/millimeter galaxies (Uematsu et al. 2024, 2025) suggests that a fraction of these galaxies may contain heavily obscured AGNs undetected in X-rays i.e., most likely, heavily CTAGNs (Fig. 2, left). Interestingly, the inferred AGN luminosity density shows a possible excess at z = 2–3 compared with that determined from X-ray surveys below 10 keV, implying their contribution to the SMBH growth may be even larger than that estimated from current X-ray surveys (Fig. 2, right). These results, however, sharply depend on the SED models used and hence should be taken with caution.

A convincing method to identify obscured AGNs is the detection of emission lines from ions with high ionization potential that can only be excited by UV to soft X-ray radiation from an AGN. In particular, those in the MIR band, such as [Ne V] 14.32 µm and [O IV] 25.89 µm, are powerful diagnostics for AGNs in heavily dust-obscured galaxies, thanks to their strong penetrating power against dust extinction.

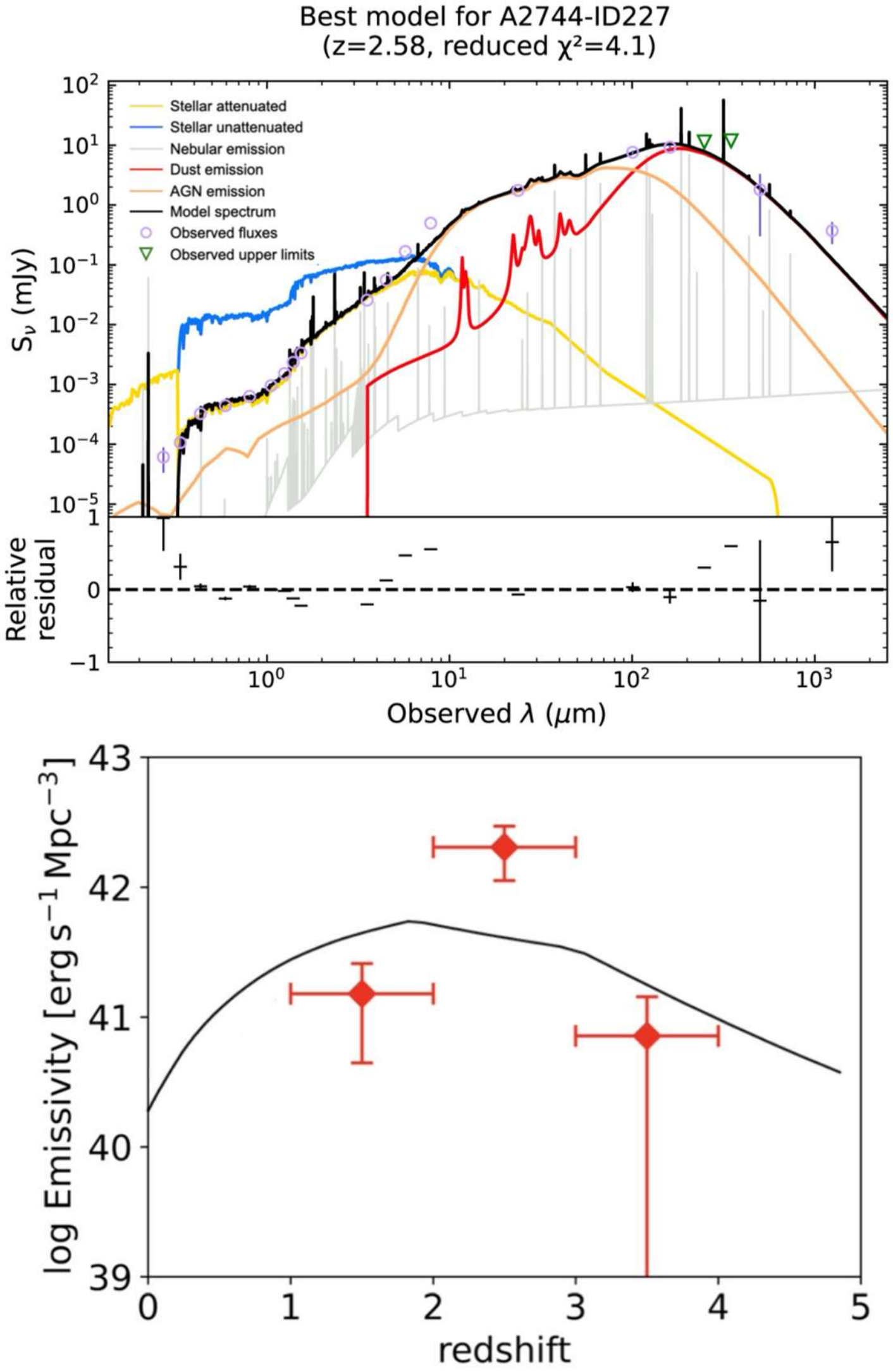

**Figure 2.** (*Left*) An example of the UV-to-millimeter SED and the best-fit model for one of the ULIRGs hosting promising obscured AGNs in the ALMA Lensing Cluster Survey (ALCS) field. (*Right*) Cosmological evolution of AGN luminosity density. The black solid line shows that determined from X-ray surveys by Ueda et al. (2014). The red points and error bars are estimated by the AGN candidates identified by the SED analysis in the ALCS field (Uematsu et al. 2024).

**Here, we propose mid-to-far IR spectroscopic observations with PRIMA/FIRESS to unambiguously detect and characterize heavily obscured AGNs in rapidly star-forming galaxies at z = 1–3.** The sample can be selected from galaxies with high star formation rates found in imaging surveys with PRIMA and/or ALMA. The most promising candidates of heavily obscured AGNs are then chosen on the basis of multiwavelength SED analysis, including X-ray data.

## Instruments and Modes Used

The FIRESS spectrometer in its pointed low-resolution mode will be used.

## Approximate Integration Time

We request ∼200 hr of exposure time with FIRESS (bands 1–4) to observe ∼50 heavily obscured AGN candidates in rapidly star-forming galaxies at z = 1–3. The exposure time is conservatively estimated to ensure >5σ detections of the [Ne V] and [O IV] lines, even in the faintest cases (∼1.3 × $10^{-19}$ W $m^{-2}$ for e.g., z = 3 ULIRGs with bolometric AGN luminosities of 1046 erg s-1; see Description of Observations).

## Special Capabilities Needed

None

## Synergies with Other Facilities

The PRIMA observation will have a strong impact on our understanding of a fundamental question in astronomy: how whole populations of galaxies and SMBHs formed over cosmic time. It is commonly accepted that SMBHs play a key role in regulating galaxy evolution via AGN feedback. In fact, recent studies of local ULIRGs (e.g., Yamada et al. 2021) indicate that obscured AGNs in late-stage mergers ubiquitously show signatures of multi-phase outflows originating from the AGNs. Therefore, to understand galaxy evolution in the context of large-scale structure, it is vital to completely locate all AGN populations including heavily obscured ones, and reveal their environments (galaxy mergers, cold streams, or isolated galaxies). In this regard, we expect strong synergy between PRIMA and current or near-future observatories at other wavelengths. To investigate the host properties of CTAGNs identified by PRIMA, observations with ALMA (morphology and gas dynamics), JWST (morphology), Subaru/HSC and ULTIMATE (photometry), and GMT or TMT (spectroscopy) will be very useful. Most of these data will be already publicly available by the time of the PRIMA operation in the 2030s and hence can be immediately utilized once the proposed PRIMA observations are performed.

## Description of Observations

The primary goal is to achieve >5σ detections of the [Ne V] and [O IV] lines from heavily obscured AGNs in star-forming galaxies. The required exposure time is estimated using a Spitzer/IRS low-resolution (R ∼ 100) spectrum of a local ULIRG UGC 5101 (z = 0.039), which hosts a CTAGN (Yamada et al. 2021, 2023). X-ray and MIR observations suggest that UGC 5101 exhibits relatively weak ionized lines compared to other local AGNs, likely due to UV line shielding by its geometrically thick torus (e.g., Oda et al. 2017; Yamada et al. 2023). A conservative exposure calculation based on this template spectrum predicts that a 4-hr observation (or 2 hr × 2 in bands 1–4) per target will achieve >5σ line detections for CTAGNs at z = 3 with a bolometric AGN luminosity of >$10^{46}$ erg $s^{-1}$, or for CTAGNs at z = 1 with >1045 erg $s^{-1}$. In both cases, the expected [Ne V] and [O IV] line flux is ≳$1.3 \times 10^{-19}$ W $m^{-2}$ (Fig. 3).

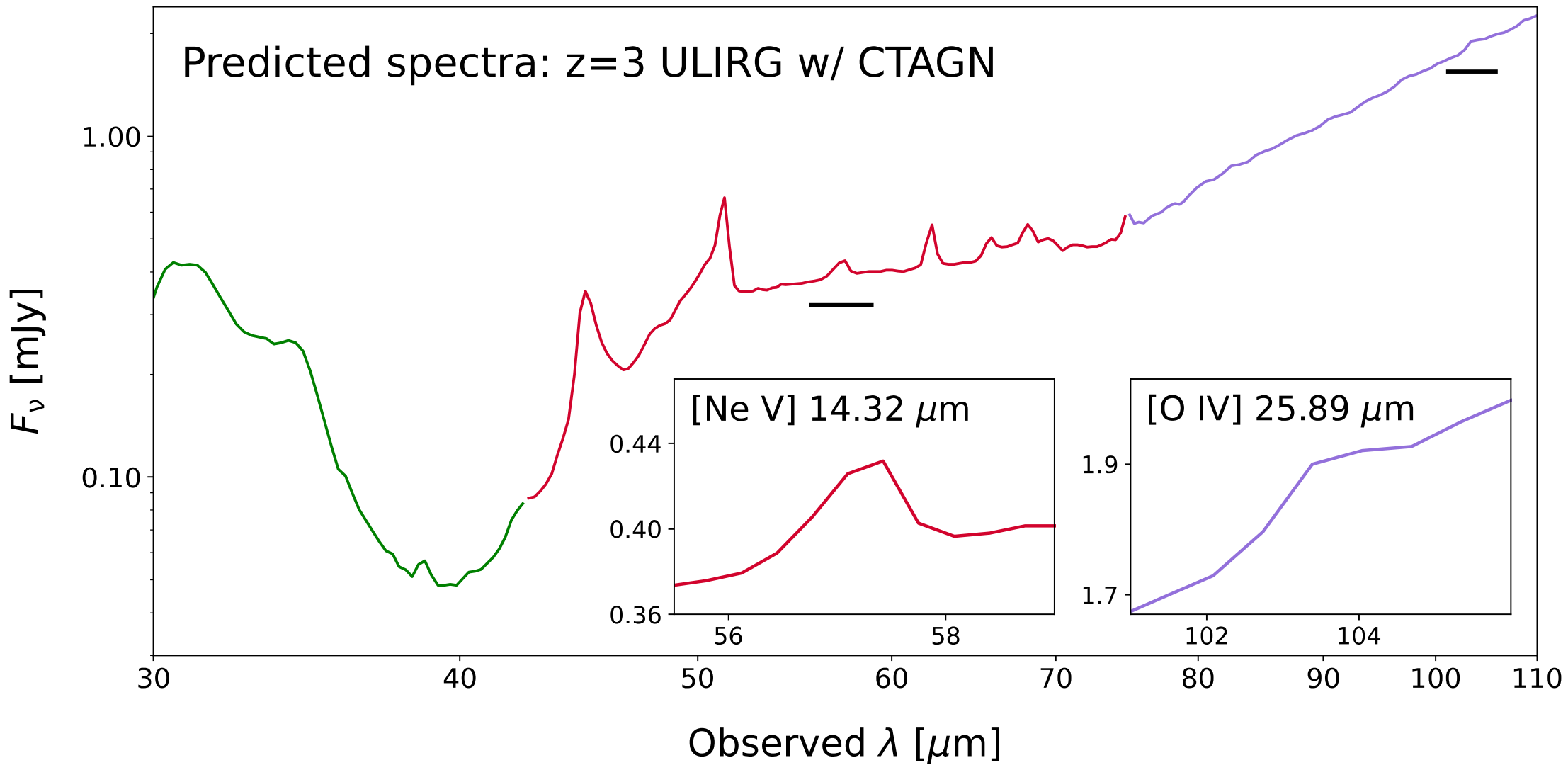

**Figure 3.** Predicted infrared spectra of a ULIRG at z = 3 hosting a CTAGN, assuming the spectral shape of UGC 5101. The template spectra were obtained from the IRSA/IRS_Enh database. The green, red, and purple segments indicate the spectral coverage of FIRESS bands 1, 2, and 3, respectively.

The targets will be selected from star-forming galaxies identified in imaging surveys with PRIMA and/or ALMA, particularly in multi-wavelength deep survey fields such as ALMA Lensing Cluster Survey (ALCS), COSMOS, SXDF, and GOODS-S. By performing multi-wavelength SED fitting, we will estimate their basic properties (e.g., star formation rates and bolometric AGN luminosities) and select a sample of approximately 50 promising CTAGN candidates at z = 1–3.

For AGNs with detected [Ne V] and [O IV] lines, we will measure their bolometric AGN luminosities. By comparing these with the upper limits of the X-ray fluxes, we can estimate the hydrogen column densities (e.g., Uematsu et al. 2024). Redshifts will be determined by combining various ionized lines, including [Ne II] and [Ne III]. Wide-band spectroscopy will resolve the complex spectra containing silicate absorption features, PAH emissions, low-ionization lines, and H2 lines (Fig. 3), allowing us to discuss the impact of AGN feedback on star-forming activity. Finally, this program will provide new insights into **the luminosity function of the most heavily obscured AGN population at z = 1–3, which have not yet been constrained in other wavelength bands.**

# 51. Uncovering the Hidden Phase of the Co-Evolution of Galaxies and Super Massive Black Holes in the Heart of the Proto-Clusters

Hideki Umehata (Nagoya University), Mariko Kubo (Kwansei Gakuin University)

Massive elliptical galaxies at the centers of local galaxy clusters host supermassive black holes (SMBHs). These most massive early-type galaxies host correspondingly massive supermassive black holes (SMBHs). But the origin of the tight bulge–SMBH mass relation remains a major mystery in astronomy. To understand this, we must study their formation in the early universe, where simulations suggest that cold gas accreted along cosmic web filaments fuels both galaxy and SMBH growth. Proto-clusters are ideal environments for such studies, hosting dusty star-forming galaxies (DSFGs) that often harbor X-ray AGNs and are thought to evolve into present-day ellipticals.

However, AGN activity in these dusty phases is difficult to observe with current instruments. Mid-infrared dust emission, which traces obscured AGNs, is key, but has been poorly constrained at high redshift. We propose to build complete infrared SEDs for DSFGs in several proto-clusters (e.g., ADF22 in SSA22 at z=3.09) using PRIMA. This enables accurate SED modeling to identify obscured AGNs, measure fundamental properties such as SMBH mass and accretion rate, connecting their growth with that of their host galaxies. This project will open a new window into understanding the emergence of galaxy–SMBH co-evolution in the early universe, revealing when and how galaxies and SMBHs assembled their masses during their peak formation epochs.

## Science Justification

In the local universe, the centers of galaxy clusters are typically occupied by massive elliptical galaxies. These most massive early-type galaxies host correspondingly massive supermassive black holes (SMBHs). Understanding the origin of the tight correlation between bulge mass and SMBH mass remains one of the most outstanding mysteries in astronomy. To address this, it is essential to observe and understand their formative phases in the early universe.

Cosmological simulations and theoretical models predict that cold gas streams flowing along intergalactic filaments supply the bulk of the material required for the growth of both galaxies and their central SMBHs (e.g., Bond et al. 1996; Springel et al. 2005; Dekel et al. 2009). During these active growth phases, galaxies are expected to be heavily obscured by dust (e.g., Alexander et al. 2008). Therefore, revealing and characterizing this dust-obscured phase of co-evolution occurring within the cosmic web is of fundamental importance. Proto-clusters, which represent massive environments where cold gas is efficiently funneled by gravity into dark matter halos to form filamentary structures, provide a unique laboratory for this purpose (Umehata et al. 2019).

Thanks to the advent of ALMA, overdensities of dusty star-forming galaxies (DSFGs) at z=2-4 have been identified (with $N_{DSFGs}$>~10, Umehata et al. 2015; 2019, Oteo et al. 2018). These galaxies are thought to rapidly quench after intense star formation, eventually evolving into passive systems. Together with their proto-cluster environments, they are believed to be the progenitors of today's massive elliptical galaxies. Intriguingly, many of these DSFGs host X-ray AGNs, showing a 2–3 order of magnitude excess in number density (or in the AGN luminosity function; e.g., Umehata et al. 2015). These bursty proto-cluster cores thus represent rare and invaluable targets for unveiling the simultaneous growth of galaxies and SMBHs during their most active phases.

However, probing AGN activity during the heavily obscured stages remains a major challenge, posing a significant gap in our understanding of early SMBH–galaxy co-evolution. In such dusty environments, measuring basic AGN properties such as SMBH mass and Eddington ratio becomes extremely difficult, as heavy extinction hampers detection in both soft X-rays and the optical, where key diagnostic lines lie.

In this context, characterizing the dust SED at (rest-frame) mid-infrared wavelengths—tracing warm-to-hot dust heated by AGN—is critically important. Currently, however, no available instrument can achieve sufficient sensitivity to detect such emission from highly obscured AGNs at high redshift. In this regard, PRIMA is poised to be a game-changer, providing coverage from 24–261 µm in the observed frame and bridging the wavelength gap between JWST/MIRI and ALMA.

We propose to construct well-sampled infrared SEDs of DSFGs in the cores of proto-clusters, including systems such as COSMOS at z=2.5 (Wang et al. 2016), ADF22 in SSA22 at z=3.1 (Umehata et al. 2015; 2019, DRC at z=4.0 (Oteo et al. 2018), COSMOS AzTEC3 at z=5.3 (Capak et al. 2011). PRIMA will fill in the critical gaps in the rest-frame near-/mid-/far-infrared SEDs. These SEDs will enable coherent modeling using multiple tools (e.g., LIGHTNING, CIGALE), allowing us to simultaneously constrain star formation and AGN components. This approach will allow us to identify obscured AGN candidates that are faint or undetected in X-ray and optical bands by fitting the full SED with both star-forming and AGN templates. Moreover, we will be able to directly recover the star formation histories (SFHs) of AGN host galaxies, shedding light on the connection between star formation and black hole accretion in the obscured AGN population within proto-clusters.

This project will open a new window into understanding the emergence of galaxy–SMBH co-evolution in the early universe. We aim to measure the SMBH-to-galaxy mass ratio and the relative growth rates in several proto-clusters at z=2-5, revealing when and how galaxies and SMBHs assembled their masses during their peak formation epochs. While QSOs are often found to host "overmassive" SMBHs relative to the local relation (Kormendy & Ho 2013), DSFGs have been suggested to host "undermassive" SMBHs (Alexander et al. 2008), potentially indicating a scenario where stellar mass growth precedes black hole growth. Both evolutionary tracks are predicted by simulations, and observations are crucial to test these models. By constructing a statistical sample and leveraging synergy with ALMA, JWST, and future facilities, we aim to constrain the triggering mechanisms of AGN activity in these dusty environments.

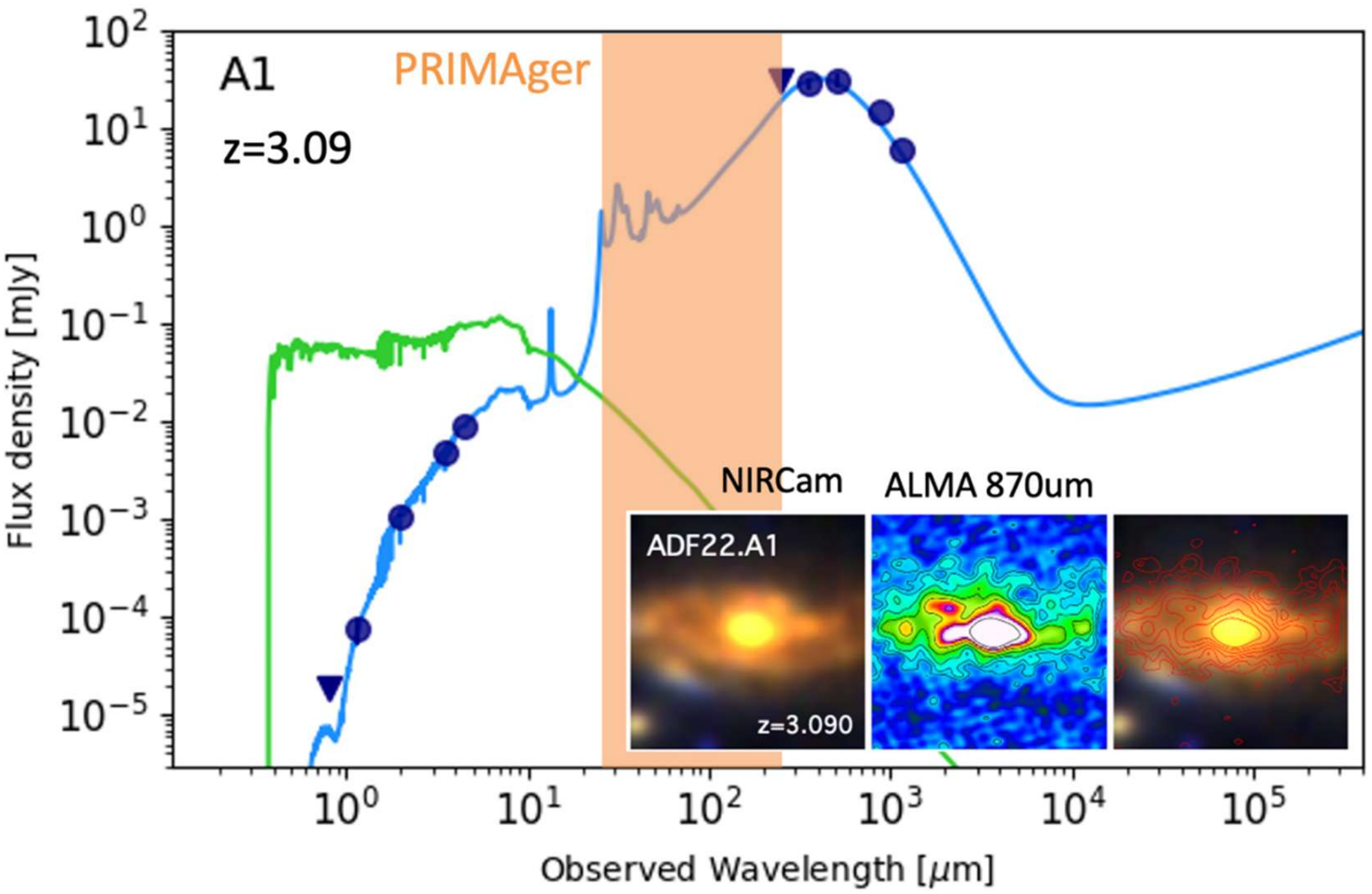

**Figure 1.** The panchromatic SED of ADF22.A1 at z=3.09, an AGN-hosting DSFG located in the core of the SSA22 proto-cluster. Heavy dust extinction significantly obscures the source, and the wavelength gap between JWST and ALMA introduces uncertainties in the characterization of the obscured AGN. (Images from Umehata et al. 2025a,c).

## Instruments and Modes Used

This observing program requires 4 maps of size 10'x10' each in both the hyperspectral and polarimeter bands of PRIMAger. Polarimetry information is not required.

## Approximate Integration Time

We aim to sample dust SEDs of proto-cluster galaxies at z=2-5 at rest-frame mid-/far-infrared wavelengths. For a proto-cluster field, we assume to obtain a 10'x10' map as recommended as a minimum area. Then the PRIMA ETC suggests the following time for observation:

PHI1_5 (38.1μm), 100 arcmin$^2$, 250μJy at 5σ, 15.5 hrs

PHI2_5 (71.7μm), 100 arcmin$^2$, 300μJy at 5σ, 9.7 hrs

PPI2 (126μm), 100 arcmin$^2$, 600μJy at 5σ, 5.1 hrs

PPI4 (235μm), 100 arcmin$^2$, 800μJy at 5σ, 9.1 hrs

Therefore, the estimated total time for PHI, PH2, PPI2, and PPI4 mapping for a proto-cluster field is 39.4 hrs. To cover the total 4 proto-cluster fields at different redshift, approximately 160 hrs are estimated.

Special capabilities needed

N/A

## Synergies with Other Facilities

We expect many synergies with other facilities. The combination of JWST/MIRI, PRIMA, and ALMA will allow us to uncover the whole picture of dust SEDs of dusty galaxies at high redshift. The high spatial resolution of JWST/NIRCam and ALMA will compensate the moderate resolution of PRIMA. A sensitive X-ray satellite is also expected to trace the X-ray emission to improve the estimates of several fundamental parameters.

## Description of Observations

We propose to observe four proto-cluster fields at z=2-5, namely, COSMOS at z=2.5 (Wang et al. 2016), ADF22 in SSA22 at z=3.1 (Umehata et al. 2015; 2019), DRC at z=4.0 (Oteo et al. 2018), COSMOS AzTEC3 at z=5.3 (Capak et al. 2011). For each proto-cluster field, we obtain a 10'x10' map using the Hyperspectral Imager (PHI1, PHI2) and two bands of Polarimetry Imager (PPI2, PPI4).

# 52. Studying the Physics of Dusty Outflows Using Hydrocarbon Dust

Andrey Vayner (IPAC/Caltech), Tanio Díaz-Santos (FORTH), Carl D. Ferkinhoff (Winona State), Peter R. M. Eisenhardt (JPL), Daniel Stern (JPL), Lee Armus (IPAC/Caltech), Brandon S. Hensley (JPL), Daniel Anglés-Alcázar (University of Connecticut), Roberto J. Assef (Universidad Diego Portales), Román Fernández Aranda (FORTH), Andrew W. Blain (University of Leicester), Hyunsung D. Jun (Northwestern College), Norm Murray (CITA/University of Toronto), Shelley Wright (UC San Diego), Chao-Wei Tsai (Beijing Normal University), Thomas Lai (IPAC/Caltech), Niranjan Chandra Roy (University of Connecticut)

Dust is crucial to the evolution of the interstellar medium (ISM), playing a critical role in the cooling and feedback processes. In dense nuclear regions of luminous galaxies, dust can be a crucial source of opacity that allows for the launching of fast-moving outflows that can affect the star-forming properties of their galaxies. To study the demographics of these outflows and to better understand the conditions of the launching mechanism, we need to study their kinematics and dynamics using spectroscopy. Measuring the kinematics of dust is challenging and relies on the detection of hydrocarbon features in absorption along our line of sight. Thus far, we know of a single case where the kinematics of dust in an outflow have been measured through the detection of 3.3 micron aromatic and 3.4 micron aliphatic hydrocarbon dust features in a luminous dust-obscured galaxy at $z = 4.6$ (Vayner et al. 2025b, submitted). With PRIMA, we would be able to perform an efficient survey to expand the sample size to many obscured quasars at the peak epoch of galaxy and SMBH growth. At $z > 2.8$, PRIMA and the FIRESS instrument will enable a study of hydrocarbon dust features, including aliphatic, aromatic, and olefinic molecules, together with silicate dust features, enabling us to study the chemistry and dynamics of dusty outflows for the first time.

## Science Justification

It is well established that feedback from active galactic nuclei plays an essential role in the evolution of massive galaxies (Mercedes-Feliz et al. 2023). Given that the majority of supermassive black hole (Delvecchio et al. 2014) and galaxy growth (Madau & Dickinson 2014) occurs at cosmic noon ($1.5<z<3$), it is critical to study galaxies hosting active galactic nuclei (AGN) at this epoch to understand how they can regulate star formation in their galaxies. Furthermore, both theoretical and observational work indicate that obscured nuclear regions of galaxies are where the energy from AGN couples most efficiently to the interstellar medium (ISM) (Costa et al. 2018). Hence, studying obscured galaxies at cosmic noon presents an excellent opportunity to investigate the evolutionary phase during which the obscured regions of galaxies undergo rapid transformation.

AGN-driven outflows are multi-phase (Richings et al. 2021), with the colder molecular phase often carrying a substantial amount of mass, energy, and momentum. Molecular outflows are often detected either through emission or absorption from small molecules ($H_2O$, OH, CO), which require a significant amount of dust to be present in the outflow. Small molecules are easily destroyed in galactic winds from shocks driven by the interaction of the outflow with the ambient media, hence dust is necessary to reform them. Furthermore, opacity from dust grains can play an important role in accelerating to velocities several times greater than the escape velocity of the inner regions of massive galaxies (Roth et al. 2012). How the dust can survive and be accelerated to large velocities is an open question that PRIMA will be critical in addressing. Furthermore, understanding the transfer of dust from the inner regions of galaxies to larger kpc scales and into the circumgalactic medium (CGM) is crucial for understanding the evolution of the interstellar medium (ISM). Understanding the transport of dust onto CGM/ICM scales can also be important for many cosmological studies.

One of the missing measurements is the kinematics and dynamics of dust in outflows, and whether they are consistent with radiation pressure as the primary driving mechanism. In the last decade, we have discovered several classes of obscured AGN at cosmic noon that host powerful, fast-moving outflows driven from extremely obscured, dense nuclear regions. Both the hot dust-obscured galaxies (HotDOGs, Eisenhardt et al. 2012) and extremely red quasars (ERQs) populations are excellent for studying the nuclear blow-out phase of quasar evolution, where the nuclear regions transform from an obscured to an unobscured phase through powerful quasar-driven outflows. Both samples contain some of the fastest-moving outflows discovered to date (Perrotta et al. 2019, Finnerty et al. 2020), and given their high Eddington ratios, high nuclear obscuration, and low number density, indicate that their phases should be short-lived.

Recent *James Webb Space Telescope* (*JWST*) observations of a HotDOG at z=4.6 (W2246-0526 ) revealed for the first time a dusty outflow driven by a heavily obscured intrinsically luminous quasar (Vayner et al. 2025b, submitted). The detection comes from the discovery of the 3.3 micron aromatic and 3.4 micron aliphatic hydrocarbon features in absorption that are blueshifted by 5250 km/s from the system redshift using the medium resolution spectroscopy mode of MIRI (Figure 1). The velocities are consistent with the scenario where the dusty outflow was driven through radiation pressure from reprocessed infrared photons near the dust sublimation radius. Surprisingly, the relative absorption strength and widths of the aromatic and aliphatic features in W2246-0526 are remarkably similar to what is seen in the Milky Way towards the galactic center Quintuplet cluster and the blue supergiant Cyg OB2-12 (Chiar et al. 2013). This potentially indicates that the outflowing dust in W2246-0526 is similar to Galactic diffuse ISM. However, with JWST, we can only detect absorption from the shortest wavelength aromatic and aliphatic hydrocarbon features, as all others either fall in bad sensitivity windows of MIRI or are completely redshifted out of the MIRI spectral range. To better grasp the composition of the dust in the outflow, we need PRIMA observations that will be able to probe additional aromatic (rest-frame 6.25, 7.7, 11.3 micron), aliphatic (6.85 micron), and olefinic (6.19 micron) features for the first time at large cosmological distances (see Figure 1). Furthermore, PRIMA will enable observations of silicate dust features, which will address the critical question of dust composition in galactic outflows.

The spectral coverage and sensitivity of FIRESS make it the ideal instrument to study these features at cosmic noon. The spectral resolution is sufficient to detect blueshifted absorption features at velocities > 500 km/s, which are expected for radiation pressure-driven outflows on dust grains in very luminous (>$10^{47}$erg/s) obscured quasars.

PRIMA, studying a large sample of obscured quasars at cosmic noon, will provide a better understanding of whether radiation pressure plays a role in the acceleration process of galactic winds. Measurements of the velocity and column density will place constraints on the driving radius of the outflow and, when compared to the luminosity of the quasar, will provide constraints on the coupling efficiency between radiation and dust in extremely obscured regions of galaxies. Observation of different hydrocarbon dust molecules will provide information about the outflowing dust composition at early epochs.

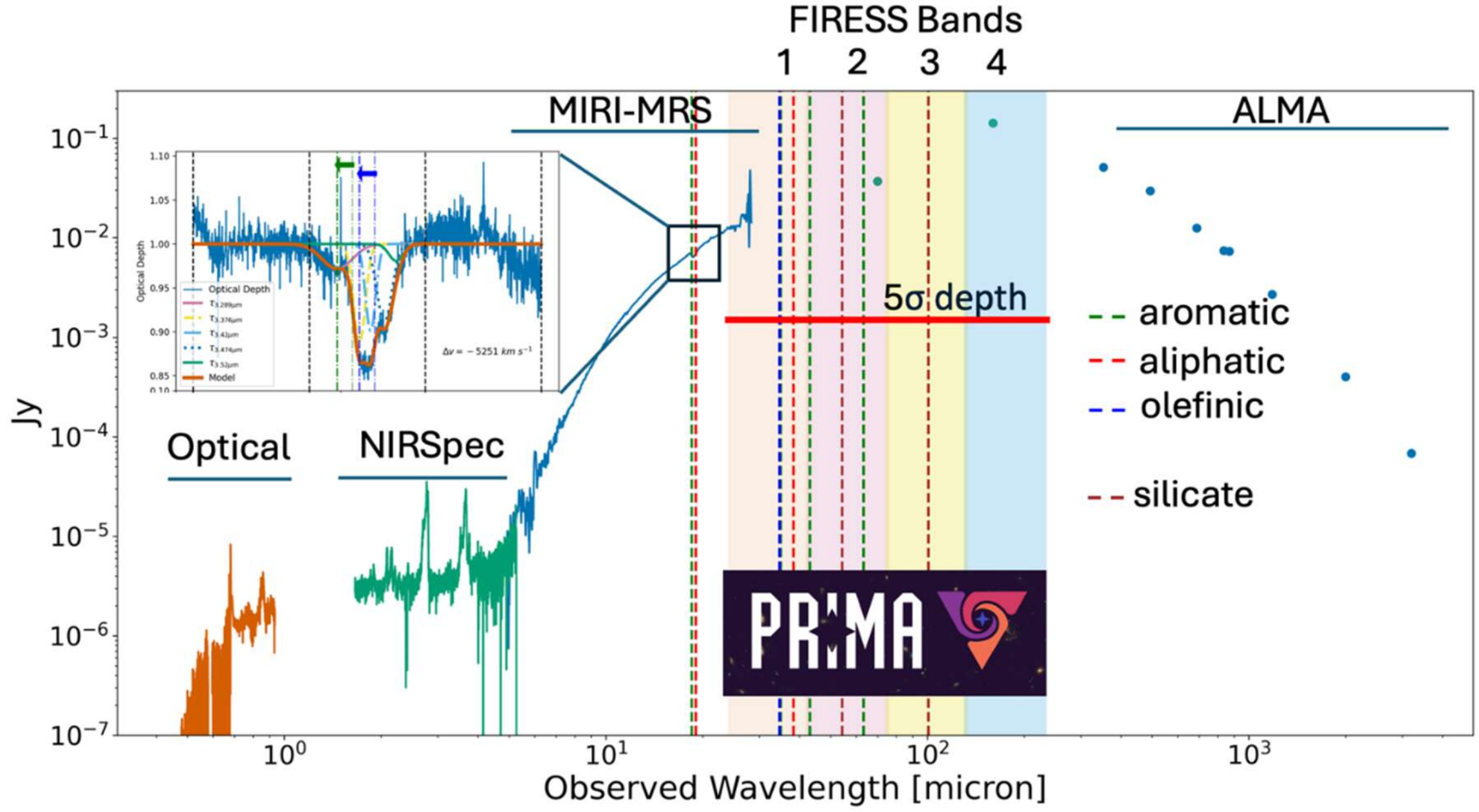

**Figure 1.** Broad-band spectral energy distribution of the luminous obscured quasar W2246-0526 at z=4.601 from observed optical to the submm. The blue spectrum shows the JWST MIRI MRS observations (Vayner et al. 2025b, submitted) with a zoomed-in insert of the normalized spectrum along with a fit to the hydrocarbon absorption features that are blueshifted from the systemic redshift by 5251 km/s, indicating a dusty outflow along our line of sight. The FIRESS low-resolution bands are highlighted in colors, with the vertical dashed lines showing hydrocarbon and silicate features that are redshifted into the different bands. The red horizontal line shows the 5σ depth reached in 0.58 hours necessary to detect a 5% absorption depth against the continuum at high significance for kinematic and dust composition analysis.

## Instruments and Modes Used

FIRESS low res mode, 1 pointing per source

## Approximate Integration Time

According to the PRIMA ETC, assuming a continuum flux density of 30 mJy at 30 micron and an absorption depth of 5% for a hydrocarbon feature requires an integration time of 0.58 hours to detect the absorption at an SNR of 5 with a spectral resolution of 500 km/s. This calculation assumes a spectral pixel size of approximately 500 km/s using the pointed low-resolution mode of FIRESS. To obtain the total wavelength coverage necessary to capture all dust features in the infrared requires a total integration time of 1.16 hours per source.

## Special Capabilities Needed

None

## Synergies with Other Facilities

JWST MIRI observations from 4.9 – 25 microns will be necessary for detecting the shortest wavelength aromatic and aliphatic hydrocarbon features for measuring the dust composition of the outflows. ALMA observations of the outflows in smaller molecules ($H_2O$, OH, CO) will be necessary for understanding how molecules can form in situ in fast-moving winds. Particularly, the wide-band upgrade to the ALMA receivers will be critical for this science case, as it will enable broader wavelength coverage to capture large velocity offsets that are currently challenging to obtain.

## Description of Observations

Spectroscopy of the infrared continuum will be critical to achieving the science goals of this program. Observations will be conducted with the low-resolution spectral mode of FIRESS. Objects will be acquired and placed in the middle of the slit, nodding the source in the slit with a total integration time of 0.58 hours per channel configuration, for a total on-source integration time of 1.16 hours. The sample will consist of 50 extremely red quasars and 50 HotDOGs to obtain critical statistics on the most obscured quasar sample at cosmic noon that hosts the fastest known outflows. The 100 sources will span a range of nuclear obscuration, quasar bolometric luminosity, and dust covering fraction to gauge the dependence of these parameters on the efficacy of launching winds through radiation pressure on dust. The total science time required to complete the program is 116 hours.

# 53. PRIMA-NEXUS: One Field for both Galaxy Evolution and Time-Domain Study

Xiurui Zhao (UIUC), Yue Shen (UIUC), Ming-Yang Zhuang (UIUC), Junyao Li (UIUC)

The North ecliptic pole EXtragalactic Unified Survey (NEXUS) is a JWST Multi-Cycle (Cycles 3–5; 368 primary hours, PI: Y. Shen) GO Treasury spectroscopic and imaging survey centered around the North Ecliptic Pole (NEP). Over three years, the program will conduct a uniform spectroscopic and photometric survey covering 400 arcmin$^2$. Given the sensitivity of JWST/NIRCam WFSS spectroscopy, all galaxies with mAB < 22.5 in F444W will have spectroscopic information thus reliable spec-z and identification across the entire 400 arcmin$^2$ field. The NEXUS field also overlaps with the Ultra-Deep Field of Euclid, offering an unprecedented optical-to-MIR spectroscopic coverage.

We propose a deep, multi-year PRIMA multi-band imaging survey of the NEXUS field, which will enable us to: 1) Achieve the first spectroscopic redshift- and classification-complete deep mid-to-far-IR contiguous survey. 2) Probe the evolution of galaxies in a large, luminosity-limited, and least-biased sample. 3) Constrain the IR luminosity functions of galaxy out to Cosmic noon. 4) Uncover the obscured AGN population and perform a least-biased census of AGN. 5) Discover and characterize a large sample of mid-IR transients. 6) Constraining the distribution of dust with different temperatures in the AGN torus by monitoring tens of AGN in five years.

## Science Justification

### Introduction

The NEXUS survey [Shen+24] aims to acquire deep (observed frame) IR spectra and imaging of all galaxies in a contiguous extragalactic field for the first time. The survey, covering about 400 arcmin$^2$, will perform NIRCam/WFSS 2.4–5µm grism spectroscopy every year for three cycles. Given the sensitivity of JWST/NIRCam slitless spectroscopy, all galaxies at mAB < 22.5 (F444W) will have reliable spectroscopic redshifts in the entire NEXUS field. The combination of uniform and broad spectroscopy and cadenced observations of NEXUS is unique among all JWST treasury programs. In addition, the NEXUS field is located in the Euclid self-calibration field, featuring the deepest 0.9–2 µm Euclid spectroscopy among all Euclid deep field [Euclid2024]. Therefore, each mAB < 22.5 galaxy in the NEXUS field will have unprecedented broadband spectroscopic coverage in 0.9µm–5µm. Furthermore, NEXUS features multiband optical to mid-IR imaging coverage by Euclid VIS/NISP and JWST NIRCam/parallel MIRI, enabling deep photometry covering 0.5–15 µm down to mAB < 28 in 400 arcmin$^2$ to produce reliable spectral energy distribution (SED) of the sources. Continuum sensitivity and wavelength coverage of JWST and Euclid are plotted in Fig. 1. These unique features make NEXUS the newest and most prominent field for tracing AGN evolution and performing time-domain studies. Mid-to-far IR observations are essential for directly probing the energy transfer in galaxies, measuring the dust surrounding SMBHs in AGN, and uncover heavily obscured AGN. PRIMA will be the best IR telescope for these objectives thanks to its exceptional band coverage, spatial resolution, sensitivity, and survey efficiency which are all crucial for detecting and characterizing a large number of galaxies in the NEXUS field and cadenced monitoring of the IR transient in the field.

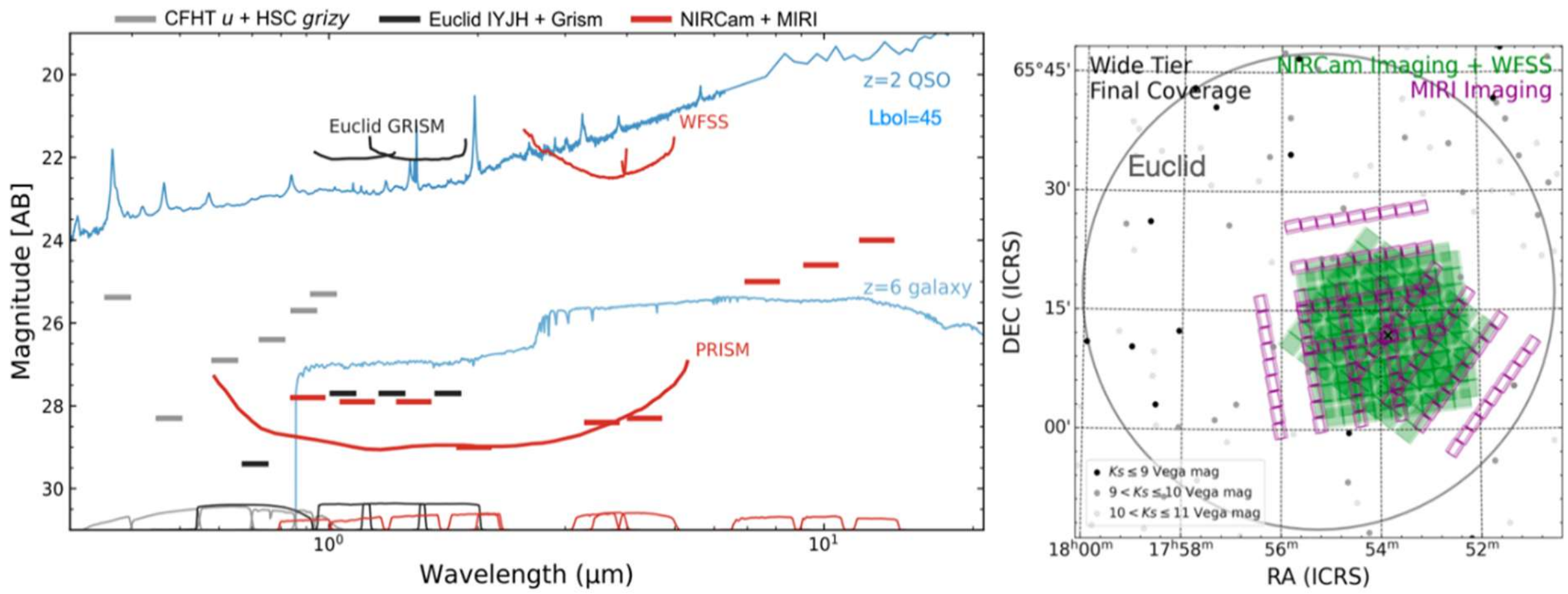

**Figure 1.** Left: Continuum sensitivity and wavelength coverage of JWST, Euclid, and ground-based telescopes (adapted from [1]). Right: JWST-NEXUS (green) layouts with Euclid UDF (gray). One *Swift*-XRT pointing (24' x 24') can cover the entire JWST-NEXUS 400 arcmin$^2$ field.

## Immediate Objectives

### 1) First Spec-z Complete Mid-IR Contiguous Survey

Various mid-IR extragalactic surveys have been conducted with different depths and area-coverages [Rieke+04, Fadda+06, Sanders+07, Davis+07, Oliver+12]. However, the lack of reliable spec-z and classifications of all mid-IR detected sources prevented these works from achieving a complete census of the mid-IR sources, thus achieving the aforementioned scientific objectives. Indeed, 50%–90% of the mid-IR detections do not have spec-z in the COSMOS [Ilbert+09], EGS [Barmby+08], SWIRE [Rowan-Robinson+08], and SERVS [Mauduit+12] fields, not to mention reliable classifications and blackhole mass measurements for AGNs. The JWST-NEXUS survey, thanks to its uniform spectroscopic survey down to deep IR fluxes, ensures that all galaxies with mAB < 22.5 at F444W will have reliable spec-z and classification in a large contiguous area. The PRIMA-NEXUS survey is designed to reach a mid-IR flux limit (∼40 µJy at 25 µm) such that the vast majority (>90%) of mid-IR sources will have counterparts with mAB < 22.5 (by comparing MIRI and NIRCam mag in JWST-JADES), and thus complete spectroscopy will already be in hand for this sample. We anticipate that the planned PRIMA-NEXUS survey will over 4,000 galaxies in the NEXUS field [Stone+24]. This ensures spec-z for all mid-IR detected sources without introducing biases against dust-rich galaxies, making PRIMA-NEXUS survey the first spec-z complete, contiguous, deep mid-IR survey. We emphasize that only with mid and far-IR data, like what PRIMA could provide, can one best constrain the star formation rate of a magnitude-complete galaxy sample with spec-z.

**2) Least Biased Galaxies Sample**

Mid-IR is able detect dust-rich galaxies, which might be difficult to discover in optical, e.g., Dust-Obscured Galaxies (DOGs), Ultra-luminous Infrared Galaxies (ULIRGs), Submillimeter Galaxies (SMGs), and obscured AGN. Therefore, Mid-IR surveys are more capable to select a more complete sample of galaxies thus measuring a least biased galaxy evolution and SMBH growth history. More intriguing, the deep optical to NIR spectroscopic and imaging data would best constrain the morphology and stellar population of galaxies, allowing us to best characterize the galaxy evolution.

**3) Uncover the Obscured AGN Population**

Obscured AGN represents a fundamental phase of SMBH growth, during which most of the BH mass is accreted [Hopkins+08]. Obscured AGN are thought to constitute a significant fraction (70%–90%) in the local Universe and the fraction is believed to increase at higher redshifts [Buchner+15, Ananna+19]. However, obscured AGN are difficult to discover in large populations due to significant obscuration and the lack of deep IR surveys. Moreover, in most previous contiguous surveys based exclusively on optical spectroscopy, type 2 AGNs with narrow emission lines at $z \gtrsim 0.5$ were often misclassified as star-forming galaxies. Therefore, the statistics and evolution across cosmic time of obscured AGN are still highly uncertain. The JWST+Euclid 0.9–5 µm spectroscopy allows reliable classification of the mid-IR sources using rest-frame optical and IR emission lines up to $z = 6$. Recent works [Lyu+24] found that broadband (including mid-to-far IR) SED fitting technique can select the most obscured AGN population, the majority of which were not recognized in deep X-ray observations. In addition, we emphasize that the future ngVLA and SKA observations will be crucial to test the effectiveness of radio diagnostics, such as radio-excess selection techniques [Bonzini+13, Smolcic+17] and radio spectral slope measurement, to unveil the most obscured AGN populations.

**4) Discover and characterize a large sample of mid-IR transients**

Time-domain astronomy is highlighted as one of the high-priority science themes by the 2020 Decadal Survey [National+21]. The NEXUS field is located in the continuous viewing zone (CVZ) of space observatories in sun-earth L2, from telescopes including JWST, Euclid [Euclid+24], WISE [Wright+10], SPHEREx [Doré+18], and The NEO Surveyor [Mainzer+23], allowing for year-long monitoring by PRIMA. Euclid will survey the field with a less than monthly cadence, ultimately making it the deepest Euclid field across the entire sky (at least 165 times more exposure than the typical wide-survey). From the ground, the Wide Field Survey Telescope [WFST, Wang+23] and the Young Supernova Experiment on the Pan-STARRS telescope [YSE, Jones+21] are both photometrically monitoring the NEXUS field in the optical with few-day cadences.

The variability of astronomical objects in mid-to-far IR is still a parameter space that is less probed. A multi-year, cadenced monitoring of a contagious field with PRIMA allows us to probe this least probed parameter space, including TDEs, AGN transients, and SNe. Any discovered mid-IR transient will be novel to the field. The event rates of different types of transients and variables will be compared with those derived in different wavelength (e.g., optical and NIR) to probe the entire SED thus the physical process of these transients and variables.

**5) Constrain dust distribution in AGN torus**

Reverberation mapping between the optical and NIR bands of AGN has been demonstrated to effectively measure the host dust distribution in the AGN torus [Yang+20, Li+23, Stone+24]. In the PRIMA-NEXUS, together with UVEX [KulKarni+21] and ground-based optical telescopes, we will monitor about 100 AGN in each epoch, allowing us to measure the medium-temperature and cold dust distribution of their tori after five-years monitoring [Treister+06]. More importantly, the SPHEREx and NEO Surveyor telescopes will monitor the NEXUS field in NIR in the following years and measure the hot dust distribution in the tori, so that we will obtain the distribution of dust with different temperature of a sizable sample of AGN torus for the first time.

## Instruments and Modes Used

PRIMAGER mapping in both bands. 400 arcmin$^2$ field, to be observed during 40 epochs over 5 years

## Approximate Integration Time

PRIMAger: in total 1000 hrs (25 h for each epoch)

## Special Capabilities Needed

None

## Synergies with Other Facilities

JWST, Euclid, SKA, ngVLA, SPHEREx, UVEX, NEO Surveyor

## Description of Observations

We plan to survey the entire 400 arcmin$^2$ of the NEXUS field in a total of 1000 hrs. The survey will be conducted in five years to have a long-based line monitoring of the AGN to measure their torus dust distribution. We plan to monitor the field 25 hrs each time with intervals separated by one week, one month, two months, allowing a high cadenced monitoring of the IR transient of the field sampling different time scales. Therefore, the monitoring includes 4 epochs in each half-year and a total of 40 epochs in 5-year. Each epoch of observation reaches 250 μJy at 25 μm in each epoch and 40 μJy after 5-year, which is still well above the confusion noise limit.

# Exoplanets, Astrobiology, and the Solar System

# 54. Measuring the Size Distribution of Inner and Outer Oort Cloud Comets

Bryce Bolin (Eureka Scientific, Oakland, CA 94602), (Affiliations): Dariusz C. Lis (Jet Propulsion Laboratory, California Institute of Technology, Pasadena, CA 91109, USA), Arielle Moullet (National Radio Astronomy Observatory, Charlottesville, VA 22903, USA)

The Oort cloud is a product of evolutionary processes in the solar system that occurred soon after its formation. Extending from 1,000 au to 500,000 au, the Oort cloud consists of ~100 billion km-scale and larger comets isotropically distributed in a spherical shell around the Sun. The Oort cloud was populated from the solar system's original debris disk during the giant planet instability phase of the solar system. Thus, it is a record of the dynamical and collision processes that were underway in the primordial disk. While the Oort cloud's inhabitants were dispersed from the inner Solar System billions of years ago, occasional passing stars cause its members to be perturbed, sending them on orbital trajectories that take them into the inner Solar System. Comets from the outer portion of the Oort cloud, >20,000 au from the Sun, can reach perihelion distances of <3 au, while comets form the inner Oort cloud, located 1,000 au to 20,000 au from the Sun, typically only get as close as ~15 au. Previously, it was difficult to observe and measure comets beyond 15 au from the Sun, making it impossible to study the properties of comets from the inner Oort cloud. However, with *PRIMA*/FIRESS, it will be possible to directly observe comets as far as 25 au, making it possible to observe and characterize inner Oort cloud comets. We propose to observe between 50 and 100 Oort cloud comets using FIRESS at wavelengths ranging from 30 microns to 70 microns to determine their radiometrically derived diameters and size distribution. We will divide our targets roughly evenly between inner and outer Oort cloud comets so that we can compare the size distribution of inner and outer Oort cloud comets, thus testing planet formation and evolutionary hypotheses describing the formation of comets and the Oort cloud's regions. Our results will have lasting value, as they will be used for the future development of planetary models that describe the formation of debris disks and will serve as a context for future searches of exoplanetary Oort clouds around other stars.

## Science Justification

### Broader Context

The Oort cloud is a reservoir of billions of comets extending from 1,000 au to 500,000 au from the Earth (Dones et al. 2015). The majority of the Oort cloud was generated when a ~20 Earth-mass disk of planetesimals was dispersed by the migration of the gas giants, sending the majority of its members into interstellar space, and a few Earth masses into what is now called the Oort cloud (Vokrouhlický et al., 2019). The Oort cloud has also experienced encounters by passing stars and galactic tides, causing its structure to become isotropically distributed as a sphere around the Sun in its outermost portions (Fouchard et al. 2017). The Oort cloud is thus a record of the early solar system's evolutionary process, as well as a record of the sun's close stellar passages and their impact on the cometary bodies in the Oort cloud. We propose to observe Oort cloud comets with *PRIMA*/FIRESS and determine the size distribution of comets coming from the Oort cloud. By studying the size distribution of comets that originate from different portions of the Oort cloud, we can test formation models that describe the formation of the Oort cloud from the dispersal of comets from the original trans-Neptunian disk and their subsequent evolution caused by galactic tides and stellar flybys.

By studying Oort cloud comets and their size distribution, we will be addressing the question from the Astro 2020 decadal survey (National Academies of Sciences 2021): "What is the range of planetary system architectures, and is the configuration of the solar system common?" and the following questions from the Planetary Science Decadal Survey (National Academies of Sciences 2023): "Q1.1 What Were the Initial Conditions in the Solar System?". We will address these questions by studying the size distribution of Oort cloud comets as a test of collisional evolution models in the original trans-Neptunian disk (Bottke et al. 2023).

Our observations will also address the Planetary Science Decadal Survey question: "Q2.6 How Did the Orbital Structure of the Trans-Neptunian Belt, the Oort Cloud, and the Scattered Disk Originate, and How Did Gravitational Interactions in the Early Outer Solar System Lead to Scattering and Ejection?". Our observations of the Oort cloud size distribution will be used as a constraint on the different processes affecting the inner and outer portions of the Oort cloud (Panel A of Figure 1), which resulted from different effects caused by scattering and interactions with the galactic tide (Rickman et al. 2008).

## Science Question

The Oort Cloud is divided into an inner region for objects with semi-major axis, a, between 1,000 au and 10,000 au, and an outer region for objects with a > 10,000 au as seen in panel A of Figure 1 (Vokrouhlický et al. 2019). The inclination, i, distribution of Oort cloud objects also differs between the inner Oort cloud, which is concentrated at ~40 degrees, whereas the outer Oort cloud is more isotropically distributed. The orbital differences between inner Oort cloud and outer Oort cloud objects also make a difference in the perihelion, q, distances they can reach. Outer Oort cloud objects can get a q of <1 au, and evenly distributed out to >20 au, whereas inner Oort cloud objects tend to have a q > 15 au, as seen in panel B of Figure 1. So far, the majority of comets observed coming from the Oort cloud are from the outer Oort cloud because it is difficult to detect comets beyond 15 au, only which a few are known with q beyond this distance (Vokrouhlický et al. 2019).

Current solar system evolutionary models predict that the Oort cloud was created when the original trans-Neptunian disk, with a mass of ~20 Earth masses, was scattered during the migration of the giant planets (Morbidelli & Nesvorný 2019). The majority of the mass of the scattered trans-Neptunian disk, which was not ejected into interstellar space, ultimately ended up in the Oort cloud, accounting for ~5% of it (Brasser & Morbidelli, 2013). Therefore, the Oort cloud represents the largest remnant of the original debris disk.

Studies of collisional models based on impact constraints from around the solar system as well as the observed size distribution of trans-Neptunian objects, indicate that the original trans-Neptunian disk before its scattering may have had a sloped structure, with at least a couple of breaks at 1 km and 5 km. For objects with a diameter < 100 to 1 km, q~-1.2, objects with a diameter 1 km to 5 km have q~-2.5, and objects with D>5 km have q~-2 (Bottke et al. 2023). However, recent evidence from ground-based observations indicates that small Oort cloud comets with diameters < 2 km may be much fewer in number compared to larger Oort cloud objects (Boe et al. 2019). The question is, if the Oort cloud and current-day trans-Neptunian belt both originated from the primordial trans-Neptunian disk, why do their size distributions look so different?

The answer could lie in the fact that the ground-based data were affected by biases caused by the obscuration of cometary nuclei due to their activity and an observational selection effect that preferentially detects comets within 5 au (Boe et al. 2021). This could have the effect of sampling only comets from the outer Oort cloud, which can have q well inside 15 au, while inner Oort cloud objects preferentially have q > 15 au (panel B of Figure 1, Vokrouhlický et al. 2019). Thus, the current observations of Oort cloud comets and size determination neglect inner Oort cloud comets. If the collisional evolution of outer Oort cloud comets differed from inner Oort cloud comets, we would only see it in the current Oort cloud comet size data, since current-day Oort cloud comet observations are biased towards outer Oort cloud comets. We will seek to address this scientific question by measuring the size distribution of both inner and outer Oort cloud comets.

## Need for PRIMA

Far-infrared observations of Oort cloud comets provide a less dust-obscured view of the comets than shorter wavelength observations (e.g., Yang et al. 2021). Observing comets in the far-infrared provides radiometrically-derived diameters (Harris, breaking albedo and diameter ambiguity, which affects visible light observations of comets (e.g. Lellouch et al. 2022).

## Interpretation Methods

We will observe known inner and outer Oort cloud comets with known orbits with FIRESS to have a sample as unbiased as possible. Inner and outer Oort cloud comets are distinguished from their orbital characteristics, with inner Oort cloud comets having semi-major axes between 1,000 and 10,000 au, and q>15, and outer Oort cloud comets having semi-major axes >10,000 au. We will use the Near-Earth Asteroid Thermal model to determine the sizes of our observed inner and outer Oort cloud comets based on the far-infrared data we obtain with FIRESS.

## Link to Testable Hypotheses

The size distribution slope of objects in the original trans-Neptunian disk can vary by as little as ~0.1 to as large as ~1.0 (Bottke et al. 2023). To test the size distribution of inner and outer Oort cloud objects with a precision that can distinguish between different slopes resulting from collisional evolution processes in the trans-Neptunian disk, we will need ~10% precision in its measurement. We will therefore need to observe ~25 objects in both the inner and outer Oort cloud categories.

We will compare the size distribution slope from our observed samples of inner and outer Cloud objects with the expected size distributions modeled from collisional simulations of the original trans-Neptunian disk (Bottke et al. 2023) as well as models that describe the scattering of the trans-Neptunian disk into the Oort cloud (Brasser & Morbidelli 2013). Our observations will be used to inform the development of future planetary formation models describing the formation of the Oort cloud, as well as indicate the extent of our solar system's Oort cloud to serve as a reference for the search of extrasolar Oort clouds (Baxter et al. 2018).

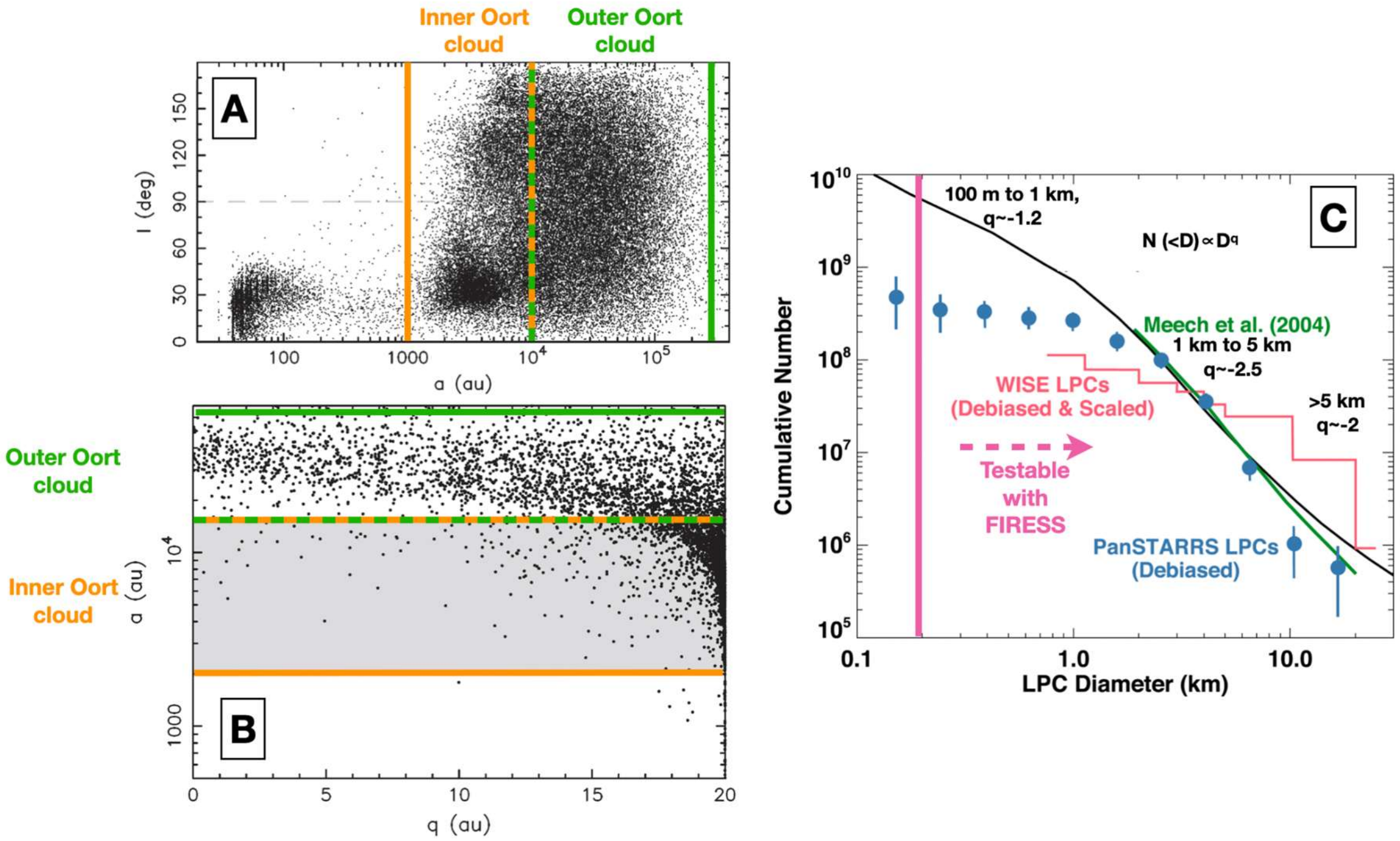

Figure 1. Observing inner and outer Oort cloud comets with *PRIMA* will provide a test of early solar system evolutionary processes for planetesimals. Panel (A): adapted from Vokrouhlický et al. 2019, the a vs i distribution of ~50,000 synthetic Oort cloud comets. The inner Oort cloud is outlined between the orange and green-orange dashed lines at 1,000 to 10,000 au. The inclination distribution of inner Oort cloud objects is skewed towards ~40 degrees. Outer Oort cloud objects are located beyond 10,000 au and have a much more uniformly distributed inclination distribution between 0 and 180 degrees. Panel (B): Adapted from Vokrouhlický et al. 2019, the q vs a distribution of Oort cloud comets. The inner and outer Oort cloud regions are indicated. Inner Oort cloud comets predominantly have q > 15 au, while outer Oort cloud comets have q more equally distributed between 0 and 20 au. Panel (C): adapted from Bottke et al. 2023, shows the modeled size distribution of the primordial trans-Neptunian disk, which has a couple of breaks at 1 km and 5 km. For objects with a diameter < 100 to 1 km, q~-1.2, objects with a diameter 1 km to 5 km have q~-2.5, and objects with D>5 km have q~-2. The size distribution of returning and outer Oort cloud comets from visible light observations is shown in blue, which has a q~-1 for objects smaller than 3 km, and a q~-2 for objects larger than 3 km.

## Instruments and Modes Used:

FIRESS point source low-resolution observations. 2 spectral setups per target.

## Approximate Integration Time

Our program requires obtaining 5-sigma detections of our comet targets between 28 microns and 64 microns. We will design our observations around an assumed single pointing time of 3.2

h. Using the FIRESS exposure time calculator in low-resolution point source mode, we can detect our targets in 3.2 h. This assumes using FIRESS's channel 1 array for comets located between ~5-11 au from the Sun, where the peak in the blackbody curve is located at the 24-43 microns covered by the array.

This calculation is done for a 200 m radius comet located at 5 au from the Sun which, using the near-Earth asteroid thermal model, has a peak black body flux at 28 microns of ~27 muJy, assuming a bond albedo of 0.015, typical for comets, an emissivity of 0.9, and a beaming parameter of 1.0. Using the FIRESS exposure time calculator, we can obtain a 5-sigma detection of a 27 µJy source at 28 microns in 3.2 h. For a 1 km radius comet located at 11 au, the peak black body flux is at 42 microns, with a flux of ~41 µJy, assuming the same parameters for bond albedo, emissivity, and beaming parameter. We can obtain a 5-sigma detection of this source in 3.2 h.

For comets located further than ~11 au, we will use the channel 2 array because their peak blackbody flux shifts to longer wavelengths than 45 microns, which is covered by chanel 2 array's coverage of 42 microns to 76 microns. Assuming a bond albedo of 0.015, an emissivity of 0.9, and a beaming parameter of 1.0, a 2km radius comet will have a peak black body flux at 49 microns of 47.8 muJy. Using the FIRESS exposure time calculator, we can obtain a 5-sigma detection in 3.2 h of observing time. Assuming the same values for bond albedo, emissivity, and beaming parameter, a 6.5 km radius comet located at 25 au from the Sun will have a peak black body flux of 73.3 muJy at 64 microns. Using the FIRESS exposure time calculator, we can obtain a 5-sigma detection of this 73.3 muJy source at 64 microns in 3.2 h.

Our targets are faint, so we must use shadow observations, repeating our observing sequences after the comet has moved out of the field to provide an accurate background subtraction. Therefore, the combined time for each of our targets will be 3.2 h for the primary observations and 3.2 h for the shadow observations, for a total of 6.4 h for each of our science targets.

## Special Capabilities Needed

Our targets will have sky-plane motions of ~10 arcsec/h. Therefore, given our 3.2 h observing blocks, we will need to use non-sideral guiding to avoid trailing losses at the 7.6 arcsec pixel scales of the FIRESS channel 1 and channel 2 arrays.

As mentioned above, we will have to use two observations for each of our targets: the primary observation of the comet and a second shadow observation of the same field as the primary comet observation, using the same exposure time and tracking rate.

## Synergies with Other Facilities

Our observations will synergize with visible and near-infrared surveys such as Rubin Observatory and the Nancy Grace Roman Space Telescope. We will combine the radiometrically-derived diameters from our FIRESS observations with the visible and near-infrared observations to determine its color and albedo (e.g., Lellouch et al. 2022, Hui et al. 2022).

## Description of Observations

Our program will target dynamically new comets from the inner and outer Oort cloud between 5 and 25 au from the Sun. We will use the FIRESS channel 1 and channel 2 arrays for our observations, which have a 4.5 arcmin field of view. We will use FIRESS's channel 1 and channel 2 arrays to observe our comet targets directly, obtaining ~5 sigma detections of >200 m to 6 km radius comets located 5 au to 25 au from the Sun. We will target specific comets with well-known orbits, using the orbital information to indicate their dynamical first passage to the planetary region of the solar system (Królikowska & Dybczyński 2017).

We will use the channel 1 array to target comets located 5 au to 11 au from the Sun where their peak black body temperature is located in the 24–43 micron range of sensitivity of the array. We will use the channel array, which covers 42–76 microns, to observe comets located 11 au to 25 au from the sun which have peak black body fluxes located at 64 microns. With the comet detections enabled by FIRESS at 24-76 microns, it will be possible to calculate the comet's diameter radiometrically using the near-Earth asteroid thermal model (Harris & Lagerros 2002).

Our targets are fainter and likely to be confused with background sources. To compensate for this, we will need to use shadow observations, which repeat the observation sequence at the same position in the sky as our targets once they have moved out of the field of view. The difference between the primary and shadow observations can remove confusing background sources, leaving the comet's source in the residual image (e.g., McKay et al. 2019).

## Acknowledgment

This research was carried out at the Jet Propulsion Laboratory, California Institute of Technology, under a contract with the National Aeronautics and Space Administration (80NM0018D0004).

# 55. Time-Critical Observations of Interstellar Objects and Impacting Asteroids with PRIMAger

Bryce Bolin (Eureka Scientific, Oakland, CA 94602), Daniel Stern (Jet Propulsion Laboratory, California Institute of Technology, Pasadena, CA 91109, USA), Mansai M. Kasliwal (Division of Physics, Mathematics, and Astronomy, California Institute of Technology, Pasadena, CA 91125, USA)

Interstellar objects (ISOs) are planetary formation byproducts ejected from extrasolar star systems. Only two are known, 1I/'Oumuamua and 2I/Borisov. Constraints are limited on ISO physical properties based on a sample size of two. Characterizing additional ISOs is desirable to constrain their physical properties and origins. The discovery of 1I and 2I occurred with little warning time, allowing for a limited window of opportunity for their study. Based on the historical discovery rate, it is expected that several ISOs will be discovered per year by surveys such as Rubin Observatory, Pan-STARRS, and ZTF, among others, in the 2030s. We propose using FIRESS to target recently discovered ISOs and determine their size distribution based on detections in the far-infrared. Our measurements and results will have legacy value in testing theories of planetesimal formation.

Earth-impacting near-Earth objects (NEOs) are a class of asteroids that possess a non-negligible chance of hitting the Earth. The destructive potential and available mitigation strategies are directly related to the size of an impacting NEO. It is expected that several ~100 m diameter NEOs with a non-negligible chance of hitting the Earth will be discovered by all-sky surveys in the 2030s. To determine their size, we propose time-critical far-infrared observations of recently discovered potentially Earth-impacting NEOs. Our observations will be able to target NEOs beyond the range of ground-based radar facilities. PRIMA will be one of the only facilities to target and observe potentially impacting NEOs in the far-infrared. Therefore, PRIMA will act as a crucial Planetary Defense asset for assessing potential NEO threats and for planning their mitigation.

## Science Justification

### Broader Context

Macroscopic interstellar objects (ISOs) encompass rogue planets (Mroz et al., 2020) and planetesimals ranging from 100 m to 1 km in scale (Moro-Martín, 2022). The latter group of ISOs is the remnant of the planet formation process, ejected from extrasolar star systems (Bannister et al., 2019; Zhang & Lin, 2020). So far, only two interstellar objects have been discovered: 1I/`Oumuamua and 2I/Borisov (Jewitt 2024). 1I was discovered by the NASA-funded Panoramic Survey Telescope and Rapid Response System (Pan-STARRS) survey in serendipitous observations during a search for hazardous asteroids (Bolin et al. 2018). 2I/Borisov was discovered by an amateur astronomer using a homemade telescope. However, serendipitous prediscovery detections of 2I were found in data taken by the Zwicky Transient Facility (Bolin et al., 2020). Recently, the Asteroid Terrestrial Last Alert System (ATLAS) discovered a third interstellar object, 3I/ATLAS (Bolin et al., 2025b).

Constraints on the physical properties of ISOs are limited to a sample size of two, with 1I and 2I having vastly different properties. 1I showed no signs of cometary activity but had a highly elongated shape evident through significant amplitude lightcurve variations (Bolin et al., 2018), while 2I and 3I possessed a comet-like appearance (Bolin et al. 2020, Bolin et al. 2025b). Different hypotheses for the origin of ISOs have been proposed, which predict different appearances and physical properties of ISOs that are detectable with ground-based observations. Theoretical models predict that ISOs originating as fragments from the tidal disruption of a larger body or formed as fragments of hydrogen ice in a planetary nebula should show little activity and highly elongated shapes such as 1I (Ćuk, 2018). Models that predict ISOs escape their home systems as distant exo-Oort cloud bodies at the gravitational boundaries of their home systems should exhibit activity similar to that of a comet, due to the colder and gentler environment from which they escaped (Peñarrubia, 2023). It is expected that up to 10's of ISOs will be discovered in the 2030s by Rubin Observatory (Schwamb et al. 2023). Therefore, it is desirable to characterize the physical properties of additional ISOs to understand their origins as more ISOs are discovered in the near future.

Impacting asteroids are a class of near-Earth objects (NEOs) that have a non-negligible chance of impacting the Earth (Milani et al. 2000). Out of ∼1 million known asteroids, there are ∼20 asteroids that are 50–100 m or larger that have a $>10^{-6}$ chance of hitting the Earth within the next 100 years (Vaduvesci et al. 2020, National Academies of Sciences 2023). The destructive potential of an asteroid is directly related to its size (Binzel 2000). Observations by ground-based radar facilities can provide measurements of an asteroid's diameter (e.g., Taylor et al. 2019). However, direct measurement of an asteroid's diameter is impossible if it is farther than a few 10s of lunar distances (>0.03 au), such as was the case for potential impactor 2024 YR4 (Bolin et al. 2025a).

Visible light observations can estimate an asteroid's diameter; however, a degeneracy between its diameter and its albedo makes determining its diameter highly uncertain to a factor of two or more (Harris & Lagerros 2002). The significant uncertainty in the estimated diameter of impacting asteroids, as determined by visible light alone, makes planning any potential mitigation procedures problematic because they depend on precise knowledge of an asteroid's size (Michel 2013). We propose using observations in the thermal infrared with FIRESS to measure the sizes of impacting asteroids from their thermal emission.

By studying interstellar objects and their size distribution, we will be addressing the question from the Astro 2020 decadal survey (National Academies of Sciences 2021):

"What is the range of planetary system architectures, and is the configuration of the solar system common?". By observing and determining the sizes of Earth-impacting asteroids, we will be addressing the planetary defense goals described in the Decadal Survey for Planetary Science and Astrobiology (National Academies of Sciences 2023)

### Science Questions

Current solar system formation models predict that more than 90% of the planetesimals accreted in the protoplanetary disk were ejected into interstellar space over the age of the solar system (Nesvorný et al., 2020). This implies that free-floating ISOs in the galaxy should be commonplace if planetesimal formation occurs in extrasolar systems, as it does in the solar system. Additionally, the detection of two known interstellar objects suggests that there are $10^{13}$ ISOs in the galaxy (Jewitt 2024). The first question this program seeks to answer is, if ISOs are evidence of macroscopic planetesimals around other stars, do the same collisional processes affecting planetesimals in the solar system also occur in extrasolar systems? Collisional evolution simulations of planetesimals suggest that planetesimals ~100 m in size should have a steep size distribution (Bottke et al. 2023) while observations of long-period comets suggest that small comets should have a shallow size distribution, implying they are fewer in number compared to larger, km-scale comets (Boe et al. 2019). FIRESS observations of ISOs will illuminate this question by accurately estimating the size distribution of ISOs.

The second question this program seeks to answer is, how much of a threat do Earth-impacting NEOs present to the Earth, if any are found? As described in the Decadal Survey for Planetary Science and Astrobiology (National Academies of Sciences 2023), it is expected that several NEOs with an impact probability of ~1% or greater will be found in the next 10 years as future all-sky surveys come online. Rapid response detection and estimation of their sizes will be necessary for assessing the destructive potential these NEOs may have, as well as providing critical physical information required for planning threat mitigation efforts (Rumpf et al., 2020).

### Need for PRIMA

Far-infrared observations of ISOs and NEOs provide accurate measurements of their radiometric flux due to thermal emission (e.g., Müeller et al. 2014). Observing ISOs and NEOs in the far-infrared provides radiometrically derived diameters, breaking the albedo and diameter ambiguity that affects visible light observations of minor bodies (Harris & Lagerros 2002). Additionally, as recent discoveries have shown (Bolin et al., 2018, 2020, 2025a, 2025b), both ISOs and impacting NEOs can have a narrow window of observability, lasting only weeks to months, necessitating their rapid follow-up with target-of-opportunity observations.

## Interpretation Methods

We will observe ISOs with ambiguous interstellar origins, eccentricities >>1, and velocity at infinity of >>0 km/s. We will observe NEOs with impact probabilities >1%. We will use the Near-Earth Asteroid Thermal model to determine the sizes of our observed NEOS and NEAs using the far-infrared data we obtain with FIRESS.

## Link to Testable Hypotheses

The size distribution of ISOs is poorly constrained, with only two known ISOs (Jewitt et al., 2020; Moro-Martín & Norman, 2022). We will constrain the size distribution of ISOs with size measurements of additional ISOs discovered in the 2030s. Our observations and measurements will test whether the size distribution of ISOs indicates a source population that is collisionally evolved in their formation environment or has undergone additional evolution in the galactic environment (Bannister et al. 2019).

To test the size distribution of ISOS at a level of precision that can distinguish between different slopes resulting from collisional and other evolutionary processes, we will need ~15–20% precision in its measurement. Therefore, we will need to observe ~5–10 ISOs to determine the ISO size distribution with enough precision to reach our science goals.

Additionally, we will observe an impacting NEO discovered in the 2030s. Our observations will provide a constraint on its size with 10% precision. This will enable an accurate assessment of its mass and other physical properties relevant to assessing impact threats (Chesley et al. 2003)

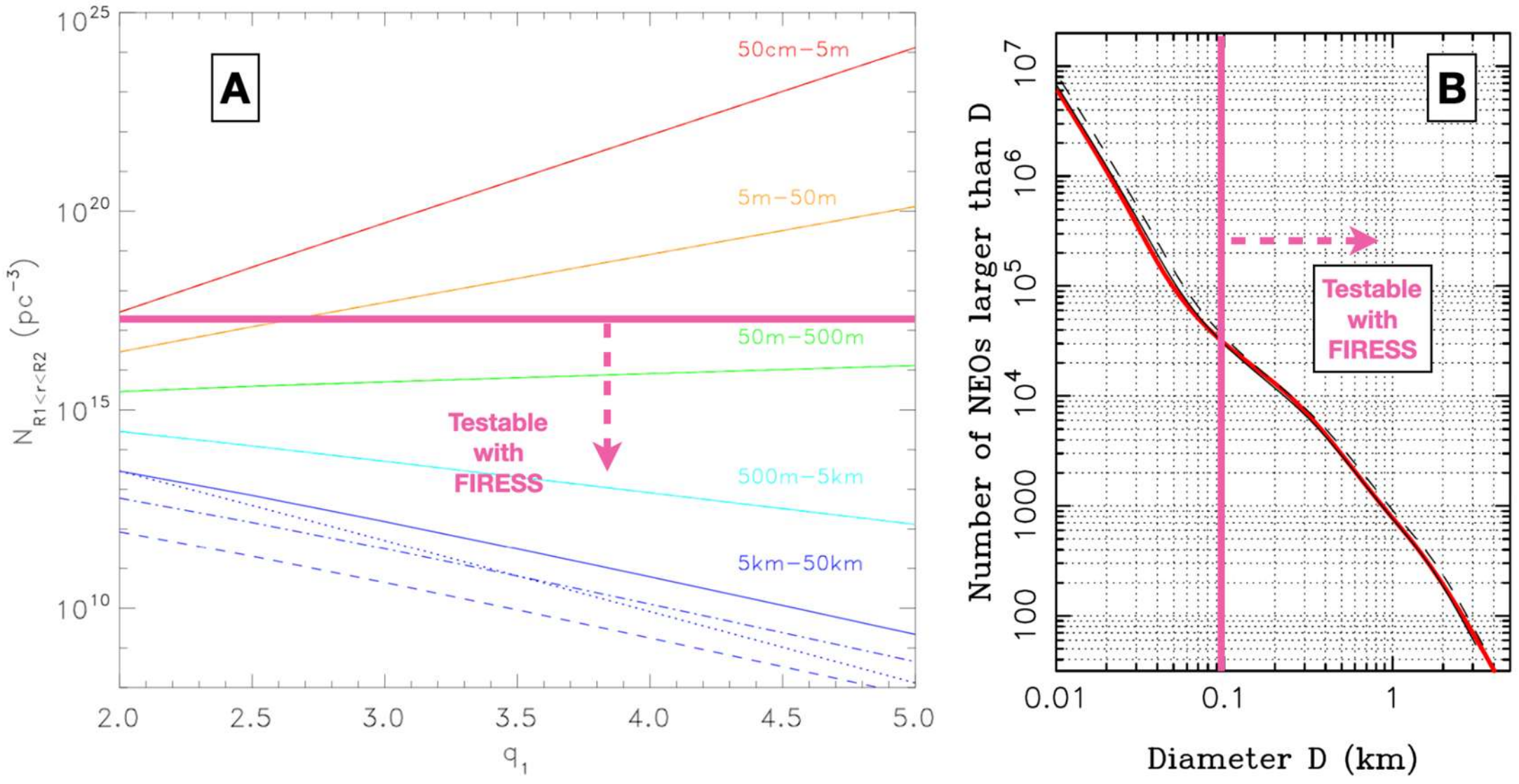

**Figure 1.** PRIMA/FIRESS is the ideal tool for measuring the sizes of interstellar objects and impacting asteroids. Panel (A): adapted from Moro-Martín & Norman 2022, ISO number density, N, vs power law index, q1. FIRESS will be able to detect interstellar objects larger than ~100 m in radius within 3 au and provide constraints on the power law of ISOs. Panel (B): Adapted from Nesvorný et al. 2024, the size distribution of NEOS from NEOMOD3 (red line). FIRESS will be able to detect NEOs as small as ~100 m in diameter within 2.3 au of the Sun.

## Instruments and Modes Used

FIRESS point source low-resolution observations.

## Approximate Integration Time

Our program requires obtaining 5-sigma detections of our ISO and NEO targets between 24 microns and 43 microns. We will design our observations around an assumed single pointing time of 3.2 h. Using the exposure time calculator for FIRESS in low resolution point source mode, we can detect our targets in 3.2 h. This assumes the use of FIRESS's channel 1 array to observe NEOs with radii of ~50 m and ISOs with radii of ~100 m, located within 2.3 au and 3 au from the Earth.

This calculation is done for a 100 m radius ISO located at 3 au from the Sun which, using the near-Earth asteroid thermal model, has a black body flux at 25 microns of ~28 muJy, the highest in the FIRESS channel 1 array, and assuming a bond albedo of 0.04, thought to be typical for ISOs (Morro-Martin et al. 2022), an emissivity of 0.9, and a beaming parameter of 1.0. Using the FIRESS exposure time calculator, we can achieve a 5-sigma detection of a 28 µJy source at 24 µm in 3.2 h. For a 50 m radius NEO located at 2.3 au from the Sun, the 24 micron flux is ~29 muJy, assuming a 0.06 bond albedo, typical for NEOs (Harris & Lagerros 2002). We can obtain a 5-sigma detection of this source in 3.2 h.

Our targets will be faint, and likely confused with background sources, so we will need to use shadow observations, repeating our observing sequences. Therefore, the combined time for each of our targets will be 3.2 h for the primary observations and 3.2 h for the shadow observations, or 6.4 h total.

## Special Capabilities Needed

Our targets have narrow windows of observability of only several weeks. Therefore, we will request that our observations be conducted in a time-critical target-of-opportunity mode as soon as possible when the PRIMA field of regard is visible.

Our targets will have sky-plane motions of ~100–200 arcsec/h. Therefore, given our 3.2 h observing blocks, we will need to use non-sidereal guiding to avoid trailing losses at the 7.6 arcsec pixel scale of the FIRESS channel 1 array.

As mentioned above, we will need to use two observations for each of our targets: the primary observation of the comet and a second shadow observation of the same field as the primary comet observation, both with the same exposure time and tracking rate, for a total of 6.4 h per target.

## Synergies with Other Facilities

Our observations will complement visible and near-infrared surveys, such as those conducted by the Rubin Observatory and the Nancy Grace Roman Space Telescope. We will combine the radiometrically derived diameters from our FIRESS observations of our targets with visible and near-infrared observations to determine their color and albedo (e.g., Harris & Lagerros 2002).

## Description of Observations

Our program will target ISOs and impact asteroids within 3 au of the Sun. We will use the FIRESS point-source mode for our observations, which has a field of view of 4.5 arcmin in width and provides coverage of wavelengths from 24 microns to 235 microns, encompassing the regions where ISOs and NEOs are brightest at 25 microns, within 3.0 au. We will use FIRESS to obtain ~5 sigma detections of ISOs and NEOs with 50–100 m radii located 2.3 au to 3.0 au from the Sun. We will target ISO and impacting NEOs with well-determined orbits, ensuring that their interstellar or Earth-crossing orbits, as well as their sky-plane position, are well-constrained (Bolin et al., 2018, 2025a, 2025b).

We will use FIRESS in low resolution point source mode to target interstellar objects and asteroids located 2 au to 3 au from the Sun. Their peak black body temperature is highest in the channel 1 range, located at wavelengths of 24–43 microns. We will use the near-Earth asteroid thermal model (Harris & Lagerros 2002) to determine the diameter of our targets from their thermal fluxes.

Due to the faintness of our targets we will need to use shadow observations, which repeat the observation sequence at the same position in the sky as our targets once they have moved out of the field of view. The difference between the primary and shadow observations can remove confusing background sources, leaving the comet's source in the residual image (e.g., McKay et al. 2019). Therefore, we will need 6.4 h to both science and shadow observations for each of our targets.

## Acknowledgement

The research was carried out at the Jet Propulsion Laboratory, California Institute of Technology, under a contract with the National Aeronautics and Space Administration (80NM0018D0004).

# 56. Exploratory Study of Bodies Within the Main Asteroid Belt

Helen Grant (INAF-IAPS), Dariusz C. Lis (Jet Propulsion Laboratory, California Institute of Technology), Arielle Moullet (National Radio Astronomical Observatory)

The origin of Earth's water and organics is a key question in Solar System science, likely linked to volatile delivery by asteroids and comets. While chondritic meteorites and their asteroidal parent bodies have been studied extensively, far-infrared (far-IR) spectral analyses remain limited. This study proposes a broad far-IR survey of Main Belt asteroids to characterize their silicate, water, and organic compositions. Using PRIMA FIRESS, it will be possible to collect and characterize high-resolution far-IR spectra of 270 asteroids previously detected by Herschel/PACS in under 60 hours. This is an exploratory study, where we aim to identify diagnostic features of silicates, sulfides, and organic ices, providing insights into asteroid formation, migration, and space weathering.

## Science Justification

The delivery of water, organics, and other volatiles to Earth is a fundamental question within Solar System science, with implications for a wide range of topics including Solar System evolution, protoplanetesimal formation, and astrobiology. Generally, since the Earth formed within the snowline and thus in a region too warm for water ice to be directly accreted it is believed that the majority of Earth’s water and other volatiles were delivered by bodies such as asteroids and comets. Chondritic meteorites, thought to originate from rocky main belt asteroids, have been extensively studied and characterized on Earth, and show evidence of notable water content as well as an abundance of organic compounds (e.g., Alexander et al., 2012; Broadley et al., 2022). Similarly, their asteroidal parent bodies have been well studied remotely in the visible, near-, and mid-infrared. However, there have been very few (if any) far-IR spectral analyses carried out on the parent bodies (or indeed the meteorites themselves) to link these two populations. As such, we are proposing a broad-scale exploratory study of a range of Main Belt asteroids to understand their silicate, water, and organic contents and compositions.

Carbonaceous chondrites (CCs), for example, are relatively water- (∼10 wt.%; Lee et al., 2023) and organic-rich (up to ~6 wt.% C; Scott & Krot, 2014), and are thought to originate from C-, B-, P-, and D-type asteroids, which reside in the outer main belt. Large scale visible studies of these asteroid types found hydrated mineral bands in ∼ 50% of asteroids, which is indicative of a high probability of measurable water in these types of bodies and provides a starting framework from which to select hydrated targets (Fornasier et al., 2014). Similarly, identification of water ice and organics on the surfaces of 65 Cybele and 24 Themis have been found in the NIR (Campins et al., 2010; Rivkin & Emery, 2010; Licandro et al., 2011). However, far-IR characterisations of these types of bodies have yet to be done.

CC parent bodies are thought to have formed in the outer SS beyond the snow line before being scattered inwards to the region of the asteroid belt as a result of giant planet formation and migration, in a process known as the Grand Tack model (e.g., Walsh et al., 2011; Scott et al., 2018). Recent analysis of material returned from the asteroid Bennu as part of the OSIRIS-ReX mission found evidence of briny, ammonia-rich fluids on the parent body similar to those predicted to still exist in the interiors of Ceres and Enceladus (Glavin et al., 2025; McCoy et al., 2025). This suggests that Bennu may indeed have formed in the outer Solar System before migrating inwards. Spectral analysis of a range of C- and B-type asteroids will help to understand whether the characteristics of Bennu are unique, or common among C+ type parent bodies.

By contrast, ordinary and enstatite chondrites (OCs and ECs respectively) originated in the inner Solar System and are thought to correlate to S-, M-, E, and V-type asteroids parent bodies (e.g., Gaffey & Gilbert, 1998; Yoshikawa et al., 2007). Studies of the least thermally metamorphosed OCs have bulk water contents up to 1 wt%, equating to matrix water contents similar to that of some CM, CR, and CI chondrites (e.g., Alexander et al., 2010; Vacher et al., 2020; Grant et al., 2023). While it is unlikely that surface water-ice exists in detectable amounts on these types of asteroid parents, it is possible that bands in the 20–100 µm range characteristic to phyllosilicates such as serpentine, saponite, and montmorillonite will be observable (Koike et al., 1982; Hofmeister and Bowey, 2006; Mutschke et al., 2008).

Asteroids are made up of a complex mixture of organic- and silicate-based phases, which can be investigated from an entirely new perspective thanks to the unique operating wavelength of PRIMA (28-240 µm). Crystalline olivines and pyroxenes have a range of features in the FIR, for example forsterite is known to have a strong transitional peak ∼34 µm (Molster et al., 2010) and a band at ∼69 µm thought to provide a useful indicator for temperature and Fe:Mg ratios (Bowey et al., 2001). Similarly, sulfides commonly found in chondrites, such as pyrite, have a range of features between ∼20-45 µm (Brusentsova et al., 2012). Furthermore, sulfides are more resistant to space weathering processes (e.g., Chaves & Thompson, 2022), so identification (or lack thereof) may help to assess the extent of weathering experienced by different bodies in the main belt. Finally, as well as water ice lines which can be found in the FIR, there are a number of organic ices, such as methanol ices which have a range of prominent features between 50-150 µm (McGuire et al., 2016). Many of these previous studies, however, focus on dusty sources so it is quite possible they will present quite differently for a whole-body point collection.

Essentially, it is uncertain exactly what we might observe with PRIMA when looking at these nearby, relatively dark bodies; however, there are a vast range of potential spectral lines to be identified which can reveal a trove of information about them that studies of meteorites on Earth simply cannot do. Far-IR flux densities of 270 main belt asteroids were obtained during observations by Herschel/PACS, which are publicly available and will be used to help determine targets (Szakáts et al., 2020).

## Instruments and Modes Used

FIRESS high-resolution point source mode.

## Approximate Integration Time

Based on Herschel/PACS observations of main belt asteroids, the targets for this study have a range of flux densities between 2–200 Jy at 100 µm (freq. $3e^{12}$ Hz; Müller et al., 2014). Thus, in high-resolution mode, R=4400 @ 100 µm, we have (2 to 200) x $1e^{-26}$ x $3e^{12}$ / 4400. This gives a range between $1.4e^{-17}$ – $1.4e^{-15}$ $Wm^{-2}$ per spectrometer channel.

Assuming a SNR=100 of on the continuum, the 5σ detection limit would be $6.8e^{-19}$ – $6.8e^{-17}$ $Wm^{-2}$ per spectrometer channel. For $6.8e^{-18}$ $Wm^{-2}$ and 2 Jy continuum, the ETC reports integration time of 0.2 hours per pointing. For $6.8e^{-16}6$ $Wm^{-2}$ and 200 Jy continuum, the corresponding time is 2.5 minutes per pointing. For full FIRESS band coverage, two pointings are required, bringing integration time per target to 5–24 minutes.

Considering the low integration time required per target, it would be possible to collect high-resolution spectra of all 270 main belt asteroids whose flux densities were measured with Herschel/PACS in only 23–108 hours.

## Special Capabilities Needed

None

## Synergies with Other Facilities

This work could be considered in collaboration with laboratory far-IR spectra of meteorites corresponding to different asteroid types. This would allow the tying together of samples that it is possible to extensively characterize here on Earth with their potential parent bodies. Additionally, JWST data of main belt comets could be considered to compare the difference between different types of bodies within the main belt (e.g., Kelley et al., 2023; Hsieh et al., 2025).

## Description of Observations

A sample of 270 main belt asteroids with previously collected flux densities will be targeted for the measurement and characterization of far-IR silicates, phyllosilicates, and water ice bands of a range of asteroid types. Comparisons will be carried out both between sources and against laboratory spectra (minerals and meteorites).

## Acknowledgment

This research was carried out at the Jet Propulsion Laboratory, California Institute of Technology, under a contract with the National Aeronautics and Space Administration (80NM0018D0004).

# 57. Size Distribution in the Solar System Kuiper Belt

Arielle Moullet (National Radio Astronomy Observatory), Joseph Masiero (Infrared Processing and Analysis Center), Noemi Pinilla Alonso (Institute of Space Science and Technology in Asturias, Universidad de Oviedo, Spain), Dariusz C. Lis (Jet Propulsion Laboratory, California Institute of Technology), Anne Verbiscer (UVA)

The size frequency distribution (SFD) of the Solar System's Kuiper-Belt objects reflects the formation history of the outer Solar System. Accessing the size regime below 100 km is important to constrain the respective role of destructive (collisions) and accretive processes that modified the initial planetesimal mass function. While optical surveys are the most effective at detecting such targets, only thermal infrared observations offer the ability to disentangle the effect of the albedo in retrieving equivalent sizes. We describe a large FIRESS targeted survey of known KBOs, completing and significantly extending earlier thermal surveys by detecting objects in size ranges mostly inaccessible to Herschel (and Spitzer), including close objects with diameters as small as 35 km. After debiasing, the retrieved SFD can be effectively matched to theoretical SFDs, discriminating against hypotheses on conditions at the time of Kuiper-Belt formation (e.g., initial mass function), as well as historical and ongoing processes.

## Science Justification

The Solar System's Kuiper Belt is considered to be both the most pristine remnant of the Early Solar nebula, and the most accessible example of the last stage of an evolved debris disk. As such, the composition of individual Kuiper Belt Objects (KBOs) can be linked to the original composition of the Outer Solar System. On the other hand, population-wide physical and dynamical characteristics are indicative of past and present processes controlling planetesimal formation and distribution across a planetary system. In particular, planetesimal accretion processes are important to understand timescales in the transition from the protoplanetary disk stage to a fully evolved planetary system.

Today's Kuiper Belt size-frequency distribution (SFD) is one of the main observables that can be compared against planetesimal evolution models in the Solar System. Theoretical models (e.g., Davis & Farinella, 1997; Kenyon and Luu, 1999; Pan & Sari, 2005) can include catastrophic and non-catastrophic collisions, collisional grinding cascades and different accretion modes and stages (e.g., streaming instability, runaway growth, pebble accretion). Note that based on the surface history of large KBOs, the leading theory is that the size distribution has not been modified by collisional and accretional effects since the surfaces of these bodies formed (Spencer et al., 2020), and the SFD must hence represent the integrated effect of historical processes up to the time of Kuiper Belt formation. The outcome of such models is usually a theoretical SFD displaying a break ('knee' or divot) in its slope at a specific size, a distinct typical size over which objects are considered of primordial size (hence representative of the initial planetesimal mass function), as well as a characteristic size under which objects are considered collisional fragments. The transition between primordial and collisionally evolved populations is likely to happen in the size regime between 10-100 km diameter (Shankman et al., 2016, Lawler et al., 2018).

The location of the SFD knee, and the slope of the SFD on either side of the slope break, are indicative of the efficiency of accreting and destructive processes over time, which are themselves linked to fundamental properties such as relative velocity (orbital inclination and eccentricity distribution) and solid bulk strength (Kenyon & Bromley, 2020; Nesvorny et al., 2019, Bottke et al, 2023).

Currently, the characterization of the SFD is mostly based on very sensitive optical photometric surveys such as the Outer Solar System Origins Survey CFHT (OSOSS, Bannister et al., 2018), which has characterized about 1000 objects with sizes down to ∼20 km diameter, and tend to show a deficit of objects below 100 km (Alexandersen & Gladman, 2014, Kavelaars et al., 2021). Impact cratering records on outer moons bring in complementary information towards smaller sizes (e.g., Nesvorny et al., 2017).

An important caveat of such surveys is that they only directly measure absolute magnitude, not size. Magnitude is only a proxy of an equivalent size based on an assumed albedo value, typically the average albedo for the target's dynamical class. But that assumption may not be solid; KBOs albedos are known to span values between 2–70%. Assuming the typical median albedo value of 10% for a given object can hence translate in a factor of 2 error in the derived diameter. Albedo-unbiased size surveys are much more limited than photometric surveys, as they typically rely on thermal measurements (infrared or mm-wave) which are challenging due to limited sensitivity in the thermal regime. The most extensive albedo-unbiased survey was achieved with Herschel through the 'TNOS are cool' large program (Vilenius et al., 2012, Muller et al. 2020). Herschel's PACS sensitivity enabled access to thermal detection on ∼130 KBOs of diameters down to 140 km diameter. **The SFD below 100 km, inaccessible by Herschel, is however much more discriminant** and informative in terms of comparison to theoretical models, as it covers the 40–100 km region where models display a slope break in the distribution (e.g., Abedin et al., 2022) due to the transition between primordial and collisionally evolved populations.

Using the exquisite far-IR point-source continuum sensitivity offered by FIRESS, which is approximately 25 times better than Herschel-PACS, one can design a Band 3 or Band 4 pointed survey of known objects. The sample can be designed to cover several dynamical classes, as well as spectral classes, for example based on the three compositional groups identified in Pinilla-Alonso et al., 2025. The survey will focus on the size regime of interest in regard to theoretical models, helping to locate the 'knee' in the SFD. Specifically, with integration time of 10 min - 4 hours on time per object, one can access thermal detections for objects down to 35 km diameter within 50 AU. The survey's immediate outcome will be photometry. Fluxes can be converted to equivalent sizes using available optical magnitudes and thermal emission models, such as NEATM (Harris et al., 1998). There is some plausible range in applicable thermal models, due to the variety of thermal and radiative properties of KBOSs surfaces. This can bring which additional uncertainty to the derived size, but in a much more moderate manner (±20%) than the uncertainty from albedo bias. In addition, the simultaneous obtention of a FIRESS band observation at shorter wavelength, though less sensitive than the Band 3 and 4 observations near the peak of thermal emission, can be used to derive a best fit of the most appropriate thermal model, through the use of a parametric variable describing the overall departure from a reference thermal model (see e.g., Vilenius et al., 2012), overall reducing the modeling uncertainty.

The derived size distribution will then be de-biased for observational limits (see Vilenius et al., 2014), and analyzed separately for distinct dynamical classes. We expect that such a survey, directly complementing the size regime investigated by Herschel and commensurate to the size regime addressed by OSSOS, can be used to:

- provide a unique, rich and low uncertainty complement to the Kuiper Belt SFD in the 35–100 km range
- strengthen the accuracy of the SFD for objects beyond 50 AU
- verify or improve the appropriateness of albedo-assumptions used in optical surveys, making the optically derived SFDs much more robust.

Together those constraints will facilitate a more meaningful comparison of the Kuiper Belt SFDs with formation models will help to understand the respective historical roles of accretion and collisions.

Note that with a pixel size (diffraction-limited) of ~13” at 100 um, some very wide binaries can be imaged separately. In addition, note that at the native resolution of FIRESS (R~100), exploratory spectra of solid surface features can be observed on the brightest objects, as described in another case of this GO Science Book (Pinilla-Alonso et al., this book).

A similar experiment could also be performed to refine the SFD of Centaurs (Bauer et al., 2013, Duffard et al. 2014) which can typically be detected at higher SNR given the smaller semi-major axis and larger thermal emission.

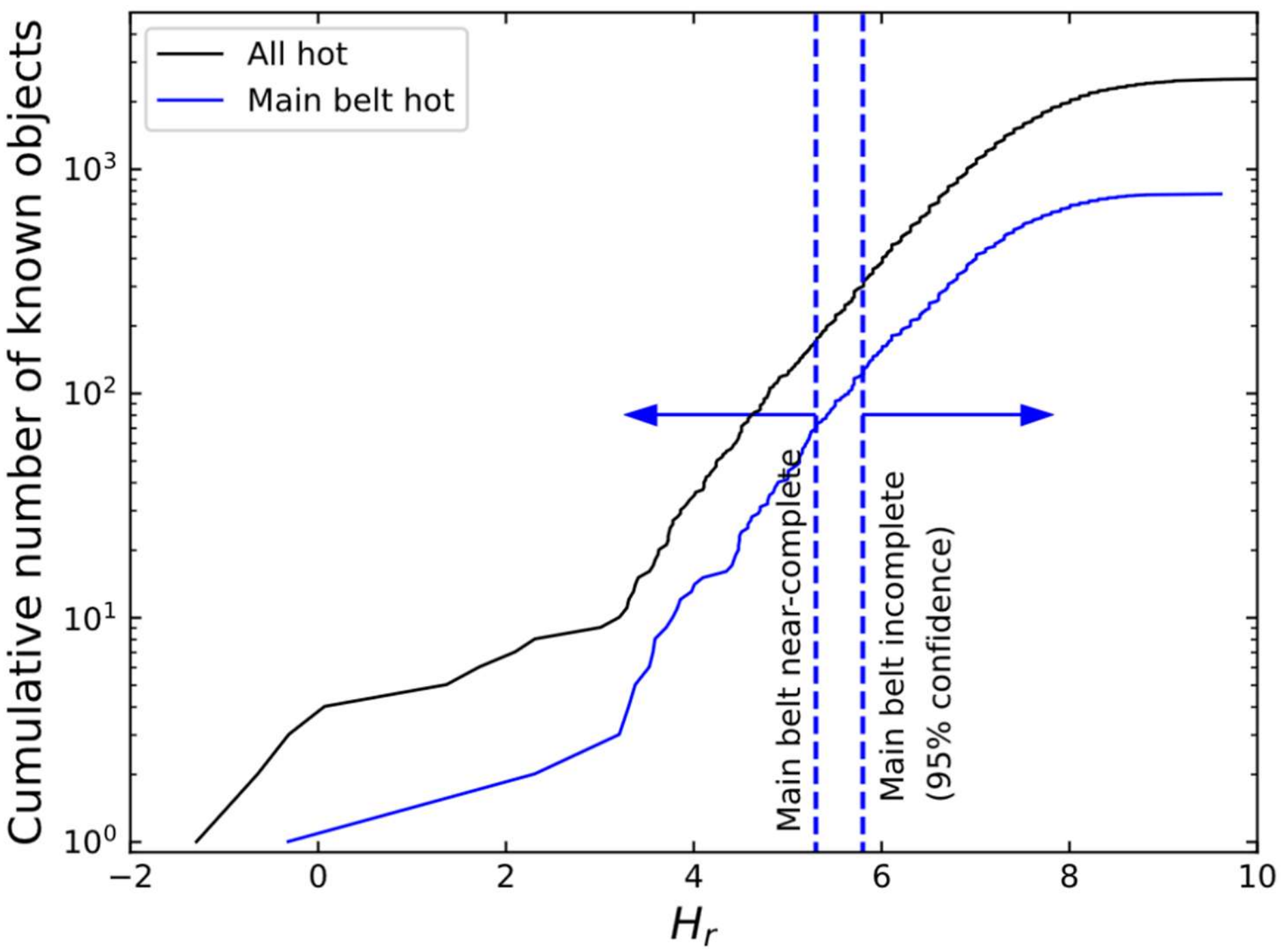

**Figure 1.** Adapted from Petit et al., 2023: Cumulative distribution of hot TNOS (black curve) based on the Minor Planet Center observations database, as a function of absolute magnitude (proxy for size). Blue is main-belt subset. Assumed geometric albedo is 0.15. The survey is almost complete for magnitudes below 5.3.

## Instruments and Modes Used

FIRESS low res mode pointing on each of the sample targets. A single spectral mode is sufficient.

## Approximate Integration Time

We require a SNR of ∼10 on the detection to be able to constrain size to a meaningful level, especially considering modeling uncertainty on the top of measurement noise.

The optimal band for detection highly depends on the body's temperature. Considering that Band 3 is close to the peak for 50 K surfaces and band 4 is close to the peak for 30 K surfaces, and the overall relative variation of the spectrum of Planck emission compared to the sensitivity spectrum, Band 3 is optimum for closer objects (~35 AU), and Band 4 for further objects (~50 AU).

By binning over the whole band 3, 1 hour of integration would yield a 10 sigma detection for a 160 uJy target. This means that, for warmer and closer targets, objects in the 35-50 km range could be detected in 1-4 h, and even shorter integration times are sufficient to achieve detection on the 50-100 km size range inaccessible to Herschel.

On the colder side, by binning over the whole band 4, 1 hour of integration would yield a 10 sigma detection for a 300 uJy target, giving access to detection of objects larger than 150 km diameter.

Deeper observations could be warranted for specific targets. Considering a sample of ~200 known KBO eligible targets, focused on closer targets with sizes inaccessible to Herschel detection but also including further targets, this leads to a total observing time of ∼400h.

## Special Capabilities Needed

All sources are Solar System sources with slow apparent motion: non-sidereal tracking is needed (less than 2”/h). The science case relies on high accuracy photometry, and special attention should be put towards flux reference accuracy.

## Synergies with Other Facilities

PRIMA will leverage the discoveries made by wide surveys (Rubin) to follow up on newly identified targets (see, e.g., Kavelaars et al., 2025).

## Description of Observations

We propose a FIRESS pointed continuum photometry survey on ~ 200 identified sources across the ecliptic, in either Band 3 or Band 4. Each target is a point source, well below 0.1” in equivalent apparent angular size. Spectral channels will be binned, to maximize continuum sensitivity.

## Acknowledgment

This research was carried out at the Jet Propulsion Laboratory, California Institute of Technology, under a contract with the National Aeronautics and Space Administration (80NM0018D0004).

# 58. Tracing Ice and Dust Across the Solar System: A Window into the Origins and Evolution of Volatiles Throughout the Universe

Noemi Pinilla-Alonso (Institute of Space Science and Technology of Asturias (ICTEA), Universidad de Oviedo, Oviedo, Spain), Moullet, A (NRAO); Brunetto, R (Université Paris-Saclay, CNRS, Institut d'Astrophysique Spatiale, Orsay, France); Emery, J (Northern Arizona University, Flagstaff, AZ 86011, USA); Verbiscer, A J(University of Virginia, Charlottesville, VA 22904, USA), Ioppolo, S (Aarhus University, Denmark).

We propose to use PRIMA, the only future facility capable of doing so, to study the distribution of ice and dust in the outer Solar System, providing essential ground truth to interpret debris disks, planetary systems, and the interstellar medium. While the James Webb Space Telescope (JWST) is unveiling a chemically rich diversity of ices on trans-Neptunian objects (TNOs), it cannot access the key spectral region needed to characterize refractory materials. Recent results combining JWST and Spitzer hint at the presence of Mg-rich silicates on TNOs, but the sample remains too limited to assess their prevalence or variability.

PRIMA's far-infrared (FIR) spectral coverage (24–240 µm) will enable the detection and quantification of silicate and organic features, assessment of crystalline vs. amorphous phases, and measurement of the dust-to-ice ratio, fundamental for reconstructing the thermal and chemical history of these bodies. Coupled with studies of the inner Solar System, particularly the asteroid belt, PRIMA will test whether a compositional continuum in minerals and organics exists, once shaped by the massive dynamical scattering of planetesimals 4.5 billion years ago. PRIMA will not simply build on JWST's discoveries; it will revolutionize our ability to trace the journey of volatiles and refractories from the interstellar medium to the solid building blocks of planetary systems.

## Science Justification

The icy bodies of the outer Solar System, particularly TNOs and some associated populations e.g. irregular moons of the giant planets and Jupiter and Neptune Trojans, are natural laboratories that preserve pristine volatiles and refractory material from the early solar nebula. Their surfaces host a complex mixture of ices ($H_2O$, $CO_2$, CO, $CH_4$, $CH_3OH$) and dust, which not only inform about local conditions during planetesimal formation (Pinilla-Alonso et al. 2025), but also establish crucial links with interstellar ices, molecular clouds, and protoplanetary disks.

Until recently, studies of ice on TNOs were largely restricted to visible and near-infrared wavelengths below 2 µm, limiting our view to the most abundant components, such as water ice (Barucci et al. 2011). Many key volatiles, including $CO_2$, $CH_4$, CO, and $CH_3OH$, have strong spectral features beyond 2 µm, often in regions strongly absorbed and obscured by Earth's atmosphere (Pinilla-Alonso et al. 2020). It is therefore not surprising, but deeply rewarding, that an exceptional instrument like the JWST is transforming this field. With its unprecedented sensitivity and spectral coverage up to 5.4 µm, JWST is unveiling a molecularly rich landscape of ices and irradiation products on TNO surfaces (Brunetto et al. 2025, Protopapa et al. 2024, Pinilla-Alonso et a. 2024, Emery et al. 2024). These discoveries, together with the detection of ices in debris disks (Xie et al. 2025) and molecular clouds (McClure et al. 2023, 2025), provide a powerful framework to trace the journey of volatiles from the interstellar medium to the building blocks of planetary systems.

However, JWST's capabilities, transformative as they are, do not extend to the study of refractory silicate material on TNOs. MIRI lacks the sensitivity to reach most TNOs and is currently limited to a few Centaurs and Trojans (Vernazza at al. 2025, Martin et al. 2023). Yet, recent studies combining JWST and Spitzer data for a small sample of TNOs and Centaurs have revealed spectral features attributed to Mg-rich crystalline silicates that resemble those seen in Jupiter Trojans, comets, and P/D-type asteroids, suggesting a shared origin as predicted by dynamical models like the Nice model (Gomes et al. 2005, Morbidelli et al. 2005, Tsiganis et al. 2005).

Importantly, the variability in their spectral contrast points to differences in porosity, grain size, and the silicate-to-ice/organics ratio. These parameters are critical to interpreting surface evolution and composition, yet the sample size remains too limited to assess whether such refractory components are ubiquitous among TNOs or restricted to a subset.

This need to better characterize the dust component gains urgency considering recent parallel breakthroughs in Solar System modeling and exoplanetary observations. Dynamical simulations of dust production and transport from the Kuiper Belt (aka trans-Neptunian belt, Corbett et al., 2025) have shown how sputtering of ice-rich grains can generate a vast, radially extended disk with compositional gradients. Strikingly, JWST IFU observations of the debris disk around HD 181327 (Xie et al., 2025) reveal exactly such a structure — including 3 µm water ice features and a radial increase in ice abundance. These two independent lines of research converge on a consistent picture in which dust evolution and ice retention are deeply intertwined. But fully validating these models requires ground truth via direct compositional measurements of Solar System bodies — something only PRIMA can offer in the FIR.

PRIMA's spectral range (28–240 µm) is uniquely suited to explore this uncharted territory. It will enable characterization of ice, silicate and organic features inaccessible to JWST (Figure 1) and quantification of the relative abundances of amorphous vs. crystalline phases, olivine/pyroxene ratios, and the dust-to-ice fraction, fundamental for reconstructing the chemical and thermal history of the outer Solar System. Figure 1 illustrates the diagnostic potential of the PRIMA spectral range to study ices, minerals, and organic materials relevant to outer Solar System bodies. In the FIR/terahertz (THz) spectral region, water ice shows distinct features associated with its amorphous and crystalline phases, allowing us to trace structural transitions driven by temperature or irradiation (panel A). Mixtures of $CO_2$ and $CH_3OH$ ices exhibit clear spectral signatures whose detectability varies with temperature and crystallization state, offering insights into thermal histories (panel B). At longer wavelengths, PRIMA captures absorption features of complex organics such as methyl formate ($HCOOCH_3$), which fall in spectral regions free from overlapping bands of dominant ices like water or $CO_2$, minimizing spectral confusion (panel C). Finally, FIR reflectance spectra of meteorites, cometary and asteroidal materials reveal key vibrational modes of hydrated and anhydrous silicates, iron sulfides (e.g., troilite), and organics that are inaccessible to JWST but lie within PRIMA's window (panel D), enabling compositional and mineralogical comparisons between laboratory analogs and small body surfaces.

PRIMA will go beyond molecular ice inventories and begin to characterize the full compositional complexity of TNOs, providing the missing piece to interpret what JWST has just started to see around other stars. In this sense, PRIMA has the potential to be transformative in building the bridge that connects the ISM and molecular clouds to proto planetary disks and debris disks as well as connecting local and extrasolar planetary systems.

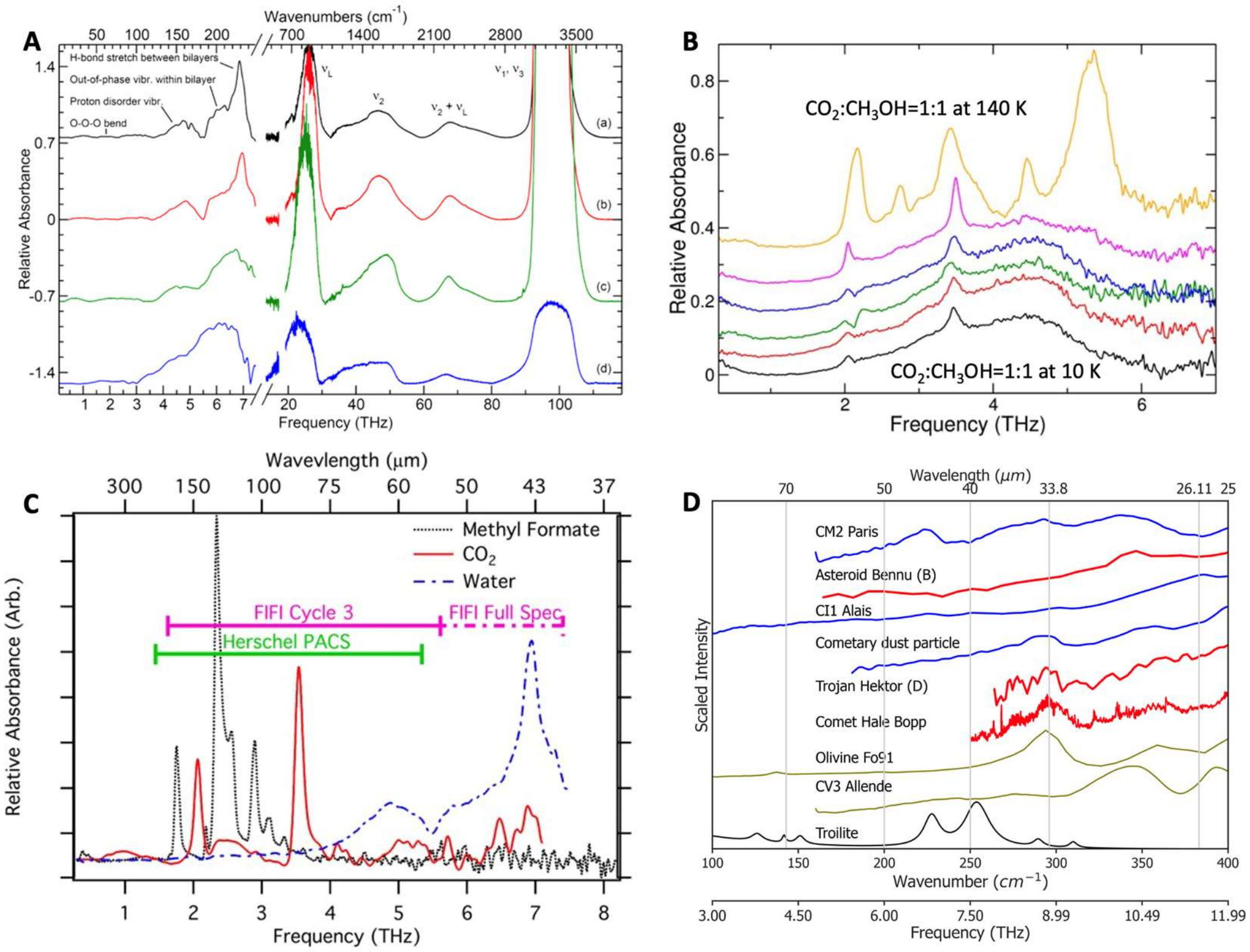

**Figure 1. Examples of ice and mineral features accessible in the PRIMA spectral range.** A: Water ice deposited at different temperatures to obtain Ih, Ic, compact amorphous, porous amorphous (a-d, respectively). B: $CO_2$ and $CH_3OH$ ice mixture 1 to 1 deposited at 10 K and heated stepwise to 140 K. C: Methyl formate (black) and other more abundant ices. D: Far-IR spectra of meteorites, cometary and asteroid materials. Panels adapted from Ioppolo et al. (2014), McGuire et al. (2016), and Brunetto et al. (2023), and references therein.

## Instruments and Modes Used

This science case needs to use FIRESS' pointed low-resolution mode, with 20 pointing and 2 spectral settings.

## Approximate Integration Time

A spectral resolution of 50-100 and the ability to detect a 10% depth per channel is necessary to identify the bands of interest. In 1 hr of observation or less, at 60 μm, it is possible to obtain a single-setting spectrum with these requirements for sources with a continuum level of 4 mJy or more. Two settings would be needed to cover the 24 – 235 μm range. Looking at the sources detected in the "TNOs Are Cool" Herschel Open Time Key Programme (Müller et al. 2009, http://public-tnosarecool.lesia.obspm.fr/Published-observations.html), about half of the sources reach that level of flux across dynamical types. With a 40 hour pilot program, one could detect about 30 of these targets and obtain their spectrum , and also sample across the 3 compositional groups identified by Pinilla-Alonso et al., 2025.

## Special Capabilities Needed

Tracking rates for non-sideral targets

## Synergies with Other Facilities

This program fosters strong synergies with ongoing and future missions to explore icy bodies and ocean worlds, including Europa Clipper, Dragonfly, and potential missions to Enceladus or Triton. It complements PRIMA's far-IR observations of debris disks by providing ground-truth from our own trans-Neptunian belt and aligns with JWST and ALMA studies of ices and dust in protoplanetary systems. The spectral data will benefit from laboratory experiments on ice and silicate analogs in the far-IR, improving compositional interpretation. It also supports wide-field surveys (e.g., Rubin, ULTRASAT, Roman) by aiding in the characterization of newly discovered comets, icy dwarf planets, and interstellar objects. Finally, it offers essential constraints for dynamical and compositional models of Solar System formation and disk evolution.

## Description of Observations

Sample justification: Past missions in the thermal IR such as Spitzer (IRAC, IRS, MIPS) and Herschel (PACS, SPIRE) have provided the most comprehensive thermal observations of TNOs and Centaurs to date. However, their thermal emission is intrinsically faint due to low surface temperatures and large heliocentric distances, requiring highly sensitive instrumentation.

Spitzer enabled the first significant detection of thermal fluxes from small icy bodies. Stansberry et al. (2004, 2008) and reported MIPS detections of ~47 TNOs and Centaurs, with follow-up studies focusing on smaller subsets or individual targets. Herschel's Open Time Key Programme "TNOs Are Cool" (Müller et al. 2009) expanded on this, providing thermal measurements for ~170 TNOs and Centaurs across all dynamical classes, leading to over 20 publications. These datasets form the foundation of our understanding of the physical properties (albedos, diameters, thermal inertias) of small bodies beyond Neptune.

In the absence of FIR reflectance spectra of TNOs and Centaurs, we use broadband flux measurements from Spitzer and Herschel to estimate optimal observing strategies with PRIMA. These fluxes allow us to simulate expected continuum levels and assess spectral detectability across the sample. Our calculations show that to detect a 10% deep absorption feature (in the order of those shown in laboratory reflectances) per spectral channel at a 5σ level in 1 hour of integration time, the target must have a continuum flux greater than ~4 mJy at 60 µm. Applying this criterion to the Herschel/Spitzer catalog, we identify approximately 80 TNOs that qualify, spanning a wide range of dynamical classes and surface compositions, and from these, we propose a pilot survey of 30 targets representative of the diversity observed in the trans-Neptunian belt at other wavelengths.

This pilot sample will not only encompass diverse dynamical populations but also span the three major compositional classes identified by Pinilla-Alonso et al. (2025): water-ice/dust-dominated, $CO_2$/CO-rich, and methanol/complex organic-bearing surfaces. If, as proposed, these classes reflect formation in distinct regions of the primordial disk, we may detect systematic differences in their mineralogical signatures. Such variations would provide evidence for radial heterogeneities in the composition of solids in the protoplanetary disk, offering a valuable link between the early Solar System and the diversity observed in other planetary systems.

## Acknowledgment

This research was carried out at the Instituto de Ciencias y Tecnologías Espaciales de Asturias, Universidad de Oviedo, Spain, thanks to the funding provided by the Ministry of Science, Innovation, and Universities (MCIU) in Spain and the State Agency for Research (AEI) for funding through the ATRAE program, project ATR2023-145683

# ISM, Star and Planet Formation

# 59. Dust Mineralogy in Protoplanetary and Debris Disks

Nicholas Ballering (Space Science Institute), Meredith MacGregor (Johns Hopkins University), Jane Huang (Columbia University), Patricia Luppe (Trinity College Dublin), Christine Chen (STScI/JHU), B. Sargent (STScI, JHU)

PRIMA offers a powerful opportunity to study the mineralogy of planet-forming solids. Far-IR spectroscopy with FIRESS can measure solid state spectral features from dust in the cold regions of protoplanetary disks and debris disks, complementing studies of warm dust mineralogy in the mid-IR with Spitzer and JWST. Far-IR features of olivine and pyroxene silicates reveal the crystallinity of dust in the outer disk, and the features are uniquely sensitive to the dust Fe/Mg ratio. PRIMA can also search for carbonate and phyllosilicate minerals—signatures of aqueous alteration. Measuring the refractory component of planet-forming material is vital to understand the diversity of exoplanet geophysics and habitability. The degree of crystallinity in the outer regions of disks indicates the efficiency of radial mixing during protoplanetary disk evolution. Comparing the mineralogy of protoplanetary dust with debris disk dust reveals the degree of grain alteration within planetesimals. Finally, comparing the mineralogy of exoplanetary systems with that of asteroids and comets reveals whether the processes that built the solar system are rare or common. PRIMA can study the mineralogy of 50 debris disks and 100 protoplanetary disks in 480 hours of integration time.

## Science Justification

Planetesimals, rocky planets, and the cores of giant planets grow from dust grains in protoplanetary disks. In more mature planetary systems, leftover planetesimals collide and release dust, generating debris disks. IR thermal emission from circumstellar dust reveals not only the dust mass and temperature but also its mineralogy via solid-state spectral features. Measuring these features can answer fundamental questions about planet composition, disk evolution, and whether our own solar system is common or unique.

While JWST and Spitzer revealed disk mineralogical features in the mid-IR, far-IR features offer unique insights into the dust composition. The far-IR traces cold dust from the outer disk compared to the warmer inner region traced in the mid-IR. The SEDs of typical debris disks peak in the far-IR while only a minority show silicate features in the 10-micron region (Ballering et al. 2014). The most prevalent silicate minerals—olivine ($Mg_{2-2x}Fe_{2x}SiO_4$) and pyroxene ($Mg_{1-x}Fe_xSiO_3$)—exhibit distinct far-IR features when in the crystalline phase (Chihara et al., 2002; Koike et al., 2003; Koike et al. 2006; Figure 1, left). These features are highly sensitive to the Fe abundance (x in the formulas above). Silica ($SiO_2$), seen in some mid-IR disk spectra (Sargent et al. 2009), have far-IR features as well (Loewenstein et al. 1973). Various other minerals that are less commonly seen in disk mid-IR spectra also show unique features in the far-IR (Figure 1, right).

These include carbonate minerals such as calcite ($CaCO_3$) and dolomite ($CaMg(CO_3)_2$), which are found in meteorites and may be the product of aqueous alteration. Carbonates were identified around evolved stars via their far-IR spectral features (Kemper et al. 2002). Phyllosilicates, also a product of aqueous alteration, have far-IR features as well (Mutschke et al. 2008).

Herschel detected silicate features in the far-IR only from disks with continuum flux density > several Jy; e.g., the beta Pic debris disk (de Vries et al., 2012) and Herbig Ae/Be type protoplanetary disks (Sturm et al., 2013). PRIMA offers the unprecedented sensitivity to measure far-IR mineralogical features of more typical protoplanetary and debris disks. In doing so, it will address four important questions regarding planet formation:

*1. What is the composition of planet-forming material?* Rocky planets are primarily made of the refractory elements Si, O, Fe, Mg, Al, and Ca. In dust grains, these elements are arranged into silicates and other minerals. Measuring grain mineralogy thus reveals the budget of refractory elements available for planet formation. This budget affects planetary geophysics and habitability. For example, the Fe abundance (measured by the 69 micron crystalline olivine feature) influences a planet's ability to generate a protective magnetic field (Noack & Lasbleis, 2020). The Mg/Si ratio – probed via the relative abundance of olivine (Mg/Si ≈ 2) and pyroxene (Mg/Si ≈ 1) – influences its ability to generate an atmosphere via outgassing (Spaargaren et al., 2020).

*2. How is material transported in protoplanetary disks?* IR spectroscopy reveals whether grains are amorphous (broad spectral features) or crystalline (narrow sharper features). PRIMA will measure the abundance of crystalline grains in the cool, outer regions of disks. Grains inherited from the ISM are amorphous (Kemper et al 2004), and crystalline grains can only form or anneal in the hot (> 1000 K) inner regions of protoplanetary disks. Crystalline grains in the outer disk must have been transported there by a large-scale mixing process such as magnetically driven winds (Arakawa et al. 2021).

*3. Are minerals altered after incorporation into planetesimals?* Dust in a protoplanetary disk aggregates into larger pebbles and planetesimals. Planetesimals that are not incorporated into planets can remain in orbit for millions or billions of years before colliding with each other and breaking apart, releasing the dust in a debris disk. Comparing the mineralogy of protoplanetary disks vs. debris disks offers a unique opportunity to study if and how grains are changed within the planetesimals by, e.g., heating from short-lived radioactive nuclides, melting, differentiation, or aqueous alteration.

*4. How common or unique is the solar system's composition?* The mineralogy of asteroids and comets in the solar system has been studied extensively using remote observations and sample return missions. Comets, which coalesced beyond the water snowline, contain a high fraction of crystalline silicates, suggesting radial mixing occurred in the solar system (e.g., Ogliore et al. 2009). Cometary olivine has a very low Fe abundance, whereas asteroid olivine is Fe-rich. A survey of disk mineralogy with PRIMA—analyzed in conjunction with mid-IR observations with Spitzer and JWST—will reveal whether the composition trends seen in the solar system are typical of most planetary systems.

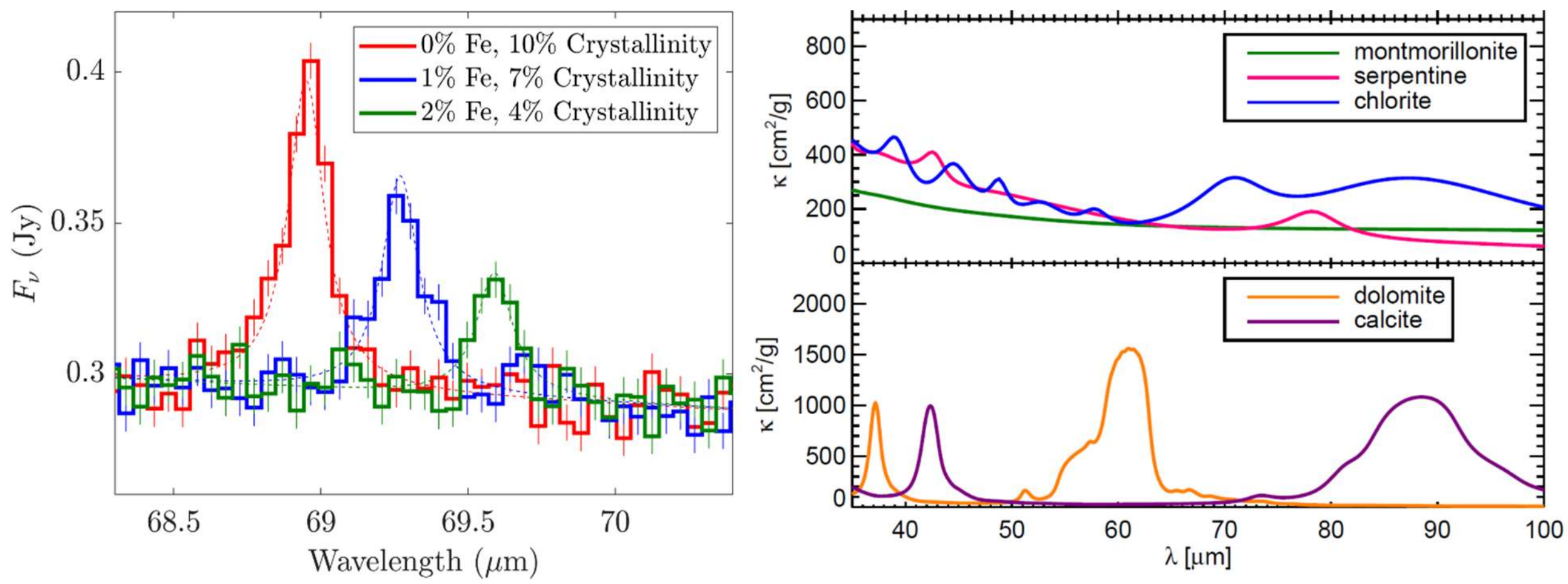

**Figure 1.** Left: simulated PRIMA observations of the forsterite 69 micron feature at three different Fe abundances (which shifts the central wavelength of the feature) and crystallinity fractions (which sets the amplitude over the continuum). This represents 4.3 hours of integration time achieving SNR = 50 per channel in spectra binned to a resolving power of R = 1400. Dotted lines are the Lorentzian profiles of the features. Right: The opacity spectra of other minerals with solid state features in the far-IR that can be measured with PRIMA (Min et al. 2016).

## Instruments and Modes Used

### FIRESS High-Resolution Pointing Mode

Mineralogical spectral features have a variety of widths and shapes across the far-IR (Figure 1). Some of the most important features, such as the 69 micron feature of crystalline olivine, are narrow enough to require the use of FIRESS in its high-resolution mode. However, binning these spectra by a factor of a few will still allow the features to be resolved while improving SNR. Most disks will be spatially unresolved with PRIMA, so pointed observations will be used.

## Approximate Integration Time

The well-characterized 69 micron feature of Mg-rich crystalline olivine (forsterite) provides a useful benchmark for setting the required SNR and integration time. This program targets a sample of disks brighter than 0.3 Jy at 70 microns (see below). The left panel of Figure 1 shows simulated SNR = 50 observations of a 0.3 Jy source. Three models are shown with crystallinity 10%, 7%, and 4%, which sets the feature strength above the continuum. For reference, the beta Pic debris disk has 4% crystallinity based on the 69 micron feature (de Vries et al., 2012). Each model also has a different Fe abundance (0%, 1%, and 2%), which shifts the central wavelength of the feature by 0.32 microns per Fe percentage (Koike et al. 2003). The features have an intrinsic Lorentzian profile with width and amplitude that vary as a function of grain temperature (Koike et al. 2006). The features shown in Figure 1 are appropriate for 50-100 K grains, typical for cold dust in debris disks and protoplanetary disks.

The FIRESS high-resolution mode offers a resolving power of R ≈ 7000 at 69 microns, which is higher than required to resolve the features and measure shifts due to small variations in the Fe abundance. The simulated observations in Figure 1 are binned by a factor of 5 to R ≈ 1400 to improve SNR. At this resolving power, the 1-$\sigma$ noise per channel to achieve SNR = 50 is $1.825 \times 10^{-19}$ W m$^{-2}$, which requires 4.3 hours of integration time according to the PRIMA ETC.

## Special Capabilities Needed

None.

## Synergies with Other Facilities

A comprehensive analysis of debris disk and protoplanetary disk mineralogy will use a combination of far-IR spectroscopy from PRIMA and mid-IR spectroscopy from Spitzer and JWST to probe both the warm and cold dust populations.

## Description of Observations

The science questions above can be addressed with a spectroscopic survey of 150 disks (50 debris disks and 100 protoplanetary disks). From the catalog of debris disks by Cotton & Song (2015), there are ≈ 50 known debris disks with 70 micron continuum flux density > 0.3 Jy. Figure 1 shows that the 69 micron feature representing 4% crystallinity or higher can be robustly measured from a 0.3 Jy disk in 4.3 hours of integration time. This is in a single spectral setting, so 8.6 hours are required to cover the full wavelength range with both settings. Brighter disks can be observed in less time, and we estimate the typical debris disk in our sample requires 2.4 hours per spectral setting (4.8 hours total). A mineralogical survey of 50 debris disks thus requires 240 hours. Note that debris disks tend to be brighter around younger and more massive stars, so a brightness-limited sample will not be fully representative across system parameters.

Protoplanetary disks are inherently much brighter in the far-IR than debris disks. A sample of 100 typical protoplanetary disks can easily be prepared with a brightness threshold of > 0.7 Jy at 70 microns based on photometric surveys (e.g., Ribas et al. 2017). According to the PRIMA ETC, obtaining SNR = 50 on a typical 1 Jy continuum source (with R = 1400) requires 1.2 hours of integration time (2.4 hours for both settings). Thus, 240 hours of observing time are required to observe the protoplanetary disk sample. The total debris disk and protoplanetary disk sample requires 480 hours of PRIMA integration time.

# 60. Protoplanetary Disk Ices

Nicholas Ballering (Space Science Institute), Kamber Schwarz (Max Planck Institute for Astronomy), Jane Huang (Columbia University), Sebastiaan Krijt (University of Exeter)

Protoplanetary disks are made of gas, dust, and ice. The ices play a critical role in delivering life-enabling volatiles to young planets. Additionally, ice mantles enhance the sticking efficiency of dust, facilitating grain growth and planet formation. PRIMA FIRESS offers a unique opportunity to measure far-IR spectral features of disk ices in emission, overcoming the challenges faced by near- and mid-IR observations of ices in absorption. In 200 hours of total exposure time, FIRESS can survey 100 disks, searching for the spectral signatures of important ice species including $H_2O$, CO, $CO_2$, $O_2$, $N_2$, $NH_3$, and HCN. This program will complement ALMA measurements of cold molecular gas in disks and JWST measurements of ice and warm gas at shorter wavelengths, providing a robust understanding of protoplanetary disk chemistry.

## Science Justification

Ices in protoplanetary disks play a critical role in the efficiency of planet formation and in the composition of material delivered to newly formed planets. Icy mantles increase the sticking efficiency of grains and facilitate the growth of dust and planetesimals (Wang et al. 2005). Ices are also a critical reservoir of the biocritical “CHON” elements that can be delivered to terrestrial planets by volatile-rich planetesimals.

Ices are observed via vibrational mode spectral signatures in the IR. To date, most ice measurements of protoplanetary disks were carried out in the near- and mid-IR. The spectral features are observed in absorption because dust grains warm enough to emit at these wavelengths are also too warm to maintain an ice mantle. In disks, thermal emission from the central star and hot inner disk provides the IR continuum on which ice absorption features are imprinted. Thus, only highly inclined disks—here the cold outer disk resides in front of the star—are suitable for ice observations. While near- and mid-IR facilities like Spitzer and JWST have successfully detected ices in several disks (Pontoppidan et al. 2005; Sturm et al. 2023, 2024), the small fraction of disks oriented edge-on to our line of sight is a fundamental limit of this technique.

Another limitation is that inferring the abundance of the ices from the absorption features is difficult due to radiative transfer effects (Ballering et al. 2021). The background continuum is not a simple point source, rather it is the extended scattered light surfaces of the disk, the geometry of which varies with wavelength. Furthermore, scattering mixes photons from ice-free regions into the same lines of sight as the icy regions, contaminating the absorption features with an unknown floor of continuum flux (Sturm et al. 2023).

Far-IR observations overcome both of these limitations. The far-IR features, which arise from lattice mode vibrations, can be studied in emission, so ices are detectable from disks of any inclination. Scattered light is negligible, so the radiative transfer is simpler, and the ice mass can be more accurately determined. FIRESS offers unprecedented sensitivity and ample spectral resolving power to study the diversity of disk ices in the far-IR.

**$H_2O$** is one of the most abundant volatiles in disks, a primary carrier of hydrogen and oxygen, and a requirement for life as we know it. $H_2O$ ice has far-IR features with shapes that distinguish between amorphous and crystalline forms, indicating its thermal history. It is the only ice species detected from disks in the far-IR to date, seen from a few bright systems with ISO (Malfait et al., 1998) and Herschel (McClure et al., 2015; Min et al., 2016).

**CO ice** is a major carrier of volatile carbon and oxygen in disks. It is seen in JWST spectra of edge-on disks, so it must be abundant and distributed far enough above the cold midplane to be detected in the scattered light disk surfaces, perhaps trapped in less-volatile ices (Bergner et al. 2024). CO ice observations will complement ALMA observations of cold CO gas that are used as a tracer of the total disk gas mass (Miotello et al. 2016). Measuring how much CO is frozen out into the ice phase will help calibrate these measurements. The most prominent far-IR feature of CO ice is centered at 199 microns (Gavdush et al. 2022).

**$CO_2$ ice** is another important carrier of carbon and oxygen that is also detected in edge-on disks with JWST. $CO_2$ is less volatile than CO, and it is non-polar, so cold $CO_2$ gas cannot be detected with ALMA. Disk chemical models predict a large fraction of CO ice may be converted into $CO_2$ (Schwarz et al. 2018), so detecting both species can test our understanding of disk chemical evolution. Far-IR features are centered at 86 and 147 microns (Gavdush et al. 2022). The left panel of Figure 1 shows a model opacity spectrum highlighting $CO_2$ and CO ice features in comparison to those from $H_2O$ ice.

**$O_2$ and $N_2$ ices** may be important reservoirs of oxygen and nitrogen in disks (the primary nitrogen carrier remains unknown). Comet 67P contains a surprisingly high abundance of $O_2$ ice, likely inherited from the solar system's natal disk (Bieler et al. 2015). With no dipole moment, these species do not emit at ALMA wavelengths in the gas phase, nor have they been detected in the ice phase with JWST. Far-IR lattice mode vibrations are the most promising method to detect $O_2$ and $N_2$ ices (Anderson et al 2024).

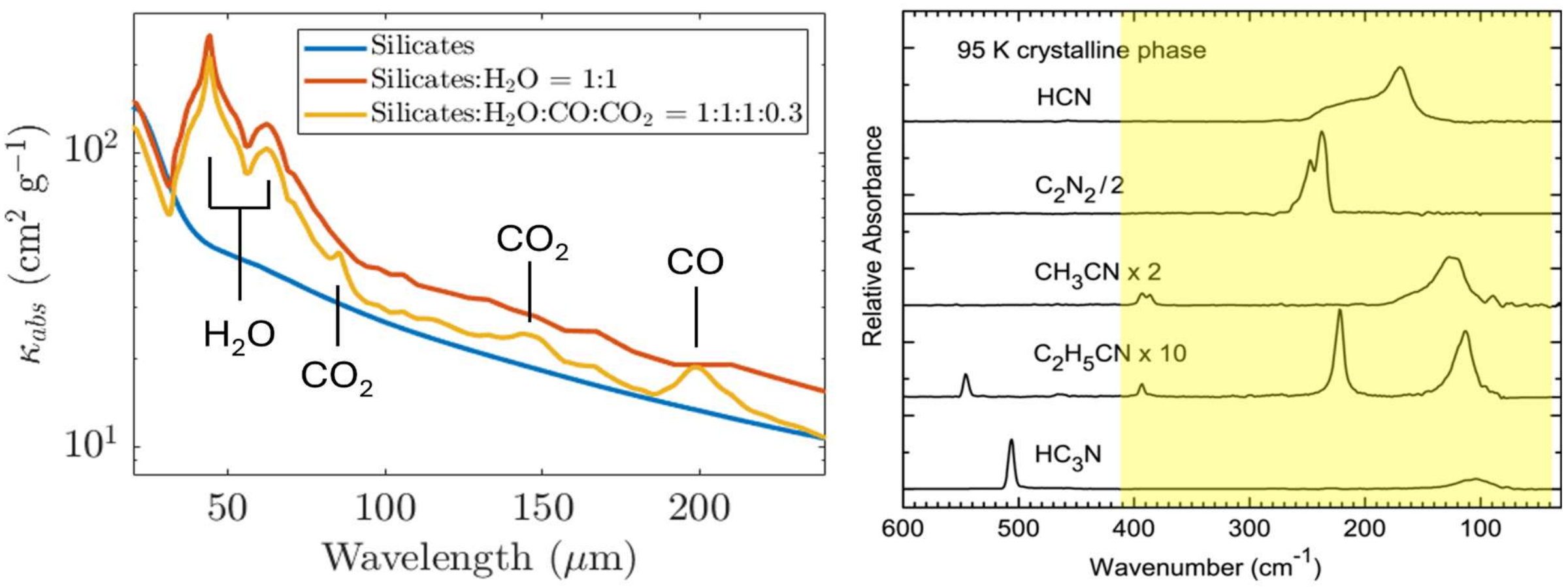

**Figure 1.** *Left:* opacity spectra of bare silicates (blue), silicates coated in $H_2O$ ice (red), and silicates coated in $H_2O$, CO, and $CO_2$ ice (orange). The most prominent spectral features of each ice species are labeled. The opacity spectra are calculated with OpTool (Dominik et al. 2021) for a grain size distribution between 0.01 and 1000 microns. References for optical constants are: silicates (amorphous olivine) from Dorschner et al. (1995), $H_2O$ ice (crystalline) from Curtis et al. (2005), and CO and $CO_2$ ices from Gavdush et al. (2022). *Right:* Laboratory-measured far-IR absorption profiles for a collection of nitrile ices (Moore et al. 2010). The yellow shaded region indicates the spectral coverage of PRIMA FIRESS.

**Other nitrogen-carrying ices** also exhibit far-IR spectral features. $NH_3$ has a broad feature centered at 30 microns (Trotta 1996). Various nitriles, especially HCN, show particularly strong features in the far-IR (Moore at el. 2010; Figure 1, right).

## Instruments and Modes Used

While many ice features are broad enough to be spectrally resolved with FIRESS in its low-res mode, protoplanetary disks are sufficiently bright to warrant observations in high-res mode. This enables complementary science to be performed at higher spectral resolution while faint ice features are extracted with improved SNR after spectral binning. Most disks are spatially unresolved with PRIMA, so pointed observations are used.

## Approximate Integration Time

The 199 micron CO ice feature provides a useful benchmark for setting the required SNR and integration time. The opacity spectrum shown in Figure 1 indicates that the height of this feature is 0.3 times the baseline continuum. However, this ratio is unlikely to hold in a disk emission spectrum because not all of the emitting dust will be coated in CO ice. A feature strength 10x weaker (3% of the continuum) provides a more conservative estimate, although the true feature strength will vary from disk-to-disk depending on the disk temperature structure and chemical makeup.

This program targets 100 disks with a typical flux density of 0.5 Jy at 200 microns (see below). For this typical disk, the amplitude of the feature is 0.015 Jy. Detecting the feature at 10-$\sigma$ significance in each spectral channel requires a 1-$\sigma$ uncertainty of 0.0015 Jy. The FWHM of the feature is approximately 15 micron, which can be spectrally resolved with 2 micron-wide channels (R = 100). The 1-$\sigma$ uncertainty on the integrated flux in each channel is then 1.124 × $10^{-18}$ W $m^{-2}$. According to the ETC, achieving this sensitivity with the FIRESS high-res mode requires 1 hour of exposure time. A second hour of exposure time is required to complete the spectrum with the other spectral setting.

## Special Capabilities Needed

None

## Synergies with Other Facilities

Icy grains detected by FIRESS in the outer regions of a disk can drift inward, eventually reaching the warm inner disk where the volatiles sublimate into the gas phase (Krijt et al. 2018). Mid-IR spectral observations with JWST are ideal for measuring this warm gas (Pontoppidan et al. 2024). Thus, combining FIRESS and JWST observations provides a holistic view of disk chemical and physical evolution. ALMA can measure the abundance and distribution of cold gas – including CO and various carbon and nitrogen carrying molecules – in the outer disk (Oberg et al. 2021), complementing FIRESS measurements of ice abundances. The radial and vertical temperature profiles of disks can be reconstructed from spatially and spectrally resolved ALMA gas observations (Law et al. 2023), providing important context for interpreting the spatially unresolved FIRESS ice measurements.

## Description of Observations

A sample of 100 protoplanetary disks can be prepared with a typical brightness of 0.5 Jy at 200 microns using photometric surveys from Herschel PACS (e.g., Ribas et al. 2017). Two hours of exposure time are required to observe the full spectral coverage per target (see above), so the program needs 200 total hours of exposure time.

# 61. Ice sublimation bursts at the water snowline with PRIMA

A. Banzatti (Texas State University, USA), N. Ballering (Space Science Institute), D. Gasman (KU Leuven), M. Audard (University of Geneva), M.J. Colmenares (University of Michigan), D. Itrich (University of Arizona), E. Dahl (California Institute of Technology), G. Perotti (Niels Bohr Institute, Max Planck Institute for Astronomy), M. Sewilo (NASA Goddard Space Flight Center and University of Maryland), C. Salyk (Vassar College), S. Krijt (University of Exeter)

The water snowline in protoplanetary disks, the sublimation/condensation surface at ~0.5-10 au around a solar-luminosity star, is considered to have a fundamental role in the formation of large planetary cores and the delivery of volatile ices to regions where super-Earths and small rocky planets are forming. However, both the location and the ice reservoir at the snowline are expected to evolve with time in ways that could speed-up or slow down planet formation. In particular, the unstable accretion phases during pre-main-sequence evolution of young stars produce variable irradiation and heating of planet-forming regions that can quickly sublimate a large part of the ice reservoir during episodic accretion outbursts. This phenomenon has been observed in both protostellar envelopes and strong early-stage FUor outbursts. Recently, by way of JWST's improved spectral resolving power, we are able to distinguish between water lines and conduct a detailed analysis of the radial distribution of water in inner disks, revealing this phenomenon in Class II phase disks as well. Yet, JWST only covers a few water lines that are sensitive to the snowline reservoir at ~170 K, lines with upper level energies of 900-1500 K. PRIMA will observe tens of lines from energies down to ~100 K, providing an unprecedented view of the cold water vapor and ice reservoirs across the snowline. In this paper, we propose to monitor ice sublimation events triggered by accretion outbursts to study processes that may have fundamental implications on planet-forming bodies near the snowline.

## Science Justification

Young stars are known to have strongly variable accretion rates that inject high-energy radiation and heat into planet-forming regions through events of episodic outbursts, where the accretion luminosity increases from factors of a few up to factors of ∼100 or more throughout the pre-main-sequence stellar evolution (Hartmann & Kenyon 1996; Audard et al. 2014; Fischer et al. 2023). The ubiquity and frequency of accretion bursts during star formation at 0.1–1 Myr suggest that burst-driven chemical evolution may constitute a fundamental ingredient in the disk chemistry that affects planet formation. Previous work supported this scenario by studying strong accretion outbursts in embedded young stellar objects that may retain long-term chemical effects in the envelope or in outer disk regions at 10–100 au (e.g., Visser et al. 2015; Cleeves et al. 2017; Rab et al. 2017; Molyarova et al. 2018; Waggoner & Cleeves 2019). Observations of Class 0/I protostars and FUor objects have indeed shown evidence in multiple objects for outburst-induced sublimation of ices leaving gas-phase molecules beyond their quiescent snowline locations (e.g., Hsieh et al. 2019; Lee et al. 2019; Jørgensen et al. 2020; Fischer et al. 2023; Calahan et al. 2024, Lee et al. 2025).

Recently, water vapor emission variability in Spitzer-IRS and JWST-MIRI spectra (Banzatti et al. 2012, Smith et al. 2025) provided the first clear evidence for a cold water vapor "burst" in low-energy lines during an accretion outburst in a Class II disk, EX Lup, the prototype of EXor outbursts (Herbig 2008) that are thought to happen later during prestellar evolution than the stronger FUor outbursts (e.g. Fischer et al. 2023). Organic emission disappeared during a strong accretion outburst in EX Lup, suggesting photodissociation of the organics by the enhanced UV radiation (Banzatti et al. 2012). The increase in water vapor line luminosity, instead, provided tentative evidence for enhanced ice sublimation caused by a recession of the snowline (Banzatti et al. 2012), as found in protostellar envelopes. JWST observations of EX Lup show unusually strong emission from cold water in comparison to other T Tauri disks, supporting the scenario of an event of increased ice sublimation during outburst and of >10 yr long freeze-out timescales in the inner disk surface (Smith et al. 2025). This demonstrates that outbursts can significantly change molecular ratios and increase the cold water reservoir, providing chemical signatures to study the recent accretion history of disks.

The thermal processing of disk regions around snowlines has significant implications for the planet formation process. The temperature conditions around the water snowline promote coagulation of icy particles (Gundlach & Blum 2014; Musiolik & Wurm 2019), while sublimation and subsequent repositioning of ices could enhance pebble growth and may trigger streaming instability (Drazkowska & Alibert 2017; Ros et al. 2019). The latter happens in a more extreme form during outbursts. Since the recovery timescales are long, these post-outburst pebble growth and ice depositioning differences may be visible in the PRIMA spectrum for decades after an outburst, and will provide insights into how outbursts and ice sublimation at the snowline may affect the seeds of planet formation (Houge & Krijt 2023; Ros & Johansen 2024).

Despite the recent results supporting the importance of accretion outburst in triggering ice sublimation, JWST only covers a few water lines that are sensitive to the snowline reservoir at ~170 K, those with upper level energies of 900-1500 K (Banzatti et al. 2023, 2025). Here we propose to use PRIMA-FIRESS to observe tens of lines from energies down to ~100 K, providing an unprecedented view of the cold water vapor and ice reservoirs across the snowline. In Figure 1, we provide a proof-of-concept simulation of how the relative change in water ice and vapor emission will be observed with PRIMA during and after an accretion outburst in a Class II disk, based on results and models from Smith et al. (2025). The water vapor models are directly based on what is provided in Smith et al. (2025) for EX Lup, assuming that we will observe either this disk or a similar one. For the ice feature, Figure 1 shows thermal emission from a disk spanning 0.1–80 au. Beyond the snowline, dust grains are 50/50 water ice and silicates (by mass), whereas grains within the snowline are pure silicates. During the outburst, the midplane snowline recedes out to 2 au while after three years the snowline returns to near its quiescent location at 0.8 au (Smith et al. 2025), and the ice feature becomes more prominent. The ice spectral features at ~44 μm and ~62 μm arise from lattice mode vibrations crystalline water ice (see also paper by N.

Ballering in this volume); if the ice were instead amorphous, the feature would peak at 46 μm (Curtis et al. 2005).

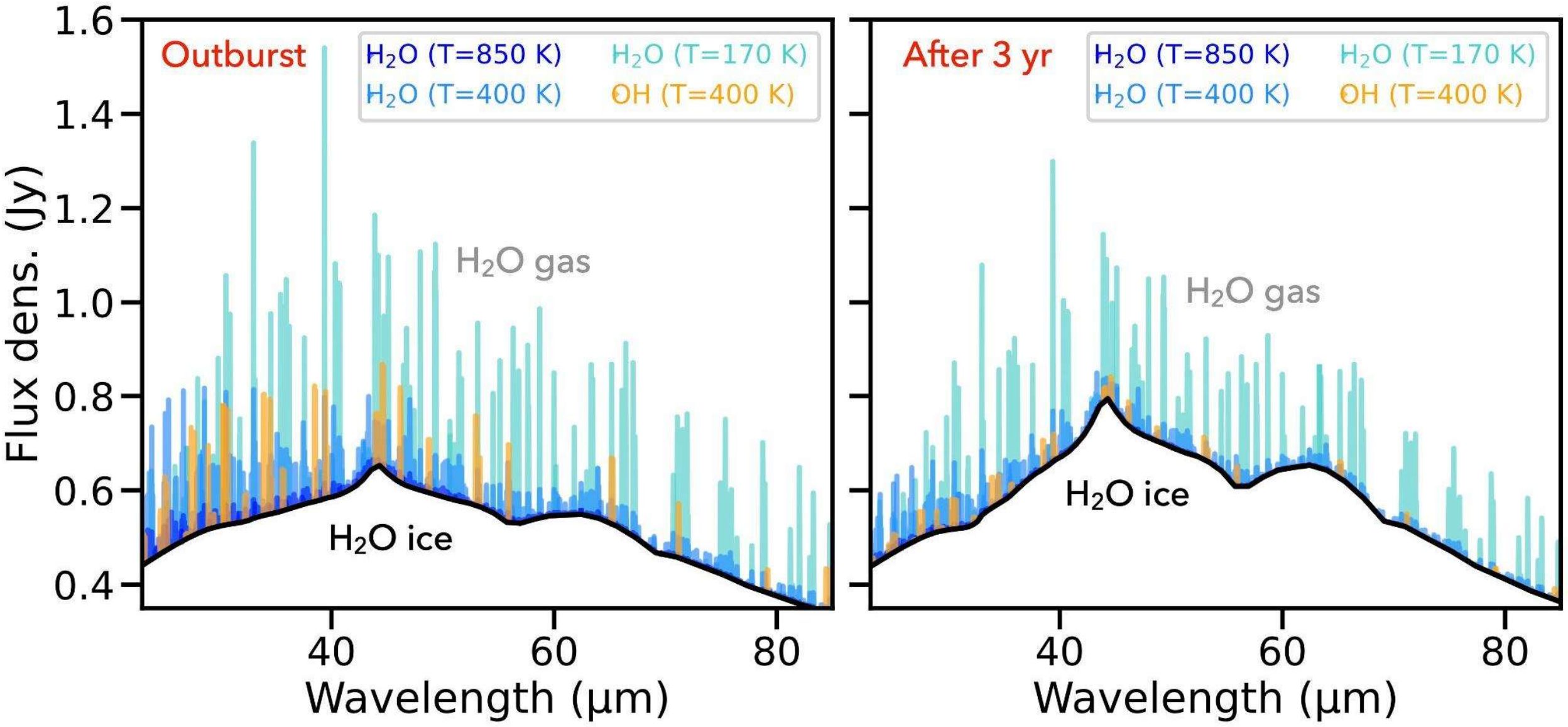

**Figure 1.** Proof-of-concept simulation of time evolution of the water ice and vapor spectrum following an accretion outburst in a Class II disk, based on Smith et al. (2025). During outburst (left), strong gas phase emission is observed on top of a reduced ice feature, with a water spectrum dominated by ~170 K water tracing ice sublimation at the snowline. Three years after outburst (right), the ice reservoir has partly built up again by freeze-out, increasing the ice feature in the spectrum. In parallel, the gas phase emission has reduced at warm temperatures (> 400 K); cold water vapor emission at ~170 K still reflects a long-lasting surface layer above the surface snowline, where the gas density is lower and the freeze-out timescales longer (Smith et al. 2025). OH lines are also expected to reflect the outburst, by being stronger when UV irradiation of the disk and water photodissociation is more intense (Banzatti et al. 2012, Tabone et al. 2024). PRIMA will likely reveal more gas lines affected by outbursts.

PRIMA will capture the relative change in both ice and vapor for the first time, enabling a test of how much of the ice reservoir at the snowline is sublimated as a function of outburst intensity and duration. We also anticipate the opportunity to spectrally resolve the location of the surface snowline from the spectrally-resolved broadening of water line widths at ~24-33 μm where the FIRESS resolving power is highest (Figure 2). The lower-energy water lines tracing the snowline are expected to be narrower during the outburst (reflecting a larger emitting area from the snowline pushed to larger radii) and broader with time after the outburst (as the snowline falls back into the quiescent location). Based on recent JWST results, it will take a few years to see this effect in the line profiles, since freeze-out is slow in the disk atmosphere (Smith et al. 2025). Over the expected lifetime of PRIMA of 5 years, we will be able to measure freeze-out timescales at the surface snowline in multiple disks and use their change in water emission to calibrate disk models of the snowline region and help determine the effects of ice sublimation events on the material available for planet formation.

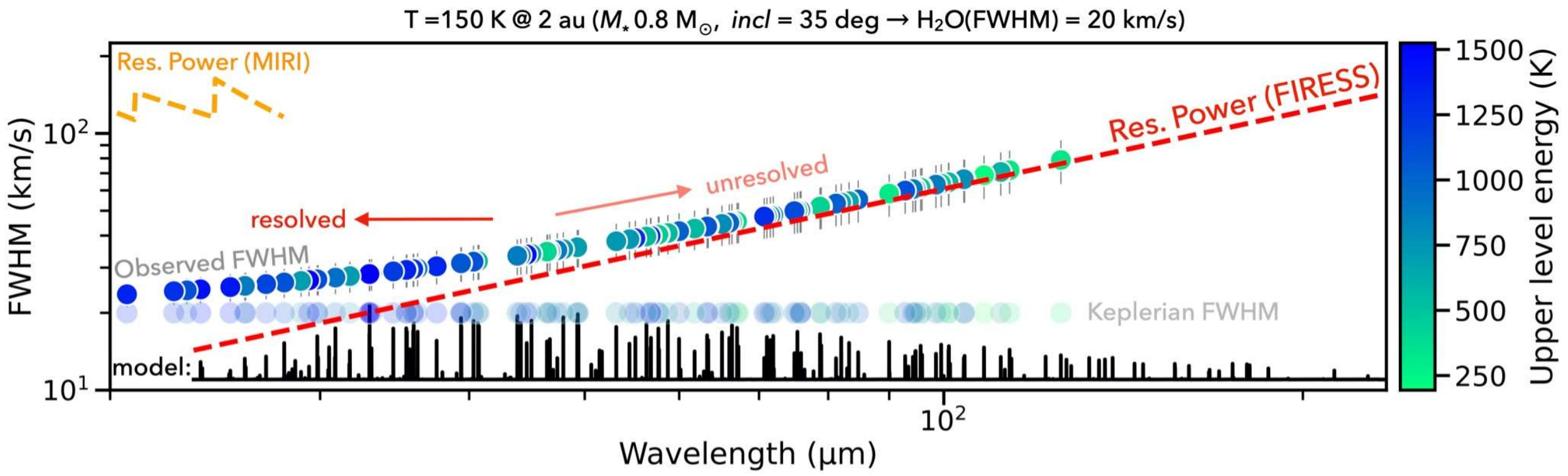

**Figure 2:** Simulation of water line broadening as observed with PRIMA FIRESS assuming the currently proposed resolving power R = 4400 × 112/$\lambda$(µm). The resolving power of MIRI is reported for comparison at the top left (from Pontoppidan et al. 2024 and Banzatti et al. 2025). Here we assume a water emission model (shown for reference in black at the bottom of the figure, from iSLAT - Jellison et al. 2024, Johnson et al. 2024) with T = 150 K and column density of $10^{17}$ cm$^{-2}$ as an annular region placed at 2 au (assuming a stellar mass of 0.8 $M_\odot$ and disk inclination of 35 deg), where the surface snowline is spectrally resolved with MIRI-MRS in the disk of GQ Lup in Banzatti et al. (2025). The colored datapoints illustrate the strongest water lines that will be observed with FIRESS, with color reflecting the upper level energy. Assuming a 20% FWHM error with FIRESS, lines at < 33 µm will be resolved at > 2$\sigma$ and should enable measuring the Keplerian radius of the snowline and its variation during and after the outburst.

## Instruments and Modes Used

FIRESS: high-resolution point source mode

## Approximate Integration Time

0.5 hours per source per epoch to detect at > 5$\sigma$ the 24 µm lines ($E_u$ ~ 1500 K, ~$10^{-17}$ W/m$^2$) using EX Lup as reference spectrum (continuum of 0.8 Jy at 24 µm), for a total of 2.5 hour per target (5 epochs, 1/yr), or ~25 hours for 10 targets. Additional time will be requested if higher cadence of recent outburst is monitored (see below).

## Special Capabilities Needed

Trigger TOO observations in case of outbursts detected during PRIMA observations (see below).

## Synergies with Other Facilities

We will coordinate with ongoing monitoring campaigns of variability in YSOs to identify outbursting sources both prior and during PRIMA observations (e.g., Contreras Peña et al. 2023, Lucas et al. 2017, 2024).

## Description of Observations

Current plan: select a sample of 10 disks by the time of PRIMA's launch based on their accretion history and recent bursts; the sample will include bursts of different intensity, which will enable to study their different effects on the snowline region. The sample will be monitored once a year, with possibly a monthly cadence for any disks that may be found to have an outburst right before PRIMA is ready for science or anytime during the first 4 years of operation (to enable monitoring for at least ~1 year after an outburst to study the fastest part of evolution in the gas phase spectrum of water).

# 62. PRIMA can measure the CO and Water Content of Exocomets

Gianni Cataldi (National Astronomical Observatory of Japan), Kenji Furuya (RIKEN), Shota Notsu (The University of Tokyo), Yao-Lun Yang (RIKEN)

Debris disks are leftover products from the planet formation process. They are analogous to the solar system's asteroid and Kuiper belts. Debris disks provide important insights into the formation and evolution of planetary systems. Mutual collisions among exocomets in debris disks are expected to release CO and water, which is photodissociated into C, O and H. Observations of the photodissociation products C and O can therefore be used to indirectly constrain the chemical composition of exocomets, in particular their CO and water content. However, only five debris disks have been detected by Herschel in [CII] or [OI] emission. We use the model by Kral et al. (2017) to construct a sample of disks to be observed in [CII] and [OI] emission with PRIMA/FIRESS. These observations are expected to significantly expand the number of disks with [CII] and [OI] detections, which is crucial to explore exocomet composition with a statistically significant sample. Comparison to the composition of solar system comets will help to set the solar system into context.

## Science Justification

### Broader Context

Debris disks consist of asteroidal and cometary bodies that continuously produce dust via mutual collisions (e.g., Hughes et al. 2018). The study of debris disks can give important insights into the origin, dynamics and evolution of extrasolar planetary systems (e.g., Wyatt 2018). This is analogous to the study of asteroids and comets in the solar system.

Gas, in the form of CO, has been observed with ALMA for about 20 debris disks (e.g., Lieman-Sifry et al. 2016; Moór et al. 2017; Cataldi et al. 2023). For systems with a large CO mass (larger than ~10-3 Earth masses), there is an ongoing debate about the origin of the gas (e.g., Cataldi et al. 2023). It could either be primordial (i.e. leftover from the protoplanetary phase) or secondary (i.e. produced, like the dust, via collisions of exocomets). On the other hand, for disks with a small CO mass (smaller than ~10-4 Earth masses), the photodissociation lifetime of CO is much shorter than the system age, implying continuous CO production, i.e. secondary gas (e.g., Matrà et al. 2017ab).

### Science Questions

Science question: In this proposal, we will focus on secondary gas, since it allows us to indirectly measure the chemical composition of the gas-producing exocomets (e.g., Kral et al. 2016, 2017; Matra et al. 2017ab). Thus, we will ignore disks with a large CO mass where the gas might be primordial.

Secondary CO gas is photodissociated by the interstellar radiation field (ISRF) on a short timescale of ~120 years (Visser et al. 2009), producing C and O. The colliding exocomets also release $H_2O$, which is typically photodissociated on even shorter timescales by stellar radiation (e.g., Cavallius et al. 2019), producing additional O and H. By observing these photodissociation products, we are able to constrain the exocomet composition in terms of CO and water content.

### Need for PRIMA

PRIMA will allow us to observe two photodissociation products of CO and $H_2O$: C (via the [CII] line at 158 µm) and O (via the [OI] line at 63 µm). These observations can constrain the amount of CO and $H_2O$ released by the exocomets. Equally important, [CII] and [OI] observations allow us to determine the physical parameters of the gas disk. This is crucial to interpret not only [CII] and [OI] data, but also CO data from ALMA (e.g., Kral et al. 2017; Matra et al. 2017b). In particular, [CII] and [OI] data allow to determine:

1. *Temperature:* Secondary gas is expected to be mainly heated by C photoionisation and cooled by [CII] 158 µm and [OI] 63 µm emission (Zagorovsky et al. 2010).
2. *Collider densities:* The expected gas densities are so low that CO and [OI] emission are likely not in LTE. Thus, the emission depends on the density of colliders (electrons and H). Electrons are mainly produced by C ionization, so observations of [CII] can determine the electron density. On the other hand, H is mainly produced by $H_2O$ photodissociation, so [OI] observations can indirectly constrain the H density.

Currently, there is no far-infrared facility that has the sensitivity required to detect [CII] and [OI] from secondary gas in debris disks. The Herschel Space Observatory detected [CII] and/or [OI] from only five disks (Brandeker et al. 2016; Donaldson et al. 2013; Riviere-Marichalar et al. 2012, 2014; Roberge et al. 2013). The improved sensitivity of PRIMA will allow detecting [CII] and/or [OI] from a significantly larger sample of disks (Figure 1).

## Interpretation Methods

To interpret the data, we will use models similar to those presented by Kral et al. (2016, 2017) and Marino et al. (2020). In those models, C and O are produced from CO photodissociation. The atomic gas viscously spreads into an accretion disk. The temperature and ionisation fraction are calculated. Finally, [CII] and [OI] fluxes are predicted using non-LTE radiative transfer. Comparison of the predicted fluxes to the data allows adjustment of the model parameters. If available, constraints from CO observations (whether detection or non-detection) by ALMA can also be included.

## Link to testable hypotheses

The secondary gas model by Kral et al. (2017) predicts [CII] and [OI] fluxes for a large sample of debris disks. The fluxes are predicted based on the fractional dust luminosity of the disks, which is a proxy for the dust (and thus gas) production rate. Our sample includes disks that are predicted to be detectable by PRIMA by the Kral et al. (2017) model, plus a few additional targets that have a CO detection from ALMA (Figure 1). In total, our sample consists of 16 targets.

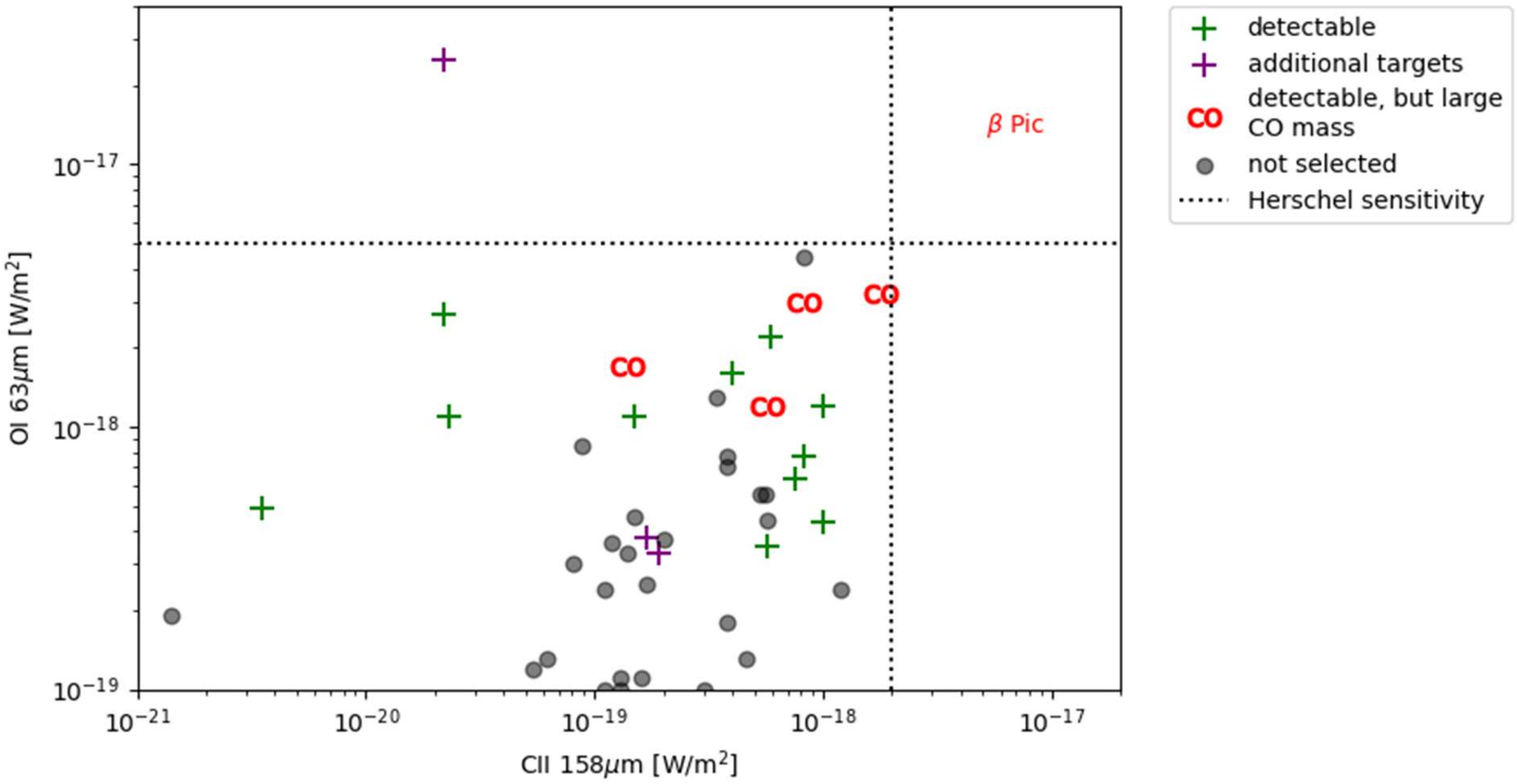

**Figure 1.** [CII] and [OI] fluxes predicted by the Kral et al. (2017) model. Our sample consists of the green crosses, which are the targets predicted to be detectable with PRIMA in [CII] and/or [OI] with an integration time of less than 5 h. Additional targets where Kral et al. (2017) predict non-detections, but have a CO detection from ALMA, are marked by purple crosses. Detectable disks with a large CO mass are excluded since their gas might be primordial (marked by red "CO"). β Pic is excluded because Herschel already detected both [CII] and [OI]. Two targets (TWA 7 and HD 109085) are not shown in the plot, but included in the sample because of their CO detections with ALMA. The black lines show the Herschel sensitivity limits.

## Instruments and Modes Used

The FIRESS spectrometer in its high-resolution mode will be used.

## Approximate Integration Time

We calculate the integration time by requesting a 5σ detection for each target. We take into account that the FIRESS sensitivity (in High-res mode) degrades with increasing continuum flux from the source.

Our sample includes four additional targets (HD 109085, HD 129590, HD 146897, HD 172555) for which Kral et al. (2017) predict non-detections in [CII] and [OI]. They were included because ALMA detected CO. We request 2 h for each of these additional targets, which is comparable to most of the other targets. One additional target (TWA 7) was not considered by Kral et al. (2017), but was detected in CO with ALMA. For this target we also request 2 h. The total integration time of this science case is about 32 h.

## Special Capabilities Needed

N/A

## Synergies with Other Facilities

There is a strong synergy with ALMA. Dust continuum observations with ALMA trace the population of exocomets, thus telling us where the gas is produced. CO observations (including non-detections) constrain the gas production rate.

## Description of Observations

To construct our sample, we selected all targets for which Kral et al. (2017) predict a [CII] or [OI] flux that can be detected with PRIMA with a maximum integration time of 5 hours. We exclude targets with a large CO mass, as their gas might be primordial. We also excluded β Pic for which Herschel already detected both [CII] and [OI]. HR 4796 and Fomalhaut are also excluded, as they were already observed by Herschel, and PRIMA might not deliver significantly deeper data because of their bright continuum flux.

We add to our sample additional targets that are predicted to be undetectable by Kral et al. (2017), but for which ALMA detected CO: HD 109085, HD 129590, HD 146897, HD 172555. Similarly, we add TWA 7, for which ALMA detected CO, but which was not considered by Kral et al. (2017).

The observational strategy is straightforward: we will use FIRESS in Pointed High-res mode. Our targets will not be resolved spectrally nor spatially. The list of our targets is given in Table 1.

**Table 1.** List of our targets. Teff is the effective temperature of the star, while f is the fractional dust luminosity of the disk (both taken from Kral et al. 2017). F60 and F160 are the continuum fluxes at 60 μm and 160 μm, respectively (Kral et al. 2017). The "5σ" columns show the 5σ sensitivity achieved with the requested integration time. The column "line" shows which lines are predicted to be detectable (it shows N/A if neither line is predicted to be detectable, but the target is included because CO was detected). The column "CO" shows whether ALMA observed CO or not, and whether it was detected or not. The requested integration time is shown in the last column. Note that for HD 138923, the predicted exposure time was very short, so we increased it to 0.4 h. For HD 172555, Herschel already detected [OI], while [CII] is predicted to be undetectable, but we still include the target in our sample since ALMA detected CO.

| name | Teff [K] | f | F60 [Jy] | F160 [Jy] | 5σ (60μm) [W/m²] | 5σ (160μm) [W/m²] | line | CO | Tint [h] |
|---|---|---|---|---|---|---|---|---|---|
| eta Tel | 9379 | 7.6e-4 | 0.47 | 0.076 | 1.2e-18 | 4.0e-19 | [CII], [OI] | non-detection | 4.8 |
| HD 38678 | 8477 | 8.9e-5 | 0.39 | 0.037 | 1.7e-18 | 5.9e-19 | [CII], [OI] | not observed | 1.9 |
| HD 61005 | 5491 | 2.2e-3 | 0.48 | 0.36 | 1.8e-18 | 1e-18 | [CII] | non-detection | 2.0 |
| HD 86087 | 9465 | 3.5e-4 | 0.69 | 0.32 | 2.4e-18 | 1.0e-18 | [CII] | not observed | 1.8 |
| HD 106036 | 8625 | 3.0e-4 | 0.032 | 0.0029 | 1.1e-18 | 9.2e-19 | [OI] | not observed | 0.58 |
| HD 131885 | 8913 | 5.4e-4 | 0.36 | 0.079 | 1.9e-18 | 7.6e-19 | [CII] | not observed | 1.3 |
| HD 138923 | 10000 | 1.1e-4 | 0.011 | 0.00097 | 1.1e-18 | 1.1e-18 | [OI] | non-detection | 0.4 |
| HD 144981 | 7538 | 2.1e-4 | 0.0079 | 0.00071 | 1.1e-18 | 1.1e-18 | [OI] | not observed | 0.4 |
| HD 164249 | 6295 | 8.7e-4 | 0.58 | 0.19 | 1.5e-18 | 5.7e-19 | [CII] | non-detection | 3.9 |
| HD 169666 | 6565 | 2.1e-4 | 0.028 | 0.0028 | 4.9e-19 | 4.2e-19 | [OI] | not observed | 2.8 |
| HD 182681 | 9621 | 2.7e-4 | 0.58 | 0.26 | 1.9e-18 | 8.3e-19 | [CII] | non-detection | 2.3 |
| TWA 7 | 4018 | 1.7e-3 | 0.092 | 0.044 | 7.0e-19 | 5.8e-19 | N/A | detection | 2.0 |
| HD 109085 | 6802 | 1.4e-5 | 0.25 | 0.21 | 1.2e-18 | 8.2e-19 | N/A | detection | 2.0 |
| HD 129590 | 5836 | 5.7e-3 | 0.36 | 0.067 | 1.5e-18 | 6.1e-19 | N/A | detection | 2.0 |
| HD 146897 | 6256 | 5.4e-3 | 0.66 | 0.1 | 2.2e-18 | 6.4e-19 | N/A | detection | 2.0 |
| HD 172555 | 7120 | 7.8e-4 | 0.32 | 0.029 | 1.4e-18 | 5.6e-19 | [OI] | detection | 2.0 |

# 63. Detecting and Characterizing Volatile Ices in Extrasolar Kuiper Belts

Christine Chen (STScI, JHU), Chen Xie (JHU), Sarah Betti (STScI), Nick Ballering (SSI), Farisa Morales (Caltech/JPL), Patricia Luppe (Trinity College Dublin), Geoff Bryden (Caltech/JPL), Meredith Macgregor (JHU), B. Sargent (STScI, JHU)

Kuiper Belt Objects in the outer Solar System are rich in water ice and other volatile ices, indicating that ices may have played an important role in the formation of the giant planets. Multiwavelength observations indicate that extrasolar planetary systems host planetesimal belts beyond the water ice snow line, analogous to the Kuiper Belt. Recent JWST NIRSpec scattered light spectroscopy of the HD 181327 and beta Pic debris disks have measured the 3 μm water ice feature, suggesting that water ice rich planetesimals may be common. PRIMA FIRESS spectroscopy will enable the detection and characterization of water and $CO_2$ ice in a large sample of debris disks. We propose a FIRESS study of 100 bright debris disks around solar-like FGK stars, already discovered using Spitzer and/or Herschel, to understand whether there are correlations between the properties of water ice and stellar spectral type and/or age.

## Science Justification

In our Solar System, water is the most abundant volatile molecule. It is found throughout the Solar System in trace quantities on the surfaces of Venus and Mars but in abundance on the surface of the Earth and in Trans Neptunian Objects (TNOs). Water ice is believed to have played an important role in the formation of Jupiter and Saturn via core-accretion. In the solar nebula, water ice doubled the available solid surface density beyond the snow line enabling the rapid accretion of solids into Earth mass cores. During the past twenty years, thousands of exoplanets have been discovered, 10-20% of which are gas giant planets. Observations of water in circumstellar disks are needed to determine whether water is a common building block in extrasolar systems and to place the water in our Solar System into context.

Debris disks are dusty disks around main sequence stars, representing the late stages of planetary system formation and evolution. Unlike protoplanetary disks, circumstellar dust and gas in debris disks are believed to be generated by collisions among planetesimals, planetary embryos, and planets. High angular resolution imaging in scattered light and thermal emission indicates that the majority of these systems possess planetesimal belts that lie beyond the water ice snow line, analogous to the Kuiper Belt (Hughes et al. 2018). Theoretical studies show that water ice is expected to survive despite sublimation and photosputtering by stellar ultraviolet photons (Kim et al. 2019). Indeed, early modeling of the scattered light and thermal emission of the debris disk around HD 181327 indicated that this system was water-ice rich (Chen et al. 2008, Lebreton et al. 2012). Far-infrared (far-IR) SED modeling of Herschel photometry for a handful of other young debris disks has also indicated the presence of icy grains (Morales et al. 2013). The exquisite sensitivity of PRIMA FIRESS will enable far-IR spectroscopic searches and characterization of solid-state water ice emission features at 42 and/or 63 μm.

Recent scattered light studies using JWST have enabled the detection and detailed characterization of volatile ices in two nearby (within 100 pc), young (~23 Myr) debris disks. Low resolution, NIRSpec IFU observations of HD 181327 have detected the broad 3 μm, solid-state, water ice feature, including a Fresnel peak at 3.1 μm, indicative of crystalline water ice (Xie et al. 2025). Similar observations of beta Pic have also detected the 3 μm water ice feature, this time indicating that the ice is amorphous. In addition, the beta Pic observations have revealed $^{12}CO_2$ and $^{13}CO_2$ frost features near 4.2 μm (Betti et al. 2025). Since beta Pic is an intermediate mass star, the discovery of water ice and $CO_2$ frost with an abundance ratio ~10:1 indicates that the planetesimals likely formed enshrouded in gas and dust during the early phases of star formation and disk evolution, similar to comets in our Solar System. While JWST has enabled the detailed characterization of ice grain properties (e.g. composition, phase - amorphous versus crystalline, size) in scattered light, JWST will only be able to study about a dozen of the brightest and most spatially extended debris disks spectroscopically.

PRIMA's outstanding sensitivity and wide wavelength coverage is expected to revolutionize the discovery and characterization of volatile ice in debris disks. To date, more than one thousand debris disks have been identified via far-IR excess using Spitzer (Chen et al. 2020) and Herschel (Matthews et al. 2014). Each object is relatively bright such that FIRESS is expected to obtain a high SNR 24 - 235 μm spectrum within an hour of observing time per target for the majority of targets. Models of the dust thermal emission inferred from scattered light observations indicate that both water ice and $CO_2$ frost may be detected in high SNR FIRESS spectra (See Figure 1). $CO_2$ frost has been even more difficult to detect than water ice in scattered light observations because $CO_2$ frost is less abundant. For example, the scattered light spectra of HD 181327 clearly show the 3 μm water ice feature but only a possible low significance detection of the 4.2 μm $CO_2$ feature that requires further confirmation (Xie et al. 2025).

PRIMA is expected to usher in an era of demographic studies in which volatile ices can not only be detected and characterized but also traced in the circumstellar environments as a function of host star spectral type and age. The JWST NIRSpec scattered light observations have revealed hints that the detectability of water ice may depend on host star spectral type. To date, three objects have been observed using the JWST NIRSpec IFU and high contrast imaging techniques: 49 Cet (A1V), beta Pic (A6V), and HD 181327 (F5V). Thus far, the depth of the 3 μm water ice feature appears anticorrelated with stellar spectral type with the latest spectral type star having the deepest feature (~30%) and the earliest type star having a very weak feature. Measurements of the coma gas production rate for Solar System TNO MU69/Arrokoth suggest that hypervolatile ices, such as CO, $N_2$, $CH_4$, Ar, and $O_2$, may have been depleted from small bodies by an age of 20 Myr (Lisse et al. 2022). This lifetime is much shorter than that of refractory H-bonded ices, such as HCN, $CH_3OH$, and $H_2O$, that are expected to survive over the lifetime of the Solar System.

The interpretation of solid-state ice features in PRIMA FIRESS spectra is expected to be straight-forward. Debris disks contain particles of a variety of sizes from sub-micron sized grains that efficiently scatter starlight to millimeter sized grains that efficiently emit at millimeter wavelengths (Hughes et al. 2018). Far-IR spectra with solid-state features can be modeled assuming thermal emission from an optically thin distribution of grains in radiative equilibrium with the star. In the simplest case, the spectrum can be approximated as the sum of solid-state features emitted by small grains and large black bodies that dominate the continuum (e.g. de Vries et al. 2012). There are already-existing libraries of optical constants for ices at far-IR wavelengths that can be used to interpret these observations (Hudgins et al. 1993).

Thus, PRIMA FIRESS observations are expected to revolutionize our understanding of the composition of planetesimal belts beyond the snow line, analogous to the Kuiper Belt, enabling detailed demographic studies of the properties and lifetimes of volatile ices.

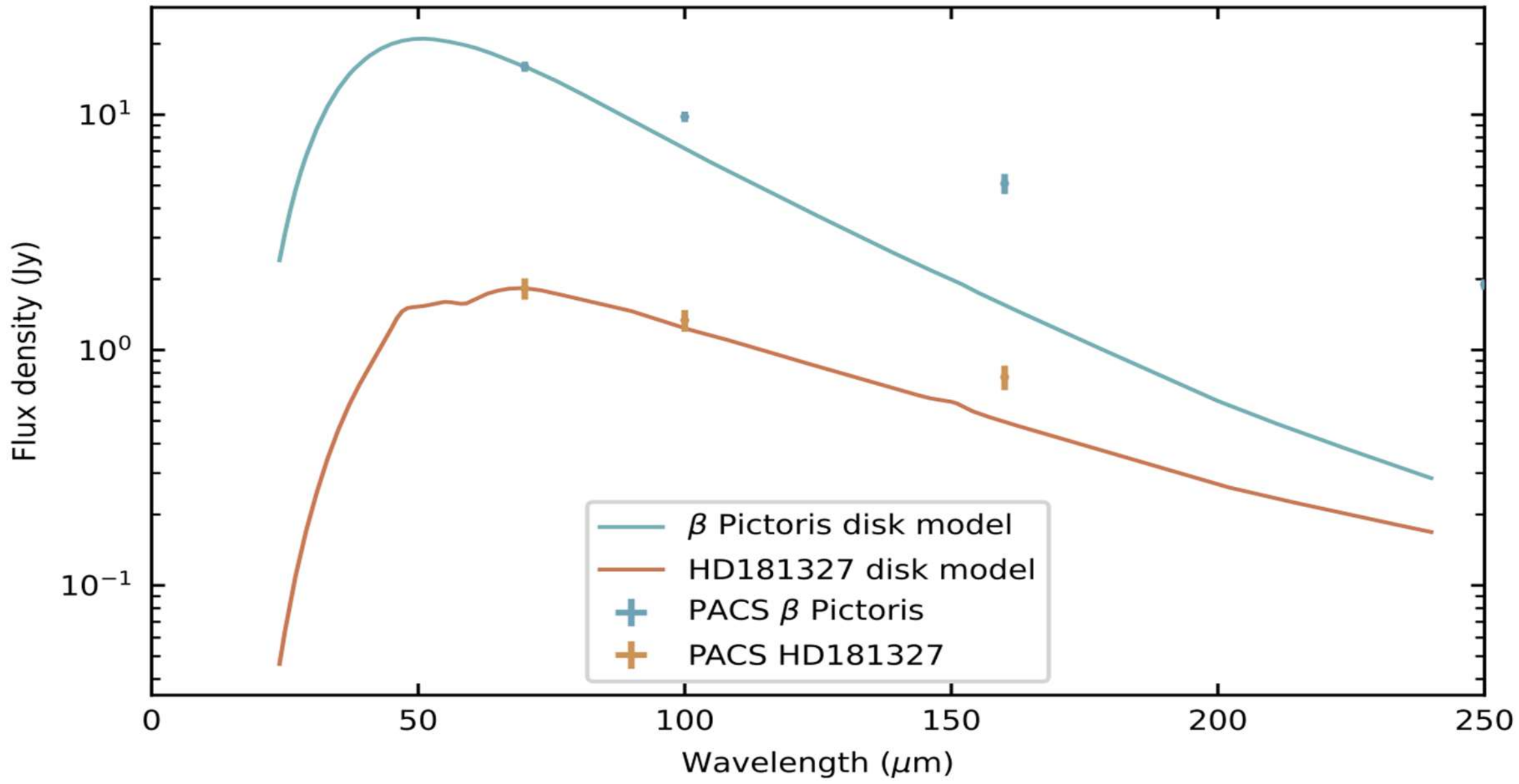

**Figure 1.** Estimated PRIMA FIRESS spectra of the debris disks around beta Pic and HD 181327 with published Herschel PACS photometry overlaid (Lebreton et al. 2012, Ballering et al. 2016), illustrating the expected detectability of 42 and 63 μm water ice and 160 μm $CO_2$ frost features. The estimated spectra were extrapolated from JWST scattered light observations that detect and characterize water ice and/or $CO_2$ frost rich grains over a portion of the disk, scaled to the 70 μm brightness.

## Instruments and Modes Used

The proposed observations require 100 pointed high-resolution observations with FIRESS. Spitzer and Herschel observations indicate that most known debris disks are sufficiently bright that they can be rapidly observed using FIRESS in high resolution mode (R~4400) even though low-resolution mode is sufficient to fulfill the science goals.

## Approximate Integration Time

Modeling of debris disk, solid-state, water-ice features (based on JWST scattered light spectroscopy) suggests that the 43 μm crystalline water ice feature will be broad with a peak flux ~10% of the continuum (see Figure 1). Therefore, far-IR spectra with a spectral resolution (R~100) and a signal to noise ratio (SNR ~ 50) should be sufficient to measure the shape of the 43 μm water ice feature well. Spitzer and Herschel identified hundreds of bright debris disks with 70 μm fluxes between 10 mJy and several Jy. For a 10 mJy source, the PRIMA ETC indicates that FIRESS in high resolution mode will detect the 43 μm continuum with a SNR~50 in approximately 1 hour of integration time, per spectroscopic setting, once the data are binned. Since the 43 μm water ice feature is in the overlap region between FIRESS Channels 1 (24-43 μm) and 2 (42-76 μm), both spectrograph settings are needed to measure the shape of the 43 μm feature well. Therefore, each target requires ~2 hours of on source integration time. For 100 sources, this is a total integration time of approximately 200 hours

The $CO_2$ ice feature is expected to be substantially fainter than the water ice feature with a line to continuum ratio of 0.5%. Therefore, spectra with very high SNR (~600) will be needed to detect $CO_2$. As a result, $CO_2$ can only be detected in very bright targets with 43 μm fluxes > 300 mJy for which it will be possible to obtain very high SNR spectroscopy.

## Special Capabilities Needed

None.

## Synergies with Other Facilities

PRIMA FIRESS will enable a demographic study of water ice in FGK debris disks for the first time to understand whether icy planetesimals are common around solar-like stars and whether these stars can retain their icy reservoirs.

Water ice has been detected in two sources using JWST NIRSpec scattered light spectroscopy thus far. For these sources, JWST and PRIMA will provide a complete spectral view from the near- to the far-IR, essential for understanding both the scattering and thermal emission properties of ice. Since the number of sources that can be studied using JWST NIRSpec is limited (~10 sources), PRIMA will significantly expand the number of debris disks studied by enabling observations of a much larger sample. The JWST NIRSpec observations (1) are limited to highly inclined debris disks that are strongly forward scattering, (2) have limited sensitivity to low surface brightness objects, and (3) are saturated near the host stars because the host stars are very bright. SPHEREx and ground-based high contrast imagers (e.g. VLT/SPHERE and Gemini/GPI) are not expected to contribute to the study of water ice in debris disks because SPHEREx lacks the angular resolution necessary to resolve debris disks from their host stars and ground-based instruments are not designed to obtain 3 μm spectroscopy because of telluric effects.

## Description of Observations

Observations of 100 debris disks around solar-like stars using the high-resolution mode of PRIMA's FIRESS spectrometer are needed to (1) determine whether water ice is common around solar-like stars and (2) trace the demographics of water and $CO_2$ ice as a function of stellar spectral type and age. Early JWST NIRSpec observations indicate that water ice may be challenging to detect around A-type stars because of their strong ultraviolet radiation. Theoretical studies of solar system KBOs indicate that volatile ices may be removed over time. Existing Spitzer and Herschel photometry indicates that the nearby solar-like debris disks are bright with 70 μm fluxes > 10 mJy.

Water ice has two solid-state features in the PRIMA FIRESS bandpass, one at 43 μm and another at 63 μm. Since the 43 μm water ice feature is in the overlap region between FIRESS Channels 1 (24-43 μm) and 2 (42-76 μm), both spectrograph settings are needed to measure the shape of the 43 μm feature well. In addition, observations over the full FIRESS wavelength range (24-261 μm) will facilitate detailed modeling of the continuum. Such modeling is critical to the interpretation of solid-state, water ice features. To detect a 43 μm water ice feature commensurate with that expected for HD 181327, each target spectrum should be observed with a minimum SNR~50 when the data are binned by a factor of 100. To detect a $CO_2$ ice feature, each target spectrum should be observed with a minimum SNR~600 when the data are binned by a factor of 100. Therefore, observations of the complete sample of 100 debris disks are expected to require ~200 hours of on-source integration time.

# 64. Measuring Interstellar Carbon Abundance and Depletion via 158 μm [CII] Absorption

Christopher Clark (Space Telescope Science Institute for the European Space Agency), Julia. C. Roman-Duval (Space Telescope Science Institute), Suzanne, C. Madden (AIM, CEA, CNRS, Universite Paris-Saclay), Marc Mertens (Max-Planck-Institut für Radioastronomie; I. Physikalisches Institut, Universität zu Köln), Claire E. Murray (Space Telescope Science Institute; Johns Hopkins University), Jürgen Stutzki (I. Physikalisches Institut, Universität zu Köln), Elizabeth Tarantino (Space Telescope Science Institute), and Kirill Tchernyshyov

Carbon plays key roles in the InterStellar Medium (ISM)—as a constituent of dust, as the carrier of the dominant far-infrared cooling line, and as a component of various important molecules. But despite this, there are very few (~19) measurements of the abundance and depletion of carbon in the diffuse ISM. As with other elements, these measurements are traditionally performed in the ultraviolet. But for carbon, ultraviolet measurements are extremely difficult, and only 19 carbon depletion values have been reported in the literature to date. This means that the fraction of interstellar carbon locked up in dust, and how this fraction varies with environment, is effectively unconstrained. Fortunately, PRIMA will enable the novel method of measuring the abundance and depletion of carbon in the diffuse Milky Way ISM, by observing absorption of the 158 μm [CII] line, using FIR-bright nearby galaxies as background continuüm sources. We identify 17 candidate sightlines where we expect PRIMA-FIRESS will be able to detect Galactic 158 μm [CII] absorption, thereby almost doubling the total number of measurements diffuse interstellar carbon abundance measurements available, and providing constraints on carbon depletion along high-extinction sightlines that will never be accessible in the ultraviolet.

## Science Justification

The fraction of interstellar carbon that is in the gas phase, compared to the fraction that is depleted into dust grains, is almost unconstrained. As with most metals, the gas-phase abundance of carbon in the neutral ISM is typically measured via UltraViolet (UV) absorption features. But in the case of carbon, the only available absorption features are either incredibly faint or super-saturated. As such, only 19 abundances for carbon in the cold ISM have been published (Sofia et al., 2011, and references therein).

A consequence of this lack of certainty is that the fraction of carbon depleted onto dust, and especially how this depletion varies with column density, is almost unknown (unlike for all other dust-refractory elements). This is illustrated starkly by Jenkins (2009), who show that the variation of carbon depletion with column density is effectively unquantified. Moreover, UV measurements of abundances – for all elements, not just carbon—become effectively impossible for sightlines with high integrated column densities, due to interstellar extinction towards the OB stars used as background sources. This prevents us from measuring depletions in many Galactic plane environments, where complex chemical evolution is no doubt occurring. Understanding how the fraction of carbon incorporated into dust grains changes with environment (especially density), and how this fraction varies with respect to other dust-forming elements, is key for improving our currently-poor knowledge of dust chemistry and its evolution (Jenkins, 2009; Roman-Duval et al., 2022).

In this science case, we summarise a new approach for measuring carbon abundance and depletion, by observing 158 µm [CII] absorption with PRIMA-FIRESS. This method is presented in full in Clark et al. (2025), where proof-of-concept using SOFIA-upGREAT is demonstrated, with a tentative detection of Galactic [CII] absorption, using FIR-bright nearby galaxies IC342 as a background continuum source. However, due do its relatively poor sensitivity, there were only 3 sightlines in the entire sky where SOFIA-uPGREAT would have been able to detect [CII] absorption in this way. PRIMA's quantum leap in sensitivity offers a way to dramatically improve our measurements of interstellar carbon abundance and depletion using this technique. We direct the reader to Clark et al. (2025) for a full description of the method, which we summarise here.

To find candidate sightlines for observation with PRIMA-FIRESS, we used the Herschel-PACS data presented by Clark et al. (2018), which provides 160 µm photometric observations of 753 nearby galaxies, with which we determined the 158 µm dust continuum brightness, by convolving to PRIMA's 20" resolution at this wavelength.

In the diffuse ISM, the vast majority of gas-phase carbon resides in C+ (Wolfire et al., 2010; Goldsmith et al., 2012); and at excitation temperatures <25 K (which is expected in the diffuse ISM), >95% of C+ is in the [CII] fine-structure line ground state. This makes [CII] an excellent tracer of carbon in the diffuse ISM. Using the HI data of HI4PI Collaboration et al. (2016) and the CO data of Planck Collaboration et al. (2014) to estimate the atomic and molecular gas column density along our candidate sightlines, we discard any candidate sightlines where molecular gas accounts for >20% of the total column, to ensure that we are sampling diffuse ISM. Carbon's transition to $C^0$ only occurs in a thin layer surrounding molecular clouds, within which carbon can also be incorporated into CO; by only considering sightlines with >80% atomic gas column, <3% of carbon should be found in these non-C+ phases, on average.

The 13' slit of PRIMA-FIRESS in band 4 (where [CII] will be observed) will measure the level of Milky Way [CII] emission towards each sightline, allowing this source of contamination to be removed.

The high-resolution mode of PRIMA-FIRESS will provide 68 km/s velocity resolution at 158 µm, meaning that the [CII] absorption line towards most candidate sightlines will be unresolved (assuming [CII] velocity width is typically 0.66x HI velocity width—Langer et al., 2014; de Blok et al., 2016). We simulate the expected peak per-channel absorption feature depth that would be observed by PRIMA-FIRESS for each of our candidate sightlines, and there are generally at depths of no more than a few percent. Fortunately, PRIMA-FIRESS is expected to be able to detect spectral features that have depth of as little as 1% the continuum level (J. Glenn, priv. comm.).

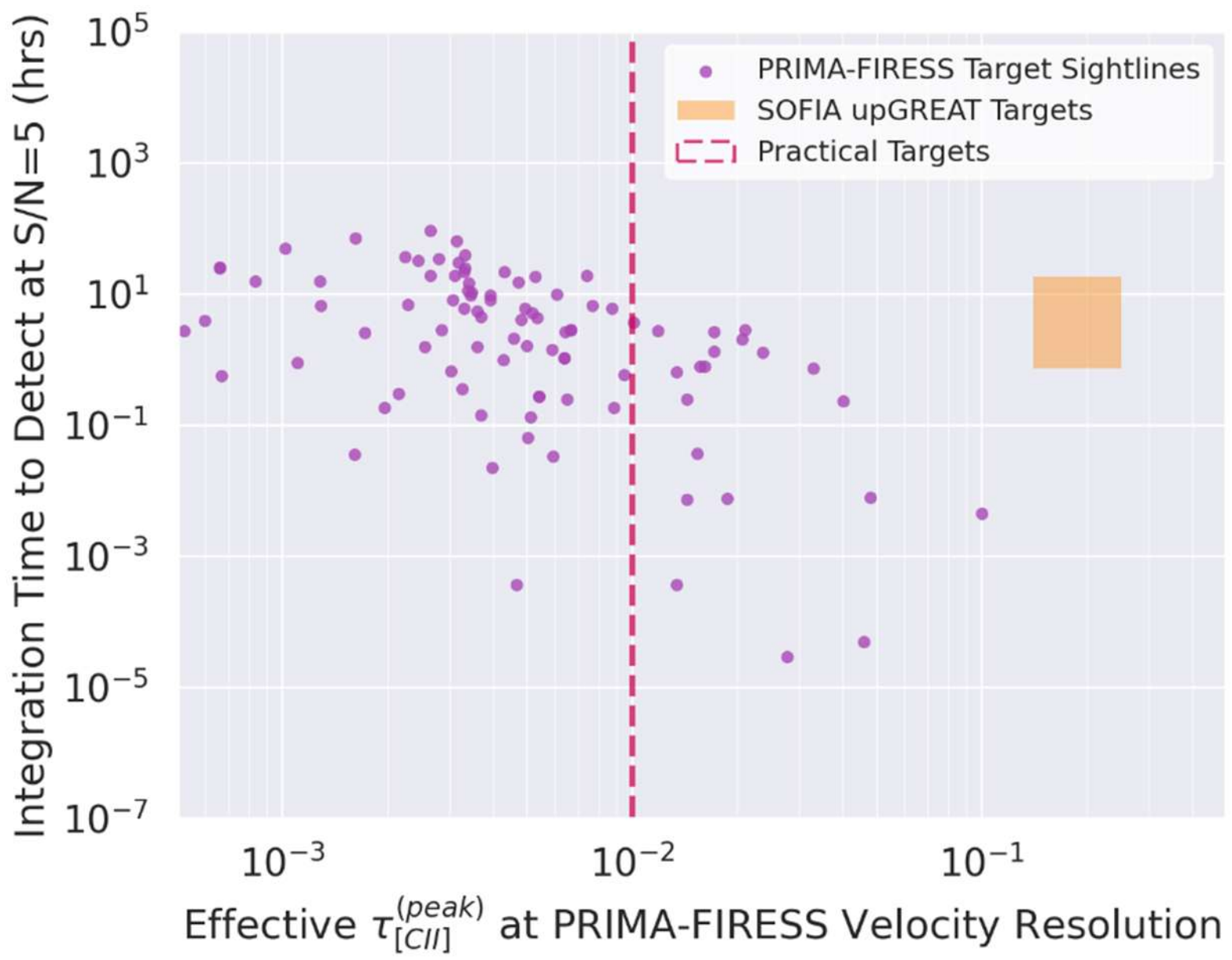

In total, we find 21 sightlines (NGC0891, NGC1266, NGC2110, NGC2146, NGC2207, NGC2273, NGC2403, NGC3256, NGC3283, NGC4945, NGC5728, NGC6300, NGC7331, NGC7714, UGC02826, ESO097-013, ESO373-008, ESO493-016, IC0010, IC0342, IC2163) where we predict PRIMA-FIRESS would observe Milky Way [CII] absorption features with >1% depth (and many more targets accessible if PRIMA-FIRESS baseline stability is allows it to observe absorption features that are <1%). The peak absorption line opacity depth observed by PRIMA for each sightline is plotted in below, along with their predicted integration times. With these 21 sightlines, PRIMA would more than double the number of measurements carbon abundance (and therefore depletion) in the Milky Way ISM, moreover sampling sightlines inaccessible in the extinction-affected UV, providing a step-change in our ability to probe this vital ISM building-block.

## Instruments and Modes Used

Pointed high-resolution FIRESS observations at 158 microns, on 21 sources

## Approximate Integration Time

To observe all 21 candidate sightlines would take about 20 hours to get to 1% of the continuum level, which is the baseline accuracy limitation . For an even more efficient pilot observing plan, we could observe the brightest 10 sightlines in less than 2 hours, to arbitrarily-high S/N (ie, limited only by instrumental systematics), as PRIMA-FIRESS should be able to detect the dust continuum from bright nearby starburst galaxies (such as NGC4945) to extremely high S/N almost instantly.

## Special Capabilities Needed

None, other than the ability for PRIMA-FIRESS to safely observe extremely bright (eg, gigajansky per steradian) continuum sources.

## Synergies with Other Facilities

Compliments the UV-based carbon depletion & extinction measurements obtainable with Hubble and HabWorlds in low-extinction sightlines.

## Description of Observations

PRIMA-FIRESS high-resolution mode observations of each candidate sightline at 158 µm, to sufficient depth to detect Wilky Way [CII] absorption. As the PRIMA-FIRESS Band 4 slit is expected to have a length of 13 arcminutes, we plan to use this to measure and subtract contamination from Milky Way [CII] emission along the line of sight. To maximise the quality of this subtraction, it may make sense to perform 2 observations for each sightline, with the slit in orthogonal directions.

For many sightlines, limiting factor for detection will be channel-to-channel baseline sensitivity, to allow detection of single-channel absorption features at a 1% level (albeit with continuum detected to arbitrarily-high S/N).

# 65. FIELDMAPS-2: Multi-scales Magnetic Field in the Bones of the Milky Way

Dr. Simon Coudé (Worcester State University/Center for Astrophysics), Prof. Ian Stephens (Worcester State University), Prof. Henrik Beuther (Max Planck Institute for Astronomy), Prof. Thiem Hoang (Korea Astronomy and Space Science Institute), Dr. Nicole Karnath (Space Science Institute), Dr. Philip Myers (Center for Astrophysics), Dr. Dylan Paré (Villanova University), Dr. Kate Pattle (University College London), Dr. Thushara Pillai (MIT), Dr. Howard Smith (Center for Astrophysics), Prof. Archana Soam (Indian Institute of Astrophysics), Dr. Le Ngoc Tram (Max Planck Institute for Astronomy)

Little is known about the role of magnetic fields on the star formation process within the dense filamentary structures found along the Galaxy's spiral arms. These dark filaments, nicknamed the "bones" of the Milky Way, are the best objects to study gas accretion in the magnetized potential of the spiral arms, as magnetic fields in diffuse media can guide the flow of gas onto filaments. The FIELDMAPS survey of dust polarization at 214 µm with SOFIA already hinted that the magnetic field was dynamically important in the densest material along the spine of bones, but the data lacked the sensitivity needed to probe the larger scale field in the more diffuse gas. Using PRIMAger's unparalleled capabilities, we will expand the FIELDMAPS survey to fully observe the far-infrared dust polarization in 18 bones and their surroundings down to a column density $N_{H_2}$ of at least $10^{21}$ cm$^{-2}$, for a total coverage of 19 square degrees in 11.9 hours. This represents a factor 27 improvement in coverage with PRIMA. We will use Histograms of Relative Orientation (HROs) to quantify the trends between the magnetic field structure and the density structures in the bones, and we will test the theoretical prediction that magnetic fields support the accretion of gas onto filaments. We will also use the Davis-Chandrasekhar-Fermi (DCF) technique to measure the plane-of-sky amplitude of the magnetic field in each bone and quantify their contribution to supporting them against gravity. Finally, the multi-wavelength data obtained with PRIMAger will enable studies of grain alignment efficiency in these clouds.

## Science Justification

Star formation primarily occurs in spiral arms where gas compresses within the magnetized spiral potential. Magnetic fields are considered a critical component for the formation of stars, as they can set the time scales for formation and can guide the flows of gas (e.g., Pattle et al. 2023). The so-called "bones" of the Milky Way (e.g., Zucker et al. 2015; 2018) are the densest velocity-coherent filaments identified in the Galactic disk, with extents > 20 pc that are mostly parallel to the plane. These elongations facilitate the study of their structure while limiting confusion along the line-of-sight. Bones may therefore be the molecular clouds best suited to study how gas collects within spiral arms. Observations thus far have been unable to fully constrain the role of magnetic fields for collecting mass in star-forming clouds, as previous and current telescopes cannot simultaneously resolve the magnetic field within molecular clouds and trace their structure at larger scales in the ISM. This is unfortunate because magnetic fields are expected to be at their most dynamically important at volume densities < $10^4$ $cm^{-3}$ (e.g., Crutcher et al. 2010; Crutcher 2012), which matches the density of the gas in and around the bones.

Interstellar dust grains are known to align with magnetic fields, which polarizes their thermal emission in the far-infrared (Andersson et al. 2015). Since dust is ubiquitous in the interstellar medium, far-infrared polarimetry is a powerful tool to probe magnetic fields in most environments within our Galaxy (Pattle & Fissel 2019). Using the HAWC+ camera on the Stratospheric Observatory for Infrared Astronomy (SOFIA), we observed the polarization signature of ten bones at 214 µm as part of the "Filaments Extremely Long and Dark: a MAgnetic Polarization Survey" (FIELDMAPS; Stephens et al. 2022; Coudé et al. 2025). Figure 1 shows the resulting plane-of-sky magnetic field structure for one of these bones, Filament 10, compared to lower resolution data from the *Planck* telescope at 850 µm. We revealed with FIELDMAPS a more complex magnetic field structure along the length of bones than seen by *Planck* or predicted by theory for filaments (Tomisaka 2015). Recent simulations hint that this complex structure may be a result of how these filaments are formed (Zhao et al. 2024), but the HAWC+ data lacks the sensitivity required to detect dust polarization in the low-density regions that would connect the cloud-scale magnetic field to larger scales.

With its unparalleled sensitivity in polarization (Dowell et al. 2024), the PRIMAger camera will probe the magnetic field structure at scales inaccessible by HAWC+ and at a much higher resolution than *Planck* (by at least a factor 11). We will use PRIMA to map the full magnetic field morphology of 18 confirmed bones and their surrounding ISM, which will provide us with the best picture of how magnetic fields interact with star-forming filaments in the spiral potential of our Galaxy. These high-resolution data of bones may also enable a direct comparison with the magnetic field structure observed in the arms of other spiral galaxies (Borlaff et al. 2023).

As part of the expanded FIELDMAPS survey with PRIMA, we will use Histograms of Relative Orientations (HROs) to quantify the trends in the orientation of magnetic fields relative to the density structures in the bones of the Milky Way (Soler et al. 2013). This will test the prediction that magnetic fields in lower density media, e.g. striations, contribute to increasing the gas accretion onto filaments (Pillsworth & Pudtritz 2024). We will complement this analysis by using the Davis-Chandrasekhar-Fermi (DCF) technique (e.g., see Pattle & Fissel 2019), combined with line surveys from radio observatories (e.g., CO, $NH_3$, $H_2CO$), to measure the magnetic field amplitudes along the spine of the bones and in the gas surrounding them. Preliminary results from HAWC+ indicate that the magnetic field in bones can resist gravitational collapse along most of their length (Stephens et al. 2022), and the data from PRIMA will establish if it is a universal characteristic of these types of star-forming filaments. Additionally, future measurements of Zeeman splitting and Faraday rotation in the radio (e.g., SKA; Heald et al., 2020) will complement the field strengths measured with PRIMA by providing additional information about the line-of-sight component of the magnetic field in bones.

Finally, the multi-wavelength polarization observations by PRIMAger will allow us to constrain the grain composition models (i.e., separate vs. composite; Tram et al. 2024), alignment mechanisms (e.g., radiative torques; Tram et al. 2024), and evolution (i.e., growth and disruption) from low to high gas densities within filaments. The largest grain sizes probed by studies of starlight polarization in filamentary structures are under 0.15 µm (Roychowdhury et al. 2025). Meanwhile, PRIMAger's observations of thermal dust polarization in denser gas will provide insights into larger grain sizes and thus supporting evidence for anisotropic grain growth effects (Hoang 2022; Tram et al. 2025). This represents a crucial advance in understanding dust physics and the role of grain geometry in the polarization efficiency observed in the far-infrared.

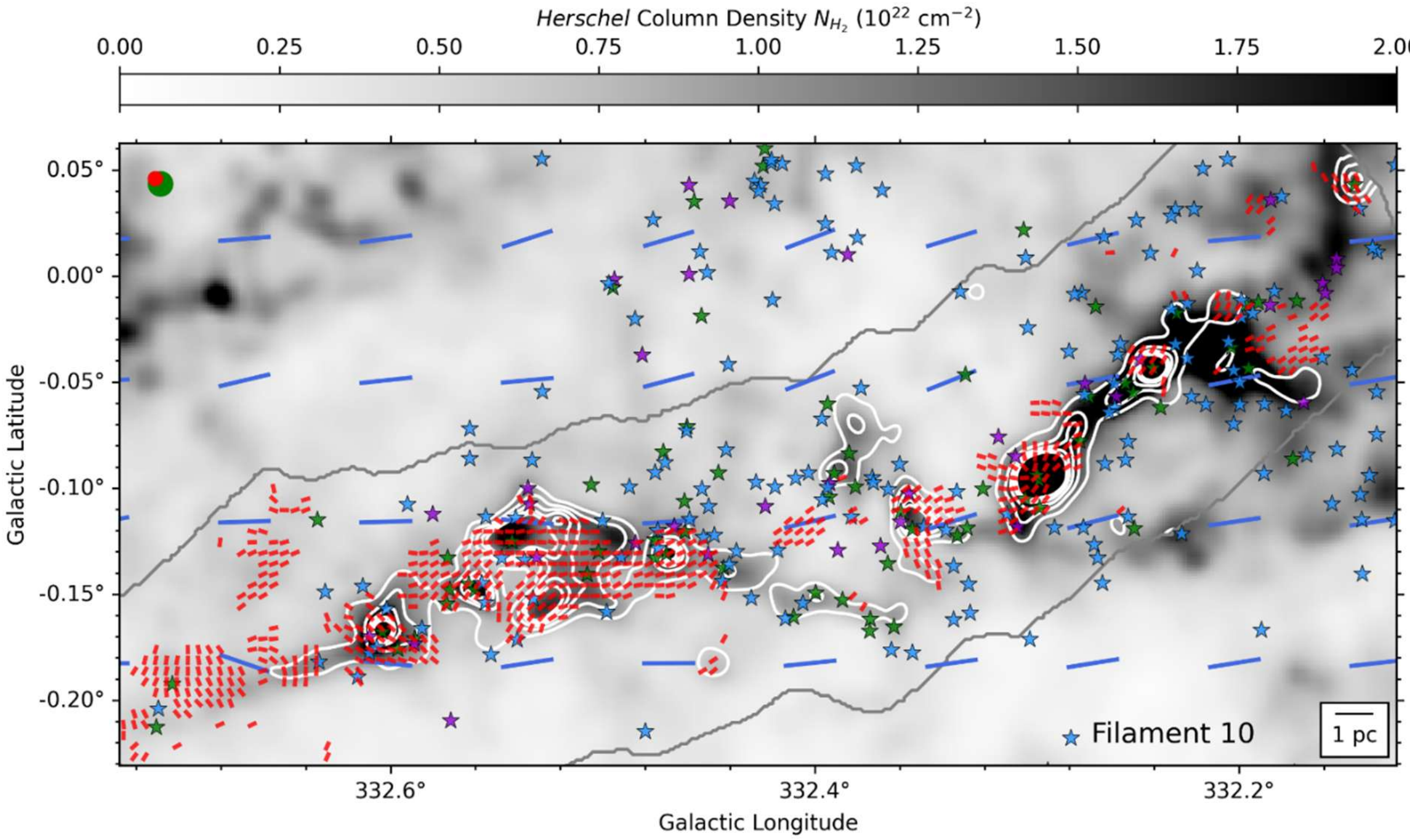

**Figure 1.** Magnetic field structure in Filament 10 ($D\sim3$ kpc) as inferred from HAWC+ 214 µm (red) and Planck 850 µm (blue) observations and plotted on the Herschel-derived $N_{H_2}$ column density map (gray scale). The white contours trace the Stokes I emission at 214 µm and the gray contour traces the $\sim0.1$ square degree area observed with HAWC+. Young Stellar Objects are identified with star symbols. The circles denote the SOFIA (red) and Herschel (green) beams. While HAWC+ was only sensitive to the densest material in the spine of the bone, PRIMAger will fully recover polarization over the entire area shown above, and beyond, at a much higher resolution than Planck (an improvement of a factor 11). Figure sourced from Coudé et al. (2025).

## Instruments and Modes Used

PRIMAger mapping mode using all four polarimeter bands to cover 19 fields (1 sq.degree each) to map 18 targets (one target needs 2 fields).

## Approximate Integration Time

According to the updated PRIMAger sensitivity tables for mapping (see also Dowell et al. 2024), the 5-sigma polarized surface brightness sensitivity over a 1 square degree map that can be achieved in 10 hours at 235 µm is 0.25 MJy/sr. Since sensitivity increases as the square root of the integration time, our targeted sensitivity of 1.0 MJy/sr will be achieved with a time of 37.5 minutes per square degree. For the full 19 square degree coverage planned for this survey, the total on-sky integration time will therefore be 11.9 hours.

## Synergies with Other Facilities

This program will have synergies with millimeter and radio observatories around the world. Specifically, this program will rely on high-resolution, in velocity space, spectroscopic data of dense gas tracers from surveys of the galactic plane (e.g., CO isotopologues with FCRAO and APEX), as well targeted follow-up observations at smaller spatial scales (e.g., $NH_3$ with GBT and $H_2CO$ with ALMA). Additionally, future capabilities to observe Faraday rotation and Zeeman line splitting in the interstellar medium (e.g., ALMA, SKA, ngVLA) will contribute the line-of-sight component of the magnetic field and will complement the PRIMA observations of the plane-of-sky field in bones.

## Description of Observations

We will observe 18 confirmed bones from Zucker et al. (2018) with PRIMAger in polarization. The original FIELDMAPS survey covered a total of $\sim 0.7$ square degrees over only 10 bones from this sample, in part due to the sensitivity constraints of HAWC+ limiting the number of targets that could be observed efficiently. In comparison, these observations will cover a 19 square-degrees area, an improvement of a factor 27 in coverage in only 11.9 hours. We target 19 fields centered on the following 18 bones and *Herschel*-identified filaments from Zucker et al. (2018): Filaments 1 to 10, Nessie, G24, G26, G28, G29, G47, G49, and G64. Due to its length on the sky, Nessie will need to be covered by two overlapping pointings.

We aim for a sensitivity of 1 MJy/sr in order to achieve a polarization detection at the 1% level for a contrast level as low as 100 MJy/sr between the low-density material surrounding the bones and the galactic emission at large scales. We expect to probe the magnetic field structure down to a column density of at least $10^{21}$ cm$^{-2}$ (see Figure 1). We based the sensitivity calculation on the *Herschel* 250 µm data as a lower limit, and the dust emission at 235 µm is expected to be the weakest for the PRIMA bands due to it peaking at shorter wavelengths. Furthermore, for an integration time of 11.9 hours, we will reach sensitivities of 1.4 MJy/sr at 172 µm, 1.88 MJy/sr at 126 µm, and 2.6 MJy/sr at 92 µm.

Although the bones themselves are elongated in Galactic longitude and have a very high aspect ratio due to their filamentary nature, extending the maps to cover a full degree in Galactic latitude will facilitate the identification of the larger scale galactic field and how it can be differentiated from the field in low density regions around the bones. As illustrated in Figure 1, the area covered by HAWC+ was limited due to observation constraints, and the detection of polarization was achieved only in the densest regions of the bones. In comparison, these PRIMA observations will have significant detection of polarization over the full area of the regions observed.

While this project is focused specifically on the dynamical importance of the magnetic field in the bones of the Milky Way, these observations with PRIMAger could be integrated into a larger survey of the galactic plane as long as the requested sensitivities are reached.

# 66. Multi-Scale Studies of the Magnetic Field of the Large and Small Magellanic Clouds

Dr. Laura Fissel (Queen's University), Kaitlyn Karpovich (Stanford University), Dr. Dennis Lee (Caltech, JPL), Dr. Dylan Paré (Villanova University), Dr. Sarah Sadavoy (Queen's University), Dr. Mehrnoosh Tahani (Stanford University)

We propose to use PRIMA's unprecedented polarization sensitivity to map magnetic fields with the PRIMAger Polarization Imager (PPI) in Bands 1-4 across the Large and Small Magellanic Clouds. With PRIMA's resolution we will be able to resolve magnetic fields on molecular cloud scales at better than 10 pc resolution in all four bands, with a best resolution of 2.6/3.2 pc for the LMC/SMC in Band 1. Our proposed survey will allow us to study the magnetic field properties of a large number of molecular clouds at the same resolution. We will also investigate how the cloud magnetic fields correlate with galactic magnetic fields, determining whether in these galaxies' clouds inherit their magnetic field morphology from the galactic magnetic field, or whether the cloud magnetic field structure is decoupled from the galactic magnetic field by local energy injection processes (such as supernovae or HII region expansion). This type of study would be extremely difficult for Milky Way molecular clouds, as most MW clouds are located near the Galactic plane where there is significant line-of-sight confusion.

We estimate that our requested survey will take 1037 hours to cover 45deg$^2$ of the LMC and SMC, but could be at the same time other proposed GO polarization studies that target the LMC and SMC (GO Science case #49 and #55 in the first PRIMA science book). The LMC and SMC are the nearest gas-rich galaxies to the Milky Way, and with PRIMA GO science these galaxies will be an excellent laboratory for the study of how galactic environment influences molecular cloud properties and their subsequent star formation.

## Science Justification

The role played by magnetic fields is a key outstanding problem in our understanding of how star formation is regulated. Understanding how magnetic fields influence the formation of stars requires multi-scale studies of how these fields influence the formation and fragmentation of dense gas in molecular clouds. Strong magnetic fields can inhibit movement of gas across field lines, while not inhibiting the motion of material parallel to the field lines, possibly leading to magnetic fields delaying the onset of star formation within molecular clouds (e.g. [3, 4, 11]).

Maps of linearly polarized emission from dust grains aligned with respect to their local magnetic field have been used to study the plane-of-sky magnetic field morphology in all phases of the interstellar medium (ISM). However, in order to understand the origin of magnetic fields within molecular clouds and how magnetic fields affect cloud formation we need to be able to directly compare the cloud fields with the surrounding galactic magnetic fields. Such observations are currently restricted to a few studies [1, 8, 7] of nearby clouds within our own galaxy. Detailed studies of magnetic fields in molecular clouds generally require observations of molecular clouds in the Milky Way. However, most clouds are located on or near the Galactic plane where it is very difficult to study the magnetic fields in molecular cloud envelopes and their connection to the local ISM magnetic field due to confusion from other polarized dust emission along the line-of-sight. Studies of magnetic fields in local galaxies have been made with the High-resolution Airborne Wideband Camera Plus instrument on the Stratospheric Observatory for Infrared Astronomy (HAWC+/SOFIA) (e.g. [6]), including many nearly face on galaxies, but these surveys were not able to resolve individual molecular clouds (scales of ≤10 pc).

With the PRobe far-infrared Mission for Astrophysics (PRIMA) we will have the sensitivity to make dust polarization observations of the nearest extragalactic regions for the study of star formation: the Large and Small Magellanic Clouds (hereafter LMC and SMC), which are the closest gas-rich galaxies to the Milky Way. Our goal is to map magnetic fields across all regions where there are molecular clouds in the LMC and SMC. At distances of 50 and 62 kpc, PRIMAger will have a resolution of 2.6/3.2 pc in PPI Band 1 for the LMC/SMC, and will be capable of resolving scales below 10 pc in all other PPI bands. Therefore, PRIMAger will resolve magnetic fields within all molecular clouds in these nearby irregular galaxies at a nearly identical linear resolution, allowing us to study how magnetic field properties correlate with local galactic environment. These observations require PRIMA's unprecedented sensitivity: HAWC+/SOFIA and ground-based polarimeters such as the James Clerk Maxwell Telescope (JCMT) and the Atacama Large Millimeter/submillimeter Array (ALMA) have so far only been able to make small polarization maps of the very brightest star-forming regions in the LMC and SMC, such as 30-Doradus.

An example of a previous polarimetric study using Planck of the LMC is shown in the left-hand panel of Figure 1 from [9]. The Planck resolution is 16 arcmin (200 pc) much lower than the sub-10 pc resolution that will be achieved with PRIMA, allowing us to improve our understanding into the spatial structure of the LMC magnetic field and the degree to which it traces distinct morphology in molecular clouds. The PRIMA resolution will in fact match resolutions achieved by HAWC+/SOFIA, which has studied particularly bright structures in the LMC, such as 30-Doradus. An example of a 89 µm study of 30-Doradus is shown in the right-hand panel of Figure 1 with a resolution of ∼3 pc [13]. This resolution will be achievable for clouds throughout the entire LMC with PRIMA PPI Band 1 observations. Our understanding of the SMC field will be improved in a similar way, allowing us to directly compare the magnetic fields of molecular clouds with the magnetic field of the ISM surrounding the clouds in both galaxies.

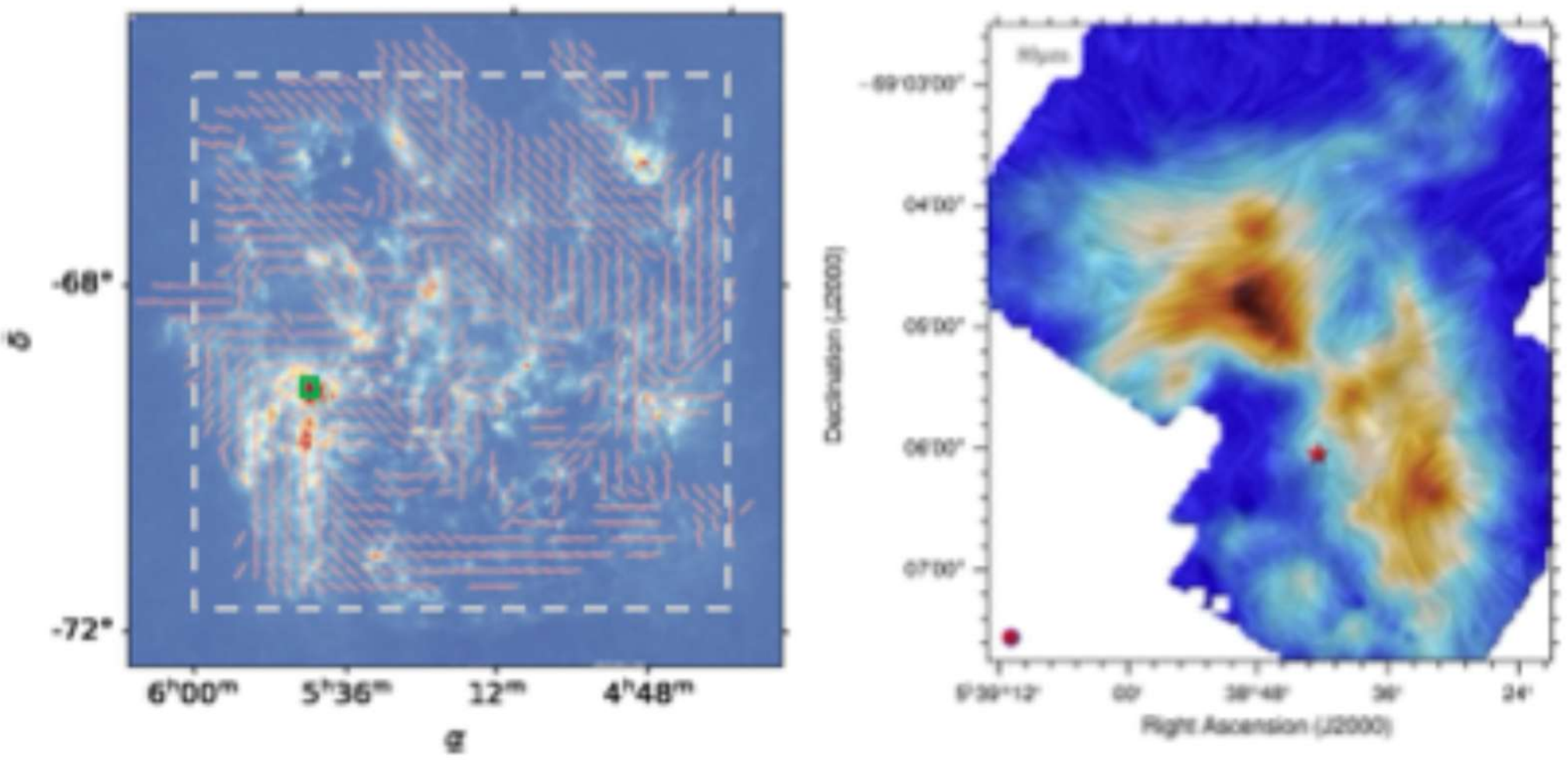

**Figure 1.** Left Panel: Planck 353 GHz inferred plane-of-sky magnetic field orientation of the LMC (after foreground subtraction) from [9]. The background map is the Herschel SPIRE 500μm from the HERITAGE survey [2]. The green rectangle shows the HAWC+ survey area of polarization observations of the brightest star forming region in the LMC: 30-Doradus (shown in the right panel), while the dashed grey region shows the area of the LMC targeted in our proposed GO program. Right panel: magnetic field inferred from HAWC+ 89μm polarization observations of 30-Doradus from [13]. Our PRIMAger survey will map magnetic fields at similar resolution over the entire LMC and SMC, directly comparing magnetic fields in molecular clouds with galactic magnetic fields.

We seek to answer the question of whether molecular clouds inherit their magnetic field from the large-scale galactic magnetic fields, as observed in some nearby Galactic molecular clouds [1, 8, 7], or whether cloud-scale magnetic fields are completely decoupled from the galactic magnetic field by local ISM physics that can alter the magnetic field morphology, such as supernovae explosions, or expanding HII regions driven by massive star clusters. Since magnetic fields can both influence and be influenced by gas motions, studying the relationship between the ISM and cloud fields will also give hints as to formation history of molecular clouds in the LMC/SMC. **The capabilities of PRIMA are uniquely suited to answering this question in a feasible observation time.**

## Instruments and Modes Used

This program requires using all four polarimetric bands of PRIMAger to create a 2.5° x 2.8° map of the SMC and a 6.2° x 6.2° map of the LMC.

## Approximate Integration Time

Our goal is to map both the magnetic field within molecular clouds and the surrounding ISM using PRIMAger's Polarimetric Imaging Mode. We base our 5-σ rms requests on archival Herschel/SPIRE data for the LMC/SMC at 100 μm, 160 μm, and 250 μm from the Herschel HERITAGE key program [2]. To detect polarization for the faintest clearly detected emission in the HERITAGE LMC/SMC we require 5-σ polarization detections Band 4 for a dust intensity of 5 MJy/sr. Planck 353 GHz maps of the LMC, shown in Figure 1, show a median polarization fraction of 3.6% [9]. Assuming 3% polarization, we require a detection of 150 kJy/sr polarized emission.

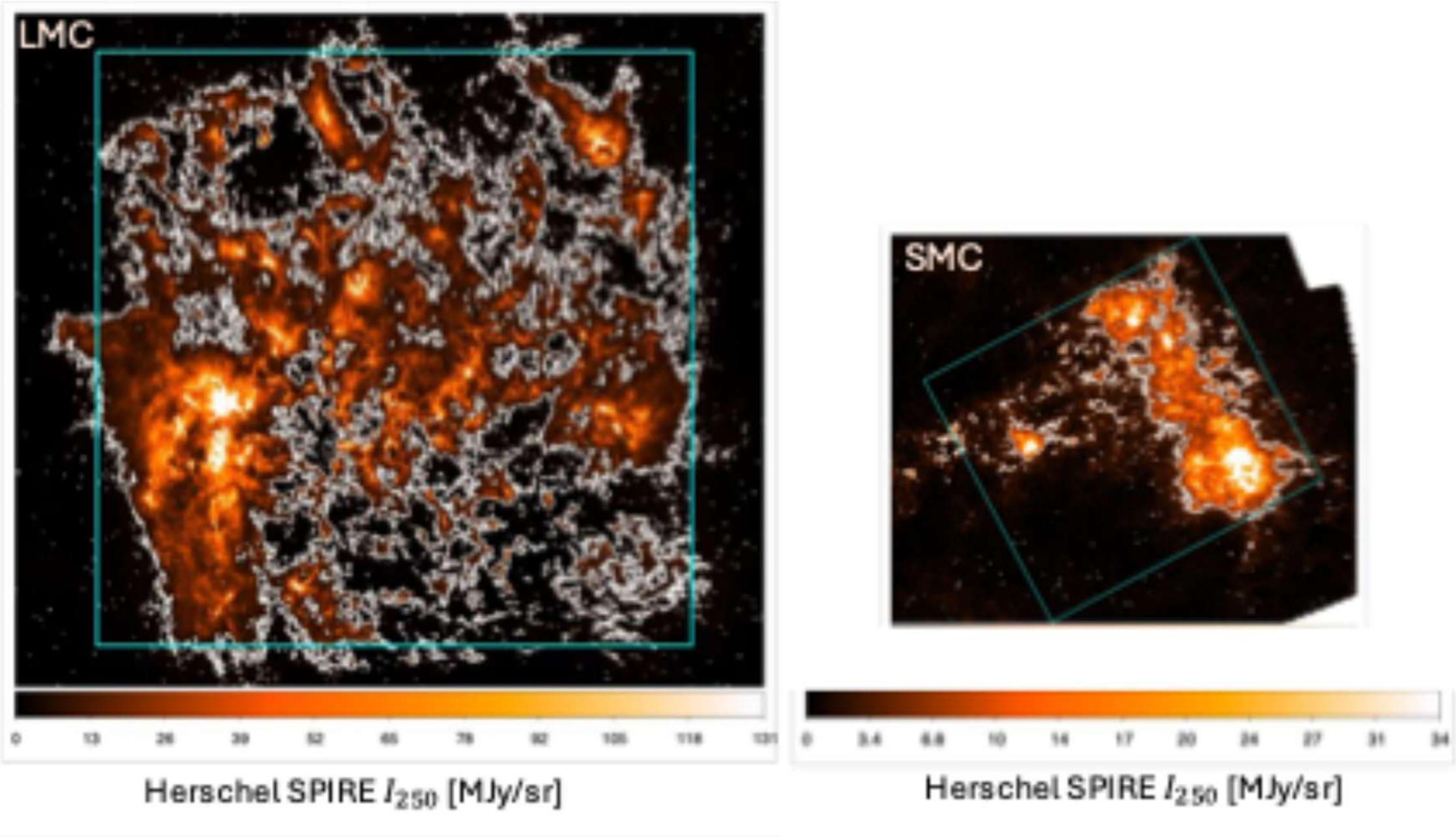

**Figure 2**. Left Panel: Herschel SPIRE 250 μm dust intensity maps of the LMC (left) and SMC (right). The cyan rectangles indicate our targeted regions for the proposed GO program. The white contour shows $I_{250}$ = 5 MJy/sr. Above this contour our survey should be able to achieve >5-σ detections of dust polarized at 3%, and >3-σ detections of 2% polarized dust.

We have estimated our approximate integration time using the PRIMA ETC for PPI Band 4. We assume a 7 square degree map for the SMC and a 38 square degree map for the LMC, for a total 45 deg$^2$ survey area. The requested survey areas are shown in Figure 2 for the LMC (left panel) and SMC (right panel). These survey areas cover the peak dust intensity regions of the LMC and SMC while also yielding coverage of the fainter, more diffuse surrounding dust. Assuming the beam area is 1.13(FWHM)$^2$, this gives a required source sensitivity of 3034 μJy. To make detect 5-σ detections with PPI Band 4 to this sensitivity we require 1030 hours according to the PRIMAger ETC. We have performed similar calculations for the other PRIMAger bands and find that the PPI Band 4 survey requires the most integration time. Therefore, we estimate a total observing time of 1030 hours to map all four bands of PRIMAger to sufficient depth over a 45 deg$^2$ area in the SMC and LMC.

**Total time:** 1030 hours for 5-σ detections of dust with a polarized intensity of 150 kJy/sr assuming the dust emission is 3% polarized.

## Special Capabilities Needed

NA

## Synergies with Other Facilities

**Ground-based sub-mm facilities:** Existing ground-based sub-mm telescopes are not yet sensitive enough to map dust polarization in the LMC and SMC, with the exception of the very brightest regions (e.g. 30-Doradus). However, these facilities are producing important dust continuum and molecular line surveys, which will complement our planned PRIMAger observations.

By the time PRIMA launches, the CCAT/Fred Young Sub-mm Telescope will be completing its planned 5-year science surveys, which includes a 100-hour 30-deg$^2$ survey of the LMC at 350 and 850 µm [10]. This survey is expected to produce 9 pc resolution maps of most molecular clouds in the LMC, but will be sensitivity limited to ∼50 pc resolution for magnetic field maps in the diffuse ISM. ALMA will also have completed its Wideband Sensitivity Upgrade, improving its mapping speed by a factor of 4, which will make it practical to use ALMA to follow-up PRIMAger polarization observations of LMC and SMC clouds at much higher resolution.

**Radio Polarimetry:** Exploring 3D magnetic field morphologies is essential for understanding the true relationship between molecular cloud and galactic magnetic fields. [8] demonstrated that the 3D magnetic fields of Orion A molecular cloud retain memory of the larger-scale Galactic magnetic fields, whereas examination of these fields only in the plane of the sky (as obtained by dust polarization) suggests otherwise. To advance this investigation, we will combine PRIMAger dust polarization observations, which trace the plane-of-sky magnetic field only, with complementary data from the Polarisation Sky Survey of the Universe's Magnetism [POSSUM; 14, 16], SPICE-RACS [12], and ultimately the full Square Kilometre Array. These observations will employ Faraday rotation measurements to infer large-scale galactic fields along the line of sight and the Faraday-based MC-BLOS technique [5, 15] to determine line-of-sight magnetic fields associated with molecular clouds.

**Synergy with other PRIMA GO Science Cases:** As the Magellanic Clouds have substantially lower metallicities than the Milky Way, they represent an ideal laboratory for exploring the dust properties of galaxies whose environment is distinct from our own. Our proposed GO Science overlaps in target area and instrument set-up with the GO Science case #49 from the first PRIMA GO Science Book (Unveiling the elusive dust properties of the diffuse ISM in nearby galaxies), and it is quite likely that both GO science programs could be achieved with the same PRIMAger survey. Since the polarization spectrum is expected to depend on various dust grain properties such as size distribution and composition, our GO science case and GO #49 complement and enhance each other. Our sample also has significant overlap with the GO science case #55 (Revealing the role of magnetic fields in the formation of giant molecular clouds and star formation in local galaxies), though we are targeting deeper polarization surveys of the LMC and SMC in order to make high resolution maps of both cloud magnetic fields and diffuse ISM fields.

## Description of Observations

To achieve the scientific objectives of this proposal PRIMA would observe the entire SMC and LMC using PRIMAger bands PPI1, PPI2, PPI3, and PPI4 with full polarimetry. **Without full polarimetric capability at these frequencies, there is no facility currently available or planned that will be able to make detailed studies of the magnetic field of the LMC and SMC spanning the magnetic field in the diffuse ISM to cloud-scale magnetic fields.**

For our observations we pick our fields of view to match the sizes of the SMC and LMC on the sky. These fields of view are shown in Figure 2. To observe this map region, we will use a scan map strategy to efficiently cover the proposed 7 and 38 square degree regions. An example of a possible PRIMA scan pattern is detailed in [17], and additional scanning strategies are currently being explored by the PRIMA team. The PRIMAger scans for each target will be obtained simultaneously, meaning that roughly 1000 hours would be required to obtain a four-color map of both the SMC and LMC (as shown above).

# 67. Surveying the Mineralogical Diversity of the ISM

Frédéric Galliano (DAp, CEA Paris-Saclay, France), Maarten Baes (Ghent University, Belgium), Léo Belloir (DAp, CEA Paris-Saclay, France), Simone Bianchi (INAF, Italy), Caroline Bot (Observatoire astronomique de Strasbourg, France), Janet Bowey (Cardiff University, UK), Jérémy Chastenet (Ghent University, Belgium), Christopher CLARK (STScI, USA), Ilse De Looze (Ghent University, Belgium), Golshan EJLALI (IPM, Tehran, Iran), Thomas Henning (MPIA, Heidelberg, Germany), Mika Juvela (Helsinki University, Finland), Hidehiro Kaneda (Nagoya University, Japan), Francisca Kemper (ICE-CSIC / ICREA / IEEC), Suzanne Madden (DAp, CEA Paris-Saclay, France), Mikako Matsuura (Cardiff University, UK), Takashi Onaka (Tokyo University, Japan), Lara Pantoni (Ghent University, Belgium), Evangelos Paspaliaris (INAF Arcetri, Italy), Monica Relaño Pastor (Granada University, Spain), Tsutomu Takeuchi (Nagoya University, Japan) Manolis Xilouris (Athens Observatory, Greece), Nathalie Ysard (IAS, Orsay, France)

PRIMA will provide continuous medium-resolution spectra in the FIR, for the first time since ISO. With its exceptional sensitivity, we should be able to detect numerous solid state features in the ISM. We plan to obtain medium resolution spectral maps of star forming regions in the *Milky* Way (MW), *Large Magellanic Cloud* (LMC) and *Small Magellanic* Cloud (SMC). Star-forming regions will provide strong enough background sources to accurately characterize the bands, and the temperature gradient as a function of distance from the central cluster will allow us to perform better decomposition. We will thus better constrain the dust composition in the nearby Universe.

## Science Justification

One of the most challenging open questions concerning the physics of the *InterStellar Medium* (ISM) is the chemical composition and structure of dust grains. We have approximate indirect constraints coming from the elemental depletions and from the detection of a few broad features. The large uncertainty of dust models however is a consequence of our ignorance of this composition and structure. For instance, Zubko *et al.* (2004) assume it is made of 4.6 % *Polycyclic Aromatic Hydrocarbons* (PAHs), 16.4 % of graphite and 79 % of amorphous silicates; Compiègne *et al.* (2011) assume it is made of 7.7 % of PAHs, 15.8 % of amorphous carbon and 76.5 % of amorphous silicates; Jones *et al.* (2017) assume it is made of 31.1 % of partially hydrogenated amorphous carbon and 68.9 % of amorphous silicates (forsterite and enstatite in equal proportions) with hydrogenated amorphous carbon mantles and Fe/FeS inclusions; Hensley & Draine (2022) assume it is made of 5 % PAHs and 95 % of a mash-up of silicate, hydrocarbons, iron and various oxides. This uncertainty is the main limitation of dust studies and of dust-based diagnostics of the physical conditions. This is also the limitation in the precision of gas-physics simulations, such as *PhotoDissociation Region* (PDR) models (Röllig *et al.* 2007).

The most straightforward way to more precisely constrain the dust composition would be to look for solid-state features. This is the only way to unambiguously identify a particular chemical composition. The *Mid-InfraRed* (MIR) range is potentially the richest domain as it is where the vibrational modes of chemical bonds are located. For instance, we know that interstellar silicates are mostly amorphous (Spoon *et al.*, 2022). However, precisely knowing this fraction and the way it varies with the physical conditions (radiation field, gas density and metallicity) would provide invaluable constraints on dust evolution. Another puzzle is that there is too much depleted oxygen in the ISM, compared to what we can put in silicates. It is thus possible that a fraction of the dust is in the form of various oxides (*e.g.* $Al_2O_3$, $CaCO_3$, *etc.*) or organic carbonates (Jones & Ysard, 2019). The bands of these compounds may have eluded previous spectroscopic surveys. Similarly, iron is heavily depleted into the grains, but we ignore what form it takes (silicates, metallic inclusions, FeS, *etc.*; *e.g.* Dwek, 2016). Finally, recent X-ray investigations of the dust chemical composition in our Galaxy suggest that Mg-rich amorphous pyroxene represents the largest fraction of dust (about 70 % on average; Psaradaki *et al.* 2023). This fraction may change with environments and amorphous pyroxenes can be studied through their MIR-to-FIR features as well.

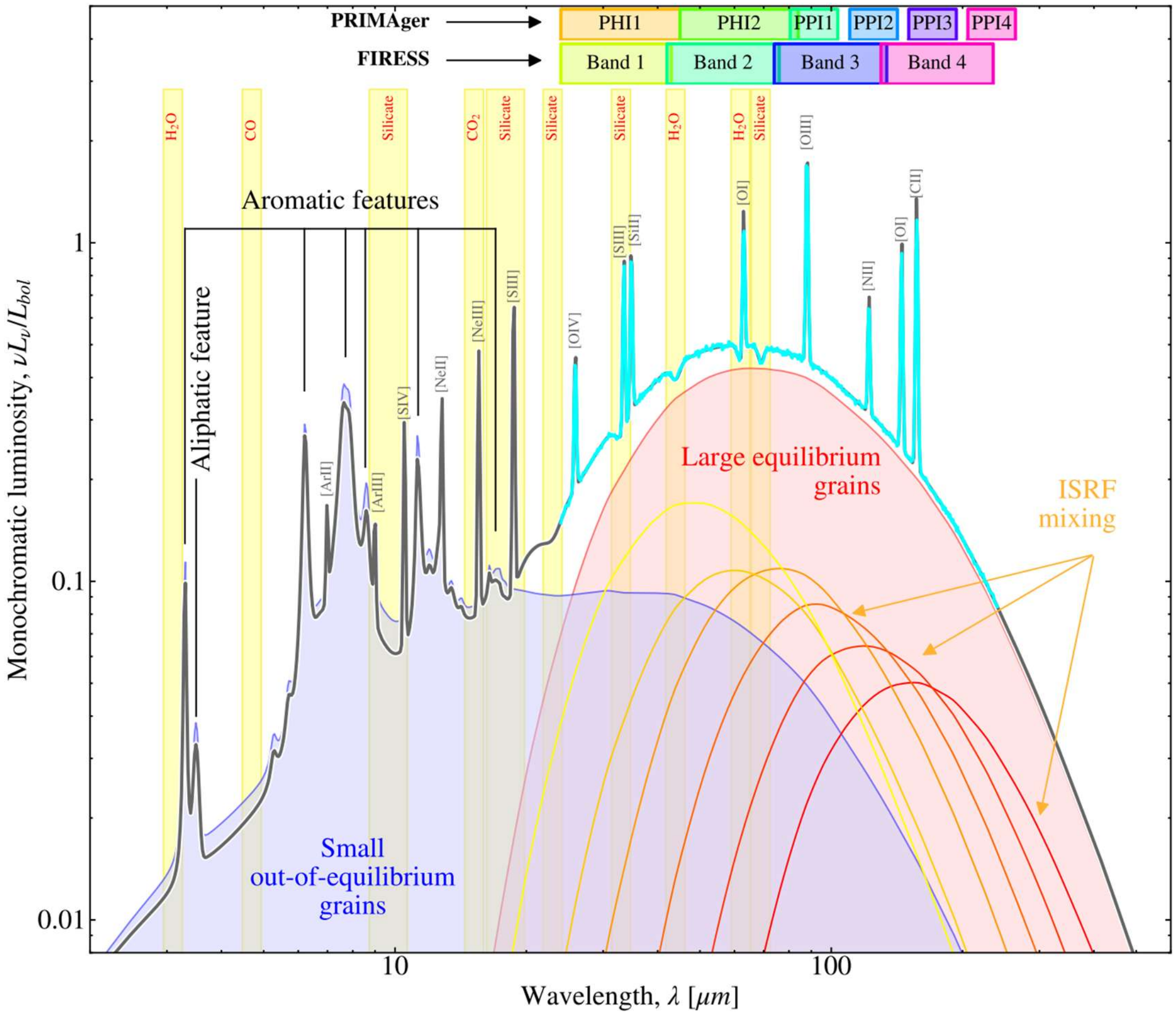

**Figure 1.** Typical SED of a star-forming region. The cyan line is a R=200 spectrum (100σ). We show a few absorption features from crystalline silicates and ices, and the brightest gas lines.

## Instruments and Modes Used

The FIRESS spectrometer in its low-resolution mapping mode will be required.

## Description of Observations and Approximate Integration Time

Low spectral resolution (*R*), high signal-to-noise ratio (S/N) observations over the whole MIR-to-FIR range would give us access to a wealth of features (Fig. 1). However, such spectra were only obtained in the past by combining ISOSWS and ISOLWS (*e.g.* Peeters *et al.* 2002), with a poor sensitivity, a poor spatial resolution and stitching problems. This is a domain where PRIMA will particularly innovate. We could map star-forming regions of different metallicities, such as in the LMC and SMC. It would be important to obtain maps in order to understand how these features vary with the distance from the central cluster. This would highlight the role of photo-processing as well as the role of mantle growth in dust evolution. We could be looking for both emission and absorption features. However, the emission features come primarily from small grains or large hot grains. They might not represent the bulk of the dust mass. This is why absorption features are preferred. This is also why we would need to target star-forming regions, so that we have a strong MIR background.

We will use PRIMA-FIRESS to perform *R*=100 spectral maps of a sample of 20 nearby extragalactic star-forming regions (≈20'×20').

20 regions could be mapped in 20 hours.

- The flux sensitivity νFν=1.7×$10^{-17}$ W/m2/beam (5σ) at 32 µm has been estimated using the dust model of Jones *et al.* (2017) with a radiation field intensity of *U*=35 and for *N*(HI)=3×$10^{21}$ H/$cm^2$. This corresponds approximately to the extended parts of the star-forming regions of the LMC.
- We require a high S/N (≈100), because the goal is not so much to detect the intensity of a feature, but to unambiguously measure it. These features are usually weak and broad, we thus need to make sure that they are not diluted in the continuum.
- The sizes are those of typical star-forming regions in the LMC/SMC (N11, N66, etc.).

Refer to Galliano et al. (2025) for more details on the sensitivity estimates.

# 68. Unveiling the Elusive Dust Properties of the Diffuse ISM of Nearby Galaxies

Frédéric Galliano (DAp, CEA Paris-Saclay, France), Maarten Baes (Ghent University, Belgium), Léo Belloir (DAp, CEA Paris-Saclay, France), Simone Bianchi (INAF, Italy), Caroline Bot (Observatoire astronomique de Strasbourg, France), Viviana Casasola (INAF, Italy), Jérémy Chastenet (Ghent University, Belgium), Christopher Clark (STScI, USA), Stavroula Katsioli (Athens Observatory, Greece), Ilse De Looze (Ghent University, Belgium), Golshan Ejlali (IPM, Tehran, Iran), Francisca Kemper (ICE-CSIC / ICREA / IEEC), Suzanne Madden (DAp, CEA Paris-Saclay, France), Mikako Matsuura (Cardiff University, UK), Takashi Onaka (Tokyo University, Japan), Lara Pantoni (Ghent University, Belgium), Evangelos Paspaliaris (INAF Arcetri, Italy), Monica Relaño Pastor (Granada University, Spain), Marc Sauvage (DAp, CEA Paris-Saclay, France), Matthew Smith (Cardiff University, UK), Tsutomu Takeuchi (Nagoya University, Japan), Manolis Xilouris (Athens Observatory, Greece), Nathalie Ysard (IRAP, Toulouse, France)

We plan to use the exceptional sensitivity of PRIMA to map the LMC and SMC with PRIMAger, down to the confusion limit. The goal is to obtain a reliable characterization of the SED of the diffuse ISM in these two systems. Combined with already existing estimates of the extinction curves and elemental depletions, this will allow us to build extragalactic dust models for these two galaxies, and thus understand the impact of metallicity on the dust properties.

## Science Justification

The constitution of interstellar dust (*i.e.* its composition, size distribution and abundance) in external galaxies is believed to differ significantly from the *Milky Way* (MW). Dust build-up and evolution indeed depend on the specific *Star Formation History* (SFH) of the galaxy. In particular, the metallicity, *Z*, which is the mass fraction of elements heavier than He, appears to be one of the most important factors (*e.g.* Galliano, 2021). This parameter, *Z*, quantifies the cumulated elemental enrichment of a system. To properly interpret observations of galaxies, we thus need to understand how the dust properties vary as a function of *Z*. Yet, contemporary dust models (*e.g.* Jones *et al.*, 2017; Hensley & Draine, 2022), that are used to provide such an interpretation, are exclusively constrained by observations of the MW, a system with a narrow *Z* range around the Solar value, $Z_\odot$. We are therefore biased by the particular properties of the MW when modeling the dust *Spectral Energy Distribution* (SED) of other galaxies. This bias especially questions our ability to accurately understand nearby dwarf galaxies and early Universe galaxies when the metal enrichment is expected to be low.

Arguably, no properly-constrained dust model of external galaxies currently exists. This is because there is a deficit of observational constraints. The mixing of physical conditions along the sightline and within the telescope beam indeed renders the SED degenerate. For instance, if we are observing a region where there is a gradient of *InterStellar Radiation Field* (ISRF), we will not be able to distinguish an overabundance of small grains from the spread due to the variation of the equilibrium temperatures of large grains (*e.g.* Figure 3 of Galliano *et al.*, 2018). This is why dust models are calibrated on observations of the diffuse *InterStellar Medium* (ISM) of the MW. The low optical depth of this medium ($A(V)\approx0.1$ for $N(H)=2\times10^{20}$ H/cm$^2$, at $Z_\odot$) ensures that the grains will be uniformly illuminated by the average ISRF. It is possible to build the observed SED of the diffuse ISM of the MW, by averaging the high Galactic latitude fluxes given by IRAS, COBE and *Planck*. This is however not yet possible in external galaxies. IRAS, COBE and *Planck* did not sufficiently resolve galaxies to allow the extraction of their diffuse ISM emission. And other observatories with a finer angular resolution, such as *Spitzer* and *Herschel*, were not sensitive enough (*Herschel* could barely go below $N(H)=10^{22}$ H/cm$^2$).

The good angular resolution and the exceptional sensitivity of PRIMA over the whole *Mid-InfraRed*-to-*Far-InfraRed* (MIR-to-FIR) window gives us a way to palliate this problem, without having to resort to feathering (*e.g.* Clark *et al.*, 2023). The two closest galaxies, the *Large and Small Magellanic Clouds* (LMC & SMC; $d_{LMC}$=50 kpc and $d_{SMC}$=60 kpc; $Z_{LMC}=1/2\ Z_\odot$ and $Z_{SMC}=1/5\ Z_\odot$; Figure 1) are ideal targets. We will be able to resolve regions of ≈10 pc size at λ=250 µm. With the addition of already estimated elemental depletions and extinction curves (*e.g.* Gordon *et al.*, 2003; Tchernyshyov *et al.*, 2015), the well characterized broad-band SED of their diffuse emission per H atom will allow us to build the first properly-constrained extragalactic dust models.

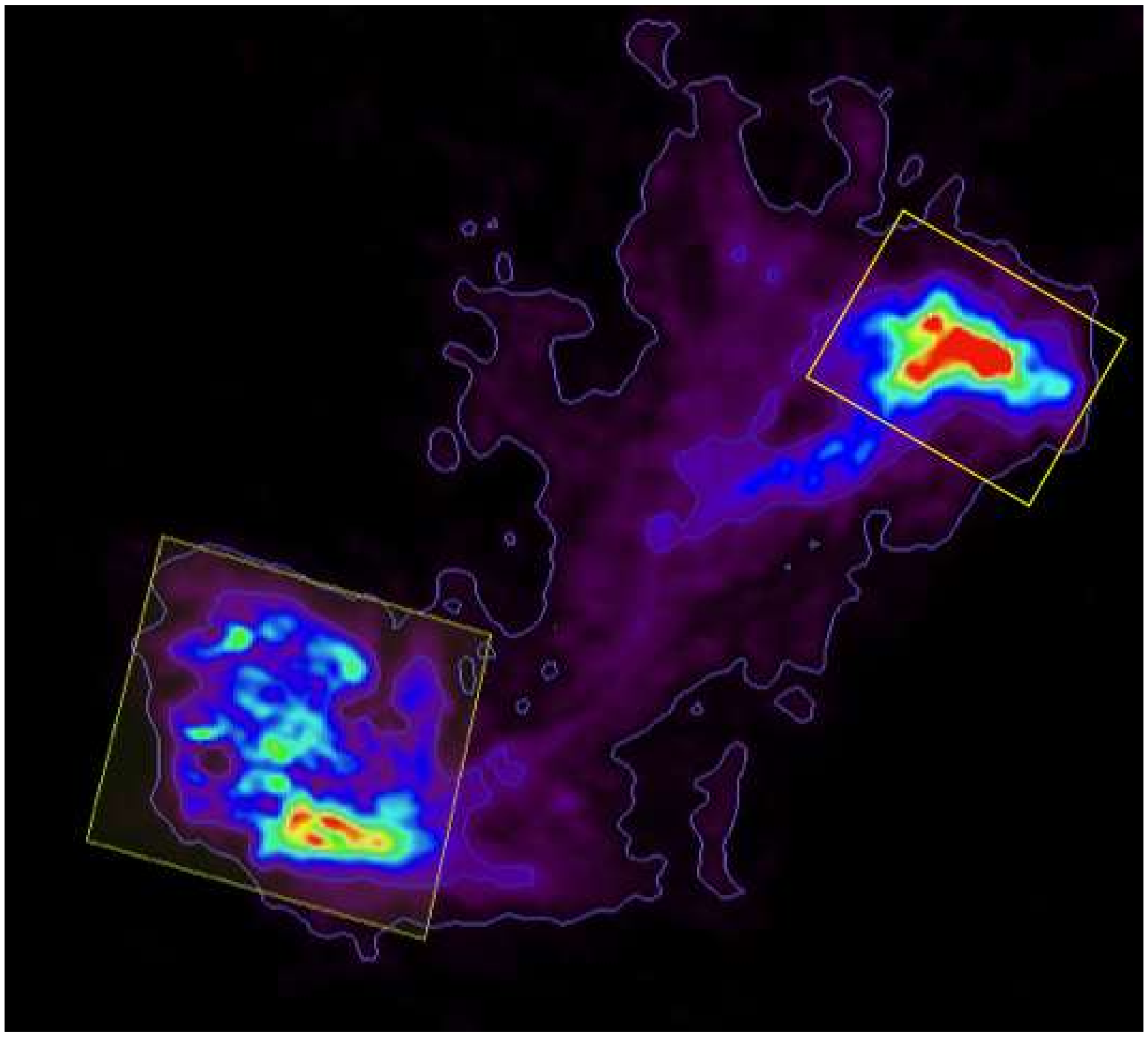

**Figure 1.** HI map of the LMC/SMC (Brüns et al., 2005). The contours correspond to N(H)=$10^{20}$ H/cm2 and N(H)=2×$10^{20}$ H/cm2. The yellow squares are areas covering most of the HI emission of both sources (LMC: 12°×12° SMC: 8°×6°).

## Instruments and Modes Used

PRIMAger mapping using all of the hyperspectral and polarimeter bands. One each of a 12°×12° map and a 0.2°×0.2° map for the LMC, and one each of a 8°×6° map and a 0.2°×0.2° map for the SMC.

## Description of Observations and Approximate Integration Time

To measure the MIR-to-FIR SED of the diffuse ISM of the LMC and SMC, we need to make deep maps of these two galaxies in all available bands. In addition, Guillet *et al.* (2018) showed how important FIR measures of the polarization fraction were to constrain the grain composition.

We will use PRIMAger, to measure both the total power and the polarization fraction.

- We will map most of the 2 galaxies with PRIMAger at the 4 long-wavelength bands.
- We will make smaller maps of a diffuse region in each galaxy, in each of the short-wavelength bands, as confusion is less problematic in the MIR.

We will make large maps of the LMC (12°×12°) and SMC (8°×6°) through all the PRIMAger bands. The flux sensitivity has been estimated using the dust model of Jones *et al.* (2017) with a ISRF intensity of $U$=1 and for $N$(H)=2×$10^{20}$ H/cm$^2$. At first approximation, the emission and the column density both scale with $Z$. We should thus scale the emission of the model by $Z/Z_\odot$. However, we are interested in the emission of an optically thin medium ($A$(V)≈0.1). This $A$(V) will be reached at a $Z_{\odot}/Z$ times higher column density than in the MW. $Z$ therefore cancels out in this estimate. Refer to Galliano *et al.* (2025) for more details on the sensitivity estimates.

This total program will thus require 63.9 hours (LMC) and 21.3 hours (SMC) large PRIMAger maps. In addition, hyperspectral imaging of a smaller region (0.2°×0.2°) of both galaxies will take twice 277 hours, and a deep polarimetry map on the same region will take an additional 3.5 hours. In total, this program will thus take around 600 hours.

**Table 1.** PRIMAger Mappings

| **Large Maps** | **Small Maps** | |
|---|---|---|
| (simultaneous observations of Hyperspectral and Polarimeter bands) | Hyperspectral band (24-84 microns) | Polarimeter band (96, 126, 172, 235 microns) |
| **LMC** | | |
| Band 1: one 12°×12° LMC map down to $\nu F_{\nu}=2.74\times10^{-16}$ W/m$^2$/beam (5σ) → 63.9 hours | One 0.2°×0.2° LMC map down to $\nu F_{\nu}=5.58\times10^{-18}$ W/m$^2$/beam (5σ) at 35 μm → 277.4 hours | Band 1: one 0.2°×0.2° LMC map down to $\nu F_{\nu}=2.74\times10^{-17}$ W/m$^2$/beam (5σ) → 1.8 hours |
| Band 2: one 12°×12° LMC map down to $\nu F_{\nu}=4.92\times10^{-16}$ W/m$^2$/beam (5σ) → 22.1 hours | | Band 2: one 0.2°×0.2° LMC map down to $\nu F_{\nu}=4.92\times10^{-17}$ W/m$^2$/beam (5σ) → 0.6 hours |
| Band 3: one 12°×12° LMC map down to $\nu F_{\nu}=5.89\times10^{-16}$ W/m$^2$/beam (5σ) → 14.5 hours | | Band 3: one 0.2°×0.2° LMC map down to $\nu F_{\nu}=5.89\times10^{-17}$ W/m$^2$/beam (5σ) → 0.4 hours |
| Band 4: one 12°×12° LMC map down to $\nu F_{\nu}=5.38\times10^{-16}$ W/m$^2$/beam (5σ) → 17.1 hours | | Band 4: one 0.2°×0.2° LMC map down to $\nu F_{\nu}=5.38\times10^{-17}$ W/m$^2$/beam (5σ) → 0.5 hours |
| **SMC** | | |
| Band 1: one 8°×6° SMC map down to $\nu F_{\nu}=2.74\times10^{-16}$ W/m$^2$/beam (5σ) → 21.3 hours | One 0.2°×0.2° SMC map down to $\nu F_{\nu}=5.58\times10^{-18}$ W/m$^2$/beam (5σ) at 35 μm → 277.4 hours | Band 1: one 0.2°×0.2° SMC map down to $\nu F_{\nu}=2.74\times10^{-17}$ W/m$^2$/beam (5σ) → 1.8 hours |
| Band 2: one 8°×6° SMC map down to $\nu F_{\nu}=4.92\times10^{-16}$ W/m$^2$/beam (5σ) → 7.4 hours | | Band 2: one 0.2°×0.2° SMC map down to $\nu F_{\nu}=4.92\times10^{-17}$ W/m$^2$/beam (5σ) → 0.6 hours |
| Band 3: one 8°×6° SMC map down to $\nu F_{\nu}=5.89\times10^{-16}$ W/m$^2$/beam (5σ) → 4.8 hours | | Band 3: one 0.2°×0.2° SMC map down to $\nu F_{\nu}=5.89\times10^{-17}$ W/m$^2$/beam (5σ) → 0.4 hours |
| Band 4: one 8°×6° SMC map down to $\nu F_{\nu}=5.38\times10^{-16}$ W/m$^2$/beam (5σ) → 5.7 hours | | Band 4: one 0.2°×0.2° SMC map down to $\nu F_{\nu}=5.38\times10^{-17}$ W/m$^2$/beam (5σ) → 0.5 hours |

## Special Capabilities Needed

None

# 69. JWST Observations of Localized Heating of the Quiescent ISM and what PRIMA can Contribute to Understanding Turbulence

Paul Goldsmith (Jet Propulsion Laboratory, California Institute of Technology), Shengzhe Wang (UCAS, Beijing), Xin Wang (UCAS, Beijing), Raphael Skalidis (Caltech), Gary Fuller (Univ. Manchester), Di Li (Tsinghua University, Beijing), Chao-Wei Tsai (NAOC, Beijing), Lile Wang (Kavli Institute, Beijing), Donghui Quan (Zhejiang Laboratory, Hangzhou), Dylan Paré (Villanova Univ.)

One of the many aspects of interstellar medium (ISM) physics that has remained elusive is turbulence. A significant fraction of the ISM energy resides in the chaotic (turbulent), superthermal motions of the gas, which play a key role in supporting molecular clouds against gravitational collapse and thus have a major influence on the rate of star formation. We have used JWST/MIRI to observe two regions in the boundary of Taurus in the 17.04 µm S(1) transition of ortho $H_2$, and very clearly detected small-scale structure on scales 1″–2.5″ (140–350 au), limited in part by signal to noise ratio considerations. The column density in the J = 3 level is typically 2e17 cm$^{-2}$. This is reasonably well-determined, but the total $H_2$ column density uncertainty is large because the 28 µm S(0) line, the lowest transition of para $H_2$, could not be observed with JWST/MIRI. While the angular resolution of PRIMA, 7.6″ at 28 µm, is significantly lower than that of JWST/MIRI, having data on the relatively low-lying S(0) transition ($\Delta E/k$ = 845 K) would be extremely valuable. The intensity measured can be compared with that integrated over an appropriate portion of a mosaic observed with JWST, giving a much better handle on the $H_2$ level populations. This is critical for modeling collisional excitation, constraining the density and temperature of the excited $H_2$, whether produced by turbulent dissipation or by some other mechanism such as shocks. The predicted 28 µm PRIMA flux is 4e-19 Wm$^{-2}$, for which we can obtain a 5σ low-resolution spectral measurement in 0.22 hr, and a 120″x120″ map in 18 hr. We propose a ∼270 hr program to survey excited $H_2$ emission from ten quiescent regions of the ISM in different environments to assess how common localized heating is, and where it occurs.

## Science Justification

### Introduction

One of the many aspects of interstellar medium (ISM) physics that has remained elusive is turbulence. Turbulence in the ISM is universal (Heyer & Brunt 2004), and a significant fraction of the ISM energy resides in the chaotic (turbulent), superthermal motions of the gas, which play a key role in supporting molecular clouds against gravitational collapse and thus have a major influence on the rate of star formation (Falceta-Gonçalves et al. 2014; Hennebelle & Falgarone 2012). It is generally felt that turbulent energy is input on large scales, by galactic differential rotation, shocks, and cloud-cloud collisions (Klessen & Hennebelle 2010). Local processes such as stellar winds and radiation can locally increase turbulence but have major effect only on relatively small fraction of the ISM. Most theoretical models have turbulent energy cascading from the large injection scale to smaller scales until the lower end of the inertial range is reached and dissipation becomes important. The scale on which dissipation occurs depends on many factors including the viscosity of the ISM gas and the magnetic field. Within the range of field strengths that plausibly permeate the ISM, the latter can affect the shocks, especially the heating of the postshock gas (LeSaffre et al. 2013). At small scales, ≦ 1e15 cm, we can have shocks generated (LeSaffre et al. 2013), or MHD vortices formed, both resulting in local heating of the gas in these so-called Turbulent Dissipation Regions, or TDRs (Falgarone, Pineau des Forêts, & Roueff 1995; Joulain et al. 1998). The presence of locally heated regions with temperatures ~ 1000 K is demanded by the presence of certain molecules, notably CH+, which have large barriers to their formation, but are widely distributed in the diffuse ISM that otherwise has temperature no higher than 100 K (Falgarone et al. 2005; Moseley et al. 2021).

The gas in TDRs includes molecular hydrogen ($H_2$), with a small fraction of atomic hydrogen depending on the density and environment of the region in question. The warm $H_2$ molecules can be observed through rovibrational transitions at infrared and far-infrared wavelengths. Emission from 3 rotational transitions of $H_2$ was observed from multiple regions around the boundary of the Taurus molecular cloud using Spitzer by Goldsmith et al. (2010). The temperature was found to be greater than 250 K from comparison of different line intensities. For sensitivity reasons, the extraction aperture used for the Spitzer data was 10.7″ x 108″, vastly larger than the expected size of TDRs. Falgarone et al. (2005) also observed $H_2$ emission from regions where UV and formation pumping should not be dominant.

To study small-scale heating on a more relevant scale, we employed JWST/MIRI medium resolution spectrometer (MRS), having 0.67″ angular resolution and spectral resolution R = 2500. We obtained 4 x 4 mosaics of the 17.04 µm S(1) $H_2$ line in two regions in the boundary of the Taurus molecular cloud, which after cropping of bad edge pixels covered an area 12.4″ x 13.2″. The lines are not resolved by MIRI/MRS, but the line flux is spread over channels surrounding and including that corresponding to the nominal wavelength. Summing over 5 channels including the line yields mean intensities of 19.6 and 11.6 MJy sr-1 for the "peak" and "edge" regions, respectively, while the rms of the background was 1.1 MJy $sr^{-1}$. We thus had highly significant detections in most pixels (Goldsmith et al. 2025).

Fig. 1 shows the intensity of the S(1) line, the lowest rotational transition of ortho-$H_2$, in the edge region. Significant small-scale structure with scale 1.0″ to 2.5″, corresponding to 140 to 250 pc is clearly visible. A characteristic intensity of 10 MJy $sr^{-1}$ corresponds to a $H_2$ column density in the J = 3 state (the upper level of the S(1) transition) of 1.6e17 $cm^{-2}$. The total column density of $H_2$ depends on the temperature and density, which is where PRIMA observations are valuable. A value of the column density that yields a reasonable geometry and is consistent with other requirements on physical conditions is $N(H_2)$ = 8e17 $cm^{-2}$.

PRIMA can observe the lowest rotational transition of para $H_2$, the S(0) line at 28.22 µm. This line has a significantly lower A-coefficient as well as lower energy (hf/k = 510 K) compared to the S(1) line of ortho $H_2$ (hf/k = 845 K). The ratio of the two lines can thus be used to constrain the density and temperature, although the ortho to para ratio (OPR) is an additional uncertainty. Different lines of evidence suggest that while the OPR can plausibly lie between 1 and 3, it is likely not far from 1. For fixed OPR, the S(0)/S(1) ratio increases at lower densities and at lower temperatures so that measurement of the ratio of the two lines can determine or at least constrain n($H_2$) and T. For accurate comparison, the JWST data will have to be convolved to the angular resolution of PRIMA.

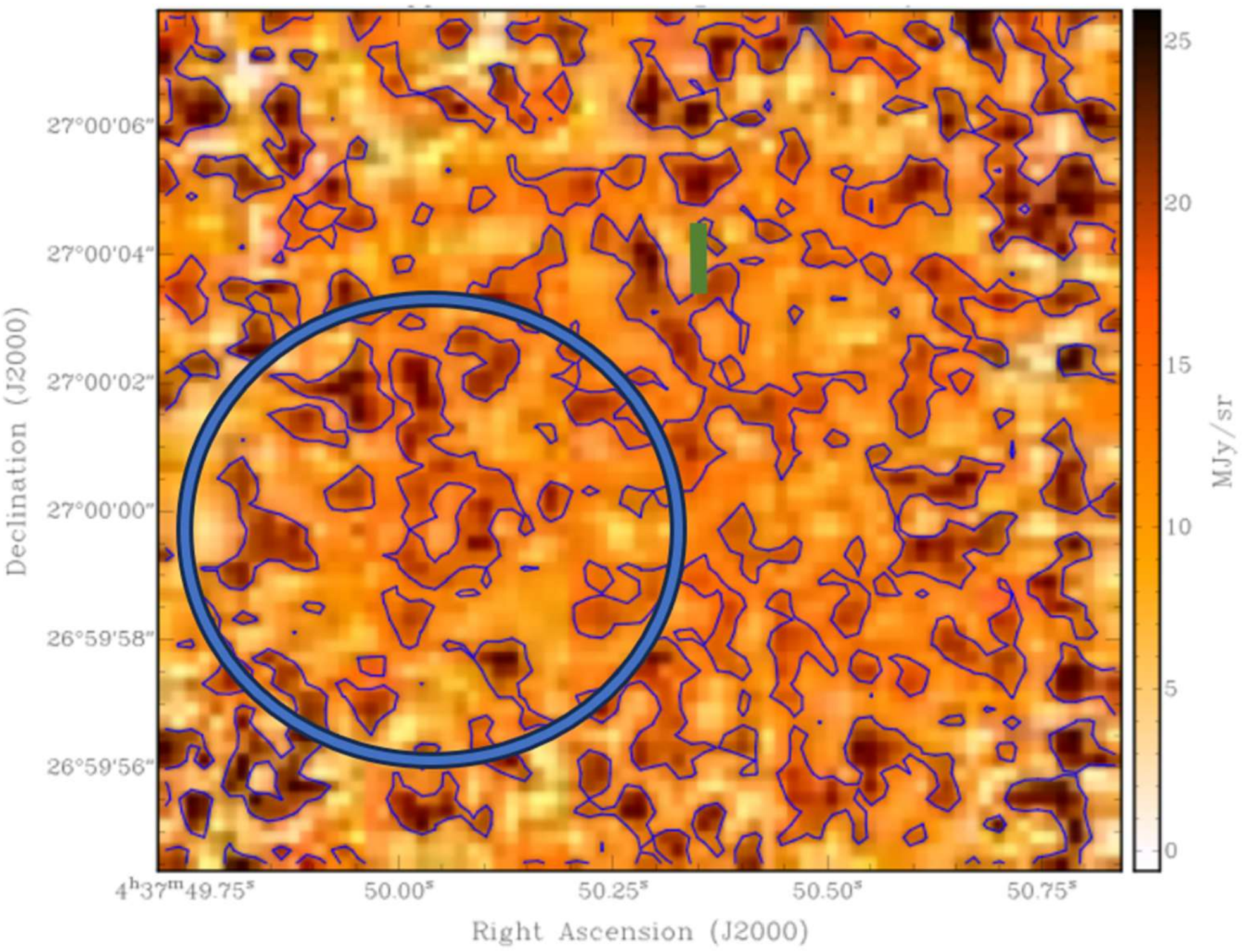

**Fig. 1.** Image of 17.04 µm $H_2$ S(1) emission from the edge region in the Taurus molecular cloud boundary obtained with MIRI/MRS instrument on JWST. The angular resolution of JWST at 17.04 µm wavelength is 0.67", corresponding to 94 au at the distance of Taurus. The green bar denotes a size of 1" corresponding to 140 au at the distance of Taurus. The blue circle denotes the 7.6" PRIMA beam at the 28.2 µm wavelength of the S(0) transition of $H_2$. Adapted from Goldsmith et al. (2015).

The S(0) line has the advantage that, as shown in Fig. 2, its intensity is relatively independent of n($H_2$) and T over the range that likely characterizes the heated condensations observed with JWST. It thus can deliver alone a reasonable estimate of N($H_2$) in the J = 2 level and so can be used for larger-scale studies of the heated material. Spitzer observations suggested that this was present in regions surrounding the Taurus molecular cloud, and other observations have detected $H_2$ rotational emission from lines of sight not associated with massive young stars or other obvious heated sources that would produce PDR regions and thus large quantities of hot $H_2$. However, we have little information on the extent and occurrence of this phenomenon, and modest-sized maps with PRIMA are a very promising way to determine these parameters.

A very large sample would clearly be desirable, but a reasonable start would be to observe 10 regions. These would be in the boundaries of molecular clouds, mostly sources with very limited star formation to avoid contamination by PDR emission. Regions with measured UV energy density would be favored, but if this information really is not available, it could possibly be obtained independently. Spitzer observations (Goldsmith et al. 2010) indicated that the rotationally excited $H_2$ emission weakened as one looked directly to the interiors of molecular clouds. This suggests that the turbulent dissipation and heating is restricted to the boundary layers of clouds. An important task is thus to study how the excited $H_2$ emission depends on position. This requires mapping the equivalent of 5 - 120″ x 120″ regions at different positions in the boundary region of the molecular cloud. A total of 15 - 120″ x 120″ region maps is requested.

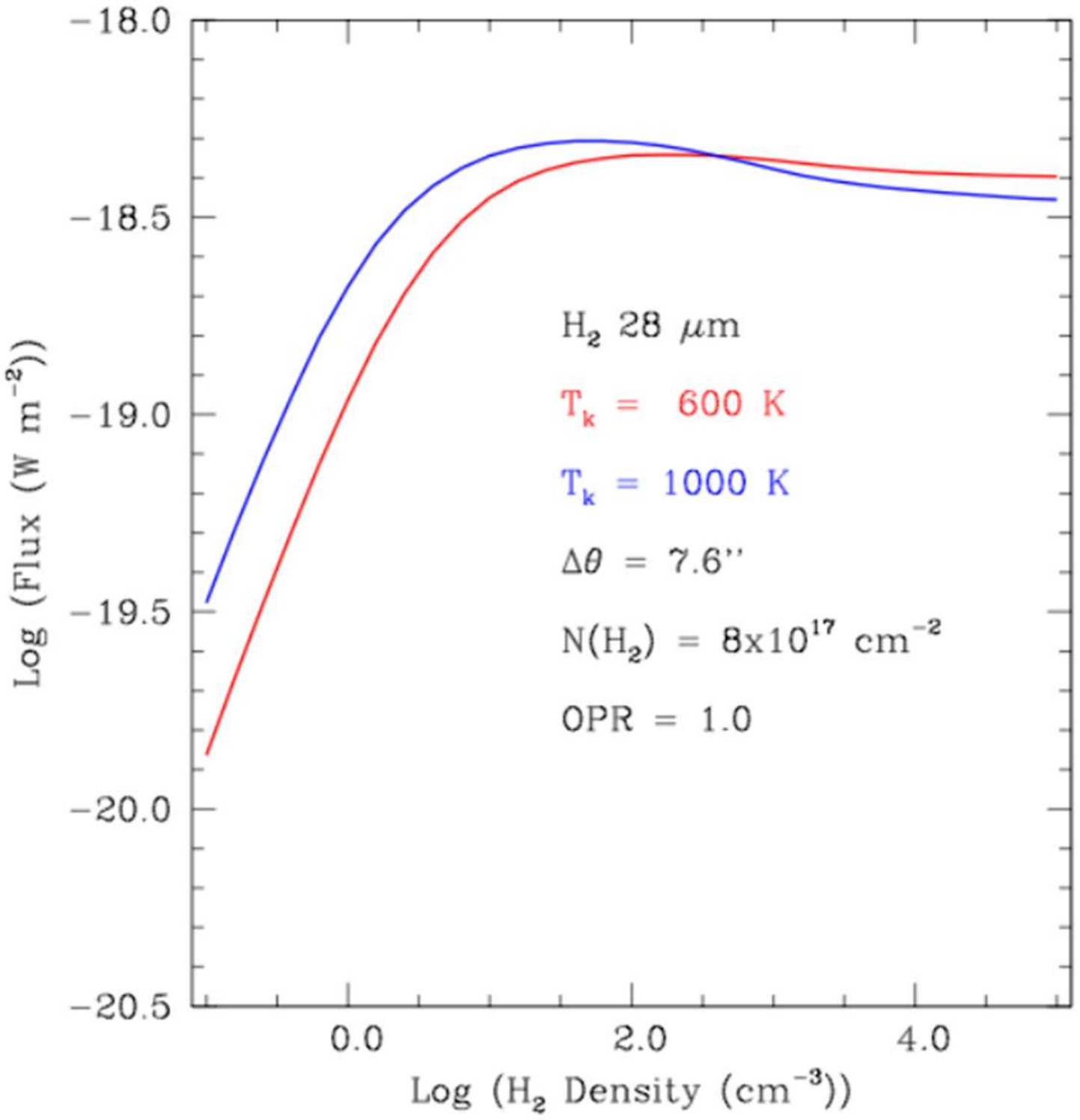

**Fig. 2.** $H_2$ S(0) 28 μm flux expected based on JWST measurements of emission from small-scale heated structures shown in Fig. 1.

## Instruments and Modes Used

120″x120″ FIRESS low resolution maps at 28 μm, one for each of 15 target regions.

## Approximate Integration Time

- Based on the ETC, a 5σ measurement of estimated 4e-19 $Wm^{-2}$ flux with PRIMA requires 0.22 hr integration time for a single pointing. A 120″ x 120″ map with ~250 spaxels requires 18 hr. To obtain a reasonable sample of regions in different environments, we propose to make 15 such maps. The total integration time will thus be ~270 hr.

## Special Capabilities Needed

None

## Synergies with Other Facilities

JWST and ALMA

## Description of Observations

We have observed two regions in the boundary of the Taurus molecular clouds with JWST/MIRI and obtained impressive images of $H_2$ 17.04 µm emission, which includes small-scale structures of 1–2.5″ size. The characteristic intensity of these condensations is 10 MJy sr$^{-1}$. The density and temperature of these regions is uncertain but based on our best estimates of Tk = 600–100 K and n($H_2$) = 200 – 400 cm$^{-3}$, as shown in Fig. 2 we predict a 28 µm flux close to 4e-19 W m$^{-2}$ for PRIMA low resolution spectroscopy. An important aspect of PRIMA observations is that, along with the larger beam size, much more extended maps can be made than are possible with JWST.

Based on the ETC, a 120″ x 120″ region can be mapped to the required 5σ sensitivity of 4e-19 W m$^{-2}$ in 18 hr. The total time required for 15 such maps is thus 270 hr.

 

# 70. PRIMA Study of a New Phase of the ISM: the Dense Warm Interstellar Medium (D-WIM)

Paul F. Goldsmith (Jet Propulsion Laboratory, California Institute of Technology), Jorge Pineda (Jet Propulsion Laboratory, California Institute of Technology), Dariusz Lis (Jet Propulsion Laboratory, California Institute of Technology)

We propose a study of the Dense Warm Interstellar Medium (D-WIM) via the two fine structure lines of ionized nitrogen ([N II]) at 122 and 205 µm. This new phase of the interstellar medium (ISM) is fully ionized gas at density 10 cm$^{-3}$ ≦ n(e) ≦ 100 cm$^{-3}$ and has effectively been identified throughout much of the Milky Way and in nearby galaxies. The unexpected conditions in the D-WIM immediately raise important questions both about this phase of the ISM and its relationship to the ISM as a whole. These include the structure of the D-WIM – what is its filling factor, distribution, and morphology? A second is – what maintains its ionization; is this a new source or can it be explained by radiation from stars? A third question is the origin of the D-WIM – what is its relationship to stars – is it the remnants of very old HII regions that were produced by clusters of massive stars? And fourth -- what is the evolution of this new phase of the ISM and does the overpressure result in expansion leading to dissipation? The PRIMA survey requires a coverage of a large fraction of the plane, plus selected regions at high galactic latitudes. A 5σ sensitivity of 2x10−9 W m$^{-2}$ sr$^{-1}$ would be highly desirable. A spectral resolution R = 10$^{2}$ would allow measuring the total intensity. Current best estimates of PRIMA sensitivity indicate that a survey covering 400 square degrees can be carried out in 1760 hours. This PRIMA survey would include other important ISM tracers including [CII] 158 µm, [OI] 63 µm and 146 µm, and higher ionization states of nitrogen and oxygen. The sensitivity requirement for study of the D-WIM is likely more stringent than for the other lines and may require special consideration for successful implementation as well as for determining the area that can be observed.

## Science Justification

Regions of ionized gas are a major component of the interstellar medium (ISM). These range from the relatively dense (with electron densities n(e) as high as 1000 cm$^{-3}$) ionized gas in the immediate vicinity of massive young stars (HII regions) to the diffuse warm ionized medium (WIM) having a density n(e) ∼0.1 cm$^{-3}$ (Reynolds 1991, Hill et al. 2008, Ferrière 2001, and references therein). Determining the properties of these regions has been an ongoing task, with the question of the relationship of these different components of the ISM being an especially difficult challenge. While a great deal of information has been gained by using radio techniques, major questions remain, especially about the lower-density ionized ISM. Densities 1–10 cm$^{-3}$, an order of magnitude higher than those of the WIM, have been suggested by observations of cm-wavelength recombination lines (Shaver 1976) and by comparison of low-frequency and high--frequency recombination lines (Anantharamaiah 1985, 1986). These regions may be the extended highly evolved HII regions and are sometimes referred to as Extended Low Density or ELD HII regions (Mezger 1978). A quite different technique to study ionized regions is to employ the fine structure lines of ionized nitrogen; there are two [NII] lines at wavelengths of 205 μm and 122 μm (e.g., Bennett et al. 1994, Oberst et al. 2006). The 14.53 eV ionization potential of atomic nitrogen means that this element will be ionized only in regions where hydrogen is also ionized and the ratio of the intensities of the two lines is an excellent probe of electron densities for 10 ≦ n(e) ≦ 1000 $cm^{-3}$ as shown in Fig. 1 (Oberst et al. 2011; Goldsmith et al. 2015).

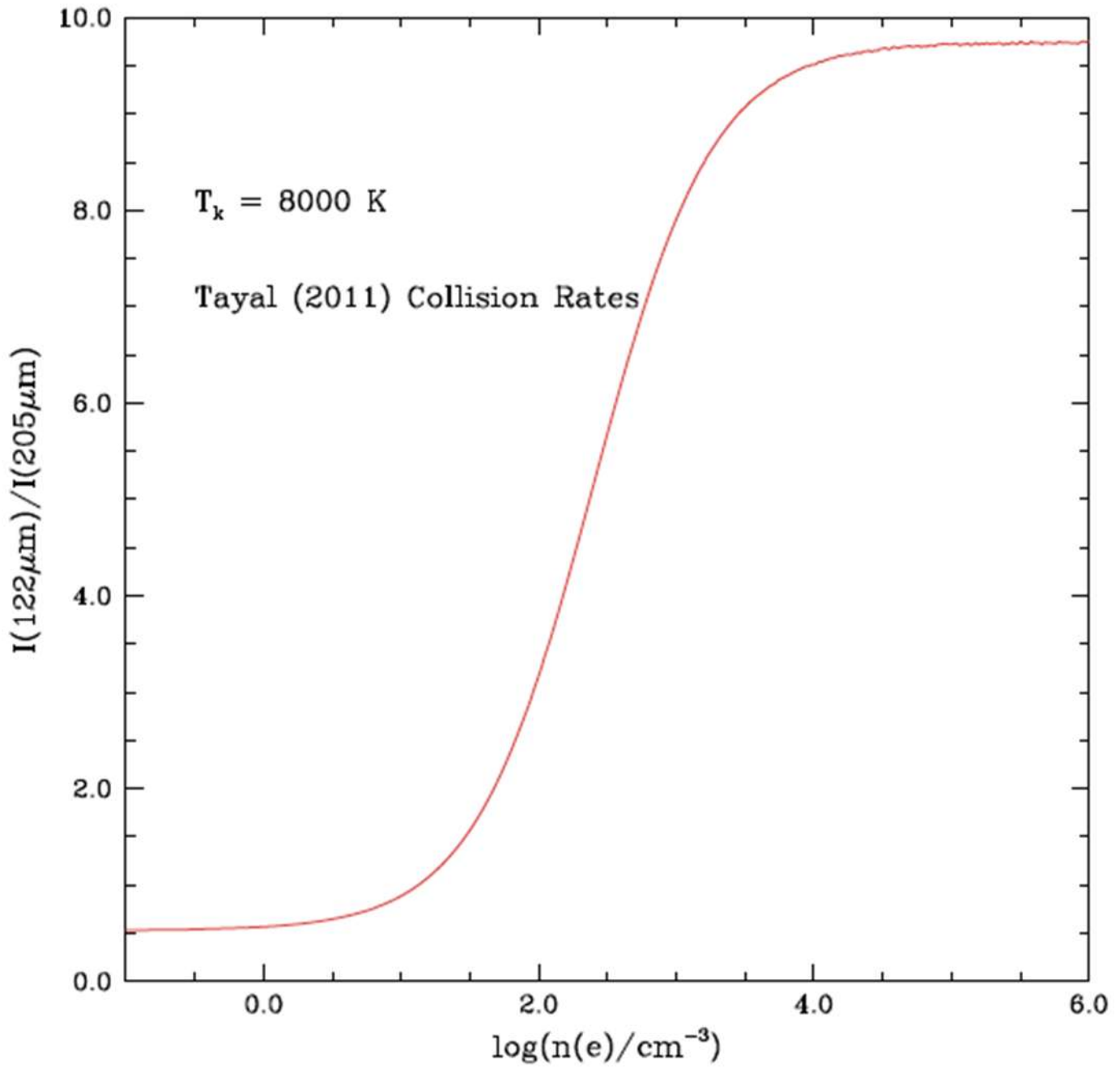

**Fig. 1.** Ratio of intensities of two [NII] fine structure lines as function of electron density. This ratio is a good measure of n(e) between 10 and 1000 $cm^{-3}$. From Goldsmith et al. (2015).

## The Dense Warm Interstellar Medium (D-WIM)

The [NII] line ratio was used by Goldsmith et al. (2015) to determine the electron density in an unbiased survey of the Galactic plane. For 116 positions in the range −60° ≦ *l* ≦ +60°, the average electron density was 29 $cm^{-3}$, with almost all lines of sight falling in the range of 10 $cm^{-3}$ to 50 $cm^{-3}$ with a few regions having n(e) up to 100 $cm^{-3}$. The presence of widespread, dense, ionized gas was a very surprising result, as its density is an order of magnitude higher than the densities of the ELD HII regions and its widespread nature makes any intimate connection with stellar-driven ionization unlikely. And if the D-WIM does not ultimately have a stellar energy source, what is providing its ionization? Compared to the WIM, the D-WIM is overpressured by 2 orders of magnitude, so it will expand. Is it being continually recreated as it diffuses and merges with the WIM? At this density, recombination is relatively rapid, and another major question is – what is the source of ionization?

Fig. 2 shows the observed [NII] fine structure line ratio and derived electron density as a function of Galactic longitude. All the above questions point to a new phase of the ISM that could imply a very large-scale impact of massive star formation but could also be indicative of a widespread additional ionization source. What is needed to pin down the properties of the D-WIM is to obtain 2D images of the [NII] lines and derived electron densities and column densities. This is what the proposed PRIMA survey will uniquely provide in two dimensions throughout the entire disk of the Milky Way.

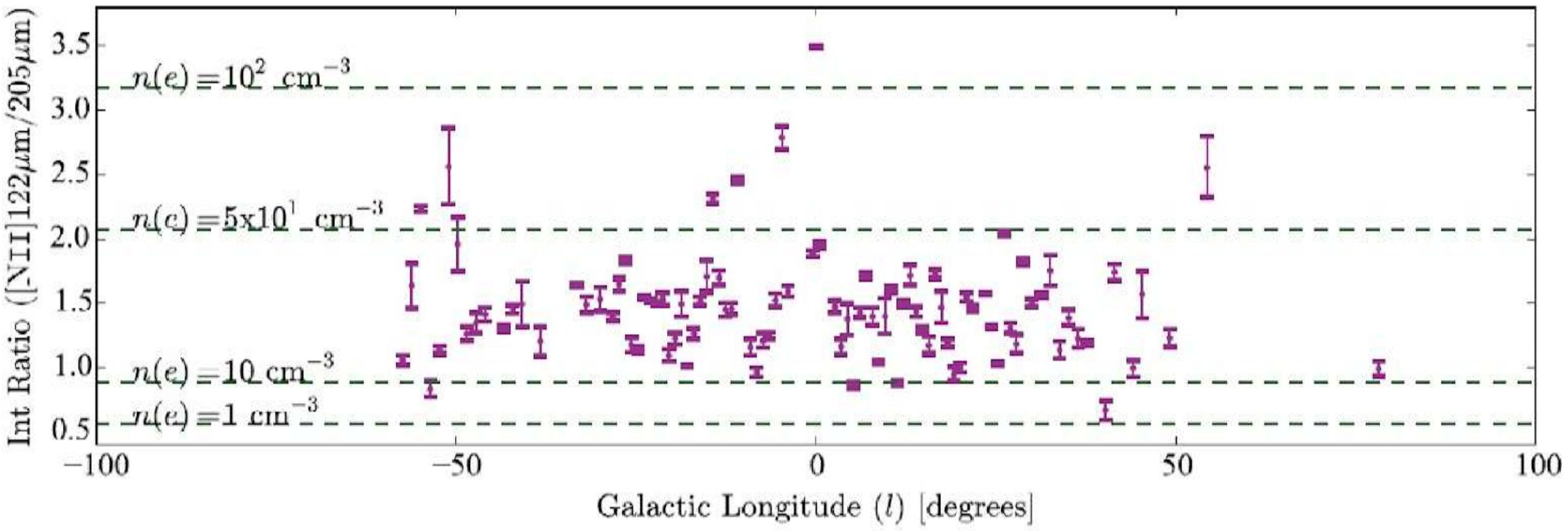

**Fig. 2.** [NII] fine structure line ratios and derived electron densities from Herschel survey (Goldsmith et al. 2015).

The extensive, n(e) ~30 $cm^{-3}$ ionized gas was called the Dense Warm Ionized Medium (or D-WIM) by Geyer & Walker (2019). These authors modeled the D-WIM as being ionized only by collisions with electrons and derived an unusually high temperature of ~19,000 K for this component of the ISM. Observations were carried out by Langer et al. (2021), and their modeling, including collisional ionization of nitrogen by proton charge exchange with protons from ionized hydrogen, suggested a much more reasonable temperature range of 3400 K to 8500 K, consistent with the line widths measured from velocity-resolved spectroscopy using the SOFIA/GREAT instrument. Observations of the same pair of [NII] fine structure lines in 21 galaxies from the KINGFISH survey yielded a range of densities from 1–300 $cm^{-3}$ (Herrera-Camus et al. 2016). This is similar to the range found in the Milky Way survey mentioned above, and even more striking, the extragalactic mean value, $\langle n(e) \rangle = 30$ $cm^{-3}$, is essentially identical to that found in the Milky Way. While there are certainly variations and gradients, this confirms on a much broader scale the existence of the D-WIM phase of the ISM. But the origin of the D-WIM remains a major puzzle. What keeps it ionized, and what is its volume filling factor? As mentioned above, obtaining additional information on the morphology of the D-WIM is thus of critical importance for understanding the ISM of a wide range of galaxies.

Pineda et al. (2019) developed a hybrid technique in which the 205 µm [NII] line was combined with radio recombination line (RRL) data. As shown by these authors, the ratio of the integrated antenna temperature of the [NII] 205 µm line to that of a RRL is dependent on the electron density (Pineda et al. 2019, equation 9). They employed this technique to obtain electron densities along 11 lines of sight using the 205 µm line data from SOFIA and RRL data from the NASA Deep Space Network and the GBT. They found electron densities between 8 and 170 $cm^{-3}$, again generally consistent with the D-WIM densities determined by Goldsmith et al. (2015). A critical point is that these LOS were chosen to **avoid known HII regions**.

This project will definitively confirm the existence of a new phase of the interstellar medium. This may be connected with the very late stages of evolution of HII regions produced by massive stars but may well point to new processes that are required to sustain this surprisingly dense, warm, ionized gas (D-WIM) filling a significant fraction of the Galactic disk. A very significant additional result will be complete imaging of the [CII] 158 µm line and the ability to determine the fraction of its emission arising in ionized regions. An extension of this project to cover a selection of nearby galaxies would be extremely valuable, but the data required may well be obtained as part of other imaging projects.

## Instruments and Modes Used

The FIRESS spectrometer in the low-res (R∼100) mapping mode will be used.

## Approximate Integration Time

1760 hr

## Special Capabilities Needed

None

## Synergies with Other Facilities

Balloon Missions (GUSTO, ASTHROS & reflights thereof)

## Description of Observations

To observe the 122 µm line we observe Bands 1 and 3 simultaneously and to observe the 205 µm line, Bands 2 and 4 simultaneously, but separate from the first observations. This means that we will be carrying out an enormously sensitive survey of the entire PRIMA spectral range, so it will cover higher ionization states of nitrogen and oxygen, and lines of S and Si in different ionization states. The [CII] line will be observable from very diffuse regions of the ISM where it has been very little studied to date.

PRIMA with a 1.8-m diameter telescope will have FWHM beam widths corresponding to individual detector pixels of ∼23″ at [NII] 205 µm and ∼13″ at [NII] 122 µm. These are comparable to or smaller than the Herschel PACS footprint, so there should not be a problem with confusion between the more diffuse ionized gas and HII regions. What is needed is well-sampled, extensive maps of a significant fraction of the Galactic plane. Exploration of off-plane emission is also important as the very large (∼7°) beams employed by Wright et al. (1991) and Bennett et al. (1994), while suggestive of extended emission, did not have adequate angular resolution or sampling to obtain definitive results. The intensity distributions of the [NII] lines observed are shown in Fig. 3 (from Goldsmith et al. 2015). The mean intensity is <log I> = $10^{-7.7}$ $Wm^{-2}sr^{-1}$ for both transitions. Given the desire to get more complete sampling of the outer galaxy as well as the high-latitude regions, a 5σ sensitivity of $2x10^{-9}$ $Wm^{-2}$ $sr^{-1}$ would be highly desirable. For the 122 µm line, the solid angle is $\sim3\times10^{-9}$ sr, yielding a desired 5σ flux of $6\times10^{-18}$ $Wm^{-2}$. For the 205 µm line, the solid angle is $\sim1\times10^{-8}$ sr, yielding a desired 5σ flux of $2\times10^{-17}$ $Wm^{-2}$.

According to the Exposure Time Calculator (ETC), for low resolution spectroscopy (R∼100) PRIMA FIRESS will map a 1 square degree area to a 5σ sensitivity of $6x10^{-18}$ $Wm^{-2}$ in 21.7 hr. for [NII] 122μm and to 2x10$^{-17}$ in 0.8 hr. for [NII] 205 μm. These two bands must be observed separately, so the total time required would be 22.5 hr per square degree. For accurate ratioing the two intensities of the two lines as well as for comparison with other tracers, we will smooth most of the map to 30″ resolution, reducing the time requirement at 122 μm by a factor of 5.6 and at 205 μm by factor 1.7, yielding time requirements of 3.9 hr and 0.5 hr./sq degree and total time requirement of 4.4 hr./sq deg.

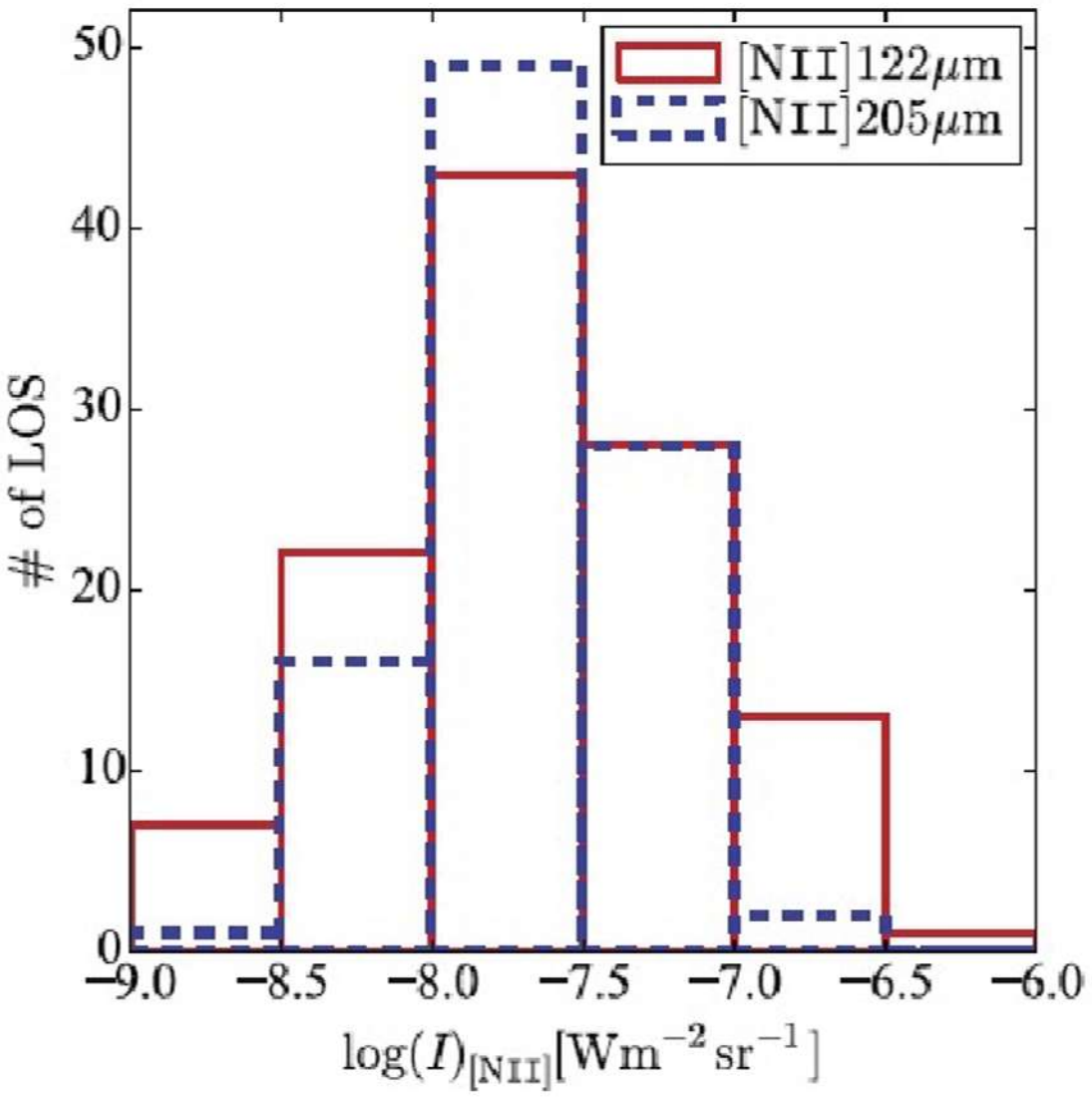

**Fig. 3.** Distribution of intensities of two [NII] fine structure lines from Galactic plane survey reported by Goldsmith et al. (2015).

The coverage of this study should as a minimum include (1) Galactic plane −90° ≦ *l* ≦ 90°; −1.0° ≦ *b* ≦ +1.0°: 360 square degrees total, and (2) Strips in the outer Galaxy and strips including high latitudes; the extent is TBD, but we want to sample at least a few degree-wide strip including the Central Molecular Zone, and another such strip at moderate longitude: 40 square degrees total. The total area to be mapped is thus 400 square degrees. This would thus take 1760 hours plus overhead and calibration time. Less areal coverage would still be acceptable and provide valuable information.

If it is possible and appropriate, the integration time could be adjusted to permit deeper integrations at larger longitudes and higher latitudes. The intensity of the [NII] lines is approximately 1/10 of that of the [CII] 158 μm line, as discussed by Goldsmith et al. (2015), making this a challenging project and one that is unlikely to be carried out by any other facility, e.g., a balloon or other suborbital mission. There is limited information on the spectral characteristics of the [NII] emission from the spectra observed in the Galactic plane by Goldsmith et al. (2015), Pineda et al. (2019), and Langer et al. (2021). In most cases, the spectral lines are single--peaked features, with line width 20 km s$^{-1}$ to 50 km s$^{-1}$. In some cases, there are multiple peaks, but generally, the [NII] emission is dominated by a single feature. Thus, low resolution PRIMA FIRESS mapping will be appropriate for this project.

 

# 71. Probing 3-D Magnetic Fields in the Star Formation Process from Clouds to Cores

Thiem Hoang (KASI, Korea), Le Ngoc Tram (Leiden), Truong Le Gia Bao (KASI), Ngan Le (KASI), Nguyen Bich Ngoc (VNSC), Diep N. Pham (VNSC), Jia-Wei Wang (EAO), Jihye Hwang (Kyushu University), Xiaohui Sun (Yunnan University, China), Kate Pattle (UCL, UK), Yue Hu (IAS)

Accurate measurements of 3D magnetic fields (B-fields) in filaments are crucial for understanding the key role that magnetic fields play in the star formation process. 3D B-fields are also crucial for modeling the propagation of cosmic rays (CRs) in molecular clouds, filaments, and cores, which play an important role in gas ionization and ambipolar diffusion during protostellar disk formation. Dust polarization induced by aligned grains only provides the B-field projected on the plane of the sky (2D B-field), but the line-of-sight component (or B-field's inclination angle) is not available. Recently, Hoang & Truong (2024) and Truong & Hoang (2025) developed a new technique to infer the B-field's inclination angle using the observed dust polarization degree with modern grain alignment theory. Through this new technique, 3D B-fields can be reconstructed by combining the 2D B-fields measured by the Davis-Chandrasekhar-Fermi (DCF) technique and the B-field's inclination angle. We propose to use the polarized dust emission mapped by Primager and innovative analysis techniques based on polarization fraction to construct 3-D structures of B-fields from molecular clouds to core scales.

## Science Justification

## Context

Understanding how stars form and evolve is the big open question in modern astrophysics. The process of star formation is complex, involving gravity, magnetic fields, turbulence, and stellar feedback. Magnetic fields are thought to play a crucial role in the formation of molecular clouds and star formation. Moreover, cosmic rays are a unique source of ionization and influence non-ideal MHD effects of protostar and disk formation. The last decade demonstrated significant progress in measuring the 2D magnetic fields projected in the plane of the sky using dust polarization from the different scales of star formation, from the ISM to molecular clouds, filaments to cores and protostellar environments. Yet, the accurate 3D structures of B-fields, which are required to test the different theories of star formation, are not yet available.

## Science Question

3D magnetic fields are also essential for understanding the formation mechanism of hub-filament systems. They may play an important role in the formation of cloud substructure and can direct the accretion of material onto star-forming filaments and hubs. Moreover, 3D B-field structures are necessary for accurate modeling of cosmic rays in molecular clouds and cores to disks (Padavoni+2013). However, key questions remain to be answered, including: (1) What are the true 3D structures of B-fields in molecular clouds, filaments, and cores? (2) What are the 3D density structures of molecular clouds and filaments? (3) How are 3D B-fields influenced by stellar feedback? (4) How do 3D B-fields influence CR propagation, gas ionization and star formation?

## Need for PRIMA

2D magnetic fields in the interstellar medium can be traced by starlight polarization and polarized thermal dust emission from non-spherical grains aligned with magnetic fields (e.g. Andersson et al. 2015). 850µm observations with Planck are low-resolution (Figure 1); while 850µm JCMT POL-2 (e.g. Pattle et al. 2017) and 50-240µm SOFIA HAWC+ observations (e.g. Pillai et al. 2020), with resolutions comparable to that of PRIMA, show the complexity of these fields. However, POL-2 and HAWC+ can only observe over very limited areas, and at the very highest surface brightnesses, due to atmospheric constraints. In particular, all ground-based telescopes inherently filter out large-scale emissions due to the need for sky background subtraction. This missing flux can introduce uncertainties in the measured polarization fraction, which is critical for accurate 3D magnetic field analysis. In contrast, space-based telescopes are essential for preserving large-scale flux and enabling reliable measurements of the polarization fraction.

## Analysis Methods

(1) We will first measure 3D B-field structures in filaments:

Step 1: Constrain the dust properties (size and shape) and the gas local properties (volume density, gas temperature, and dust temperature). This step is complementary to Tram et al.

Step 2: Constrain the B-field's inclination angles using the observed polarization and theoretical polarization model obtained by Dustpol-py.

Step 3: Derive 3D B-fields by combining the B-field's inclination angle with the POS B-fields measured by the DCF method by Pattle et al. to derive full 3D B-fields (morphology and strength).

(2) We will investigate the key roles of 3D B-fields on filament formation and evolution. We calculate the actual magnetic energy using the full 3D B-fields and compare it with gravity and turbulence at multiple scales.

(3) Study the role of 3D B-fields on CR transport and gas ionization. We will analyze the correlation between 3D B-fields and cosmic ray ionization measured on multi-scales from filaments to cores.

## Link to Testable Hypotheses

(1) Do weak and turbulent B-fields influence the CR propagation into MCs/filaments/cores and reduce the gas ionization? (2) The bending of B-fields is expected to help drive the gas inflow into the filament hub. We will analyze the correlation between the 3D B-fields and kinematics of the hub-filament systems (3) We will investigate proposed relationships between magnetic fields and (proto)stellar feedback.

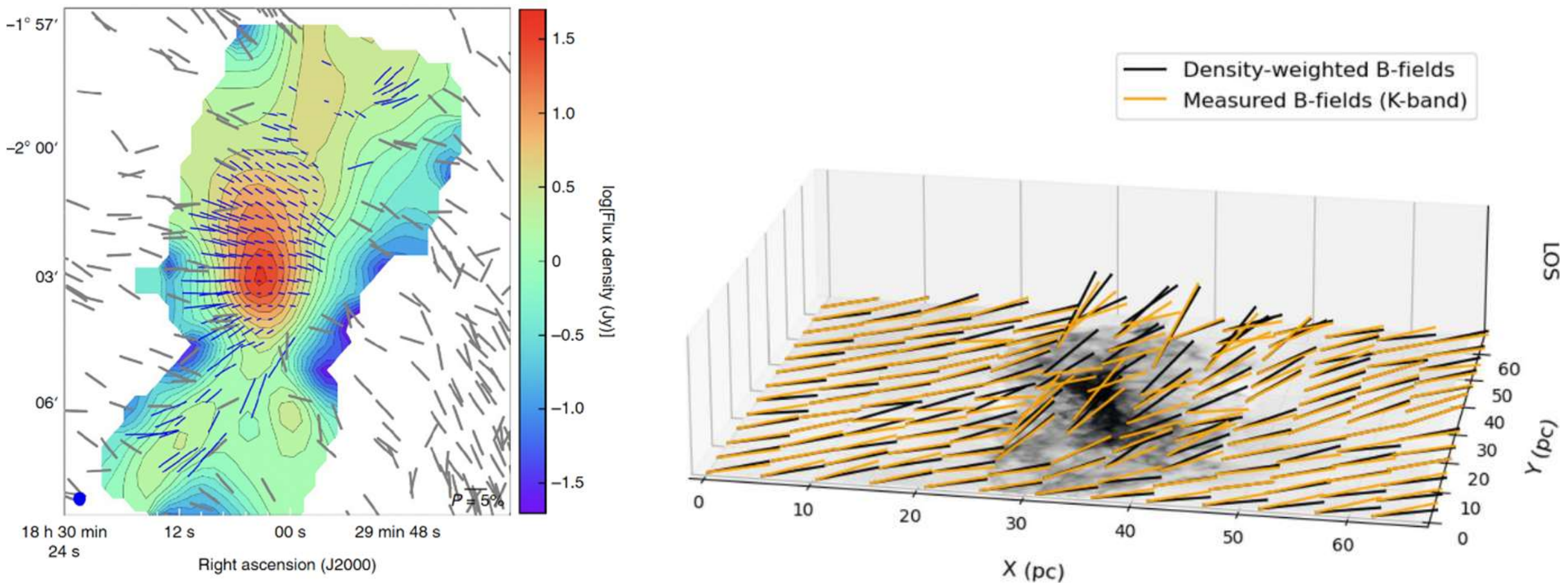

**Figure 1.** 2D B-fields from dust polarization toward the filament G11.11-012 by SOFIA/HAWC+. Right: 3D B-fields from the new techniques in Truong & Hoang (2025)

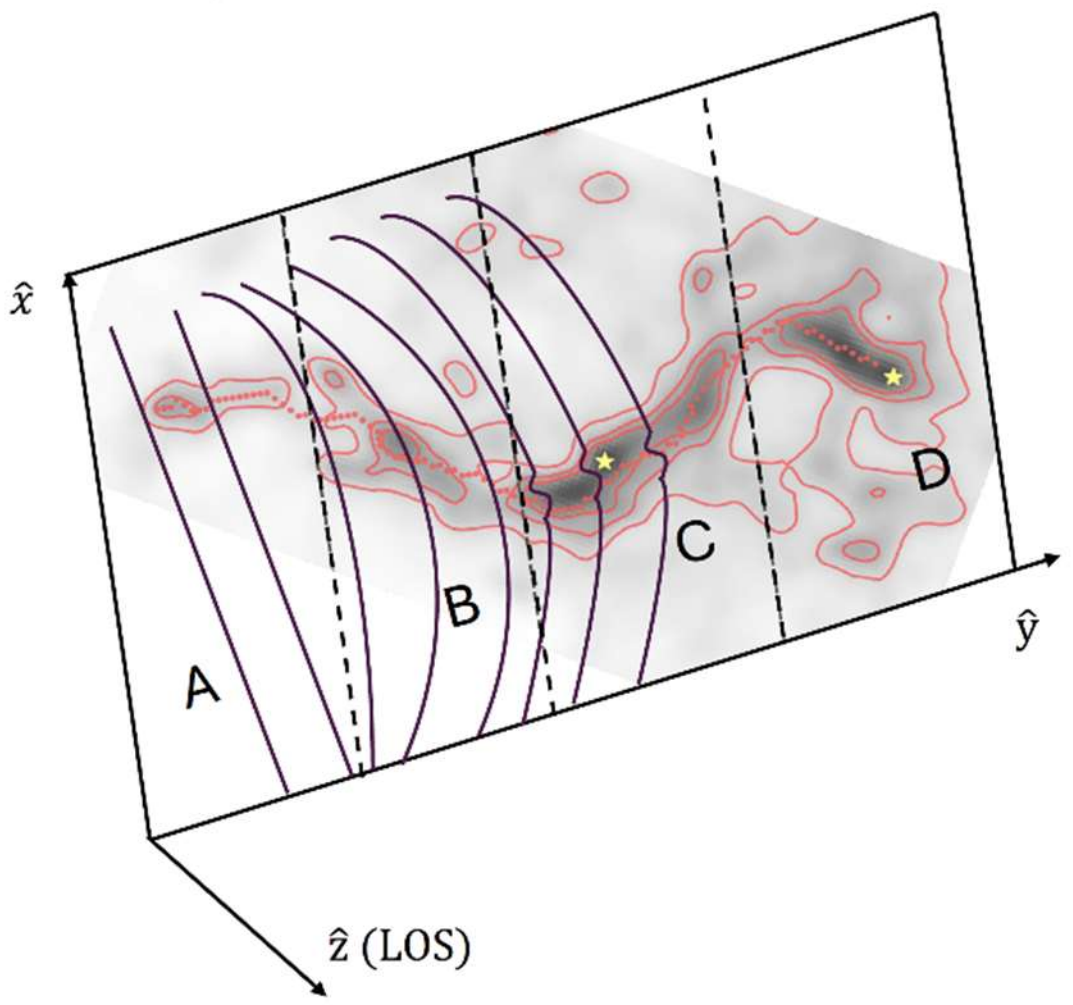

**Figure 2.** Illustrations of 3D B-fields of the massive filament G11.11-012 derived from the starlight polarization by Chen et al. (2023) at K-band (Truong et al. in preparation)

## Instruments and Modes Used

PRIMAGer polarimetric maps, totaling 160 square degrees

## Approximate Integration Time

We propose polarization observations with PRIMAger toward a sample of **10** targets of high-mass star-forming regions (SFRs), located at distances ranging from ~500 pc to ~6 kpc. Each region has an area of ~1-9 square degrees, except for the two IRDC filaments with a smaller size of ~25'x25' (~500 - 700 square arcminutes).

We use *Herschel* data at 70 µm, 160 µm, and 250 µm to estimate the required observing time for PRIMAGer in two bands, PPI1 (91 µm), (172 µm), and PPI4 (235 µm), respectively. For PPI3 (126 µm), we adopt the average intensity from 70 and 160 µm *Herschel/*PACS bands as a proxy.

Since our targets are high-mass star-forming regions, their continuum emission at longer wavelengths is expected to be relatively strong. In the outermost region of the faintest target (S106), we find that the minimum intensity in band PPI4 is roughly 40 MJy/sr.

In this band (PPI4), with the beamsize of 27.6", assuming the 5% polarization level in outer regions, a total-intensity limit of 40 MJy/sr (~560 mJy/beam) would correspond to a polarized intensity of 2 MJy/sr (~20 mJy/beam). Using the PRIMAger Exposure Time Calculator, achieving a 5σ sensitivity in PPI4 requires an observation speed of ~0.5 hour/deg². Our targets cover about 25 deg², implying a total of ~13 hours for all targets.

For some areas, we will add deeper observations.

For the band PPI3, we estimate observing time for a smaller area of 10'x30' in Cygnus-X, which covers the DR21 filament. We find that it would take ~ 22.0 hours to observe this target, if the intensity in this band is ~0.66 mJy/beam and 20% polarization level is assumed. We plan to make observations for one-third (⅓ ) of the targets, but focus on their central region of the brightest targets. This would require ~65 hours in total.

For PPI1, we assume the brightness intensity at 91 µm to be approximately similar to that at 70 µm from *Herschel*/PACS. We propose to focus on the densest and hottest region of the targets. For example, in DR21, we will observe the DR21 main region with a size of 5'x5', which requires ~20.1 hours. We plan to make observations for one-third (⅓) of the targets with the brightest center, which approximately costs ~60 hours for observing time.

Total time is about 140 hours

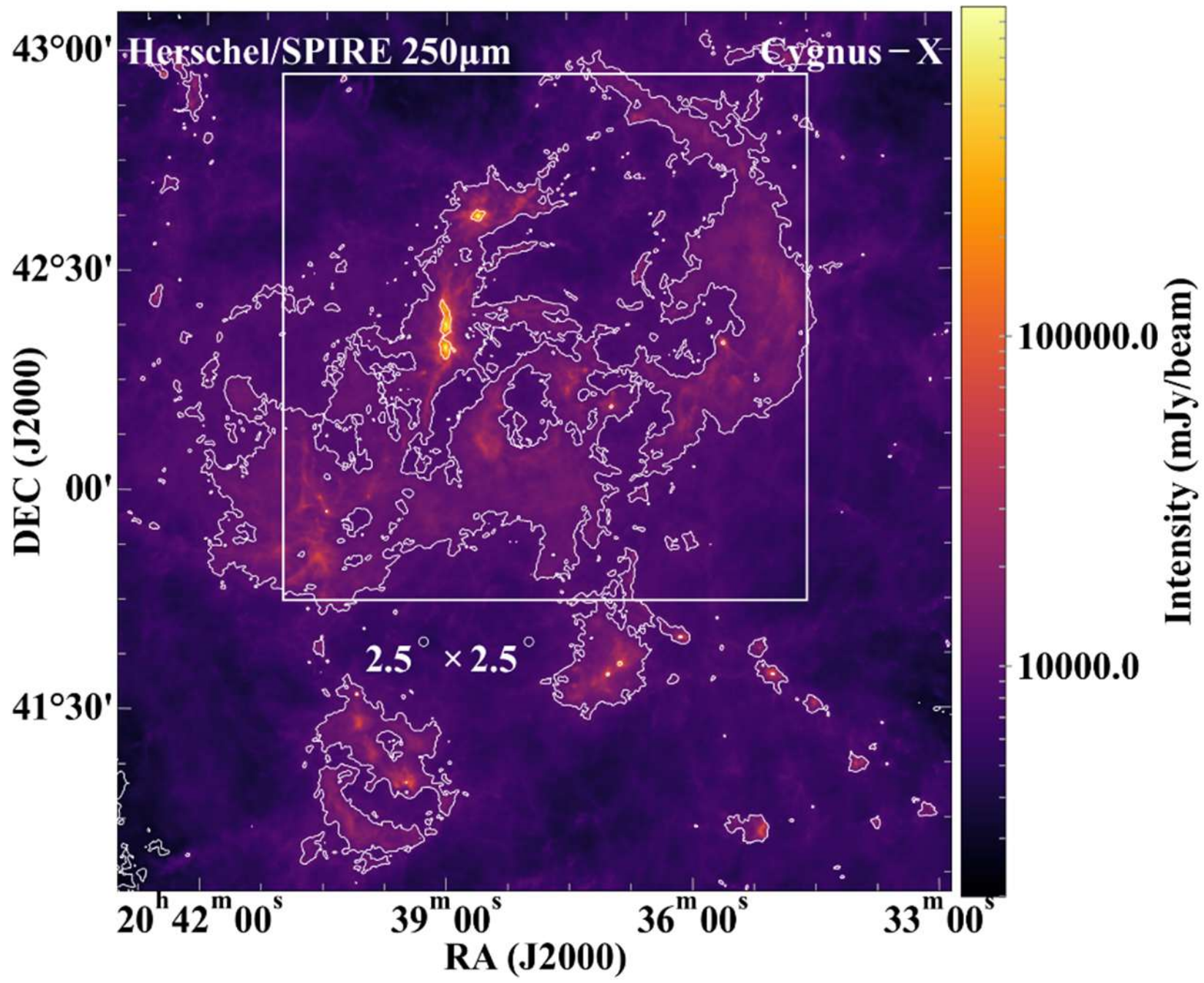

**Figure 3.** *Herschel* map at 250 µm in the Cygnus-X region, representing the expected emission map observed with PRIMAger in band PPI4. White contours show the intensity at levels of (2e3,3e3,1e4, 1e5) mJy/beam. The white rectangle with a size of 2.5x2.5 deg shows the area to be observed in band PPI4.

## Special Capabilities Needed

None

## Synergies with Other Techniques

Faraday rotation and Zeeman measurements will be used to combine with dust polarization to derive 3D magnetic fields.

## Description of Observations

The selected targets are associated with high-mass star-forming activities at various evolutionary stages, ranging from Infrared Dark Cloud (IRDC) filaments to dense high-mass SFRs, and further to more evolved H II regions.

Number of Targets: 10 (target highlighted in boldface as representative of the class)

IRDC: G16.96+0.27, G34.43

High-mass SFRs: NGC 6334, Cygnus X North, G28, Mon R2, W3, W4

HII regions: S106, M17

## Description of Data Analysis and Modeling

We will use the numerical codes Dustpol-Py to model dust polarization (starlight and thermal emission polarization) and constrain dust properties. We will utilize our new POLARIS code to perform synthetic continuum total emission polarization observations by post-processing multi-scale MHD simulations from AREPO, thereby comparing synthetic total emission and polarization with observations. Whenever available, Zeeman and Faraday rotational measurements (whenever available) will be used to check the B-field's inclination angle obtained from dust polarization.

# 72. From Dust to Planetesimals/Comets: Probing Radial Mixing and Mineral Evolution in Planet-Forming Disks with PRIMA

Mitsuhiko Honda (Okayama University of Science), Ryo Tazaki (University of Tokyo), Hiroki Chihara (Osaka Sangyo University), Shogo Tachibana (University of Tokyo), Aki Takigawa (University of Tokyo), Shota Notsu (University of Tokyo), Hanako Enomoto (University of Tokyo)

Understanding the transformation of interstellar dust into planetesimals and comets is a cornerstone of planetary system formation studies. In particular, the processes that govern the mineralogical evolution and radial transport of solids in protoplanetary disks remain poorly constrained. Pristine amorphous silicate dust exists in protoplanetary disks, while crystalline silicate dust, which may be formed in the hot inner region by thermal annealing, is also often found in the disks, showing the diversity of thermal and mixing histories of the disks. Solar System comets also show crystalline silicate components and suggest large-scale mixing of thermally processed materials, but it is unclear whether this is a universal phenomenon or a peculiarity of our own system. To address this, it is critical to know the silicate dust composition at cold outer comet forming regions of the disk, only traceable by far-infrared spectroscopy. Here we propose using PRIMA's mid- to far-infrared spectroscopic capabilities to conduct sensitive surveys of solid-state features such as forsterite, enstatite, and water ice across statistically significant samples of disks.

These observations will trace the radial distribution and thermal history of dust and volatiles, providing insight into the efficiency of radial mixing and the formation conditions of planetesimals. By linking disk dust compositions with those inferred for extrasolar debris and comets, we can test key hypotheses about the ubiquity of these processes. The resulting data will form a crucial bridge between disk chemistry and the architectures of mature planetary systems, ultimately informing our understanding of planet formation and the emergence of habitable environments.

## Science Justification

The initial stages of planet formation involve the growth of submicron-sized interstellar dust grains into kilometer-sized planetesimals within gas-rich disks. While dust constitutes only ~1% of the disk's mass, it is critical for the formation of solid planetary bodies (Birnstiel et al. 2016). The mineralogical and volatile composition of these grains evolves through heating, irradiation, collisions, and radial transport (Nittler & Ciesla 2016). Observations of Solar System comets reveal the presence of crystalline silicates—materials requiring high-temperature processing—implying substantial radial mixing from the hot inner regions to the cold outer disk (Crovisier et al. 1997; Wooden et al. 2000).

Yet it remains an open question whether this mixing is universal in protoplanetary disks or unique to our Solar System. The far-infrared spectral features of silicates and water ice provide a direct window into these processes, but current facilities lack the sensitivity and spectral coverage to observe them in large, diverse samples of disks.

**We will address three central questions:**

(1) Is radial mixing of thermally processed dust from inner to outer disk regions a common feature across planetary systems?

(2) How does the composition of solids in disks change from primordial to debris stages?

(3) Can we link the dust composition in young disks with the observed properties of comets and extrasolar planetesimals?

These questions remain largely unanswered due to the limited sensitivity of existing facilities in the far-infrared domain, where many key solid-state features are found.

To answer these questions, highly sensitive mid- to far-infrared spectroscopy is essential to probe colder disk regions where icy planetesimals form. PRIMA is uniquely suited to address these questions due to its capability to conduct sensitive, high-resolution infrared spectroscopy across mid- to far-IR wavelengths. The PRIMA/FIRESS is capable of detecting the 45 µm water ice and 69 µm crystalline forsterite features in disks as faint as T Tauri stars and brown dwarfs. This level of sensitivity enables large statistical surveys of disks across evolutionary stages, capturing the thermal history and radial distribution of dust and ice. This capability is especially important because JWST lacks sufficient sensitivity beyond 28 µm, and ALMA, though powerful, cannot access the far-infrared solid-state features.

By measuring the spectral features of crystalline silicates (e.g., forsterite 69 µm) and water ice (e.g., 45 µm) toward various disks, PRIMA will allow us to:

- Determine the degree of dust processing and annealing in different radial zones (e.g., Maaskant et al. 2015). Figure 1 illustrates the importance of forsterite 69 µm as the tracer of cold crystalline silicate grains.
- Trace the presence of cold water ice in outer disk regions, which informs the formation conditions of icy planetesimals (Maldoni et al. 1999). Note that the degree of mixing of water ice and silicate affects the appearance of forsterite 69 µm profile (Figure 2), thus modeling ice and dust components all together is required.
- Constrain radial mixing efficiency through the detection of high-temperature materials at large disk radii (e.g. van Boekel et al. 2004), offering constraints on turbulence and transport mechanisms. Furthermore, with the statistical samples, we can answer the universality of the significant radial mixing that happened in Solar system formation.
- Combining these observations with sub-mm data from ALMA (e.g., CO line distributions, dust mass) and PRIMA's molecular line observations, especially water vapor with various transitions, will offer a comprehensive picture of disk chemistry & evolution.

By combining spectroscopic observations with disk modeling, PRIMA will transform our understanding of how dust evolves into planetesimals—and whether the path followed in our Solar System is a cosmic norm or an exception.

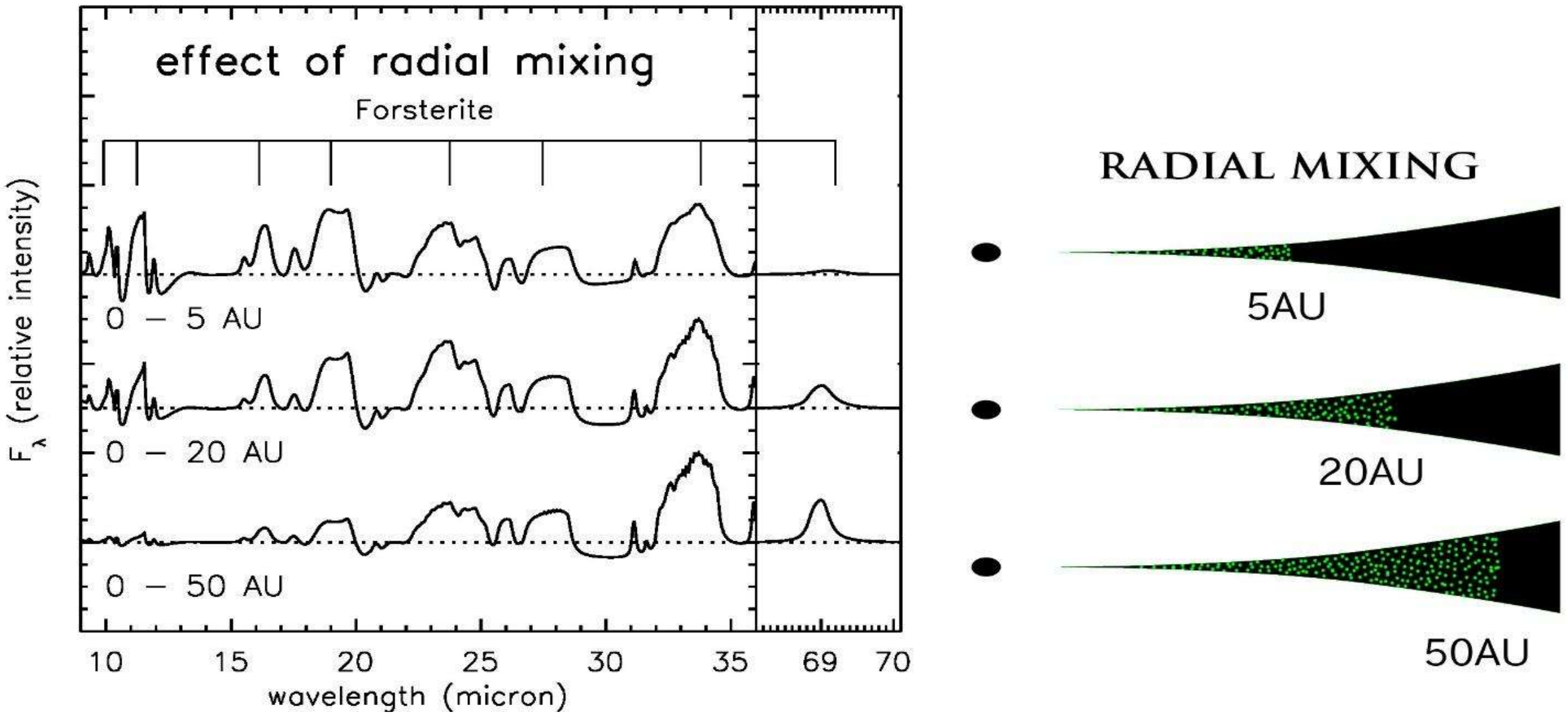

**Figure 1.** The effect of radial mixing of forsterite (crystalline silicate) grains in the disk on the spectra. When forsterite resides only within 5 au (top), the 69 μm forsterite feature is very weak. On the other hand, when forsterite appears within the 0–50 au radius of the disk (bottom), a strong 69 μm feature appears (Kamp et al. 2021, Maaskant et al. 2015).

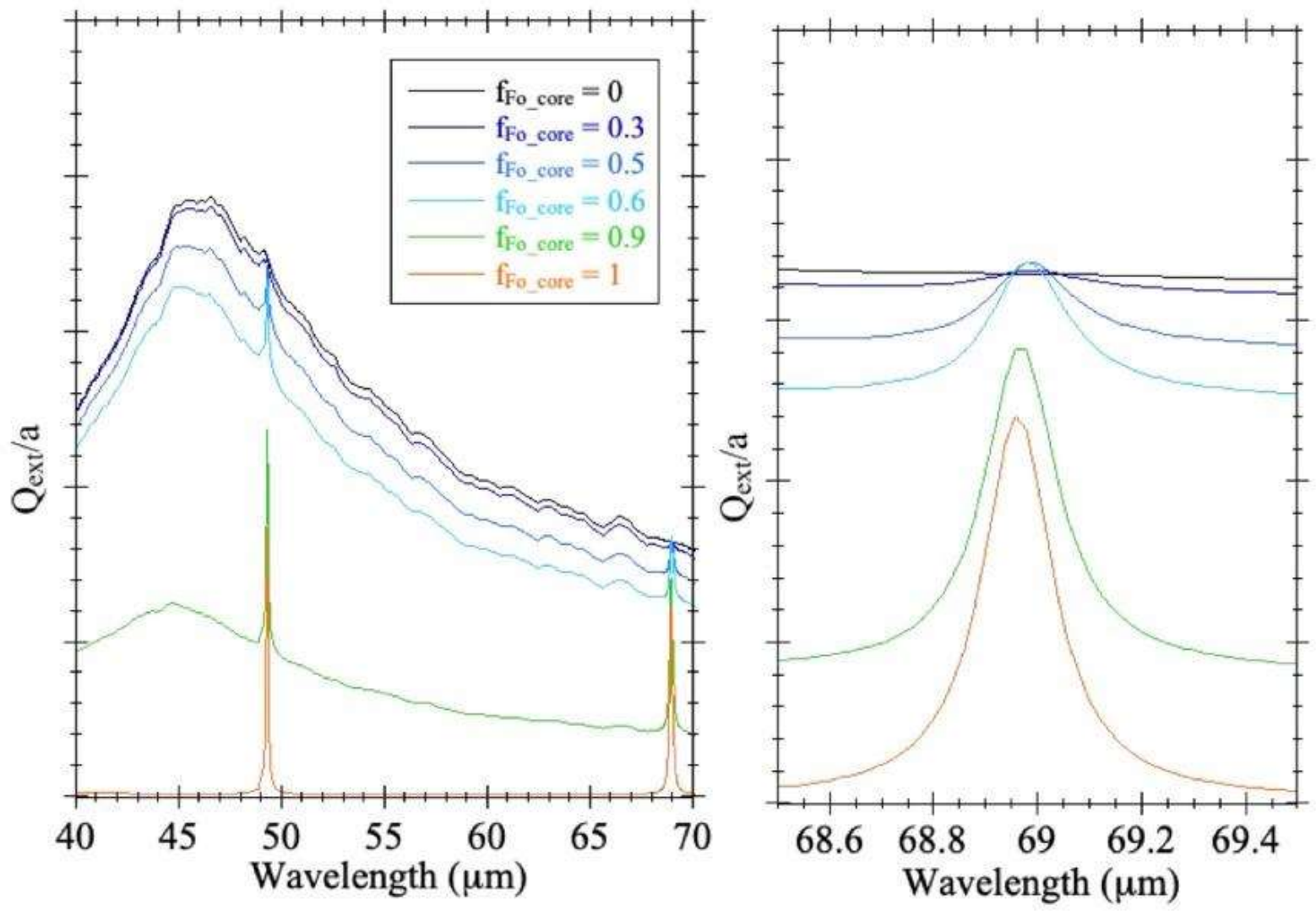

**Figure 2.** Calculated extinction coefficients of forsterite dust covered with amorphous H2O ice mantle at ~100 K (left) and the zoom at 69 μm (right). fFo_core is the size ratio of the forsterite core to the whole grain. Increasing ice mantle fraction broadens the width of the forsterite 69μm feature accompanied by a small peak shift. (based on data from Suto et al. 2006, Schmitt & Trotta 1991)

## Instruments and Modes Used

This science case will use the FIRESS spectrometer in the low-resolution mode for 300 point-source targets.

## Approximate Integration Time

Although the flux density of the target varies significantly depending on the type of object and wavelengths of interest, here we assume the typical flux density of 1 Jy at 70 µm and required S/N of 100 for 5% feature to continuum ration at 69um forsterite feature (5 sigma detection), which requires a few minutes for integration time for PRIMA/FIRESS Low resolution spectroscopy mode. According to the ETC, we assume 5 minutes for each source, and we would like to propose ~300 targets in this program. Therefore, the total integration time requested will be ~25 hours (300 targets x 5 min = 1500 min = 25 hours). Note that no overheads included in this estimation.

Highly accurate calibration will be a more important requirement for this study. Because of the strong continuum expected underneath the features, the relative calibration is required to be better than 1%.

## Special Capabilities Needed

High accuracy in the spectral response of FIRESS (better than 1%) is required.

## Synergies with Other Facilities

To understand the disk properties such as structure, composition, geometry, and so on, multi wavelength studies using JWST, ALMA, and other ground-based facilities are strongly favorable.

## Description of Observations

The main targeted features are the crystalline forsterite 69 µm feature and the (crystalline) water ice ~44 µm feature. As the expected features are weak relative to the underlying continuum even for the crystalline feature, we need the S/N > 100 for the continuum component. Thanks to the unprecedented sensitivity of PRIMA, this requirement is easily achieved for most of the target, even low-mass T Tauri stars and brown dwarfs. To make the statistical discussion, we propose to make observations of ~300 targets from nearby star forming regions (e.g. Taurus, Ophiuchus, Chameleon)

# 73. Tallying the Fuels of Star Formation in the outer Galaxy

Arshia Maria Jacob (University of Cologne), Dariusz C. Lis (Jet Propulsion Laboratory, California Institute of Technology)

Once thought inhospitable to star formation, the outer regions of the Milky Way and other galaxies—unlike the spiral-arm-dominated inner disks—are now recognised as active star-forming environments. This paradigm shift has been driven by multi-wavelength observations from instruments such as *Chandra*, *Spitzer*, ALMA, and JWST, whose sensitivity and resolution have revealed star-forming clusters and individual young stellar objects in these low-density regions. Despite these advances, the amount and distribution of material available for star formation in galaxy outskirts remain poorly constrained. PRIMA, with its unique access to far-infrared lines of key atomic and molecular gas species, offers a vital opportunity to map and characterise this elusive reservoir. In particular, the cold neutral medium (CNM)—a critical component of the galactic gas cycle and a precursor to molecular and star-forming gas—remains poorly understood in terms of its large-scale distribution, especially beyond the solar circle.

## Science Justification

Our understanding of star formation begins with the compression of gas in giant molecular clouds, leading to the formation of dense molecular cores. These cores collapse into protostars, which evolve through accretion and the development of circumstellar disks (Zinnecker & Yorke 2007). While this represents the standard framework, environmental conditions can strongly influence the physical processes involved and potentially alter key outcomes, such as the initial mass function (IMF). The efficiency with which molecular clouds form stars is thought to depend on factors such as density, temperature, and chemical abundances—particularly that of neutral hydrogen in its atomic and molecular forms, or the cold neutral medium (CNM) of the interstellar medium (e.g., Evans 1999).

Due to their lower metallicities (Rudolph et al. 1997), reduced average gas temperatures (Mead & Kutner 1988), diminished cosmic-ray fluxes, and lower mass surface densities, the outer regions of the Milky Way and other galaxies have long been considered less efficient at forming stars than their spiral-arm-rich inner counterparts. However, near-infrared observations by the Subaru Telescope and JWST—capable of penetrating dense dust lanes—have revealed a surprising abundance of young stellar objects and star-forming clusters in these outer regions, challenging this long-standing view.

With low metallicities resembling conditions in the early universe and dynamics shaped by bar-driven inflows, the outskirts of galaxies present a unique laboratory for studying star formation in environments distinct from inner galactic regions. These conditions offer a valuable opportunity to refine our understanding of star formation under near-primordial conditions.

### The Distribution of the Cold Neutral Medium

The significance of galaxy outskirts in disk evolution and star formation has been recognised for over 60 years, dating back to early extragalactic H I studies, when extended atomic hydrogen beyond optical disks was first observed (e.g., Höglund & Roberts 1965). More recently, building on this foundation, Dickey et al. (2022), using data from the Australian Square Kilometre Array Pathfinder (ASKAP), showed that in the inner Milky Way, the CNM as traced by HI has a vertical scale height similar to that of molecular gas. In contrast, in the outer Galaxy, the CNM and WNM are well mixed, maintaining a nearly constant CNM/WNM ratio out to galactocentric radii of at least 40 kpc (see Fig. 1). However, molecular line studies have not yet probed these outer regions, leaving the nature and distribution of the CNM—as traced by molecular material—largely unknown. Although far-IR observations with the IRAS satellite and large-scale maps of low-J CO lines have offered valuable insights into the molecular gas content of the outer Galaxy (Dame et al. 2001, Colombo et al. 2021), they either do not match the spatial extent of H I maps or may be limited as tracers in these regions, where a significant fraction of the molecular gas is likely CO-dark and thus not detectable through CO emission alone. For instance, Digel et al. (1994) found several molecular clouds beyond 18 kpc; however, they are all much less CO-bright (although of comparable mass) than Orion viewed at a similar distance and resolution. This limitation is underscored by the recent serendipitous discovery by Busch et al. (2021) of extremely broad (~150 km $s^{-1}$), faint (<10 mK), and widespread 1665 and 1667 MHz ground-state thermal OH emission in the second quadrant of the outer Galaxy (R> 8 kpc), using the Green Bank Telescope (see Fig. 1). Their findings suggest the presence of a vertically extended (−200 pc < z < 200 pc), diffuse molecular gas disk in the outer Galaxy—material that has remained undetected in previous all-sky $^{12}CO$ surveys.

This discovery raises a critical problem: if a significant reservoir of molecular gas is not traced by traditional CO observations, our current understanding of the molecular gas distribution—and thus the star formation potential—in the outer Galaxy is incomplete. Using alternative tracers and conducting comprehensive surveys of the outer galaxy are essential steps toward accurately mapping this hidden component and refining models of galactic star formation.

## Need for PRIMA

PRIMA, with its access to multiple CO-dark and molecular line tracers such as $C^+$, CH, and OH (Pineda et al. 2023, Rugel et al. 2018, Jacob et al. 2019, 2023), enables simultaneous observations across diverse phases of the interstellar medium (ISM). These tracers—often detected in absorption—are crucial for disentangling the contributions from distinct ISM components and enable galactic tomography through their unique velocity structures. Targeting the outer Galaxy offers key advantages: observations are free from the kinematic distance ambiguities that affect the inner Galaxy (Roman-Duval et al. 2009; Wienen et al. 2015), and source confusion is significantly reduced due to the lower density of molecular clouds. Moreover, PRIMA's enhanced sensitivity and high spatial resolution at far-infrared (FIR) wavelengths will allow for a significantly more comprehensive and efficient survey of the molecular gas reservoir in the outer Galaxy—surpassing what is currently achievable through time-intensive radio observations. These capabilities position PRIMA as an essential tool for uncovering the full extent of the molecular

ISM and refining our models of star formation in extreme, low-metallicity environments analogous to those in the early universe.

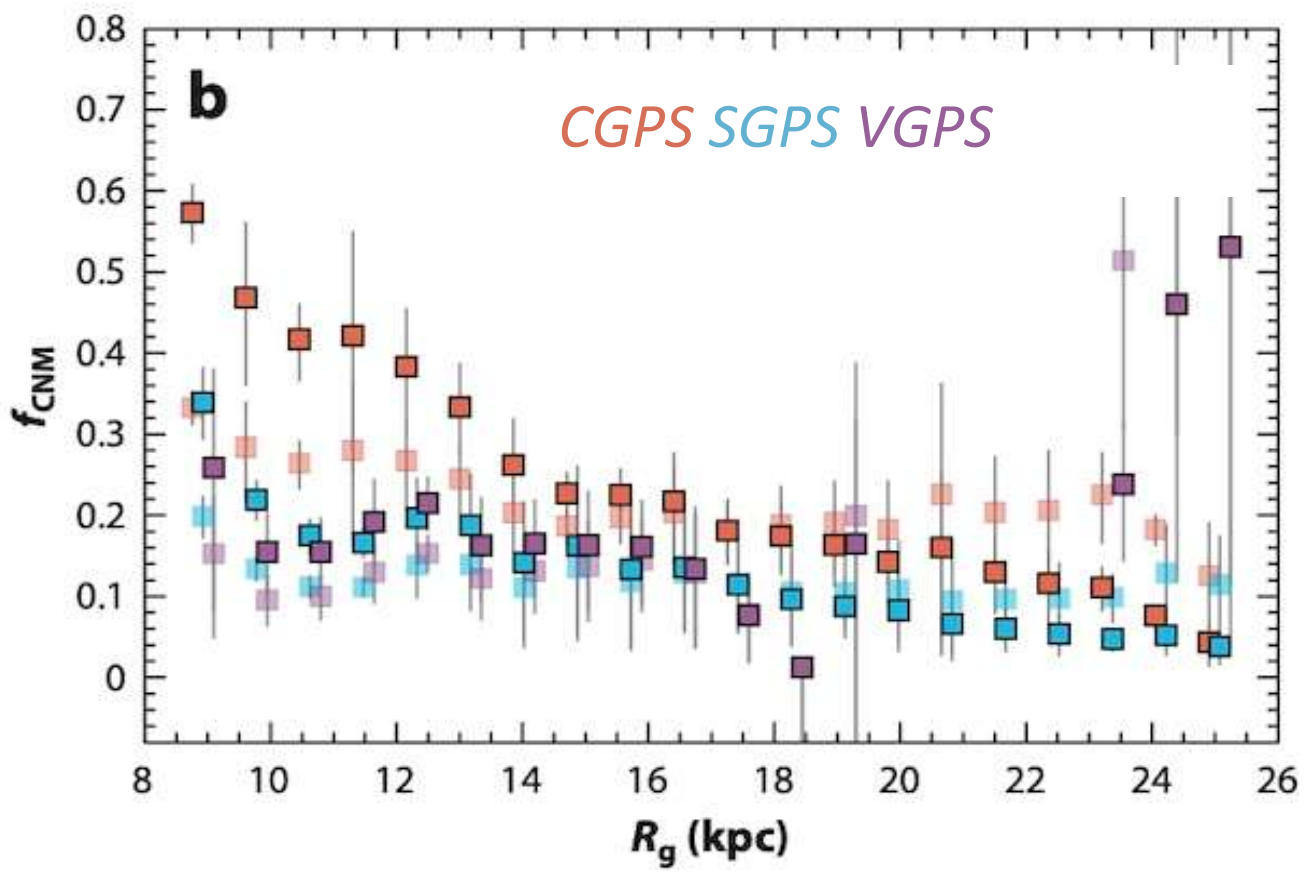

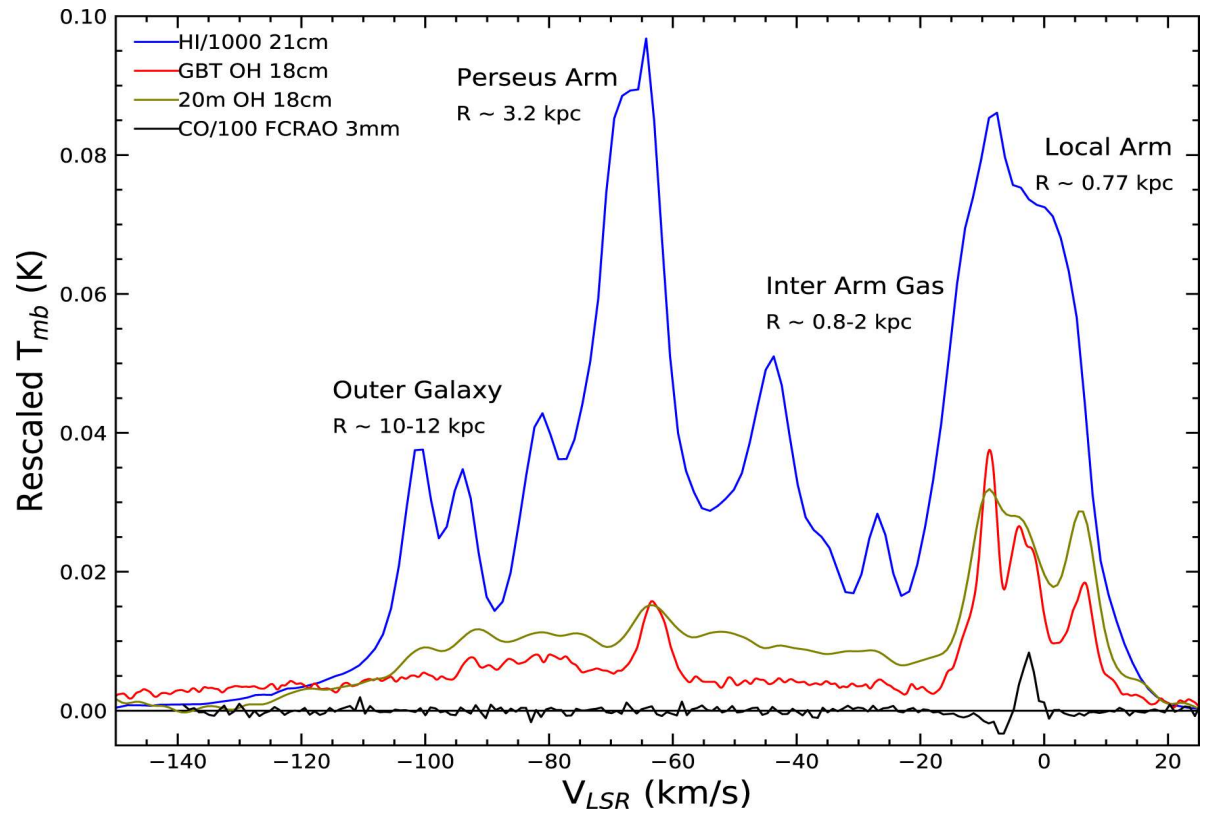

**Figure 1.** *Left:* The implied CNM fraction from HI surveys, $f_{CNM}$ =Ts/⟨Ts⟩, as a function of Galactic radius. The CNM fraction shown here is dependent on the assumed cold gas temperature. Taken from McClure-Griffiths et al. 2023. *Right*: The observed 1667 and 1665 MHz OH broad emission toward the outer Galaxy ($l$ = 108°, $b$ = 3°) in red and dark green, CO emission in black and HI emission in blue. Taken from Busch et al. 2021.

## Instruments and Modes Used

This case will use the FIRESS R~100 mapping over 1 deg x 1 deg.

## Approximate Integration Time

The key lines of interest (CH 149 µm, CII 158 µm, and NII 205 µm) are all in PRIMA FTS Band 4. To achieve the required sensitivity $2.16 \times 10^{-18}$ $Wm^{-2}$ (5σ) across a map of size 1 deg × 1 deg area, results in a total time of 107 hours (Urquhart et al. 2025). The sensitivity criteria is made by scaling on previous CO observations made using the APEX 12m sub-mm telescope under the OGHReS survey.

## Special Capabilities Needed

No

## Synergies with Other Facilities

The main objectives of this survey are to map the distribution of molecular gas in the outer Galaxy and to investigate star formation under very different conditions from those found in the inner Galaxy and Galactic centre region. When combined with dedicated follow-up observations of key star formation tracers such as $HCO^+$ using instruments like ALMA (Braine et al. 2023), PRIMA data can be used to investigate the relationship between dense gas and star formation. On kilo parsec scales, a strong correlation is observed between dense gas mass and the star formation rate (SFR), but whether this relationship holds in the low-density, low-metallicity environments of the outer Galaxy remains an open question. These observations are especially timely, as they will also inform the interpretation of future studies with instruments like CCAT, which aim to probe star formation in high-redshift galaxies under similarly extreme conditions.

## Description of Observations

Observations will be carried out using the PRIMA FIRESS Spectrometer in the low-resolution FTS mode, which provide sufficient resolution to probe the main line of interest in this study—[CII] 158 µm.

## Acknowledgment

Part of this research was carried out at the Jet Propulsion Laboratory, California Institute of Technology, under a contract with the National Aeronautics and Space Administration (80NM0018D0004).

# 74. Hyperspectral Imaging of the Magellanic Clouds with PRIMA

Olivia Jones (UK Astronomy Technology Center) Kathleen Kraemer (Boston College), Peter Scicluna (University of Hertfordshire), Beth Sargent (STScI/JHU), Sundar Srinivasan (IRyA-UNAM)

As stars evolve and die, both low and intermediate-mass asymptotic giant branch (AGB) stars and high-mass, red supergiants (RSGs), and supernova explosions inject material from their stellar interior into the interstellar medium (ISM). This drives the chemical enrichment of galaxies (see Schneider & Maiolino 2024 for a recent review). If we wish to achieve a self-consistent understanding of galaxy evolution, we must thus understand how stars evolve and die on a galactic scale, where distances are known, and we can link the stellar content to the physical conditions. This will also enable us to compare and test stellar, dust, and chemical evolution models with a high degree of precision. Previously, dust-injection rates for the Magellanic Clouds have been computed using mid-IR fluxes, which are sensitive to warm (100 K) dust. Yet intriguing evolved stars have been detected in the Magellanic Clouds at sub-mm wavelengths (Seale et al. 2014; Jones et al. 2015b). Stars emitting at these wavelengths must have significant cold dust (T < 50 K) reservoirs; as such, they trace a unique population of mass-losing sources. Surveying the Magellanic Clouds with the PRIMA Hyperspectral imager will measure the emission from evolved stars longwards of 24 microns. Constraining this emission will both separate very red evolved stars from young sources in the process of formation and quantify dust production by metal-poor evolved stars and supernova remnants, by constraining their far-IR emission

## Science Justification

SN ejecta and AGB winds are generally considered to be the primary source of dust in galaxies with their relative contribution depending on various factors, for instance AGB stars may start to dominate dust production in a galaxy after only a few 100 Myr or this process may take more than 1 Gyr depending on the stellar initial mass function (IMF; e.g, Schneider & Maiolino 2024 and references therein). Massive stars, e.g., RSGs (van Loon et al. 2005), Wolf–Rayet stars (Lau et al. 2022), and Luminous Blue Variables can also produce vast quantities of dust on short timescales, comparable to the expected dust masses from core-collapse SNe, after reverse shock. However, most of this dust is expected to be destroyed by the subsequent SN explosion. These processes and the relative contribution between different sources of dust are still poorly understood, and there is a large variety of models and assumptions, which can result in completely different implications for early galaxy formation and the evolution of baryonic matter.

Infrared surveys of nearby galaxies and their stellar populations are the perfect mechanism for providing tighter observational constraints on the enrichment of the ISM, as a wide variety of sources eject their nucleosynthetic material through dust-driven winds or as ejecta dust in the case of core-collapse supernovae.

The Large and Small Magellanic Clouds at a distance of 50 and 60 kpc, respectively, have metallicity of ~½ and ⅕ solar. These low-metallicity, nearby interacting galaxies have been extensively surveyed from the UV to the far-infrared. This has resulted in a well-characterized stellar population. Thus providing both a galaxy-wide view and simultaneously allowing the detailed study of individual objects and their interaction with the local environment.

The Spitzer Surveying the Agents of Galaxy Evolution (SAGE; Meixner et al. 2006; Gordon et al. 2011) and *Herschel* Inventory of The Agents of Galaxy Evolution (HERITAGE; Meixner et al. 2010, 2013) legacy surveys of the Magellanic Clouds operating in the mid- and far-infrared, respectively, first detected this emission from cool dust originating from individual objects on a galaxy-wide scale. Here, AGB stars, RSGs, Wolf–Rayet stars, post–AGB stars, planetary nebulae, and supernovae (remnants) are exceptionally bright at infrared and sub-millimeter wavelengths as a result of their thermal emission and characteristic solid-state features due to resonances in the dust grains. Using mid-IR photometry, several works have characterised the dust budget of the Magellanic Clouds (e.g., Matsuura et al. 2009; Boyer et al. 2011; Riebel et al. 2012; Matsuura, Woods & Owen 2013; Srinivasan et al. 2009; Boyer et al. 2012; Srinivasan et al. 2016) they detected over 9700 carbon-rich and oxygen-rich Asymptotic Giant Branch stars, which are important sources of carbonaceous and silicate grains, and quantified their dust production rates. The most extreme AGB stars are detectable at > 8 $\mu$m and are thought to dominate the mass return to the ISM, producing more than 75% of the dust but constitute only 3% of the total AGB population by number (Boyer et al. 2012, Srinivasan et al. 2016, Nanni et al. 2018). Furthermore, there were circa. 100 – 300 very dusty objects with a rising spectral energy distribution (Boyer et al. 2011, Srinivarsan et al. 2016) that may potentially be evolved stars; however, the lack of well-sampled photometry at these longer wavelengths makes it difficult to separate these evolved sources from young stars in the process of formation. To do so would either require spectroscopy or photometry from an instrument like PRIMAGER's hyperspectral imager, which would enable the peak of the dust emission from very red evolved stars undergoing a superwind to be constrained, and cleanly separated from cool young objects in the process of formation, which are still expected to have a rising SED over this wavelength regime. Measuring the slope of the SED at these wavelengths would also characterize the dust grain size and temperature.

The HERITAGE Band-Matched catalog (Seale et al., 2014) covers the 3.6–500 $\mu$m range and contains over 42,000 unique sources; these represent the dustiest populations of sources in the Magellanic Clouds. From this catalogue, Jones et al (2015b) conclusively identified 35 evolved stars and stellar end products that are bright in the far-infrared. These sources are likely to be among the dustiest evolved objects in the Magellanic Clouds and originate from both intermediate and high mass stars. There are a further ~500 evolved star candidates in the HERITAGE point source catalogs of the Magellanic Clouds, which may contribute to the stellar dust return to the ISM. Determining if this emission originates from the circumstellar envelope or if it is due to swept-up ISM dust requires in-depth observation of the far-IR emission, which PRIMA will provide. In doing so, these observations will fully characterise the stellar dust return to the ISM and the mass-loss history of these stars.

Previous surveys of the Magellanic Clouds connecting the mid to far infrared were limited by sensitivity, detecting only the brightest sources. With PRIMA, we can efficiently map the bar, wings, and the Magellanic Bridge in a uniform manner, detecting dust production from objects to > 1 x $10^{-9}$ Msun/year. PRIMA can provide both the wavelength coverage and the spatial resolution required to untangle the cold dust emission from evolved stars and young stellar objects, which can have considerable overlap in colour space at 8 to 24 microns. The ~5 times improvement in point source sensitivity and three times better angular resolution at 70 microns will permit a complete census of evolved stars with mass loss rates greater than 1 x $10^{-9}$ Msun/year, while simultaneously detecting Stage 0 and I YSOs to less than 5 Msun. In doing so, we will provide a template for mass-loss and dust production as a function of metallicity, and a foundation for interpreting observations of high-redshift galaxies (z~8), with a significant amount of dust. Dust in these galaxies must have been produced within the first 1.5 billion years after the Big Bang, with both SN ejecta (initially) and AGB winds (35 Myr after their formation) considered primary contributors to the dust production in the early Universe.

## Instruments and Modes Used

This science case will use 3 PRIMAger maps.

## Approximate Integration Time

We wish to map the Large and Small Magellanic Clouds with the Hyperspectral Imager. This will require three maps to completely image both galaxies, including the Magellanic Bridge. Using the galaxy sizes listed above, chosen to match the survey area of the Spitzer SAGE and SAGE-SMC, and the Herschel HERITAGE survey of the Magellanic Clouds, the PRIMA Exposure Time Calculator returns a total of 491.4 hours for a Survey Depth (5 σ) 4.9 mJy at 70 microns.

## Special Capabilities Needed

Ability to map large areas with the same position angle on the sky so as not to introduce any gaps in the mosaic.

## Synergies with Other Facilities

Euclid, JWST, ALMA, Spitzer, Herschel, WISE, Roman, SKA, AtLast, ELT, Rubin.

The SMC and LMC have been surveyed at many wavelengths, from HI 21 cm and CO (e.g., NANTEN, Fukui et al. 1999) at the longest wavelengths, to optical and (e.g., MCPS,

Zaritsky et al. 2002, 2004) and near-IR (2MASS) surveys. Warm dust emission from the mid-IR was surveyed extensively with Spitzer, such as SAGE (LMC; Meixner et al. 2006), S3MC (Bolatto et al. 2006), SAGE-SMC (Gordon et al. 2011), SAGE-Spec (Kemper et al. 2010), and SMC-Last (Mizuno et al. 2022, Kuchar et al. 2024).

The PRIMA survey will complement these.

## Description of Observations

We propose a uniform photometric survey of the Magellanic Clouds with the PRIMA Hyperspectral Imager. This will image the LMC and SMC in 12 filters from 24 to 84 microns to a 5σ sensitivity of 4.9mJy at 70 microns. In doing so, we can also map the Magellanic Clouds with longer wavelength imaging with PRIMA polarimetry bands at 96, 126, 172, and 235 microns if the scan pattern is carefully chosen. This will produce a point source inventory of these galaxies in the far-infrared with an angular resolution comparable to that of Spitzer-MIPS 24 at 65 microns. This photometric coverage is necessary to understand these galaxies' complex stellar populations. Both evolved stars and Young stellar objects can emit light at these wavelengths, with PRIMA tracing the earliest and coldest stages of star formation (Class 0 and Class 1 YSOs) and cold dust associated with evolved stars and supernova remnants. At the distance of the Magellanic Clouds, we will detect (per beam, 5σ) the emission from individual evolved stars and YSOs. Plus, the integrated emission from compact young clusters. These far-infrared observations are important to constrain the dust mass-loss rates of evolved stars (e.g., Boyer et al 2010, A&A, 518, L142). The depth available by the PRIMA Hyperspectral Imager is 11-20 times fainter than what was possible with Spitzer MIPS, reaching fainter dust emission and lower mass objects (~<5 Msun).

# 75. Spectro-Polarimetry Survey of Local, Low-Mass Lynds Clouds: Tracing Isolated Cloud Scale Magnetic Fields at the Onset of Star Formation

Dr. Janik Karoly (University College London), Dr. Kate Pattle (University College London), Dr. Simon Coudé (Worcester State University / Center for Astrophysics), Dr. Dylan Michelson Paré (Villanova University), Dr. Archana Soam (Indian Institute for Astrophysics), Dr. Isabelle Ristorcelli (Institut de Recherche en Astrophysique et Planétologie), Dr. David Eden (Bath University)

The role of magnetic fields in star formation is still not entirely understood from observations. Star formation which is magnetically regulated is theorized to begin with a magnetically-dominated (magnetically sub-critical) molecular cloud envelope with cores forming within the molecular cloud which then become matter-dominated (magnetically super-critical). The size and density scales of this transition from sub- to super-critical is not known, including if there is a single value or if it varies across environments. We propose to observe the magnetic field in eight molecular clouds with PRIMAger to investigate this transition. These eight sources are relatively isolated, low-mass molecular clouds/cores and have submillimeter polarimetry observations which trace the magnetic fields of the dense core. PRIMAger is uniquely positioned and the only instrument which can observe the transition in these molecular clouds because of its ability to trace the extended, diffuse and warmer dust components in the molecular cloud which is where the transition is expected to occur. Being able to build up a sample of characteristics of transitions in magnetic criticality will help to answer the question when/where it occurs and if there is a single value when this transition occurs in star formation. The multiple bands of PRIMAger will allow us to trace different dust densities and temperatures as well, providing a self-consistent comparison. PRIMAger's ability to observe the magnetic field in the most diffuse material down to the embedded cores with high resolution for nearby sources (~0.01pc for a distance of 150pc) makes it crucial in understanding what role magnetic fields play in star formation.

## Science Justification

Due to the bright nature and relative abundance of filaments, great strides have been made recently in understanding how star formation proceeds in filamentary molecular clouds (see Hacar et al. 2023 and references therein). This work includes observations of magnetic fields made using both Planck (e.g. Soler 2019) and the JCMT (e.g. Arzoumanian et al. 2021), tracing filamentary magnetic fields of both nearby and far-away star forming regions, respectively. Meanwhile, nearby low-mass molecular clouds provide a unique environment in which to study star formation since they can be quite isolated with a diffuse large-scale molecular cloud envelope and dense core(s).

Many of these low-mass molecular clouds have been observed in the sub-millimeter, where the densest regions of the cloud are traced and the magnetic field at the dense core scale is recovered. Crutcher (2004) predicts that overall, star formation in these molecular clouds proceeds from a magnetically sub-critical envelope into a super-critical core. Magnetically sub-critical means that the magnetic pressure support is enough to prevent further gravitational collapse. On the other hand, super-criticality means the magnetic pressure is no longer sufficient to support against collapse. While observations generally support this sub- to super-critical transition with increasing density (Crutcher 2012, Pattle et al. 2023), both the spatial and density scale when this transition occurs is not clear, if there even is a single value. For example, in L1544 (Ching et al. 2022), one of the proposed targets in this study, they found the cloud to be magnetically supercriticality occurs earlier than expected. **Understanding this transition and when/where it may occur will greatly contribute to the understanding of what roles magnetic fields play in star formation. Tracing this transition from sub- to super-critical is something that only PRIMAger can do with the multi-band polarimetry and ability to observe extended structures.**

Herschel observations at 250μm reveal the warmer and more diffuse dust structure of some of these isolated molecular clouds (see Figure 1). PRIMAger at its longest wavelength of 235μm would trace similar dust populations and as wavelength decreases, it will trace even warmer and more diffuse populations in the other bands. These dust structures will be the envelopes of the molecular clouds where we expect sub-critical values. All the proposed targets have observations at 850μm with ~14” resolution using SCUBA-2/POL-2 on the James Clerk Maxwell Telescope (JCMT) which trace the dense cores embedded in the warmer envelope. These cores have been found to cover a range of sub- and super-critical regimes.

Polarized dust emission can be used to infer the magnetic field direction assuming dust grains are spun up by Radiative Alignment Torques (RATs) and align with the local magnetic field (Andersson et al. 2015). We will use the Davis-Chandrasekhar-Fermi (DCF) method to calculate the magnetic field strength (Davis 1951, Chandrasekhar & Fermi 1953). The DCF method is the current best estimate for plane-of-sky magnetic field strength derived from dust polarimetry (Pattle et al, 2023).

In addition, the transition of magnetic field morphology across different scales is not well understood. In some cases the molecular cloud or core magnetic field resembles the large-scale magnetic field, often traced by Planck or starlight polarization. These PRIMAger observations will allows us to compare existing Planck observations, which have a resolution of 5’ and therefore will sufficiently sample the proposed one square degree field of view, with the 10-28” resolution PRIMAger observations to determine how the structure of the magnetic field varies from the largest scales (and most diffuse medium) down to the cloud and even core scales, tracing the slightly warmer dust envelope of the cores.

PRIMAger is unique in its ability to observe these core regions out to the most diffuse envelope, therefore enabling us to probe for a transition spatial scale and density of when star formation switches from magnetically dominated (sub-critical) to gravity dominated (super-critical). With the multiband capabilities, we can see if this changes in different dust populations, whether it be different temperatures or densities. In this survey we will observe eight molecular clouds, chosen because they have existing POL-2 observations which trace the magnetic field and dynamics at the core scale and because they are nearby (all less than 170pc) so PRIMAger can resolve magnetic fields at the sub-parsec scale.

Figure 1 shows the 250 µm emission in the one square degree field of view for each of the targets. Black contours denote the dense core locations observed with JCMT/SCUBA-2. L1498 is a core where Lee, Myers & Tafalla (2001) found signs of infall motions and suggest that the core is pre-protostellar and on the verge of collapse. L1517B consists of a series of quiescent filaments and starless cores and Hacar & Tafalla (2011) suggested that the main dense core (L1517B) formed out of subsonic filaments and may be at a very early evolutionary stage still. L1512 is an isolated starless core which is chemically evolved and older than 1.4 Myr (Lin et al. 2020). The magnetic field of the core was also observed in Lin et al. (2024), finding a magnetic field strength of 18 µG. L1544 is one of the strongest infall candidates among prestellar cores and shows clear signs of gravitational collapse (Caselli et al., 2002). It is most likely in the process of forming a star and gravity has begun to dominate the dynamics. L183 has tentatively been identified as an infall candidate based on very significant blue excess in its asymmetric, double-peaked CS spectra (Lee, Myers & Tafalla, 2001). The magnetic field has previously been observed within the main dense cores (Karoly et al. 2020) and they were found to be magnetically sub-critical. L43 is a partially evolved molecular cloud with a dense, submillimeter-bright core and two low-mass protostars. The magnetic field in the core is unique from the large-scale magnetic field and there is an observed sub- to super-critical transition from the edge inward (Karoly et al. 2023). L1495 is a series of filaments in a triangle configuration and 8 starless cores (Ward Thompson et al. 2023). The filaments are not observable at 850 µm and the understanding of the magnetic field within them will be important in understanding the evolution of the cores. L1527 is the only source among these which has an embedded stellar object (an embedded Class 0/I protostar; van't Hoff et al., 2023) which also drives a large outflow detected by JWST and in CO. It is also situated in the least isolated environment (see Figure 1).

L43

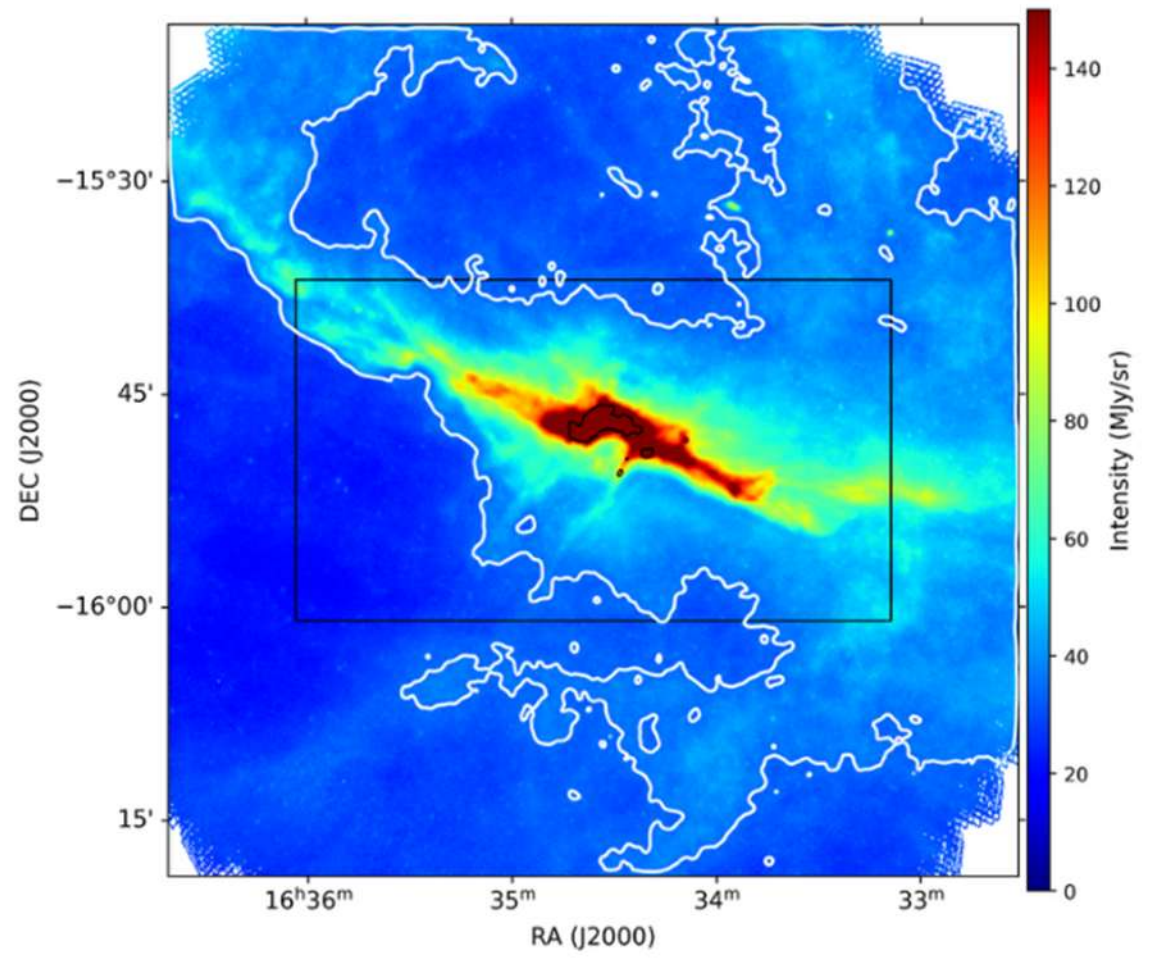

L183

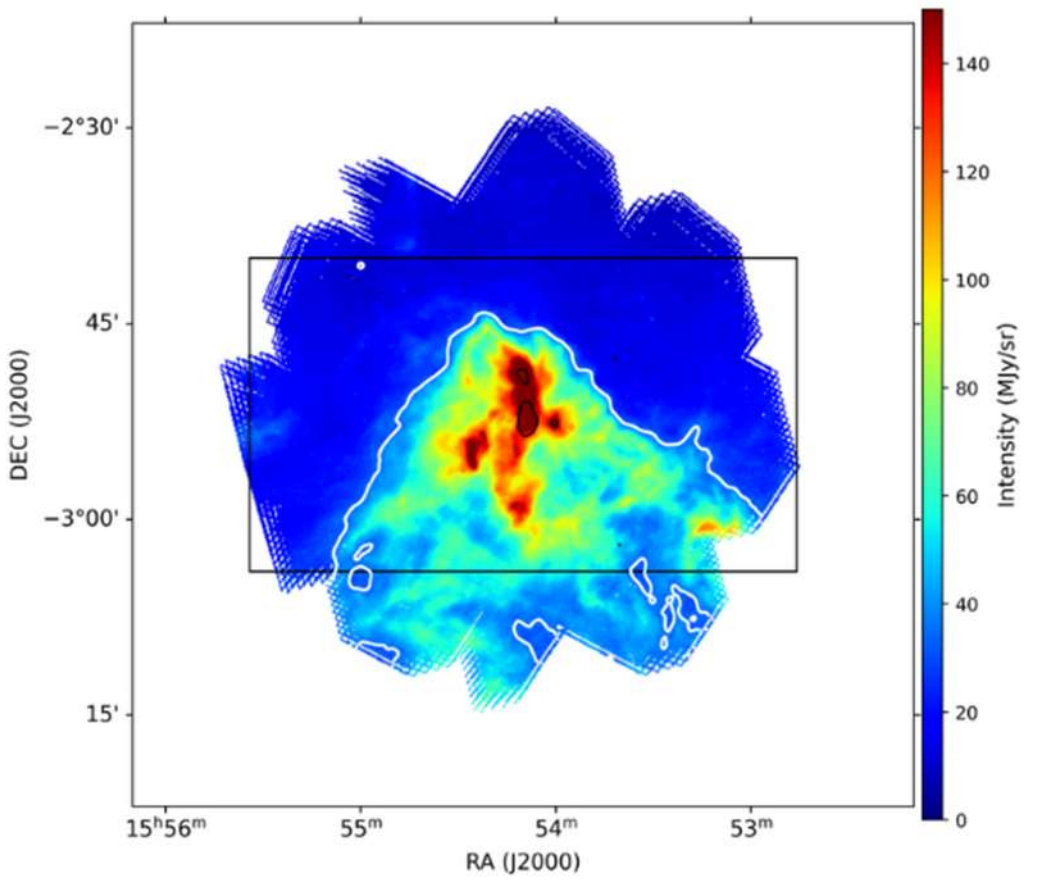

L1495

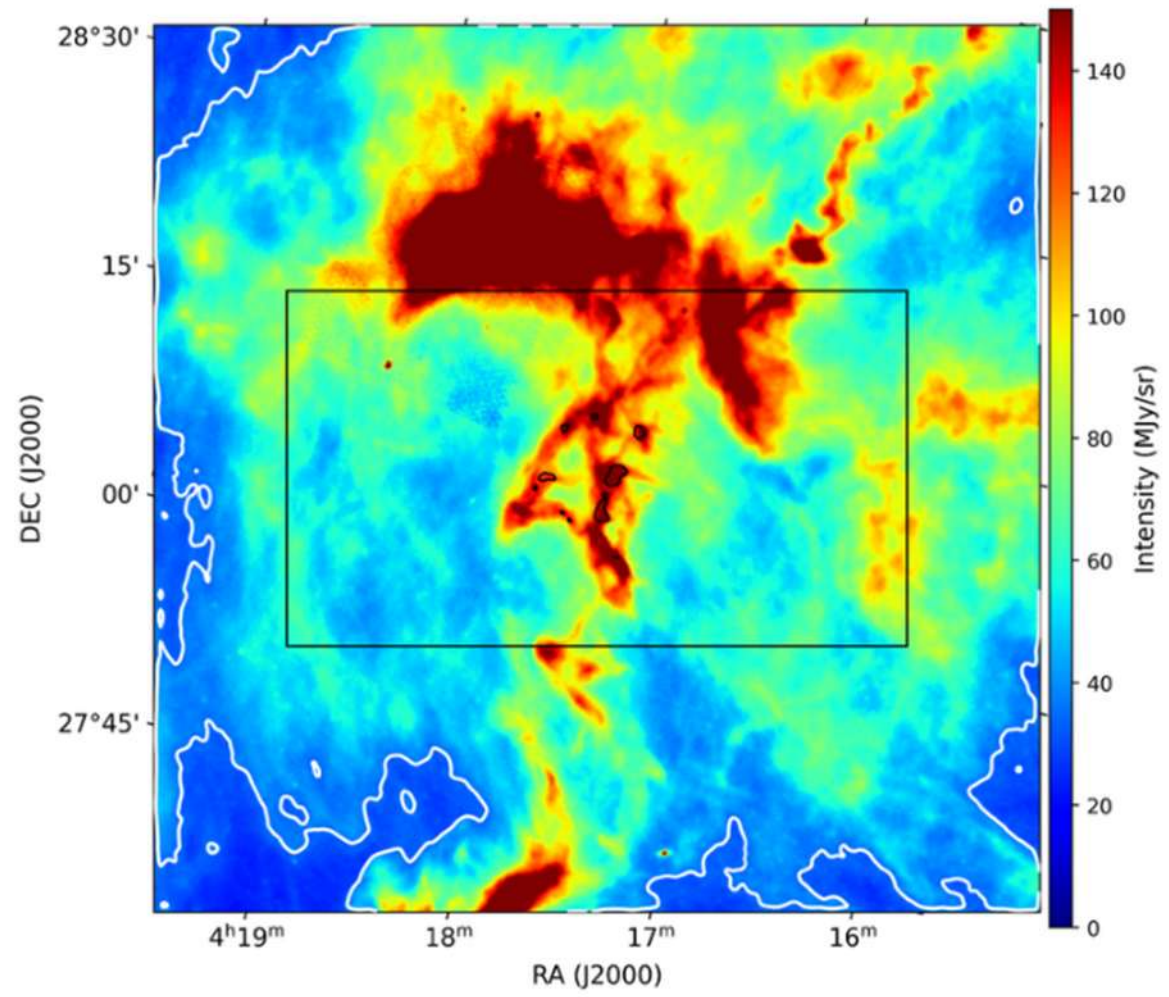

L1498

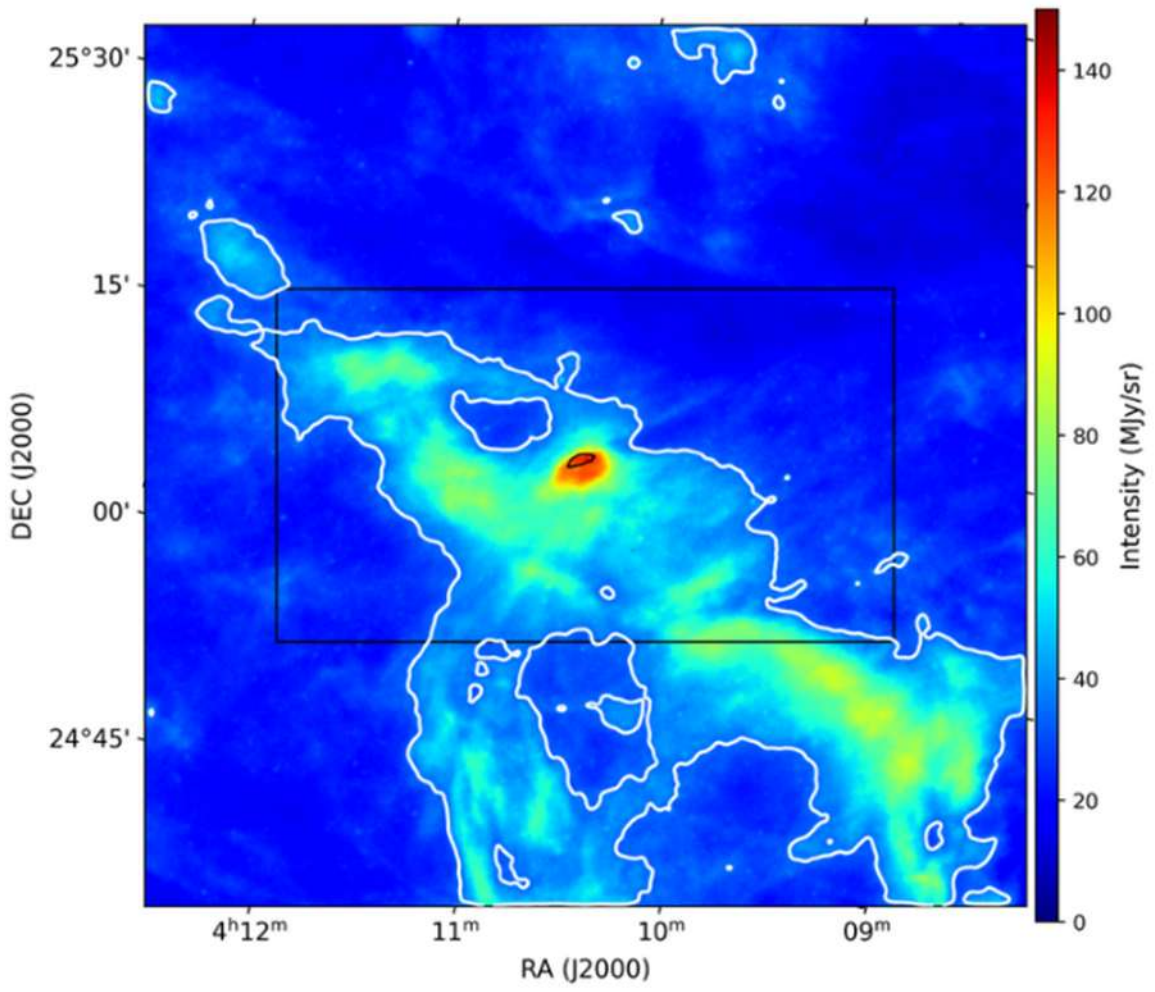

L1512

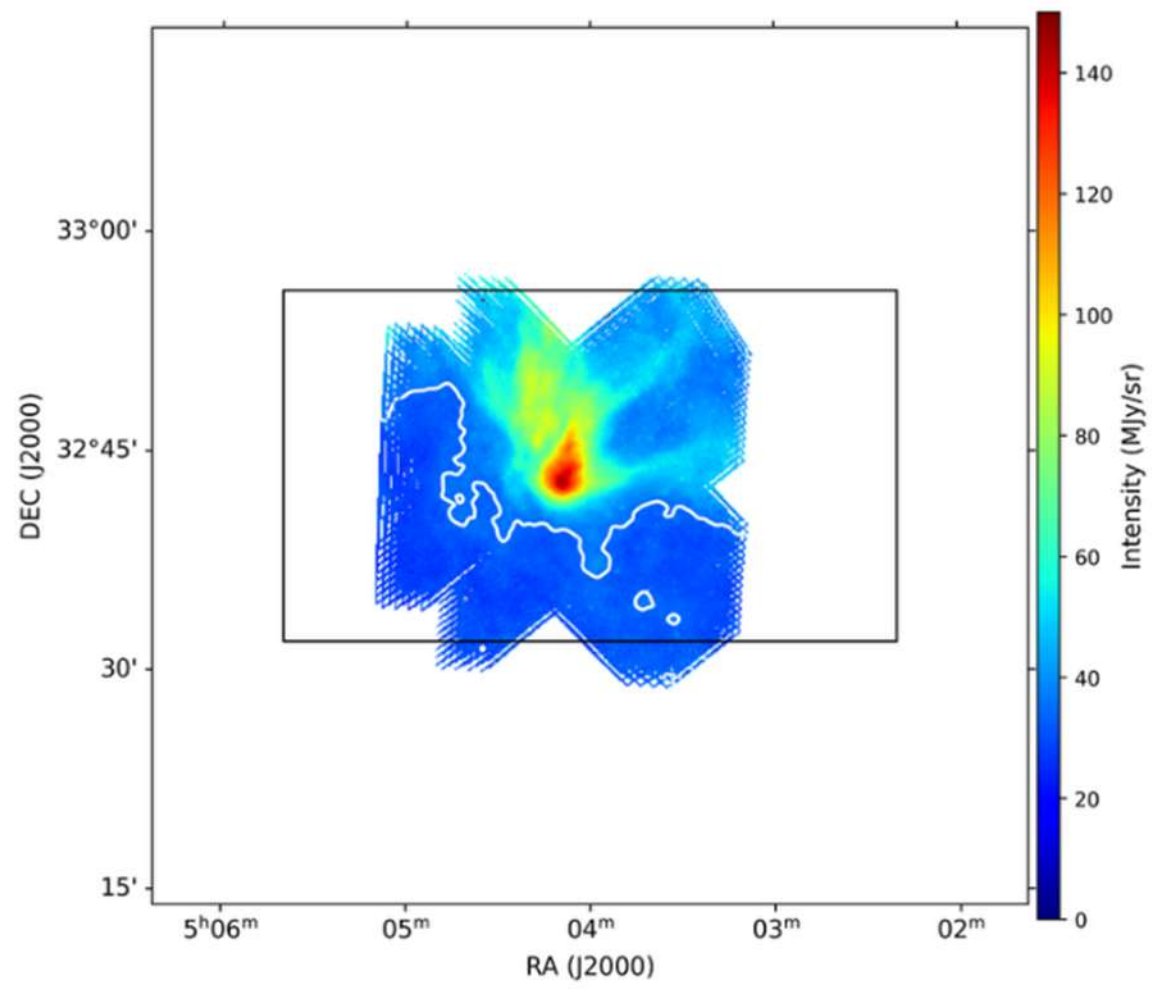

L1517B

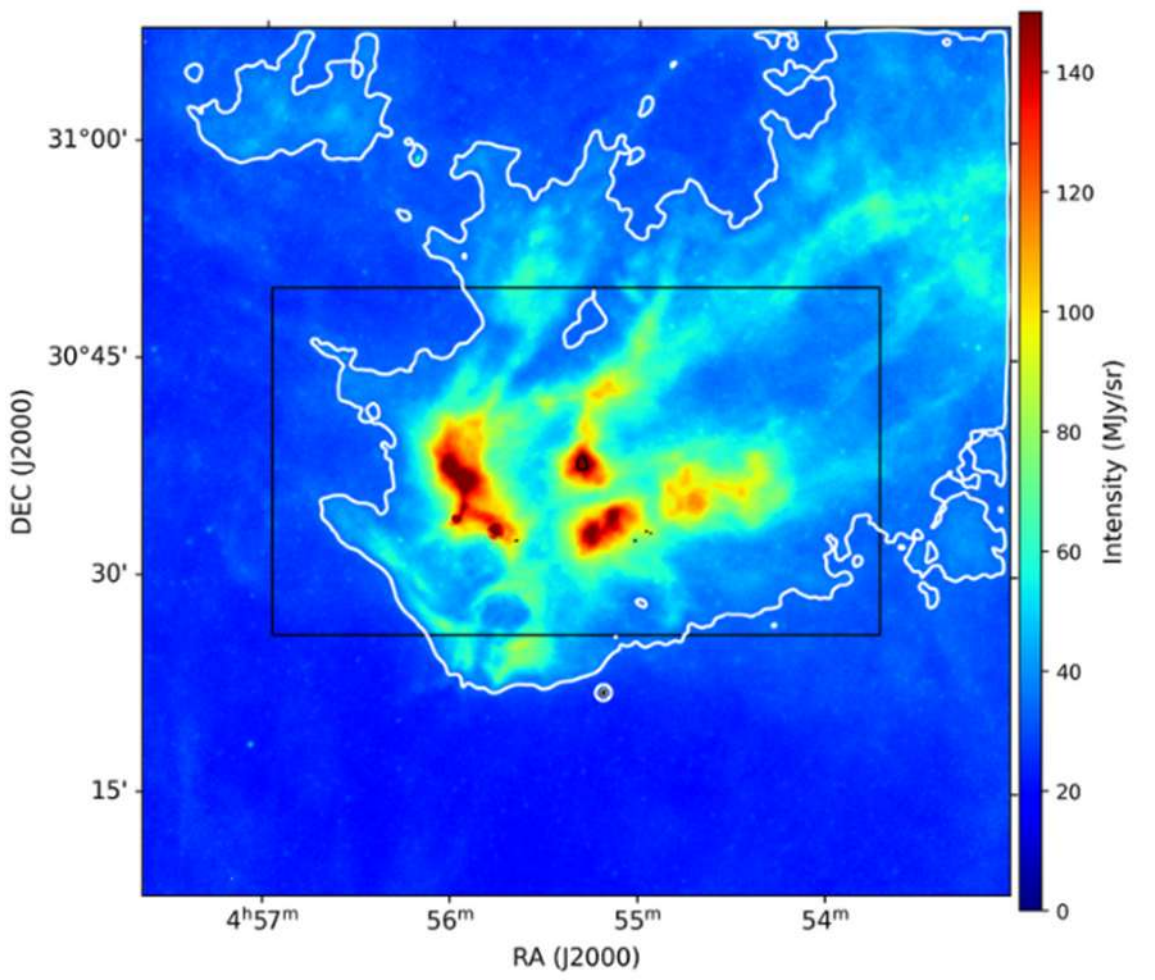

L1527 L1544

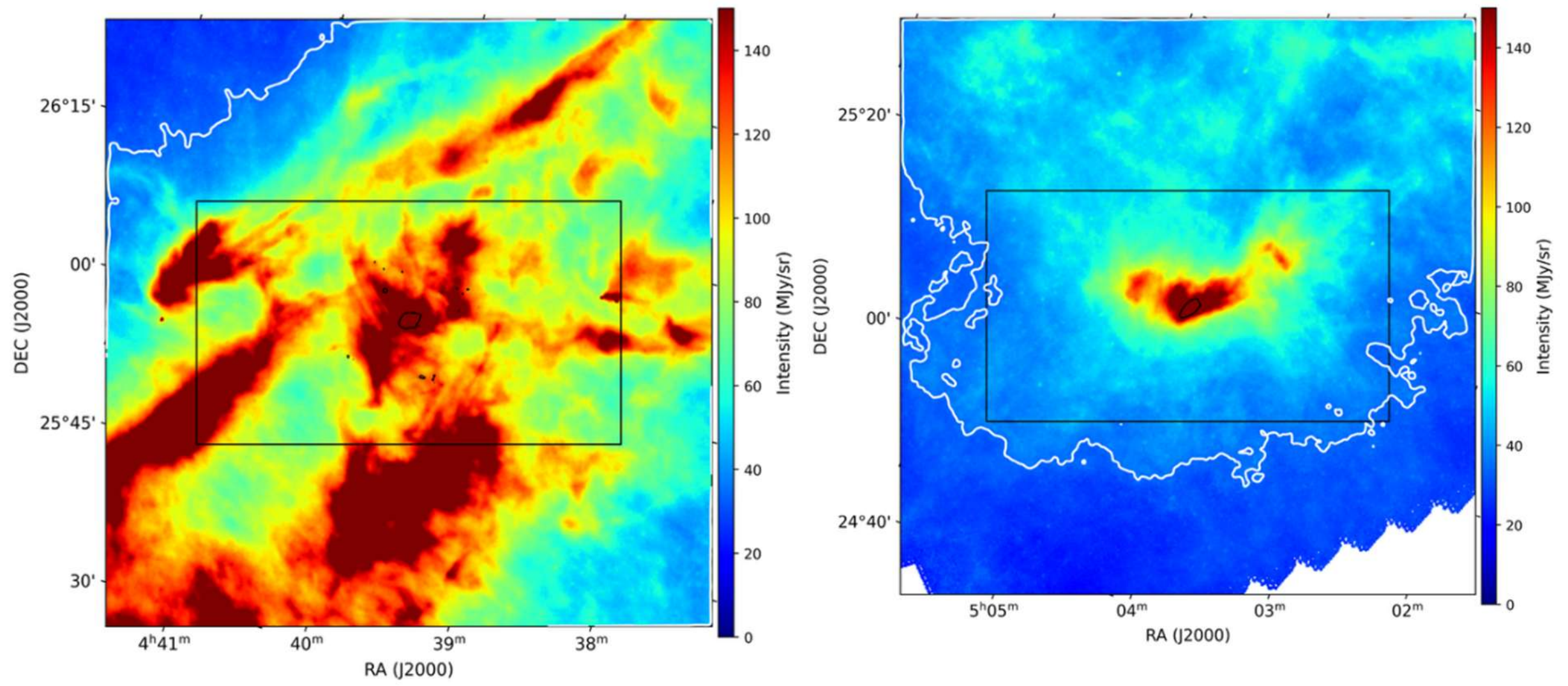

**Figure 1.** Herschel 250 µm maps downloaded from the ESA Herschel archive. The white contours show a flux level of 35 MJy/steradian in each source. Black contours show the 850 µm dust emission intensity from JCMT/SCUBA-2/POL-2 observations when available. These contours trace the dense cores where magnetic fields have already been observed by JCMT/POL-2. The black box shows the proposed FOV of PRIMAger of 42x24 arcminutes (Dowell et al. 2024).

## Instruments and Modes Used

This science case will use 8 PRIMAger maps over 1° x 1°.

## Approximate Integration Time

We start by assuming that PRIMAger Band 4 235 µm observations are similar to Herschel SPIRE 250 µm observations. Figure 1 shows a white 250 µm contour for each source at 35 MJy/sr. This contour encompasses all the diffuse material. If we assume that diffuse material will be polarized at 20% (the maximum ISM polarization level, Planck Collaboration 2015), that gives a polarized intensity value of 7 MJy/sr. To achieve a 5-sigma detection, the mapping at 235µm would need to reach a sensitivity of 1.4 MJy/sr which corresponds to a value of $1.235 \times 10^4$ µJy. Entering this into the PRIMA ETC for Band 4 and choosing a survey area of 8 square degrees, 1 square degree maps of the 8 sources, gives a time estimate of for a non-polarized source.1 hours. For dust temperatures in the range of 20-40 K, the blackbody peaks at 73 and 145 µm respectively, so for the shorter wavelengths we would expect larger fluxes and therefore the time requirement at 235 µm is our limiting factor due to the ability to observe all four bands at once. We do check this with the ETC where we scale the fluxes at 250µm assuming a dust temperature of 30 K and beta of 1.7. Using the same method as above for calculating required sensitivities gives $7.1 \times 10^4$, $1.06 \times 10^5$ and $1.04 \times 10^5$ µJy for 172, 126 and 96 µm respectively. These give < 1 hour integration time in the ETC. If we assume a more conservative 10% polarization for the 35 MJy/sr material, the required 235 $\mu$m observation time for a 5-sigma detection is 44 hours.

**PRIMAger times:** 11.1 (96 $\mu$m), 11.1 (126 $\mu$m), 11.1 (172 $\mu$m), 11.1 (235 $\mu$m) hours. However, the ETC mentions that "Depth in polarized intensity is √2 times larger". The exposure time should assume use 12.35/√2 = 8.73 mJy which gives 22.1 hours.

**Total time:** 22.1 hours

## Special Capabilities Needed

None

## Synergies with Other Facilities

PRIMA will focus on the infrared to FIR of the electromagnetic spectrum and will observe the extended, warm, diffuse dust population of molecular clouds.

Single dish observations which typically observe in the submillimeter will complement PRIMA with observations of the dense, cold dust. This includes current facilities like the JCMT, IRAM, LMT and many others. For example, where observations do overlap with JCMT/SCUBA-2/POL-2, we would have 6+ data points for investigating dust properties and 3D magnetic fields. There are also future planned facilities such as AtLAST which is a proposed 50m single dish telescope observing in the millimeter. This telescope would most likely have multi-band polarimetry capabilities. These single dish telescopes also provide kinematic observations which can be used to characterize turbulence in these molecular clouds. In addition, the Fred Young Submillimeter Telescope (FYST, CCAT-Prime Collaboration 2023) will have observations at 350, 730, 850, 1000, and 1400 $\mu$m which will additionally complement the goal of studying how the polarization and magnetic field varies as a function of wavelength, potentially enabling better understanding of ISM dust and 3D magnetic fields.

Interferometers such as ALMA, SMA and NOEMA can observe the small-scale B-field, probing scales even smaller than those with single dish telescopes. ALMA has a planned upgrade and will continue to be useful across the decades.

The Square Kilometer Array (SKA) should be sensitive to observe the Zeeman effect which provides direct information and the magnetic field strength and line-of-sight component. Zeeman observations are often done in the densities which will be observed by PRIMAger since dense cores have a weak signal and current observatories rarely achieve the required sensitivity.

## Description of Observations

To achieve the scientific objectives of this proposal PRIMA would observe the selected Lynds molecular clouds using all four PRIMAger bands with full polarimetry. There is no facility previously available, currently available or planned that will be able to study specifically the magnetic field of the diffuse regions in these molecular clouds, especially at the resolution and sensitivity of PRIMAger.

For these PRIMAger observations we require only a single square degree field of view. This field of view encompasses the entirety of these low-mass molecular clouds (see Figure 1). This will also allow us to compare existing Planck observations, which have a resolution of 5’ and therefore will still sufficiently sample the one square degree, with the 10-28” resolution PRIMAger observations to determine how the structure of the magnetic field varies from the largest scales (and most diffuse medium) down to the intermediate scales probed by PRIMA.

To observe this map region, we will use a scan map strategy to efficiently cover the proposed one square degree region. An example of a possible PRIMA scan pattern is detailed in Dowell et al. (2024), and additional scanning strategies are currently being explored by the PRIMA team. As seen in Figure 1, some of these structures are not as large as the proposed square degree or even the field of view of PRIMAger. Dowell et al. (2024) suggested a beam-steering mirror mapping mode for structures smaller than the instrument field of view.

# 76. A Census of Dust Production from "Gap" Transients with PRIMA/FIRESS

Ryan Lau (NOIRLab), Jacob Jencson (Caltech/IPAC), Kishalay De (Columbia U), Viraj Karambelkar (Caltech), Mansi Kasliwal (Caltech), Sam Rose (Caltech), Tom Steinmetz (CAMK Torun)

Dust grains are a key component of the ISM. However, the origin of dust remains uncertain. Although core-collapse supernovae (ccSNe) have proven to be strong candidates as rapid dust factories, the destructive nature of SN shocks raises questions on the survival of SN-formed dust. The key question of this science case is to investigate other (non-SN) sources of dust, which is in line with PRIMA's core science theme 3: *How do interstellar dust and metals form and build up in galaxies over cosmic time?*

The role of newly emerging classes of IR-luminous "gap" transients as significant dust producers has been largely unexplored due to mid- and far-IR observational limitations. With JWST, the mid-IR emission from ~15 gap transients will be observed to investigate their dust production in an approved Cycle 4 program. Such observations will set the stage for single-pointing PRIMA/FIRESS low-resolution (R~10 binned) follow up in the far-IR where cooler and more massive dust components may hide. Upcoming wide-field time-domain surveys telescopes such as the Vera C. Rubin Observatory and the Nancy Grace Roman Space Telescope will importantly enable the discovery of new gap transients. The primary goal of this science is to determine whether gap transients are significant sources of dust compared to ccSNe.

## Science Justification

### Background and Broader Context

Dust grains are the seeds of star and planet formation, play a key role in the physical and chemical conditions of the interstellar medium (ISM), and are direct tracers of the chemical enrichment of the Universe. Despite the importance of dust, its origin remains uncertain. Given the presence and abundance of dust observed in some of the earliest galaxies formed with the first than 1 Gyr of cosmic time (e.g., Witstok et al. 2023), dust formation must be linked to the brief but extreme lives and deaths of massive stars. The metal-rich ejecta of core-collapse supernovae (ccSNe) are naturally considered dominant dust sources and may form up to ~1 $M_\odot$ of dust per event (Schneider & Maiolino 2024). However, given the destructive nature of SN shocks, ccSNe may be net dust destroyers (e.g., Temim et al. 2015; Kirchschlager et al. 2024). The key science question of this science case is in line with PRIMA's core science theme 3: *How do interstellar dust and metals form and build up in galaxies over cosmic time?*

New classes of IR-luminous transients with eruptive mass loss have emerged from recent time-domain surveys and are reshaping our understanding of massive star evolution (e.g., Kasliwal et al. 2017). These IR-dominated transients, occupying the luminosity "gap" between novae and supernovae, show substantial dust production (Jencson 2020; see Figure 1) and occur at rates that are significant fractions of ccSN rates. For example, luminous red novae (LRNe), which are a class of gap transients that are associated with stellar mergers, exhibit occurrence rates ~50% that of ccSNe (Karambelkar et al. 2023). The role of gap transients as dust producers, however, currently remains largely unexplored due to mid- and far-infrared observational limitations (λ ≳ 5 μm; Rose et al. 2025). An approved JWST Cycle 4 Survey program (GO 7040) will capture the mid-IR emission from ~15 gap transients out to ~20 μm to investigate their dust production. The JWST survey program will set the stage for PRIMA/FIRESS follow up in the far-IR where cooler and more massive dust components may hide.

**IR-Luminous "Gap" Transients.** Gap transients fall into three main classes (Jencson 2020) that are distinguished by their physical origins and observational signatures as follows:

**Luminous Red Novae (LRNe)**: Stellar mergers or common-envelope ejections, characterized by relatively low luminosities and ejecta velocities (Kulkarni et al. 2007; Tylenda et al. 2011; Karambelkar et al. 2023).

**Intermediate-Luminosity Red Transients (ILRTs)**: Transients thought to result from the electron-capture SNe of super-AGB stars, with light curves that resemble scaled-down ccSNe (Thompson et al. 2009; Rose et al. 2025).

**SN Impostors / LBVs**: Non-terminal, giant eruptions from luminous blue variables (LBVs), exhibiting SN-like luminosities but bluer colors and different profiles (Smith et al. 2011). For simplicity, we refer to these as LBVs throughout.

Our science goal is to determine whether gap transients are significant sources of dust compared to ccSNe.

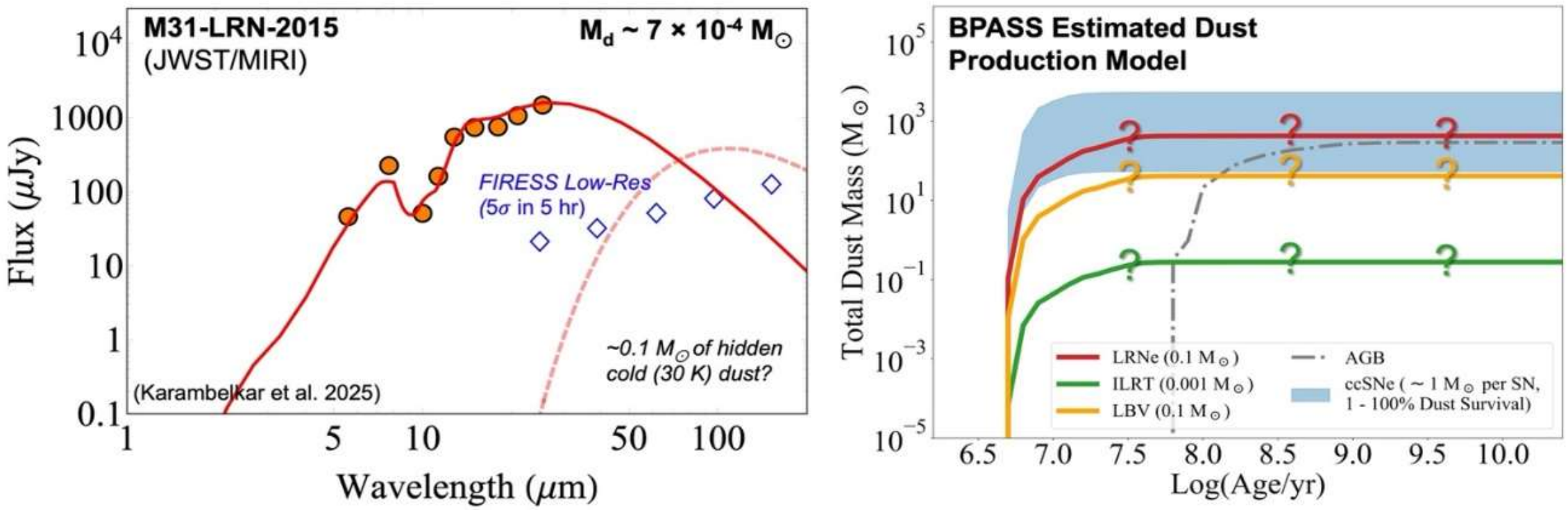

**Figure 1. (Left)** DUSTY model fit to JWST/MIRI photometry of M31-LRN-2015 (Karambelkar et al. 2025) overlaid with PRIMA/FIRESS Low-Res (R~10) sensitivities and an example of a ~0.1 $M_\odot$ component of "hidden" cold dust that PRIMA would be able to probe. **(Right)** BPASS dust-production model of ccSNe assuming 1 - 100% ejecta dust survival, AGB stars, and gap transients with adopted guesses for dust production: 0.1 $M_\odot$ per LRN, 0.001 $M_\odot$ per ILRT, and 0.1 $M_\odot$ per LBV.

**Dust Mass in Gap Transients.** Mid- and far-IR photometry provide vital insight into thermal dust produced by gap transients. Dust masses can be derived from infrared SEDs using JWST/MIRI + PRIMA/FIRESS observations in four filters spanning 5 - ~200 μm. PRIMA would notably be sensitive to cooler (~30 K) dust components that may be more massive than warmer components detected by JWST (Figure 1, Left). A DUSTY model fit to JWST/MIRI Imaging of M31-LRN–2015 (Karambelkar et al. 2025) demonstrates the utility of PRIMA/FIRESS for capturing the wavelength range needed to detect a far-IR emitting dust component. As much as ~0.1 $M_\odot$ of cold dust – an order of magnitude more than what was measured from the dust detected by JWST from M31-LRN-2015 (Karambelkar et al. 2025) – could be "hidden" in the far-IR.

To model dust production across stellar populations, Binary Populations and Spectral Synthesis (BPASS; Eldridge et al. 2017) models can be used and will build on ccSNe dust production models from Lau et al. (2020). An example of such a model is shown in Figure 1 (Right). Gap transient classes can be integrated into this framework based on their observed rates—approximately 80% (LRNe), 5% (ILRTs), and 5% (LBVs) of the ccSNe rate (Graur et al. 2017; Karambelkar et al. 2023).

Gap transients may be key contributors to cosmic dust production, potentially filling the deficit left by ccSNe. Upcoming wide-field time-domain surveys telescopes such as the Vera C. Rubin Observatory and the Nancy Grace Roman Space Telescope will importantly enable the discovery of new gap transients. PRIMA/FIRESS will uniquely capture the far-IR emission from potentially hidden cool dust that builds on our investigation of gap transients with JWST. The pursuit of this science case will place critical constraints on stellar dust sources and shed light on explosive mass loss during the final stages of massive star evolution.

## Approximate Integration Time

Based on the PRIMA ETC, each gap transient would take ~5 hr of integration time with FIRESS Low-Res (R~10) to achieve a sufficient SNR (>5σ) to characterize the dust mass in the far-infrared.

## Special Capabilities Needed

No special capabilities needed.

## Synergies with Other Facilities

The proposed science case has synergies with JWST for full IR SED coverage. There are also notable synergies with upcoming wide-field time-domain survey telescopes such as the Vera C. Rubin Observatory and the Nancy Grace Roman Space Telescope.

## Description of Observations

The observational strategy for conducting a far-IR census of gap transients is straightforward and entails single-pointing PRIMA/FIRESS low-resolution spectroscopy (in a synthesized R ~ 10 bin) of known gap transients. Observations in this mode provide the most sensitive configuration for obtaining the far-IR (~25 - 200 µm) SED from cool dust emission formed by gap transients and will enable dust-model fitting with radiative transfer codes such as DUSTY.

# 77. Mapping Magnetic Field Strength from Molecular Cloud Envelopes to Dense Cores and Filaments

Dariusz C. Lis (Jet Propulsion Laboratory, California Institute of Technology), Paul F. Goldsmith (Jet Propulsion Laboratory, California Institute of Technology), Simon Coudé (Worcester State University), Ivana Bešlić (Observatoire de Paris), Maryvonne Gerin (Observatoire de Paris), Jérôme Pety (Institut de Radioastronomie Millimétrique), Brandon S. Hensley (Jet Propulsion Laboratory, California Institute of Technology), Nathalie Ysard (Université Paris-Saclay and Université de Toulouse), Karine Demyk (Université de Toulouse)

Magnetic fields play a key role in supporting molecular clouds against gravitational collapse and consequently in reducing the rate of star formation far below that expected based on free-fall collapse. Low-spatial-resolution *Planck* dust polarization maps revealed a dramatic change in the alignment between the magnetic field and dense gas distribution, from parallel in the diffuse regions to perpendicular in dense star-forming supercritical filaments and ridges. While the plane-of-the-sky magnetic field geometry can be studied with far-infrared (FIR) polarimetry alone, determination of the magnetic field strength additionally requires velocity-resolved spectral line imaging to determine the turbulent velocity dispersion.

We describe a multi-wavelength FIR—radio project intended to be a precursor to a future effort aimed at mapping magnetic fields in molecular clouds over tens of square degrees, for which PRIMA would be a key element. We focus on the RCW 36 region in Vela, where 89 and 214 µm dust polarimetry and high-resolution [C II] spectroscopic data are available from SOFIA/HAWC+ and up-GREAT observations. The archival FIR observations are combined with radio observations of ammonia inversion lines at 1.2 cm obtained using the 70 m Canberra antenna of NASA's Deep Space Network that trace dense molecular gas.

We demonstrate that 89 µm dust polarimetry combined with high-resolution [CII] spectroscopy can be used to map the magnetic field strength in low-density molecular cloud envelopes down to $A_V$=0.1 mag. Longer wavelength 214 µm dust polarimetry combined with velocity-resolved observations of ammonia inversion lines, with the DSN in the southern hemisphere or GBT in the northern hemisphere, can then be used to trace the evolution of the magnetic field in the transition region from the envelope gas to dense filaments and cores above $A_V$=1 mag.

## Science Justification

The question of what controls the star formation efficiency in molecular clouds has long been at the center of star formation research. Early studies (e.g., Zuckerman & Evans 1974) showed that if all the gas within dense interstellar clouds were to collapse freely under self-gravity, the star formation rate in the Milky Way would be two orders of magnitude higher than the observed rate of 2 $M_{\odot} yr^{-1}$ (Robitaille & Whitney 2010). Theories proposed to explain such a low star formation rate invoke the presence of turbulence or magnetic fields supporting interstellar clouds against gravitational collapse. In addition to preventing the gas from collapsing, the turbulence and the magnetic field can also bolster the star formation processes. For example, the coupling between the magnetic field and the neutral gas can allow parts of low-density clouds to fragment and initiate star formation (Fiedler & Mouschovias 1993).

As discussed in Bešlić et al. (2024), it has been difficult to test specific star-formation models observationally. Despite the efforts made over the past 40 yr to characterize interstellar turbulence, we still struggle to understand the different energy sources that contribute to the observed line-of-sight gas velocity dispersion. In addition, measurements of magnetic fields are observationally extremely challenging, as they typically involve relatively weak signals, both for Zeeman gas line splitting (Crutcher 2012) and for dust polarization measurements (see Pattle et al. 2019, for a recent review).

Determination of the magnetic field strength is necessary to ascertain which regions in molecular clouds are sub/super-critical and may consequently collapse to form stars. The Davis-Chandrasekhar-Fermi (DCF) method (Davis 1951; Chandrasekhar & Fermi 1953) and the Skalidis-Tassis (ST) method (Skalidis & Tassis 2021) provide a way of calculating the plane-of-the-sky B-field strength using changes in the direction of the magnetic field direction provided by FIR dust polarimetry, together with information about gas density and turbulence from velocity-resolved spectroscopy of atomic and molecular lines. The current white paper focuses on *identifying optimal tracers for measuring magnetic field strength over extended areas, from low-density molecular cloud envelopes to dense star-forming filaments and cores.*

### Magnetic Fields in Molecular Cloud Envelopes

The IRAM-30m ORION-B Large Program (Pety et al. 2017) provided some of the largest images of molecular line emission (5 square degrees or ~20 pc across the Orion B molecular cloud), including tracers such as CO, $HCO^+$, HCN, and CS, as well as their optically thin isotopologues. However, extending such observations to even wider areas and more diffuse regions is very challenging due to the weakness of the lines. The 158 µm fine structure line of [CII] is well established as the brightest FIR line that can be mapped at high velocity-resolution over extended areas. An additional advantage of the [CII] line is that it is considered a good tracer of the "CO-dark gas," where hydrogen is molecular, but molecular tracers, such as CO, are photo-dissociated. Such gas has been suggested to contain as much as 50% of the total mass of molecular clouds.

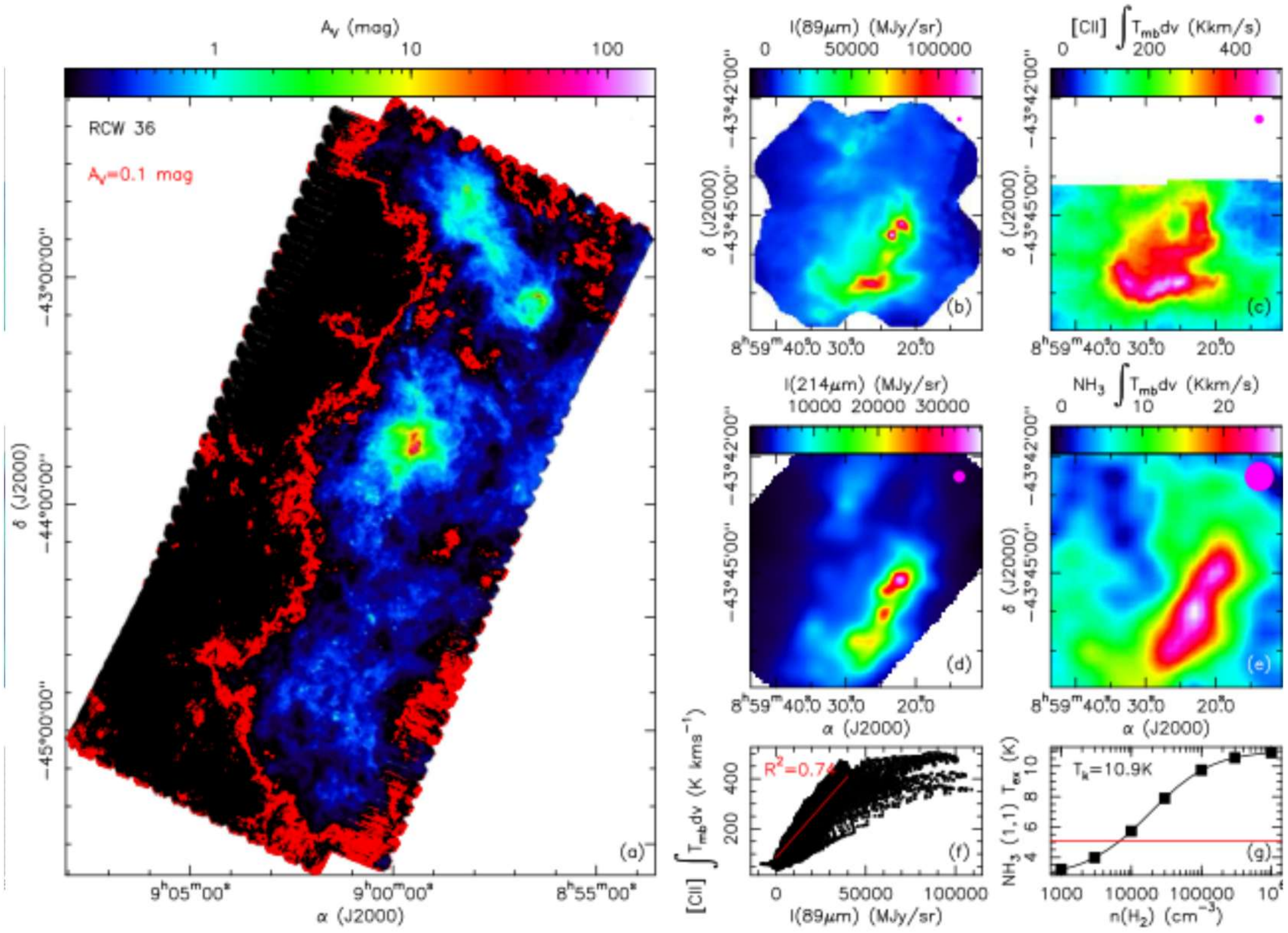

**Figure 1.** (a) Extinction map of the RCW36 region in Vela based on PACS 160 µm dust continuum map. The red contour corresponds to $A_V$=0.1 mag. (b and d) SOFIA/HAWC+ 89 and 215 µm images of the Vela C molecular cloud (Stokes I). (c) SOFIA/up-GREAT moment 0 image of the [CII] emission in the same region. (e) DSS-43 moment 0 map of the (1,1) ammonia inversion line. Images are plotted at their native resolution with the FWHM beam sizes showed by the magenta circles in the upper right corners. (f) Pixel-by-pixel correlations of the [CII] emission with the 89 µm continuum intensity. The red line is a linear fit to the points with 89 µm intensities below 40,000 MJy/sr. The non-zero intercept is owing to the filtering of low-level extended emission in the HAWC+ images, which are affected by the correlated noise from the Earth's atmosphere. (f) Excitation temperature of the ammonia (1,1) line as a function of density for a fixed kinetic temperature of 10.9 K.

We use archival SOFIA HAWC+ and up-GREAT observations of the RCW 36 region in Vela as a proof of concept. Figure 1 (a) shows the visual extinction map of the region based on observations of the 160 µm dust continuum emission with *Herschel*/PACS. Our requirement is to map the magnetic field geometry and strength, as well as the gas velocity field in molecular cloud envelopes down to a visual extinction of 0.1 mag (H nuclei column density of $2.2\times10^{20}$ cm$^{-2}$), the region within the red contour. The total area mapped with PACS in RCW 36 is 4.7 sq. deg, of which 2.8 sq. deg. (60%) is above our extinction cutoff $A_V$=0.1 mag, which corresponds to a mass surface density of ~1.8 M$_\odot$pc$^{-2}$, a factor of ~70 lower than the extinction threshold for star formation (Heiderman et al. 2010; Lada et al. 2010). The total mass of this region is estimated at $4.5\times10^4$ M$_\odot$. To determine the magnetic field strength requires measurements of the polarized FIR dust continuum emission (using PRIMAger) and velocity-resolved [CII] observations (from a balloon or an Explorer-class mission equipped with a 158 µm heterodyne array capable of at least 1 kms$^{-1}$ velocity resolution).

Panels (b) – (c) show SOFIA images of the 89 µm continuum emission (Stokes I), [CII] integrated line intensity, and 215 µm continuum across the region. The [CII] morphology closely resembles that of the 89 µm continuum emission, showing extended low-level emission originating in the cloud envelope. It differs significantly from the 214 µm continuum emission, which traces primarily the dense star-forming ridge at the center of the cloud. Panel (f) shows a pixel-by-pixel correlation of the [CII] emission with the 89 µm continuum intensity. A good correlation ($R^2$=0.74) is seen up to 89 µm intensities of about 50,000 MJy/sr, which flattens at higher intensities. This is consistent with chemical models, which show that ionized carbon is only abundant in the outer layers of the cloud with $A_V$~1.5 mag, transitioning to neutral carbon, CO, and other carbon-bearing species in better shielded regions. The hydrogen in the [C II] emitting region is mostly, but not fully molecular, as [C II] is also a good tracer of atomic gas. For standard ISM conditions, the transition between $H_2$ and H is about $A_V \simeq 0.3$ mag (see, e.g., Liszt & Gerin 2023). The details depend on the gas density and the UV field strength, but typical gas temperatures in this region are 50 – 150 K, which explains the good correlation with the 89 µm continuum intensity.

Multi-wavelength dust continuum observations can be used to determine the dust temperature and hydrogen column density distribution across the cloud envelope from SED fits. The gas volume densities can then be estimated using, e.g., the inverse modelling method recently developed by Orkisz & Kainulainen (2024). Since the [CII] emission is sensitive to pressure, independent constraints for the gas density can be obtained from the [CII] line strength combined with temperature estimates derived, e.g., from the dust SED. The only remaining parameter required to compute the magnetic field strength is the turbulent velocity dispersion, which can be determined from the [CII] line width. In the case of RCW 36, the derived magnetic field strength is in the range 100 – 800 µG. The dense filament traced by the 214 µm dust continuum emission is supercritical, while the lower-density envelope gas is sub-critical (Coudé et al. 2025).

The observing time requirements for the FIR dust continuum polarimetry and [CII] line observations of the envelope regions are discussed below.

### Magnetic Fields in Dense Cores and Ridges

While [CII] is an excellent tracer of the lower-density envelope material, including the CO-dark gas, the $C^+$ abundance drops rapidly at visual extinctions $\gtrsim$1.5 mag, as described above. A different molecular gas tracer is thus required in higher extinction regions.

We suggest that radio observations of the inversion of lines of ammonia at ~1.2 cm wavelength can provide the information needed for the magnetic field strength measurements in dense gas. Figure 1 (e) shows a map of the $NH_3$ (1,1) inversion line in the RCW 36 region obtained with the 70-m antenna of the NASA Deep Space Network at Canberra Australia (DSS-43). $NH_3$ emission is detected over an extended region, showing a morphology similar to the 214 µm dust continuum emission. The advantage of the ammonia inversion lines over other high-density tracers is that the intensities of the (2,2) and (1,1) lines, which can be observed simultaneously using the DSS-43 K-band spectrometer, provide a measure of the gas kinetic temperature (Lis et al. 2025). The excitation temperature can be determined independently from the hyperfine-structure line ratios. Since collisional excitation rates are available, the gas density can then be estimated (Figure 1, panel g).

We set a limit of $A_V$ >1 mag as a boundary of the region that must be mapped in ammonia, to provide some overlap with [C II] observations. *Herschel* observations of nitrogen hydrides demonstrated the presence of ammonia in diffuse gas down to $A_V \simeq 1$ mag (Person et al. 2012). In RCW 36, such regions correspond to a relatively small area of ~360 sq. arcm, or ~3.5% of the area with $A_V$ > 0.1 mag traced by [CII].

The combination of the FIR polarimetry from PRIMAger, with [CII] and $NH_3$ velocity-resolved imaging gives *all the information necessary* to study the evolution of the magnetic field geometry and *strength* in the transition region from the molecular cloud envelopes to dense cores and filaments using the DCF or ST methods. We demonstrate below that such observations will be feasible in the PRIMA era.

## Instruments and Modes Used

This program proposes to map an approximately 4.7 sq. deg region of the sky, such as the RCW 36 region using the polarimeter band of PRIMAger.

## Approximate Integration Time

### PRIMAger Observations

We consider as a proof of concept a study of the magnetic field geometry and strength in a molecular cloud of the size comparable to the map of RCW36 shown in Figure 1(a), with a total area of 4.7 sq. deg. We want to map the whole region down to $A_V$=0.1 mag. We use the standard formulae to convert the FIR continuum flux to $A_V$ (see, e.g., Lis et al. 2024 and references therein). We assume the grain emissivity coefficient $Q(\lambda)=7.5\times10^{-4}$ 125µm/λ in the PRIMA wavelength range (Hildebrand 1983), a uniform dust temperature of 35 K as given by SED fits to the multi-wavelength Herschel PACS and SPIRE data in the RCW36 region, and the conversion factor $N_H=2.21\times10^{21}$ cm$^{-2}$ $A_V$(mag) (Güver & Ozel 2009). The resulting continuum intensities corresponding to $A_V$=0.1 mag are 55.1, 43.6, and 23.8 MJy/sr at 90, 160, and 235 µm, respectively.

We require, conservatively, a 5σ PRIMAger detection of 1% polarized intensity, 0.55 and 0.24 MJy/sr, at 90 and 235 µm, respectively. The 1% limit corresponds approximately to 5$^{th}$ percentile in Planck full-sky polarization maps (see Fig. E.1 in Planck Collaboration et al. 2020), while Ashton et al. 2018 found a median value of 3.2% for the polarization fraction in a translucent cloud based on a combination of Planck and BLASTPol data.

From the PRIMAger specification web page:

(https://prima.ipac.caltech.edu/page/instruments#primager), the surface brightness sensitivity (5σ, 10 h, 1 sq. deg.) for the polarimetry imager is 0.65 and 0.25 MJy/sr at the two wavelengths, respectively. The total observing time (including overheads) to cover a 4.7 sq. deg. area is thus $4.7\times(0.65/0.55)^2\times10$ h = 66 h at 90 µm and $4.7\times(0.25/0.24)^2\times10$ h = 51 h at 235 µm. The two bands can be observed simultaneously. With a single cloud requiring ~70 h of observing time, a significant sample of ten clouds of comparable size can be observed in a 700 h PRIMA GO observing program.

## Acknowledgment

This research was carried out at the Jet Propulsion Laboratory, California Institute of Technology, under a contract with the National Aeronautics and Space Administration (80NM0018D0004). 

## Special Capabilities Needed

None.

## Synergies with Other Facilities

To map the strength of the magnetic field across molecular cloud envelopes, velocity-resolved observations of [CII] at 158 µm are required. These can be provided by space-borne or balloon platforms (e.g., a future FIR MIDEX mission or an ASTHROS-like balloon mission for pilot studies of smaller regions).

Radio frequency observations of the ammonia inversion lines can be provided by the DSN antenna at Canberra, Australia, for southern hemisphere sources and the Green Bank Telescope for the northern hemisphere sources.

### [CII] Observations

For [CII] observations we consider a multi-beam heterodyne FIR receiver placed on a balloon or a space-borne platform (e.g., ASTHROS-like balloon mission or a future FIR MIDEX mission).

We consider the same extinction threshold $A_V$=0.1 mag, corresponding to an $H_2$ column density of $1.1\times10^{20}$ cm$^{-2}$. Assuming a $C^+$ fractional abundance of $2.5\times10^{-4}$ relative to $H_2$ as suggested by PDR models leads to a $C^+$ column density of 2.8 $\times10^{16}$ cm$^{-2}$. The corresponding [CII] line intensity can be computed using the formulae from Goldsmith et al. (2012). Assuming a gas temperature of 50 K and an $H_2$ density of 1000 cm$^{-3}$ in the envelope gas, we derive an integrated [CII] line intensity of 0.64 K kms$^{-1}$, which for a line width of 3 kms$^{-1}$ corresponds to a peak brightness temperature of 0.20 K, or an antenna temperature of 0.14 K, assuming a 70% main beam efficiency.

For the observing time calculation, we assume a single side band (SSB) receiver temperature of 2000 K and a velocity resolution of 1 kms$^{-1}$. For position switched observations, the ON+OFF observing time needed for a 7σ detection of the [CII] line is 1.75 h per point. This allows ~15% accuracy in the fractional line width estimate. With a 90 cm diameter telescope (the same beam size of 0.75’ as the DSN Canberra antenna at the ammonia line frequency), the number of individual pointings to cover a 2.8 sq. deg. area is $1.67\times10^4$. With a modest 7-pixel array that could be built today, the total time required is 5.7 months. However, ten years from now, a larger array with 32-pixels can be envisioned, reducing the total time to 1.25 months. An observing program including ten clouds of comparable size can thus fit with a 3-year MIDEX primary mission. Smaller areas can be mapped from a balloon as a pilot study.

### Ammonia Observations

Based on pilot observations carried out using the NASA Deep Space Network DSS-43 antenna (Fig. 1e), we estimate that about 40 h of the telescope time is required to map a 10’×10’ area down to the sensitivity required to measure accurately the line width and the central line velocity. The total observing time required per cloud is thus of order 160 hours, making the observations feasible.

## Description of Observations

The whole cloud would first be mapped using PRIMAger. The multi-wavelength Stokes-I images would then be used to compute the dust temperature, $H_2$ column density, volume density, and visual extinction maps. Areas with visual extinctions $A_V$>0.1 mag and $A_V$>1 mag would be identified for follow-up observations of the [CII] and ammonia line emission, respectively.

# 78. Mapping the Cold Dusty Reservoirs of Extra-Solar Planetary Systems at Birth

Joshua B. Lovell (Center for Astrophysics | Harvard & Smithsonian), Nicholas Ballering (Space Science Institute); Mark Booth (UK Astronomy Technology Centre); Katie Crotts (STScI); Patricia Luppe (Trinity College Dublin); Meredith MacGregor (Johns Hopkins University); Brenda Matthews (Herzberg, National Research Council of Canada); Karl Stapelfeldt (NASA/JPL/Caltech); David J. Wilner (Center for Astrophysics | Harvard & Smithsonian)

Debris disks are the cold, dusty, outer reservoirs of planetary systems. Through mutual planetesimal collisions, debris disks trace where planetesimal belts formed, and allow insights as to how planetary systems evolve over Myr–Gyr timescales. At their very earliest stages when debris disks can be observed, that is, following the dispersal of the bulk protoplanetary disk, there remains a significant lack of observations, limiting our understanding of the birth population of these disks, and their immediate evolution towards the stellar main-sequence. This stage of evolution is of critical importance for the evolution of planetary systems, as it marks the period over which mature extra–Solar planetary systems come into existence. We propose to use PRIMA to map a complete birth population of debris disks. We show by example the utility of such a survey in the Taurus star forming region, which is both well-characterised, nearby, and host to hundreds of class III YSOs. YSOs at this stage have only recently dispersed their protoplanetary disks, and host the youngest population of debris disks. We will utilise the unprecedented imaging sensitivity of PRIMA's PRIMAger instrument in the far–infrared, where these disks peak in emission, to construct spectral energy distributions of all known class III YSOs from 24–235-microns (utilizing Gaia, 2MASS and WISE photometry for optical and near-infrared), to i) determine the presence of young debris disks and their occurrence rates, and ii) investigate the population of disks, by modelling their dust temperatures/radii. From these data, we will investigate whether the steady-state collisional cascade evolution holds; whilst this model canonically describes the evolution of main-sequence debris disks, we will test whether the earliest stages hold, for example, if enhanced collision rates from recently-formed terrestrial planets contribute significantly to the observed disk emission.

## Science Justification

In the nearby galaxy, recent Gaia studies of nearby star-forming regions have shown the presence of over 10,000 young stellar objects (YSOs; i.e., within 350pc, see e.g., Luhman 2022). YSOs evolve rapidly – from protostellar sources with envelopes and thick disks, to naked T-Tauri stars – with number counts showing this YSO population is dominated by the most evolved 'class III' YSOs.

Class III YSOs are identified based on their lack of near-/mid-infrared excess emission above the stellar photosphere, physically attributed to these having already dispersed the bulk of their warm inner dust (Williams & Cieza 2011). Dust observed at earlier YSO stages is seen in large quantities, and provides the bulk reservoir from which large solid bodies – planetesimals and planets/planet cores – are built; therefore, the lack of near-/mid-infrared excess emission at the class III stage implies that planetesimals must have formed by this stage (Wyatt+15). Class III stars therefore offer a unique view on the planetary system components beyond the bulk dispersal of protoplanetary disks, as they advance towards the stellar main-sequence.

Collisions between planetesimals in dense planetesimal belts are the primary source of dust observed in extra-Solar debris disks, disks at 10s of au, analogous to the Solar System's Kuiper Belt (Wyatt+08; Matthews+14; Hughes+18). Observations with Spitzer, Herschel and ALMA of the most massive debris disks demonstrate these evolve via a 'steady-state' collisional cascade (Wyatt+07; Sibthorpe+18; Matra+25). Collisional cascade models imply that dust in planetary systems is continually replenished as large bodies successively grind down to smaller pebbles and dust, replenishing these at rates enabling their detection over >Gyr–timescales.

Whilst the steady state collisional cascade model is well-understood towards debris disks around main-sequence stars, there have been only a handful of investigations to understand the demographics and evolution of the birth population of debris disks. Despite this, the epoch of evolution as stars enter the main-sequence can be chaotic for planetary systems; following the dispersal of the primordial gas disk, all giant planet formation has ceased, and giant impacts (between rocky bodies) become more frequent, marking the phase of terrestrial planet growth (Morbidelli+12; Su+19). Whether the steady state model for disk evolution holds over this pre-main sequence epoch remains unknown. By studying a relatively small sample of class III YSOs with ALMA, Lovell+ (2021a) showed that the steady state model offers a valid interpretation of their disks. However, with so few detections of class III stars (<20 total; Michel+21), the census of this disk population is almost completely unknown (complete to less than ~1%) and thus constraints on how debris disks evolve over their first few Myr remain a mystery. Competing ideas, such as delayed stirring models, can be uniquely tested during this epoch (Kenyon & Bromley 2004).

What is critically missing is a survey instrument capable of detecting faint excess dust emission to enable a deeper census of faint excess dust emission, and produce a complete picture of disk evolution, including at the youngest ages. Whilst ALMA and the SMA have provided important information on the presence of small samples of disks (upcoming work of Lovell/Andrews; Michel+21), these instruments are inefficient at mapping the dust emission from the 100s–1000s of known class III stars, in part owing to their sensitivity to cold, optically thin dust which peaks between 50–200 microns. Few targets at this evolutionary stage were targeted by Herschel, however as highlighted by Lovell et al. (2021), far-infrared data is crucial in constraining the properties of disks around class III YSOs. In this context, PRIMA opens up brand-new possibilities to detect emission from these faint disks, at the wavelengths where their emission peaks.

We show in Fig.1 the landscape that PRIMA opens up for unbiased studies of debris-disk hosting stars out to 150pc (typical distance to nearby star-forming regions), by considering the fractional disk luminosity as a function of temperature of a 40au disk around a $1L_{Sun}$ star. To the same distance of 150pc, whereas Herschel could have detected only the very brightest debris disks (i.e., those with fractional luminosities, $f=L_{disk}/L_{star}$>1e-3, where to first order, f defines the total cross-sectional area of dust in the disk; see eq.4, Wyatt+08), PRIMA is sensitive to those with f>1e-4. Overall, this demonstrates PRIMA's ability to detect statistically large populations of debris disks in nearby star-forming regions.

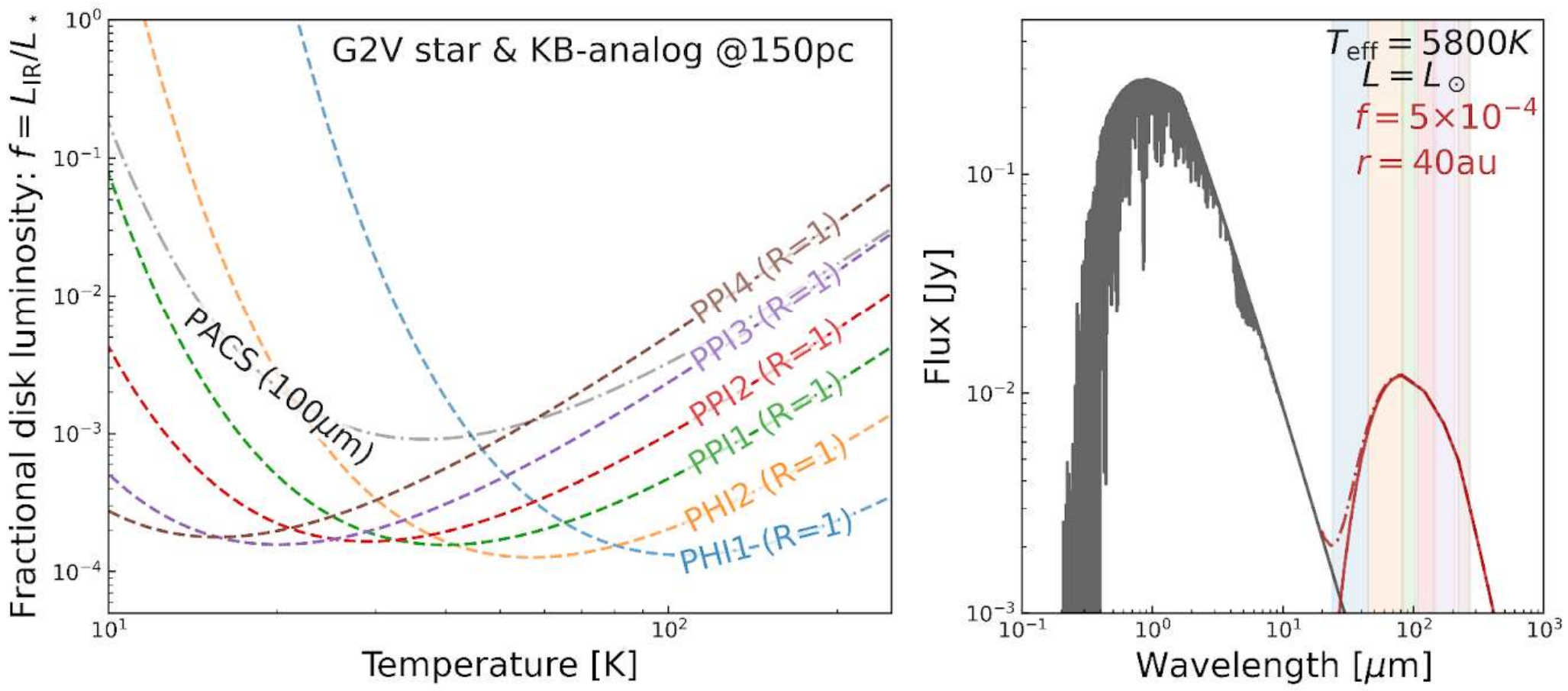

**Fig.1.** Left: Fractional disk luminosity vs. dust temperature, showing the detection curves for the PRIMA's PHI (Hyperspectral Imager) and PPI (Polarimetry Imager). Right: A simulated mid-G star and a disk at 150pc (with the median f of the REASONS survey; Matrà+2025), highlighting the PRIMAger wavebands where disk emission clearly dominates over that of the star, for which emission would be strongly detected (>5σ).

We propose a survey of a complete, nearby star-forming region. In this science case, we base our analysis on the Taurus star-forming region, which hosts approx. 350 class III stars (and over 600 YSOs in total), and spans an area approx. 200-square degrees (though, we note that in principle this type of survey could equally be conducted in other comparably large, similar age and distance star-forming regions, such as Lupus, Ophiuchus). Although such a large area would be prohibitively expensive to map debris disks in at mid-infrared or sub-millimeter wavelengths (e.g. with ALMA, SMA or JWST), PRIMA's mapping efficiency means this can be done in a fraction of the time to equivalent fractional luminosity sensitivities. To obtain sensitivity to disks and study a complete birth population of a large sample of debris disks, a 250-hour survey is required (see integration time below). We note that with Herschel, unbiased surveys detected debris disks towards ~20–30% of nearby main-sequence stars (Thureau+14; Sibthorpe+18). It is expected that as debris disks evolve, their fractional luminosities decrease, and so by observing a complete disk population at t=0, we will collect essential data to verify and/or constrain disk evolution models. Our survey provides very good sensitivity to debris disks at 150pc, and we estimate that many dozens of new debris disks will be discovered. We note this number is likely a lower-limit, depending on the collisional status of these targets, i.e., if class III YSO debris disks undergo giant impacts more frequently than main-sequence debris disks, these will be even brighter. Overall, this science proposal therefore addresses a key question highlighted by NASA's Decadal Review – 'What is the range of planetary system architectures, and is the configuration of the solar system common?'

We will obtain far-infrared photometry with PRIMA's Hyperspectral Imager and Polarimetry Imager, binned into a 6 (or more) wavebands, PHI1, PHI2, PPI1, PPI2, PPI3, and PPI4, that will enable accurate measurements of the total flux of the disk and star from 24–235 µm. In combination with the known stellar emission of these targets from Gaia, 2MASS and WISE, we will produce broadband spectral energy distributions and fit stellar and disk models, to i) confirm the presence of cold dust excesses in the far-infrared, ii) model the dust radii/temperatures associated with any detections, and iii) model the population in the context of collisional evolution, to investigate how the birth population of debris disks compares to main-sequence populations. For free, we will also measure for faint hot (>100K) dust with the PHI1 band, that may be indicative of dust at just a few au, in the inner terrestrial planet-forming zones.

## Instruments and Modes Used

The program will use PRIMAger in the hyperspectral and polarimeter dual-mode observation mode. 200 square degrees should be covered, all data binned to R=1.

## Approximate Integration Time

No overheads/slew time included. We estimate the sensitivity based on the ability to measure the dust spectrum of a median REASONS disk (f=5e-4; Matrà+25) in 6 PRIMAger bands, as 250 hours total, utilising a dual observing mode; integrating simultaneously with the Hyperspectral Imager and the Polarimetry Imager. This estimate yields the sensitivity curves shown in Fig. 1, and is based on the ETC (late May 2025), achieving 5-sigma sensitivities (with R=1) in the 6 bands of: 1.6 mJy, 2.8 mJy, 5 mJy, 7 mJy, 10 mJy, 13 mJy, (PHI1, PHI2, PPI1, PPI2, PPI3 and PPI4 respectively) . Although the data is observed with R=8-10 (PHI) and R=4 (PPI), these will be translated into the 6 data points for the construction of SEDs, though we note that bright sources will require less spectral binning, and thus more accurate modelling of their dust properties.

## Special Capabilities Needed

N/A

## Synergies with Other Facilities

Detected disk candidates will form a high-priority target list for high-resolution images with ALMA to resolve their dust (and possible gas) disks at (sub)-millimeter wavelengths. These targets plausibly form the population of gas-rich/CO-rich debris disks, and therefore will be high-priority candidates for JWST MIRI MRS imaging of their mid-infrared spectra to investigate their compositions (e.g., silicate features, gas emission lines), and evolutionary status (e.g., atomic lines that may indicate their dispersal status). AtLAST, a proposed 50-metre single dish (sub)-millimeter telescope offers further utility, and at comparable spatial resolution, could simultaneously map the same sky areas as PRIMA. Indeed, a science paper for AtLAST specifically highlighted potential science opportunities for studying the youngest debris disks in nearby star-forming regions (Klaassen+2024).

## Description of Observations

We propose to map the near-200 square degrees of the Taurus Star Forming region. Our observations do not need to be conducted simultaneously. We will avoid any ultra-bright IR sources to avoid reaching dynamic range limits of the instrument, understood to be in the range 1000–10,000 (i.e., by providing a list of coordinates designated as `keep-clear' zones). We will use the PRIMAger Hyperspectral Imager and Polarimetry Imager in a dual-observation mode; integrating with both instruments simultaneously as the selected region is scanned.

# 79. Probing Debris Disks Around the Nearest M Dwarfs with PRIMA

Patricia Luppe (Trinity College Dublin), Joshua B. Lovell (Center for Astrophysics | Harvard & Smithsonian), Meredith A. MacGregor (Johns Hopkins University), David J. Wilner (Center for Astrophysics | Harvard & Smithsonian), Nicholas Ballering (Space Science Institute), Gianni Cataldi (National Astronomical Observatory of Japan), Katie Crotts (STScI), Mark Booth (UK Astronomy Technology Centre)

Debris disks offer insights into the evolution of planetary systems, tracing the ongoing collisional activity of remnant planetesimals. While many *Herschel* and ALMA surveys have established the disk occurrence rates around A-, F-, G-, and K-type stars, the statistics for M dwarfs remain limited. The *Herschel* DEBRIS survey was the largest targeted survey to observe M dwarfs so far, which observed the closest 94 M dwarfs. However, it only detected infrared excesses around 2% of them. This low detection rate doesn't mean that these disks do not exist but rather shows the limitations of *Herschel*'s sensitivity.

With PRIMAger, we have the opportunity to significantly increase the known number of M dwarf debris disks. With its enhanced far-infrared (far-IR) sensitivity, it can probe much fainter dust luminosities and colder disks. By reobserving the same M dwarf sample targeted by the *Herschel* DEBRIS survey, PRIMAger can provide a direct comparison to previous surveys and give us a better understanding of M dwarf planetary systems.

The results can further be used to test whether the steady-state collisional evolution model holds for M dwarf debris disks, as well as to investigate whether the model is applicable to very cool and faint disks.

## Science Justification

To understand planetary systems, it is essential to not only look at the planets but also at the smaller bodies and structures within these systems. Debris disks are second-generation dust disks, formed by collisions of planetesimals, which tell us about the late stages of planet formation. These disks are mostly detected through mid- and far-IR excess emission, where the dust spectrum peaks, which is stellar radiation, re-radiated from dust grains at longer wavelengths. Because the dust in these disks would get blown away very quickly, the existence of such disks tells us that the dust must be replenished by collisions, and so tells us about the dynamics in the systems.

Despite the large number of terrestrial planets that have been detected around M dwarfs (Mulders et al. 2015, Dressing & Charbonneau 2015), detections of debris disks around these stars are rare. Large M dwarf surveys only show detection rates of around 2% (Lestrade et al. 2012, 2025), while debris disks around FGK stars show rates of about 17% (Sibthorpe et al. 2018) and around A stars of about 24 % (Thureau et al. 2014). Because M dwarfs are much cooler and have very low luminosities, we expect their debris disks to be much cooler than those around AFGK-type stars. Luppe et al. (2020) showed that debris disks around M dwarf stars may be as common as around other stars, but not detectable with current telescopes.

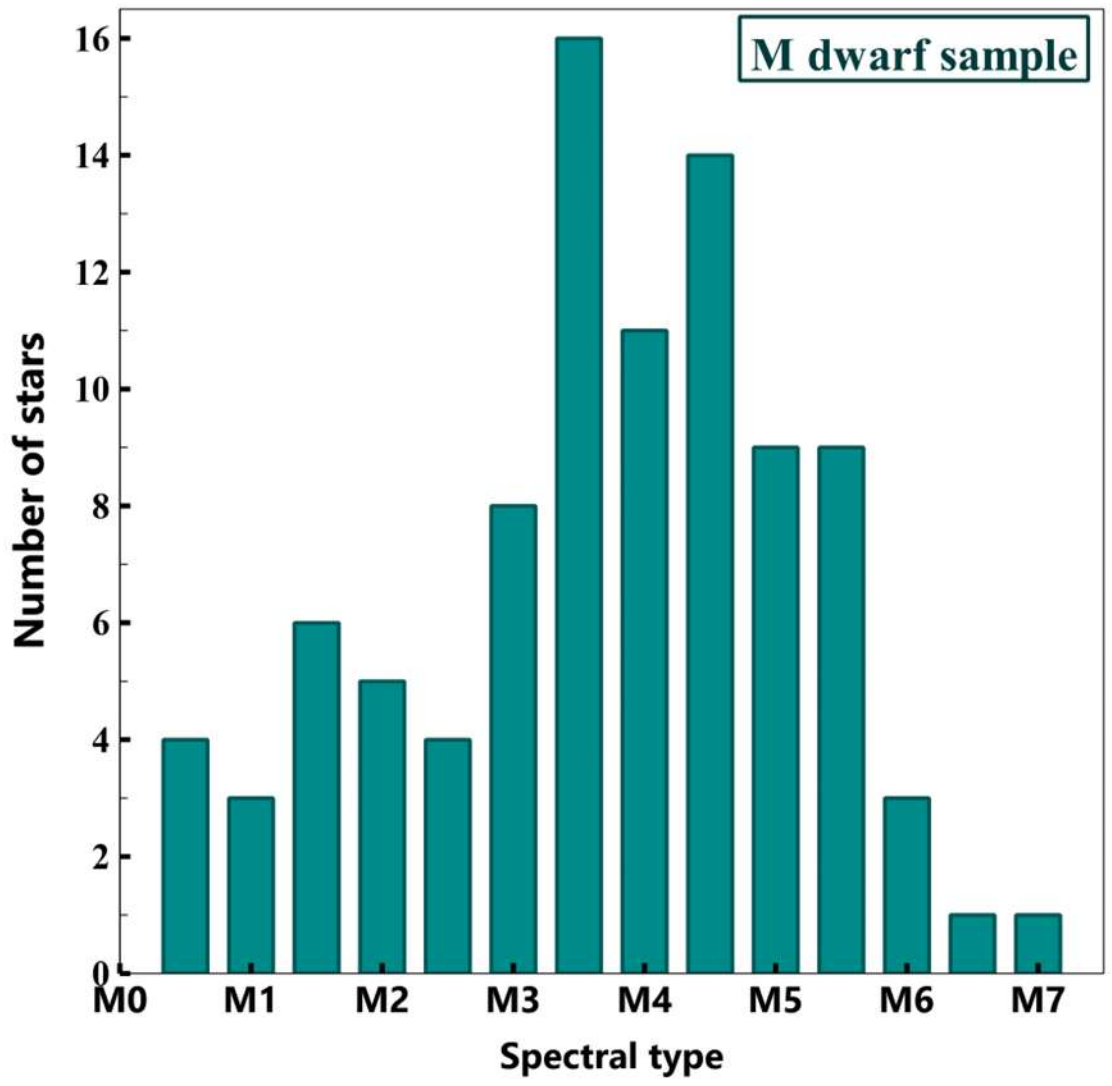

Fig. 1: Number of stars to be observed for each spectral type

The biggest survey to search for debris disks around M dwarfs was the Herschel DEBRIS survey (Matthews, B. C. et al 2010), which observed the nearest 94 M dwarfs (Lestrade et al. 2025), see Fig. 1. We propose to re-observe these M dwarfs with PRIMAger to have a direct comparison with the AFGK disks from the *Herschel* DEBRIS.

In the field of debris disks, we describe the brightness of a disk by comparing it to the luminosity of the star. We call this fractional luminosity $f_d = L_{disk} / L_{star}$.

In Fig. 2, we compare the sensitivity of the *Herschel* DEBRIS survey (PACS 100µm and PACS 160µm) with the sensitivities of the different PRIMAger bands (PPI1 to 4) for a 5-hour observation of the M5 dwarf. We plot the 68 disks detected by Herschel around AFGK stars and appropriately scale them to a distance of 3 pc and the temperature of an M5 dwarf (see Luppe et al. 2020 for scaling details). These disks are shown as green dots in the plot. By applying this scaling for all stars in our sample (for different spectral types and distances), we find that over two-thirds of these disks would be detectable by PRIMA. This would result in an overall M dwarf disk detection rate of 14.5%, which is comparable to the detection rate of the *Herschel* DEBRIS FGK stars. Because of PRIMAger's broad wavelength coverage, it wouldn't just detect the disks but also characterize their SEDs.

For stars where we don't find disks, it still gives us information about upper limits for disk luminosity and dust masses, giving us constraints for future observations and debris disk theory.

Because disks around M dwarfs are colder, the PPI bands are the ones most suitable and cover large parts of the cold and faint disk range.

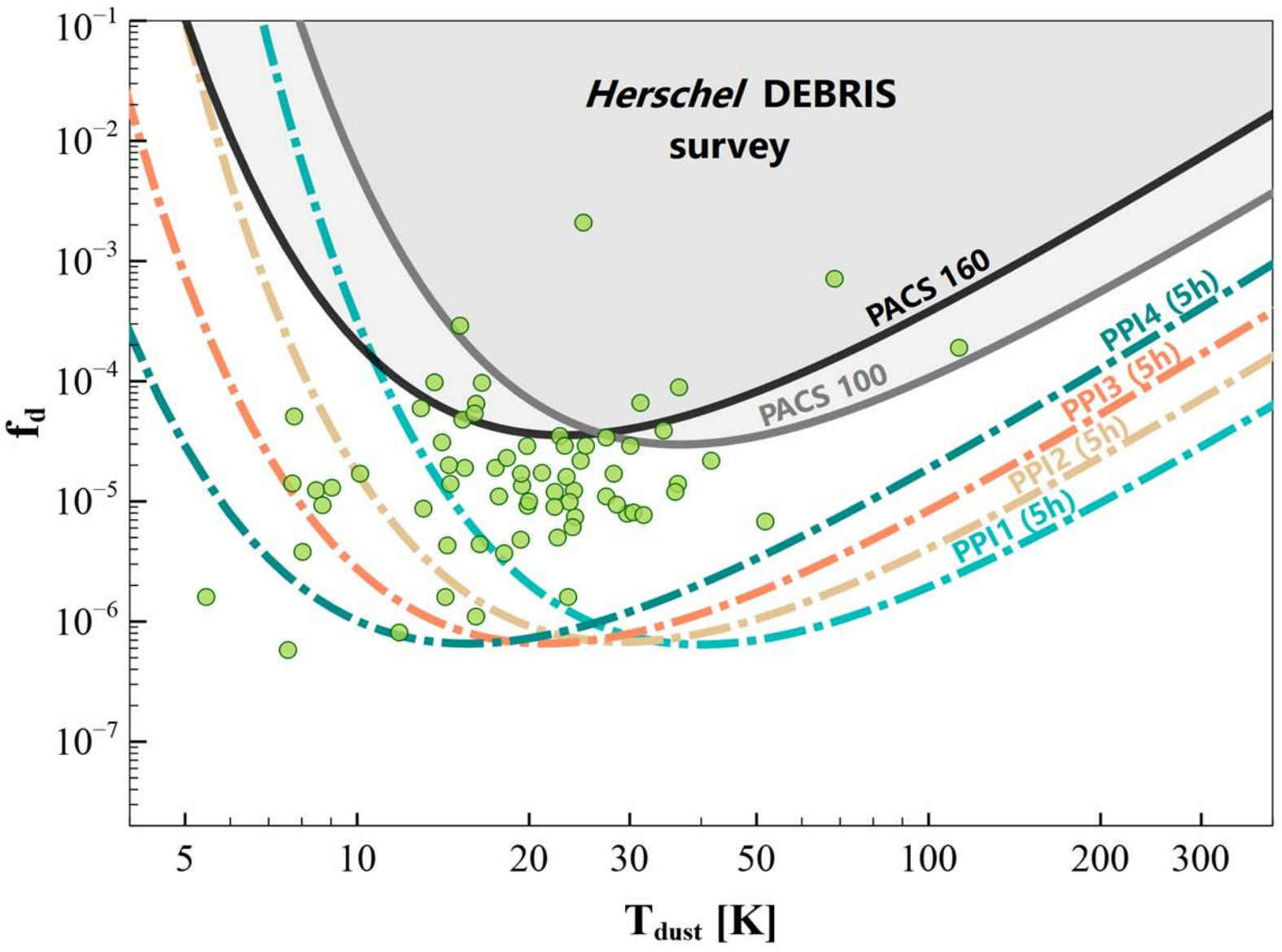

Fig. 2: Detection limits of PRIMAger and the Herschel DEBRIS survey for an M5 dwarf at about 3 pc distance. With an observation time of 5 hours for PPI1 (92μm), PPI2 (126μm), PPI3 (172μm) and PPI4 (235μm). The green dots represent 68 disks detected around AFGK stars by the DEBRIS survey and are scaled to a 3pc distance and a stellar temperature of an M5 dwarf. Sensitivities are calculated with the ETC.

## Need for PRIMA

Far infrared observations are the best opportunity to detect cold dust disks. Because of its unprecedented sensitivity, PRIMA would be the ideal instrument for an M dwarf debris disk survey. The sensitivity of Herschel PACS was enough to detect cold disks around AFGK stars, but we need PRIMA to find these disks around M dwarfs.

Our results will provide important data for testing the steady-state collisional evolution model of Wyatt et al. (2007) on M dwarf debris disks. It has been well-tested for AFGK stars, however, due to the small number of known M dwarf debris disks, it was not yet possible to test its validity for low mass star disks. Because all stars in the sample are field stars, their ages are estimated to be distributed between ~100 Myr and ~10 Gyr (Lestrade et al. 2025). We will therefore be able to test the model not only for stars with lower masses than before, but also for the long-term predictions, due to the large age range.

We are aware that for unresolved detections, signals could also be mimicked by background galaxies, as we might hit the confusion limit (Bethermin et al. 2024). Methods to get past the confusion limits and distinguish between galaxies and debris disks are, e.g. additional FIRESS observations to characterize spectral lines, characterizing the SEDs, re-observing high proper motion sources or using methods to model it out (Donnellan et al. 2004).

## Instruments and Modes Used

This observing program requires 10’x10’ PRIMAger PPI1-4 band maps towards 94 stars (polarimetry information is not needed).

## Approximate Integration Time

When scaling the disk temperature and fractional luminosity for the Herschel DEBRIS disks regarding the M dwarfs in our sample, we estimated M dwarf debris disk fractional luminosities to range from 2e-3 to 6e-7 for the brightest and faintest disks, respectively. We will observe a sample of the closest 94 M dwarfs, all located within a distance of 10 pc. For the PRIMAger Polarimetric Imager a 5-hour observation leads to a sensitivity of 343 $\mu$Jy for PPI4 at 235$\mu$m and a total observation time of 470 hours. The sensitivities were estimated using the PRIMA ETC (May 2025) and achieve 5-sigma sensitivities. We show the sensitivity curves in Fig. 2 for an M5 dwarf at a distance of approximately 3 pc. For our overall estimated detection rate, we accounted for the different distances and spectral types in our sample, as the sensitivity curves change with stellar luminosity and distance.

## Special Capabilities Needed

None

## Synergies with Other Facilities

Newly discovered M dwarf debris disks will provide us with new targets for follow-up high-resolution observations. In the (sub-)mm range, ALMA (Atacama Large Millimeter Array) could provide information about the dust. An ALMA upgrade (ALMA2040) and the possible future single-dish telescope AtLAST (Atacama Large Aperture Submillimeter Telescope) could give us more information about large grains and identify disk asymmetries or gaps that could be induced by planet-disk interactions.

Follow-up observations with JWST can reveal small dust grain distributions in the systems and detect high-mass planets.

For future missions searching for Earth-like planets, like the Habitable Worlds Observatory,

characterizing the nearby M dwarfs can help, because the dust in the system has an influence on the detectability of planets. At the same time, the structure of the dust can reveal the influence of planets in the system (Pearce et al. 2022, Imaz Blanco et al. 2023).

## Description of Observations

We will observe a sample of 94 M dwarf stars with the PRIMAger Polarimetric Imager. All stars lie within a distance of 10 pc, and their spectral types range from M0.5 to M7, see Fig. 2. Each star will be observed for 5 hours with a 10’x10’ PRIMAger survey area. Of the 94 M dwarfs, two are already known to have debris disks (Lestrade et al. 2012, Kennedy et al. 2013).

# 80. Comparing the Carbon-to-Oxygen Ratio of Debris Disks to Exoplanets

Meredith MacGregor (Johns Hopkins University), Christine Chen (Space Telescope Science Institute), Giani Cataldi (NAOJ), Patricia Luppe (Trinity College Dublin), Jay Chittidi (Johns Hopkins University) and others

Debris disks are the end stage of planet formation. Within these remnant disks, comets and asteroids collide to produce small dust grains and gas. As a result, the detailed composition of the dust and gas are expected to reflect the composition of the underlying, larger bodies. We will use PRIMA to observe both the [OI] 63 µm and [CII] 158 µm lines to determine the C:O ratio of circumstellar gas in 35 debris disks that have either been detected in CO emission by ALMA or show evidence for exocomets and thus high levels of collisions. Measurements of the C:O ratio are especially relevant today because JWST is poised to measure the C:O ratios of hundreds of exoplanetary atmospheres during the next decade. The results of previous studies suggest a dichotomy in the measured values between massive directly imaged and smaller transiting exoplanets. This study with PRIMA will allow us to make the first comparisons between a statistically significant sample of disk environments and planetary results to inform our understanding of planetary composition and formation mechanisms.

## Science Justification

### Broader Context

Given that the bulk gas in debris disks has dissipated (>10 Myr, Hughes et al. 2018), the circumstellar dust and gas in these systems are believed to be generated by collisions among planets, planetary embryos, and planetesimals. As a result, the detailed composition of the dust and gas are expected to reflect the composition of the underlying, larger bodies (Youngblood et al. 2021, Kral et al. 2017) contrary to protoplanetary disks in which the composition of the gas is complementary to that of the dust (Oberg et al. 2011). The Carbon-to-Oxygen (C:O) ratio of debris disk gas (and by extension that of the parent bodies) is expected to provide key constraints on the bulk chemistry of exoplanetary systems. In our Solar System, the terrestrial planets are Oxygen-rich, however, modeling indicates that rocky exoplanets could also be Carbon-rich with diamond cores (Allen-Sutter et al. 2020). Although hundreds of debris disks have been discovered thus far, via infrared excesses, the detailed composition of the gas and dust and therefore the larger bodies remains enigmatic (Chen et al. 2020).

Measurements of the C:O ratio are especially relevant today because JWST is poised to measure the C:O ratios of hundreds of exoplanetary atmospheres during the next decade. The results of previous studies suggest a dichotomy with massive directly imaged planets having a C:O ratio approximately solar (~0.5) and transiting exoplanets having a wide variety in C:O ratio from 0.3 - 1.6 (Hoch et al. 2023). The directly imaged planet 12.7 $M_{Jup}$ planet $\beta$ Pic b with a semi-major axis of 10 au has a stellar C:O ratio (GRAVITY Collaboration 2020), while the 1.76 $M_{Jup}$ transiting exoplanet WASP-77 A b with a semi-major axis of 0.024 au has a super stellar C:O ratio (Line et al. 2021, Reggiani et al. 2022).

The absence or presence of gas during the formation process is the primary difference between super-Earth and mini-Neptune planets. Formation of gaseous mini-Neptunes requires the presence of an adequate gas reservoir to accrete material from. On the other hand, rocky super-Earth formation requires gas dispersal prior to the accumulation of sufficiently massive rocky cores (Venturini & Helled 2013). The solar C:O ratio of more massive planets may indicate that they formed via accretion of large amounts of icy planetesimals following envelope accretion while the large variation in the C:O ratio of lower mass planets may indicate that they formed via a wider variety of pathways, including gas accretion beyond the water ice snowline. Bulk planetesimal C:O ratios are critical, along with those of protoplanetary disks (Anderson et al. 2021), to place exoplanetary atmosphere measurements into context.

Abundance measurements of debris disk gas have provided an alternate pathway to constrain the detailed composition of the circumstellar material. Ultraviolet, absorption line spectroscopy using HST has detected narrow absorption from cold atomic carbon and other species in a handful of young debris disks (Youngblood et al. 2021; Ren et al. 2023). A combined FUSE and HST inventory of the circumstellar gas around the archetypal debris disk Beta Pic found a gaseous C:Fe ratio 16x the solar value (Roberge et al. 2006), which could be the result of inefficient removal of carbon by weak radiation pressure (Fernández et al. 2006) and/or a carbon enrichment in the underlying planetesimal bodies. Unfortunately, the number of systems that can be studied with HST is severely limited to stellar host stars that are nearby (within 100 pc) and early-type (A0 or earlier). Stars that are more distant and/or cooler do not emit sufficient ultraviolet continuum emission against which to detect carbon and oxygen gas absorption.

## Science Question and Approach

We will use PRIMA to observe both the [OI] 63 μm and [CII] 158 μm lines. These observations will allow us to determine the C:O ratio of circumstellar gas in 35 debris disks (Figure 1, left) to constrain the C:O ratio of the parent exo-asteroids, comets, and Kuiper Belt Objects (KBOs). These targets have either been detected in CO emission by ALMA (Figure 1, right) or show evidence for exocomets and thus high levels of collisions. This study with PRIMA will allow us to make the first comparisons between a statistically significant sample of disk environments and planetary results to inform our understanding of planetary composition and formation mechanisms.

Debris disks were originally thought to be gas-free. However, the number with detected gas has grown substantially over the last decade due largely to detections of CO emission using ALMA (Marino et al. 2020). There are two possible scenarios for how this gas is produced: (1) gas remains from the primordial phase or (2) gas is regenerated due to collisions between larger bodies. The primordial scenario would be surprising, since gas is expected to dissipate on much shorter timescales. However, self-shielding can extend the CO lifetime if densities are high enough (Kospal et al. 2013, Nakatani et al. 2021, Nakatani et al. 2023). The secondary scenario is generally favored for most disks, yet some systems have high enough gas masses that the implied collision rates are exceptionally high (e.g., Cataldi et al. 2023). Our proposed PRIMA program has the potential to shed new light on this debate by measuring the relative abundances of [OI], [CII], and CO.

### Need for PRIMA

No other past or currently proposed FIR facility has the sensitivity necessary to measure the [OI] and [CII] emission in a statistically significant sample of debris disks. Indeed, during Herschel's lifetime, PACS measured [OI] and/or [CII] emission around only a handful of debris disks (Riviere-Marichalar et al. 2012, Roberge et al. 2013, Donaldson et al. 2013). This PRIMA GO program will make the first comparisons between a statistically significant sample of disk environments and planetary results to inform our understanding of planet formation mechanisms.

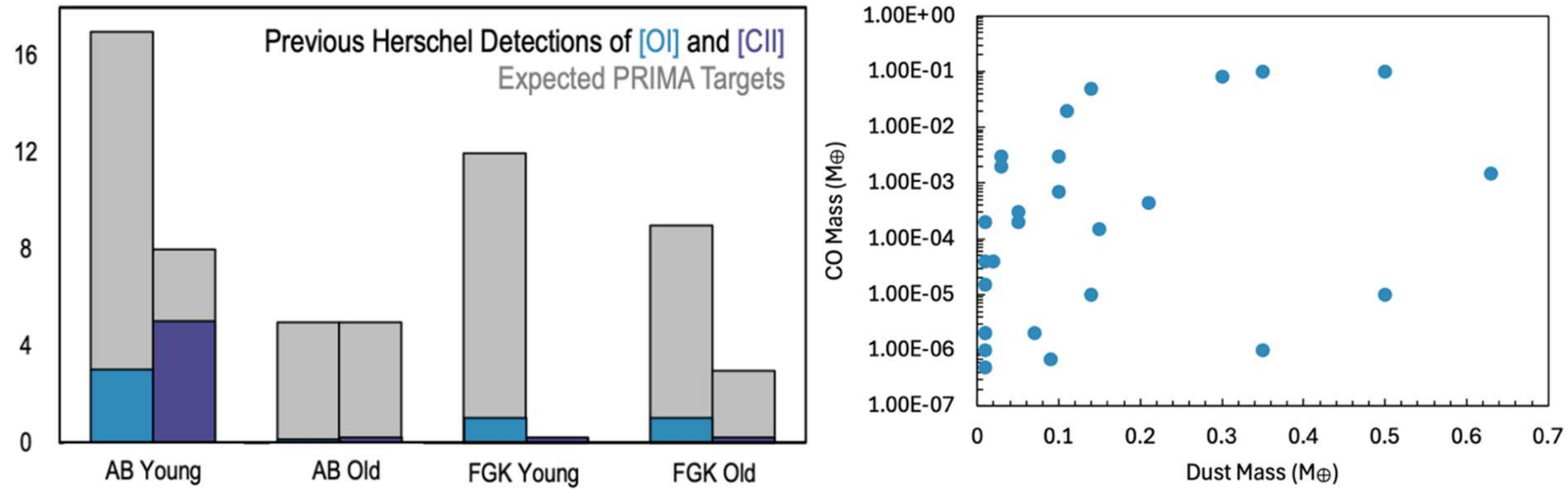

**Figure 1.** (Left) Previous facilities have detected [OI] and [CII] emission from only a handful of debris disks. Here, 'old' systems are older than 50 Myr. (Right) The sample of debris disks with CO masses measured by ALMA.

## Instruments and Modes Used

This observing program requests observations of 35 disks using FIRESS' pointed high-resolution mode.

## Approximate Integration Time

The estimated [OI] fluxes range from 1.0 x $10^{-16}$ W $m^{-2}$ to 1.0 x $10^{-18}$ W $m^{-2}$ for the brightest and faintest sources in our sample, respectively. Using the PRIMA ETC for FIRESS, we estimate that roughly 0.5 – 1 hours will be needed for a 5$\sigma$ detection of each target. We roughly double this estimate to obtain [CII] fluxes as well. Therefore, the total time for this program is roughly 60 hours excluding overheads.

## Special Capabilities Needed

None

## Synergies with Other Facilities

We have defined our sample to include all debris disks with previously measured CO and/or [CI] emission by the Atacama Large Millimeter/submillimeter Array (ALMA). We will examine our sample for correlations between the [OI], [CII], [CI], CO, and dust masses when possible, to shed light on the origin of gas in debris disks. 49 Ceti, β Pictoris, and HD 172555 also have previous detections of [OI] from Herschel.

## Description of Observations

We will observe both [OI] and [CII] emission from a sample of 35 debris disks (Figure 1) using the high-resolution mode of PRIMA's FIRESS spectrometer. Of the 35 target sources, 28 have previous detections of CO gas from ALMA. The remaining 7 targets are inferred to have exocomets either through IR variability or transit detections. Although these sources do not yet have gas detections, the inferred presence of exocomets indicates a high level of collisions and thus likely high gas abundance. The table below gives relevant information on all the targets in the sample.

We use [OI] and [CII] total flux measurements to constrain the C:O ratio. As discussed above, the total observing time is calculated to ensure at least a 5$\sigma$ detection of each target in both lines. To accurately determine the total flux in each line, we need to know the radial location of the emitting gas. We assume that [OI] and [CII] share the same radial distribution, as is expected if both are produced through outgassing during planetesimal collisions. For the majority of the sample drawn from ALMA CO detections, the dust and gas location are previously known. For targets where [OI] and/or [CII] are strongly detected, we can use tomographic mapping to determine the radial location and test these assumptions. Herschel previously detected [OI] from 3 targets, which we will use as an important check on our derived fluxes. It also important to note that the critical density is very high of [OI], which means we will most likely detect it in NLTE. Under these conditions, the oxygen mass is strongly dependent on the density of colliders (Kral et al. 2017). To get the total carbon mass, knowledge of the ionization fraction is required. Given these constraints, we expect the uncertainty on the C:O ratio to scale with detection significance; the constraints will be strongest for the brightest targets.

ALMA searches to date have not detected any gas species in debris disks besides CO and [CI] and therefore have not provided constraints on the gas-phase C:O ratio. If the origin of gas in debris disks is cometary, we expect large amounts of water and thus an overabundance of oxygen atoms to carbon ions by a factor of 10. This program with PRIMA to measure [OI] and [CII] will provide the most direct constraint on the C:O ratio in debris disks for the largest sample and will help distinguish between cometary and primordial gas formation scenarios.

**Table 1.** PRIMA C/O Debris Disk Targets

| Name | RA | DEC | SpT | ALMA CO? | Exocomets? |
|---|---|---|---|---|---|
| **HD 131488** | 14 55 08.03 | -41 07 13.40 | A1V | Y | |
| **HD 141569** | 15 49 57.75 | -03 55 16.34 | A2Vek | Y | |
| **HD 131835** | 14 56 54.47 | -35 41 43.66 | A2IV | Y | |
| **HD 21997** | 03 31 53.65 | -25 36 50.94 | A3IV/V | Y | Y |
| **HD 121617** | 13 57 41.13 | -47 00 34.25 | A1V | Y | |
| **HD 121191** | 13 55 18.88 | -53 31 42.71 | A5IV/V | Y | |
| **HD 36546** | 05 33 30.76 | 24 37 43.73 | B8 | Y | |
| **HD 156623** | 17 20 50.62 | -45 25 15.00 | A0V | Y | Y |
| **HD 32297** | 05 02 27.44 | 07 27 39.68 | A0V | Y | Y |
| **HD 138813** | 15 35 16.11 | -25 44 02.99 | A0V | Y | |
| **HD 9672** | 01 34 37.78 | -15 40 34.90 | A1V | Y | |
| **Beta Pic** | 05 47 17.09 | -51 03 59.41 | A5V | Y | Y |
| **HD 110058** | 12 39 46.20 | -49 11 55.55 | A0V | Y | |
| **HD 146897** | 16 19 29.24 | -21 24 13.27 | F2/F3V | Y | |
| **49 Ceti** | 01 34 37.78 | -15 40 34.90 | A1V | Y | Y |
| **Fomalhaut** | 22 57 39.05 | -29 37 20.05 | A0V | Y | |
| **Eta Corvi** | 12 32 04.23 | -16 11 45.62 | F2V | Y | |
| **HD 172555** | 18 45 26.90 | -64 52 16.53 | A7V | Y | |
| **HD 48370** | 06 43 01.02 | -02 53 19.32 | K0V | Y | |
| **HD 95086** | 10 57 03.02 | -68 40 02.45 | A8III | Y | |
| **HD 181327** | 19 22 58.94 | -54 32 16.98 | F6V | Y | |
| **HD 5267** | 00 54 35.22 | 19 11 18.32 | A1Vn | Y | |
| **HD 158352** | 17 28 49.65 | 00 19 50.25 | A8Vp | Y | |
| **HD 183324** | 19 29 00.99 | 01 57 01.62 | A0IVp | Y | |
| **Eta Tel** | 19 22 51.21 | -54 25 26.15 | A0Vn | Y | |
| **HR 4796** | 12 36 01.03 | -39 52 10.22 | A0V | Y | |
| **HD 61005** | 07 35 47.46 | -32 12 14.04 | G8Vk | Y | |
| **HD 110411** | 12 41 53.06 | 10 14 08.25 | A0Va | Y | |
| **HD 166191** | 18 10 30.34 | -23 34 00.27 | F3/5V | Y | |
| **HR 6864** | 18 22 00.14 | -28 25 47.90 | A3III | | Y |
| **HD 37306** | 05 37 08.77 | -11 46 31.90 | A1V | | Y |
| **HR 3685** | 09 13 11.98 | -69 43 01.90 | A1III | | Y |
| **HD 85905** | 09 54 31.82 | -22 29 14.90 | A1IVn | | Y |
| **HD 145964** | 16 14 28.88 | -21 06 27.50 | B9V | | Y |
| **HD 42111** | 06 08 57.90 | 02 29 58.90 | B9V | | Y |

Note: Targets are ordered by inferred CO mass from largest (top) to smallest (bottom).

# 81. Variability and Transient Debris Dust with PRIMA

Jonathan P. Marshall (Academia Sinica Institute of Astronomy and Astrophysics), Mark Booth (UK Astronomy Technology Centre), Joshua B. Lovell (Center for Astrophysics | Harvard & Smithsonian), Nicholas Ballering (Space Science Institute), Patricia Luppe (Trinity College Dublin), Christine Chen (STScI, JHU), Meredith MacGregor (Johns Hopkins University), Katie Crotts (STScI)

Debris disks around main sequence stars are composed of dusty and icy bodies, from kilometer-size planetesimals to micron-size dust, created in belts around their host stars. These belts are analogous to the Solar system's Asteroid and Edgeworth-Kuiper belts, and are commonly identified by detection of excess emission at infrared wavelengths. Bright asteroid belt analogues, known as 'extreme debris disks' have been interpreted as evidence of catastrophic collisions occurring near the habitable zone of their host stars. These systems may exhibit monotonic or quasiperiodic variability as the hot dust is removed or replenished. A handful of debris disk systems also exhibit evidence for the existence of star-grazing planetesimal populations, exocomets, flung onto eccentric orbits far from their birthplace and can be identified by occultation of the stellar photosphere. Here we propose far-infrared characterisation of fifty extreme debris disks with PRIMA, measuring their far-infrared spectral slopes, and searching for cool, outer belts as counterparts to the hot inner emission. These will provide important context for the origin, delivery, and aftermath of dusty material to the habitable zone around these stars, provide constraints on the mostly unknown architectures of these systems, and place these extreme systems in the wider ensemble of debris disks.

## Science Justification

Debris disks are formed from belts of planetesimals and their collisional debris. They are commonly identified through detection of excess emission from a star at infrared to millimetre wavelengths (Wyatt 2008), with fractional luminosities ($L_{disk}/L_{star}$) below $10^{-3}$ and sizes between 10s to 100s au (Hughes+ 2018). These structures are more massive, larger analogues to the Asteroid and Edgeworth-Kuiper belts of the Solar System (Horner+ 2020). Cool debris belts (EKB analogues) are found around 15-30% of AFGK-type stars (Sibthorpe+ 2018, Thureau+ 2014), but warm (AB analogues) are rarer and found both in conjunction with cool belts and in isolation (e.g. Kennedy + Wyatt 2012, Fujiwara+ 2013). The current evolutionary model for debris belts suggests a steady-state grinding down of bodies in a collisional cascade such that disk mass and brightness decline over time (Najita+ 2022).

Some bright, young debris disks exhibit spectroscopic or photometric absorption at the < 1% level associated with exocomets - planetesimals passing perilously close to their host stars (Strom+

2020). For these systems, the planetesimals do not just reside in belts but were flung from their point of formation to a fiery demise (e.g Rodet + Lai 2024).

A handful of warm debris disks are so bright that their evolution cannot fit the steady state paradigm. With fractional luminosities up to a few percent, these systems are known as 'extreme debris disks' (EDDs; Su+ 2019, Moor+ 2024). Such systems fall into two broad categories; young systems believed to be undergoing the final stages of rocky planet assemblage, or much older systems where destabilization of a planet-moon system may have resulted in a catastrophic collision.

A third kind of transient system has been identified wherein the star is occulted by a dense and extended cloud of dust which may or may not exhibit (mid-)infrared excess. The archetype is Boyajian's star (KIC8462852, Boyajian+ 2016), and a handful of subsequent events matching these criteria have been identified (Schmidt 2020). Surveys such as ASASSN identify optical dimming which can be combined with (NEO)WISE to characterize potential dust emission from these stars. For example, a recent extreme event around ASASSN-21qj was attributed to either an exocometary breakup close to the star (Marshall+ 2023, or a planet-planet collision at a few au (Kenworthy+ 2023).

In total there exists a population of ~50 debris disks that might be considered EDDs (including debris disks with exocomets, WISE W3 [12 µm] bright debris disks, and transient excesses). The occurrence, evolution, and persistence of the extreme excesses generated through catastrophic collisions provides a natural point of leverage to examine planet formation and dynamical evolution through planetesimals and their debris.

With PRIMA we can address several outstanding questions:

**What is the spectral slope of the warm emission?** This constrains the size distribution of debris, and the material strength of the original colliding bodies. Results from Spitzer suggest the size distribution is steeper for warm belts than cold belts (Mittal+ 2015), which can be more precisely determined with this analysis.

**Do EDDs have cool outer belts?** A handful of EDDs have evidence of cool EKB analogues. Determining the architectures of planetesimal belts within EDDs constrains the potential origins of the dust-producing bodies. Absent, or systematically fainter, EKB-analogues would point to in-situ dust production rather than external delivery as the main cause of EDDs. A substantial number of EDDs have been identified post-Herschel, such that current far-infrared observations of these systems are incomplete.

**What is the brightness distribution of those cool belts?** With PRIMA we can detect cool belts around EDDs down to levels comparable to faint Herschel-detected disks at $L_{disk}/L_{star} > 10^{-5}$ at a distance of 100 pc. If these cool belts are systematically brighter around EDDs, then the collisional excitation could be system-wide. With the population available (50), the time-evolution of the disk brightness can be accounted for in the detection statistics. An all-sky survey with PRIMA would increase the number of EDDs enabling more robust interpretation from a larger population.

The PRIMA instrument is thus uniquely capable of addressing the origins and evolution of EDDs. Tracing the Rayleigh-Jeans tail of dust emission to far-infrared wavelengths constrains the dust size distribution. This spans not just emission from the known, warm component, but also fully characterising potential cool belts around these stars. JWST lacks the wavelength range required and ALMA is too insensitive to warm dust to characterize such a population in the (sub)millimetre.

## Instruments and Modes Used

PRIMAGer maps of minimum size, in both bands. One map for each target

## Approximate Integration Time

Estimates presented here based on calculations using the online PRIMA ETC.

The integration time required for a 100 sq. arcmin map is 9 hrs to achieve a 5-σ rms of 300 μJy at 70 μm using PRIMAger, sufficient to detect a $L_{disk}/L_{star} = 10^{-5}$ debris disk at 100 pc. For 50 targets the program would require 450 hrs of integration time.

## Special Capabilities Needed

To characterize the variability of these sources across timescales relevant to dust evolution (years to decades) the same targets should be regularly observed such that the time request presented here is per epoch of observation (i.e. 570 hrs/year).

## Synergies with Other Facilities

The WISE and NEOWISE catalogues provide an all-sky survey to mid-infrared wavelengths for the identification of EDDs (WISE) and hot transient excess emission (NEOWISE). In particular the NEOWISE catalogue (W1 and W2, 3.6 and 4.5 μm photometry) has yet to be systematically searched for such systems that would be prime targets for this program.

Spitzer/IRS has a legacy of several hundred debris disk observations including sources with silicate features for characterization and composition analysis.

JWST NIRSpec + MIRI observation of 5 to 25 μm spectrum reveals the 10 and 20 μm silicate features, carbonaceous and water features. A subset of 11 EDDs have already been observed in a Cycle 2 program (PID #3189). Prior to PRIMA many potential targets could be characterized with JWST as a prelude to this program.

SPHEREX will perform an all-sky near-infrared spectroscopic survey up to 5 μm complementary to previous (NEO)WISE photometric surveys for the detection and characterization of EDDs and transient hot dust

ALMA observations constrain the architectures of cool belts around ~ 100 debris disk host stars at millimetre wavelengths, providing context for the cool excess emission detected by PRIMA. EDDs may have substantial CO gas masses from exocomet disruption, revealed at mm wavelengths (e.g. HD 172555, Schneiderman+ 2021).

## Description of Observations

PRIMAger: We aim to detect cool debris disks around EDD stars down to a level consistent with faint debris disks detected by Herschel ($L_{disk}/L_{star}$ ~ $10^{-5}$). At a representative distance of 100 pc, the 70 µm flux density of such a disk would be 240 µJy. With a minimum map size of 100 sq. arcmin and a 5-σ limit of 600 µJy (at R=10, to be rebinned to R=2, 270 µJy) this would take 9 hrs (per target) with PRIMAger.

# 82. PRIMA to Test Dust Alignment Theories and Measure the Properties of Dust Grains in the Dense ISM and Star-Forming Cores

Anaëlle Maury (ICE/CSIC, Spain & CEA Astrophysics dpt, France), V. Le Gouellec (ICE/CSIC, Spain), Le Ngoc Tram (Leiden Observatory, Netherlands), V. Guillet (IAS, France), T. Hoang (KASI), L. Testi (Univ. Bologna, Italy), N. Ysard (IRAP, France), S. Coudé (CfA, USA), D. Paré (Villanova U, USA), A. Soam (IIA, India), I. Ristorcelli (IRAP, France), D. Johnstone (NRC Herzberg Astronomy and Astrophysics, Canada)

Dust plays a key role in the chemistry and the physics of the interstellar medium (ISM) , yet its composition and growth during early star formation remain poorly constrained. Observing the total and polarized submm dust emission, Planck, BLASTPOL and ALMA have shown that the standard dust models clearly need to be revised. Further multi-wavelength observations are crucial to sample the polarized spectral energy distribution (SED) in the Wien regime (in the short-wavelength spectrum for the thermal emission from objects) where dust models can be efficiently discriminated, providing key insights on the dust properties and dust evolution and alignment processes, from the irradiated regions of the ISM to planet-forming disk scales.

We propose to use PRIMAger to sample polarized dust emission in both diffuse and dense protostellar environments, testing dust alignment models and constraining grain sizes at the onset of star and planet formation. PRIMAger will allow us to get access to local variations of the total and polarized SED, which can now be confronted against predictions from synthetic observations produced from state-of-the-art MHD models combined with radiative transfer models, to solve some long-standing issues regarding the nature and evolution of dust grains in the ISM and in star-forming cores.

## Science Justification

Dust is the key ingredient to form planetary systems around young stellar objects through its evolution from sub-micronic ISM dust particles to macroscopic objects. Moreover, dust is responsible for coupling the magnetic field to star-forming material, and the grain size distribution plays a role in setting the efficiency of magnetic fields to regulate, for example, the properties of disks around solar-type protostars. Detailed knowledge on dust properties, including their evolution from the diffuse ISM to the densest phases where stars and planets form, is thus essential. Planck observations of the Galactic diffuse and cold ISM have revealed discrepancies between the dust masses predicted from dust models and those derived from the optical, as well as local variations in the diffuse ISM where empirical dust models such as Draine & Li (2007) are usually calibrated (Planck Collaboration Int. XI 2014).

While multi-wavelength dust thermal emission is used to estimate masses in dense astrophysical structures, polarized dust emission is routinely used to trace the magnetic fields in astrophysical structures. In the current paradigm, dust polarization is produced by grains aligned with local magnetic fields, as their initial momentum is set by the local radiation field (Radiative Alignment Torques, RATs, Draine & Weingartner 1997, Hoang & Lazarian 2009, 2016). Complementing the total dust emission, Planck and BLASTPOL sampled the polarized dust SEDs at submm wavelengths, providing constraints on the nature of grains like their compositions and emissivities (Planck collaboration XXI & XXII, 2015; Planck Collaboration XI 2018). Local variations of the polarization fraction can be attributed to fluctuations on the structure of the B-field along the line of sight (Planck Collaboration XII 2018), to the evolution of grain alignment efficiency (Andersson et al. 2015, Hoang et al. 2021) and the grain shape's elongation due to anisotropic grain growth (Hoang et al. 2022). The shape of the polarized SED is a direct signature of the variations in the grain physics itself (Smith et al. 2000, Fanciullo et al. 2017, Fanciullo et al. 2022, Le Gouellec et al. 2023b, Tram et al. 2025). Exploring this aspect, Guillet et al. (2018) showed in particular that the flat profile of **the polarization fraction from the IR to the sub-mm depends on the chemical composition of the grains responsible for the observed polarization**. More especially, it depends on the amount of aligned carbon grains compared to silicate ones (Fig. 1). While a model with carbon grains aligned and silicates with carbon inclusions seems to better reproduce the polarization observed. In spite of these promising results, many questions remain on the nature of the various grain populations. In particular, Planck is not sensitive to the emission from warm dust, and the temperature of the aligned grain population thus remains mostly unconstrained. While new dust models and modeling tools are currently updated to include this polarized component of the emission (Jones et al. 2017, Hensley & Draine 2023), they predict that **observing the polarization fraction of dust thermal emission in the Wien regime provides unique probes of the dust composition**. As ALMA, NOEMA and other telescopes now provide polarization datasets that are widely used to probe the geometry of B-fields, and hence their role, in dense environments forming stars, understanding the exact origin of the dust polarized emission, its dependency with wavelength and with the local radiation field becomes crucial to interpret polarimetric observations and constrain dust properties.

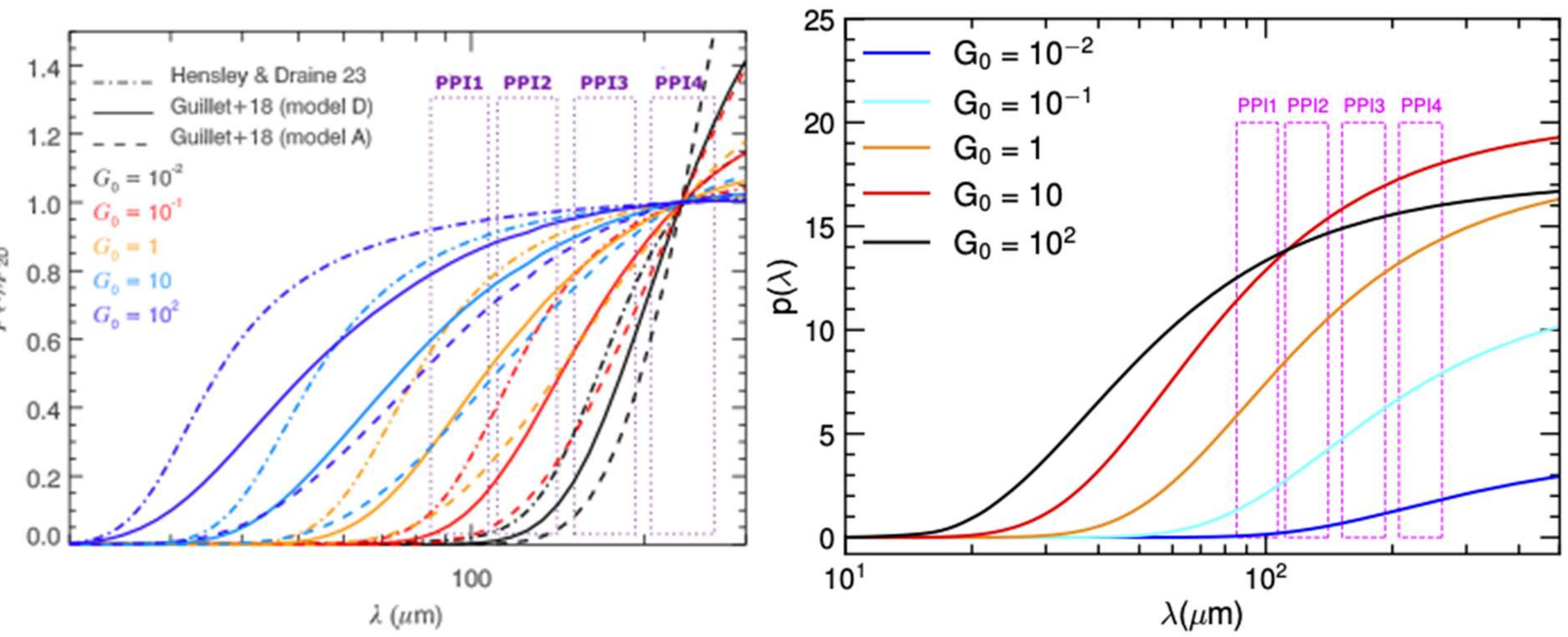

**Figure 1**. **Left**: Dust polarization fraction, p = P/I (normalized to the P/I of PRIMAger band 2), as a function of wavelength, for two dust models of Guillet et al. (2018) and the Astrodust model by Hensley & Draine (2023). Model A includes silicates grains that are aligned and carbon grains that are not. Both are aligned in model D, with carbon inclusions (6% in volume) in the silicate matrix. The vertical boxes show the PRIMA bands sampling the Wien regime where the polarized emission evolves dramatically depending on the dust model considered. The figure also illustrates how the polarization fraction is also expected to evolve under a range of irradiation conditions, with values $G_0$ typical from diffuse ISM to dense gas. - **Right**: Models from grain alignment with RATs also predict that not only the shape but the absolute percentage of the polarization from thermal dust emission varies for different $G_0$, with a fraction higher for higher $G_0$ at a specific wavelength. Note that, when radiation $G_0$>10, rotation disruption occurs, leading to a reduction in the polarization fraction for wavelengths greater than 100 $\mu$m (black line).

Recent mm observations find protostellar dust with **emissivity indices β < 1 and very high polarization fractions at mm wavelengths,** two independent possible signatures of significant grain growth in dense star-forming regions toward protostellar envelopes (see Fig. 2, Galametz et al. 2019, Valdivia et al. 2019). These findings, along with unexplained emissivities in filaments (Lowe et al. 2022), highlight the need for new constraints on dust properties in the dense ISM, testing the early phases of dust evolution, potentially building large dust grains before the material even reaches the disk scales. Moreover, characterizing large grains in protostellar envelopes is vital for understanding both dust evolution toward planets and the coupling of magnetic fields to circumstellar material (see Zhao et al. 2016 and the review by Maury et al. 2022).

Recently, it has been shown that **polarized dust emission can be a powerful tool tracer to measure the sizes of the largest dust grains in dense environments**: indeed, since the RAT mechanism only allows to align grains with sizes comparable to the wavelength of the incoming photons, in high-density environments where most of the protostellar radiation is quickly reprocessed in infrared / far-infrared photons, the resolved study of the dust polarized emission might be a unique window to put significant constraints on the underlying grain size distribution. Observations of the dust polarization fractions at millimeter wavelengths compared to models suggest protostellar environments may be favorable to develop, early on, a grain size distribution containing a significant fraction of large grains, with sizes 10μm (Valdivia et al. 2019, Le Gouellec et al. 2019, see the right panel of Fig. 2). However, in dense but highly irradiated regions, a fraction of large grains may be aligned with the radiation direction instead of the magnetic field (Lazarian & Hoang 2007, Pattle et al. 2021). Observing the polarized emission from both dense cores where dust grains align with the magnetic fields and highly irradiated environments where dust grains may in turn align with radiation, is thus a powerful tool not only to measure dust properties, but also to inform us about the grain alignment mechanisms at play in different environments.

Observing both dense cores and highly irradiated regions is key to testing both grain alignment mechanisms and measuring dust properties. The effect of changes in the dust grain size distribution on the thermal dust emission from infrared to millimeter wavelengths in dense environments, either as a consequence of grain growth or radiative torque disruption (RAT-D; see Hoang 2020 for a review), are yet to be explored. However, the lack of FIR data, needed to probe grains 50–200 µm in size, limits our understanding of early grain growth and alignment in dense environments. Hence, not only are all the first steps of grain growth missed but also uncertainties remain on the respective roles of RATs and radiative alignment in the dense ISM. This is a key domain of grain sizes, where dust grains are expected to decouple from the gas thus promoting faster growth: we currently are blind to these dust grains in the pristine dense environments that are the cradle of star and planet formation processes, which is a significant shortcoming of our current understanding of planet formation.

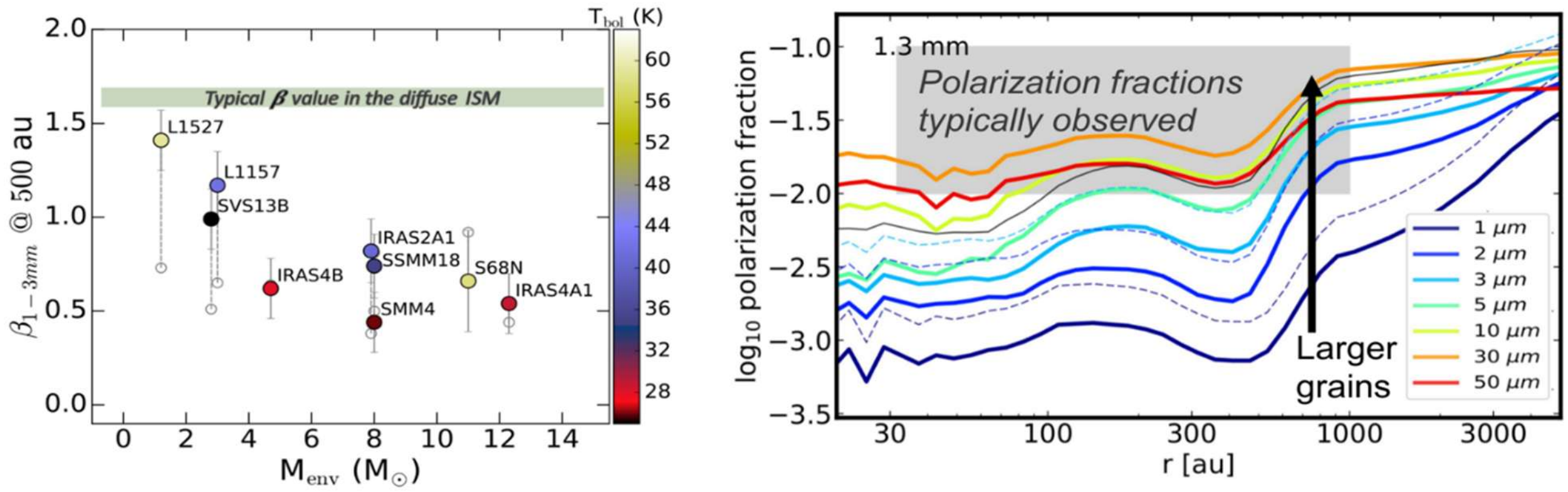

**Figure 2.** Left: Observed millimeter dust emissivity index in a sample of young protostars (Galametz et al. 2019): all protostars show lower values than the progenitor diffuse ISM at radii < 1000 au. Right: Modeled radial profiles of the polarization fraction from the 1.3 mm dust emission from RATs and MHD models of protostellar cores (Valdivia et al. 2019). Large (> 10 µm) dust grains are required to match the observations of the polarized emission (grey area), however the lack of far-infrared wavelengths prevents a more refined characterization of the size of these dust grains.

The presence and formation mechanisms of these large dust grains in protostellar envelopes are vigorously debated in the community, and testing whether they are present already at large scales in filaments and in the outer layers of protostellar cores would also put stringent constraints on the scenario and physical processes responsible for the first steps of grain coagulation. Not only this characterization of dust properties is key to set constraints on the first steps of dust evolution towards planetary systems in dense environments, but also because dust grain properties are key to couple the magnetic field to the circumstellar material and their sizes for example may directly influence the size of disks formed around protostars, for example.

## Observables

The spectral dependence of the polarization fraction (sampled with Planck and BLASTPOL) is relatively flat in the submm range (Fig. 1), which complicates its use to disentangle between the contributions of the various (silicate and carbon) dust populations. On the contrary, dust models predict measurable variations of the polarization fraction with the dust composition in the PRIMA wavelength regime (especially in Bands 1 and 2 at 90-100 µm, see Fig. 1). Using PRIMA in polarization will allow to test the various models of dust grains by measuring the polarization fractions of dust grains in the Wien domain. For example, as shown in Figure 1, PRIMA could discriminate very different dust models with large grains (1-micron sized dark dust as in Siebenmorgen 2023 or the THEMIS-II model by Ysard et al. 2024), models with two dust populations (Guillet et al. 2018) and new models with only one component of composite grains as in Hensley & Draine 2023.

Sampling the polarized SED in different FIR bands will constrain the optical properties, temperature and composition of the dust population. Dust models also predict dramatic variations of the polarization fraction with the local radiation field intensity, which also depends on the dust mixture (Fig. 1): probing the polarization fraction in different $G_0$ (and thus dust temperature) regimes could be another additional test of dust models, from diffuse (high $G_0$) to dense (low $G_0$) environments.

Moreover, the fraction of polarized dust emission P at far-infrared wavelengths can be a powerful tool to investigate the very early stages of grain growth in dense, shielded environments where the local radiation field is well understood. Obtaining polarization fractions of the dust emission in several bands at far-infrared wavelengths with PRIMA would put unique constraints on the maximum size of dust grains ranging from a few dozens to a few hundreds of microns. This kind of work would not be possible without detailed modeling of the physical conditions in protostellar envelopes (temperature, density), which became possible only recently thanks to the development of radiative transfer tools and analytical models of aligned grains (see for example works by Valdivia et al. 2019, Le Gouellec et al. 2023a, Chau Giang et al. 2023).

## Need for PRIMA

The good spatial resolution and sensitivity of PRIMA to image the continuum emission and its polarized component will allow to directly investigate the grain properties, both in regions of the ISM showing large $G_0$ variations, and in star-forming regions where dust temperatures and radiation fields are known from ancillary data (e.g., ALMA, VLA, JWST), thus providing invaluable information on the size of the largest grains present in these protostellar cores.

The high fluxes of star-forming structures in the far-IR bands make them easily detectable at the PRIMA sensitivities: PRIMAger will be a unique tool to test state-of-the-art dust models and firmly establish grain properties at the onset of the star and planet formation processes.

## Interpretation Methods

Analyzing the local variations of the total and polarized dust SED in the far infrared obtained with PRIMA will allow direct tests constraining the nature of the aligned grains. The observational constraints can now be confronted to predictions from synthetic observations produced from state-of-the-art MHD models combined with radiative transfer models including different alignment mechanisms and dust models (see e.g. Le Gouellec et al. 2020, Valdivia et al. 2022 and Chau Giang et al. 2023 for example).vAnalyzing the variation of the polarization fraction (p) with the dust temperature (Td) inferred from SED will help test the key p-Td prediction from the RAT paradigm (Lee et al. 2020, Tram et al. 2021). We will also analyze the correlation of beta with Td to understand the influence of dust evolution and radiation field on the measured beta from total SED. If the change in beta is coincident with the change in the p-Td, it is evident for the effect of RAT-D. The observed p-Td will be confronted to predictions of dust polarization models.

## Instruments and Modes Used

PRIMAger large maps: in polarimetric band

PDRs: 10 maps of 60x60arcmin

Protostars: 5 maps of 60x60 arcmin

## Special Capabilities Needed

None

## Description of Observations and Approximate Integration Time

To describe how such observations could be carried out, we used as prototype the NGC1579 (500-700 pc, Andrews & Wolk 2008) Hii region, illuminated by the early-type B star LkHα101. NGC1579 also possesses a regular B-field morphology on large scale that is crucial if we want to disentangle the dust composition and alignment from variations linked with a non-uniformity in the B-field orientation. This region shows large range of local radiation field intensities, as $G_0$ varies from 200 to 7000. We used Herschel maps at 100, 160 and 250 µm to estimate the required sensitivities, since they are the closest to the PRIMAger bands we ought to use in this science case. Moreover, their spatial resolutions are similar (from 9” to 18” for 100 and 250 µm respectively). As an example, to detect the polarized dust emission at a few percent level one would require a typical 5-sigma 8.5 MJy/sr (0.2 mJy per squared arcsec) in polarized emission in the Band PPI1, which is critical for this science case. PRIMAger provides a typical 5-sigma in 10hours of 0.65 MJy/sr for mapping a 1 squared degree sky area (Ciesla et al. 2024) in Band 1, in Large Mapping mode including all estimated overheads. Note that this estimate should also account for the two orthogonal scan passes needed to calibrate and remove the 1/f noise of the detectors. Hence, a typical region such as NGC1579 would need around 1 hour of integration time for its polarized SED to be investigated with PRIMAger. Fig. 3 illustrates the surface area covered in such a time for the NGC1579 region. Assuming a handful of such regions to be targeted, a program testing dust alignment models in highly irradiated regions would require around 10 hours.

PRIMAGER observations to measure the polarized SEDs of the dust in the dense protostellar cores and characterize the dust properties, such as the maximum grain size, would target the densest parts of filaments and their protostars, cold structures much less bright than PDRs, especially below 100 microns. The time required to map their polarized emission has been estimated in a similar fashion, using the far-IR flux densities from the Herschel/SPIRE observations at 250 µm in the protostars of the Aquila/Serpens South cluster (see Maury et al. 2011), and assuming a conservative 3% polarized fraction. Note that a non-detection at such level in PPI1 would allow to put stringent constraint on the maximum grain size producing the polarization observed at longer wavelengths with ground-based facilities. Obtaining the polarized dust SEDs in a nearby star-forming region, such as Serpens, would allow to sample about 40 protostars and would require about 30 hours with PRIMAger. Mapping all protostars in the 3 closest clouds at d < 300 pc would therefore require around 100 hours of PRIMAger.

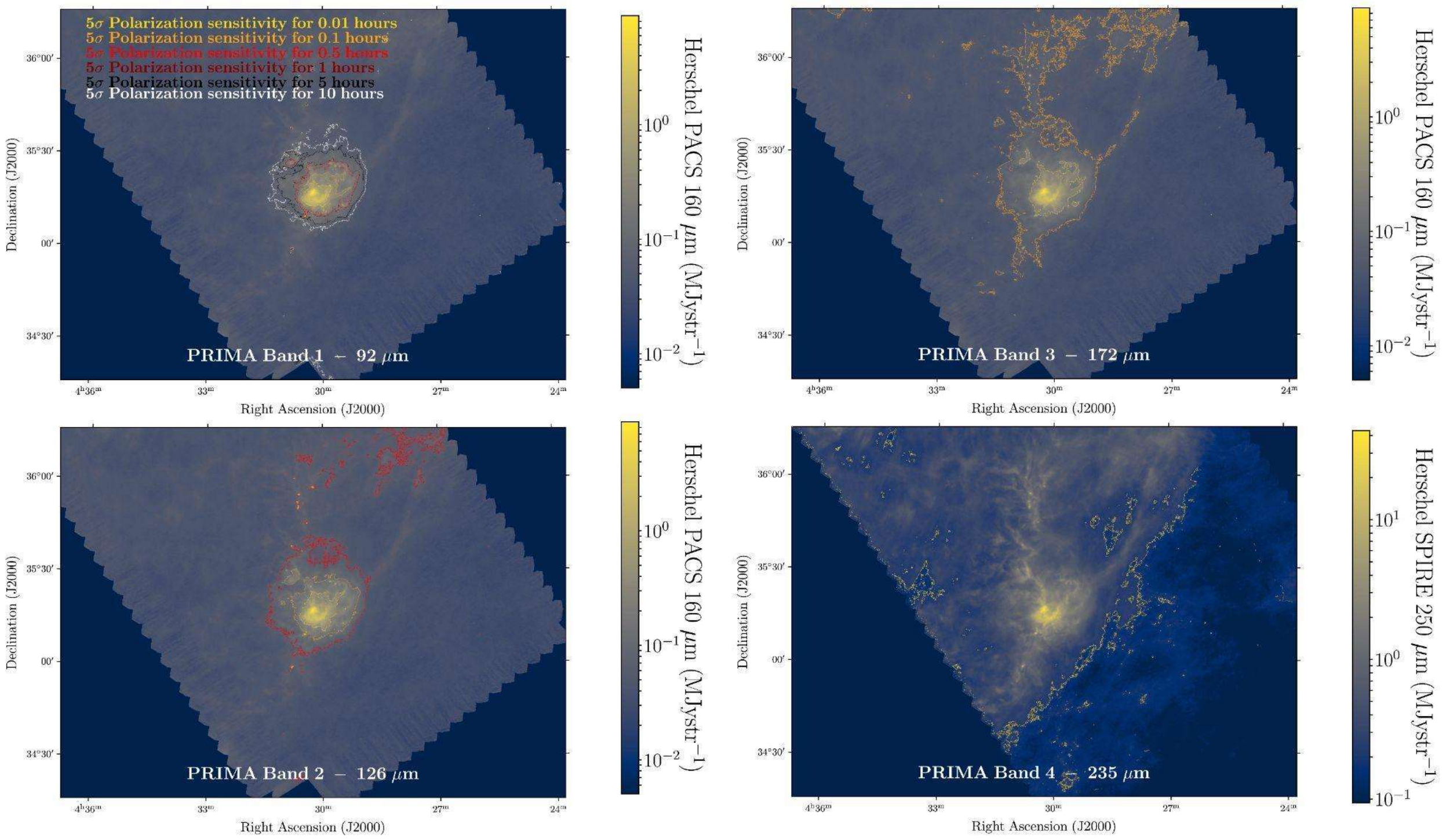

**Figure 3.** Estimation of sensitivity in polarized dust emission for the four bands of PRIMAGER for the NGC1579 region. In each panel, the Herschel total dust intensity is shown in color scale. The contours indicate the PRIMA 5σ sensitivity in polarization for the corresponding integration time, where we assumed a polarization fraction of 30%, a surface covered of 1 squared degrees, and used the Herschel dust temperature maps to interpolate the dust thermal emission at the wavelengths of PRIMAGER.

## Synergies with Other Facilities

Ancillary datasets: AKARI, Herschel and Planck. Synergy with ALMA, NOEMA, JWST and SOFIA/HAWC+.

# 83. A Baseline Spectroscopic Survey of Protostars within 500 pc: Accretion Driven Feedback in Evolving and Outbursting Protostars

Tom Megeath (U. of Toledo), Rob Gutermuth (UMass Amherst), Savio Oliveira (U. of Toledo), Sam Federman (INAF-OACN), Yao-Lun Yang (RIKEN), Wafa Zakri (Jazan University), Libby Banks (U. of Toledo), Mae Higgens (U. of Toledo)

Accretion-driven outflows play a fundamental role in the evolution of galaxies, shaping the IMF and regulating star formation on molecular cloud scales. Far-IR spectra taken with PRIMA can simultaneously observe diagnostics of accretion (using the far-IR continuum) and outflow (using lines), providing a unique capability for exploring the connection between accretion luminosity and outflow mechanical luminosity for the well-characterized sample of 743 protostars within 500 pc of the Sun. We propose to measure how the properties of outflows depend on the far-IR luminosities and evolutionary stages of protostars using lines of CO augmented by [OI], $H_2O$, and $H_2$. These data will measure how the mechanical luminosities of outflows depend on the accretion luminosities of the protostars, and promise to establish a direct and almost instantaneous connection between accretion and outflow. This survey will extend previous work with Herschel Space Observatory to a 27x larger sample with a 10x increase in sensitivity.

We further propose that these observations be taken early in the mission, so that they serve as a baseline sample for protostars showing large changes in far-IR luminosity in PRIMAger monitoring surveys. FIRESS measurements of outbursting protostars identified by PRIMAger can be compared to the baseline spectra to directly measure changes in the outflows during the outburst. Such outbursts may play an important, if not dominant, role in the mass assembly of stars. These data can show, for the first time, how the mechanical luminosity of outflows changes during outbursts. Since most outbursts happen for deeply embedded sources, and the changes in the outflow will first be apparent in the inner, highly extinguished regions of protostars, far-IR diagnostics are essential for detecting changes in the outflows in response to bursts and fluctuations in the accretion rate.

In total, this will be an unparalleled survey to establish the connection between accretion-driven feedback to the accretion luminosities, evolution, and variability of protostars over the rich sample of protostars in the nearest 500 pc.

## Science Justification

Accretion-driven outflows from low to intermediate-mass protostars shape the initial mass function (IMF) and regulate star formation. These outflows, driven by collimated jets and wider-angle winds launched by accretion disks, eject infalling gas and lower the masses of the nascent stars, thereby determining the peak of the IMF (e.g., Guszejnov+2022). On scales of molecular clouds, outflows reduce the rate of star formation and regulate star formation (Federrath+2015, Pokhrel+2021, Grúdic+2022)

It is of fundamental importance to understand the dependence of this feedback on the luminosities and evolutionary stages of protostars. Far-IR observations can simultaneously measure the luminosity of a protostar—which can be used to estimate its mass accretion rate (Fischer+2017, Fischer+2024)—and the far-IR shock tracers of the outflow - a measure of the outflow mechanical luminosity (Figure 1, Manoj+2016). Far-IR observations are much less affected by the extinction that obscures the inner regions of protostars, even from mid-IR observations with JWST (Federman+2024; Narang+2024).

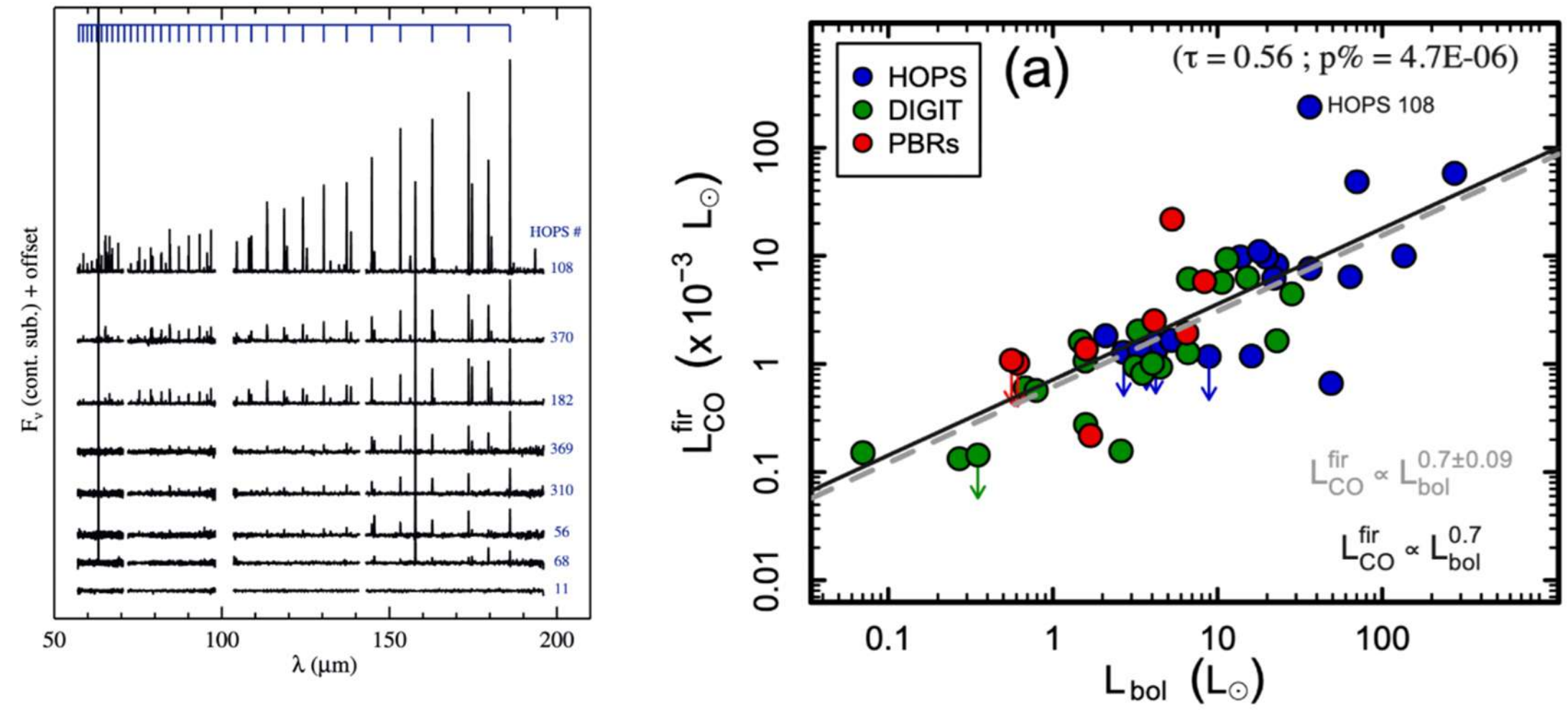

**Figure 1.** Left: spectra from the Herschel PACS spectrometer that show the prominence of the far-IR CO rotational ladder (Manoj+2013). Lines of [OI] and $H_2O$ are also present. Right: the far-IR luminosities of CO vs. the bolometric luminosities for 21 protostars. These data show a strong correlation, which is interpreted as the mechanical luminosity of the outflow growing with the mass accretion rate of the protostars (Manoj+2016).

Herschel observations show a correlation between the bolometric luminosities of protostars and the mechanical luminosities of their outflows, but find no significant correlation with the evolution of the protostars (Figure 1). These data, however, are dominated by the youngest protostars with luminosities over 1 Lsun. With the higher sensitivity of FIRESS, we will extend the work of Manoj+2016 to the total sample of 743 well-characterized protostars within 500 pc (Table 3, Furlan+2016, Pokhrel+2023). These will provide the most extensive study of the correlation between accretion-generated bolometric luminosities and the outflow mechanical luminosities over the full range of luminosities and evolutionary stages present in the sample. We will further extend the analysis to lines of [OI], $H_2$, and $H_2O$ as a function of both protostellar luminosity and evolution (Furlan+2016, Fischer+2017, Federman+2023). In contrast to millimeter lines, which show flow properties typically averaged over thousands of years, these lines are dominated by inner, highly obscured regions of the outflow near the protostar, providing a measure of the outflow averaged over timescales <100 years (Manoj+2016). Thus, they provide a more instantaneous measure of outflow that can be simultaneously connected to the accretion rates estimated from the far-IR luminosity.

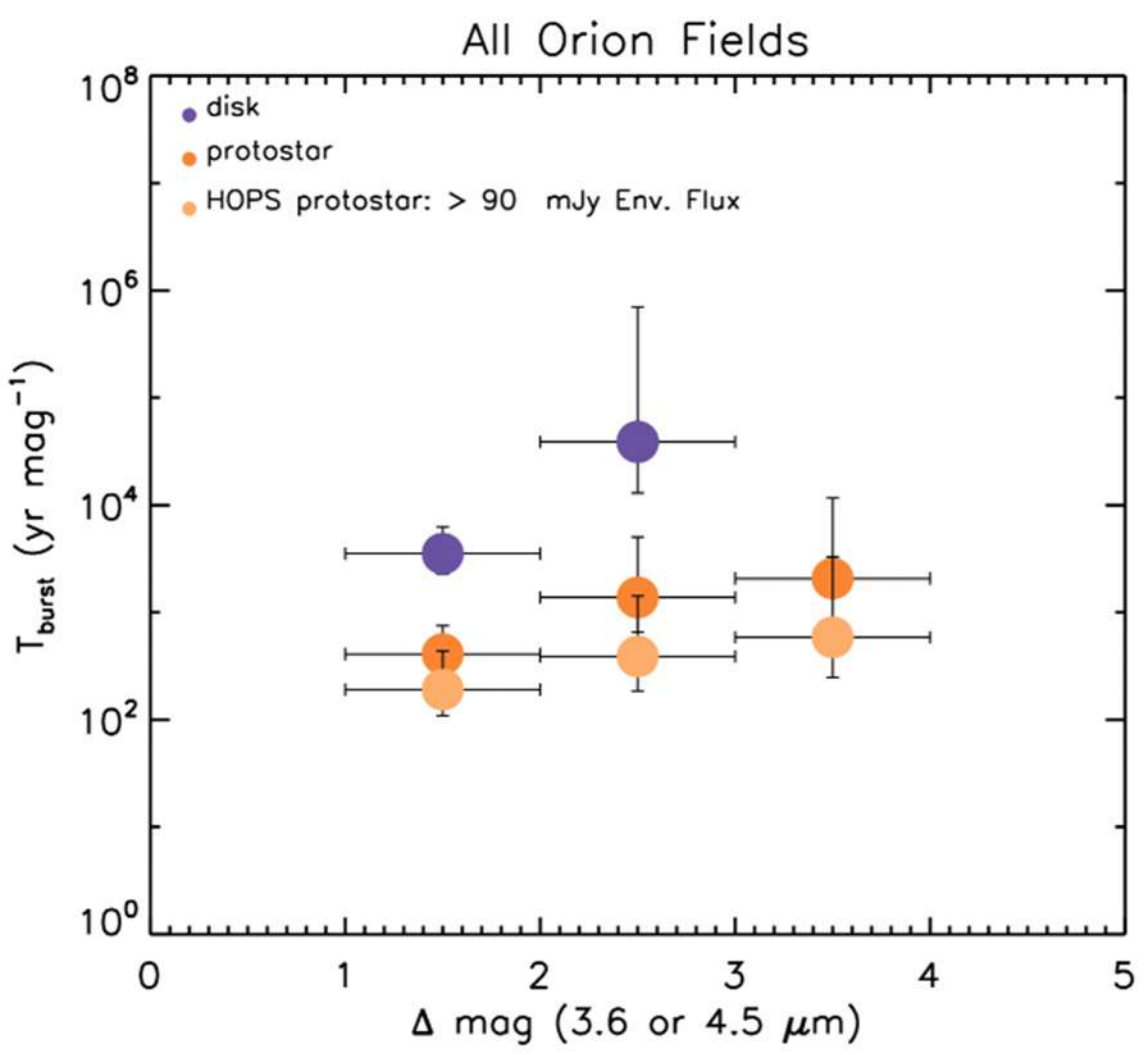

**Figure 2.** The typical interval between outbursts as a function of the evolution of young stellar objects. These show shorter time intervals (i.e., more rapid outbursts) as we go from pre-ms stars with disks to all protostars to the most deeply embedded protostar. The latter may undergo outbursts every few hundred years. Zakri in prep.

The accretion rates of young stars vary substantially over timescales of years and decades. A significant fraction of the masses of stars is thought to be accreted during outbursts where the accretion increases by up to a factor of 10 over a decade or more (Zakri+2022). Far-IR photometric monitoring surveys with PRIMAger will measure variations in the accretion rate for protostars (Fischer+2024; Battersby GO). For YSOs shown to undergo outbursts in the far-IR monitoring surveys, we will obtain spectroscopic measurements of those sources with the identical spectral coverage and resolution as the baseline data. Through comparisons to the baseline spectra, we will directly measure how the outbursts influence outflows. The most deeply embedded protostars, those still in their primary mass assembly phase, that show the most frequent outbursts (Zakri+2022, Figure 2). Far-IR observations are required to trace the outflow in the inner regions of these highly obscured protostellar systems, where the effect of the outburst on the outflow will first be apparent. Using simultaneous measurement of the far-IR continuum flux (i.e., protostar luminosity) and the far-IR CO, we will obtain the most direct connection between outburst and outflow to date. We estimate that several outbursts will be detected per year.

These data will provide a unique view of the connection between accretion and outflow. This will provide a deeper understanding of feedback from protostars and provide constraints on the mechanisms driving bursts, an essential step towards an understanding of both star formation and the effect bursts have on planet formation.

## Instruments and Modes Used

The FIRESS spectrometer in its pointed high-resolution mode will be used.

## Approximate Integration Time

We will observe spectra with the high-resolution point source mode to get the far-IR spectra. To determine our signal-to-noise, we used the flux of the CO 14-13 line in Manoj+2016. We adopted a flux of 0.46 x $10^{-17}$ W $m^{-2}$, approximately a factor of 10 lower than the sensitivity limit of the Herschel observations. We also adopted a continuum flux of 2 Jy, which was similar to the low luminosity source IRAS 16253 in Ophiuchus. This gave a time of 0.5 hours per setting. In total, we expect 1 hour per source using both settings.

Table 1. Protostar Sample

(these are available from the HOPS and eHOPS samples available on the IRSA website).

| **Orion Complex** | **330 protostars** | **380-450 pc** |
|---|---|---|
| **Auriga Cloud** | 41 protostars | 470 pc |
| **Cepheus Flare Complex** | 21 protostars | 350 pc |
| **Chameleon** | 13 protostars | 150-200 pc |
| **Corona Australis** | 11 protostars | 150-200 pc |
| **Lupus** | 5 protostars | 150-200 pc |
| **Ophiuchus** | 51 protostars | 140 pc |
| **Perseus** | 82 protostars | 300 pc |
| **Pipe Nebula** | 16 protostars | 150-200 pc |
| **Aquila** | 172 protostars | 436 pc |

## Special Capabilities Needed

We must observe all of these targets as early as possible in the PRIMA mission. The legacy dataset would then serve as a reference for repeated FIRESS spectra of any transients detected in PRIMAger photometric monitoring.

## Synergies with Other Facilities

these observations may be complemented by visible light to near-IR spectra and photometry to search for signatures of the accretion from the inner disk for more evolved sources. NEO surveyor will cover many of the selected star forming regions and can provide complementary time domain photometry in the mid-IR.

## Description of Observations

We would observe all targets with both FIRESS modes to obtain full 24-235 µm spectra of each source. Our integration times of 1 hr per mode per source represent a typical value, but since all of our targets are Herschel and Spitzer-identified, we have detailed SEDs available (Furlan+2016; Pokhrel+2023) to tune the integration times to the source. Importantly, these observations should be performed early in the PRIMA mission to set a baseline for each protostar that can be compared against when a substantial change is detected in the source's luminosity during the PRIMAger monitoring.

# 84. PRIMAGAL, a full Galactic Plane Survey with PRIMAger: The Role of Magnetic Field in the Formation, Fragmentation and Demise of Dense star-forming Filamentary Clouds

Sergio Molinari (INAF-IAPS, Rome), J. D. Soler, V.-M. Pelkonen, A. Nucara, E. Schisano, A. Traficante, D. Elia, C. Mininni, M. Benedettini, S. Pezzuto (INAF-IAPS, Rome), C. Battersby (Univ. Connecticut), H. Beuther (MPIA, Heidelberg), D. Johnstone (HAA-NRC, Canada), R. Paladini (IPAC-Caltech), X.H. Sun (Yunnan University), D. Paré (Villanova University), A. Saydjari, T. Dharmawardena (Princeton), T. Hoang (KASI), S. Coudé (SSC, NASA Ames), D. Eden (University of Bath), D. Lis (JPL), B. Svoboda (NRAO), J. Pineda (MPE, Munich), M. Audard (University of Geneva), M. Beltràn (INAF-OAA, Arcetri), R. Klessen (Univ. Heidelberg), P. Hennebelle (CEA, Saclay), L. Testi (Università di Bologna), F. van der Tak (SRON, Groningen), and G. White (Open University)

We illustrate the concept of PRIMAGAL, a proposed survey of polarized dust emission with the PRIMAger instrument in the Milky Way Galactic Plane to quantify the role of magnetic fields in the formation, evolution and demise of dense molecular clouds and filaments, and assemble the ultimate SED atlas of dense cluster-forming clumps. This survey will determine the strength and orientation of magnetic fields towards several thousands of filamentary clouds in a wide range of linear masses, column densities, evolution, star-formation rates and efficiencies, and Galactic environments, addressing for the first time in a statistically significant fashion the role that magnetic fields play in shaping the formation, evolution and fragmentation of dense ISM filaments down to a minimum scale of 0.4 pc up to 8 kpc distance from the Sun. In parallel to polarization mapping, the possibility to use PRIMAger's hyperspectral channels will unlock free access to narrow-band photometric imaging in the so far unexplored 24-70 µm wavelength range. This will allow us to assemble, in synergy with already existing surveys with Spitzer and Herschel (among others), the complete and ultimate atlas of Spectral Energy Distributions (SEDs) of dense clumps in the Galaxy, constraining the physical and evolutionary parameters of hundreds of thousands of these cluster-forming structures. A full Galactic Plane survey within $|b| \leq 1^\circ$ (a total of 720 sq. deg.) with a 5 σ sensitivity of $\sim 1.5$ MJy/sr in 183 µm polarized emission, can be executed by PRIMAger in ∼400 hours.

## Science Justification

Star formation is a complex multi-scale process in which a variety of physical agents are at work to shape clouds of diffuse Interstellar Medium (ISM) into dense tens-of-parsecs filamentary structures that fragment into dense and massive parsec-scale clumps. These are sites where further fragmentation produces clusters of denser thousands-au cores, the seeds where individual stars finally form.

Early evidence from Zeeman-splitting line observations (Crutcher et al. 2010) suggested that while in relatively diffuse ISM the strength of the magnetic field is constant with gas density, it showed a rise with increasing density close to the criticality regime in terms of mass/flux ratio, after a threshold value roughly located at $10^2 \leq n_H \leq 10^3$ $cm^{-3}$. This possibly hinted at different roles played by B at different stages of the formation process of dense structures. Planck 353 GHz polarization maps showed a fundamental statistical relationship between cloud structure and magnetic field orientation in 10 nearby (d <500pc) molecular clouds (Soler et al. 2016). In the surroundings of dense filamentary clouds the magnetic field lines tend to be oriented perpendicular to the backbone axis of clouds, and parallel to the faint filamentary pattern observed in their low column density environment. The orientation changes to parallel the closer one gets to the denser cloud regions. Simulations indicated that clouds only show this change in relative orientation when their magnetic fields are dynamically important; that is, when the magnetic energy density is equal to or larger than the turbulent kinetic energy density (Soler et al. 2013). The spatial resolution of the Planck maps in polarization (~5′ in the Galactic Plane), however, was not sufficient to map magnetic fields across the transition region where magnetic field may become subdominant with respect to gravity. To understand how to role of B changes with clouds density and morphology it is crucial to probe polarization down at sub-arcminute level.

An important leap in this direction was offered by facilities that allowed such resolution either because of large telescope apertures, such as Pol-2 at the JCMT, or because of shorter wavelengths of observations, like HAWC+ on board SOFIA Observatory. Polarization observations with these two instruments showed (e.g. Pillai et al. 2020, Liu et al. 2018) that the magnetic field may change back to aligning parallel to dense filaments as field lines are altered by gravitational collapse.

The PRIMAger instrument (Ciesla et al. 2024) on board of the PRIMA observatory (Glenn et al. 2024) offers the unique and unprecedented combination of sensitivity, spatial resolution and mapping speed to effectively map infrared polarized emission over the entire Galactic Plane, from the scale of entire large filamentary complexes down to the dense core scale. The PRIMAger's beam FWHM between 11′′ and 28′′ in its 90-230μm spectral range will resolve linear scales down to ~0.35 pc up to ~8 kpc heliocentric distance (i.e., the distance to the Galactic Center). Initial experiments with post-processing of numerical MHD simulations of an entire galaxy (an improved version of Tress et al. 2024, including a barred Galactic potential) to produce synthetic maps of Stokes Q and U parameters show that PRIMAger will indeed be a game-changer (Molinari et al. 2025). Figure 1 shows a Line Integral Convolution (LIC) representation of the direction of magnetic fields from the post-processed numerical simulations as reconstructed at the Planck (left) and the PRIMAger (right) spatial resolution. The higher spatial resolution of PRIMAger gives access to an extended range of spatial frequencies that allow filtering the large scale emission. The map on the right in Figure 1 has been obtained by filtering out low spatial frequencies (corresponding to scales larger than 20 arcmin in the post-processed Stokes Q and U maps.

By deploying PRIMAger's capabilities on the Galactic Plane we will measure magnetization properties of entire molecular filaments/clumps structures, and correlate the magnetization with local star formation rates, efficiencies and cloud/clump fragmentation degree derived from far-IR continuum and spectroscopic large-scale surveys at comparable resolutions such as Hi-GAL (Molinari et al. 2016, Elia et al. 2021), GLIMPSE (Benjamin et al. 2005), MIPSGAL (Carey et al. 2009), SEDIGISM (Duarte-Cabral et al. 2021), FUGIN (Umemoto et al. 2017), and others.

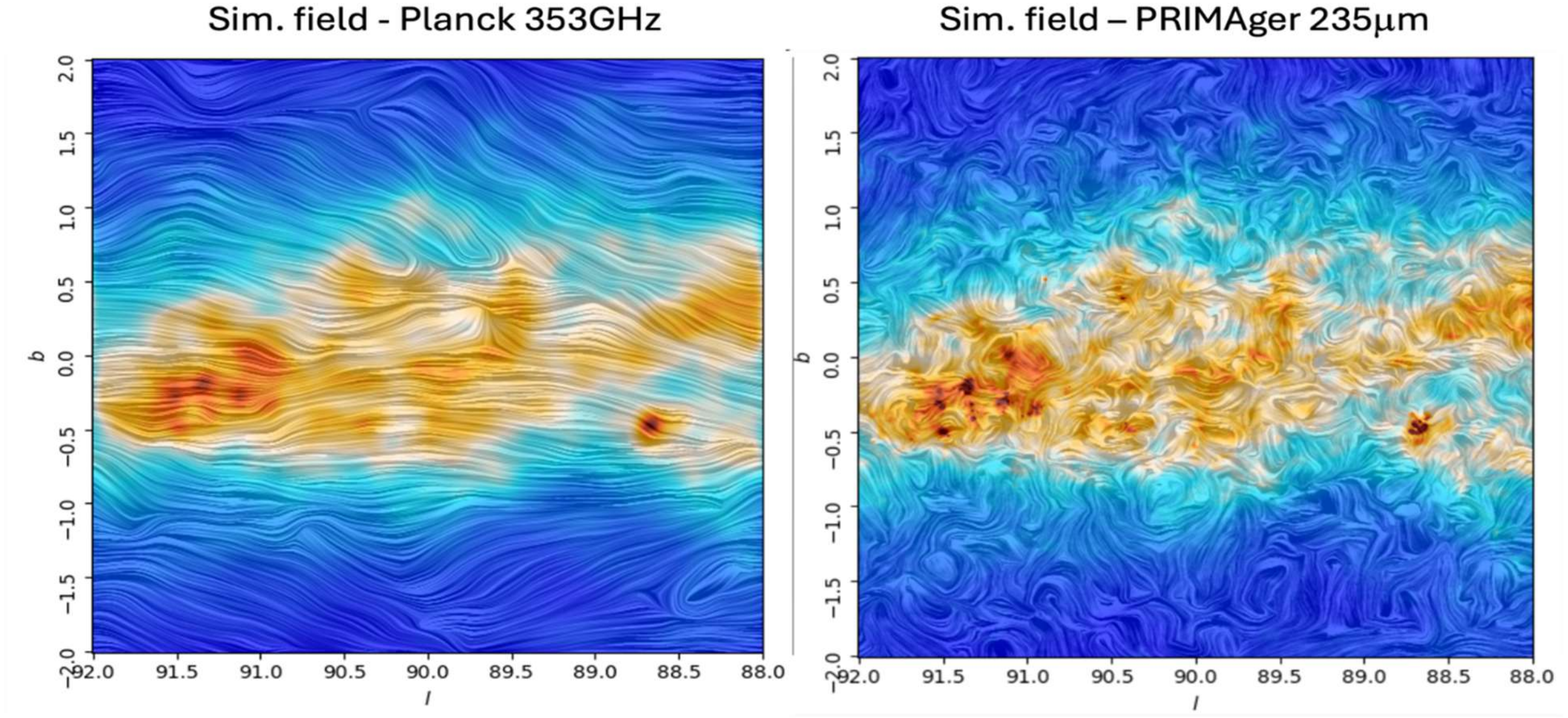

**Figure 1.** Direction of the magnetic field obtained from the post-processing of MHD numerical simulations (Tress et al. 2024) at the Planck spatial resolution (left panel), and as it would be seen with PRIMAger large-scale mapping at 235μm (right panel).

We will enable for the first time the development of a quantitative statistical picture of the role of the magnetic fields in modulating cloud and star formation efficiencies in our Galaxy thanks to the great diversity in density, star formation activity levels, turbulence properties, evolutionary stages and environmental conditions that the observed star-forming filamentary structures span in our Galaxy (Schisano et al. 2020). The variance offered by the thousands of filaments that will be mapped offers a continuum of situations, from structures with low column densities and virtually no fragmentation into dense clumps (or clumps in a relatively early stage with L/M≤1), where we will be able to map magnetic fields during the filament assembly process, to very dense, supercritical and fragmented structures where we will map the properties of magnetic fields against the rates and efficiencies of the ongoing star formation process.

Dust grain alignment efficiency and the inclination angle of the magnetic field can be constrained by exploring the dependence of the polarization fraction (p) on the local dispersion (S) in the field orientation angle, and the gas column density (N) using the method called *pNS* relations (Fissel et al. 2016, King et al. 2019). Plane-of-sky magnetic field strength component can then be estimated using the Polarization Dispersion Analysis technique (PDA, Houde et al. 2009), a refinement of the Davis-Chandrasekhar-Fermi method (Davis 1951, Chandrasekhar & Fermi 1953). It uses large numbers of independent polarization vectors to study dispersion of the magnetic field direction vs. scale. Combined with the velocity information obtained from existing and forthcoming molecular line observations at similar spatial resolutions, and field inclination angle estimates from the *pNS* relations, the PDA will provide estimates of the magnetic field strength, turbulence power spectrum, and energy balance between magnetic pressure, turbulence, and self-gravity. Finally, the PDA-derived magnetic field properties can be tested by the analysis of the coupling between the density and magnetic fields with the Histogram of Relative Orientations (HRO) technique (Soler et al. 2013) and directional analysis metrics like the projected Rayleigh statistic (Jow et al. 2018, Soler 2019), which by quantifying the relative orientation between cloud density structures and the magnetic field constrains the magnetic to turbulence energy density ratio. HRO magnetization estimates are independent of those made using the PDA technique, thus providing a complementary constraint on the magnetic energy density.

In addition to the polarimetric mapping, PRIMAger offers the possibility to acquire, in parallel, imaging data from the hyperspectral bands. These complementary mapping acquired essentially for free at R=10 between 24 and 80μm will allow to obtain complete census of compact sources in a yet uncharted spectral range, complementing ancillary mapping at similar spatial resolution at longer wavelength from Herschel and single dish mm surveys, and shorter by Spitzer (e.g. Elia et al. 2021). It will then be possible to compile the most complete atlas od SEDs from dense clumps in the Galactic Plane with the complete coverage from 1μm to 1mm, allowing to optimally constrain models of dense clumps (e.g. Molinari et al. 2019). As sensitivity and then observing time for the survey are driven by the polarimetric mapping (see below), the point source sensitivity that will be possible to reach in the hyperspectral bands will likely be between 5 and 20mJy in the 24-84μm range. This does not particularly improve with respect to the state of the art for the mid-IR in terms of sensitivity, but it will give us for the first time access to the 24-80μm regime.

## Instruments and Modes Used

This PRIMAGAL survey maps the entire 360° Galactic Plane in a |b|≤1° strip, for a total of 720 sq. deg., with PRIMAger. The data can be acquired by assembling 2°x2° tiles. Both hyperspectral and polarimeter bands will be used, where the polarimetric information in the polarimeter bands are critically important.

## Approximate Integration Time

The filamentary structures detected in the Herschel Hi-GAL maps (Schisano et al. 2020) have a typical brightness from few tens to thousands of MJy/sr at 250μm, that corresponds into a range of mean $N_H$ columns densities between a few $10^{20}$ and a few $10^{23}$ $cm^{-2}$. By adopting 50 MJy/sr as a working figure for detection of continuum surface brightness we make sure that we will be able to map structures characterized by $A_V$<1, and hence allow us to investigate the formation of high-density filaments from low-density molecular gas. Since we would like to detect a 50 MJy/sr minimum brightness at least 3% of the continuum as polarized signal to trace depolarization across the filaments (as observed, e.g., for nearby Gould Belt filaments) our target detection is 1.5 MJy/sr at a 5σ significance. With the PRIMAger instrument (https://prima.ipac.caltech.edu/page/instruments) a brightness of 0.35 MJy/sr in the 165μm band can be reached over a mapped area of 1 sq. deg. at a 5σ level in 10 hours in Large Mapping mode including all estimated overheads. Note that this estimate should also account for at least two orthogonal scan passes needed to calibrate and remove the 1/f noise of the detectors. Our required sensitivity of 1.5 MJy/sr at 5σ will then be reached in ~0.55 hours for a one sq. deg. map. Mapping the entire Galactic Plane in a |b|≤ 1° strip, a 720 sq. deg. total area, requires ~400 hours total observing time.

## Special Capabilities Needed

N/A

## Synergies with Other Facilities

Data from this survey will be analyzed in the context of extensive existing base of ancillary photometric and spectroscopic data from the near-IR to the radio (see above). In terms of future facilities, the most important synergy will be with the SKAO, and with its precursor MeerKAT. In particular there are Galactic Plane surveys in preparation with both facilities that would cover the southern Galactic Plane at sensitivities higher than existing radio continuum surveys between 1 and 15GHz at arcesec and subarcsec resolution. A most important goal of these future radio surveys will be to quantify the feedback (radiative and dynamical) from massive forming stars into the ISM, that together with gravity and turbulence (addressed in IR-submm continuum and spectroscopic surveys) is a most important agent that modulates the rate and efficiency of star formation. With the MeerKAT/SKAO quantifying feedback, and PRIMAGAL quantifying the role of magnetic fields, we will have the complete framework at hand.

## Description of Observations

We envisage a similar mapping strategy as the one adopted for the Herschel large scale mapping mode, given that the focal plane assembly of the polarimetric arrays is similar to the Herschel/SPIRE arrays. Continuous bustrophedic satellite scanning ensures that the same position in the sky is seen by several different pixels, therefore providing the optimal redundancy needed to properly calibrate and eliminate instrumental drifts. The entire Galactic Plane can be covered by assembling together 2°x2° square tiles that are covered with two mutually orthogonal scanning passes. This proved an ideal strategy with Herschel to beat the 1/f noise due to slow detector drifts.

Since the two hyperspectral channels arrays are located about 16 arcminutes away in the focal plane from the polarimetric arrays, a slightly oversized (10%) mapping tile would ensure that the same 2°x2° area is effectively covered for polarimetric mapping as well as for the 24-84μm continuum mapping.

# 85. Locating Water Snowline Positions in Protoplanetary Disks by Analyzing Keplerian Rotation Profiles of Water Lines

Shota Notsu (Department of Earth and Planetary Sciences, The University of Tokyo, Japan), Hideko Nomura (National Astronomical Observatory of Japan, Japan)

Understanding the water vapor and ice distribution, including the water snowline position, in planet-forming disks will provide information on the origin of water in planetary systems, including our Solar System. Water can be delivered to rocky planets, including Earth, through icy pebbles, icy planetesimals, comets, and asteroids, supporting the emergence of life on these planets. Water gas abundance is high ($\sim10^{-4}$) within the water snowline, and it is relatively high ($\sim10^{-8}$–$10^{-7}$) in the hot surface layer and the photo-desorbed region of the outer disk, compared to its value ($\sim10^{-12}$) in the regions outside the water snowline near the equatorial plane (e.g., Walsh et al. 2012).

As one of PI Sciences, PRIMA will characterize the main reservoirs of warm and cold water vapor and the far-infrared crystalline and amorphous water ice features for around 200 protoplanetary disks, giving unique access to the time evolution of the water throughout the entire disk. In PRIMA, the spectral resolution of high-resolution mode with Fourier transform module (FTM) is $R\sim\lambda/\Delta\lambda\sim$ 4400 * (112 μm/λ), and it is R>10,000 at λ<40 μm and R~20,000 at the shortest wavelength coverage (λ~24 μm) of PRIMA/FIRESS. Thus, PRIMA will expect to resolve line profiles and/or measure the line widths with Keplerian rotation in the shortest wavelength water lines, especially λ<40 μm. By obtaining water line profiles and constraining the positions of the water snowline for around 200 disks within PI Science survey, PRIMA will trace the time evolution of the water snowlines in the planet-forming disks which divide the regions between rocky and gas giant planet formation (see also Notsu et al. 2016; 2017; Kamp et al. 2021; Inoue et al. 2023).

We note that the cooperative observations (such as same target objects) with PRIMA and GREX-PLUS will be useful. GREX-PLUS will expect to obtain high spectral resolution spectra (R~30,000) towards 100 disks for λ=10–18 μm, where several candidate water lines such as the 17.75 μm line are included (Inoue et al. 2023).

## Science Justification

Understanding the water vapor and ice distribution, including the water snowline position, in planet-forming disks will provide information on the origin of water in planetary systems, including our Solar System. Water can be delivered to rocky planets, including Earth, through icy pebbles, icy planetesimals, comets, and asteroids, supporting the emergence of life on these planets. Water gas abundance is high ($\sim10^{-4}$) within the water snowline, and it is relatively high ($\sim10^{-8}$–$10^{-7}$) in the hot surface layer and the photo-desorbed region of the outer disk, compared to its value ($\sim10^{-12}$) in the regions outside the water snowline near the equatorial plane (e.g., Walsh et al. 2012).

Recently, spatially resolved water line emission within the water snowline have been observed for younger (Class 0-I) protostellar disks and envelopes with e.g., ALMA (e.g., Harsono et al. 2020, Tobin et al. 2023; Faccini et al. 2024). In addition, hot and warm water gas from innermost region of the disks have been detected with JWST, and these observations suggested that the warm water line emission are stronger in compact disks, suggesting that there is indeed a higher inward mass flux of icy pebbles in compact disks (e.g., Banzatti et al. 2023, 2025; Gasman et al. 2025).

As one of PI Sciences, PRIMA will characterize the main reservoirs of warm and cold water vapor and the far-infrared crystalline and amorphous water ice features for around 200 protoplanetary disks, giving unique access to the time evolution of the water throughout the entire disk. In PRIMA, the spectral resolution of high-resolution mode with Fourier transform module (FTM) is $R\sim\lambda/\Delta\lambda\sim$ 4400 * (112 μm/λ), and it is R>10,000 at λ<40 μm and R$\sim$20,000 at the shortest wavelength coverage (λ$\sim$24 μm) of PRIMA/FIRESS. Thus, PRIMA will expect to resolve line profiles and/or measure the line widths with Keplerian rotation in the shortest wavelength water lines, especially λ<40 μm. By obtaining water line profiles and constraining the positions of the water snowline for around 200 protoplanetary disks within PI Science survey, PRIMA will trace the time evolution of the water snowlines in the planet-forming disks which divide the regions between rocky and gas giant planet formation (see Figure 1 and Notsu et al. 2016; 2017; Kamp et al. 2021; Inoue et al. 2023).

Sophisticated physical and chemical disk models (e.g., Notsu et al. 2016, 2017) suggested that water lines with small Einstein A coefficients ($\sim10^{-6}$–$10^{-3}$ $s^{-1}$) and relatively high upper state energies ($\sim$1000 K) are dominated by emission from the hot midplane region inside the water snowline, and therefore through analyzing their profiles the position of the water snowline can be located. They also discussed the possibility to conduct line survey observations to locate the water snowline positions with future mid-infrared, far-infrared, and sub-millimeter observations (see Figure 1). The 25.25 μm line, which satisfies the condition to trace the water snowline, is the best candidate transition in the wavelength range of PRIMA/FIRESS, since the spectral resolution is highest (R$\sim$20,000). The longer wavelength water lines such as the 33.83 μm and 37.98 μm water lines are similarly suited, but the spectral resolution even with the FTS is lower (R $\sim$ 10,000). Thus, these lines will not be able to resolve the double peak structure of this emission lines sufficiently but will enable us to measure the line widths. This will give us additional information on the location of the snowline (see also Kamp et al. 2021 and Banzatti et al. 2025). Although not discussed enough in PI Science, water line profile analyses can be done within the same dataset as PI science.

These water lines cannot be accessed from ground-based telescopes due to telluric absorption. The sensitivity of the Spitzer Space Telescope was not high enough to detect these lines (Antonellini et al. 2015; 2016; Antonellini 2016). JWST will be able to detect the 25.25 μm lines, but its highest spectral resolution (R $\sim$ 3,000) is not high enough to resolve the line profiles.

However, JWST observations will be key in measuring line strengths (see e.g., Banzatti et al. 2023, 2025), a key input for the planning of future observations by PRIMA. Suitable water lines to trace snowlines with ALMA require long integration times to detect the line profiles in Class II disks (Notsu et al. 2018; 2019) and thus it will not be possible to study the time evolution of the location of water snowlines across large samples of disks.

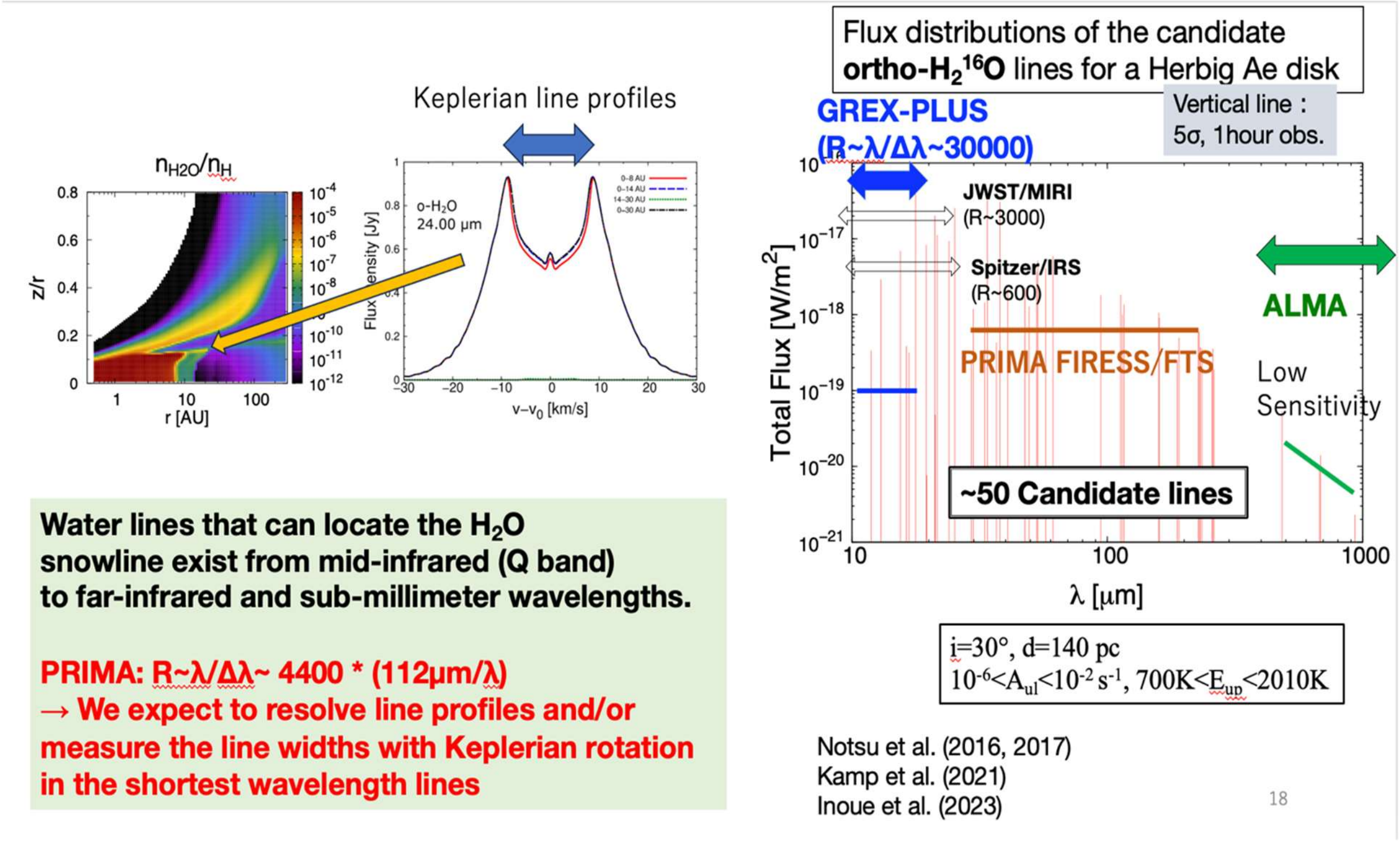

Fig. 1

We note that the cooperative observations (such as same target objects) with PRIMA and GREX-PLUS will be useful. GREX-PLUS (Galaxy Reionization EXplorer and PLanetary Universe Spectrometer) is a mission candidate for a JAXA's strategic L-class mission to be launched in the 2030s (Inoue et al. 2023). Its primary sciences are two-fold: galaxy formation and evolution and planetary system formation and evolution. GREX-PLUS will expect to obtain high spectral resolution spectra (R∼30,000) towards around 100 disks for $\lambda=10^{-18}$ μm, where several candidate water lines such as the 17.75 μm line are included (Inoue et al. 2023).

## Instruments and Modes Used

FIRESS high-resolution point source mode

## Approximate Integration time

The detection of the 25.25 μm, 33.83 μm, and 37.98 μm water lines for Herbig Ae disks with enough sensitivity to locate the snowline positions requires only short integration times (∼0.6, 0.4, and 0.3 hours, respectively) with PRIMA/FIRESS if the disks are located at the distance of the Taurus Molecular Clouds (∼140 pc). In this estimation, we adopted the modeled line fluxes from Notsu et al. (2017). T Tauri disks in nearby star-forming regions (<300 pc) will require typically

~1-10 hours exposure times to achieve the required sensitivity for a reliable snowline estimate. For a statistically significant sample of ~50 T Tauri disks and ~50 Herbig Ae disks, the project needs around 300 hours.

## Special Capabilities Needed

None

## Synergies with Other Facilities

ALMA

- Several Herbig Ae disks and younger T Tauri disks in nearest star forming regions (<150pc)
- Obs. Time per object: >10 hours per source (much longer than GREX-PLUS and PRIMA)
- Higher spectral resolution (R>100,000)

GREX-PLUS (2030s~)

- Higher sensitivity than ALMA and PRIMA for water line detection
- Several candidate ortho- and para-water lines (e.g., 17.75 µm, 17.64 µm (JWST detected),16.24 µm lines)
- Surveys of the water snowline positions for
  - bright T Tauri disks (>$10^{-20}$ W $m^{-2}$) in nearest star forming regions (<150pc) : ~190 min. per source
  - many Herbig Ae disks (>$10^{-18}$ W $m^{-2}$) in nearest star forming regions (<150pc): <30 min. per source
  - Herbig Ae disks in Orion and other massive star forming regions (~400pc, >$10^{-19}$ W $m^{-2}$): <1 hours per source

JWST/MIRI (R~3000)

- We will not be able to resolve the line profiles sufficiently.
- We can make target selections for PRIMA and GREX-PLUS using JWST spectra.

The cooperative observations (such as same target objects) with PRIMA and GREX-PLUS will be useful. GREX-PLUS will expect to obtain high spectral resolution spectra (R~30,000) towards around 100 disks for $\lambda=10^{-18}$ µm, where several candidate water lines such as 17.75 µm line are included. By using both GREX-PLUS and PRIMA, we would like to obtain more numbers of water lines and to constrain the inner water gas distribution and the water snowline positions more precisely.

## Description of Observations

FIRESS high-resolution point source spectra must be obtained on each target. Some targets from the PI disk survey may be used for the project.

# 86. Search for Missing Oxygen in the Interstellar Medium

Takashi Onaka (University of Tokyo), Itsuki Sakon (University of Tokyo), Takashi Shimonishi (Niigata University), Mitsuhiko Honda (Okayama University of Science)

The recent study of interstellar elemental depletion poses an important problem in the interstellar matter that at least a quarter of the total oxygen (~160 ppm relative to hydrogen) is not accounted for in any known form in the translucent or dense interstellar medium (ISM). The study of the 3 μm absorption feature of water ice suggests that one fifth of the missing oxygen may reside in 3 μm-size water ice grains. Recent near-infrared (NIR) spectroscopy indicates the ubiquitous presence of water ice in star-forming regions, further supporting a possibility that a significant fraction of oxygen may hide in even larger water ice grains in the dense ISM. However, the 3 μm feature becomes complex and weak for grains larger than 3 μm, and thus the NIR spectroscopy is not the best means to study large ice grains reliably. Here we show that sensitive observations of the far-infrared (FIR) features of water ice at 44 and 62 μm enable us to constrain the amount of crystalline water ice grains up to 5 μm or even larger accurately. Oxygen is one of the key elements of the ISM chemistry, and [O I] 63 μm is a dominant cooling line in the neutral ISM. The understanding of the actual form of the missing oxygen in the ISM is crucial for the study of the ISM and star-formation. To detect the FIR features of the crystalline water ice over the expected strong continuum, a sensitive FIR spectrograph represented by PRIMA/FIRESS is indispensable for this study. Since the feature is broad, the low resolution of R~130 is sufficient, but accurate relative calibration better than 1% is required.

## Science Justification

The study of elemental depletion provides unique, quantitative information on the relative abundance of the major elements in gas and solid phases and strongly constrains the model of dust grains in the ISM (e.g., Draine 2003). The recent improvement of the analysis of the abundances of the Sun and B-stars removes the uncertainties in the reference abundance (Asplund et al. 2009; Nieva & Przybilla 2012). Together with the improved reference abundance, a detailed study of elemental depletion observations, however, poses a new, significant problem of the missing oxygen in the ISM. Jenkins (2009) and Whittet (2010) show that at least a quarter of the total oxygen (~160 ppm relative to hydrogen) is not accounted for in any known form in the translucent or dense ISM with the density larger than 7 $cm^{-3}$, while recent X-ray spectroscopy suggests that there may not be missing oxygen in the less dense, diffuse ISM (Psaradaki et al. 2023). Jenkins (2009) proposes that the missing oxygen may be hidden in large water ice grains because water ice is only oxygen-bearing species that can be in the solid phase without conflicting the abundance constraint and because the strong absorption band at 3 μm of water ice becomes weak and complex as the grain size increases and thus can be easily elusive in the 3 μm spectroscopy. Poteet et al. (2015) investigated this possibility in detail towards the well-studied line-of-sight to ζOph using the 3 μm spectrum, concluding that even including contributions from 2.8 μm-size water ice they detected and other species, there remains two thirds of the missing oxygen (~100 ppm), which is not accounted for.

However, the 3 μm feature becomes weak and complex for grains larger than 3 μm, and it is difficult to study the presence of micron-sized grains in the NIR. High-energy resolution X-ray observations will offer unique, independent data on the elemental abundance both in gas and solid phases in the ISM (e.g., Psaradaki et al. 2023). However, large grains (> 1 μm) also become optically thick at the oxygen K-edge and the X-ray spectroscopy is not sensitive to large ice grains either. FIR features do not have this problem unless the grains are too large, e.g., larger than 10 μm. The FIR spectroscopy provides a unique means to study the presence of large ice grains and their amount in the dense ISM unambiguously.

Figure 1 shows the results of simple simulations. We assume a mixture of water ice and carbon dust and calculate the average refractive indexes using the effective medium theory of Maxwell Garnett (Bohren & Huffman 1983). We adopt the refractive indexes by Bertie et al. (1969) for crystalline ice and those for amorphous ice by Hudgins et al. (1993) at 10 K for 2.5–200 μm. We assume the refractive indexes of BE carbon for the carbon inclusion (Zubko et al. 1996). The results with the silicate inclusion do not differ significantly. We fix the volume fraction of carbon inclusion as 0.1, following the value used in Poteet et al. (2015) of 0.15. Figure 1a clearly shows that the 3 μm extinction feature of water ice per mass shows complicated features and becomes very weak for grains larger than 2 μm (e.g., Dartois et al. 2024). On the contrary, Figure 1b indicates that the absorption cross-section per mass at 44 μm of crystalline water ice does not significantly change even for the grain of 5 μm and the feature is visible even for that of 10 μm, demonstrating the effectiveness the FIR feature in detecting large water ice grains compared to the NIR feature. Figure 1c shows the predicted emission from the ISM with and without ice grains (red and black) for the crystalline water ice of 5 μm under the radiation field intensity relative to the solar value, $U$ = 600, using the DustEM code (Compiègne et al. 2011). We assume that the missing oxygen of 160 ppm all resides in the water ice grains. Note that the vertical scale is expanded to emphasize the difference in the spectral shape. The 44 μm feature is only ~1% over the linear baseline. For the amorphous ice, the feature becomes weaker even with $U$ = 1000 (Figure 1d). The detectability of amorphous ice increases if we observe a target with much higher $U$.

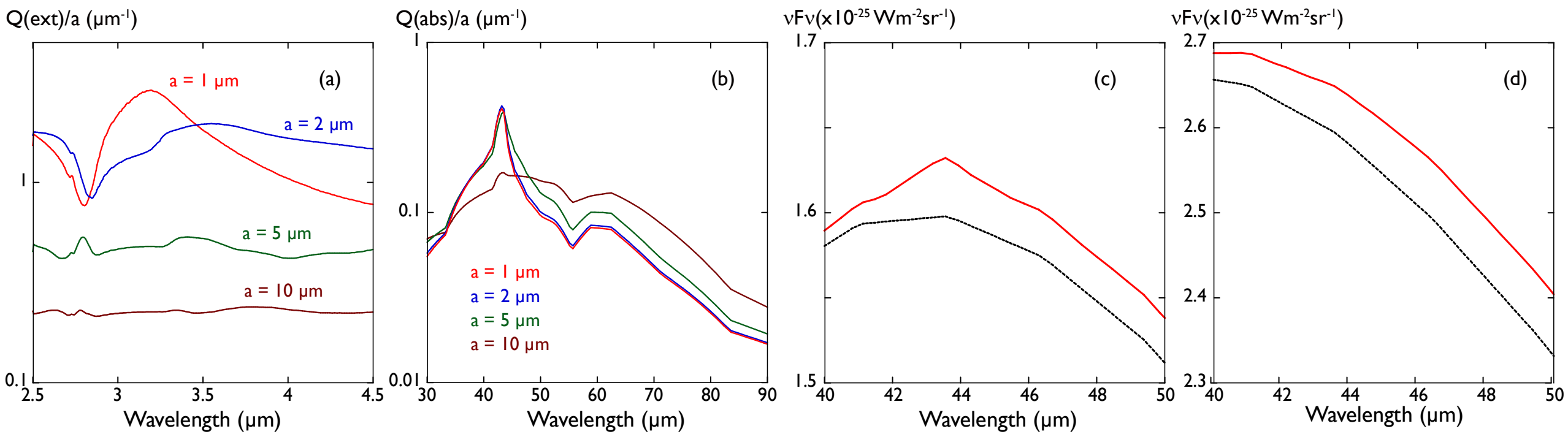

**Figure 1.** (a) Extinction cross-section per mass at 3 μm and (b) absorption cross-section per mass at 44 μm for a mixture of crystalline water ice and carbon dust for 1, 2, 5, and 10 μm size. (c) Expected FIR emission with (red) and without (black) crystalline water ice grains of 5 μm with *U* = 600 per hydrogen. (d) Expected FIR emission with and without amorphous water ice grains of 2 μm with *U* = 1000 per hydrogen.

NIR spectroscopy of star-forming regions with AKARI reveals the relatively ubiquitous presence of the broad, shallow 3 μm absorption feature of water ice (Mori et al. 2014), supporting the possibility that the missing oxygen may in fact be sequestered in large water ice grains. The 3 μm spectroscopy also suggests the presence of crystalline water ice (Boogert et al. 2015) and large icy grains (Terada & Tokunaga 2012) in various objects. Formation mechanisms of crystalline water ice in the ISM have been discussed (Jenniskens & Blake 1996). A recent laboratory experiment of liquid-like behavior of amorphous ice at 65-150K under UV radiation suggests that crystallization of ice could proceed in a relatively short time scale in the ISM (Tachibana et al. 2017). A significant fraction of water ice could be crystallized in the ISM. If oxygen hides in large icy grains, FIR spectroscopy is the only means to put a stringent upper limit on the total mass oxygen in them unambiguously. It should also be added that [O I] 63 μm emission is a dominant coolant in the dense ISM and oxygen is one of the important elements in the ISM chemistry (Hollenbach & Tielens 1997). It is crucial to correctly understand the relative abundance in the gas and solid phase of oxygen. FIR spectroscopy offers a unique opportunity to study oxygen in both phases. Since the water ice feature is faint and the underlying continuum emission is strong, we need a very sensitive, well-calibrated FIR spectrometer for this study.

## Instruments and Modes Used

The proposed observations will be carried out by using the FIRESS spectrometer in the low-resolution mode. Each target will be mapped with a typical area of ~15 × 15 arcmin$^2$.

## Approximate Integration Time

In this study, ~10 targets will be selected and FIRESS low-resolution spectral map for ~15 × 15 arcmin$^2$ will be obtained for each target. To obtain the spectral map data with the signal-to-noise ratio of 100 for emission with 2 × $10^{-16}$ W m$^{-2}$ at around 40 μm (see the “Description of Observations” section about the justification of this requirement), the integration time of 1.6 hours for each target and band is needed. To cover the wavelength range of 24 – 74 μm, the required integration time is 3.2 hours per target, i.e., 32 hours for 10 targets.

## Special Capabilities Needed

High accuracy in the spectral response of FIRESS (better than 1%) is required.

## Synergies with Other Facilities

High-energy resolution spectroscopy in the X-ray, e.g., XRISM and Athena, will provide complementary information on the depletion of elements. Radio observations of CO will also be important to estimate the total oxygen in the ISM.

## Description of Observations

The crystalline water ice has three lattice vibration modes in the FIR at 44, 52, and 62 µm. The 44 µm band is the strongest. However, because of the lack of sensitive spectrometer in this spectral range, only a few detections have been reported so far; one in emission in an OH/IR star (Omont et al. 1990) and a few in absorption in embedded infrared sources (Dartois et al. 1998). The 62 µm band has been observed in several more sources in emission in planet-forming regions and in a young star (McClure et al. 2012, 2015; Min et al. 2016). It has also been detected in a shock region (Molinari et al. 1999). For the present study, we will observe dense, but warm ISM regions. At 44 µm, the dust emissivity is still proportional to the mass of the crystalline ice as far as the size is smaller than 5 µm and the fraction of the inclusions (silicate or carbon) is small. The size of ice grains larger than 5 µm can be estimated by the band profile (Figure 1b). Once the temperature of ice grains is well determined, the mass of the crystalline ice can be constrained accurately, and we can put a strict upper limit of oxygen hiding in crystalline grains up to ~5 µm size. The temperature of the dust grains that emit the continuum can be estimated from its shape. Although it may be similar, the ice grains could have a different temperature from the grains that emit continuum. The grain temperature depends on the size and the fraction and composition of the inclusions. We need to estimate the temperature of ice grains independently. Detection of both 44 and 62 µm features is thus crucial.

Figures 1c and d show the FIR emission of typical interstellar clouds per hydrogen atom. With $U = 600$, the temperature of the ice grain becomes $T = 30.8$ K. Since the features in question is at 44 and 62 µm, higher T and thus higher U environments will significantly enhance the detectability of the feature (e.g., Kamp et al. 2018, 2021). Figures 1c and d present rather conservative cases in this regard. We will select about 10 dedicated targets for this study carefully based on the past FIR surveys of Spitzer, AKARI, and Herschel observations, for which the feature at 62 µm can also be detected and a signal-to-noise ratio larger than 100 is expected within the integration time of several hours. Since the targets will be diffuse regions, a small map (< 15 arcmin) will be useful to remove any artifacts and to study spatial variations of the spectrum. Here, we assume that the HI column density of the target is $1 \times 10^{20}$ cm$^{-2}$. With the pixel size of 7.6", the estimated flux at around 40 µm becomes about $2 \times 10^{-16}$ W m$^{-2}$. For a small map of 15' × 15', the signal-to-noise ratio of 100 can be achieved in 1.6 hr for one band. Since we need to use band 1 and band 2 to cover the range 24–74 µm, in total we estimate the integration time of 3.2 hr per target.

As indicated in Figures 1c and d, the expected feature is very weak relative to the underlying continuum even for the crystalline feature, which is the primary target of the present study. For amorphous water ice, the 44 μm feature becomes weak and broad, and the 62 μm feature is almost invisible. For the detection of amorphous water ice feature at 44 μm, therefore, we need accurate knowledge of the shape of the underlying continuum emission. For this purpose, FIR spectra taken in other programs for various environments, which include more diffuse regions, can be very beneficial for the accurate estimate of the shape of the underlying continuum emission. They will also be used to increase the reliability of the present study and the sample size. FIR spectra of planet-forming regions, where the ice features have been detected, will also provide the useful actual crystalline ice features in the ISM.

# 87. Gaining New Insights Into the Icy, Dusty Regions of the Galaxy with Combined SPHEREx and PRIMA Observations

Roberta Paladini (Caltech-IPAC), Matthew Ashby, Gary Melnick, Joseph Hora, Volker Tolls, Jaeyeong Kim (Center for Astrophysics, Harvard & Smithsonian), Michael Werner (JPL), Carey Lisse (John Hopkins University)

The cold, dense inner regions of molecular clouds are the sites of star and planet formation. Measuring the content of ices and dust in these environments is essential for constraining the initial conditions of this important process.

The SPHEREx space mission—which successfully launched on March 11th 2025—is observing the full sky every six months. In the context of the Ices Investigation, one of the three main goals of the mission, SPHEREx is generating an unprecedented statistically-significant spectroscopic NIR (0.75 – 5 micron) database for nearly ten million carefully-selected Galactic targets located towards the line-of-sight of star forming regions at different stages of evolution. The SPHEREx spectral range covers key absorption spectral features—such as the 3.0 micron $H_2O$, the 4.27 micron $CO_2$ and the 4.67 micron CO features—that directly probe the amount and composition of (predominantly) amorphous ice mantles in these regions. In addition, the shape of these ice absorption features, as well as the observed NIR continuum, contain important information about the size and composition of dust grains. PRIMA is uniquely positioned to provide essential complementary information to SPHEREx to characterize the Milky Way's icy, dusty regions.

The proposed PRIMA observations will target a sub-sample of the nearly ten million SPHEREx Ices sources.

Low-res (R ~ 100) FIRESS observations in Band-1 (24 – 43 micron) and Band-2 (from 42 to 76 micron) will probe the 44 micron and 62 micron (mostly) crystalline $H_2O$ ice emission features and, in combination with the SPHEREx NIR absorption spectroscopy data—that trace (mostly amorphous) H2O, $CO_2$ and CO ice - they will allow us to set very tight constraints on the physical properties (amount, composition and state) of ice mantles. In addition, the FIRESS measurements will provide a robust estimation of the MIR—FIR continuum. This continuum, combined with the SPHEREx data at shorter wavelengths, will allow us to reconstruct the slope of the NIR – FIR dust extinction curve, for the densest and more extincted lines of sight, or the shape of the dust thermal emission spectrum, for the less dense and extincted lines of sight. In both cases, it will encode information on the dust grains optical properties, such as grain size and composition;

PRIMAger observations, both in intensity and polarization, in the PP1 (92 micron), PP2 (126 micron), PP3 (172 micron), PP4 (235 micron) bands. The PRIMAger FIR broadband photometry, combined with FIRESS spectroscopy, will allow the definitive determination of the continuum, all the while providing, thanks to its polarimetric capabilities, unprecedented independent information on the size of the grains. Last but not least, the PRIMAger polarization measurements will shed light on the behavior of magnetic fields which, together with gravity and turbulence, regulate the star/planet formation process.

## Science Justification

### Broader Context

In recent years the exoplanet community has been dedicating an increasing amount of effort towards finding planets in the so-called *habitable* zone, i.e., the distance from the host star at which water can exist in liquid form on the surface of a planet. A testimony of this is represented by the current and future observations that are and will take place using sophisticated coronographs on-board JWST, Roman and HWO. A parallel undertaking aims at the characterization of planet atmospheres and at constraining favorable conditions for the formation of molecules essential for life, such as amino acids. Understanding how this process unfolds implies understanding how the main chemical building blocks—such as the $H_2O$, $CO_2$, CO molecules, which are precursors to complex organic molecules (COMs)—are synthetized.

One important discovery made in recent years (e.g., Boogert et al. 2011; 2013) is that these key simple molecules are mostly locked into ice, i.e., in amount far exceeding that in the gas form. Indeed, the interior of molecular clouds, where the bulk of star and planet formation occurs and external photo-desorbing (and photo-dissociating) FUV photons are well attenuated, is characterized by low temperatures and high densities (e.g., $T_d$ < 15 K, $A_V$ ~ 2 - 3 mag). These physical conditions have been shown (e.g., Whittet et al. 2003) to be ideal for the formation of amorphous ice mantles on dust grains. These ice mantles are dominated by $H_2O$ in the early stages, followed by other species, such as CO, $CO_2$, $CH_3OH$.

The path of ice formation is intimately related to the evolution of dust grains. The addition of ice coatings on the grains increases cohesion and their sticking efficiency, which promotes growth through grain coagulation (Ormel et al. 2009; Juvela & Ysard 2011; Jones et al. 2016). Some of these particles will eventually survive the star formation process and coagulate into larger bodies, including comets, planetesimals, and ultimately planets.

Achieving a full understanding on the coupling of ices and dust grains and on how they evolve, independently or in a correlated fashion, is therefore central to inform theories of star and planet formation.

## Science Questions Addressed by PRIMA Observations

The specific questions that the combined SPHEREx-PRIMA measurements will address are:

- Does the total ice reservoir (column density) increase with density (i.e., $A_V$)?
- If there is a trend between the ice reservoir (column density) with $A_V$, is the trend uniform across ice species and ice state (amorphous and crystalline)?
- Do dust grains undergo continuous growth throughout the star formation process, i.e. does grain size linearly increase with evolutionary stage of a molecular cloud cores (from pre-stellar to proto-stellar)?

- Is there a correlation between the increase of ice reservoir (column density) and the increase of dust grain sizes?
- Is the detection of amorphous vs. crystalline ice (i.e., absorption vs. emission ice features) correlated with the environmental properties (e.g., density, radiation field hardness and intensity, etc.)
- Is the evolution of ice mantles and dust grains correlated with the behavior (orientation and intensity) of magnetic fields?

## The Need for PRIMA

PRIMA is uniquely positioned to vastly complement the SPHEREx 0.75-5.0 micron spectra and answer the questions posed above. The PRIMA FIRESS instrument extends the SPHEREx spectroscopic coverage to MIR and FIR wavelengths, allowing us to detect additional ice features as well as to trace the slope of the dust extinction curve (for denser and more extincted lines of sight) or the shape of the dust thermal emission spectrum (for less dense and extincted lines of sight). The PRIMAger instrument further extends the wavelength coverage of FIRESS to the longest FIR photometric bands, enabling the definitive determination of the continuum, all the while providing, thanks to its polarimetric capabilities, unprecedented independent information on the size of the grains. The PRIMAger polarization measurements will also shed light on the behavior of magnetic fields which, together with gravity and turbulence, regulate the star/planet formation process.

## Observables

The observables for the PRIMA observations are:

### Ice Emission Features

- 44 and 62 micron crystalline H2O ice emission features

### The Shape of the Continuum

- The densest, most extincted lines of sight, once the stellar photosphere is subtracted out, what is left is the extinction curve for the intervening dust grains; for the less dense and extincted lines of sight, this is a measure of dust thermal emission

### Polarization

- Intensity and geometry of the observed polarization signal

## Link to Testable Hypothesis

The combined SPHEREx and PRIMA observations will provide, in the star and planet formation context, a complete picture on the amount and distribution of both amorphous (SPHEREx) and crystalline (PRIMA) $H_2O$ ice mantles, and on dust grain size and composition. This information will be obtained as following:

**Ice Emission Features**

The detection of the 44 and 62 micron emission features is considered a diagnostics for crystalline $H_2O$ ice, in contrast to amorphous ice, commonly detected through absorption features. The appearance of these features suggests ice temperatures approximately above 110 K, as this is when amorphous ice transitions to crystalline ice (Smith et al. 1994). In addition, Onaka et al. (2025) have shown that FIR spectroscopy is much more effective than NIR spectroscopy in detecting very large grains (i.e, >> than the average size of ISM-grains, ~ 0.1-0.2 micron), as the FIR 44- and 62-micron feature profiles are proportional to grains as large as 5 micron.

### The Dust Extinction Curve

The predicted effect of larger-than-ISM dust grains on both the NIR and MIR extinction curve is to cause a *flattening* of the slope beyond 1-2 micron. Indeed, a confirmation of this behavior has started emerging from the limited number of dense lines of sight observed, mostly using broadband photometry (e.g., Indebetouw et al. 2005; Chapman et al. 2009; McClure et al. 2009; Nishiyama et al. 2009; Lim & Tan 2014; Nogueras-Lara et al. 2019).

## Dust Thermal Emission

the shape of the dust thermal emission spectrum allows the accurate determination of the dust spectral emissivity index ($\beta$), which dictates the behavior of the Rayleigh-Jeans slope of the observed SED and is sensitive to dust composition, porosity and size, with several models predicting , in particular, a significant increase in the values of $\beta$ for larger-than-ISM grains (Ossenkopf & Henning 1994, Koehler et al. 2012, 2015).

## Polarization

The shape of the polarized dust SED and, in particular, the measured polarization fraction have a strong dependence on the grain physics, such as size and composition (Smith et al. 2000, Fanciullo et al. 2017, Guillet et al. 2018). Therefore, polarization measurements provide a robust estimation of grain properties, which can be cross-examined with those derived independently from either the spectral emissivity or the slope of the extinction curve, as well as from the analysis of the spectroscopic features.

## Instruments and Modes Used

FIRESS low res mode, 30 pointings per cloud PRIMAGer minimum size maps in both bands

## Approximate total Integration Time

85 hrs

## FIRESS

The sensitivity estimate is based on the observations by McClure et al. (2015) of VW Cha, i.e. the faintest target in this study's sample, and on the PRIMA Survey Calculator. We make the conservative assumption that the intensity of the fainter (compared to the 44-micron line) 62-micron line is of the order 5e-14 W m-2. Therefore, in order to detect the line with a S/N ~100, the required integration time per source is 5 mins. Note that the continuum level may exceed the saturation limits. Given that we aim at observing a total of 300 sources (30 sources per cloud), the **total estimated observing time is 25 hrs.**

PRIMAger: the sensitivity estimate is based on the observations by McClure et al. (2015) of the protoplanetary disks VW Cha, GQ Lup, Do Tau and Haro 6-13, combined with the PRIMA Survey Calculator. We assume that the average continuum at 92 microns is of the order of 3e-14 W m-2. Adopting a polarization fraction of 1% and a S/N ~10, the estimated time to map a region 15' X 15' centered on a source is 0.2 hr, totaling an integration time of just 60 hrs for a subset of 300 sources.

## Special Capabilities Needed

N/A

## Synergies with Other Facilities

The SPHEREx space mission—which successfully launched on March 11$^{th}$ 2025—is observing the full sky every six months. In the context of the Ices Investigation, one of the three main goals of the mission, SPHEREx is generating an unprecedented statistically-significant spectroscopic NIR (0.75 – 5 micron) database for nearly ten million carefully-selected Galactic targets located towards the line-of-sight of star forming regions at different stages of evolution. The SPHEREx spectral range covers key (absorption) spectral features—such as the 3.0 micron $H_2O$, the 4.27 micron $CO_2$ and the 4.67 micron CO features— that directly probe the amount and composition of ice mantles in these regions. At the same time, the shape of these ice absorption features, as well as the observed NIR continuum, contain important information about the size of dust grains. The full SPHEREx spectral database for the Ices sources is expected to be available at the completion of the fourth all-sky survey, roughly by ~2027.

## Description of Observations

### Target Selection

Within the context of the SPHEREx Ices Investigation, one of the three main science objectives of the SPHEREx mission, a list of targets likely to exhibit ice-absorption features has been developed prior to launch, to provide spectra for a wide variety of objects and in sufficient numbers to trace the history of organic molecules from their time of formation within molecular clouds to their incorporation in planetary systems. This list takes the name of SPHEREx target List of ICE Sources (SPLICES, Ashby et al. 2023). It contains 8.6 X 10 6 objects brighter than W2 ∼ 12 Vega mag over much of the sky, principally within a broad strip along the Galactic plane, but also within high-latitude molecular clouds and even the Magellanic Clouds.

The proposed PRIMA observations will target a sub-sample of SPLICES sources located in nearby (~< 1 kpc) molecular clouds. The decision to restrict the selection to background stars behind nearby clouds assures that the line-of-sight does not intercept multiple astrophysical components. In total, 10 nearby molecular clouds are considered for the PRIMA observations, comprising low- to high-mass star forming regions. These are: Chamaeleon, Lupus, Mon R2, Ophiucus, Orion A, Orion B, Perseus, Rosette, Serpens/AqR, Taurus. For each cloud, we will draw from the SPLICES catalog 30 high-S/N sources distributed in 5 density bins ($A_V$ ~ 5 to 25 mag), for a total of 300 sources.

### Observations

Low-res (R ~ 100) **FIRESS observations** in Band-1 (24 - 43 micron) and Band-2 (from 42 to 76 micron). These observations will probe the 44 micron and 62 micron (mostly) crystalline $H_2O$ ice emission features and, in combination with the SPHEREx NIR absorption spectroscopy data—that trace (mostly amorphous) H2O, $CO_2$ and CO ice – they will allow us to set very tight constraints on the physical properties (amount, composition and state) of ice mantles. In addition, the FIRESS measurements will provide a robust estimation of the MIR—FIR continuum. This continuum, combined with the SPHEREx data at shorter wavelengths, will allow us to reconstruct the slope of the NIR – FIR dust extinction curve, for the densest and more extincted lines of sight, or the shape of the dust thermal emission spectrum, for the less dense and extincted lines of sight. In both cases, it will encode information on the dust grains optical properties, such as grain size and composition.

**PRIMAger observations**, both in intensity and polarization, in the PP1 (92 micron), PP2 (126 micron) , PP3 (172 micron), PP4 (235 micron) bands. The PRIMAger FIR broadband photometry, combined with FIRESS spectroscopy, will allow the definitive determination of the continuum, all the while providing, thanks to its polarimetric capabilities, unprecedented independent information on the size of the grains. Last but not least, the PRIMAger polarization measurements will shed light on the behavior of magnetic fields which, together with gravity and turbulence, regulate the star/planet formation process.

# 88. Spectro-Polarimetry of the Dust Population in the Central Molecular Zone: Magnetic Field Structure and Strengths

Dr. Dylan Michelson Paré (Villanova University), Dr. David T. Chuss (Villanova University), Dr. Natalie Butterfield (National Radio Astronomy Observatory), Dr. Mark Morris (University of California – Los Angeles), Kaitlyn Karpovich (Stanford University), Dr. Janik Karoly (University College London), Dr. Simon Coudé (Worcester State University / Center for Astrophysics), Dr. Thiem Hoang (Korea Astronomy and Space Science Institute)

The Central Molecular Zone (CMZ) of the Galactic Center (GC) Milky Way contains a substantial fraction of the molecular mass of the Galaxy ($10^7$ $M_{sun}$) yet exhibits a far lower star formation efficiency (SFE) than expected given the high densities found in this region (a star formation rate of <0.1 $M_{\odot}$/yr). There are multiple possible explanations for the depressed SFE in the CMZ, like feedback, strong turbulence, and magnetic fields. It is currently unclear which of these mechanisms is the dominant inhibitor of star formation in the CMZ. It is important to understand the star formation process in the extreme environment of the CMZ because it exhibits properties similar to those observed at redshifts of z∼2, the time of peak cosmic star formation in the universe. It is possible to discriminate between the different SFE inhibiting mechanisms through a combination of high-spatial resolution magnetic field mapping and estimates of the magnetic field strength. Furthermore, studying the properties of the magnetic field over a range of wavelengths will also enable investigations of how the magnetic field varies along the line of sight when traced by different dust populations. The PRobe far-Infrared Mission for Astrophysics (PRIMA) will be uniquely capable of providing all these capabilities. Polarimetric observations conducted from 92–235 µm with the PRIMAger instrument will probe the cool and warm dust populations in the GC, allowing us to assess how the magnetic field varies with different local conditions within the CMZ. Furthermore, the PRIMA FIRESS spectrometer will enable studies of spectral lines that can be used to characterize the turbulence in the CMZ. Studies of the turbulence will enable estimates of the CMZ magnetic field strength through the application of methods like the Davis-Chandrasekhar-Fermi (DCF) method. The combination of these data products will make it possible to determine the importance of the magnetic field in the CMZ to determine whether the magnetic field is the primary mechanism inhibiting star formation in the region.

## Science Justification

The molecular cloud population in the central 200 pc region of the Milky Way, known as the Central Molecular Zone (CMZ), contains a substantial percentage of the molecular mass of the Milky Way (4% or $10^7$ M$_\odot$, Barnes et al. 2017). The CMZ possesses greater molecular temperatures and densities (e.g. Mills et al. 2018), elevated turbulence (Henshaw et al. 2019), and stronger magnetic field strengths (Pillai et al. 2015) than what is observed in the Galactic disk. The conditions of the CMZ, coupled with its relative proximity to Earth (∼8 kpc away, Gravity Collaboration 2019) makes it an important local analog to the conditions of the universe that were present during the time of peak cosmic star formation (z∼2, Kruijssen et al. 2013). The CMZ can therefore be observed to obtain insight into the physics of the cosmic history of star formation to greater detail than is currently possible using extragalactic observations.

The fraction of the Galactic star formation rate (SFR) in the CMZ is roughly the same as the fraction of molecular mass of the Galaxy present in the region (3–6%, e.g. Rickert et al. 2019). However, this rate is an order of magnitude lower than would be expected given the high densities (e.g. $10^7$ cm$^{-3}$, Mills et al. 2018) found in the CMZ. There are multiple possible explanations for the low SFR in the CMZ, which include the strong magnetic field of the CMZ, the strength and compressibility of turbulence, and that the CMZ could be in a period of inactivity between episodic bursts of star formation (Krumholz & Kruijssen 2015). **Determining which of these factors is inhibiting star formation in the GC will expand our understanding of the properties of star formation in extreme environments like the CMZ and the universe during the time of peak cosmic star formation.**

The PRobe Far-Infrared Mission for Astrophysics (PRIMA) will be uniquely suited to helping uncover the nature of what is regulating star formation in the CMZ. The PRIMA PRIMAger Imager will be capable of studying the dust polarization from ∼90–250 µm with a resolution ranging from ∼10–28″. This wavelength range enables a study of how the magnetic field geometry in the CMZ varies in the cool (T≅22 K) and the warm (T≅41 K) dust. These dust populations are known to be distinct in the CMZ, as can be seen in the background grayscale images in Figure 1. Although the magnetic field of the cool dust has been mapped throughout the CMZ using SOFIA/HAWC+ (Butterfield et al. 2024; Paré et al. 2024; Yang et al. 2025), there is only very sparse coverage of the magnetic field in the warm dust. The PRIMAger range of wavelengths will therefore greatly enhance our understanding of how the magnetic field behaves in the different dust populations of the CMZ.

Furthermore, the PRIMA FIRESS spectrometer is ideally situated for studying spectral lines that probe the turbulent velocities within the CMZ clouds. The determination of these turbulent velocities is an important input for estimations of the magnetic field strength using methods like the Davis-Chandrasekhar-Fermi (DCF) method (Davis 1951; Chandrasekhar & Fermi 1953). This method has been successfully used to derive magnetic field strengths in individual CMZ molecular clouds before (e.g. Guerra et al. 2023). The FIRESS frequency range covers the H2 line at 28 µm, at band 1. At this wavelength, using FIRESS in high-res mode will yield a velocity resolution of ∼15 km/s, sufficient to resolve the turbulence in the clouds in the extreme GC region. This study of the turbulence can then be used to estimate magnetic field strengths using DCF. Furthermore, the results obtained from FIRESS can be compared to previous spectral line observations of the GC, such as the ALMA CMZ Exploration Survey (ACES) observations of the region.

The combination of magnetic field morphology mapping and strength estimation is of critical importance for our understanding of the role of the field in regulating star formation. Paré et al.

(2025) determined that shear plays an important role in the CMZ, though the impact of shear varies on a cloud-by-cloud basis. This analysis was made by inspecting the relative angle between the magnetic field and the column density using statistical methods like the Histogram of Relative Orientation (HRO, Soler et al. 2013) and the Projected Rayleigh Statistic (PRS, Jow et al. 2018). The 4-band nature of the PRIMAger observations will make it possible to conduct this alignment analysis over a range of frequencies to determine whether the alignment changes with dust temperature. The studies of relative alignment can then be compared to the magnetic field strength estimates derived from the FIRESS observations to determine whether there is a connection between field strengths and shear.

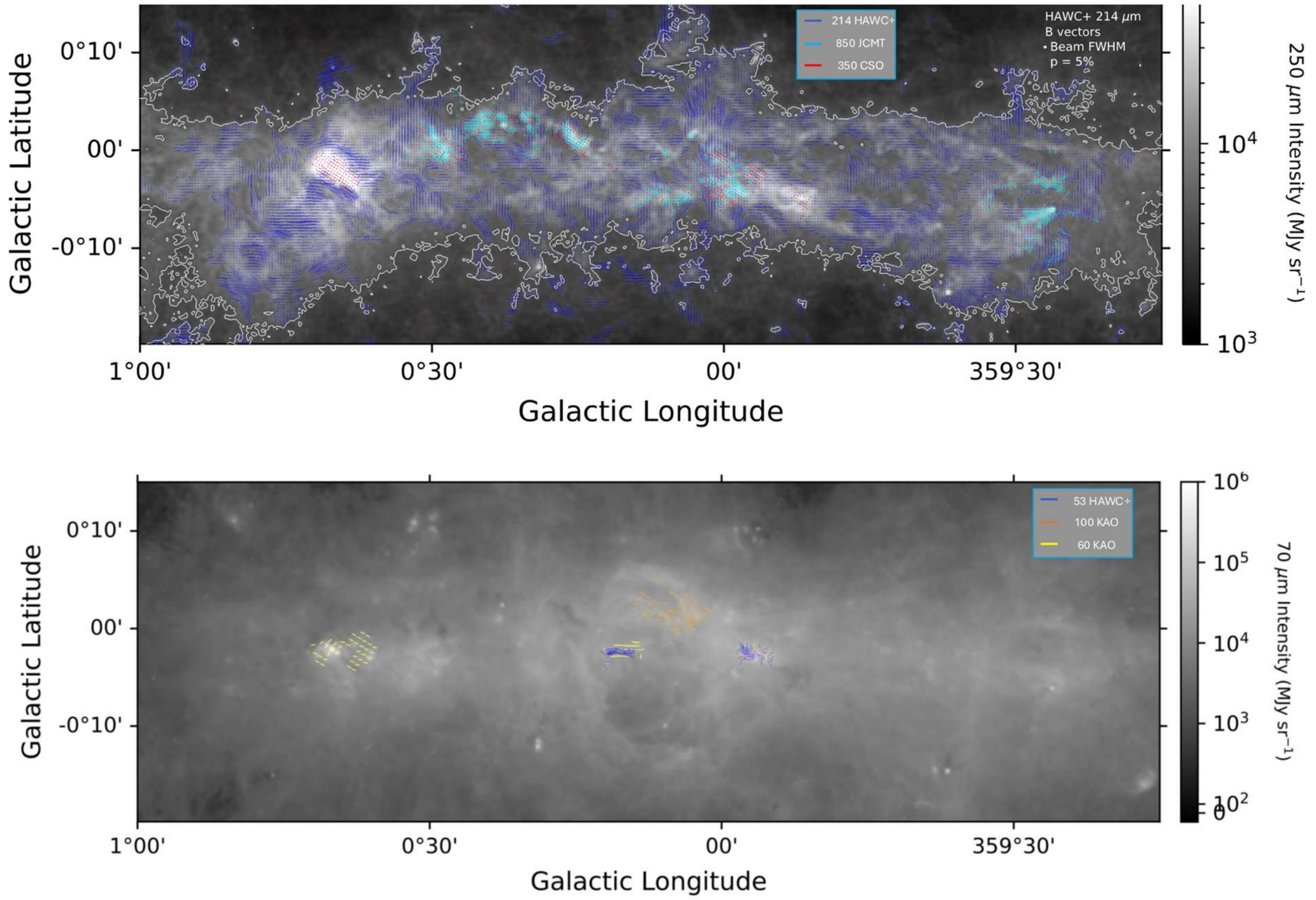

**Figure 1.** The complexity of the magnetic field in the CMZ. Upper panel: the cool dust emission is shown as a grayscale background as observed by Herschel/SPIRE at 250 μm (Molinari et al. 2011). Overlayed colored dashes represent magnetic field orientations obtained from previous polarimetric observations of the CMZ cool dust from the CSO at 350 μm (red, Dotson et al. 2010), JCMT at 850 μm (cyan, Liu et al. 2023), and SOFIA/HAWC+ at 214 μm (Paré et al. 2024). Lower panel: the warm dust emission is shown as a grayscale background as observed by Herschel/SPIRE at 70 μm. Overlayed colored dashes represent magnetic field orientations obtained from previous polarimetric observations of the CMZ warm dust from the KAO at 60 and 100 μm (yellow and orange, Dotson et al. 2010) and SOFIA/HAWC+ at 53 μm.

## Instruments and Modes Used

- PRIMAger mapping, polarimeter bands at 96, 126, 172 and 235μm with polarimetric capability. One large map (4 deg x 7 deg) for each band.
- FIRESS high-res map for band 1. One large map (90' x 30').

## Approximate Integration Time

We estimate our approximate integration time using the PRIMA ETC. For the four PRIMAger bands we assume a 28 square degree map size and determine our 5σ rms required by scaling from Herschel/SPIRE 250 μm and 70 μm observations of the GC where we assume the cool dust at 250 μm has a temperature of T = 41 K and β = 1.67 and we assume the warm dust at 70 μm has a temperature of T = 22 K and β = 1.67. These sensitivities equate to values of: 1.06 x $10^4$ μJy at 96 μm, 2.60 x $10^4$ μJy at 126 μm, 4.08 x $10^4$ μJy at 172 μm, and 8.50 x $10^4$ μJy at 235 μm (these are the input values for the PRIMA ETC to determine the required PRIMAger times). This map size has the advantage of observing the regions of the Galaxy surrounding the CMZ, possibly providing insight into how dust flows into the CMZ from the Galactic disk. Furthermore, the 7° latitude range of the map will enable an in-depth of polar spurs that may be outflowing from the GC.

For FIRESS we estimate the time by assuming a 2700 square arcminute map size that is focused on the brighter molecular emission concentrated in the central 200 pc region of the GC. We will use the high-resolution mapping mode for band 1, which is not represented in the ETC. We instead estimate the time required by assuming it takes FIRESS 1 hour to map 4.8 square arcminutes to a sensitivity of $\sim10^{-17}$ W $m^{-2}$. Scaling this time to the size of our proposed FIRESS FOV results in a total FIRESS time of 562.5 hours.

- PRIMAger times: 7.8 (96 μm), 2.9 (126 μm), 1.9 (172 μm), 0.8 (235 μm) hours
- FIRESS time: 562.5 hours
- Total time: 570.3 hours

## Special Capabilities Needed

The FIRESS component of this proposal is contingent on being able to perform high-resolution mapping (R∼10,000 for 25 – 35 μm). The low-resolution mapping mode is simply not able to provide the velocity resolution necessary for sensitivity to the turbulence (R∼100). We therefore advocate for the inclusion of the FIRESS high-res mapping mode, as having this option will uncover critically important information on the turbulence of the clouds in the CMZ and in the Galaxy as a whole.

## Synergies with Other Facilities

These PRIMA observations would synergize with recent, existing, and forthcoming instruments. For example, the observation specifications of PRIMA detailed here would synergize well with the Atacama Large Millimeter/sub-millimeter Array (ALMA). ALMA is able to provide arcsecond and sub-arcsecond resolution capabilities that would probe protostellar core and envelop spatial scales in the GC, whereas PRIMA is sensitive to molecular cloud scales. The behavior of the magnetic field can be compared at these different scales (in conjunction with MHD modeling) to determine whether the magnetic field or turbulence dominates in CMZ clouds. This CMZ PRIMA project will also synergize with next-generation radio instruments like the next generation Very Large Array (ngVLA) and Square Kilometer Array (SKA). For example, the ngVLA and SKA will primarily operate at cm wavelengths, a wavelength range that will not be sensitive to the far-infrared dust emission. ngVLA observations can therefore be compared to the μm PRIMA observations to study connections between the CMZ clouds and the radio non-thermal structures observed in the GC. Similar work has previously been performed using VLA and SOFIA/HAWC+ observations (Paré et al. 2025), but the combined analysis of PRIMA and ngVLA observations will represent a significant improvement in sensitivity to what is possible with current and recent observatories.

The PRIMA observations detailed here can also be used with the Wide-field Survey (WFS) that will be conducted using the Fred Young Submillimeter Telescope (FYST, CCAT-Prime Collaboration 2023). The frequencies of the WFS survey (350, 730, 850, 1000, and 1400 µm) will complement the goal of studying how the polarization varies as a function of wavelength. In complex regions, the polarization can be used to distinguish between magnetic fields originating from dust with different temperatures. In simpler regions, the polarization spectrum can test models of dust alignment (Lee et al. 2024).

PRIMA provides unique and crucial capabilities in both resolution and frequency space compared to other instruments that will be operating during PRIMA's lifetime. PRIMA's capabilities will be invaluable for furthering our understanding of the role of magnetic fields and the nature of dust in the CMZ. Work of this form will enhance our understanding of Galactic nuclear regions and the mechanisms inhibiting star formation more generally.

## Description of Observations

To achieve the scientific objectives of this proposal PRIMA would observe the entire CMZ using PRIMAger bands PPI1, PPI2, PPI3, and PPI4 with full polarimetry. Full polarimetry with all four of these bands will be critical to obtain polarimetric observations of both the cool and warm dust populations. **Without full polarimetric capability at these frequencies, there is no facility currently available or planned that will be able to study the magnetic field of the warm dust component in the CMZ, making it impossible to study the role the magnetic field plays in the full CMZ dust population.**

For our PRIMAger observations we will match the field of view of the 240 µm PILOT observations of the GC (4° x 7°, Mangilli et al. 2019). This field of view encompasses the entirety of the CMZ and the surrounding Galactic region (like the base of the polar spurs and the inflowing material along the dust lanes connecting the CMZ with the Galactic Disk). Furthermore, we will be able to compare the 2′ resolution PRIMA observations with the 10–28″ resolution PRIMAger observations to determine how the structure of the magnetic field varies over this range of resolutions. To observe this map region, we will use a scan map strategy to efficiently cover the proposed 28 square degree region. An example of a possible PRIMA scan pattern is detailed in Dowell et al. (2024), and additional scanning strategies are currently being explored by the PRIMA team. The PRIMAger scans for this project will be obtained simultaneously, meaning that only 7.8 hours are required to obtain a four-color map of the CMZ using PRIMAger (as shown above).

In addition to the PRIMAger mapping, we will also create a map using the FIRESS high-resolution mapping mode. The map size for this observation will focus on the bright, high density molecular clouds in the central 200 pc region of the Galaxy. We will use a 2700 square arcminute map size for FIRESS to study the H2 line at 28 µm. This observation will have to be obtained separate from the PRIMAger observations, since the two instruments are not designed to be operated simultaneously.

# 89. The Distribution of Elemental Abundances in the Milky Way

Jorge L. Pineda (Jet Propulsion Laboratory, California Institute of Technology), Shinji Horiuchi (CSIRO & NASA DSCC), Loren Anderson (W. Va. University), Paul F. Goldsmith (JPL)

The distribution of elemental abundances within the disks of galaxies provides essential observational data that can be used to develop and refine models of galaxy formation and evolution. Currently, only the Milky Way allows for detailed, high-resolution studies of the interstellar medium (ISM) and star formation, which provide a deeper understanding of the elemental abundance distribution. Traditionally, abundance distributions in the Milky Way have been determined by observing H II regions with optical lines. Due to dust extinction, optical studies are limited to the solar neighborhood and are unable to probe the inner Galaxy where most star formation takes place. Longer-wavelength observations — including far-infrared (FIR) fine-structure lines, radio continuum, and hydrogen radio recombination lines (RRLs) — can be used to determine elemental abundances without being obscured by dust. PRIMA/FIRESS will revolutionize the study of elemental abundances in the Milky Way by providing sensitive maps of numerous FIR spectral lines across a wide range of environments. These observations will allow us to determine electron densities and N/O ratios across the Milky Way. By incorporating RRLs from the Green Bank Observatory (GBO) and the 70 m antenna at NASA's Canberra Deep Space Communications Complex (DSS-43), we can also determine the nitrogen abundance distribution (N/H) over a wide range of Galactic environments. Enhanced distance knowledge from GAIA and VLBI observations will enable us to learn about azimuthal distributions, the vertical distribution of metals in the Milky Way, and the nature of the flattening of the N distribution in the inner 4 kpc of the galaxy (Pineda et al. 2024).

In this white paper we discuss the potential of PRIMA/FIRESS to advance our understanding of the distribution of elemental abundances in the Milky Way. We describe the FIR fine structure line observations that FIRESS on PRIMA will enable and a potential survey of HII regions to be carried out with PRIMA. Observations of the distribution of elemental abundances in the Milky Way will be important for the interpretation of extra-galactic abundance determinations in the optical and FIR with PRIMA.

## Science Justification

As galaxies grow, inflows from the circum-galactic (CGM) medium reach the disks of galaxies, cooling and forming dense molecular clouds where star formation takes place. As star formation occurs, newly formed massive stars emit copious amounts of far-ultraviolet (FUV) photons that ionize their surroundings, drive stellar winds, and eventually explode as supernovae, slowing down further star formation and driving outflows of gas through the ISM back into the CGM. Understanding the interplay between inflows, star formation, stellar feedback, and outflows, is the key for a full characterization of galaxy growth and evolution.

The distribution of elemental abundances within the disks of galaxies offers essential observational data that can be used to develop and refine models of galaxy formation and evolution. Elements such as C, N, and O are products of “primary" and "secondary" processes in massive and intermediate mass stars (Johnson et al. 2019) and therefore their abundances are related to the types of stars, star formation rate, and star formation history at a given location, with each element having a different enrichment timescale. Studies of large samples of galaxies (e.g. Belfiore et al. 2017) show that galaxies show negative elemental abundance gradients in their disks that steepen as a function of stellar mass. In the high stellar mass end, an O/H flattening is observed in the central parts of galaxies. In addition, the O/H distribution is observed to flatten in the outer parts of galaxies. Abundance gradients in galaxies are typically interpreted as the result of "inside-out" formation, in which the disks of galaxies form by gas accretion with a rate that is faster in the inner Galaxy compared with the outer Galaxy (Larson1976, Matteucci et al 1989, Boissier et al. 1999, Pilkington et al. 2012). However, other mechanisms, such as variable star formation efficiency and/or radial flows can better explain the observed slope in the Milky Way (Palla et al. 2020, Pineda et al. 2024). Numerical simulations predict that a signature of radial flow of metals can be seen in the azimuthal distribution of abundances (Orr et al 2023), with observations suggesting that indeed the azimuthal distribution of metals in the Milky Way is not uniform (Balser et al. 2011). The O/H flattening in the center of galaxies is interpreted as these regions reaching a "metallicity equilibrium", in which metal production is balanced by metal consumption by star formation expulsion by outflows (Peng et al. 2014, Weinberg et al 2017). However, the flattening in the Nitrogen abundance observed in the Milky Way's center is more likely due to radial flows induced by the stellar bar potential (Pineda et al. 2024).

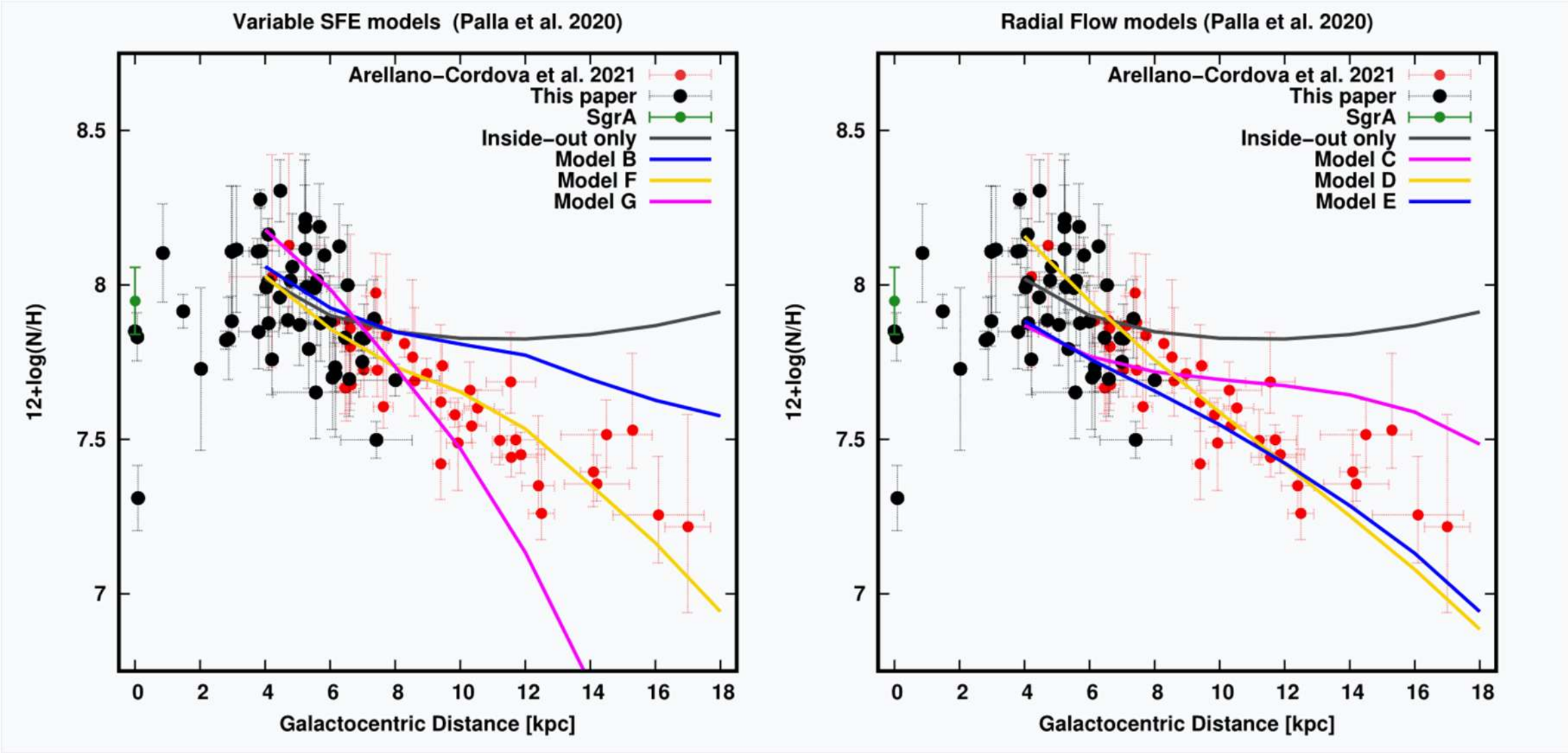

**Figure 1.** Far--infrared spectroscopy and radio recombination line observations enable the measurement of Nitrogen elemental abundances in the inner galaxy which is obscured by dust in optical wavelengths. This data set is complemented with optical measurements in the outer Galaxy (Arellano-Cordova et al. 2021) to obtain measurements of the N/H distribution in the inner 18 kpc of the Milky Way. Such measurements can be used to constrain models of the formation of the Milky Way (Pineda et al. 2024)

It is important to test whether the properties of abundance distributions observed in nearby galaxies also apply to the Milky Way. Detailed studies of the properties of the interstellar medium and star formation at high spatial resolution are currently only possible in the Milky Way. Therefore, by studying the elemental abundance distribution in the Milky Way, we can obtain a deeper insight into the nature of these distributions, which in turn can be used to interpret the results obtained over large samples of unresolved, external galaxies. The abundance distributions in the Milky Way have been traditionally obtained observing HII regions with optical lines in a variety of environments. These observations are mostly focused on nearby HII regions, where dust extinction obscures these optical lines moderately, and have been used to infer that the abundance of Oxygen and Nitrogen increase with Galactocentric distance from the outer Galaxy inward to Rgal= 6 kpc (Esteban et al. 2018). Due to increased dust extinction, optical studies are unable to probe the inner Galaxy, where most of the star formation takes place, and which is thought to have formed during the early stages of the Milky Way's evolution. Therefore, these observations have been unable to test the flattening in the O/H distribution and a possible increase of the N/O ratio observed in the central parts of external galaxies with similar stellar masses as the Milky Way's. In addition, by studying the vertical distribution of metals we can learn about how metals are distributed into the ISM by stellar feedback, and whether they can be deposited into the CGM.

Longer-wavelength observations—including FIR fine-structure lines, radio continuum, and hydrogen RRLs—can be used to determine elemental abundances without being obscured by dust. PRIMA/FIRESS will revolutionize the study of elemental abundances in the Milky Way by providing sensitive maps of numerous far-infrared spectral lines across a wide range of environments. These observations will allow us to determine electron densities and N/O ratios across the Milky Way. By incorporating RRLs from the 100m Green Bank Telescope and the 70 m antenna at NASA's Canberra Deep Space Communications Complex, we can incorporate information about the column density of H+, which in turn enables us to determine the distribution of the nitrogen abundance relative to hydrogen in a wide range of Galactic environments. Enhanced distance knowledge from GAIA and VLBI observations will enable us to learn about azimuthal distributions, the vertical distribution of metals in the Milky Way, and the nature of the flattening of the Nitrogen distribution in the inner 4 kpc of the Galaxy (Pineda et al. 2024).

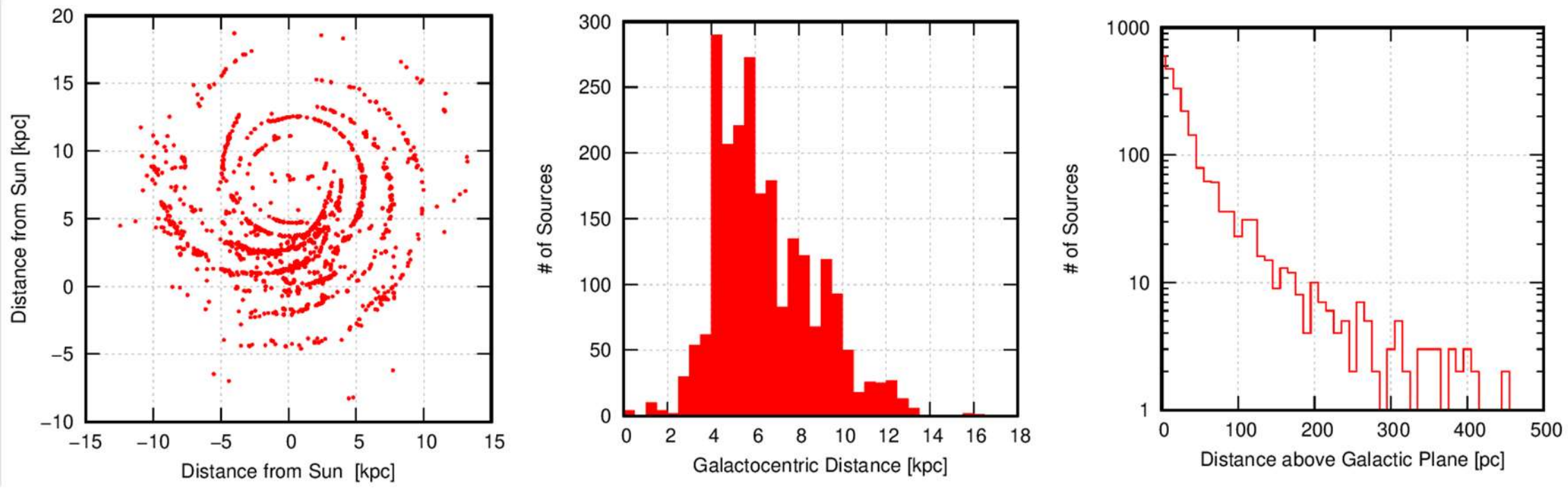

**Figure 2.** The unprecedented sensitivity, mapping speed, and spectral coverage of PRIMA/FIRESS make a survey of all known HII regions in the Milky Way feasible and will provide a complete picture of the distribution of elemental abundances. In the left panel we show the distribution of HII regions across the disk (distances obtained with the Reid et al. model). In the middle and central panels we show histograms of HII regions as a function of distance to the Galactic center and the vertical distance from the plane of the disk.

## Proposed PRIMA Observations

The FIRESS instrument on PRIMA will provide sensitive observations of the mid- to far-infrared range (24 µm–235 µm) and therefore will provide maps of several key FIR fine-structure lines that provide fundamental diagnostics of the physical conditions of the interstellar medium and that can be used to probe the distribution of metals in H II regions (Simpson et al. 2004). These lines include the fine structure lines of singly ionized Nitrogen, [N II] 122 µm and 205 µm, that trace low ionization gas, and their ratio is a sensitive probe of volume density (Goldsmith et al. 2015; Figure 3 right panel). In addition, higher ionization state lines, such as [N III] 57 µm, and [O III] 88 µm and 52 µm, can be used to determine the N/O ratio in H II regions, and electron densities via the [O III] 52 µm/88 µm ratio (Figure 3 right panel). Together with observations of radio continuum or Hydrogen recombination lines, with facilities such as the NASA DSN DSS-43 antenna in Canberra Australia and the Green Bank Observatory, the nitrogen abundance can be determined (Pineda et al. 2024).

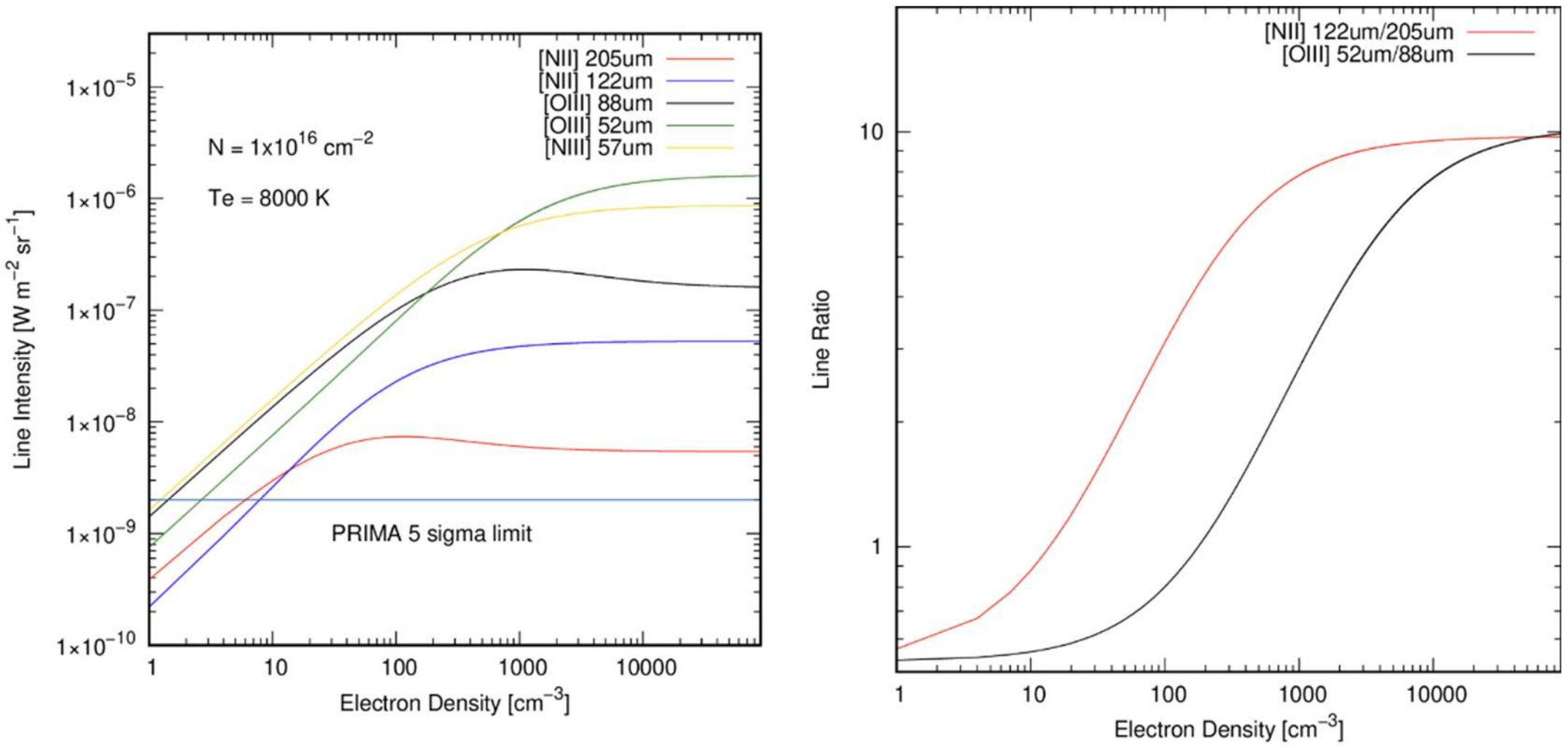

**Figure 3.** (Left panel) Predicted line intensity of $N^+$,$N^{++}$, and $O^{++}$ fine structure lines as a function of electron density. We show as a straight line the 5 sigma sensitivity limit proposed of the PRIMA/FIRESS instrument. (Right panel) Ratio of the [NII ] 122μm/205μm and [OIII ] 52μm/88μm lines as a function of electron density.

To determine the sensitivity needed for FIR fine structure lines to be detected with high signal-to-noise ratio, we calculated in the left panel of Figure 3 the expected intensities of the $N^+$, $N^{++}$, and $O^{++}$ lines as a function of electron density for a column density of these species of $1x10^{16}$ $cm^{-2}$ and T=8000K. (Figure 3). As we can see, a sensitivity limit of $2x10^{-9}$ W $m^{-2}$ $sr^{-1}$ allows us to detect these lines with SNR larger than 5. (Note that intensities scale linearly with assumed column density.) For Nitrogen, such a sensitivity will enable us to detect Nitrogen abundances that are 0.1 Solar for $H^+$ column densities, $N(H^+)$, above $2x10^{21}cm^{-2}$, 0.5 Solar for $N(H^+)$ above $4x10^{20}$ $cm^{-2}$, and Solar for $N(H^+)$ above $2x10^{20}$ $cm^{-2}$.

HII regions in the Milky Way have been identified and catalogued in the Milky Way disk with multi-wavelength surveys. The WISE catalog presented by Anderson et al. 2014 provides positions and sizes of over 2000 HII regions distributed across the Galactic plane. Distances to these HII regions can be determined using Maser parallaxes and/or proper motions from GAIA. Reid et al. (2019) presented a model of the spiral structure of the Milky Way based on 200 trigonometric parallaxes of masers associated with massive star forming regions. Distances measured from maser trigonometric parallaxes have the advantage that they do not rely on any assumption on the kinematics of the Galaxy. The Reid et al. (2019) model enables us to estimate distances to sources using their Galactic coordinates and LSR velocity. In Figure 1 we show the Azimuthal, radial, and vertical distribution of the WISE HII region catalog in the Milky Way. As we can see, determining elemental abundances in the HII regions in this catalog will PRIMA/FIRESS will enable us to have a complete sample of the distribution of elemental abundances in the Milky Way. The total area in the sky that is covered by HII regions (excluding those with complex velocity structure and unknown LSR velocity) in the WISE catalog is 111 square degrees.

## Instruments and Modes Used

FIRESS Spectrometer Low-res (R~100) Mapping. Total area to map 110 square degrees.

## Approximate Integration Time

FIRESS Spectrometer Low-res (R~100) Mapping: 161h

## Special Capabilities Needed

None

## Synergies with Other Facilities

Radio Telescopes (NASA DSN, GBT, VLA, and ngVLA).

## Description of Observations

To accomplish the goal of obtaining a complete picture of the distribution of elemental abundances across the Milky Way, a large scale survey of HII regions distributed in the Galactic disk is needed. The unprecedented sensitivity and mapping speed provided by PRIMA/FIRESS makes it possible to image all HII regions in the WISE catalog. For the sensitivity described above, 5 sigma flux of $2x10^{-9}$ W $m^{-2}$ $sr^{-1}$ ,the Exposure Time Calculator (ETC) suggests a 1.5h per square degree integration time 30" pixel at 57μm. Therefore, for a survey covering 110 square degrees, a total of 161h of observing time is required.

## Acknowledgment

This research was carried out at the Jet Propulsion Laboratory, California Institute of Technology, under a contract with the National Aeronautics and Space Administration (80NM0018D0004).

# 90. A Census of Dust Production from Nearby Supernovae with PRIMA/FIRESS

Sam Rose (Caltech), Kishalay De (Columbia), Jacob Jencson (Caltech/IPAC), Viraj Karambelkar (Caltech), Mansi Kasliwal (Caltech), Ryan Lau (NOIRLab), Maggie Li (Caltech), Samaporn Tinyanont (NARIT)

Improving observations of high redshift galaxies over the last decade have revealed large dust reservoirs very early in cosmic history. These large dust masses challenge existing theories of dust formation and require large amounts of dust to be produced by fast-evolving massive stars, primarily within the ejecta of supernovae. The bulk of dust formation in supernovae is expected to occur decades after the explosion, when the ejecta and forming dust is cold and emitting primarily in the far-infrared (FIR). With the limited sensitivities of previous FIR missions like Herschel, only Galactic supernova remnants and the extremely nearby SN 1987A have been observed in the FIR at late times. PRIMA/FIRESS will offer an unprecedented view of dust formation in supernovae, increasing the number of supernovae with late-time dust mass measurements by at least two orders of magnitude. Understanding dust formation by supernovae in the local Universe will improve our understanding of early-Universe dust formation and address PRIMA's core science theme 3: *How do interstellar dust and metals form and build up in galaxies over cosmic time?*

## Science Justification

In the last decade there has been growing recognition of the ubiquity of dust across cosmic time and in a variety of environments, even while our understanding of how that dust is formed has remained incomplete. It is well known that young stars in the local universe are formed in giant clouds of dense gas and dust within galaxies (Zavala et al. 2021). Newly formed stars are observed to be enshrouded in dusty disks within which small particles of dust can begin to stick together, grow, and eventually form planets (Dipierro et al. 2015). The most fundamental building block of rocky planets, like our own Earth, are these small astronomical dust particles. To understand our own cosmic origins, it is necessary to understand how this dust forms.

Dust is formed in large quantities early in cosmic history. Sub-mm observations with ALMA of distant gravitationally lensed galaxies (e.g., Laporte et al. 2017), and recent studies by the James Webb Space Telescope (JWST, e.g., Witstok et al. 2023) have shown that dust is formed in significant quantities for a variety of galaxy masses even at very early times (within 600 million years of the Big Bang).

While low-mass asymptotic giant branch (AGB) stars can form significant quantities of dust in the local Universe, dust which forms very early in cosmic history must come from faster evolving progenitors. Supernovae, which are produced very soon after star formation, are likely significant contributors to the cosmic dust budget at early times (e.g., Schneider & Maiolino 2024). It has been shown that large amounts of dust can form in the ejecta of supernovae (e.g., Wesson et al.

2015, Shahbandeh et al. 2023), but supernovae may also destroy large quantities of dust when they explode (e.g., Priestley et al. 2021). In particular, the reverse shock may destroy dust that is newly formed in the ejecta of the supernova (e.g., Micelotta et al. 2016). SN 1987A is likely the most well-studied example owing to its proximity (e.g., Wooden et al. 1993, Dwek et al. 2010, Matsuura et al. 2011, Matsuura et al. 2015, Dwek et al. 2019), and there have been many efforts to model dust formation in supernova ejecta (e.g., Sarangi & Cherchneff 2015, Sarangi et al. 2018, Sluder et al. 2018, Brooker et al. 2022). In the last 3 years JWST has opened up a new frontier in the study of dust formation in supernovae explosions. Late-time mid-IR (MIR) observations of many different supernova types at a variety of post-explosion times have allowed for direct measurements of dust mass, and spectroscopy has helped constrain dust composition (e.g., Shahbandeh et al. 2023, Tinyanont et al. 2025).

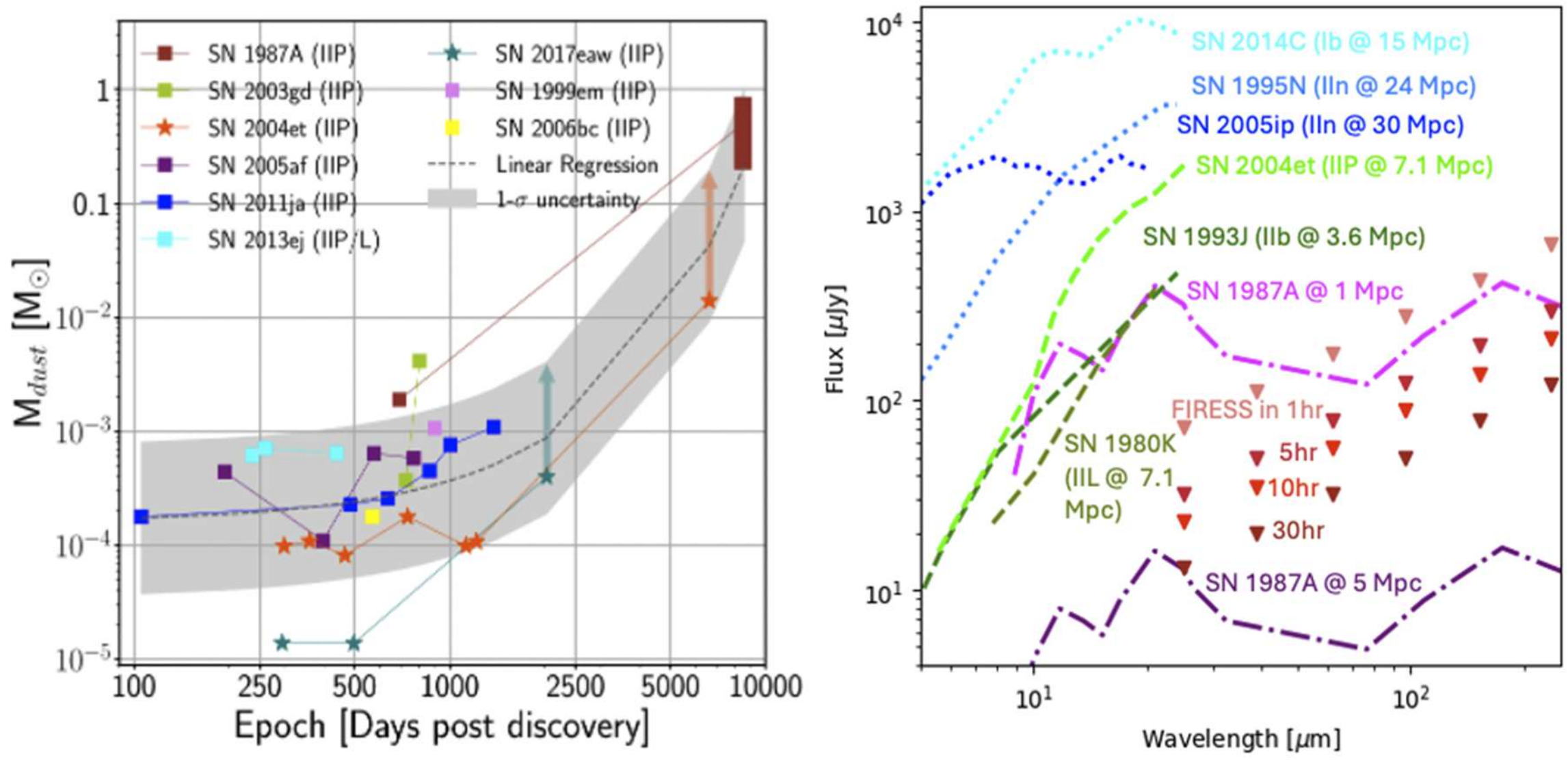

**Figure 1.** (Left) Dust mass measurements for a variety of supernovae as a function of time post-discovery from Shahbandeh et al. 2023. The total mass of dust observed increases with time, but at late times the bulk of the dust is cold and emitting in the FIR. (Right) MIR photometry of 6 nearby supernovae as well as MIR and FIR photometry for SN 1987A scaled to 1 and 5 Mpc (SN 1980K with Spitzer/IRAC at 28 yrs from Sugerman et al. 2012; SN 1987A with Herschel/PACS&SPIRE at 25 yrs from Matsuura et al. 2015; SN 2004et with JWST/MIRI at 18 yrs from Shahbandeh et al. 2023; SN 1995N with Spitzer/IRAC&MIPS at 14yrs from Wesson et al. 2023; SN 1993J with JWST/MIRI at 30yrs from Szalai et al. 2025; SN 2014C with JWST/MIRI at 9 yrs from Tinyanont et al. 2025; SN 2005ip with JWST/MIRI at 17yrs from Shahbandeh et al. 2024). Supernovae plotted in blue (SN 2014C, SN 1995N, and SN 2005ip) showed signatures of CSM interaction while supernovae plotted in green were, at least initially, non-interacting (SN 2004et, SN 1993J, and SN 1980K). While non-interacting at early times, late-time observations of SN 2004et do show signatures of CSM interaction. The continuum sensitivity of FIRESS with 6 wavelength bins in 1hr, 5hr, 10hr, and 30hr of integration time is shown by red triangles. PRIMA/FIRESS ought to be sensitive to interacting supernovae out to 50 Mpc in less than 1 hour of integration time per target and to at least some non-interacting supernovae (especially those which start to interact at late times) out to 25 Mpc in less than 5 hours of integration time. The FIR probed by PRIMA/FIRESS contains the bulk of the cold dust SED.

On the timescales of years to decades, MIR observations of nearby supernovae from Spitzer, and now from JWST, have shown that dust mass increases with time. However, at later times, as the ejecta expands and cools, the dust around these supernovae is cold and consequently the peak of the SED is at longer wavelengths than currently accessible. Therefore, observing the full dust mass, as measured for SN 1987A (∼0.1 solar masses of dust), is only possible with long wavelength observations (from Herschel and ALMA, Matsuura et al. 2015). In order to fully characterize the dust produced by supernovae explosions longer wavelength observations are necessary. The FIRESS instrument on PRIMA will offer orders of magnitude improvement on the depths provided by Herschel, as well as coverage of the peak of the cold dust SED currently missed by sub-mm ground-based instruments, like ALMA. This increase in depth will allow for FIR observations of dusty supernovae out to 50 Mpc for interacting supernovae (which are brighter due to the power provided by shock interaction with a dense CSM), and out to 25 Mpc for at least some subset of non-interacting core-collapse supernovae (see Figure 1). With 1 hour single pointing FIRESS observations for the ∼35 interacting supernovae, which have been observed in the last 40 years within 50 Mpc, and 5 hour single pointing FIRESS observations for the ∼80 core-collapse supernovae within 25 Mpc over the same time frame (of which at least some subset should be detectable), PRIMA is capable of increasing the number of supernovae which have dust masses measured out to late times in the FIR by 100 times in only 435 hours of exposure time.

## Instruments and Modes Used

FIRESS pointed low-resolution

## Approximate Integration Time

Based on the PRIMA ETC, each interacting supernovae would take less than 1 hr of integration time with FIRESS Low-Res (synthesized R∼10 bins) to achieve a sufficient SNR (>5σ) to characterize the dust mass in the far-infrared. Deeper observations are required for non-interacting supernovae which lack the additional power provided by the interaction such that each target would require ∼5 hr of integration time. The proposed depths obtained from the ETC are shown in Figure 1.

## Special Capabilities Needed

None

## Synergies with Other Facilities

The proposed science case has synergies with JWST for full IR SED coverage. Several of these nearby supernovae have observations from JWST already. FIR observations with PRIMA in the next several years will probe the ongoing formation of dust.

## Description of Observations

The observational strategy for conducting a far-IR census of nearby supernovae is to obtain single-pointing PRIMA/FIRESS low-resolution spectroscopy (in synthesized R ∼ 10 bins) of known supernovae within 50 Mpc (interacting) or 25 Mpc (non-interacting). Observations in this mode probe the far-IR (∼25–200 µm) SED from cool dust formed in the supernovae ejecta and will enable dust-model fitting with radiative transfer codes such as DUSTY to obtain information about total mass, composition, and grain size distribution.

As we are interested in supernovae at late times (at least 5 years), all targets for the program are already known. In the last 40 years there have been about 80 core-collapse supernovae within 25 Mpc and 35 interacting core-collapse supernovae within 50 Mpc. It is important to note that a supernova can begin to interact at late-times even if it was not initially observed to interact (e.g., Myers et al. 2024) such that even some of the "non-interacting" supernovae may be brighter at late times than expected which is why we choose a more permissive distance cut of 25 Mpc even though 1987A-like objects would be completely undetectable at that distance.

# 91. Measurement of Water Isotopologue Ratios: a Probe of Planet Formation Cosmochemistry

Colette Salyk (Vassar College), Danny Gasman (KU Leuven)

Solar system materials show mass-independent variations in oxygen isotopes proposed to arise from isotope-selective photodissociation. These variations allow oxygen isotopes, particularly in water, to be used as tracers of a material's parent body. It is not yet clear to what extent this phenomenon may occur in other planet-forming regions. Confirmation of this process outside of the solar nebula would enable more detailed testing of the phenomenon at the time of its occurrence, open the door to using oxygen isotopes as origins tracers in exoplanetary systems, and contextualize the process in our own Solar System. Therefore, we propose here to measure the abundance of $H_2^{18}O$ in nearby planet-forming regions, and determine whether it has an enhanced (relative to ISM) abundance consistent with predictions from isotope-selective photodissociation. Detection of the second most common isotopologue of water, $H_2^{18}O$, remains at the limit of detectability for JWST MIRI-MRS, but is more readily detectable with PRIMA-FIRESS, which is sensitive to cooler, more spatially extended water vapor. A set of strong $H_2^{18}O$ emission lines (plus the suite of $H_2^{16}O$ lines that appear throughout the FIRESS wavelength range) will be observed in a representative sample of 5 nearby protoplanetary disks to test this scenario. Models of the $H_2^{16}O$ and $H_2^{18}O$ will be used to measure the $H_2^{16}O/H_2^{18}O$ and test for differences relative to ISM values.

## Science Justification

### Does isotope-selective photodissociation occur in protoplanetary disks?

In solar system materials, oxygen isotope ratios follow a mass-independent fractionation trend (in a three-isotope plot) believed to originate in the solar nebula itself. Oxygen isotope ratios are thus a fingerprint pointing back to a material's radial origin in the nebula (Clayton et al. 1973, McKeegan 2011, Genda 2016). Since most fractionation processes are mass-dependent, new theories needed to be devised to explain these observations; the origin of this fractionation trend is proposed to be isotope-selective photodissociation of CO, in which CO molecules bearing heavier isotopes of oxygen ($C^{18}O$ and $C^{17}O$) are more readily dissociated than $C^{16}O$, which effectively self-shields (Thiemens & Heidenreich 1983, Yurimoto & Kuramoto 2004, Lyons & Young 2005). The heavier oxygen atoms are then free to enrich other molecules, like $H_2O$, with heavier oxygen.

It is not yet clear to what extent this process takes place beyond the solar system. Confirmation of the action of isotope-selective photodissociation in protoplanetary disks would solidify the leading hypothesis for the observed mass-independent fractionation trend, and allow for more detailed study of its physical processes. In addition, confirmation that radial gradients in oxygen isotopes universally develop in planet-forming environments would open the door to using oxygen isotopes as radial origin tracers in exoplanetary systems.

Some studies of Carbon-bearing species in the outer regions of disks support the action of isotope-selective photodissociation (Smith et al. 2009, Hily-Blant et al. 2017, 2019, Yoshida et al. 2022, Bergin et al. 2024). However, it is not yet known whether water is affected by this process as it was in the solar system. JWST's MIRI-MRS spectrograph is sensitive to disk regions expected to have isotopically-enhanced water vapor (Calahan et al. 2022), but the most common isotopologue of water, $H_2^{18}O$, appears to lie right at the limit of detectability for that facility (Temmink et al. 2024, Salyk et al. 2025, in preparation). In contrast, longer-wavelength PRIMA observations can detect emission from lower energy transitions that probe cooler, more spatially extended water vapor reservoirs. With nominal disk models, and ISM-like abundances, $H_2^{18}O$ emission should be readily detectable in nearby protoplanetary disks with PRIMA. If, instead, the water vapor is enhanced in heavy isotopes as predicted, the emission will be even easier to detect. The potential enhancement is also expected to be radially dependent (Calahan et al. 2022), and this prediction can therefore be tested by comparing PRIMA and JWST observations.

Thus, PRIMA will be able to answer the question: does isotope-selective photodissociation, leading to heavy water, occur in protoplanetary disks?

In addition to providing constraints on the prevalence of isotope-selective photodissociation, detections (and even non-detections) of $H_2^{18}O$ will provide improved constraints on overall water vapor column densities in protoplanetary disk atmospheres. Many observed infrared $H_2^{16}O$ emission lines in protoplanetary disks are optically thick (e.g., Carr et al. 2004, Salyk et al. 2008, Banzatti et al. 2023, Temmink et al. 2024) and the emitting layer may only show 1-10% of the total column density (e.g., Walsh et al. 2015, Bosman et al. 2022), preventing the determination of the total water abundance. Less common isotopologues, such as $H_2^{18}O$, are helpful to constrain the $H_2^{16}O$ abundance in the disk.

## Methods

FIRESS spectra of nearby protoplanetary disks are expected to include hundreds of detectable $H_2^{16}O$ emission lines. These emission lines can be modeled readily with existing modeling frameworks (e.g. Romero-Mirza et al. 2024, Pontoppidan et al. 2024, Temmink et al. 2024) to determine the column density, temperature, and emitting location of the water vapor reservoirs. FIRESS observations, which probe cool and colder water vapor reservoirs, can also be supplemented with the already hundreds of JWST MIRI-MRS spectra of the warmer reservoirs of disks, to improve constraints on the distribution of the primary istopologue of water.

Once the water vapor is properly modeled, these reservoirs can be assumed to also contain $H_2^{18}O$. Models can be produced with a range of $H_2^{16}O/H_2^{18}O$ ratios, and varied to find the best fit to observed $H_2^{18}O$ emission lines. The best-fit $H_2^{16}O/H_2^{18}O$ ratios can be compared with ISM values, and with those expected from isotope-selective dissociation.

## Testable Hypothesis and Expected Outcomes

### Hypothesis

Selective photodissociation does/(does not) affect water vapor in protoplanetary disks

**Expected outcome**

Observed $H_2^{16}O$ / $H_2^{18}O$ ratios are equivalent to/(are smaller than) ISM values

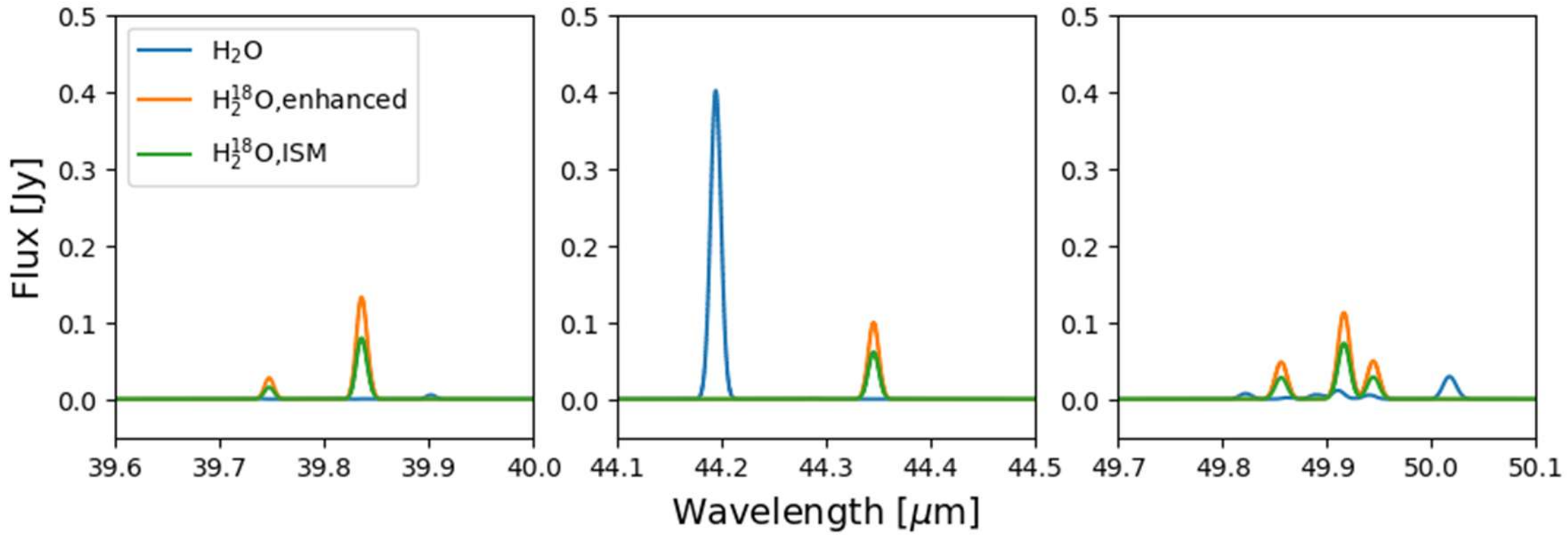

**Figure 1.** $H_2^{16}O$ and $H_2^{18}O$ emission model predictions highlighting 3 strong, isolated $H_2^{18}O$ emission lines in the PRIMA FIRESS wavelength range. Green curves show the expectation assuming an ISM $H_2^{18}O$ abundance relative to $H_2^{16}O$; orange curves show an enhanced $H_2^{18}O$ abundance caused by selective photodissociation (a factor of 1.8 difference; Calahan et al. 2022).

## Instruments and Modes Used

FIRESS: high-resolution point source mode

## Approximate Integration Time

25.3 hours per pointing

## Special Capabilities Needed

None

## Synergies with Other Facilities

Targets with strong water emission, most likely to have detectable $H_2^{18}O$ emission, can be readily selected from prior JWST MIRI-MRS protoplanetary disk samples (e.g. MINDS; Henning et al. 2024 and JDISCS; Pontoppidan et al. 2024.) JWST MIRI-MRS data will also be used to assist in constraining $H_2^{16}O$ emission models.

## Description of Observations

We propose to observe a representative sample of nearby (d~150 pc) protoplanetary disks to detect $H_2^{18}O$ and determine whether $H_2^{18}O$ has been enhanced by selective photodissociation. Targets with the strongest expected $H_2^{18}O$ line emission will be selected via modeling of existing JWST MIRI-MRS spectra.

Based on models derived from past mid-to far-IR data (Blevins et al. 2016), the strongest $H_2^{18}O$ lines lie near 40-50 microns and three lines in particular are isolated from prominent $H_2^{16}O$ emission – see Figure 1. These lines have fluxes of order 1e-18 $W/m^{-2}$. Assuming a nominal point source continuum of 0.5 Jy, a 5$\sigma$ detection of the shortest wavelength $H_2^{18}O$ feature in Figure 1 requires 13.8 hours per target to detect in Band 1. A 5$\sigma$ detection of the 44 micron line would require 11.5 hours per target to detect in Band 2, and would also include a detection of the 49.9 micron line.

# 92. PRIM(All): A PRIMA All-Sky, Polarized, Dust Map and Point Source Catalog

Andrew K. Saydjari (Princeton), Ritoban Basu-Thakur (JPL/Caltech), Robert Benjamin (Wisconsin), Arjun Dey (NOIRLab), Thavisha Dharmawardena (NYU), Bruce Draine (Princeton), Thomas Essinger-Hileman (NASA/GSFC), Richard M. Feder (UCB/LBNL), Douglas Finkbeiner (CfA), Laura Fissel (Queen's), Alyssa Goodman (CfA), Brandon S. Hensley (JPL/Caltech), David W. Hogg (NYU), Eric Koch (NRAO), Dustin Lang (Perimeter), Robert Lupton (Princeton), Joan Najita (NOIRLab), Giles Novak (Northwestern), Theo O'Neill (CfA), Kate Pattle (UCL), Josh Peek (STScI), Edward Schlafly (STScI), Jason Wright (Penn State), Rongpu Zhou (LBNL), Catherine Zucker (CfA)

Interstellar Milky Way dust is a pervasive observational challenge, but also a vital window for understanding the Galaxy. PRIMAger's FIR coverage from 92–235 µm perfectly straddles the black body peak for interstellar dust, leading to robust constraints on dust temperature. Thus, a PRIMA, all-sky, polarized, 4-band, 3-epoch survey with simultaneous all-sky hyperspectral imaging (R8) is ideal for converting FIR emission to dust extinction maps. Legacy all-sky FIR dust maps have had a transformative impact on the community (Schlegel et al. 1998, SFD98, 13,570 citations, COBE+IRAS), by providing constraints on the structure of the interstellar medium traced by Galactic dust and enabling foreground removal for studies of the circumgalactic medium, galaxy evolution, and cosmology. However, the low angular resolution, temperature sensitivity, and point-source contamination of SFD are now limiting factors in both foreground removal and uniform, all-sky studies of the interstellar medium. An all-sky PRIMAger survey, which is feasible in 15% of GO time, will revolutionize 2D emission-based dust maps. In just 5000 h, PRIMA will obtain an all-sky dust map with 34x the angular resolution of SFD98, while providing new constraints on magnetic fields via polarized dust emission resulting from dust grain alignment. These observations will produce a catalog of point-sources 63x deeper than the IRAS all-sky catalog, which will synergistically reduce point-source contamination in the dust map and provide polarized flux measurements that will guide exposure times for GO programs. These sensitive, high-resolution measurements of dust polarization, extinction, and temperature will facilitate a broad range of science cases including studies of feedback, star formation, the interstellar radiation field, and Galactic scale magnetic fields all while producing multi-epoch, legacy data products with lasting community impact for decades to come.

## Science Justification

### Motivation

Dust maps are multidisciplinary data products used by many subfields of astronomy, so generating a high-quality dust map will significantly broaden PRIMA's impact on the astrophysical community. Current 2D emission dust maps are limited in angular resolution, sensitivity, and by (point source) contamination (Figure 3). Many astronomers only use dust maps for "extinction correction," to remove the effect of dust as a foreground. In cosmology, such corrections are crucial for measuring the selection function of cosmological tracers, photometric redshifts of galaxies, and calibration of standard candles (e.g., SN 1a). In stellar astrophysics, extinction correction is important for precision measurement of stellar parameters (e.g., log(g)), necessary for probing either stellar astrophysics or host-planet correlations (e.g., TEMPO, Soares-Furtado et al. 2024). Careful extinction correction is also crucial for spectral and photometric flux calibration. High angular resolution maps of the Galactic "cirrus" are important for removing foreground (Milky Way) dust emission when studying low surface brightness features in the circumgalactic medium of nearby galaxies (e.g., Dragonfly; Liu et al. 2025).

Our motivation in proposing a PRIMA all-sky survey and dust map go far beyond extinction correction, though. Below, we highlight a handful of field-defining improvements that PRIMA's all-sky dust map will enable.

**PRIMA will distinguish variations between dust composition and the ISRF:** PRIMA's wavelength coverage (Figure 1) vastly improves sensitivity to the shape of the Galactic interstellar dust SED, which is generally modeled as a black body modified by a frequency dependent (power-law) opacity.

$$I_\nu \propto \left(\frac{\nu}{\nu_0}\right)^\beta B_\nu(T)$$

PRIMA will thus enable better constraints on the dust SED parameters: β, the frequency power-law exponent, and T, the temperature. This formulation, where tradeoffs between β and T can cause degeneracy, remains a key limitation with Planck data, where different pipelines (Commander 2016 vs. GNILC 2018) produced drastically different β-T distributions because they were primarily relying upon bands in the Rayleigh-Jeans tail of the dust blackbody (Planck Col. et al. 2016, Planck Col. et al. 2020). With PRIMA, we will have robust measurements of β to directly test the previously reported R(V)-β correlation (Schlafly et al. 2016, Zhang & Green 2025), which has implications for dust composition and its variation throughout the Milky Way (Zelko & Finkbeiner 2020).

Additionally, improved constraints on dust T are essential to self-consistently model the interstellar radiation field (ISRF; e.g., Hα sky in 3D, McCallum et al. 2025) informing how stellar feedback from massive stars reshapes its local environment. Constraining this stellar feedback is an essential piece to understand the fundamental properties of the ISM, particularly for simulations ranging from cosmological scales to MHD fluids seeking to describe ISM physics from first principles (e.g., Federrath et al. 2021). These careful measurements of dust β-T can also be used to disentangle starless and star-forming cores by the contributions of internal and external heating.

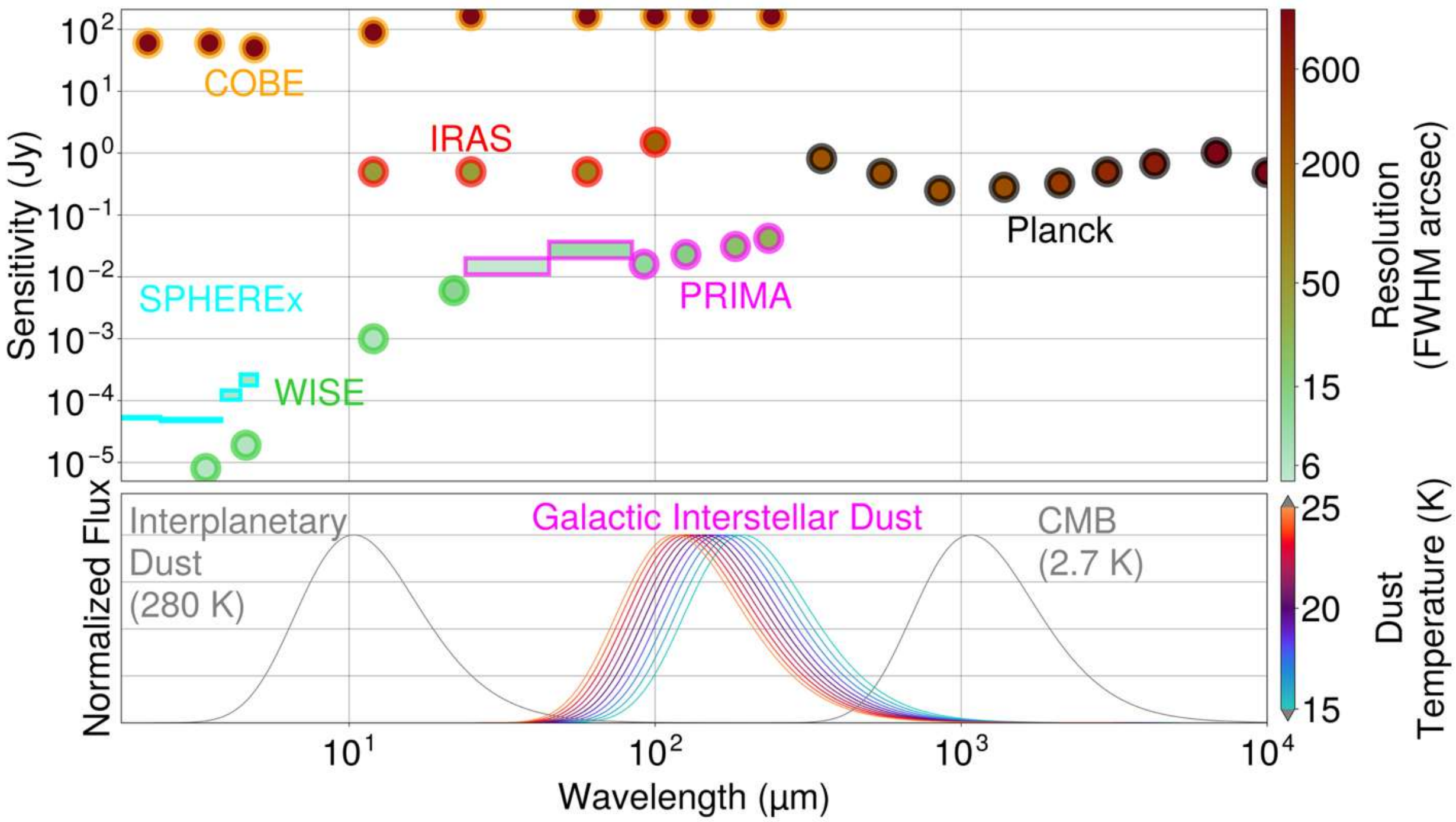

**Figure 1.** (*Top*) Summary of (5σ point source) sensitivity, wavelength coverage, and angular resolution of previous all-sky infrared surveys for comparison to PRIMA. (*Bottom*) Black body curves at representative temperature for interplanetary dust and the CMB for reference. Black bodies of temperatures typical for Galactic interstellar dust (15–25 K) are shown in color. Note that PRIMA will provide the best all-sky sensitivity over the full wavelength range of interstellar dust emission, as compared to Planck which only covered the Rayleigh-Jeans tail of the dust emission.

**PRIMA will reveal the fine-grained structure of molecular clouds across the Milky Way:** FIR dust emission is one of the most powerful tracers of the environments where stars form, with previous facilities like Herschel and Planck informing much of our understanding of the internal structure of molecular clouds (e.g. filaments and cores; Andre et al. 2010, Zari et al. 2016, Molinari et al. 2010). Compared to Herschel's limited area coverage and Planck's relatively poor sensitivity and resolution, PRIMA's all-sky 2D dust emission maps will provide a near-complete inventory of cloud masses at unprecedented dynamic range with implications for our understanding of physical processes (gravity, turbulence; Chen et al. 2017) shaping cloud structure.

Moving beyond 2D, 3D dust maps have recently used high-stellar densities to achieve angular resolutions (Zucker, Saydjari, & Speagle et al. 2025) rivaling the highest angular resolution FIR 2D dust observations (e.g., Herschel/SPIRE, Griffin et al. 2010). However, the resolution of 3D dust maps is fundamentally limited by the (stellar) source density of extinction measurements. Joint processing of FIR 2D emission maps with stellar extinctions is a unique route to vastly **increase the angular resolution** of 3D dust maps.

The resulting high-resolution 3D dust maps will dramatically expand the spatial scales and environments over which we can study the statistical properties of the ISM, such as filament widths, cloud mass and curvature, bubbles, and hierarchical substructure. These measurements are essential to understanding how and where star formation is initiated in 3D: whether in isolated pockets or interconnected cloud complexes, whether magnetically aligned or feedback-triggered, and how these conditions vary with the Galactic environment. PRIMA is the only facility that can provide high resolution, all-sky dust mapping to fill this need. These 3D capabilities will also enable recovery of cloud properties in previously inaccessible environments across the Milky Way, where confusion (e.g. superposition of multiple clouds at different distances in the Galactic plane) limited previous studies.

**PRIMA will provide a high-resolution map of the Galactic magnetic field:** Many basic questions remain on the role of the magnetic field in shaping the ISM, from MHD turbulence in diffuse ISM filaments to the formation of molecular clouds and the eventual collapse during star formation. Planck all-sky polarization maps revealed the Galactic magnetic field and reshaped our understanding of the magnetic field's role in the diffuse ISM (Soler et al. 2016), but tracing finer scales across ISM phases remains limited by Planck's angular resolution. Finer scale polarization mapping from SOFIA (e.g., FIREPLACE, Paré et al. 2024) and balloon experiments like BLASTPol (Fissel et al. 2016) provide the needed improvement in spatial resolution, but only offer maps of small areas of the sky. PRIMA will provide Planck's all-sky coverage with the high angular-resolution previously limited to dedicated efforts like BLASTPol and SOFIA, thus delivering an unprecedented high-resolution view of the Galactic magnetic field (e.g. Unger et al. 2024). Based on previous Herschel observations, we predict that our PRIMAger all-sky polarization survey will be able to map the magnetic fields of all visible Milky Way molecular clouds, with at least 10 times better resolution than Planck, for all dust with AV> 3 (i.e., we will be sensitive to magnetic fields in the low-density envelopes of clouds, including thousands of filamentary structures identified with the Herschel HiGAL survey, Schisano et al. 2020). These sensitivity estimates make the conservative assumption that all dust in molecular clouds is 2% polarized; however, low column density dust often shows significantly higher fractional polarization levels (e.g., Planck Collaboration, 2020b), so we will likely be sensitive to polarization at even lower dust columns.

PRIMA all-sky polarization mapping will enable measurements of the changes in the optical extinction curve that result from dust grain alignment with the local magnetic field (R(V)\~0.2, Hensley 2025). At low dust densities (high Galactic latitude), FIR polarized dust emission as a result of this alignment is likely the highest signal-to-noise measurement of these extinction curve variations, which has implications for extragalactic extinction corrections. Practically, we can also leverage this polarization as an additional axis to improve the subtraction of light from interplanetary (zodiacal) dust emission (Ganga et al. 2021) and point sources.

**All-sky PRIMA dust maps will improve cosmology surveys:** Precise extinction measurements are crucial for reliable observations of ultra-large-scale galaxy clustering, as probed by surveys such as DESI, SPHEREx, and Rubin. Important forecasting work by Huterer et al. (2013) showed that photometric calibrations need to be understood at the 0.1% (1 mmag) level for next generation precision cosmology experiments that rely on galaxy clustering. They specifically call out systematics (3 mmag) in SFD98 identified using "standard crayons" by Peek & Graves 2010 as a major limitation on those photometric calibrations, and one without a clear route to a solution.

A recent explicit example of the limitations of SFD98 for cosmology is Zhou et al. 2024, where systematics and the low angular resolution resulted in a systematic floor to the precision of DESI's cosmological measurement introduced in the selection function of the cosmological tracers. Using a combined analysis of DESI imaging and spectroscopy, the authors derived corrections relative to SFD98 with an RMS at the level of 13 mmag. For context, the authors show that a 10 mmag change in E(B-V) caused a change in the expected target density of emission line galaxies (ELGs) of 10%, which can significantly impact cosmological parameter estimation (e.g. $f_{NL}$, σ8), and is far worse than the forecasted precision requirements from Huterer et al. 2013.

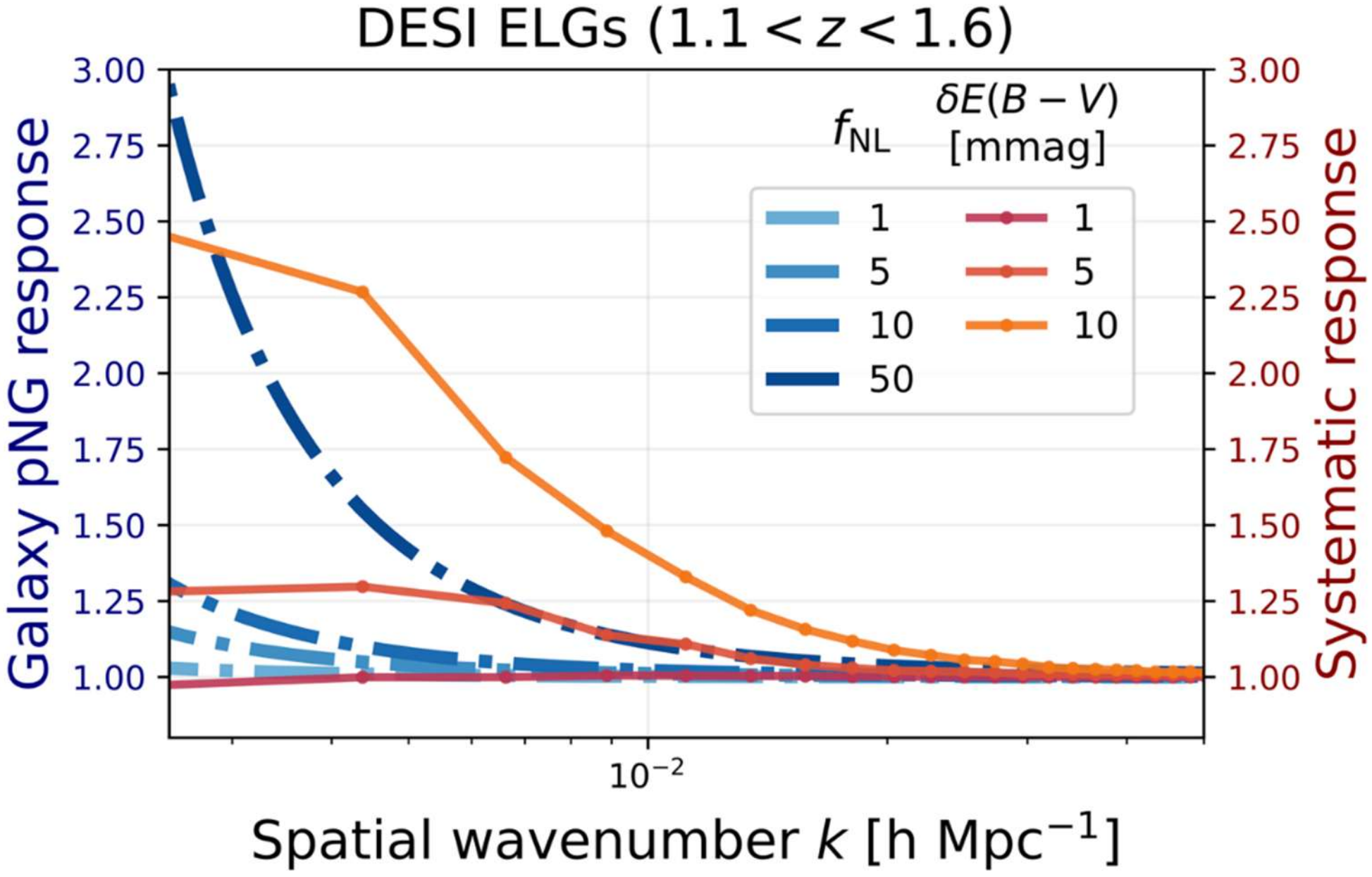

**Figure 2.** Comparison of fractional galaxy power spectrum response to $f_{NL}$ and systematic biases introduced by extinction errors. Without reducing the extinction errors in SFD to the levels achievable with PRIMA, next generation cosmology surveys will struggle to achieve σ($f_{NL}$) 1 sensitivity in tests of inflationary physics.

We quantitatively demonstrate the impact of extinction systematics on cosmological parameter inference in Figure 2, which compares the fractional galaxy power spectrum response to local primordial non-Gaussianity (pNG) — parameterized by different values of $f_{NL}$ (dashed curves) — with the systematic bias introduced by extinction errors, simulated using DESI ELG EZmocks (solid curves). Following Huterer et al. (2013), we emulate extinction errors via contamination fields that match the angular power spectrum of the SFD map, scaled to various RMS levels. While ELG selections are more sensitive to dust extinction than other galaxy tracers, Figure 2 highlights that, without reducing reddening systematics to the 1 mmag level, extinction-driven biases can mimic the scale-dependent signature of primordial non-Gaussianity and thus compromise the ability of next-generation surveys to achieve σ($f_{NL}$) 1 sensitivity in tests of inflationary physics.

In addition, PRIMA's improved constraints on dust parameters (β and T) will improve modeling of dust foregrounds, yielding cleaner measurements of the CMB and CMB spectral distortions. Uncertainty on the complexity of the dust foreground will be a fundamental limitation for every planned and future CMB experiment (e.g., Simons Observatory, South Pole Observatory, CMB-S4, LiteBIRD, PICO).

**PRIMA will resolve a significant fraction of the Cosmic Infrared Background (CIB):** Existing measurements with Spitzer/MIPS and Herschel suggest that >50% of the CIB intensity can be resolved into sources above 10 mJy (Dole et al. 2006, Béthermin et al. 2012, Oliver et al. 2012), and a plethora of techniques exist that can facilitate effective source de-blending and separation from structured foregrounds (Schlafly 2021, Saydjari et al. 2022, Feder et al. 2023), including those that specifically utilize the polarization capabilities of PRIMA (Donnellan et al. 2024). A full sky catalog derived from PRIMA would measure >$10^7$ galaxies, which enables a host of galaxy evolution and cross-correlation studies with existing CIB tracers.

CIB maps derived from PRIMA are valuable for “de-contamination” of existing FIR-based extinction maps (e.g., Chiang 2023), which is needed for accurate measurements of cosmic dust and global star formation history (Chiang et al. 2025). Thermal dust emission from the CIB has been identified as a major limiting foreground in studies of large-scale structure, including those from measurements of CMB secondary anisotropies (McCarthy et al. 2024, Schaan et al. 2018) and FIR line intensity mapping, e.g., [CII] with CONCERTO (Van Cuyck et al. 2023). PRIMA full-sky CIB maps would lead to significant improvements over current mitigations in these domains, which are primarily limited to Planck data.

## Technical Justification

**Why All-Sky:** We have presented several science justifications that benefit most strongly from either coverage of high Galactic latitudes (extinction correction for cosmology/extragalactic studies, CIB), the Galactic plane (Galactic magnetic fields, fine structure of molecular clouds), or the entire sky (point source catalog for targeting, dust temperature map for ISRF modeling, dust spectral index for composition variation modeling). Together, we think these cases form a compelling argument for a large and early allocation of time for a homogeneous all-sky survey.

**PRIMA Dust Temperature Coverage:** Making a dust map from infrared emission requires sensitivity to blackbody spectra at the temperatures of Galactic dust, typically 15–25 K. SFD98 used the 100 µm and 240 µm COBE/DIRBE bands (smoothed to 1.3° FWHM) to constrain the dust temperature. The 25 µm COBE/DIRBE band was used as a template for removing emission from interplanetary (zodiacal) dust at ∼280 K. This was then combined with higher sensitivity and angular resolution IRAS 100 µm data to produce the final map. The final SFD98 data product is thus an extinction map with 6.1 arcmin angular resolution and 0.4 mmag extinction precision, but only 3–10 mmag extinction accuracy due to systematics (see above). Further, its conversion from FIR emission to dust extinction relies on dust temperatures from much lower angular resolution (1.3°) and (44x) lower sensitivity DIRBE data, which unfortunately limits the effective angular resolution and extinction precision/accuracy of SFD98.

In the decades since IRAS (1983) and COBE (1989), we have not had a mission capable of making an all-sky FIR map at wavelengths near the peak of Galactic dust emission. However, blackbody peak emission falls within the bandwidth of PRIMA polarimetric bands for dust at temperatures of 11–36 K, which is well matched to typical Galactic dust temperatures of 15–25 K. Additionally, the PRIMA hyperspectral bands are sensitive down to 24 µm, providing sensitivity to emission from interplanetary dust so that it can be removed/separated from emission from Galactic dust. Thus, PRIMAger’s wavelength coverage makes PRIMA a perfect dust mapper, well-positioned to convert FIR emission to interstellar dust density (Figure 1).

**Dust Map Trade Space:** Figure 3 shows the extinction precision–angular resolution–exposure time trade space for an all-sky dust map with PRIMA, estimated using the 92 µm band, and assuming 18.4 mmag of E(B-V) per MJy/sr at 100 µm found by SFD98. Because most “diffuse” maps trade off angular resolution for smoothing data in order to achieve higher signal-to-noise ratio, we present this space as extinction precision in mmag of SFD E(B-V) versus pixel scale in arcsec. Colored level sets trace this trade off for a constant number of hours allocated to this PRIMA all-sky GO program, with our recommendation of 5000 h highlighted in red (15% of GO time).

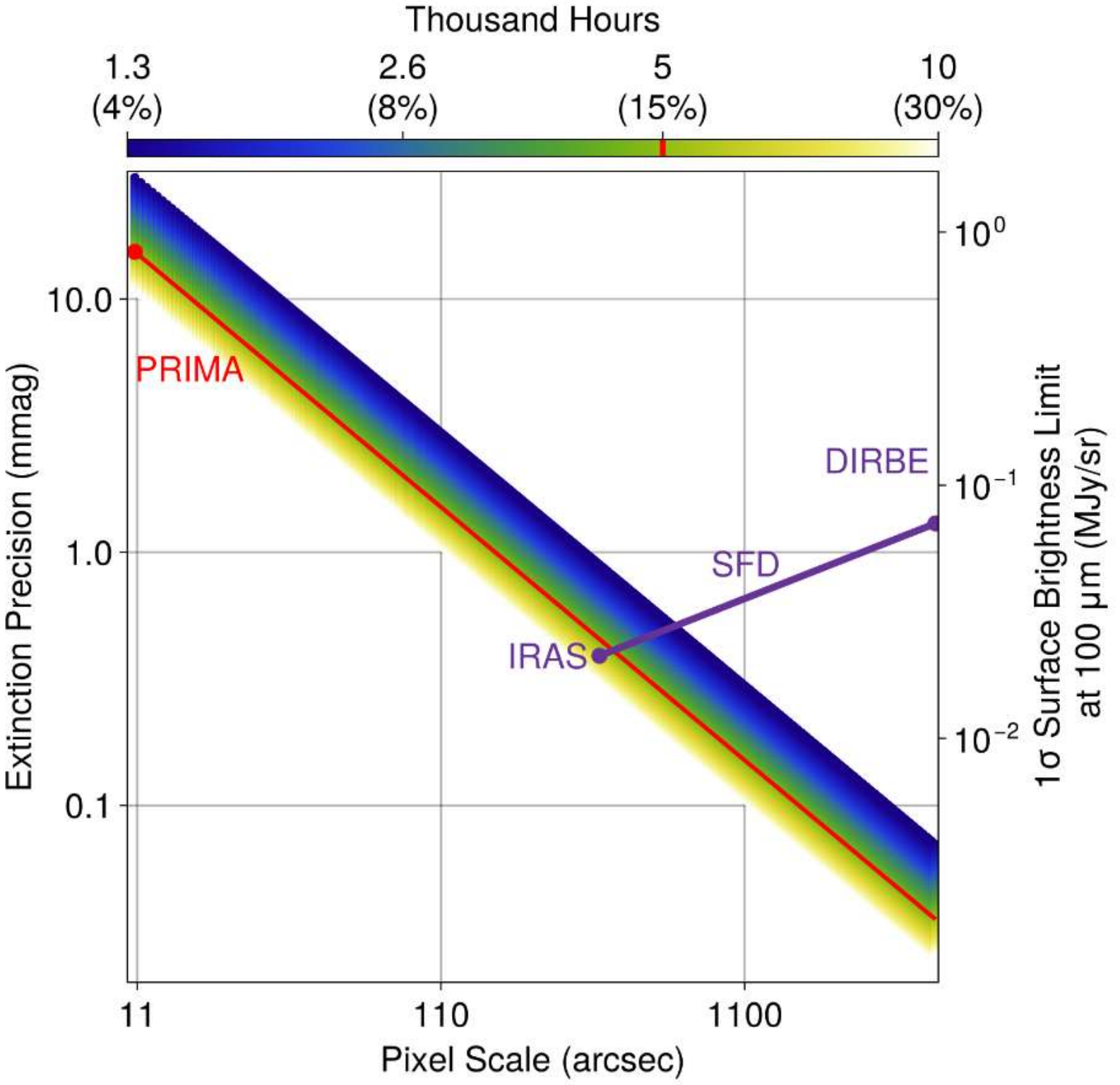

**Figure 3.** Trade space for PRIMA dust map extinction precision, smoothed resolution, and GO program time allocation. Recommendation of 15% highlighted in red. Time ranges from minimum time for an all-sky epoch with PRIMAger polarimetry (4%) to 30%.

Scatter points indicate the sensitivity at 100 μm and the highest angular resolution of PRIMA and the SFD98 datasets (IRAS, DIRBE). Clearly, a PRIMA all-sky survey provides orders of magnitude (34x, 433x) increase in angular resolution. At the same angular resolution as IRAS, the PRIMA all-sky survey will have comparable extinction precision to IRAS and 38x the sensitivity of DIRBE. Further, by simultaneously observing in all 4 polarimetric bands, a PRIMA all-sky survey will finally break us free of the awkward situation where the dust extinction and temperature measurements are at different angular resolutions and sensitivities, as they are in SFD98.

**Point Source Catalog:** This all-sky survey will also produce a point source catalog 63x deeper than the IRAS PSC, with polarization information (Table 1). The 5σ-limits for point sources are shown for both the polarimetric and hyperspectral bands in Table 2. This deeper point source catalog will be crucial for reducing the point source contamination compared to SFD98, which is currently limited by these systematics. Extrapolations from 50σ (2% photometry) detections in AKARI predict this catalog would contain 0.5–10 million sources, compared to 250,000 in the IRAS PSC (Wright et al. 2023). This all-sky point source catalog, if prioritized early in the mission, would enable serendipitous discovery of new FIR, polarized sources of interest (e.g. YSOs, debris disks, protostars) down to 16 mJy, providing targets for additional GO time follow-up studies. The hyperspectral observations would enable detailed SED modeling for sources and studies of variations in the FIR dust SED, similar to recent successes by Gaia XP and SPHEREx (Korngut et al. 2018, Green et al. 2024, Zhang & Green 2025).

Table 1. Comparison of angular resolution, point source and surface brightness limits, and dust precision for 100 µm all-sky surveys.

| Mission @100 $\mu m$ | 5σ Point Source (Jy) | Angular Resolution (arcsec) | 1σ Surface Brightness (MJy/sr) | | E(B-V) mmag | |
|---|---|---|---|---|---|---|
| | | | Native | 6.1’ | Native | 6.1’ |
| DIRBE | 165. | 4680. | 0.07 | 0.93 | 1.3 | 17.1 |
| IRAS | 1.5 | 366. | 0.021 | 0.021 | 0.39 | 0.39 |
| PRIMA | 0.016 | 10.8 | 0.8 | 0.029 | 15.4 | 0.45 |

Table 2. 5σ-limits for point and extended sources per band for the 4-band, all-sky polarimeter survey and hyperspectral imager survey with PRIMAger. [Limits on polarized intensity and polarized surface brightness are shown in square brackets.]

| Band | Center Wavelength (µm) | 5σ Limit [Polarized] | | | |
|---|---|---|---|---|---|
| | | Point Source (mJy) | | Surface Brightness (MJy/sr) | |
| PPI1 | 92 | 16 | [23] | 4 | [6] |
| PPI2 | 126 | 23 | [33] | 3 | [4] |
| PPI3 | 183 | 31 | [42] | 2 | [3] |
| PPI4 | 235 | 42 | [59] | 2 | [2] |
| PHI1 | 24—45 | 11—20 | | 15—6 | |
| PHI2 | 45—84 | 20—37 | | 7—5 | |

**GO Synergies:** This point source catalog would provide both accurate fluxes and polarization fractions, which would inform the exposure time calculations for other GO programs. Many of the polarization-focused GO programs currently use conservative estimates for this calculation, which may significantly overestimate the needed exposure time. As a result, while a substantial time investment, the PRIMA all-sky survey will likely recover some of this time investment by improving those estimates. Several smaller GO programs identified in the PRIMA GO V1 Handbook (Moullet et al. 2023) would reach their required sensitivities under the all-sky survey (#4, #63, #70). In addition to those science cases already identified, we anticipate that a PRIMA all-sky survey will enable population-level statistical studies of point-source classes and investigations of diffuse / low-surface brightness objects with arcmin angular scales. The 3 epoch cadence for the polarimetric band all-sky coverage will provide synergies with transient studies.

## Instruments and Modes Used

PRIMAger will be used for mapping the whole sky (41253 square degree) parallel with the polarimeter and the hyperspectral bands. 1 epoch in hyperspectral band, 3 in polarimetry band.

## Approximate Integration Time

5000 h (15% of 75% of 5 years, the GO time)

*based on https://prima.ipac.caltech.edu/page/instruments (2025/05/18), which includes estimated overheads.

## Special Capabilities Needed

PRIMA will be an amazing mission for dust mapping. Here we try to outline a few considerations that are specific to all-sky surveys and dust maps and try to propose feasible modifications to survey strategies and calibrations to make PRIMA even better for dust mapping.

**Low-Frequency Noise:** Making a dust map, or any map of extended structures with large angular size, requires careful handling of the “backgrounds.” This motivates a cadence of calibrations using the internal relative calibration source sufficient to achieve less than 0.5 kJy/sr drift in gain between observations. Making a PRIMA dust map would also be significantly easier if correlated, low-frequency (“1/f”) noise were rejected using “dark detectors,” rather than by removing all low-frequency spatial scales in the images. The KIDs for these “dark detectors” are co-fabricated with their optically-coupled kilopixel counterparts, however their coupling is purposefully limited, e.g., by not having a matched lens. This is common practice in many detector arrays and is done in general for stability monitoring. For the proposed effort they would serve as references to capture common mode 1/f noise unrelated to any FIR excitation. The addition of ``dark detectors’’ to the instrument would thus significantly strengthen the instrument capabilities for dust mapping.

**Zodiacal Light:** It will be important to be able to carefully model and remove the emission from zodiacal dust, which contributes significantly to the observed FIR emission, but negligibly to optical extinction. At 100 µm, the zodiacal dust is only 10 MJy/sr in the ecliptic plane (SFD98), so the simultaneous observations with the PRIMAger hyperspectral bands (24–84 µm) are important for capturing the emission peak of the zodiacal dust. The intensity of the zodiacal light is also very sensitive to the dust temperature, and thus the solar elongation angle of the observations. It will be useful to design a survey strategy that incorporates some observations at the full range of Solar elongations (as much as possible within the field of regard) to map out these temperature variations, but spends most of its time observing at Solar elongations near 90°. Modeling zodiacal dust emission also requires that PRIMA process and release images with reasonable relative zero point calibration before coaddition, especially before coaddition across observations taken with different Solar elongations (i.e., calibrated single exposures).

**Absolute Zeropoint:** A method of calibrating an absolute zeropoint to better than 0.05 MJy/sr at 100 µm (1 mmag in E(B-V)), would strengthen this science case. Almost all extinction corrections to date are tied to SFD98, which inherits its zero point from the COBE/DIRBE internal reference to a “cold-load” (Boggess et al. 1992) and the zero-point calibration during zodiacal light subtraction using the Leiden-Dwingeloo 21-cm HI map (Hartmann & Burton 1997). An absolute zeropoint from PRIMA would represent an opportunity to significantly reduce the uncertainty and potential systematics in extinction correction zero-points, which is especially important at high Galactic latitudes and extragalactic work. Frequent observations of an internal reference after each retuning of the Kinetic Inductance Detector (KIDs) on sky and a rigorous, laboratory calibration campaign is a possible solution.

KID responsivity (i.e. resonator frequency shift as a function of optical power) does not fundamentally drift, except for potential sub-Hz low frequency noise. Systematic laboratory calibration with a cold black body can thus be pursued in lieu of an absolute in-flight temperature reference. Such calibrations can be used to infer on-sky optical loading from the KID response. This process could be sufficient to achieve the target absolute zeropoint precision if systematics are properly accounted for. The two main systematics are discussed below, and neither are prohibitive. The detector and instrument teams can pursue these with existing testing platforms and a well-defined work plan to deliver on these sensitivity metrics.

First, the optical transfer function of a lab system is different from, though arguably much simpler than, the complete optical pathway of the full telescope. We can thoroughly measure the optical transfer function in the laboratory with a Fourier Transform Spectrometer and use basic inverse computation methods to calibrate the optical power, per spectral channel. This effort for a laboratory-based blackbody-based calibration has been done by the JPL KID team (Day et al. 2024) and can be repeated on the flight-grade detectors with their flight-grade filters. A separate calibration of telescope optics before integration and testing is already planned. Thus, it is feasible to connect these two calibrations to achieve an absolute zeropoint calibration, if they are held to metrics tied to this science requirement.

Second, magnetic flux from the environment and the ADR cooler can become trapped and affect the superconducting properties of KIDs. The amplitude of this effect can be quantified in laboratory testing, with detectors in flight-like packaging. In this test, we would measure the variation in KID response under different magnitudes of externally applied magnetic field. The observed fractional frequency shift will quantify the KIDs' magnetic field susceptibility. We can also repeat the aforementioned black-body calibrations after cryogenic cycling in the presence of known magnetic fields to study the calibration repeatability. Such efforts are standard in CMB and other detector development programs. Practically, these tests will provide guidance on the level of magnetic field shielding required to make this systematic quantifiably subdominant compared to the required absolute zeropoint precision.

## Synergies with Other Facilities

The PRIMA all-sky dust map will enable precise extinction corrections that will improve photometric redshift estimations at the core of key science projects in the LSST at the Vera C. Rubin Observatory, Euclid, and the Nancy Grace Roman Space Telescope. The resulting extinction maps will also be invaluable for understanding and accounting for the foreground dust extinction and emission affecting cosmological tracer samples used by wide-field spectroscopic surveys (e.g. DESI, 4MOST, PFS), photometric surveys (e.g. Euclid, LSST), and planned/future CMB experiments (e.g., Simons Observatory, South Pole Observatory, CMB-S4, LiteBIRD, PICO). The fluxes from the point source catalog will be invaluable FIR points for broad wavelength SED modeling in conjunction with other all-sky photometric catalogs from ROSAT, eROSITA, GALEX, UVEX, Gaia, 2MASS, Roman, SPHEREx, WISE, and the VLA.

While the hyperspectral imager should be sufficient to constrain and remove contamination from interplanetary dust, SPHEREx imaging would provide a helpful additional constraint on the blue end of the interplanetary dust SED. Measurements of HI from DSA-2000 and ASKAP will provide important independent (cosmic infrared background free) measurements of the ISM column densities, which we will use for calibration, zeropointing, and for studies of variations in the gas-to-dust ratio.

## Description of Observations

We propose dividing the total exposure time into 3 epochs of all-sky PRIMAger polarimetric imaging, spatially offset to combine to 1 full-sky pass with the hyperspectral imager.

**Cadence:** Ideally, the first polarimetric all-sky pass would be completed early in the mission for synergies with targeting for other GO programs. Breaking up the total all-sky exposure time into three epochs, has the significant benefit of enabling transient studies and improving relative calibrations, while only slightly complicating the zodiacal light modeling.

**Scanning Strategy:** The best calibrations are “self”-calibrations— calibrating observations with one part of the instrument against another. This has been borne out time-and-again, such as with the calibration of SDSS (Padmanabhan et al. 2008; Finkbeiner et al. 2016), GALPHA-HI DR2 (Peek et al. 2018), DECaPS2 (Saydjari et al. 2023), and APOGEE (Saydjari et al. 2025), using methods sometimes called “ubercalibration.” Observational strategies that are uniform (such as regular tiling) are anathema to self-calibration, leading to (paradoxically) non-uniform photometric precision across the field of view and an order of magnitude decrease in the average photometric precision (e.g. Holmes et al. 2012, 1.8 mmag versus 12.7 mmag uncertainties for non-uniform versus uniform tilings).

Thus, the careful addition of non-uniform offsets and rotation angles to the survey strategy has the potential to increase the photometric precision of the survey by an order of magnitude. This argues against 10° x 10° tiling or sweeping the sky in great circles between the Ecliptic poles, which is also inefficient and non-uniform in depth, despite the fact that these are the simplest survey strategies. However, the elements of non-uniformity need to be traded off against the orientation direction and spatial offset (between epochs) requirements for the hyper-spectral imager as well as solar elongation considerations discussed above.

**Maximum Mapping Rate:** For an all-sky survey, it is relevant to consider the limitations on the maximum mapping rate, which sets a lower bound on the number of hours one can allocate to get a single all-sky epoch. We will estimate this using the “blurring limit,” when one is slewing so fast that high-frequency spatial structure is lost because of spatial averaging over the readout. A preliminary estimate for that rate (for the most blurring-sensitive band at 24 μm) is when the observatory is moving at 7 arcmin / second.

For a survey with the hyperspectral imager, movement of the telescope in the spectral direction would then yield a maximum mapping rate of 1.5 arcmin x 7 arcmin / second, or 10.5 sq. deg. /

hour (conservative, using narrower PH1). This means, the minimum time to map an epoch of the full sky in the hyperspectral bands is 3900 hours. The wider footprint of the polarimetric bands yields a faster mapping rate of 4.6 arcmin x 7 arcmin / second, or 32.2 sq. deg. / hour (conservative, using narrower PPI4). This means, the minimum time to map an epoch of the full sky in the polarimetric bands (with 27% sky coverage in the hyperspectral bands) is 1300 hours. Both of these maximum mapping rate estimates are larger than our recommended rate of 8.2 sq. deg / hour, illustrating that we can do both an all-sky polarimetric and hyperspectral survey in the recommended 5000 h.

# 93. Investigating the Role of Magnetic Fields in the Formation and Evolution of Striations in Interstellar Clouds with PRIMA

Raphael Skalidis (Caltech & JPL, USA), Konstantinos Tassis (University of Crete & IA-FORTH, Greece), Aris Tritsis (EPFL, Switzerland), Paul F. Goldsmith (JPL, USA)

The importance of magnetic fields in star formation is a longstanding open question. Observations indicate that magnetic fields influence the gas properties and kinematics of diffuse interstellar medium regions, but ambiguities persist in some models. Striations—linear and magnetized low column density structures found in the outskirts of molecular clouds—are ideal testbeds to investigate the role of magnetic fields in the early phases of star formation. PRIMAger offers us a unique view of the magnetic field properties of striations by observing the polarized thermal emission of striations. It will require around 59 hours to observe the striations of three nearby molecular clouds – the Polaris Flare, Musca, and Taurus—at bands 3 (172µm) and 4 (235µm) of PRIMAger. In this article, we update the estimated fluxes and integration times reported in volume 1 of the G.O. book. We refer to the peer-reviewed article by Skalidis et al. (2025), published in the JATIS PRIMA paper collection, for a detailed discussion of the proposed theoretical models for magnetized striations, flux estimates derived using PRIMA, and the expected scientific outcomes.

## Science Justification

The star formation rate of our Galaxy (1.3 Solar masses per year) is almost two orders of magnitude lower than what is expected from the available gas mass reservoir (Zuckerman and Evans 1974). Star formation theories suggest that interstellar medium (ISM) magnetic fields play a key role in suppressing the formation of stars (Mouschovias 1976). If we consider flux-freezing conditions—which accurately apply to molecular clouds with ionization fractions of the order of $10^{-4}$—the compression of the field lines due to gravitational contraction would significantly enhance the magnetic energy density, hence preventing the cloud from collapsing. A pathway to forming stars comes from ambipolar diffusion, which allows neutral particles to slide from the bulk motions of ions, which directly interact with magnetic fields, and form local centers of gravity (Mouschovias et al. 2011). However, the competition of magnetic fields with other pressure terms, like radiation from feedback, obscures the exact role of magnetic fields in star formation models (e.g., Suin et al. 2025).

Dust polarization observations over the last decade have shown that magnetic fields influence the gas distributions and kinematics in the ISM. This is evident in the strong correlations between HI and molecular column density structures with the ISM magnetic field orientation (e.g., Clark et al. 2014, Skalidis et al. 2022). Additionally, both the kinematics and column densities show anisotropies with respect to the local mean orientation of the magnetic field, which is consistent with models of strongly magnetized turbulence (Heyer et al. 2008). The Planck Collaboration concluded that turbulence in ISM molecular clouds is sub- to trans- Alfvenic, which indicates dynamically important magnetic fields in the ISM (Planck Collaboration 2016).

One of the most striking correlations between the magnetic field orientation and gas-dust structures in the ISM was established in the diffuse parts of the Taurus molecular cloud. Goldsmith et al. (2008) discovered linear column density structures, called striations, which are well aligned with the magnetic field and vary quasi-periodically across the magnetic field lines (Fig. 1). The observed properties are consistent with the propagation of compressible magnetohydrodynamic (MHD) waves (Tritsis & Tassis 2016). Striations are in some cases linked to dense filaments with protostars and are thus considered a necessary element to form stars (Palmeirim et al. 2013). Therefore, striations are ideal testbeds to investigate the role of magnetic fields in the early phases of star formation.

PRIMA is uniquely equipped to study the magnetized striations. PRIMAger's sensitivity and angular resolution would probe the small-scale fluctuations of striations in both translucent and dense molecular clouds. The polarization data from PRIMAger would allow us to test key predictions of the linear MHD wave theory and calibrate methods for estimating the strength of the magnetic field that use the dispersion of polarization angles (Skalidis & Tassis 2021, Skalidis et al. 2021) and the dispersion relation of linear MHD waves (Tritsis et al. 2018).

Below we present the main requirements of the proposed survey and report the updated fluxes and integration times of the corresponding science case presented in Volume 1 of the G.O. book. A detailed discussion of the science motivation, theory, and flux calculations is included in the peer-reviewed article of the PRIMA paper collection by the JATIS journal (Skalidis et al. 2025).

## Instruments and Modes Used

PRIMAger mapping, polarimeter bands at 172 μm and 235 μm with polarimetric measurements. One large map (1 deg2) for each of the three targets in the sample.

## Approximate Integration Time

| | | | Integration time for a 10σ detection (hours) | | Polarization Fluxes (MJy/sr) | |
|---|---|---|---|---|---|---|
| **Cloud** | **R.A.** | **Dec.** | **172 μm** | **235 μm** | **172 μm** | **235 μm** |
| Polaris Flare | 05:46:19 | +87:57:12 | 114.9 | 98.2 | 0.15 | 0.12 |
| Musca | 12:26:24 | -71:59:48 | 1.4 | 0.7 | 1.36 | 1.36 |
| Taurus | 04:54:31 | +27:00:44 | 3.7 | 2.2 | 0.83 | 0.78 |

The integration times in the table are without smoothing.

## Special Capabilities Needed

None

## Synergies with Other Facilities

Synergies between the polarization data from PRIMAger and ground-based radio observatories are required for the completion of this program.

## Description of Observations

We propose to observe with PRIMAger the striations of three nearby molecular clouds (Fig. 1): the Polaris Flare, Taurus, and Musca. We have constrained the averaged magnetic field properties of these regions in past works (Taurus: Chapman et al. 2011, Tritsis et al. 2018, Polaris Flare: Panopoulou et al. 2016, Skalidis et al. 2023, and Musca: Tritsis & Tassis 2018). The target regions are ~1 sq. degrees each. The updated estimated average polarization fluxes with the corresponding integration times are given in the table below, where values are extracted from Skalidis et al. (2025). We expect ~20% intrinsic variance in the obtained fluxes. The reported integration times are based on the ETC to obtain 10σ measurements. In the Polaris Flare, we plan to smooth out the data and combine measurements from bands 3 and 4 to enhance the signal-to-noise ratio. In Taurus and Musca, high-resolution measurements are required, and thus we will employ the native resolution of bands 3 and 4. With the averaging procedure, the total integration time of the proposed program becomes ~59 hours.

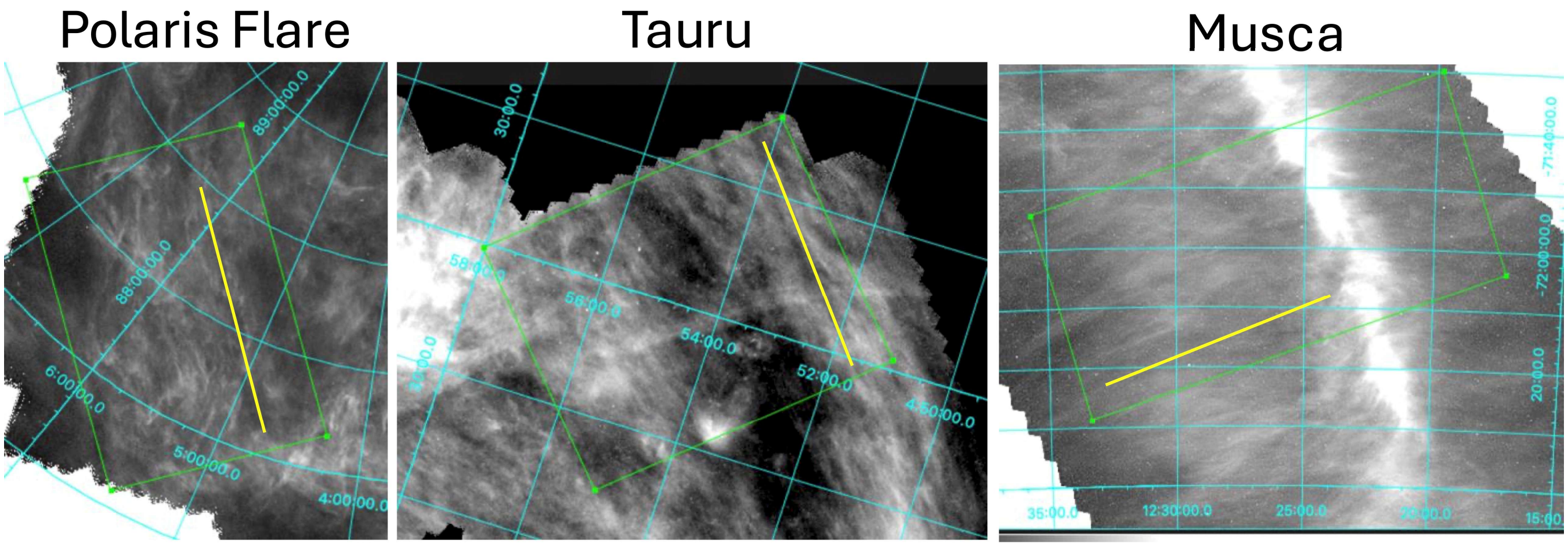

**Figure 1.** The green rectangles correspond to the proposed target region towards the three clouds with prominent striations. The background maps correspond to the total dust intensity at 350μm observed with the SPIRE instrument onboard the Herschel observatory. The yellow segments show the orientation of the striated structures, which are aligned with the plane-of-the-sky magnetic field morphology. The center coordinates of each regions are shown in the table. The area of each region is approximately 1 square degree.

# 94. Charting Circumstellar Disk Evolution through a PRIMA All-Sky Survey

Karl Stapelfeldt (NASA/JPL/Caltech, on sabbatical at IPAG / Grenoble), Nicholas Ballering (Space Science Institute), Mark Booth (UK Astronomy Technology Centre), Geoffrey Bryden (NASA/JPL/Caltech), Christine Chen (STScI, JHU), Joshua B. Lovell (Center for Astrophysics | Harvard & Smithsonian), Patricia Luppe (Trinity College Dublin), Meredith A. MacGregor (Johns Hopkins University), Jonathan P. Marshall (Academia Sinica Institute of Astronomy and Astrophysics), Brenda Matthews (Herzberg Astronomy & Astrophysics, NRC Canada), Tiffany Meshkat (IPAC/Caltech), Farisa Y. Morales (NASA/JPL/Caltech), Marion Villenave (IPAG / Grenoble), David Wilner (Center for Astrophysics | Harvard & Smithsonian)

An all-sky survey with PRIMAger can serve many diverse goals in astrophysics. Stellar infrared excess emission traces the presence of circumstellar dust, which can take the form of shells, disks, and envelopes. While the 2010 WISE all-sky survey has characterized the population of stars with warm circumstellar dust across mid-infrared wavelengths, no comparable all-sky survey has been done in the far-infrared since the IRAS satellite in the 1980s. The cool dust emitting at 50 microns and beyond is a key tracer of the evolution of planetary systems, from protoplanetary disks to debris disks and to the mass loss processes of evolved stars. With orders of magnitude greater sensitivity than IRAS and broad spectral coverage, a PRIMA all-sky survey can explore trends in circumstellar dust luminosity and temperature versus stellar age, spectral type, metallicity, host star multiplicity, cluster environment, and the presence of planets. In conjunction with the WISE survey data, PRIMA's sample size will be large enough to reveal rare objects such as dusty brown dwarfs, edge-on protoplanetary disks, disks transitioning between the protoplanetary and debris phases, and disks that have undergone a major recent planetesimal collision releasing dust. A PRIMA far-IR all-sky survey will reveal the full populations of disks in various evolutionary states and around various host objects, seeding follow-up work with other facilities and with PRIMA itself.

## Science Justification

Circumstellar disks are found through the mid- and far-infrared excess emission of their host stars, emission that is produced by surrounding dust grains that absorb starlight and re-radiate it at longer wavelengths. Protoplanetary disks have very strong infrared excesses that can be orders of magnitude brighter than the stellar photospheres at these wavelengths, while debris disks are seen down to excess emission levels as little as 20% above the stellar photosphere. The Spitzer and Herschel missions mapped nearby star-forming regions to characterize the protoplanetary disk population, as well as making pointed surveys of main sequence star samples to study the incidence of debris disks. These surveys covered many thousands of stars in broad photometric bands, and a smaller sample via low-resolution spectroscopy, but were biased in favor of pre-selected targets. The WISE mission escaped this limitation by surveying the whole sky but could only detect warm disks with excess emission at 22 micron and shorter wavelengths. The nearby population of cooler disks emitting in the far-IR, which Spitzer results (Trilling et al. 2008) showed to be many times larger than the population of warm debris disks WISE could detect, remains inadequately surveyed and characterized.

A key question for disk evolution is to understand the transition from an optically-thick, gas-rich, relatively massive protoplanetary disk populated with ISM-like grains to an optically thin, gas-poor, and dust-dominated debris disk around a main sequence star. A class of "transition disk" objects has been identified by the community over the past several decades, based on central holes in their radial dust distributions. But other than their central holes, these objects have the same general properties as protoplanetary disks: small dust grains, large masses, and abundant gas that continues to accrete onto the star. A new approach is needed to identify the true transition disk objects, the systems where young planets first emerge from disk extinction into direct view.

A PRIMAger all-sky survey will find the full population of objects intermediate between protoplanetary and debris disks and fill out their spectral energy distributions. The basis for identifying them will be the use of the disk *fractional infrared luminosity* L_d/L_star to identify these transition objects (Wyatt et al. 2015). The debris disk community has long-used this parameter—the ratio of the infrared excess luminosity to the stellar luminosity—to characterize a disk's evolutionary state. The protoplanetary disk community has not, instead using the slope of the system's spectral energy distribution to characterize it, which cannot be compared directly to L_d/L_star. Yet L_d/L_star is a more useful parameterization because it separates the disk emission from the stellar emission and is an integral quantity tracing the total solid angle occupied by dust around the star. By collecting the missing far-IR data points not covered by WISE, the PRIMA all-sky survey will enable accurate and consistent L_d/L_star values to be derived for tens of thousands of dusty stars. Debris disks have L_d/L_star values at or below a few times 0.1%, while protoplanetary disks (including the so-called "transition disks") have L_d/L_star values ~10%. Relatively few transition objects with L_d/L_star values near 1% are known (the so-called "Peter Pan" disks for example, Silverberg et al. 2020). As the direct descendants of protoplanetary disks, they are expected to be dispersed from their formation regions, in areas not surveyed by Spitzer or Herschel. Do these "luminosity transition" disks generally have inner holes or not? The PRIMA all-sky survey will find this crucial missing link population in disk evolution, enabling a range of imaging and spectroscopic follow-up across the electromagnetic spectrum, and shining a light on the processes of disk dissipation, dynamical evolution, and the properties of young planets.

A PRIMA all-sky survey can be sensitive to faint debris disks down to the levels of Ld/Lstar of a few x $10^{6}$, or about 10x that of the solar system's Kuiper Belt, and thus help place our solar system in context with other planetary systems. To achieve this, far-infrared sensitivity down to the level of the stellar photosphere infrared excesses is needed. The Herschel DUNES survey (Eiroa et al.

2013) pursued this strategy for a sample of 133 FGK stars, detecting 25—a sample not large enough to robustly demonstrate trends in disk properties versus their host star properties. PRIMA should be capable of the needed sensitivity in a much larger target sample, which will 1) Allow the debris disk luminosity function to be placed on a much sounder statistical basis; 2) establish trends in dust temperature versus spectral type that will reveal whether disk incidence favors a specific radial temperature zone or one defined by dynamical timescales; and 3) indicate which nearby planetary systems are likely to have dusty habitable zones, as indicated by its correlation with the presence of Kuiper Belts (Ertel et al. 2020), and which will impede direct imaging and spectroscopy of Earth analogs.

## Instruments and Modes Used

PRIMAger PPI and PHI, all-sky survey

## Approximate Integration Time

The entire sky can be covered in six months as PRIMA scans along lines of ecliptic longitude. Six months corresponds to 5000 hours of integration time, as proposed in the all-sky survey design in the GO book 2 white paper by Saydjari et al [PRIMA Team: Add cross-reference]. For this survey area in this integration time, the achieved 5 sigma sensitivities are 16, 23, 30, and 41 mJy in the polarimetric bands at 92, 126, 172, and 235 microns respectively (neglecting source confusion). The simultaneously obtained hyperspectral maps at shorter wavelengths would be over-resolved spectrally for the needs of this program; the data will be rebinned to R= 4 to make four effective photometry bands centered at 28, 38, 53, and 72 microns respectively. This should provide about a factor of two better sensitivity per band than reported in the ETC for the hyperspectral mode, or 6, 8, 12, and 16 mJy (5 sigma) in the four synthesized hyperspectral bands. The 28-micron sensitivity would be comparable to the WISE survey's 22 micron point source sensitivity of 6 mJy (5 sigma; Wright et al. 2010), while the 92-micron sensitivity would be comparable to that of the Herschel DEBRIS survey (10 mJy, 5-sigma at 100 microns; Lestrade et al. 2025). This shows that the PRIMA all-sky survey would be well-matched for variability studies vs. the WISE survey in the small wavelength-region where they overlap, and to the prior Herschel survey (which was for select targets, not all-sky).

These sensitivities correspond to detecting the photosphere of a star with apparent K magnitudes of 7.4, 6.4, 5.2, and 4.2 at 28, 38, 53, and 72 microns respectively. To assess the reach of such a survey, we performed an analysis of the Hipparcos catalog. Using the B-V color of each star to map to its V-K color, the number of stars bright enough for a S/N = 5 detection of the photosphere was tallied. To restrict the sample to dwarfs and sub-giants, a star was rejected if its absolute V magnitude exceeded that of a main sequence dwarf, with the same B-V color, by more than 0.4 mag. The results are tabulated in the table below.

**Table 1.** Number of Hipparcos main sequence stars, broken down by spectral type, where the photosphere could be detected with 5$\sigma$ significance in the PRIMA all-sky survey described here, at the specified wavelengths.

| | A | F | G | K | M | Total |
|---|---|---|---|---|---|---|
| 53 μm | 169 | 273 | 100 | 56 | 12 | 610 |
| 72 μm | 50 | 73 | 26 | 17 | 1 | 167 |
| 92 μm | 22 | 25 | 10 | 8 | 0 | 65 |

If the disk is spatially resolved, then the sensitivity could be worse than the values above, which assume a sensitivity appropriate to point sources. The detection of the stellar photosphere at the Gaia/Hipparcos position will strongly mitigate source confusion as an issue for the SED beyond 100 microns.

## Special Capabilities Needed

N/A

## Synergies with Other Facilities

The PRIMA all-sky survey is designed to complement the 2010 WISE survey. Together they will provide the full SED coverage of mid-infrared and far-infrared excess needed to characterize the presence and evolution of circumstellar matter from the pre- to post- main sequence. The disks with 1% fractional luminosity found in this survey will be excellent candidates for near-/mid-infrared high contrast imaging follow-up (ELT) to discover and characterize young planets just emerging from their surrounding dust—as well as ALMA & JWST studies of their gas content, and ALMA mapping the distribution of their large grain populations.

## Description of Observations

All-sky survey in 6 months. Scanning along lines of ecliptic longitude, perpendicular to the sun direction. Observations will be made in all PRIMAger bands simultaneously, with the large survey area resulting in all sources being covered at all wavelengths.

## Acknowledgment

This research was carried out at the Jet Propulsion Laboratory, California Institute of Technology, under a contract with the National Aeronautics and Space Administration (80NM0018D0004).

# 95. The PRIMA View on Hydrocarbons in Planet-Forming Disks

Benoît Tabone (IAS, CNRS, Paris-Saclay University), P. Estève (IAS, Paris-Saclay University), M. J. Colmenares (University of Michigan), M-A. Martin-Drumel (ISMO, CNRS, Paris-Saclay University) M. Zannese (IAS, Paris-Saclay University), E. Habart (IAS, Paris-Saclay University), E. F. van Dishoeck (Leiden Observatory), A. M. Arabhavi (Kapteyn Astronomical Institute, University of Groningen), T. Henning (Max-Planck-Institut für Astronomie), J. Huang (Columbia University, Department of Astronomy), E. Pacetti (INAF-IAPS, Rome), G. Perotti (Niels Bohr Institute, Max-Planck-Institut für Astronomie), I. Kamp (Kapteyn Astronomical Institute, University of Groningen), K. Zhang (UW-Madison)

The atmospheric characterization of planets constitutes a new avenue to constrain their formation history since planets inherit their composition from the disk. One of the essential goals for the disk community is then to observationally constrain the partition of chemical elements across the disk and as a function of time. Ground-based (sub-)millimeter instruments and JWST provide unique constraints on the partitioning of elements in disks. Yet, the current facilities miss the critical region where the majority of giant planets are expected to form, located between $\simeq$ 1-10 au. Moreover, the emission of some of the key carbon and nitrogen carriers can only be observed in the far-IR. Today, the absence of a far-IR instrument limits our ability to interpret the atmospheric composition of planets. Here, we demonstrate the potential of PRIMA to probe the gas-phase carbon content of disks. We show that the emission of small hydrides like CH and $CH_2$ provides complementary constraints on the emission of the cold outer disk. In addition, the bending modes of at least $C_4H_2$ and $C_3$ constitute promising cases to constrain the C/O and C/H in the giant planet-forming regions (1-10 au) in synergy with HD and $H_2O$ lines covered by the PI science.

## Science Justification

### Context

Planets form, migrate, and obtain their elemental composition in disks orbiting young stars (a few Myrs). With JWST, ARIEL, and ground-based instruments (e.g., VLT-CRIRES, ELT-METIS) the diversity of exoplanets will be scrutinized from a new vantage point: their elemental compositions with the objective of using these constraints to trace back the history of planet formation. In this context, one of the goals of the disk community is to determine the radial distribution of elements as a function of time.

The *Herschel* Space Observatory and ground-based millimeter facilities have started the revolution by showing that the outer disks are likely depleted in oxygen and carbon (Bergin+2010, Miotello+2017) with elevated C/O (> 1) in the majority of the disks (Mitello+2019). Since 2022, JWST has provided an unprecedented view on the innermost regions of disks by accessing the mid-infrared emission of the main oxygen and carbon carriers. Focusing on Sun-like stars, disks appear on average bright in oxygen-bearing species pointing towards low gas-phase C/O inside of $\simeq$ 1 au. Therefore, disks seem to exhibit a radial gradient in the gas-phase C/O ratio between their outermost regions (≥20 au) and their innermost regions (≤ 1 au). However, where the transition between C-dominated and O-dominated regions is located remains an open question.

Today, the absence of sensitive far-IR instruments severely limits the characterization of disks questioning our ability to interpret the atmospheric composition of exoplanets. The FIRESS spectrometer onboard PRIMA will fill this critical gap. HD emission will provide the total hydrogen content in the outer and inner disk whereas $H_2O$ lines will trace the distribution of oxygen in intermediate regions where the majority of gas giants form (1-10 au). The access to the far-IR emission of ammonia will also complement the nitrogen budget in disks (Bergner, PRIMA GO book 1). Here, we demonstrate the potential of PRIMA to access the emission of carbon-bearing species to complement the carbon budget in the outer disk and peer into the gas-phase carbon in the intermediate regions (1-10 au).

## Results

### Missing C-carrier in the outer disk: rotational emission of hydrides

The far-IR domain covers the rotational lines of small hydrides that cannot be observed from the ground. Figure 1a shows a synthetic PRIMA-FIRESS spectrum predicted by the DALI thermochemical model (Bruderer+2012) for a typical disk around a 1 $L_\odot$ bright T Tauri star. The disk model is very similar to that of Bosman+2022 and Tabone +2024, but the chemical network includes hydrocarbons with up to 5 carbon atoms using both the UMIST and KIDA databases (Esteve+ in prep.). CH and $CH_2$ lines are predicted to be detectable by FIRESS for gas-phase C/O ratios above 1, whereas the line fluxes drop by more than an order of magnitude for C/O <1. FIRESS will cover the full rotational ladder of CH and $CH_2$ opening the possibility of detailed excitation studies. The brightest lines are not expected to be kinematically resolved, but the excitation of CH and $CH_2$ along with the constraints on the physical structure of the disks obtained by SED modeling and spatially resolved ALMA observations of CO will allow us to robustly trace back the emitting region of these hydrides. In general, we find that the detectable emission of CH and $CH_2$ originates from 10-50 au offering a synergy with ALMA and NOEMA observations of other small hydrocarbons like $C_2H$ and c-$C_3H_2$.

The CH+ rotational lines constitute another promising science case for PRIMA. *Herschel*-PACS detected $CH^+$ far-IR emission only for Herbig disks with cavities of about 5 au, highlighting the need for irradiated cavity walls to trigger FUV-driven ion chemistry (Thi+2011, Fedele+2013). This is well in line with the PACS and now JWST-NIRSPEC detections of $CH^+$ in interstellar PDRs (Parikka+2017, Zannese+2025) and the MIRI detection of $CH_3^+$ in TW Hya (Henning+2024). In contrast, $CH^+$ lines are not expected to be bright in full disks, unless they are exposed to external UV. In general, simultaneous observations of $CH^+$, CH, and $CH_2$ emission will provide fundamental constraints on the carbon chemistry under dense and irradiated conditions with a synergy with the studies of PDRs (see Zannese+ in this Volume).

**Gas-phase carbon in giant planet-forming regions: bending modes of hydrocarbons**

To access the intermediate regions where giant planets are thought to form, one could use rotationally excited lines of the hydrocarbons detected by ALMA. However, these lines are likely too weak as shown in Figure 1a for the case of $C_2H$. As demonstrated by JWST, the bending modes of the hydrocarbons give unique access to the volatile carbon content in the hot inner regions (≥ 300 K, ≤1 au). The far-IR regime is poor in bright rovibrational bands of hydrocarbons. For example, $C_2H_2$ has a band around 80 µm, but it is predicted to be orders of magnitude too weak to be detectable. However, of the most abundant hydrocarbons, diacetylene ($C_4H_2$), $C_3H_4$, and tricarbon ($C_3$) have strong emission bands in the far-IR. $C_4H_2$ is already detected by JWST around a Sun-like star (Colmenares+2024) via its 16 µm feature. The unique contribution of PRIMA will be to map the $C_4H_2$ to larger radial distances since the 45 µm feature corresponds to lower energy levels. The targeted spectral resolution of FIRESS of 30 km/s at 45 µm will be crucial to constrain the emitting region of $C_4H_2$. This will indeed allow us to measure the width of the individual lines on either side of the Q-branch (see Figure 1b) and infer their Keplerian broadening following deconvolution, as already done with MIRI-MRS (Banzatti+2025). $C_3H_4$, which is also detected by JWST around some very low-mass stars (Arabhavi+2024, Kanwar+2024), constitutes another promising species for PRIMA thanks to its bending mode around 30 m. The synergy between JWST and PRIMA observations of at least these two species will allow us to determine the radial extent of hydrocarbon enrichment.

$C_3$ is a special molecule since it possesses a very low-energy bending mode at 63.9 cm$^{-1}$ (156 µm, Martin-Drumel+2023) already detected in dense and warm star-forming regions by ISO and *Herschel* (Cernicharo & Goicoechea 2000, Mookerjea+2010).

The first excited state will likely trace $C_3$ in the cold outer disk while more excited bands will unveil the building up of carbon chains in the warm regions.

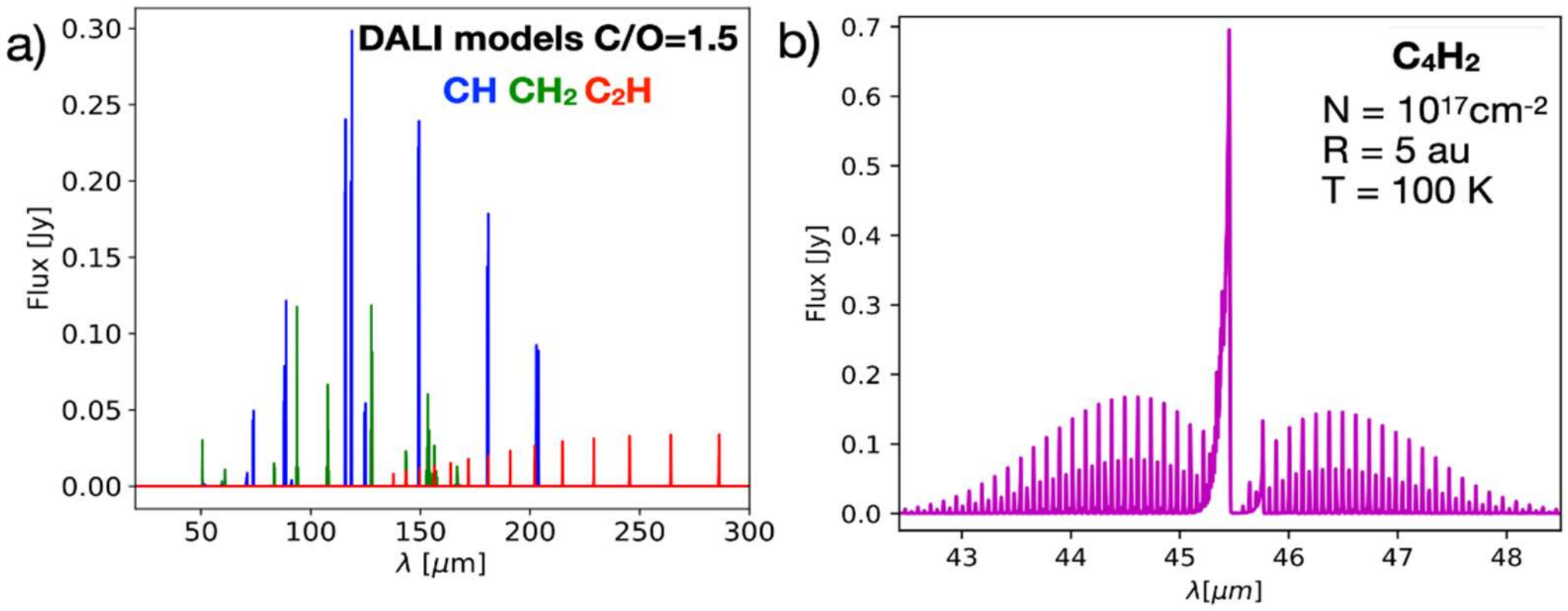

**Figure 1.** Synthetic spectra of hydrocarbons at PRIMA-FIRESS spectral resolution. a) DALI thermochemical model predictions for small hydrides, providing key insights into carbon chemistry in cold regions. The emission of $C_2H$ is also added. b) A slab model of $C_4H_2$ illustrating the potential of low-lying bending modes of hydrocarbons to probe the warm gas located in the formation zone of giant planets.

## Instruments and Modes Used

FIRESS in the high-resolution point source mode covering Band1+3 and Band 2+4 to access the emission of at least $C_4H_2$, CH, $CH_2$, and $C_3$.

## Approximate Integration Time

5h per source and spectral setting to reach a 5-sigma sensitivity of 6e-19 W $m^{-2}$ around 180 µm and 2e-18 W $m^{-2}$ around 46 µm assuming a continuum emission of 0.5 and 1 Jy respectively. With this sensitivity, we will detect 5 to 10 lines of CH and $CH_2$ and the Q-branches of $C_3$ and $C_4H_2$.

## Special Capabilities Needed

None.

## Synergies with Other Facilities

The (sub)millimeter facilities access the cold outer regions of disks (≥ 20 au) through the observation of rotational lines of small hydrocarbons. JWST reveals the warm inner regions (≤ 1 au) via the ro-vibrational emission from multiple carbon-bearing species rotational lines of OH and $H_2O$. In this context, PRIMA will uniquely complement these facilities by filling the 1-10 au observational gap and access species emitting from the outer disks but not observable from the ground.

## Description of Observations

We propose to use the FIRESS instrument in the high-resolution mode to observe a sample of 6 well-characterized disks around T Tauri disks ($M_{\odot}$=0.5-1 $M_{\odot}$), with existing JWST/MIRI-MRS and ALMA (or NOEMA) observations. We will target disks with bright hydrocarbons in their outer regions (typical of most T Tauri disks). This will ensure that hydrocarbons are present over extended spatial scales. To study the gradients in elemental abundances across the disks, three sources will have prominent mid-IR features of oxygen-bearing species ($H_2O$, $CO_2$) and three sources will exhibit bright hydrocarbon features with the MIRI-MRS detection of at least two hydrocarbons (e.g., $C_2H_2$ and $C_4H_2$). With 10 hours of integration time per source (60 hours total), we will be able to detect lines of CH, $CH_2$, $C_3$, and $C_4H_2$ with a 5-sigma sensitivity.

# 96. Refractory Minerals in the Outflow from the Massive Young Stellar Object Orion Source I—Toward Understanding the Formation of the First Solids in the Solar System

Shogo Tachibana (Univ. of Tokyo), Yao-Lun Yang (RIKEN), Aki Takigawa (Univ. of Tokyo), Tomoya Hirota (NAOJ), Nami Sakai (RIKEN)

An outflow from a high mass YSO (young stellar object) candidate contains refractory metal-bearing molecules such as AlO, SiO, and SiS. The limited spatial distribution of AlO in the outflow near the base indicates recondensation of AlO to form Al-bearing refractory minerals. We here propose spectroscopic observation of the outflow from Orion Source I to detect refractory minerals such as melilite for the first time around YSOs. This observation aims to address two key scientific questions: (1) What are the solid-gas chemistry and dynamics of the outflow from Orion Source I? and (2) How did the first solids (CAIs) in the Solar System form? Refractory minerals in the outflow will be a useful tracer of the chemistry and dynamics of the outflow and will provide us with insight into the formation process of the first solids in the Solar System, which mainly consist of refractory minerals condensed from gas. We will use PRIMA/FIRESS to obtain 24-100 μm high-resolution spectra, probing the absorption features due to the Al-bearing silicate dust. The high-resolution mode mitigates the brightness of the source.

## Science Justification

The Orion Kleimann–Low (KL) nebula, the nearest massive star formation site (~400 pc; Grossschedl et al. 2018), harbors a candidate high-mass YSO (young stellar object), Source I, which has a hot circumstellar rotating gas disk emanating a wind of $H_2O$, AlO, SiO, SiS, SO, and $SO_2$ (e.g., Hirota et al. 2017; Ginsburg et al. 2018; Tachibana et al. 2019; Wright et al. 2020, 2024). Emissions of salt molecules (NaCl and KCl) have also been observed near the surface of a protoplanetary disk or within the disk (Ginsberg et al. 2019; Wright et al. 2020, 2024), which may be excited by accretion shock of materials in the outflow onto the disk (Wright et al. 2024). The findings of refractory metal-bearing molecules such as AlO, SiO, and SiS indicate high-temperature destruction of silicate and oxide dust and high-temperature gas chemistry involving oxygen and sulfur (Tachibana et al. 2019; Wright et al. 2024).

Tachibana et al. (2019) showed spatially resolved distributions of aluminum monoxide (AlO) emission lines at 497 and 650 GHz in the rotating outflow of Orion Source I (Fig. 1). Wright et al. (2020) also showed similar spatial distributions of AlO emission lines at 229 and 344 GHz in the outflow. Observation by Wright et al. (2020) was made with 4-times higher angular resolution than Tachibana et al. (2019), and it has revealed that the AlO emission peaks are located farther downstream than the $H_2O$ emission. This spatial offset suggests two possibilities on the origin of AlO molecules (Wright et al. 2020): (1) AlO molecules detected in the gas phase may have been produced by oxidation of aluminum-bearing dust grains with oxygen released via $H_2O$ dissociation or (2) AlO may have been released further out in the outflow than $H_2O$.

Both observations (Tachibana et al. 2019; Wright et al. 2020) showed the spatial distributions of AlO emissions limited near the base of the outflow in spite of their low excitation temperatures (~167 and ~281 K for 497 and 650 GHz emissions, respectively), compared to the broader distribution of $Si^{18}O$ emission with the emission temperature of 128 K in the outflow (Hirota et al. 2017). Such limited distributions of AlO emissions indicate that AlO is not present in the gas phase in the outer part of the outflow lobes away from the disk surface, most likely due to recondensation of AlO as refractory mineral dust, as discussed for the spatially limited distribution of AlO, consistent with the distribution of dust, in outflows of an oxygen-rich M-type AGB star W Hya (Takigawa et al. 2017).

Refractory minerals, which require AlO to form, include corundum ($Al_2O_3$), hibonite ($CaAl_{12}O_{19}$), spinel ($MgAl_2O_4$), melilite ($Ca_2Al_2SiO_7$-$Ca_2MgSi_2O_7$), and anorthite ($CaAl_2Si_2O_8$). These minerals are found in primitive meteorites, called chondrites. Ca- and Al-rich inclusions (CAIs) in chondrites consist of such refractory minerals and are the oldest solid objects in the Solar System (e.g., Connelly et al. 2012). The chemistry, mineralogy, and petrology of CAIs strongly suggest their high-temperature origin in the very early stage of disk evolution, and some CAIs consisting mainly of Al-rich melilite are thought to have condensed directly in the high-temperature gas phase (e.g., MacPherson 2014). However, it has been under debate for years in the fields of meteoritics and cosmochemistry where and how CAIs formed in the early Solar System.

We propose a mid-infrared spectroscopic observation of the outflow from Orion Source I to achieve the first detection of Al-containing refractory minerals (e.g., melilite) in a young stellar object (YSO). This observation aims to address two key scientific questions:

**(1) What are the solid-gas chemistry and dynamics of the outflow from Orion Source I?**

The mineralogy of dust grains reflects the physicochemical conditions in the outflow and serves as a powerful tracer of chemical and dynamical processes, combining with radio observations of metal-bearing gases,

**(2) How did the first solids (CAIs) in the Solar System form?**

Observing refractory dust in the outflow provides a unique opportunity to investigate processes analogous to those that formed the first solids in the Solar System.

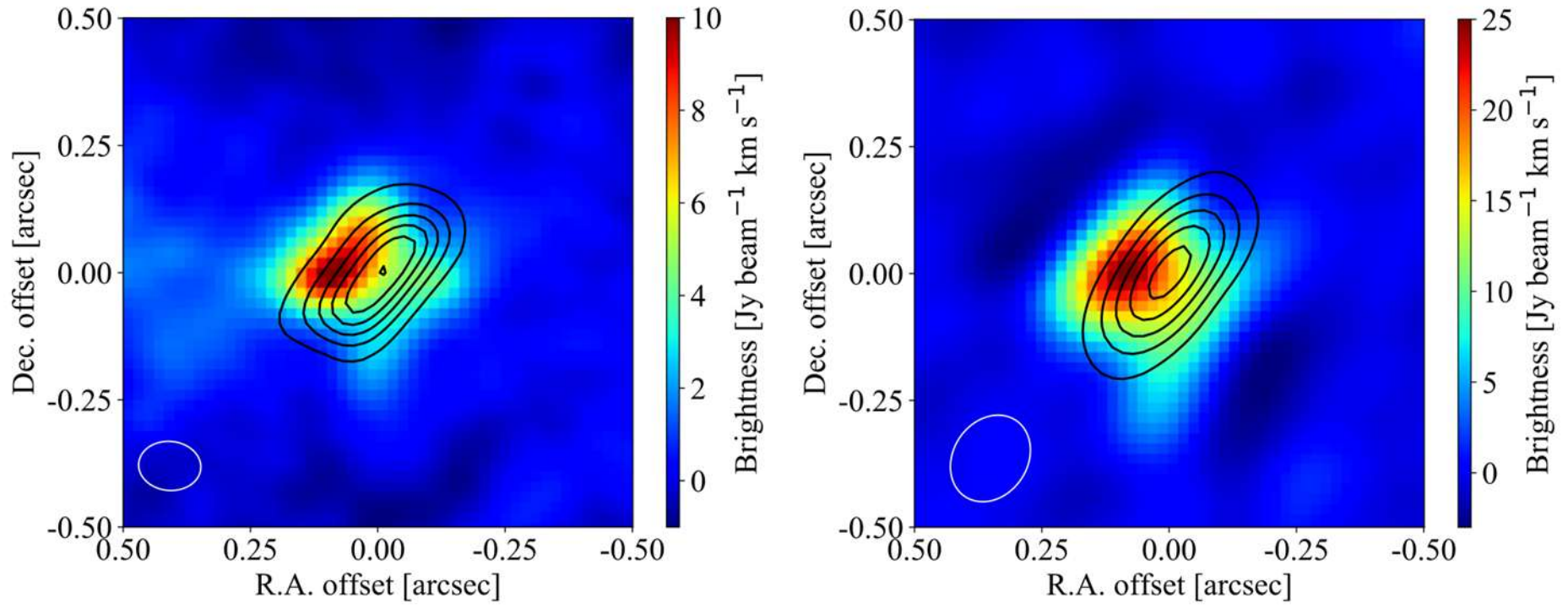

**Figure 1.** AlO emissions in the outflow from Orion Source I at 497 (left) and 650 GHz (right) (Tachibana et al. 2019)

The feasibility of infrared observation of refractory minerals in the 10-micron band was discussed by Posch et al. (2007), who considered the case where refractory minerals are mixed with crystalline silicates such as olivine and pyroxene. They concluded that, with the typical abundance ratio of CAIs in chondrites (<5%), it would be difficult to observe refractory minerals in the 10-micron band due to multiple intense peaks of olivine and pyroxene.

The distributions of AlO and SiO molecules in the outflow of Orion Source I suggest that the formation of refractory Al-bearing minerals (found in CAIs) predate the formation of silicates after the launch from the disk. Therefore, we expect to observe refractory minerals in the outflow without significant contribution of silicates such as olivine and pyroxene. Melilite has absorption peaks, distinct from those of olivine and pyroxene, in the wavelength range of PRIMA/FIRESS (Chihara et al. 2007; Fig. 2), and it is a major refractory mineral present in CAIs condensed from gas. Therefore, melilite is one of the good target minerals to search with PRIMA around Orion Source I.

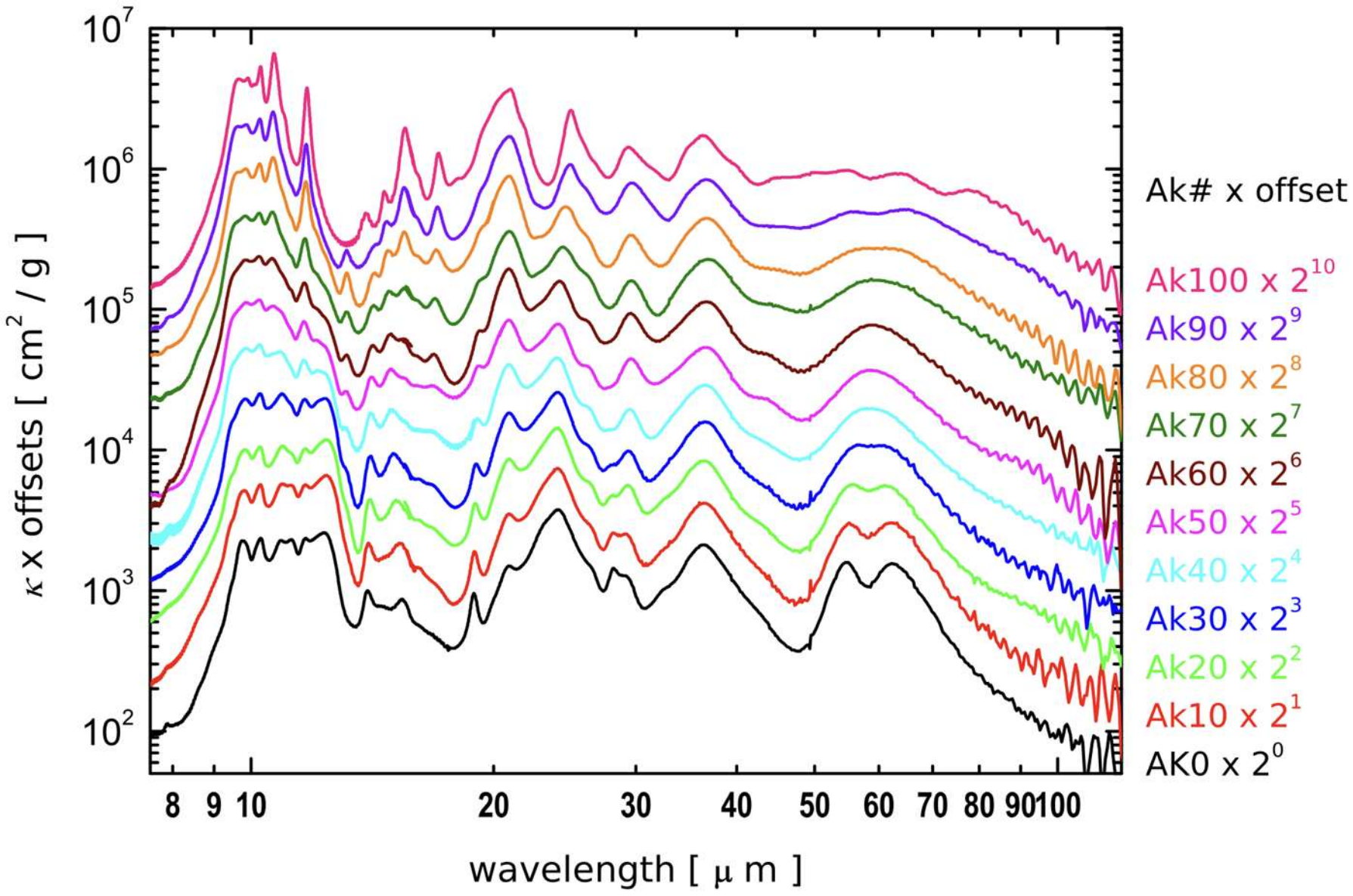

**Figure 2.** Mass absorption coefficients of melilite ($Ca_2Al_2SiO_7$-$Ca_2MgSi_2O_7$) (Chihara et al. 2007). Ak# in the figure represents the molar fraction of the $Ca_2MgSi_2O_7$ component.

## Instruments and Modes Used

The FIRESS it its pointed high-resolution mode is required.

## Approximate Integration Time

From Fig. 2, the peak opacity of dust mixtures with various levels of Al-inclusion is around 1000 $cm^2\ g^{-1}$, but the difference between each mixture can be resolved with a sensitivity of 100 $cm^2\ g^{-1}$. To estimate the column density of the Al-bearing dust, we start from visual extinction. In this region, Scandariato et al. (2011) find an $A_V$ ranging from 6 to >30. If we take an $A_V$ of 10 as a reference, we get a hydrogen column density of $1.4 \times 10^{22}$ $cm^{-2}$ based on the extinction law in Weingartner & Draine (2001) with $R_V = 5.5$. Then, we further assume a gas-to-dust mass ratio of 100 and a mean molecular weight of 2.8 (Evans et al. 2022), resulting in a dust mass column density of 6.5 x $10^{-4}$ g $cm^{-2}$.

Taking an opacity difference of 10 $cm^2$ $g^{-1}$ and a continuum flux of $10^2$ Jy derived from Okumura et al. (2011) at 20 µm, we get a signal strength of 1.9 x $10^{-17}$ W $m^{-2}$ per element of spectral resolution in the high-resolution mode of FIRESS. Using a continuum flux of $10^2$ Jy at 30 µm, the exposure time calculator indicates an exposure time of 12 hours for the estimated signal strength per pointing. This observation requires two pointings to cover the entire spectrum of the Al-bearing silicate dust. The total time will be 24 hours.

## Special Capabilities Needed

No special capability is needed.

## Synergies with Other Facilities

ALMA observations of SiO and AlO are critical for source characterization and indicate suitable candidates for this PRIMA study. At this moment, Orion Source I is the only source where the presence of Al-bearing silicate dust is hinted at from the distributions of the SiO and AlO emission. Thus, ALMA surveys of refractory species are critical to explore suitable candidates beyond Orion SrcI. In addition to observations, laboratory experiments on dust sublimation, condensation, and optical properties are also highly complementary for interpreting the observed absorption features.

## Description of Observations

The features of Al-bearing silicate extend from the shortest wavelength coverage of PRIMA to 100 µm. This program will use PRIMA to cover 24-100 µm range with FIRESS at low-resolution. While the low-resolution mode is sufficient to resolve the spectrum of Al-bearing silicate dust, we choose the high-resolution mode to mitigate the brightness of the source. We will use Band 1-3 of FIRESS, thus, effectively getting the full FIRESS spectrum of PRIMA coverage with two pointings. At the moment, only Orion SrcI has gaseous AlO detection together with the gaseous SiO detection, which warrants the search of Al-bearing silicate dust. We anticipate a few more detections of AlO may be reported in the next 5 years given the recent discovery of refractory species in the disks of high-mass protostars (Ginsburg et al. 2019; Wright et al. 2024).

# 97. Tracing Cloud Formation and Evolution Through 3D Magnetic Field Mapping

Mehrnoosh Tahani (Stanford University, University of South Carolina), Laura Fissel (Queen's University), Enrique Lopez Rodriguez (University of South Carolina), Kate Pattle (University College London)

We propose to use the unprecedented polarization sensitivity of PRIMA's PRIMAger Polarization Imager (PPI) and its high resolution in Band 1 (92 µm) to map magnetic fields across two contrasting molecular cloud environments: the well-studied Perseus cloud and the isolated Musca filament. This comparative study will leverage our existing VLA radio observations that provide line-of-sight magnetic field component of the Perseus cloud, along with upcoming POSSUM survey results for Musca, to construct the first detailed 3D magnetic field vector maps at sub-parsec resolution. Perseus, with its known formation history through interstellar structure (e.g., bubble) interactions, will reveal how magnetic fields evolve during active star formation phases, while Musca, an isolated filament with lower star-formation activity, will show magnetic field morphology in early evolutionary stages. With PRIMA's resolution of 0.01 pc for Perseus and < 0.01 pc for Musca, we will resolve magnetic field structures at scales necessary for understanding cloud fragmentation and star formation efficiency, and the roles that magnetic fields play in these processes. Our survey will require approximately 1438 hours to cover 42 deg$^2$ of Perseus and 12 deg$^2$ of Musca, providing the first comprehensive view of how environment and evolutionary state influence magnetic field evolution in molecular clouds and how magnetic fields regulate cloud formation, fragmentation, and star formation.

## Science Justification

**Unveiling Cloud History Through 3D Magnetic Fields:** Magnetic fields play significant roles in molecular cloud formation, evolution, and star formation (Pattle et al. 2023; Planck Collaboration et al. 2016). However, understanding their true impact requires mapping three-dimensional (3D) field morphologies across multiple scales. Recent studies by Tahani et al. (2022a,b) mapped 3D interstellar magnetic field vectors for the first time, revealing the full orientation and direction of these vectors and providing the only complete 3D interstellar field reconstructions to date. Using these 3D vectors in 3D space (6D maps), they demonstrated how field morphology can reveal the formation history of molecular clouds, as an archaeological record of past interactions with galactic structures.

**Perseus – Probing Magnetic Field Evolution During Active Star Formation:** The 3D field morphology of the Perseus cloud revealed a formation scenario inconsistent with the previously proposed Per-Tau supershell alone (Bialy et al. 2021). The reconstructed magnetic field structure predicted the existence of an additional interacting structure, which was subsequently confirmed by kinematic observations (Kounkel et al. 2022). This success demonstrates the transformative power of 3D magnetic field studies for understanding cloud formation and evolution mechanisms.

These 3D reconstructions combined radio observations for rotation measure (Tahani et al. 2018) with far-infrared observations by Planck (Tahani et al. 2019, 2022a,b; Tahani 2022). Due to resolution and sensitivity limitations of both datasets, the resulting 3D magnetic field vectors provide only coarse spatial information (> 1 pc). To understand the detailed roles that magnetic fields play in cloud evolution, we need to improve the resolution of these 3D fields to sub-parsec scales. PRIMA's unprecedented resolution and sensitivity, combined with currently available and upcoming radio observations, make this advancement possible.

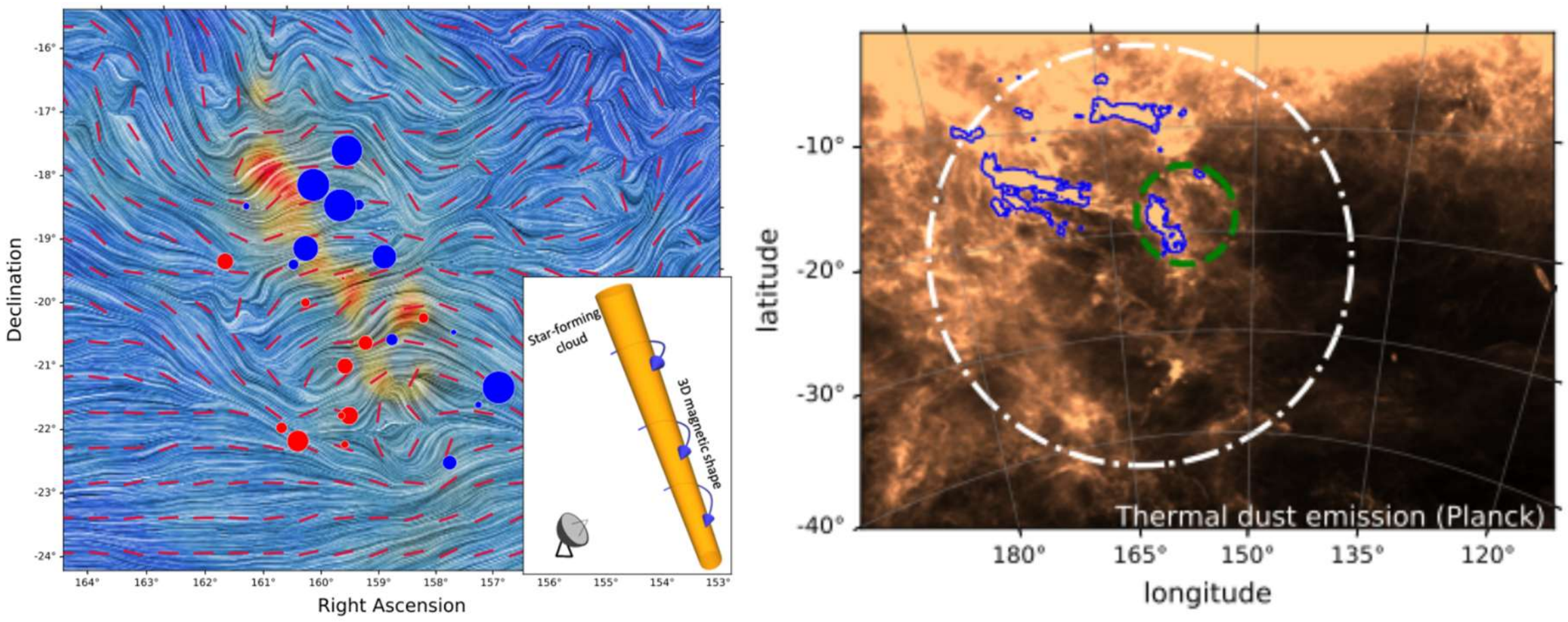

**Figure 1.** Left: Perseus molecular cloud's line-of-sight and 3D magnetic fields. Blue/red circles: line-of-sight magnetic fields toward/away from us. Red/drapery lines: Planck-observed plane-of-sky fields. Background: visual extinction map (Kainulainen et al. 2009). Right: Bubbles shaping Perseus (blue contour inside the green circle). Per-Tau (white) and newly identified structure (green) influence cloud formation. Background: thermal dust emission.

Perseus represents an ideal laboratory for studying magnetic field evolution during active star formation. Our existing VLA L-band observations provide a high-density rotation measure map of the Perseus cloud, improving the cloud's Faraday rotation measure source density by a factor of several compared to previous studies. Additionally, upcoming rotation measure maps from the Polarisation Sky Survey of the Universe's Magnetism (POSSUM; Jung et al. 2024; Gaensler et al. 2025), SPICE-RACS (Thomson et al. 2023), and ultimately the full Square Kilometre Array will further improve line-of-sight magnetic field resolution and source density. Current plane-of-sky observations from Planck, however, lack the resolution to reveal magnetic field evolution at scales where gravity or turbulence may dominate cloud dynamics. PRIMA's capabilities will enable us to map 3D magnetic field morphology at sub-parsec scales both in the cloud environment and inside molecular clouds, trace magnetic field evolution from quiescent to active star-forming regions with varying stellar ages, and quantify magnetic field-density relationships with unprecedented detail.

**Musca – Magnetic Fields in Early Stages of Cloud Evolution:** The Musca filament offers significantly different properties compared to the Perseus cloud. Musca is an isolated, quiescent molecular cloud in the early stages of evolution (Evans et al. 2014; Cox et al. 2016). Located at ~170 pc distance (Zucker et al. 2021) with minimal active star formation, Musca represents magnetic field morphology largely unperturbed by stellar feedback (see Figure 2) with relatively high observed polarization fraction through dust polarization (Planck Collaboration et al. 2015; Ngoc et al. 2021). This contrasting environment is important for understanding several key aspects of magnetic field physics in molecular clouds, including: a) examining how magnetic fields influence filament formation in isolation, b) exploring the relationship between magnetic fields and cloud/filament fragmentation efficiency, c) investigating whether 3D magnetic field morphology differs fundamentally between bubble-compressed clouds like Perseus and isolated clouds like Musca, d) exploring the role of magnetic fields in setting initial conditions for star formation, and e) unveiling cloud-formation history through 3D magnetic field mapping. These insights will be necessary when combined with observations of a larger sample of clouds and their 3D field morphologies.

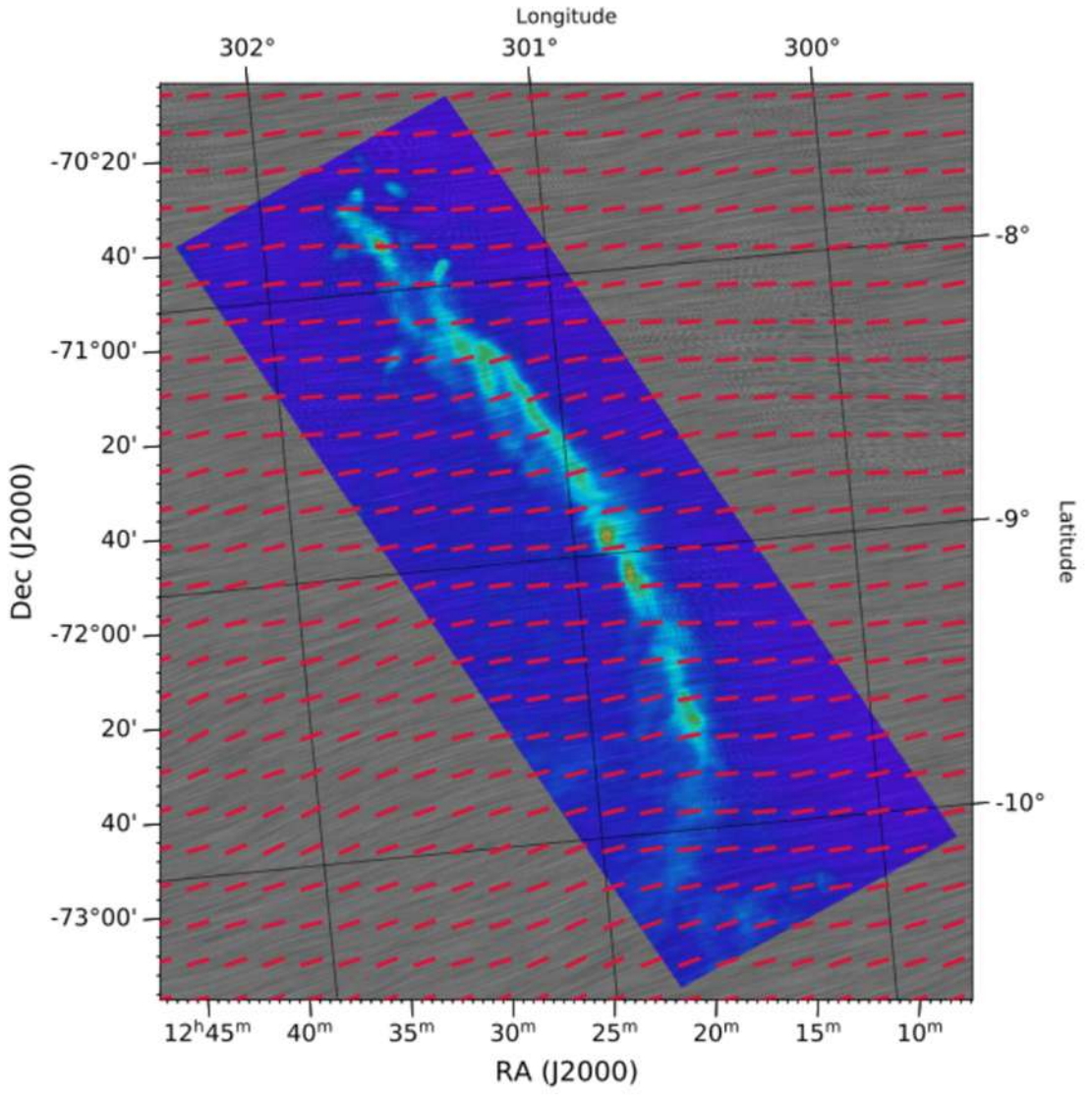

**Figure 2.** Herschel (background color) and Planck observations of Musca. The red and drapery lines show the plane-of-sky magnetic fields.

**A Framework for Understanding Cloud Formation and Evolution:** The 3D magnetic field morphologies serve as unique tracers for deciphering the formation and evolutionary history of molecular clouds. This comparative study of Perseus and Musca will establish a framework for understanding how magnetic fields regulate molecular cloud evolution across different galactic environments and evolutionary stages. By combining PRIMA's sub-parsec resolution observations with radio-derived line-of-sight magnetic fields (Tahani et al. 2024), we will quantify for the first time how environmental factors such as bubble interactions versus isolation, and evolutionary states from active to quiescent star formation, influence magnetic field morphology, strength, and coupling with gas dynamics. This approach will enable us to explore cloud formation scenarios in different galactic environments and examine the role these scenarios play in the co-evolution of clouds and their magnetic fields. PRIMA's high sensitivity and spatial resolution will allow us to trace magnetic field evolution from large-scale galactic fields through cloud formation to gravitational collapse onset. This examination will provide observational constraints for magnetohydrodynamic simulations and theoretical models of star formation. This dual-cloud approach will serve as a foundation for future systematic studies of the broader molecular cloud population.

## Instruments and Modes Used

Not provided

## Approximate Integration Time

Our sensitivity requirements are driven by the need to detect polarized emission from diffuse cloud material and map magnetic field morphology at scales comparable to the radio observations.

**Perseus:** We target 5-σ polarization detections for dust intensities of 20 MJy/sr, based on Herschel Gould Belt observations, assuming 3% polarization efficiency. This requires detection of 600 kJy/sr polarized emission across our 6°× 7° (42 deg$^2$) survey area. Using the PRIMA ETC for PPI Band 1 (92 µm), this requires 972.6 hours of exposure time.

**Musca:** For the same sensitivity target of 20 MJy/sr dust intensity with 3% polarization fraction, we require detection of 600 kJy/sr polarized emission. The 4° × 3° (12 deg$^2$) survey area covering the main filament and surrounding regions requires 277.8 hours exposure time.

**Total time:** 1250.4 hours exposure time. Including 15% overhead, the total requested time is 1438 hours.

## Special Capabilities Needed

None

## Synergies with Other Facilities

**Radio Observations:** Our study leverages existing and upcoming radio facilities to provide the line-of-sight magnetic field component required for 3D reconstruction. For Perseus, we will utilize completed VLA L-band observations providing high-density rotation measure data across the cloud (VLA/24A-376, VLA/20A-165, VLA/19B-053). These observations significantly improve source density compared to previous studies. For Musca, we will utilize upcoming rotation measure maps from the Polarisation Sky Survey of the Universe's Magnetism (POSSUM), SPICE-RACS, and ultimately the full Square Kilometre Array to obtain the first line-of-sight magnetic field measurements of this isolated filament. These studies incorporate the MC-BLOS technique (Tahani et al. 2018, 2024), which uses Faraday rotation of background point sources combined with extinction maps and chemical evolution modeling to determine line-of-sight magnetic fields of molecular clouds. Additionally, targeted Zeeman splitting observations of OH and HI lines will provide independent magnetic field strength measurements to complement our Faraday rotation results.

**Optical Polarimetry:** Complementary plane-of-sky stellar polarization observations from optical polarimetry surveys such as Dragonfly Polarimetry (https://dragonflypol.github.io/DragonflyPol/) will enable detailed mapping of foreground and background magnetic field contributions, essential for constructing accurate 3D magnetic field vectors of both clouds. These optical observations will help disentangle the complex line-of-sight structure by providing independent constraints on magnetic field orientations at different distances along the sight lines. Combined with Gaia stellar parallax measurements, this multi-wavelength approach will allow us to construct a complete 3D picture of magnetic field morphology from larger cloud scales down to individual star-forming cores and provide unprecedented insight into the magnetic field environment both within and surrounding the molecular clouds.

**Sub-mm Polarimetry:** Musca also will be mapped by the CCAT/FYST 6-m telescope in linear polarization at 350 µm, 850 µm, and 1.1 mm using the PrimeCam instrument (CCAT-Prime Collaboration et al. 2023). With this PRIMAger + CCAT multiscale data we will be able to search for regions where there are differences in the inferred magnetic field direction as a function of wavelength. These differences could indicate that the magnetic field morphology changes for dust grain populations of different temperatures along the same sightline through the cloud. This will allow us to probe magnetic field orientation as a function of dust temperature, complementing our 3D field measurements. As ancillary science, with CCAT+ PRIMAger we will measure the dust polarization spectrum of Musca from 96 µm to 1.1 mm. We will compare our measurements with existing polarization spectrum predictions from different dust models to constrain the composition of the dust grain populations in Musca (Guillet et al. 2018; Hensley & Draine 2023).

## Survey Design

Our observational strategy employs PRIMA's PPI Band 1 (92 µm) to map magnetic field morphology of the Perseus (6° × 7°) and Musca (4° × 3°) clouds. These observations will provide sub-parsec resolution plane-of-sky magnetic fields of both the cloud complexes and their surrounding environments. We will combine PRIMA's plane-of-sky magnetic field measurements with our radio-derived line-of-sight components using established 3D reconstruction techniques. This approach will deliver the first comprehensive 3D magnetic field maps at sub-parsec scales for both clouds. These 3D maps serve as novel tracers of cloud formation and evolution (Tahani et al. 2022b,a) and will enable us to assess how environment and evolutionary state influence magnetic field and cloud co-evolution and star formation regulation. This study will help establish a framework for future systematic studies of the broader molecular cloud population.

# 98. An Unbiased PRIMA Survey for Debris Disks around Nearby Stars

David J. Wilner (Center for Astrophysics | Harvard & Smithsonian), Joshua B. Lovell (Center for Astrophysics | Harvard & Smithsonian), Mark Booth (UK Astronomy Technology Centre), Meredith MacGregor (Johns Hopkins University), Patricia Luppe (Trinity College Dublin), Nicholas Ballering (Space Science Institute), Karl Stapelfeldt (NASA/JPL/Caltech), Brenda Matthews (Herzberg, National Research Council of Canada), Tiffany Meshkat (IPAC/Caltech), Katie Crotts (STScI)

We propose a complete, volume limited survey of Sun-like stars (FGK type) within 10 pc for far-infrared dust emission from debris disks. By reaching levels sufficient to reveal true analogs of the Kuiper Belt for the first time, these observations will place the properties of our Kuiper Belt into broader context. This unbiased PRIMA survey will build on the 17% detection rate of Herschel for debris disks around nearby FGK stars. The resulting improved debris disk demographics will provide the basis for a test of whether or not current steady-state collisional evolution models remain viable at fainter levels, reaching below the currently estimated median fractional luminosity, and also enable an assessment of models that invoke dynamical depletion from outer planetary migration.

## Science Justification

### Broader Context

## Debris Disks probe planetary system formation and evolution.

Debris disks are the relics of planet formation around main-sequence stars that are made detectable through dust replenished by collisions of planetesimals, the extrasolar analogs of Kuiper Belt Objects. Their structure encodes important information about planetary system formation and dynamical history, just as the Kuiper Belt in our Solar System is believed to have been shaped by the outward migration of Neptune at an early epoch, clearing away much of its initial mass, scattering some KBOs to orbits with high eccentricities and inclinations, and trapping others in mean motion resonances. Debris disks can also affect a planet's habitability, e.g. by bombarding planets with water-rich comets or destructive impacts, and may have implications for the development of life.

Debris disks are generally first identified at far-infrared wavelengths, where their dust emission spectrum peaks at high contrast with the central star. However, despite decades of improving far-infrared observations from IRAS to ISO, AKARI, Spitzer and Herschel, only the dustiest debris disks have been found so far. For Sun-like stars – spectral types FGK – a volume limited census to approximately 20 pc at far-infrared wavelengths is largely complete, and 47/275 = 17% of those systems show detectable emission at a limiting fractional luminosity, $f = L_{disk}/L_{star}$, of about $5 \times 10^{-6}$ (Sibthorpe et al. 2018), where the fractional luminosity provides a proxy for the total cross sectional area of dust. Applying completeness corrections to adjust for stellar and disk properties results in an incidence of about 28%.

These observations of nearby FGK stars are consistent with a steady state model in which each star is born with a planetesimal belt that undergoes collisional evolution, although this interpretation must be tempered by the fact that it requires substantial extrapolations from the small fraction of systems that are detected. An accounting of debris disks at lower fractional luminosities is needed to determine if the debris disk demographics remain consistent with passive collisional evolution. For our Kuiper Belt, Neptune's outward migration at an early epoch is a critical factor in its current low mass (Malhotra 1993). The primordial Kuiper Belt, with mass about 35 Earth masses, would be readily detected nearby, while the depleted post-migration Kuiper Belt would not (Booth et al. 2009).

## Science Question

## How typical is the Kuiper Belt in our Solar System?

The far-infrared emission from our Kuiper Belt today is an order of magnitude below the detection threshold for Herschel in the local volume. The dust spectrum peaks at 40-50 microns but at less than 1% of the photospheric flux at these wavelengths, with a corresponding fractional luminosity of about $5 \times 10^{-7}$ (see Vitense et al. 2012 and Poppe et al. 2019). A population survey of the 70 FGK stars within 10 pc that achieves Kuiper Belt detection levels would show if the properties of our Kuiper Belt are typical, and perhaps also if depletion from planetary perturbations is a common feature of debris disk evolution.

## Need for PRIMA

A telescope with improved sensitivity to dust emission at far-infrared wavelengths is needed to detect debris disks at lower fractional luminosities than seen in previous surveys, and to reveal true analogs of our Kuiper Belt around nearby stars.

Figure 1 (left) shows curves for the minimum detectable values of fractional luminosity for a G2V star at 10 pc ($L_{star} = 1\ L_{Sun}$, $T_{eff} = 5800$ K) as a function of $T_{dust}$ (and equivalent blackbody radius) for Herschel/PACS at 100 microns, together with projections for the PRIMA instruments (assuming the sensitivity limits can be achieved) As Figure 1 makes clear, PRIMA has the potential to access a new regime in this parameter space. The best band for detection of Kuiper Belt analogs will be PHI2 from 45-84 microns, as explained below. These observations will also include PHI1 from 24-45 microns, which will be useful to detect/characterize warmer dust.

Because the disk dust signal must be distinguished from the central star, raw sensitivity is only one of the technical requirements for detection. This issue has been previously explored in the context of the SPICA mission concept—a cold, high sensitivity, single dish telescope similar to PRIMA—focusing on absolute calibration uncertainties for unresolved disks, accurate point spread function subtraction for resolved disks, as well as the impact of confusion from background galaxies (Kamp et al. 2021).

For the typical 10% calibration uncertainties of previous far-infrared telescopes, spatially unresolved Kuiper Belt analogs are simply not accessible with single band photometric observations, even if they can achieve sufficiently high sensitivity with deep integrations.

Saturation from the stellar photosphere emission must be avoided, too. For debris disks that can be resolved spatially, a limiting factor becomes knowledge of the point spread function, with stability requirements at the $10^{-4}$ level of the peak in the relevant region of the wings for accurate star subtraction to reveal the extended disk emission. At 10 pc distance, the classical Kuiper Belt with radii 42 to 47 au, subtends a diameter of about 9 arcseconds, (slightly) larger than the PRIMA PHI2 beam. Closer debris disks will have a larger angular extent, but generally still subtend less than 2 PRIMA PHI2 beams.

One way to partly overcome the observational issues for unresolved or marginally resolved debris disks will be to make use of spectral information, since the excess emission from cold dust that peaks in the far-infrared has a very different *shape* from the (Rayleigh-Jeans) stellar photosphere spectrum that peaks in the optical. Figure 1 (right) shows spectra of a G2V star and Kuiper Belt analog at 10 pc, highlighting the contrast in their spectra. Observations at different wavelengths will need precise common calibration, as for PRIMAger that will deliver sub-bands at R=10 (PHI2).

To the extent that debris disks are marginally resolved or unresolved, background confusion will remain an issue, as dusty background galaxies have the potential to mimic a faint disk signal. For example, the nominal PRIMA confusion limit for PPI1 is about 5 mJy (Bethermin et al. 2024), comparable to or exceeding a nearby Kuiper Belt analog. Any disk candidates identified through far-infrared spectral excess emission would need to be vetted for the presence of background confusion using additional observations, e.g. obtaining spectral line detections with FIRESS that would discriminate gas-rich galaxies from gas-poor debris disks (, or through observations at higher angular resolution at infrared, millimeter, or radio wavelengths, and applying deblending methods (Donnellan et al. 2004). For some nearby stars with sufficiently high proper motions, multi-epoch observations could provide a check as the system moves with respect to the background.

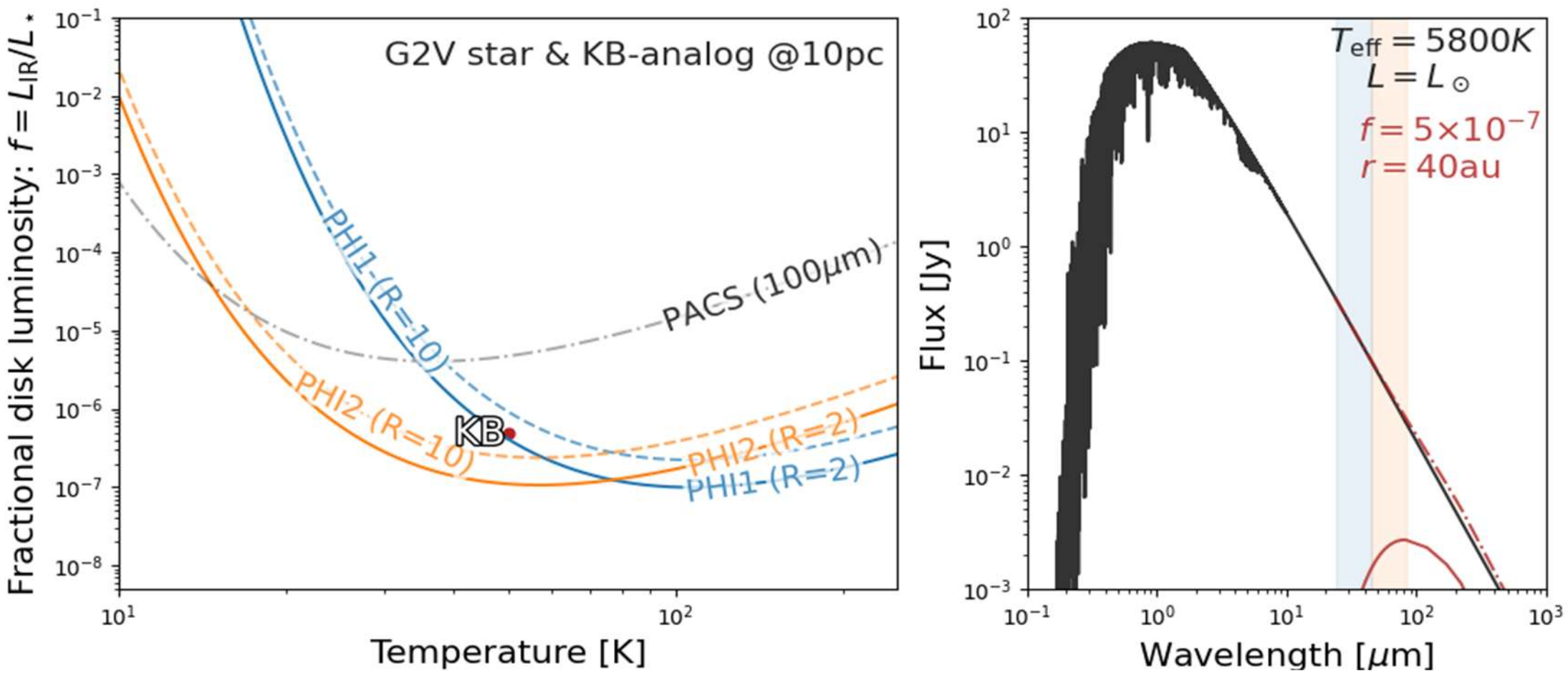

**Figure 1.** (left) PRIMA sensitivity curves (2 hours) for debris belt fractional luminosity ($f = L_{disk}/L_{star}$) as a function of dust temperature for PHI1 and PHI2 at R=10 and R=2 (top) compared to Herschel/PACS at 100 microns. PRIMA provides a substantial improvement on Herschel, sufficient to reach true Kuiper Belt analogs around a sample of nearby stars for the first time. (right) The stellar spectrum (black), Kuiper Belt analog spectrum (solid red), and the sum (dot-dashed red). A Kuiper Belt is faint compared to the stellar photosphere, and detections will benefit from leveraging PRIMA's spatial and spectral resolution.

## Interpretation Methods

The PRIMA far-infrared observations provide direct detections, or upper limits, on the presence of cool dust emission associated with each target star at a level much lower than previously available. The excess emission (at all available wavelengths) will be analyzed with a standard steady-state collisional evolution model that assumes a narrow belt at birth around each star drawn from a distribution of radii or masses informed from observations of protoplanetary disks (Wyatt 2007). The fit of this model, its parameters, and departures in the data will allow an assessment of the viability of simple, passive collisional evolution, as well as the role of planetary perturbations in dynamically depleting exo-Kuiper Belts. These results will place the properties of our Kuiper Belt in the context of other Sun-like stars.

## Instruments and Modes Used

PRIMAGER small maps in Hyperspectral Band

## Approximate Integration Time

Using the ETC for PHI2 (45-84 microns) and the smallest map size, 10' x 10', gives a 5 sigma limit of 1.2 mJy in 2 hours, sufficient to detect a Kuiper Belt analog around a G2 star at a distance of 10 pc. The observations will also include PHI1 (24-45 microns), which will also provide useful constraints on warmer dust. A survey of all 64 FGK stars within 10 pc to this limit will require 128 hours.

## Special Capabilities Needed

This project requires the ability to do accurate PSF subtraction and/or precise relative calibration of spectral sub-bands in order to leverage both spatial and spectral information to distinguish faint debris disk emission from bright stellar photospheres and from faint background galaxies.

## Synergies with Other Facilities

The discovery of new debris disks through this survey will provide new targets for high resolution follow-up with facilities such as JWST that can determine the distribution of small grains in the system, identify the presence of ices, and detect any high mass planets in the systems. In the (sub-)mm, deep observations with single-dish telescopes such as the current Large Millimeter Telescope and future Atacama Large Aperture Submillimeter Telescope will be able to determine the sub-mm slope of the SED, providing constraints on the dust collisional and optical properties. High resolution follow-up with an upgraded Atacama Large Millimeter Array has the potential to map the distribution of large grains in the system, which can highlight imprints of planet-disk interactions. Characterizing these nearest stars will inform future surveys for Earth-like planets with facilities like the Habitable Worlds Observatory as the amount of dust in the system impacts the detectability of planets.

## Description of Observations

This sample consists of the 64 FGK stars (8 F, 18 G, 38 K) within 10 pc (Gaia, Reyle et al 2021). Each star will be observed with a 10’x10’ PRIMAger map for 2 hours, 5 sigma depth of 1.2 mJy with PHI2, to maximize sensitivity to dust emission spectra at Kuiper Belt temperatures and provide spectral information to distinguish dust emission from the stellar photosphere. Accurate PSF subtraction is also highly desirable to spatially separate the marginally resolved debris disk dust emission from the stellar photosphere. A handful of these stars have already known debris disks, which will be included in this unbiased, uniform sample and offer a control for comparison with previous observations.

# 99. Surveying Atomic Outflows in Protostars Across Mass Spectrum

Yao-Lun Yang (RIKEN), Alessio Caratti o Garatti (INAF-OACN), Robert Gutermuth (UMass Amherst), Daniel Harsono (National TsingHua University), Dominika Itrich (University of Arizona), Doug Johnstone (HAA, NRC Canada), S. Thomas Megeath (University of Toledo), Marta Sewilo (NASA Goddard Space Flight Center and University of Maryland), Łukasz Tychoniec (Leiden Observatory, Leiden University)

Protostellar outflows are signposts of ongoing star formation across the universe and across mass. Outflows play a crucial role in star and planet formation by removing excess angular momentum, regulating accretion onto protostars, and by dissipating gas and dust from disks. Protostellar outflows are observed to have fast collimated component and a wide angle part. In addition, they are typically diverse in terms of chemical composition that may be linked to their launching region and the planet forming disks. Despite their importance in star and planet formation, protostellar outflow origin and evolution can be better understood via deep spectral surveys in far-infrared wavelengths. The atomic component - particularly traced by forbidden [O I] emission - remains underexplored due to observational challenges from the ground. This study outlines the significance of velocity-resolved [O I] and [C II] far-infrared observations in identifying shocked atomic gas and disentangling it from photodissociation regions (PDRs). Existing data from SOFIA provide preliminary evidence for high-velocity atomic outflows from low- and high-mass protostars, yet are limited by sensitivity and resolution. The proposed PRIMA/FIRESS survey, with its superior sensitivity and sufficient spectral resolution, will enable systematic detection and kinematic characterization of atomic outflows across a large protostar sample. By kinematically resolving the line profiles of atomic and ionized lines, such as [O I] and [C II], and other key tracers, such as CO, $H_2O$, the survey aims to constrain mass loss rates, momentum transfer, and the oxygen budget in protostellar environments. This work will fill a critical gap in the atomic component of outflows, completing our understanding of outflow and accretion mechanisms and their role in star and planet formation.

## Science Justification

In star and planet formation, outflows play the essential role in mediating the exchange of mass and angular momentum from the envelope through the disk and onto the forming star. Thus, outflows are intimately connected with the star and planet formation process. At the early stage of star formation, observational characterization of the accretion processes is hindered by high extinction from the dusty envelope surrounding the forming protostar. Thus, because of its close relation, outflows become the powerful probe of accretion as well as the feedback to the surrounding quiescent gas, which could regulate the ongoing star formation activity. Most observations of outflows probe the molecular and ionized components, using facilities such as ALMA, HST, JWST, and ground-based near-infrared observatories. A common picture emerges with collimated fast ionized jets at the center of wide-angle slow molecular outflows. However, the ionized component, presumably comes from the deep in the potential well near the star and therefore cannot remove significant angular momentum, the molecular component may well be influenced by entrained material from the outflow boundary with the infalling envelope - thus the atomic component is likely the best bit to probe to understand the role of the outflow is liberating angular momentum from the outer disk - where it poses the biggest problem for star formation.

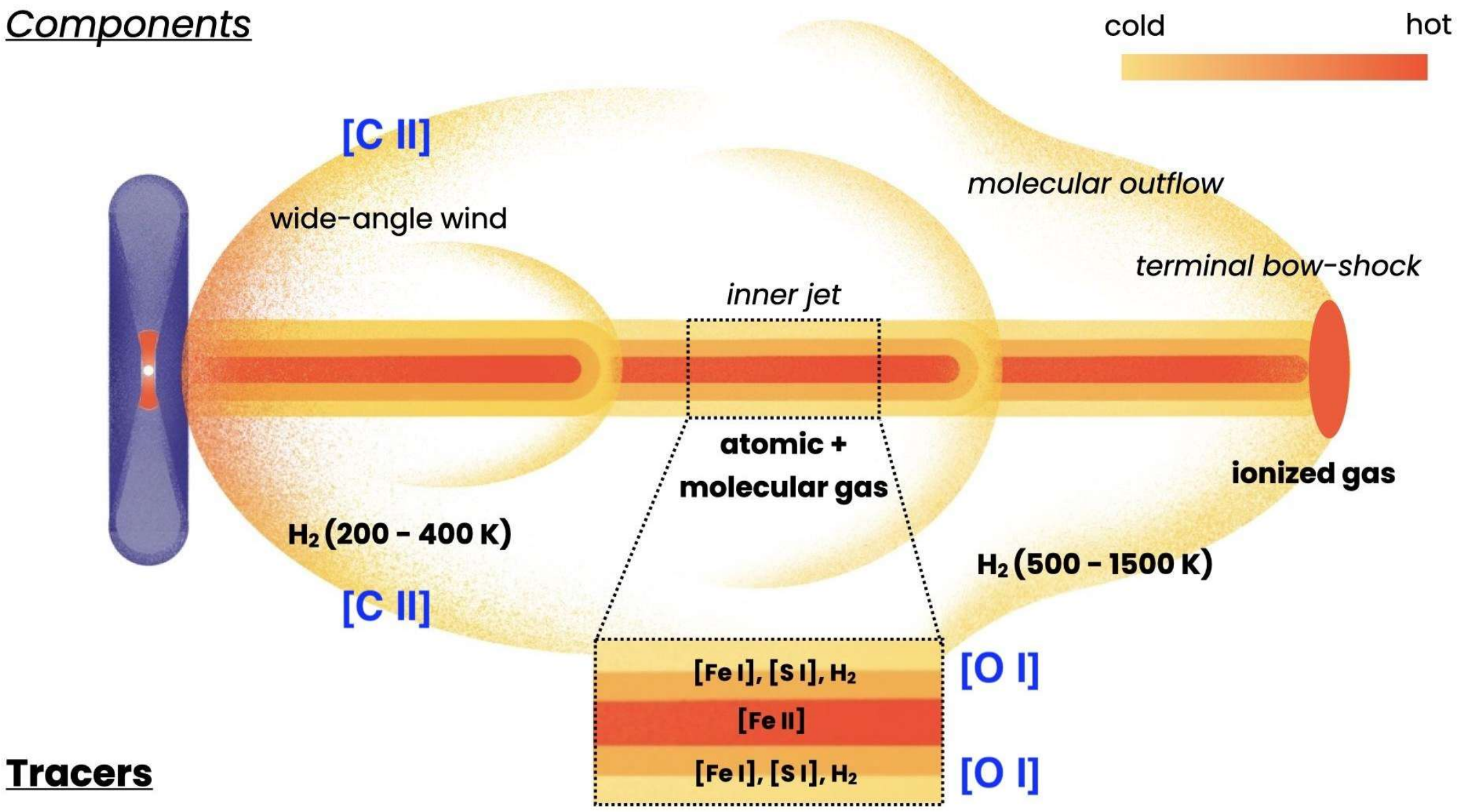

**Figure 1.** Outflow structure (Caratti o Garatti et al. 2024). The components traced by [O I] and [C II] are labelled.

Theories have predicted that the emission of forbidden atomic oxygen traces the shocked gas in outflowing winds, allowing us to constrain the physical properties of the missing atomic outflows. Hollenbach & McKee (1989; hereafter HM89) shows that, in a J-shock model, the dominant gas coolant [O I] 63 µm line flux is proportional to the mass flux in the shock, which can be related to the mass outflow rate. However, atomic oxygen can also be present in photodominated regions (PDRs), confusing a direct association between the [O I] flux and shocks. To address this confusion, the HM89 model shows that the [C II] 158 µm is more sensitive to PDR; thus, the flux ratio between the [O I] 63 µm and [C II] 158 µm lines can serve as an indicator to discriminate between J-shocks and PDR. Although this model provides a simple approach to estimate the mass outflow rate, a critical parameter in understanding the accretion process and, ultimately, the rate at which the protoplanetary disk is dispersed, it has many inherited assumptions and simplifications that require observational tests. We find only a weak correlation between [O I] luminosities and the mass outflow rate inferred from CO emission with Herschel (Green et al. 2013; Karska et al. 2013), underlining the need for a direct measurement of outflow rate in the atomic component.

Velocity-resolved observations of [O I] and [C II] lines will distinguish the shocked atomic gas in the outflow, providing a direct measurement of atomic outflow kinematics. Observations with sufficient velocity resolution (Δv < a few 10 km s$^{-1}$) at far-infrared wavelengths were only available by using SOFIA GREAT/upGREAT, although the Herschel Space Observatory extensively surveyed the unresolved fluxes of both [O I] and [C II] lines (Green et al. 2016; Karska et al. 2018). In some cases, the [O I] line is marginally resolved, suggesting the presence of high-velocity components, and shows velocity structure consistent with an outflow (Lee et al. 2014). But a definitive measurement still relies on isolating the high-velocity component of [O I]. For SOFIA, this measurement was extremely difficult due to the limited sensitivity of the telescope. The peak intensity of shocked atomic gas becomes lower when the line width gets broader in stronger shocks, making it less likely to be detected. Nonetheless, a few observations resolved the [O I] line profiles in both high- and low-mass protostars, giving us a preliminary test of using [O I] as a shock diagnostic.

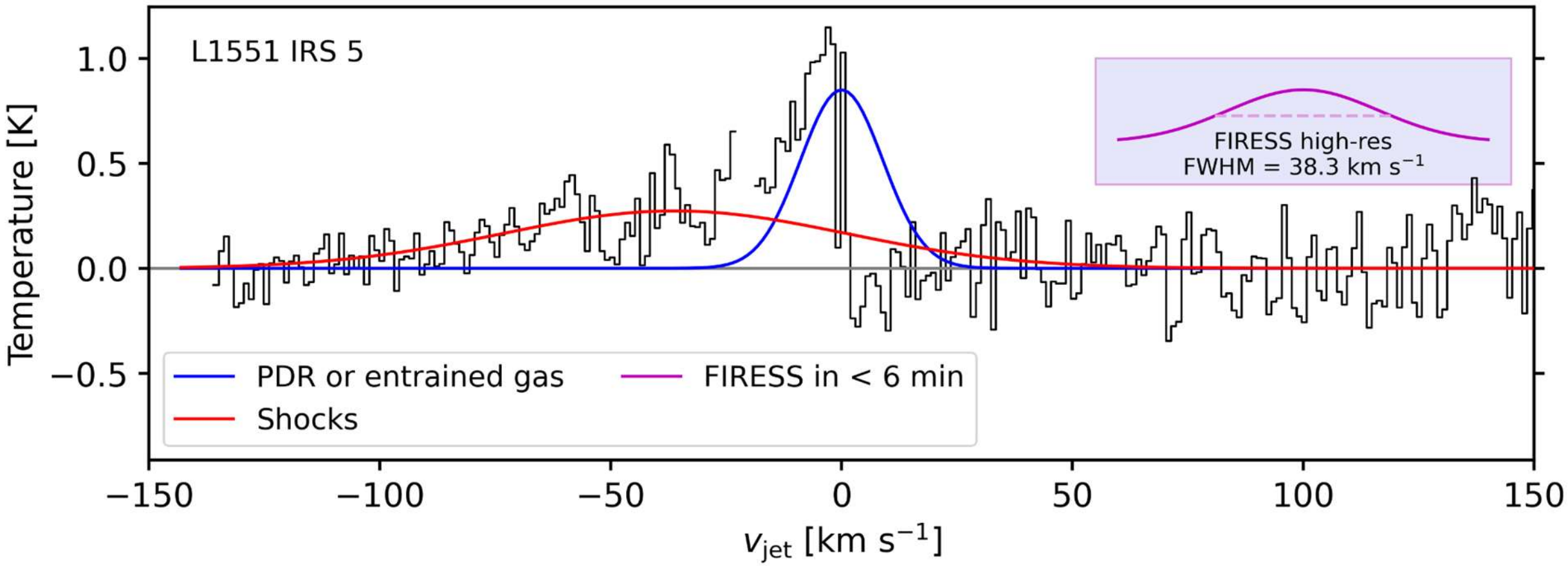

Figure 2. SOFIA/upGREAT spectrum of [O I] 63 μm line in a Class I protostar, L1551 IRS 5 (Yang et al. 2022). The high-velocty shock-tracing component is highlighted by the red line. The inset purple line profile shows the spectral resolution of FIRESS at 63 μm, 38.3 km s$^{-1}$, and its strength can be observed within a minimal integration time of PRIMA (< 6 minutes).

In high- and intermediate-mass protostars, the [O I] lines show complex line profiles with substantial absorption at the line center of the [O I] 63 μm line due to the foreground gas (Leurini et al. 2015; Gusdorf et al. 2017; Schneider et al. 2018, 2021; Karska et al. 2025). Observing the less populated [O I] transition at 145 μm can address this optical depth effect. Only two Class 0/I low-mass protostars have resolved [O I] line profiles from SOFIA. In the shock knot of NGC 1333 IRAS 4A, Kristensen et al. (2017a) find a high-velocity [O I] component up to ∼50 km s$^{-1}$, tracing the shock previously detected in CO and $H_2O$. In L1551 IRS 5, a high-velocity [O I] component was detected and interpreted as the atomic outflow (Yang et al. 2022; Figure 2). The consistent mass loss rate probed by molecular (CO), atomic ([O I]), and ionized ([Fe II]) outflows supports the momentum-conserved scenario of outflows, where the momentum simply transports between different outflow components. Moreover, as the protostars evolve, atomic jets can become the main component of the outflow, where atomic oxygen serves as one of the main driving agents (Nisini et al. 2015; Benedettini et al. 2017). Despite detailed measurements of a few atomic outflows, the limited number of sources with spectrally resolved [O I] emission prevents us from building a complete picture of outflows in low-mass star formation.

## Experimental Design and Interpretation

PRIMA/FIRESS provides the necessary velocity resolution and sensitivity to enable a systematic survey of atomic outflow in low- and high-mass protostars. The velocity resolution of FIRESS at 63 μm is ~35 km $s^{-1}$, sufficient to isolate the high-velocity component detected in SOFIA observations. Direct measurements of this high-velocity atomic oxygen gas can immediately estimate the kinematics of atomic outflows and further compare with molecular and ionized outflows surveyed by ALMA and JWST. With the simultaneous measurements of the [O I] 145 μm line, we can constrain the excitation of the atomic oxygen, deriving its column density and hence the mass outflow rate of the [O I] outflow. With assumptions of chemical abundances, we can obtain the mass outflow rate of the entire atomic outflow. The mass outflow rate is an important constraint for models of outflow launching mechanisms and the underlying accretion models (Watson et al. 2016). With the [O I] velocity, we can derive the momentum carried by [O I] outflows and compare it with the momenta in other outflow components, such as CO, HI, and [Fe II]. This comparison tests a fundamental picture where the most energetic ionized outflow drives lower energetic outflows by simply transporting its momentum.

The [C II] 158 μm line serves as the indicator of PDR. Comparisons between the line profiles of [C II] and [O I] will distinguish shocks from PDR. Therefore, any PDR contribution to the [OI] line profile can be characterized and excluded from the dynamical analysis of atomic outflows. The PDR component can be further modeled using, for example, the Meudon PDR code (Le Petit et al. 2006) or the PDR toolbox (Pound & Wolfire 2011).

## Auxiliary Science

While this survey targets [O I] and [C II] lines using FIRESS Band 2 and 4, other emission lines will be observed simultaneously, including high-J CO, $H_2O$, as well as forbidden lines, such as [OIII], [NII], and [CI]. The high-J CO and $H_2O$ lines often originate from shocks in the outflows or on the outflow cavity walls (Mottram et al. 2014; Kristensen et al. 2017a; Karska et al. 2018; Yang et al. 2018). The atomic and ionized forbidden lines also trace high-velocity outflow components, constraining the ionization as well as the velocity and density in shocks.

This survey also provides an opportunity to characterize the oxygen budget in outflows. The coverage of [OIII], which was only accessible by SOFIA with much less spectral resolution, allows us to probe the ionized outflow to complete the chemical budget of oxygen. Together with CO, $H_2O$, and [O I], we can include all the major oxygen carriers in protostellar outflows.

## Need for PRIMA

PRIMA/FIRESS provides the ideal capabilities of superior sensitivity and sufficient spectral resolution to carry out a large survey of atomic outflows in low- and high-mass protostars. FIRESS has a spectral resolution of ~35 km $s^{-1}$ at 63 μm, which can resolve the broad component detected in L1551 IRS 5 by SOFIA. The sensitivity gain is the game-changer to enable this survey. PRIMA/FIRESS can detect the broad component in L1551 IRS 5, which was detected with 2.5σ by SOFIA, with 5σ using a minimal integration; it can achieve 10 times better sensitivity with 0.7 hours of integration time. Using PRIMA/FIRESS, we can efficiently survey the [OI] line at 63 μm on 50–100 Class 0/I protostars with less than 100 hours, increasing the current sample by a factor of >50. Leveraging the synergy of CO outflows surveyed by ALMA and ionized jets mapped by JWST, PRIMA observations of atomic outflows provide the missing piece in the puzzle of understanding protostellar outflows in low-mass star formation.

## Instruments and Modes Used

FIRESS pointed high resolution mode, 1 pointing per target

## Approximate Integration Time

Based on Yang et al. (2022), the peak intensity of the high-velocity [O I] component in L1551 IRS 5 is 0.3 K. The line width is 87.5 km $s^{-1}$. FIRESS has a resolution of ~35 km $s^{-1}$ at 63 µm. Thus, we estimate the line flux within one third of the line width, which is $7.1 \times 10^{-16}$ W $m^{-2}$. We use the continuum flux in Yang et al. (2018), 210 Jy. From the Exposure Time Calculator, FIRESS high-resolution mode can achieve a 5σ detection of this line flux within a minimum integration of 2.5 minutes. To explore the limit of FIRESS, we find that it can detect this flux with 100σ after an integration of 1.9 hours. From the COPS-SPIRE survey (Yang et al. 2018), which surveyed 26 low-mass Class 0/I protostars, the lowest [O I] line flux is 3% of that found in the reference source, L1551 IRS 5. Assuming the intensity of the high-velocity [O I] component scales with the total [O I] line flux, we can still achieve a 3σ detection with an integration time of up to 1.9 hours.

The two [O I] lines, 63 µm and 145 µm, can be observed in Bands 2 and 4 simultaneously. The same integration time of 1.9 hours results in a 5σ detection of an [O I] 145 µm line flux of $1.7 \times 10^{-17}$ W $m^{-2}$, about two times better than that at 63 µm. However, from the COPS-SPIRE survey, the [O I] 145 µm line flux is typically 7% of the flux of the 63 µm line. Detecting the weakest [O I] 145 µm line with 3σ following the approach described above requires an integration time of 33.3 hours. However, such high sensitivity may not be necessary because, in the weak [O I] sources, the optical depth effect becomes negligible. Thus, detecting the [O I] 63 µm line alone (with 1.9 hours) may be sufficient. To have a statistically significant sample of protostars from low- to high-mass, we aim to observe 100 relatively isolated sources in the nearby star-forming regions (<500 pc for low-mass and <2 kpc for high-mass). Assuming that 10% of the sample requires high sensitivity for the [O I] 145 µm line, the entire survey requires 504 hours (90 sources with 1.9 hours and 10 sources with 33.3 hours).

In summary, using PRIMA/FIRESS, we can efficiently survey 100 Class 0/I protostars with ~200 hours if only targeting the [OI] 63 um line, increasing the current sample by a factor of ~20. Including the few sources requiring high S/N observations of the [OI] 145 µm line, which address the optical depth effect on the 63 µm line, leads to a total time of ~500 hours.

## Special Capabilities Needed

- None

## Synergies with Other Facilities

The atomic outflows probed by PRIMA are complemented by JWST and ALMA observations. JWST probes high-velocity jets via [Fe II] and shocks in outflow cavities via H2, testing the relation between the ionized and atomic jets (Watson et al. 2016; van Dishoeck et al. 2025). In comparison, ALMA is probing some of the shocks - in SO and SiO - as well as the cold gas through CO at high spatial resolution and therefore mapping the shape of the outflows and probing the underlying flow and the entrainment of material.

## Description of Observations

This program aims to resolve the [O I] emission to measure mass outflow rates across protostellar masses. Targeting protostars across the mass spectrum allows us to prove the outflow launching mechanism as well as the inferred accretion process from low- to high-mass, addressing a fundamental question on whether we can understand the high-mass protostars based on what we learn from their low-mass counterparts.

The source selection will be based on the protostars with known infrared fluxes with either Spitzer or Herschel. For low-mass sources, we can select from the Spitzer c2d survey (Evans et al. 2009), the HOPS (Furlan et al. 2016), and the eHOPS survey (Pokhrel et al. 2023). For intermediate- and high-mass sources, we can select sources from the ATLASGAL survey (Gutermuth & Heyer 2015; Heyer et al. 2016). These sources need to be either isolated or substantially dominated by a single source, so that the [O I] line flux can be unambiguously associated with an accretion rate. Given the spatial resolution of PRIMA, sources within 500 pc would be ideal. For high-mass sources, we need to relax this limit to be within 2 kpc to have a sufficient number of sources.

The two [O I] lines, 63 µm and 145 µm, can be observed simultaneously using FIRESS Band 2 and 4, respectively. The [C II] 158 µm line will be observed in Band 4 as well, together with the [OIII] 52 µm line at Band 2. Given the standard integration time of 1.9 hours, we will observe the [C II] 158 µm with a 5σ sensitivity of $1.6\times10^{-17}$ W m$^{-2}$, which is also about S/N∼5 for the weakest [C II] line in Green et al. (2016). Thus, only one pointing is required for each source. We will use the high-resolution mode of FIRESS so that the high-velocity [O I] component can be resolved.

# 100. Observation of Hydrides in Photodissociation Regions

Marion Zannese (IAS, Orsay), Emilie Habart (IAS, Orsay), Benoît Tabone (IAS, Orsay)

Massive stars play a critical role in shaping the interstellar medium (ISM) through stellar feedback, primarily via radiation. Their extreme ultraviolet (EUV) radiation ionizes gas, forming H II regions, while far-ultraviolet (FUV) radiation creates photodissociation regions (PDRs) with complex chemical and thermal structures. Hydrides like $CH^+$, OH, and HD are key tracers of the physical and chemical processes in the molecular zones of these irradiated regions. Observations with instruments like *Herschel* and the *James Webb Space Telescope* (JWST) have revealed out-of-equilibrium excitation mechanisms, such as formation pumping, for these hydrides, which are sensitive indicators of conditions like gas temperature and density and which complete the range of precision astrochemistry tools. Here, we demonstrate how the instrument FIRESS of PRIMA (24–235 µm) will advance this field by detecting hydride rotational transitions emissions in various PDRs, providing broader insight into their physical structure and chemistry triggered by FUV photons. We propose to focus on emblematic PDRs in the giant Orion molecular clouds (i.e., the Orion Bar and the Horsehead Nebula) and in the Cepheus cloud associations (i.e., NGC 7023) with the FIRESS Spectrometer to cover a wide range of physical conditions (UV, density).

## Science Justification

In our Galaxy and distant star-forming galaxies, massive stars are major sources of radiative and mechanical energy injection in the interstellar medium. This stellar feedback affects the structure, thermal balance, chemistry and ionization state of the ISM, and therefore has a direct impact on the star-formation rate of galaxies and their evolution. The radiation emitted by massive stars is one of the principal mechanism of stellar feedback. Indeed, massive stars emit extreme ultraviolet photons (EUV, i.e. E > 13.6 eV) which ionize the gas thereby creating H ii regions, which expand and disperse the gas and dust around them. They also emit lower energy photons (Far UV, hereafter FUV, E < 13.6 eV) which can penetrate slightly inside the surrounding molecular cloud, creating a photodissociation region (PDR) at the edge. Inside the PDR, the FUV photon flux is gradually extincted by dust grain and gas line absorption which leads to strong temperature gradients, and a layered structure with different chemical compositions.

Hydrides, such as $CH^+$, OH and HD, are key molecules to study physico-chemical processes in strongly irradiated environments and the chemical origin of the molecular species. These hydrides have been observed in several PDRs at the edge of molecular clouds (e.g. Orion Bar, NGC 7023) with *Herschel* via their rotation emission (Goicoechea et al. 2011, Nagy et al. 2013, Parikka et al. 2017, Joblin et al. 2018) and the JWST for their rovibrational emission (Zannese et al. 2025, Misselt et al. in prep). JWST detections of rovibrational emission were also obtained towards PDR in the photoevaporated wind of externally irradiated protoplanetary disks (Berne et al. 2024, Zannese et al. 2024). The analysis of their emission by JWST and *Herschel* revealed out-of-equilibrium excitation processes such as chemical pumping (e.g. $C^+ + H_2(v,J) = CH^+(v',J') + H$, Godard et al. 2013, Zannese et al. 2025 ; $O + H_2(v,J) = OH+(v',J') + H$, Zannese et al. 2024) or prompt emission following the photodissociation of water ($H_2O + h\nu = OH(v,J) + H$, Zannese et al. 2024). The observation of highly excited rotating OH attests that water is efficiently destroyed by UV radiation. The formation pumping processes have been proved to be robust diagnostics of physical parameters, such as gas temperature and gas density, and the warm UV-induced chemistry. It has been shown that the presence of $H_2$ excited by UV photons results in very active chemistry, forming $CH^+$ and in a second step $CH_3^+$, a key molecular ion detected for the first time in space (Berne et al. 2023, Zannese et al. 2025). These unprecedented results on the warm ISM chemistry (very different from cold chemistry), internationally recognized, are now stimulating laboratory experiments (in particular on $CH_2^+$ which remains undetected for the moment).

$CH^+$ and OH have also been detected toward isolated protoplanetary transitional disks with *Herschel* (Meeus et al. 2013) and JWST (Henning et al. 2024). In the layers of the disks which are directly exposed to the stellar radiation field, the reaction like $C^+ + H_2(v,J) = CH^+(v',J') + H$ is expected to be dominant. Despite its importance in the thermal balance of the gas and in the determination of primeval planetary atmospheres in irradiated regions, the warm UV-induced chemistry remains poorly constrained. This is why it is imperative to study hydrids in PDRs.

PRIMA, observing between 24 and 235 $\mu$m, will detect the pure rotational lines of these hydrides (see Figure 1). Its wavelength range and high sensitivity will allow the detection of higher excited levels of $CH^+$, OH and HD compared to *Herschel* (e.g., see Figure 2 where it is visible the intensity of the more excited lines is weaker) and the detection of these hydrides in low-excited PDRs, such as the Horsehead Nebula (e.g. see Figure 3 where line intensities decrease with UV radiation and pressure). The systematic detection of the emission of these hydrides will allow us to constrain of the physical and chemical conditions. That is why, we propose to observe three various PDRs such as the Orion Bar, NGC7023 and the Horsehead Nebula, with the FIRESS spectrometer.

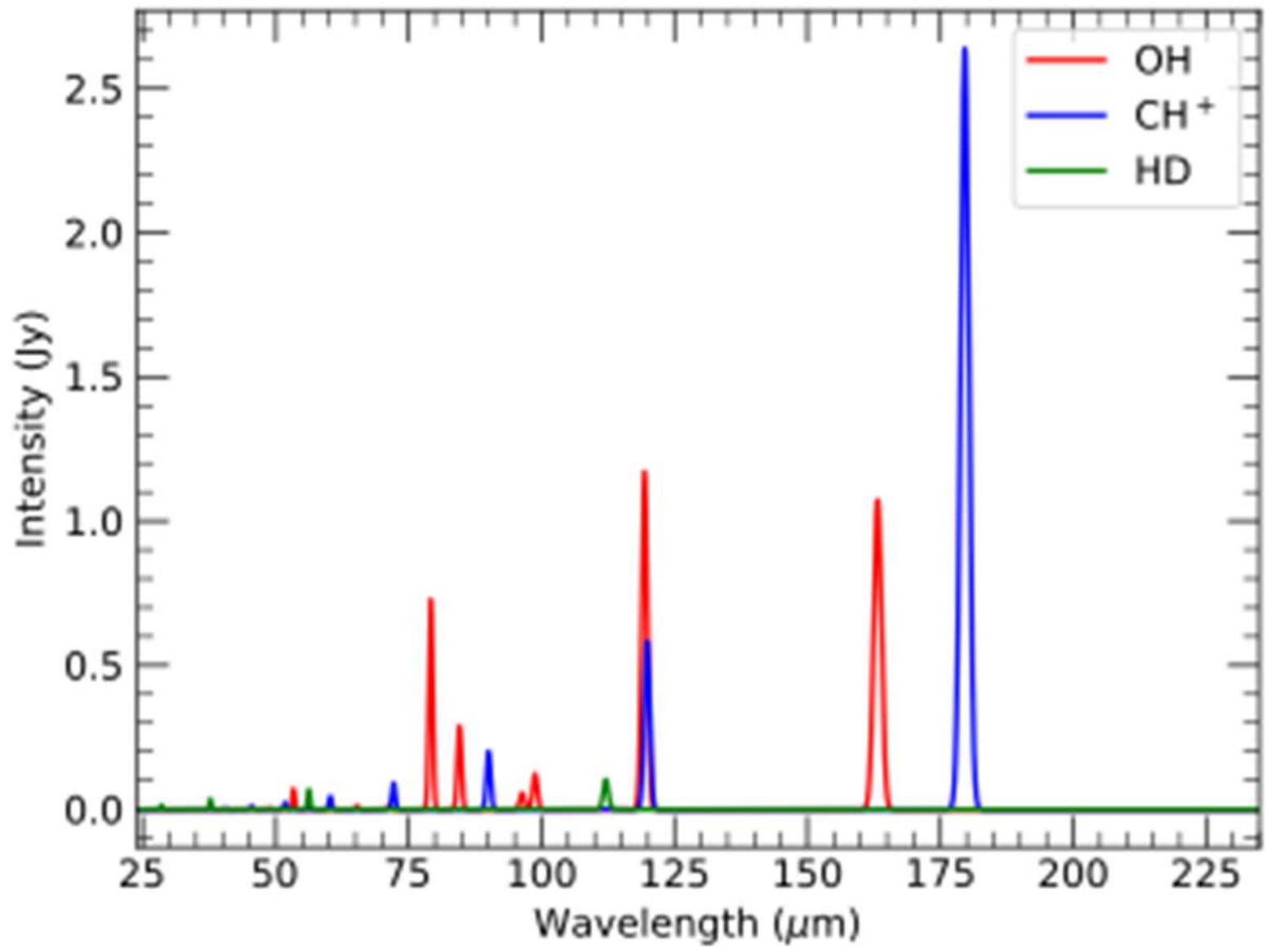

**Figure 1.** PRIMA simulated spectra of hydrides produced with a Meudon PDR Code model at P = $10^8$ K cm$^{-3}$ and $G_0$ = $10^4$.

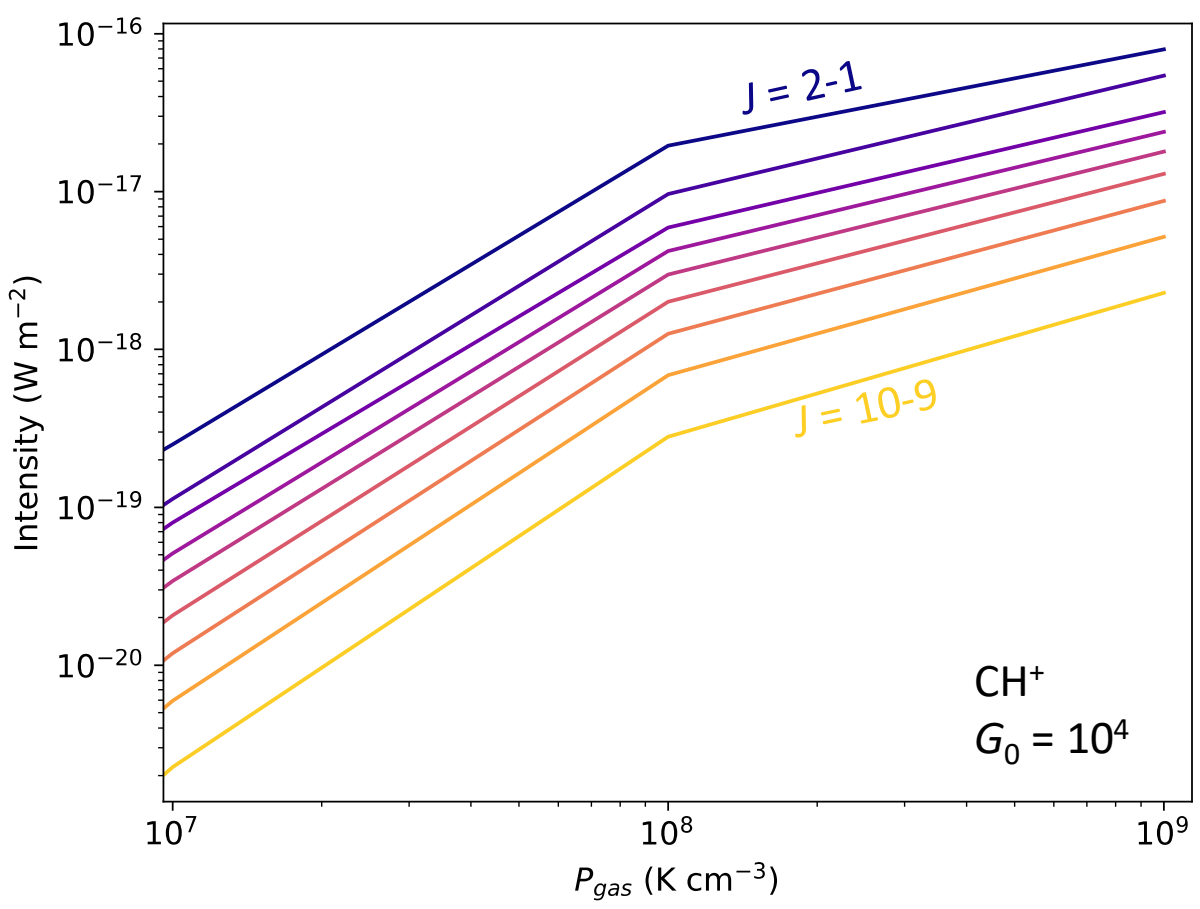

**Figure 2.** Intensity of $CH^+$ rotational lines detectable with PRIMA in function of the thermal pressure considering $G_0 = 10^4$. These intensities are underestimate as the models does not take into account chemical pumping.

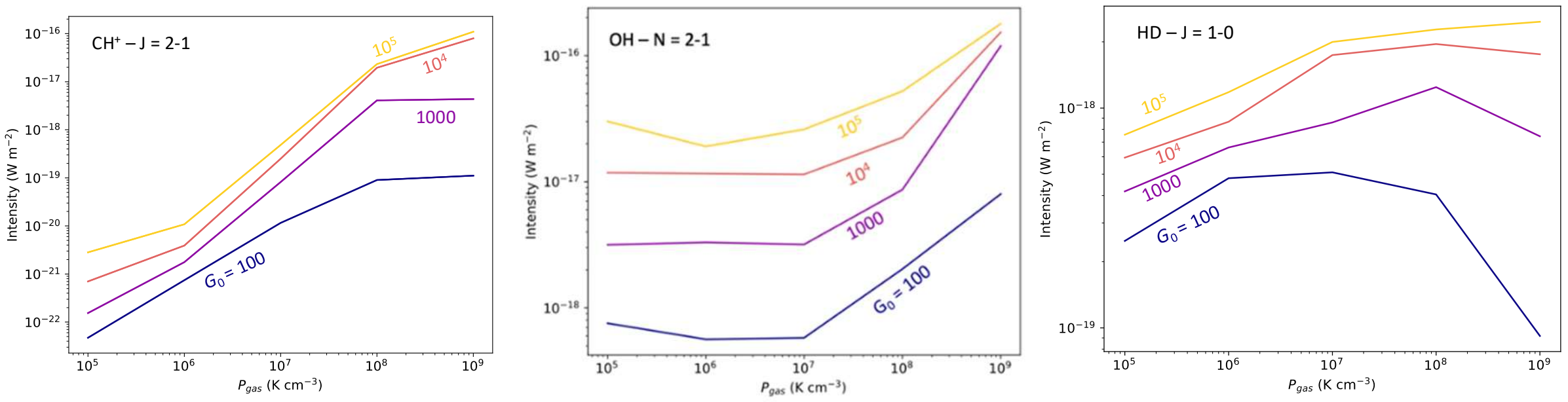

**Figure 3.** Intensity of $CH^+$, OH, HD line coming from the lowest levels detectable by PRIMA as a function of thermal pressure and UV field intensity.

## Instruments and Modes Used

FIRESS: Low-resolution map mode R~120-150 on the full spectral range to cover 5′ maps on three PDRs.

## Approximate Integration Time

10h (only band 1-3) for the Horshead Nebula ($P_{gas}$ ~5x$10^6$ K cm$^{-3}$, $G_0$ ~100) to detect the OH and HD lines. This would allow detecting the first rotational lines of HD J=1-0 - 112 $\mu$m (5x$10^{-19}$ W m$^{-2}$) and OH J=2-1 - 120 $\mu$m (6x$10^{-19}$ W m$^{-2}$).

10h (full spectral range: 2x5h) per source to detect highly excited lines of OH, $CH^+$ and HD in NGC7023 ($P_{gas}$ ~ 5x$10^7$ K cm$^{-3}$, $G_0$ ~ 1000) and the Orion Bar ($P_{gas}$ ~$10^8$ K cm$^{-3}$, $G_0$ ~$10^4$). In NGC7023, we would be able to detect OH N=3-2 - 84 $\mu$m (5x$10^{-19}$ W m$^{-2}$) ; $CH^+$ J=4-3 - 90 $\mu$m (6x$10^{-19}$ W m$^{-2}$) and HD J=3-2 - 38 $\mu$m (9x$10^{-19}$ W m$^{-2}$). In the Orion Bar, we would be able to detect OH N=3-2 - 84 $\mu$m (5x$10^{-18}$ W m$^{-2}$) ; $CH^+$ J=7-6 - 51 $\mu$m (1x$10^{-18}$ W m$^{-2}$) and HD J=4-3 - 28 $\mu$m (4x$10^{-18}$ W m$^{-2}$). Nevertheless, we emphasize that the detection of these lines might be limited by instrumental effects due to the low contrast line to continuum. With the low-resolution mode and considering the high continuum of PDRs, we expect this contrast to be possibly below 1% for some of the lines in the three PDRs. The possibility of detection of the faintest lines will be determined by the PRIMA instrumental capacity.

## Special Capabilities Needed

N/A

## Synergies with Other Facilities

These observations will allow synergies with the already existing observations of the JWST of these three PDRs. The JWST has allowed the detection of the rovibrational transition of these hydrides, unveiling important out-of-equilibrium effects. Observation of the rotational transitions of these species will complement our understanding of the out-of-equilibrium processes and their importance for pure rotational levels. The detection of rotational levels will also constrain the total abundance of these hydrides, which cannot be accessed through their rovibrational emission detected by the JWST. Indeed, the abundance in highly excited states represents only a small fraction of the total abundance, two orders of magnitude lower than that estimated with *Herschel*. Coupled analysis of the emission of hydrides with JWST and PRIMA will thus provide a more global view on the physics and chemistry at play in photodissociation regions.

## Description of Observations

We propose to use the FIRESS instrument in and low-resolution mode to map 3 emblematic PDRs (Orion Bar, Horsehead Nebula, NGC 7023), with existing JWST and *Herschel* observations. The strategy will be to make maps for different transitions in the full spectral range in order to detect the strong emission zones of the hydrids.

With 10 hours of integration time per source (30 hours total), we will be able to detect hydrides lines in all PDRs and some highly excited rotational transitions in the highly-excited Orion Bar.

We underline that PRIMA with FWHM beam widths ~7-25′′ will not be able to spatially resolve the thin warm molecular layers of ~1′′ where hydrides emit mostly (e.g., Zannese et al. 2025) but PRIMA is the only instrument than can detect their key rotational lines and complementary observations at higher angular resolution (JWST, ALMA) allow us to constrain the structure of the emission zone.

# 101. Characterizing the Interplay between Galactic Star Formation and Ionization Feedback with PRIMA

Annie Zavagno (LAM, Marseille, France), Russeil, Delphine (LAM, France), Traficante, Alessio (IAPS-INAF, Italy), Molinari, Sergio (IAPS-INAF, Italy), Battersby, Cara (University of Connecticut, USA), Di Francesco, James (University of Victoria, Canada), Schneider, Nicola (University of Cologne, Germany), André, Philippe (CEA-Saclay, France), Arzoumanian, Doris (Institute for Advanced Study, Kyushu University, Japan), ZHANG, Siju (Universidad de Chile, Departamento de Astronomía, Chile), SUIN, Paolo (LAM, France), Figueira, Miguel (Max Planck Institut, Germany & NCNR, Poland), Yadav, Ram Kesh, SAMAL, Manash Ranjan (PRL, India), RAWAT Vineet (PRL, India), André, Philippe (AIM-CEA, France), Mattern, Michael (AIM-CEA, France), Berthelot, Loris (LIS & LAM, France), Liu, Hong-Li (Yunnan University, China), Sadavoy, Sarah (Queen's University, Canada), Nozari, Parisa (Queen's University, Canada), Epinat, Benoît (CFHT, Hawaï)

We propose using PRIMA to study the impact of the early feedback of high-mass stars (e.g. photoionisation and winds) on the properties of star formation in our Galaxy. There are more than 8,000 ionised (HII) regions in our galactic plane, which have a significant impact on shaping the interstellar medium and the future history of star formation.

We intend to map the surroundings of a series of Galactic HII regions at different stages of evolution in order to characterise ongoing star formation in their associated photo-dissociation regions (PDRs). The high-sensitivity far-infrared data obtained with PRIMA will enable us to sample star formation activity down to low masses. Combined with multi-wavelength data and state-of-the-art, high-resolution numerical simulations, this programme aims to study the impact of early feedback from high-mass stars on star formation properties and star formation laws (SFR, SFE versus $\Sigma_{gas}$). A Galactic plane survey combining imaging and polarisation with PRIMA should be considered to maximise the scientific return, as demonstrated by previous infrared surveys such as IRAS, Spitzer-GLIMPSE and Hi-GAL. The PRIMAGAL survey, as proposed by Molinari et al. (2025), could be one such survey.

## Science Justification

Massive O and B stars ($M_{Star} > 8\ M_{\odot}$) have a profound impact on their surroundings through radiative feedback. Recent images of nearby galaxies, such as the Phantom galaxy (NGC 628), taken by the James Webb Space Telescope (JWST) with its Mid-Infrared Instrument (MIRI), demonstrate the significance of this impact in shaping the surrounding molecular medium (Kim et al. 2023). However, the physics of this feedback are not well understood. In particular, the effect that this feedback might have on the properties of future stars is a matter of much debate. Does this feedback have a constructive impact, favouring the formation of new stars, or a destructive impact, dispersing the gas and halting further star formation? While numerical simulations tend to conclude that this feedback is destructive (Walch et al. 2013; Geen et al. 2015), observations indicate the opposite, favouring the formation of a new generation of high-mass stars (Deharveng et al. 2010).

In many Galactic ionised (HII) regions, the photodissociation region (PDR) surrounding the HII region is observed to be a dense layer of gas and dust in which new stars, including high-mass stars, form. The supersonic expansion of the ionised gas in the surrounding medium creates a shock wave that travels ahead of the ionisation front, enabling the formation of a dense layer between the two fronts. This layer becomes unstable under its own gravity and fragments, forming dense clumps in the PDR where a new generation of stars forms.

Star formation occurs in dense clumps (0.1 pc in size) that fragment into cores (about 0.01 pc in size). These clumps and their associated cores are hosted by filaments. The process of star formation, from the assembly of the interstellar medium into filamentary structures to the formation of cores, is still highly debated.

The impact of the physical conditions (e.g. level of turbulence, magnetic field) of the original medium on the star formation process also remains puzzling, even though recent studies show that these conditions clearly affect the properties of future star formation (Zhang et al. 2020, 2021). In particular, the compression exerted on the surrounding medium by an expanding HII region can increase the local density, favouring the formation of a new generation of stars (Zavagno et al. 2020). This phenomenon has been observed in numerous galactic (Palmerim et al. 2017) and extragalactic (Bernard et al. 2016) star-forming regions. Young stars are observed at the edges of HII regions, towards their associated dense PDRs, within the filamentary structures that comprise them.

We propose using PRIMA and its suite of instruments to study how high-mass stars affect star formation in our galaxy. The questions we want to address are:

* How does compression from incoming stellar radiation (from ionising stars forming HII regions) influence the properties of filaments and clumps in PDRs around HII regions?

* How do the properties of filaments and clumps differ in zones that are directly irradiated (facing the incoming radiation) and zones that are less irradiated?

* How do the evolving physical conditions (density, temperature, turbulence and magnetic field) impact star formation evolution in the PDR?

We will compare the results of the PRIMA observations with state-of-the-art, high-resolution magnetohydrodynamic simulations of star formation (Verliat et al. 2022, Suin et al. 2024, 2025) and dedicated PDR models (Tiwari et al. 2022, Pound & Wolfire 2023).

## Need for PRIMA

Thanks to its high sensitivity, efficient imaging power, and polarimetric capability, PRIMA is particularly well suited to study the evolution of physical conditions in Galactic star-forming regions.

Active star-forming regions in the Galactic plane are bright in the infrared (from 30 to 2000 MJy⁄sr at 8 μm and from 20 to 1000 MJy⁄sr at 24 μm as observed with *Spitzer* and from 50 to 3500 MJy⁄sr at 250 μm as observed with *Herschel* (Pilbratt et al. 2010). In addition, star-forming regions such as the W3-W4-W5 complex are extended (about 9 deg x 3 deg in this case) and will require large maps to fully understand the star formation process there (see Figure 1).

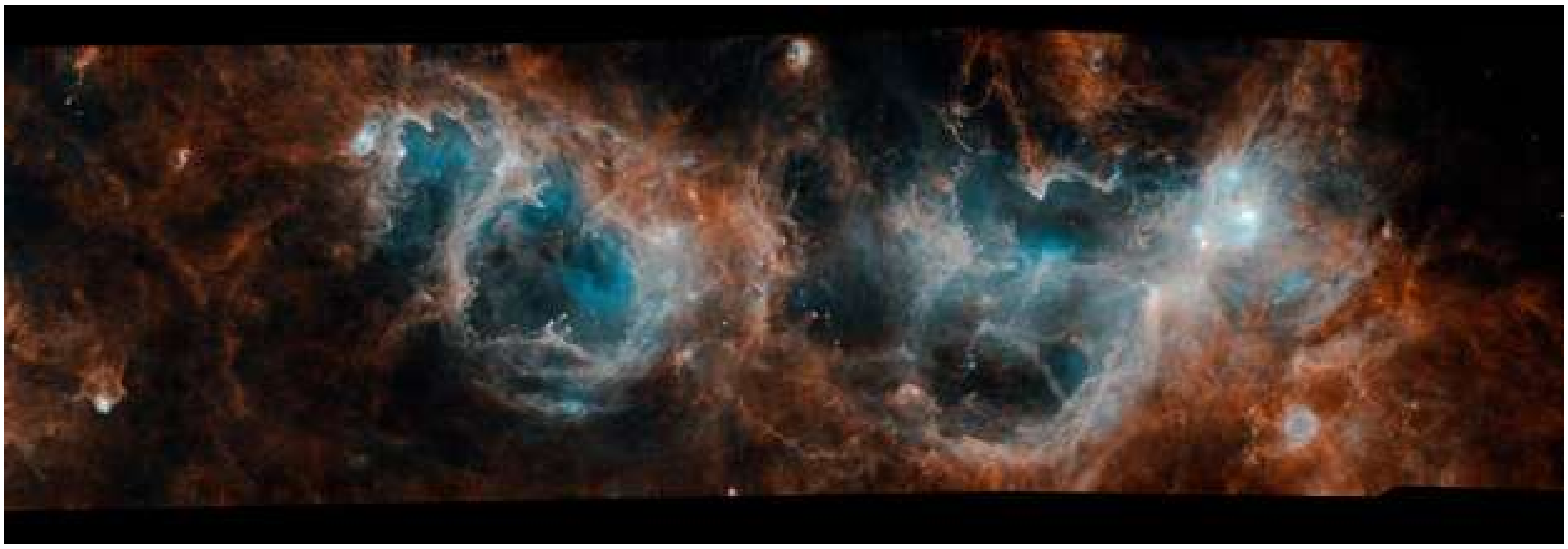

**Figure 1.** The W3-W4-W5 star forming complex as seen by *Herschel*. This two-colour image combines Herschel observations at 70 microns (cyan) and 100 microns (orange), and spans about 8.4° by 2.9°; north is up and east to the left. The HII regions appear in blue and are responsible for compressing the surrounding molecular gas in which new stars form. Copyright: ESA/Herschel/NASA/JPL-Caltech, CC BY-SA 3.0 IGO; Acknowledgment: R. Hurt (JPL-Caltech).

HII regions located above the Galactic plane are often large star-forming complexes that will benefit from PRIMA's high mapping efficiency. PRIMAger will achieve a surface brightness sensitivity (total power I, background subtracted) in the hyperspectral imaging bands PHI1 and PHI2 of about 1 and 0.7 MJy⁄sr (5σ, 10 h, 1 square degree), respectively (see Table 2 in Ciesla et al. 2025).

This will allow the detection of the lowest mass/most embedded young stellar objects in the field for a new estimate of the star formation rate and efficiency in the clouds under (and away from) the influence of radiative feedback. PRIMA's combination of photometric imaging, polarimetry, and far-infrared spectroscopy can address the important open questions such as the time evolution of the feedback effect of high-mass stars on low to high-density regions, and the role of the magnetic field in this evolution. The many Galactic H II regions already observed in the optical and infrared will be observed with PRIMA to cover a wide range of geometries (compact, bipolar, extended, as seen in Figure 1) that are directly related to the physical conditions of the medium in which they form and to their stage of evolution. PRIMA will have the sensitivity and mapping capabilities to cover a large area (>1 square degree), allowing the necessary statistics to be achieved in terms of the Galactic regions covered to envisage quantifying the impact of radiative feedback on star formation.

## Imaging

The infrared domain (25 to 300 μm) covered by the PRIMAger Hyperspectral Imager (PHI) is particularly well suited to study the star formation process over a wide range of densities, including the low-density medium ($N(H_2)$~$10^{20}$ $cm^{-2}$).

PRIMA's high-efficiency mapping provides such an opportunity and will be invaluable for studying star formation in large Galactic complexes.

Thanks to its high sensitivity and high mapping efficiency, PRIMAger will be able to map Galactic HII regions that sample different physical conditions (column density, temperature, turbulence level, and magnetic field) in their surroundings. The surface brightness sensitivity that can be achieved in a reasonable time (total power I) in the hyperspectral imaging bands PHI1 and PHI2 of 4.5 and 2.5 MJy/sr (5σ, 10 h, 1 sq deg) will allow the detection of the low-mass/most embedded young stellar objects (with typical fluxes of 1 mJy at 30 μm and 2 mJy at 70 μm) present in the field to obtain an unbiased census of the young stars/clump population. The sensitivity of PRIMAger will be key to a new estimate of the SFR and star formation efficiency and to discuss these quantities under (and without) the influence of ionization feedback.

## Polarimetry

The PRIMAger Polarimetric Instrument (PPI) will allow us to understand how the evolution of magnetic field properties (orientation, intensity) affects the star formation process and how this evolution is related to the radiative feedback over a wide range of densities. This is particularly important in regions such as the hub-filament systems, where the feeding of fresh material by the filaments allows star formation in the central cluster to continue and be modified as a function of time.

The typical surface brightness sensitivity (polarized intensity P) between 0.65 MJy/sr in PPI1 and 0.25 MJy/sr in PPI4 (5σ, 10 h, 1 sq deg, see Ciesla et al. 2025) will allow 5 to 10σ detection in the low-density regions ($N(H_2)$~$10^{20}$ $cm^{-2}$), where the polarization fraction is expected to be around 1%, and a higher signal to noise ratio in regions where the polarization fraction increases up to 8%.

## PRIMA FIRESS

Several positions towards Galactic HII regions have been observed with the *Herschel*-SPIRE-FTS (Griffin et al. 2010), showing a rich diversity in the physical conditions that are linked to the evolution stage of the media, from purely ionized to the molecular medium, that can be star-forming or not (Rodon et al. 2015)

The PRIMA FIRESS spectral range is ideal for observing the forbidden lines such as [O I] 63 μm, [O III] 88 μm [Si II] 35 μm, and [C II] 158 μm, which trace the PDR and characterize the physical conditions, including shocks and outflows, in star-forming regions using PDR models (Tiwari et al. 2022, Pound & Wolfire, 2023). Both low- and high-resolution spectra can be obtained at key positions that sample the variety of conditions observed in star forming regions, such as the ionized, neutral, and molecular medium, with or without ongoing star formation. Looking at the evolution of physical conditions from low- to high-density regions affected (and unaffected) by ionization feedback will be of particular interest to quantify the role of feedback. The FIRESS spectra will nicely complement the many *Herschel*-PACS and SPIRE-FTS spectra obtained on Galactic HII regions and their associated PDR. New spectroscopic results from the JWST will also be very important to get an overview of the physical processes at work in Galactic PDRs. Regions observed with *Herschel* show bright lines and continuum (both in the PACS and SPIRE bands). Using the current estimate of $7x10^{-19}$ W/m$^2$ in line emission (5σ, 1 h sensitivity), a FIRESS spectrum covering the range 24 to 235 µm can be obtained in 2 h (Bradford et al. 2025).

## Instruments and Modes Used

PRIMAGer mapping in both bands, using either large maps (1°x1° or larger), or pointed observations of 100-1000 regions

Pointed FIRESS low-res observations, 4 pointings per HII region.

## Approximate Integration Time

A Galactic plane survey with PRIMAger would be ideal to integrate this program.

The PRIMAGAL survey, proposed by S. Molinari et al. (2025), is a four-band polarisation survey of the galactic plane with |b| ≤ 1° (covering a total area of 720 square degrees). This can be executed by PRIMAger in approximately 1200 hours, including all mapping and instrument overhead. Imaging is also obtained in parallel. This time estimate does not include the FIRESS spectroscopy.

## Special Capabilities Needed

None

## Synergies with Other Facilities

There are many surveys of the Galactic plane. Multi-scale and multi-wavelength observations at different resolutions are available, including optical, near-infrared, mid-infrared, far-infrared Herschel and ALMA millimetre data. This means that the PRIMA data will complement existing data very well, enabling an analysis of the star formation process with an unprecedented level of sensitivity. This is a unique feature of PRIMA in this wavelength range. A survey of the Galactic plane is also proposed using the SKA, which would allow for a detailed study of the ionised gas within HII regions (Traficante et al. 2025).

Please note that a suite of dedicated numerical simulations of early (pre-supernova) radiative feedback is also available, which will enable a thorough understanding of the underlying physical mechanisms involved in star formation under feedback (Suin et al. 2024, 2025, see Figure 2).

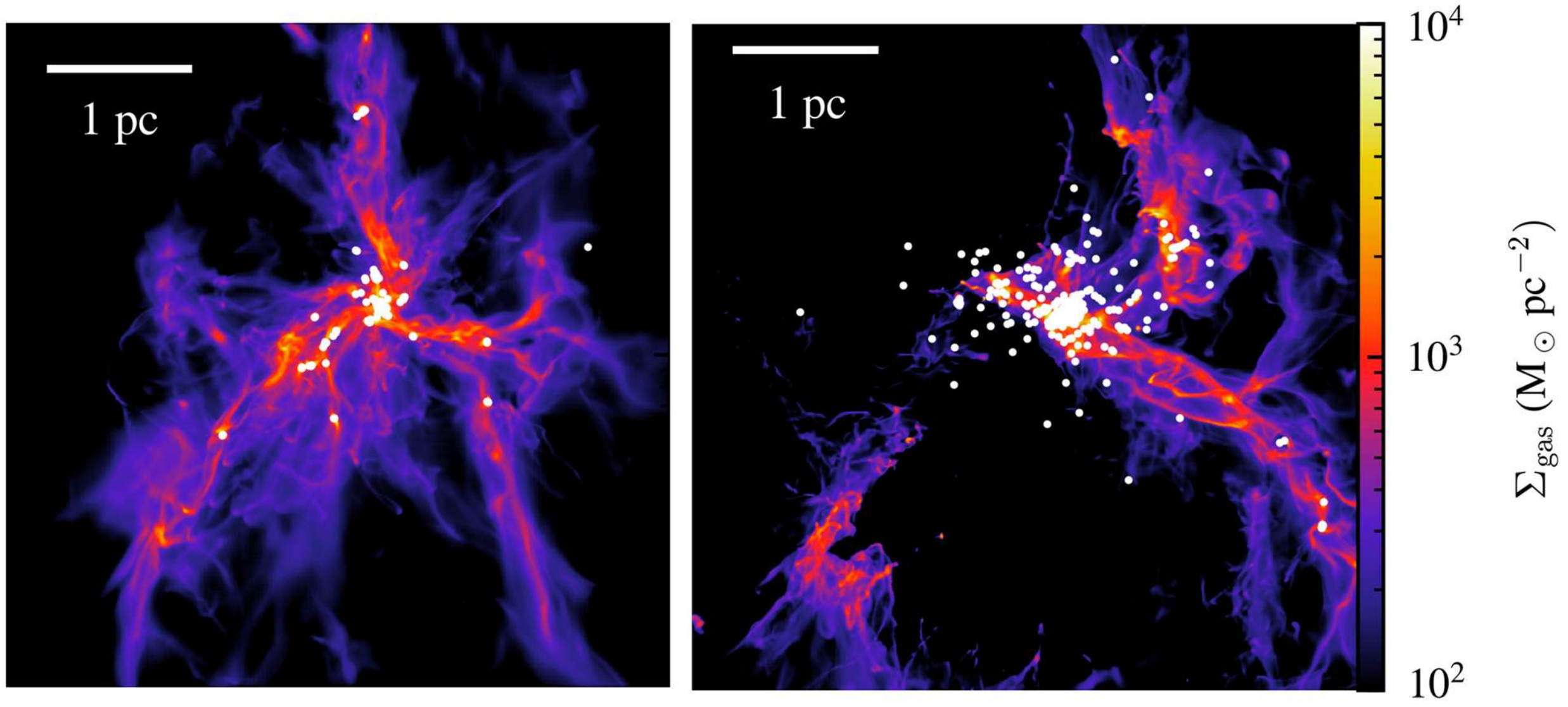

**Figure 2.** Numerical simulation showing the density projections along the z-axis for two snapshots of the simulation including the effect of feedback from a 100 $M_\odot$ star and protostellar jets from young stars on a $10^4$ $M_\odot$ molecular cloud at 2.3 Myr (left) and 3.15 Myr (right) (figure taken from Suin et al. 2024).

## Description of Observations

The most effective approach for this programme would be to incorporate the scientific objectives into a survey of the Galactic plane (or a portion thereof), as suggested by Molinari et al. (2025). This would maximise the scientific value of the survey, as was the case with earlier space-based infrared missions, such as IRAS, ISO, Spitzer and Herschel. Adding spectroscopic measurements with FIRESS will clearly add value to the programme, including determining the physical conditions in the observed regions.

# 102. Hidden Sulfur: Tracing Interstellar Metal-sulfides with PRIMA

Shaoshan Zeng (RIKEN, Japan), Izaskun Jiménez-Serra (Centro de Astrobiología, Spain), Yao-Lung Yang (RIKEN, Japan), Ange le Taillard (Centro de Astrobiología, Spain), Marta Rey-Montejo (Centro de Astrobiología, Spain), Laura Colzi (Centro de Astrobiología, Spain), Nami Sakai (RIEKN, Japan), Asunción Fuente (Centro de Astrobiología, Spain), Yuki Kimura (Hokkaido University)

Sulfur's significant depletion in the gas phase of the interstellar medi (ISM) remains a long-standing astrochemical mystery. While some sulfur is found in gas-phase and icy molecules, these account for less than 1% of the expected cosmic abundance, suggesting a substantial solid-phase reservoir. Meteoritic and cometary studies point to refractory metal-sulfides, such as iron sulfide (FeS) and magnesium sulfide (MgS), as major sulfur carriers in early solar system materials. Recent gas-phase detections of MgS and NaS in a shocked Galactic Centre cloud hint a solid-phase origin, likely from sputtered dust grains. However, direct observational evidence of FeS and MgS in interstellar dust has been elusive due to instrumental limitations. The upcoming PRIMA mission, with its FIRESS spectrometer, offers a unique opportunity to address this. FIRESS's far-infrared coverage (24–235 µm) encompasses diagnostic absorption features of FeS and MgS, particularly between 25–80 µm, which are inaccessible to current instruments like JWST. This program aims to identify and quantify FeS and MgS in interstellar dust grains towards various young stellar object (YSO) sources. By doing so, we will investigate whether these metal-sulfides are the dominant solid-phase sulfur carriers in the ISM and explore their formation and persistence conditions across different environments. Simulated profiles confirm FIRESS's capability to detect these features. This study will test the hypotheses that FeS and MgS are the primary solid-phase sulfur reservoirs, that their IR spectral features are reliable diagnostics of sulfur content, and that their abundance varies with environmental conditions. The sensitive detection of metal-sulfides by PRIMA will be a transformative step towards resolving the sulfur depletion problem and enhancing our understanding of interstellar dust composition and astrochemical evolution.

## Science Justification

Sulfur is an essential element in astrochemical processes and prebiotic chemistry, yet its observed gas-phase abundance in dense regions of the interstellar medium (ISM) is significantly lower than expected from solar abundances (S/H~1.5x10-5; Asplund et al. 2009). This discrepancy, known as the "sulfur depletion problem," indicates that a large fraction of sulfur is removed from the gas phase, but the identity of the solid-phase sulfur reservoir remains uncertain. While sulfur-bearing species such as $SO_2$ and OCS have been detected in ices in dense molecular clouds and star-forming regions (Geballe et al. 1985; Palumbo et al. 1995; Boogert et al. 1997; Rocha et al. 2024), more recently observations with James Webb Space Telescope (JWST) revealed that $H_2S$ is not detected in interstellar ices (McClure et al. 2023). However, the detected abundances are too low to account for the missing sulfur. In some cases, the contribution is less than 1% of the total expected sulfur budget, prompting the need to explore alternative solid-phase carriers.

Mineralogical studies of meteorites and cometary samples suggest that sulfur is commonly locked in refractory materials like metal-sulfides such as iron sulfide (FeS) and magnesium sulfide (MgS). Observations from the Rosetta mission and samples returned from asteroid Ryugu confirm that a substantial fraction of sulfur exists in non-volatile form, consistent with sulfide minerals (Calmonte et al. 2016; Yoshimura et al. 2023). These findings from the early solar system imply that metal-sulfides formed and persisted in the dense ISM prior to incorporation into planetary bodies. The recent detection of MgS and NaS in the gas phase within the molecular cloud G+0.693-0.027 in the Galactic Centre (Rey-Montejo et al. 2024), whose chemistry is believed to be produced by a large-scale shock driven by a cloud-cloud collision (Zeng et al. 2020), supports the hypothesis that such compounds are sputtered off dust grains by shocks, hinting at a solid-phase origin. However, direct observational evidence of FeS and MgS in dust remains elusive, largely due to the limitations of current instrumentation in sensitivity and wavelength coverage.

The PRIMA mission, equipped with the FIRESS spectrometer, offers a breakthrough opportunity to address this problem. The far-infrared range covered by FIRESS (24–235 µm) includes the diagnostic absorption features of FeS and MgS, particularly between 25–80 µm (Hony et al. 2002; Kimura et al. 2005). In particular, the 30 µm emission feature, long associated with evolved carbon-rich stars and planetary nebulae, is a prominent mid-IR signature mainly attributed to MgS. However, it has never been observed in the ISM since the absorption feature of MgS and FeS are weak and required high sensitivity observations (Hofmeister et al. 2003). These features are inaccessible to JWST due to its limited wavelength range (≤28 µm) and significantly lower sensitivity beyond 20 µm. In contrast, FIRESS's low-resolution mode ($R > 85$) provides sufficient spectral resolution and sensitivity to detect the relatively weak but distinct IR bands of metal-sulfides.

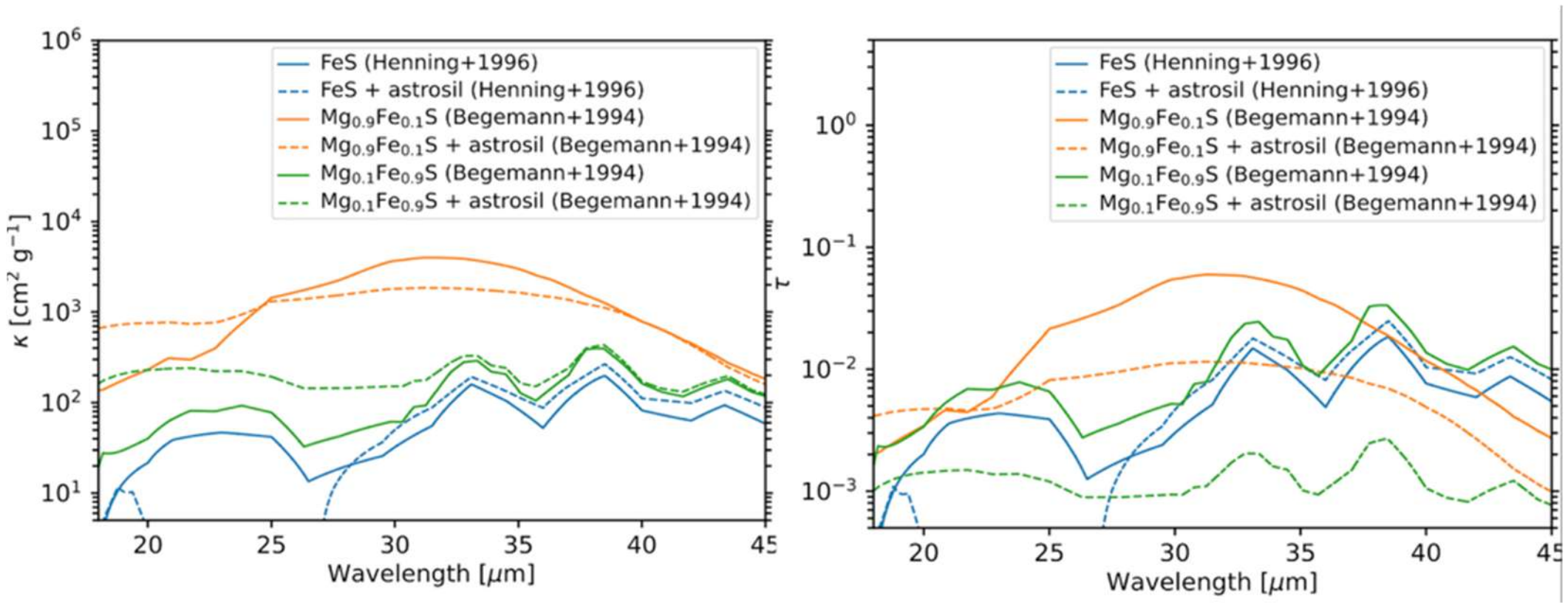

**Figure 1.** Left panel: Synthesised dust opacity profiles of 100% FeS (blue), of 100% $Mg_{0.1}Fe_{0.9}S$ (green), and 100% $Mg_{0.9}Fe_{0.1}S$ (orange) for pure samples (solid) and with astrosilicates mixtures (dashed) calculated using the dust model of Draine et al. (2003). Middle panel: optical depth for FeS and MgS molecular bands expected between 20 and 50 µm. The opacity and optical depth profiles have been smoothed to the expected resolving power of R>85 of the FIRESS low-resolution mode. Right panel: the synthesized absorbance profile.

The science objective of this program is to identify and quantify the presence of FeS and MgS in interstellar dust grains towards a collection of young stellar object (YSO) sources located in different environments, where sulfur elemental abundance has recently been constrained (Fuente et al. 2023). By doing so, we aim to determine whether these compounds represent the dominant sink of sulfur in the ISM and under what physical conditions they form and persist. Simulated profiles (Figure 1) for metal-sulfides and mixtures with silicates, to reflect realistic interstellar grain composition, indicate that these features are resolvable with FIRESS. We obtained the profiles with the optool software (Dominik et al. 2021) with a built-in model of FeS (Henning & Stognienko, 1996), and by using the JPDOC database for the optical constants of $Mg_xFe_{1-x}S$ (Begemann et al. 1994). Based on expected column densities and assumed sulfur depletion levels (e.g., 50% of sulfur in FeS and 5% in MgS), our models predict detectable absorption towards typical YSO source. The models incorporate realistic grain size distributions and mixing with astrosilicates, enabling robust identification of target features even under complex grain compositions. Measurements across various environments will help determine the ubiquity and abundance of FeS and MgS in the ISM.

This study tests several key hypotheses: (1) that FeS and MgS are the main solid-phase reservoirs of sulfur in dense molecular clouds; (2) that the IR spectral features of these compounds can be reliably used as diagnostics of sulfur content in dust; and (3) that the abundance and distribution of these materials vary with environment, indicating dynamic sulfur chemistry in star-forming regions. Moreover, the results will set constraints on competing hypotheses such as the presence of sulfur allotropes (e.g., $S_8$), which our analysis shows are not detectable with PRIMA within feasible integration times (>500 hours). Therefore, metal-sulfides are the most promising and accessible solid-phase sulfur candidates for PRIMA's observational capabilities. By enabling the first sensitive detection of metal-sulfides in interstellar dust, PRIMA will offer a transformative step toward resolving the sulfur depletion puzzle and advancing our understanding of interstellar dust composition, astrochemical evolution, and the origins of sulfur-bearing material in planetary systems.

## Approximate Integration Time

To demonstrate detectability, we present a case study using BHR 71, a typical ~10L⊙ low-mass protostar (Yang et al. 2017). Considering the observed continuum flux is ~3 Jy at 30 µm, the MgS absorption band at this wavelength yields an expected intensity of ~30 mJy while the FeS features are weaker, at ~3 mJy. Detecting the FeS feature with high sensitivity, S/N = 100, requires 2.5 hours using FIRESS in low-resolution mode. These short integration times make it feasible to survey a variety of environments, enabling exploration of how metal-sulfide abundance varies with physical conditions. Furthermore, predicted detection limits indicate that MgS can be observed in sources with continuum flux as faint as 15 mJy, and FeS in sources ≥ 150 mJy.

For each target, we will use Band 1 and 2 to probe the MgS and FeS, requiring two pointings. For a sample size of ~20, the total required observation time is 100 hours.

## Special Capabilities Needed

No special capability is needed

## Synergies with Other Facilities

Complementary JWST observations of sulfur-bearing ices (such as $SO_2$ and OCS) and complex organics in the mid-infrared will help distinguish between volatile and refractory sulfur reservoirs.

Highly sensitive ALMA observations of gas-phase sulfur-bearing molecules in the millimeter/submillimeter regime will not only contribute to a comprehensive sulfur budget when combined with PRIMA data, but also enable joint studies of gas–dust chemical interactions in protostellar environments and molecular clouds.

## Description of Observations

While FeS exhibits multiple distinct narrow features between 30 and 50 µm, MgS displays a broad feature at ~30 µm. These features will be resolved using the low-resolution mode of the FIRESS instrument with resolving power R>85, which corresponds to spectral resolutions between 0.35 – 0.5 µm within the target wavelength range. Observations will be conducted towards selected star-forming regions that provide strong continuum emission ideal for detecting the weak absorption features.

Suitable targets will be chosen from star-forming regions with known gas-phase surful abundances and will include embedded protostars with brightness levels comparable to our reference source, BHR71. Fuente et al. (2023) surveyed nine filaments in nearby star-forming regions (Taurus, Perseus, and Orion) to determine sulfur abundance, making these regions excellent candidates for studying refractory sulfur content. In each filament, we will select 2-3 embedded protostars, compiling a sample of approximately 20 sources. Both Band 1 and Band 2 will be used to fully cover the MgS and FeS features, requiring two pointings per source.

# 103. Probing Warm Gas Atmospheres of Protoplanetary Disks with CO Ladders

Ke Zhang (UW-Madison), Colette Salyk (Vassar College), Emma Dahl (JPL), Benoît Tabone (IAS, CNRS)

Understanding the physical and chemical structures of planet-forming disks is key to constraining the conditions of planet formation. Protoplanetary disks are known to have strong radial and vertical gradients, making characterizing the disk gas temperature structure crucial for us to interpret the observed line fluxes and infer the underlying density and chemical abundance structures in planet-forming disks. The pure rotational lines of CO, the so-called CO ladder, are an ideal "thermometer" to characterize gas temperature structures in disks. Previous Herschel observations detected high-J (J > 10) rotational CO lines in a few Herbig and T Tauri disks, and models have shown that these lines trace gas (>100 K) in the warm disk atmosphere between 10–100 au. With its superior far-IR sensitivity, PRIMA offers the opportunity to systematically survey CO ladders across a large sample of nearby Herbig and T Tauri disks. This will enable important constraints on the thermal structure of disk atmospheres, advancing our understanding of conditions of planet-forming environments.

## Science Justification

Accurate characterization of the physical and chemical structures of planet-forming disks is essential for our understanding of planet-formation conditions. In disk atmospheres, the gas temperature is expected to be much higher than the dust temperature (e.g., Kamp & Dullemond, 2004, Najita et al. 2011, Bruderer et al. 2012), as the gas can be heated by UV and X-ray radiation and the collisions between gas and dust grains are not sufficiently frequent to equilibrate the temperature between gas and dust in that low density environment. The exact temperature structure in a typical disk atmosphere is still largely unconstrained. The pure rotational lines of optically thick CO line emission (CO ladder) provide a useful thermometer to measure disk temperature structure. Low-J (J≤6) CO lines have been readily detected by ALMA, and in some disks, spatially resolved images (beam size of 10–50 au) have been used to reconstruct the disk's 2D temperature structure (e.g., Pinte et al. 2018, Law et al. 2021, Galloway-Sprietsma et al. 2025). However, the low J CO lines are mostly sensitive to outer disk regions (50–200 au), and vertical regions with Z/R <0.3. The temperature structure of the higher atmosphere between 10–50 au is still largely unconstrained.

High-J (J>10) CO lines have been detected in ∼10 Herbig disks and 3 T Tauri disks (Sturm et al. 2010, van Kempen et al. 2010, Meeus et al. 2012, 2013, Fedele et al. 2013). The absolute fluxes and shape of the CO ladder provide important constraints on the absolute gas temperature and how the gas temperature changes with radius in the disk atmosphere. Thermo-chemical models of the CO ladder in the HD 100546 disk indicated that its disk atmosphere (Z/R >0.3) has $T_{gas}$ >> $T_{dust}$ in the 10–100 au region (Bruderer et al. 2012). Using a small sample of CO ladders detected in Herbig disks, Fedele et al. 2016 showed that disk atmospheric temperature structures of Herbig disks appear to be diverse. The warm atmospheric temperature structure of T Tauri disks is still largely uncharted due to the lack of detections of multiple high-J CO lines.

As a sensitive far-IR facility, PRIMA provides an exciting opportunity to measure CO ladders in a large number of disks that are otherwise inaccessible with current observatories. Fedele et al. 2016 provided predictions of CO ladder fluxes of Herbig disks based on their thermo-chemical models, and showed that the CO ladder line fluxes are most sensitive to flaring angle, gas-to-dust mass ratio, scale height, and dust settling conditions. Here we add new DALI thermo-chemical models for T Tauri disks to estimate the detectability of CO ladders. Figure 1 shows the CO line emitting regions in two fiducial cases of T Tauri disks and Herbig disks. It shows that CO ladder fluxes are highly sensitive to the flaring condition of the disk. For nearby Herbig disks (<400 pc), we expect CO ladder lines ($J_{up}$ from 12–28) can be detected in both flared and flat conditions. For nearby T Tauri disks (<200 pc), the CO ladder lines can be detected in flared disks but would be challenging for flat disks.

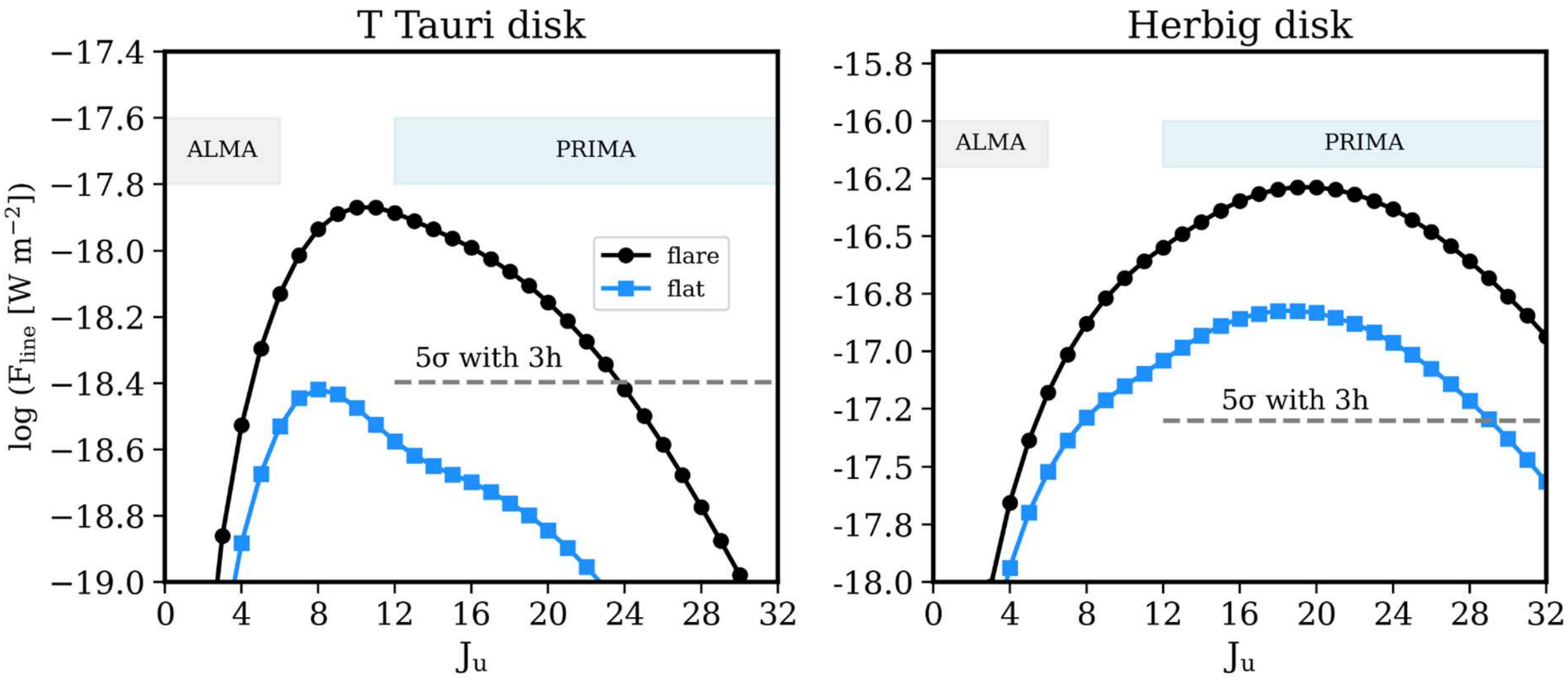

**Figure 1.** CO ladder fluxes from DALI thermo-chemical models for fiducial T Tauri and Herbig disk models at 150 pc. The T Tauri disk models are for $M_{disk}$ = 1e-3 $M_{sun}$, $L_{star}$ = 1$L_{sun}$, gas-to-dust mass ratio of 100, and Rc = 60 AU. The flared case has H/R = 0.09x(R/Rc)^0.25, and the flat case H/R = 0.05x(R/Rc)^0.1. The Herbig disk models are for $M_{disk}$ = 1e-2 $M_{sun}$, $L_{star}$ = 15$L_{sun}$, g2d=100, and Rc = 120 AU. The flared case has H/R = 0.1x(R/Rc)^0.25, and the flat case H/R = 0.05x(R/Rc)^0.1. The 5σ detection limit is estimated at 100 micrometers with a continuum flux of 70 mJy for T Tauri models, and 15 Jy for Herbig models.

PRIMA can detect CO ladder absolute fluxes and the shape of CO ladder to characterize the radial gas temperature structure in the disk atmosphere. The results are important for us to understand the conditions in the high atmospheric region, which is critical for disk wind and photoevaporation conditions, both of which can affect the material available for planet formation.

High-J $^{13}CO$ line fluxes are expected to be 10x lower than $^{12}CO$ lines, based on two detections with Herschel (Fedele et al. 2016). With PRIMA's sensitivity, we expect $^{13}CO$ and $C^{18}O$ to be detectable in Herbig disks, but would be much harder to detect in T Tauri disks except for the most flaring ones. The physical structure obtained from $^{12}CO$ is important to interpret the other racers like $H_2O$, N-bearing species, and hydrocarbons. CO is the main O and C carrier, and therefore, the optically thinner $^{13}CO$ and $C^{18}O$ isotopologue lines will provide additional information to construct vertical temperature structures in disk atmospheres.

## Instruments and Modes Used

FIRESS Spectrometer, High-resolution mode for bands 2, 3, and 4, single pointing

## Approximate Integration Time

Herbig disks within 400 pc typically exhibit 100 µm continuum fluxes between 1–50 Jy, while T Tauri disks within 200 pc generally range between 50–200 mJy at the same wavelength.

Assuming a representative 100 µm continuum flux of 15 Jy for Herbig sources, we estimate a 5σ line sensitivity of $5 \times 10^{-18}$ W m$^{-2}$ with a 3-hour integration. This sensitivity is sufficient to detect approximately 15 high-J CO lines in our Herbig disk model. Given the typical flux range of Herbig disks, we anticipate requiring 1–5 hours per source, allowing a survey of ~50 Herbig disks to be completed within a total of 150 hours.

For T Tauri disks, we assume a 100 µm continuum flux of 70 mJy, yielding a 5σ detection limit of $5 \times 10^{-19}$ W m$^{-2}$ with a 3-hour integration. This sensitivity is adequate to detect ~10 CO lines in the flared T Tauri disk model, but insufficient for detection in flatter disk models. We therefore anticipate requiring 3–5 hours per source and a total of 50 hours to survey ~10–15 highly flared T Tauri disks within 200 pc.

## Special Capabilities Needed

None

## Synergies with Other Facilities

Combine with CO low J lines measured by ALMA to constrain the thermal structure of the disk atmospheres between 10–500 au.

## Description of Observations

These observations use the FIRESS Spectrometer in high-resolution mode. The line emissions are spatially and spectrally unresolved, and therefore just require a single pointing per source. The PI science case of far-IR water line observations will cover the same wavelength range for CO ladders, and therefore, the CO ladder lines can be obtained for free for PI science targets. For additional sources, we propose to observe 50 nearby Herbig disks within 400 pc and 10–15 flared T Tauri disks (based on initial lower-resolution far-IR photometry studies) to measure CO ladders.

# Stars, the Sun and Stellar Populations

# 104. How Stars Assemble their Mass: Uncovering the Primary Mode of Stellar Accretion with PRIMA

Cara Battersby (UConn), Henrik Beuther (MPIA), Adam Ginsburg (University of Florida), Doug Johnstone (NRC-Herzberg), Rachel Lee (UConn), Klaus Pontoppidan (STSCI), Marc Audard (University of Geneva), Alessio Caratti o Garatti (INAF-OACN), Michael M. Dunham (Middlebury College), Robert Gutermuth (UMass Amherst), Dominika Itrich (University of Arizona), Jeong-Eun Lee (Seoul National University), Dylan Paré (Villanova University), Marta Sewilo (NASA Goddard Space Flight Center, University of Maryland), Yao-Lun Yang (RIKEN)

The mass of a star predominately determines its physical characteristics, life cycle, and ultimate fate. Decades of research have uncovered a common initial stellar mass distribution, however, the processes by which stars assemble their mass remain largely undetermined and may have a significant effect on the resulting star and its planetary system. Several lines of evidence suggest that forming stars, independently of their mass and age, can undergo substantial episodic mass accretion events that are short compared to the formation time; however, the predominance of this episodic mode of accretion is unknown. Do episodic accretion events dominate the buildup of a star's mass ($\Sigma M_{burst} \gtrsim 50\% M_{star}$) or are they less consequential? With PRIMA, we can unequivocally answer this question.

Only in the far-infrared (far-IR) is a protostar's luminosity directly linked with its mass accretion rate. While mid-IR and submm observations have demonstrated widespread and substantial luminosity variations indicative of both large and small accretion events towards protostars, emission at these wavelengths does not directly correlate with the magnitude of the accretion events. Through Monte Carlo simulations, we demonstrate that repeated observations of 2,000 Galactic protostars with PRIMA allow us to detect a sufficient number of large accretion bursts to unambiguously determine the primary mode of stellar mass accretion.

Protostars are highly clustered, and assuming a minimum map size of 3° x 1°, mapping 2,000 known Galactic protostars within ~2 kpc can be achieved through observing a total of 60 sq deg. over 13 star-forming regions. We propose to regularly monitor these protostars on cadences from two weeks to the entire five-year PRIMA mission lifetime and compare over 25 years against archival Herschel Space Observatory (and other relevant) data. We suggest that the first monitoring observations should be conducted with FIRESS FTM to enable meaningful target of opportunity (ToO) high-resolution spectral line follow-up for outbursting targets. Additionally, we anticipate that stacking PRIMAger data over all of the monitoring observations will provide ample polarization sensitivity across these key star-forming regions to enable exciting magnetic field science.

## Science Justification

### The mass of a star is very important but how this mass is accumulated is largely unknown.

The mass of a star determines nearly every aspect of its life and death. Its mass determines what elements it will produce in its lifetime, how those elements will ultimately be re-distributed, and the feedback effect that the star will have on its environment. The lifetime and final fate of a star is mostly dependent upon its mass and how that mass accretes may have a significant effect on the resulting star and its planetary system.

Countless studies have focused on determining the initial mass function (IMF) of stars and whether this distribution varies across environments in the universe (e.g. Motte et al. 2022, Damian et al. 2021, Offner et al. 2014). However, the detailed processes by which a star assembles its mass are largely shrouded in mystery. Since planet formation begins during the protostellar stage, the stellar mass assembly process will also affect the forming planets (e.g. Cridland et al. 2022, Tychoniec et al. 2020). What is the role of secular accretion[3] versus stochastic accretion bursts (e.g. Dunham et al. 2014)? How common are small ($\lesssim$ factor of 2 change in luminosity) versus large (factor of ~100) accretion bursts and how long does each last? How do such episodic bursts affect the chemical composition of the circumstellar disk and forming planets? PRIMA will enable us to answer these key questions and unambiguously determine the dominant accretion processes for forming stars.

### Mass accretion events are of great consequence, but parameter space is massively under-sampled.

A decade of observations from the mid-IR (e.g., Park et al. 2021; Zakari et al. 2022) to sub-mm (e.g., Mairs et al. 2024, Lee et al. 2021) have established that protostellar variability is common. These protostellar luminosity variations (ranging from a factor of 2 to 100) are the result of mass accretion events (Audard et al. 2014; Hartmann et al. 2016, Fischer et al. 2023). Classically, it was thought that protostars smoothly accrete through their protostellar disks via secular accretion from the surrounding molecular cloud (Stahler et al. 1980). However, several lines of evidence suggest that episodic mass accretion events play an important role in their mass assembly, independently of their mass or evolutionary stage.

The idea that episodic accretion events may be highly consequential was originally motivated by the luminosity problem (e.g., Kenyon et al. 1990), where mass accretion rates inferred from protostellar luminosities and integrated over the protostellar lifetime yield final stellar masses that are incompatible with the observed IMF. This is easily explained if mass accretion onto a protostar changes with time. The luminosity problem combined with observed protostellar variability suggests a solution involving stochastic accretion bursts (such that a large amount of time is spent at low accretion rates and a small amount of time is spent at high accretion rates; Dunham & Vorobyov 2012).

[3] The term “secular accretion” refers to very long-term changes in the accretion rate that are not due to distinct burst-like phenomena but are instead due to changes in the evolutionary state. If the accretion rate increases because the central mass has grown or if it decreases because the core density is decreasing with time, that's a secular change.

Additional evidence that variable accretion is common and important in the star formation process can be found throughout astrophysics. For example, the highly variable knotted morphology of protostellar outflows (e.g., Plunkett et al. 2015, Frank et al. 2014) suggests time-variable protostellar accretion. There is chemical evidence for episodic accretion, including water snowlines at larger orbital diameters than would be expected from measured protostellar luminosities (e.g. Lee et al. 2019). A few dozen extreme mass accretion events have been directly detected through time domain studies of protostars (e.g., Strom & Strom 1993; Caratti o Garatti et al. 2011; Safron et al. 2015; Audard et al. 2014; Lee et al. 2025).

The current state of the field and observations to date are summarized in the review by Fischer et al. (2023). We know that protostellar luminosities can vary on every timescale that has been observed, from a few weeks (Billot et al. 2012) to many decades (Kenyon et al. 2000). Thus far, observations are consistent with a continuum between these two extremes, but parameter space is massively undersampled. Current data defy any simple classification system. It additionally remains unknown if all protostars undergo large amplitude, stochastic accretion changes, or if slow, small-amplitude, secular-driven accretion changes can also resolve tensions between theoretical and observed protostellar luminosities (Dunham et al. 2014; Fischer et al. 2023).

**The far-IR is the only wavelength regime that reliably traces mass accretion in protostars. We need PRIMA to understand how stars gain their mass.**

The far-IR is the only wavelength regime capable of accurately tracing mass accretion onto forming protostars. The UV is not accessible due to the heavy extinction surrounding forming stars. While outbursts can and have been tracked at the mid-IR and sub-mm wavelengths, the measured luminosity at these wavelengths only loosely correlates with the amount of mass accreted. In Fischer et al. 2024 (see also Figure 1), we demonstrate that by measuring far-IR luminosity changes over time one can reliably and with relatively little scatter infer the true mass accretion magnitude of the event. At mid-IR wavelengths, optical depth (extinction) causes high uncertainty and depending on the orientation of the disk, outflow, and envelope, the mass accretion estimate can be off by a large factor. At sub-mm wavelengths, the optically thin envelope emits freely but the brightness depends instead on the dust temperature which varies with distance from the protostar and responds in a non-linear way to changes in the source brightness.

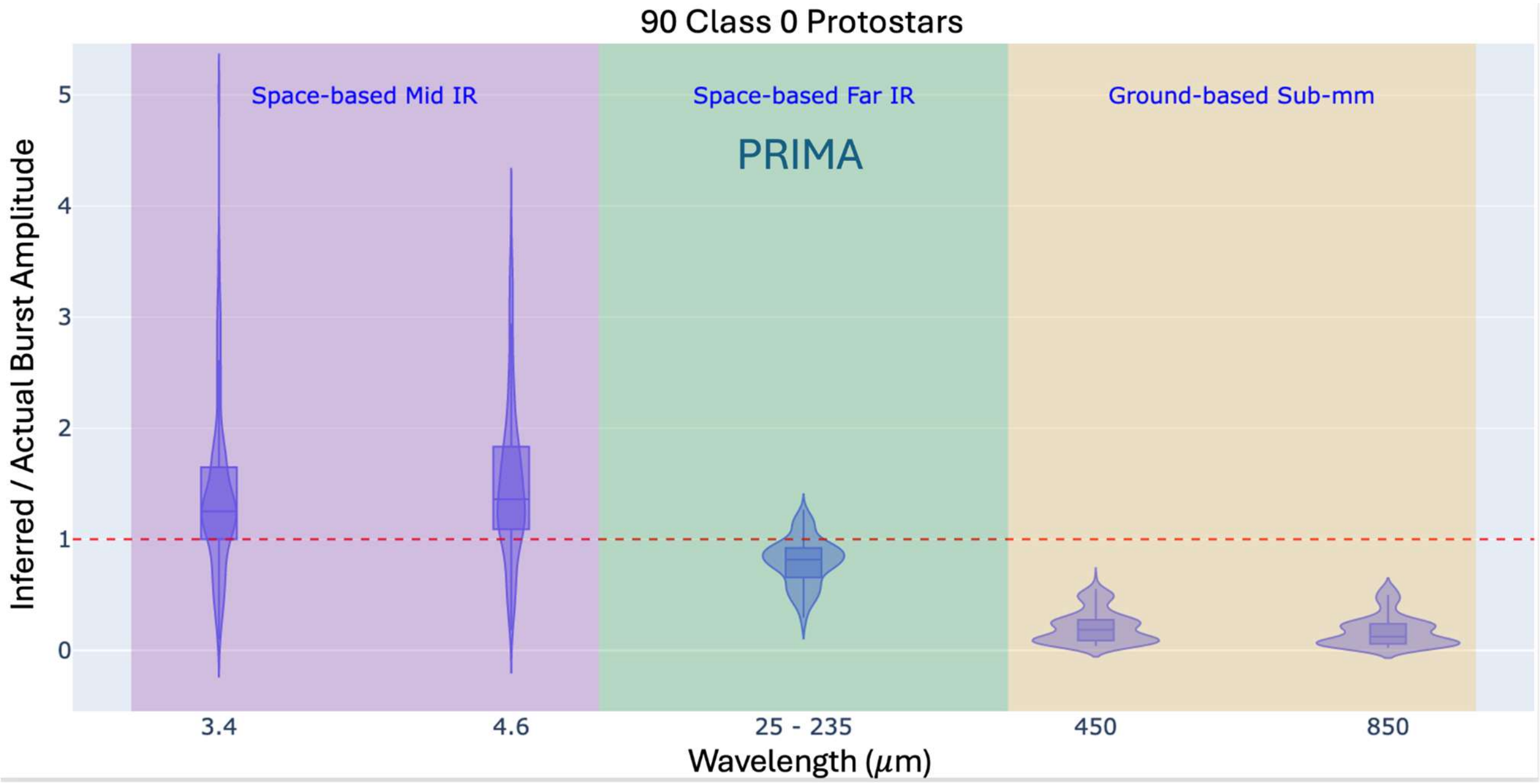

**Figure 1.** The far-IR is the only observable wavelength regime sensitive to the amount of mass being accreted onto a protostar during an accretion event (from Fischer et al. 2024). This figure uses detailed protostellar models towards 90 Class 0 protostars in Orion (Furlan et al. 2016) with a range of burst amplitudes (from 10 to 100) to plot the inferred burst amplitude versus the true (modeled) burst amplitude. In the mid-IR, there is a high amount of uncertainty due to extinction and disk, envelope, and outflow orientation and it is not possible to infer the true burst amplitude. In the sub-mm wavelength regime, the optically thin envelope emits freely but the brightness depends on the dust temperature which varies with distance from the protostar and responds in a non-linear way to changes in the source brightness. However, in the far-IR with PRIMA, we can unambiguously trace the protostellar mass accretion rate.

### PRIMA observations of 2,000 protostars are required to unambiguously determine the primary mode of protostellar mass assembly.

With PRIMA, we aim to unambiguously determine whether or not the majority of a protostar's mass is assembled via large, stochastic mass accretion events or via slow, secular accretion changes; in other words, is $\sum M_{burst} \gtrsim 50\%\ M_{star}$? As a limiting case, we assume that a protostar gains its mass primarily in rare, but large, episodic accretion events, which sets the limit of the number of protostars required for observations since these are the rarest events. Specifically, we take the case of extreme (FU Ori) bursts whereby the stellar luminosity and accretion rate increases by a factor of 100x over quiescence and lasts for a century or more. In order for such bursts to account for 50% of the final stellar mass, the protostar must undergo such bursts 1% of the time (Lee et al., in prep.). We consider this the limiting case, because the rarity of these bursts makes them the most challenging observationally. In any other case, where smaller, more common bursts play such an important role in mass assembly, we will naturally detect a sufficient number of bursts since they occur more often.

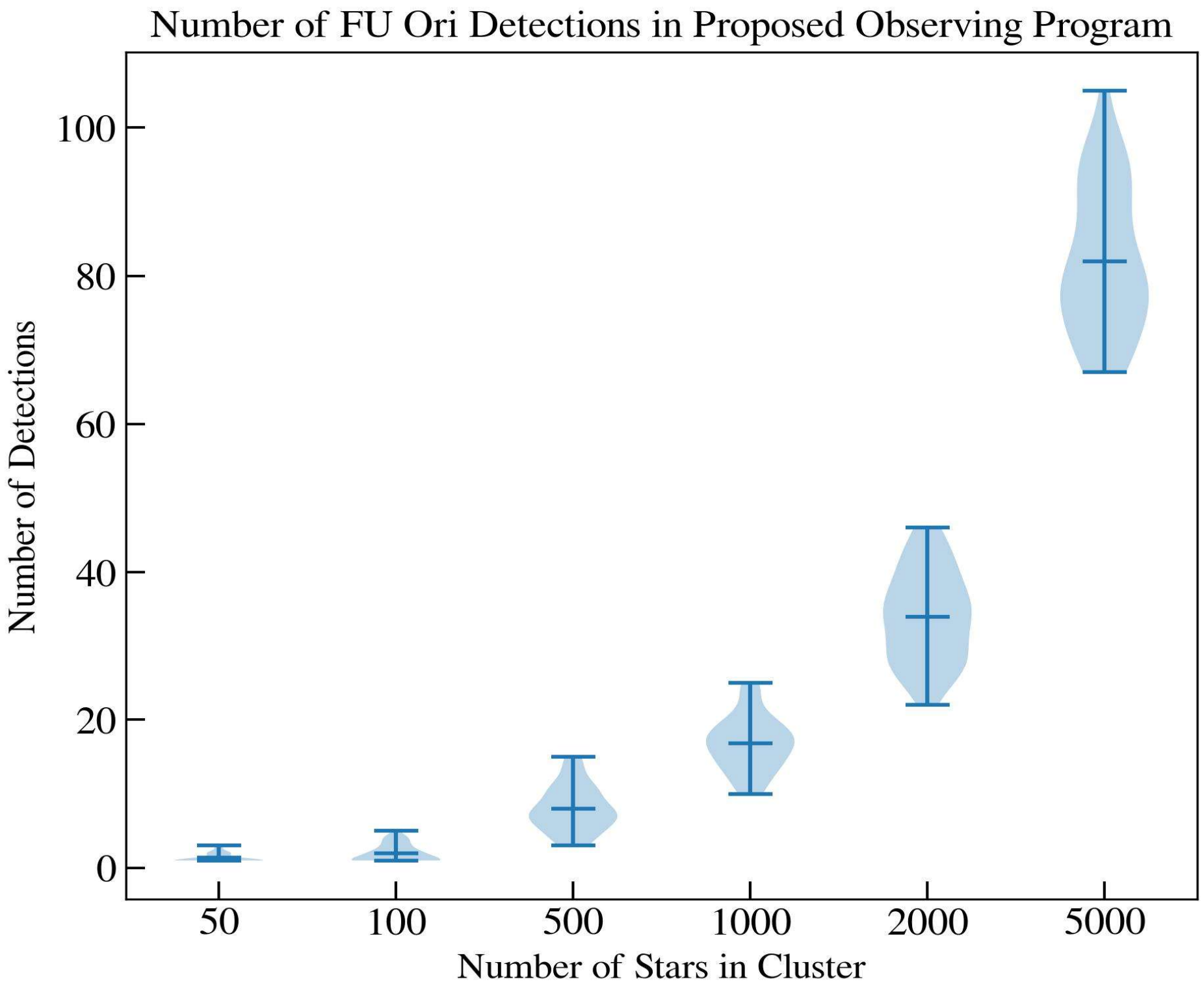

**Figure 2.** Monte Carlo simulations from Lee et al. (in prep.) show that we require observations of 2000 protostars to reliably and unambiguously determine the primary mode of protostellar mass assembly. This figure shows the number of stars that we would detect undergoing a 100x burst in each "PRIMA lifetime survey" (including the 25 year comparison with archival Herschel data) for different protostellar cluster sizes. The burst frequency was set at 1% as described in Lee et al. (in prep.) and we determine that the cluster size of 2000 is necessary and sufficient to reliably detect sufficient bursts in our repeated simulations.

We run simple Monte Carlo simulations of protostellar clusters with various burst magnitudes, lifetimes, and probabilities that are then "observed" with the proposed PRIMA monitoring campaign (Lee et al., in prep.). We find that in this limiting case, where a protostar gains most of its mass through large episodic accretion bursts, we need to observe 2,000 protostars, covering an extensive range of protostellar masses, in order to be able to detect a sufficient number of rare, large bursts. We demonstrate this in Figure 2, which shows (for a fixed input burst fraction of 1% ; Lee et al. in prep.) the expected number of large (100x) burst detections in a PRIMA lifetime survey ranging from 50 to 5000 objects. We determine that observing 2,000 protostars is necessary and sufficient in order for PRIMA to unambiguously decide the primary mass assembly mode for protostars.

Understanding how stars assemble their mass is of fundamental importance in astrophysics. Many essential open questions about star and planet formation have remained unanswered due to previous observational limitations. With our ambitious large survey with PRIMA, we can unambiguously determine the main mode of stellar mass assembly. This survey is planned as a community collaboration to optimize the science return, including Target of Opportunity Observations (ToOs), a complementary spectral line survey and post-outburst follow-up, as well as a complementary polarization survey.

## Instruments and Modes Used

For the main science case, mapped total bolometric luminosity is the required measurement, and this could be achieved using PRIMAger PPI or PHI mapping or FIRESS low-resolution mapping. Possible initial observations with FIRESS FTM high-res mode.

## Approximate Integration Time

### Main Science Case

10 observing epochs with PRIMAger or FIRESS

The main science case (protostellar luminosity monitoring) requires a measurement of total bolometric luminosity over the mapped area. This could be performed with PRIMAger or FIRESS (low-res or FTM). Assuming that the PRIMA mapping speed is about 1 sq. deg. per hour, then each target epoch of 60 sq. degs will require 60 hours. We plan to obtain 10 epochs for a total of about 600 hours of science time. Estimates with the ETC based on typical source fluxes indicate the telescope sensitivity is not the limiting factor since protostars are relatively bright, rather we would be limited by mapping speed.

### Bonus ToO Science Case

First epoch with FIRESS FTM plus follow-up

In the Bonus ToO science case, we request the first epoch of observations occur with FIRESS FTM to allow the opportunity for ToO spectral line follow-up of outburst events. We estimate that the time is the same for high-resolution mapping with FIRESS. So, if we do the full first epoch of observations with FIRESS that would be approximately 60 hours. Additionally triggered ToO observations would probably be for an individual field, ranging from about 1 to 5 sq. deg. in size or about 1 to 5 hours for follow-up. Please see GO Science Case by T. Megeath which describes this specific case in more detail.

### Bonus Polarization Science Case

PRIMAger polarization maps of all epochs

In the Bonus polarization science case, we use all four-bands of PRIMAger at every epoch to observe the protostars with sensitivity to polarization. We then stack the polarimetric images of each region at the end to achieve a polarization map of each region.

### Summary

- Main science case: 600 hours with PRIMAger or FIRESS (low-res or FTM)
- Bonus ToO science case: 60 hours with FIRESS FTM plus follow-up.
- Bonus polarization science case: The same as the main science case but all 600 hours are observed with PRIMAger polarization-sensitive observations. Stack the observations to achieve sensitivity polarized emission maps of each star-forming region.

## Special Capabilities Needed

### Main Science Case

We require protostar observations on cadences from 2 weeks to 5 years (start and end of mission). The short cadence observations should occur at the start of the mission so they can be repeated if necessary. We require a wide dynamic range to capture both dim protostars and bright outbursts in the same field (factor of $10^4$ - $10^5$ likely sufficient).

### Bonus ToO Science Case

FIRESS high-resolution mapping towards all targets in the first epoch. Triggered ToO follow-up observations with FIRESS high-resolution mapping towards outbursting sources. We estimate that a timescale of 2 weeks should be sufficient for follow-up due to the nature of these sources.

## Synergies with Other Facilities

This observing program is bolstered by comparison with archival data from the Herschel Space Observatory which has observations at far-IR wavelengths towards our proposed targets about 25 years before PRIMA's expected launch. In addition, we will compare with other archival data, such as Spitzer MIPS at 70 or 25 micron where available.

We plan to coordinate with ground-based and space-based facilities for complementary protostellar outburst monitoring. Coordinating observations with SPHEREx if still operating in the 2030s would be highly synergistic. We could trigger a ToO towards an active burst in multiple facilities to more fully map out the protostellar mass assembly process. We aim to coordinate with relevant observing programs and new facilities, for example, the ALMA 2030 upgrade as well as ngVLA and the SKA. We would connect with the maser monitoring project (https://www.masermonitoring.com) as well as the monitoring program planned for TAO (The University of Tokyo Atacama Observatory). TAO/MIMIZUKU can observe 24-38 micron and its monitoring program targeting bright protostellar sources will lend critical insights on the variability timescale of similar sources.

## Description of Observations

Main science case: We will observe the nearest ~2,000 protostars (including both low- to high-mass protostars), within 2 kpc, over the five year PRIMA mission lifetime. This is essential, since the most powerful outbursts are also the most rare, and we need to maximize our time baseline. The targets are listed in Tables 1 and 2. We will monitor the protostars ten times on cadences from 2 weeks to the full 5-year mission lifetime. Since the protostars are highly clustered, we will observe them in small maps of 1-3 sq. deg. in size. The main science case can be accomplished in any instrumental mode (PRIMAger, FIRESS low-res or FTM).

Bonus ToO science case: We will observe the 2,000 protostars in the first epoch using FIRESS FTM to get baseline spectra for each target. We would then propose ToO follow-up observations for outbursting targets and coordinate with other facilities. Please see GO Science Case by T. Megeath which describes this specific case in more detail.

Bonus polarization science case: We will observe all epochs with PRIMAger sensitive to polarization. While a single map alone may not have sufficient sensitivity to detect dust polarization toward these star-forming regions, integrating all ten region visits will produce interesting dust polarization maps of these star-forming regions.

### Cadence

The shortest timescale on which we expect protostellar variation is about two weeks, the time it takes light to travel across a 6” beam at the distance of Orion, a typical region in our proposed survey (Johnstone et al. 2013). We plan to observe all the protostars in small maps as soon as possible when PRIMA begins science observations, then to follow up on the targets again two weeks later. Additional observations, save the final one, can happen on a somewhat flexible schedule, with the caveat that we request to have the shortest cadence observations at the start of the mission. The first observations may yield unexpected variability which would lend key insights into the accretion process onto stars and may result in an altered observing program.

The final PRIMA protostellar observations should occur near the end of the mission lifetime, about five years after the first observations. We also plan to compare our observations with archival Herschel data to look for protostellar variability on timescales up to about 25 years, as well as with Spitzer MIPS archival data. The rarest type of protostellar outbursts are also the most powerful, increasing the protostars luminosity by a factor of about 100 and lasting for about a century. These are extremely rare and in order to unambiguously determine the importance of these outbursts, we need to maximize our time baseline as much as possible.

### Mapping Strategy

Since protostars are highly clustered, we plan to map each region in the smallest map possible (down to about 1° x 1° to be the most efficient), but for the purpose of this GO science case, we assume a minimum map size of 3° x 1°). The protostars are bright so the limiting factor is the mapping speed, not the sensitivity required.

In the Bonus ToO science case, we map the first epoch with FIRESS FTM and request ToO follow-up for outbursting events. The ToO observations may be maps or pointed observations.

In the Bonus polarization science case, we map all epochs with PRIMAger sensitive to polarization.

### Target of Opportunity Observations (ToOs)

In the Bonus ToO science case we do a baseline observation with FIRESS FTM in the first observing epoch see GO Science Case by T. Megeath which describes this specific case in more detail). Then, during the first major[4] detected outburst(s), we will trigger additional observations of that protostar / those protostars to observe: 1) how the outburst luminosity changes with time and 2) how the spectral features change with time. The ToO will be with PRIMA FIRESS FTM as well as being coordinated with relevant ground- and space-based facilities. For these follow-up ToO observations to be most useful, we require an initial spectral scan of all the targets with FIRESS at the start of the PRIMA mission to establish the pre-burst spectrum.

[4] TBD based on modeling efforts in the years to come what is meant by major and when we should trigger additional observations, but likely at least a 3-5x increase in luminosity.

### Bonus Polarization Science Case

Inclusion of the PRIMAger hyper-spectral band as well as polarization observations were highly regarded by community members. By integrating all ten region visits, we expect to detect substantial dust polarization signal toward each of the molecular clouds (protostars polarized generally at 10% level), fueling new and interesting studies of magnetic field properties towards these regions, e.g. determining whether ISM-scale magnetic fields play a role in regulating burst timescales. This determination can be made by studying how ordered the magnetic field is local to the protostars and whether the order of the magnetic field varies with time and in conjunction with burst events. This analysis can be done observationally and with the incorporation of model predictions.

### Targets

We plan to observe the following regions in individual maps (the smallest we can make, 1° x 1° is the most efficient, but our estimates assume a minimum map size of 3° x 1°). We have two categories of objects, those within 1 kpc and those out to about 2 kpc. Within 1 kpc, we have about 1000 protostars (mostly low-mass) over approximately 44 sq. deg. The farther out targets yield about 1500 additional protostars (including high-mass protostars) over between 10-16 sq deg, for a total estimated observing area of about 60 sq. deg to yield at least 2,000 Class 0 protostars.

Targets within 1 kpc:

| **Region** | **# of protostars** | **Reference** | **Map size (sq deg)** | **Typical source separation (pc)1** | **Distance (kpc)** | **Physical PRIMA resolution (pc)2** |
|---|---|---|---|---|---|---|
| Perseus | 111 | Dunham et al. (2015) | 6 (one 2x1 + one 2x2) | 0.05 | 0.25 | 0.009 |
| Auriga/CMC | 43 | Dunham et al. (2015) | 5 (5x1) | 0.05 | 0.45 | 0.016 |
| Orion | 319 | Furlan et al. (2016) | 10 (two 3x1 + one 4x1) | 0.05 | 0.42 | 0.015 |
| Mon R2 | 188 | Gutermuth et al. (2011) | 6 (3x2) | 0.05 | 0.83 | 0.029 |
| Ophiuchus | 71 | Dunham et al. (2015) | 6 (3x2) | 0.05 | 0.13 | 0.005 |
| Aquila | 148 | Dunham et al. (2015) | 6 (3x2) | 0.05 | 0.26 | 0.009 |
| Serpens | 52 | Dunham et al. (2015) | 0 (included in Aquila) | 0.05 | 0.43 | 0.015 |
| Cepheus | 29 | Dunham et al. (2015) | 3 (three 1x1) | 0.05 | 0.32 | 0.011 |
| IC 5146 | 38 | Dunham et al. (2015) | 2 (2x1) | 0.05 | 0.95 | 0.033 |
| Total | 999 YSOs | | within about 44 sq. degs. | | | |

[1] Figure 2 of Gutermuth et al. (2009) suggests that adopting a 0.05 pc source separation for all regions would be a reasonable way to proceed.

[2] at 70 micron

Targets within 2 kpc:

| Region | # of protostars | Reference | Map size (sq deg) | Typical source separation (pc) | Distance (kpc) | Physical PRIMA resolution (pc) |
|---|---|---|---|---|---|---|
| Cygnus X | 200 class 0<br>670 class I | Beerer et al. 2010 | 4 | clustered | 1.5 kpc | 0.052 pc |
| Carina | 247 class 0/I | Povich et al. 2011 | 2.25 | clustered | 2.3kpc | 0.081 pc |
| M17 | 406 YSOs | Povich et al. 2009 | 1.7 | clustered | 2.1kpc | 0.074pc |
| W3 | 592 class 0/I | Rivera-Ingraham et al. 2011 | 1.5 | clustered | 2.0kpc | 0.070 pc |
| Total | 1445 - 2115 YSOs | | about 10-16 sq. deg. when accounting for map shape | | | |

We acknowledge our beloved colleague Will Fischer who made major contributions to this PRIMA proposed observing program and field of protostar research before he passed away on April 16, 2024: https://www.mariettatimes.com/obituaries/2024/04/william-jack-fischer/

# 105. Characterizing Cold Dust Around White Dwarfs with PRIMA

Amy Bonsor (Institute of Astronomy, University of Cambridge), Jay Farihi (Dept of Physics & Astronomy, University College London), Mark Wyatt (Institute of Astronomy, University of Cambridge), Christine Chen (Space Telescope Science Institute, John Hopkins University)

The planetary systems orbiting, sometimes transiting, and often polluting white dwarfs provide a unique window into the bulk composition and geochemical assembly of exoplanets that is not available around other types of stars. Closely orbiting circumstellar dust is often observed in those stars externally polluted by heavy elements, and detected at wavelengths available to the *Spitzer* and *WISE* missions, but to date the parent body reservoirs necessary to supply this dust have not been detected. These reservoirs should be analogous to belts of debris seen in the solar and extrasolar planetary systems, namely main-belt and Kuer-belt analogs, and the PRIMA mission is well suited to detect these for the first time. Here we describe a model that predicts long-term dust production in these cold debris belts around white dwarfs, and propose a PRIMA survey of 50 young and luminous targets where there is sufficient sensitivity to detect these. The model predicts cold debris belts surviving from the main-sequence can be detected around 7-8 of these white dwarfs in-situ, but it is likely that cold dust supplying the inner discs can be detected in more cases. This survey will for the first time trace the fate of outer planetary systems as the star evolves to become a white dwarf.

## Science Justification

Almost all planet-host stars known to date will end their lives as white dwarfs. While stellar evolution is dramatic and exoplanetary systems out to a few au are likely cleared (Veras et al. 2016), the effects on the outer regions of are relatively benign (Bonsor et al. 2011; Smallwood et al. 2018). Mass loss leads orbits to expand by the ratio of the main-sequence to white dwarf mass (Jeans 1924), a factor that is typically two to three If we take the solar system as a benchmark, white dwarf planetary systems are likely characterized by cold, outer planetesimal belts and planets orbiting at tens of au.

These outer planetary systems have never been directly observed, although two attempts have been made with *Herschel* and ALMA of the prototype and nearest and brightest example, G29-38 (Farihi et al. 2014). Our knowledge of white dwarf planetary systems comes instead from the inner regions, where observations of emitting dust, gas, and transiting debris probe the accretion of material in close proximity to the star (e.g. Farihi 2016). White dwarfs are unique laboratories for exoplanetary science (e.g., Jura & Young 2014; Buchan et al. 2024). Owing to enormous surface gravities, their spectra should contain only hydrogen or helium, with heavier elements sinking out of sight on timescales of days to millions of years. The accretion of planetary material even from bodies as small as kilometres in size (the accretion of one human per second is detectable with the most sensitive observations), shows up as heavy elements in their spectra (Koester et al. 2014; Ould Rouis 2025). Photospheric characterization of white dwarf spectra uniquely provide the bulk composition of planetary material (e.g. Zuckerman et al. 2007; Gänsicke et al. 2012). One of the biggest open questions is the origin of this material, where possibilities in include main—or Kuiper-belt analogs, and destabilized exo-moons that might create a new debris field during the post-main sequence (Payne et al. 2017; Akiba et al. 2024). The star-grazing orbits necessary to pass within the stellar Roche limit require such an intact reservoir, as well as major planets to dynamically perturb their orbits.

Emission from hot dust in the near infrared connects the orbiting planetary debris with the metals observed in the photosphere. All the dust observed to date through its thermal emission is found within ~100 stellar radii of the white dwarf at temperatures comparable to the sublimation temperatures for silicate dust (1,000-1,600K). The dust emission is detected photometrically, and in a handful of cases spectroscopically, with most detections at 3 and 4 µm with *Spitzer* or *WISE* and a handful of objects having longer wavelength detections at 24 µm with *Spitzer*. In fact, 25 metal polluted white dwarfs have such 24 µm observations, but with few actual detections.

While the hot dust traces material that is closest to the star and in the long-term process of being accreted onto the stellar photosphere, significant questions remain regarding the nearly certain link with an outer, surviving planetary system. Exactly how is the dust supplied from the outer planetary system? Tidal disruption models suggest multiple populations of dust, including between the inner hot component and outer cool component (e.g. Malamud & Perets 2020). The growing handful of white dwarfs with transiting debris on orbital periods of hours to days to hundreds of days highlight the likely diversity of orbiting dust streams and reservoirs within these systems (e.g. Guidry et al. 2021).

## Science Question

### Are White Dwarfs orbited by cold, outer Kuiper-belts? What are their properties?

While models predict that white dwarfs should be orbited by a population of cold, outer planetary belts and cold, outer planets, there are currently no observations capable of probing these regions of their planetary systems. Exoplanets are essentially undetectable around white dwarfs using radial velocity or transit searches—the best hope may be direct imaging or astrometric detection via *Gaia* astrometry (e.g. Rogers et al. 2024)

Although ALMA*, Herschel, Spitzer* can probe the brightest Kuiper-belt analogues around main-sequence stars, by the white dwarf phase, a combination of collisional grinding and a faint host-star means that these dust discs are intrinsically orders of magnitude fainter. In a similar manner to the solar system Kuiper belt, belts orbiting white dwarfs are modeled to lie below current detection thresholds (Bonsor et al, 2010). PRIMA is sufficiently sensitive to detect these belts, particularly around the hottest and most luminous (youngest) white dwarfs.

## Science Question

### Does the population of cold, outer Kuiper-belts orbiting white dwarfs trace their evolution from the main sequence?

Kuiper-belt analogues should survive stellar evolution to the white dwarf phase. While heating and drag may remove the small dust during the giant phases, these are quickly replenished. Bonsor et al., 2010 predict that 15% of cold, outer belts orbiting young (<10Myr) white dwarfs (L_star>0.5L_solar) should have 70 µm emission brighter than 0.02mJy within 100pc. The proposed PRIMA survey can confirm if discs white dwarf discs evolved as predicted by the aforementioned model. Any discrepancies would be indicative of the physical processes at play, which can include dynamical instabilities within their planetary systems and extreme thermal processing during epochs of high luminosity. In particular, comparison between the properties of these discs and their main-sequence counterparts can probe the unseen population of large planetesimals orbiting white dwarfs that must supply the discs observed to date and atmospheric pollution.

## Science Question

### How is planetary material supplied to the inner regions of the white dwarf system?

Both photospheric metals and the inner, hot dust must originate from the colder, outer planetary system, but it remains an open question exactly how these are supplied. This PRIMA survey has the potential to resolve this question by detecting any dust orbiting between the innermost discs that are known, and any cold, outer belts. This dust could be a result of the tidal disruption process, which can create subsantial streams of dust spanning a wide range of orbits but all sharing the same periastra, and this is likely linked to the longer-term transiting events detected to date (Vanderbosch et al. 2020; Brouwers et al. 2022).

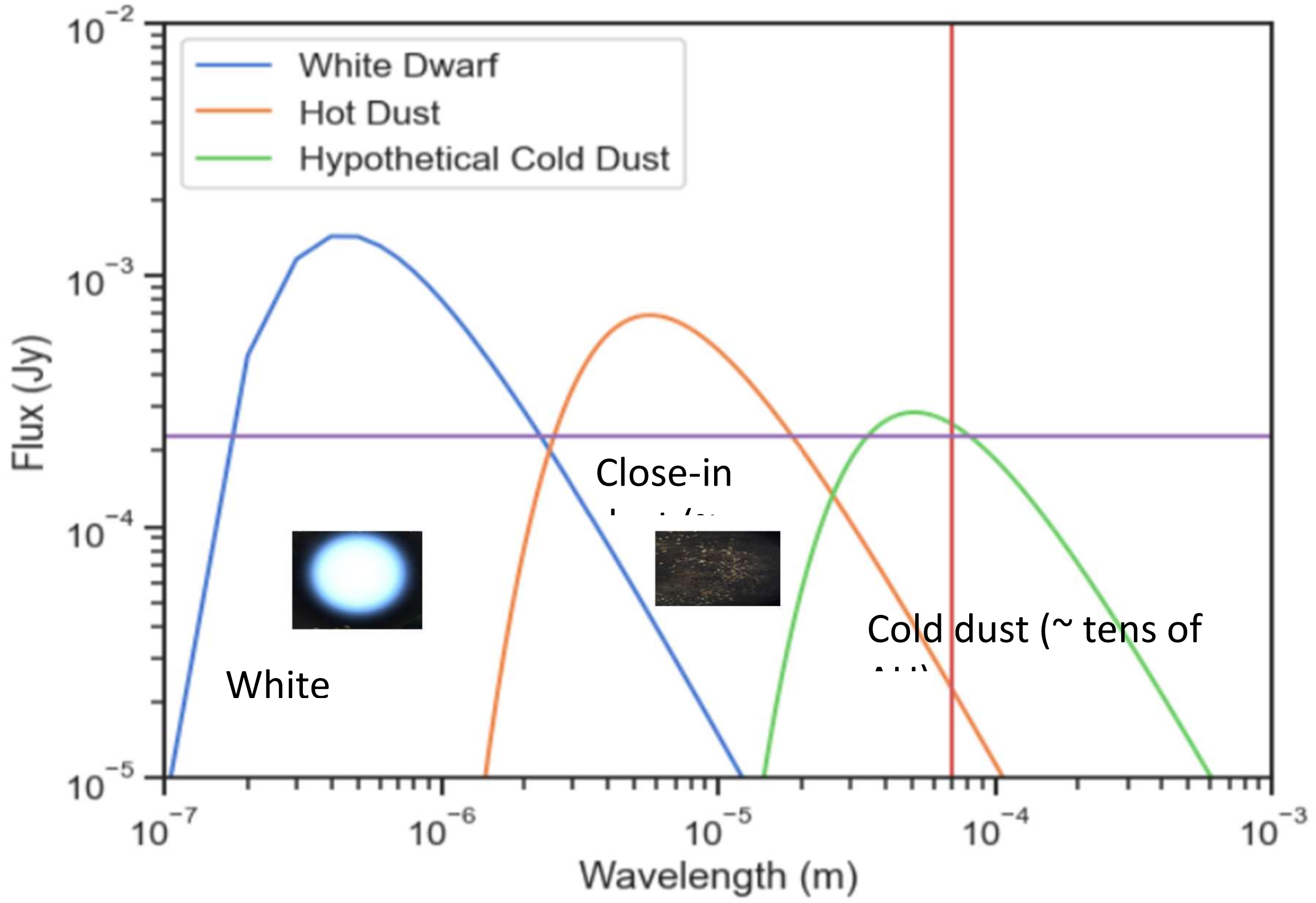

**Figure 1.** Toy model to illustrate that PRIMA could detect hypothetical cold, outer dust belts. Plotted is the flux in Jansky from a white dwarf with a hot dust component at 900K (modelled on GD 16 above). An additional cold component at 100K, which lies just above the PRIMA sensitivity of 0.23mJy (purple horizontal line) at 70 µm (red vertical line).

## Description of the Observations

### The Survey

The proposed PRIMA survey has the best chance to detect the cold, outer debris belts orbiting white dwarfs by observing a sample of 50 white dwarfs within 200pc and with L_star ~ L_sun. The model prediction is 7-8 of these target should have detectable cold outer belts at fluxes greater than 0.03mJy at 70 µm. Each system with a detected cold, outer belt will have an enormous impact on the field, opening up new avenues for the characterization of white dwarf planetary systems, connections to their main-sequence counterparts, dynamical and evolutionary modeling of the parent bodies and their debris. It woudl also provide the currently missing link between the photospheric metals and the source regions of the parent bodies; where they were formed and their likely chemistry at formation. The low-resolution spectra from this PRIMA survey will characterize these outer belts, determining their radial location and extent.

## Instruments and Modes Used

FIRESS low res pointed spectroscopy, one pointing per source, 2 spectral setups

## Need for PRIMA

To date, attempts to detect outer dust belts around white dwarfs have been limited to *Spitzer* 24 µm observations with a sensitivity limit of roughly 0.1mJy (Farihi et al. 2016), where a similar limit is possible with ALMA band 7 (1.3 mm; Farihi et al. 2014). However, these wavelengths do not probe the peak of emission from dust with temperature near 50K, which instead peaks close to 60 µm. Many extrasolar Kuiper belts have dust temperatures near 100K, and if these cold belts expand by a factor of a few owing to post-main sequence mass loss, it is expected their dust will be in the range of 30-70K for sufficiently bright white dwarfs. Only PRIMA can probe these dust temperatures and corresponding orbital radii in this range, owing to its unique spectral range.

## Approximate Integration Time

3.6 hours per target yields a sensitivity of 0.08mJy at 24 µm and 0.23mJy at 70 µm at R=100. In order to reduce the sensitivity, we will bin the data down to a resolution of R=2 and a sensitivity of 0.03mJy at 70 µm. This survey will be a volume-limited sample, focusing on the 50 white dwarfs with luminosities above solar within 200pc. This sample size is selected to reflect the best chances of detecting the cold, outer belts (Bonsor & Wyatt 2010). Total integration time: 2 x 3.6 x 50 = 360 hours

## Synergies with Other Facilities

Discovery of new debris disks through this survey will provide new targets for high resolution follow-up with facilities such as JWST that will characterize the interplay between the inner dust and cold, outer belts, probing composition and population of small grains. In the (sub-)mm, deep observations with the Atacama Large Millimeter Array for the brightest discs, will have the potential to map the larger grains and the disc surface morphology in these systems, based on the knowledge that the dust is detectable.

# 106. Determination of Forsterite Grain Temperatures in Circumstellar Environments Using Narrow Emission Bands at 49 µm and 69 µm

Janet E. Bowey (Cardiff UniversityOlivia C. Jones (UK Astronomy Technology Center)

Infrared emission bands due to forsterite (the regularly identified crystalline silicate in the mid- to far-infrared) are observed in circumstellar regions around evolved stars, young stellar objects, protoplanetary discs and debris discs in the Milky-Way and beyond. Systematic changes in the peak wavelengths and widths of forsterite bands near 49 and 69 µm with temperature can be used to estimate grain temperatures from emission features in circumstellar environments. Grain temperatures of the optically thin component are indicative of their location relative to optically thick emission from dust in a circumstellar disc, envelope or outflow. This information is used to constrain the system morphology, to aid the interpretation of images and could constrain chemical studies. Time-series observations in rapidly evolving environments including SN, AGB stars and eruptive variables might also prove instructive. PRIMA will be the first observatory able to provide data for the 49-µm peak and will detect the 69-µm peak in more environments than was possible with Herschel or ISO.

## Science Justification

Infrared emission bands due to forsterite (the most regularly identified crystalline silicate in the mid- to far-infrared) are observed in circumstellar regions around evolved stars, young stellar objects, protoplanetary discs and debris discs in the Milky-Way and beyond. Laboratory data by Bowey et al. (2001) showed bands near 49 and 69 µm in the spectrum of crystalline forsterite, $Mg_2SiO_4$, sharpened and shifted to shorter wavelengths when the temperature was reduced from 295 K to 4K. Systematic changes in the peak wavelengths and widths of features in the laboratory data with temperature (Figure 1) have been utilized to estimate the grain temperatures of forsterite grains responsible for optically thin 69 µm emission in circumstellar environments (Bowey et al. 2002, Sturm et al. 2013, Blommaert et al 2014). PRIMA will be the first observatory able to provide data for the 49-µm peak.

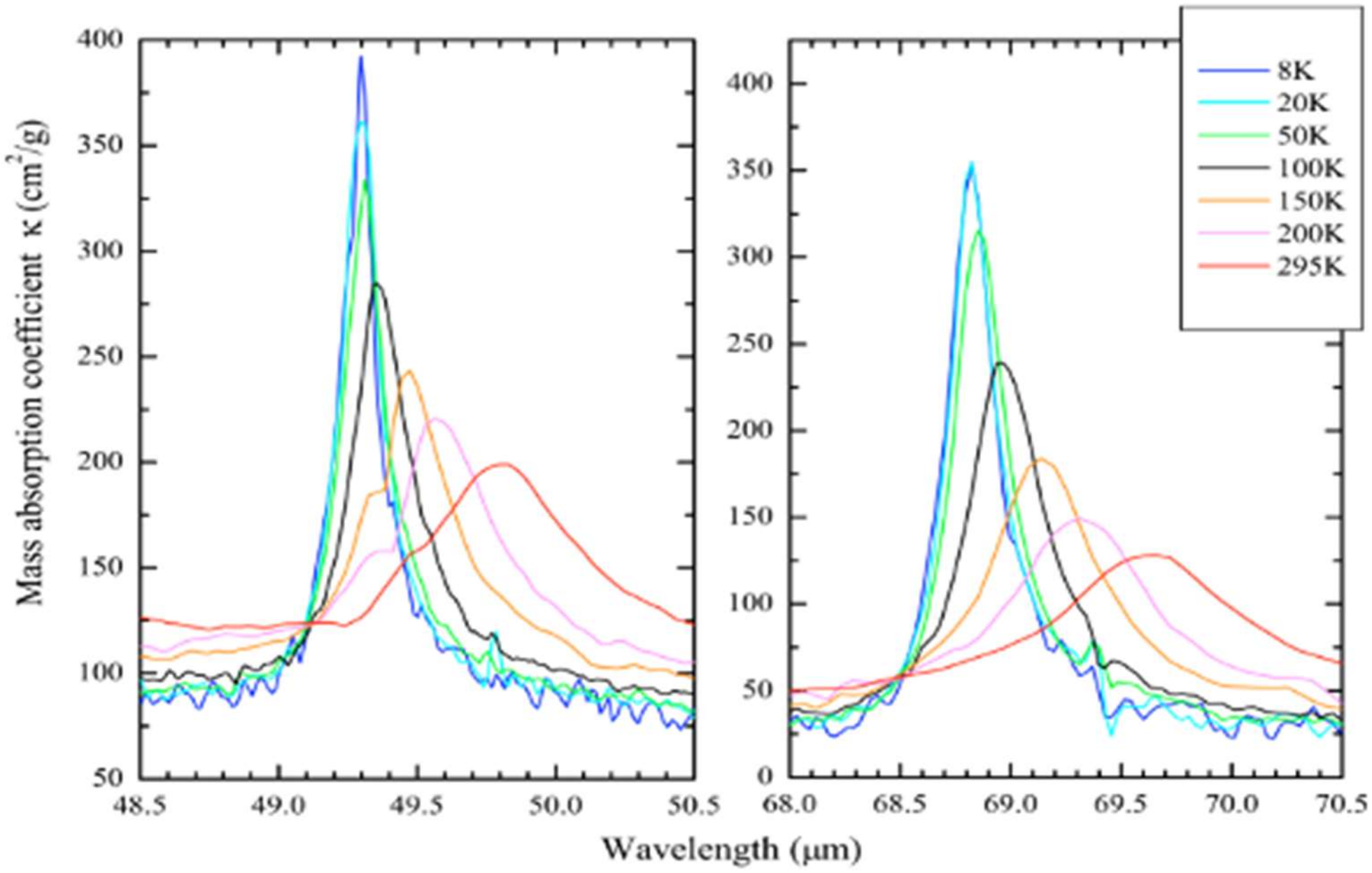

**Figure 1.** Effect of temperature on the 49- and 69-micron bands of forsterite (figure from Koike et al. 2006)

These temperature shifts constrain radiative transfer models (e.g. Maaskant et al. 2015) because dust temperatures are usually estimated from the broader continuum or the ratios of the 23 µm and 69 µm forsterite band strengths (I23/I69 ratio). Grain temperatures of the optically thin component are indicative of their location relative to optically thick emission from dust in a circumstellar disc, envelope or outflow. This information is used to constrain the system morphology and may aid the interpretation of images. For example, Maaskant et al. found that forsterite in flat disks surrounding Herbig Ae/Be stars is located in the inner few AU and that the detection rate for the 69 µm feature is higher for objects with lower millimetre luminosity. This may indicate that as disks evolve toward lower masses, optical-depth effects or increased production of forsterite or transportation of forsterite into cold regions of the outer disk enhance the strength of the 69 µm feature. Temperature data can also be utilized in chemical studies of grain condensation from the gas-phase and dust evolution. Time-series observations in rapidly evolving environments including SN, AGB stars and eruptive variables might also prove instructive because grain temperatures will vary as the dust is transported from hotter to colder regions, or vice versa.

## Science Questions

- How frequently are the features observed?
- Where are they most common?
- Is there any correlation with the orientation of the source?
- Is there any correlation with its location – metal-rich environment, metal-poor environment in the MW or beyond.
- What is the temperature of the forsterite dust?

- Are there several optically thin temperature components (indicated by broadened bands)?
- Are temperatures deduced from the 49- and 69-µm bands in each source the same?

## Need for PRIMA

PRIMA is required to conduct the study because these wavelengths are inaccessible from ground-based observatories. Infrared features at shorter wavelengths exhibit smaller temperature changes which cannot be disentangled from the effect of grain composition or other molecular bands within circumstellar envelopes. At the lowest temperatures the 49- and 69-µm bands are 0.15 and 0.3 µm wide, respectively, indicating they could be detected at R∼770, and R∼230. However, the lower resolutions limit the accuracy of width measurements and introduce confusion with gas-phase emission. Hence, we require the high-resolution (R∼4400 at 112µm) spectrometer.

The other constraint is the contrast between the continuum flux and the optically thin emission because source lists would be tailored to match the overall flux sensitivity. Detected PACs/Herschel (e.g., the PNe in Figure 2) and LWS/ISO peak strengths were typically 3% to 8% of the continuum with outliers at 1% and 29% so we request a contrast sensitivity of 1%. Blommaert et al. detected the band in 8/15 oxygen-rich AGB OH/IR stars, 11/18 post-AGB and low-mass evolved stars and 3/8 massive evolved stars. Maaskant et al. (2015) found the band in 6/23 Herbig YSOs.

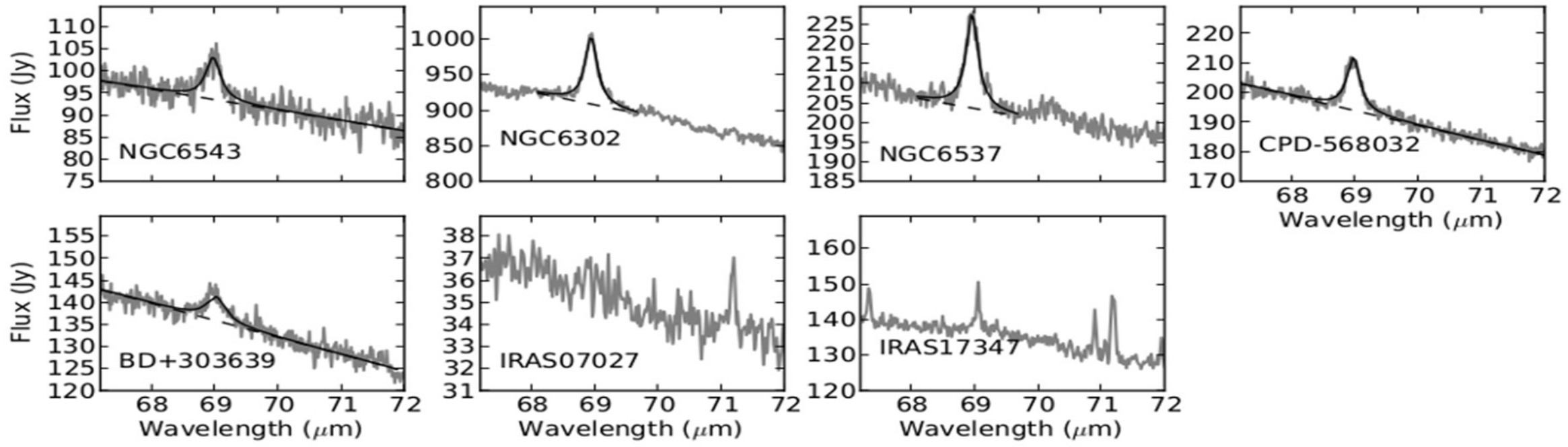

**Figure 2.** Spectra of PNe from Blommaert et al. 2014. Detected 69 µm features are fitted with Lorentzian profiles. IRAS 07027 and IRAS 17347 are OH maser sources which contain no forsterite band at 69 µm but might contain gas lines.

## Interpretation Methods

The peaks are fitted with Lorentzian profiles superimposed with low order polynomials to represent the continua. Results are then compared with those for the laboratory data (see Figure 3, reproduced from Blommaert et al.).

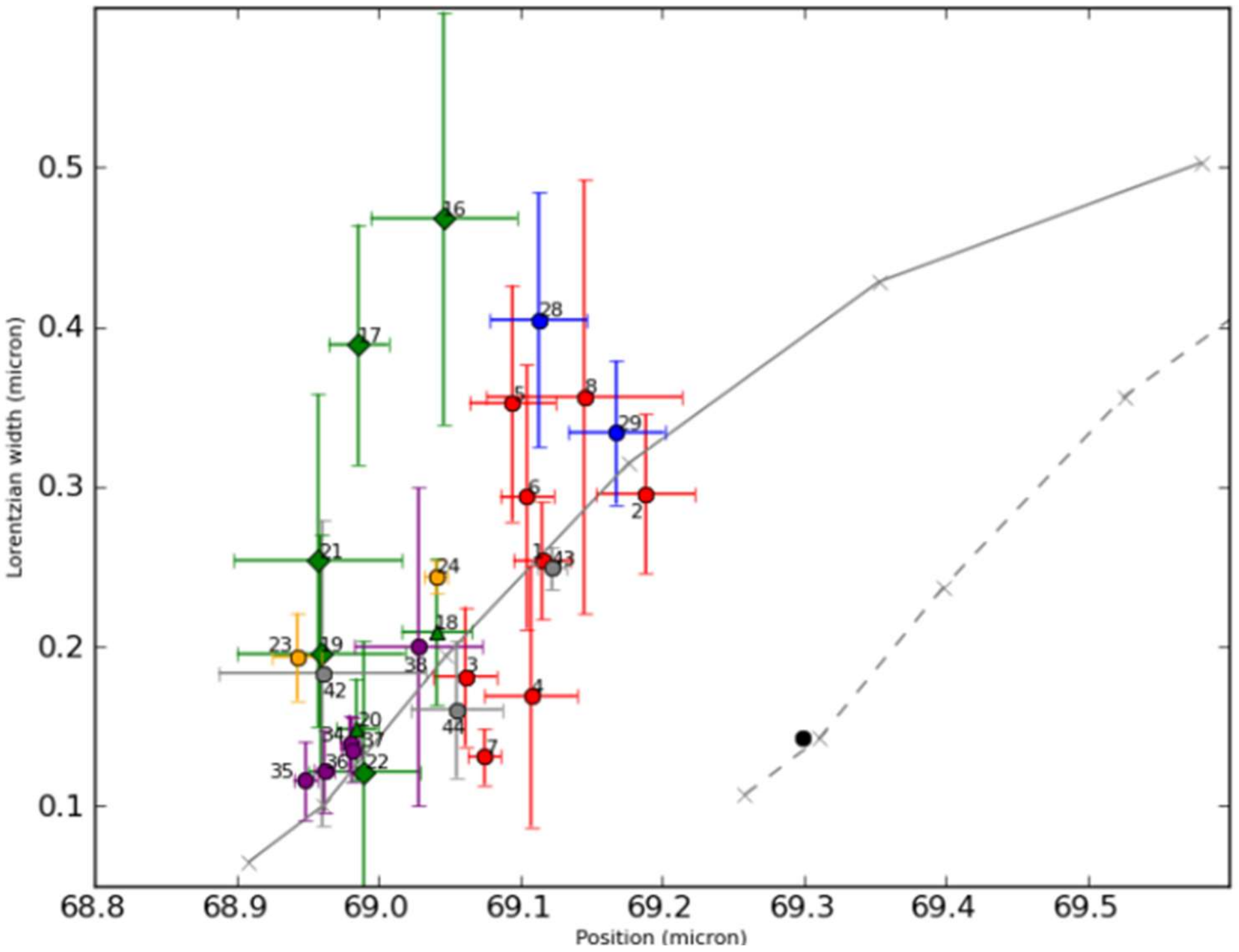

**Figure 3.** Position and width of all detected 69 μm bands. In red are the OH/IR stars, in green are the post-AGB stars, in blue the post-AGB stars with circumbinary disks, in yellow, two other likely disk sources, in purple the PNs, and in grey the massive, evolved stars. In contrast to the detected 69 μm bands of the evolved stars, the position and width of the 69 μm band of the disk around the young main-sequence star β Pictoris is shown in black. In solid and dashed grey are the width and position of 69 μm bands using lab measurements of Koike et al. (2003) and Suro et al. (2006). The solid grey is for crystalline olivine without iron, and the dashed grey is with 1% iron. The crosses on the curve indicate the temperatures going from 50, 100, 150, 200, and 295 K with increasing wavelength. The dashed grey curve is created using the extrapolation of de Vries et al. (2012).

## Instruments and Moades Used

FIRESS high res, 1 pointing per source

## Approximate Integration Time

On the faintest sources (~0.5 mJy continuum), a 1% contrast line at 69 μm can be detected in 10 hours of integration. We can target sources with continuum fluxes spanning from 0.5 mJy – 1000 Jy continuum at 69 μm, requiring typically under 10 min of integration. We estimate that a meaningful sample of 30 sources would hence require about 30 h integration'. About 30 hours of observations could be sufficient to observe ~30 targets of different brightnesses.

## Special Capabilities Needed

None

## Synergies with Other Facilities

Crystalline silicates are prevalent in the mid-infrared, where they clearly appear as sharp resonance features in the spectra from ~11 microns. Amorphous silicates features are present at 10 and 20 microns. Thus, there are strong synergies with mid-infrared spectroscopic instruments such as MIRI onboard JWST, and possibly MATISSE at ESO which due to the atmospheric windows can observe the 10 µm amorphous silicate feature of bright sources.

## Description of Observations

To precisely measure the crystalline forsterite grains, we require pointed observations with the high-resolution spectrograph on the FIRESS Spectrometer. Our sources span a large range of fluxes, and to detect the crystalline silicates we require a contrast of at least 1% above the continuum, Furthermore, to address the science cases regarding the frequency and correlation of the 49 µm and 69 µm forsterite grains in circumstellar environments, a statistically significant sample covering a range of sources and luminosities is also required. We estimate that a medium of 30 sources is required to achieve our science objectives. Due to the relative rarity of these sources, ideally this would be coupled with a high slew speed and pointing accuracy, to reduce overheads when observing a large number of targets across the sky.

About 30 hours of observations could be sufficient to observe ~30 targets of different brightnesses.

# 107. Investigating the Anomalous Chemistry and Complex Structures in the Circumstellar Envelopes Surrounding UV/X-ray Emitting AGB Stars

Jaime Alonso-Hernández (CAB, CSIC-INTA), Carmen Sánchez Contreras (CAB, CSIC-INTA), Marcelino Agúndez (IFF, CSIC), Raghvendra Sahai (JPL/Caltech), José Pablo Fonfría (IFF, CSIC), Luis Velilla-Prieto (IFF, CSIC), Guillermo Quintana-Lacaci (IFF, CSIC), and José Cernicharo (IFF, CSIC)

Despite advances over the last decades, it is still poorly understood which are the mechanisms that drastically transform the roughly spherical circumstellar envelopes (CSEs) around Asymptotic Giant Branch (AGB) stars into the large variety of aspherical shapes observed in Planetary Nebulae (PNe). Nevertheless, it is generally accepted that stellar binarity and multiplicity play an important role in shaping the stellar winds.

This proposed study aims to increase our knowledge on the formation of anomalous chemistry and structures that deviate from those typically found in AGB CSEs. We will focus on a sample of AGB stars that, unlike the majority of their class, show ultraviolet excesses and, in some cases, X-ray emission. These AGB stars are binary candidates in which the high-energy radiation is likely related with active accretion processes. Under this context, the interaction with stellar companions can produce the development of the structures observed in Planetary Nebulae. Furthermore, The presence of high-energy radiation is able to impact the CSE chemistry, leading to molecular abundances and degrees of ionization that differ significantly from those observed in most of AGB CSEs.

PRIMA will provide unprecedented observational capabilities in the far-infrared (FIR), allowing the study of these systems from a double perspective combining deep imaging and high-resolution spectra covering a large wavelength range (24-235µm). This will provide key information to characterize the dynamics, molecular gas chemistry and mineralogy of the CSEs surrounding these AGB binary candidates from the cold external regions to the warmer their interiors.

## Science Justification

### Broader Context

When low and intermediate mass stars (i.e., between 1-8 solar masses) reach the AGB evolutionary phase, they experience intense mass-loss through stellar winds, leading to the formation of dusty and roughly spherical CSEs (see Höfner & Olofsson 2018). During the following evolutionary stages (known as post-AGB) CSEs suffer a metamorphosis and change into the complex shapes that are commonly found in PNe. The shaping processes that produce this dramatic transformation are not well understood yet. Nevertheless, it is widely accepted that binarity should play a crucial role (Balick & Frank 2002).

The advancement of high-resolution FIR spectrometers unveiled the presence of a very rich spectra in AGBs, post-AGBs and PNe (see Ramos-Medina et al. 2018a, 2018b; Nicolaes et al. 2018; da Silva Santos et al. 2019) composed of atomic lines (from neutral atoms and ions such as OI, NII and CII), rotational and vibrational molecular lines (from molecules such as CO, SiO and $H_2O$) as well as solid state features (from minerals such as forsterite).

One interesting discovery was the unexpected detection of water lines in C-rich envelopes (see Melnick et al. 2001, Decin et al. 2010) due to the fact that in these environments most oxygen should be locked into CO with little left to form other O-bearing molecules. Still there is no consensus on the underlying disequilibrium mechanism, but photochemistry (see e.g., Agúndez et al. 2010) and shock-induced chemistry (see e.g., Cherchneff 2012) are postulated as the most promising ones.

On the other hand, previous FIR images (with angular resolutions of a few arcseconds) unveiled the presence of large scale structures (with angular sizes ranging from tens to hundreds of arcseconds) surrounding some AGB stars (see e.g., Cox et al. 2012). These structures mainly formed by warm (~10-100K) dust are particularly suited to be detected in the FIR (~30-300μm) through their thermal emission. A few of these extended structures display highly complex shapes (e.g., spiral and arcs) and require the presence of stellar companions to be formed (see Mayer et al. 2011, 2013). Moreover, other binary-induced structures (e.g., disks and bipolar outflows) have been found at the innermost regions of their CSEs with high-angular resolution observations (e.g., Decin et al. 2020).

In the last decades we have witnessed the discovery of several AGB stars with intense ultraviolet (UV) excess and X-ray emission, indicative of accretion onto a companion star (see Sahai et al. 2008, 2015 and Ortiz & Guerrero 2021). Recently, it has been shown that UV/X-ray driven chemistry in these sources can be very effective and produce a significant enhancement of certain molecules (e.g., SiO, HNC, $HCO^+$, see Van de Sande & Millar 2022, Alonso-Hernández et al. 2025).

## Science Questions

These AGB binary candidates are ideal laboratories to explore the effect of UV/X-ray driven chemistry in AGB circumstellar envelopes. This science case aims to achieve a better understanding of the physical processes that shape the stellar winds during the transformation of CSEs to PNe by a dedicated study of a statistically significant sample of candidate binary AGB stars, selected based on their high ultraviolet (UV) excess and X-ray emission. This science case includes the following scientific topics:

- The characterization of the effect of UV/X-ray emission on the circumstellar chemistry and the abundance enhancement of certain key molecules.
- The estimation of the degree of ionization produced by the high-energy radiation.
- The identification of large scale spatial structures in their CSEs.

- The characterization of the main physical parameters at largest scales.

### Need for PRIMA

High-resolution (up to R∼20,000) observations with FIRESS will enable detailed studies of a large amount of physical processes in the warm, inner regions of CSEs surrounding AGB binary candidates. These regions can be probed via spectral lines, solid-state features, and dust continuum emission. High-excitation molecular lines are essential for characterizing these inner layers and understanding the influence of internal UV and X-ray radiation. FIRESS will uniquely provide simultaneous access to dozens of spectral lines, with the spectral resolution, sensitivity, and wavelength coverage required to disentangle molecular and atomic/ionized species.

PRIMAger Hyperspectral Imager will complement these observations with deep, multi-band maps of extended CSE structures. This will enable direct identification and characterization of large-scale envelope structures (e.g., spirals, arcs, and detached shells, see Cox et al. 2012). Its wide field of view, high sensitivity, and spectral resolution will deliver spatially resolved SEDs, providing key constraints on the dust temperature, composition, and density distribution in the outer envelope regions.

## Interpretation Methods

We will combine spectral line identification with radiative transfer modeling (of both line and continuum emission) to characterize the physical and chemical properties of the gas and dust components in the envelopes. CO rotational transitions will be used to estimate density and temperature gradients (see Ramos-Medina et al. 2018b; da Silva Santos et al. 2019). High-J molecular transitions of various species will be used to derive abundance profiles in the warm (∼100-1000 K) inner regions ($r<10^{16}$ cm), where UV/X-ray-driven chemistry is most effective (see Fig. 1). These observational results will be compared with predictions from chemical kinetics models that include internal UV and X-ray irradiation (see Van de Sande & Millar 2022; Alonso-Hernández et al. 2025, respectively). The analysis will focus on classic tracers detectable in the FIR (e.g., CO, HCN and SiO), molecules sensitive to UV/X-ray radiation, including species with enhanced but not fully understood abundances (e.g., $H_2O$, OH, and $NH_3$), which are primarily observed in the FIR, as well as other UV/X-ray tracers (e.g., $CO^+$, $HCO^+$ and $N_2H^+$). Furthermore, FIRESS will enable a better identification of additional molecular lines and bands in the FIR range.

We will also estimate the degree of ionization using atomic lines sensitive to high-energy radiation (see Saberi et al. 2018). While lines such as [O I] 63 µm, [N II] 122 µm, and [C II] 158 µm are commonly seen in pPNe/PNe, they are typically absent in AGB CSEs (see da Silva Santos et al. 2019). In addition, PRIMA's broad wavelength coverage enables detailed modeling of the spectral energy distribution (SED), allowing a more accurate estimation of dust temperature, composition, density, and grain sizes, which are degenerated when using only photometric data (e.g., Ysard et al. 2018). The FIR spectral shape may also reveal anomalous structures formed in the innermost regions (see Wiegert et al. 2020).

PRIMAger maps will allow the identification of large-scale structures following standard procedures (see Cox et al. 2012). Additionally, the provided spectral resolution throughout the different photometric channels (R∼10) will allow spatially-resolved SED modeling and the determination of dust temperature and density along the structures.

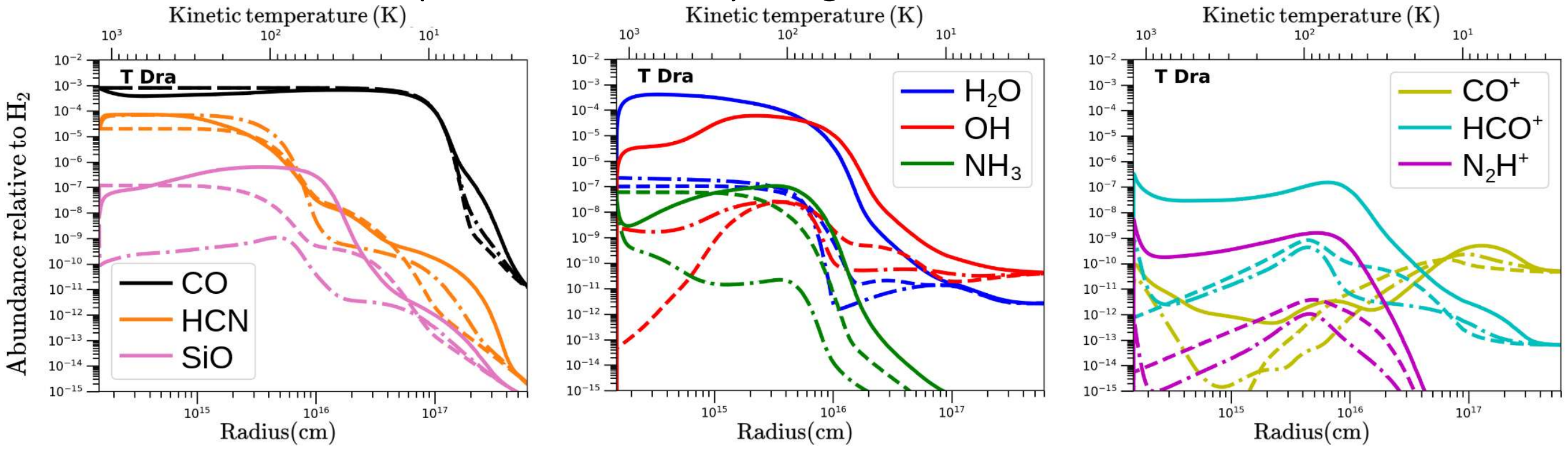

**Figure 1.** Model predictions of the radial abundance distribution of certain key molecules based on the chemical kinetics model presented in Alonso-Hernández et al. 2025 for a porous envelope in the AGB star T Dra. Left: CO (black), HCN (orange) and SiO (pink). Center: $H_2O$ (blue), OH (red) and $NH_3$ (green). Right: $CO^+$ (olive), $HCO^+$ (cyan) and $N_2H^+$ (purple). Solid lines represent the model including internal UV and X-ray emission, dash-dotted lines represent the model including only internal UV emission and dashed lines represent the model without internal UV and X-ray emission.

## Link to Testable Hypotheses

One of the goals of this study is to delve onto the warm, inner regions of CSEs surrounding AGB binary candidates, where UV and X-ray radiation from stellar companions is expected to drive unique chemical and ionization processes. FIRESS will enable us to characterize these inner zones through high-excitation molecular lines. In addition, PRIMAger will map the large-scale dusty structures formed during the AGB phase, revealing the imprint of binary interactions on envelope morphology. By combining FIRESS and PRIMAger data, we will obtain a comprehensive view of the structure and evolution of AGB CSEs. A sample of ∼20 targets (10 O-rich and 10 C-rich) will provide sufficient diversity for comparative studies and help assess how binarity influences both chemistry and envelope shaping.

## Instruments and Modes Used

This science case will require FIRESS high-resolution spectroscopic observations and PRIMAger hyperspectral imaging observations for a sample of 20 AGB stars. FIRESS observations should cover the entire FIRESS spectral coverage by taking two exposures for each target, while the PRIMAger observations should obtain 5’×5’ map for each target.

## Approximate Integration Time

We propose to perform FIRESS high-resolution point source observations with exposure times of 2.2 hours in bands 1 and 3 to achieve a sensitivity of $1\times10^{-17}$W/m² at 30µm assuming 5 Jy continuum flux, and 1.4 hours in bands 2 and 4 to achieve a sensitivity of $2\times10^{-18}$W/m² at 70µm assuming 0.5 Jy continuum flux. These observations will provide deep FIR spectra, allowing the detection and characterization of a large variety of spectral observables. For a sample of 20 sources, the total integration time will result on 72 hours.

We propose to perform PRIMAger mapping on square regions of 5’×5’ (based on the results from Cox et al. 2012) with a sensitivity of 500 μJy (5σ limit) to exploit PRIMAger dynamic range (up to $10^3$-$10^4$) based on the expected stellar fluxes at 30μm (5 Jy). The expected integration time is 3.2 hours for each map according to the PRIMA exposure time calculator. These deep maps will allow the detection of large scale structures. For a sample of 20 sources, the total integration time will result on 64 hours.

## Special Capabilities Needed

This science case requires a wide dynamic range (∼$10^4$) of PRIMAger to capture the high contrast between bright AGB stars and their faint circumstellar environments avoiding image distortion. Furthermore, high accuracy in the flux density measurements of FIRESS will provide accurate intensity ratios between the different emission lines and a reliable spectral shape of the continuum.

## Synergies with Other Facilities

Observations from high-angular resolution interferometers will provide essential complementary information. (Sub)millimeter facilities (e.g., ALMA, NOEMA, and SMA) can map low-excitation molecular line emission and help to constrain the physical conditions and molecular abundances (e.g., CO, HCN, CN, $H_2O$, OH, $HCO^+$) in the intermediate and outer layers of the envelope. Similarly, optical and near-IR interferometers (e.g., VLTI, CHARA) offer crucial insights into the spatial distribution of dust and the development of complex structures near the stellar surface.

In addition, archival data from past, current, and upcoming infrared missions (e.g., ISO, Herschel, Spitzer, SOFIA, JWST, SPHEREx) will extend the spectral coverage for both continuum and line emission. This broader wavelength baseline will enhance radiative transfer modeling and allow for the inclusion of a wider set of molecular transitions covering a broader range of excitation conditions.

## Description of Observations

Brightest AGB stars will be excluded from this study due to FIRESS saturation limits. Therefore, the sample will be selected from fainter and more distant sources. However, it is not discarded a risk of (partial) saturation in FIRESS band 1 depending on the stellar flux variability on the infrared, related with stellar pulsations and poorly constrained at these wavelengths (e.g., Monnier et al. 1998; Hron et al. 1998).

Both FIRESS and PRIMAger observations can be carried out using standard observational procedures, with no need for time-critical scheduling. Exposure times can be adjusted to match the flux levels of individual targets or revised based on future updates to instrument performance. No special calibration strategies are required.

A potential follow-up project including multi-epoch observations would be valuable to assess spectral variability across different pulsation phases and UV/X-ray intensities. Such monitoring could provide insight into temperature changes in the innermost envelope regions, particularly where FIR tracers are sensitive to warm gas and dust.

# 108. Understanding the Mineralogy in Evolved Stars

Francisca Kemper (ICE-CSIC / ICREA / IEEC), B. Sargent (STScI / JHU), Olivia Jones (UK-ATC)

We propose to do a spectroscopic survey of dust-producing evolved stars to quantify the mineralogy of the stellar dust production and compare it with the interstellar dust properties, in our own Milky Way and external galaxies. The survey will be done with the FIRESS instrument on PRIMA, using the pointing observations in low-resolution (R~90-130) mode. We will target dusty evolved stars in the Local Group, covering a range of metallicities, galaxy properties, but also stellar dust mass-loss rates and ZAMS masses. A sample of a few hundred objects will be targeted to obtain statistically relevant results. Specifically, we will be looking to measure the crystalline silicate content in the stellar ejecta, which may be used as a tracer of recent star formation in external galaxies, and hope to detect more instances of circumstellar diopside, water ice and carbonates.

## Science Justification

The interstellar medium of galaxies is continuously replenished with the ejecta from post-main-sequence stars. These ejecta are often enriched with the products of nucleosynthesis and are an important driver for galactic chemical evolution. The conditions in the stellar winds are also conducive for the condensation of solid state particles (dust), and, in the case of Asymptotic Giant Branch (AGB) stars, the condensation of dust is even required to accelerate the wind and drive the (enhanced) mass loss. Thus, post-main-sequence stars can also directly contribute dust particles to the interstellar dust reservoir, rather than just contributing the elements for later dust assembly in molecular clouds. Indeed, the presence in meteorites of presolar grains with an isotopic signature pointing to an AGB or other non-solar origin (Clayton & Nittler, 2004), proves that dust grains formed by post-main-sequence stars not only get deposited into the ISM directly, but some even survive up until the point of incorporation in a planetary system in formation.

There is some question, however, to what extent the dust in the interstellar medium is being processed after being ejected by post-main-sequence stars. Considering that the residence timescale of dust in the ISM, based on elemental depletions, is 3 x $10^9$ yr, and that the erosion life time of dust under supernova shock processing is 4 x $10^8$ yr for refractory elements like Si, Mg and Fe, and 2 x $10^8$ yr for C, we expect that around 11% of the silicate and 6% of the carbon dust in the ISM is of stellar origin (Jones & Nuth, 2011). The rest is recondensed or regrown in the interstellar medium. For the Large Magellanic Cloud (LMC), the erosion life times due to supernova processing are a factor of ~10 shorter (Temim et al. 2015), and consequently, even smaller fractions of original stellar dust in the ISM remain. Indeed, the composition of freshly produced stellar dust does not match with the composition of interstellar dust in the LMC (Kemper, 2015), and it has also been established that the total dust production by evolved stars does not match the dust mass measured in the interstellar medium (Srinivasan et al. 2016), in this case for the Small Magellanic Cloud (SMC).

Specifically, the crystalline fraction of silicates in the interstellar medium is observed to be very low (<2 %; Kemper et al. 2004, 2005), while reports quantifying the crystalline fraction of silicates in evolved stars seem to suggest a fraction of ~10%, consistent with our understanding of the resident and dust destruction time scales in the ISM. However, a systematic measurement of the crystalline fraction in the circumstellar shell of evolved stars is lacking.

Although the silicates in the interstellar medium of our Milky Way are amorphous, this is not the case for all galaxies. We now have the capacity of detecting crystalline silicates in external galaxies, with Spoon et al. (2022) reporting them in around 800 galaxies, almost a quarter of the sample observed with Spitzer. Because crystalline silicates require temperatures higher than what is available in the ISM to form, it is thought that these crystalline silicates must be of stellar origin, and can trace recent star formation activity. To put a strong constraint on the importance of this process, we need to obtain a systematic inventory of the mineralogy of dust around post-main-sequence stars, including its composition and crystalline fraction.

The PRIMA Far-Infrared Enhanced Survey Spectrometer (FIRESSS) will operate from 24-235 micron. In this wavelength range crystalline silicates forsterite and enstatite have strong resonances at 33 and 43 micron, respectively, with forsterite having an additional weaker feature at 69 micron (Molster et al. 2010). In this wavelength range, amorphous silicates do not have any spectral features. Assuming the continuum at these wavelengths is due to amorphous silicates at the same temperature, and the dust is optically thin, the crystalline fractions can be calculated using the feature over continuum ratio, as is for instance demonstrated for NGC 6302 in Figure 1. The 69 micron feature is found to be a sensitive tracer for the Fe-content in the lattice, thus further probing the conditions at the time of dust formation and crystallization. Additional species with resonances in the FIRESS spectral range include Ca-bearing crystalline silicates, such as diopside (Koike et al. 2000); carbonates (see Figure 1, below); and crystalline water ice (Hoogzaad et al. 2002); all of these detected around at least one evolved star. Due to their specific mineralogy, these species trace physical conditions in the stellar outflows. The FIRESS low resolution mode, with a spectral resolution of R~90-130 matches very well with the spectral resolution of the available optical properties.

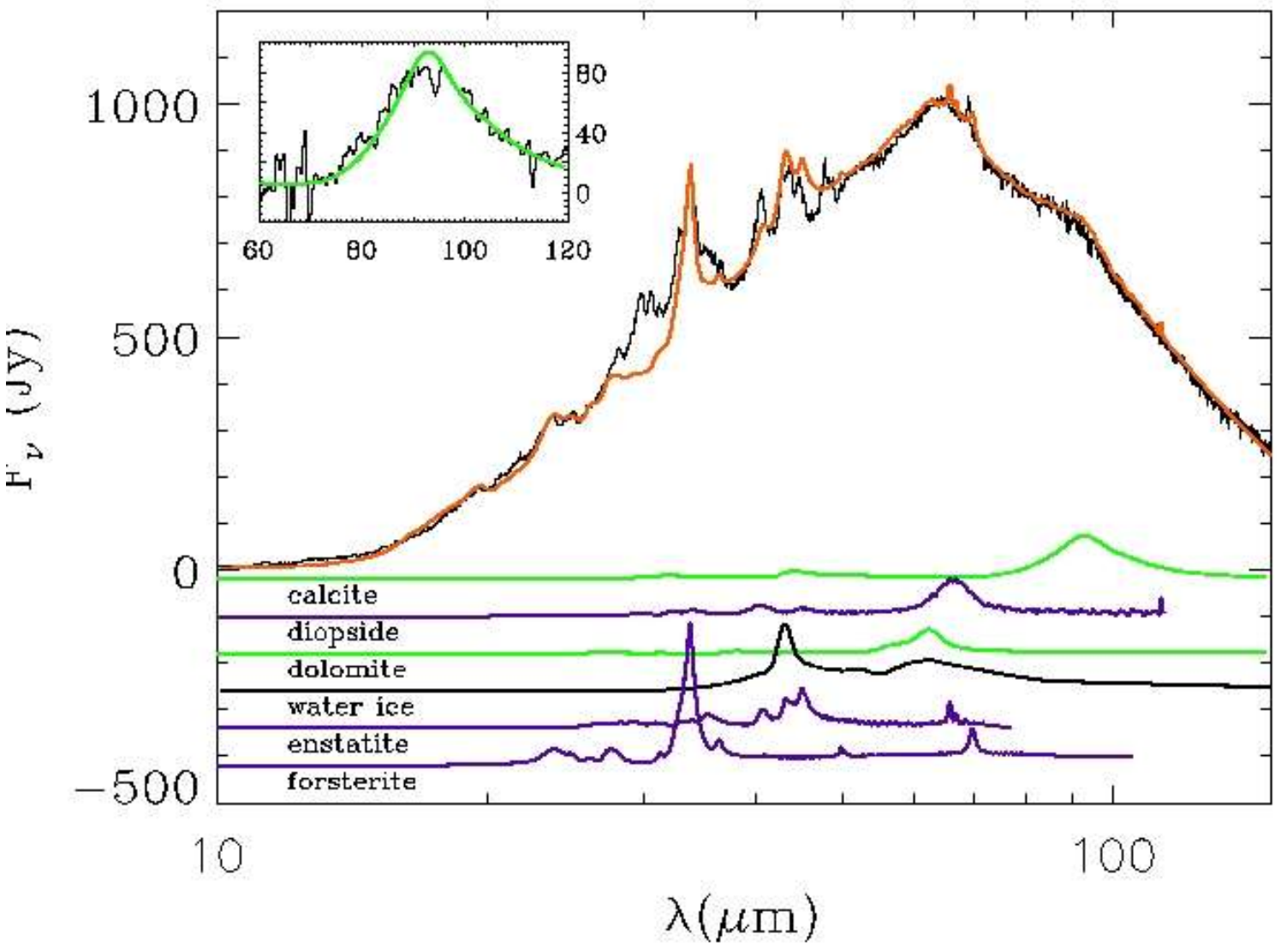

**Figure 1.** ISO SWS/LWS spectrum of Planetary Nebulae NGC 6302, along with a fit showing thermal dust emission from crystalline silicates, diopside, water ice and carbonates (Kemper et al. 2002).

## Instruments and Modes Used

FIRESS low resolution pointed observations, one for each of the 1250 sources

## Approximate Integration Time

All 1250 sources are bright enough that a spectrum can be obtained in the minimum integration time of 5 minutes per setting, according to the ETC. Two settings per source are required to get a full spectrum, yielding 10 minutes per source. We thus require a total of 208 hours of integration time, see also the section

## Synergies with Other Facilities

The spectra can be complemented with existing spectroscopy from e.g. Spitzer and JWST (and, in case of Galactic objects, from ISO) in the 8-24 micron range. In this range both crystalline and amorphous silicates have spectral features, and this will provide additional constraints on the dust composition. However, it is important to stress that the longer wavelengths (>24 micron) accessible with PRIMA, provide more stringent and conclusive constraints on the dust composition. This is because amorphous silicates have no features in this wavelength range, and because the the crystalline silicates and other oxygen-rich minerals have spectral features that are more characteristic for individual species, allowing for a more fine-grained determination of the mineralogy.

## Description of Observations

Targets will be selected from the IRAC and MIPS point source catalogues currently available for the LMC (Meixner et al. 2006), SMC (Gordon et al. 2011), and other Local Group galaxies (Khan et al. 2015). Within the Milky Way, the GLIMPSE survey may be used (Benjamin et al. 2003). It has been demonstrated that the reddest AGB stars are the most dust-producing ones (Srinivasan et al. 2016) and therefore the most relevant to quantify the properties of dust injected into the ISM, but we will ensure to include lower mass-loss rate objects. Furthermore, we will target higher-mass post-main-sequence stars which are also thought to contribute significantly, such as LBVs (Agliozzo et al. 2021), as well as supergiant B[e] stars which may return significant amounts of dust to the interstellar medium as well (Kraus et al. 2019). For most non-Galactic AGB stars, a single point source observation will be sufficient, but it may be possible to resolve some of the Galactic AGB stars with a small map observing mode.

Based on the above-mentioned catalogues, we estimate that the reddest AGB stars, which are the most significant dust producers, have a flux density of ~10-15 W/m2 at 100 micron, at the distance for M31. Therefore, it is possible to survey the composition of these extreme dust producers in all Local Group galaxies, by obtaining a full spectrum for each object in 10 minutes per source, the minimum integration time according to the PRIMA exposure time calculator.

The SMC contains about 300 “extreme” AGB stars, which dominate the dust production (Srinivasan et al. 2016). In the SMC these sources will be mostly carbon-rich, due to the low metallicity, but in higher metallicity environments the fraction of silicate-rich sources is higher. Furthermore, Agliozzo et al. (2021) argue that LBVs, which are silicate-rich, are also major dust sources in the Magellanic Clouds. The sources can be selected based on color selection criteria. Envisioning a survey in which we select ~50 sources per major dwarf galaxy (assuming ten representative dwarf galaxies), and ~250 sources in the Milky Way, M31 and M33, it is possible to do a spectroscopic survey of AGB stars and dust-producing massive stars over the entire Local Group covering a range of metallicities, masses, and other properties, in 208 hours.

# 109. Exploring the Properties of Stellar Flares in the Far-Infrared with PRIMA

Meredith MacGregor (Johns Hopkins University), Rachel Osten (STScI)

M dwarfs are the most common stars in the Galaxy and have a high frequency of Earth-sized planets at an equilibrium temperature that places them in the habitable zone, making them favored targets of upcoming missions to detect and characterize exoplanets. In recent years, M dwarf flaring has been studied at near-infrared (NIR) and millimeter wavelengths with JWST and ALMA, respectively. At NIR wavelengths, blackbody emission dominates, while optically thin synchrotron or gyrosynchrotron are thought to produce the millimeter wavelength emission. To date, M dwarf flares remain unexplored in the FIR. Monitoring with PRIMA would fill this gap and complete our understanding of the flare spectrum at IR through millimeter wavelengths. Many potential targets host exoplanets whose atmospheres will be studied extensively by current and future observatories, so this is a unique opportunity for PRIMA to not only contribute to our understanding stellar physics but also of planetary habitability. In addition to dedicated monitoring, an all-sky survey with PRIMA is likely to detect numerous incidental stellar transients from both M dwarfs and RS CVn binaries.

## Science Justification

### Broader Context

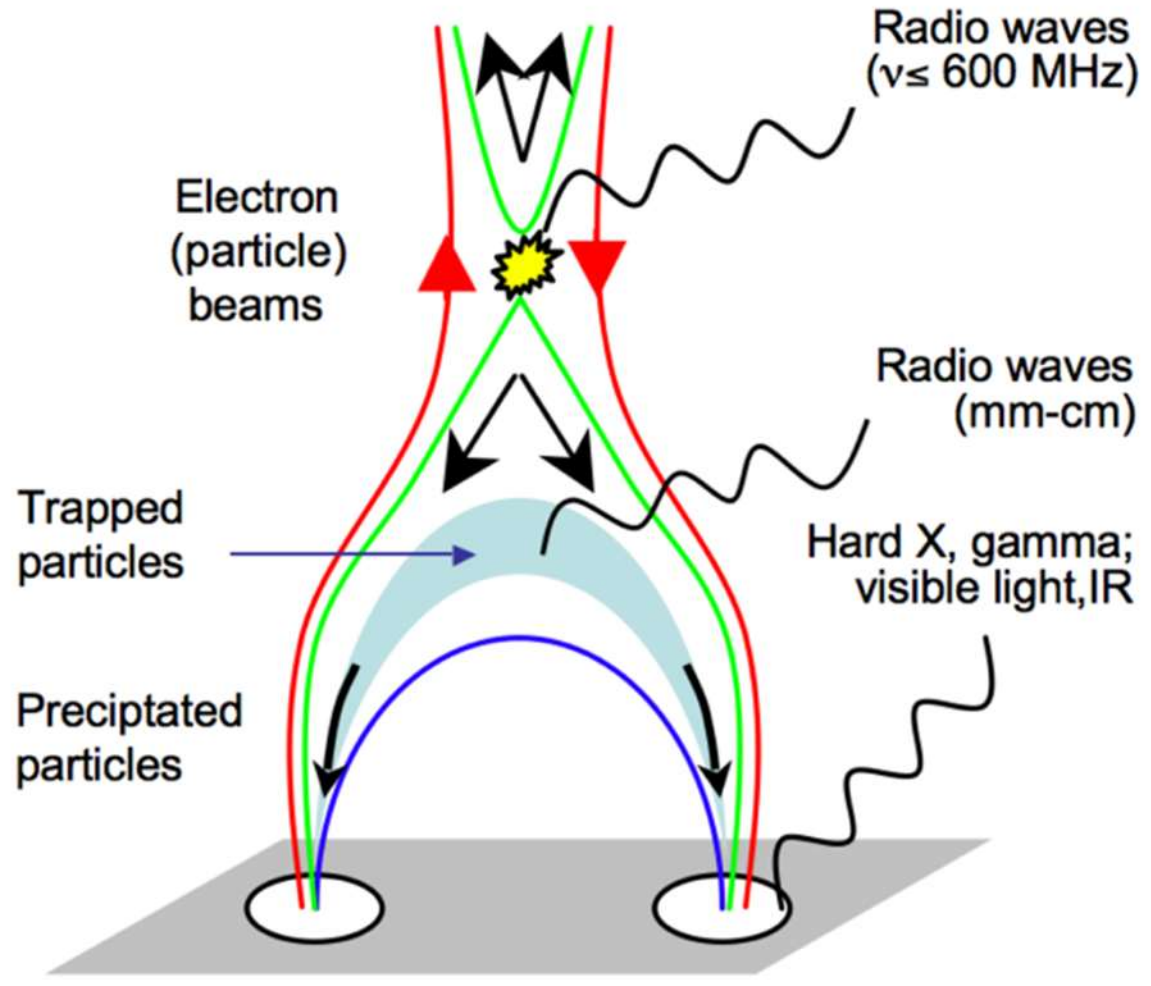

**Figure 1.** Different wavelengths of light trace different emission mechanisms (from Klein & Dalla 2017).

M dwarfs are the most common stars in the galaxy (Henry et al. 2006) and have a high frequency of Earth-sized planets at an equilibrium temperature that places them in the habitable zone, where liquid water could be stable on their surfaces (Dressing & Charbonneau 2015). As a result, these stars are favored targets of current upcoming missions to detect and characterize exoplanets including JWST, Pandora, and HWO. However, M dwarfs exhibit high levels of stellar activity and flaring throughout their entire lifetimes (e.g., Schneider & Shkolnik 2018; France et al. 2016). Over time, large flares can deplete a planet's atmosphere of water and ozone, raising questions about the habitability of planets orbiting these stars (e.g., Segura et al., 2010; Tilley et al., 2019). At the same time, planets orbiting M dwarfs may require frequent flaring to produce enough UV fluence for abiogenesis (Ranjan et al. 2017; Rimmer et al. 2018).

Stellar flares release energy across the entire electromagnetic spectrum. Magnetic reconnection accelerates electrons along magnetic field lines, producing non-thermal synchrotron radio and bremsstrahlung hard X-ray emission (Figure 1). Electrons impact and heat ambient gas, resulting in prompt visible/UV flares from the chromosphere and longer-duration soft X-ray flares from the corona.

Millimeter flares have been detected with ALMA in Band 6 from six M dwarfs (MacGregor et al. 2018; MacGregor et al. 2020, Burton et al. 2025). As a group, these flares have distinctive characteristics, namely short durations (< 30 sec), falling spectral indices with frequency ($\alpha < -1.5$), and linear polarization ($|Q/I| > 0.1$) that challenge our understanding of stellar flaring mechanisms. Figure 2 displays a remarkable flare from Proxima Centauri, our closest extra-solar stellar system, seen at millimeter wavelengths as well as at optical and ultraviolet (FUV) wavelengths. In contrast, solar flares measured at comparable wavelengths (Krucker et al. 2013) are roughly 10-100x less luminous and often characterized by rising (positive) spectral indices ($\alpha$>2). If millimeter flares result from optically thin synchrotron emission, these events must occur in a low-density environment of only modest magnetic field strength (100s – 1000 G). While flares may be brighter at longer radio wavelengths, synchrotron emission frequently becomes optically thick in this regime (Tristan et al. 2025). Flaring from solar-neighborhood magnetically active stars has also been detected by ACT (Guns et al. 2021) and SPT (Naess et al. 2021; Tandoi et al. 2024) pointing to the ubiquity of particle acceleration in magnetic reconnection events.

At NIR wavelengths (1 – 3.5 μm), continuum emission dominates flare emission, assuming a 9000 K blackbody (Fuhrmeister et al. 2008). Until JWST, flare detections at these wavelengths were limited. Multiple flares have now been detected from TRAPPIST-1 with effective temperatures below 5300 K, lower than optical wavelengths (Howard et al. 2023). To date, no flares have been detected at longer wavelengths into the FIR. Observations at these missing wavelengths would connect the blackbody emission observed in the NIR to the rising synchrotron spectrum observed in the millimeter filling in a key gap in the flare spectrum.

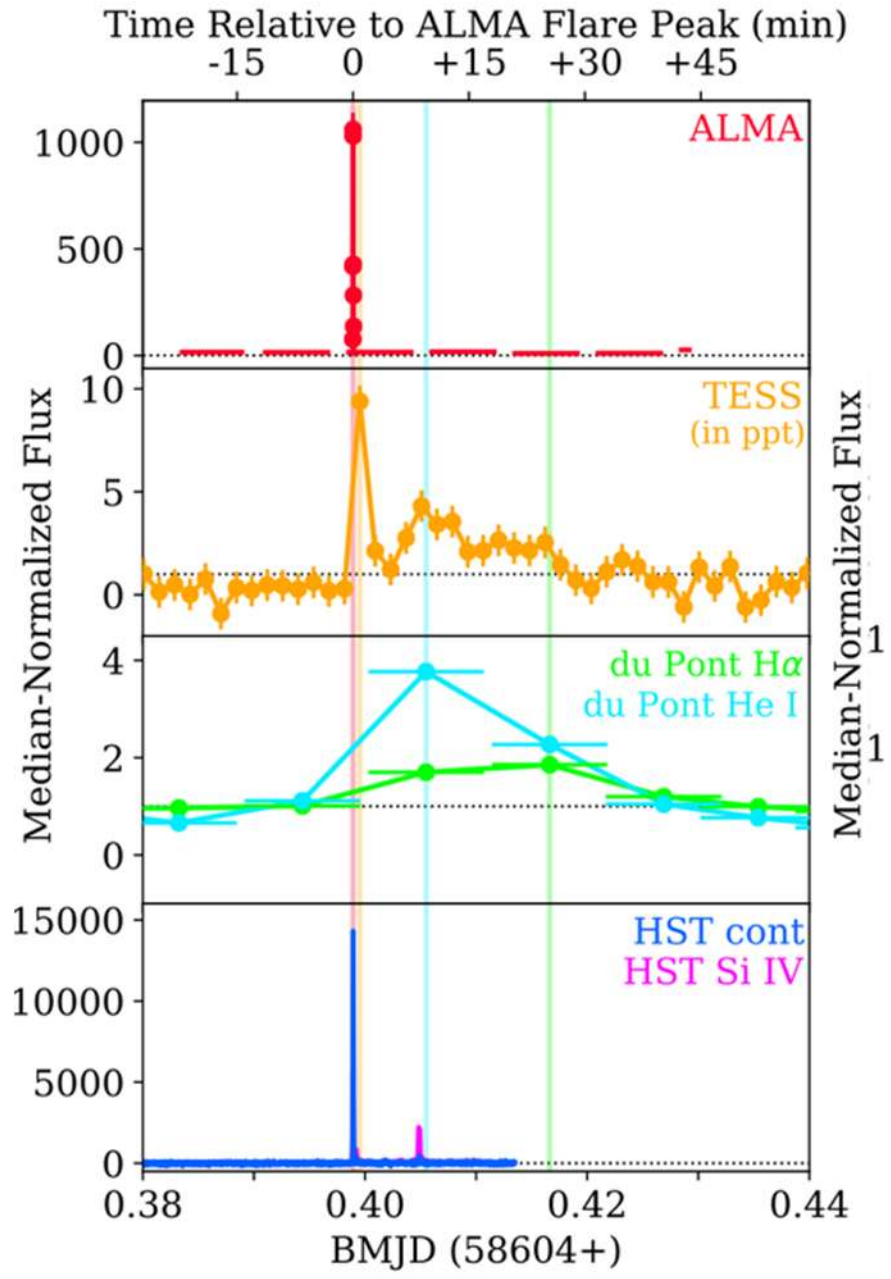

**Figure 2.** A flaring event observed from Proxima Centauri shows a tight correlation between UV and millimeter emission (from MacGregor et al., 2021).

It is worth noting that binary stars are another likely source of flares at these wavelengths (e.g., Beasley & Bastian 1998; Brown & Brown 2006). In a tidally locked binary system where spin and orbital angular momenta are coupled, stars maintain fast rotation with age and even spin up. RS CVn- (including at least one evolved star) and BY Dra-type (consisting of two main sequence stars, typically K- or M- spectral type) binary systems exhibit enhanced levels of magnetic activity for Gyrs, including frequent and extreme flares. RS CVn HR 1099 was observed with Herschel. No obvious flaring events were detected, but there is evidence for a possible excess above the expected photospheric flux.

### Science Question and Approach

We will use PRIMA to study flare rates, energies, and properties in the FIR to determine the dominant source of emission at these wavelengths filling in a gap between existing studies in the millimeter and NIR. The first approach to tackle this question is through dedicated monitoring campaigns targeting active stars or stellar systems. Numerous existing facilities including ALMA, JWST, TESS, and HST have carried out similar monitoring campaigns of exoplanet-hosting M dwarfs including Proxima Centauri, TRAPPIST-1, and AU Mic. New PRIMA observations could leverage these existing datasets to provide flaring constraints in a new wavelength regime. Many of these exoplanet systems are targets of campaigns to characterize planetary atmospheres, so this is a unique opportunity for PRIMA to not only contribute to our understanding stellar physics but also of planetary habitability.

There is also the possibility to detect flaring events by chance in other large area surveys. An all-sky survey is proposed by Saydjari et al. in GO Volume 2. In six months or roughly 5000 hours, PRIMA would scan along lines of ecliptic longitude to cover the entire sky. As demonstrated by CMB all-sky surveys at millimeter wavelengths, we expect that such a survey would likely find at least 10s – 100 flaring stars.

### Need for PRIMA

The importance of studying stellar and planetary coevolution is highlighted in the Astro2020 decadal survey and has motivated many current missions and observing programs. Since no other observatory current or planned covers the FIR, PRIMA is the only observatory that can carry out this science.

## Instruments and Modes Used

This observing program requires 20 pointed low-res observations with FIRESS.

## Approximate Integration Time

There are two approaches to take to study stellar flares: (1) dedicated monitoring campaigns and (2) chance detection through large area surveys. Given typical M dwarf flare rates, monitoring for 10 hours per target will likely detect sufficient flares to constrain the flare frequency distribution (FFD). To complete a sample of 20 nearby M dwarf flare stars yields a total program duration of 200 hours. We also strongly endorse PRIMA all-sky surveys proposed in both GO Volumes 1 and 2, since they will likely yield a large number of incidental flare detections.

## Special Capabilities Needed

This program would benefit from time constrained observations to enable coordination with other facilities including ALMA, JWST, and HST. In addition, stellar flares often have short durations requiring fast cadence observations (1 – 2 min or less) in order to resolve the shape of the light curve.

## Synergies with Other Facilities

In recent years M dwarf flares have been studied frequently at millimeter wavelengths with ALMA and CMB telescopes such as ACT and SPT. The wavelengths likely trace optically thin synchrotron or gyrosynchrotron emission (MacGregor et al. 2020). Flare are now also being studied with JWST at NIR wavelengths (Howard et al. 2023) where hot blackbody emission is expected to dominate. FIR emission from stellar flares remains essentially unexplored. Observations with PRIMA would fill in this missing piece and constrain the flare spectrum across the infrared and submillimeter.

## Description of Observations

A targeted monitoring campaign to study stellar flares at FIR wavelengths does not require high spatial or spectral resolution. The goal is simply to measure the total flux density as a function of time. With JWST, spectral lines have been detected during flares so we use the low-resolution mode of PRIMA's FIRESS instrument to enable spectroscopy. However, we note that PRIMAger could also be used for monitoring although there is no need for mapping. As discussed above, any all-sky surveys that make use of PRIMAger are likely to yield incidental stellar transient detections as foregrounds.

The key observational constraint is total on-source observing time. We need to monitor targets for long enough to build up a sample of flares with a range of energies. Many potential targets have already been studied with other facilities, which inform the time request here. ALMA monitored Proxima Centauri for 40 hours (MacGregor et al. 2021), which has greatly improved statistics on the frequency of millimeter flares from M dwarfs by creating the first stellar FFD at millimeter wavelengths (Burton et al. 2025). Observations were carried out over 11 nights (roughly 6 hours per night with overheads), and multiple flares were detected on all nights. The smallest easily detectable flares have flux densities ~10 mJy and occur multiple times per day. Medium flares with flux densities of 20-50 mJy occur roughly 1-2x per day. The largest flares with flux densities >50 mJy (FUV energies >$10^{30}$ erg) occur in ~50% of the observing blocks. Thus, we expect that 10 hours on-source is likely enough time to sample the FFD of typical M dwarf flare stars.

# 110. Why are Far-infrared Observations Important for Supernovae and Supernova Remnants?

Mikako Matsuura (Cardiff University), Ilse De Looze (Ghent University), Ori Fox (STScI), Ciska Kemper (E-CSIC / ICREA / IEEC), Florian Kirchschlager (Ghent University), Dan Milisavljevic (University of Purdue), Armin Rest (STScI), Laurance Sabin (UNAM), Arka Sarangi (University of Copenhagen), Roger Wesson (Cardiff University)

Supernovae (SNe) are critical engines that drive the evolution of the interstellar medium (ISM) in galaxies: SNe are considered to be the dominant source of chemical enrichment, a key source of dust creation, and the origin of energetic shocks and outflows, which destroy dust present in the ISM. Moreover, an instant burst energy from a SN can illuminate the surrounding ISM, so that SNe can work as a lighthouse that enables us to investigate the ISM gas and dust in distant galaxies, which normally may not be detected without SN illumination.

Yet, it is not well determined how much of the elements are synthesized or condensed into dust within the SNe/SNRs and how much of the freshly formed SNe dust and existing ISM dust is destroyed by SNe and SNRs. With high-sensitivity imaging and spectroscopic capabilities, PRIMA will be able to measure the true amount of dust mass formed in SNe, unveiling the dust compositions within.

## Science Justification

PRIMA's far-infrared (FIR) imaging capabilities will detect dust emission, and atomic and molecular line emissions from SNe and SNRs in the Milky Way, local group galaxies and galaxies up to 5 Mpc. With these new exploitations of wavelengths and distant galaxies, PRIMA will solve the following questions, relevant to 'What are the role of SNe for ISM evolution of galaxies?'

- Are SNe important sources of dust in cosmic history? Can supernovae form the first dust of the Universe?
- What kind of dust grains are produced by supernovae? Are the grain properties (grain compositions and grain sizes) the same as those in the interstellar medium?
- How do supernovae cool down to become supernova remnants?
- What are the dust properties of the interstellar media in distant galaxies?

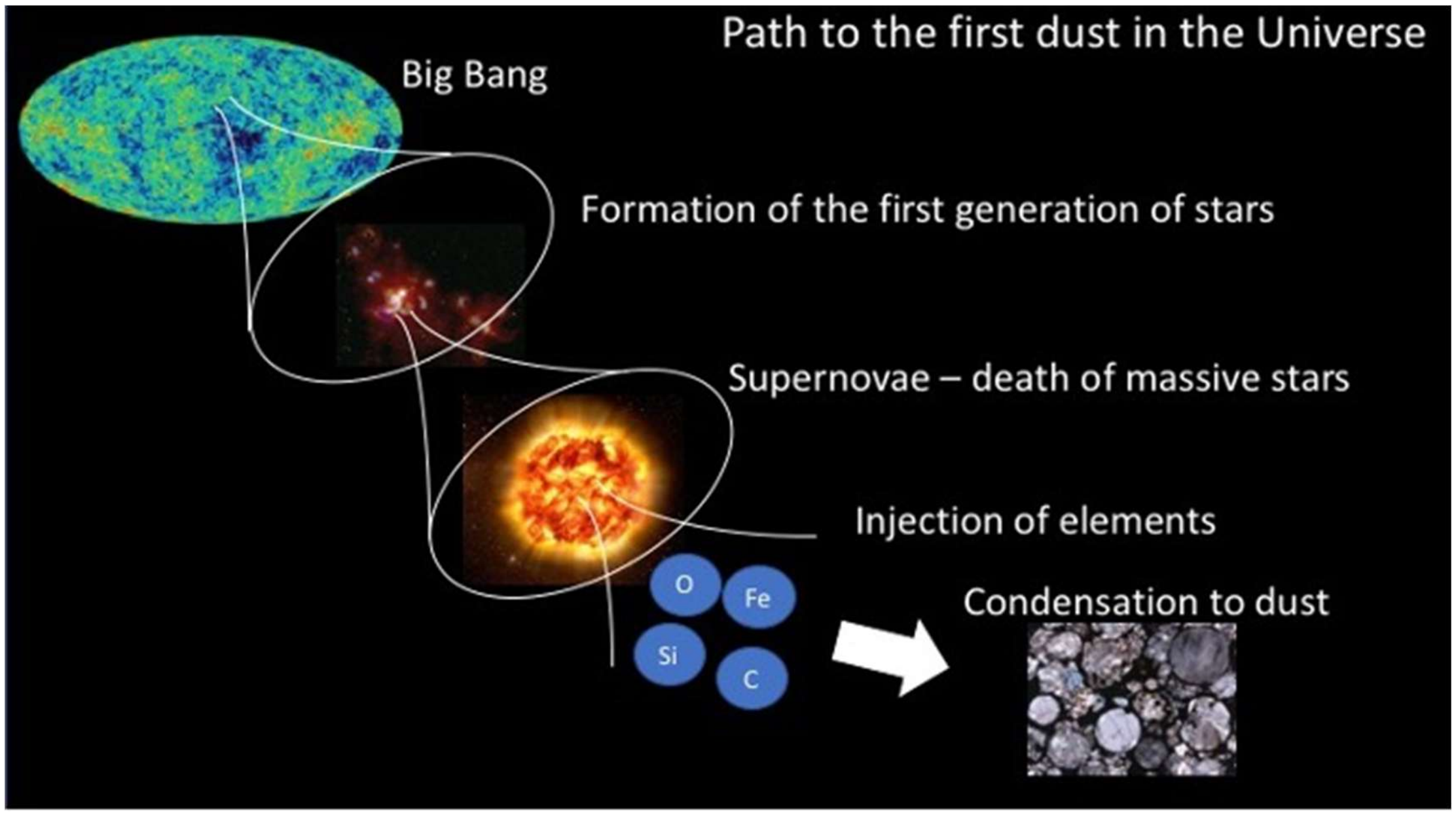

**Figure 1.** The pathway to form the first dust in the universe: after the Big Bang, the first generation of stars are formed, and high mass stars die as supernovae. These SNe synthesize elements, and some of refractory elements condense into dust grains.

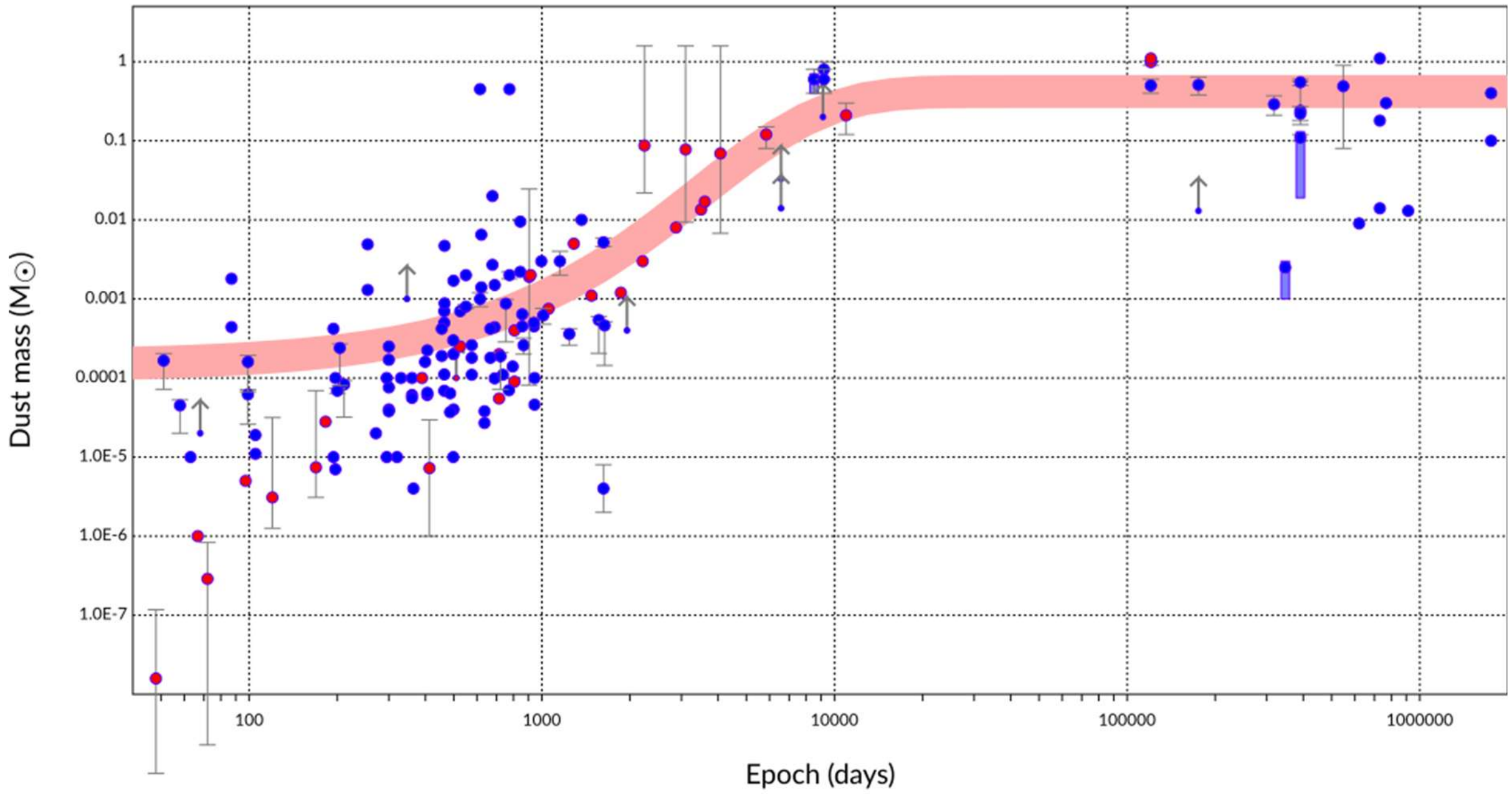

**Figure 2.** Dust masses measured in SNe, as a function of epoch since the SN explosion (Revised from Niculescu-Duvaz et al. 2022; https://nebulousresearch.org/dustmasses/)

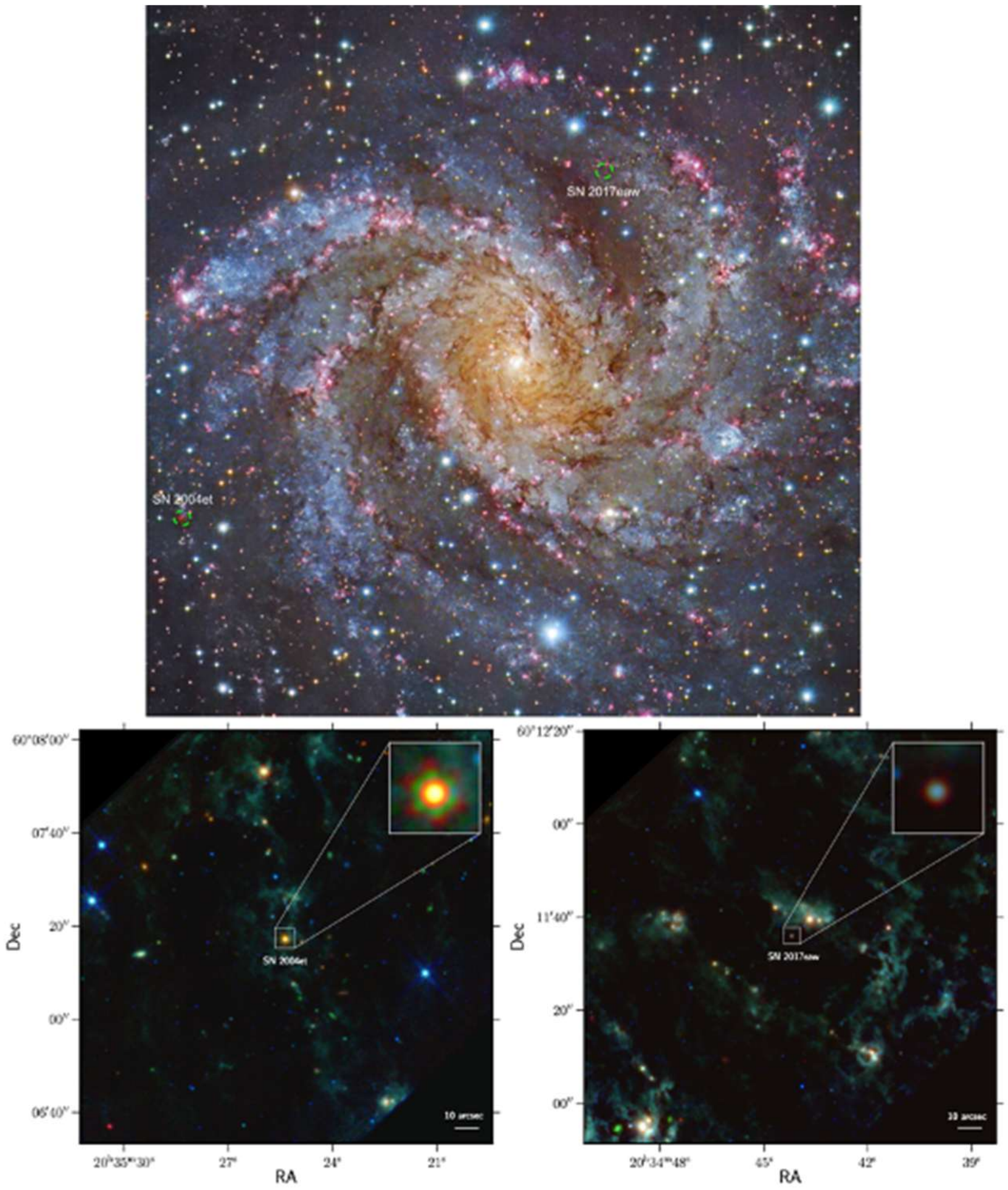

**Figure 3.** Optical images (top) of the galaxy NGC 6946 with two SNe, and JWST MIRI images (bottom) of SN 2004et and SN 2017eaw (Shahbandeh et al. 2023)

## [1.1] Dust in Supernovae

### [1.1.1] Freshly formed dust: what is the true amount of SN dust mass?

It has been proposed that SNe formed the first dust in the Universe, and they are the key source of dust from high-redshift to the Milky Way (e.g. Dwek 1998; Nozawa et al. 2003). Massive stars (>10 Msun) evolve quickly and die as SNe with a large quantity of heavy elements being ejected from a single star (Fig 1). If some of these heavy elements can condense into dust grains, SNe can fill the ISM of a galaxy with dust quickly (Laporte et al. 2017; Nanni et al. 2020). Theory predicts that if a SN can produce on average 0.1-1.0 Msun of dust per SN, the dust in high redshift galaxies can be accounted for by SN dust (Dwek & Cherchneff 2011).

The issue of SN dust is the discrepancy of the inferred dust masses in young (<2 years) SNe and older (>10 years) SN remnants. Fig 2 demonstrated the reported dust masses. The reported dust masses of young SNe tend to be in an order of $10^{-6}$ to $10^{-3}$ solar masses, while higher dust masses (0.01-0.5 solar masses; e.g. Sugerman et al. 2006) have been reported in older SNe and SN remnants (Gomez et al. 2012, Matsuura et al. 2011, De Looze et al. 2017, Cigan et al. 2019, Chawner et al. 2019). The reasons for this discrepancy are undetermined: either dust masses slowly increase in time over 10 years (Wesson et al. 2015), or dust is optically thick in early days (Dwek et al. 2019), so that inferred dust masses are underestimated. The largest problem is lack of dust measurements in later phases. Young SNe have hotter dust, hence, their SED peaks at near- and mid-infrared, which previously covered by Spitzer, Gemini, and now JWST. While SNe are getting older, gas and dust in SNe gradually cool so that the peak of the dust SEDs shift towards FIR wavelengths. Historically, there were only a few FIR instruments which are capable of detecting colder and older SN dust, with only a handful of measurements with Herschel. JWST started detecting dust in ~18 years old SNe (Shahbandeh et al. 2023) but is still limited to the warm dust (>100K). Hence, in order to evaluate the true amount of dust mass formed in SNe, FIR observations are essential, and the high-sensitivity of PRIMA will provide a unique opportunity.

**[1.1.2] What types of dust grains are formed?**

Although it is getting clearer that SNe do form dust, what types of dust grains being formed remain as an unresolved question. Currently, only limited examples are known: Cas A and G54.1+0.3 from Spitzer (Rho et al. 2008; Temim et al. 2017; Rho et al. 2018) and potentially, more analysis will follow with JWST/MIRI. As seen in Fig 4, dust found in SNRs are far from those found in cosmic silicate dust grains, including $MgSiO_3$, FeO, $SiO_2$, FeS, $Al_2O_3$. These dust grains typically have features beyond 25 μm, and PRIMA will be able to identify the characteristics of SN dust.

PRIMA, which has capabilities to detect numerous SNe, will open opportunities to investigate what types of dust are formed in which types of SNe.

Typically, elemental yields of SNe have the core mass dependence, which would result in different dust compositions (Sarangi et al. 2015), and this will be a new research field to be opened by PRIMA.

The formation of molecules, molecular clusters, and their transition to solid grains of dust, control the chemical budget of SNe from their nebular phase to the remnants. Based on the type of progenitor, theories predict the dominance of C-rich (expected in smaller progenitor stars) or O-rich dust components (expected in larger progenitors). Moreover, the timescales of dust formation for individual species differs based on the densities of the clumps within the ejecta, which had been developed by Rayleigh-Taylor instabilities immediately after SN explosion. The cooling rates of the clumps are defined by elements present in those clumps, affecting dust compositions formed in clumps.

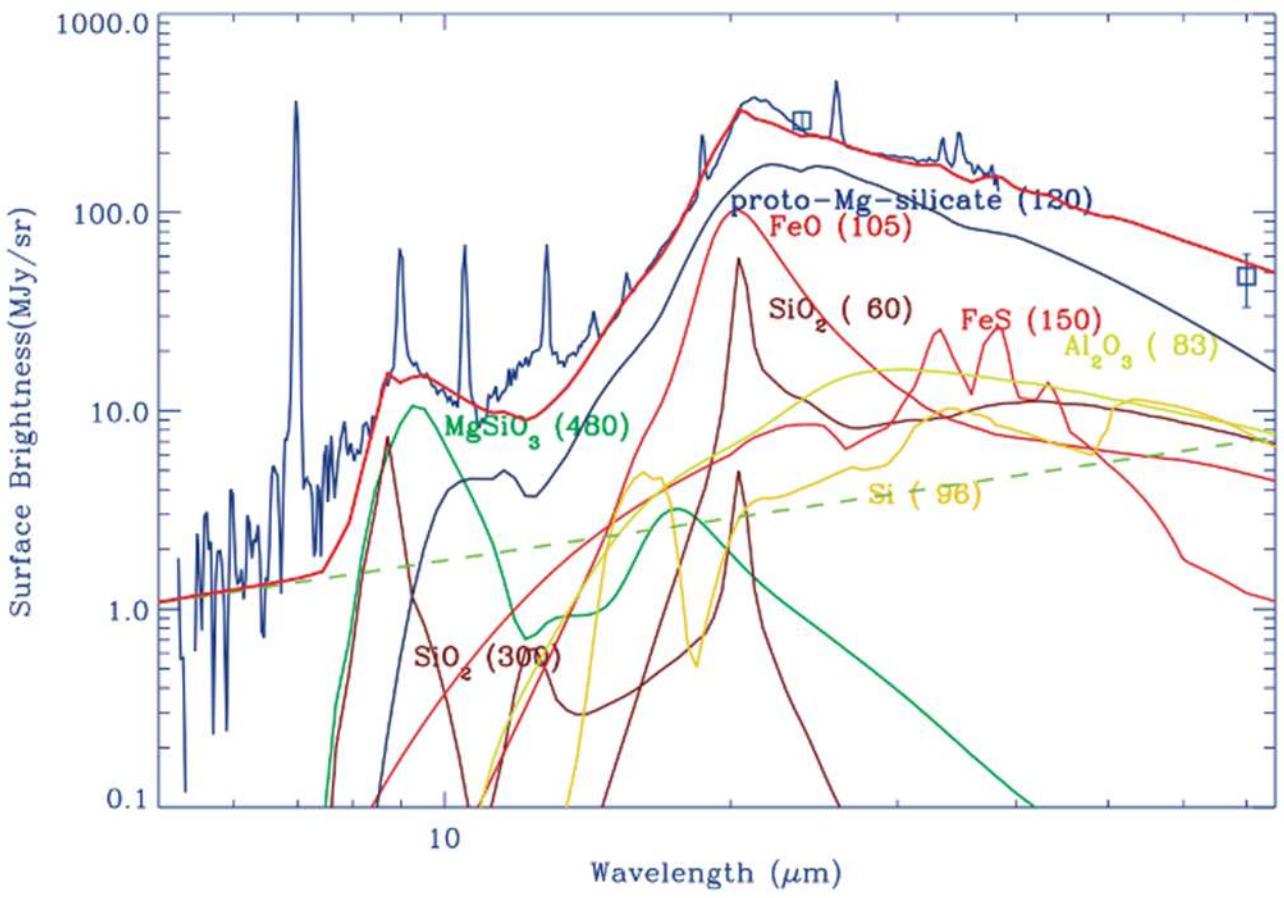

**Figure 4.** Dust identified in Cas A (Rho et al. 2008)

### [1.1.3] SNe illuminating surrounding Circumstellar and interstellar dust

The interaction of the SN shock with circumstellar environments, which has been formed by the mass loss from the progenitor star, influence the physical and chemical evolution of the SN profusely. In interaction-dominated SNe, typically donated as type IIn SNe, dust is assumed to form in the post-shock gas (behind the SN blast wave) with surrounding circumstellar material.

### [1.1.4] Survival of Supernova and ISM dust: How much of dust being destroyed?

Florian: While supernovae can be significant producers of dust, a large fraction of the dust can potentially be destroyed by the reverse shock. Moreover, the forward shock can trigger the destruction of interstellar dust grains. The net dust survival rate is crucial for determining whether SNe can contribute significantly to the dust budget in the ISM.

Theoretical studies about the destruction rate of SN formed grains in the SNR (e.g. Silvia et al. 2010, Bocchio et al. 2016, Slavin et al. 2020) or destruction of pre-existing ISM grains by the forward shock (e.g. Bocchio et al. 2014, Slavin et et al. 2015, Hu et al. 2019) vary significantly. The dust survival rates of these studies span a broad range between 0 and 99 % which is due to different initial conditions in the SNR, considering different dust processes, and different initial grain size distributions and materials.

Highly resolved MHD simulations have improved our understanding of shocked clumps in the SNR and shocked regions in the ISM. We know that the gas densities, gas-to-dust mass ratios, magnetic field strengths and grain sizes are the main quantities that impact the dust survival rates. Taking all these aspects into account and considering the combined effects of gas sputtering and grain-grain collisions, we are able to better constrain the surviving dust masses (e.g. Kirchschlager et al. 2019, 2022, 2023). However, due to a lack of observational data, key parameters like the grain size or clump gas densities are uncertain, which makes the dust survival rates unclear. PRIMA's wavelength coverage, resolution and sensitivity can help to derive these key parameters and thus shed light on the dust survival rates.

### [1.2] From SNe to SN remnants: cooling of gas

The explosion energy of $10^{51}$ erg is gradually lost, and SNe can cool down to become SN remnants. In very early phases, the majority of the SNe energy is emitted at UV and optical, but once it reaches to nebular phase after a few years, FIR atomic lines are predicted to be the main cooling in SNe and SNRs. The dominant cooling lines are predicted to be [Fe II] 25.99 μm, [Si I] at 68.47 μm and [O I] 63.19 μm (Jerkstrand et al. 2015). However, the detections of FIR lines are very limited and found in SNR Kes 75 with its pulsar age of 450±50 years (Temin et al. 2019) and Cas A (Rho et al. 2008). Hence, detecting these FIR dominant cooling lines are crucial to test SNe/SNR physics, and PRIMA is the only telescope that can tackle this physics.

About a few months after the SN explosion, part of the ejecta gas is getting cooler, and molecules are starting to form. Molecules have been proposed to be responsible for the rapid cooling of the gas (Jerkstrand et al. 2015). On top, molecules are indispensable chemical constituents to form dust grains in chemical reactions (Sarangi et al. 2015), Detection of molecules in extragalactic SNe, a few months to a few years after the explosion, will provide the critical evidence for determining the pathways of ejecta gas cooling process, and dust formation.

In order to truly describe the evolution of SNe to SN remnants, and evaluate their impact on surrounding ISM, detecting these cooling lines are important. With PRIMA's sensitivity, these lines could be detected in SNRs in a few Mpc.

## Instruments and Modes Used

The proposed observations require both PRIMAger hyperspectral bands as well as FIRESS in low-resolution mapping mode to create a 25 square arcmin map towards CAS A.

## Approximate Integration Time

Two complementary observing programs are proposed:

1. Observe five newly discovered extragalactic SNe with PRIMAger hyper spectral bands (time estimate of 100min to 190 hours depending on brightness and distance)
2. Observe CasA SNR, as an example of nearby SNRs (Crab, SN 1987A etc), with FIRESS in 3 bands in a 25 sq arc minute map to the 0.5e-17 W m-2 level. This time estimate is 54 hours.

For Observation #1, SN brightnesses have substantial diversity; hence, we state only a guideline brightness. We estimate an approximate integration time required of about 190 hours.

Brightness estimates for dust-forming supernovae after 1 yr and 10 yrs - SN1987A at a distance of 50 kpc had a flux at 10μm of about 4Jy after 415 days and 12Jy after 615 days (Wooden et al), while dust evolution models by Wesson et al. (2015) predict the peak of the SED shift to FIR at day 1153, with an estimated flux of about 1Jy at 50 μm. Wesson et al. estimated that the SED would peak at around 100μm, with a flux of 0.2-0.45Jy after 10 years. SN2004et, meanwhile, at a distance of about 7.75Mpc, has an SED peaking at about 15μm, with a flux of 0.75mJy, at age 18 yrs (Shahbandeh et al 2023). Hence, there are brighter SNe at FIR, too. For an approximate estimate, 1 microJy for 1 sigma at 84 micron is used for a pessimistic scenario of an extra-galactic (5-10 Mpc in distance) SN. Assuming 5 microJy 5-sigma (pessimistic scenario), it would take 18.7 hours at 84 micron per SN. Having 2 bands require 38 hours per SN per observing. Repeating this process for five times would cost 190 hours. For an optimistic case, 10 min per band (overhead limited). Taking 2 bands for five times with a few months' interval between would be about 100 min total.

Observation #2 is a spectroscopic survey of CAS A with FIRESS which requires a total time of 54 hours.

Herschel PACS showed 170 Jy for the total flux at 70μm. Assuming that flux spread 100 arcsec radius circular aperture, it would be 1 Jy per 1 pixel for PRIMA FIRESS. That would approximate to 0.5e-17 W $m^{-2}$ level. In order to cover 25 square arcmin survey area, it requires about 18 hours for 1e-19 W $m^{-2}$. Having three bands would require 54 hours.

## Special Capabilities Needed

Monitoring up to the whole mission lifetime might be required, if appropriate targets are found. Young SNe but distant SNe monitoring could last a few years with a few months' interval, while older (>5 years old) SNe monitoring would require the whole mission lifetime with a year to two-year interval.

Reasonably quick response time (one months), after ToO trigger

## Synergies with Other Facilities

JWST and ALMA

## Description of Observations

There are two complementary observing strategies.

Observation #1: One is to detect newly discovered SNe. First, PRIMA will observe the SNe detection sites after discovery and keep monitoring with imaging. Once detected and bright enough in FIR, SNe will be observed with spectra. These strategies will be used to study dust formation and line cooling.

Observation #2: The second strategy is to observe historically known SNRs. These PRIMA observations include Cas A, Crab, SN 1987A, and monitor dust formation and destruction science with Herschel time. The observations require imaging and spectroscopy with a large or small mapping.

# 111. PRIMA's View on Clumping in Massive Star Winds

Francisco Najarro (Centro de Astrobiología, (CAB), CSIC-INTA, Spain), María del Mar Rubio Díez (Centro de Astrobiología, (CAB), CSIC-INTA, Spain), Miriam García (Centro de Astrobiología, (CAB), CSIC-INTA, Spain)

Massive stars play a fundamental role throughout the evolution of the Universe. Through their radiation-driven winds, they significantly impact the dynamics and energetics of the interstellar medium. Among them, OB stars are the most extensively studied. The mass-loss rates of these luminous stars are typically inferred from various diagnostics: ultraviolet P Cygni line profiles, H-alpha emission, and radio and far-infrared (FIR) continuum emission.

How these rates, and to which extent, may be overestimated by the presence of small-scale density inhomogeneities (clumping) in the stellar winds is still an open question. This can only be fully addressed through a consistent analysis of all available diagnostics—probing different regions of the wind— by means of state-of-the-art model atmospheres, which will ultimately yield reliable mass-loss rates. For most of nearby Galactic massive stars, multi-wavelength data are available in all but one critical observational component: FIR diagnostics of free-free emission. This wavelength region is very sensitive to clumping properties at intermediate wind heights. Therefore, we propose to use the unique capabilities of PRIMA instruments to study the winds of a selected sample of 50 Galactic OB stars. With the FIRESS instrument we will obtain spectra for the brightest objects, while PRIMAGER will secure photometric measurements for the fainter ones. These observations will provide the missing information needed to characterize clumping throughout the entire outflow, enhancing our understanding of wind physics, and derive accurate mass-loss rates.

## Science Justification

Massive stars play a central role at all stages of the Universe, from re-ionization of the early Universe (e.g., Haiman and Loeb 1997, Amorín et al. 2017) to the formation of massive black holes and gravitational waves (e.g., Abbott et al. 2020). These stars exhibit mass-loss rates ($\dot{M}$) of $10^{-8}$–$10^{-4}$ $M_\odot$ $yr^{-1}$ and terminal wind velocities up to 2000 km $s^{-1}$. Through their radiative output, stellar winds and deaths in explosive events, massive stars rule galactic evolution. They enrich the interstellar medium (ISM), inject momentum, and influence star formation through compression or disruption of molecular clouds (e.g., Palacios et al. 2005, Prialnik et al. 2000, Rodríguez-Puebla et al. 2016)

Radiatively driven winds from OB stars are key mechanisms in this feedback process and mass-loss is considered the dominant process shaping their evolution and fate (e.g., Chiosi and Maeder 1986, Kippenhahn et al. 2012, Björklund et al. 2021). Traditional models assumed smooth outflows. However, nowadays it is well accepted that significant inhomogeneities—"clumping"— are present within these winds. Clumping, parameterized by the clumping factor (f_cl, Owocki et al. 1988), alters diagnostics sensitive to density-squared processes (e.g., recombination lines, free-free emission), leading to "true" mass-loss rates that may have been overestimated by up to a factor of ten (e.g., Fullerton et al. 2006, Cohen et al. 2010, Sundqvist et al. 2010).

Thus, characterizing clumping and inhomogeneities of radiation-driven winds is one major frontier in the field of massive stars, since it critically impacts the overall mass lost to winds during stellar evolution. Results will impact the current paradigm of whether the winds rule the evolution of massive stars (Meynet et al. 1994) or if their role is almost irrelevant in the context of other processes, like eruptions and binary interactions (Smith, 2014).

A full characterization of clumping and its structure can only be carried out through multiwavelength studies mapping different radial zones of the wind.

## 2. Current Limitations in Mass-Loss Diagnostics

Several spectral diagnostics are used to infer $\dot{M}$ and f_cl, each probing different regions of the stellar wind:

### UV P-Cygni Lines

Sensitive to ion column density but suffer from saturation in strong winds and dependence on X-ray ionization. P V lines suggested dramatic clumping, but newer 2D models with optically thick clumps revise $\dot{M}$ upward, revealing major uncertainties.

### Hα Emission

A key tracer of inner wind density. Models assuming smooth winds matched theoretical predictions, but later studies show clumping can cause $\dot{M}$ to be overestimated by ~2.5×. Hα is limited to regions <2 $R_\star$.

### Radio and FIR Continuum

Free-free emission in the radio and FIR bands is highly sensitive to $\rho^2$ and provides insight into outer and intermediate wind regions. These diagnostics are ideal for constraining clumping across the full wind but have so far lacked data in the critical intermediate zone.

Combining these diagnostics enables a full radial characterization of clumping throughout the stellar wind. However, the lack of FIR observational data, which best map between ~3–10 $R_\star$ leaves the clumping structure in this region largely unconstrained.

## 3. PRIMA FIRESS and PRIMAGER: A Unique Opportunity

We propose to use the unique capabilities of PRIMA instruments to study the winds of a selected sample of ~50 Galactic OB stars. With the FIRESS instrument we will obtain spectra for the brightest objects, while PRIMAGER will secure photometric measurements for the fainter ones. FIR fluxes depend on the wind's clumping properties in this region. Variations in f_cl by factors of 5–10 produce detectable changes (>2×) in predicted fluxes throughout the full PRIMA wavelength range. These measurements will:

- Constrain the wind clumping structure in the intermediate wind zones,

- Improve accuracy of mass-loss rate estimates,
- Allow robust, multi-wavelength (UV to radio) modeling using advanced NLTE atmosphere codes that include X-ray and clumping effects.

## 4. Goals and Impact

Our ultimate goal is a consistent, physically motivated derivation of both mass-loss rates and clumping factors throughout the entire wind. This will:

Provide insight into the origin of wind clumping, likely linked to the instability of line-driven winds:

- Validate or revise theoretical predictions for radial clumping stratification.
- Establish more accurate mass-loss recipes to be implemented in evolutionary codes
- Enable precise modeling of OB star evolution and feedback.

Previous efforts (e.g., Najarro et al. 2008; Puls et al. 2006) highlighted the need for intermediate wind diagnostics, especially for stars with strong winds, where f_cl varies significantly with radius. A first step was carried out with HERSCHEL-PACS observations (Rubio-Díez et al. 2022), though only upper limits could be obtained for a significant number of sources at longer wavelengths. Our proposed PRIMA FIRESS and PRIMAGER data will provide full FIR coverage, filling this critical gap and serving this way as a cornerstone for future stellar wind studies.

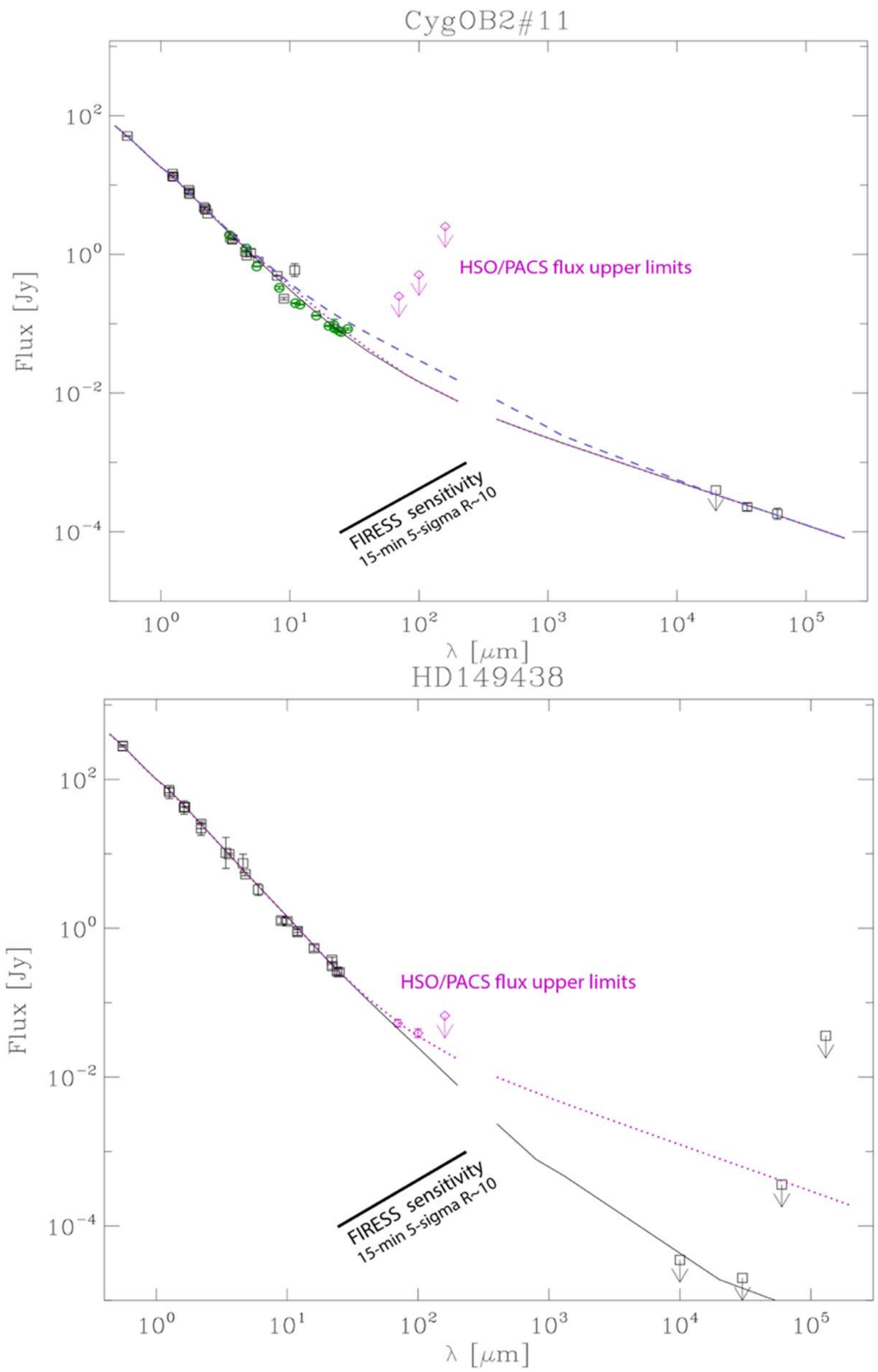

**Figure 1**. Observed and best-fit fluxes vs. wavelength for the O5 Supergiant Cygnus 11 (top) and for the B2 dwarf Tau Sco (bottom). Figure adapted from (Rubio-Díez et al. 2022).

## Instruments and Modes Used

This science case will use 50 FIRESS R~100 pointings and 50 PRIMAger pointings.

## Approximate Integration Time

Based on PRIMA ETCs we estimated that for FIRESS the observing time required to achieve our scientific goals in low point resolution mode is 25 min/source for the weaker sources while 13 mins/source should suffice for the brighter sources of the sample. For PRIMAger, we will use the hyperspectral imager mode for which we expect a somewhat higher sensitivity than FIRESS.

## Synergies with Other Facilities

ALMA, JWST, VLT (XShooter), etc.

## Description of Observations

We seek accurate determinations of the SED at FIR wavelengths of a sample of 50 massive OB stars. The luminosity of our sample at this spectral range requires the use of FIRESS and PRIMAger, which are much more sensitive than previous FIR instruments such as PACS on board of Herschel Space Telescope.

Our two main selection criteria for our targets are (i) that the sample should consist of stars for which we have extensive complementary data at other wavelengths, and (ii) the sample should give representative coverage of spectral types O4–B8. Feasibility of FIRESS and PRIMAger observations provides further criteria: targets should have a reasonably high predicted flux at 84 and 235 µm, a minimum of 1 mJy at the longest wavelength.

To select the targets, we will use available catalogs of well-studied O and B stars (IACOB, Fullerton et al. 2006, Najarro et al. 2011) and the investigation by Puls et al. (2006), Najarro et al. (2008) and Rubio-Díez et al. (2022).

To estimate fluxes, we use the stellar and wind parameters listed in the literature for O. As confirmed by more-sophisticated models, the FIR continuum emission (24 to 235 µm) can be estimated simply by adding the contribution of the stellar wind to a Planck function representing the photospheric emission. The wind contribution is calculated from the Lamers & Waters (1984) model, assuming the smallest amount of clumping consistent with present observational constraints. The true amount of clumping may be larger, which would result in higher fluxes. Additionally, we will use previous flux observations of OB stars at 70, 100 and 160 µm with Herschel/PACS.

As all targets are point sources, a single point source observation will be enough. For the brightest sources we will use FIRESS spectrometer in low-resolution point source mode (R~90-130), whereas for the fainter ones PRIMAger small maps (5’× 5’) in both hyperspectral bands (R~10) will be selected.

Based on PRIMA ETC we estimated that for FIRESS the observing time required to achieve our scientific goals in low-resolution point source mode is 25 min/source for the weaker sources while 13 mins/source should suffice for the brighter sources of the sample. For PRIMAger, we will use the hyperspectral imager mode for which we expect a somewhat higher sensitivity than FIRESS. Thus, we expect to cover the whole sample with roughly 20 hours (overheads excluded).

# 112. Exploring the Magnetized Circumstellar Environment of Evolved Stars Through Dust Polarization Analysis

Laurence Sabin (Institute of Astronomy-UNAM, Mexico), Edgar Ramirez (Tecnologico de Monterrey, Guadalajara, Mexico), Abraham Luna (INAOE, Mexico), Omar Serrano (INAOE, Mexico), Jesús Toalá (IRyA-UNAM, Mexico), Alejandro Garcia (INAOE, Mexico), Margaret Meixner (USRA)

The role of magnetic fields in evolved stars, from both low- and high-mass progenitors, remains a relatively underexplored area of research. Magnetic fields are known to play a critical role in stellar formation and evolution, significantly impacting the circumstellar environment. Most evolved systems—such as (pre)Planetary Nebulae, Symbiotic Stars (Symb), Wolf-Rayet stars (WR), and Supernova Remnants (SNRs)—exhibit strong mass loss, producing large amounts of dust and gas that enrich the Interstellar Medium (ISM). Key open questions include how magnetic fields interact with stellar winds and surrounding material, how they influence stellar evolution, what their strength and geometry are, and how they shape the distribution of material ejected from the star. Dust polarization measurements provide a powerful means to probe and characterize magnetic fields in these stars, revealing not only the field structure and strength but also the properties of the dust itself. PRIMA will be ideally suited for this task, enabling sensitive, multi-wavelength, and multi-scale far-infrared polarization observations and thus delivering critical new insights into magnetic processes during the final stages of stellar evolution.

## Science Justification

### Broader Context

Unlike the star formation phase, the role of magnetic fields in later stages of stellar evolution is not as well understood or even studied in some cases. This is particularly true for evolved stars that arise from both low-mass (1-8 solar masses) and high-mass (>8 solar masses) progenitors.

These low-mass descendants, such as (pre-)Planetary Nebulae, Symbiotic Stars, and type Ia supernovae, as well as the high-mass descendants such as Wolf-Rayet nebulae, and Supernova remnants with a core-collapsed origin, are characterized by intense mass-loss episodes induced by stellar winds or explosive events.

These episodes release substantial quantities of gas, carbonaceous- and/or silicate-type dust (e.g., ~0.001-0.2 Solar Masses in young PNe and WR nebulae, Rubio et al., 2020; Jiménez-Hernández et al., 2020), energy, momentum, and processed material into the Interstellar Medium (ISM), contributing to galactic chemical enrichment.

Additionally, magnetic activity is believed to play a crucial role in the generation and modulation of X-ray emission in these evolved systems, either through magnetically confined winds, shocks, or reconnection events.

The best solution is to explore magnetic fields through dust polarization analysis to advance our understanding of the role of magnetism in the formation and evolution of evolved stars and their interaction with the ISM.

This technique is based on the principle of aligning spinning, non-spherical paramagnetic dust grains with magnetic field lines. The dust emission is then polarized, as all the grains have their long axis perpendicular to the field lines.

The two more popular methods of dust alignment are by radiative torques (Dolginov & Mytrophanov, 1976; Draine & Weingartner, 1996) and by mechanical alignment (Lazarian, 2007). Radiative alignment torques (RATs) are considered a more efficient process, as they involve (isotropic) radiation and thus allow grains (with sizes similar to the wavelength) sufficient time to rotate and become fully aligned with the field. In contrast, Mechanical Alignment occurs when dust grains are subjected to a gas flow that imparts a torque, aligning the grains with respect to the flow direction or the magnetic field.

Instruments like PRIMAGer, with its polarimetric imaging capability, enable multi-wavelength and multi-scale polarization analysis in the mid- to far-infrared, leading to a significant advance in our understanding of magnetism in evolved stars. This potentially reshapes our models of their final evolutionary stages, their X-ray emission mechanisms, and their feedback into the interstellar medium (ISM).

## Science Questions

This project primarily aims to understand the impact of magnetic fields in the final stages of stellar evolution (in evolved stars) for both low- and high-mass progenitors through the study of polarized dust emission and absorption. Specifically, to address the following questions:

**How do magnetic fields interact with stellar winds and the circumstellar environment?**

*(e.g., comparison of magnetic vs. thermal energy, influence on mass loss, and shaping of outflows)*

**How do magnetic fields influence the late stages of stellar evolution, including their role in shaping asymmetries, launching collimated outflows, and regulating mass ejection?**

What are the strengths, structure, and topology of magnetic fields in evolved stars? *(e.g., dipolar, toroidal, or tangled configurations)*

**How do magnetic fields contribute to or regulate X-ray emission in evolved stars?**

**How do magnetic field strengths and dust grain properties vary between low-mass and high-mass progenitors?**

**How do magnetic fields affect dust grain polarization and the resulting thermal emission in evolved stars?**

**What information about dust composition and distribution is obtained from scattering, absorption, and emission processes?**

**How can multi-wavelength polarimetric observations with instruments like PRIMAGer improve our understanding of magnetic phenomena in late stellar evolution?**

By gathering and analyzing observational data from a variety of evolved stars, the project will provide a better understanding of the properties of dust and magnetic fields in evolved stars. This will provide first-hand, valuable insights into the late stages of stellar evolution that can be compared to existing (MHD) models.

## Need for PRIMA

With SOFIA no longer operational since 2022, the mid-to-far-infrared (MIR-FIR) wavelength range lacks adequate instrumentation from both ground and space. The PRIMA telescope, equipped with the PRIMAger instrument, will address this critical gap by delivering unprecedented polarimetric capabilities in the far-infrared (FIR) range (25-300 µm).

The key justifications for PRIMA are:

The analysis of extended nebulae around evolved stars: The Dusty circumstellar structures (∼30″-5′ in radius)observed in AGB stars and Planetary Nebulae require wide-field FIR imaging to study the spherical-to-axisymmetric symmetry transition, linked to magnetic fields and stellar winds. In the case of Supernovae Remnants, their extended nebulae (2-7′ radius) with dust and hot gas exhibit around 5-30% FIR polarization, critical for tracing magnetic fields and nucleosynthesis. PRIMAger will map these structures without the interstellar contamination affecting lower-resolution instruments.

The large field of view and angular resolution: PRIMAger will deliver FIR polarization maps with sub-arcminute resolution and a tens-of-square-degree field of view, which is essential for capturing interactions between circumstellar envelopes and the interstellar medium (ISM) or galactic magnetic fields. For example, the PRIMAGAL survey (Molinari et al. 2025) will map the Galactic plane to quantify the role of magnetic fields in stellar filament formation, analyzing thousands of clouds with varying linear masses and densities.

The limitations of interferometry: Interferometric instruments will likely filter out extended emission (>1″) due to limited *uv*-plane coverage, making them unsuitable for large structures such as AGB envelopes or SNRs. PRIMAger avoids this issue by combining high-sensitivity direct imaging (cryogenic detectors at 4.5 K) and integrated polarimetry, empirically calibrated using ∼1% polarization fractions observed in local galaxies.

The opportunity for unique science: PRIMAger features a variable linear filter (25-80 µm, R ∼10) and four polarized broadbands (92-235 µm), allowing for simultaneous studies of dust chemistry and magnetic alignment. The loss of FIR access (28–1600 µm) with SOFIA's retirement makes PRIMA urgent for continuing research on dust formation in SNe, which is crucial for understanding the evolution of the early universe.
In summary, PRIMA will provide unique capabilities to study the role of magnetic fields in stellar evolution, filling a critical observational void in the post-SOFIA era.

## Interpretation Methods

As previously noted, the determination of the magnetic field geometry with PRIMA will be primarily based on the analysis of dust polarization, governed by the alignment of spinning, non-spherical paramagnetic dust grains via Radiative Alignment Torques (RATs). However, it is essential to consider the possible contributions from other polarization mechanisms, such as scattering and dichroic absorption, which may also significantly affect the observed signal. In environments associated with evolved stars, the dominant polarization mechanisms are typically synchrotron emission, dust emission and absorption (through dichroism), and scattering. Among these, only scattering can occur without a magnetic field. In contrast, magnetic fields can be crucial for aligning dust grains, which are responsible for the polarized emission or absorption observed in the FIR.

We will apply multi-wavelength analysis techniques to disentangle the various polarization mechanisms in the PRIMA data. These include methods such as those employed by Chastenet et al. (2022) for SOFIA/HAWC+ observations, which exploit the wavelength dependence of polarization efficiency across the NIR to FIR to separate the contributions of different physical origins. We will also incorporate morphological and model-based diagnostics, such as those used by Jones (2000) in the study of M82, to interpret the thermal emission patterns of aligned, non-spherical dust grains. Finally, we intend to combine the PRIMA observations with data from other facilities to achieve a sizable wavelength coverage of the sources (see Figs. 1 & 2).

By combining observational and theoretical approaches, we aim to reliably isolate the polarization signal associated with magnetic field alignment and distinguish it from other competing or contaminating effects, thus ensuring robust magnetic field mapping with PRIMA.

## Link to Testable Hypotheses

The dust emission and polarimetric data obtained with PRIMA will serve as key observational constraints for magneto-hydrodynamical (MHD) simulations, specifically designed to reflect the diversity of circumstellar and interstellar environments around evolved stars. These simulations will be tailored to reproduce the observed morphologies, polarization patterns, and dust properties, enabling a deeper understanding of the interplay between magnetic fields, stellar winds, and dust dynamics in shaping nebular structures. In the GO BOOK 1 (L. Sabin et al.), we described the case study for (Pre-)PNe, Wolf-Rayet nebulae, and SNRs.

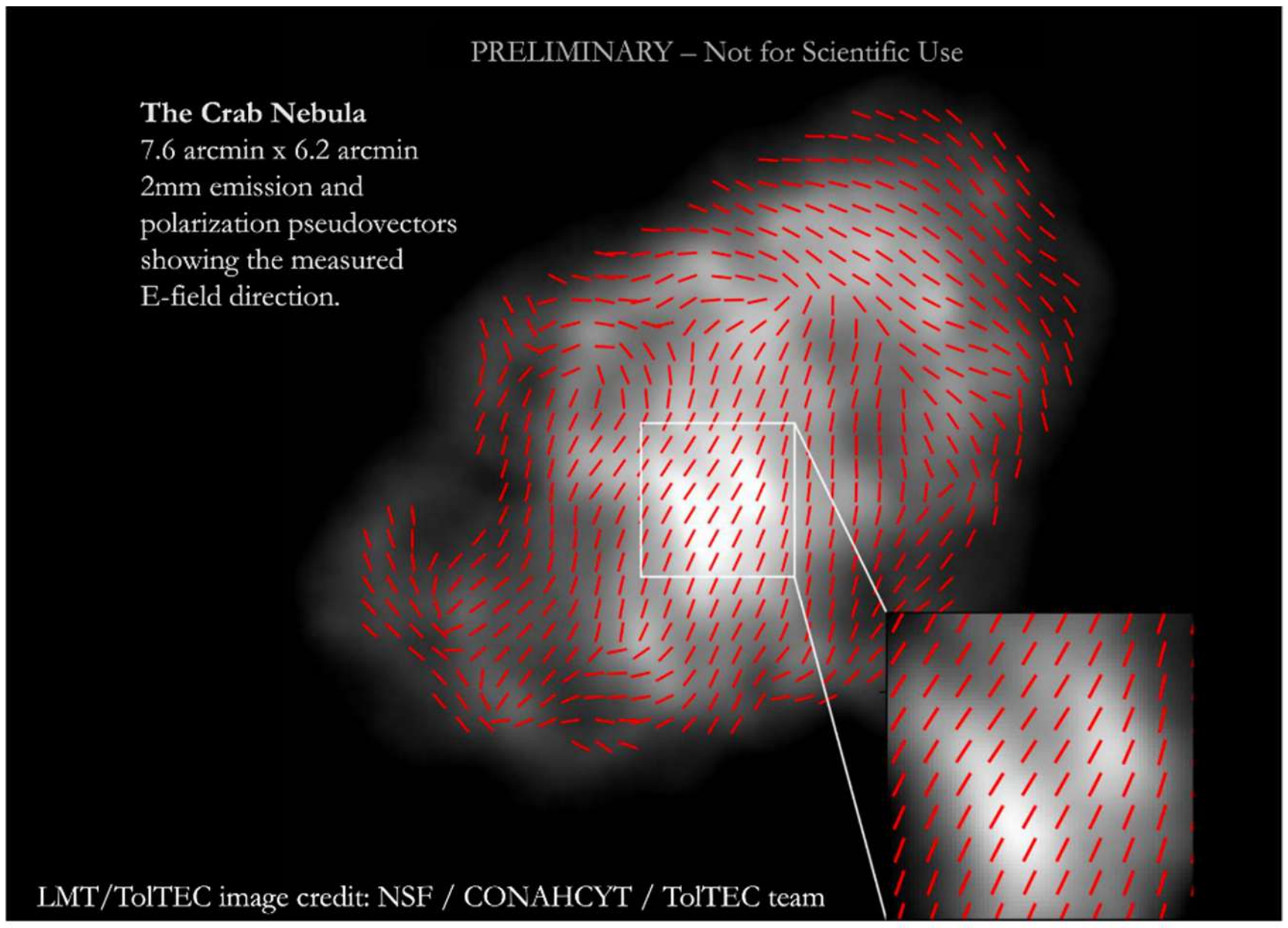

**Figure 1.** 2 mm total intensity map of the Crab Nebula (M1) with overlaid polarization vectors. Observed by LMT-TolTEC.

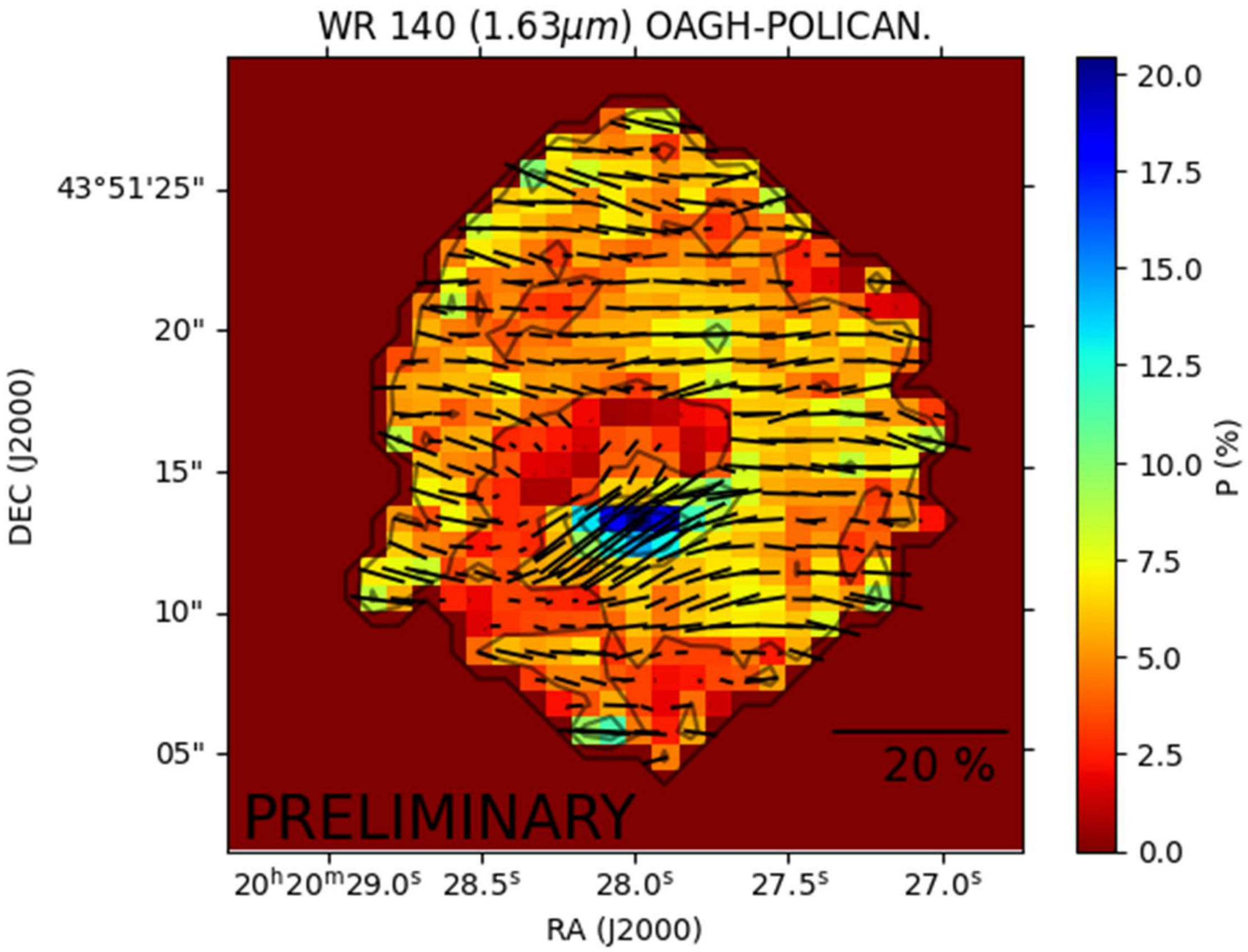

**Figure 2.** 1.63 μm polarization map of WR 140 with overlaid polarization vectors. The color wedge indicates the degree of polarization (%). Preliminary results obtained with OAGH-POLICAN.

## Approximate Integration Time

According to the PRIMA's new ETC calculation, the worst case of point source sensitivity in polarized flux at Band 2A (92µm) is 300 µJy for a 10-hour observation over an area of 10'x10', so it's necessary to make an individual estimation for each scientific case. An estimation of time using this case for the different kinds of sources is described as follows:

**SNRs:** The brightest case for SNRs is M1. It is approximately 7'x5', with local polarized flux reported at 150 GHz of approximately 14 Jy (Ritacco et al. 2018). With a relatively high flux at mid-infrared wavelengths (De Looze et al. 2019, Chastenet et al. 2022) a lower survey observational time (1.8 hour over an area of 10'x10' with a sensitivity of 1 mJy on the polarized emission). We select small-sized SNRs of different types (II, IIb, Ia-Ib, Ic) with polarized fluxes larger than 10 mJy. Such objects would need a minimal integration time, of about 10 minutes. Four objects with the characteristics above sum to a subtotal of 3 hours of integration time.

**Low-intermediate mass evolved stars:** We will use the sample from the THROES catalog (Ramos-Medina et al. 2018), which includes various objects already detected and measured with Herschel/PACS, resulting in 121 targets. The total flux ranges from ∼2 Jy to ∼0.5 Jy (92– 235 µm) for the faintest targets up to ∼1000 Jy to ∼250 Jy (92–235 µm) for the brightest. Assuming polarized fluxes of 250 mJy and an average size estimation of 10' x 10' arcmin, our estimated time for these cases sums to a subtotal of approximately 2 hours of observing time for around 10 different pointings.

**WRs:** These sources are bright (>1 Jy) in polarized flux and small (<2×2) arcmin$^2$. We can use the case of WR140, which is an object with an average flux of 1 Jy in every PRIMAger wavelength. For five selected cases (NGC 6888, NGC 2359, M1-67, WR16, WR140), we estimated 5 hours of observing time if an observation of 10'x10' is used.

In summary, the total observational time required for all the aforementioned projects is approximately 20 hours to complete the estimated objectives for each scientific case. We are focusing on observations at 92 µm. However, the 235 µm case represents the case with the lowest survey depth of all the PRIMAger wavelengths. According to the PRIMA survey calculation, with 0.7 hours of observational time over an area of 10'x10', we estimate an observational depth of 2 mJy at 235 µm.

## Special Capabilities Needed

None

## Synergies with Other Facilities

The TolTEC camera has a total of 7716 detectors distributed in the 1.1, 1.4, and 2.0 mm bands. Each element is a Lumped Element Kinetic Inductance Detector (LEKID), sensitive to the linear polarization of the incoming radiation and additionally has a half-wave plate to recover the polarization in all three bands. Four general science areas focused on by TolTEC align with PRIMA's science objectives, particularly the "The Fields in Filaments Legacy Survey", which is designed to probe the distribution of magnetic fields in filaments as traced by the polarization of dust. Just now, TolTEC is in the commissioning stage (http://toltec.astro.umass.edu/) (see Figure 1)

The ALMA interferometer completes PRIMA's wavelength coverage in the submm and mm range with sensitive imaging from 84 to 950 GHz (3 mm to 320 µm), allowing an angular resolution down to 0.015" (at 300 GHz) and full polarization capabilities. It will therefore be possible to map the sources near their central engine and retrieve the magnetic field's information at this location.

The Next Generation Very Large Array (ngVLA), although still a project, will serve as a valuable complement to the PRIMA data in the 1.2–116 GHz range. Set to be 10 times more sensitive than ALMA and to reach mas resolution. Its polarization capabilities will allow tracing the magnetic field even closer to the central structure and then perform multi-scale/resolution analysis of the magnetic field and dust.

The full-intensity, high-resolution images obtained from the JWST are a valuable supplement to the PRIMA data, enabling the clear identification of the magnetic field topology with respect to the nebula and the completion of the dust analysis using models.

## Description of Observations

Polarimetry requires strict control and precision of the angles; therefore, it is recommended to use the raster observation mode at different angles to eliminate correlation effects per trace in the coadded map. Mappings must consider observing a "sky" at least equivalent in area to the observed object. For example, for M1 (7×5 sqr.arcmin), the minimal PRIMAGer map of 10x10 sqr.arcmin is sufficient to get enough data from the object and sky. Several calibrators are needed, such as a gain calibrator (e.g., 0730-116), a passband calibrator (e.g., 3C 84), and a flux calibrator: generally, a planet (e.g., Mars/Uranus). But these can be adjusted as in all cases we will use the closest calibrators from the sources. Those can be shared with other programs.

# 113. Probing Mass-Loss During the Final Dying Gasps of Ordinary Stars

Raghvendra Sahai (Jet Propulsion Laboratory, California Institute of Technology), Toshiya Ueta (University of Denver)

Most stars in the Universe that evolve in a Hubble time or less (i.e., with main-sequence masses of ~1-8 Msun) go through their death throes as Asymptotic Giant Branch (AGB) stars, ejecting half or more of their total mass in the form of nucleosynthetically- enriched material into the interstellar medium (ISM) via dusty, isotropic, slowly- expanding winds. The heavy mass-loss dramatically alters the course of stellar evolution, seeds the ISM for future generations of stars and habitable planets, and drives galactic chemical evolution. Yet this crucial stellar evolutionary process is very poorly understood, especially the observed dramatic transformation of the geometry of mass-ejection from roughly isotropic to a rich diversity of bipolar, multipolar and elliptical shapes during the final death throes of these stars which convert them to planetary nebula. The current consensus is that a variety of strong binary interactions are responsible, but in order to pin down theoretical models for this transformation, a knowledge of the mass ejected during this transition phase (~$10^3$ yr) as a function of morphology is needed. PRIMA, with its unprecedented imaging and spectroscopic capability in the far-IR, is needed to probe these ejecta. We therefore propose a sensitive far-IR line survey of a large sample of young planetary nebulae spanning the full range of observed morphologies, using the FIRESS instrument.

## Science Justification

### Broader Context

Most stars in the Universe that evolve in a Hubble time or less (i.e., with main-sequence masses, m(MS) ~1-8 Msun) go through their death throes as Asymptotic Giant Branch (AGB) stars, ejecting half or more of their total mass in the form of nucleosynthetically-enriched material into the interstellar medium (ISM) via dusty, isotropic, slowly-expanding (Vexp~10-20 km/s) winds at rates of dM/dt~1e-7 to 1e-4 Msun/yr (e.g., Olofsson 2008). The heavy mass-loss dramatically alters the course of stellar evolution, seeds the ISM for future generations of stars and habitable planets, and drives galactic chemical evolution (e.g., increase in metallicity in the early Universe). Yet this crucial stellar evolutionary process is very poorly understood, especially the observed dramatic transformation of the geometry of mass-ejection from roughly isotropic to a rich diversity of bipolar, multipolar, and elliptical shapes during the final death throes of these stars, which convert them to planetary nebula. There is increasing observational evidence that binary interactions may play a crucial role in the late evolution and deaths of such stars (Sahai & Trauger 1998, Sahai+2011 [SMV11], Decin+2020). For example, HST imaging surveys of planetary nebulae or PNe (the end-products of AGB evolution) show that although AGB stars eject mass roughly isotropically over most of their lives, most young PNe have extreme and varied shapes—very few are round (SMV11). Theoretical studies show that strong binary interactions during the late AGB phase—such as common envelope (CE) evolution (e.g., Blackman & Lucchini 2014), are likely needed to produce high-velocity jets that can sculpt (from the inside-out) the mass ejected during the final death throes of AGB stars and result in the observed departures from spherical symmetry (Sahai & Trauger 1998). Mass ejection on the relatively short (~1e3 yr) time-scale that is associated with these very last death throes of AGB stars, sometimes referred to as the "superwind" phase, is best probed by studying young PNe. Such studies are crucial for understanding the physics of strong binary interactions (a process of fundamental and wide-ranging importance for stellar astrophysics) that result in the varied shapes of PNe. A fundamental question that such a study of young PNe can address is whether binarity also impacts the very onset of the star's death phase.

## Science Question

The very last death throes of AGB stars that transform (most) of them into aspherical PNe (with WDs at their centers), hold the key to understanding the binary interaction mechanisms that are believed to either cause (as, e.g., Common Envelope Evolution or CEE, Grazing Envelope Evolution or GEE, resulting in the ejection of most or part of the stellar envelope) and/or influence (as, e.g., formation of a jet-powering accretion-disk due to Bondi-Hoyle or Roche-lobe Overflow) this transformation. The diversity of morphologies (Fig. 1), which mostly includes bipolar, multipolar, and elliptical shapes (with a ubiquitous presence of point-symmetry) and only a small fraction of round ones, is well cataloged.

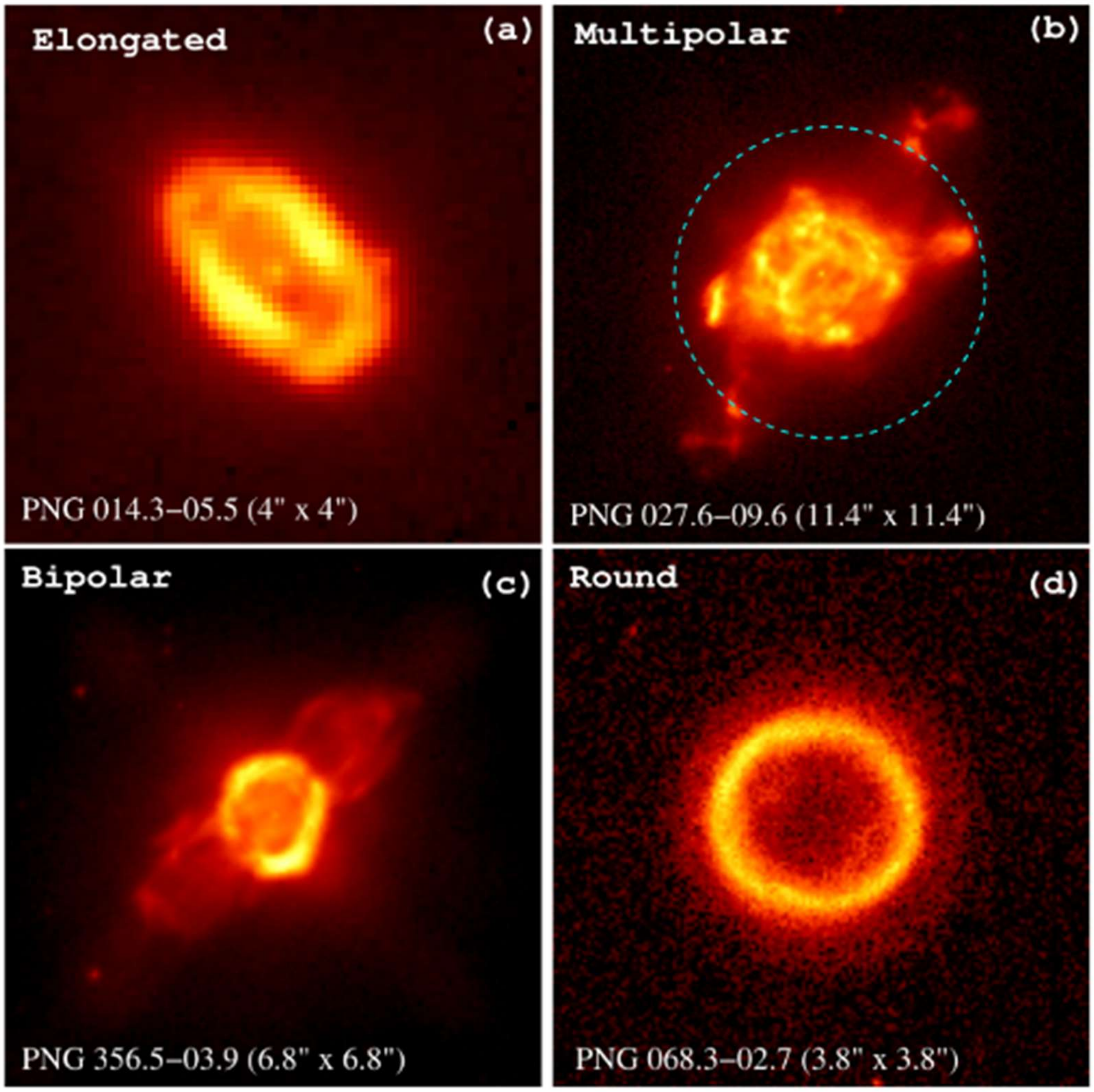

**Figure 1.** HST H-alpha images of PNe belonging to 4 major primary morphological classes. The dashed circle in panel b shows a 6.5 arcsec diameter circle for scale (adapted from SMV11).

A knowledge of the mass ejected during this transition phase as a function of morphology in young PNe is crucial for pinning down the nature of the binary interactions (e.g., in contrast to the other binary interactions, CEE will eject a lot of mass, ~few x 0.1 Msun, in a relatively short time), but is lacking. This is because most of this ejecta mass is expected to be neutral and warm, for which the primary gas cooling lines (e.g., [OI]63 micron, [CII]158 micron) lie in the far-IR region – but only very few young PNe have been observed in these lines (e.g., with ISO: Liu+2001, or PACS: Ramos-Medina+2018; Fig. 2).

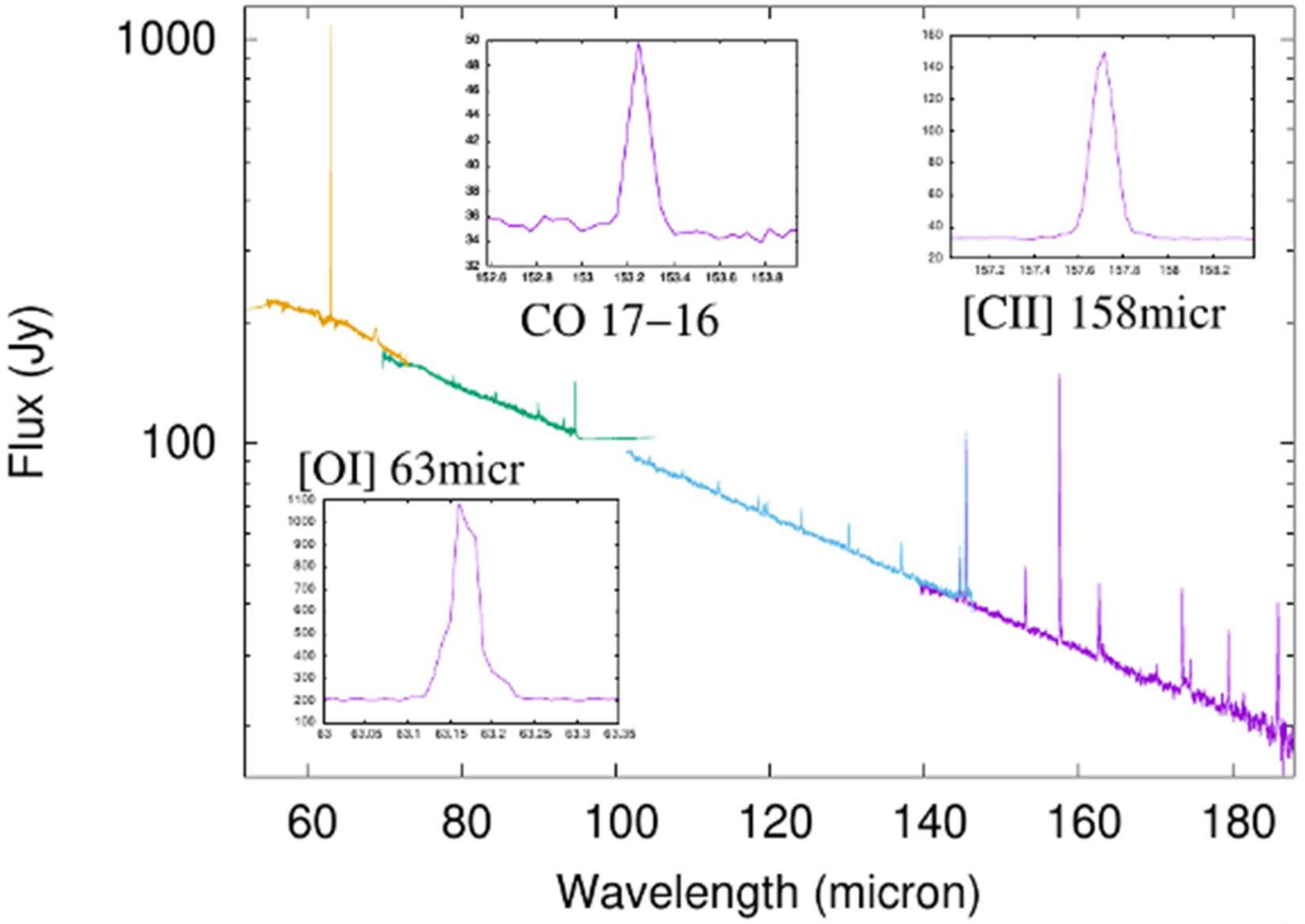

**Figure 2.** Typical spectrum of a young PN (CPD-56d8032) showing fine-structure and CO lines, taken using PACS/Herschel (data from Ramos-Medina+2018). A PACS survey of a statistical sample of such PNe, as a function of morphological type, will enable the determination of the mass ejected during the final death throes of AGB stars as they transition into PNe, and help constrain theoretical models of binary interactions that are believed to cause this transition.

## Instruments and Modes Used

FIRESS pointed high-res observations of 150 targets

For each FIRESS high-res pointing, we aim for a 5-sigma sensitivity of ~1e-16 (0.3e-16) W/m^2 for lines at 63 (158) micron, with a corresponding underlying continuum of 100 (50) Jy, which will take ~0.2 hr to cover all 4 bands. 150 targets will require a total of 30 hours

## Special Capabilities Needed

None

## Synergies with Other Facilities

Millimeter-wave observations of molecular lines such as those from the low-J transitions of CO, 13CO, C17O, C18O, HCN, HC3N, CN, HCO+, using interferometric facilities such as ALMA, NOEMA, SMA, will (i) provide important kinematic information on the mass outflows in our targets, as well as probe structures with much higher angular resolution (~0.1 arcsec) than PRIMA; (ii) probe the chemistry and key isotope ratios such as 12C/13C, 16O/17O, and 17O/18O that are affected by nucleosynthesis during post-main-sequence evolution. JCMT will probe the extended molecular H2 haloes and their structure (which can indirectly reveal the presence of unseen close companions), as well as compact dust clouds around the central stars that may have resulted from the grinding of planetesimals or outgassing from comets left over from the main-sequence phase (e.g., Sahai+2023, 2025; De Marco+2022).

## Description of Observations

We will carry out a FIRESS survey of a statistical sample of young PNe in multiple far-infrared lines that are diagnostic of the density and temperature in the PDR (e.g., [OI]63, 146 micron & [CII]158 micron) and molecular region (two CO lines). Young PNe generally have small angular sizes (typically < 10 arcsec) (Fig. 1), so a single-pointing per target will be adequate to capture all of the flux, as well as provide information on the emission extent (in the unlikely case that the emission is extended). The faintest lines of importance observed in a few targets observed with PACS are ~1e-16 W/m2. We therefore aim for a 5-sigma sensitivity of ~1e-16 (0.3e-16) W/m2 per pointing at 63 (158) micron, which together with an underlying continuum , typically of ~100 (50) Jy (using the PRIMA ETC) will take ~0.2 hr, using Bands 2, 3, and 4 to cover our main lines of interest. Thus, our target list of ~150 young PNe taken from compilations of HST-imaged young PNe (SMV11, Guzman-Ramirez+2011, Stanghellini+2016), most of which have Spitzer spectra (Stanghellini+2012) revealing their dust chemistry, will require 30 hr of integration time. The sample size is dictated by the requirement of obtaining statistical samples of the 7 different primary morphological classes (bipolar, collimated-lobe-pair, multipolar, elliptical, spiral, round, and irregular: SMV11).

## Acknowledgment

This research was carried out at the Jet Propulsion Laboratory, California Institute of Technology, under a contract with the National Aeronautics and Space Administration. 

# 114. The Extraordinary Deaths of Ordinary Stars: Probing their Full Mass-Loss Histories

Raghvendra Sahai (Jet Propulsion Laboratory, California Institute of Technology)

Most stars in the Universe that evolve in a Hubble time or less (i.e., with main-sequence masses of ~1-8 Msun) go through their death throes as Asymptotic Giant Branch (AGB) stars, ejecting half or more of their total mass in the form of nucleosynthetically-enriched material into the interstellar medium (ISM) via dusty, isotropic, slowly-expanding winds. The heavy mass-loss dramatically alters the course of stellar evolution, seeds the ISM for future generations of stars and habitable planets, and drives galactic chemical evolution. Yet this crucial stellar evolutionary process is very poorly understood because the full mass-loss histories of such stars over long time scale is very poorly understood. Observations that can probe mass-loss from intermediate to long time-scales (~10^3 to 10^5 yr), for a statistical sample of AGB stars, are needed. Only PRIMA, with its unprecedented imaging and spectroscopic capability in the far-IR and high sensitivity to low surface brightness structures, can carry out such observations. Using FIRESS, we propose a sensitive far-IR imaging survey of thermal dust emission and line emission from the extended regions of a statistical sample of AGB stars to understand, for the first time, the long-term mass-loss histories and thus estimate the total ejected mass during the AGB phase.

## Science Justification

### Broader Context

Most stars in the Universe that evolve in a Hubble time or less i.e., with main-sequence masses, m(MS) ~1-8 Msun, go through their death throes as Asymptotic Giant Branch (AGB) stars, ejecting half or more of their total mass in the form of nucleosynthetically-enriched material into the interstellar medium (ISM) via dusty, isotropic, slowly-expanding (Vexp~10-20 km/s) winds at rates of dM/dt~1e-7 to 1e-4 Msun/yr (e.g., Olofsson 2008). The heavy mass-loss dramatically alters the course of stellar evolution, seeds the ISM for future generations of stars and habitable planets, and drives galactic chemical evolution (e.g., increase in metallicity in the early Universe). Yet this crucial stellar evolutionary process is very poorly understood because the full mass-loss histories of stars, with m(MS)~1-8 Msun as a function of m(MS) and metallicity, are very poorly known. The mass-ejection may occur explosively, e.g., as a result from a strong binary interaction, such as common-envelope evolution or CEE (as e.g., in the Boomerang Nebula: Sahai et al. 2013) or grazing-envelope-evolution or GEE (Soker 2015). Such interactions are an important subclass of the diversity of explosive phenomena observed across the electromagnetic spectrum (e.g., Red Novae, Red Transients). The ejected mass matter interacts with the ISM, contributing to its turbulent structure and energy content.

## Science Question

### Probing the Long-Term Mass-Loss History of AGB Stars

Determining the full history of heavy mass-loss in AGB stars remains a major observational challenge. The time-scales over which AGB stars lose most of their mass (prior to beginning their post-AGB journey) can be very long (>~1e5 yr for an average dM/dt ~ 1e-5 Msun/yr). Hence, we do not know, in general, how long AGB stars have been undergoing dense mass-loss (i.e, at rates >~10^-7 Msun/yr), if the mass-loss rates change by significant *factors, and if so, whether this happens in a periodic, episodic or irregular manner over long time-scales.* A dramatic example of such a change is the strong, episodic increase that has been found (from observations of millimeter-wave CO line emission) for a few C-rich (C/O>1) stars by the detection of a detached, geometrically-thin expanding circumstellar shell (Olofsson+1996), inferred to have resulted from a thermal pulse associated with He-burning and 3rd dredge-up nucleosynthesis around the stellar core (this process produces much of the C in the Universe and converts the normally O-rich (C/O<1) stellar atmosphere to one that is C-rich (C/O>1). Millimeter-wave CO line emission from gas and scattered light from dust in the circumstellar envelope (CSE)—the standard tracers of the mass-loss history from ground-based observations – are limited typically to time-scales of <6,500 yr for a typical expansion velocity of 10 km/s and mass-loss rate <10^-5 Msun/yr, because, at relatively large radii (> 3.5x10^17 cm) in the CSE, CO is photodissociated by the interstellar UV field (e.g., Ramstedt et al. 2020), and dust scattered light becomes undetectable because it becomes too faint. Observations of atomic hydrogen (HI) from the wind, generally resulting from the photodissociation of H2 in the molecular wind, are usually strongly confused by Galactic emission (e.g., Matthews et al. 2015).

Hence, the total amount of mass ejected into the ISM, M(ejecta) (which depends on the envelope's outer extent e.g., M(ejecta)~r(out) (for a constant mass-loss rate at a constant expansion velocity) remains unknown; as a result the progenitor mass is also unknown. For example, in the case of IRC+10216, the best-studied mass-losing AGB star, the CSE seen in CO or scattered light extends to about 200'' (3.5 x 10^17 cm), and the inferred M(ejecta) is ~0.15 Msun, a small fraction of what this star has had to have lost, given its late evolutionary phase. The unexpected discovery of a very extended astrosphere (resulting from the shock interaction of the stellar wind with the ISM) in far-UV emission around IRC+10216 (Sahai & Chronopoulos 2011), led to a first estimate of the total duration of heavy mass-loss, t(hml)~69,000 yr, and total ejected mass, M(ej,tot)~1.4 Msun, a value consistent with the theoretical expectation that C-rich stars have progenitor masses of ~2 Msun (and that IRC+10216 is close to transitioning to the post-AGB phase).

Far-IR emission from dust in the CSE provides a method of tracing the mass-loss history over very long time-scales. Previous far-IR missions, beginning with IRAS, followed by ISO, Akari and Herschel detected very extended emission from CSEs, at radii much larger than probed by standard tracers in a few objects, either from detached shells, and/or from the astrospheres (e.g., Waters+1994, Izumiura+1996, Ueta+2006). A dedicated far-IR survey at 70 and 160 micron of a sample of 78 evolved stars (mostly AGB stars and a few red supergiants), using Herschel's PACS instrument revealed astrospheres in ~40% of the sample (Cox et al. 2012: Cox12; Fig. 1). About 20% of the sample, comprising of C-rich or S-type (C/O~1) stars (which experience thermal pulses), showed circular rings. The astrosphere radius, Rc, and width provide a natural estimate of the time-scale for the full duration of AGB mass-loss.

Even *for intermediate mass-loss time-scales* that can be probed in CO mm-wave lines, far-IR imaging may offer a distinct advantage. For CO, the combination of interferometric and total-power (to recover interferometric flux losses) observations needed to image areas of (say) 5 x 5 sq. arcmin, with high (~few arcsec) angular resolution, is very time-consuming (several hours/target) even for relatively bright sources and prohibitive for fainter ones. In contrast, striking discoveries of unexpected morphological structures (inferred to result from the presence of a binary companion) via far-IR imaging could be made with PACS imaging at 70 and 160 micron, e.g., the presence of 2 perfectly diametrically opposed spiral-shaped arms in the symbiotic star R Aqr and possible spiral structures in the S-star W Aql (Mayer+2013) with significantly smaller exposure times.

*In summary, in spite of the above studies, many questions remain unanswered.* For example, for most AGB stars, the values for M(ej,tot) derived from the far-IR studies (as well as the far-UV) are generally far below those required to bring these stars close to the post-AGB evolutionary phase (Maercker+2022, Sahai & Stenger 2023). Interestingly, within the small sample of stars for which M(ej,tot) could be estimated (~30), the C-rich and S-type stars typically show significantly higher values of t(hml) and M(ej,tot), which is expected since O-rich stars evolve into C-rich stars. So it is a puzzle as to why there are also O-rich (C-rich) stars with M(ej,tot) comparable to the highest (lowest) ones found for C-rich (O-rich) stars. In addition, some stars show shock structures that do not look like bow-shocks, but have “eye” shapes – the origin of such is not understood (the simplest explanation, that the mass-loss ejection properties are different in two opposing hemispheres, is unlikely.) Others, like W Hya, show exceedingly complex mass-loss histories via their FUV emission (Fig. 2); hints of these complexities are seen in the PACS images of W Hya, and in many other stars in the Cox12 study. In the current sample of AGB stars with extended rings, attributed to a thermal pulse, surprisingly only one or two show two rings (Mecina+2014), instead of just one, although AGB stars are expected to undergo a large number before their post-AGB journey. This may be due to sensitivity (the outer ring is much fainter than the inner one) and/or because of the star’s initial mass (since the interpulse timescale is smaller for stars with higher initial masses, e.g., ~2.8e4 yr for 3.6 Msun).

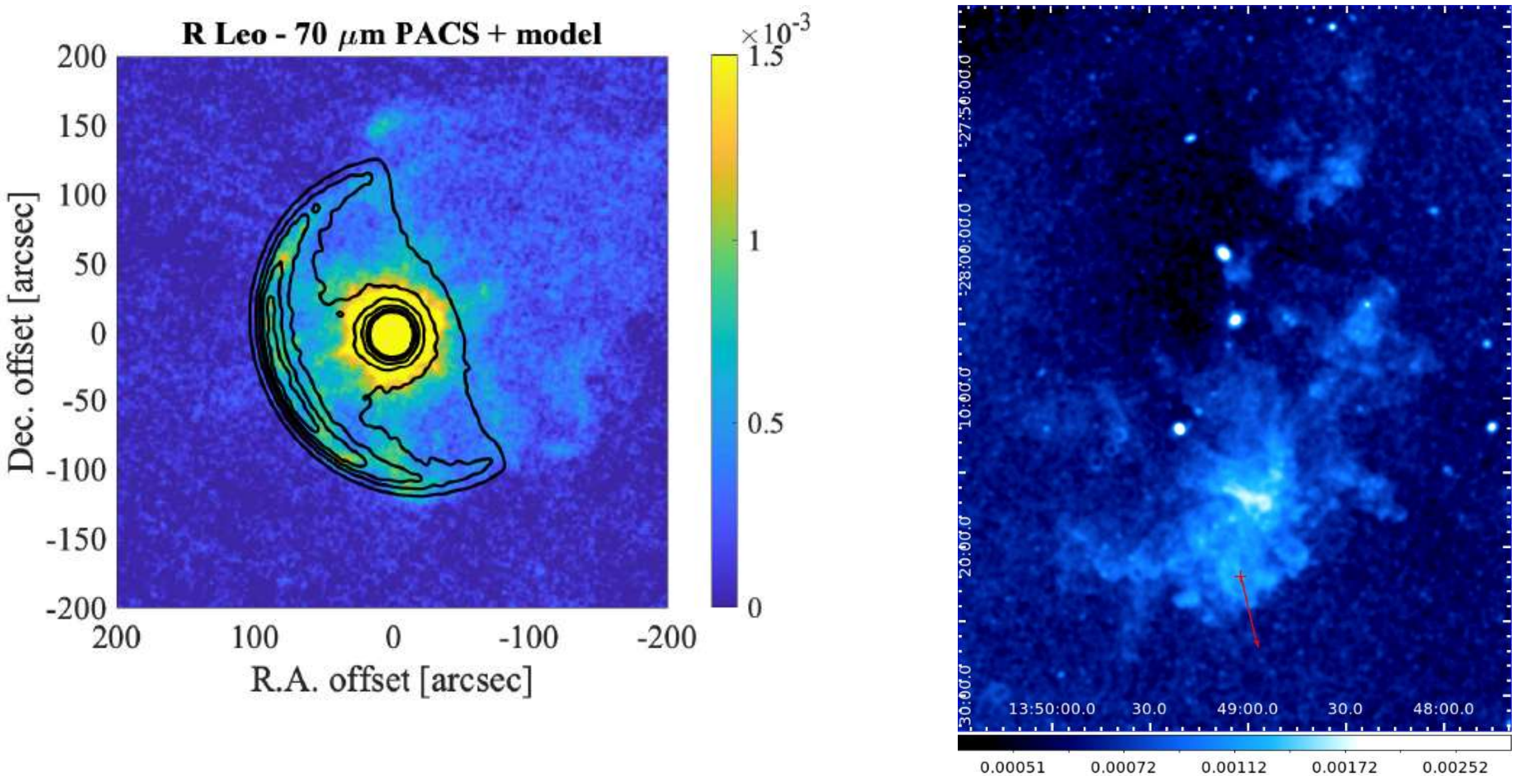

**Figure 1.** The astrosphere of the O-rich AGB star, R Leo, as seen at 70 micron with PACS/Herschel (color scale), together with a model (contours) for analysing the shock interaction of the AGB wind with the ISM (adapted form Mercker+2022) to estimate the long-term mass-loss history. A PRIMA far-IR imaging survey will enable us to similarly estimate the mass-loss histories of a statistical population of AGB stars on ~10^5 yr time-scales

Finally, the potential contribution of line emission (e.g., [OI] 63 μm, [CII] 158 μm) from shocked gas in the astrospheres remains completely unknown, but may be significant as indicated by numerical simulations (Mohamed+2012). *A much more sensitive survey of a larger sample (e.g., with PRIMA) can help address the above questions and issues.*

## Instruments and Modes Used

FIRESS low-resolution mapping mode (a 28 x 28 arcmin^2 map for each target)

## Approximate Integration Time

70 hrs

## Special Capabilities Needed

None

## Synergies with Other Facilities

Millimeter-wave observations of molecular lines such as those from the low-J transitions of CO, 13CO, C17O, C18O, HCN, HC3N, CN, HCO+, using interferometric facilities such as ALMA, NOEMA, SMA, will (i) provide important kinematic information on the mass outflows in our targets, as well as probe structures with much higher angular resolution (~0.1 arcsec) than PRIMA; (ii) probe the chemistry and key isotope ratios such as 12C/13C, 16O/17O, and 17O/18O that are affected by nucleosynthesis during post-main-sequence evolution.

## Description of Observations

We will carry out a PRIMA survey of a statistical sample of AGB stars that is a factor ~100 deeper in sensitivity than what has been achieved so far. Our survey has strong synergies with surveys of mass-loss from AGB stars that have been (or are being, or will be) carried out with ALMA (e.g., DEATHSTAR: Ramstedt+2020, ATOMIUM: Decin+2020) and JCMT (NESS: Scicluna+2022). The dust temperature in the shocked wind in astrospheres is found to be typically 30-50K, hence the far-IR wavelength range is well-suited for probing the astrosphere dust continuum emission. We will aim for detecting a continuum surface brightness of ~0.1 MJy/sr in the 50-200 micron wavelength range, which is a factor 100 deeper than the faintest sources detected in Cox12. We will determine the contribution of any line emission in different bands down to about 1% of the wide-band emission seen in the PACS 70 and 160 micron filters (i.e., unresolved line flux ~0.4e-16 W/m^2 for a line-width of ~1 km/s emitting over a 0.1 arcmin^2 area). We will use FIRESS to make small, low spectral resolution maps in both bands of the far-IR continuum and line emission. Since the FOVs of bands 1+3 and bands 2+4 are separated by about ~3.3 arcmin, we will map 28 x 28 arcmin^2 regions per target to obtain full spectral coverage over a FOV of ~20 x 20 arcmin^2. In order to achieve a 5 sigma detection of such line emission in the ~50-200 micron range, we estimate, using the PRIMA/FIRESS ETC, that we will need 0.35 hr per source, or a total of 70 hrs for a survey of the nearest 200 AGB stars (extracted from the compilation of about 18,400 AGB stars by Suh 2021). The sample size is dictated by the requirement of obtaining statistical samples of the 4 different morphological classes of extended emission sources (fermata, eyes, rings, irregular: Cox12), and the 3 different abundance types, and conservatively assuming a detection rate that, although expected to be higher, may be the same as in Cox12, i.e., 40%.

## Acknowledgment

This research was carried out at the Jet Propulsion Laboratory, California Institute of Technology, under a contract with the National Aeronautics and Space Administration. 

# 115. The Impact of the PRIM(ALL) All-sky Survey on Studies of Dusty Evolved Stars

Peter Scicluna (University of Hertfordshire), Fonteini Lykou (Konkoly Observatory), Sundar Srinivasan (IRyA-UNAM)

Asymptotic giant branch stars and red supergiants are key sources of dust and metals, and amongst the brightest infrared sources in galaxies. Despite decades of study, we have yet to fully understand key questions, such as how their mass-loss evolves over time, what the properties of the dust they produce are, and how these questions depend on the properties of the stars. PRIMA presents an opportunity to answer these questions, partly through an all-sky survey which would reveal the far-infrared properties of tens or hundreds of thousands of Galactic evolved stars. While large populations can be studied in nearby galaxies, the Galactic population is critical for two key reasons: firstly, only in the Galaxy can we spatially resolve the envelopes, and hence map time evolution; secondly, only in the Galaxy can we access the full detail of the other properties of the stars. By detecting a significant fraction of the total population of Galactic evolved stars, an all-sky survey with PRIMAger's polarimetric mode would reveal the relationships between dust production, mass loss, stellar properties and metallicity, as well as resolving the envelopes of large samples of nearby evolved stars, from which mass-loss histories and the processing of newly-formed dust can be derived. In addition, it will prove an excellent resource to select sources for follow-up observations, both with PRIMA and numerous ground- and space-based observatories.

## Science Justification

**The dusty outflows of evolved stars—including asymptotic giant branch (AGB) stars and red supergiants (RSGs)—are major contributors to the chemical evolution of galaxies and the production of dust.** While the most-recently formed dust is hot, it cools in the outer outflow, and its radiation shifts from the mid-infrared to longer wavelengths as the dominant radiation source shifts from the central star to the interstellar radiation field. Studies with Herschel, IRAS and AKARI as well as a wide variety of large ground-based sub-mm telescopes have exploited this to study evolved stars in the solar neighbourhood, such as the statistics of mass loss, the time-evolution of mass-loss in individual stars, and the properties of the dust in their outflows (e.g. Jura & Kleinmann 1989; Ueta et al., 2019; Cox et al., 2012; Kerschbaum et al., 2010; Maerker et al., 2018; Dharmawardena et al., 2018; Scicluna et al., 2022).

An all-sky survey with PRIMAger's long-wavelength bands - such as that proposed in the PRIMA science book volume 1 by Wright et al—would extend these questions throughout the Galaxy, and answer a wide range of science questions regarding stellar evolution and the chemical enrichment of galaxies. These questions include, but are not limited to:

1. What is the total dust production by Galactic AGB stars?

2. How does mass loss evolve over time?
3. How is dust processed in the outflows of evolved stars and as it mixes with the interstellar medium?
4. What are the long-wavelength properties of dust produced by AGB stars, and do they change as the stars evolve or dust is processed?
5. What is the relationship between dust, mass-loss and pulsations?
6. How does dust production depend on metallicity?

All of these questions could be answered in whole or part by an all-sky survey in the 100-micron range, either through the catalogues and images the survey provides, or by providing the fluxes and images required for targeted follow-up.

The last all-sky surveys to cover the ~100 micron range were IRAS and AKARI, whose catalogues reached point-source depths of 1.5 and 0.55 Jy/beam at 100 and 90 micron respectively, with angular resolutions of ~2' and 39". **In just 5000 hours, PRIMA could achieve a point-source depth of 16 mJy/beam (Saydjari et al) at a resolution of 10.8", 93 times deeper and 11 times higher resolution than the IRAS point-source catalogue at similar wavelength.** This would be sufficient to detect more than 90% of AGB stars with significant dust production ($> 3 \times 10^{-9}\ M_{\odot} yr^{-1}$, see Figure 1) closer than the Galactic Centre, and >90% of Galactic evolved stars with extreme dust production rates $> 10^{-7}$ up to a distance of 30 kpc. At an increased survey depth of ~1 mJy/beam (as originally proposed by Wright et al), >90% of AGB stars with dust-production rates $> 3 \times 10^{-9}\ M_{\odot} yr^{-1}$ and half with dust-production rates $> 10^{-10}$ would be detected to distances of 30 kpc, approaching a complete census of **all** dust-producing AGB stars in the Galaxy - hundreds of thousands of AGB stars alone. This ignores the impact of confusion (likely to be significant in the direction of the Galactic Centre); nevertheless a PRIMAger all-sky survey would detect at least many tens of thousands of Galactic AGB stars. The improved angular resolution compared to IRAS and AKARI will be key to minimising confusion from other compact Galactic sources, and extended objects including background galaxies, star-forming regions and Galactic cirrus.

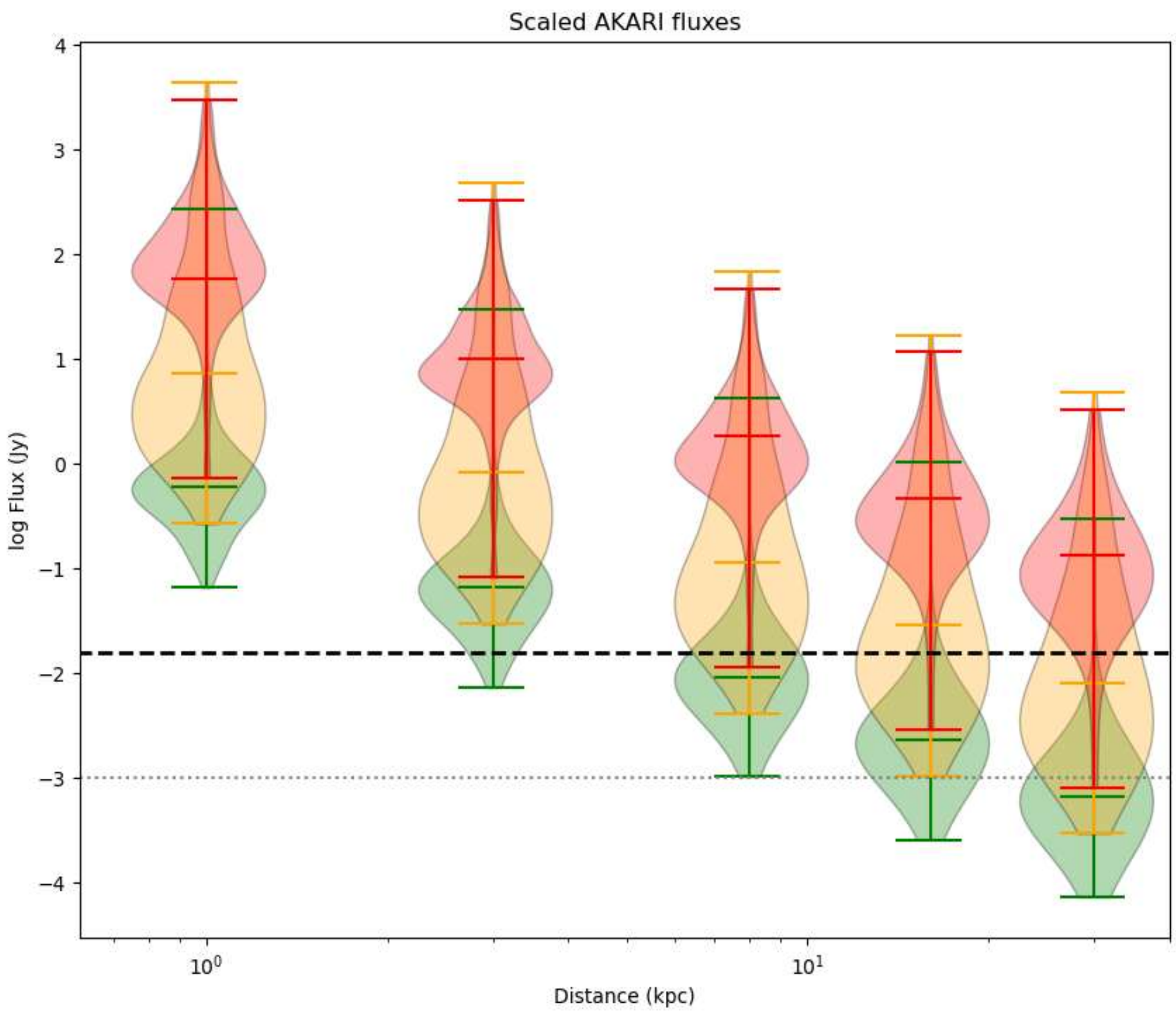

**Figure 1.** Distribution of 90 µm (AKARI WIDE-S) fluxes of Solar Neighbourhood AGB stars from the Nearby Evolved Stars Survey (Scicluna et al., 2022), scaled to different distances. Different colours indicate different bins in present-day dust-production rate – red: $> 10^{-7} M_{\odot} yr^{-1}$; orange: $3 \times 10^{-9} \rightarrow 10^{-7} M_{\odot} yr^{-1}$ ; green: $10^{-10} \rightarrow 3 \times 10^{-9} M_{\odot} yr^{-1}$. The lines correspond to different 5$\boldsymbol{\sigma}$ detection limits for a PRIMA survey – the proposed 16 mJy limit is shown as a dashed line, and the deeper 1 mJy limit as a dotted line.

The majority of these detections would be distant, unresolved sources. The sheer size of the resulting sample would enable vastly improved statistics on the total Galactic dust production, while also being large enough to explore the dependence on Galactocentric radius. This could be used as a proxy for metallicity, complementing JWST observations of low-metallicity dwarf galaxies, and potential PRIMA observations of the Magellanic Clouds (Jones et al.) to understand how the emissivity of dust changes with metallicity.

Nearby sources will be resolved, and PRIMA will detect arcminute-scale emission from thousands of individual AGB stars in multiple bands. These images will be perfect to probe both the time-evolution of dust production (and hence mass loss) and how the properties of the dust evolve over time for individual stars. Previous works (e.g. Dharmawardena et al., 2018) have shown how multiband imaging can be used to trace mass-loss history and the evolution of dust properties, and the sensitivity of PRIMA would make this feasible for large samples. While the all-sky data may not be sensitive enough for all possible targets, it could be key to identify targets for follow-up with deeper observations similar to those proposed in the Vol 1 science cases by Ueta et al. and Sahai et al.

We anticipate that an all-sky survey with the polarimetry bands will also cover roughly a quarter of the sky with simultaneous hyperspectral imaging observations. These data will provide further constraints on the composition and distribution of dust in the envelopes of evolved stars.

## Approximate Integration Time

Other science cases (e.g. Saydjari et al.) consider the observing-time estimates in detail, including sizes of scans, number of scans required, etc, and we defer to them for the full calculations. They show that a 5-σ depth of 16 mJy can be achieved across the whole sky in 5000 hours.

## Special Capabilities Needed

None

## Synergies with Other Facilities

SKA/ngVLA/DSA-2000: Imaging of evolved stars in HI has long proven a technical challenge, but new facilities are poised to change that. PRIMA's unprecedented sensitivity to dust will match the capabilities of the SKA and ngVLA to detect neutral gas in evolved-star outflows, providing the first direct and unbiased gas-to-dust ratio estimates. An all-sky survey will be the most effective route to choose targets for HI observations.

Roman: Time-series imaging of the Galactic Plane with the proposed Early-definition Galactic Plane Survey and Galactic Bulge Time-Domain Survey will identify AGB stars on the far side of the Galactic Centre through their long-period variability.

SPHEREx will perform an all-sky spectroscopic survey covering the wavelength range in which the stellar spectra of AGB stars peak. Given the depth of the survey, it is likely that SPHEREx will have a similar yield of Galactic AGB stars to PRIMALL—i.e., approximately all of them—at least in unconfused and unsaturated regions of the Galaxy. These spectra will be perfect for characterising the central stars, complementing PRIMA's unprecedented power to characterise their dusty outflows.

Sub-millimetre line and continuum observations with e.g. ALMA, JCMT, or in the future AtLAST will provide complementary constraints on the dust properties and outflows of AGB stars. PRIMA's higher frequencies provide similar angular resolution to the JCMT at 650 – 1300 micron and to AtLAST at 1.8 – 4.5 mm (ALMA bands 2, 3 and 4). This synergy can extend studies of dust properties to longer wavelengths, where lab data have larger discrepancies. Single-dish telescopes can also provide complementary maps of molecular-line emission, e.g. CO, HCN, for nearby sources which reveal the distribution of gas and how the gas-to-dust ratio changes in the outflows. Meanwhile, ALMA can perform efficient follow-up on more distant targets, probing longer wavelengths and higher angular resolutions both in lines and continuum.

## Description of Observations

We propose an all-sky imaging survey using the PRIMAger polarimetric bands, observing the whole sky to a depth of 16 mJy (5-σ) in the 92 micron band, while also covering a large fraction of the sky in parallel in the hyperspectral imaging bands. This will produce a point-source catalogue of dusty Galactic sources from young stellar objects to evolved stars nearly 100 times deeper and 10 times higher resolution than IRAS at the same wavelength. With the potential to detect tens or hundreds of thousands of evolved stars in 4 bands, PRIMA will not only efficiently disentangle YSOs and evolved stars, it will probe the properties of dust throughout large samples of evolved stars and provide key input for deeper follow-up studies with PRIMA on key samples.

# 116. Time-Domain Studies of Stars and Stellar Remnants with PRIMA

Peter Scicluna (University of Hertfordshire), Sundar Srinivasan (IRyA-UNAM), Mikako Matsuura (Cardiff University)

We propose a high-resolution spectroscopic monitoring survey of nearby evolved stars and stellar outbursts with FIRESS. By detecting the variation in large numbers of highly-excited molecular transitions, this survey will reveal the dependence of wind acceleration and dust formation on the stellar variability. Single-epoch observations have been shown to be possible with Herschel, which observed small samples. PRIMA's vastly improved sensitivity will enable large, statistically-relevant samples to be observed with a modest time investment. We will be able to probe the time variation of the density, temperature, and velocity profiles over a large range of physical scales. These observations will improve our understanding of the interaction between shocks and dust formation in the acceleration zone and provide critical constraints to hydrodynamical models.

## Science Justification

The last decade has seen a revolution in time-domain science across the entire EM spectrum. PRIMA stands to bring time-domain science to the far-infrared, through its combination of high sensitivity and broad wavelength coverage. Here we consider 2 science cases of stars and circumstellar material evolving and varying on human timescales.

### Key Science

**[1] Tracing pulsation-induced shocks in the wind-acceleration zone of AGB stars**

AGB mass loss is believed to be initiated by radial pulsations which generate shocks to levitate the atmosphere and form molecules (and, further outward, dust) in the circumstellar envelope (CSE). While hydrodynamical models that incorporate a treatment of the conditions in the wind acceleration zone (e.g., Bladh et al. 2019) are able to reproduce the general behavior, the models lack calibration of several important parameters, including the time variation of the number density of key molecular species and the velocity profile in the CSE. A clearer understanding can come from multi-epoch observations that trace the evolution of the inner CSE as a function of the pulsation phase.

High-J molecular rotational transitions in the far-infrared offer one way to probe the conditions in the acceleration zone of AGB CSEs. A related Vol 2 science case (Scicluna et al., "Tracing mass-loss, wind acceleration and dust formation with FIRESS") discusses the possibility of a spectroscopic survey of a large (~560 sources) number of AGB stars in the Solar Neighborhood to obtain integrated luminosities of high-J lines. We will refer to this in the rest of this case as the "single-epoch science case", as here we propose a multi-epoch monitoring survey of a subsample of the sources mentioned in the aforementioned case.

The FIRESS instrument on PRIMA will be able to detect molecular lines with very high excitation energy from the wind-acceleration zone of evolved stars. Doing this even at single epochs for a large sample of AGB stars would already be an advance on Herschel results. However, PRIMA would actually be sensitive enough to perform time-series observations in this range, revealing not only how the acceleration and excitation changes between stars, but how it responds to the pulsations of the central star itself! Changes in line strengths over time will reveal how the density, temperature and velocity of gas within the wind-acceleration zone changes in response to stellar pulsations. This is key to understanding how momentum is transferred from gas on ballistic trajectories before dust forms, to accelerating and eventually unbound gas after dust has formed. This will answer key questions such as how shocks propagate from the stellar pulsations into the extended envelope, and how those shocks impact the acceleration of the outflow. Models suggest dust forms in the wake of these shocks, which then accelerate the wind, which time-domain spectroscopy with FIRESS may be able to observe in action.

Given expansion speeds of ~5-10 km/s, we expect the lines to be at best marginally resolved. The integrated line luminosities of the large number of lines expected in the wavelength range covered by FIRESS will, however, provide reasonable constraints on the velocity structure in the envelope over a large physical scale.

These data may also provide serendipitous detections of atomic and ionised lines, e.g. [CII], [OI], and prominent dust features (e.g. Forsterite @ 69 micron). If so, we will also be able to investigate their evolution over the pulsation cycle.

**[2] Detecting FIR variability in novae and stellar outbursts**

It is getting clear that recurrent novae and stellar outbursts emit circumstellar dust emission at far infrared wavelengths. They also emit numerous molecular lines, such as CO, SiO and H2O (Exter et al. 2016). The energy peaks of both dust and molecular lines fall onto far-infrared wavelengths, hence, they are important coolants of the circumstellar envelope. Indeed, Exter et al. (2016) detected time variation in outburst of the Galactic luminous red nova, V838 Mon. Currently, this is the only nova or outburst whose far-infrared variability has been observed. Therefore, further observations with PRIMA will be required, if the far-infrared detection of outbursts are common in these objects.

PRIMA can make a survey of novae and outbursts, characterising the temperature and density structure and identifying variations in these for the first time. This will include T CrB, the 84-year recurrent nova, which consists of a red giant and white dwarf. Mass transfer from the red giant to the white dwarf is believed to fuel the nova explosion, and PRIMA may potentially be able to detect remnants of the mass transfer.

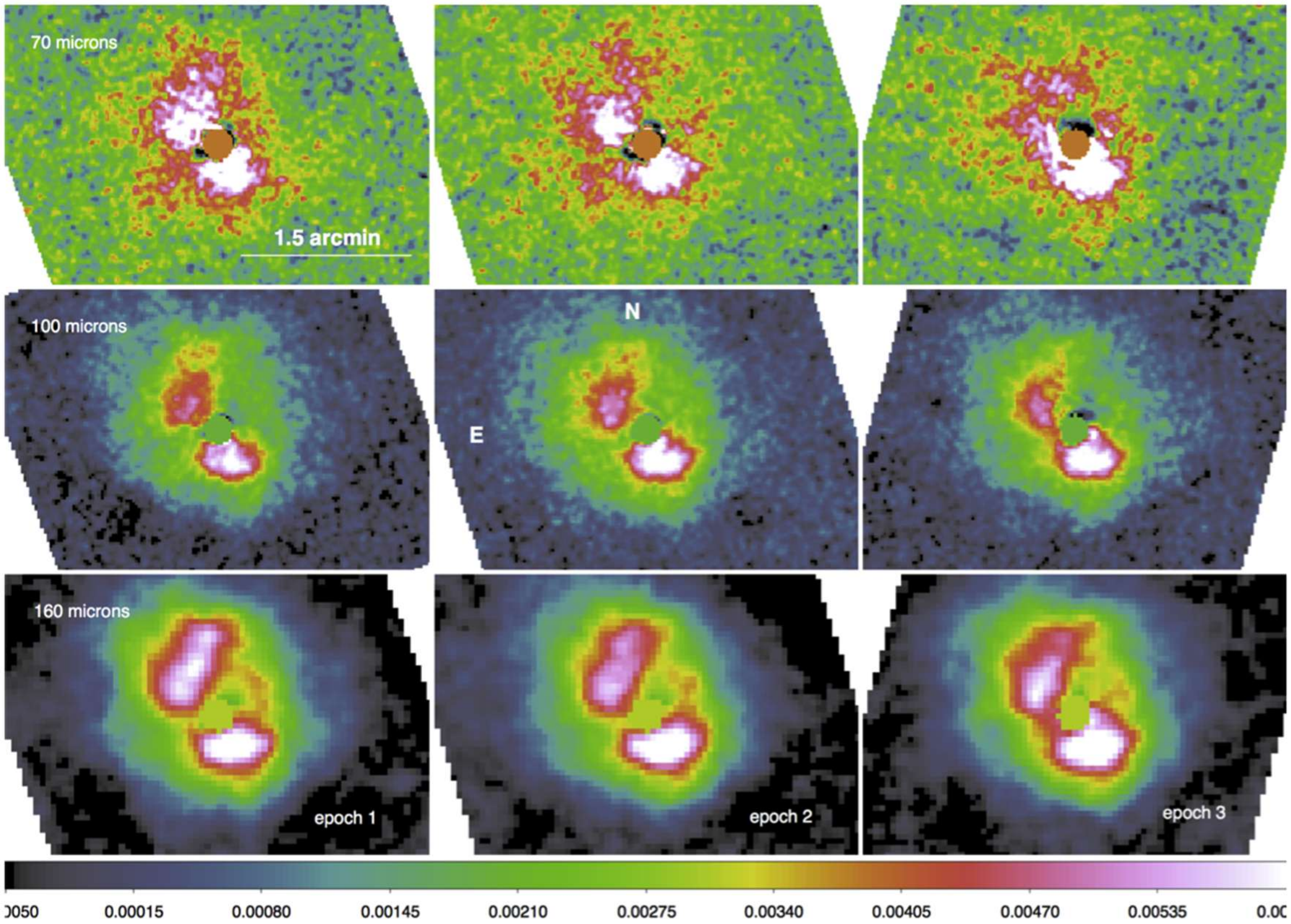

**Figure 1.** Herschel/PACS imaging of V838 Mon at 3 epochs, showing the evolution of the flux and morphology of the envelope (Exter et al. 2016).

## Instruments and Modes Used

This spectroscopic monitoring program will be carried out by using the FIRESS spectrometer in the high-resolution mode. The full spectral range of FIRESS should be covered to observe important molecular emission lines. In this program, typically 10 observations per AGB star are needed for a sample of roughly 50 stars.

## Approximate Integration Time

For the monitoring of high-J lines in AGB stars, as in the single-epoch science case, we feed the line and continuum fluxes from Nicolaes et al (2018) into the PRIMA ETC to find that the most important lines (e.g. ground-state rotational transitions of CO, $H_2O$, HCN) can be detected in less than the minimum full-scan time for a typical source. For a sample of sources, we therefore expect to need no more than 30 minutes per source per epoch. Therefore 1 hour per source per epoch is needed to cover the full spectral range of FIRESS. For ten epochs, this comes to 10 hours per source. We can therefore perform time-series observations for 50 sources from the single-epoch sample with a total time request of 500 hours over ten epochs. Given that the sources in the NESS sample have periods ranging from 30 days to 2 years, the observations will have to be spread over at least as much time. For the outburst science, taking stellar outburst V 838 Mon as an example, which is the unique Herschel observations of a transient, the CO lines are typically $10^{-18}$ – $10^{-17}$ W $m^{-2}$. This object was resolved, hence, requesting 4 square arcmin low-resolution map. At 200 micron, this requires 0.38 hours. Because it requires two separate sets of bands, this requires 0.76 hours per observing epoch. Assuming 5 outbursts over 5 epochs for each target, the total required time would be approximately 25 hours.

## Special Capabilities Needed

Time-series observations require time-constrained scheduling to control the cadence, either to ensure that observations have approximately equal sampling over a given period, or to space observations e.g. "no closer than X days apart". Finally, some time-series observations may be most effective if triggered like target-of-opportunity observations; this may be particularly relevant to observations of novae and of AGB stars with periods significantly longer than one observing period. This would also benefit from the opportunity to propose for time that can be spread over multiple observing cycles.

To prevent smearing in time, each observation (i.e. one epoch) and any associated calibrations may need to be executed as a single block without interruption. However, as these are short observations this is unlikely to be an issue.

AGB stars are long-period variables, so for maximum scientific gain, we require the flux calibration and baseline to be stable over very long time periods - then the absolute line and continuum fluxes can be used for precision science. If this cannot be achieved, relative calibrations (e.g. line-to-continuum ratios) can still be used to study variation but with reduced scientific gains.

## Synergies with Other Facilities

Between Gaia DR4, LSST, ZTF, NEOWISE and NEOSurveyor, optical to mid-infrared light curves are or will be available for all potential targets. These will enable efficient source selection while providing the required information on periods and amplitudes.

Optical and NIR time-series spectra will probe changes in the dust-formation layer and, with R ~ 100,000, can resolve the velocity profiles of shocked gas over time and will therefore complement the PRIMA observations discussed here.

Ground based interferometers, including ALMA, and in future the SKA and ngVLA, have the capability to resolve (spectrally and spatially) high-excitation transitions of other species, particularly rotational lines from vibrationally-excited states of molecules, and dust precursor species (e.g. citations). This complements the high-abundance, unresolved tracers that FIRESS is sensitive to.

The low-J data obtained from ground-based single-dish and interferometric facilities can be combined with the PRIMA data to obtain a comprehensive view of the envelope's physical structure and dynamics. The combined data will probe the radial temperature, density, and velocity profile out to ~10^4 au.

## Description of Observations

For evolved stars, targets will be selected based on the line strengths in single-epoch spectra, ensuring that line variability can be detected, and optical variability data (from e.g. OGLE, Gaia and other surveys). By stratifying the sample selection across variability periods and amplitudes, as well as mass-loss rate and chemical type, robust trends in line variability can be extracted. For each source, roughly 10 epochs will be observed for the full spectral range of FIRESS, with roughly even sampling across one pulsation period, ensuring the different stages of the wind-acceleration zone's response is sampled.

# 117. Tracing Mass-Loss, Wind Acceleration and Dust Formation with FIRESS

Peter Scicluna (University of Hertfordshire), Sundar Srinivasan (IRyA-UNAM), Beth Sargent (STScI/JHU)

We propose a high-resolution spectroscopic survey of nearby evolved stars with FIRESS. By detecting large numbers of highly-excited molecular transitions, this survey will reveal the dependence of wind acceleration and dust formation on the properties of the stars. Herschel proved such observations are straightforward, observing small samples, but PRIMA will enable large, statistically-relevant samples to be observed with a modest time investment.

## Science Justification

Evolved stars, including asymptotic giant branch (AGB) stars and red supergiants, drive the chemical evolution of the Galaxy; their mass loss returns the products of nucleosynthesis to the ISM, providing fuel for future star formation (Höfner & Olofsson, 2018). However, key aspects of this process remain poorly understood, particularly the mechanisms behind the formation of dust grains, and how the outflow is initiated and accelerated to escape velocity (e.g. Weigert et al., 2024). Building on work with Herschel, PRIMA will revolutionise our understanding by—for the first time—enabling these processes to be observed in statistically-relevant samples of evolved stars.

High-J rotational line emission in the far-infrared has been studied for both Galactic (Nicolaes et al. 2018) and extragalactic objects (Matsuura et al. 2016). Matsuura et al. 2016 obtained detections for a number of lines in the CO ladder (11 → 10 to 15 → 14) for the Large Magellanic Cloud red supergiant IRAS 05280–6910, as well as one CO transition for the OH/IR star WOH G64. All the lines were unresolved. For the red supergiant, they fit the observed integrated line fluxes, enabling them to estimate the mass-loss rate. Nicolaes et al. 2018 obtained unresolved line spectra for many molecules for ~40 Galactic AGB stars. These data included CO transitions up to 45 → 44, corresponding to much higher excitation temperatures and therefore probing the wind acceleration zone for these stars.

Such high-J observations can therefore not only provide the mass-loss rate but will also reveal the temperature and density gradients in the acceleration zone. In addition, if the lines are marginally resolved, we can also infer the velocity gradient.

Using the FIRESS high-resolution mode, rotational transitions of highly-excited molecules in the wind acceleration zone can be efficiently detected. For example, CO transitions from J= 12 → 11 up to 95→94 are inside FIRESS' coverage, while $H_2O$ alone may have hundreds of detected lines in a typical FIRESS observation (Nicolaes et al., 2018). The emission from lines with such high excitation has significant contributions from the wind-acceleration zone, and in many cases are dominated by it (De Beck et al., 2010). These lines reveal the temperature and density in this critical region, as well as the evolution of the outflow velocity and chemical abundances. While FIRESS will only be able to, at best, marginally resolve the lines, the sheer number of detectable lines will ensure that a statistical picture of the wind acceleration is clear, both for the overall population and for individual sources.

In addition to these strong lines from abundant species, FIRESS has the potential to detect lines from a wide range of molecules, including potential dust precursors, e.g. SiO and CCH. Recent ALMA observations (e.g. Decin et al., 2017) have painted a confused picture of dust condensation, with presumed dust precursors (e.g. AlO) showing gas-phase emission on much larger scales than the dust-formation radius, indicating low condensation efficiency. By sampling large numbers of lines from multiple species in a single observation, PRIMA may be able to shed new light on this question even without spatially-resolving the lines. By exploiting the differences in excitation, lines can be traced to different radii, as with the wind acceleration zone. Moreover, FIRESS spectra will provide continuum shapes and fluxes for all the sources, revealing the slope of the dust emissivity in great detail.

These observations will also provide serendipitous constraints on weaker lines from less abundant molecules, allowing a comprehensive chemical inventory to be taken for large samples. Moreover, these observations will also cover different tracers, such as [CII] and [OI] emission lines, which may be excited by interstellar UV photons in the outer reaches of the outflows, and will provide strong constraints on the 69 micron crystalline forsterite feature seen in O-rich stars and the unidentified broad 30 micron feature in C-rich stars.

Herschel was only able to observe 40 evolved stars with heterogeneous spectral coverage, but typically detected several hundred lines per source with integrated line fluxes ranging from $2.10^{-18} – 10^{-14}$ W $m^{-2}$ (Nicolaes et al., 2018). PRIMA could provide data of similar or higher quality *in just 1 hour per source*, covering the entire FIRESS wavelength range, enabling observations of samples an order of magnitude larger. This, and the large sample size, would enable a statistical exploration of the differences between AGB stars of different chemical types and between high- and low-mass O-rich sources, as well as differentiating red supergiants from AGB stars.

The Nearby Evolved Stars Survey (NESS; Scicluna et al. 2022) sample is ideal for the proposed observations; the NESS targets consist of ~780 AGB stars within ~5 kpc of the Solar Neighborhood. Of these, 560 have 65 µm fluxes below the FIRESS saturation limit of 300 Jy. In comparison to other existing surveys of nearby stars, the NESS sample has a relatively larger fraction of stars with lower mass-loss rates. Figure 1 shows that this sample explores a larger range of colors (and therefore dust temperatures) and mass-loss rates than the one studied by Nicolaes et al. The increased number of sources with such low mass-loss rates could enable us to pinpoint the stage where the mechanism driving mass loss shifts from pulsations to dust-driven winds.

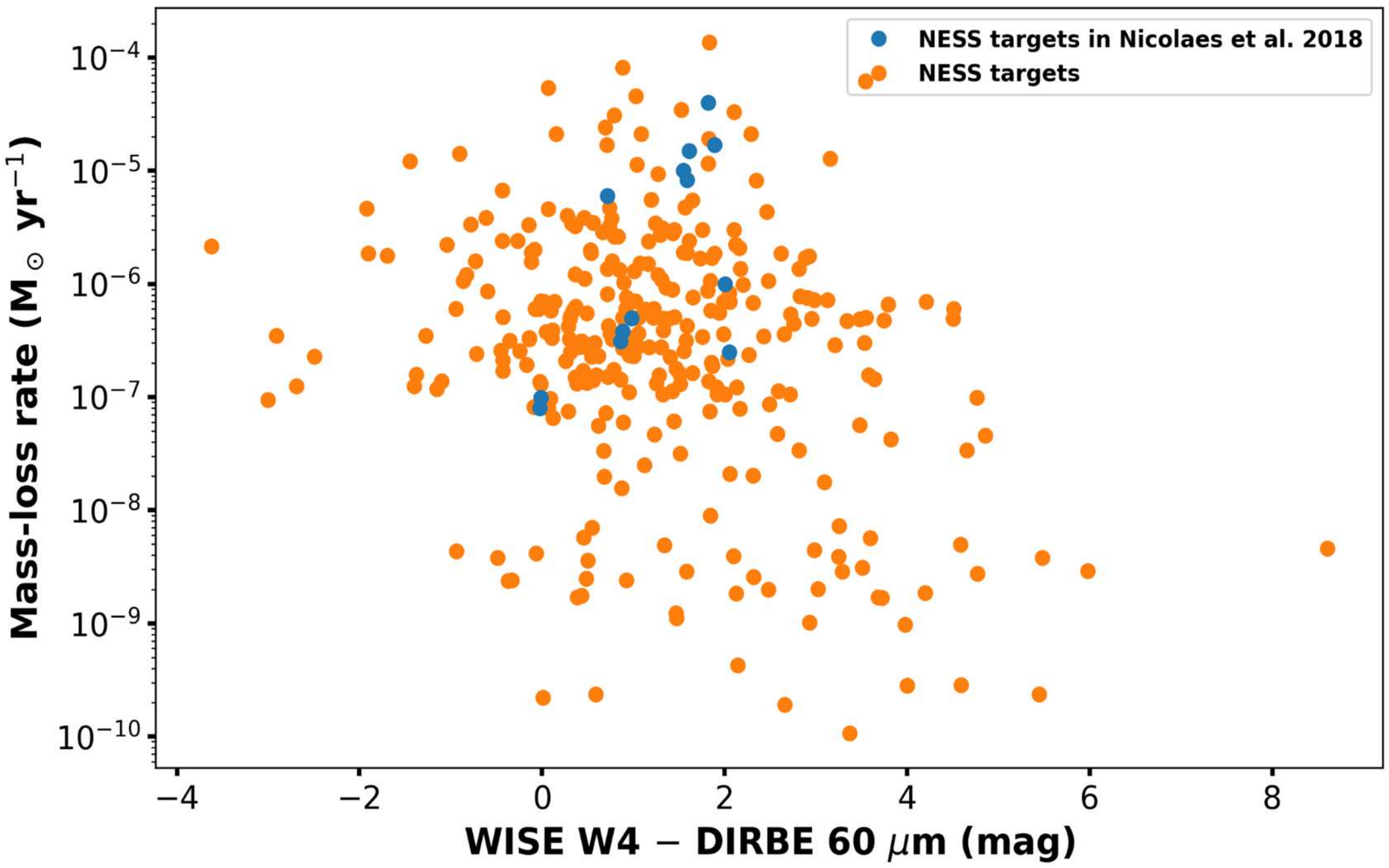

**Figure 1.** Mass-loss rate (from dust-production rates assuming a gas:dust ratio of 200) as a function of infrared color for NESS targets with 60 µm fluxes below the FIRESS saturation limit of 200 Jy (orange) compared to those NESS sources that are in the Nicolaes et al. 2018 sample (blue).

A large fraction of the NESS targets have already been studied at sub-mm and radio wavelengths. When combined with PRIMA observations, the data will allow us to probe the velocity profile over the entire CSE.

## Instruments and Modes Used

FIRESS high-res mode pointings, on each of up to 560 targets

## Approximate Integration Time

For line and continuum fluxes reported for AGB stars in Nicolaes et al (2018), the PRIMA ETC shows that the most important lines (e.g. ground-state rotational transitions of CO, $H_2O$, HCN) can be detected in less than the minimum full-scan time for a typical source. For a sample of sources, we therefore expect to need no more than 30 minutes per source per spectral setting, or one hour for the full spectral range.

There are 560 sources in the NESS sample with 65 µm fluxes below the saturation limit of 300 Jy. Of these, 113 have 24 µm fluxes below the saturation limit of 18 Jy. For the full sample, we will get CO transitions for J=12-11 up to at least 40-39. In addition, we will obtain J=41-40 up to (potentially) 95-94 for the sources with 24 µm fluxes below the saturation limit.For the full sample, we would expect about 600 h of integration.

## Special Capabilities Needed

None

## Synergies with Other Facilities

Large ground-based optical/IR telescopes now feature high-throughput spectrographs with R > 100,000, capable of resolving the velocity profiles of e.g. CO ro-vibrational band emission, complementing the large numbers of rotational lines that PRIMA can detect. Ground based interferometers, including ALMA, and in future the SKA and ngVLA, have the capability to resolve (spectrally and spatially) high-excitation transitions of other species, particularly rotational lines from vibrationally-excited states of molecules, and dust precursor species (e.g. citations). This complements the high-abundance, unresolved tracers that FIRESS is sensitive to.

## Description of Observations

We propose a homogeneous survey of far-infrared molecular lines in Solar Neighbourhood AGB stars using FIRESS. This will detect molecular lines from species including CO, $H_2O$, and their isotopologues across the entire 24-236 micron range producing a catalogue of line intensities for identified and unidentified species. Using non-LTE models, this will probe the mass-loss rates, temperature, density and velocity profiles of the inner outflows of these stars, which is only possible thanks to the large number of detectable lines and the wide range of conditions they are sensitive to. Thanks to the sensitivity of PRIMA, a large, statistically relevant sample can be observed for a modest time investment, providing a statistical understanding on how wind acceleration depends on the properties of the central star. Serendipitously, these observations may provide complementary constraints on dust-precursor molecules and the dust-formation process, as well as the properties of gas in the outer envelope, while the slope of the continuum emission will constrain the wavelength-dependence of the dust properties.

# 118. Magnetic Field in Young Supernova Remnants (SNRs)

Xiaohui Sun (Yunnan University)

Core-collapse supernova (SN) is an important dust factory and is likely the primary dust source in the early Universe. Compelling evidence has been found that there exist cold dust components in young SN remnants (SNRs), such as Cas A and the Crab Nebula, based on polarization observations at 154 μm and 850 μm. However, there is yet a systematic polarization observation of young SNRs in the Galaxy. We compile a sample of young SNRs with ages less than about 2 kyr and with far infrared total intensity emission over the wavelength range of 70-250 μm detected by Herschel, and propose to image them with PRIMAger at 92, 126, 172, and 235 μm, aiming to detect polarized emission for the first time at these wavelengths. This will help us understand the dust production and components in young SNRs and will also help us understand the magnetic fields in SNRs in combination with low-frequency radio observations.

## Science Justification

### Background and science questions

Core-collapse supernovae (SNe) are considered major sources of dust production, contributing significantly to the abundant dust observed in the early universe (e.g. Sarangi, Matsuura & Micelotta 2018). Despite this, the exact quantity of dust an SN can generate and the nature of its dust components are still uncertain. Studying young supernova remnants (SNRs) is crucial to resolving these uncertainties.

Recent studies have focused on two supernova remnants (SNRs): Cassiopeia A (Cas A) and Crab Nebula. Using the SCUBA array on the JCMT, observations of Cas A, approximately 350 years old, identified a cold dust component between 2 and 4 $M_{\odot}$ at around 18 K for the first time in a young SNR (Dunne et al. 2003). Polarization studies of Cas A using SCUPOL on the JCMT measured a dust polarization fraction of nearly 30% at 850 $\mu$m (Dunne et al. 2009). Additionally, the SOFIA HAWC+ detected up to a 20% polarization fraction at 154 $\mu$m in a more limited region of Cas A (Rho et al. 2023). Regarding the Crab Nebula, which is about 1000 years old, Herschel's findings also confirmed a cool dust component of approximately 0.24 $M_{\odot}$ and 28 K for silicate grains and about 0.11 $M_{\odot}$ at 34 K for carbon grains (Gomez et al. 2012). Though De Looze et al. (2019) reported a reduced mass for the Crab Nebula's cool dust, they observed a notable condensation efficiency, indicating that SNe might be significant dust contributors. More recently, polarized emissions at 89 $\mu$m and 154 $\mu$m have been detected, suggesting a cold dust component with a mass near 0.1 $M_{\odot}$ for silicate grains (Chastenet et al. 2022).

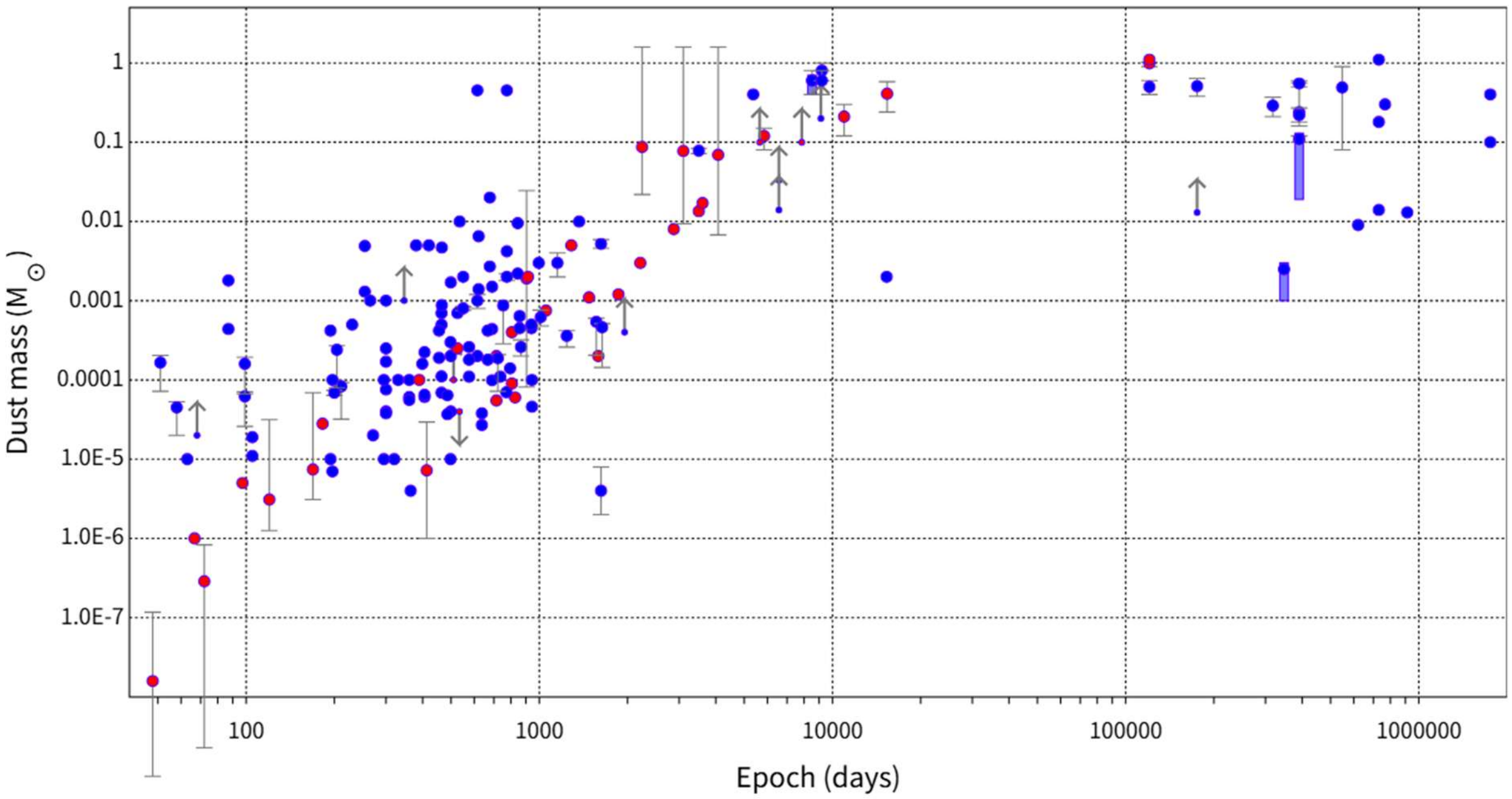

**Figure 1.** Dust mass versus SN epoch compiled by Dr. Roger Wesson (https://nebulousresearch.org/dustmass).

The compilation by Dr. Roger Wesson of dust mass versus ages is shown in Fig. 1. Dust mass ranges from about 0.01 to about 1 $M_{\odot}$ for young SNRs with ages from several hundred to several thousand years. How polarized observations will determine the dust production and components and will help understand the magnetic field in SNRs are the questions to be addressed in this proposal.

## Need for PRIMA

In contrast to early-stage SNe, dust in young SNRs cools, peaking in emission at wavelength of about 100 $\mu$m (e.g. Priestely et al. 2022). However, far-infrared emission interstellar medium (ISM) dust contaminates SNR dust emission. Stronger magnetic field in SNRs causes tighter alignment of dust with magnetic field, resulting in a higher polarization fraction. Therefore, polarization is crucial to differentiating SNR dust emission from ISM dust emission. For SNRs, the polarization fraction can reach up to about 30%, whereas it is only about a few percent in ISM (e.g. Rho et al. 2023).

Previous dust polarization observations were mainly conducted at 850 $\mu$m. At this wavelength, polarization from synchrotron emission is still comparable with dust polarization. In contrast, at 92-235 $\mu$m, the synchrotron emission is negligible.

Therefore, PRIMA is well suited to observe these young SNRs. We propose to observe a sample of young SNRs in the Milky Galaxy with PRIMAger at 92-235 $\mu$m to obtain the polarization images for the first time for most of them.

## Expected Science

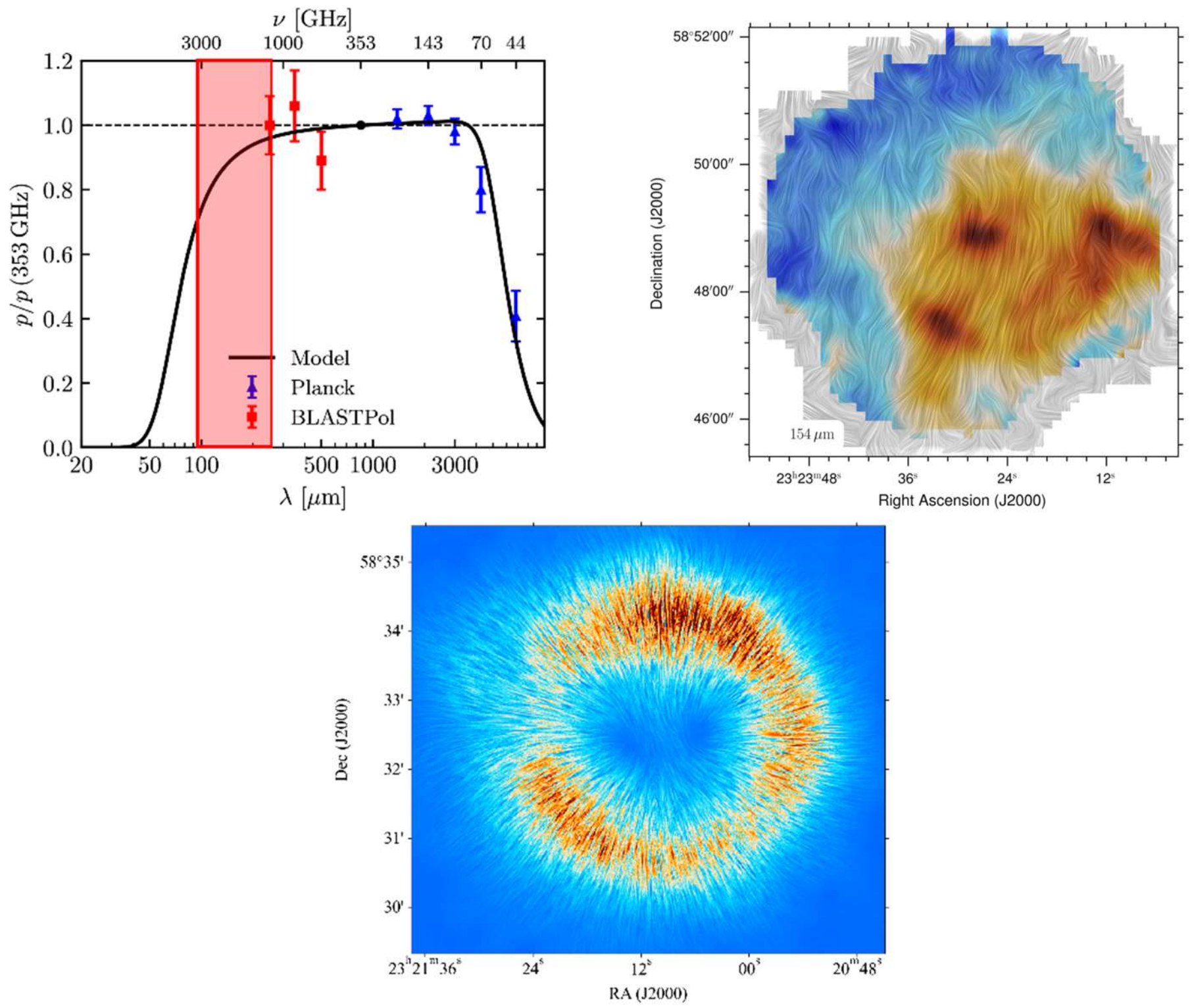

**Figure 2.** Left: Dust polarization fraction normalized to 353 GHz (Hensley & Draine 2023). The shaded rectangle indicates the PRIMA wavelength range. Middle: Magnetic field of Cas A from the polarization observations at 154 $\mu$m (Rho et al. 2023). Right: Magnetic field of Cas A from radio observations at 32 GHz (Reich 2002).

What are the dust compositions in SNRs? Dust is typically modeled with two components: carbonaceous and silicate, and the latter contributes more to polarization because of alignment with the magnetic field. Recently, Hensley & Draine (2023) proposed a one-component model with uniform compositions of carbon-rich and silicon-rich grains, termed astrodust. Based on this model, a nearly uniform fractional polarization from far-infrared to submm can be reproduced. However, there is a lack of polarization observations in the range 92-235 $\mu$m to verify this model (Fig. 2, left panel). At this range, the polarization fraction is expected to deviate from that at 353 GHz. By observing a large sample of SNRs, we will be able to examine the stardust model in SNRs in comparison with that in ISM.

How does dust form regarding the SNR evolution? The SNR sample spans ages of SNRs from hundreds to thousands of years. This corresponds to various stages of the evolution of SNR reverse shocks that process and destroy dust. As reviewed by Micelotta, Matsuura & Sarangi (2018), the surviving dust mass depends on age, which is similar for the first several hundred years and diverges afterward. The dust mass measured from the observations of SNRs of varying ages therefore constrains the models of dust process in SNRs.

What is the magnetic field in SNRs traced by dust? The magnetic field in SNRs is frequently studied using radio polarization observations of synchrotron emission. However, Faraday rotation by the magnetized medium inside and in front of the SNRs modulates polarized emission, which makes the measurement of magnetic field challenging. Polarization observations at 92-235 $\mu$m experience virtually no Faraday rotation and thus trace the intrinsic magnetic field. As an example, the magnetic field traced by dust polarization at 154 $\mu$m and radio polarization at 32 GHz is shown in Fig. 2 (middle and right panel). There is clear discrepancy between the fields, which can be attributed to Faraday rotation or magnetic field from different locations in SNRs. The combination of these observations will gain us a better understanding of magnetic fields in SNRs.

## Instruments and Modes Used

| PRIMAger Imager | | |
|---|---|---|
| Mapping details | Hyperspectral band (24-84 microns; R=8-10) | Polarimeter band (96, 126, 172, 235 microns; R=4) |
| Map size: 10 arcmin x 10 arcmin<br>Number of maps: 16 | Not necessary. | Require all 4 bands!<br>R=4 |
| If you selected Polarimeter Band, do you need polarimetry information? Yes | | |

*note that the minimum usable map size is 10'x10'. For maps under <15'x15' (small maps mode), having both bands requires two distinct observations.

## Approximate Integration Time

We selected 11 SNRs with ages less than about 2 kyr and with far infrared emission detected by Herschel from the catalog by Chawner et al. (2020): G11.2-0.3, G15.9+0.2, G21.5-0.9, G27.4+0.0, G29.7-0.3, G34.7-0.4, G43.3-0.2, G54.1+0.3, G332.4-0.4, G349.7+0.2, and G350.1-0.3. We also included Cas A, 3C 58, and Crab Nebula, all with strong far infrared emission. For comparison, we added Tycho and Kepler, which are from Type I SN rather than core-collapse SN. Previous observations indicated that they both contain cold dust components. In total, there are 16 SNRs in our sample.

Assuming an average diameter of about 10 arcmin (Chawner et al. 2020MNRAS.493.2706C) the **10 x 10 arcmin^2** map will cover the supernova remnant. The beam size is 27.6 arcsec or 0.46 arcmin at 235 $\mu$m. Using the formula from Planck Coll. et al. (2011), we derive a mean flux S(mJy) = B (MJy/sr) x A (arcmin^2) x 84.616 = 0.07 * 0.46 * 0.46 * 84.616 = **1.25 mJy.** With these values, the required time is 3.7 h based on the PRIMA's ETC.

The total exposure time for the 16 maps with sizes of 100 arcmin$^2$ would thus amount to 59.2 h.

## Special Capabilities Needed

N/A.

## Synergies with Other Facilities

N/A.

## Description of Observations

We require a small-map mode with a map size of 10 arcmin x 10 arcmin. The scanning shall ensure that a uniform polarization surface brightness sensitivity is achieved across the map and the scanning effects are minimal.

# 119. Enabling Spatially-Resolved Diagnostics of Low-Temperature Dusty Plasma Regions of Planetary Nebulae in the Solar Neighborhood

Toshiya Ueta (University of Denver), Raghvendra Sahai (Jet Propulsion Laboratory, California Institute of Technology)

We propose a far-infrared (far-IR) spectroscopic survey of planetary nebulae (PNe) using the next-generation space observatory PRIMA, building upon the legacy of *Herschel*. *Herschel* uniquely enabled spatially resolved studies of ionized, atomic, and molecular gas components in PNe through far-IR spectral mapping—an advance over the spatially integrated analyses of previous missions like *ISO*. However, Herschel's spectral mapping lacked data in critical lines necessary to close diagnostic calculations such as [O III] at 52 microns and [N II] at 205 μm. With PRIMA's enhanced sensitivity and mapping capabilities, we aim at systematically tracing variations in line and continuum emission across nebular structures, thereby unveiling the spatial distribution of electron density, temperature, and elemental abundances. These diagnostics will offer new insights into the mass-loss history of PN progenitors, the propagation of ionization fronts, and the role of dust-obscured regions in shaping nebular evolution. Compared to optical diagnostics, far-IR line ratios are less sensitive to electron temperature and largely unaffected by extinction, allowing robust analyses even in dusty environments. Moreover, generally low-temperature excitations would allow probing of outer, lower-temperature regions in which most of the circumstellar mass is expected to reside. PRIMA's spectral-imaging survey will thus provide critical constraints on the chemical enrichment and energetics of PNe, with broad implications for multi-wavelength studies of late stellar evolution.

## Science Justification

### Broader Context

The planetary nebula (PN) phase marks the last throes of stellar evolution for low to intermediate initial mass stars (of about 0.8-8 Msun). During this phase, the circumstellar envelope of gas and dust, which is created by mass loss in the preceding asymptotic giant branch (AGB) and post-AGB phases, undergoes a dramatic transformation (i.e., ionization, photo-dissociation, and dynamical shaping) caused by the fast wind and the intense radiation from the central star and by the less powerful but often significant interstellar radiation field coming from the surrounding interstellar space. As a consequence, a wide variety of underlying physical conditions are showcased within PNe, from fully ionized hot plasma to dusty cold atomic/molecular clouds, which exist (at least to first order) in a stratified manner around the central star. Therefore, PNe provide excellent astrophysical laboratories to test theories of stellar evolution as well as theories of gas-dust dynamical processes in interacting stellar winds that can also interact with the surrounding interstellar medium (ISM).

### Science Question

While PN investigations have been traditionally done through diagnostics of optical emission lines, PNe are bright sources at a wide range of wavelengths from the radio through the UV, and in some cases, even in the X-ray. Investigations using far-IR radiation are especially critical to comprehend PNe as complex physical systems in their entirety, because a large fraction of the nebula mass resides outside the central ionized region. However, according to the mass budget estimates based on the UV to mid-IR photometric survey of the Magellanic Clouds, the amount of circumstellar dust grains has been severely underestimated: only about 3% of the ISM dust grains is accounted for in the warm component of the circumstellar envelopes. What this implies is that the most extended cold regions of the circumstellar envelope could contain this "missing mass" component, which can only be detected in the wavelength ranges in the far-IR and longer.

## Need for PRIMA

Our previous attempts with Herschel found that the bulk of the circumstellar mass resided beyond the ionized region of planetary nebulae in the form of atomic/molecular gas but did not reach sufficient sensitivities to probe this colder wind-ISM interface region. With spectral images at various diagnostic lines in the PRIMA range augmented with other line maps, the physio-chemical conditions will be spatially resolved in a sample of extended planetary nebulae and shell-ISM interaction cases in the Solar neighborhood. We can then understand how emission arises especially in some exotic molecular radials and at the shell-ISM interface as well as individual and bulk/statistical abundances to be compared with predictions from existing PDR models. Herschel unfortunately did not deliver data in some critical lines to do diagnostics, such as [O III] 52 μm and [N II] 205 μm. PRIMA's FIRESS offers enough sensitivity and spectral coverage to allow fully closed plasma diagnostics with far-IR lines detected by the instrument.

## Interpretation Methods

Far-IR fine-structure line ratios, especially [O III] 52/88 μm and [N II] 122/205 μm, are relatively insensitive to the electron temperature (Te) and provide excellent diagnostics of the photo-ionized regions (PIRs). Meanwhile, low-excitation and atomic line maps at [C II] 158 μm and [O I] 63 and 146 μm are typically used to probe the physical conditions of the photo-dissociation regions (PDRs). Using a full complement of far-IR diagnostic line maps that spatially resolve extended structures, we will be able to see how the colder, low-excitation "missing mass" components are distributed in the interface between the ISM and the warmer circumstellar shell for the first time.

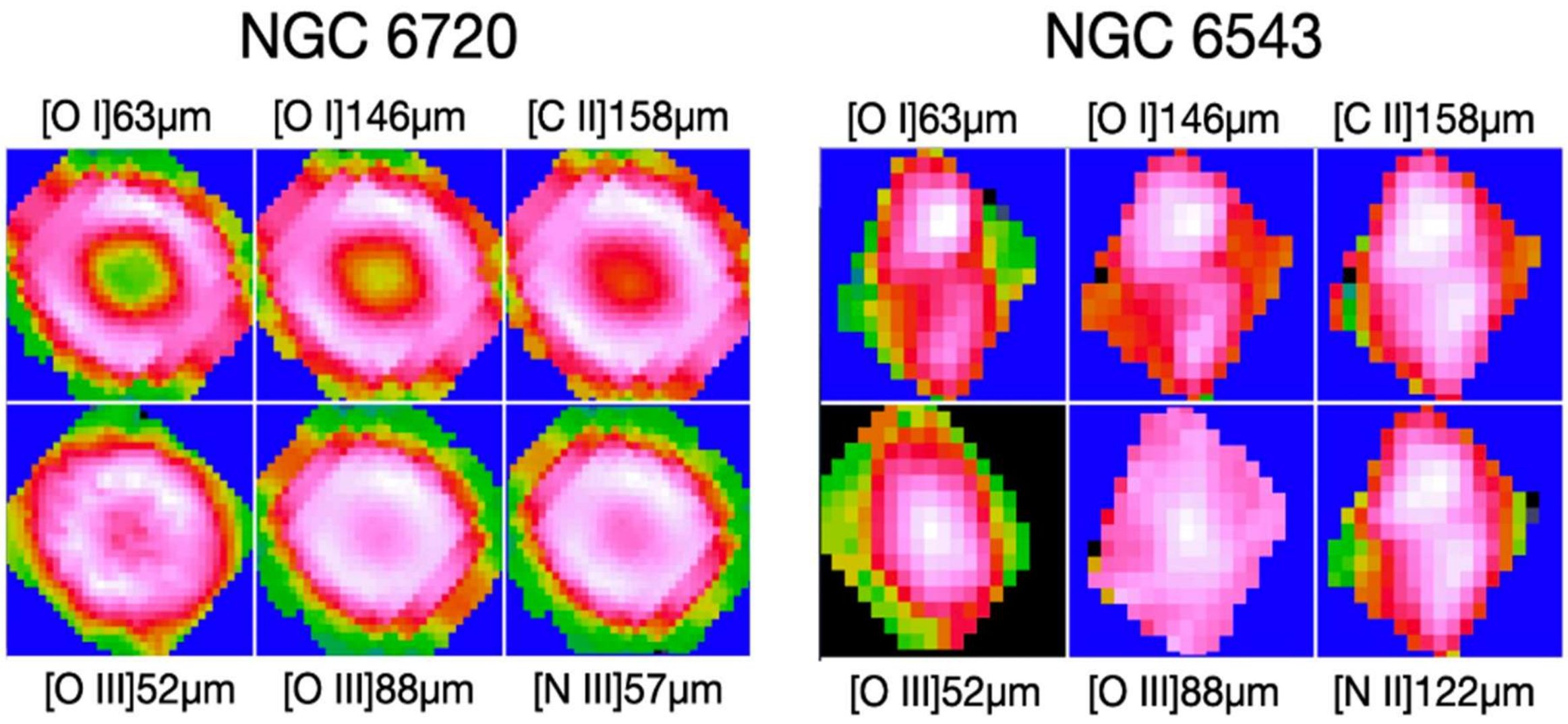

Figure 1: Examples of PACS spectral maps: **[Left]** NGC 6720 (Ring Nebula, roughly 90″ × 90″) and **[Right]** NGC 6543 (Cat's Eye, roughly 50″ × 50″). The top frame maps represent the PDR, while the bottom frame maps represent the PIR: the bottom-right frame shows the [N III] 57 $\mu$m map that represent high-excitation PDR for NGC 6720 and the [N II] 122 $\mu$m map that represent low-excitation PDR for NGC 6543. In both cases, the emission morphology of the PDR and PIR is distinct, suggesting to yield distinct ($n_e$, $T_e$) and ($n_H$, $T_H$) distributions after full plasma diagnostics performed as outlined above. The blue-edge [O III] 52 $\mu$m map (the bottom-left frame) has not been integrated over the line width - exhibiting a single wavelength bin map of the spectral cube.

## Instruments and Modes Used

This observing program requests 10'x10' low-res maps with FIRESS towards about two dozen targets.

## Special Capabilities Needed

N/A

## Approximate Integration Time

Based on Herschel/PACS spectral mapping, we aim at achieving 5 sigma detection at 5e-18 W/m$^2$. Using the PRIMA Exposure Time Calculator for FIRESS, we need 2.88 hrs to cover a 10' x 10' field at the target sensitivity at the blue end of the spectral coverage (with the same exposure time, we gain x3 sensitivity at the red end of the spectral coverage). To observe about two dozen sources known to be extended in the far-IR from the previous Herschel/PACS observations, we need about 70 hrs.

## Description of Observations

Default scan-mapping of a 10′ × 10′ field will do for most of the cases. Some target sources are smaller than this field size. However, we will adopt this field size to map targets in all 4 bands simultaneously, instead of separating mapping in two spectral coverage blocks.

## Acknowledgement

The research was carried out at the Jet Propulsion Laboratory, California Institute of Technology, under a contract with the National Aeronautics and Space Administration (80NM0018D0004).

# 120. Observational Constraints on Thermal Pulses at the Interface Regions Between the Circumstellar Shells and the Interstellar Space in a Sample of Solar-Neighbor Evolved Stars

Toshiya Ueta (University of Denver), Raghvendra Sahai (Jet Propulsion Laboratory, California Institute of Technology)

The life cycle of matter in the Universe begins with stellar nucleosynthesis. Various chemical elements synthesized deep within stars must be dredged up to the surface and dispersed into the circumstellar environments via stellar mass loss. The circumstellar shell forms as the direct product of this process, retaining the pristine records of mass loss and showcasing subsequent physico-chemical processes. Our studies are aimed at providing empirical quantitative evidence of mass loss thermal pulses that have not yet been established from first principles of physics. The past far-IR opportunities enabled us to glimpse at the interface region between the circumstellar shell and the interstellar medium, where the bulk of the circumstellar mass was found at the coldest temperatures. We will use PRIMA to gather quantitative evidence that is directly comparable with existing theories of thermal pulses for the first time. Specifically, we will spatially resolve multiple shell structures resulting from thermal pulses during the late stages of post-main-sequence evolution to directly measure the inter-pulse spacing and timescale as empirical constraints. Also, we will spectrally resolve thermal pulses to perform full-fledged atomic and molecular line diagnostics to look into the degree of gas-dust coupling that is the cornerstone of dust-driven stellar mass loss. These investigations will be done with a set of about two dozen evolved stars with extended shells in the Solar neighborhood—all resolvable with PRIMA—that were detected in the previous far-IR opportunities.

## Science Justification

### Broader Context

Stellar mass loss is one of the critical processes of stellar evolution that has significant implications for the rest of the cosmos, from the chemical yield of individual stars, the ISM metallicity, to the dust yield in the cosmological context. While existing stellar evolution codes are equipped with various mass loss prescriptions that would yield results that are generally consistent with observations in a statistical sense, specifics are still unknown as mass loss has not been theoretically formulated from first principles and observational clues of this physically important process of mass loss have not been clearly established as observables.

### Science Question

We aim at addressing the following two specific issues surrounding stellar mass loss in terms of providing quantitative observational constraints on existing theories/models:

**(1) The geometry and timescale of thermal pulses during the extreme AGB phase**

During the latest stages of evolution during the asymptotic giant branch phase, the rate of mass loss is known to be enhanced by thermal pulses so that the bulk of the stellar envelope can be lost to the circumstellar space by the end of the AGB phase. Also, the geometry of mass loss is expected to change dramatically around this phase. These thermal pulses are theoretically predicted, and mass loss ejecta from a single thermal pulse event have been observationally seen. However, in no case have multiple shells of thermal pulse origin been spatially resolved, while unresolved cases have been detected (Figure 1). We will detect and spatially resolve these multiple shells to observationally constrain the time interval between such shells of thermal pulse origin to be compared with existing model predictions. Also, hyperspectral mapping in thermal dust continuum can probe time-lapse of the geometrical transformation as bands at longer wavelengths see the spatial distribution in the further past.

**(2) Spatially resolved gas-dust coupling in the shell-ISM interface regions**

Another critical issue in the theoretical framework of dust-driven stellar winds is the gas-dust coupling. Robust gas-dust coupling is paramount for the successful launching and acceleration of dust-driven stellar winds. It ensures that the momentum gained by the dust from stellar radiation is efficiently transferred to the gas, enabling a substantial mass outflow from the star. Without strong coupling, the wind would be weak or non-existent. Given that momentum transfer in dust-driven stellar winds relies on the drag between dust and gas, which inherently implies a slight spatial differentiation due to their differing initial responses to radiation pressure and subsequent interactions, observing the spatial separation of dust and gas components becomes significantly meaningful. We aim at providing such crucial observational insights into the efficiency of the dust-gas coupling, the dynamics of momentum transfer within the wind, and ultimately, a more comprehensive understanding of the wind's driving mechanisms and overall structure.

## Need for PRIMA

PRIMA would have sufficient resolution both spatially and spectrally, and especially would have sufficient extended-source sensitivities, all of which were not afforded by the previous far-IR facilities. To probe the coldest extended shell-ISM interface regions, access to the far-IR spectral range is essential. As demonstrated in Figure 1, the success of this program depends on achieving an optimal balance between spatial resolution and detector sensitivity. If the resolution is too high, individual detector elements may not collect sufficient surface brightness for detection and reliable measurements. Conversely, if the resolution is too low, dust shells may remain spatially unresolved and intervals of mass loss events not measured.

## Interpretation Methods

Thermal pulses in stellar winds are expected to manifest themselves as semi-periodic surface brightness modulations in circumstellar shells extended to a few x 100” and more. Spacing between brightness peaks would yield the timescale of thermal pulses, given the expansion velocity and distance.

### Link to Testable Hypotheses

Thermal pulses were theoretically found in the 1980s and have been the hallmarks of dust-driven mass loss. Yet direct observational evidence has been elusive.

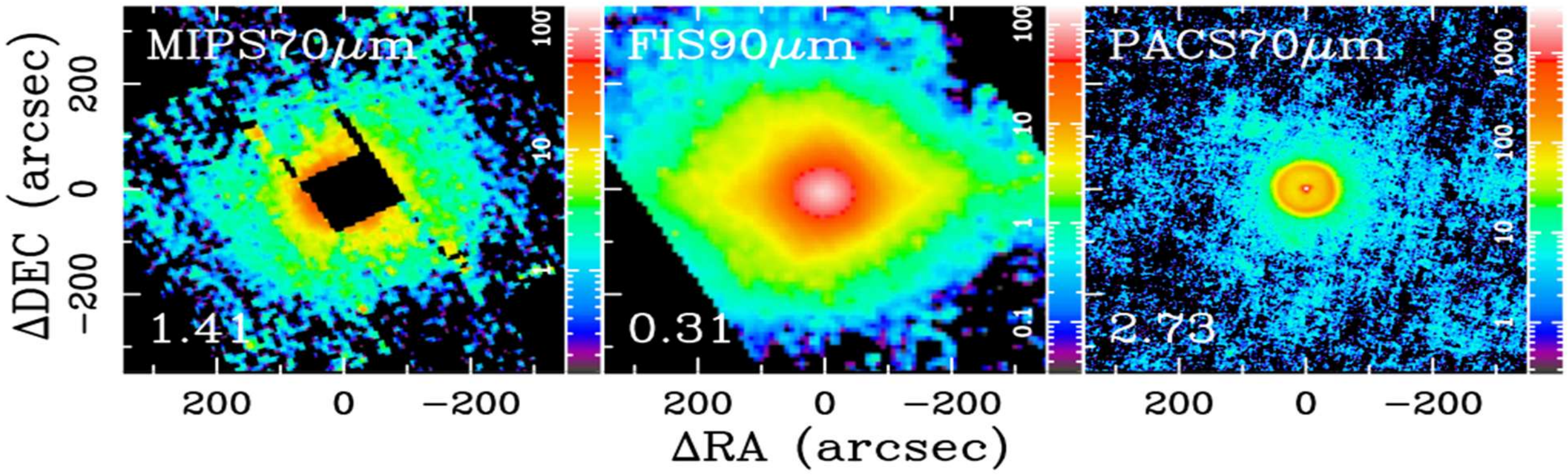

**Figure 1.** Far-IR maps of the extended circumstellar shell of U Ant taken with Spitzer at 70μm (left; Ueta et al. 2007), AKARI at 90μm (middle; Ueta et al. 2017, 2019), and Herschel at 70μm (right; Kerschbaum et al. 2010). Each color wedge shows the surface brightness in MJy/sr. The detection limit in MJy/sr is indicated by the value given at the lower left corner. The Herschel map (right) clearly resolved the detached shell caused by the last thermal pulse, but did not detect any previous detached shells. The Spitzer map (left) showed hints of the presence of more extended shells around the detection limit. The AKARI map (middle) achieved the sub-MJy/sr sensitivity needed to detect the extended shell caused by the thermal pulse before last. However, the spatial resolution was not enough to resolve the detached shell generated by the previous thermal pulse. PRIMAger will provide both spatial resolution and sensitivity to detect both of the detached thermal pulse dust shells, allowing the direct measurement of the inter-pulse time scale. FIRESS mapping is aimed at detecting the corresponding molecular/atomic gas shell to investigate if there is any spatial offset between dust/gas shells, indicative of drag between them.

## Instruments and Modes Used

PRIMAGer imaging in the polarimetry band (92 and 126 μm). Minimal map size for each target. FIRESS low res 10’x10’ maps for each target.

## Approximate Integration Time

### PRIMAger Spectral Imaging

Based on the existing most sensitive far-IR images of evolved star dust shells taken with Spitzer and AKARI, we aim at achieving 5σ/pix with 0.3 MJy/sr at 92 μm (Band 1) and 0.2 MJy/sr at 126 μm (Band 2). According to the PRIMA Exposure Time Calculator, we would need 1.2 hrs in Band 1 and 1.6 hrs in Band 2 to map a 100 square-arcmin field that is large enough to capture extended circumstellar shells. For a sample of about two dozen extended targets, the whole 2-band mapping would take about 70 hrs.

### FIRESS Spectral Imaging

Based on the existing Herschel/PACS spectral images of planetary nebulae in far-IR atomic lines, the detection limit is typically about $1 \times 10^{-17}$ W/m$^2$. If we want to aim at detecting such line emission at 5σ when mapping a region of 10’x10’, the PRIMA Exposure Time Calculator indicates 0.72 hrs for band 1 and 2, 0.28 hrs for band 3, and 0.11 hrs for band 4. Thus, we need 1.44 hrs to map a 10’x10’ region in all 4 bands. For a sample of about two dozen extended targets, the whole spectral mapping would take about 35 hr.

## Special Capabilities Needed

None

## Synergies with Other Facilities

Ground-based optical IFU observations to get optical forbidden line maps. ALMA spectral mapping for the molecular component.

## Description of Observations

In both thermal dust emission PRIMAger/PPI and FIRESS imaging, simple scan-mapping of a 10’ × 10’ field will do for most of the cases. Because of the nature of our target sources (emission is typically centrally concentrated), if there is a mapping strategy to increase redundancy (hence S/N) in the outer part of the extended structure would be nice.

## Acknowledgement

The research was carried out at the Jet Propulsion Laboratory, California Institute of Technology, under a contract with the National Aeronautics and Space Administration (80NM0018D0004).